# Deep Underground Science and Engineering Laboratory – Preliminary Design Report

Kevin T. Lesko[1, §], Steven Acheson[1], Jose Alonso[4, †], Paul Bauer[3], Yuen-Dat Chan[1, 2], William Chinowsky[1, †], Steve Dangermond[3], Jason A. Detwiler[1, 2], Syd De Vries[1, 2], Richard DiGennaro[1, 2, †], Elizabeth Exter[1, 2], Felix B. Fernandez[1], Elizabeth L. Freer[3], Murdock G. D. Gilchriese[1, 2], Azriel Goldschmidt[1, 2], Ben Grammann[3], William Griffing[1, †], Bill Harlan[4], Wick C. Haxton[1, 2], Michael Headley[3], Jaret Heise[4], Zbigniew Hladysz[3], Dianna Jacobs[1, 2], Michael Johnson[4], Richard Kadel[1, 2], Robert Kaufman[3], Greg King[4], Robert Lanou[6, †], Alberto Lemut[1, 2], Zoltan Ligeti[2], Steve Marks[1, 2], Ryan D. Martin[1, 2], John Matthesen[3], Brendan Matthew[4], Warren Matthews[3], Randall McConnell[4], William McElroy[4], Deborah Meyer[4], Margaret Norris[5], David Plate[1, 2, †], Kem E. Robinson[1, 2], William Roggenthen[3], Rohit Salve[1, 2], Ben Sayler[5], John Scheetz[4], Jim Tarpinian[1], David Taylor[3], David Vardiman[3], Ron Wheeler[4], Joshua Willhite[3], and James Yeck[1, ‡]

[1] University of California at Berkeley, Berkeley, California 94720
[2] Lawrence Berkeley National Laboratory, Berkeley, California 94720
[3] South Dakota School of Mines & Technology, Rapid City, South Dakota 57701
[4] South Dakota Science and Technology Authority, Lead, South Dakota 57754
[5] Black Hills State University, Spearfish, South Dakota 57799
[6] Brown University, Providence, Rhode Island 02912

[§] Corresponding Author
[†] retired
[‡] Permanent Address: University of Wisconsin

Abstract

*The DUSEL Project has produced the Preliminary Design of the Deep Underground Science and Engineering Laboratory (DUSEL) at the rehabilitated former Homestake mine in South Dakota. The design satisfies the requirements of the project Readiness Stage detailed in the National Science Foundation's Large Facilities Manual. The Facility design calls for, on the surface, two new buildings—one a visitor and education center, the other an experiment assembly hall— and multiple repurposed existing buildings. To support underground research activities, the design includes two laboratory modules and additional spaces at a level 4,850 feet underground for physics, biology, engineering, and Earth science experiments. On the same level, the design includes a Department of Energy-shepherded Large Cavity supporting the Long Baseline Neutrino Experiment. At the 7,400-feet level, the design incorporates one laboratory module and additional spaces for physics and Earth science efforts. All underground areas will be connected by a distributed-access network extending from the surface down to a depth of 2.25 km and extending over 30 km. With input from some 25 science and engineering collaborations, the Project has designed critical experimental space and infrastructure needs, including space for a suite of multidisciplinary experiments in a laboratory whose projected life span is at least 30 years. From these experiments, a critical suite of experiments is outlined, whose construction will be funded along with the*

*facility. The putative users, with funding independent of the Project, are advancing their designs and pursuing research and development on detector technology. The Facility design permits expansion and evolution, as may be driven by future science requirements, and enables participation by other agencies. The design leverages South Dakota's substantial investment in facility infrastructure, risk retirement, and operation of its Sanford Laboratory at Homestake. The Project is planning education and outreach programs, and has initiated efforts to establish regional partnerships with underserved populations—regional American Indian and rural populations. The DUSEL Project enhances South Dakota and regional university participation in world-leading research and proposes major construction activities for these and other South Dakota research entities. The competition-sensitive nature of the estimates contained in Volume 2 of the Preliminary Design Report resulted in this Volume being redacted from this distribution of the Report.*

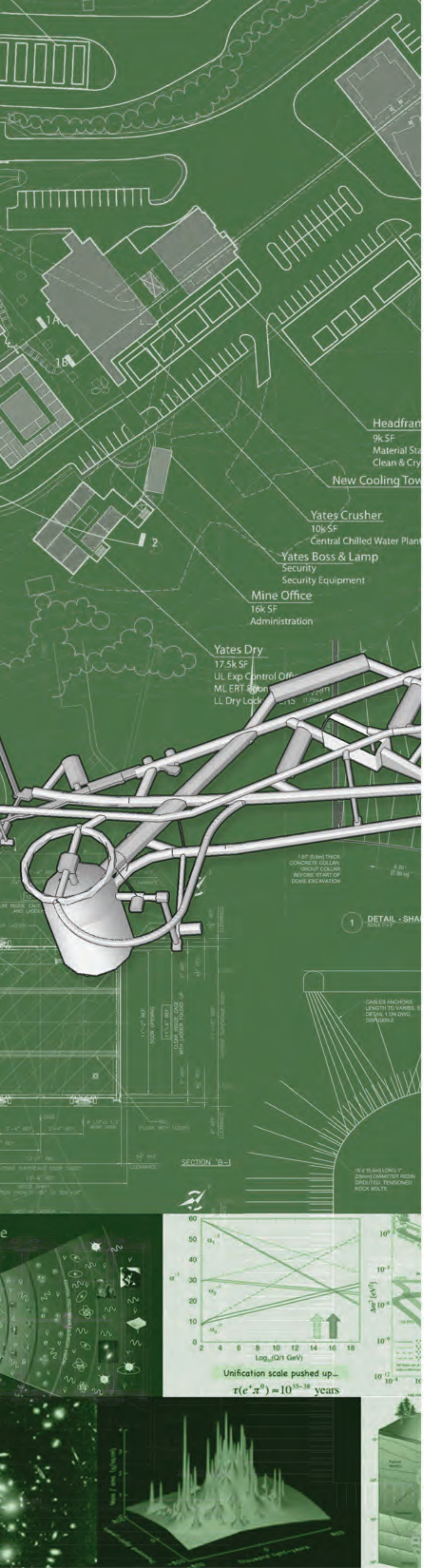

# Preliminary Design Report

May 2011

DUSEL
Deep Underground
Science and
Engineering Laboratory

This page intentionally left blank

## Table of Contents

# List of Figures

## Volume 1

## Volume 2

## Volume 3

## Volume 4

## Volume 5

## Volume 6

## Volume 7

## Volume 8

## Volume 9

## Volume 10

# List of Tables

## Volume 1

## Volume 2

## Volume 3

## Volume 4

## Volume 5

## Volume 7

## Volume 8

## Volume 9

# List of Acronyms

| | |
|---|---|
| $(C_0)$ | uniaxial compressive strength |
| $(T_0)$ | average tensile strength |
| °C | degrees Celsius |
| °F | degrees Fahrenheit |
| 1TGe | 1 Tonne Germanium |
| 2-D | two-dimensional |
| 3-D | three-dimensional |
| µm | micrometer |
| AARM | Assay and Acquisition of Radiopure Materials |
| ABA | Architectural Barriers Act |
| AC | Actual Cost; alternating current |
| ACAMS | Access Control and Alarm Monitoring System |
| ACBM | asbestos-containing building material |
| ACWP | Actual Cost of Work Performed, a.k.a. AC |
| AD | Associate Director |
| ADA | Americans with Disabilities Act |
| ADAAG | Americans with Disabilities Act Accessibilities Guidelines for Buildings and Facilities |
| AED | Automatic External Defibrillators |
| AEG | American Society of Engineering Geologists |
| AGB | asymptotic giant branch |
| AGV | automated guided vehicle |
| AHJ | Authority Having Jurisdiction |
| AHU | air handling unit |
| AISES | American Indian Science and Engineering Society |
| ALD | Associate Laboratory Director |
| ANL | Argonne National Laboratory |
| ANSI | American National Standards Institute |
| AoR | Area of Refuge |
| APA | Anode Plane Assembly |
| APTS | Asset/Personal Tracking System |
| ARD | Acid Rock Drainage |
| ArDM | Argon Dark Matter |
| ARIN | American Registry of Internet Numbers |
| ANSI | American National Standards Institute |
| ARMA | American Rock Mechanics Association |
| ASCE | American Society of Civil Engineers |
| ASLA | American Society of Landscape Architects |
| ASME | American Society of Mechanical Engineers |
| ASTC | Association of Science and Technology Centers |
| atm | atmosphere |
| ATV | all-terrain vehicle; acoustic televiewer |
| B&M | Blackstone and McMaster |
| B–L | baryon number minus lepton number |
| BAC | Budget At Completion |
| BBN | Big Bang nucleosynthesis |
| BCCP | Berkeley Center for Cosmological Physics |
| BCWP | Budgeted Cost of Work Performed, a.k.a. EV |
| BCWS | Budgeted Cost of Work Scheduled, a.k.a. PV |
| BER | Biological and Environmental Research |
| BES | Basic Energy Sciences |
| BGE | biology, geology, and engineering |
| BH | borehole |
| BHP | Black Hills Power |
| BHSU | Black Hills State University |
| BMS | Building Management System |
| BNL | Brookhaven National Laboratory |
| BOD | Basis of Design |
| BOE | Basis of Estimate |
| CAC | Cultural Advisory Committee |
| Caltech | California Institute of Technology |
| CAM | control account manager |
| CBB | Contract Budget Baseline |
| CCB | Configuration Control Board |
| CCC | Command and Control Center |
| CCD | charge-couple device |
| CCR | Configuration Change Request |
| CCS | Carbon dioxide Capture and Storage |
| CD-0 | Critical Decision-0 (DOE) |
| CD-1 | Critical Decision-1 (DOE) |
| CDE | communications distribution enclosure |
| CDMS | Cryogenic Dark Matter Search |
| CDR | Conceptual Design Report; communications distribution room |
| CE | communications enclosure |
| CERN | European Organization for Nuclear Research |
| cfm | cubic feet per minute |
| CFO | Chief Financial Officer |
| CFR | Code of Federal Regulations |
| CHW | chilled water |
| CI | Cyberinfrastructure; Construction Institute |
| CIAC | Cyberinfrastructure Advisory Committee |
| CIDR | classless inter-domain routing |
| CLEAN | Cryogenic Low Energy Astrophysics with Noble gases |
| cm | centimeter |
| CM | Construction Manager |
| CMMI | Capability Maturity Model Integrated |
| CMP | Contract Management Plan |
| CNGS | CERN Neutrinos to Gran Sasso |
| CNO | carbon-nitrogen-oxygen |
| CO | carbon monoxide |
| Co-PI | Co-Principal Investigator |
| COR/COTR | contracting officer’s technical representative |
| COUPP | Chicagoland Observatory for Underground Particle Physics |
| CP | Charge and Parity |
| CPI | Cost Performance Index |
| CPR | Cost Performance Report |
| CR | communications room |
| CRAC | computer room air conditioning |
| CSIRO | Commonwealth Scientific and Industrial Research Organization |
| CSM | Colorado School of Mines |
| CSO | Chief Science Officer |
| CTF | Counting Test Facility |
| CUBED | Center for Ultralow-Background Experiments at DUSEL |
| CUORE | Cryogenic Underground Observatory for Rare Events |
| CV | cost variance |
| DAEdALUS | Decay at Rest Experiment for $\delta_{CP}$ studies At the Laboratory for Underground Science |
| DAMA/LIBRA | DArk MAtter/Large sodium Iodide Bulk for RAre processes |
| DAQ | data acquisition |
| DAWD | days after well was drilled |
| dBA | decibels, A-weighted |
| DC | direct current |
| DDC | Direct Digital Control |
| DEDC | DUSEL Experiment Development and Coordination |
| DEM | distinct element method |
| DHA | David Heil & Associates |
| DHCP | Dynamic Host Configuration Protocol |
| DIANA | Dakota Ion Accelerators for Nuclear Astrophysics |
| DKA | Dangermond Keane Architecture |
| DLL | Deep-Level Laboratory |
| DLM | Davis Laboratory Module |
| DM | dark matter |
| DMR | Deferred Maintenance Rehabilitation |
| DMTPC | Dark Matter Time Projection Chamber |
| DNS | Domain Name System |
| DOE | Department of Energy |
| DOORS | Dynamic Object-Oriented Requirements System |
| DPD | Deputy Project Director |
| DST | Distributed Strain and Temperature |
| DTA | Davis Transition Area |
| DTS | Distributed Temperature Sensing |
| DUGL | Deep Underground Gravity Laboratory |
| DULBCF | DUSEL Low Background Counting Facility |
| DuRA | DUSEL Research Association |
| DUSEL | Deep Underground Science and Engineering Laboratory |
| DX | direct expansion; air-cooled direct expansion unit |
| DZ | disturbed zone; damage zone |
| EA | Environmental Assessment |
| EAC | Estimate At Completion; Education Advisory Committee |
| E&O | Education and Outreach |
| EDM | electrical discharge machining |
| EEAC | External Experimental Advisory Committee |
| EGB | Education Governing Board |
| EH&S | Environment, Health, and Safety |
| EHSD | Environmental Health and Safety Director |
| EHSOC | Environmental Health and Safety Oversight Committee |
| EIP | Early Implementation Program; Experiment Implementation Policy |
| EIRD | External Interface Requirements Document |
| EIS | Environmental Impact Statement |
| EMI | electromagnetic interference |
| EPA | Environmental Protection Agency |
| EPS | Experimental Planning Statement |
| EPSCoR | Experimental Program to Stimulate Competitive Research |
| ERP | Emergency Response Plan |
| ERT | Emergency Response Team |
| Esnet | Energy Sciences Network |
| ESR | Excavation Support Ratio |
| ESS | electronic security system |
| ETC | Estimate To Complete |
| EV | Earned Value |
| EVMS | Earned Value Management System |
| EVMT | Earned Value Management Technique |
| EXO | Enriched Xenon Observatory |
| FAARM | Facility for Assay and Acquisition of Radiopure Materials |
| FACU | fire alarm control unit |
| FAQ | frequently asked questions |
| FAR | Federal Acquisition Regulation |
| FAS | fire alarm system |
| FBI | Federal Bureau of Investigation |
| FBG | fiber Bragg grating |
| FDM | finite difference method |
| FDR | Final Design Report; Final Design Review |
| FEA | Finite Element Analysis |
| FEMP | Federal Energy Management Program |
| FEM | finite element method |
| Fermilab | Fermi National Accelerator Laboratory |
| FF | fracture frequency |
| FHA | Fire Hazard Analysis |
| FLS | Fire-Life-Safety |
| FLUKA | FLUktuierende KAskade |
| FMS | Facility Management System |
| FNAL | Fermi National Accelerator Laboratory |
| FOB | fine-ore bin |
| FRGP | Front Range GigaPop |
| FTE | full-time equivalent |
| FV | fiducial volume |
| FY | fiscal year |
| G1 | Generation One |
| G2 | Generation Two |
| G3 | Generation Three |
| GAAP | generally accepted accounting principles |
| GAC | Geotechnical Advisory Committee |
| Gbit | Gigabit |
| Gbps | Gigabits per second |
| GBR | Geotechnical Baseline Report |
| GCS | geologic carbon sequestration |
| Gd | gadolinium |
| GE | General Electric |
| GEANT | GEometry ANd Tracking |
| GEAR UP | Gaining Early Awareness and Readiness for University Programs |
| GEIA | Government Electronic Industries Alliance |
| GEODM | Germanium Observatory for Dark Matter |
| GEOX™ | GEologic Optical eXtensometer and Tiltmeter |
| GERDA | GERmanium Detector Array |
| GES | Geotechnical Engineering Services |
| GFCI | ground fault circuit interrupter |
| gpm | gallons per minute |
| GPN | Great Plains Network |
| GPS | global positioning system |
| GSA | General Services Agreement; General Services Administration |
| gsf | gross square feet |
| GSI | Geologic Strength Index |
| GUT | Grand Unified Theory |
| GXe | xenon, gas phase |
| HAR | Hazard Analysis Report |
| HARCC | Homestake Adams Research and Cultural Center |
| HAZCOM | Hazard Communication |
| HAZWOPER | HAZardous Waste OPerations and Emergency Response |
| HD | high definition |
| HEP | High Energy Physics |
| HI | hollow inclusion |
| HID | High-Intensity Discharge |
| HMI | human/machine interface |
| HMC | Homestake Mining Company |
| hp | horsepower |
| HPGe | high-purity germanium |
| HPR | Highly Protected Risk |
| HTTP | Hyper Text Transfer Protocol |
| HTTPS | Hyper Text Transfer Protocol Secure |
| HUD | Housing and Urban Development (U.S. Department of) |
| HVAC | Heating, Ventilation, and Air-Conditioning |
| HV | high voltage |
| IAB | Infrastructure Advisory Board |
| IBC | International Building Code |
| ICARUS | Imaging Cosmic and Rare Underground Signals |
| ICC | International Code Council |
| ICD | Interface Control Document |
| ICDP | International Continental Scientific Drilling Program |
| ICPMS | inductively coupled plasma mass spectrometry |
| ICT | information and communications technology |
| ID | identification |
| IDS | intrusion detection system |
| IEC | International Electrical Code |
| IEEE | Institute of Electrical and Electronics Engineers |
| IETF | Internet Engineering Task Force |
| IFC | International Fire Code |
| IH | inverted hierarchy |
| IMB | Irvine, Michigan, Brookhaven experiment |
| IODP | Integrated Ocean Drilling Program |
| ION | Inter-Operability Network |
| IP | Internet Protocol |
| IPC | International Plumbing Code |
| IPS | Integrated Project Schedule |
| IPT | Integrated Product Team |
| IRC | Internal Review Committee |
| IRD | Interface Requirements Document |
| ISE | Integrated Suite of Experiments |
| ISE IRD | Integrated Suite of Experiments Interface Requirements Document |
| ISM | Integrated Safety Management |
| IT | Information Technology |
| iZIP | Interdigitated Z-dependent Ionization- and Phonon-mediated |
| Ja | joint alteration |
| JHA | Job Hazard Analysis |
| Jn | joint set number |
| Jr | joint roughness number |
| Jw | joint water reduction factor |
| JOG | Joint Oversight Group |
| K2K | KEK (Koh-Ene-Ken, an abbreviation for a Japanese name of the National Laboratory for High Energy Physics) to Kamioka experiment |
| km | kilometer |
| KPP | key performance parameter |
| kT | kilotonne |
| kV | kilovolt |
| kW | kilowatt |
| L | Level |
| LAB | linear alkylbenzene |
| LAN | local area network |
| LAr | liquid argon |
| LArTPC | Liquid Argon Time Projection Chamber |
| LBNE | Long Baseline Neutrino Experiment |
| LBNL | Lawrence Berkeley National Laboratory |
| LBP | lead-based paint |
| LC | Large Cavity |
| LC-1 | Large Cavity 1 |
| LCAB | Large Cavity Advisory Board |
| LED | light-emitting diode |
| LEED | Leadership in Energy & Environmental Design |
| LEPC | Lawrence County Emergency Management Commission |
| LGC | WBS element for the LBNE Facilities |
| LHC | Large Hadron Collider |
| LHD | load-haul-dump |
| LIDAR | Light Detection and Ranging |
| LIGO | Laser Interferometer Gravitational-wave Observatory |
| LLC | Limited Liability Company |
| LM | laboratory module |
| LM-1 | Laboratory Module 1 on the 4850L |
| LM-2 | Laboratory Module 2 on the 4850L |
| LMA | large mixing angle |
| LMD-1 | Laboratory Module 1 on the 7400L |
| LN | liquid nitrogen |
| LNGS | Laboratori Nazionali del Gran Sasso |
| LOE | level of effort |
| LOI | Letter of Intent |
| lpm | liters per minute |
| LRDP | Long Range Development Plan |
| LTE | long-term evolution |
| LUCI | Laboratory for Underground $CO_2$ Investigations |
| LUNA | Laboratory for Underground Nuclear Astrophysics |
| LUX | Large Underground Xenon |
| LXe | liquid xenon |
| LZD | LUX-ZEPLIN DUSEL |
| m | meter |
| $m^2$ | square meters |
| $m^3$ | cubic meters |
| M&M | measurement and monitoring |
| MAX | Multiton Argon and Xenon |
| Mbps | megabits per second |
| MC | Monte Carlo |
| MCM | thousand circular mils |
| MCR | main communication room |
| MDU | Montana-Dakota Utilities |
| MEP | mechanical, electrical, plumbing |
| MER | mechanical and electrical room |
| MERV | minimum efficiency reporting value |
| MG | motor-generator |
| MINERvA | Main INjector ExpeRiment for v-A |
| MicroBooNE | Micro-Booster Neutrino Experiment |
| MiniBooNE | Mini Booster Neutrino Experiment |
| MiniCLEAN | Mini Cryogenic Low Energy Astrophysics with Noble gases |
| MINOS | Main Injector Neutrino Oscillation Search |
| MK | McCarthy Kiewit |
| MLL | Mid-Level Laboratory |
| mm | millimeter |
| MOU | Memorandum of Understanding |
| MPa | Mega Pascal |
| MPLS | Multi Protocol Label Switching |
| MPS | Mathematical and Physical Sciences |
| MR | Management Reserve |
| MREFC | Major Research Equipment and Facilities Construction |
| MRTG | Multi Router Traffic Grapher |
| MSDS | Material Safety Data Sheet |
| MSHA | Mine Safety and Health Administration |
| MSU | Michigan State University |
| MT | megaton |
| MUL | Mobile Underground Laboratory |
| MUTCD | Manual of Uniform Traffic Control Devices |
| MVA | mega volt ampere |
| MW | megawatt |
| mwe | meters-water-equivalent |
| NAA | neutron activation analysis |
| NAG | net acid generation |
| NASA | National Aeronautics and Space Administration |
| NAT | Network Address Translation |
| NDT | non-destructive testing |
| NEC | National Electric Code |
| NEXT | Neutrino Experiment with a Xenon TPC |
| NEPA | National Environmental Policy Act |
| NERSC | National Energy Research Scientific Computing Center |
| NESC | National Electric Safety Code |
| NFPA | National Fire Protection Association |
| NGI-Q | Tunneling Quality Index |
| NH | normal mass hierarchy |
| NHPA | National Historic Preservation Act |
| NIC | network interface card |
| NOC | network operations center |
| NPDES | National Pollutant Discharge Elimination System |
| NRHP | National Register of Historic Places |
| NSB | National Science Board |
| NSBP | National Society of Black Physicists |
| NSHP | National Society of Hispanic Physicists |
| NSF | National Science Foundation |
| NSU | Northern State University |
| NTN | Northern Tier Network |
| NTP | Network Time Protocol |
| NuMI | Neutrinos at the Main Injector |
| NUSL | National Underground Science Lab |
| NVMS | Network Video Management System |
| NVR | network video recorder |
| O&M | Operations and Maintenance |
| OBS | Organizational Breakdown Structure |
| OCIP | Owner's Controlled Insurance Program |
| ODH | oxygen deficiency hazard |
| OHEP | Office of High Energy Physics |
| OHN | occupational health nurse |
| OLI | Oppenheim Lewis Inc. |
| OLR | other levels and ramps |
| OMB | Office of Management and Budget |
| OMS | Outage Management System |
| ONP | Office of Nuclear Physics |
| OPERA | Oscillation Project with Emulsion-tRacking Apparatus |
| ORP | oxidation-reduction potential |
| OSG | Open Science Grid |
| OSCARS | On-demand Secure Circuits and Advance Reservation System |
| OSHA | Occupational Safety and Health Administration |
| OSTP | Office of Science and Technology Policy |
| OTV | optical televiewer |
| P5 | Particle Physics Project Prioritization Panel |
| P&P | policies and procedures |
| PAC | Program Advisory Committee |
| PAN | Physics of Atomic Nuclei |
| PASAG | Particle Astrophysics Scientific Assessment Group |
| PBS | Project Breakdown Structure |
| PC | personal computer |
| PCB | polychlorinated biphenyl |
| PCM | Project Controls Manager |
| PDA | Property Donation Agreement |
| PDR | Preliminary Design Report |
| PE | photoelectron |
| PEP | Project Execution Plan |
| PFA | peak friction angle |
| PGA | peak ground acceleration |
| PHA | Preliminary Hazard Analysis |
| PHY | NSF Division of Physics |
| PI | Principal Investigator |
| PICASSO | Project In CAnada to search for SuperSymmetric Objects |
| PI/ED | Principal Investigator/Executive Director |
| PI/LD | Principal Investigator/Laboratory Director |
| PLC | Programmable Logic Controller |
| PMB | Performance Measurement Baseline |
| PMCS | Project Management Control System |
| PMT | photomultiplier tube |
| PO | Project Office |
| POD | Project and Operations Director |
| PODS | Petrology, Ore Deposits, and Structure |
| POT | protons on target |
| PP | Planning Package |
| PPE | personal protective equipment |
| ppt | parts per trillion |
| PPV | peak particle velocity |
| psi | pounds per square inch |
| PTFE | polytetrafluoroethylene |
| PTP | precision time protocol |
| PTZ | pan-tilt-zoom |
| PV | Planned Value |
| QA | quality assurance |
| QAM | Quality Assurance Manager |
| QASP | Quality Assurance Surveillance Plan |
| QC | quality control |
| QUPID | QUartz Photon Intensifying Detector |
| RAM | Responsibility Assignment Matrix |
| R&D | research and development |
| R&E | research and education |
| RBC | rotating biological contactor |
| RBS | Resource Breakdown Structure |
| RCRA | Resource Conservation and Recovery Act |
| RCS | radio communications system |
| REED | Research, Education, and Economic Development |
| RETC | Rapid Excavation and Tunneling Conference |
| RFID | radio frequency identification |
| RFP | Request for Proposal |
| RG | red giant |
| RH | relative humidity |
| RHEL | Red Hat Enterprise Linux |
| RMP | Risk Management Plan |
| RMR | Rock Mass Rating |
| RMT | Risk Management Team |
| RO | reverse osmosis |
| ROI | region of interest |
| RQD | Rock Quality Designation |
| R&RA | Research and Related Activities |
| RU | Regis University |
| S1 | Solicitation 1 (NSF) |
| S2 | Solicitation 2 (NSF) |
| SACNAS | Society for Advancement of Chicanos and Native Americans in Science |
| SCADA | supervisory control and data acquisition |
| SCBA | Self-Contained Breathing Apparatus |
| SCCC | Satellite Command and Control Center |
| SCS | structured cabling system |
| SCSE | Sanford Center for Science Education |
| SD | spin dependent; South Dakota |
| SD DENR | South Dakota Department of Environment and Natural Resources |
| SD GEARUP | South Dakota Gaining Early Awareness and Readiness for Undergraduate Programs |
| SDN | South Dakota Network |
| SD SHPO | South Dakota State Historic Preservation Office |
| SDSM&T | South Dakota School of Mines and Technology |
| SDSTA | South Dakota Science and Technology Authority |
| SDSU | South Dakota State University |
| SE | Systems Engineering |
| SEI | Software Engineering Institute |
| SEMP | Systems Engineering Management Plan |
| SHM | structural health monitoring |
| SHPO | State Historic Preservation Office |
| SI | spin independent |
| SIP | Systems Integration Plan |
| SLAC | Stanford Linear Accelerator Center |
| SME/AIME | Society of Mining Engineers/American Institute of Mining, Metallurgical and Petroleum Engineers |
| SMTP | Simple Mail Transfer Protocol |
| SOP | Standard Operating Procedure |
| SOW | Statement of Work |
| SPCC | Spill Prevention Control and Countermeasure |
| SPD | Scientific Programs Director |
| SPI | Schedule Performance Index |
| SR | stress ratio; strength ratio |
| SRF | stress reduction factor |
| SRN | supernova relic neutrinos |
| SSC | Superconducting Super Collider |
| SSI | Sustainable Sites Initiative |
| SSM | standard solar model |
| STEM | Science, Technology, Engineering, and Mathematics |
| Sub-PI | Sub-Principal Investigator |
| Super-K | Super-Kamiokande |
| SUSY | supersymmetry |
| SV | schedule variance |
| SWOT | strengths, weaknesses, opportunities, and threats |
| SWP | Safe Work Permit |
| TBD | to be determined |
| Tbps | terabits per second |
| TBR | to be reviewed/refined/revised |
| TBX | Generic designation that refers to either a "to be reviewed/refined/revised" (TBR) or a "to be determined" (TBD) requirement |
| TCP/IP | Transmission Control Protocol/Internet Protocol |
| TDS | total dissolved solids |
| THMCB | Thermal-Hydrological-Mechanical-Chemical-Biological |
| TPC | time projection chamber |
| TPM | Time Projection Modules |
| Tr1 | Tertiary rhyolites |
| TVSS | transient voltage surge suppression |
| T2KK | Tokai to Kamioka and Korea |
| UC Berkeley | University of California at Berkeley |
| UCD | University of California at Davis |
| UCS | uniaxial compressive strength |
| UGI | Underground Infrastructure |
| UGL | Underground Laboratory |
| UNC | University of North Carolina at Chapel Hill |
| UND | University of Notre Dame |
| UPS | uninterruptible power source |
| USD | University of South Dakota |
| USGBC | U.S. Green Building Council |
| USGS | United States Geological Survey |
| USO | User Support Office |
| UTK | University of Tennessee, Knoxville |
| UTRC | Underground Technology Research Council |
| UU | University of Utah |
| V | volt |
| VCRM | Verification Cross-Reference Matrix |
| vDUSEL | Virtual DUSEL |
| VE | Value Engineering |
| VEMP | Value Engineering Management Plan |
| VFD | variable frequency drive |
| VLAN | Virtual Local Area Network |
| VMS | virtual matrix switches |
| VOC | volatile organic compounds |
| VOIP | Voice Over Internet Protocol |
| VRT | virgin rock temperature |
| VSS | Video Surveillance System |
| VUV | vacuum ultraviolet |
| WAN | wide area network |
| WARP | WIMP Argon Programme |
| WBS | Work Breakdown Structure |
| WCD | water Cherenkov detector |
| WCE | water Cherenkov equivalent |
| WCh | water Cherenkov |
| WIMAX | Worldwide Interoperability for Microwave Access |
| WIMP | Weakly Interacting Massive Particle |
| WIPP | Waste Isolation Pilot Plant |
| WMU | West Michigan University |
| WWTP | Waste Water Treatment Plant |
| WWW | World Wide Web |
| ZEPLIN | ZonEd Proportional scintillation in LIquid Noble gases |

# List of Appendices

## Volume 1

## Volume 2

## Volume 4

## Volume 5

## Volume 6

## Volume 7

## Volume 8

## Volume 9

## Volume 10

## Executive Summary

The DUSEL Project has produced the Preliminary Design of the Deep Underground Science and Engineering Laboratory (DUSEL) at the rehabilitated former Homestake mine in South Dakota. The design satisfies the requirements of the project *Readiness Stage* detailed in the National Science Foundation's (NSF's) *Large Facilities Manual* (NSF 10-012). The Facility design calls for, on the surface, two new buildings—one a visitor and education center, the other an experiment assembly hall—and multiple repurposed existing buildings. To support underground research activities, the design includes two laboratory modules and additional spaces at a level 4,850 feet underground for physics, biology, engineering, and Earth science experiments. On the same level, the design includes a Department of Energy (DOE)-shepherded Large Cavity supporting the Long Baseline Neutrino Experiment (LBNE). At the 7,400-feet level, the design incorporates one laboratory module and additional spaces for physics and Earth science efforts. All underground areas will be connected by a distributed-access network extending from the surface down to a depth of 2.25 km and extending over 30 km. With input from some 25 science and engineering collaborations, the Project has designed critical experimental space and infrastructure needs, including space for a suite of multidisciplinary experiments in a laboratory whose projected life span is at least 30 years. From these experiments, a critical suite of experiments is outlined, whose construction will be funded along with the facility. The putative users, with funding independent of the Project, are advancing their designs and pursuing research and development on detector technology. The Facility design permits expansion and evolution, as may be driven by future science requirements, and enables participation by other agencies. The design leverages South Dakota's substantial investment in facility infrastructure, risk retirement, and operation of its Sanford Laboratory at Homestake. The management structure planned under the Preliminary Design Report encourages both participation in and growth of DUSEL's construction and science programs by universities in South Dakota and around the region.

## Intellectual Merit/Broader Impacts

DUSEL's intellectual merit and broader impacts are closely connected, coupled to user research and to education and public outreach. The Project has designed an education and public outreach center, is planning programs, and has initiated efforts to establish regional partnerships with underserved populations—regional American Indian and rural populations. The DUSEL Project enhances South Dakota and regional university participation in world-leading research and proposes major construction activities for these and other South Dakota research entities.

DUSEL's physics efforts will advance knowledge of and understanding of dark matter; neutrinoless double-beta decay; leptonic violations of fundamental symmetries; neutrino properties; proton decay lifetime; and element formation in stellar interiors. Results from Earth science and engineering research will advance understanding of the processes that shape the surface of the Earth, knowledge of subsurface life, and improvements in carbon sequestration technology.

DUSEL presents NSF and DOE an opportunity for interagency cooperation on major science projects already started with the existing DUSEL Joint Oversight Group. DUSEL'S innovative monitoring devices and excavation techniques may be useful to commercial mining excavation and civil construction.

Addressing underground activities hazards, the Project, with the South Dakota Science and Technology Authority, has crafted Environment, Health, and Safety programs to ensure an early rollout of many elements of the Integrated Safety Management system of general use for any underground work.

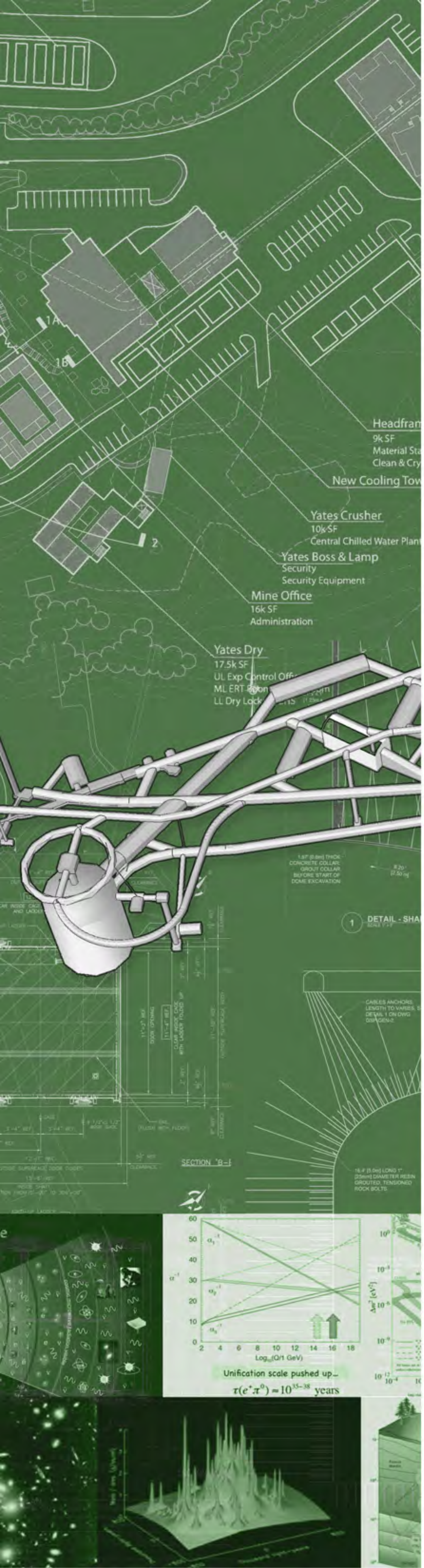

# Preliminary Design Report

May 2011

# Volume 1:
# Project Overview

DUSEL
Deep Underground
Science and
Engineering Laboratory

# Project Overview

## Volume 1

### 1.1 Introduction

This Preliminary Design Report (PDR) summarizes the design of the National Science Foundation's (NSF's) Deep Underground Science and Engineering Laboratory (DUSEL). The design incorporates and enables both a multidisciplinary suite of world-class experiments and the Facility to support these experiments. All of the envisioned experiments share the requirement of using underground space to conduct their activities. The experiments are drawn from nuclear and high energy physics, Earth sciences, biology, and engineering. In addition to synergistic research opportunities, the research and engineering applications will participate in a comprehensive education and public outreach program.

The scientific and engineering program hosted in DUSEL will be of the highest caliber, with major discovery potential not in one but in five or six different experiments. Within DUSEL's suite of experiments lies the potential for revolutionizing our understanding of the physical universe, significantly extending the theories describing our universe, and pursuing multiple research topics beyond our current understanding. The discoveries sought by DUSEL's suite of critical experiments will transform our understanding of the underpinning of physics, biology, and Earth science. Within the physics experiments are at least four experiments, each worthy of the highest academic honors—be it identifying the mysterious dark matter that makes up ~25% of the universe; unambiguously observing neutrinoless double beta—one of nature's rarest decays—which would establish the particle-antiparticle nature of the neutrino; observing proton decay and establishing the ultimate instability of matter and the theories that describe the ultimate fate of matter; and completing our understanding of neutrino oscillations and perhaps through the observation of violation of fundamental symmetries (Charge and Parity) helping to establish the origins of the matter-antimatter asymmetry in the universe. Within the Earth sciences the goals are to isolate and examine rare life-forms in the subterranean environment; to make significant steps in completing our understanding of the tree of life, not just on the surface but extending our understanding to the limits of life; to attack problems of the highest importance to our society's economic well-being, including problems of carbon sequestration, understanding large-scale subterranean excavation, and the interactions of the processes in the subsurface; and to examine the complex interactions in the underground from the nanometer scale to the kilometer scale involving thermal, mechanical, hydrological, biological, and chemical processes. Each of these endeavors is capable of transforming our understanding of the universe, rewriting the textbooks and curricula across the world's universities. Individually, these endeavors would establish true scientific leadership to the U.S. science and engineering programs. As an integrated suite of experiments in a single laboratory, DUSEL provides an unequalled opportunity to transform our understanding of the physical universe and propel the United States into a world leadership across multiple disciplines, enabling the nation to benefit from multiple scientific revolutions and the associated societal impacts. These experiments are presented in Figure 1.1-1.

The Project's overarching goal is to develop an enduring international underground laboratory with a best-in-world-class scientific program of research, education, and outreach and to do so as quickly and as cost-efficiently as is consistent with the highest level of safety.

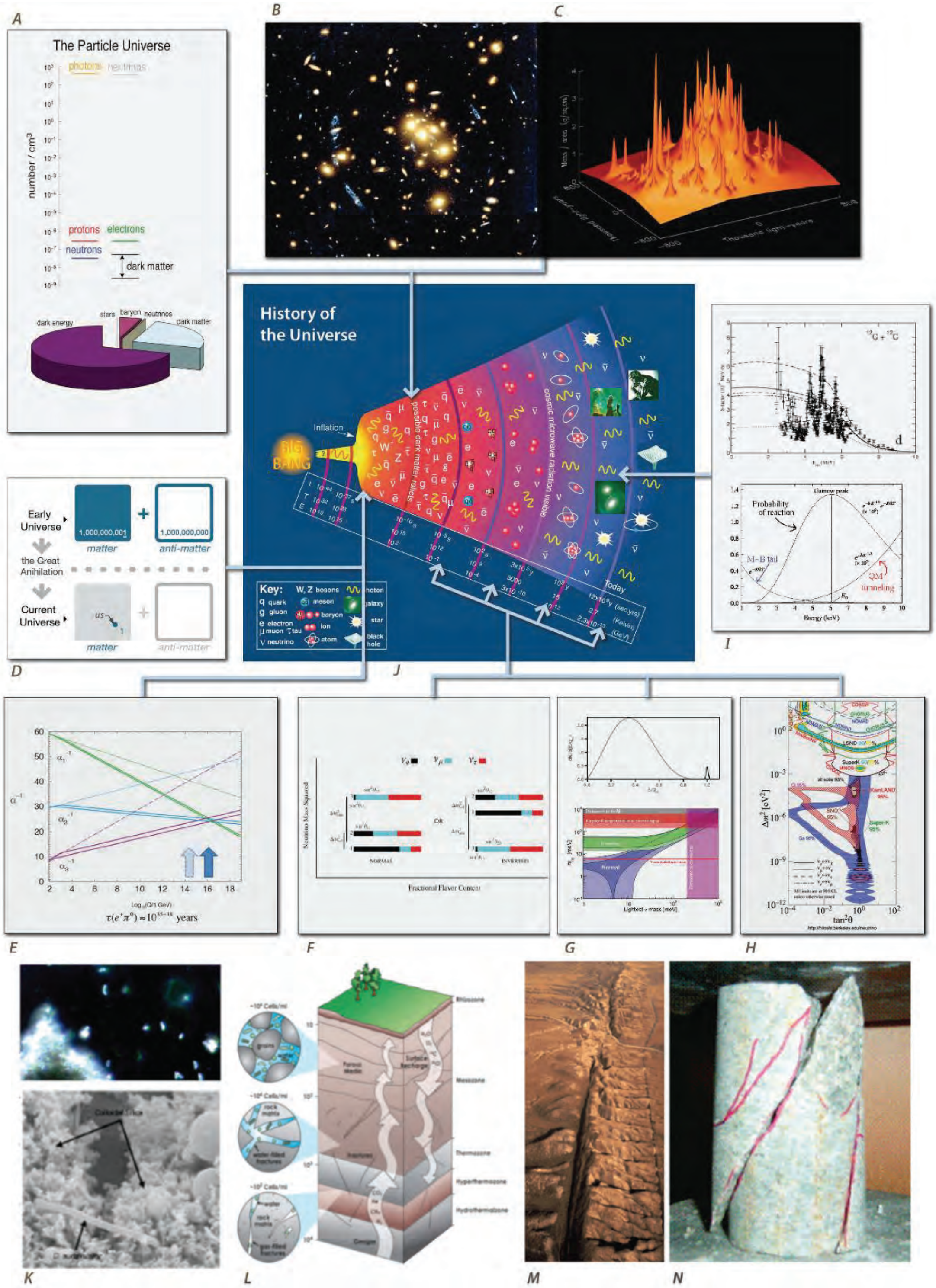

**Figure 1.1-1** These panels present the discovery potential for DUSEL in physics and Earth sciences. DUSEL proposes to support experiments in Dark Matter (A, B, and C), Neutrinoless Double Beta Decay (F and G), Nuclear Astrophysics (I), Proton Decay (E), and Long Baseline Neutrinos (D, F, and H). This panel maps the significance of these experiments to the understanding of our physical universe and the history of the universe (J). The Facility will host a diverse program in Earth sciences, including geomicrobiology (K), fault rupture (M and N), excavation monitoring, coupled processes (L), and seismic monitoring arrays (M). This panel represents how the Earth science facilities will pursue these disciplines, making use of the DUSEL Facility's access to an extensive subsurface facility. [Graphics provided by DKA; Particle Data Group, LBNL; Deep Science Report[1]; and Hitoshi Murayama, UC Berkeley]

Initial ideas for a dedicated underground research laboratory were developed and proposed in the 1950s and 1960s. Following significant scientific discoveries in the 1990s and early 2000s, NSF launched a rigorous process to evaluate the scientific potential for a dedicated deep-underground laboratory and to establish a structured approach for evaluating potential sites. The process was aimed at establishing the general requirements for DUSEL, updating elements of the scientific case for the most compelling experiments, and establishing elements of the Conceptual Design for the Facility. NSF developed a series of solicitations to evaluate and document the scientific needs for an underground laboratory and to develop Facility requirements. The first solicitation (S1) defined the site-independent assessment of underground science and explored Facility requirements. This effort is summarized in the NSF publication *Deep Science.*[1] A variety of collaborations, including the site of the former Homestake Gold Mine in Lead, South Dakota, received support to develop general Conceptual Designs exploring a wide spectrum of environments, access options, and organizational arrangements. The S1 solicitation was followed by Solicitations 2 to 4, where the second solicitation (S2) enabled collaborations throughout North America to propose their sites for consideration, the third solicitation (S3) provided the down-select to the Homestake site, and the fourth solicitation (S4) provided funding to individual collaborations to develop designs for their experiments.

The Homestake site was selected for design development following initial conceptual development supported under S2 and evaluation by NSF expert review panels. The evaluation included both visits to the proposed sites and reverse site visits by the collaborations to NSF. The development of the Preliminary Design was created with cooperative agreements between NSF and the University of California at Berkeley (UC Berkeley) following the third solicitation.

This design proposes the former Homestake Gold Mine as the site for DUSEL. The UC Berkeley team is a collaborative effort working closely with South Dakota government and university entities, in particular the South Dakota School of Mines and Technology (SDSM&T). The state of South Dakota established the South Dakota Science and Technology Authority (SDSTA) to facilitate the development of Homestake DUSEL as well as to advance higher education and technology activities in the region.

The SDSTA received title to the Homestake site in 2006 from the Barrick Gold Corporation (Barrick) after the 2003 closure of the mining Facility. The closure of the Homestake Facility was conducted and documented under a comprehensive Mine Closure Plan monitored by the Environmental Protection Agency and others. Reports documenting compliance with the Closure Plan are included in supporting material.[2] The conditions of the site donation are detailed in the Property Donation Agreement (PDA) between Barrick and the SDSTA. The property donation includes 186 surface acres with about 65 buildings, and mineral rights to over 7,000 subsurface acres. The site includes nearly 600 km of existing drifts, ramps, tunnels, and shafts, although the design of the DUSEL Facility limits the actual footprint to a much smaller area to reduce operating and construction costs. The Surface Campus includes the buildings and infrastructure to support surface and underground operations, including hoisting facilities for the (existing) Yates and Ross Shafts extending from the surface (absolute elevation about 5,300 feet above sea level) to the 4850L (feet below Yates Shaft ground level). The deeper campus extending from the 4850L to the 8000L is accessed through the #6 Winze and a new winze to be installed between the 7400L and the 4850L. The vision of the proposed laboratory is shown in Figure 1.1-2, and illustrates the access shafts and the proposed research campuses. Figure 1.1-3 is a cross section of the Facility and presents the approximately 60 levels in the Facility and the additional raises, ramps, and existing infrastructure.

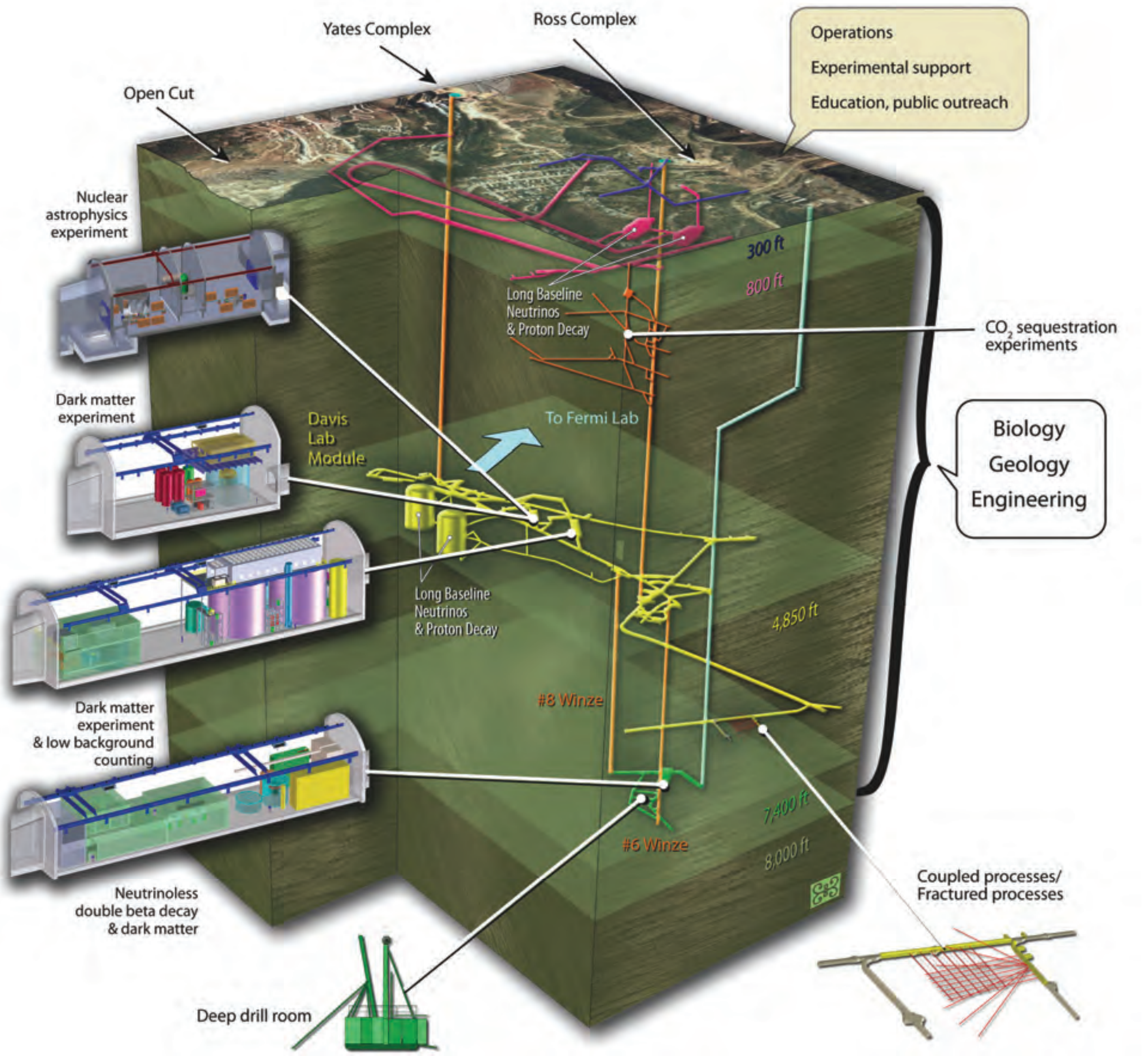

**Figure 1.1-2** Graphic representation of the Deep Underground Science and Engineering Laboratory, illustrating the different research campuses and concepts for the experimental deployments. [Graphic provided by Zina Deretsky]

The SDSTA, using a U.S. Department of Housing and Urban Development (HUD) grant as well as state and private philanthropic funding, has stabilized the site, re-established access to the underground, and re-established pumping of the accumulated water from the underground, including treatment and purification of the water. The rehabilitation efforts began in mid-2007. The initial Ross Shaft rehabilitation and pump refurbishment reached the level of the accumulated water mid-2009. The water, which had reached 4,529 feet below the collar in August 2008, has been subsequently pumped below the 5,331-foot level by January 1, 2011. Disposal of the water from the underground meets all applicable rules, regulations, and permit requirements. The impacts of flooding the 4850L have been mitigated. Significant multiple infrastructure and safety enhancements were identified and installed, including removing much of the unused mining infrastructure in the Yates and Ross Shafts, and designing and installing safety-related infrastructure. The Ross and Yates Shaft access and hoisting infrastructure has been upgraded to a level appropriate for the Final Design, maintaining the Facility and providing safe access from the surface to the 5000L. The Davis Laboratory Module (DLM), which housed the 2002 Nobel Prize-winning solar neutrino research of Dr. Ray Davis, has been expanded and additional infrastructure installed to support physics experiments. Near the DLM, a new hall (Davis Transition Area [DTA]) (135 feet x 50 feet) has been excavated to support several physics experiments. A collection of biology, geology, and engineering collaborations has initiated research at the Sanford Lab, an underground laboratory at the 4850L that was created and supported by the SDSTA using state-controlled funds. Working with SDSTA personnel, the DUSEL Project has crafted, tested, and refined an appropriate Environment, Health, and Safety (EH&S) program of Integrated Safety Management. The Davis Campus and additional space in the underground will be used for hosting science throughout the Major Research Equipment and Facility Construction (MREFC) construction period and into Facility operations.

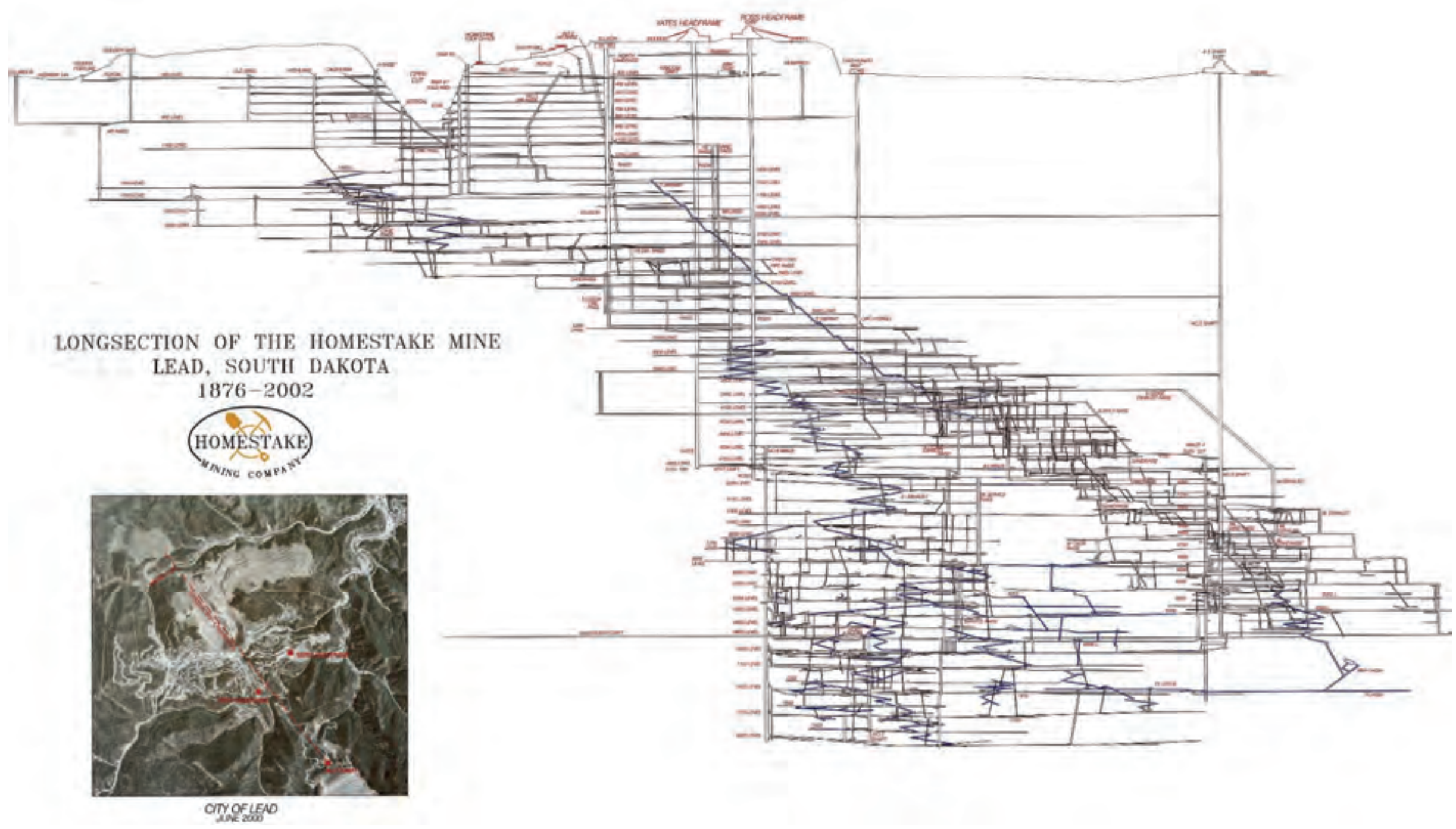

**Figure 1.1-3** The long section of the former Homestake Gold Mine. This figure illustrates the 60 underground levels extending to greater than 8,000 feet below ground. The location of cross section is indicated in the inset along a NW to SE plane. The projection extends for 5.2 km along this plane. [Graphic provided by Homestake Mining Company]

The starting conditions for the DUSEL Final Design effort are defined by 1) the SDSTA reentry and rehabilitation efforts; 2) infrastructure improvements and completion of deferred maintenance by SDSTA staff and the DUSEL Project team; 3) the documented assessment, inspection, and analysis of the existing Sanford Laboratory buildings and infrastructure accomplished by SDSTA and DUSEL teams; 4) DUSEL design efforts creating the Conceptual and Preliminary Designs; and 5) those additional design efforts, critical safety system refurbishment, and upgrades proposed in FY 2011 with the Transitional Funding Proposal.

Working with over 25 collaborations (consisting of more than 700 scientists and engineers), the Project has developed facility requirements crafted from a suite of critical, multidisciplinary experiments. These requirements have in turn been used to develop the design of research spaces, laboratories, and associated infrastructure.

The current design presents the Facility at the Preliminary Design level (about 25-30% construction ready), identifies a suite of compelling and transformational experiments to be hosted in the Facility, and gives estimates of the operational requirements for this dramatic world-class research laboratory.

## 1.2 Project Overview

Responding to community interest and scientific opportunities, NSF established a process to evaluate, design, engineer, and potentially construct a deep underground laboratory. DUSEL will be constructed with NSF support from the MREFC account. The DUSEL proposal embodies a world-class facility closely coupled with a multidisciplinary suite of transformational experiments and applications. UC Berkeley is leading the design of the DUSEL Facility and the integration of a suite of experiments into the Facility. DUSEL involves significant participation by multiple science agencies in the United States, notably NSF and Department of Energy (DOE). The DOE (Office of High Energy Physics [OHEP] and Office of Nuclear Physics [ONP]) and NSF have established a Joint Oversight Group (JOG) to define and manage their cooperative participation in the DUSEL Facility and its suite of experiments. The agency participation in DUSEL will be formalized in the Memorandum of Understanding (MOU) between NSF and DOE. It is anticipated that in the future, the JOG's scope would be expanded to encompass the biology, geology, and engineering efforts as well.

The joint agency participation in the experimental program is modeled on the successful NSF-DOE partnership for the Large Hadron Collider program at the European Council for Nuclear Research (CERN). For each discipline, the joint participation will be defined and managed within the JOG. Joint agency participation in DUSEL and its suite of experiments will distribute the construction responsibilities among the agencies while assuring full and unrestricted access to the scientific data and shared responsibility for managing the scientific analysis and publication of all results.

The DUSEL design team is acquiring and managing facility requirements gathered from approximately two dozen DOE- and NSF-supported collaborations. The NSF-supported collaborations, funded with the fourth solicitation (S4), are listed in Table 1.2-1. A dedicated DUSEL Facility design is being developed to host a suite of world-leading experiments. The experimental disciplines span high energy and nuclear physics, biology, engineering, and Earth sciences. In many cases and disciplines, technology choices are yet to be made and collaborations are still developing. Consequently, the Facility design is being developed without a specific experiment or collaboration having been selected, with the exception of DOE's Long Baseline Neutrino Experiment (LBNE), which received Critical Decision 0 (CD-0) by DOE in January 2010. In several examples, efforts of multiple collaborations are being supported to pursue alternative technology options. The selection of specific collaborations and technologies will be made as the experimental designs advance and mature and as essential R&D activities are completed. The Facility design is capable of hosting any of the technologies or collaborations, and the design team will continue to interact and liaise with the collaborations as their designs mature. The DUSEL Project refers to the collection of experiments used as the basis of design for the Facility, therefore, as the "generic suite of experiments," stressing that the final selections have not been made while maintaining the greatest potential for discovery and major revolutions in a variety of scientific fields. In some cases, the experimental selection process will extend beyond the anticipated Facility construction start and will conclude with these experiments being installed in the last years of the DUSEL Facility construction project.

Community input and interest in the suite of experiments was informed by workshops, letters of interest, long-range plans, National Academy studies, and interagency studies. Additional guidance from funding agencies, the DUSEL Program Advisory Committee (PAC), and funding agency advisory panels[3] indicates that within the physics experiments are four essential experimental pillars—Long Baseline Neutrino Experiment, Proton Decay, Neutrinoless Double-Beta Decay, and Dark Matter Searches.

DUSEL's proposed suite of experiments includes additional well-motivated physics experiments and a selection of multidisciplinary uses drawn from biology, geology, and engineering (BGE) collaborations (see Volume 3, *Science and Engineering Research Program*). The DUSEL proposal includes a well-integrated Education and Public Outreach component (see Volume 4, *Education and Public Outreach*).

The suite of experiments used to guide the Facility design (Table 1.2-1) consists of:

- **Dark Matter Searches**: one or more Generation Three (G3) experiments, and G2 experimental and Facility support
- **Neutrinoless Double Beta Decay Searches**: one G2 experiment as well as facility support for a G1 effort
- **Biology, Geology, and Engineering Experiments**: support for multiple experiments
- **LBNE and Proton Decay Searches**: support for the program defined by DOE's LBNE project and its CD-0 (~200 kT of water Cherenkov equivalent detector mass)
- **Nuclear Astrophysics Experiments**: support for phased nuclear astrophysics accelerator-based experimental program
- **Low-Background Counting and Material Assay Efforts**: support for advanced low-background assay and materials selection

The DUSEL Facility, presented in Volume 5, *Facility Preliminary Design,* is being designed to support this suite of critical experiments and consists of:

- A **Surface Campus** capable of supporting staging and pre-assembly of experiments, and later operations of experiments, and the Education and Public Outreach efforts
- An underground research campus at the 4850L, the **Mid-Level Laboratory**, with two laboratory modules; one Large Cavity, stewarded by the DOE's LBNE project; and use of the existing Sanford Laboratory's **Davis Laboratory Module** and **Davis Transition Area**. BGE experiments would be supported on the 4850L.
- An underground research area at the 7400L, the **Deep-Level Laboratory**, with one new laboratory module. BGE experiments would be supported on the 7400L.
- BGE experiments would also be supported using the Facility-wide network of ramps, drifts, and shafts, the **Other Levels and Ramps**.
- Potential locations for additional LBNE detector sites are being evaluated at several underground locations in collaboration with the DOE's LBNE project.

The Facility described above to support this suite of experiments provides EH&S systems; utilities, including ventilation, air conditioning, power, communications, water; dual access and egress for research campuses; basic laboratory outfitting; and stabilized underground spaces. Approximately 1.5 M tonnes of rock will be excavated, producing ~35,000 $m^2$ of new underground laboratory and associated support space, not including the existing system of drifts and ramps extending over 30 km.

| Collaboration | Principal Investigator | Institution |
|---|---|---|
| **Physics** | | |
| **Dark Matter** | | |
| MAX | Galbiati | Princeton University |
| LZ20 | Shutt | Case Western Reserve University |
| GEODM | Golwala | Caltech |
| COUPP | Collar | University of Chicago |
| **Neutrinoless Double Beta Decay** | | |
| EXO | Gratta | Stanford University |
| GE1T | Wilkerson | University of North Carolina |
| **Long Baseline Neutrinos and Proton Decay** | | |
| LBNE Collaboration | Svoboda | University of California at Davis |
| LBNE Project | Strait | Fermilab |
| **Nuclear Astrophysics** | | |
| DIANA | Wiescher | University of Notre Dame |
| **Advanced Assay and Ultrapure Materials** | | |
| FAARM | Cushman | University of Minnesota |
| **Biology, Geology, and Engineering Experiments** | | |
| Transparent Earth | Glaser | UC Berkeley |
| Fiber-Optic Array | Wang | University of Wisconsin |
| Fault Rupture | Germanovich | Georgia Tech |
| Coupled Processes (THMC) | Sonnenthal | UC Berkeley and Lawrence Berkeley National Laboratory |
| EcoHydrology | Boutt | University of Massachusetts |
| Excavation Monitoring | Bobet | Purdue University |

**Table 1.2-1** S4-supported collaborations providing support to DUSEL.

The Facility design includes a Large Cavity accessed from the 4850L. Because of its size and depth, this cavity represents a technically challenging aspect of the experimental facility design. The DUSEL Project, therefore, advanced significant Preliminary Design elements for the Large Cavity. DOE is responsible for completing the design, management, and construction of this cavity. The Large Cavity costs are not included in the DUSEL MREFC proposal. The DUSEL schedule reflects the DOE-led efforts to construct the cavity in concert with the DUSEL Facility. There is a contribution to the LBNE effort included within the science partition of the proposed MREFC funding and it is defined in the DUSEL agency stewardship model.

In parallel with the development of the Facility design, the Project has created an estimate for operations and maintenance for the Facility and the experimental programs. This estimate was created with input from the original Homestake Mining Company operations and maintenance requirements, SDSTA activities at Sanford Lab, and analyses of other underground research facilities, including Gran Sasso, SNOLab, Neutrinos at the Main Injector (NuMI), Waste Isolation Pilot Plant (WIPP), and Kamioka.

To coordinate and manage Project activities across multiple partners (UC Berkeley and the South Dakota entities of SDSTA and the SDSM&T), UC Berkeley and the Project have taken steps to create and empower a Limited Liability Company in South Dakota (DUSEL LLC) to oversee the Final Design, Construction, and Operations on the behalf of UC Berkeley. The Regents of the University of California and the Governor of South Dakota would appoint the Board of Directors of the DUSEL LLC. The formation of the DUSEL LLC is not a requirement for the advancement of DUSEL's design, and current Project participants would be able to maintain the present organization should the DUSEL LLC development require additional work.

The details of the DUSEL Project Execution Plan and Project Management Systems are discussed in detail in Volumes 7 and 8, respectively. The DUSEL Project is following standard large-project protocols to execute the design and construction projects as well as to manage the overall Project. The DUSEL Project reports to the Vice Chancellor for Research at UC Berkeley and benefits from strong leadership and support from the UC Berkeley campus. The Project collaborates with South Dakota entities, including the SDSM&T and the SDSTA. The Project receives significant benefits from a close relationship with Lawrence Berkeley National Laboratory (LBNL) for engineering, scientific, and project management staff as well as for project management systems.

In advance of the DUSEL LLC-managed activities, the Project and SDSTA have established a plan and taken steps to merge activities and staff into a single organization. The SDSTA's Sanford Laboratory mission is to facilitate the creation of DUSEL. This includes supporting basic operations of the Homestake site to enable safe access to the underground and to stabilize the Facility, thus enabling the design and planning for the DUSEL Facility and its suite of experiments. The Project schedule calls for the complete integration of activities within the DUSEL Project, with the exception of operation of the Sanford Laboratory scientific program within the Davis Campus. These activities would continue to be managed by SDSTA, and would ultimately become integrated into DUSEL in the later stages of the DUSEL Final Design. SDSTA would continue to serve as landlord for the Facility and would hold title to the surface and underground real estate, ensuring the conditions established in the Barrick PDA are met and requirements of philanthropic gifts are satisfied, Notably the SDSTA has received a $70 million donation from T. Denny Sanford to facilitate the creation of the Sanford Laboratory and the associated Sanford Center for Science Education (see Volume 4).

The total DUSEL MREFC-funded Project will request $875 million (FY 2010$). The Facility cost is estimated at $575 million, including a ~35% project-wide contingency on the facility elements. The NSF MREFC budget for the suite of experiments is $300 million (FY 2010$), which includes a $125 million contribution for LBNE. Additional support for DUSEL science programs would be provided by DOE, according to the stewardship model of the experimental program, and by non-U.S. or other sources.

The DUSEL MREFC-funded construction would span eight years beginning in FY 2014, but would not be technically limited to this start date. This schedule begins with major facility construction activities and introduces support for the science programs as they are ready for construction funding and are phased to match beneficial occupancy of the Facility. The MREFC-funded facility construction would start and complete before the MREFC-funded experiment construction would be complete. The schedule reflects approximately 18 months of schedule contingency.

Within the DUSEL MREFC budget, funds for the Facility construction will be partitioned from those supporting DUSEL's scientific program. Allocations for specific science construction projects will be confirmed and codified by the JOG. The funds supporting scientific instrumentation will be made available as the experiments are ready for construction and pass the necessary review requirements. The DUSEL Project proposes that an initial additional partitioning of the science support component of the MREFC funding be imposed early in the process to preserve funding for those experiments installed later in the construction period. The size of the scientific program and the partitioning of the $300 million contribution to the science programs must be informed by participating agencies' evaluation process. The results of these processes will be formalized by the JOG. To facilitate the establishment of the MREFC budget, the DUSEL Project and the scientific collaborations have assembled approximate cost ranges and schedules for the science programs. Additional effort will be required before these estimates would be usable by the DOE O-413.3A process; however, they are sufficient for establishing the MREFC budget. Changes to the MREFC budget partition functions would require consent of the JOG.

Table 1.2-2 presents the assembled cost ranges as well as the MREFC budget assignments for the research efforts that make up the DUSEL scientific scope.

| Science Goal | Total Estimated Experimental Cost Range* ($M) | Proposed MREFC Budget Contribution ($M) | Number of Deployments |
|---|---|---|---|
| Dark Matter (per experiment) | 80 - 100 | 175 | ≥ 1 |
| 0νββ (per experiment) | 220 - 300 | | 1 |
| Bio/Geo/Eng | 60 - 180 | | multiple |
| Nuclear Astrophysics Facility | 30 - 45 | | 1 |
| Adv. Low-Background & Assay | 2 - 15 | | 1 |
| Long Baseline Neutrinos & Proton Decay‡ | 785 - 1065 | 125 | 200 kT WCE |

*These cost ranges are not to be confused with or substituted for DOE CD estimates.

‡LBNE CD-0 range includes beam, near detector, and far detectors; this range includes MREFC-budget contribution.

**Table 1.2-2** The DUSEL scientific program, estimated experimental cost ranges, proposed MREFC-only budget contributions to each of these efforts, and the estimated number of experimental deployments within DUSEL. Cost ranges include a rough estimation of contingency at ~50% level. Estimated costs are in FY 2010 million dollars.

## 1.3 Project Evolution

The development of the DUSEL Facility follows the steps presented in the NSF *Large Facilities Manual*.[4] With an award to UC Berkeley in 2005 (PHY0528103) resulting from the submission to NSF Solicitation 05-506, elements of the Conceptual Design were developed and summarized in an initial report, *The Deep Underground Science and Engineering Laboratory at Homestake: Conceptual Design Report*, dated January 9, 2007.[2] This Conceptual Design Report (CDR) served as the basis for the selection of the Homestake site and the UC Berkeley team by the comprehensive peer-review process, involving both site visits and reverse site visits. These concepts were subsequently developed with Cooperative Agreements (CA) PHY0717003, PHY0938228, PHY0940801, and PHY1059670 (NSF Solicitation 06-614) between NSF and UC Berkeley.[5] The DUSEL Project exited the Conceptual Design phase and entered the Readiness stage, as defined in the *Large Facilities Manual*, in 2009, following the annual review of the Project and receiving funding of Cooperative Agreement, CA2. This document represents the culmination of these efforts to create and document the Preliminary Design of the DUSEL Facility and its suite of critical, world-class experiments.

The DUSEL Project has undergone frequent and comprehensive reviews by NSF-organized review panels. DOE has frequently sent observers to Project reviews. In addition to the comprehensive annual review of Project progress, NSF has established focused progress reviews to analyze a subset of the Project's activities. Several focused reviews have concentrated on safety systems and infrastructure for both DUSEL and SDSTA at Sanford Laboratory activities. Although the SDSTA effort originally held independent reviews of its efforts at Sanford Laboratory, at this point the reviews and advisory committees are fully integrated with DUSEL reviews. Reports from these committees have been incorporated into the review process.

The Project has established a comprehensive annual internal review as well as establishing advisory and consultation committees and boards:

**Internal Review Committee** (IRC) regularly reviews the Project progress, proposals, and major reports. The IRC meets annually.

**Large Cavity Advisory Board** (LCAB) provides expert advice on design and construction of large-scale excavations.

**Infrastructure Advisory Board** (IAB) provides expert advice on the necessary infrastructure to support the operations, including hoists, conveyances, and safety systems.

**Environmental Health and Safety Oversight Committee** (EHSOC) provides UC Berkeley's Vice Chancellor for Research with high-level advice and an assessment of the Project's progress in creating the necessary EH&S systems and programs. The EHSOC was initiated in 2010 and replaced the earlier Homestake EH&S advisory committee. The EHSOC held its first full meeting in August 2010.

**Cultural Advisory Committee** (CAC) provides insight and recommendations on engaging rural and underserved populations, as well as suggesting paths to maximize regional participation in DUSEL.

**Cyberinfrastructure Advisory Committee** (CIAC) provides expert advice on cyberinfrastructure and data management and transportation.

**DUSEL Program Advisory Committee** (PAC) provides the Vice Chancellor for Research at UC Berkeley with scientific guidance and advice concerning DUSEL scientific missions. The PAC held its first meeting in July 2010 and reviewed the proposed generic suite of experiments used to develop the Facility design.

**Education Advisory Committee** (EAC) provides expert advice on education and public outreach activities associated with DUSEL and the Sanford Center for Science Education.

The Project carefully tracks and responds to recommendations and findings from the reviews and from the advisory boards and committees. The review database is included in the documentation supplied to the NSF-organized and internal-review processes.

During DUSEL's Conceptual Design phase, a nascent scientific advocacy group, the DUSEL Experimental Development Committee, was established with independent funding to the committee from NSF. As the scientific community expanded and developed, a formal users' committee, the DUSEL Research Association (DuRA), was established, along with an executive management committee for the users. DuRA provides DUSEL scientific users with an additional mechanism to communicate with DUSEL Project management and for the DUSEL Project to regularly communicate to the growing user community.

## 1.4 Scientific Programs and Integrated Suite of Experiments Overview

The scientific goals for DUSEL span physics, biology, Earth sciences, and engineering. DUSEL will be a multidisciplinary dedicated research facility providing support for scientific research, engineering, and education efforts that benefit from access to the underground site. The relevance of DUSEL in achieving a broad spectrum of scientific goals is documented in National Academy and agency reports[3] and the NSF publication *Deep Science*.[1] More recently, the 2008 Particle Physics Project Prioritization Panel (P5) report emphasized the importance of DUSEL to the domestic high energy physics community.[6] Late in 2009, the high energy physics Particle Astrophysics Scientific Assessment Group (PASAG) affirmed the importance of direct dark-matter detection as a core mission for the particle physics community.[7] The LBNE, a major element of the DUSEL Facility and a major focus for the design efforts, received CD-0 by DOE in January 2010. The Project will capitalize on the significant opportunities for education and public outreach that this suite of transformational science creates. The Project is integrating education and outreach activities throughout the planning for DUSEL. The educational and scientific uses have been well documented.[2,5] Education and outreach efforts receive significant support from the state of South Dakota and philanthropic donors, and are included in the proposed DUSEL MREFC construction budget.

The overall scientific motivations for DUSEL were originally summarized in the CDR.[2] Briefly summarized below are the key elements of the potential scientific program at DUSEL. The experimental programs described are anticipated to be elements of the Integrated Suite of Experiments (ISE), to be implemented concurrently with the construction of the DUSEL Facility. The relevance of this program has grown in the years since this suite was proposed in the CDR and has been reaffirmed in the NSF guidance given to the Project late in 2009. The experiments presented in this section reflect the Project's evaluation of the active community participation in creating the ISE, as demonstrated in part by the response to the NSF DUSEL S4 Program Solicitation[8] and subsequent S4 awards.[9] Anticipated ongoing activities in all of the areas are described in Volume 3, *Science and Engineering Research Program*.

### 1.4.1 Physics and Astrophysics Research

Physics experiments will dominate research at DUSEL with respect to cost and size of experiments, size of collaborations, and facility infrastructure requirements. These experiments predominantly share the criterion of requiring extraordinary shielding from cosmic rays and other sources of background, including naturally occurring radioactivity. They require well-equipped underground laboratories to mount and operate the experiments, and well-regulated environments. Many of the proposed experiments anticipate operating for years and in some cases for decades. Our Facility Preliminary Design groups the experiments into research campuses to share resources and infrastructure and achieve a commensurate reduction in construction and operating costs. The principal physics underground campuses are located at the 4850L and 7400L. The LBNE project is considering a shallower deployment of the liquid argon detector at the 800L.

### 1.4.2 Direct Detection of Dark Matter

There is compelling evidence that most of the matter in the universe consists of non-Standard Model particles subject to gravitational forces. This nonluminous material directly influences large-scale cosmology, galactic formation, and evolution, and provides convincing evidence for new physics beyond the Standard Model. Experiments seeking direct detection of dark matter have made impressive advances in sensitivity in the past five years, pursuing multiple technologies and techniques focused on detecting nuclear recoils following collisions between the dark matter and detector nuclei. The technologies frequently develop discrimination between these recoil signals and other backgrounds, notably internal and external radioactivity. Collaborations in this area require well-controlled environments to deploy more massive experiments. Our interactions with this community indicate we should anticipate significant phased deployments at the 4850L and 7400L. These experiments may involve targets of cryogenic noble liquids or solid-state detectors, water or other shields, and various background-rejection technologies, which will require careful integration into the DUSEL Project to ensure that deployment and safety issues are addressed. We anticipate dark-matter experiments to be among the first physics experiments deployed in DUSEL. Dark-matter experiments represent the largest number of collaborations seeking to make use of DUSEL, as represented by a total of four S4 awards. The Project is following these collaborations as well as several additional direct-detection efforts. The Project anticipates hosting at least one G3 dark-matter experiment in the ISE as well as potential G2 experiments, hosted in the Davis Campus, in advance of the completion of the new laboratory modules. The G3 dark-matter experiments are necessary to complement the Large Hadron Collider Experiments in seeking to identify dark matter.

### 1.4.3 Neutrinoless Double-Beta Decay Searches

Much of DUSEL's physics mission focuses on completing our understanding of neutrino properties. While oscillation experiments have presented compelling evidence that neutrinos oscillate between massive families, there remain significant challenges to completing our understanding of the neutrino. The issues remaining to be addressed include: the absolute neutrino mass, the ordering of the three neutrino families (mass hierarchy), the full mixing matrix describing the oscillations among the three families, and possible charge and parity (CP) symmetry violating phases and possible Majorana phases in the neutrino mixing matrix. The experiments searching for neutrinoless double-beta decay will address three of these outstanding questions: those of absolute neutrino mass, mass hierarchy, and Majorana phases. When coupled with other experiments measuring these properties, especially long-baseline neutrino experiments, even neutrinoless double-beta decay null results are valuable. Two collaborations have expressed interest in DUSEL for ~1 tonne scale detectors. Both are interested in the 7400L, while one is initiating research at the 4850L (in the Davis Campus) as part of the Sanford Laboratory early science program. The Project anticipates hosting one large detector effort in neutrinoless double-beta decay.

Both dark-matter and neutrinoless double-beta decay experiments may have sensitivity to solar and supernova neutrinos, presenting opportunities for additional research goals.

### 1.4.4 Nuclear Astrophysics Experiments

The Big Bang and subsequent expansion created the elements hydrogen, helium, and lithium. All additional elements were created by nuclear synthesis reactions within stars. While the theoretical framework of nuclear synthesis was developed nearly 50 years ago, the precise determination of the details of the stellar proton-proton fusion chain and the stellar carbon-nitrogen-oxygen cycle has just begun. Despite the high stellar temperatures at the sites of these reactions, most of these reactions' energies are far below the Coulomb barrier, resulting in challenging rates for laboratory measurements. A collaboration proposes to develop both a light-ion and heavy-ion accelerator at the 4850L and to begin an ambitious campaign to measure dozens of reaction rates essential for understanding the details of the reactions responsible for the creation of elements essential for our own existence as well as possibly holding the key to understanding supernova explosions.

Operation of small, low-energy accelerators underground will require careful attention to integration of the infrastructure requirements, including safety, access, and shielding to ensure noninterference with neighboring experiments. The 4850L provides adequate shielding to support the full experimental program. The Mid-Level Laboratory Campus will be able to support the development of a nuclear astrophysics facility.

### 1.4.5 Large Detector Research—Long Baseline Neutrino Experiment and Proton Decay Searches

Atmospheric, solar, and reactor neutrino experiments have required the first modification of the very successful Standard Model within the framework of the known particles. To fully understand the neutrino mixing matrix with the known neutrino species, to establish mass hierarchy, to test the unitarity of the neutrino mixing matrix, and to probe for possible leptonic CP violation will require substantial investments in a program consisting of massive detectors (fiducial masses of hundreds of kT), intense beams of neutrinos produced at an accelerator (beam intensities approaching several MW), and an appropriate long baseline between the detectors and the beam source. This long baseline is required to introduce adequate material for the neutrino beams to acquire the matter-dependent oscillation phases and break the accidental degeneracies in the three-neutrino oscillation equations. DUSEL is well situated to provide these large cavities, with appropriate rock characteristics, good access to the proposed site at the 4850L, a designated rock-disposal site, and the appropriate distance (about 1,300 km) from Fermi National Accelerator Laboratory (Fermilab). The same detectors will be used for a compelling program of searching for proton decay and detection of astronomical neutrinos (solar, atmospheric, and supernova).

The LBNE project is investigating multiple options for realizing the very large detectors required, including combinations of water Cherenkov detectors and/or liquid argon detectors. The feasibility of the proposed cavities to house water Cherenkov detectors is being extensively studied by the DUSEL team, in partnership with the LBNE project. These large cavities will require thorough geotechnical and site investigations and a careful design approach for the Large Cavity excavation and rock stabilization. The large scale of these detectors requires careful planning to ensure that detector construction issues are adequately addressed, in particular the transportation of huge volumes of detector media and components. The facility requirements for the liquid argon option are primarily being studied by the LBNE project with support from the DUSEL team. The water Cherenkov detectors would be accessed from the 4850L. The liquid argon detectors could be located at the 4850L or at shallower depths such as the 800L. DOE's LBNE project is led by Fermilab.

### 1.4.6 Physics Experimental R&D and Future Uses

The Homestake DUSEL site offers many opportunities for innovative research. These experiments seek to exploit access to large-scale vertical shafts and large, well-equipped underground laboratories that can be adapted for a variety of experiments. The spectrum of experimental topics is broad, and includes dark-matter detectors with directionality, low-energy astronomical neutrinos, neutron antineutron oscillations, tests of gravity, gravity wave detection, atom interferometry, and atmospheric sciences, including aerosol and cloud formation. Several physics communities, including dark-matter and neutrino experiments, are creating experimental road maps that span several decades of research with increasing sensitivity and/or precision.

### 1.4.7 Bioscience Research

DUSEL provides a unique window into subsurface life using the existing drifts, shafts, and workings. In contrast to the physics experiments that rely upon the rock overburden for shielding, the Earth sciences—including geomicrobiology, geosciences, and engineering—investigate the rock itself and the dynamic processes operative in the rock volume. The DUSEL Facility will allow answers to questions such as identifying the source of energy for subsurface populations, the energy budget that sustains this life, and how life has adapted to the constraints of energy availability, temperature, and perhaps even pressure. The limits of life in the underground as well as the search for new life and, perhaps, even the development of life will involve examination of the diversity and distribution of microbial life forms from the surface to ~16,000 feet below ground, where rock temperatures are estimated to be ~120°C. This research will focus on the interactions among the microbes, the host rock, and associated fluids, and will provide a unique research opportunity to probe and understand the relationships among all the parts of the tree of life.

A dedicated deep-drilling facility will be required to aseptically core to great depths and at elevated temperatures. The geomicrobiological research will also involve bioprospecting throughout the Facility as well as closely affiliated geochemical investigations of the rock matrix and adjacent fluids. DUSEL proposes to create a facility at the 7400L to support the geomicrobiology drilling efforts.

### 1.4.8 Geosciences Research

Rock fracturing is a fundamental factor in the movement of fluids in the subsurface. The effect of fracture networks, however, is a function of scale, and DUSEL will provide an opportunity to gain important insights into this fluid flow as well as the associated seismic events on a wide range of scales. The coupling of thermal, hydrological, chemical, biological, and mechanical processes will validate models of permeability and transport in the fracture networks and rock deformation that play such a critical role in human interaction with the subsurface. The long-term, variable-scale access experiments will be dominated by investigations of the rock mechanics, heat flow, and the in situ coupled processes. The research will involve both run-of-the-facility investigations, taking advantage of the access to the ~35 $km^3$ represented by the Facility for hydrological, biological, chemical, and geophysical "prospecting," as well as purpose-built experimental deployments. Additional research efforts, and support functions for the research efforts, will take advantage of additional existing spaces, including other levels, ramps, and shafts in the Facility.

### 1.4.9 Ground Truth and Engineering Research

A need to understand the mechanical properties of rock, the response of rock to human activity, the extrapolation of rock properties between sparse sampling points (boreholes), and the development of seismic imaging technologies is complementary with the ambitious underground construction program, especially with regard to the creation of the extremely large cavities that are of great importance to the physics research. These cavities will be at depths that are unprecedented and will create a natural synergism among the engineering, geosciences, and physics research. Not only will these excavations support decades of physics research, they will also provide opportunities to greatly advance techniques for excavation and monitoring, which will be required to understand the "health and status" of the Facility. Ground truth and engineering research will span the Facility and will develop a number of purpose-built experimental facilities at a variety of depths. Some of these research facilities may be shared with geosciences and geomicrobiology efforts.

### 1.4.10 Research on Questions of Societal Importance

DUSEL provides a large-scale, three-dimensional laboratory to research and develop geothermal energy extraction and carbon sequestration. Although inappropriate for large-scale energy extraction or carbon sequestration, the Facility will provide a laboratory setting to examine the effects of chemical alteration and the examination of the mechanisms governing the flow of critical carbon dioxide. The $CO_2$ sequestration collaboration is pursuing a custom-built vertical shaft extending from the surface to the 1700L to host these studies.

## 1.5 Early Scientific Programs

An early science program has been established at SDSTA's Sanford Laboratory.[10] State and private funds have been secured to create and initially operate this facility. Support for the experiments and collaborations has been secured from NSF and DOE base programs, National Aeronautics and Space Administration (NASA) research and federal Experimental Program to Stimulate Competitive Research (EPSCoR) programs, as well as from state and university sources. The scientific scope and scale of these science projects and experiments are presently divided into two distinct categories: 1) sites with a fixed-footprint laboratory space specifically constructed or refurbished and 2) those having a very small footprint in existing drifts. The former group includes two physics experiments—the Large Underground Xenon (LUX) experiment,[11] searching for direct detection of dark-matter particles; and the MAJORANA DEMONSTRATOR experiment,[12] to develop technology and to advance searches for neutrinoless double-beta decay. The latter group includes approximately 10 small biological, geological, and engineering efforts principally investigating suitable locations and environments for future experiments. Several geology and engineering experiments have installed instruments and are collecting data.

The LUX experiment is a large two-phase liquid/gas xenon dark-matter detector with water shield, to be installed in the refurbished Davis Campus. LUX is preparing to assemble and test its detector components, including a significant water shield, in a refurbished facility aboveground. The MAJORANA DEMONSTRATOR experiment requires the use of electroformed copper made underground to achieve low backgrounds. Copper will be produced in the next few years in a dedicated clean-room area in a refurbished shop near the Ross Shaft. The MAJORANA DEMONSTRATOR experiment would be assembled and installed at the Davis Campus when refurbishment of the campus is completed in the last quarter of 2011.

DUSEL design and engineering efforts directly benefit from the Sanford Laboratory science program. The DUSEL Project has hired and shared experienced staff from the SDSTA who are well trained in underground activities, including design, engineering, and operations. The health and safety program and operations procedures required for DUSEL are being created and refined based on the smaller-scale Sanford Laboratory effort. The Project is applying Sanford Laboratory's interface and installation experiences to improve its procedures and prepare for DUSEL's much larger and more complex experiments.

## 1.6 Facility Overview

The DUSEL Facility is a custom-designed requirements-driven research laboratory. Physical and operating requirements are obtained and managed by the DUSEL Project from interactions with approximately two dozen scientific collaborations. The Facility design satisfies these requirements for the generic suite of experiments. Figure 1.1-2, a graphic representation of the Facility, indicates locations of interest for the various scientific efforts. Many of these have received support from NSF through the S4 solicitation. Several collaborations, notably the LBNE project, have received DOE funding.

The DUSEL Complex will be grouped into a variety of campuses distributed within the Homestake property, descending from the surface down to the Deep-Level Campus at the 7400L, as presented in Figure 1.6-1. The Project has performed assessments of the existing, accessible SDSTA infrastructure provided through the property donation from Barrick, including buildings, utilities-distribution systems, conveyance and hoisting systems, underground excavations, and environmental systems including the water-pumping and -treatment facilities. SDSTA has invested in significant facility enhancements, including utility distribution, water-pumping facilities, and new custom-built excavations for the Sanford Laboratory early science program. Where feasible and cost effective, existing systems are incorporated into the DUSEL design. The Sanford Laboratory design and planning efforts were integrated with DUSEL's to reduce the risk of subsequent interference or mandatory replacement of Sanford Laboratory systems by the DUSEL Project. The Barrick-donated infrastructure, along with the SDSTA improvements, represents a sizable contribution to the ultimate DUSEL Facility and significant reduction in the total estimated DUSEL cost.

**Surface Campus**

The Surface Campus houses DUSEL administration; provides office, staging, preparation, and support space for the scientific collaborations; hosts the education and outreach efforts and the Sanford Center for Science Education (SCSE); and supports Facility operation, including conveyance, safety, and environmental functions. Most of the existing 65 surface structures that transferred to the SDSTA through the Barrick Property Donation Agreement have been assessed. These buildings total in excess of 253,000 gross square feet (gsf). The DUSEL design proposes continued-use buildings at the Ross and Yates Campuses and Waste Water Treatment Plant. Additional new structures are proposed, principally at the Yates Campus. Other structures are proposed for demolition and site restoration as outlined in Table 1.6-1. The proposed Surface Campus is presented in Figure 1.6-2.

**4850L Mid-Level Campus**

The 4850L is a major campus spanning nearly 1 km distance between the access shafts. This level provides an average of ~4100 meters water equivalent (mwe) of shielding from cosmic rays. The purpose-built laboratory modules, Large Cavity (the Large Cavity is managed by the LBNE collaboration for the DOE, and would be accessed from this Campus) and basic access to a major fixed BGE site are provided at the 4850L. Figure 1.6-3 presents the 4850L principal features. Table 1.6-2 presents key parameters of the 4850L Campus. The Ross Shaft, extending from the surface to the 5000L, is maintained for construction and maintenance access as well as routing utilities from the surface. The Yates Shaft will be overhauled to provide customized access from the surface to the 4850L for scientific users and material. Access on the 4850L will be facilitated by enlarging existing drifts. Significant additional excavation is included in the design to support experiment-related utilities (e.g., mechanical, electrical, piping, HVAC, communications) and safety systems and Areas of Refuge (AoRs).

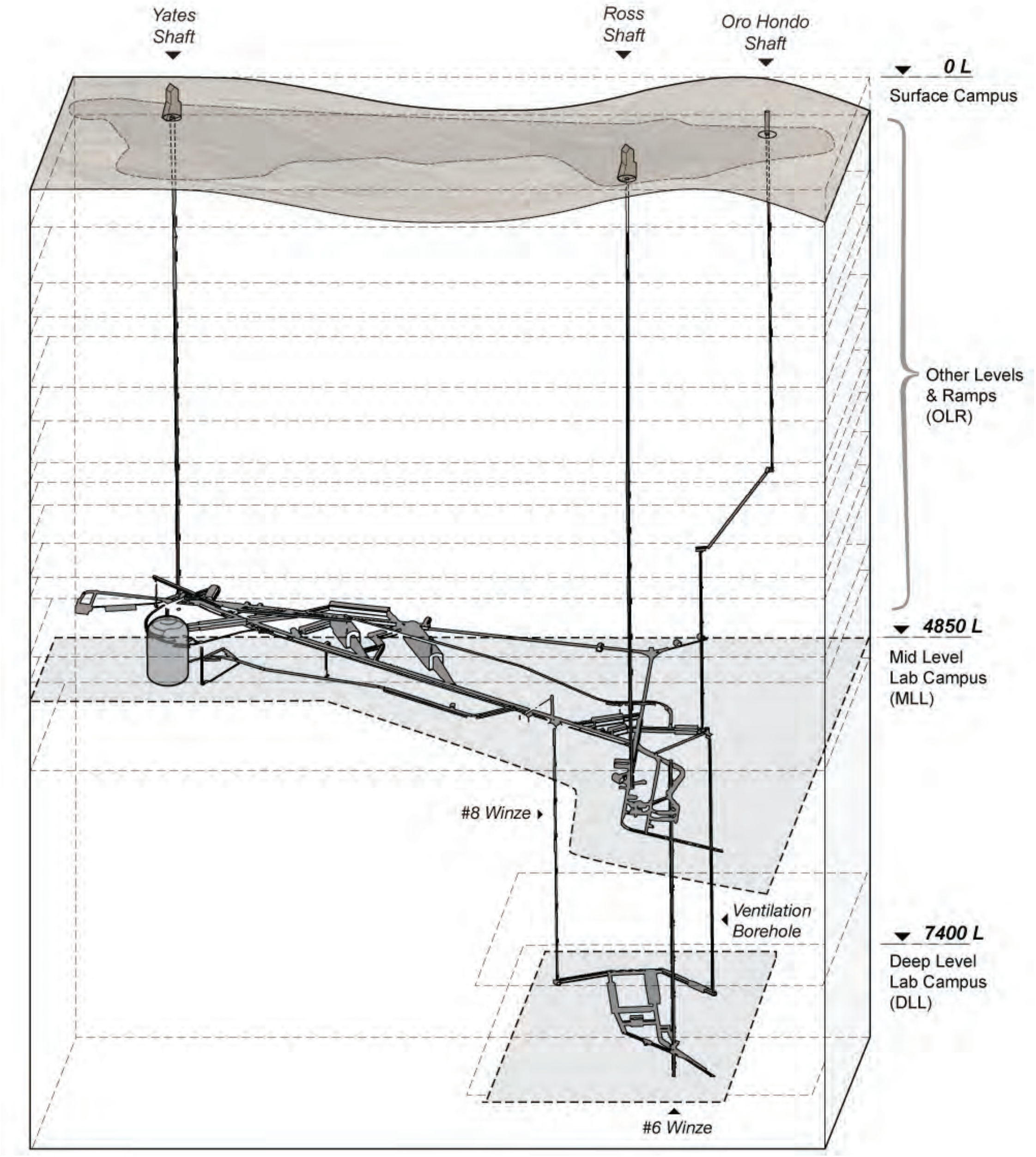

**Figure 1.6-1** DUSEL Complex. Shown are the major elements of the Facility, extending from the Surface Campus to the Mid-Level Laboratory Campus at the 4850L and the Deep-Level Campus at the 7400L. [DKA]

| Surface Facility Configuration | Gross Square Feet |
|---|---:|
| Existing Structures—Yates and Ross Campuses | 253,000 |
| Structures Proposed for Removal | 48,000 |
| Structures Proposed for Adaptive Reuse | 205,000 |
| New Structures | 42,000 |
| Total DUSEL Proposed Structures | 247,000 |

**Table 1.6-1** DUSEL Surface Facility, including adaptive reuse of existing structures and new construction.

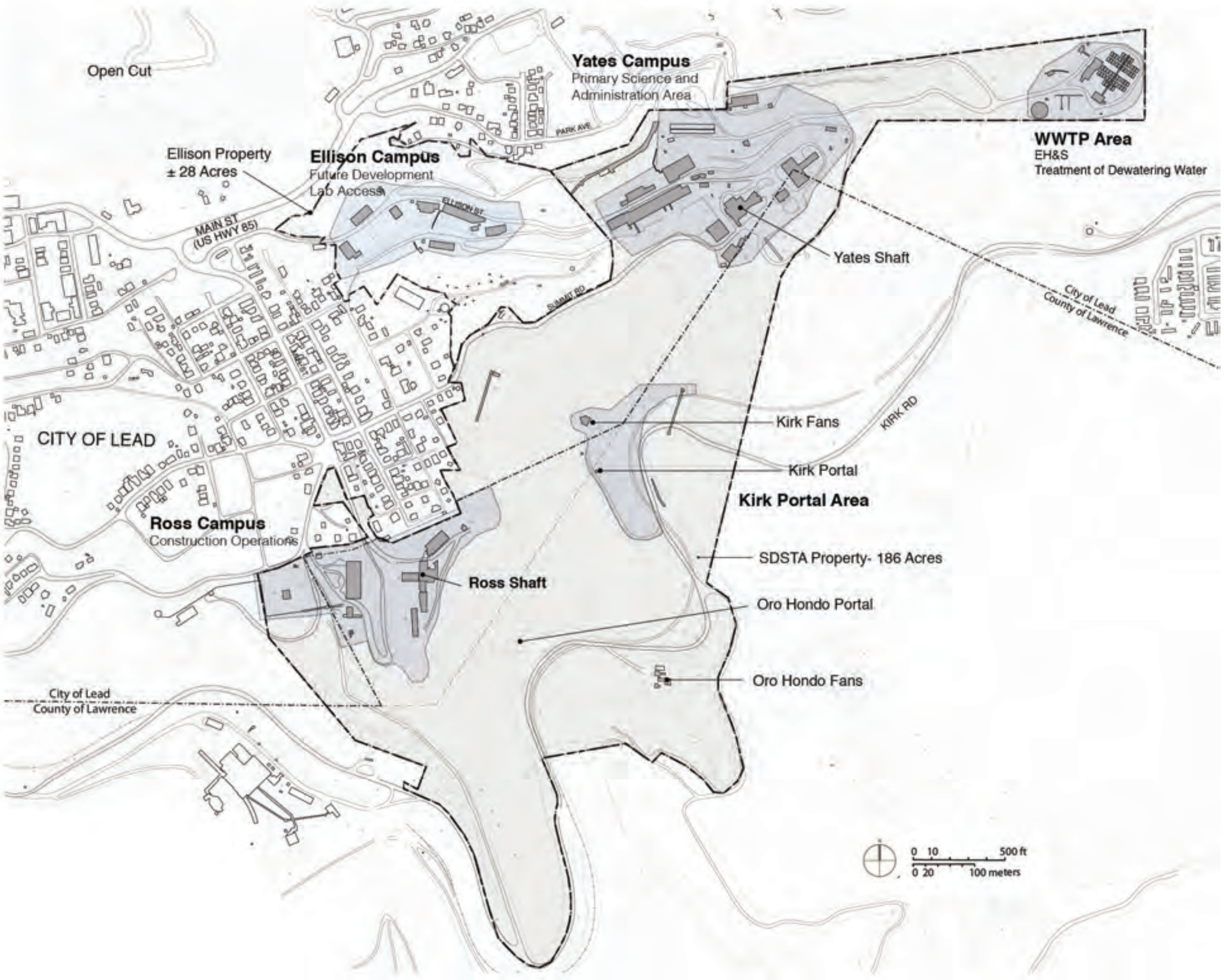

**Figure 1.6-2** DUSEL's Surface Campus. Shown are the existing and proposed new surface structures. [DKA]

The Mid-Level Laboratory (MLL) Campus is designed to support experiments from DUSEL's transformational suite of experiments, including: G3 dark-matter search experiments, a nuclear astrophysics facility, at least one LBNE Large Cavity with options for additional large cavities or liquid argon modules, and low-background counting facility. A fixed BGE research laboratory would be located southwest of the Ross Shaft near an existing series of drifts. Also shown in Figure 1.6-3 are the DLM and the new excavation, DTA, hosting the LUX and MAJORANA DEMONSTRATOR experiments. These spaces would be repurposed from their current early science experiments to house additional DUSEL experiments and provide an option for hosting experiments much earlier in the DUSEL MREFC-funded construction phase. The existing network of drifts on the 4850L is extensive. Additional BGE uses will likely be deployed within this network for experiments with lower facility impact or requiring less-frequent access and occupancy.

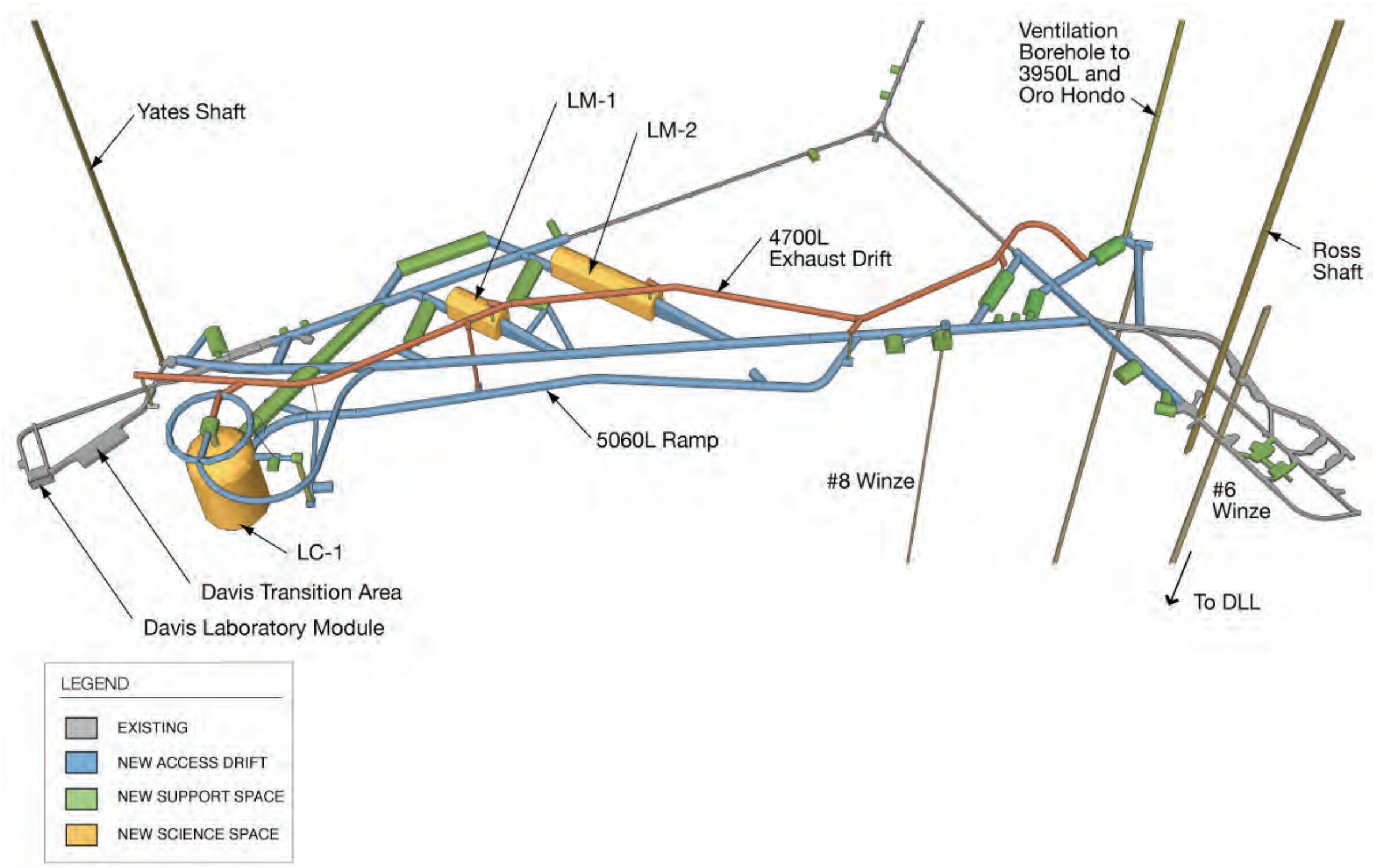

**Figure 1.6-3** DUSEL's Mid-Level Laboratory. Laboratory Modules 1 and 2 (LM-1, LM-2) are included in the DUSEL MREFC budget. The Large Cavity is a DOE-stewarded project within the DUSEL Facility. The Davis Laboratory Module (DLM) and the new Sanford Laboratory excavation Davis Transition Area (DTA) initially hosting the LUX and MAJORANA DEMONSTRATOR projects are also shown and will be available for use by experiments associated with the MREFC budget. [DKA]

| Experiment Space | Width (m) | Height (m) | Length (m) | Floor Area ($m^2$) | Finished Volume ($m^3$) |
|---|---|---|---|---|---|
| LM-1 | 20 | 24 | 50 | 1,000 | 22,495 |
| LM-2 | 20 | 24 | 100 | 2,000 | 44,990 |
| LC-1 (DOE) | - | 83 | 55 (dia) | 2,376 | 185,947 |
| DLM | 11 | 13 | 17 | 187 | 2,431 |
| DTA | 16 | 5 | 43 | 688 | 3,440 |

**Table 1.6-2** Critical 4850L Campus parameters.

The LBNE collaboration is investigating options to achieve their full detector fiducial mass of 200 kT water Cherenkov equivalent target defined by their DOE CD-0 mission-need statement. These options include additional large cavities at the 4850L, or laboratory modules at the 4850L or 800L.

### 7400L Deep-Level Campus

The Deep-Level Laboratory (DLL) at the 7400L is reserved for users requiring great depth to achieve extensive shielding from cosmic rays or to obtain access to the deep subsurface for BGE studies subjected to extreme physical conditions. The 7400L provides ~6400 mwe of shielding. A purpose-built laboratory module and additional research space is provided at the 7400L, presented in Figure 1.6-4, and the critical parameters in Table 1.6-3. Access to the 7400L is provided through the existing #6 Winze, with secondary egress provided by a new winze from the 4850L. A second new borehole from the 7400L upward to the 4850L will provide positive ventilation for the 7400L experimental areas.

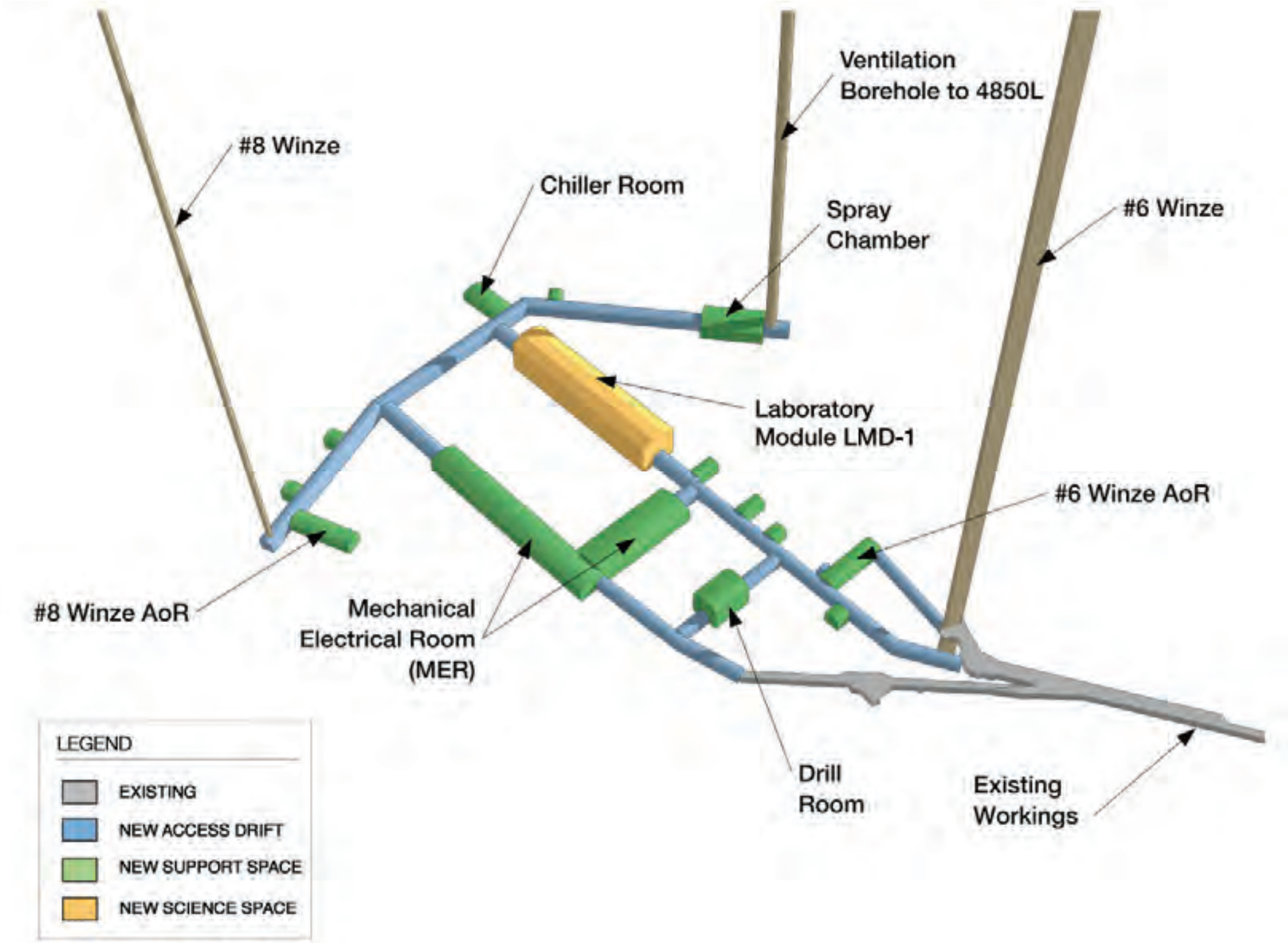

**Figure 1.6-4** DUSEL's Deep-Level Laboratory Campus. [DKA]

| Experiment Space | Width (m) | Height (m) | Length (m) | Floor Area ($m^2$) | Finished Volume ($m^3$) |
|---|---|---|---|---|---|
| LM-1 (DLL) | 15 | 15 | 75 | 1,125 | 14,288 |
| Drill Room | 11 | 11 | 16 | 176 | 1,644 |

**Table 1.6-3** Critical Deep-Level Campus parameters.

### Other Levels and Ramps

The Facility provides access from the surface to levels approximately 8,000 feet below local ground level, although the 7400L is anticipated to be the lowest routinely accessible level for science. Sixty existing levels are vertically spaced approximately every 150 feet. The DUSEL design is primarily focused in the region bounded by the Ross and Yates Shafts and extending below the 4850L via the #6 Winze. Many of the levels are connected by a system of ramps, shafts, and raises, presented in Figure 1.1-3. Some of these systems are proposed for use in DUSEL to provide operations support for pumping accumulated groundwater and providing ventilation to the underground. The far-reaching system of ramps and levels proposed for Facility operation and additional scientific research is presented in Figure 1.6-5; this system provides access to much of Homestake's subterranean 35 $km^3$ rock mass.

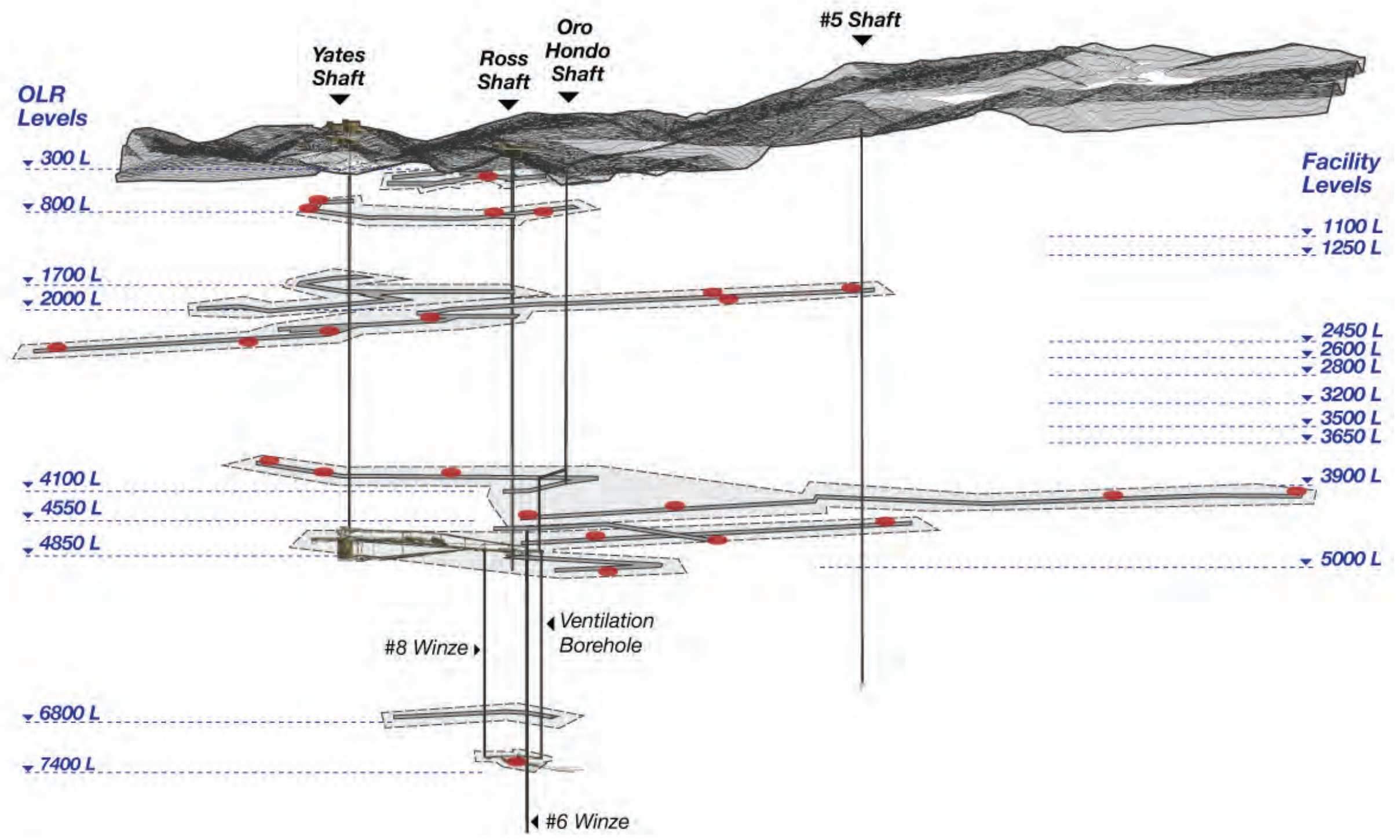

**Figure 1.6-5** The use of Other Levels and Ramps (OLR) within the Facility for biology, geology, and engineering research spans much of the Facility, as shown by the levels proposed by many of the BGE experiments. Approximate OLR experiment locations are noted in red. [DKA]

In total, nearly 30 km of drifts, ramps, and raises would be maintained as part of DUSEL, with half of this total being used to maintain the Facility and shared with experimental uses, and the remaining indicated levels to provide access for these scientific uses probing the rock mass. Additional levels include 300L, 800L, 1700L, 2000L, 4100L, 4550L, 6800L, and 7400L, as well as ramps connecting some of these levels.

One experiment proposing to investigate $CO_2$ sequestration technologies would make use of multiple levels to access their purpose-built raise, shown in Figure 1.6-6, extending from the surface to the 1700L.

### 800L LBNE Laboratory Module Option

DOE's LBNE project is pursuing options to host detector modules at shallower levels for technology choices to be made in the future. The LBNE project has developed a conceptual design of a liquid argon detector module at the 800L also presented in Figure 1.6-6.

### DUSEL in International Context

Physics experiments over the past ~20 years have highlighted major discoveries from underground experiments and led to major revolutions in our understanding of the physical universe. These discoveries have principally come from Japan's Kamioka Facility, from Europe's Gran Sasso Facility, and from Canada's Sudbury Neutrino Observatory. These discoveries have spawned renewed international interest in underground research in a variety of fields, placing increasing pressure on the existing facilities and triggering rigorous competition for the sparse available underground research space, especially at great depths. Most of the world's facilities are heavily subscribed and a majority of these provide space at depths now considered to be too shallow for most of DUSEL's suite of experiments. The suite of experiments considered for deployment in DUSEL are, with one or two exceptions, uniquely aligned with DUSEL and are not considering alternative underground facilities.

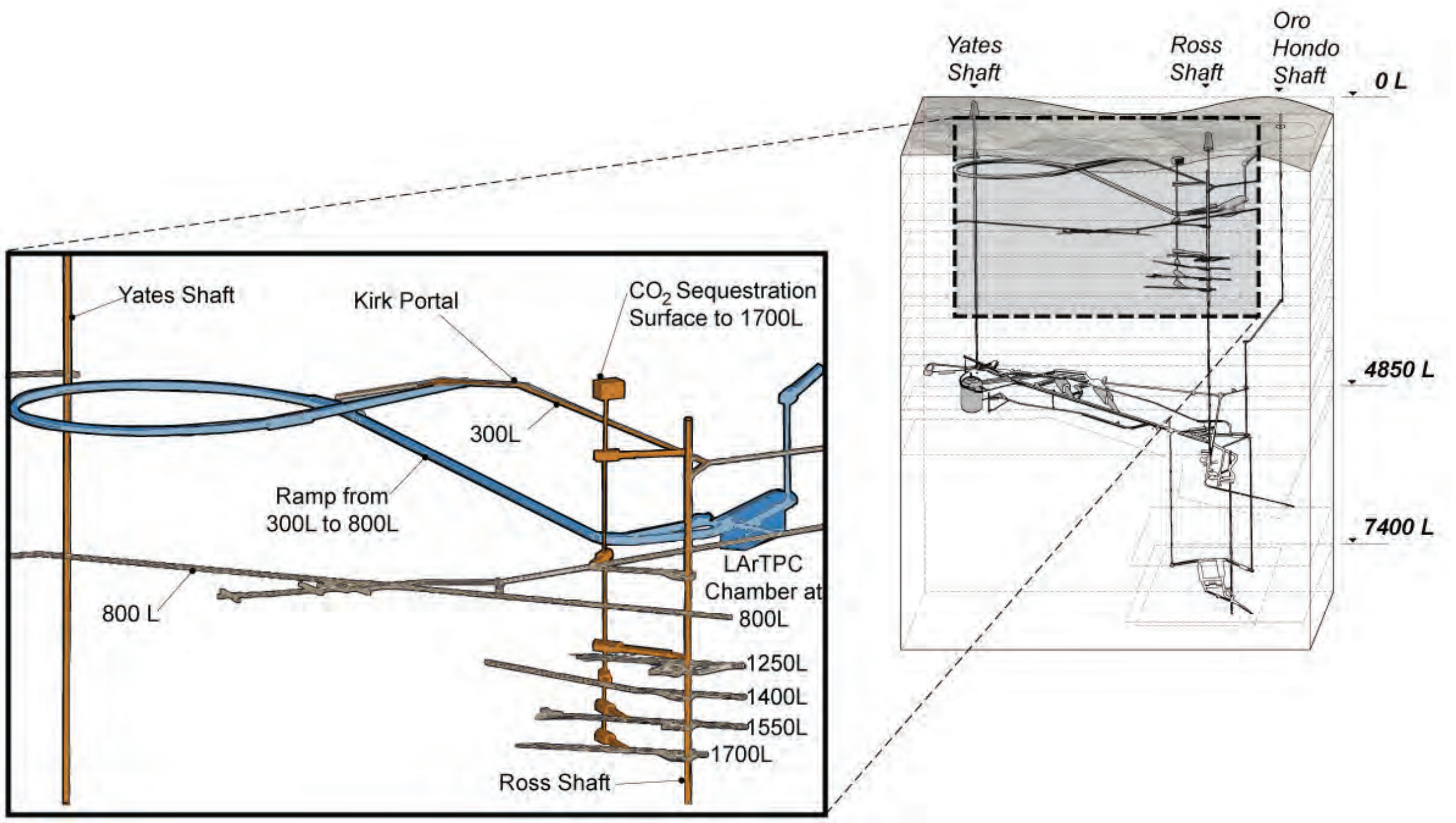

**Figure 1.6-6** LBNE Conceptual Design of liquid argon detector module at the 800L and the adjacent $CO_2$ Sequestration (LUCI) site. [DKA]

The Earth science communities have had access to essentially no dedicated, long-term basic research facilities worldwide. And no existing facilities provide the access and availability to such a large block of the Earth's crust for dedicated research.

Figure 1.6-7 presents a volumetric comparison of the existing and proposed underground research facilities worldwide. DUSEL would more than double the world's inventory of underground research space and provide access to a wide variety of depths as appropriate for DUSEL's multidisciplinary research programs. Access to DUSEL's research campuses would provide much-needed space for the coming decades of underground research and facilitate potential major scientific discoveries in several scientific disciplines.

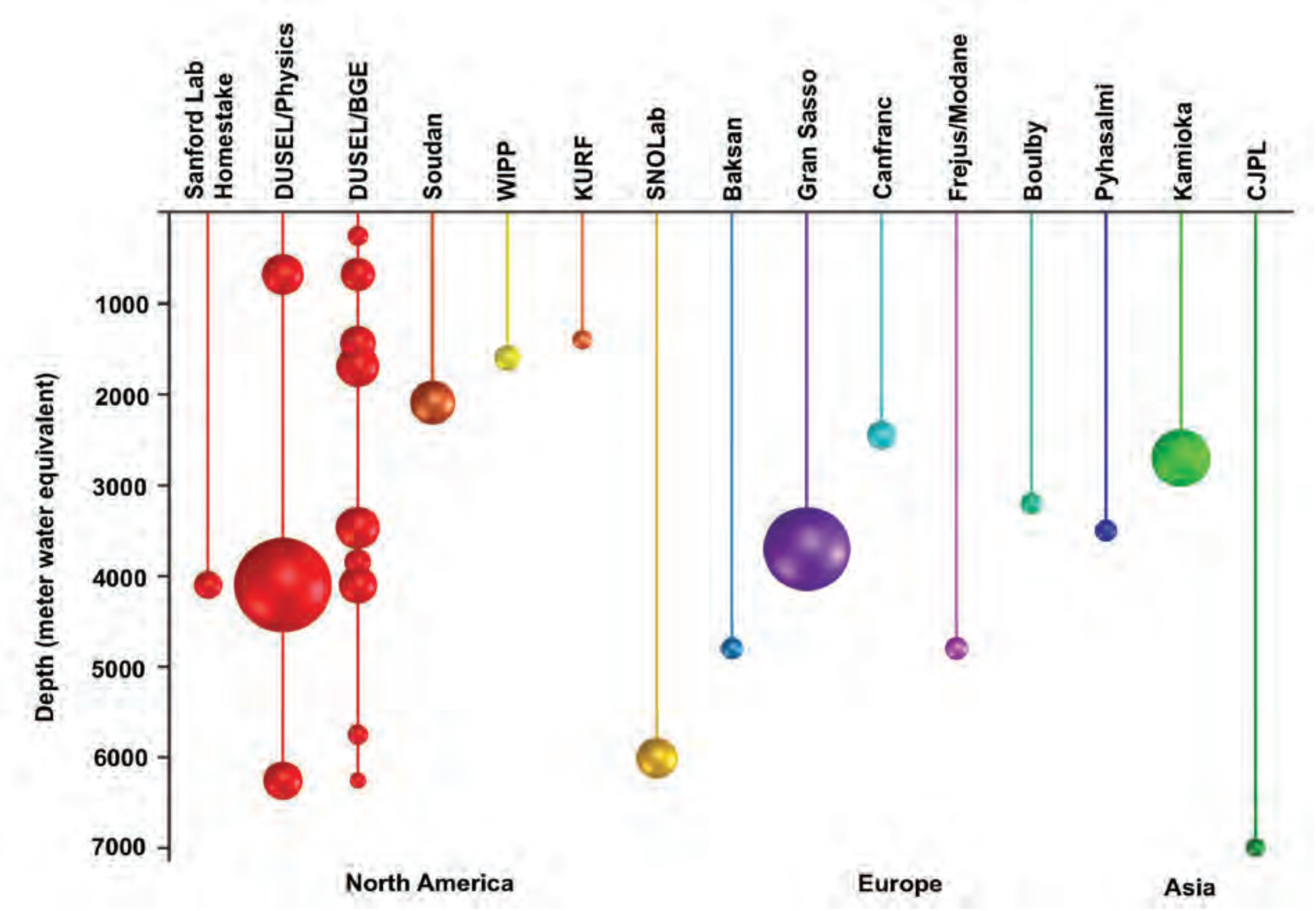

**Figure 1.6-7** Volumetric comparison of the world's existing and proposed underground research facilities. The vertical scale is the depth of the facilities in meters-water-equivalent. The size of the plotted symbols scale with the volume of the research facility. For the DUSEL design, we have assumed one large water Cherenkov detector at the 4850L and one liquid argon detector at the 800L. The 4850L and 7400L Laboratory Modules are presented. [Dave Plate, DUSEL]

## 1.7 Project Systems Overview

The Project Management Control System (PMCS) provides the Project with the systems necessary to manage the cost and schedule of the DUSEL Project. At the heart of the PMCS is the Work Breakdown Structure (WBS) and associated dictionary that define the Project's scope. From the WBS and associated cost and schedule estimates, the detailed Integrated Project Schedule and cost estimate are developed from the contracted architectural and engineering contractor's work-packages. The Project has established the performance measurement baseline and will track accrued value and manage this to the baseline schedule using an ANSI-compliant Earned Value Management System (EVMS). The Project has exercised the use of the EVMS during the development of the Preliminary Design. The DUSEL design is controlled and managed with a Configuration Control Board as presented in Volume 8, *Project Management Control*.

The DUSEL Project MREFC-funded Baseline is presented in Figure 1.7 along with major Level 1 milestones for the MREFC-funded construction Project. A more complete list of Level 1 milestones is presented in Table 1.7.

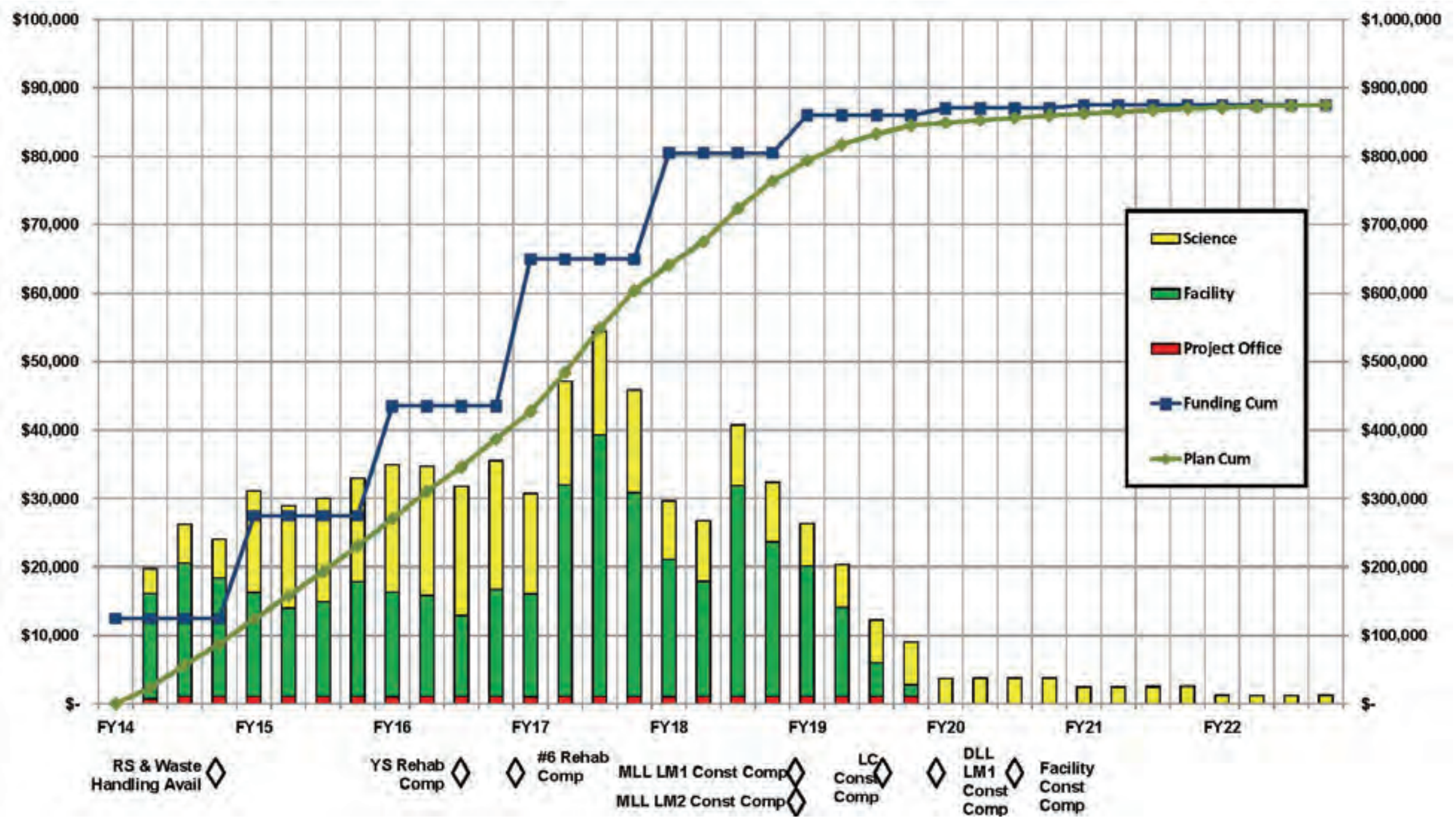

**Figure 1.7** DUSEL's MREFC Construction Budget Baseline. The quarterly budgets for the Science, Facility, and Project Office are indicated on the vertical scale at left, while the Funding and Plan Cumulative amounts are indicated on the vertical scale at right. Several major project milestones are indicated across the top of the figure. Budget values shown in FY 2010 thousand dollars including Management Reserve.

| Level 1 External—Major Milestones | |
|---|---|
| **Milestone Description** | **Date** |
| S3 Award Complete | 15-Sep-08 |
| S4 Awards Announced | 4-Aug-09 |
| NSF Approval of CA2 Funds to Complete PDR Efforts | 24-Sep-09 |
| LBNE CD-0 DOE Approval | 8-Jan-10 |
| NSF Release of additional funds to complete PDR efforts | 12-Jan-10 |
| PDR 30% A&E Cost Estimates & Schedule Complete | 7-Apr-10 |
| PDR 60% A/E & CM Cost Estimates, Reconciled Cost Estimates & Schedules Complete | 4-Aug-10 |
| PDR 90% A/E & CM Cost Estimates, Reconciled Cost Estimates & Schedules Complete | 7-Oct-10 |
| PDR 100% A/E & CM Cost Estimates, Reconciled Cost Estimates & Schedules Complete | 22-Nov-10 |
| NSF Annual Review Complete: Spring 2011 | 18-Apr-11 |
| Preliminary Design Report (PDR) submitted to NSF | 29-Apr-11 |
| NSF Start PDR Baseline Review by National Science Board | 2-May-11 |
| NSF Approval of DUSEL Funding Proposal for R&RA-funded Final Design 2012-2013 | 16-Jun-11 |
| NSF Approval of DUSEL Funding Proposal for R&RA-funded Operations 2012-2013 | 16-Jun-11 |
| Final Design - Contract Award | 2-Feb-12 |
| NSF Annual Review Complete: Spring 2012 | 19-Apr-12 |
| LBNE CD-1 DOE Review Complete | 6-Jul-12 |
| LBNE CD-1 DOE Approval | 6-Sep-12 |
| Final Design - 60% | 17-Sep-12 |
| MREFC Construction Funding - NSF Authorized | 1-Oct-12 |
| Final Design - 90% | 14-Feb-13 |
| Final Design Review (FDR) Complete | 1-Apr-13 |
| NSF Annual Review Complete: Spring 2013 | 18-Apr-13 |
| Construction Bid Package - Prepare | 14-May-13 |
| Final Design Review (FDR) Approval | 3-Jun-13 |
| NSF Approval of DUSEL Funding Proposal for R&RA-funded Operations 2014-2022 | 18-Jun-13 |
| Final Design - 95% | 23-Jul-13 |
| Construction Bid Package - Release | 13-Aug-13 |
| Construction Bid Package - Award Contract | 18-Nov-13 |
| Ross Shaft Rehabilitation Complete | 31-Jan-14 |
| MREFC Construction Funding - NSF Released - Start On Site Work | 3-Feb-14 |
| NSF Annual Review Complete: Spring 2014 | 18-Apr-14 |
| Ross Shaft and Waste Handling Available | 2-Aug-14 |
| NSF Annual Review Complete: Spring 2015 | 20-Apr-15 |
| DLL Final Design - 60% | 24-Dec-15 |
| DLL Final Design Review (FDR) Complete | 28-Jan-16 |
| Yates Shaft Full Ventilation Available | 4-Feb-16 |
| DLL Final Design - 90% | 1-Mar-16 |
| DLL Construction Bid Package - Prepare | 29-Mar-16 |

| Level 1 External—Major Milestones | |
|---|---|
| **Milestone Description** | **Date** |
| DLL Final Design Review (FDR) Approval | 31-Mar-16 |
| NSF Annual Review Complete: Spring 2016 | 19-Apr-16 |
| Yates Shaft Rehabilitation Complete | 1-Jun-16 |
| DLL Final Design - 95% | 13-Jun-16 |
| DLL Construction Bid Package - Release | 13-Jun-16 |
| DLL Construction Bid Package - Award Contract | 15-Jul-16 |
| #6 Winze Rehabilitation Complete | 26-Oct-16 |
| NSF Annual Review Complete: Spring 2017 | 20-Apr-17 |
| NSF Annual Review Complete: Spring 2018 | 19-Apr-18 |
| MLL Lab Module 2 Construction Complete (ready for researcher fitout) | 15-Nov-18 |
| MLL Lab Module 1 Construction Complete (ready for researcher fitout) | 16-Nov-18 |
| NSF Annual Review Complete: Spring 2019 | 18-Apr-19 |
| LGC Large Cavity 1 Construction Complete (ready for LBNE fitout) | 31-Jul-19 |
| DLL Lab Module 1 Construction Complete (ready for researcher fitout) | 9-Dec-19 |
| NSF Annual Review Complete: Spring 2020 | 17-Apr-20 |
| MREFC-funded Facility Construction Complete | 27-Jul-20 |
| NSF Annual Review Complete: Spring 2021 | 20-Apr-21 |
| NSF Approval of DUSEL Funding Proposal for R&RA-funded Operations 2022- | 18-Jun-21 |
| NSF Annual Review Complete: Spring 2022 | 20-Apr-22 |
| MREFC Experiments Construction Complete | 29-Mar-24 |

**Table 1.7** DUSEL Level 1 milestones. Milestones reflect late finish dates that include Schedule Reserve.

## 1.8 Systems Engineering and Integration Overview

The DUSEL Project has assembled an experienced team to implement a rigorous systems engineering process warranted by the scale and scope of this Project. With its large stakeholder pool and Project staff, the DUSEL Project will be successful with a disciplined and deliberate approach to system development. The DUSEL Systems Engineering (SE) team has established requirements baselines, put in place a robust configuration management system, and drafted the Project plans that will ensure continued program integration as the Project team advances into the Final Design phase. As the crosscutting technical organization that interacts with EH&S, Facility, and Science departments, the SE team is able to assist senior management to optimize the program.

Systems Engineering uses many tools, including the following to accomplish these goals:

**Risk Registry.** Systems Engineering works with all members of the Project team to identify, classify, and assess Project risks. Risk management is a continual process, and periodic reviews of current and new risks are conducted along with development and monitoring of risk mitigation plans.

**Requirements and Interface Documents.** The requirements and interface document structure provides the ability to identify how stakeholder needs are translated into technical requirements along a mechanism to control interfaces between the Facility, Science, and external entities. The disciplined development and review process ensures that all aspects of the Project are aligned and in agreement with the baseline requirement set.

**Configuration Management System.** A project the size and scope of DUSEL requires a structured process to establish and control an approved baseline (typically for documents and/or written procedures) as configuration-controlled items.

**System Verification.** This process uses the requirements and interfaces developed for the program and coordinates verification plans and procedures to ensure a successful implementation of the requirements and that user needs are satisfied.

**Value Engineering and Trade Studies.** The Value Engineering and Trade Study processes are utilized particularly by the Facility and Science divisions to develop data to support cost, schedule, or design changes required to make the Project more efficient, align the design with Project requirements, or cost estimates with allocated budgets.

## 1.9 Business Systems and Staffing Overview

DUSEL Senior Management has analyzed the staffing requirements for each phase of the Project, spanning Preliminary Design through the ongoing operations of the science programs and the DUSEL Facility.

The DUSEL Project employed ~55 full-time staff during the Preliminary Design phase. During the transition from Preliminary Design to Final Design, the Project staff will merge with SDSTA's Sanford Laboratory staff, and together will number ~160 full-time and 20 part-time staff. The combined organization is shown at a summary level in Figure 1.9 and the full Project organizational chart is included in Appendix 1.A.

The Business Systems for the DUSEL Project have relied to date on systems resident at UC Berkeley and the SDSM&T. The SDSTA has established independent business systems fulfilling the requirements for its operations within the state of South Dakota. The anticipated formation of the DUSEL LLC prior to Final Design will unite the Project stakeholders into one organization with consolidated Business Systems and will have the capacity to support the MREFC-funded Construction and laboratory Operations.

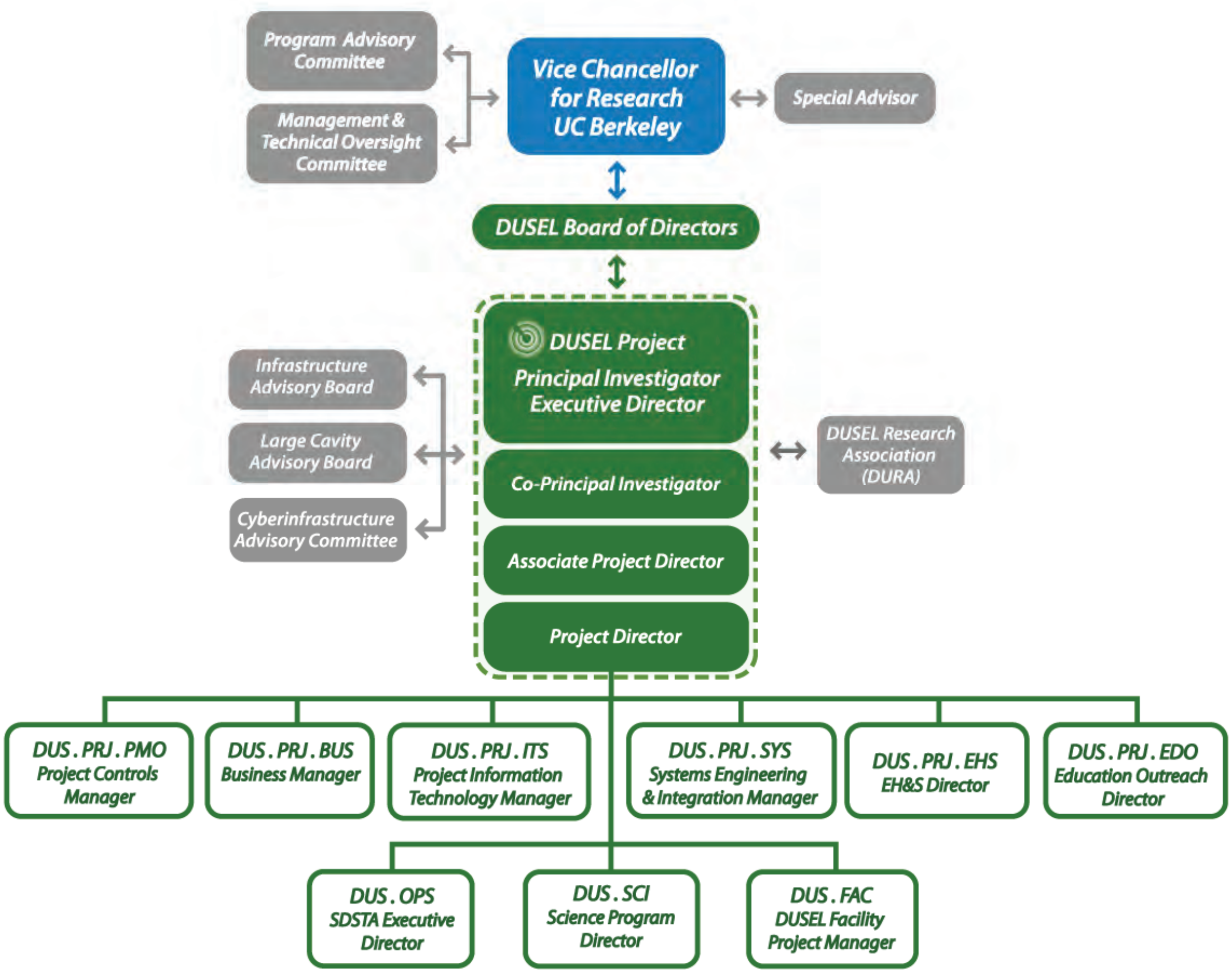

**Figure 1.9** The summary level DUSEL organization chart.

## 1.10 Integrated Environment, Health, and Safety Management Overview

Safety is the highest priority of the DUSEL Project. The primary objective is to systematically integrate excellence in environment, health, and safety (EH&S) into the management and work practices of all activities at all levels so that DUSEL's mission is achieved while protecting the public, employees, contractors, and the environment. The Integrated Safety Management plan combines the efforts of SDSTA's Sanford Laboratory and the DUSEL Project into a single, integrated organization. This goal is accomplished by ensuring that the overall management of EH&S functions and activities is an integral part of the mission accomplishment. As a function of this integration, it is important to recognize that "safety" refers to the reduction or elimination of all hazards, including hazards to health and environment.

DUSEL and SDSTA employees and users are required to conduct all work and operations in a safe and environmentally sound manner. These policies apply to all employees, users, visiting scientists, contractors, and their subcontractors. All are expected to fully comply with all procedures, instructions, and directives contained in the *EH&S Manual* (discussed in Volume 6, *Integrated Environment, Health, and Safety Management*) in order to reduce or eliminate hazards in the workplace. It is DUSEL's policy to integrate EH&S protection into all aspects of work, using the principles and core functions of the Integrated Safety Management system and to seek improvements in management and performance at every opportunity. Every person on site is responsible for EH&S and is accountable for performing all activities in a safe and environmentally responsible manner.

The basic EH&S policies and procedures have been established at Sanford Laboratory in advance of the DUSEL construction Project. As the program matures, the practice of subjecting any changes to these policies and procedures, or the creation of new ones, to a Project-wide review will ensure that hazard-control systems within a specific functional area do not conflict with controls established in other functional areas.

- Departments are allowed to tailor the DUSEL (institutional) EH&S Program to meet the needs of their organizations when possible. They participate in the development of institutional programs, and considerable effort is made to obtain universal buy-in as DUSEL programs are developed to reduce implementation variations.
- The implementation of a successful EH&S program cannot be imposed from above. The work will be conducted safely and with minimal environmental impact only if workers are involved in the process of planning the work to identify potential hazards associated with work activities.
- The foundation of Integrated Safety Management is line responsibility; i.e., the line organization must have the authority and responsibility, and be held accountable for integrating EH&S into, and as a part of, all the work it does.
- DUSEL policies and procedures identify the EH&S responsibilities of all employees. They further call out the roles and responsibilities of management, various safety personnel, other staff with special EH&S responsibilities, contractors, and visiting scientists performing work at or otherwise using the facilities.
- Clear and unambiguous lines of authority and responsibility for assuring successful EH&S performance have been established at all organization levels. Key management and EH&S positions are appointed by the EH&S Director, based on knowledge of the skills needed for the position and the competency of the candidates. The key to balancing priorities is to ensure that those who make the decisions are authorized to do so and that they have accurate information about the nature of the work, the hazards, and appropriate controls.

- Resources must be effectively allocated to address EH&S, programmatic, and operational considerations.

Engineering and administrative controls tailored to the work being performed are in place and discussed in the *EH&S Manual* to prevent and mitigate EH&S hazards.

Before work is performed by the Project, hazards are identified and analyzed so that appropriate controls can be developed. Hazard analyses are performed at the facility level and at the project level for major projects as well as at the activity level by employees, visiting scientists, and contractors. The complexity and formality of the hazard identification process and subsequent development of work controls are tailored to the conditions and work activity. Similarly, individual hazard analyses are tailored to the specific conditions and nature of the work. In addition, the *EH&S Manual* identifies the requirements and training necessary to ensure that personnel conducting these reviews are qualified.

## 1.11 Education and Public Outreach Overview

The Sanford Center for Science Education (SCSE) will be the education and outreach arm of DUSEL,[13] and will be fully integrated into the operations of the DUSEL Facility and science programs. SCSE is receiving significant philanthropic funding from T. Denny Sanford, and aims to exceed typical expectations for education and outreach at a national scientific research facility. This science education center will provide innovative programs, expand educational opportunities for a wide array of students, and generally enhance the efforts of DUSEL. The Homestake location provides a unique environment in which to structure education and outreach programs—using the area's geology and ecology, its history, and its native people.

The SCSE will feature engaging, hands-on, and highly interactive science activities with components geared to a wide array of audiences, from tourists to school groups to educators to scientists. The somewhat remote location necessitates development of a compelling off-site interpretive program as well, to complement what is envisioned on-site and to offer equally captivating experiences to explore DUSEL science through virtual experiences and interactions. The key messages of the SCSE program will be drawn from the scientific research programs and discoveries taking place at DUSEL.

Due to the historical importance of the Black Hills area to Lakota culture and also because of the low representation of American Indians in all disciplines of science and engineering, people of American Indian descent represent an especially important audience. Historically under-represented groups throughout the region, especially within the physical sciences and engineering, also include women and girls, and rural populations. DUSEL is deeply committed to serving all these audiences.

## 1.12 Beneficial Impact on South Dakota's Universities, Regional Universities, and Research Organizations

The DUSEL Project has established a major goal of achieving significant and lasting benefits to universities in both South Dakota and the region. These universities have demonstrated a high level of interest in the scientific and engineering opportunities represented by DUSEL, and are geographically well situated to participate in the DUSEL Project through its multiple phases, including design, construction, and laboratory operations. Several institutions have invested significant resources in both Sanford Laboratory and DUSEL.

This participation was facilitated by establishing the DUSEL South Dakota Project Office at the SDSM&T. Furthermore, the education and public outreach efforts are spearheaded by Black Hills State University (BHSU) faculty. The SDSTA plays an essential role as landlord for the Facility site, establishing and operating the early science program, maintaining and operating the Facility prior to the formation of the joint operating entity (the DUSEL LLC), maintaining the PDA with Barrick, and receiving and overseeing philanthropic donations. The DUSEL LLC recognizes and integrates these efforts into an effective management organization during the Final Design and maintains this through Construction and Operations.

During the Operations phase, the DUSEL Project has defined scientific research staff to assist with the suite of experiments. This staff will maintain research positions supported within the DUSEL Project and will have the opportunity for joint faculty positions within the South Dakota and regional universities. This will afford the research staff the benefits of academic posts and laboratory research positions. Furthermore, these positions will significantly enhance the universities' participation in DUSEL's experimental programs. To facilitate this interaction, UC Berkeley and SDSM&T plan to establish complementary centers for underground science and engineering at each of the institutions. These multidisciplinary research centers will promote underground research areas and will coordinate and provide postdoctoral and graduate student assistance to the experimental efforts. The cooperative underground science and engineering centers approach will ensure that regional, scientific, and engineering objectives involving research positions and educational opportunities will be achieved.

## 1.13 DUSEL Operations Estimate

The DUSEL Preliminary Design includes estimates of the staff support to develop the DUSEL Final Design, develop DUSEL's suite of experiments, construct the Facility and its experiments, and ultimately operate the Facility and support an ongoing R&D effort. Consistent with the NSF *Large Facilities Manual,* the DUSEL Project will undergo a series of transitions as it prepares for the Steady-state Operations phase, as described in Volume 10, *Operations Plans*. The projected staffing profile is presented in Figure 1.13. The operations estimates are presented in Volume 2, *Cost, Schedule, and Staffing*.

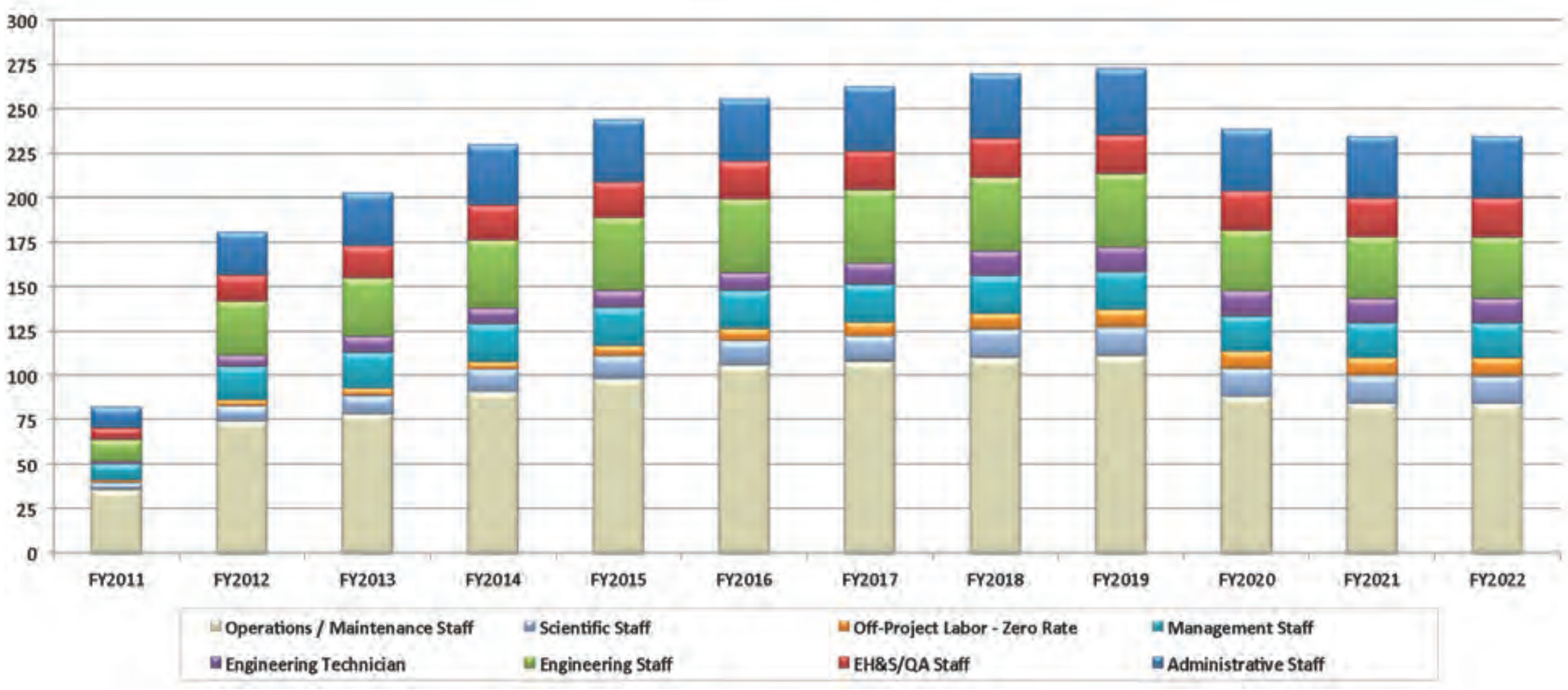

**Figure 1.13** The operations, technical, and scientific staff populations. The details of the staffing profiles are discussed in Volumes 2 and 10. (FY 2011 represents six months' staffing to correspond with the planned start of the Transition phase in April 2011.)

## 1.14 Preliminary Design Report Document Organization

The DUSEL Preliminary Design Report (PDR) is presented in 10 volumes. These volumes present the design of the DUSEL Facility and its integrated suite of experiments at the level of completeness defined in the *Large Facilities Manual*.[4]

**Volume 1** provides an overview of the DUSEL Project.

**Volume 2** presents the baseline cost, schedule, and staffing assessments for the Project from Final Design through Operations phases.

DUSEL's Science and Engineering Research Program is presented in **Volume 3**. These programs constitute the DUSEL suite of transformational experiments. The experiments span multiple efforts in physics, Earth science, biology, and engineering.

The DUSEL Education and Public Outreach efforts and plans, which are integrated with the DUSEL Project and its experimental programs, are presented in **Volume 4**.

The Preliminary Design for the construction of the DUSEL Facility is presented in **Volume 5**.

The Integrated Environment, Health, and Safety Management is presented in **Volume 6**.

The DUSEL Project Execution Plan is presented in **Volume 7** including topics of management and organization, risk management and assessment, and Project-management systems.

**Volume 8** summarizes the associated Project management control systems, including Project controls, acquisition strategy, and procurement plans.

The Systems Engineering plans and implementation are presented in **Volume 9**.

DUSEL will become a major, multidisciplinary research facility, and the plans for operating the Facility and managing the science programs are presented in **Volumes 2, 7,** and **10.** In addition, **Volume 10** summarizes the transitions between each of the Project phases and provides a description of the level of effort for the staffing for each phase. While this PDR provides a thorough presentation of the Facility and its science programs, the complete presentation requires the use of extensive **Appendices**, presented in the Table of Appendices.

## Volume 1 References

This page intentionally left blank

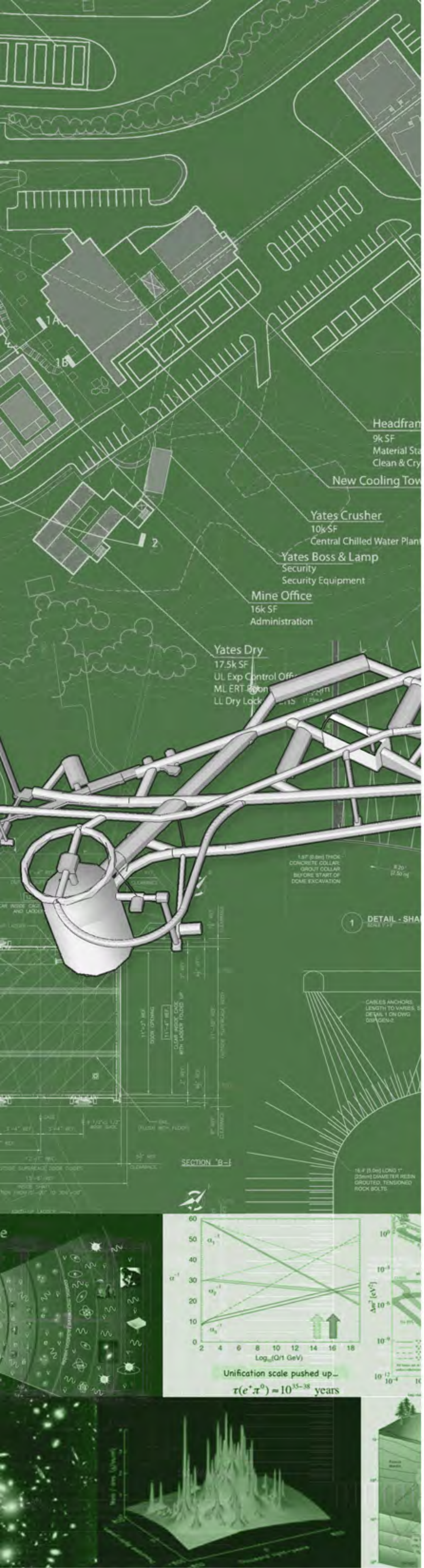

# Preliminary Design Report

May 2011

# Volume 3:
## Science and Engineering Research Program

DUSEL
Deep Underground Science and Engineering Laboratory

This page intentionally left blank

# Science and Engineering Research Program

## Volume 3

### 3.1 Introduction

Integrating the scientific goals and the experiments designed to meet these goals is an essential driver of the DUSEL Facility design. A brief overview of the scientific goals for experiments at DUSEL is presented in Chapter 3.2; a detailed review being beyond the scope of this volume. A short overview of the Facility design is given in Section 3.3.2. More detailed descriptions are in other volumes of this Preliminary Design Report, in particular Volume 5, Facility Preliminary Design. The candidate experiments that were used to establish the requirements for the DUSEL Facility are described in detail in Chapter 3.3. The Early Scientific Program, under way at Sanford Laboratory, is described in Chapter 3.4. Chapters 3.5 to 3.9 describe the process for establishing Facility design requirements based on the candidate experiments, the experience at Sanford Laboratory, and other criteria. The essential requirements of the Facility are also presented in these sections. Finally, preliminary research-program planning is described in Chapter 3.10.

### 3.2 Overview and Scientific Motivation

The deep underground, low-background environment at DUSEL is required to confront experimentally some of the most critical issues in fundamental physics and cosmology. DUSEL will be a unique facility not only in which to address questions in fundamental physics, astrophysics, and cosmology but also simultaneously to engage in cutting-edge research in underground biology, geosciences, and construction engineering. In the sections below, very short introductions to the scientific goals for experiments projected to operate at DUSEL are given. Specific examples of experiments proposed for a DUSEL are presented in Chapter 3.3.

#### 3.2.1 Physics

The Standard Model of elementary particles, though successful in describing the basic structure of elementary particle components and their interactions, is incomplete. There is uncertainty about the as-yet unobserved Higgs boson and associated mechanism of electroweak symmetry breaking, and its charge and parity (CP) symmetries violation effects are entirely too weak to account for the baryon asymmetry of the universe. An outstanding and fundamental question is how to accommodate neutrino masses. It is not yet known whether neutrinos are *Dirac* fermions (as are quarks) or *Majorana* fermions. In the latter case, lepton number conservation, a fundamental standard-model rule, is violated. The most promising way to probe this is by searching for neutrinoless double-beta decay ($0\nu\beta\beta$), a process only allowed if neutrinos are *Majorana* particles. The theoretical expectations for $0\nu\beta\beta$ decay depend sensitively on other neutrino parameters as well, most notably on the mass scale and the ordering (hierarchy) of the neutrino masses.

While the existence of dark matter is well established by cosmological observations on a wide range of scales, its nature is unknown, a major puzzle of nature. Speculation abounds. Indeed, well-motivated theoretical considerations suggest strongly that new physics will appear at the TeV energy scale. Theoretical models typically contain new particles, often including an electrically neutral, stable, Weakly Interacting Massive Particle (WIMP). An example is supersymmetry (SUSY), which is a new symmetry principle, and necessarily implies extending the Standard Model to include a new set of particles. The lightest neutral SUSY particle, the neutralino, is thought to be stable. It is a suggestive and nontrivial coincidence that the TeV scale and typical weak interaction cross sections are just right, so that such a particle could be a thermal relic of the early universe and account for the observed dark-matter abundance. If this is correct, it may be possible to produce such a particle in proton-proton collisions at the Large Hadron Collider (LHC), and detect it in experiments there. The LHC alone, however, cannot establish its identity as dark matter; other complementary techniques will be required. The identity of dark matter in the universe is a central question in particle physics and cosmology. Knowing more about the properties of dark matter will, therefore, have profound and broad implications on fundamental particle physics as well as astrophysics. The large-scale dark-matter experiments that will be possible at DUSEL are likely to be critical to establishing and understanding the nature of dark matter.

The uniquely quantum-mechanical phenomenon that neutrinos of different flavors oscillate into one another and must therefore have non-zero masses has been established. Precision neutrino-oscillation experiments are essential to improve knowledge of the oscillation parameters and to search for CP violation in the neutrino sector. CP violation might be key to understanding the still incomprehensible baryon asymmetry in the universe. Over the past decade, substantial progress has been made toward determining neutrino properties. However, the *Dirac* or *Majorana* nature of neutrinos is as yet unknown, which can only be probed by more sensitive neutrinoless double-beta decay experiments. The opportunity to provide spaces both for the needed large-scale neutrinoless double-beta search experiments as well as neutrino-oscillation experiments is a feature of the DUSEL Facility.

Are protons absolutely stable? There is currently no evidence for proton decay after decades of experimental research: The proton lifetime is greater than $1x10^{34}$ years, about $10^{24}$ times the lifetime of the universe. The detection of proton decay would have profound implications for understanding of the universe. The very large detectors needed to advance determination of neutrino-oscillation parameters would also extend substantially the search for proton decay.

Nuclear astrophysics is concerned with the origin of elements in stars and stellar explosions and informed by measurements of nuclear processes. This research is at the intersection of nuclear physics, astrophysics, cosmology, and observational astronomy. Measurements of the very low reaction cross sections, a critical aspect of increased understanding, can only be done with accelerators deep underground to achieve acceptably low backgrounds.

#### 3.2.1.1 Dark Matter[1]

A broad range of experimental observations tells us that our universe consists of ~73% dark energy, ~23% nonbaryonic dark matter, and ~4% baryons. Of fundamental importance to cosmology, astrophysics, and elementary particle physics, the nature of dark matter awaits elucidation. It is likely some new form of matter pointing to physics beyond the Standard Model.

Among the many candidates suggested but as yet undetected, one of the most compelling possibilities is that the dark matter is composed of WIMPs that were produced moments after the Big Bang from collisions of ordinary matter. WIMPs denote a general class of particles produced in the hot early universe. They drop out of equilibrium when the temperature becomes less than their mass, so that they can no longer be pair-produced. Their density today is inversely proportional to their annihilation rate; to represent ~23% of the critical density, their annihilation cross section must be typical of electroweak-scale interactions, hinting at new physics outside the Standard Model at the TeV scale. The SUSY neutralino is a natural dark-matter candidate.

WIMPs from the galactic halo are detected by their scattering on atomic nuclei in terrestrial detectors. Because the energy of these nuclear recoils is about 10 keV, where electromagnetic backgrounds can dominate by many orders of magnitude, the technical challenge is to combine low-radioactivity materials and environments with rejection of electron recoil events to keep spurious signals at bay. These detectors must be sited in deep-underground laboratories to shield them from cosmic-ray-induced backgrounds. Examples of the broad range of techniques that have been developed to address these challenges are detailed in Chapter 3.3. At present, the upper limit on the WIMP-nucleon cross section approaches $10^{-44}$ $cm^2$, well into the region of parameter space where SUSY particles could account for the dark matter (illustrated in Figure 3.2.1.1). The next 2 to 3 orders of magnitude represent a particularly rich region of electroweak-scale physics. The combination of what is to be learned from direct searches and astrophysical observations, combined with accelerator-based experiments, is truly profound.

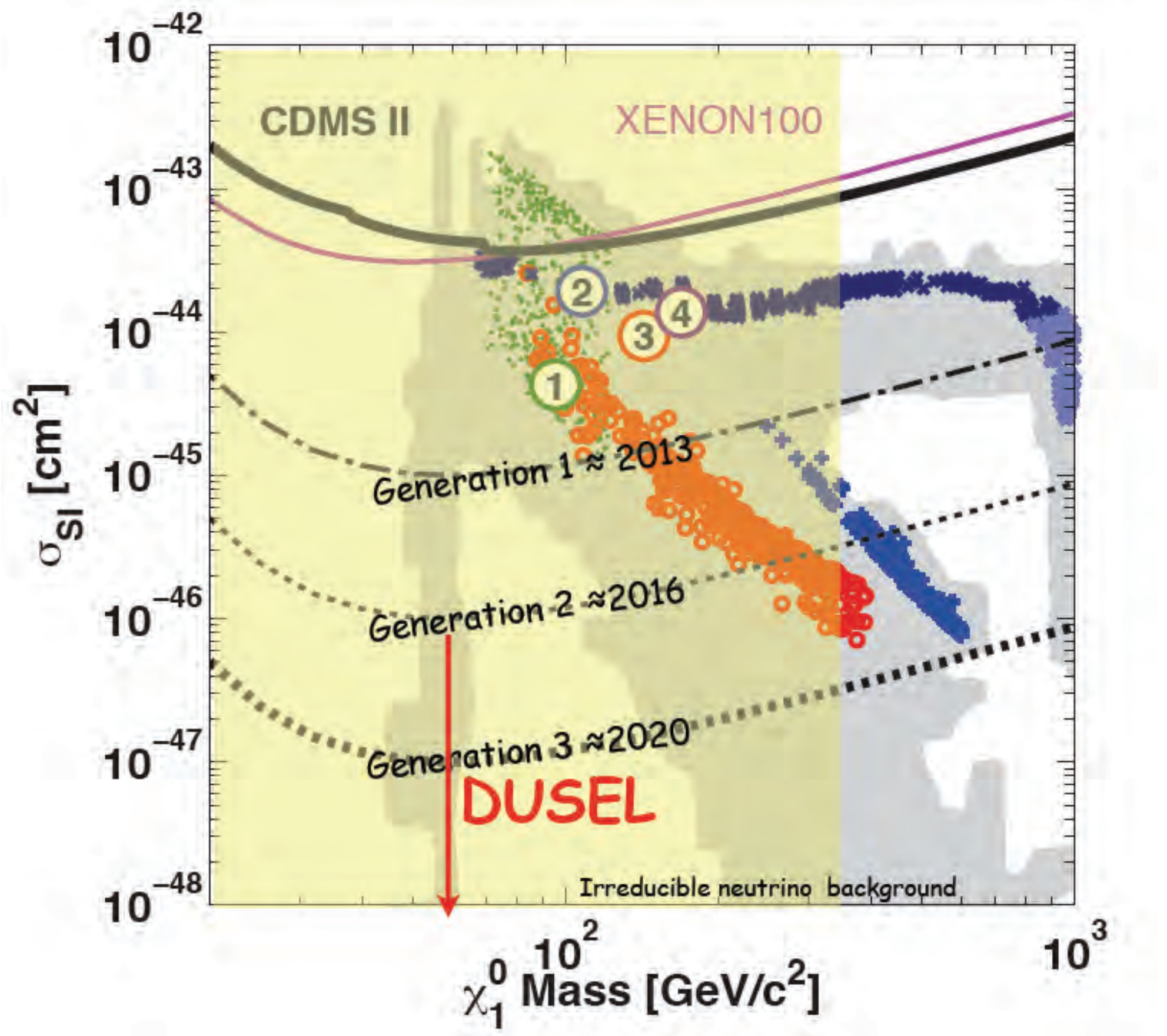

**Figure 3.2.1.1** WIMP (lightest neutralino) current limits (solid lines) from CDMS II and XENON100 and sensitivity and goals for the next three years (dashed-dotted line), G2 (dotted line) experiments (with results in 2016), and G3 (heavy dotted) experiments (yielding results in ≈2020). The shaded regions represent expectation for minimum supersymmetry models. The points indicate the various regions (green = "bulk," dark blue = "focus point," red = "co-annihilation," blue = "Higgs funnel") of an oversimplified mSUGRA model (A=0, μ>0). Numbers are the benchmarks of Baltz et al. (2006).[2] The yellow region is accessible to the LHC. Experiments in DUSEL will target primarily the Generation 3 (G3) sensitivity level.[3] [Courtesy U.S. dark matter working group]

### 3.2.1.2 Neutrinoless Double-Beta Decay[4]

The consequences of the discovery of neutrino oscillation are far reaching. The existence of finite yet peculiarly small neutrino masses and nearly maximal flavor mixing provide a picture of the lepton sector that is very different from that of the quarks. Indeed, clues to the mechanisms underlying elementary particle masses may emerge from the study of neutrinos, and physics phenomena at very large energy scales—well beyond what is possible with conceivable accelerators—may become accessible through some form of the *seesaw mechanism*[5] The magnitude of the neutrino mass scale may first become accessible through the observation of neutrinoless double-beta ($0\nu\beta\beta$) decay. That would simultaneously be the first example of a lepton-number violating process and demonstrate that two-component *Majorana* particles exist in nature. The lepton-violating nature of *Majorana* neutrinos could provide a means for generating the observed matter/antimatter asymmetry in the universe through *leptogenesis.*[6]

In conventional, two-neutrino, double-beta decay, the Z of a nucleus changes by two, emitting two electrons and two neutrinos. Only the two electrons are experimentally observable. Because some of the energy is carried away by the neutrinos, the energy sum of the two electrons follows a continuous spectrum. If neutrinos are *Majorana* particles, a second decay mode is allowed in which a single neutrino appears as a virtual particle, and no neutrinos appear in the final state so that the electrons carry away the entire available kinetic energy. In a calorimetric experiment, this neutrinoless mode is detectable as a mono-energetic peak at the endpoint of the two-neutrino mode spectrum (see Figure 3.2.1.2-1). The decay rate scales with the *Majorana* mass of the neutrino. The detection of a neutrinoless double-beta decay peak would therefore not only confirm the *Majorana* nature of the neutrino, but also simultaneously give a measure of its absolute mass.

Conventional double-beta decay has been observed in a number of nuclei, with measured half-lives of $10^{20}$ years or longer. Present limits on neutrinoless double-beta decay half lives exceed $10^{25}$ years. Hence, experiments that aim to reliably detect $0\nu\beta\beta$ decay in a finite time require:

- Large amounts of source material
- Sufficient energy resolution to identify the $0\nu\beta\beta$ peak
- Ultralow backgrounds in the region of the $0\nu\beta\beta$ peak

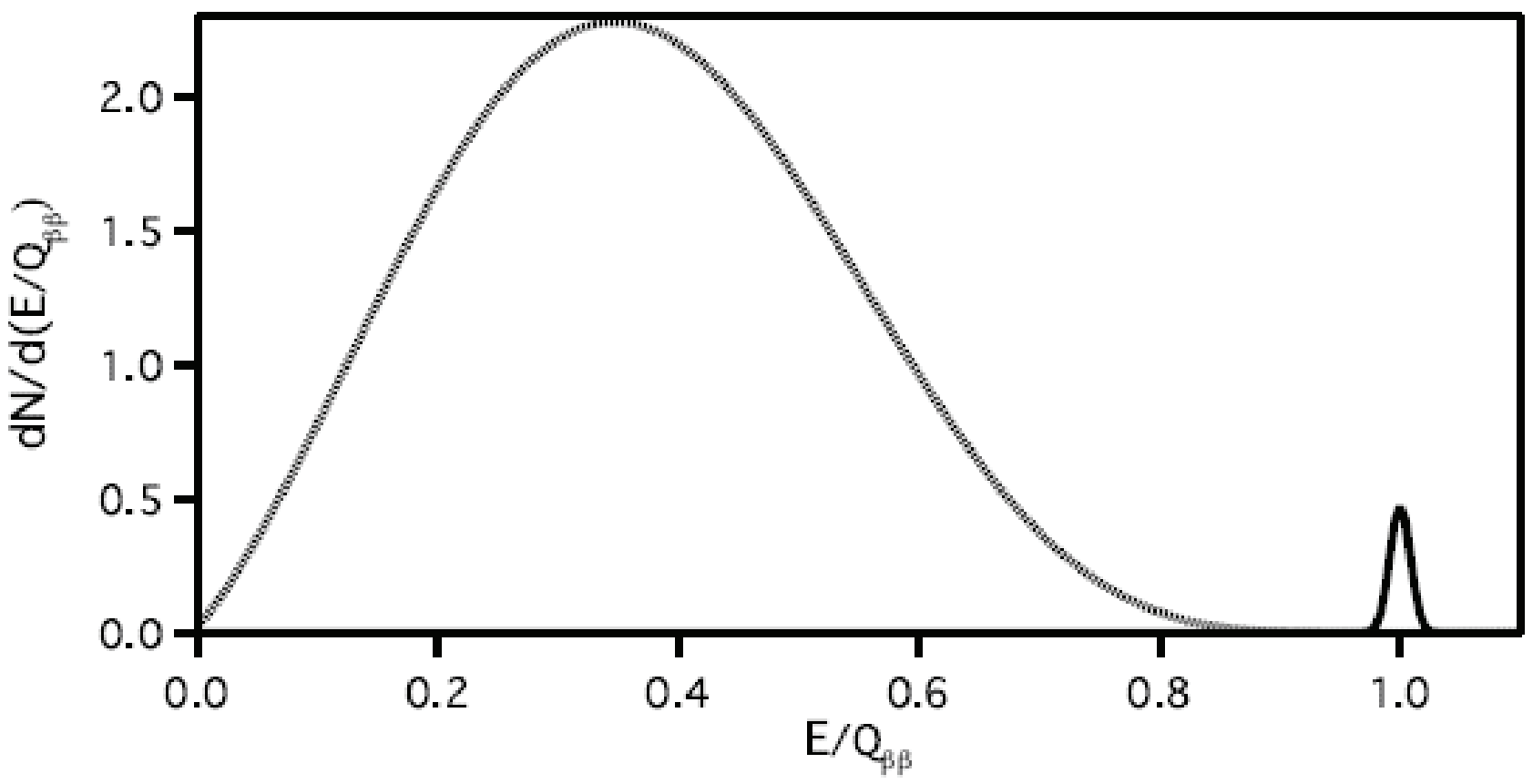

**Figure 3.2.1.2-1** A schematic double-beta decay energy spectrum, as would be measured by a calorimetric detector. The small peak from $0\nu\beta\beta$ decay is shown here with arbitrary height beside the background from the two-neutrino mode. [Courtesy 1TGe collaboration]

The half-life of 0νββ decay $T_{½}^{0\nu}$ depends on a phase-space factor $G^{0\nu}$, a nuclear matrix element $M^{0\nu}$, and an effective double-beta decay neutrino mass $m_{\beta\beta}$:[7]

$$T_{1/2}^{0\upsilon} = \left(G^{0\upsilon}\left|M^{0\upsilon}\right|^2 m_{\beta\beta}^2\right)^{-1}$$

$$m_{\beta\beta} \equiv U_{e1}^2 m_1 + U_{e2}^2 m_2 e^{i\phi_2} + U_{e3}^2 m_3 e^{i\phi_3}$$

where $U_{\alpha i}$ are elements of the neutrino mixing matrix (see Section 3.2.13), $m_i$ are the neutrino masses, and $\phi_i$ are *Majorana* phases. The parameter space for which 0νββ decay is allowed depends on the neutrino mass hierarchy, as seen in Figure 3.2.1.2-2. The phase space factor increases with the $Q$ of the decay, varying from $10^{-25}$-$10^{-26}$ years$^{-1}$ eV$^{-2}$, depending on the isotope. The nuclear matrix element is also isotope dependent, with different calculations giving values ranging from ~1-5. Reaching $m_{\beta\beta}$ sensitivity corresponding to the atmospheric mass splitting (~50 meV) requires half-life sensitivities above $10^{27}$ years; covering the inverted hierarchy parameter space for $m_{\beta\beta}$ requires sensitivity to half-life greater than $10^{28}$ years.

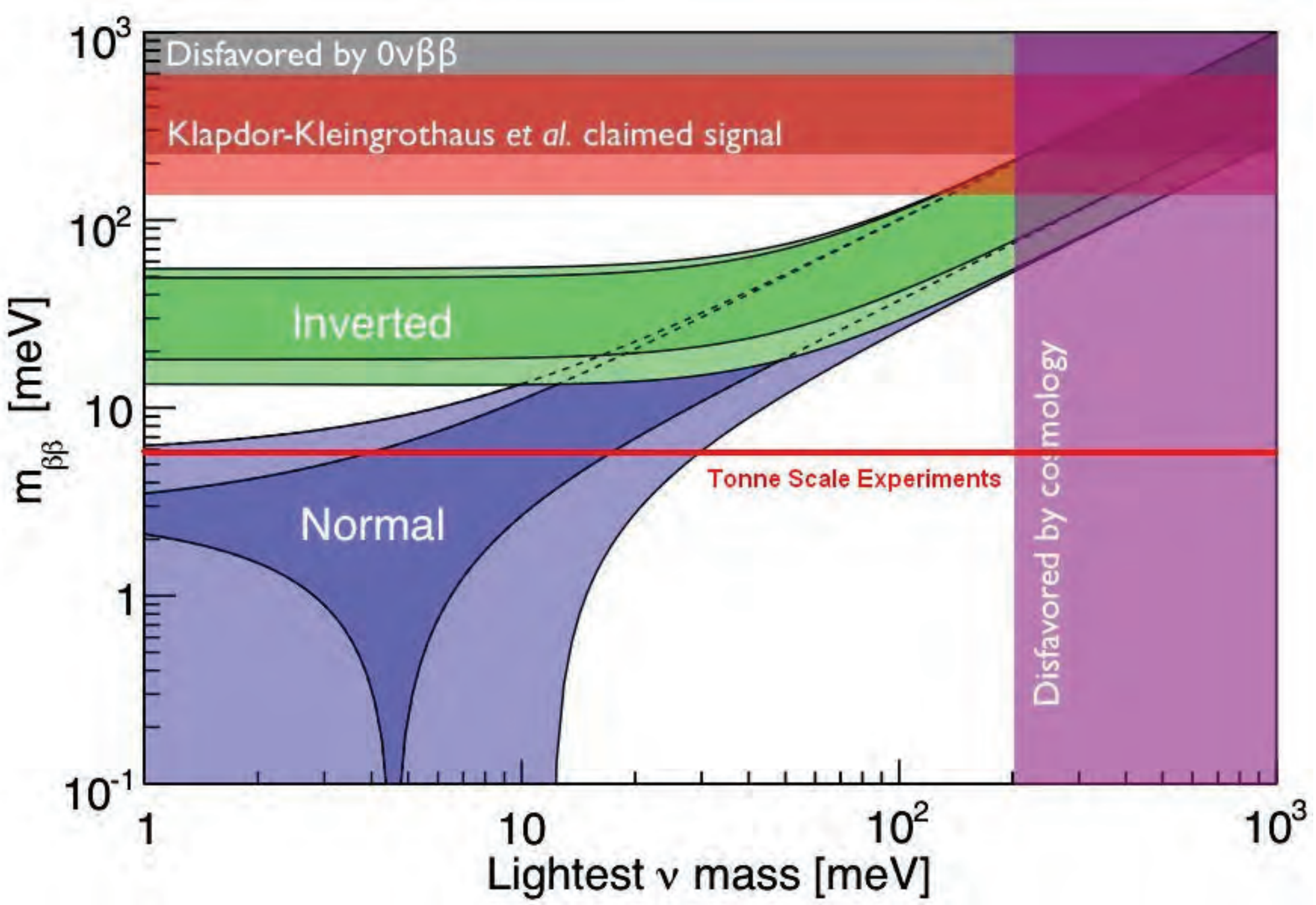

**Figure 3.2.1.2-2** Phase space for neutrinoless double-beta decay shown in terms of the lightest neutrino mass (constrained by cosmology) and the effective double-beta decay neutrino mass (constrained by current 0νββ-decay experiments). Tonne-scale experiments could extend the $m_{\beta\beta}$ sensitivity down to the red line. Also shown is a recent controversial claimed observation of a 0νββ signal,[8] which will be tested by the generation of experiments prior to the tonne-scale ones proposed for DUSEL. [Jason Detwiler, DUSEL]

Figure 3.2.1.2-3 illustrates how the sensitivity to $m_{\beta\beta}$ depends on the exposure and the background level of an experiment. Although Figure 3.2.1.2-3 is drawn using values of $G^{0\nu}$ and $M^{0\nu}$ calculated for $^{76}$Ge, the situation is qualitatively the same for any ββ isotope.[9] In the absence of background, reaching the extreme

half-life sensitivity required to definitively test the inverted mass hierarchy will require counting tonnes of isotope for multiple years. The presence of even a few background counts in the analysis region of interest (ROI) for the 0νββ peak search causes rapid degradation of the sensitivity. Hence, experiments that aim to reliably detect 0νββ decay in a finite time require:

- Large amounts of source material
- Ultra-low background in the vicinity of the 0νββ peak

Since most ββ isotopes have a small natural abundance, it is typically advantageous to build 0νββ decay detectors using isotopically enriched material.

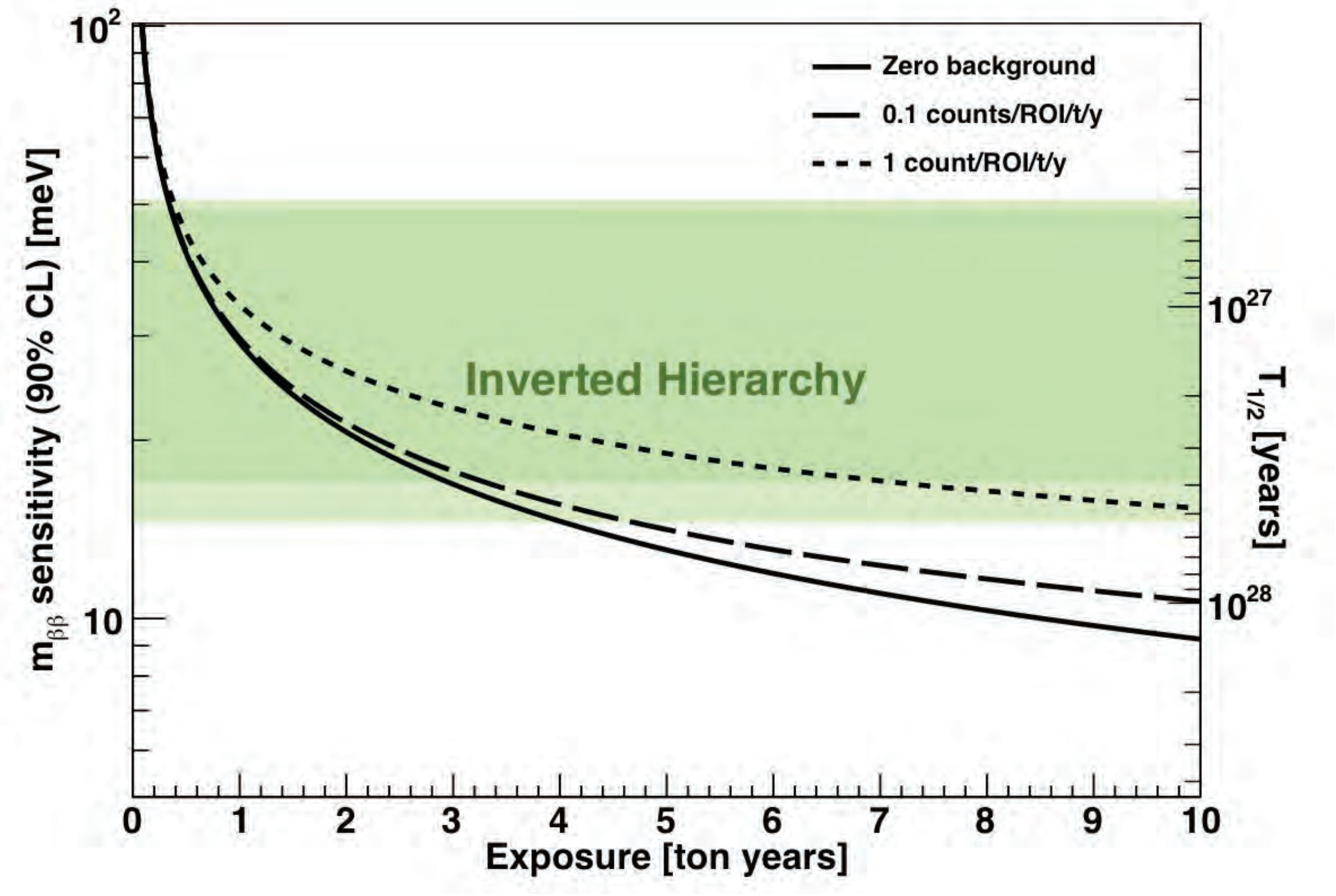

**Figure 3.2.1.2-3** Sensitivity to $m_{\beta\beta}$ as a function of exposure (mass x counting time) under different backgrounds scenarios for $^{76}Ge$. The background rate is expressed as a counting rate per unit mass in the analysis ROI for the 0νββ peak search. The shaded region corresponds to the allowed values of $m_{\beta\beta}$ for inverted hierarchical masses in the limit $m_1 \rightarrow 0$. [Jason Detwiler, DUSEL]

Although enrichment can be a major cost driver, the cost and complexity of the required infrastructure can be greatly reduced because less total mass can be employed to achieve the same results. Further, enrichment can reduce the background per unit active mass, leading to substantial improvements in sensitivity.

Achieving ultra-low backgrounds requires also that the detector be constructed of radiopure materials using clean techniques. Some materials must be shielded during transport or even produced underground to reduce long-lived cosmogenic spallation products to acceptable levels. External radiation must be attenuated with thick, graded shielding. These experiments are also extremely sensitive to cosmogenic activity, especially inelastic interactions with high-energy spallation neutrons. These considerations drive the need to build large-scale experiments deep underground in an ultraclean environment.

### 3.2.1.3 Neutrino Oscillations[10]

Results from the past decade reveal that the three known types of neutrinos (e,μ,τ) have nonzero mass, mix with one another, and oscillate between generations. These facts point to physics beyond the Standard Model. Measuring the mass and other properties of neutrinos is fundamental to understanding the deeper underlying theory and will profoundly shape understanding of the evolution of the universe.

The concept of neutrino oscillations was first put forward by Bruno Pontecorvo in 1957,[11] considering the possibility of $\nu \rightarrow \overline{\nu}$ oscillations. Neutrino oscillations imply that neutrinos have small masses and mix between states. Consider two neutrinos oscillating (adequate for solar- and atmospheric-neutrino oscillations). The probability of flavor change, for a neutrino of energy *E* detected a distance *L* from the source, is given by:

$$P(\nu_a \rightarrow \nu_b) = \sin^2 2\theta_{12} \sin^2(1.27\Delta m_{12}^2 \frac{L}{E})$$

where $\Delta m^2$ is in $eV^2$, *L* is in km, and *E* is in GeV. Thus neutrino oscillations at a suitable range of (*L/E*) can give information on $\Delta m^2$ and the mixing angle.

In 1998, the Super-Kamiokande Collaboration announced “Evidence for Oscillation of Atmospheric Neutrinos” based on the observation of a deficit of muon neutrinos as a function of the distance traversed through the Earth.[12] The data were consistent with two-flavor $\nu_\mu \rightarrow \nu_\tau$ oscillations with $sin^2 2\theta_{12} > 0.82$ and $5\times10^{-4} < \Delta m_{12}^2 < 6\times10^{-3}$ at 90% confidence level. This result, along with the solar neutrino deficit and subsequent results from the SNO experiment,[13] led to the solid conclusion that neutrino oscillations had indeed been observed, and in fact with two distinct $\Delta m^2$. The KamLAND experiment[14] confirmed oscillations at the “solar” mass difference via $\overline{\nu_e}$ disappearance from reactors. The KEK-to-Super-Kamiokande (K2K) and Main Injector Neutrino Oscillation Search (MINOS) accelerator long-baseline experiments[15,16] confirmed the oscillation phenomena at the “atmospheric” mass difference via $\nu_\mu$-disappearance. The ongoing Oscillation Project with Emulsion-tRacking Apparatus (OPERA[17]) experiment in Gran Sasso, using the CERN Neutrinos to Grand Sasso (CNGS)[18] neutrino beam, is attempting to measure oscillations in the $\nu_\mu \rightarrow \nu_\tau$ appearance channel.

There are known three neutrino flavors and mass states related by a 3 × 3 unitary matrix, *U*:

$$\begin{pmatrix} \nu_e \\ \nu_\mu \\ \nu_\tau \end{pmatrix} = \begin{pmatrix} U_{e1} & U_{e2} & U_{e3} \\ U_{\mu 1} & U_{\mu 2} & U_{\mu 3} \\ U_{\tau 1} & U_{\tau 2} & U_{\tau 3} \end{pmatrix} \begin{pmatrix} \nu_1 \\ \nu_2 \\ \nu_3 \end{pmatrix}$$

where *U* is defined, by convention (neglecting possible *Majorana* phases, which do not affect oscillations experiments), with three mixing angles, $\theta_{12}$, $\theta_{23}$, $\theta_{13}$, and a phase $\delta$:

$$U = \begin{pmatrix} \cos\theta_{12}\cos\theta_{13} & \sin\theta_{12}\cos\theta_{13} & \sin\theta_{13}e^{-i\delta} \\ -\cos\theta_{23}\sin\theta_{12} - \sin\theta_{13}\sin\theta_{23}\cos\theta_{12}e^{i\delta} & \cos\theta_{23}\cos\theta_{12} - \sin\theta_{13}\sin\theta_{23}\sin\theta_{12}e^{i\delta} & \sin\theta_{23}\cos\theta_{13} \\ \sin\theta_{23}\sin\theta_{12} - \sin\theta_{13}\cos\theta_{23}\cos\theta_{12}e^{i\delta} & -\sin\theta_{23}\cos\theta_{12} - \sin\theta_{13}\cos\theta_{23}\sin\theta_{12}e^{i\delta} & \cos\theta_{23}\cos\theta_{13} \end{pmatrix}$$

The absolute value of the neutrino mass scale, though known to be very small, is unknown, as is the hierarchical ordering of the mass states—though it has been determined from the solar neutrino data that $m_1 < m_2$. There are two hierarchy possibilities:

$$m_1 < m_2 < m_3$$

or

$$m_3 < m_1 < m_2,$$

the "normal" (NH) and the "inverted" mass hierarchy (IH)—see Figure 3.2.1.3. Distinguishing these has important implications for models of neutrino mass. It is also interesting to note that a consequence of an inverted hierarchy is that it predicts a larger rate for neutrinoless double-beta decay than does the normal hierarchy. The predicted rate in the IH is in fact within the reach of the next-generation neutrinoless double-beta decay experiments.

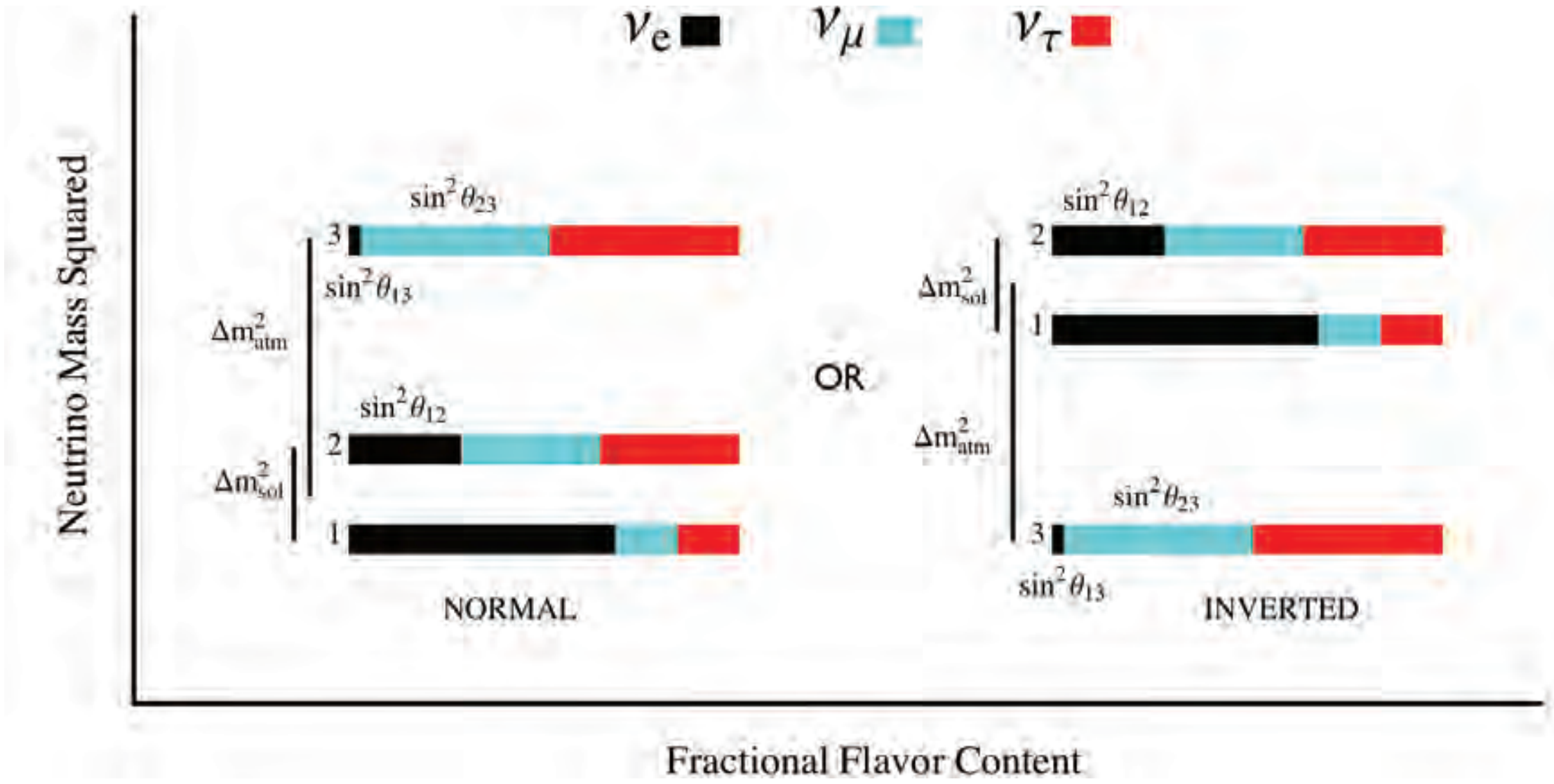

**Figure 3.2.1.3** Mass and mixing for the two arrangements of the hierarchy Illustration showing the flavor content of the three neutrino mass states for the NORMAL and INVERTED hierarchy.

The equations that give the probability of flavor transitions among the three neutrino flavors (e, μ, τ) now become more complicated. Further, should the neutrinos propagate in matter, not vacuum, the oscillation probability for $\nu_\mu \rightarrow \nu_e$ is modified by the presence of electrons in the intervening matter. The resulting "matter effect" enhances the oscillation probability for neutrinos and suppresses the probability for antineutrinos if the mass hierarchy is normal. The effect is opposite in the inverted hierarchy. The magnitude of this matter effect is governed by the amount of matter penetrated and their energy. If $L$ is sufficiently long and $E$ sufficiently high, it is possible to distinguish the NH from the IH by measuring the oscillation probabilities for neutrinos and antineutrinos.

The transition probability for muon neutrinos in matter is:

$$P(\nu_\mu \rightarrow \nu_e) \approx \sin^2 2\theta_{13} T_1 + \alpha \sin 2\theta_{13} T_2 + \alpha \sin 2\theta_{13} T_3 + \alpha^2 T_4$$

where $\alpha$ represents the small ratio $\Delta m^2_{21} / \Delta m^2_{31} \sim 1/30$, and $T_1$-$T_4$ are given by:

$$T_1 = \sin^2\theta_{23}\frac{\sin^2[(1-x)\Delta]}{(1-x)^2}$$

$$T_2 = \sin\delta\sin 2\theta_{12}\sin 2\theta_{23}\sin\Delta\frac{\sin(x\Delta)}{x}\frac{\sin[(1-x)\Delta]}{(1-x)}$$

$$T_3 = \cos\delta\sin 2\theta_{12}\sin 2\theta_{23}\cos\Delta\frac{\sin(x\Delta)}{x}\frac{\sin[(1-x)\Delta]}{(1-x)}$$

$$T_4 = \cos^2\theta_{23}\sin^2 2\theta_{12}\frac{\sin^2(x\Delta)}{x^2}$$

Here $\Delta = \Delta m_{31}^2 L/4E$ is the oscillation phase of a neutrino of energy $E$ over a path length $L$; $x = 2\sqrt{2}G_F N_e E/\Delta m_{31}^2$ introduces the matter effects, $G_F$ is the Fermi constant, $N_e$ represents the average number density of electrons along the neutrino path that competes with the CP violating phase $\delta$. The term $T_1$ ($T_4$) corresponds to the atmospheric (solar) oscillation, and $T_2$ ($T_3$) is the CP violating (conserving) term. Values for the various mixing angles and two mass differences are shown in Table 3.2.1.3. For antineutrinos, substitute $x \rightarrow -x$, and $\sin\delta \rightarrow -\sin\delta$. The matter effects compete directly with CP violation via the $T_2$ and $T_3$ terms above. Given the typical energy of the Fermi National Accelerator Laboratory (Fermilab)-DUSEL neutrino beam of ~0.1-5GeV, and a path length of $L \geq 1000$ km, the matter effects can be used to disentangle the mass hierarchy as inverted or normal.

| | |
|---|---|
| $\Delta m_{12}^2$ | $7.59 \pm 0.02 \times 10^{-5} eV^2$ |
| $\Delta m_{23}^2$ | $2.43 \pm 0.13 \times 10^{-3} eV^2$ |
| $\sin^2 2\theta_{12}$ | $0.87 \pm 0.03$ |
| $\sin^2 2\theta_{23}$ | $> 0.92$ |
| $\sin^2 2\theta_{13}$ | $< 0.19$ (90% CL) |

**Table 3.2.1.3** Current knowledge[10] of three mixing angles and two $\Delta m^2$.

To date, only an upper limit on the value of the third mixing angle $\theta_{13}$ is known. It should be noted that if $\theta_{13}$ were identical to 0, there would be no $\nu_e$ content in the $\nu_3$ mass state and there would be no CP violation possible in the neutrino sector. The dependence on $\theta_{13}$ of the experimentally observable CP asymmetry in $\nu_\mu \rightarrow \nu_e$ is not linear.[19] At very small values of $\theta_{13}$, the CP asymmetry is diluted. For very large $\theta_{13}$, the CP asymmetry would have been rather small; however, the current experimental bound indicates this is not the case: $\theta_{13}$ is relatively small. This situation poses a challenge: The small value of $\theta_{13}$ already indicates that the event rates will be small and therefore will require very large beam intensities and detector mass. On the other hand, the asymmetry is expected to be rather large (~30%), observable above background, as long as $\theta_{13}$ is not "too" small.

### 3.2.1.4 Proton Decay

The search for proton decay tests the apparent but unexplained conservation of baryon number. In the Standard Model, baryon and lepton number conservation is basically a consequence of the quarks and leptons being organized in separate multiplets. Grand Unified Theories (GUTs) organize the quarks and leptons into combined multiplets and allow for the conversion of quarks into leptons by the exchange of a force-carrying particle. Indirect evidence for such unification is the predicted meeting of the strengths of the three forces, electromagnetism, weak, and strong, to occur at scales of $10^{14}$ to $10^{16}$ GeV. These energy scales are inaccessible to accelerators. Assuming the mass of the force-carrying particle is at these scales, a proton or bound neutron is then predicted to decay to leptons and mesons, but at very slow rates, commensurate with observed stability of matter. Establishing this picture of the violation of baryon-number conservation would have profound implications for understanding of cosmology and particle physics.

When GUTs were first developed in the 1970s, early theories, for example, SU(5), predicted lifetimes as small as $10^{29}$ to $10^{30}$ years with a principal decay mode $p \rightarrow e^{+} + \pi^{0}$. This prompted first-generation searches in kiloton-scale experiments such as Soudan, Frejus, Kamiokande, and Irvine-Michigan-Brookhaven (IMB).[20] No signals were observed in these experiments for a large number of decay modes, ruling out the minimal SU(5) theory and severely constraining and limiting the development of GUTs. Nevertheless, revised theories were developed that evaded the experimental constraints. In particular, GUTs based on SUSY were developed that disallowed the decay $p \rightarrow e^{+}\pi^{0}$, but predicted new principal modes such as $p \rightarrow \bar{\nu} K^{+}$.

If the scale of SUSY is just above the electroweak scale (where it could be observed at LHC), very rapid proton decay is predicted. A new conservation principle (such as R-parity) is needed to avoid this rapid decay. Decay modes could still occur via exchange of supersymmetric particles in dimension five operators.[21] The lifetime of the proton then depends on the mass of the supersymmetric exchange particles and the strength of coupling to "normal" particles. There is therefore great uncertainty in the theoretical predictions.

Following the round of first-generation searches, the 22.5 kT Super-Kamiokande experiment became operational in 1996, and it continues to integrate exposure time. At present, the best limit[22] on the classic mode, $p \rightarrow e^{+} + \pi^{0}$ comes from a 0.17 megaton-years exposure of Super-Kamiokande, expected to improve by a factor of ~3 by 2030. The detection efficiency of 45% is dominated by final-state $\pi^{0}$ absorption or charge-exchange in the nucleus (the efficiency for LAr suffers from the same problem), and the expected background is $1.63\,^{+0.42}_{-0.33}\,(stat)\,^{+0.45}_{-0.51}\,(sys)$ events/Mton-yr.[23] The decay mode $p \rightarrow K^{+}\ \bar{\nu}$ is experimentally more difficult in water Cherenkov detectors due to the unobservable neutrino and the fact that the kaon daughter is below the Cherenkov threshold. The present limit from Super-Kamiokande is the result of combining several channels, the most sensitive of which is $K^{+} \rightarrow \mu\nu$ accompanied by a de-excitation signature from the remnant $^{15}$N nucleus. Monte Carlo studies suggest that this mode should remain background-free for the foreseeable future. To date, no significant signature has been observed in numerous modes. The limits placed on the key modes $e^{+}\pi^{0}$ and $\bar{\nu} K^{+}$ are currently at a 90% confidence level:

$$\tau / B(p \rightarrow K^{+}\nu) > 3.3 \times 10^{33} \text{ years.}$$

$$\tau / B(p \rightarrow e^{+}\pi^{0}) > 1 \times 10^{34} \text{ years}$$

Figure 3.2.1.4 illustrates these limits, along with the predicted lifetime ranges allowed by a wide variety of GUTs. The Super-Kamiokande lifetime limits achieved serve as severe constraints and strict guidance for the development of further theories.

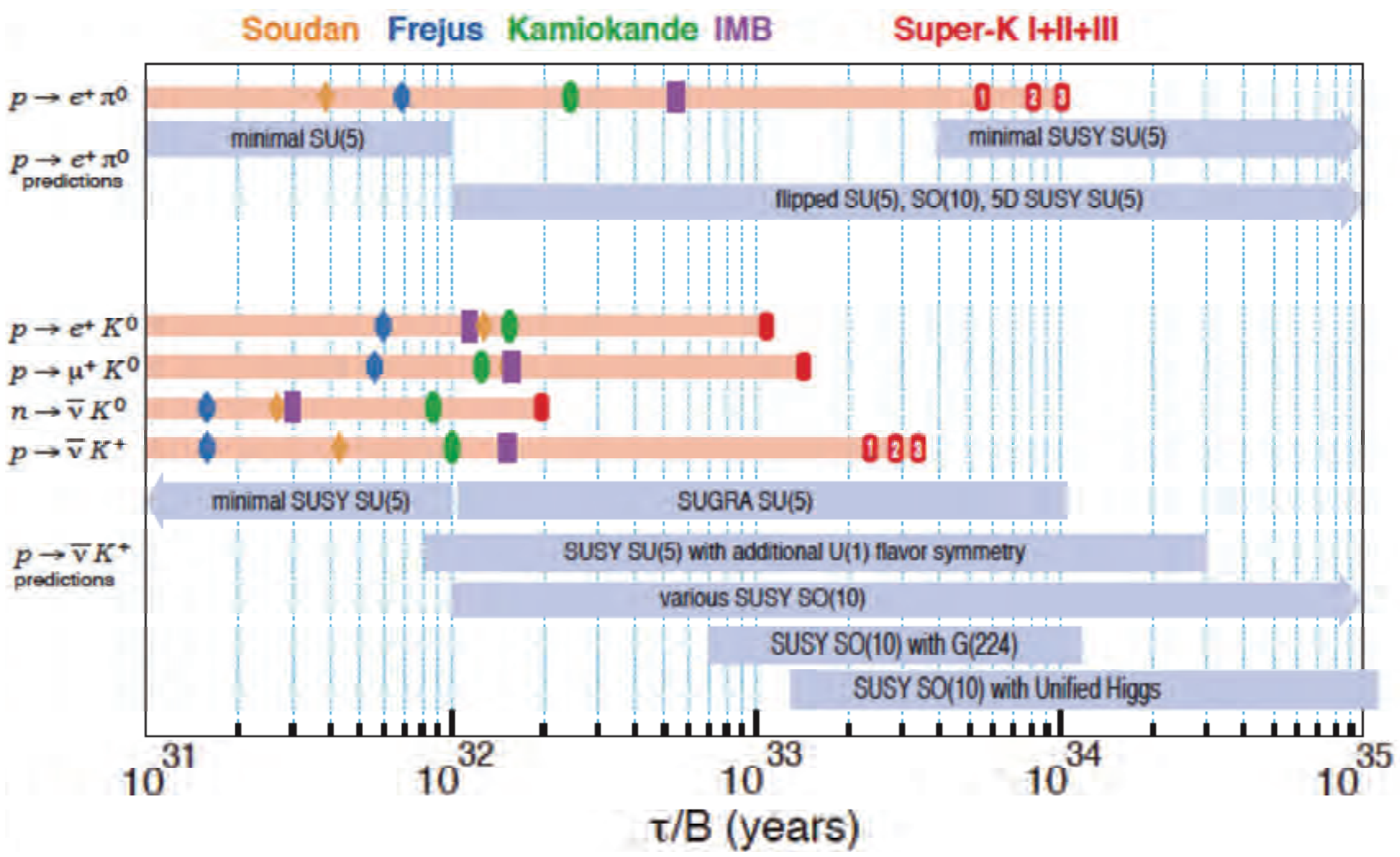

**Figure 3.2.1.4** Comparison of present limits with the range of proton and bound nucleon lifetimes supported by various GUTs. The top section shows the limits and predictions for $p \rightarrow e^+\pi^0$, which are typically favored by the exchange of a heavy force-carrying particle. The bottom section shows the predictions for $p \rightarrow \nu\, K^+$, which is typically favored by SUSY. In addition to the experimental limit of this mode, the limits for other modes favored by SUSY, which have comparable lifetime predictions, are included. [Courtesy LBNE]

An assessment of early 21st century theories shows that several predict proton decay, but lifetime predictions are not precise and typically vary over two or three orders of magnitude.

Experimental searches now need to be prepared to search for very long lifetimes; very massive detectors and very long exposure times are needed. GUTs based on a minimal supersymmetric model predict a unification energy of about $10^{16}$ GeV, pushing the partial lifetime in the $e^+\pi^0$, channel to $10^{36}$ years or so—more than two orders of magnitude beyond present experimental limits. However, some of these models do predict a partial lifetime of order $10^{34}$ years in the mode p → $\bar{\nu}K^+$.

### 3.2.1.5 Supernovae

A core-collapse of a supernova in our galaxy or a nearby galaxy may provide a wealth of information via its neutrino signal.[24] About 99% of the supernova's energy is released in an initial neutrino burst that lasts a few tens of seconds, expelling about half the neutrinos in the first second. Supernova neutrino energies range in the few tens of MeV, and their luminosity is divided approximately equally among the three neutrino flavors and between neutrinos and antineutrinos. The following science topics would be addressed by observing a high-statistics core-collapse neutrino signal:

**The properties of neutrinos.** In particular, neutrino oscillations in the core can provide information on oscillation parameters, mass hierarchy, and $\theta_{13}$, possibly down to very small values inaccessible to accelerator-base experiments. To get meaningful results, the systematics of the supernova models and neutrino transport out of the dense supernova need to be well understood.

**The astrophysics of core collapse.** The time, energy, and flavor distribution of the detected neutrinos will give valuable information on the explosion mechanism, accretion, neutron star cooling, and possible transitions to quark matter or to a black hole.[25]

**Early alert.** Because the neutrinos emerge promptly after core collapse—in contrast to the electromagnetic radiation, which must beat its way out of the stellar envelope—an observed neutrino signal can provide a prompt supernova alert.[26] This would allow astronomers to find the supernova in early light turn-on stages, which may yield information about the progenitor and its environment, and possibly allow coincident gravity waves.

The observation of 19 neutrino events in two water Cherenkov detectors for SN1987A in the large Magellanic Cloud (55 kpc) confirmed the baseline model of core collapse, but left many questions that will be unanswered until the next supernova neutrino detection.[27]

Core-collapse supernova explosions throughout the history of the universe left behind a diffuse background of neutrinos, which should be detectable on Earth. While these supernova relic neutrinos (SRN) undoubtedly permeate the universe, they have thus far evaded detection. The flux and spectrum of these neutrinos contain information about the rate of supernova explosions (and consequently the star

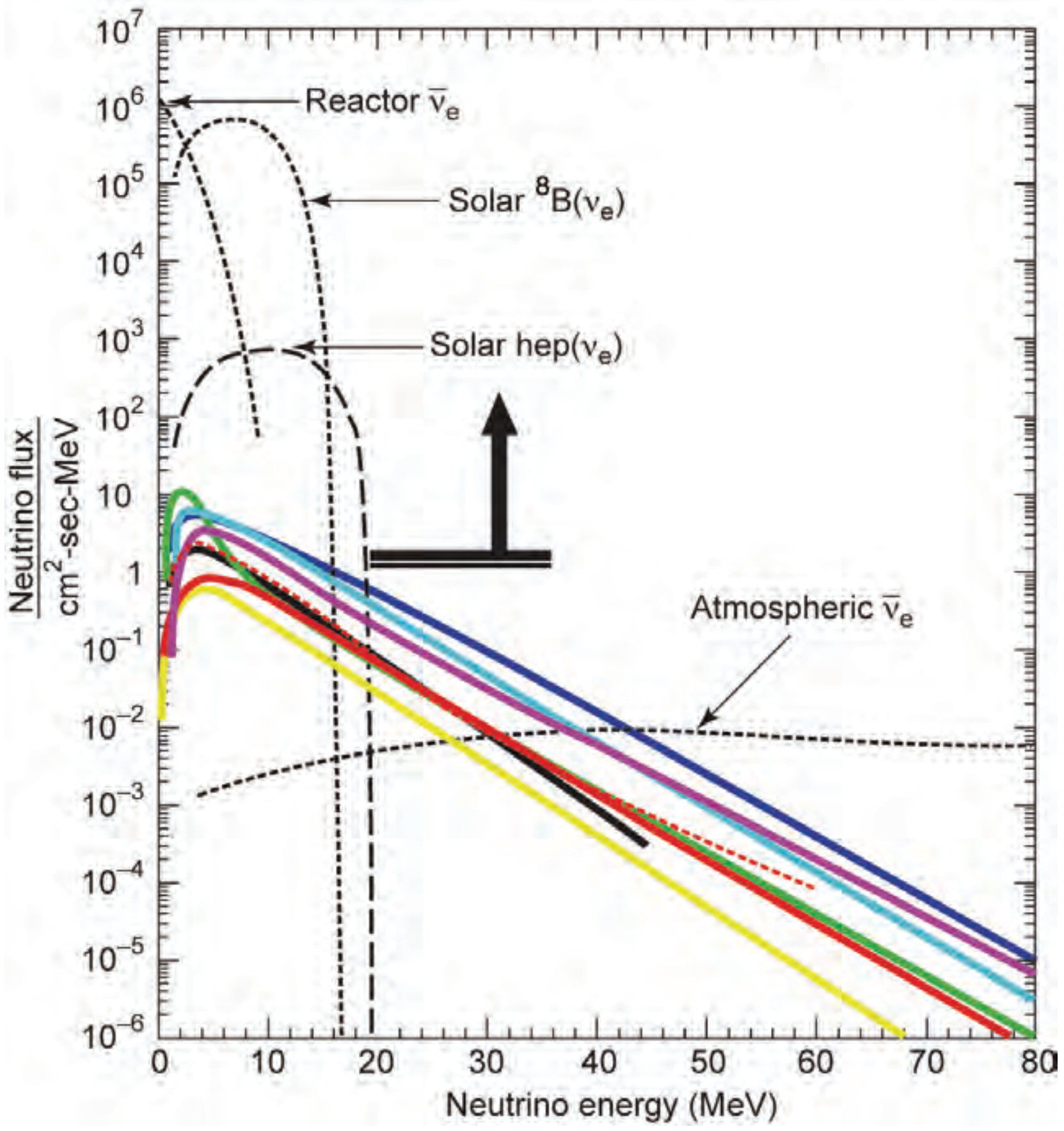

**Figure 3.2.1.5** Various theoretical expectations for diffuse neutrinos from relic supernovae. The reactor neutrino background is shown at the Super-K site. The bold line indicates the Super-K flux limit at 1.2 events/cm$^2$/sec with $E_\mu$>19 GeV. At DUSEL, the reactor background will be much smaller. [Courtesy Super-K collaboration]

formation rate) in the past. Though valuable information can be obtained from a nearby supernova burst, that is a very rare event and one may not be detected within the next few decades, so that detection of the relic neutrinos is especially important. To date, Super-Kamiokande[28] has set the best limit on their flux: flux($\nu_e$) < 1.2 $cm^{-2}s^{-1}$ for $E_\nu$ > 19.3 MeV. This limit is tantalizingly close to the majority of modern theoretical predictions. Figure 3.2.1.5 shows the predictions for the flux of SRN from several published models and the Super-K limit.

### 3.2.1.6 Other Neutrino Science

A large-volume neutrino detector can be used for more precise measurements of solar and atmospheric neutrino properties. Generally speaking, all of the current Super-K measurements can be repeated with higher precision, since the detector mass is an order of magnitude higher, and the rate of cosmic rays is an order of magnitude lower.

Atmospheric neutrinos can be used to determine some of the neutrino oscillation parameters, including $sin^2 2\theta_{13}$, the mass hierarchy, and the CP-violating phase $\delta_{CP}$. Generally, the study with atmospheric neutrinos will not be as sensitive as an experiment with a neutrino beam, but atmospheric neutrinos will allow the possibility of revealing different physics. This is because the atmospheric neutrino sample covers five orders of magnitude in neutrino energy and three orders of magnitude in baseline, including long paths through matter. The atmospheric neutrino flux is a mixture of muon and electron neutrinos and antineutrinos. One expects 14,000 atmospheric neutrino interactions per 100 kT of detector mass per year. Most atmospheric-neutrino events occur at neutrino energies under 1 GeV, where both water Cherenkov and LAr (LAr) perform well. A LAr detector may have significant capability to identify neutrino versus antineutrino by observing the recoil proton present in charged current neutrino scattering. This large sample of neutrino interactions allows for a comparison of the neutrino oscillation framework under different observational conditions.

### 3.2.1.7 Laboratory Nuclear Astrophysics

The past decade has seen extraordinary strides in our ability to probe the universe in different wavelengths and over a wide range of redshifts. While the underlying goals of observational programs may be astrophysics or cosmology, the radiation we observe typically comes from the nuclear and atomic processes that control the evolution of astrophysical objects. Consequently, without a precise understanding of this associated microphysics, we can neither interpret the observations we are making nor understand their astrophysical implications. Classic examples come from big bang nucleosynthesis (BBN) and the solar neutrino problem. The former became our first precise tool for cosmology, determining the baryon-to-photon ratio, only because laboratory astrophysicists had carefully measured the nuclear cross sections needed to build a theoretically sound BBN network. In the case of solar neutrinos, the first neutrino detector would not have been built without the laboratory astrophysics discovery that a key nuclear reaction was 1,000 times stronger than expected, leading to detectable higher-energy solar neutrinos. Similarly, the discovery of neutrino oscillations was driven by three decades of laboratory astrophysics that effectively determined the solar core temperature to 1%, thereby making an astrophysics solution to the solar neutrino problem highly unlikely.

These examples also demonstrate the critical importance of continually improving our laboratory astrophysics capabilities. More precise astrophysical observations are useful only if we have the capacity to interpret those observations. Anomalies that are discovered—two relevant examples are the BBN Li abundance and the discrepancies between helioseismology and recent measurements of metal abundances in the solar photosphere—can be traced to possible new physics only if the uncertainties in standard physics are adequately controlled.

Laboratory nuclear astrophysics is concerned with measuring the nuclear processes, including charged-particle-, neutron-, and weak-interaction-induced reactions that govern the big bang, stars, and various catastrophic or transient events seen in astrophysics. The advancement of this field allows us to better understand cosmological and stellar nucleosynthesis, the associated chemical enrichment of our galaxy, the energy production and lifetimes of stars, the mechanisms responsible for catastrophic events such as thermonuclear and core-collapse supernovae, the origin of the gamma rays, cosmic rays, and astrophysical neutrinos observed from Earth, the properties of neutron stars, and the gravitational wave forms from events like neutron star mergers. The tools of the field include very-low-energy accelerators that can measure cross sections at or near those characterizing the cores of ordinary stars; radioactive beams, which can be used to study properties of the exotic nuclei, far from the valley of stability, that are found in the billion-degree plasmas of core-collapse supernovae; reactors and spallation facilities, which produce intense neutron sources for studying the nucleosynthesis that occurs in red giants and other stars in more advanced stages of evolution; low-energy accelerators for studying weak and electromagnetic transitions important to neutrino and gamma-ray production, including surrogate reactions for measuring the nuclear responses that govern inclusive neutrino interactions; and a variety of specialized experiments at heavy-ion and electron facilities to characterize properties of nuclear matter important to the nuclear equation of state.

The envisioned laboratory nuclear astrophysics program at DUSEL would be based on next-generation low-energy accelerators. DUSEL's exceptional overburden will provide a low-background environment for the program, allowing measurements to be extended to very low energies, where counting rates could be typically one event per month.  By permitting measurements near or in the Gamow peak—the critical energy where reactions in stars take place—the DUSEL Facility will greatly reduce the uncertainties that would otherwise be introduced by the need for theoretical extrapolations to low energies. Three specific examples are given below.

**Solar neutrino sources and the metallicity of the sun**. Over the past two decades a remarkable program to probe the interior of our sun has been developed, based on two probes, solar neutrinos, and helioseismology. Recently, because of new analyses of photospheric absorption lines that have impacted estimates of key solar metal abundances, discrepancies have arisen between the interior sound speeds determined from helioseismology and the standard solar model (SSM). Solar neutrinos are also impacted, as the abundance uncertainty produces changes of 20% in the high-energy solar neutrino flux measured by SNO and Super-Kamiokande. This metallicity discrepancy has led to the re-examination of one of the assumptions of the SSM, that the sun was initially fully mixed and thus homogeneous when it began main-sequence burning. There is a mechanism that could invalidate this assumption—late in the evolution of the solar system, the growth of planets extracted large quantities of metal from the last 5% of solar system gas—and there are controversial claims in astrophysics that metallicity anomalies seen in "solar twin" systems are correlated with properties associated with planets. Were this claim to hold up, the metallicity profile of host stars could become an important clue in planet hunting.

There are two potential DUSEL connections. First, the most direct way to determine whether the sun is homogeneous is to directly measure the metal content of the solar core. This can be done to an accuracy of about 10% by measuring the carbon-nitrogen-oxygen (CNO) contribution to solar neutrinos, but at much greater depths than Gran Sasso to eliminate a critical background. Such a measure would determine the core abundance of C and N. Two candidate locations providing adequate depth are SNOLab and DUSEL. Second, the key uncertainty in interpreting the experiment is the uncertain nuclear physics of the CNO cycle. While the nuclear physics uncertainties have been greatly reduced recently by Gran Sasso's Laboratory for Underground Nuclear Astrophysics (LUNA) collaboration, further improvement is needed as nuclear uncertainties remain dominant. Improved measurements could be made at the Dakota Ion Accelerators for Nuclear Astrophysics (DIANA) Facility, proposed for DUSEL (see Section 3.3.6).

**Carbon-based nucleosynthesis**. The broad energy range that will be covered by the proposed DUSEL nuclear astrophysics Facility is important to studies of α–induced reactions on light and medium-mass nuclei. Alpha-capture reactions govern stellar helium burning in red giant (RG) stars and asymptotic giant branch (AGB) stars. They are also critically important to the first generation of massive, metal-free stars. Such stars lacked the C, N, and O needed to catalyze CNO cycle burning, and thus must have first synthesized carbon via the triple-alpha process on primordial helium. Other particularly critical reactions are $^{12}C(\alpha,\gamma)^{16}O$ and $^{16}O(\alpha,\gamma)^{20}Ne$ that define the $^{12}C/^{16}O$ abundance ratio in the subsequent carbon-burning stage, when processes such as $^{12}C+^{12}C$ and $^{16}O+^{12}C$ take place. The interplay of these reactions influence the carbon-oxygen distribution in the post-helium burning or post-carbon burning matter in supernova progenitors, and they also dictate the light element abundance distribution in white dwarf matter. Because of the uncertainties in the nuclear physics, the structure of the progenitor star's carbon and oxygen zones remains one of the key variables in the nucleosynthesis accompanying a core-collapse supernova. These uncertainties also affect the ignition and burning conditions for both nova and type Ia supernova explosions.[29] The modeling of type Ia supernovae is drawing great attention because of the need to assess systematic uncertainties in their use as "standard candles" in cosmology.

**Neutron sources for the production of trans-Fe elements in stars**. The heavy elements above Fe were produced by neutron capture through one of two mechanisms. Some were created far from the valley of stability through the r-process, which requires short-lived explosive conditions characteristic of core-collapse supernovae or neutron-star mergers. The remaining elements were produced in the s-process, which requires more quiescent conditions characteristic of red giants or other stars burning in hydrostatic equilibrium. The s-process nucleosynthesis of heavy elements in stars requires a steady but significant source of neutrons. The reactions $^{13}C(\alpha,n)^{16}O$, $^{17}O(\alpha,n)^{20}Ne$, and $^{22}Ne(\alpha,n)^{25}Mg$ are considered the most likely sources for neutron production in a variety of stellar helium-burning[30,31,32] and carbon-burning environments.[33] The low-energy cross sections of these reactions determine the neutron flux during core helium burning[32] and shell carbon burning of massive stars[33] as well as during inter-shell helium burning in AGB stars.[30] The DUSEL nuclear astrophysics program would be able to determine the reaction rates for candidate neutron sources over the relevant energy range. Standard evaluated databases now show large differences in recommended astrophysical rates for reactions like $^{13}C(\alpha,n)^{16}O$ and $^{17}O(\alpha,n)^{20}Ne$ due to the absence of good data in the critical 10-40 keV range.

### 3.2.2 Biology, Geosciences, and Engineering—Goals and Motivation

The DUSEL Facility will provide access to a large volume of a dedicated underground laboratory for a variety of experiments in biology, geosciences, and underground construction engineering. A wide range of critical scientific questions and topics could be addressed in this underground laboratory including:

- What controls the distribution and evolution of subsurface life?
- How deeply does life extend into the Earth?

- How do fluid chemistry, rock mechanical properties, and microbes evolve in fractured crust under changing temperature, stress, and flow?
- Basic processes relevant to $CO_2$ sequestration
- Mechanical response of rock masses on length scales from cm to km and timescales from milliseconds to years
- Advanced seismic and electromagnetic probes and monitoring; precision seismology
- Advanced engineering design for large cavities and other underground construction and monitoring over many years, leading to advances in the design of underground cavities
- The nucleation, propagation, and fluid conductivity of faults and fractures on significant scales

The several cubic kilometers of crystalline rock within the DUSEL Facility provide a diverse and varied environment for Earth science investigations. The large extent of the developed underground access ensures that a variety of conditions and features can be identified that support diverse and rich biological, geoscience, and geoengineering investigations. The mechanical quality of the rock, in general, reduces the maintenance costs for providing long-term access to the experiments, resulting in lower operational costs than might be expected in other geologic environments at similar depths. The multidisciplinary biology, geology, and engineering (BGE) research typically focuses on issues that are of immediate societal importance such as carbon sequestration, faulting and fracturing in the subsurface, resource recovery, new life forms, and underground construction. In general, this research facilitates studies that explore geologic and life processes and how they interact in the deep-subsurface environment. The NSF-supported BGE collaborations, as currently envisioned, lay the groundwork for a long-lasting program in this type of research that is important to questions as to how humans interact with the subsurface environment.

#### 3.2.2.1 Geologic Carbon Sequestration

Geologic carbon sequestration (GCS) is part of the Carbon dioxide Capture and Storage (CCS) process in which $CO_2$ is injected into deep geologic formations. This strategy is being evaluated as a national and global priority for its potential to help mitigate climate change. Fossil-fuel-based energy sources are estimated to emit some 30 gigatonnes $CO_2$ per year.[34] If this technology is to be adopted as a widespread approach, issues such as the geochemical stability of the host reservoirs in the presence of large amounts of $CO_2$, well cement stability, and an understanding of the flow behavior of $CO_2$ at critical temperatures and pressures are important. This approach to long-term climate-change mitigation will be successful only if the injected $CO_2$ remains in the intended storage regions. Because of the lower density of $CO_2$ at all conditions in deep subsurface reservoirs, there is a tendency for injected $CO_2$ to leak upward out of intended storage reservoirs. Leaked $CO_2$ could pose environmental hazards, impact subsurface resources, or discharge into the atmosphere, negating the climate-change-mitigation objective and causing potential loss of carbon storage credit. Advancement of GCS requires a sound understanding of the processes controlling $CO_2$ storage, trapping, and migration in the subsurface environment.

#### 3.2.2.2 Deformation of Large Underground Rock Masses

Deformation of the Earth occurs on spatial scales from atomic to global, and timescales from instantaneous to millions of years, and involves loads such as the sudden dislocation of a rock burst, Earth tides, and the slow accumulation of tectonic strain. In DUSEL and other large underground facilities, an overarching problem is to predict how the rock mass will respond to different forces in a subsurface environment composed of a complex material that is both heterogeneous and anisotropic and filled with fluid. Predicting the rock response to different types of deformation is important to the safety and longevity of infrastructure on the surface and underground. Many current techniques for measuring deformation are restricted to a small subset of the broad scales of movement in the Earth's crust. DUSEL presents the special opportunity to access—via many kilometers of drifts, winzes, and boreholes—several cubic kilometers of rock mass of different lithologies for several decades while both natural and anthropogenic loads are imposed. This is of particular importance to the development of our understanding of long-term strain mechanisms, which will benefit greatly by the extended life envisioned for this laboratory. By creating a large underground network for strain monitoring, this experiment will be a bridge between small-scale and large-scale deformation-monitoring techniques.

#### 3.2.2.3 Coupled Thermal-Hydrological-Mechanical-Chemical-Biological (THMCB) Processes

Most natural and engineered Earth-system processes involve strong coupling of thermal, chemical, mechanical, and sometimes biological processes in rocks that are heterogeneous at a wide range of spatial scales. One of the most pervasive processes in the Earth's crust is that of fluids (primarily water, but also $CO_2$, hydrocarbons, volcanic gases, etc.) flowing through fractured, heated rock under stress. Although rocks and fluids can sometimes be analyzed for physical and chemical properties, it is very difficult to create quantitative numerical models based on fundamental physics and chemistry that capture the dynamic changes as they take place. Initial conditions and history are known roughly at best, and the boundary conditions have likely varied over time as well. Recognition of the importance of how these related processes can affect the behavior of rock in the subsurface has been a major step forward. By actively varying and controlling these parameters, their relative importance in the natural system can be evaluated and prioritized systematically. A fundamental understanding of these processes can be of great use in practical applications such as geothermal energy recovery, carbon sequestration projects, hydrocarbon development, waste disposal, and even the evaluation of groundwater resources.

#### 3.2.2.4 Ecohydrology Studies of Deep Fractured Rocks

The next and perhaps final frontier of ecosystem discovery may lie within the vast, unexplored inner space of continents. The deep subsurface has recently been recognized as an ecosystem that can profoundly influence the way the origin and early evolution of life on Earth is viewed, the search for novel life forms and enzymes, and approaches to future energy production. The assumption is that like surface ecosystems, deep-subsurface ecosystems involve complex interactions between life and environmental processes, such as the transport and availability of chemicals and energy, and the extent and distribution of settings that provide suitable habitats. But the ecology of the deep subsurface has yet to be defined because it is much more difficult to make observations at depth than it is in the familiar surroundings of Earth's surface. Although the deep subsurface comprises a significant fraction of the living carbon on our planet, it is the most poorly understood ecosystem. DUSEL represents a historic opportunity for the controlled exploration of a novel rock-hosted ecosystem that spans the subsurface biosphere from its top at the base of the photosphere to its bottom at the abiotic fringe.

#### 3.2.2.5 Underground Cavity Design

As the requirements for energy, waste storage, and resources grow, so do the challenges that engineers face to meet societal needs. Tunnels are being built at depths of more than 5,000 feet; mining for resources and drilling for oil extraction often reaches depths of tens of thousands of feet, where rock pressures and temperatures reach record magnitudes that challenge our knowledge of rock behavior. The search for new sources of energy is turning toward geothermal energy, where mechanics, temperature, fluid flow, and chemistry must be carefully integrated. Although the engineering for rock behavior at shallower depths is well advanced, the science and engineering required to address these and other problems at deeper depths is not yet sufficiently mature. Finding solutions for these questions is one of the great challenges of this century, and also one of the greatest opportunities. DUSEL provides the means, because of its size, depth, and duration of service, to advance the field of rock mechanics and specifically to scale laboratory results to field observations.

### 3.2.2.6 Fracture Processes

Fractures and fluids influence just about all of the mechanical processes in the Earth's crust, but many aspects of these processes remain poorly understood, in large part because of a scarcity of controlled field experiments at appropriate scales. Advancing the understanding of faulting and fracturing processes is critically important to many fields of Earth science and engineering, including seismology, resource recovery, environmental remediation, economic and structural geology, and disposal of radioactive wastes and carbon dioxide in the deep subsurface. In particular, the understanding of processes of earthquake triggering, fracture nucleation, and propagation will be improved by field-scale experiments on controlled fault initiation. The ability to examine faults through great vertical depths offers the opportunity to study how fluids can move through fault zones as well as to understand potential effects on the biological component that may be specific to the fracture surfaces. Experiments involving fluid-rock reactions coupled with microbial transport in the deep subsurface have implications ranging from the evolution of rock properties to geochemical cycles to how life evolved on Earth.

### 3.2.2.7 Subsurface Imaging and Sensing

Astronomers and astrophysicists can image deep into space to study the formation of the universe, creation and death of stars, collision between stars, and the dynamics surrounding black holes—yet geologists and engineers often find it difficult to image and study processes even tens of meters into the Earth surface at sufficiently desirable resolution. The geophysical systems that can be installed at DUSEL cover a very broad spectrum of applications ranging from improving the short-range, high-resolution imaging of microdeformation such as might be associated with the excavation of the laboratories or the controlled fracturing studies, to high-resolution teleseismic arrays that may provide much better resolution of Earth structure through the use of three-dimensional arrays. This would lead to a research effort analogous to that of using the Hubble Space Telescope, imaging into, not away from, the Earth. This will develop and refine measurement methodologies and science required to image the Earth at multiple scales, combine a number of different physical measurement and inversion methodologies to provide complementary information, and strong constraints, for the necessary inversion solutions. This methodology, perhaps using multiple and complimentary types of geophysical techniques such as seismic and electromagnetic, will bring images into sharper focus and will help provide deeper scientific understanding of geological and engineering processes and behavior in the context of a highly stressed geological environment. Geophysical imaging will characterize heterogeneity and anisotropy in the rock mass and illuminate complex phenomena such as the rearrangement of in situ stresses due to excavation. The research will also examine rock damage process precursors and onset of tremors.

## 3.3 DUSEL Research Program

The overall scientific goals and motivation for the DUSEL Facility have been summarized in Chapter 3.2. An overview of the candidate experiments and the process that has been used to determine critical requirements of the Facility is provided here. A short summary of the characteristics of the Facility follows; details are in Volume 5, *Facility Preliminary Design*. What follows below is a description of potential experiments that address the major scientific goals of DUSEL with an emphasis on the characteristics of the experiments that guide the requirements and design of the DUSEL Facility.

The historical development of the scientific program consists of an Early Science program that will be followed by the Integrated Suite of Experiments (ISE) as part of the DUSEL Major Research Equipment and Facilities Construction (MREFC) funded Project. A scientific program at the South Dakota Science and Technology Authority (SDSTA) Sanford Laboratory was initiated in late 2005 through a call for letters of interest from the scientific community. This program precedes the MREFC-funded Construction and the operations and the Facility development for this program are funded through private and South Dakota state sources. Chapter 3.4 describes the ongoing experiments that make up the initial science effort at Sanford Laboratory. These projects are pursuing dark-matter detection, neutrinoless double-beta decay and ultrapure crystal growth as well as endeavors in BGE. Information from these experiments, along with ongoing experiments at other underground laboratories, has helped define a generic ISE, which is described in Chapter 3.5. The DUSEL requirements for the generic ISE and the resulting DUSEL design requirements to accommodate the ISE are summarized in Chapters 3.6-3.9.

### 3.3.1 Candidate Experiments

Primary input to the scientific requirements that has been used to guide the DUSEL Facility design is from the successful awardees from the NSF DUSEL S4 solicitation. This solicitation, *Development of Technical Designs for Potential Candidates for the DUSEL Suite of Experiments*, was issued in 2008, and the S4 awards were made in fall of 2009. S4 funds provide support for potential candidates for the ISE at DUSEL. Selection for an S4 award constitutes a commitment only to research and development of the design of specific experiments, not for inclusion in the integrated suite. Similarly, future proposals that were not awarded S4 funds are not excluded from the suite of experiments. The S4 awardees are listed in Table 3.3.1.

The Long Baseline Neutrino Experiment (LBNE) Project (see Section 3.3.5) was formed in 2008 and received CD-0 approval from the Department of Energy in January 2009. An S4 award for the development of a water Cherenkov detector is a substantial component of the LBNE project. Large LAr detectors are also under study. The very large detectors needed at the DUSEL Facility to fulfill the scientific mission of LBNE are critical determinants of many aspects of the DUSEL design.

| Experiment Type | Experiment | Principal Investigator |
|---|---|---|
| Dark Matter | MAX | Galbiati (Princeton) |
| Dark Matter | LZD | Shutt (Case Western) |
| Dark Matter | GEODM | Golwala (Caltech) |
| Dark Matter | COUPP | Collar (Chicago) |
| $0\nu\beta\beta$ | EXO | Gratta (Stanford) |
| $0\nu\beta\beta$ | 1TGe | Wilkerson (N. Carolina) |
| Long baseline $\nu$, proton decay | LBNE- Water Cherenkov | Svoboda (UC Davis) |
| Nuclear astrophysics | DIANA | Wiescher (Notre Dame) |
| Low Background Assay | FAARM | Cushman (Minnesota) |
| Bio/Geo/Eng | Transparent Earth | Glaser (UC Berkeley) |
| Bio/Geo/Eng | Fiber Optic Array | Wang (Wisconsin) |
| Bio/Geo/Eng | Fault Rupture | Germanovich (G'a Tech) |
| Bio/Geo/Eng | THMCB (coupled processes) | Sonnenthal (UC Berkeley) |
| Bio/Geo/Eng | $CO_2$ Sequestration | Peters (Princeton) |
| Bio/Geo/Eng | Ecohydrology | Boutt (U. Mass.) |
| Bio/Geo/Eng | Monitoring | Bobet (Purdue) |

**Table 3.3.1** NSF S4 solicitation awardees.

A broad program of principally physics experiments at underground sites is under way in Asia, Canada, Europe, Russia, and the United States. The DUSEL Facility design has been informed by these experiments and their associated underground facilities. In particular, we have benefitted greatly from the assistance of colleagues at the Laboratori Nazionali del Gran Sasso (LNGS) in Italy and SNOLab in Canada in establishing the DUSEL Facility design.

### 3.3.2 Overview of the DUSEL Facility

A very brief summary of the planned DUSEL Facility as it relates to housing experiments is presented in this section to provide background information for the description of experiments and requirements in subsequent sections of this volume.

A 3D underground view of the Facility, including key options for LBNE, is shown in Figure 3.3.2-1.

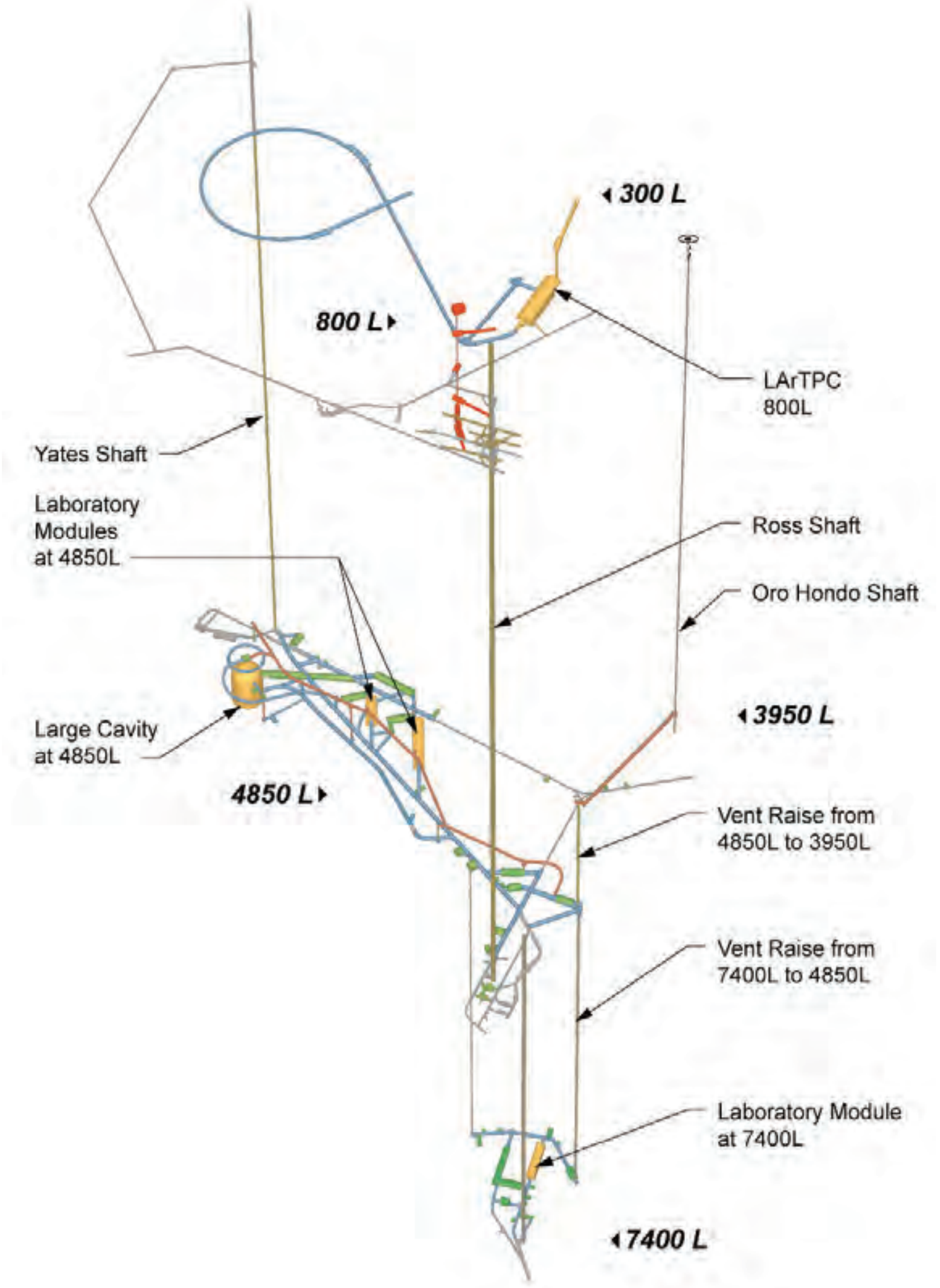

**Figure 3.3.2-1** A 3D view of the major elements of the planned DUSEL Facility. [DKA]

The principal components of the baseline Facility to accommodate the scientific program are:

- A Surface Campus to support operation of the Facility, its scientific program, and education and outreach
- A Mid-Level Laboratory (MLL) campus at the 4850L
- One large cavity (LC-1) for LBNE at 4850L. The option of locating a LAr detector complex at the 800L for LBNE is also shown.
- A Deep-Level Laboratory (DLL) campus at the 7400L
- Other levels and ramps at a variety of depths to support the operation of the Facility, including support of physics experiments and locations primarily for experiments in BGE (Figure 3.3.2-2)
- Shafts for access (Yates and Ross Shafts) and for ventilation (Oro Hondo) and internal access from the MLL to the DLL via the #6 and #8 Winze

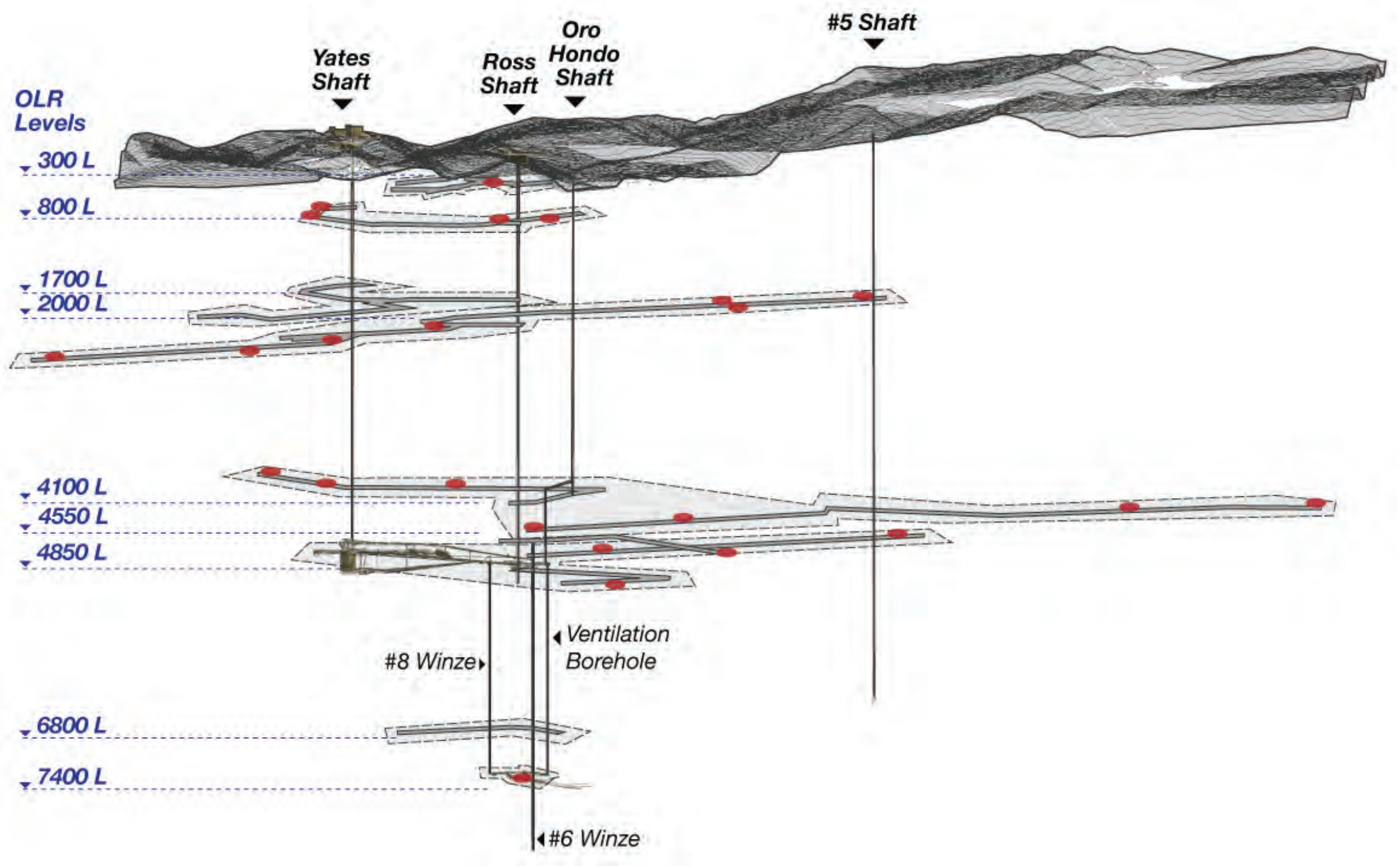

**Figure 3.3.2-2** Summary of other levels and ramps to be developed for the potential DUSEL experimental program with approximate science locations noted in red. [DKA]

## 3.3.3 Dark Matter Experiments

### 3.3.3.1 Overview of Dark Matter Experiments

For more than two decades, experimental groups have sought to detect halo Weakly Interacting Massive Particles (WIMPs) in the laboratory to solve the dark-matter mystery (the experimental evidence for the existence of dark matter and the science motivation of WIMPs as a dark-matter candidate are summarized in Section 3.2.1.1). With the notable exception of the DArk MAtter/Large sodium Iodide Bulk for RAre processes (DAMA/LIBRA) experiment that reported[35] a signal positive result in the form of an annual modulation in the signal rate in NaI scintillator crystals consistent with halo WIMP interactions, all other efforts have, so far, have not confirmed these results. This apparent inconsistency could be due to either a more complex dark-matter particle sector than single heavy WIMPs or to unknown experimental artifacts in the DAMA/LIBRA setup. In any case, further experimentation is warranted.

The pursuit of a WIMP signal has led to the development of a wide range of detection techniques, each with unique strengths and challenges and applicable to different target materials. The current best limits for the WIMP spin-independent (SI) interaction cross section (see Figure 3.2.1.1) of a few times $10^{-44}$ $cm^2$/nucleon (at ~70 $GeV/c^2$ WIMP mass) come from Cryogenic Dark Matter Search (CDMSII) and XENON100 experiments.[36] These used 4 kg of germanium detectors with ionization and phonon sensing at 50 mK temperatures and 40 kg of xenon in a dual-phase time projection chamber (TPC), respectively. The best limits for the detection of WIMPs for spin-dependent (SD) couplings at the level of $10^{-37}$ $cm^2$ come from the Chicagoland Observatory for Underground Particle Physics (COUPP) and the Project in CAnada to Search for SuperSymmetric Objects (PICASSO).[37] These used a bubble chamber filled with superheated $CF_3I$ liquid and a bubble detector with Freon droplets in a gel, respectively. These direct SD limits are substantially less stringent than for spin-independent (SI) couplings, as they do not benefit from the $A^2$ enhancement factor of coherent scattering. Note that more stringent limits from SuperKamiokande[38] and IceCube[39] exist from indirect WIMP detection.

Additional exposure and extensions of the above experiments and other experiments/targets/techniques—such as Large Underground Xenon (LUX), Mini Cryogenic Low Energy Astrophysics with Noble gases (MiniCLEAN), Argon Dark Matter (ArDM), the WIMP Argon Programme (WARP), DarkSide, and XMASS—typically with masses in the few hundreds of kilograms, are expected to reach $10^{-45}$ $cm^2$ sensitivities for SI couplings within the next one to three years. This class of experiments is known as Generation 1 (G1). Generation 2 (G2) experiments, with sensitivities at or better than $10^{-46}$ $cm^2$ and masses in the 100 kg to tonne range, are expected to come online around 2013, with full results available in ≈2014-2016. XENON1T, LZS, SuperCDMS at SNOLAB, DEAP 3600, DarkSide-G2, and COUPP 500 kg are some of these planned G2 detectors.[40]

With each significant increase in sensitivity come additional challenges, uncertainties, and risks from background events and their suppression. Total background rates must be kept near zero (after discrimination cuts are applied) while the volume and mass increase by factors of ~10. Thus, there is at this moment uncertainty as to what technique/s will successfully achieve the intended sensitivities by the time the first dark-matter experiments are installed in DUSEL.

Irreducible backgrounds from solar, atmospheric, and diffuse supernovae neutrinos begin to contribute below WIMP cross sections of $\sim 10^{-48}$ $cm^2$, and will likely determine the ultimate reach of direct dark-matter counting searches.

As noted in Section 3.2.1.1, an important part of the theoretically well-motivated WIMP parameter space will probably already have been probed by the time DUSEL dark-matter experiments are in operation. If no WIMPs are observed in the pre-DUSEL era, initial DUSEL dark-matter experiments will have a chance to discover dark matter. On the other hand, should WIMPs be discovered in a pre-DUSEL experiment, DUSEL dark-matter experiments will obtain large samples of WIMP events that will confirm the discovery and provide important information about the WIMP nature, its interactions, and its mass.

If WIMPs are discovered in a direct-detection experiment, the ultimate cross-check will be confirming their galactic origin by observing secondary signatures related to the motion of the Earth and solar system as well as the WIMP velocity distribution. Directional WIMP detectors, possibly large volumes of gas imagers currently in the R&D phase, could be deployed in the future at DUSEL for this purpose.

WIMPs could also be observed, directly produced, in accelerator experiments at the LHC at CERN. Finding the same mass derived from direct detection as reconstructed for LHC events would show unambiguously that the particles produced at the LHC are stable over the age of the universe and would fully justify the use of the relic abundance as a constraint on theoretical parameters.

The planned construction completions of the MLL and DLL laboratories are beyond the planned time scale of G2 experiments. Dark-matter experiments at DUSEL will target sensitivities in the $10^{-47}$-$10^{-48}$ $cm^2$ /nucleon range (at 70 $GeV/c^2$ WIMP mass) until reaching the ultimate sensitivity set by irreducible neutrino backgrounds. These G3 experiments require multi-tonne mass targets with large and often sophisticated shield and veto systems against natural radioactivity in the surrounding materials, against residual cosmic ray muon fluxes, and against high-energy neutrons from cosmic ray muon spallation.

The four proposed G3 experiments for DUSEL that received S4 funding are:

- COUPP, 16 tonnes of $CF_3I$ in bubble chambers
- GEODM, 1.5 tonnes of Ge detectors with ionization and athermal phonon sensors at mK temperatures
- LUX-ZEPLIN DUSEL (LZD), 20 tonnes of Xe in a dual phase TPC
- Multiton Argon and Xenon (MAX), two detectors, each a dual-phase TPC with:
    - 20 tonnes of depleted liquid Ar
    - 6 tonnes of liquid Xe

Note that the noble-liquid TPCs proposed by the LZD and MAX collaboration are similar in structure and scope. The LZD and MAX collaborations have started a preliminary series of discussions, including members of the CLEAN collaboration (another potential G3 experiment described below), to evaluate converging toward a proposal for a single experiment at DUSEL. This is in line with the recommendation of the NSF S4 Review Panel (July 2010): “The MAX and LZD collaborations are encouraged to continue working toward a single DUSEL experiment.”

All these large and challenging G3 experiments proposed for DUSEL represent expansions over pre-DUSEL experiments. As such, they will benefit from the techniques and know-how developed by their respective predecessors, thus reducing the risk and ensuring their optimal designs. In the larger worldwide context, other G3 experiments are in conceptual and/or planning stages in Canada, Europe, Japan, and China on a similar timescale as DUSEL’s initial experiments.

### 3.3.3.2 Candidate Experiments

#### 3.3.3.2.1 COUPP

COUPP employs ultraclean (and thus indefinitely metastable) $CF_3I$ bubble chambers sensitive to WIMP-induced nuclear recoils. Due to the presence of both fluorine and iodine, this target is sensitive to both SD and coherent SI WIMP couplings. The signature from a WIMP interaction is a single bubble induced by the large energy loss density of the recoil of a nucleus. Neutrons can be efficiently differentiated from WIMP interactions thanks to their large probability of scattering elastically that often produces multiple bubbles from nuclear recoils in the detector volume. Electrons (and consequentially also gammas) are intrinsically rejected at the $\sim 10^{-13}$ level because their energy-loss density is below the threshold required for bubble formation. This important and unique intrinsic rejection is achieved by deliberate selection of the temperature and pressure of the bubble chamber. Alpha particles, on the other hand, can produce bubbles. Such events can be identified and rejected, at a level yet to be determined for COUPP modules, by their distinct acoustic signal pattern. To mitigate this background, the uranium and thorium contents of the target material must be reduced to $<10^{-15}$ g/g through purification techniques (similar purity levels have been achieved before by neutrino experiments using liquid scintillators) and alpha-particle-induced bubbles must be discriminated from nuclear recoils at the $10^{-3}$ level by their distinct acoustic signatures.

Large target masses can be easily monitored due to the built-in amplification that the macroscopic phase transition provides. Bubble formation can be detected by monitoring for a pressure rise, the acoustic signal accompanying the initial stage of the nucleation, and/or by video motion detection. A real-time software comparison of video frames taken every ~10 ms triggers the recompression of the chamber before the bubbles grow much larger than a few mm. The use of multiple cameras allows a spatial granularity of about 1 mm.

The proposed experiment for DUSEL consists of 32 identical bubble chambers with 500 kg of target material each—a total of 16 tonnes. The modules are immersed in a single 20 m x 12 m x 7 m water tank that provides both shielding against external neutrons and functions as a temperature bath to keep the bubble chambers at their ~40 °Celsius set point. The operating pressure is in the range 0-50 psi when the chambers are in the expanded, superheated state and increases to 200 psi in the compressed state.

Figure 3.3.3.2.1 shows the proposed layout and configuration for the COUPP experiment in the 7400L laboratory. A minimum water shield thickness of 2 m at that depth is required to reduce the fast neutron background from cosmic-ray muon interactions (see Section 3.3.10). An alternate deployment in the 4850L laboratory would require either ~3 m additional water shielding on all sides or installation of an active neutron tagger in the shield volume, or some combination of both.

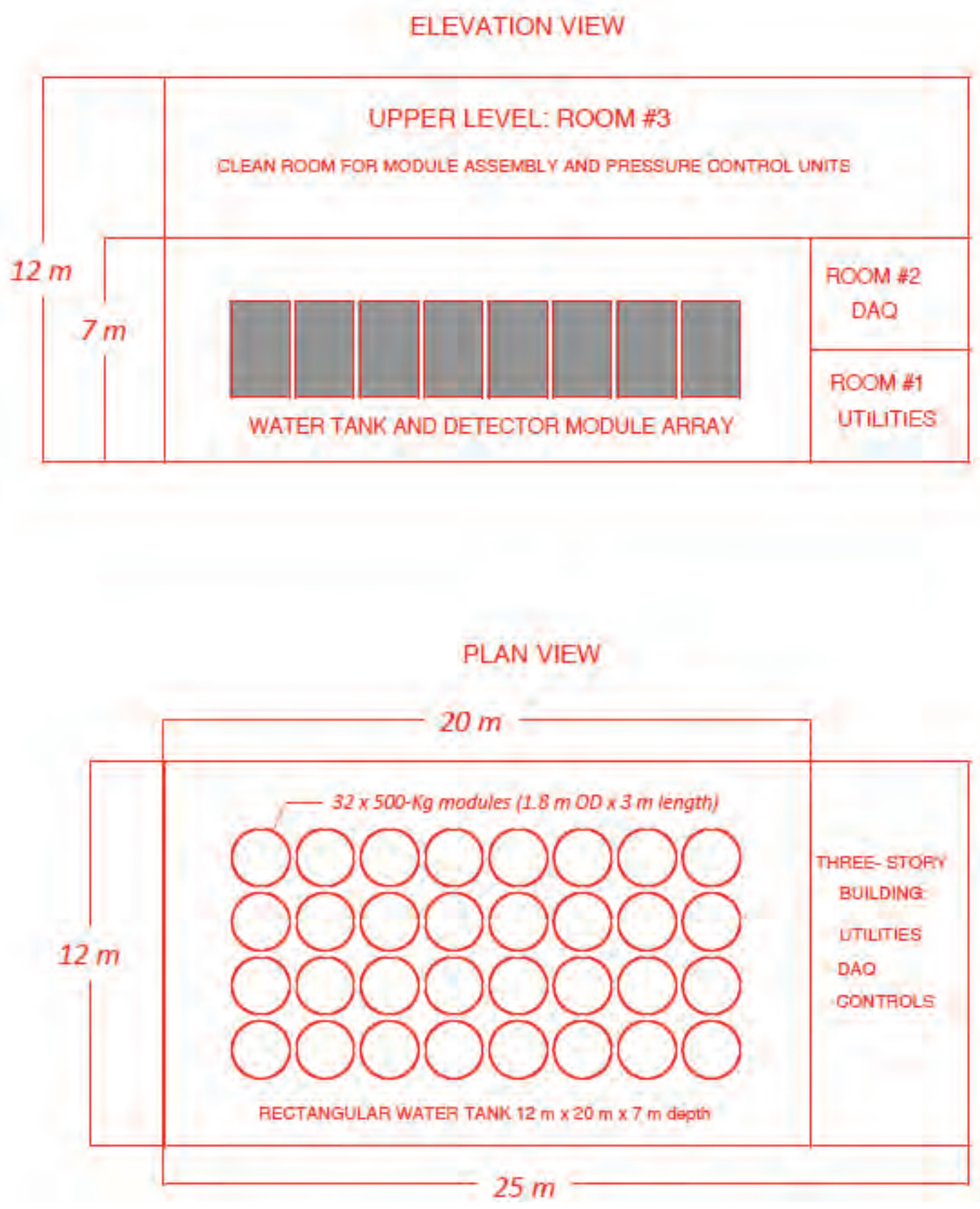

**Figure 3.3.3.2.1** COUPP layout at the 7400L laboratory. The 16 tonnes of active target are contained in 32 identical 500 kg bubble chambers. The bubble chambers are immersed in a rectangular water tank held at ~40 C. [Courtesy COUPP collaboration]

### 3.3.3.2.2 GEODM

The Germanium Observatory for Dark Matter (GEODM) uses interdigitated Z-dependent Ionization- and Phonon-mediated (iZIP) detectors. iZIPs employ simultaneous athermal phonon and ionization measurements in germanium substrates operated at approximately 40 mK temperature to reject electromagnetic backgrounds (photons and electrons). Such particles scatter off electrons in the detectors, while WIMPs (and neutrons) scatter off nuclei. In iZIPs, recoiling electrons are more ionizing and thus can be discriminated against using the ratio of ionization to recoil energy. For recoils within the first few micrometers of the detector surface (primarily caused by low-energy electrons), ionization yield provides less discrimination than for bulk events, but such near-surface events can be rejected via the asymmetric ionization signal observed in ionization electrodes on opposite faces of the detector. Further, events near the surface have faster phonon signals than do bulk events, providing additional discrimination

The target mass of 1.5 tonnes will consist of 300 germanium crystals 15 cm in diameter and 5 cm thick, each weighing 5.1 kg. Each detector will use a phonon sensor segmented into 32 elements to obtain detailed x-y-z position information. The phonon signal is measured using photolithographed superconducting phonon absorbers and transition-edge sensors. The detectors will employ a multiplexed phonon readout and a low-power ionization readout to accommodate the large phonon channel count and the large number of detectors. An integrated cryogenic system and shield provide a 40 mK environment for the detector mass (cold volume $\approx$1 $m^3$), shielding the detectors from radiogenic photon and neutron backgrounds, and allowing for calibration using insertable photon and neutron sources. The baseline shield design is of copper and polyethylene. Specialized production techniques such as electroforming will be used to obtain low-activity cryostat and shield stock materials. The experiment will be housed in a clean room facility, with radon abatement active when detectors are exposed.

Figure 3.3.3.2.2 shows the proposed layout and configuration for the GEODM experiment in the 7400L laboratory. The central 5.6 m diameter, 5.6 m tall cylinder contains the 100 tonnes of copper and 125 tonnes of polyethylene of the passive shield and the cryostat that houses the 1.5 tonnes of germanium crystals. The dilution refrigerator is sited immediately outside the shield. An adjacent Class 100 clean room will be used for detector assembly. A 20-tonne crane is currently planned to be located inside the 2.3 $m^3$ Class 10,000 clean room that surrounds the detector, shield, and detector assembly areas. Alternatives to use the laboratory module (LM) crane will be studied in conjunction with meeting cleanliness requirements.

With a larger and active (instrumented) shield and at increased risk of background events from neutrons and long-lived isotopes from cosmic ray muon activation (see Section 3.3.10), GEODM could also be sited in the 4850L laboratory. For this shallower site, and to further suppress the fast neutrons from spallation, an additional thick water shield will surround the detector. The inside shield/cryostat is not directly immersed in the water.

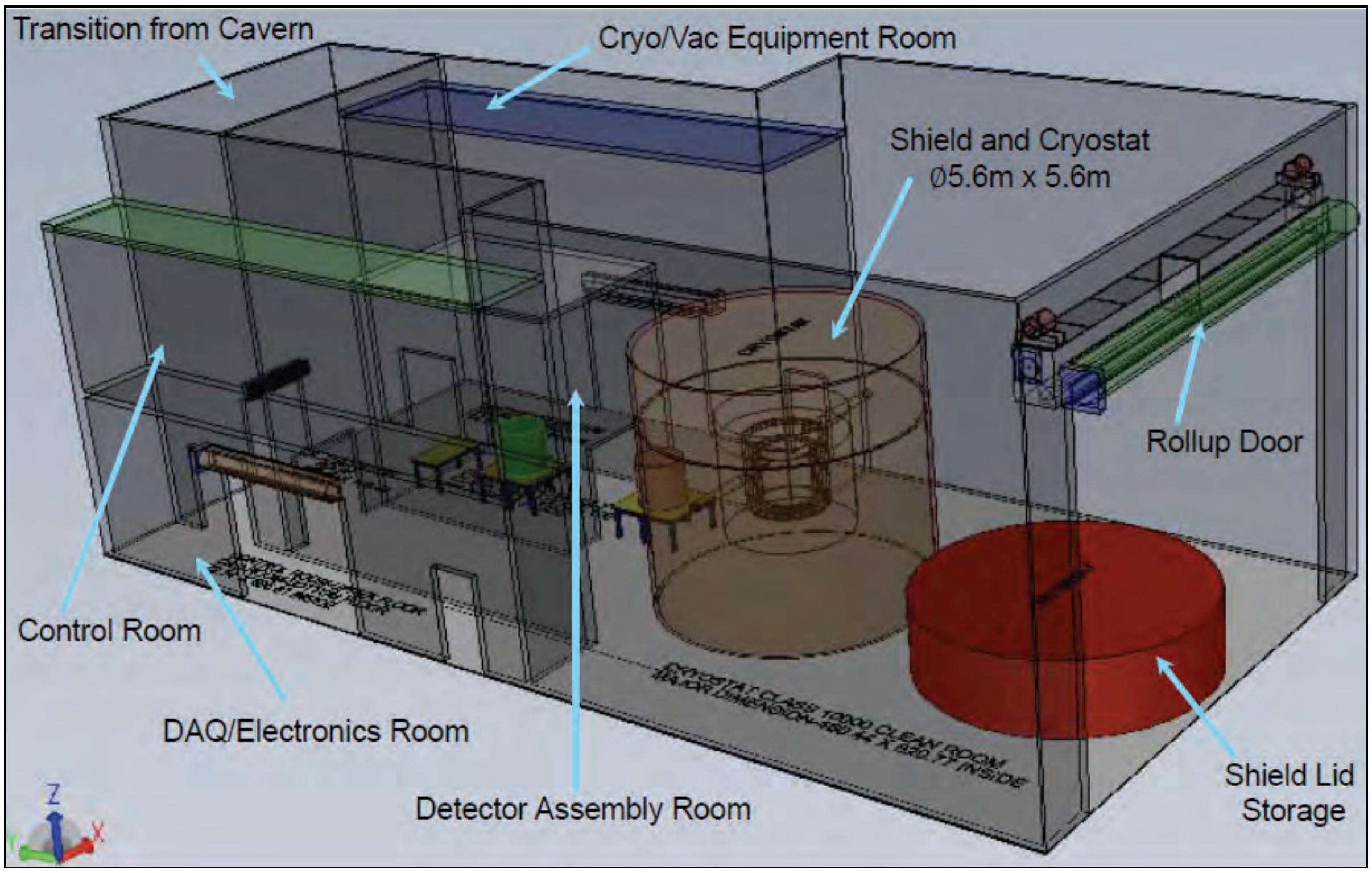

**Figure 3.3.3.2.2** GEODM layout at the 7400L laboratory with a 25 (L) x 12 (W) x 10 (H) m envelope. [Courtesy GEODM collaboration]

### 3.3.3.2.3 LZD

The LUX-ZEPLIN DUSEL (LZD) detector, a dual-phase xenon time projection chamber (TPC), is to be located inside a large-scale liquid scintillator and water shield. Particle interactions in a central cylindrical active region of liquid xenon (LXe) create both prompt scintillation signals (S1) and ionization electrons, the ratio of which is governed by recombination occurring along the recoiling particle's track. An applied electric field extracts a fraction of the electrons from the event site and drifts them up to the liquid surface, where a stronger field extracts them into the gas phase. The electrons are accelerated in the gas by another field and create a large secondary proportional scintillation signal (S2) before being collected on an anode grid. Both S1 and S2 signals are measured with two arrays of photomultiplier tubes (PMTs), the first located below the fully active LXe region and the second just above the gaseous S2-creation region. The ratio of the S1to S2 signals provides discrimination between electron and nuclear recoil. The S2 signal recorded in the top array determines the x-y position of the event with ~1-cm accuracy. The drift time—(difference between arrival of S1 and S2)—measures the event depth to an accuracy also of ~1 cm.

The 20 tonnes of LXe in LZD will operate at ~175 K in an all-Ti cryostat surrounded by a 1 m thick liquid scintillator to tag neutrons and gammas which, in turn, is placed in an ultrahigh-purity water shield, instrumented with large PMTs to also act as a Cherenkov muon veto. A system of thermosyphons between the detector and a liquid nitrogen (LN) reservoir above the water tank cools the apparatus. The 450 PMTs used in the LXe volume of LZD will be constructed using ultralow-background 7.5 cm metal cans with total activities less than 1 mBq/PMT. The walls of the 200 cm diameter and 200 cm tall cylindrical active volume are lined with a segmented set of polytetrafluoroethylene (PTFE) panels whose high reflectivity improves the light-collection efficiency in this geometry. A set of grids and field-shaping rings creates the fields that drift electrons. The primary drift field requires 100 kV applied to the bottom (cathode) grid.

The high Z and density of the 20 tonnes of LXe provide strong self-shielding. This significantly suppresses the event rates of neutrons and gammas (with mean free paths of ~10 cm and 5.5 cm, respectively) that could compete with the dark-matter signal within the fiducial region. The background particles predominantly deposit energy well above the ~10-keV WIMP range, and/or have multiple vertices that are largely rejected by the ~1 $cm^3$ 3D imaging of the TPC. Low-energy single-scatter rates of neutrons and gammas in the central 14 tonnes are, respectively, at least $10^3$ and $10^6$ times lower than event rates in the full 20 tonnes. In addition, electron recoil backgrounds can be discriminated against using the ratio of ionization to prompt scintillation (S2/S1), which provides a further factor of 200-1000 discrimination since electron recoils have less-dense tracks and higher S2/S1 (less recombination) than denser nuclear tracks from WIMPs (and neutrons).

An important internal-background-events source from $^{85}Kr$, a beta emitter present in commercial xenon, will be reduced to a level of ~0.05 ppt Kr/Xe by a chromatographic separation system whose throughput goal is ~50-100 kg/day.

Figure 3.3.3.2.3 shows the proposed layout at the 4850L laboratory. The 12 m height by 12 m diameter water tank contains the cryogenic liquid scintillator vessel that surrounds the liquid xenon vessel. To provide a safe containment in case of prolonged loss of cooling, the full mass of boiled xenon can be captured in four large tanks containing passive charcoal.

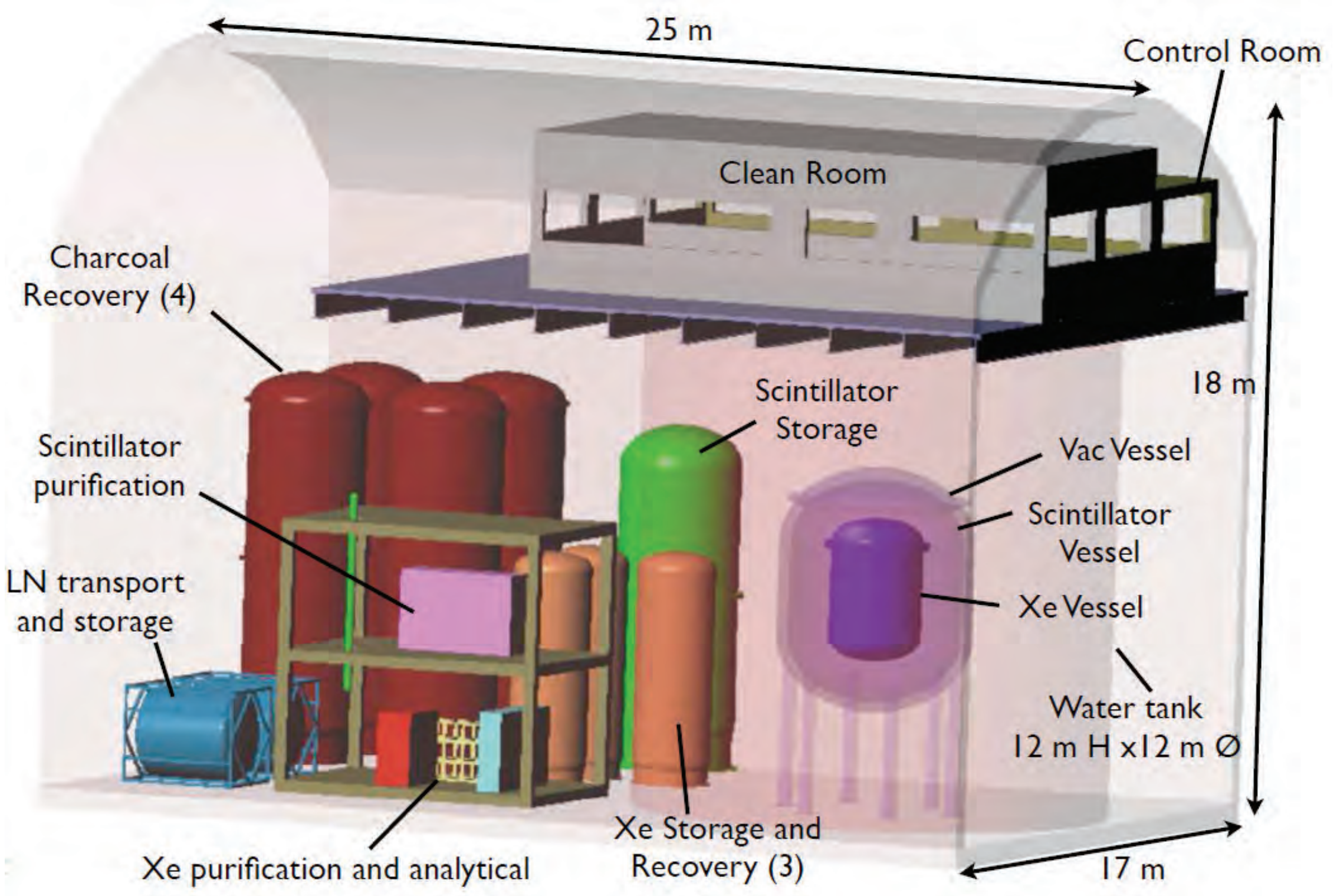

**Figure 3.3.3.2.3** LZD layout at the 4850L laboratory with a 25 (L) x 17 (W) x 18 (H) m envelope. The envelope of the LZD layout is shown, not the envelope of the LM in which it might be housed. [Courtesy LZD collaboration]

### 3.3.3.2.4 MAX

The Multiton Argon and Xenon (MAX) detectors are dual-phase TPCs, one filled with 20 tonnes of argon depleted of the radioactive $^{39}Ar$ isotope, and the other with 6 tonnes of xenon. Each detector is immersed in respective room-temperature liquid scintillator vessels that, in turn, are placed inside two large water tanks. The use of two target materials with matching sensitivity would serve to confirm a putative WIMP signal by verifying the expected $A^2$ dependence of the WIMP spin-independent coherent scattering rate and by verifying the consistency of the WIMP mass deduced from the two recoil spectra. It is also due to the $A^2$ dependence of the interaction rate that the argon TPC target mass is significantly larger than the xenon TPC. The depleted argon and xenon detectors feature many common core technologies and subsystems.

The principles of detection, fiducialization, and discrimination (using the S2/S1 ratio) between nuclear and electron recoils for both TPCs are the same as described in the previous section.

To sense the S1 and S2 scintillation light while minimizing the radioactive contamination in the inner volume of the TPCs, the MAX group intends to use a newly developed ultralow-activity Quartz Photon Intensifier Detector (QUPID) with activities significantly smaller than the lowest-activity PMTs. This will help improve light collection, position sensitivity, and lower the energy threshold.

The 6-tonne xenon TPC operation principle and the purity requirements are very similar to those of LZD. Cooling is achieved using a system of pulse tube refrigerators. $^{85}Kr$ will be removed by cryogenic distillation, validated with gas chromatography and with an Atom Trap Trace Analysis System, currently under development within the XENON program.

For the argon TPC, an additional discrimination tool is the pulse shape of the S1 primary scintillation signal. This discrimination stems from a very large decay-time difference between the two excimer states (singlet and triplet) responsible for the emission of the vacuum ultraviolet (VUV) scintillation light, which are populated differently by low- and high-density tracks. It should provide an additional $10^8$ rejection of beta- and gamma-induced backgrounds.

One of the main background sources in large argon detectors is $^{39}Ar$, a β emitter produced in the atmosphere by cosmic rays. The specific activity of $^{39}Ar$ (Q=565 keV, τ=388 yr) in conventional argon supplies is ~1 Bq/kg of atmospheric argon, corresponding to a concentration of $^{39}Ar/Ar=8\times10^{-16}$. Even though for $^{39}Ar$ in argon the S1 pulse shape discrimination is strong enough to discriminate against the $^{39}Ar$ activity, the 20-tonne unsegmented-detector event pile-up demands the reduction of the $^{39}Ar$ fraction. For this reason, the target material will be obtained from recently discovered underground sources of argon depleted in $^{39}Ar$. A common cryogenic distillation plant will purify the underground argon into detector-grade argon and will reduce the Kr contamination in xenon below the part-per-trillion (ppt) level.

Figures 3.3.3.2.4-1 and 3.3.3.2.4-2 show the layout and configuration of the two MAX detectors and ancillary equipment at the 4850L laboratory. The water tanks are 16 m in height by 16 m in diameter. The clean-room area above the water tanks, with a crane to lower the detectors into the scintillator vessels, is shared by both detectors. The current MAX layout exceeds the simple available envelope to allow full bridge crane access in one of the LMs at 4850L with a crown height of 24 m. Monorail crane access would be maintained in the proposed layout but with small clearance. Additional design work is needed to understand the consequences of the proposed layout. The feasibility of increasing the height (and other dimensions) of an MLL LM is under study (see Chapter 3.6).

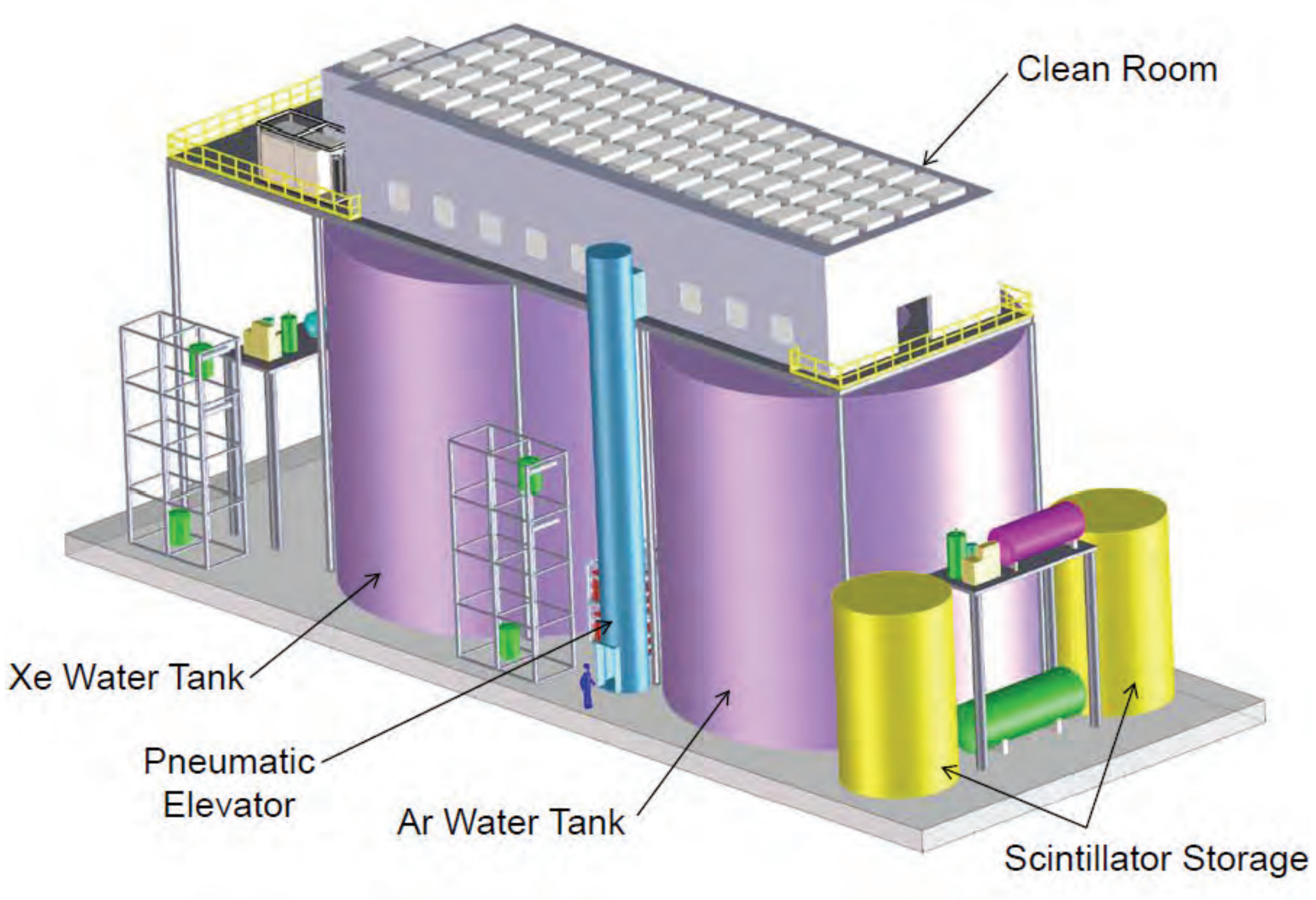

**Figure 3.3.3.2.4-1** Layout of MAX, with its two water tanks and respective depleted argon and xenon detectors at the MLL laboratory in a 50 (L) x 17 (W) x 22 (H) m envelope. [Courtesy MAX collaboration and David Taylor, DUSEL]

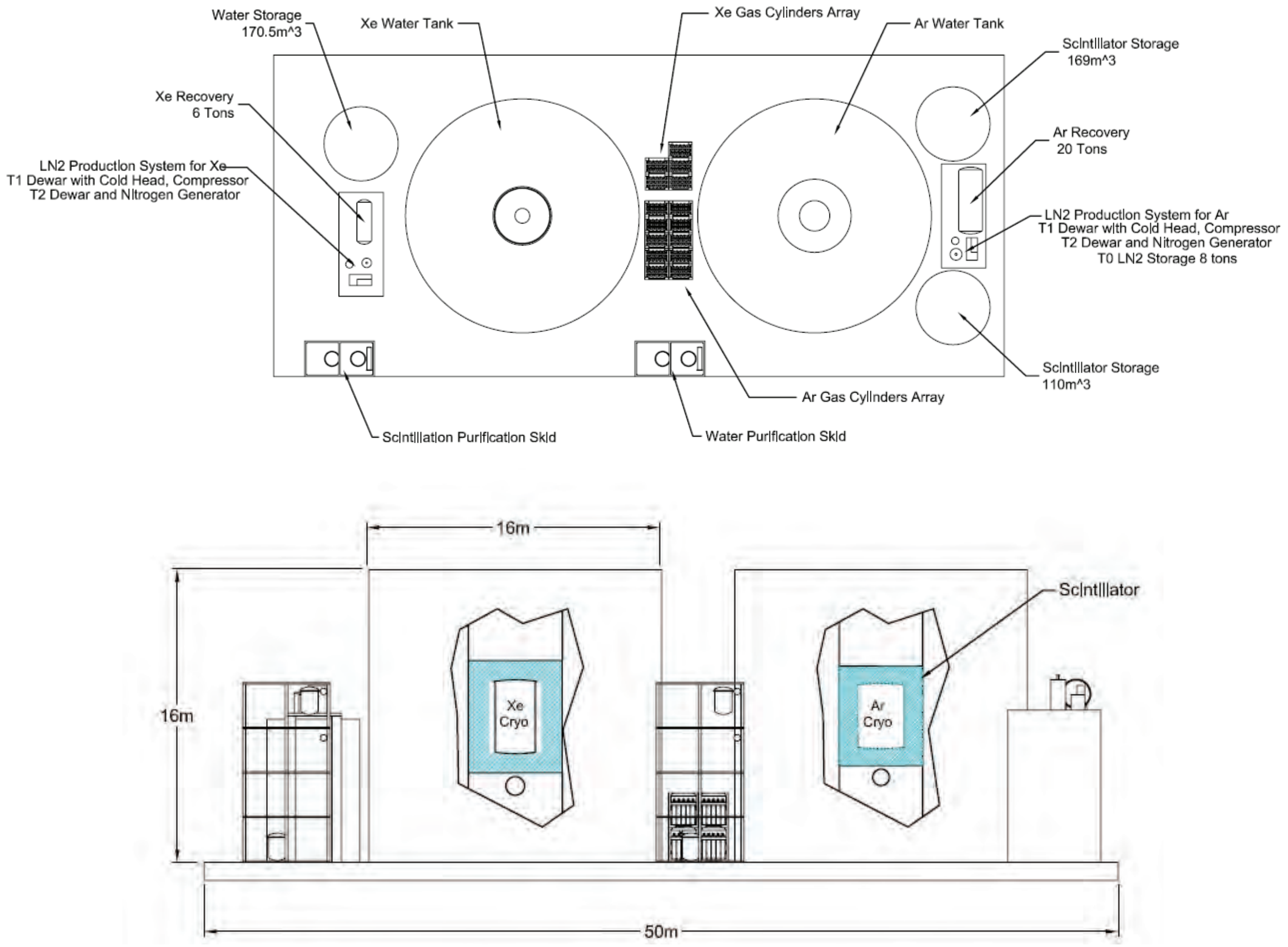

**Figure 3.3.3.2.4-2** MAX configuration at the MLL. [Courtesy MAX collaboration]

### 3.3.3.3 Other Experiments

#### 3.3.3.3.1 CLEAN

The Cryogenic Low Energy Astrophysics with Noble gases (CLEAN) team is studying a single-phase 50-tonne (10-tonne fiducial) LAr detector, which could also be filled with 10 tonnes of liquid neon.[41] The scintillation light produced by interactions in the target is sensed by an array of PMTs on the envelope area of the active volume. Shielding is provided by water. Discrimination against background is based on the combination of the scintillation pulse shape and the radial position reconstruction. Considerable precautions need to be taken to prevent radon contamination on the outer surface of the target. Currently the team envisions a 15 m diameter x 15 m high water tank.

#### 3.3.3.3.2 DMTPC

The Dark Matter Time Projection Chamber (DMTPC) is a proposed apparatus to detect the direction of WIMP recoils.[42] The WIMP detector is a TPC, consisting of a target gas volume in a strong electric field (1 MV/m). A nuclear recoil in the target volume as it loses energy will ionize the gas; a low-pressure gas is used to extend the ranges of these ionization tracks to a few millimeters for typical WIMP-induced recoil energies (~100 keV). The detector uses CF4 as a target gas, which allows detection via scintillation photons from the electron avalanche, as well as sensitivity to spin-dependent interactions. $^{4}$He is added to increase the fast neutron interaction cross section, allowing an in situ neutron background measurement. The team has also experimented with adding xenon, which would allow a search for excited dark matter with a cubic meter device. A charge-couple device (CCD) camera images the scintillation light caused by the recoiling nucleus to determine its direction of travel.

This technique, which has the potential of confirming the galactic-halo source of a WIMP nuclear recoil signal, is in its development phase. As such, for DUSEL, a 1-2 $m^3$ detector for R&D in a 10 m long by 4 m wide and 3 m height area at the 4850L laboratory could be accommodated.

### 3.3.3.4 Experiment Requirements

DUSEL Project staff obtained experiment requirements for the four S4-funded proposed experiments by phone interviews and face-to-face meetings with the teams and by direct entry into a requirements database by experimental-group members. Tables 3.3.3.4-1 through 3.3.3.4-4 show the current state of the main experiment requirements with significant impact on the DUSEL design.

| Requirement | Value/Description | Comment/Justification |
|---|---:|---|
| **Layout** | | |
| Depth | 7400L | |
| Footprint | 25 m L x 12 m W | |
| Height [m] | 12 | Constrained value |
| Floor Load [kPa] | 100 | |
| **Utilities** | | |
| Power [kW] | 100 | Heaters/chillers to keep water temperature at 40 °C |
| Standby Power [kW] | 5 | Large thermal mass, in emergency-mode only monitoring required |
| Chilled Water [kW] | 0 | |
| Waste Heat to Air [kW] | 100 | |
| Purified Water [$m^3$] | 1680 | Entire volume of water shield |
| Potable Water [lpm] | Nominal use | |
| Compressed Air | Nominal use | |
| Network | 1 Gb/s | |
| **Environment** | | |
| Temp. Min [ºC] | 20 | Nominal, may be tighter in clean rooms |
| Temp. Max [ºC] | 25 | |
| Humidity Min [%] | 20 | Nominal, may be tighter in clean rooms |
| Humidity Max [%] | 50 | |
| Rn Background [$Bq/m^3$] | TBD | Experiment provides Rn scrubbing area |
| **Crane** | | |
| Max. Load [Short Tonne] | 5 T | Inside COUPP clean room |

| Requirement | Value/Description | Comment/Justification |
|---|---|---|
| **Occupancy** | | |
| Peak Installation Occupancy [count] | 24 | |
| Installation Duration [months] | 24 | |
| Peak Commissioning Occupancy [count] | 5 | |
| Commissioning Duration [months] | 9 | |
| Peak Operation Occupancy [count] | 10 | |
| Operation Duration [months] | 120 | |
| **Cryogens** | | |
| LN Storage | 500 L | |
| LN Consumption | 100 L/day boil-off | Gas buffer on top of water shield. |
| **Major Hazards (Other Than Cryogens)** | | |
| Water Flood Hazard | 1680 $m^3$ | |
| Health Hazard | 16 tonnes of $CF_3I$, a low-vapor pressure fire-extinguishing compound | Cardiac sensitization at exposures of >0.2% for >1 minute. If one module/tank breaks >0.2% in laboratory. |
| **Assay and Storage** | | |
| Assay Needs | Nominal | |
| Underground Storage | None | |

**Table 3.3.3.4-1** COUPP Experiment requirements.

| Requirement | Value/Description | Comment/Justification |
|---|---|---|
| **Layout** | | |
| Depth | 7400L | Alternate design developed for 4850L |
| Footprint | 25 m L x 12 m W (7400L) | |
| Height [m] | 10 (7400L) | |
| Floor Load [kPa] | 100 | |
| **Utilities** | | |
| Power [kW] | 200 | |
| Standby Power [kW] | 75 | 70 kW to keep system cold;5 kW for monitoring |
| Chilled Water [kW] | 160 | |
| Waste Heat to Air [kW] | 40 | |
| Purified Water [$m^3$] | 140 | Entire volume of water shield |
| Potable Water [lpm] | Nominal use | |
| Compressed Air | Nominal use | |
| Network | 10 Gb/s | |

| Requirement | Value/Description | Comment/Justification |
|---|---|---|
| **Environment** | | |
| Temp. Min [ºC] | 20 | Nominal, may be tighter in clean rooms |
| Temp. Max [ºC] | 25 | |
| Humidity Min [%] | 20 | Nominal, may be tighter in clean rooms |
| Humidity Max [%] | 50 | |
| Rn Background [$Bq/m^3$] | TBD | Experiment provides Rn scrubbing area |
| **Crane** | | |
| Max. Load [Short Tonne] | 20 T | Internal experiment clean room crane nominally used |
| **Occupancy** | | |
| Peak Installation Occupancy [count] | 24 | |
| Installation Duration [months] | 24 | |
| Peak Commissioning Occupancy [count] | 5 | |
| Commissioning Duration [months] | 9 | |
| Peak Operation Occupancy [count] | 10 | |
| Operation Duration [months] | 120 | |
| **Cryogens** | | |
| LN Storage | Storage dewars, tens of liters in cold traps | |
| LN Consumption | Few L/day boil-off | |
| LHe Storage | Several hundred liters | |
| LHe Consumption | 2 L/day | |
| **Major Hazards (Other Than Cryogens)** | | |
| Fire Hazard | ~6 $m^3$ of solid plastic scintillator | Plastic scintillator veto |
| **Assay and Storage** | | |
| Assay Needs | Nominal | |
| Underground Storage | None | |

**Table 3.3.3.4-2** GEODM Experiment requirements.

| Requirement | Value/Description | Comment/Justification |
|---|---|---|
| **Layout** | | |
| Depth | 4850L | |
| Footprint | 25 m L x 17 m W | |
| Height [m] | 18 | |
| Floor Load [kPa] | 200 | |
| **Utilities** | | |
| Power [kW] | 220 | |
| Standby Power [kW] | 50 | Enough LN to maintain cold for 12 hours (in the absence of this emergency power) |
| Chilled Water [kW] | 50 | |
| Waste Heat to Air [kW] | 170 | |
| Purified Water [$m^3$] | 1400 | Entire volume of water shield |
| Potable Water [lpm] | Nominal use | |
| Compressed Air | Nominal use | |
| Network | 1 Gb/s | |
| **Environment** | | |
| Temp. Min [ºC] | 20 | Nominal, may be tighter in clean rooms |
| Temp. Max [ºC] | 25 | |
| Humidity Min [%] | 20 | Nominal, may be tighter in clean rooms |
| Humidity Max [%] | 50 | |
| Rn Background [Bq/$m^3$] | TBD | Experiment provides Rn scrubbing area |
| **Crane** | | |
| Max. Load [Short Tonne] | 10 T | Outside clean room (penetrates into clean room) |
| **Occupancy** | | |
| Peak Installation Occupancy [count] | 24 | |
| Installation Duration [months] | 24 | |
| Peak Commissioning Occupancy [count] | 5 | |
| Commissioning Duration [months] | 9 | |
| Peak Operation Occupancy [count] | 10 | |
| Operation Duration [months] | 120 | |
| **Cryogens** | | |
| LN Storage | 3000 L | |
| LN Consumption | 200-500 L/day | For cooling, are generated in situ |
| LXe | 20 T | |

| Requirement | Value/Description | Comment/Justification |
|---|---|---|
| **Major Hazards (Other Than Cryogens)** | | |
| Water Flood Hazard | 1400 $m^3$ | |
| Oxygen deficiency | Large volumes of cryogenic liquids, 20 tonnes LXe, and few tonnes LN | Approach is fail-safe recovery of xenon, vessel engineering to prevent LXe mixing with water. |
| Fire hazard | 100 tonnes of organic scintillator (possibly isohexane) and 100 tonnes of charcoal | Charcoal is contained in stainless steel vessel |
| **Assay and Storage** | | |
| Assay Needs | Nominal | |
| Underground Storage | None | |

**Table 3.3.3.4-3** LZD Experiment requirements.

| Requirement | Value/Description | Comment/Justification |
|---|---|---|
| **Layout** | | |
| Depth | 4850L | |
| Footprint | 50 m L x 17 m W | |
| Height [m] | 22 | |
| Floor Load [kPa] | 200 | |
| **Utilities** | | |
| Power [kW] | 260 | |
| Standby Power [kW] | 44 | |
| Chilled Water [kW] | 70 | |
| Waste Heat to Air [kW] | 190 | |
| Purified Water [$m^3$] | 6500 | Entire volume of both water shields |
| Potable Water [lpm] | Nominal use | |
| Compressed Air | Nominal use | |
| Network | 1 Gb/s | |
| **Environment** | | |
| Temp. Min [ºC] | 20 | Nominal, may be tighter in clean rooms |
| Temp. Max [ºC] | 25 | |
| Humidity Min [%] | 20 | Nominal, may be tighter in clean rooms |
| Humidity Max [%] | 50 | |
| Rn Background [Bq/$m^3$] | TBD | Experiment provides Rn scrubbing area |
| **Crane** | | |
| Max. Load [Short Tonne] | 40 T | Centerline Crane |
| Max. Load [Short Tonne] | 10 T | 2 flat-beam cranes |

| Requirement | Value/Description | Comment/Justification |
|---|---|---|
| **Occupancy** | | |
| Peak Installation Occupancy [count] | 24 | |
| Installation Duration [months] | 24 | |
| Peak Commissioning Occupancy [count] | 5 | |
| Commissioning Duration [months] | 9 | |
| Peak Operation Occupancy [count] | 10 | |
| Operation Duration [months] | 120 | |
| **Cryogens** | | |
| LN Storage | 9 T | |
| LN Consumption | | |
| LXe Storage | 6 T | |
| LAr Storage | 20 T | |
| **Major Hazards (Other Than Cryogens)** | | |
| Water Flood Hazard | 6500 $m^3$ | |
| Oxygen deficiency | Large volumes of cryogenic liquids, 20 tonnes LAr, 6 tonnes LXe, and 9 tonnes LN | Approach is double-walled cryostat with leak detection in intermediate vacuum region. If break detected, rapid drain of warm-side liquid (water or scintillator) and full recovery of the cryogen liquid. |
| Fire hazard | 200 tonnes of organic scintillator | |
| **Assay and Storage** | | |
| Assay Needs | Nominal | |
| Underground Storage | 100 $m^2$ | |

**Table 3.3.3.4-4** MAX Experiment requirements.

For the DMTPC proposed R&D area, basic requirements were captured in the database; they are expected to have no impact (besides the area required) on the Facility interface.

#### 3.3.3.5 Schedule

For the large dark-matter G3 experiments expected for DUSEL, about three to four years of procurement, construction, and preparation will be needed before starting installation underground.[43] Therefore, for a timely deployment in DUSEL, selection among the G3 candidates would need to be made about four years before the scheduled completion of the respective 4850L LM(s) and/or the DLL LM. Additional discussion of the selection process and timescale is in Chapter 3.10.

### 3.3.4 Neutrinoless Double-Beta Decay Experiments

#### 3.3.4.1 Candidate Experiments and Requirements[44]

Two experiments have received S4 funding from the NSF to develop proposals to be located in DUSEL. The 1 Tonne Germanium (1TGe) experiment would look for $0\nu\beta\beta$ decay in approximately 1 tonne of high-purity germanium (HPGe) detectors enriched in $^{76}$Ge. The Enriched Xenon Observatory (EXO) experiment plans to search for 0νββ decay in a TPC containing 1 to 10 tonnes of $^{136}$Xe. These experiments will be sensitive to effective *Majorana* neutrino masses of about 10-20 meV after 5+ years of running. Both experiments are investigating multiple configurations for the detector design: 1TGe is evaluating passive lead shielding versus a large liquid-phase active shield, while EXO is developing designs for both a liquid-phase and a gas-phase Xe TPC. Both experiments are assuming occupancy at the 7400L. 1TGe and EXO are discussed in detail in the following sections.

In addition to 1TGe and EXO, several other groups have expressed interest in the future use of DUSEL. These include current members of the Cryogenic Underground Observatory for Rare Events (CUORE)[45] and the Neutrino Experiment with a Xenon TPC (NEXT) collaborations.[46] CUORE is an Italian-led group that is constructing a 750 kg array of (unenriched) $TeO_2$ bolometers in the Gran Sasso National Laboratory to search for the 0νββ decay of $^{130}$Te. The NEXT collaboration is performing R&D toward a 100 kg enriched xenon high-pressure gaseous TPC design to be installed in the Canfranc Underground Laboratory, Spain.[47] Both collaborations have mentioned DUSEL as a potential location for future experiments deploying larger masses of ββ-decay isotopes.

Requirements of a 0νββ experiment at DUSEL have been set based on Conceptual Designs of the 1TGe and EXO experiments. A facility that is designed to accommodate either 1TGe or EXO (or both) would be suitable for other neutrinoless double-beta decay experiments as well.

#### 3.3.4.1.1 1TGe

The 1TGe apparatus would consist of an array (~1000) of HPGe diodes. Several advantages of HPGe detectors are:

- They are a well-established technology.
- Enrichment from the natural abundance of 7.6% up to 86% in active isotope has been demonstrated.
- The source is the detector (minimizing mass and space requirements).
- They have excellent energy resolution (less that 0.2% at 2039 keV, the endpoint of the $^{76}$Ge double-beta decay spectrum).
- They are extremely radiopure (providing a substantial reduction in radioactive backgrounds).

Exploring the critical design features of a tonne-scale $^{76}$Ge detector is the topic of current R&D by the MAJORANA (U.S.-led effort)[48] and GERmanium Detector Array (GERDA) (European-led effort)[49] collaborations, which are currently constructing $^{76}$Ge-based detector systems. The construction and successful operation of the MAJORANA DEMONSTRATOR (at Sanford Laboratory) and GERDA (at LNGS) detectors will not only address the key technical challenges of the 1TGe experiment, but will also definitively test the claim of an observed 0νββ decay signal in $^{76}$Ge by Klapdor-Kleingrothaus, et al.[50]

The MAJORANA DEMONSTRATOR, a 40-60 kg array of novel p-type point contact HPGe detectors, is scheduled for operation of its first module (20 kg of $^{nat}$Ge) in 2013 in the Davis Transition Area (DTA) at the 4850L of Sanford Laboratory, as described in Chapter 3.4. The DEMONSTRATOR uses a conventional vacuum-cryostat design in which electroformed copper is the primary structural component. The cryostat is surrounded by passive and active bulk shielding composed of electroformed copper, lead, polyethylene (as a neutron moderator), and a plastic scintillator muon veto. The DEMONSTRATOR's HPGe detectors offer background rejection capabilities that are optimal for double-beta decay searches, and its scalable, modular design readily permits future upgrades. Furthermore, the Ge crystals can be reconfigured into different cryostat or shielding designs should such a need be identified.

The GERDA experiment, scheduled to begin taking first data in 2011 using 18 kg of enriched $^{76}$Ge detectors at LNGS, is investigating a novel approach of immersing the Ge diodes directly in a liquid cryogen. In the GERDA approach, the degree of background shielding can be scaled easily, and if a scintillating cryogen (e.g., liquid Ar) is used, the scintillation would provide an additional tag for external background rejection.

The two collaborations have signed a letter of intent to join together to construct a tonne-scale Ge detector array with a *Majorana* mass sensitivity below 50 meV using the best techniques demonstrated in the current phase. The 1TGe collaboration consists of the MAJORANA collaboration along with scientists from the Max Planck Institute in Munich, most of who are also members of GERDA. The objective of the 1TGe collaboration is to develop a realistic design of a 1-tonne $^{76}$Ge detector, using the knowledge and experience gained from the MAJORANA DEMONSTRATOR and GERDA experiments.

The goal of the 1TGe Project is to develop the basic elements of a Preliminary Design for an experiment that could operate at DUSEL. If required for design development, specific R&D will be conducted to mitigate high-risk technical elements of the Project and to provide a realistic schedule for risk retirement. In particular, R&D will be conducted to better understand the recycling options that need to be implemented to maximize the use of the enriched $^{76}$Ge material. The collaboration will also build upon R&D that is already under way as part of the MAJORANA DEMONSTRATOR and GERDA Projects. The final implementation of the 1TGe experiment will only be determined after the data from the MAJORANA DEMONSTRATOR and GERDA experiments have been analyzed and technical choices have been made. The 1TGe collaboration has been interacting with the DUSEL Project Team to define a set of requirements for both possible implementations of the shield—Cu/Pb shield or LAr shield:

> **The Cu/Pb shield option.** The copper-lead shield would be based on the MAJORANA DEMONSTRATOR design, which is scalable in a straightforward manner. A crucial need for this design is a large amount of electroformed copper grown underground to avoid activation by cosmic rays. One of the goals of the DEMONSTRATOR is to show that electroformed copper can be successfully grown and processed underground, to result in the extremely low levels of radioactivity (less than 1ppt U/Th) that are required for the experiment.

> **The LAr shield option.** The LAr shield would be based on the configuration of the GERDA experiment. LAr scintillates and can be instrumented for use as an active shield. The volume of LAr could be contained within a large water shield to further attenuate neutrons and provide additional active veto capability against muons. The 1TGe collaboration is also exploring similar options in which the detectors are enclosed in a vacuum cryostat that is immersed in a large LAr, water, or liquid scintillator shield.

Results from the DEMONSTRATOR and GERDA experiments will help in the understanding of the size of shielding required for a tonne-scale experiment that uses this design.

#### 3.3.4.1.1.1 Depth

The 1TGe collaboration has submitted a detailed justification for location at the 7400L of DUSEL—to reduce backgrounds. For the experiment to be successful, background levels of less than 1 count/ tonne/year in a 4 keV region of interest around the 2039-keV $0\nu\beta\beta$ peak are required. The background budget is essentially consumed by the materials that compose the experiment; it requires that backgrounds from cosmic rays be negligible. The reduction in muon-induced neutrons is about a factor of 20 when going from the 4850L to the 7400L. They conclude that locating at the 4850L carries large risks, especially due to the uncertainties calculating neutron production rates. They comment that locating the experiment at the 4850L would require substantial simulations and use of the R&D data from the MAJORANA DEMONSTRATOR to establish if there exists a shielding configuration that could still produce the desired reduction in cosmic ray-induced backgrounds. If this were the case, it presumably would look like a larger version of the LAr shield design. Further discussion of the arguments for the 7400L is given in Section 3.3.10.

#### 3.3.4.1.1.2 Layouts

The 1TGe experiment has provided DUSEL with conceptual layouts for the 7400L LM for both shield configurations (see Figures 3.3.4.1.1.2-1 to 3.3.4.1.1.2-2). The Cu/Pb layout is based on scaling up the DEMONSTRATOR experiment; and the LAr shield layout is based on scaling up the GERDA experiment. Neither layout fits in the current LM footprint guideline that was provided by the DUSEL Project Team to the collaboration; the Cu/Pb version is too long (41 m compared with 25 m guideline) and the LAr version is too wide (by 4 m) and too tall (by 2 m). The Cu/Pb layout could be scaled to better match the constraint, and some of the space (gowning area, control and break rooms, etc.) could be shared with another experiment in the LM. The LAr-shield would require a larger LM if the shielding requirements remain unchanged; this has prompted a Trade Study (Section 3.8.5) examining the costs of redimensioning the 7400L LM. As of late 2010, neither of the R&D versions of the experiment has acquired physics data, and these design options should be considered still conceptual. In all likelihood, the final implementation of the tonne-scale experiment will look quite different from either of the prototypes, having been informed by simulations of the possible shielding configurations for the detector and construction and operation of the prototypes.

No layout has been provided for the copper electroforming facility. However, it is expected that the copper electroforming will be done using the facilities installed for the DEMONSTRATOR in the Sanford Laboratory (4850L). The DEMONSTRATOR will use 16 baths to grow the parts it needs. Based on preliminary estimates, the addition of four baths (of slightly larger diameter) will be sufficient to satisfy the needs for the 1TGe experiment. A plan for the evolution of the DEMONSTRATOR laboratory space will be developed once better estimates of the achievable copper growth rates become available.

In addition to electroforming space, the 1TGe experiment will require a clean underground machine shop. The machine shop that is planned for the MAJORANA DEMONSTRATOR will most likely be too small to accommodate the needs of the 1 tonne experiment and the space will likely have to be expanded.

It should be noted that the underground machine shop and electroforming facilities do not need to be as deep as the experiment. Either could be installed at shallower levels in DUSEL (300L, for example).

Finally, the Center for Ultralow-Background Experiments at DUSEL (CUBED) collaboration (Chapter 3.4) is investigating the underground production of the HPGe detectors—zone refining, crystal growing, and detector fabrication.[51] CUBED would also require space underground. As the decay of cosmogenic isotopes produced in the Ge detectors while they are aboveground presents a potentially significant source of background to 1TGe, underground Ge detector fabrication would benefit the 1TGe experiment.

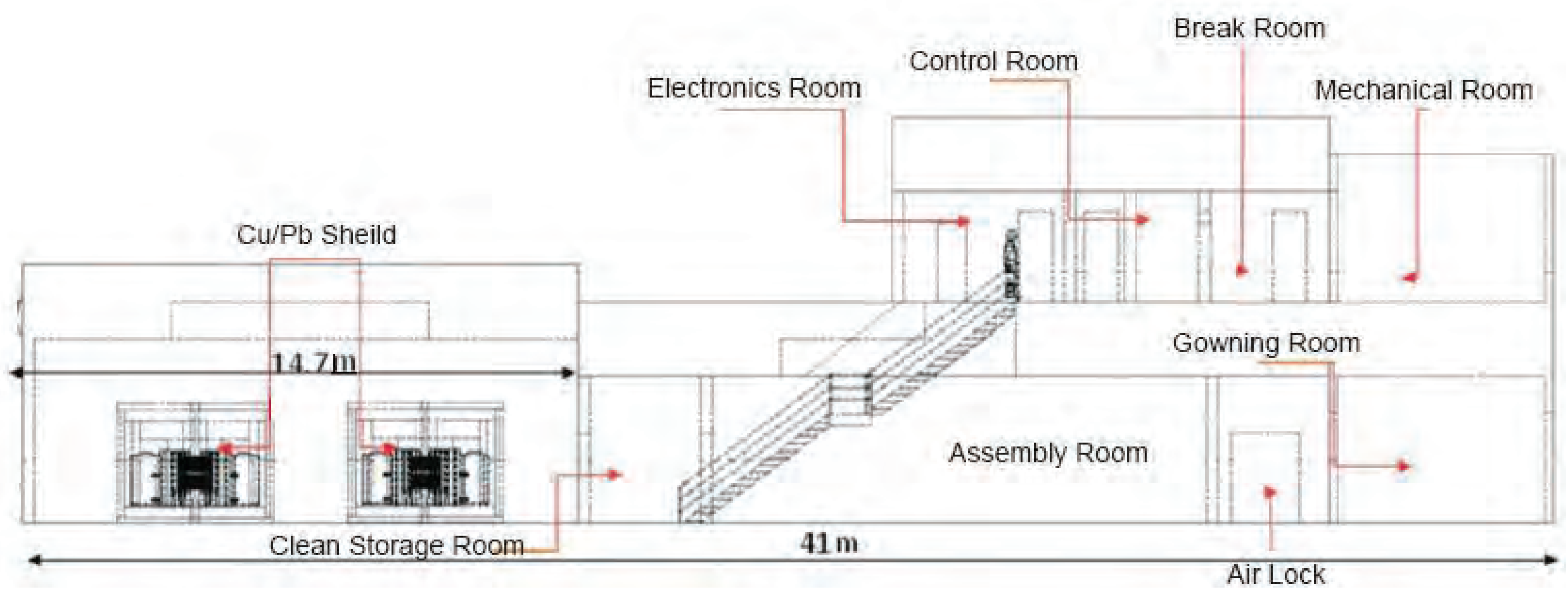

**Figure 3.3.4.1.1.2-1** Side view of Cu/Pb shield layout for the 1TGe experiment. The footprint of this design has a length of 41 m, a height of 11 m, and a width of 11.6 m. [Courtesy 1TGe collaboration]

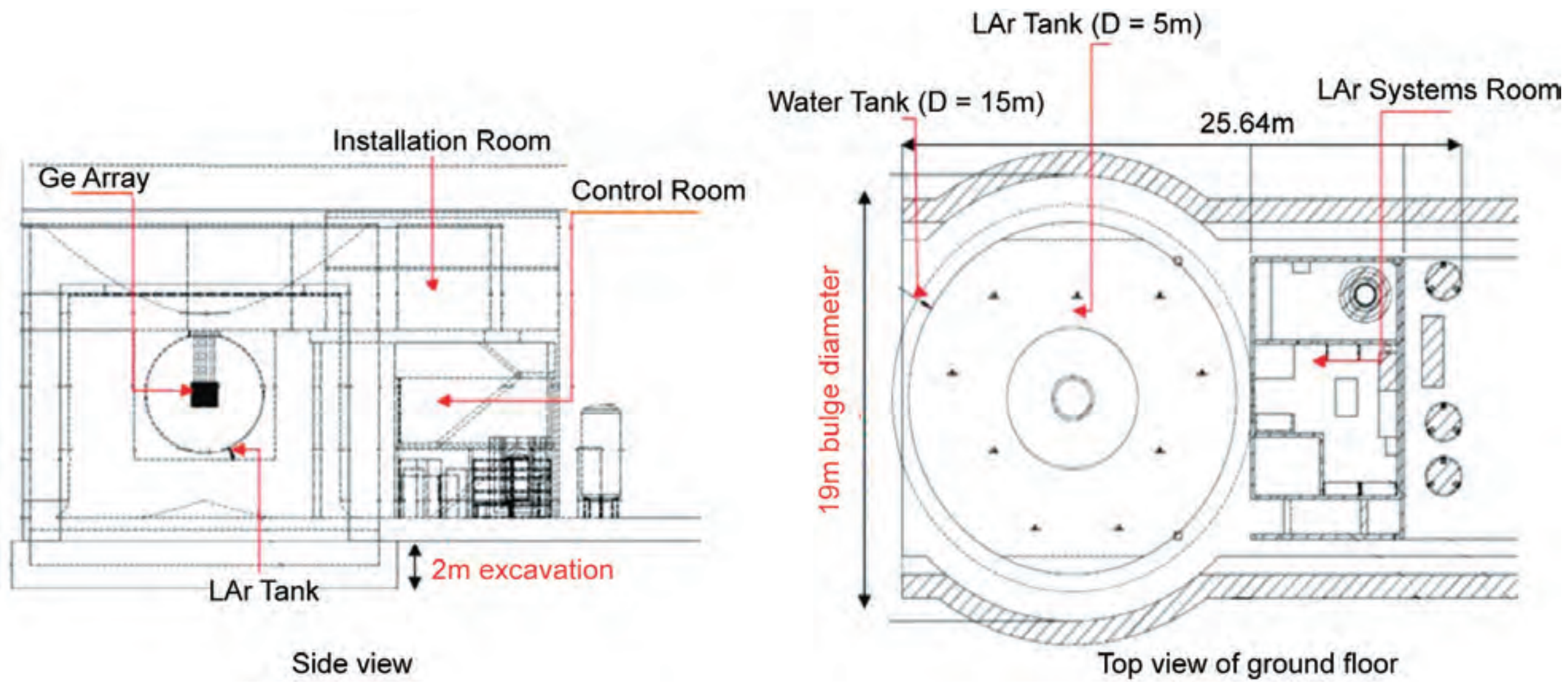

**Figure 3.3.4.1.1.2-2** Side and top view of the layout for the 1TGe LAr shield option. [Courtesy 1TGe collaboration]

### 3.3.4.1.2 The EXO Experiment

In 2000, the Enriched Xenon Observatory (EXO) collaboration started a program to push the sensitivity for $0\nu\beta\beta$ decay using $^{136}$Xe as a source and detector. $^{136}$Xe is particularly appropriate for a very large experiment:

- The use of a material in the form of a liquid or gas allows for easy transfer of the enriched isotope from one detector to another. In addition, the possibility of using the material either in gas or in liquid phase, with complementary properties, opens a broad set of possibilities for a program of experiments in which a large fraction of the cost is the isotopic enrichment. Indeed, the possibility of using different types of detectors is an integral part of the EXO program.

- $^{136}$Xe, being a gas at standard temperature and pressure (STP) and hence easy to process in ultracentrifuges, is particularly economical to enrich from the natural fraction of 8.9%.
- Xenon can be used at the same time as a source and as a homogeneous detector, either in gas phase (GXe) or in liquid phase (LXe). In either case, Compton scatterings of photons (the main source of background) can be readily distinguished from "single site" events produced in the ββ decay. The energy resolution in an LXe or GXe TPC is known to be sufficient to separate the standard-model two-neutrino (2νββ) decay from the 0νββ decay for *Majorana* mass sensitivities below 10 meV.
- In a large detector, the xenon can be continuously extracted and repurified, if necessary, during the lifetime of the experiment. This is important, as the background requirements for G-3 ββ-decay experiments are so extreme that the final detector is really the only device with sufficient sensitivity to verify the purity of the source.
- Xenon, being a noble element, is particularly easy to purify from all chemically active elements. Contaminations of $^{85}$Kr (a fission fragment injected into the atmosphere by nuclear reactors) and $^{42}$Ar (bred in atmospheric nuclear testing) produce decays with low Q-value not relevant for the 0νββ decay mode. In addition, ultracentrifugation greatly reduces the contamination of these substantially lighter isotopes.
- No long-lived isotopes of Xe exist. Hence, after a short "cooldown" period underground and chemical purification, no contamination should remain in the gas or liquid.
- The $^{136}$Xe Q-value of 2457.8 keV is among the largest of the candidate double-beta isotopes
- The ββ decay of $^{136}$Xe produces a barium ion ($^{136}Ba^{2+}$) that can in principle be detected by optical spectroscopy on the ion $Ba^{+}$, using the shelving technique.[52] The possibility of "tagging" the chemical species of the final state of the decay would provide a new variable to be used for background suppression. This technique, only applicable to the case of xenon, could drastically improve the quality of ββ decay detection and make extremely large experiments possible.

The EXO collaboration manages a diverse program, including the construction of a large detector (EXO-200); R&D on LXe and GXe technologies; an isotope-enrichment program; and barium-tagging R&D employing techniques from atomic, molecular, and optical (AMO) physics and radioactive beam physics. As of late 2010, the EXO-200 detector is in the final phase of preparation for data-taking at the Waste Isolation Pilot Plant (WIPP) near Carlsbad, New Mexico. It is expected that EXO-200 will have a relatively short technical run with 150 kg of natural Xe, followed by a run of three to five years with xenon enriched to 80% in the isotope 136 (the enrichment grade chosen by the collaboration). The technical run of the largest liquid xenon TPC ever built will provide a wealth of information essential for the design of a multitonne EXO, while the subsequent physics run is expected to measure the 2νββ decay in $^{136}$Xe and substantially improve the sensitivity to *Majorana* neutrino masses through the 0νββ decay mode.

### 3.3.4.1.2.1 Depth

The EXO baseline assumes that the experiment will be located at the 7400L of DUSEL. The EXO proponents are evaluating the depth requirement for a multitonne EXO detector. This requires balancing the technical risk of backgrounds larger than expected with the practical inconvenience of the deeper site and of the smaller cavity that will likely be available. EXO will be measuring the background rate induced by cosmic radiation in the upcoming runs of EXO-200 located about 1590 meters-water-equivalent (mwe) underground at the WIPP site, having started in late 2010. The data may be used as a basis for the extrapolation of backgrounds to larger depths. Further discussion of the arguments for the 7400L is given in Section 3.3.10.

### 3.3.4.1.2.2 Layouts

The EXO baseline design for an installation at DUSEL includes a LXe TPC, using between 1 and 10 tonnes of xenon, enriched to 80% in the isotope 136. As for EXO-200, the TPC will have charge and scintillation readout, the latter used both to provide a start time for the drift time measurement and to improve the energy resolution. The detector's physical properties will be similar to those of EXO-200, and the light readout will probably employ large-area avalanche photodiodes of similar nature to those used in EXO-200. The detector will localize candidate events in real time and trigger a mechanical system capable of inserting a grabber tip in the LXe and bringing it to a distance of ~1 cm from the decay site in tens of seconds. It is assumed that the ion trap and the optics related to the fluorescence detection will be built out of conventional (i.e., not low-radioactivity) components and will be housed in a special laboratory outside the detector shield. Depending on the total volume sought, two or more double drift spaces (each with a central cathode at high voltage) will be required. The vessel containing the LXe will be cooled and shielded with the same HFE7000 fluid used in EXO-200, contained in a vacuum-insulated cylindrical cryostat with its axis vertical.

Because of the large volume of the full EXO detector and the thick shielding required by the higher radioactivity of hard rock compared to salt at WIPP, it is assumed that the shielding will, for the most part, be water. The cryostat will be housed in a radio-quiet chamber, surrounded in all directions but the top by ~5 m of water. A water tank of this size can accommodate a TPC containing 1 to 10 tonnes of LXe without a significant change in footprint. After the installation of the cryostat, the volume above it will be outfitted as a Class 100 clean room, similar in quality to the innermost part of the EXO-200 clean rooms. The top of the shielding is at present thought to be made of 50-cm thick lead to provide a relatively thin layer to be penetrated by the ion grabber probe. The laboratory housing the ion trap and the optics will be located above the top shield. A concept of the detector within one of the DUSEL standard laboratories is shown in Figures 3.3.4.1.2.2-1 and 3.3.4.1.2.2-2. An additional advantage of the water shielding is the possibility of instrumenting it with PMTs to use it as a cheap and hermetic cosmic-ray veto detector. The current configuration of the 7400L LM is too narrow to accommodate EXO's water tank configuration. A Trade Study quantifying the costs of resizing the LM is described in Section 3.8.5.

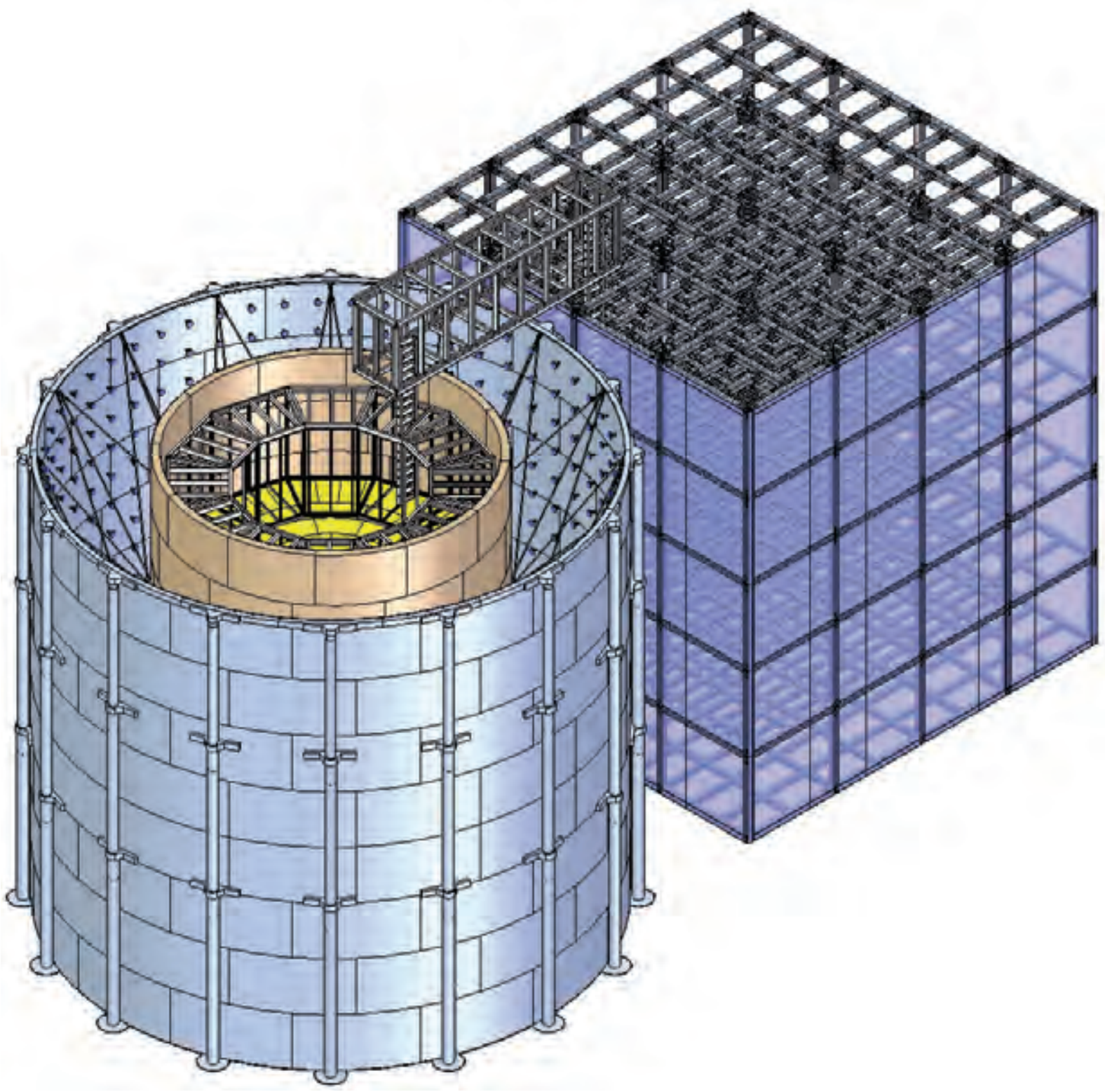

**Figure 3.3.4.1.2.2-1** 3D view of the EXO baseline design for a 1 to 10 tonne LXe TPC surrounded by a water Cherenkov veto. Next to the tank is a five-story support building containing clean rooms, offices, Xe/cryogen/refrigerant and water handling, UPS/electrical utilities, and a machine shop. The water tank has a 17 m diameter and a height of 15.5 m (not including the walkway linking to support building). The support building has a 15 m height and a 14 m x 14 m square footprint. [Courtesy EXO collaboration]

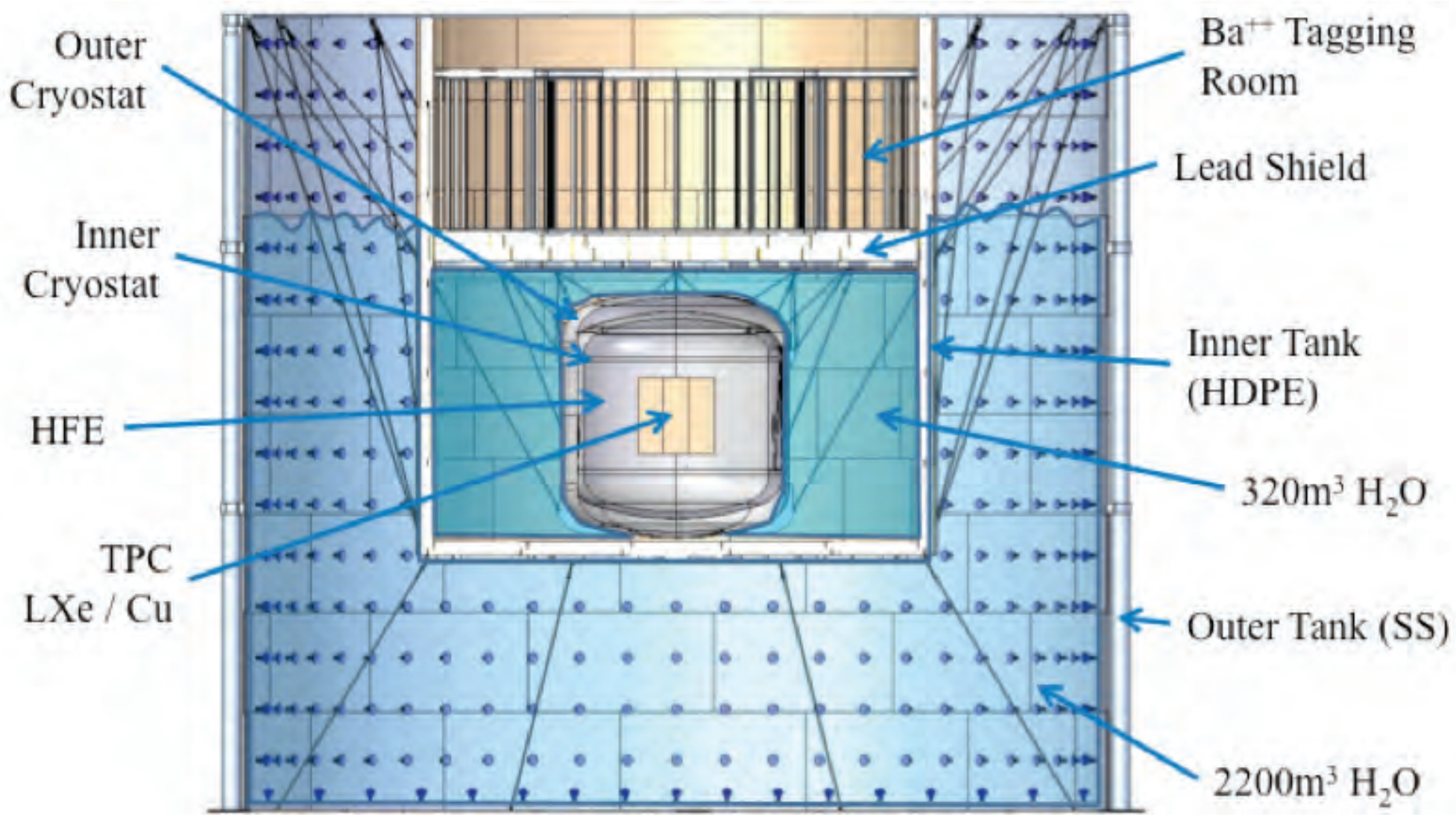

**Figure 3.3.4.1.2.2-2** Schematic of the EXO detector. The water tank has a 17 m outer diameter and a height of 15.5 m. [Courtesy EXO collaboration]

### 3.3.4.1.2.3 GXe Option

While a high-pressure gas-phase TPC was used in the early conceptual description of EXO, it was later decided that because of the finite resources, priority would be given to a liquid phase 200 kg prototype. At the same time, R&D on high-pressure GXe versions of the detector continued and is in progress within the EXO collaboration. The tradeoff between the GXe and the LXe options can be summarized as follows. A GXe detector is expected to be superior in terms of energy resolution and topology. Depending on the density, topological information should result in better background rejection and push the requirement for Ba tagging to higher fiducial masses. At sufficiently low density, the $\beta\beta$ correlation may become measurable, providing insights on the underlying physics. On the other hand, for a very large detector, low density implies very large size and, at the extreme, unmanageable costs. At a pressure of 10 atm, a chamber in the shape of a square cylinder filled with 3 tonnes of Xe would have an active diameter and length of ~4.4 m. An LXe detector would be substantially smaller and trade a large pressure vessel for some modest cryogenics. The large size of a GXe detector increases the channel count and dimensions of the readout wires or other gas-gain structures and of the photon shielding. In addition, a larger detector requires a larger amount of clean materials for its construction. Structural engineering issues related to the very large pressure vessel are also a concern. Ba tagging would require rather different techniques in the two cases, and at the present state of R&D, it is unclear which technique would be simpler and more effective.

Nevertheless, R&D on a GXe $0\nu\beta\beta$ detector is an important investment, whatever the technology choice for EXO, because of the flexible nature of $^{136}$Xe for $\beta\beta$ searches. Should a positive signal be found in EXO, using technology accepted for the detector, a cross-check measurement with a very different detector will probably be desirable, making the advancement of both LXe and GXe detectors very appropriate.

## 3.3.4.2 Experiment Requirements

The 1TGe and EXO collaborations have provided a list of requirements they would need in DUSEL in order to operate successfully. These are summarized in Tables 3.3.4.2-1 to 3.3.4.2-3. In most cases, for 1TGe, a Cu/Pb shield experimental configuration was assumed, while for EXO, a liquid-phase TPC configuration was assumed. A few comments are given after the tables.

| Requirement | Value/Description | Comment/Justification |
|---|---|---|
| **Layout** | | |
| Footprint | 41 m L x 11.6 m W | For Cu/Pb shield at 7400L |
| Height | 11 | For Cu/Pb shield at 7400L |
| Footprint | 25m L x 19m W | For LAr shield at 7400L, includes 2 m of extra width in module |
| Height [m] | 14.5 | For LAr shield at 7400L, includes 2 m excavation to provide overhead clearance |
| Floor Load [kPa] | 24 | |
| **Utilities** | | |
| Power [kW] | 70 | |
| Standby Power [kW] | 70 | |
| Chilled Water [kW] | 0 | |
| Waste Heat to Air [kW] | 70 | |
| Purified Water [$m^3$] | 2600 | Water shield volume |
| Potable Water [lpm] | Nominal | |
| Compressed Air | | Used for lifting in Cu/Pb configuration, can be brought down in bottles |
| Network | 1 Gb/s | |
| **Environment** | | |
| Temp. Min [ºC] | 20 | |
| Temp. Max [ºC] | 25 | |
| Humidity Min [%] | 20 | |
| Humidity Max [%] | 50 | |
| Rn Background [$Bq/m^3$] | 3 | Experiment to provide Rn scrubbing area |
| **Crane** | | |
| Max. Load [Short Tonne] | 5.5 | |
| **Occupancy** | | |
| Peak Installation Occupancy [count] | 20 | Estimate |
| Installation Duration [months] | 36 | |
| Peak Commissioning Occupancy [count] | 12 | Estimate |
| Commissioning Duration [months] | 24 | Begins year 2 of construction |
| Peak Operation Occupancy [count] | 6 | Estimate |
| Operation Duration [months] | >60 | |
| **Cryogens** | | |
| LAr Storage | 21,000 L | 1T Ge GERDA style. 30 T used in GERDA |
| LAr Consumption | NA | |
| LN Storage | 10,000 L | Assume 5 days storage—tied to consumption rate |
| LN Consumption | 2,000 L/day | Based on 2 L/day per 1,000 detectors |

| Requirement | Value/Description | Comment/Justification |
|---|---|---|
| **Major Hazards (Other Than Cryogens)** | | |
| Oxygen Deficiency Hazard (ODH) | From cryogens | |
| Failure in LAr Vessel | Rapid boil-off of LAr could present ODH hazard | |
| Chemical Hazards | Chemicals used in electroforming lab, issues will be addressed by the MAJORANA DEMONSTRATOR at Sanford Laboratory | |
| Water Flood Hazard | 2600 $m^3$ | |
| **Assay and Storage** | | |
| Assay Needs | Nominal | |
| Storage | 5 m x 6 m x 3 m at 4850 | Needed for crystal storage |

**Table 3.3.4.2-1** 1TGe experiment requirements.

| Support Area at 4850L Requirements (Electroforming) | |
|---|---|
| Dimensions (W x L x H) | 14 m x 9 m x 3 m electroforming lab at 4850L, will reuse current MAJORANA facility<br>6m x 10m x 3m cleaning and passivation lab, will reuse current MAJORANA facility |
| Network Connection Bandwidth | 1 Gb/s |
| Telephone Connection | Yes |
| Temperature Min-Max | 20-25 (°C) |
| Humidity | 15-60% |
| Purified Water | 250 gal/month |
| Maximum Radon Activity | 1 (Bq/$m^3$), experiment to provide Rn exclusion area |
| Access Control | Yes |
| Clean Room | Electroforming to have Class 100 area |

**Table 3.3.4.2-2** 1TGe electroforming requirements.

| Requirement | Value/Description | Comment/Justification |
|---|---|---|
| **Layout** | | |
| Footprint | 32 m L x 17 m W | Based on 17 diameter tank and 14 m x 14 m x15 m support building |
| Height [m] | 17.5 | Based on a 2 m height from a walkway on top of 15.5 m tank |
| Floor Load [kPa] | 1670 | Lead shield option |
| **Utilities** | | |
| Power [kW] | 300 | |
| Standby Power [kW] | 60 | |
| Chilled Water [kW] | 250 | For refrigeration system |
| Waste Heat to Air [kW] | 50 | |
| Purified Water [$m^3$] | 3500 | Water shield volume |

| Requirement | Value/Description | Comment/Justification |
|---|---|---|
| Potable Water [lpm] | Nominal | |
| Compressed Air | Nominal | |
| Network | 1 Gb/s | |
| **Environment** | | |
| Temp. Min [ºC] | 20 | |
| Temp. Max [ºC] | 25 | |
| Humidity Min [%] | 20 | |
| Humidity Max [%] | 50 | |
| Rn Background [$Bq/m^3$] | 1 | Experiment to provide Rn scrubbing area |
| **Crane** | | |
| Max. Load [Short Tonne] | 20 | |
| **Occupancy** | | |
| Peak Installation Occupancy [count] | 20 | Estimate |
| Installation Duration [months] | 42 | |
| Peak Commissioning Occupancy [count] | 15 | Estimate |
| Commissioning Duration [months] | 12 | |
| Peak Operation Occupancy [count] | 6 | Estimate |
| Operation Duration [months] | >60 | |
| **Cryogens** | | |
| LXe Storage | 3400L | 10 T |
| LXe Consumption | N/A | |
| **Major Hazards (Other Than Cryogens)** | | |
| Pressure Vessel | For GXe | |
| Lasers | Lasers used in Ba-tagging room (minor hazard) | |
| Chemical hazards | Chemicals used during construction (minor hazard) | |
| **Assay and Storage** | | |
| Assay Needs | Nominal | |
| Storage | TBD | |

**Table 3.3.4.2-3** EXO experiment requirements.

The LAr option for 1TGe would require significant additional power for cryo-coolers and refrigeration, but this has not been estimated at this time. Neither group's requirements include power for clean room HVAC. UPS power would be needed primarily to control shutdown of detector electronics, and to keep the cryogens cold. This is critical for EXO, whose enriched xenon cryogen represents the bulk of the cost of the experiment. The same would be the case for 1TGe if the LAr were the shield to use $^{39}$Ar-depleted argon. For both experiments, sufficient amounts of $LN_2$ can be stored underground to keep the cryogens cold during an emergency to alleviate the risk of venting.

The 1TGe electroforming laboratory would use a significant amount of purified water; this could be provided from the LBNE water-purification plant or other sources. For EXO and for the 1TGe LAr option, the water-tank shield would need ~1 kT of pure water and a purification plant to recirculate it.

Ventilation will also have to take into account exhaust from $LN_2$ boil-off as well as chemicals from the electroforming laboratory (see *Major Hazards*, below).

**Cryogen Needs**

Substantial amounts of cryogen will have to be transported to the 7400L for the 1TGe experiment to keep the HPGe diodes cold. The #6 Winze limits the size of transport dewars that can be used and can be an issue.

EXO will use refrigeration for its LXe and hence has limited need for additional liquid cryogens.

**Other Requirements**

Compressed air will be needed for the machine shops and, for 1TGe, the electroforming operation. The Cu/Pb shield option for 1TGe additionally requires compressed air for lifting and moving around part of the shield; these needs are, however, temporary, and could be accommodated by bringing in compressed air in tanks.

Both experiments will require IT support (data line, phone, intercom). 1TGe requires a >100 Mbps data line, while EXO requires >10 Gbps to handle very high calibration-data rates. The 1TGe electroforming laboratory will also require IT support to implement slow controls for the process.

**Major Hazards**

The main hazards involved in the 1TGe operation are those associated with the use of large amounts of cryogen (oxygen depletion). Of most concern would be a failure of the vessel containing the LAr in the liquid cryogen option of 1TGe. Such a failure would result in a thermal coupling of the LAr to the large water tank, and would induce a rapid boil-off of the LAr. The GERDA experiment has developed a scheme to rapidly evacuate the water tank should such a failure occur.

Chemical hazards associated with 1TGe electroforming should not be a serious concern, as relatively small quantities would be involved in a spill. These hazards will be addressed as part of the scope of the MAJORANA DEMONSTRATOR.

Laser hazards associated with the EXO barium tagging can be handled with standard controls. High-voltage hazards will be similar to those encountered with very low-current, high-voltage power supplies employed in large drift chambers.

### 3.3.4.3 Assembly Study

A study was performed to examine aspects of material transport from the Surface to the 7400L LM for assembly and installation of EXO. Although specific to the EXO experiment because of its high level of complexity, the study's conclusions apply generically to any experiment of similar size at the deep level.

The route to the 7400L LM is from the Surface to the 4850L using the Yates Shaft, transporting the materials to the #6 Winze for lowering to the DLL. Table 3.3.4.3-1 shows the size and load capacities that were assumed at the time of this study (completed mid-2010) for both shaft cages; these values differ slightly from the baseline design (Yates width = 3.4 m and Yates length = 3.7 m) described in Volume 5,

*Facility Preliminary Design*. The #6 Winze is the limiting factor for size and payload for transport from the 4850L to the 7400L. Experimental hardware must be designed to be transported in planned load geometries and/or by weight to be assembled by bolting or—in the case of cryostats, water tanks, and structures—welding into the required subassemblies.

| Access | Width [m] | Height [m] | Length [m] | Max. Payload [kg] |
|---|---|---|---|---|
| Yates Super-cage | 3.2 | 3.5 | 3.8 | 18,100 |
| #6 Winze cage | 1.4 | 2.1 | 3.7 | 5,400 |

**Table 3.3.4.3-1** Lift cage size and load capacity.

The Trade Study also considered the use of a rehabilitated ramp system for transport from the 4850L to the 7400L. Rehabilitation of the ramp system was subsequently removed from the DUSEL Project baseline on the basis of cost considerations. Although the ramp system is no longer a possibility, the comparison with the use of the #6 Winze is documented here for reference.

The result of the study is a Work Breakdown Structure (WBS)-based list of planned load designs and quantities for each subassembly requiring in-module fabrication. The overall transport timeline can then be estimated when coordinated with winze schedules and transit times. A preliminary comparison of the required number of trips using the #6 Winze or the ramp system is shown in Table 3.3.4.3-2, below. For example, trips for the detector reduce from 294 to 125 by using the ramps. Assumptions used for this table are as follows:

1. Ramp transport allows planned loads of larger-size construction material pieces. This results in fewer trips, with the additional benefit of possible reduced fabrication time, for example, in weld assembly of subassemblies.
2. Ramp transport would also be used for larger surface-assembled subassemblies such as the TPC. This would allow fabrication and assembly of more critical hardware in a more controlled environment.
3. Ramp travel is likely to be the rougher of the two modes of transport so, if there is a choice, delicate items like PMTs would be transported in the #6 Winze.
4. Incidental fabrication items and possible day-to-day needs could be transported in the #6 Winze, which would be a faster “delivery” system.

| Component | Total Number of Trips | |
|---|---|---|
| EXO Detector | **#6 Winze only** | **Ramp only** |
| Shielding Assembly | 237 | 86 |
| Barium Tagging | 20 | 10 |
| Cryostat | 16 | 8 |
| TPC | 1 | 1 |
| PMTs | 20 | 20 |
| Total | 294 | 125 |
| Walkway | 20 | 10 |
| Support Building | 256 | 100 |
| **Total** | 570 | 235 |

**Table 3.3.4.3-2** Trips to the 7400L for #6 Winze or ramp.

At present, there are a considerable number of perceived challenges for installation of experiment hardware in the 7400L LM. Many of these may be mitigated during the evolution of the present EXO Conceptual Design to the Preliminary and Final Designs.

Installation of the experiment will require a well-thought-out process plan developed as an integral part of experiment equipment design, transport planning, and constrained-space fabrication. Transport and staging must include not only experiment hardware but also fabrication equipment and fixtures. The fabrication equipment and fixtures will have to be set up prior to start of fabrication or stored nearby for use as needed.

Module crane hook height may be close to EXO's containment water tank's 17-m height, possibly limiting over-the-wall installation of contained equipment. Consequently, fabrication of contained hardware may need to take place inside the partially built inner and outer water tank walls or fabricated and subassembled outside the outer wall and lifted into place and supported in position. The completion of the inner and outer water tanks would be a closing-up fabrication process. Figure 3.3.4.1.2.2-1 shows the detector baseline design.

## 3.3.4.4 Schedule

### 1TGe

The MAJORANA DEMONSTRATOR is expected to start taking data in 2013 and to run until at least 2015, possibly into 2017. The GERDA experiment started commissioning in 2010 and is expected to run into 2015. It is expected that plans for a Preliminary Design of 1TGe could start in 2015 and lead to a Final Design within three years. Procurement (particularly for the enriched Ge) could start before the Final Design is complete. Electroforming operations could start before beneficial occupancy at the 7400L and it is expected that construction could begin in line with beneficial occupancy at the 7400L. For a Cu/Pb-shield option, the modular design would allow for a subset of the array to begin operating within the first two years of construction. The full construction of the experiment is expected to take about three years; the entire array would then be operated for at least five years.

**EXO**

The EXO-200 experiment started commissioning in 2010 and is expected to run into 2015. The design of the "full" EXO apparatus is anticipated to require three years to complete, with construction activities beginning as early as one year before beneficial occupancy in DUSEL. Fabrication, procurement, assembly, and installation are expected to take 3.5 years after beneficial occupancy. Initial operation will follow a short commissioning period expected to last about one year after final construction. The detector will then be operated for a minimum of five years.

### 3.3.5 Neutrino Oscillations and Proton Decay

In its 2008 report, the Particle Physics Project Prioritization Panel (P5) recommended a world-class neutrino physics program as a core component of the U.S. particle physics program.[53] Included in this report is the long-term vision of a large detector in DUSEL, a high-intensity neutrino source at Fermi National Accelerator Laboratory (Fermilab) and associated neutrino beam from Fermilab to DUSEL.

On January 8, 2010, the Department of Energy (DOE) approved the "Mission Need" for a new Long Baseline Neutrino Experiment (LBNE) that would enable this world-class program. DOE has provided engineering and design funds for the LBNE project, and the NSF has awarded an S4 grant for the water Cherenkov detector R&D. An active scientific collaboration of ~250 physicists from more than 50 institutions has been formed to participate in the research. Together, DOE and the National Science Foundation (NSF) have indicated their intention to fund two 100 kT water Cherenkov "equivalent" detectors, and have encouraged foreign participation for a third. Here, "equivalent" is used because Monte Carlo (MC) estimates have indicated a LAr (LAr) detector has a higher efficiency for neutrino oscillation physics: A smaller-mass LAr detector may have the equivalent performance of a larger water Cherenkov detector.

The goal of the neutrino part of the program is to make precise measurements of neutrino oscillation parameters, search for CP violation in the neutrino sector, determine the neutrino mass hierarchy (normal or inverted), and measure the value of the third mixing angle—see Chapter 3.2.

Neutrino interaction cross sections are very small. Detectors must be very massive—tens of kT. Therefore, the reference configurations of the LBNE with a large neutrino detector located deep enough underground (or with an explicit, very efficient muon veto at shallower depths) to remove cosmogenic background have a natural synergy with research on proton decay as well as the detection of neutrinos from astrophysical sources.

Additionally, the large-mass detectors can detect supernovae from as far away as the Andromeda Galaxy, and according to current theories, may have the sensitivity to see the diffuse neutrino flux from relic supernovae above background.

#### 3.3.5.1 Measuring $\nu_e$ Appearance in LBNE[54]

Plots of the $\nu_\mu \rightarrow \nu_e$ probability versus energy for a 1,300-km baseline from Fermilab to DUSEL are shown in Figure 3.3.5.1. The four plots show the results for antineutrinos in the right column, and neutrinos in the left column. The figure shows an example of the rate of (unoscillated) $\nu_\mu$ charged current events (black histogram, left vertical scale) versus $\log_{10}$(energy[GeV]) from a wide band beam from Fermilab on a 100 kT water Cherenkov detector at DUSEL after 10 x $10^{21}$ protons on target (POT). Approximately 20,000 charged current events would be collected for this exposure, were there no oscillations. The colored curves show the probability of $\nu_e$ appearance events for $sin^2 2\theta_{13} = 0.04$, and three different values of $\delta_{CP}$, and one curve (light blue) where $sin^2 2\theta_{13} \equiv 0$. The rate of appearance events can be obtained by multiplying the charged current event rate by the appearance probability. This calculation does not account for detector efficiency due to event selection cuts. Comparing the colored curves for $\nu_e$ appearance, one observes that to differentiate the various curves for different values of $\delta_{CP}$ it is important that the incident neutrino beam cover a substantial portion of the oscillations between 0.5 and 5 GeV. At lower energies, the oscillation probability starts increasing due to the effect of the solar term; at the same time, the effect of matter is substantially decreased. It is experimentally very challenging to obtain a sufficient number of events below 1 GeV; however, that observation can put important constraints on all of the oscillation parameters, including the solar parameters.

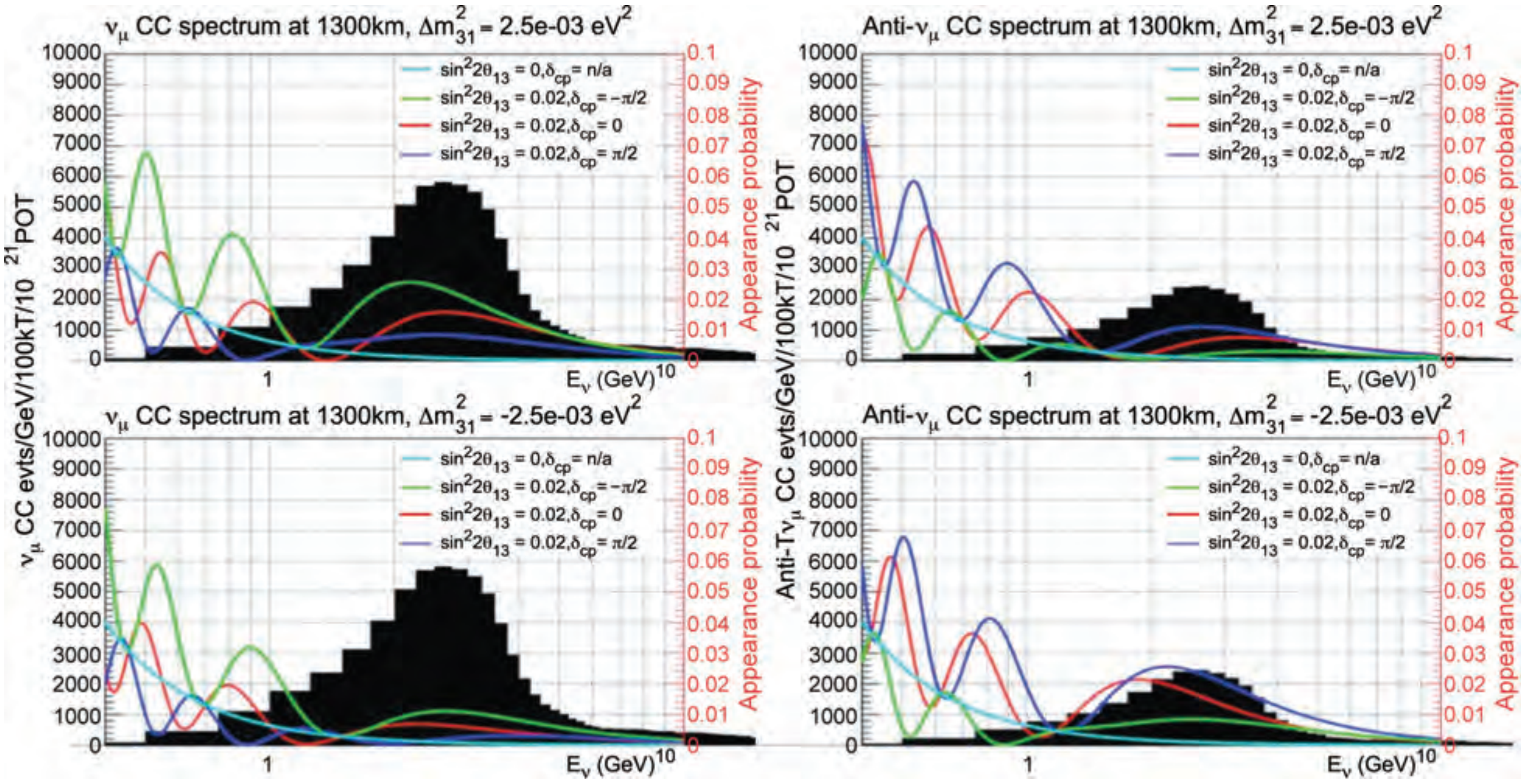

**Figure 3.3.5.1** This figure (top left) shows an example of the rate of (unoscillated) $\nu_\mu$ charged current events (black histogram, left vertical scale) versus $\log_{10}$(energy(GeV)) from a wide band beam from Fermilab on a 100 kT water Cherenkov detector at DUSEL after 10 x $10^{21}$ POT, normal hierarchy. Approximately 20,000 charged current events would be collected for this exposure without oscillations. Plotted in color is the probability of $\nu_e$ events from oscillations (right vertical scale), assuming $\sin^2 2\theta_{13} = 0.04$, and one line (light blue) for $\sin^2 2\theta_{13} = 0$. The top right figure is the same as the left, except for antineutrinos. The two bottom figures are the same as the corresponding pictures in the top row but representing results for the inverted mass hierarchy. [Courtesy LBNE project]

### 3.3.5.2 Sensitivity Reach in LBNE

#### 3.3.5.2.1 $\theta_{13}$, Mass Hierarchy and $\delta_{CP}$

This section describes the sensitivity for measuring $\theta_{13}$, determining the mass hierarchy and measuring the CP phase for LBNE. Figure 3.3.5.2.1-1 shows the measurement ellipses and sensitivity limits for neutrino oscillation parameters. These limits are calculated for a total exposure of 60 x $10^{20}$ POT for neutrino running and equivalent antineutrino running. Because the live time of the experiment is ~100 sec per year (10 μs pulse × ~$10^7$ pulses/year), the background from cosmic ray events is negligible. At the 1000 mwe depth, the number of cosmic ray events in a 100 kT water Cherenkov detector is about equal to the number of charged current events (see Table 3.3.5.2.1-1) and can be eliminated from topological considerations. At the 4850L (4290 mwe), the number of cosmic ray events is negligible. LAr detectors, because of their higher granularity, could operate closer to the surface for this physics measurement; however, full simulations are needed to determine the minimum acceptable depth.

| Rate(Hz) | In-time cosmic/yr | Depth (mwe) |
|---|---|---|
| 500 kHz | $5 \times 10^7$ | 0 |
| 3 kHz | 300,000 | 265 |
| 400 Hz | 40,000 | 880 |
| 5 Hz | 500 | 2300 |
| 1.3 Hz | 130 | 2960 |
| 0.60 Hz | 60 | 3490 |
| 0.26 Hz | 26 | 3620 |
| 0.09 Hz | 9 | 4290 |

**Table 3.3.5.2.1-1** The rate[55] of cosmic ray muons in a 50-m height/diameter detector assuming a $\cos^2\theta$ distribution (there will be a small correction at the deepest levels). The second column is the number of cosmic rays in 10-microsecond-long pulses for $10^7$ pulses, corresponding to approximately one year of running, versus depth in mwe. The 4850L is equivalent to 4290 mwe.

Figure 3.3.5.2.1-1(a) shows the one- and two-sigma measurement ellipses for $\delta_{CP}$ and $sin^2 2\theta_{13}$, as compared to the MC input value (“×”). Note that the error ellipses are roughly the same size independent of the value of $sin^2 2\theta_{13}$. This is because as the value of $sin^2 2\theta_{13}$ increases, the number of events increases, but the asymmetry decreases. These effects roughly cancel, so the sizes of the error ellipses do not change as a function of $sin^2 2\theta_{13}$.

Figure 3.3.5.2.1-1(b) shows the three- and five-sigma sensitivity limits for $sin^2 2\theta_{13} \neq 0$. This plot is made by calculating the number of events seen at the detector for each value of $sin^2 2\theta_{13}$ and $\delta_{CP}$, and comparing this number to the result where $sin^2 2\theta_{13}=0$, for all $\delta_{CP}$. One set of scatter plots is made for each of the two mass hierarchies, and the three- and five-sigma exclusion limits are drawn. The three- or five-sigma limit is taken as the minimal value of $sin^2 2\theta_{13}$ that excludes both mass hierarchies at the desired level. Figure 3.3.5.2.1-1(c) is for the three- and five-sigma exclusion of the mass hierarchy, and is calculated in an analogous way to (b). Finally, (d) shows the exclusion plots for $\delta_{CP}$ as a function of $sin^2 2\theta_{13}$. The two-lobed structure results, as there is no CP violation for $\delta_{CP}$ equal to zero or π. For comparative purposes, the sensitivity is usually quoted as the minimum value of $sin^2 2\theta_{13}$ that excludes 50% of the $\delta_{CP}$ axis ($0 \leq \delta_{CP} \leq 2\pi$). In the example shown, this occurs at $sin^2 2\theta_{13} = 0.01$. Figure 3.3.5.2.1-2 shows the same limits for a 50-kT LAr detector.

Table 3.3.5.2.1-2 shows a comparison of the sensitivity of the experiment for several different exposures and configurations. The results shown in Figures 3.3.5.2.1-1 and -2 correspond to the last two lines of Table 3.3.5.2.1-2. As can be seen by comparing these lines, for the purposes of measuring neutrino oscillation parameters, a 50 kT LAr detector is roughly equivalent to a 300 kT water Cherenkov detector.

Finally, Figure 3.3.5.2.1-3 shows the sensitivity of water Cherenkov and LAr detectors as a function of exposure, compared with two proposals from the past: a large LAr detector (100 kT) at Ash River in the United States (with an upgraded NuMI beam from Fermilab), and the T2KK proposal, with large water

Cherenkov detectors at Kamioka and in Korea. At this time, there are no proposals for neutrino oscillation experiments at nearly the same advanced stage of sophistication as LBNE.

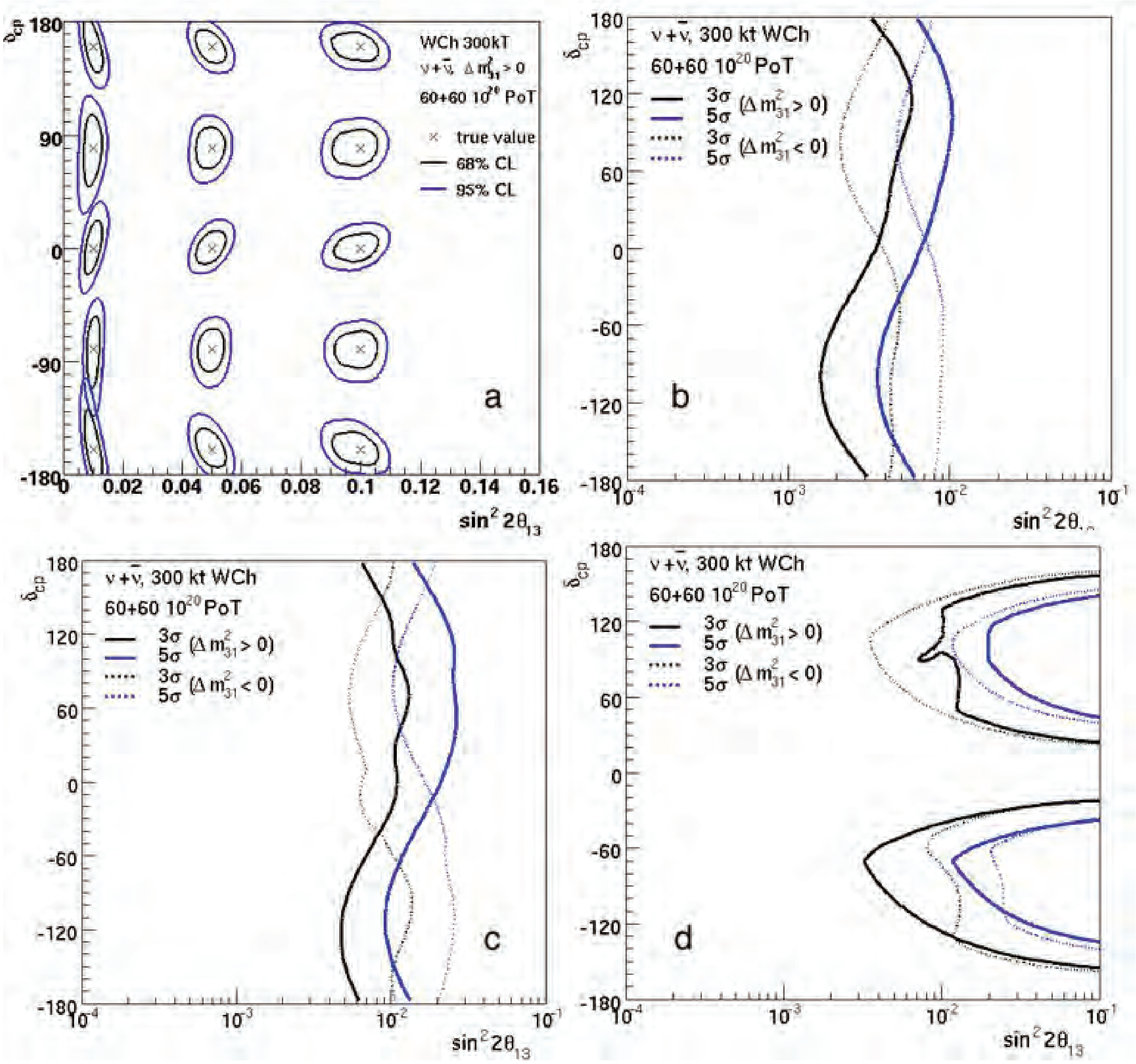

**Figure 3.3.5.2.1-1** Discovery and sensitivity limits for neutrino oscillation parameters. a) One- and two-sigma measurements limits for $\delta_{CP}$ and $sin^2 2\theta_{13}$ for a water Cherenkov detector. The “x” indicates the MC input value, and the one- (black) and two- (blue) sigma error ellipses are shown. b) The three- and five-sigma sensitivity limits for $sin^2 2\theta_{13} \neq 0$. The three-sigma limits for both mass hierarchies are indicated. The solid set of lines is for the normal hierarchy, and the dotted set is for the inverted hierarchy. The three-sigma limit for either solution is indicated. c) Three- and five-sigma limits for measuring the mass hierarchy. The solid and dotted lines have the same meaning as in b). d) The three- and five-sigma exclusion limits for $\delta_{CP}$ as a function of $sin^2 2\theta_{13}$ for both mass hierarchies. The line indicates where 50% of the $\delta_{CP}$ axis is excluded. The exposure is 60 ($\nu_\mu$) + 60 ( $\overline{\nu_\mu}$) x $10^{20}$ POT for a 300 kT water Cherenkov detector. The limits on $\delta_{CP}$ for a LAr detector are similar—see Figure 3.3.5.2.1-2. [Courtesy LBNE project]

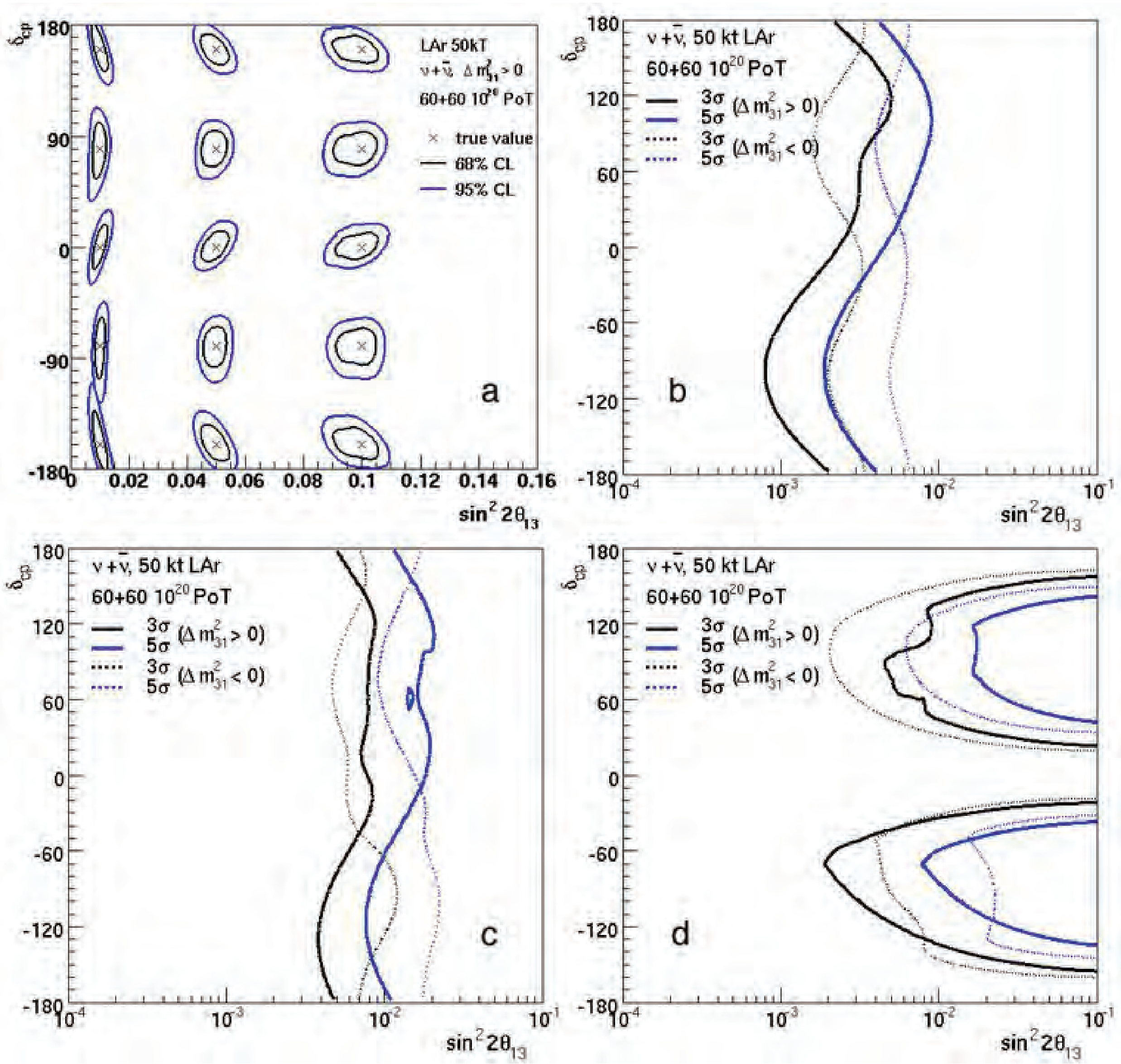

**Figure 3.3.5.2.1-2** Sensitivity plots for a 50 kT LAr detector. The plots a-d, for a LAr detector, are constructed in the same fashion as in Figure 3.3.5.2.1-1. [Courtesy LBNE project]

| Detector Size (kT) | POT ($x10^{20}$) @120 GeV (1 MW = $10^{21}$ POT/yr) | Years $\nu$ $\bar{\nu}$ | 3 σ Sensitivity, Minimum value of $sin^2 2\theta_{13}$ | | |
|---|---|---|---|---|---|
| | | | $\sin^2 2\theta_{13} \neq 0$ | Mass Hierarchy | CPV (50% of $\delta_{CP}$ coverage) |
| H20 100 (Bishai) | 30+30 | 3+3 | 0.014 | 0.031 | >0.1 |
| H20 300 (Bishai) | 30+30 | 3+3 | 0.008 | 0.017 | 0.025 |
| H20 600 (Bishai) | 30+30 | 3+3 | 0.005 | 0.012 | 0.012 |
| H20 300 (Bishai) | 60+60 | 3+3 | 0.005 | 0.012 | 0.012 |
| LAr 50 (Dierckxsens) | 60+60 | 3+3 | 0.005 | 0.011 | 0.010 |

**Table 3.3.5.2.1-2** Examples of the sensitivity to neutrino oscillation parameters for a variety of detector and beam configurations. The results shown in Figure 3.3.5.2.1-1 and -2 correspond to the last two lines of this table. As can be seen by comparing the last two lines, a 300 kT water Cherenkov detector has approximately the same sensitivity as a 50 kT LAr detector for these measurements. [Courtesy LBNE project]

| Item | 100 kT LAr | WWB LAr | WWB WCh | T2KK WCh |
|---|---|---|---|---|
| POT/yr $x10^{20}$ ( $\nu$ ) | 10 | 22.5 | 22.5 | 52 |
| POT/yr $x10^{20}$ ( $\bar{\nu}$ ) | 10 | 45 | 45 | 52 |
| Yrs $\nu$ + $\bar{\nu}$ | 3+3 | 5+5 | 5+5 | 5+5 |
| Power (MW) | 1.13 | 1 ( $\nu$ ) 2 ( $\bar{\nu}$ ) | 1 ( $\nu$ ) 2 ( $\bar{\nu}$ ) | 4 |
| Baseline (km) | 810 (Ash River) | 1290 | 1290 | 295 + 1050 |
| Mass (kT) | 100 | 100 | 300 | 270 + 270 |
| Duty cycle | 0.54 | 0.54 | 0.54 | .32 |
| Exposure: (Mt-MW- $10^7$ s) | 1.15 | 2.55 | 7.65 | 17.85 |

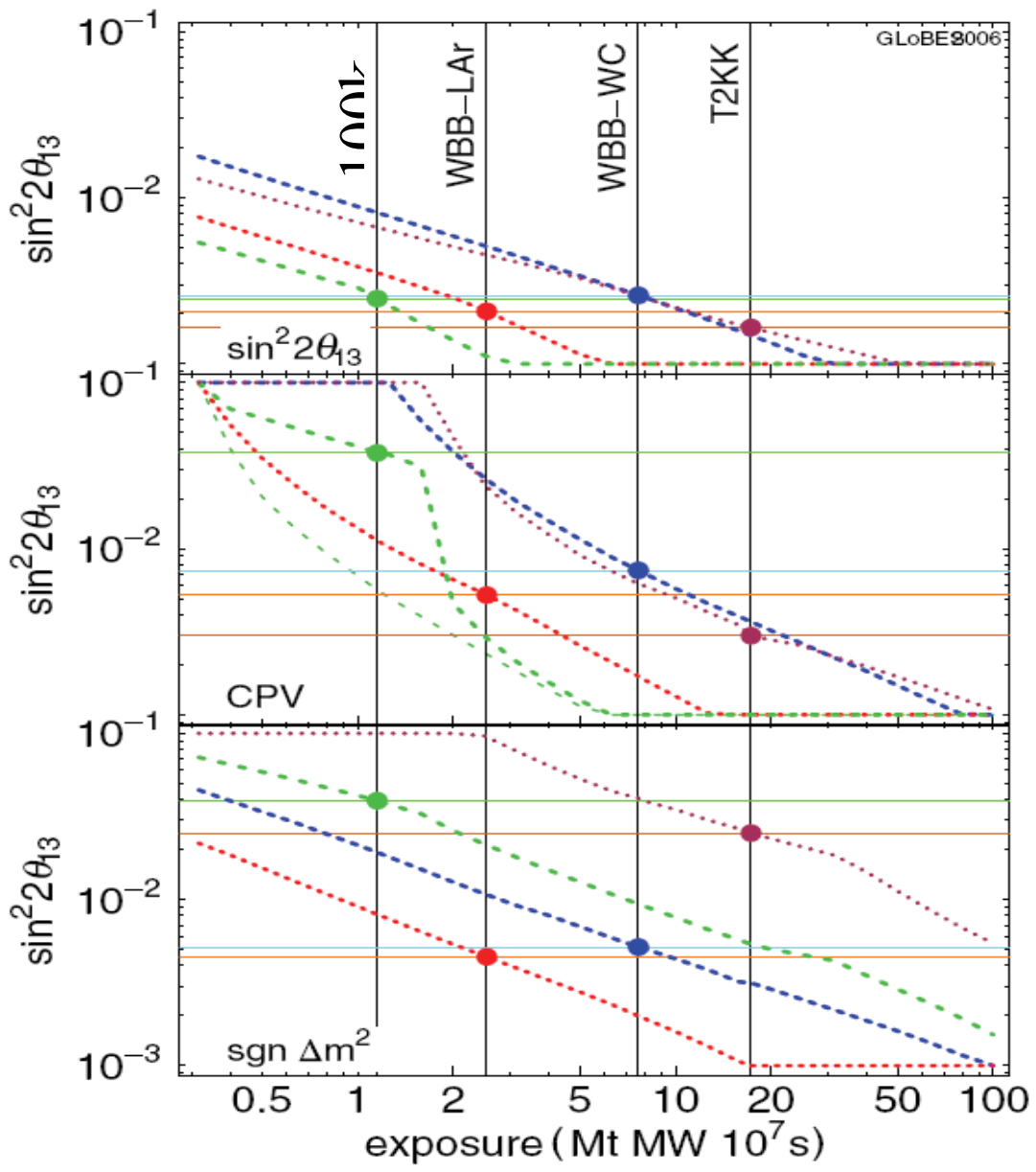

**Figure 3.3.5.2.1-3** Sensitivity[56] for various detectors as a function of exposure, measured in terms of MT of target mass, MW of beam power, and $10^7$ seconds of live time. The parameters of the four proposals shown are listed in the table on the left, and their 3σ sensitivity is plotted on the right. The three sets of curves are for (top to bottom) $\sin^2 2\theta_{13} \neq 0$, $\delta_{CP}$, and the sign of the mass hierarchy. [Courtesy LBNE project]

### 3.3.5.3 Non-accelerator Physics and LBNE

The following sections discuss the capability of the proposed LBNE detectors to extend the search for proton decay and detection of other than LBNE neutrinos

#### 3.3.5.3.1 Detector Performance

Table 3.3.5.3.1 displays the efficiencies for two modes of proton decay detection for the water Cherenkov and LAr detectors and the estimated background for each detector and decay channel. The efficiency for the $p \rightarrow e^{+}\pi^{0}$ channel is equal in the two detector technologies and is dominated by $\pi^{0}$ absorption on the nucleus. The background for the water Cherenkov detector is estimated from the Super-K exposure.

In $p \rightarrow \nu K^{+}$ the efficiencies of the two technologies are quite different. In the water Cherenkov detector the charged kaon, being below Cherenkov threshold in water, is invisible. This mode is predominantly

| | Water Cherenkov | | Liquid Argon TPC | |
|---|---|---|---|---|
| | Efficiency | Background | Efficiency | Background |
| $p \rightarrow \pi^{0} e^{+}$ | 45% | 0.2 | 45% | 0.1 |
| $p \rightarrow \nu K^{+}$ | 14% | 0.6 | 97% | 0.1 |

**Table 3.3.5.3.1** Summary of efficiency and background for the two decay modes for water Cherenkov detector and LAr. Background is quoted in terms of events per 100 kT/yr. The background for water Cherenkov is evaluated from Super-K experience and data for depth similar to or greater than Super-K (~2300 mwe). For LAr, the background is evaluated for ~300-800 mwe with large uncertainties. [Courtesy LBNE project]

detected via the coincidence of a gamma cascade from the remnant $^{15}N$ nucleus, and the detection of the $\mu^{+}$ and $e^{+}$ daughters from the $K$ decay. In the LAr detector, the kaon and its entire decay chain are visible, and the efficiency is estimated to be significantly higher, but background estimates are less certain because of the lack of experimental experience with a LAr detector, and the lack of a full simulation of a veto for cosmic background at shallow levels (300L-800L).

#### 3.3.5.3.2 Sensitivity versus Time

To determine the sensitivity as a function of time, for a given detector mass (or evolution of mass) the needed inputs are: 1) signal detection efficiency, 2) background rate, 3) exposure starting date, 4) detector mass as a function of time, and 5) live-time assumptions. For a third-generation search such as LBNE, one needs to compare the potential reach to the integrated exposure that could be achieved by Super-K on the same time scale. Assuming no candidate events are found, by 2020 the Super-K exposure could be approaching 0.5 Mtonne-years, resulting in a limit on the proton lifetime of $\geq 2 \times 10^{34}$ years. Figure 3.3.5.3.2 shows the evolution of sensitivity, beginning in 2020, for several possible LBNE far detector configurations at DUSEL.

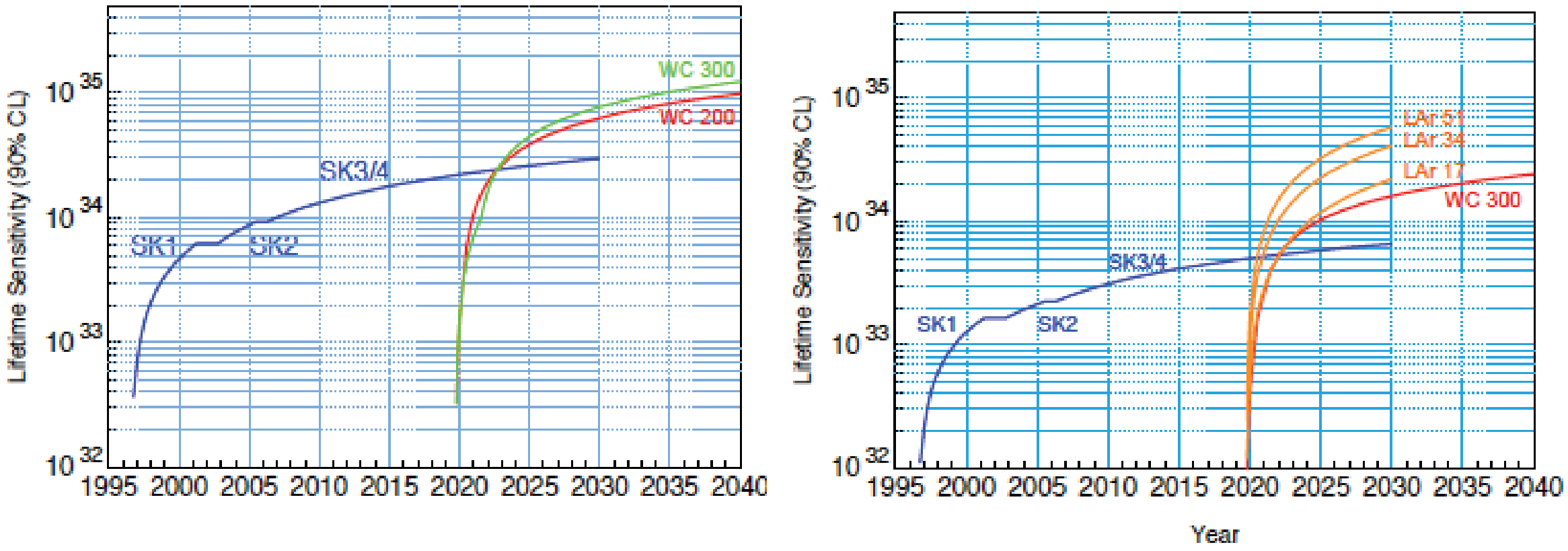

**Figure 3.3.5.3.2** Proton lifetime sensitivity for two different decay modes. (Left) Proton lifetime sensitivity for the $p \rightarrow e^{+} + \pi^{0}$ mode for Super-K (SK1-4), a 200 (WC200) and 300 (WC300) kT water Cherenkov detector. For this decay channel, a LAr detector must have the same mass as a water Cherenkov detector for the same sensitivity. (Right) Proton lifetime sensitivity for the decay $p \rightarrow \bar{\nu} K^{+}$ for the Super-K experiment, and a water Cherenkov detector of 300 kT or LAr detectors from 17 kT to 51 kT fiducial mass. [Courtesy LBNE project]

### 3.3.5.3.3 Conclusion—Proton Decay

There are two potential "game changers" in the search for proton decay. The first is a discovery of supersymmetry (SUSY) particles at the Large Hadron Collider (LHC). This would provide extremely strong motivation to search for the modes involving kaons, in particular $\bar{\nu}K^{+}$. Second would be the emergence of candidate events in Super-K. This would clearly motivate a confirmation in a larger detector. With a 300 kT water Cherenkov detector, the limits on the $p \rightarrow e^{+} + \pi^{0}$ increase by an order of magnitude. For the $p \rightarrow \bar{\nu} K^{+}$ decay, a 300 kT water Cherenkov detector improves the Super-K limits by a factor of two to three over 10 years, and a 50 kT LAr detector by a factor of eight based on the assumption that it would have higher kaon efficiency. The addition of gadolinium to water could allow tagging of backgrounds that might appear with long-term running.

### 3.3.5.3.4 Galactic Supernova Bursts

A supernova burst in the Milky Way galaxy[57] would produce a huge signal in an LBNE water Cherenkov detector, as shown in Table 3.3.5.3.4. Further, a 300 kT water Cherenkov detector is sensitive to supernovae in the Andromeda Galaxy, with about 10 events in a 30-second interval.

| Event type | Expected number of Events |
|---|---|
| Charged Current $\bar{\nu}_e$ | 60,000 |
| Neutral Current $\nu_x$ | 3,000 |
| Elastic Scattering $\nu_e$ | 3,000 |

**Table 3.3.5.3.4** Rate of observed neutrino events in a 300 kT water Cherenkov detector for a supernova 10 kpc distant. [Courtesy LBNE]

Note that the elastic scattering events indicate the direction of the supernova and allow comparison of the $\nu_e$ and $\bar{\nu}_e$ flux. Both inverse beta decay and elastic scattering cross sections are known to better than 1% in this energy range on water (hydrogen), and the observed positron energy for inverse beta decay is a near-exact mirror of the spectrum of the parent neutrino flux. For elastic scattering, the outgoing electron energy is a simple convolution of the parent $\nu_e$ spectrum. Thus, flavor-resolved spectra can be extracted.

### 3.3.5.3.5 Diffuse Neutrinos from Relic Supernova

Though supernova relic neutrino (SRN) models vary, according to one widely accepted modern analysis,[57] a 300 kT water Cherenkov detector located deep underground would permit a sensitivity limit of about 0.02 $cm^{-2}$ $sec^{-1}$ (see Figure 3.3.5.3.5). The addition of Gd to the water would allow LBNE to tag the inverse beta decay events using the final state neutrons. This tagging will lower background from atmospheric neutrinos and other sources. As can be seen by the figure, the sensitivity limit of the water Cherenkov detector exceeds all theoretical predictions, and should be able to distinguish between several of them. These data would undoubtedly stimulate new theoretical (and perhaps even experimental) developments in the neutrino and cosmology communities.

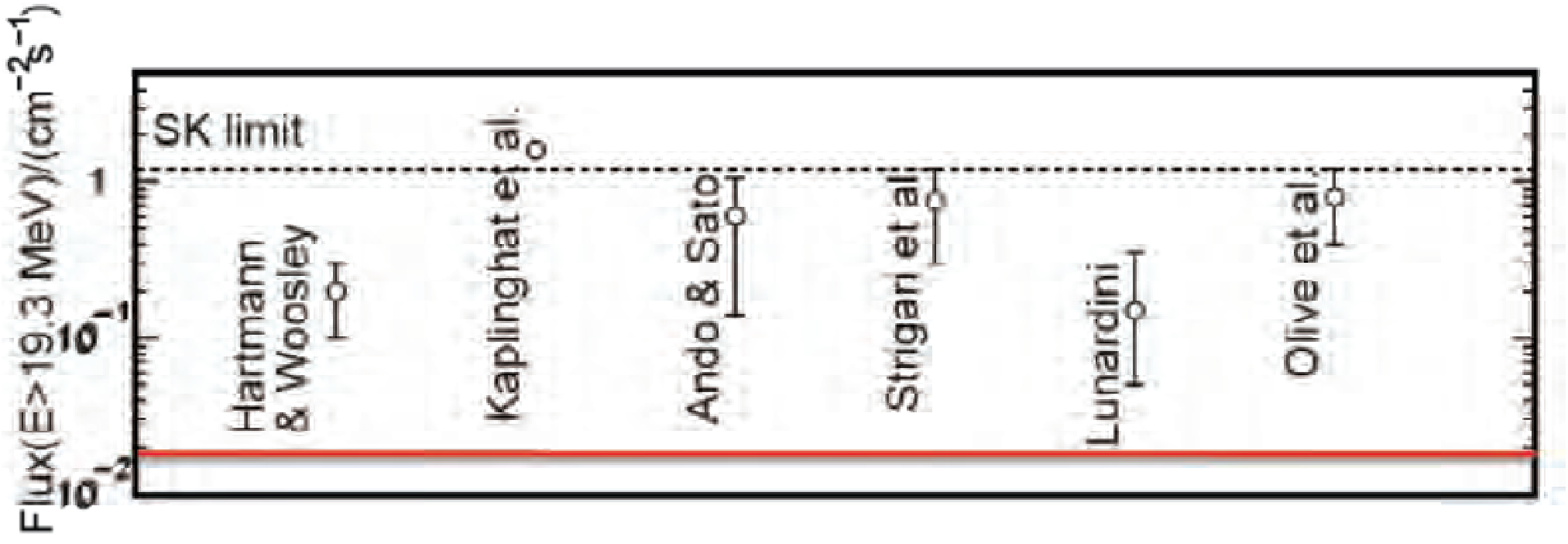

**Figure 3.3.5.3.5** Comparison of the Super-K limit[58] with theoretical estimates[57] for the diffuse flux of neutrinos from relic supernovae. The solid (red) line at the bottom represents the expected sensitivity of a 300 kT water Cherenkov detector with a threshold of 15.5 MeV, compared with the Super-K threshold of 19 MeV. [Courtesy LBNE project]

### 3.3.5.3.6 Other Neutrino Science

As an example of future studies of solar and/or atmospheric neutrinos, consider the predicted day-night asymmetry of neutrinos from the sun. These neutrinos pass through the dense core of the Earth at night, and the difference between the forward scattering amplitude of $\nu_e$s and the other flavors leads to a flavor transformation similar to that which occurs within the solar interior. As the beam from the sun arrives at the Earth, it is nearly a pure $\nu_2$ state and therefore its flavor content is only ⅓ $\nu_e$. The flavor transformation within the Earth thus leads to a net gain in $\nu_e$ content—the sun "shines brighter" in $\nu_e$s at night than during the day. A measurement of the day-night asymmetry can take several forms. At its simplest, an integral asymmetry measurement can be made:

$$A = \frac{2(\phi_{\nu_e}^{night} - \phi_{\nu_e}^{day})}{(\phi_{\nu_e}^{night} + \phi_{\nu_e}^{day})}$$

Currently, the measurements by the Super-Kamiokande and SNO collaborations have found $A = 0.021 \pm 0.02 +^{0.013}_{-0.012}$ [15] and $A = -0.037 \pm 0.063 \pm 0.032$,[59] respectively, each within 1σ of $A = 0$ when both statistical and systematic error estimates are included. For a 300 kT water Cherenkov detector, the event rate in the detector is roughly 130/day, and consequently the statistical precision on this asymmetry after a year should be significant: ~0.005, depending on the achievable analysis energy threshold. For the current best-fit large mixing angle (LMA) parameters, the integral asymmetry is expected to be near 0.02. More sophisticated analyses, involving fits to the energy and zenith-angle dependent survival probabilities, have already provided noticeably better measurements of the asymmetries in both Super-K and SNO, and could be applied to a larger detector as well. The day-night asymmetry is expected to manifest itself at very low energies (<5 MeV in scattered electron energy). This measurement might require additional photosensors to be installed.

Generally, the study with atmospheric neutrinos will not be as sensitive as that of the LBNE, but atmospheric neutrinos will allow the possibility of revealing different physics. This is because the atmospheric neutrino sample covers five orders of magnitude in neutrino energy and three orders of magnitude in baseline, including long paths through matter. The atmospheric neutrino flux is a mixture of muon and electron neutrinos and antineutrinos. One expects 14,000 atmospheric neutrino interactions per 100 kT of detector mass per year. The majority of the atmospheric neutrino events occur at neutrino energies below 1 GeV, where both water Cherenkov and LAr perform well. A LAr detector may have significant capability to identify neutrino versus antineutrino by observing the recoil proton present in charged current neutrino scattering. This large sample of neutrino interactions allows for a comparison of the neutrino oscillation framework under different conditions than those presented by the long-baseline neutrino beam experiment.

### 3.3.5.4 Detector Depth Requirements

In 2008, the LBNE science collaboration made a detailed study of the depth requirements for the main physics topics of interest with large detectors—*The Depth Document*.[55] The topics considered were accelerator-generated neutrinos, supernovae, solar and atmospheric neutrinos, and nucleon decay. The requirement on the depth of the detector is guided by the rate of the desired signals and the rate of backgrounds from cosmic rays over a very wide range of energies, from solar neutrino energies of 5 MeV to high energies in the range of hundreds of GeV. Table 3.3.5.4 shows the overburden required for different physics processes for both technology options. Since this study was carried out, placement of the LAr detector at a relatively shallow depth such as 800 feet overburden has become a serious option, and inclusion of a veto around the detector may be successful to reject background at the shallower depth for proton decay and other non-accelerator physics.

| Physics | Water | Argon |
|---|---:|---:|
| Long-Baseline Accelerator | 1,000 | 0-1,000 |
| $p \to K^+ \nu$ | >3,000 | >3,000 |
| Day/Night $^8$B Solar $\nu$ | ~4,300 | ~4,300 |
| Supernova Burst | 3,500 | 3,500 |
| Relic Supernova | 4,300 | >2,500 |
| Atmospheric $\nu$ | 2,400 | 2,400 |

**Table 3.3.5.4** Depth requirements in mwe for different measurements and the two detector technologies being considered.

### 3.3.5.5 The LBNE Project

The LBNE project will include an intense neutrino source pointing toward a distant large detector and a much smaller detector located close to the source. The far detector must be a long distance (>1,000 km) from the neutrino source to increase sensitivity to neutrino oscillation parameters. A nearby detector close to the neutrino source is necessary to measure the initial composition of the beam.

LBNE's target scope is to build a neutrino facility that uses a proton beam to produce a beam of neutrinos directed toward near and far detectors.

The preferred alternative for the neutrino source and near detector site is Fermilab, as it has already developed the expertise for construction of neutrino beams as part of the Neutrinos at the Main Injector/Main Injector Neutrino Oscillation Search (NuMI/MINOS) Project. The 700 kW upgrade of the Fermilab proton source, a component of the current NOνA[60] Project, offers a platform from which to launch a new neutrino beam for a long-baseline detector. DUSEL is located at an optimum distance of 1,290 km from Fermilab to detect neutrino oscillations. The LBNE science collaboration currently consists of more than 250 scientists and engineers from 54 institutions that have come together to carry out an experiment using the facilities currently being designed for the LBNE project at Fermilab and DUSEL. The collaborators come from universities and national laboratories, both from the United States and around the world. The collaboration encourages and anticipates further international participation.

The collaboration is in the process of putting together a science report to evaluate the scientific sensitivities and costs for potential variants of the experiment—what type of beam and detectors to use. The collaboration's science report will provide input to the LBNE project's Conceptual Design.

#### 3.3.5.5.1 The LBNE Beamline

The components of the LBNE neutrino beamline are designed to take a proton beam extracted from the Fermilab main injector and transport it to a target area, in which a neutrino beam is generated and aimed toward the far detectors. The neutrino beam will have sufficient intensity and a specific energy spectrum to meet the physics goals of the LBNE oscillation experiment.

The primary proton beam is extracted from the main injector at the same location where the beam is extracted for the presently active NuMI beam to MINOS. A short distance from the main injector extraction enclosure, the LBNE primary proton beam will be directed along a trajectory pointed west toward the DUSEL site.

The LBNE primary beam uses only conventional magnets with an optics design based on the Fermilab main injector. The magnets are designed to transport the beam to the target with very low losses and an energy range of 60 to 120 GeV. Although the NuMI (and NOνA) beam operates at 120 GeV, the lower energy of 60 GeV may be preferred in some scenarios, depending on the level of background processes seen at the far detectors. To reach the far detectors, the generated neutrino beam must be aimed downward into the Earth at an angle of approximately 5.6 degrees, or 10% slope relative to the surface. Figures 3.3.5.5.1-1 and 3.3.5.5.1-2 show a plan view and a cross-sectional view of the beamline, respectively.

The primary proton beam is directed at an actively cooled target, whose interaction products are focused by a set of two horns. The focus provided by the horns maximizes the number of pions that can emit a neutrino in the direction of the far detectors. The pion decay volume in LBNE is a circular cross-section pipe, 4 m in diameter and 250 m long, with its axis pointing toward the far detectors. The preferred design has this pipe filled with air, although an alternative design uses a helium-filled pipe.

Non-interacting protons (~15%) and the non-decayed pions (or kaons) are absorbed in a specially designed aluminum and steel pile protecting the rock from beam-activated nuclides. The absorber occupies an excavated enclosure at the end of the decay pipe.

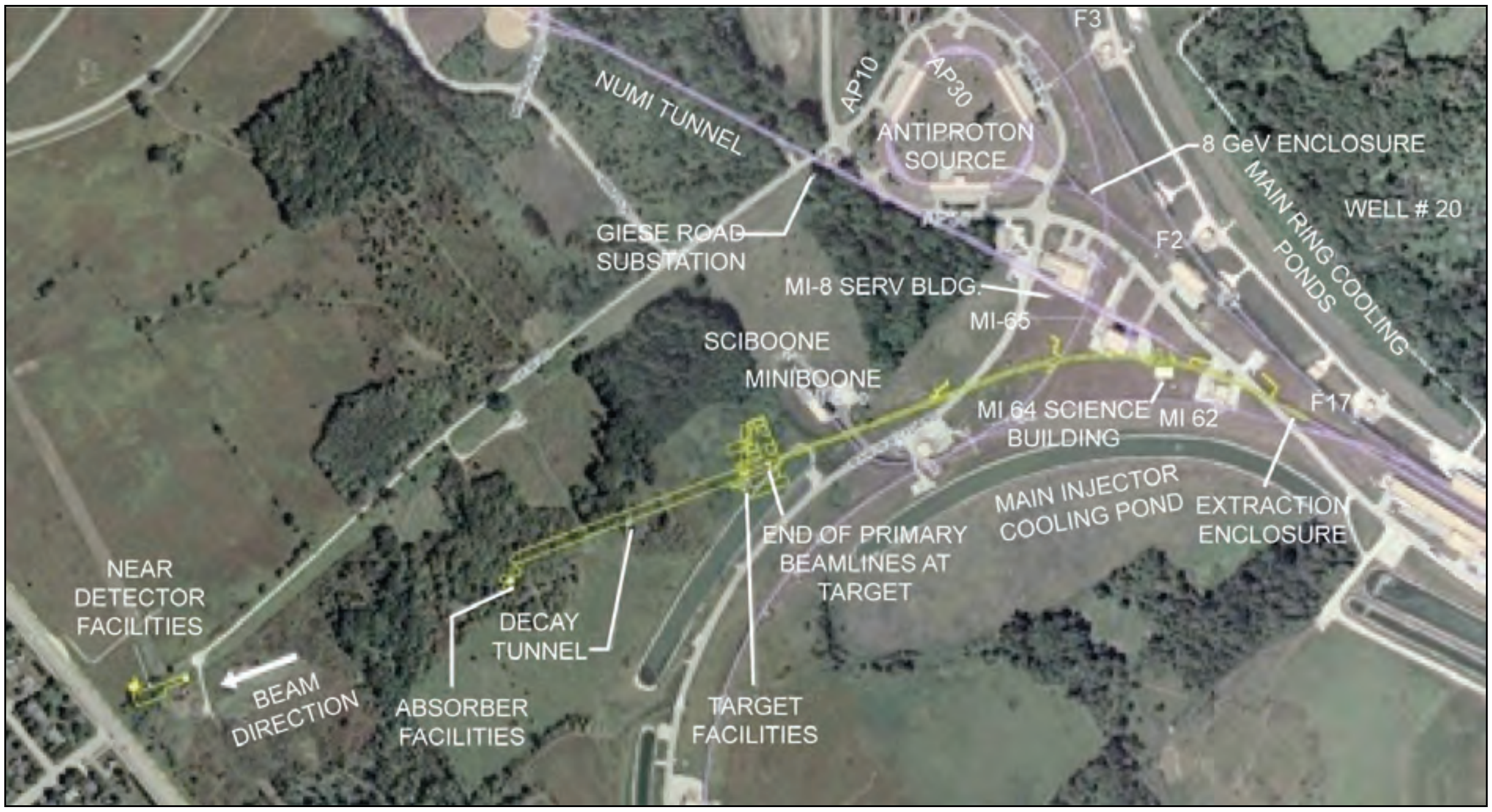

**Figure 3.3.5.5.1-1** Aerial view of the Fermilab site showing the LBNE beamline (yellow) from the main injector to a target hall near the center of the figure. This is followed by a 4-m-diameter decay pipe pointed at DUSEL. The near detector is at the lower left side of the figure close to the Fermilab west property line. [Courtesy LBNE project]

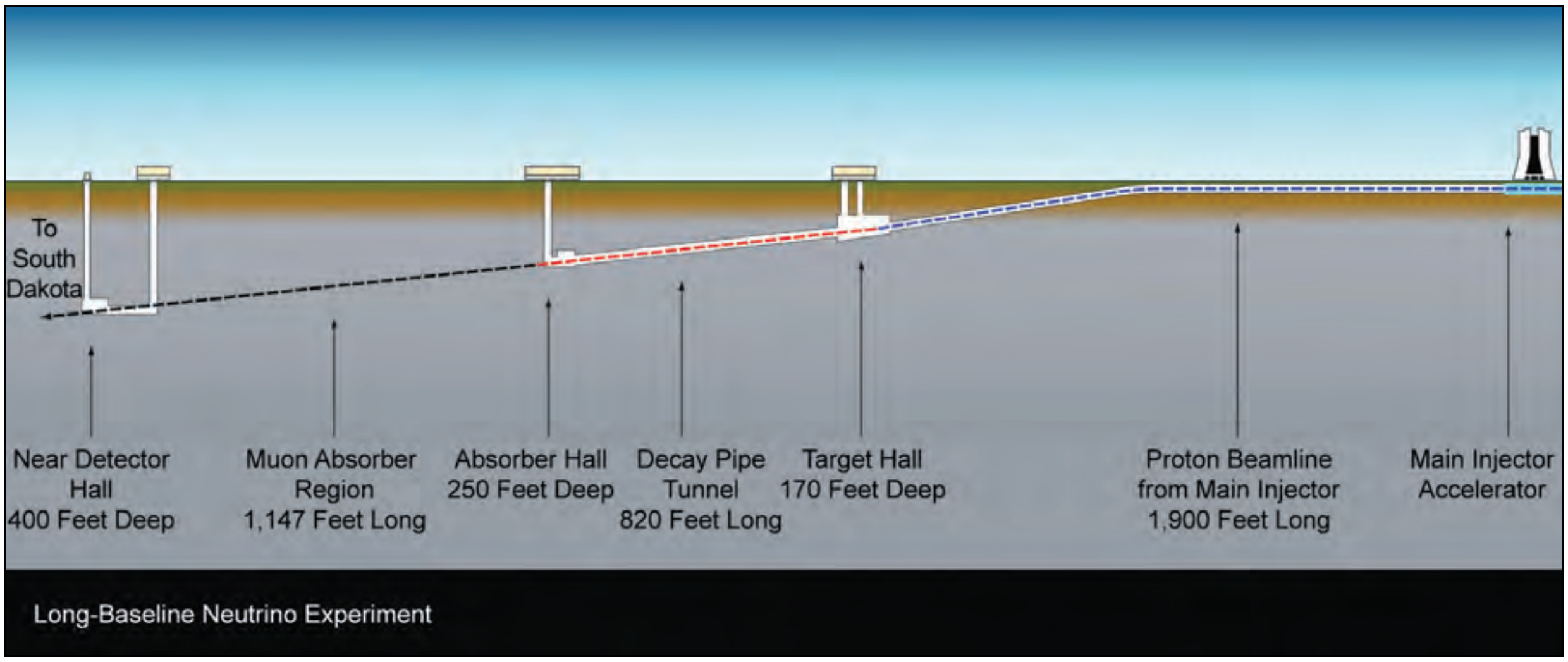

**Figure 3.3.5.5.1-2** Vertical cut through the Fermilab site showing (right to left) the downward slope of the beamline, the target hall, decay pipe, and near detector complex. [Courtesy LBNE project]

### 3.3.5.5.2 The LBNE Near Detectors

The purpose of the near detector complex is to make measurements that are needed or useful for LBNE's long-baseline oscillation physics analysis. Because the neutrino flux at the near detectors is greater than the flux at the far detectors, the near detectors can be much smaller than the far detectors and still be much more precise in their measurements.

The near detector complex has two primary goals: to measure the neutrino flux coming from the target before the neutrinos oscillate, and to measure the rate of background processes that might contaminate electron neutrino oscillations. The flux of muon neutrinos, electron neutrinos, and antineutrinos must be measured with very high precision. Whatever target material is chosen for the far detectors—hydrogen and oxygen (a water Cherenkov detector) or argon (LAr detector)—must be included in the near detectors so that accurate measurements of the fluxes and interactions with the target nuclei can be made.

Much is still unknown about how exactly the near detectors will be designed, and many parameters for the detectors must still be optimized. For DOE Critical Decison-1 (CD-1), a cost and schedule range will be generated for several options. A fine-grained tracker is necessary to study backgrounds in detail. Designs being considered for that tracker are a scintillating tracker, like Main INjector ExpeRiment for v-A (MINERvA),[61] and a straw-tube tracker with transition radiation detectors. A Liquid Argon Time Projection Chamber (LArTPC) will be a necessary component of the near detectors if that technology is chosen for the far detectors. There are two options for an LArTPC for the near detectors: using Micro-Booster Neutrino Experiment (MicroBooNE)[62] or building a smaller magnetized LArTPC. To measure the flux, Michel decay detectors could be used, with the option of placing them in the alcoves or in the absorber. A threshold Cherenkov counter is also necessary for measuring the post-absorber muon flux—both its absolute rate and energy spectrum.

### 3.3.5.5.3 The Far Detector Technology

Three alternatives are being considered for the far detector configuration: two modules of water Cherenkov, two modules of LAr, or one of each. Figure 3.3.5.5.3 illustrates events from both technologies.

Some virtues of water Cherenkov as a technology for massive detectors are the relatively low cost, relative simplicity of design, ease of operation, and extensive operations experience. The active target

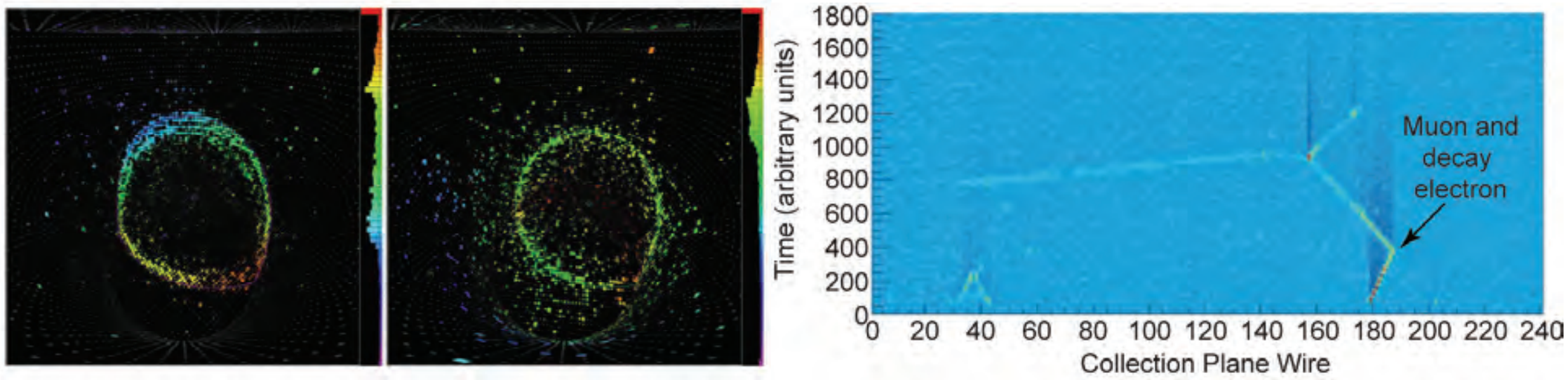

**Figure 3.3.5.5.3** (Left) Response of a water Cherenkov detector to a muon track. Each dot represents a PMT, and the dot size indicates the number of photoelectrons per PMT. The colors represent the signal's time of arrival, with a maximum time separation of ~150 ns. (Center) A Cherenkov ring from an electron passing through a water Cherenkov detector. The electron ring is "fuzzier" than the muon ring because of scattering of the electron as it passes through the water. [Courtesy of Super-K Collaboration] (Right) Muon decay in the ArgoNeuT LAr detector. [Courtesy LBNE project]

medium is water, a very abundant, very cheap, easy-to-handle source for the target material with which to build the massive detectors. The wall of the water container is instrumented with photomultiplier tubes (PMTs) whose signals are read out with well-understood electronics, which includes charge to digital converters and time to digital converters. The PMT readouts are then used to analyze the arrival time and the number of photons produced by the Cherenkov radiation of charged-particle tracks in the water and detected by the PMTs to reconstruct vertex, direction, and energy of the track.

LArTPC is a newer technology under development for future beam-based neutrino research. This technology promises precise event reconstruction and particle identification, as well as potential scalability to large detectors. Preliminary simulations have indicated that LAr detectors perform with higher efficiency and better background rejection than water Cherenkov detectors. It has been suggested that a 50 kT LAr detector would have similar performance to a 300 kT water Cherenkov detector for some physics processes, e.g., neutrino oscillation physics. If this is accurate, one can build smaller LAr detectors to achieve similar results to larger water Cherenkov detectors for neutrino oscillation physics. Smaller detectors would require less excavated volume, which would potentially reduce costs. The advantages of smaller size are offset by the need for a cryogenic vessel and cryogenic liquid transport, the large channel count, ~1 M channels per 17 kT detector—and safety issues related to the large cryogenic liquid inventory. Significant additional work is required to demonstrate the performance ratio and cost comparison between the two technologies.

The LBNE project will develop Conceptual Designs, cost estimates, and construction schedules for both the water Cherenkov and the LAr detectors. For both technologies, it is assumed that the desired detector mass required to achieve the science goals of the experiment will need to be reached by modular construction, and a reference detector module has been specified for each. For water Cherenkov, the reference detector has a fiducial mass of 100 kT or a total mass of about 130 kT. Water Cherenkov detector modules will be sited at the 4850L. For LAr, the reference detector will have a fiducial mass of about 17 kT, or a total mass of about 25 kT. The preferred depth for LAr is the 800L.

### 3.3.5.6 A Water Cherenkov Detector for LBNE

#### 3.3.5.6.1 Detector Elements

The size of a water Cherenkov detector is determined by three factors. First is the maximum transverse dimension allowed by the rock properties and appropriate ground support. Second is the maximum depth of the water, currently limited by the pressure tolerance of the PMTs to ~60 m depth. Finally, the maximum transverse distance between any two PMTs is limited by the clarity of the water to ~80-100 m. The good uniformity of rock stress in the horizontal plane leads to the current reference design of a cylindrical cavity, with a water diameter and depth of 53 m and 60 m, respectively. The total mass of water is 130 kT, and with a 2.5-m fiducial cut around the boundaries (including PMTs and their mounting), the total fiducial mass is 100 kT. Studies are under way to understand if different cavity geometry is more cost effective than the current reference design for a fixed total mass. Studies include a larger-diameter cavity, a different shape to minimize excavation cost, and increasing the cavity depth by a suitable PMT enclosure.

The DUSEL baseline design (Chapter 5.7) describes in detail one large cavity for a water Cherenkov detector. Preliminary concepts for multiple large cavities were investigated (see Chapter 5.7) but are not included in the DUSEL baseline design.

The total fiducial mass is best achieved with multiple detectors because:

- It would be technically challenging to excavate a single cavity for 200 kT or more.
- Excavating multiple chambers in parallel is considerably faster than excavating a single enormous chamber, even if that were technically feasible.
- Construction of these multiple modules can be phased as funding evolves.
- Multiple chambers permit at least one detector to be active all the time. While one detector is taking data, any other single detector can be running calibrations, or be down for scheduled maintenance or occasional nonscheduled maintenance. By staggering scheduled calibrations and maintenance between detectors, only one detector will be "off the air" at any given time. This is essential for supernova detection, for example.
- Each of the detector modules can be optimized for different scientific investigations while still maintaining sensitivity to the basic items listed above. For example, one of the detector modules can be instrumented for low-energy sensitivity with an increased photocathode detection area and/or Gd loading or with a veto region to allow a lower trigger threshold.

The DUSEL LBNE water Cherenkov design consists of a large excavated cavity in a very strong and stable rock formation, lined with a smooth, watertight liner and then filled with extremely pure water. Each PMT will be connected via single cable carrying both high voltage (HV) and signal to readout electronics above the water.

An extensive water-purification plant has been designed to fill the detector in about three months, and to repurify one volume of water in about one month. The system includes features to remove the heat from the water, as shown in Table 3.3.5.6.1-1. Adding insulation on the deck at the 4850L can significantly reduce the thermal load on the water-cooling system.

| Item | Value |
|---|---:|
| Ambient rock temperature | 33.4 °C at 4850L |
| Water temperature | 13 ± 2 °C |
| Heat influx from rock | 53 kW |
| Heat influx from PMTs | 15 kW |
| Heat influx from dome (no insulation) | 33 kW |
| Estimated total heat inflow | ~100 kW |

**Table 3.3.5.6.1-1** Temperatures and thermal flux in the water Cherenkov reference design.

**Excavation and Liner.** Table 3.3.5.6.1-2 lists the physics and safety requirements for the water Cherenkov detector. Various design assumptions and the dimensions of the chamber are shown in Tables 3.3.5.6.1-3 and 3.3.5.6.1-4, respectively. The excavation diameter is determined by the strength of the rock that encloses the cavity. For our reference design, LBNE has assumed a 55-m diameter excavation, as shown in Figure 3.3.5.6.1-1.

| Item | Value | Reason |
|---|---|---|
| Total volume of WCh detector (assumes foreign participation for 100 kT WCh equivalent) | ≥300 kT | Neutrino oscillation parameter measurement, proton decay |
| Minimum fiducial volume (FV) per cavity | ≥100 kT | Scientific competiveness |
| Maximum distance between cavities | 5 km perpendicular to neutrino beam | Neutrino beam opening angle |
| Maximum depth of water above lowest PMT | 60 m | PMT collapse under water pressure |
| Depth of cavity below surface | 4850L | Cosmogenic background to proton decay |
| Maximum distance between any two PMTs | 80 m | Water attenuation length at peak PMT wavelength sensitivity when convoluted with Cherenkov light spectrum |
| FV cut | 2 m from PMT photocathode: top, bottom, and sides | Cosmic ray rejection |
| Cavity lifetime | ≥30 years | Proton decay, neutrino oscillation parameter measurement |
| Egress | Dual egress from drifts and cavities during all phases of construction and operation | Personnel safety |
| Water management | Inherently protect the remainder of the DUSEL Facility from catastrophic failure of the water containment | Personnel, equipment, or Facility safety |
| Water temperature | 13 °C ± 2 °C | Reduce biological growth |
| Temperature and humidity of dome air | ~22 °C ± 4 °C, 40% ±10% RH | Standard DUSEL ventilation air temperature |
| Code requirements | DUSEL Doc. EHS-29-200-L5-01 | Personnel, equipment, or Facility safety |

**Table 3.3.5.6.1-2** Requirements for the water Cherenkov detector.

| Assumption | Value |
|---|---|
| FV | 100 kT |
| Shape, based on Homestake rock properties and Super-K experience | Right cylinder |
| Minimum diameter of water | 53 m |
| Buffer size, PMT +FV cut (radially and axially, based on Super-K experience) | 2.5 m |
| Allowance for drainage and liner (sides and bottom) | ~1 m |
| Allowance for freeboard and deck (top only) | 2 m combined |
| Top of deck | Level with 4850L |

**Table 3.3.5.6.1-3** Assumptions used to design the water Cherenkov detector excavation volume.

| Item | Specification |
|---|---|
| Shape | An unobstructed right cylindrical volume free of rock outcroppings or ground support * |
| Free diameter | 55 m* |
| Top of right cylindrical | Level with the 4850L |
| Free height from (possibly virtual) flat floor to 4850L | 64.3 m* |
| FV | 48 m diameter by 55.3 m high, 100 kT FV |
| Lifetime | >30 years |
| Water temperature | ~13 °C ± 2 |
| Dome air temperature & relative humidity | ~ 22 °C ± 4 °C, 40% ± 10% RH |

* Follows from assumptions

**Table 3.3.5.6.1-4** Water Cherenkov cavity specifications.

The main function of the watertight liner is to provide an absolute barrier between the highly purified water (ASTM Type 1, ultrapurified water) in the detector and any underground water that might seep into the excavations.

Two liner concepts have been explored: a liner that is directly mounted on the inner rock of the excavation or a self-supporting structure independent of the rock. The liner attached to the rock has been defined as the reference design (for the DUSEL Preliminary Design Report) as it maximizes the fiducial volume for a constant excavation size, and is cheaper to build (Figure 3.3.5.6.1-2).

**Photodetectors.** Recent Cherenkov detectors have used photomultipliers of various diameters, ranging from 20 cm to 50 cm. LBNE is focused on tubes in the 25-30-cm diameter range. This range seems to provide the maximum signal-per-unit cost, considerably reduced risk of implosion compared with the 50-cm diameter tubes used by Super-Kamiokande, and a reasonable number of tubes per module. The base design calls for 50,000 PMTs (with high quantum efficiency photo-cathodes) per 100-kT fiducial-mass module. This design corresponds to photocathode coverage sufficient to collect 5% of light emitted by events in the fiducial volume. However, this number will ultimately be adjusted as more information is obtained about PMT performance achievable by the various potential vendors and how it relates to cost.

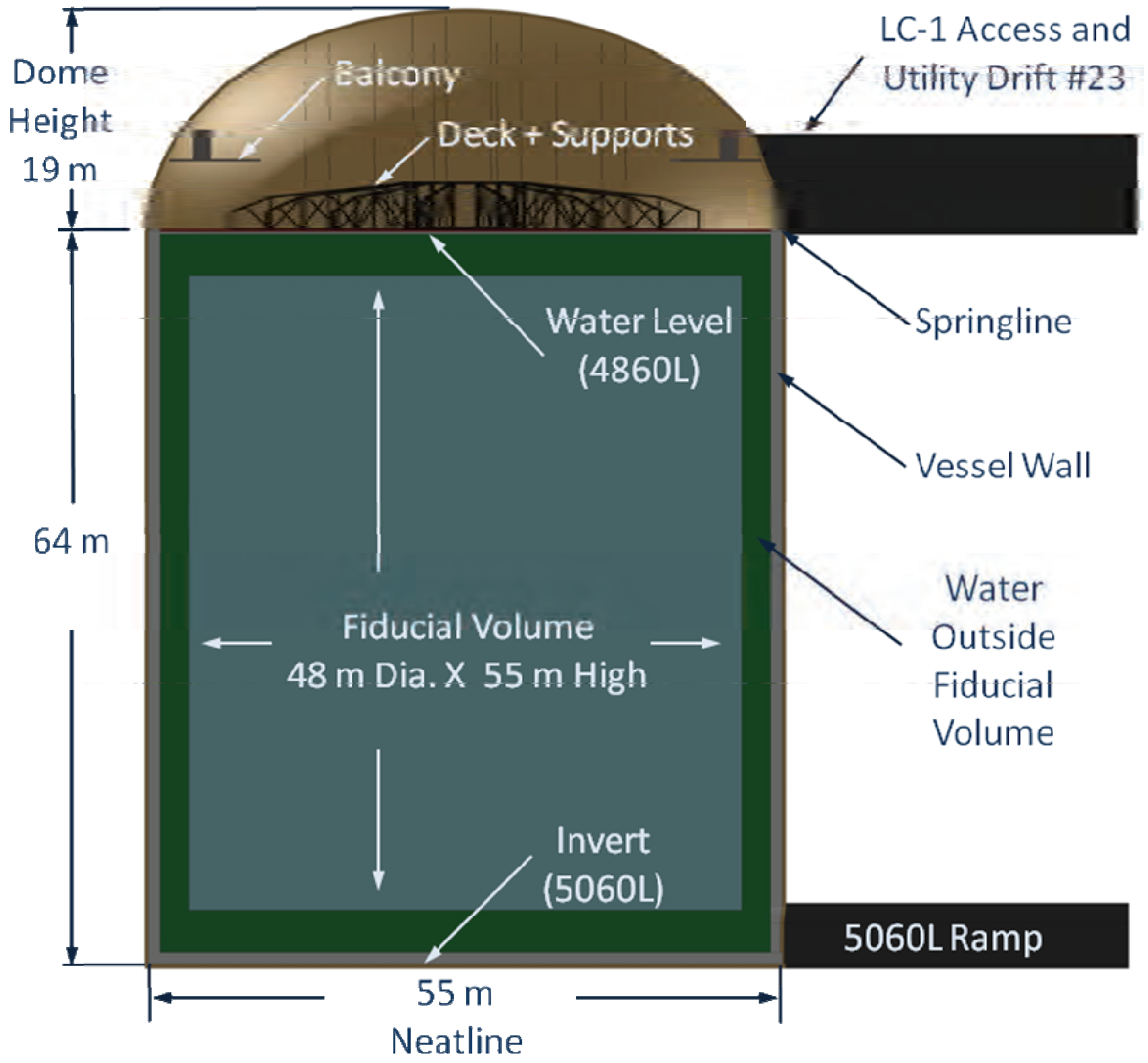

**Figure 3.3.5.6.1-1** Drawing, corresponding to the specifications of Table 3.3.5.6.1-4, showing the cross section through the water Cherenkov detector for the reference design. The dimensions are in meters. [Dave Taylor, DUSEL]

**Photomultiplier mounting and signal cables.** The natural mounting structure for the PMTs is the watertight liner. Using this liner avoids the construction of a separate costly and space-consuming framework while providing a strong, rigid, and stable support. The most attractive approach is to mount a number of tubes onto one frame and then attach that frame to the liner. The number of tubes per frame will depend on the tube spacing and a reasonable weight and size frame. Typical numbers of tubes per frame unit are six to nine. Ideally, the tube mount will avoid torques on the tube due to the buoyancy of the spherical section and the long power/signal cable at the end of the tube.

One of the critical issues is to minimize the risk of tube implosion and to prevent propagation of the implosion to adjacent tubes in case one tube does implode. There are a number of approaches to a solution, ranging from total tube enclosure, encasement in a shock-wave-dampening shield, to a shock-wave deflector between tubes. The choice of approach will depend on the results of various implosion shock-wave studies now under way and on the pressure resistance of the tubes of the various vendors**.**

**Readout electronics.** There are two possible approaches to the readout electronics. One is to locate electronics for batches of tubes (e.g., 16 tubes per batch) underwater, adjacent to the tubes. This reduces the cable length between tube and readout electronics but makes access to the electronics and maintenance extremely difficult.

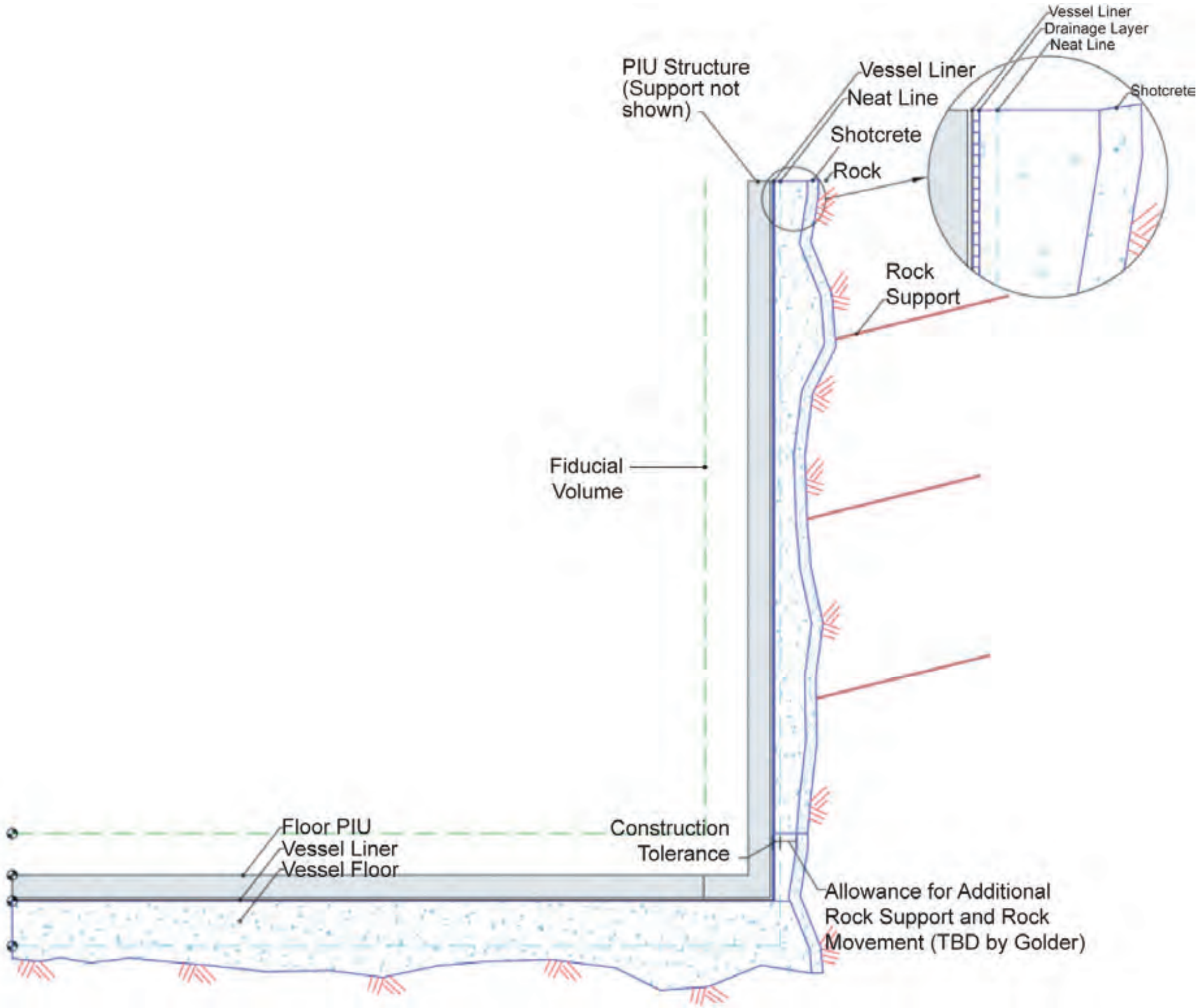

**Figure 3.3.5.6.1-2** Preferred vessel design integrated with rock. [Courtesy LBNE project]

The other approach is to locate the electronics on a deck directly above the water detector and link each tube to the electronics with a cable. The most distant tubes—those in the center of the bottom of the detector—will require a cable of 80-100 m. In previous detectors, all tube-to-readout electronics cables were of equal length. Using the equal-length cable approach for this detector will generate several thousand kilometers of cable slack. The storage of that amount of cable can become a major space issue. The reference design for the detector assumes equal-length cables. This option remains under study. A black sheet separates the active volume of the water forward of the PMTs from the annular volume that contains the PMT supports. This will prevent reflections from the supports from creating false hits in the detector. This annular volume behind the PMTs is being considered as a “thin veto” for cosmic rays or through-going particles. The LBNE project is studying whether this would allow using more of the fiducial volume for the LBNE physics by allowing us to tag rock interactions versus contained muons originating from ν's in the beam.

**Water fill, recycling, and cooling.** The surface-water processing system has been designed to purify about 1,000 liters/minute of fill water, resulting in a total fill time of several months. In addition, the system will recycle the detector water through a repurification system at a rate about five times higher. Figure 3.3.5.6.1-3 shows a 3D-CAD isometric drawing of the repurification plant located on the 4850L near the detector. The purification requirements for this detector are within the normal range of commercial systems, similar to those of previous water Cherenkov detectors.

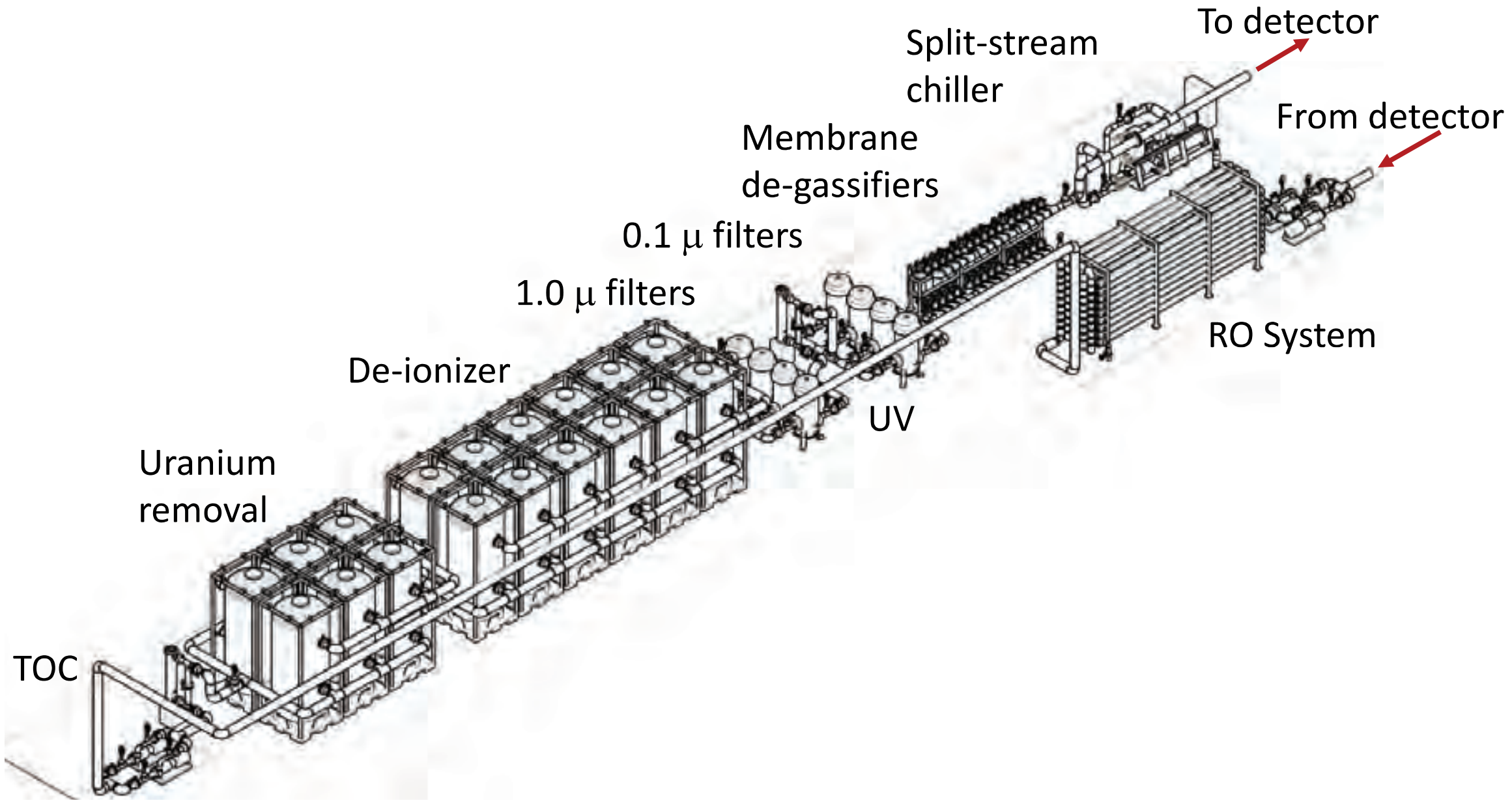

**Figure 3.3.5.6.1-3** Drawing of ~4500 lpm water recirculation system. [Courtesy LBNE project]

The planned temperature of the water is about 13°C. This will reduce the photomultiplier electronic noise and inhibit biological growth in the detector. This water temperature requires both that the initial fill water be cooled from its surface temperature and that the recycled water be cooled to remove the thermal energy due to heat flow from the rock and heat input from the photomultiplier bases. The internal rock temperature prior to excavation at the 4850L is about 33°C. As the excavation proceeds and increasing cavity surface area is exposed to air, the near-surface rock will cool, with the surface rock approaching the ventilation air temperature. Reasonable estimates are that by the time the detector is filled with water, the heat flow from the rock will be between 50 and 100 kW. The photomultiplier bases are likely to add another 10-20% to this heat flow (see Table 3.3.5.6.1-1). A significant heat flow into the detector will come through the top surface from the room air contact. Insulation at this surface may be necessary.

**Veto counters.** Although the cosmic-ray-muon flux is very low at the depth of this detector, about four muons per day per $m^2$, the large aperture of each module, about $2 \times 10^3$ $m^2$, will still result in a significant muon flux through each 100 kT module—approximately 0.1 Hz, the muons very peaked in the vertical direction. Thus, a veto counter placed directly above the top set of photomultiplier tubes and directly beneath the top deck of the detector can veto a significant fraction of the incident cosmic-ray muons.

**Fiducial volume definition.** The PMT mounting structure will include a black-light shield at the equatorial plane of the PMTs, that is, at the largest diameter of these tubes. That light barrier will separate any Cherenkov light generated in the outer annular region of the detector from light that is generated in the central detector cylinder. The present plan is to then define the fiducial-volume limit to be 2 meters radially inward from this light barrier. Since this is a software definition, it can be dynamically varied once the detector is in operation and events are being reconstructed. It is even possible to define different fiducial volumes for different signals.

**Calibrations and monitoring.** The energy scale and linearity, energy resolution, directional dependencies of energy scale and resolution, and the stability of the energy calibrations must all be well understood in order to achieve the physics goals. The goal for energy scale uncertainty is 2% or better in all energy regions (MeV to GeV). The energy calibration can be accomplished by a combination of naturally occurring events inside the detector as well as dedicated sources deployed at various locations inside the detector volume. Cosmic-ray muons can be used in the energy range of hundreds of MeV to several GeV. For low-energy calibration, radioactive gamma and beta sources, a low-energy linac (5-16 MeV) as well as Michel electrons can be used as sources. A novel laser-wake electron accelerator is being considered for high-energy calibration in the range 100 MeV to 1 GeV.

A centrally located LED diffuser ball will be used for timing and charge calibration. The PMT timing should be calibrated to better than 1 ns over a pulse height range of 1-1000 photoelectron (PE). After charge calibration, the uncertainty in number of PE in each PMT over the range of 1-1000 PE should be <10%.

The water transparency needs to be continuously monitored. An attenuation length of 100 m or more must be measured to 5-10%. Muons or LEDs inside the detector could be used. Alternatively, the attenuation length could be measured for samples of water in an external system, specially designed or commercially available.

Other environmental variables that will require monitoring include water temperature, the flow rate and pattern of water circulation, water level, pH, resistivity, total dissolved solids, radon, magnetic fields, and biologics.

**Computing.** The computing requirements can be divided into three categories: online, offline, and infrastructure. Online computing includes processing the raw data received from data acquisition (DAQ) system, run control, and detector monitoring.

The LBNE offline computing group will take care of simulations, reconstruction software, and official production data processing. Infrastructure supports the efforts of the online and offline groups by providing a software framework and data archive, and assuring adequate hardware and network connectivity.

**Installation and integration.** Coordination of the construction of the chamber and the installation of the detector components will be challenging. The laboratory hoist system imposes constraints on the mass and volume of materials that can be brought underground. The underground detector-staging area and detector-chamber entrance require careful coordination and sequencing of the lowering of detector components. There are concerns about interactions of dust associated with chamber excavation, installation of the chamber liner, drilling of holes for the photomultiplier mounts, and the protection of the "clean" components of the detector, PMTs, electronics, and water-handling system. Fortunately, simulations of the installation sequence and process can easily be carried out and optimized. A clean surface-staging area will be required. It will also be necessary to specify the degree of cleanliness required by each of the detector components and what restrictions these requirements impose on the installation process.

#### 3.3.5.6.2 Alternatives and Options

Electron antineutrinos are products of supernovae, and background rejection at the low-energy cutoff imposed by solar neutrinos can be improved via the double coincidence of inverse beta decay reaction: $\overline{\nu}_e + p^+ \rightarrow n + e^+$, with the prompt positron detected through its Cherenkov radiation, and the neutron via delayed capture (~30 μs) on gadolinium (probably in the form of gadolinium sulfate, ~0.1% Gd by weight) in the water, releasing a cascade of photons with total energy ~8 MeV of energy of which ~6 MeV is detected by the PMTs. The successful detection of inverse beta decay depends on the Gd doping of the water. Neutron capture on a proton only releases a single 2.2 MeV photon, which is not detectable in the water Cherenkov detector.

There are two main technical issues regarding the Gd doping of the water:

> **Water recirculation and purification.** The recycling of the water through the purification system will remove the Gd salt, so the system will have to remove the Gd before the water reaches the purifier and then redissolve the Gd after purification. Also, there may be light attenuation due to Gd and questions about materials of construction that are in contact with the Gd. Although Gd is not in the baseline detector design, it would be desirable to avoid materials in the detector construction that interact with the Gd salts to preserve this option for the future.
>
> **Larger photocathode coverage.** Based on Super-Kamiokande's experience, LBNE believes at least 20% photocathode coverage with normal quantum efficiency tubes is desirable to realize the major physics goals of the experiment. However, the possibilities for low-energy physics (e.g., solar and supernova neutrinos) would be greatly enhanced by increased photocathode coverage.

**Increased fiducial volume.** The single most critical parameter of the DUSEL water Cherenkov detector array is the total fiducial volume. There are two issues here. One is the maximum volume of rock that can be safely excavated at reasonable cost. The second is the maximum number of such modules that can be constructed. The goal is to get as close to 300-kT fiducial mass as possible. For example, an increase of 10 m in the excavated diameter of a chamber results in a 44% increase in fiducial mass for that chamber. Since the excavations for these detectors are already the largest deep-underground excavations, this question must be approached with considerable caution. A study by Golder Associates on alternative cavity designs indicates that maximum allowed cavity diameter from geotechnical considerations is 65 m. Mailbox-style cavities with flat walls are discouraged. A 65-m diameter cavity would allow a 150 kT fiducial volume detector as a single, right, vertical cylinder, meeting all other technical specifications for the detector (maximum PMT depth in water, transmission length underwater, etc). Current and planned R&D on the PMT implosion, including PMT enclosures to protect them from implosions of adjacent PMTs, could allow the depth of the water to increase, permitting 65-m diameter cavities as large as 200 kT fiducial mass. The Golder alternative cavity shape document is currently being considered by the collaboration to understand how to best optimize the cavity shape and size, the distribution and coverage of PMTs in the fiducial volume and value engineering studies to minimize the overall cost for the equivalent physics reach. See Chapter 5.7, *Large Cavity for the Long Baseline Neutrino Experiment,* for further details on this alternative.

#### 3.3.5.6.3 Overall Underground Layout and Facility Resource Requirements

The design of the large cavity and related spaces at the 4850L and 5060L are presented in Chapter 5.7. Explicit space has been provided to accommodate the water-purification plant, including maintenance considerations. Apart from the large cavity (LC), there is about 1,300 m$^2$ of other space on the 4850L related to LBNE activities. Figure 3.3.5.6.3 is an isometric view of the dome area, showing the cable penetrations and electronics racks and magnetic compensation for the Earth's magnetic field. The magnetic field compensation maximizes the PMT efficiency and also reduces the asymmetries in the detector efficiency. If the Earth's field is not compensated, there is a 10-15% efficiency loss per PMT.

A utility drift holds the electrical room for the LC, the HVAC room, the control room, and calibration storage room. A radon-abatement system (or LN2 plant) for suppressing radon between the deck and the water is also included in the design.

On the 5060L, sumps are needed to collect native and detector water, and pumps to recirculate the water to the purification system. In the event Gd is added to the water, there will be a small Gd recovery plant at this level to remove Gd from water leaking from the detector volume**.** Table 3.3.5.6.3 lists the facilities requirements for one water Cherenkov detector.

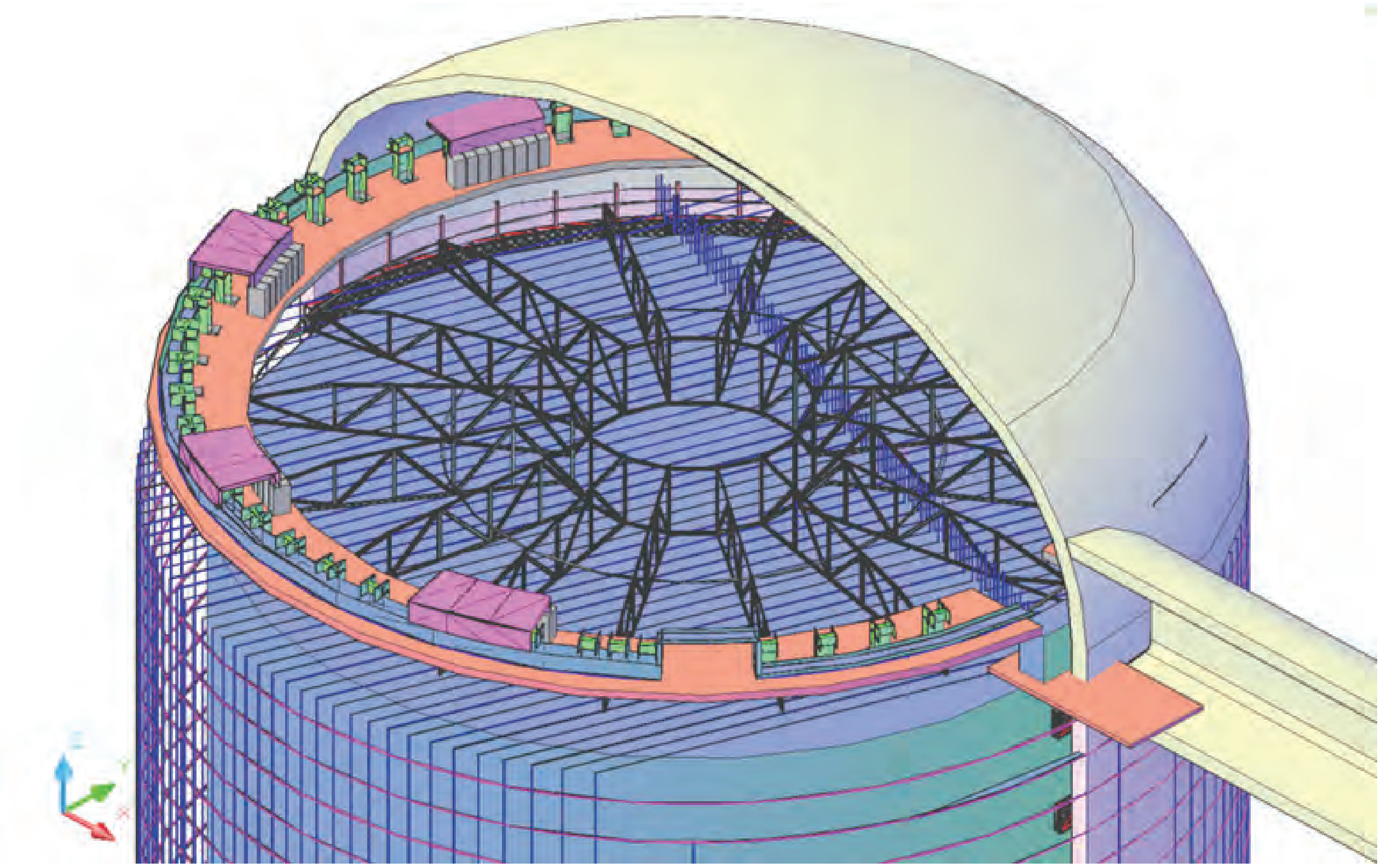

**Figure 3.3.5.6.3** Isometric view of the dome of the LC1. There is a "balcony" (red) around the perimeter of the cavity about 3 m above the 4850L. The balcony supports the cable penetrations around the perimeter (green) and the eight sets of electronics racks. The deck is supported by large trusses that are in turn mostly supported from the rock above the dome. The total load on the rock above the dome is ~ 600 T. The trusses are oriented so as not to obscure the entrance from the utility drift, the calibration drift, or calibration ports in the deck itself. The narrow, circumferential black lines and the black lines running down the side of the cylindrical volume represent the magnetic compensation cables for canceling the Earth's magnetic field within the cavity. [Courtesy LBNE project]

| Requirement | Value/Description | Comment/Justification |
|---|---|---|
| **Layout** | | |
| Depth | 4850L at deck top, base at 5060 | |
| Footprint [$m^2$] | 2206 | 53-m diameter water, excavation is to 55 m, area is for one LC only, without utility or $H_2O$ purification areas |
| Height [m] | 83 | Top of dome at 83 m above bottom neat line; spring line at 4850L, 64 m above bottom neat line |
| Floor Load [kPa] | 1667 | |
| Total Underground Area, including LC1 [$m^2$] | 4267 | |
| Total Surface area [$m^2$] | 3181 | |
| **Utilities** | | |
| Power [kW] | 2248 | Underground power only—does not include surface power requirements. Power for 1 LC only. |
| Standby Power [kW] | 110 | Sump pump + partial control system. Power for 1 LC |
| Chilled Water [kW] | 1411 | Water system pumps are assumed to reject 85% of heat into the sump water, 15% heat to HVAC. Heat for one LC. |
| Waste Heat to Air [kW] | 691 | Heat for 1 LC |
| Purified Water [$m^3$] | 138,000 | LBNE is responsible for this system, quantity is for 1 LC |
| Surface Power [kW] | 771 | |
| Industrial Water | 600 gallons / minute on the surface | To feed the surface water purification system |
| Potable Water | Not defined | |
| Compressed Air | Up to 380 cfm (650 $m^3$/hr) | Current Facility requirement |
| Network [Gb/s] | 10 | 1 dedicated line per cavity |
| **Environment** | | |
| Temp. Min [ºC] | 18 | |
| Temp. Max [ºC] | 25 | |
| Humidity Min [%] | 30 | |
| Humidity Max [%] | 50 | Guidance given that humidity is key requirement with 50% as target and minimum to be 30% with temperature secondary consideration |
| Rn Background [Bq/$m^3$] | TBD | |
| **Occupancy** | | |
| Peak Installation Occupancy [count] | 50 | |
| Installation Duration [months] | ~24 | |
| Peak Commissioning Occupancy [count] | TBD | |
| Commissioning Duration [months] | 6 | |

| Requirement | Value/Description | Comment/Justification |
|---|---|---|
| Peak Calibration Occupancy [count] | 5 | |
| Average Calibration Occupancy [count] | 2 | |
| Calibration Duration | 1 day/month | |
| Operation Duration [months] | >360 | |
| **Cryogens** | | |
| LN Storage | 100 kg dewar | For calibration |
| LN Consumption | 100 liter/week | |
| Transportation Quantity per dewar [kg] | TBD | |
| Transportation Frequency [shipments per week] | 1 | |
| **Major Hazards (Other than Cryogens)** | | |
| | Flooding, ultrahigh-purity water is toxic, gadolinium is possible additive that might be considered toxic, falls, drowning, rock collapse, fire, electrocution | |
| **Assay and Storage** | | |
| Assay Needs | TBD | |
| Underground Storage | NA | |

**Table 3.3.5.6.3** Facility requirements for the water Cherenkov detector.

## 3.3.5.7 The Liquid Argon Detector

### 3.3.5.7.1 Liquid Argon Technology

The very large water Cherenkov detector included in the conceptual design for LBNE is an extension of current technology, particularly of Super-Kamiokande. This section describes a different technology: Liquid Argon Time Projection Chamber (LArTPC). While at least one 600 T LArTPC exists (Imaging Cosmic and Rare Underground Signals [ICARUS]),[63] it does not have the same level of operational experience as Super–Kamiokande. A LArTPC features notable strengths. First, it enables detailed, reconstructed images of neutrino-scattering events, which leads to a high level of background rejection. Second, it allows for comparatively localized event topologies, which lets the detector simultaneously measure multiple events from, for example, cosmic rays, and distinguishes between them. More specifically, LArTPC technology provides:

- Highly accurate differentiation of electrons vs. photons by high-resolution measurements of electromagnetic shower development in the vicinity of the interaction vertex
- High-resolution reconstruction of the recoil hadronic shower, including nuclear debris
- Excellent sensitivity to low-energy hadrons that are below Cherenkov threshold in water

Current understanding indicates that a LArTPC detector can be located at a moderate depth (~800 feet) and still achieve sufficient rejection of cosmic-ray-induced background, even for non-beam-event related studies, such as the searches for proton decay and supernova neutrinos. In addition, LAr pattern-recognition capabilities make this technology more efficient. The physics observational capabilities of a LAr detector for neutrino oscillation studies may be comparable to those of a much larger water Cherenkov detector, which provides less-detailed information about events. Although the exact equivalence factor between LAr and water Cherenkov technologies depends on specific event topologies, simulation studies suggest that a LAr detector has equivalent physics reach (for neutrino oscillation studies), to a water Cherenkov detector approximately six times larger in mass.

While LArTPC technology is promising because of its likely high spatial resolution and excellent measurement of deposited ionization along isolated tracks, it clearly requires development and operational experience. It is also important to assess the scalability of this technology to masses of as much as $10^5$ tonnes that may be required for future neutrino experiments. Currently, only simulations are available that show LBNE can do $\nu K^+$ proton decay channel at 800 feet and trigger the LAr detector for supernovae or proton decay. There is no information at all on spallation product backgrounds. No large-scale LAr detector has yet given actual numbers on these important issues.

#### 3.3.5.7.2 A LArTPC Implementation for LBNE: The LAr20 Detector

The *LBNE Conceptual Design Report* will describe a particular LArTPC implementation, the LAr20 Detector. The LAr20 conceptualization has a total mass of 25 kT and a fiducial mass of 16.7 kT. High-purity LAr serves as both the neutrino target and the tracking medium for the particles produced in the interaction. The overall dimensions of the active volume are 15.0 m wide (in X) by 14.0 m high (in Y) by 71.1 m long (in Z, the beam direction).

The LAr20 Detector will identify neutrino events through the observation of the outgoing charged particles resulting from neutrino interactions in the LAr. A uniform electric field in the LAr volume will cause ionization electrons produced by the passage of these charged particles to drift to three wire planes. The electric potentials of the three wire planes will be arranged such that the electrons will pass through the first two planes, and be collected on the third. The passage of electrons through the first two planes will produce induced bipolar signals on those wires. The deposition of electrons on the third plane will produce negative unipolar pulses. “Cold” electronics within the LAr20 vessel will amplify the signals on each wire and continuously digitize the amplified waveforms at 2 MHz. The proposed LAr20 wire pitch in all planes is 3 mm; therefore, LAr20 position resolutions will be at the millimeter scale. The trajectory of particles in the detector will be reconstructed from the known wire positions and the arrival times of electron signals on the wires, combined with the time the interaction took place in the detector. The amplitude of the ionization electron signals measures the energy loss of the particles, which enables an estimate of their momentum and particle type.

The main features of the LAr20 Detector are shown in Figure 3.3.5.7.2. Selected parameters are given in Table 3.3.5.7.2. Major sources of power and cooling are listed in this table. However, the design of the LAr detector lags with respect to the water Cherenkov detector: Many of the detailed parameters and impact on the DUSEL facility have not yet been developed.

| Parameter | Value | Unit | Note |
|---|---|---|---|
| Wire Spacing | ~3 | mm | Typical |
| No. of Wire Planes | 3 | | Wire orientations per module |
| Stereo Angle | 0, +45, -45 | deg | |
| Drift Distance | 2.47 | m | |
| Wires per Module | 5520 | | |
| Readout Channels | 30 | | 128 times multiplexed |
| Module Size | 5 x 7 x 2.5 | $m^3$ | (X x Y x Z) |
| UV Wire Length | 9.9 | m | |
| UV Wire Capacitance | 238 | pf | In LAr |
| Electric Field | 500 | V/cm | X direction |
| Max. Drift Voltage | ~125 | kV | |
| Maximum Drift Time | ~1.5 | ms | |
| Module Volume | 87 | $m^3$ | |
| Module Active Mass | 0.122 | kT | |
| X,Y Fiducial Cut | 0.3 | m | |
| Z Fiducial Cut | 1.5 | m | End modules |
| Module Fiducial Volume | 73 | $m^3$ | |
| Module Fiducial Mass | 0.102 | kT | |
| Number of Modules | 3 x 2 x 28 = 168 | | (X x Y x Z) |
| Total Fiducial Mass | 16.4 | kT | |
| Total Readout Wires | 645120 | | |
| Total Readout Channels | 336 | | 2 optical fibers per APA |
| Cryostat Dimensions | 16 x 16 x 74 | $m^3$ | (X x Y x Z) [15 x 14 x 71.1 $m^3$, active volume only] |
| Total Mass LAr | 25 | kT | |
| Cryostat Insulation Thickness | ~1 | m | |
| Cryostat Insulation Heat Loss | 36 | kW | |
| Max Recirculation Rate | 163 | $m^3$/hr | |
| LAr Volume Turnover | 5 | days | |
| LN2 Refrigeration Plant Capacity | 59 | kW | ARUP conceptual report |
| LN2 Storage Dewar | 50 | $m^3$ | ARUP conceptual report |
| LN2 Backup Capacity | 40 | hours | ARUP conceptual report |

**Table 3.3.5.7.2** Selected parameters of the LAr detector reference design. The X direction is horizontal and perpendicular to the beam, Y is vertical, and Z is horizontal near the beam direction. This design is the preferred current design of different concepts.

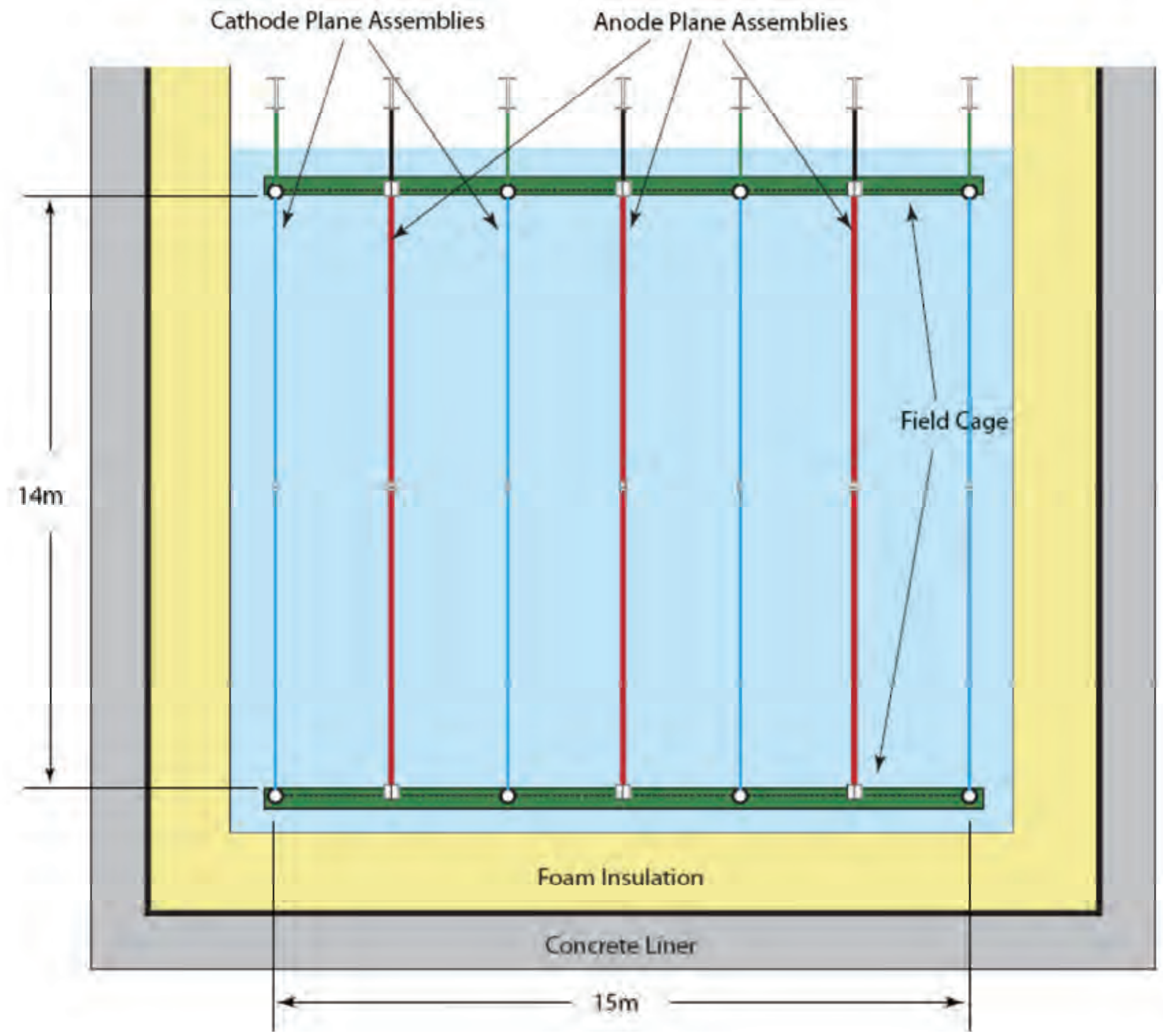

**Figure 3.3.5.7.2** Cross section through the LAr cryostat, indicating the foam liner, field cages, anode plane assemblies, and cathode plane assembles. [Courtesy LBNE project]

Figure 3.3.5.7.2 shows the overall layout of the LAr Time Projection Modules (TPM). Each TPM will consist of two cathode planes and a central Anode Plane Assembly (APA) of the same size. The maximum drift distance will be 2.5 m. The TPMs will be arranged to share cathode planes, so that in the "x" direction, there will be four cathode planes interleaved by three APAs to form three TPMs. Thus, in all, the LAr20 Detector will have 168 APAs and 224 cathode planes.

The LAr20 cathode planes will be held at an electric potential of ~ -125 kV to create an electric field of 500 V/cm between the cathode and anode planes. This field will produce an electron drift velocity in the LAr of 1.6 mm/µs. Each APA will contain four planes of wires in a wrapped configuration. The wire planes are: the grid plane, induction plane 1, induction plane 2, and collection plane. The purpose of the grid plane is solely to improve the effectiveness of induction plane 1; it will not be instrumented with readout electronics. In total, LAr20 will have 654,065 readout wires and 282,240 grid wires.

A "field cage" constructed of copper-plated circuit-board material will surround each row of cathode planes and APA. The purpose of the field cage is to shape the electric field in the LAr to ensure uniform electron-drift trajectories near the detector edges. A resistor chain between the cathodes and anodes will establish the electric potential of the field.

A major feature of LAr20 will be the use of cryogenic, or "cold," electronics. Signals from each wire channel will be amplified, shaped, and digitized by the analog section of integrated electronics, mounted directly on the APA and within the cryogenic volume. The digital section of these same electronics will provide zero-suppression and 128-fold multiplexing. Multiplexed signals will be routed by optical fibers through cryogenic feedthroughs, located at the top of the cryostat. A local computer cluster will provide triggering and event selection. Data will be stored at the detector location and also transported to Fermilab and other collaborating institutions for archival storage and offline analysis.

A single cryostat will house all the TPMs, containing the LAr at a temperature of 87 K and insulating it from external heat. Conceptual design studies resulting in this report suggest that the optimal choice for cryostat design is a single "membrane cryostat." The most notable feature of a membrane cryostat is the use of a thin metallic liner to contain the liquid argon. The metallic liner will be constructed of 1.2 mm-thick stainless steel, corrugated in both directions to enable thermal expansion and contraction. The liner will attach to insulation units constructed of plywood boxes filled with polyurethane foam. The plywood will be "marine grade," typically used for the construction of boats. The hydrostatic load of the liquid argon will transfer through the liner and insulation to the walls of the cavity, resulting in a highly efficient use of the excavated cavity volume. A secondary liner will provide an annular space for argon gas purges. A tertiary liner will prevent groundwater infiltration. Membrane cryostats have been used for ocean transport and onshore storage of liquefied natural gas for several decades.

The LAr in the membrane cryostat will be cooled by a cryogenics system located primarily on the surface. Insulated cryogenic piping will connect the surface refrigeration plant with the underground cryostat and LAr purification system that must be located adjacent to the cryostat. A surface location simplifies installation and maintenance and minimizes oxygen deficiency hazard (ODH) because of the enhanced possibilities for air circulation and venting. Redundant LAr pumps, located inside the cryostat, will be used to recirculate the LAr through the purification system.

Drifting electrons over several meters requires minimizing electronegative contaminants in the LAr. Water vapor and oxygen are the major sources of electronegative contamination. The maximum design electron-drift time is 1.54 ms. The design electron lifetime is 1.4 ms. The design equivalent-$O_2$-contamination is 214 ppt. Contaminants will be removed by recirculating LAr through molecular sieves and copper filters. Purity monitors at the filter outlets will monitor their effectiveness. Argon flow will be diverted to a second set of filters when the first set is saturated. Circulating a 95% argon-5% hydrogen gas mixture through them at elevated temperature will regenerate saturated filters. Argon gas boil-off from the top of the detector will be reliquefied by a condenser and purified before it is returned to the cryostat.

The detector electronics will be configured to enable both continuous and triggered data acquisition. A trigger may be initiated by either a beam-spill signal from Fermilab or a signal from a scintillation light-detection system. A beam spill will trigger data acquisition from the entire detector. Upon initiation of a photon-detector trigger, only wires in the vicinity of the source of the scintillation light will be read out. Signals from some processes of physics interest (e.g., relic supernovae) may be below threshold for the light-detection system. Continuous readout of the detector will enable the study of such processes, given sufficient computing resources to store and offline-analyze the large amount of data that will be generated in a continuous readout mode.

### 3.3.5.7.3 LAr20 Detector Location

The depth for the LAr20 Detector location represents a trade-off between physics background and detector size and cost. For a fixed-dollar budget, LBNE must choose between a larger detector at shallower depth (due to lower cost per tonne), or a smaller one placed deeper underground. The former offers higher potential measurement capability due to size, whereas the latter offers minimization of cosmic-ray-induced backgrounds due to depth. For detector optimization, the layout of the Homestake Mine suggests strong consideration of three possible depths:

- **300 feet.** At this shallow depth, approximately horizontal access is possible by constructing tunnels from the canyon east of the Yates Shaft. However, cosmic ray background, including from both the east and west sides, is likely to complicate non-neutrino beam measurements.
- **800 feet.** At this moderate depth, cosmic ray background is reduced from that at the 300-feet level by about a factor of 10, by both increased vertical overburden and a depth profile that in percentage terms is more flat and less like a mountain. Adding an active shield to the LAr20 Detector at this depth may provide physics sensitivity for a range of beam and non-beam experiments competitive with the same detector at 4,850 feet. Primary access to the LAr20 Detector Laboratory would be through a decline road tunnel, thus mostly isolating LAr20 from shaft contention with other DUSEL activities.
- **4,850 feet.** This location at DUSEL's most active level would significantly limit the rate of background events and remove the necessity for an active shield. The LBNE executive board concluded that the ~$100 million necessary to install the LAr detector at the 4850L was not justified by the physics benefit. It will not be pursued further.

The Conceptual Design process concluded that the most favorable depth for LAr20 is the 800L. Compared with the 4850L, the preferred 800L simplifies both access and the cryogenics system design, and reduces the possibility of shaft-access contention with DUSEL and its other experiments.

Figure 3.3.5.7.3-1 shows possible locations of the LAr surface facilities with respect to the DUSEL campus structures. The main access would be via an adit near the Kirk Portal that descends via a 12% grade to the 800L in a spiral. Two LAr cavities could be located not far from the Ross Shaft at this level. Figure 3.3.5.7.3-2 is an isometric view of the cavities and ramp system. Two cavities are shown, but one or two could be constructed, depending on the technology choice for the experiment. Finally, Figure 3.3.5.7.3-3 shows a cross section, approximately perpendicular to the beam direction, through a single LAr cavity. The membrane cryostat is located below the normal access/working level, and the ~1-m-thick insulating cryostat walls are supported from the rock. Figure 3.3.5.7.3-4 is an isometric view of a cross section through the membrane cryostat.

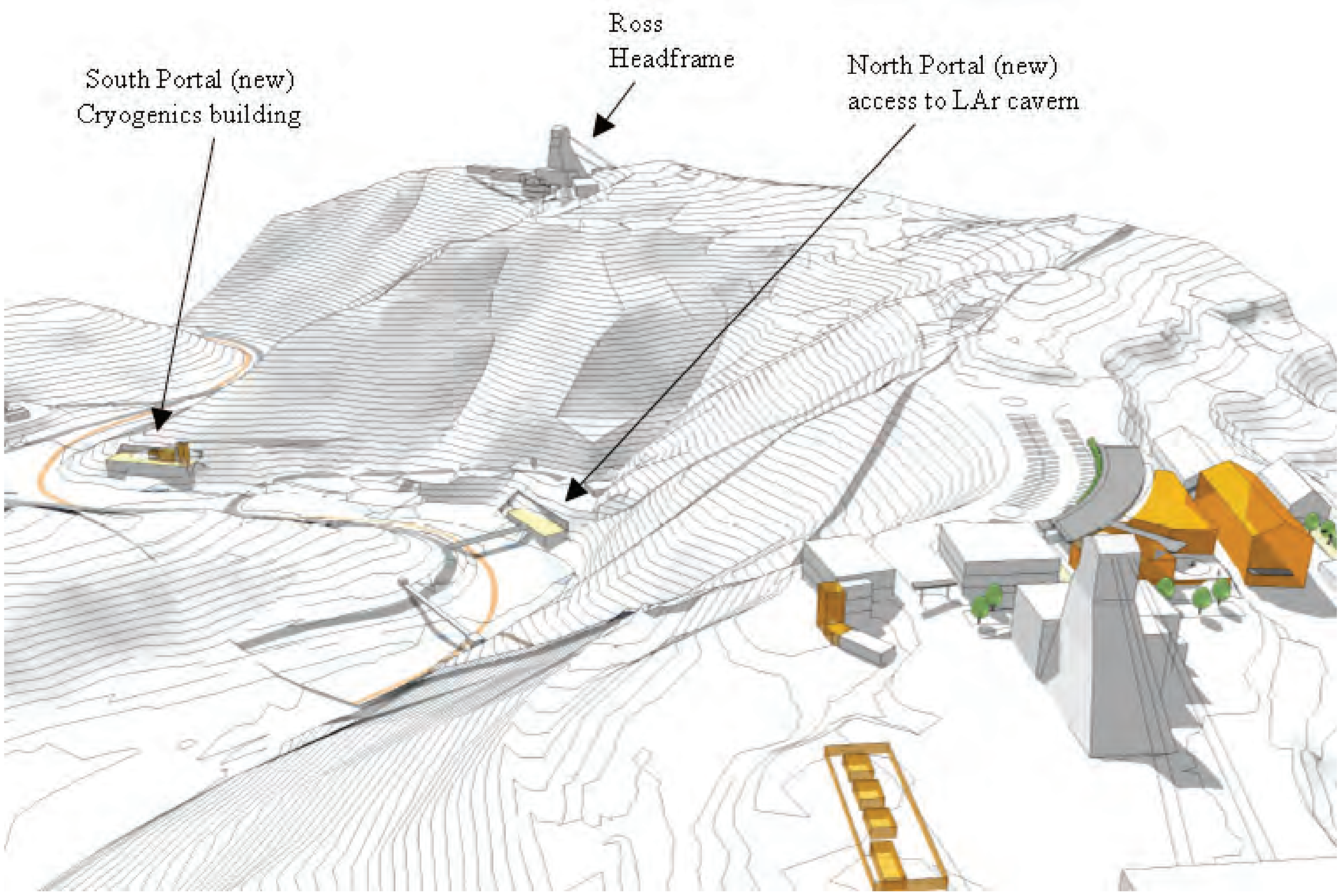

**Figure 3.3.5.7.3-1** Isometric view of the DUSEL campus showing a possible layout of the LAr surface buildings on the DUSEL Campus. The cavities are accessed by a spiral ramp system with a portal at the Kirk fans (300L) that connects to the 800L, and from there to the Ross Shaft for secondary. [HDR]

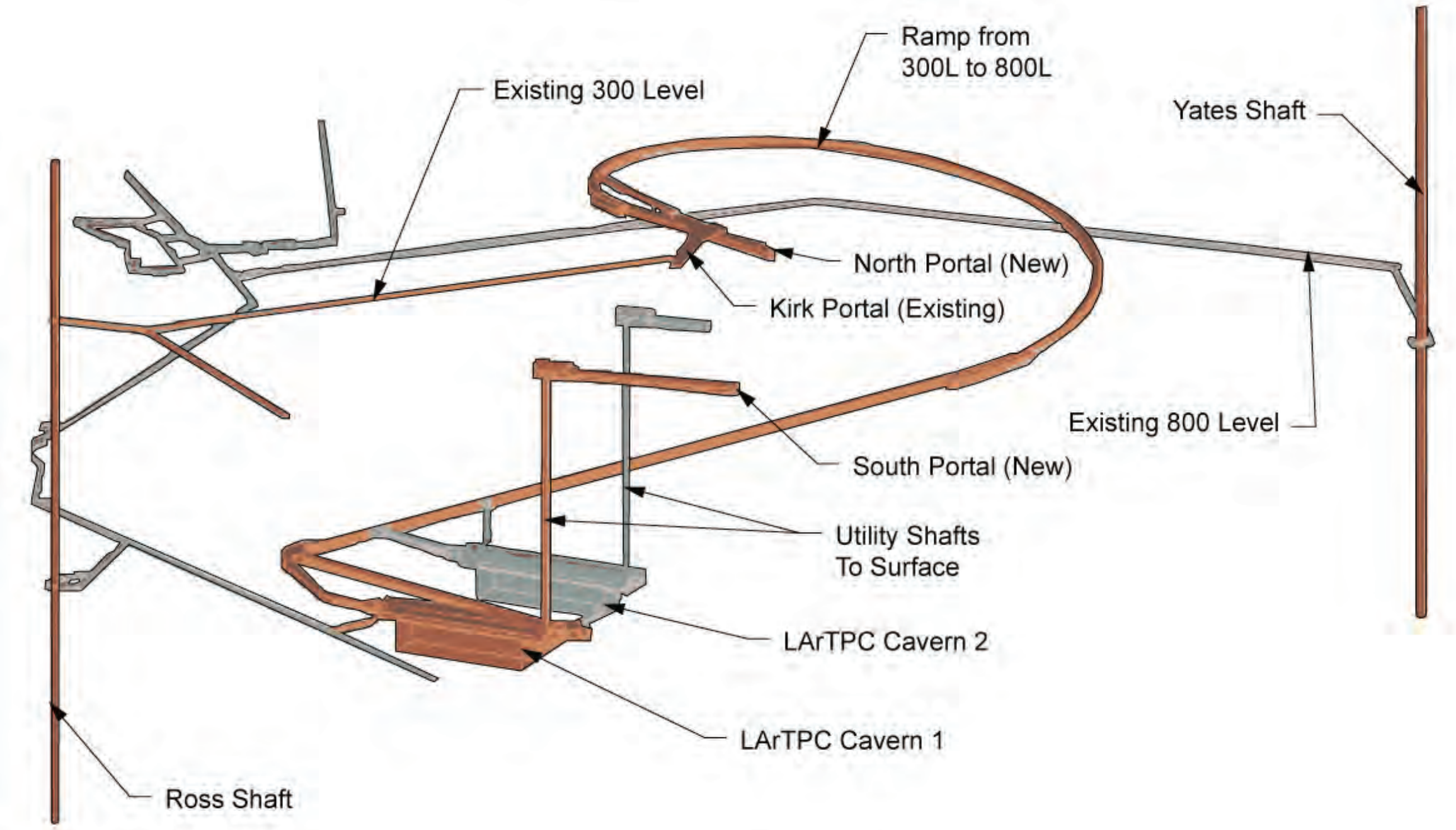

**Figure 3.3.5.7.3-2** Isometric view of two cavities for LAr detectors. Shafts from the cavities connect to new drifts at the 300L for utilities and venting. The ramps to the upper (right) side of the cavities are for normal experimental access. The ramps connecting to the bottom of the cavities are for construction access and would be plugged prior to operations. The connection to the 800L allows the Ross Shaft to act as an emergency access. [DKA]

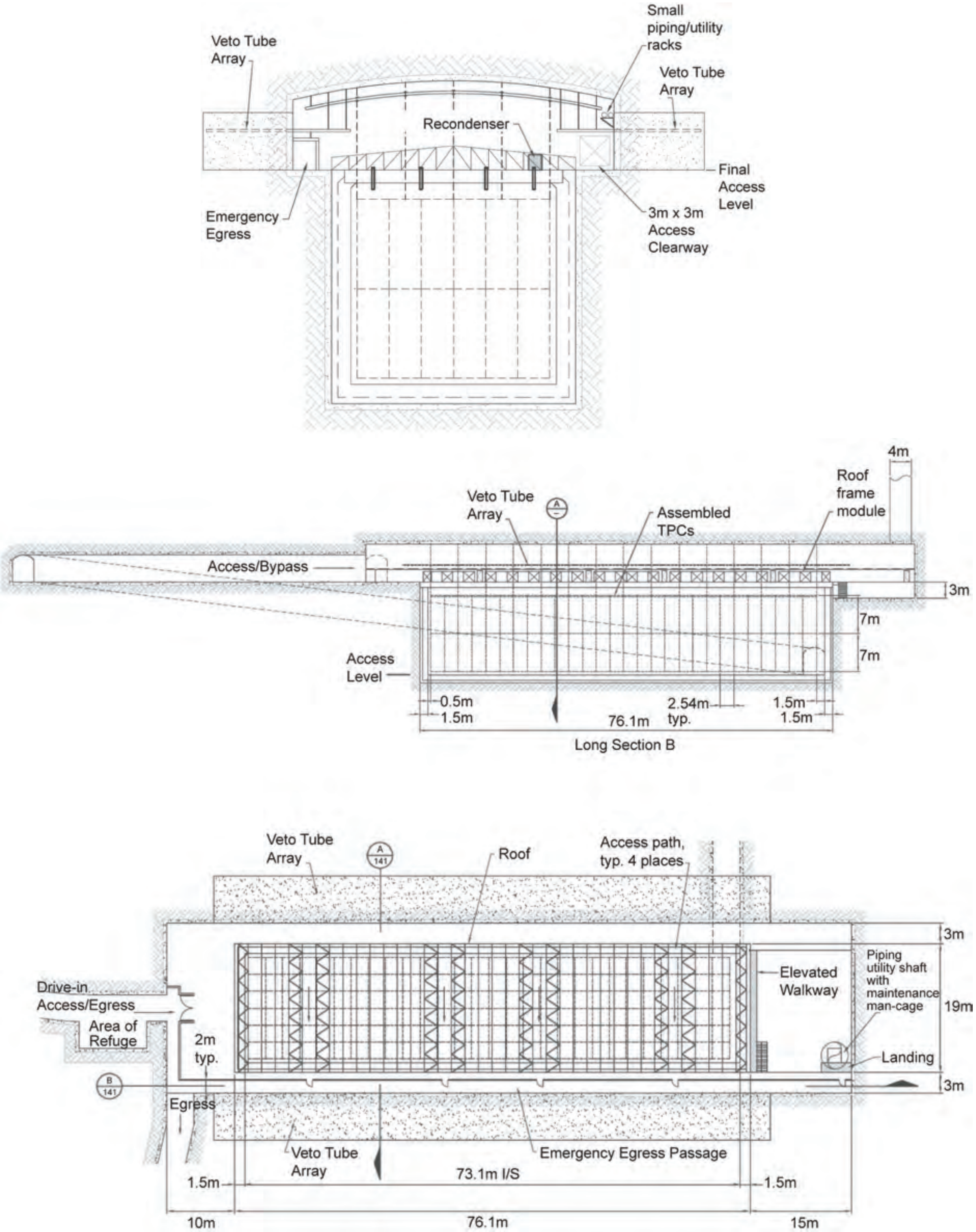

**Figure 3.3.5.7.3-3** (Top) Cross section through a LAr cavities, approximately perpendicular to the beamline. The detector is in a pit below the normal access level (i.e., working level) of the cavity, and the membrane cryostat walls are supported from the rock. (Middle) Long section of the LAr cavity, showing the main access at floor level, the vent borehole on the right side, and construction mucking drift (dotted diagonal lines). (Bottom) Plan view of a LAr detector, showing the top of the cryostat and the emergency-egress pathways. An emergency-access corridor along one long side of the cavity (lower edge of cavity in this view) with doors every ~20 m allows sheltered evacuation of workers to the ramp system in event of a catastrophic failure of the LAr containment. [Courtesy LBNE project]

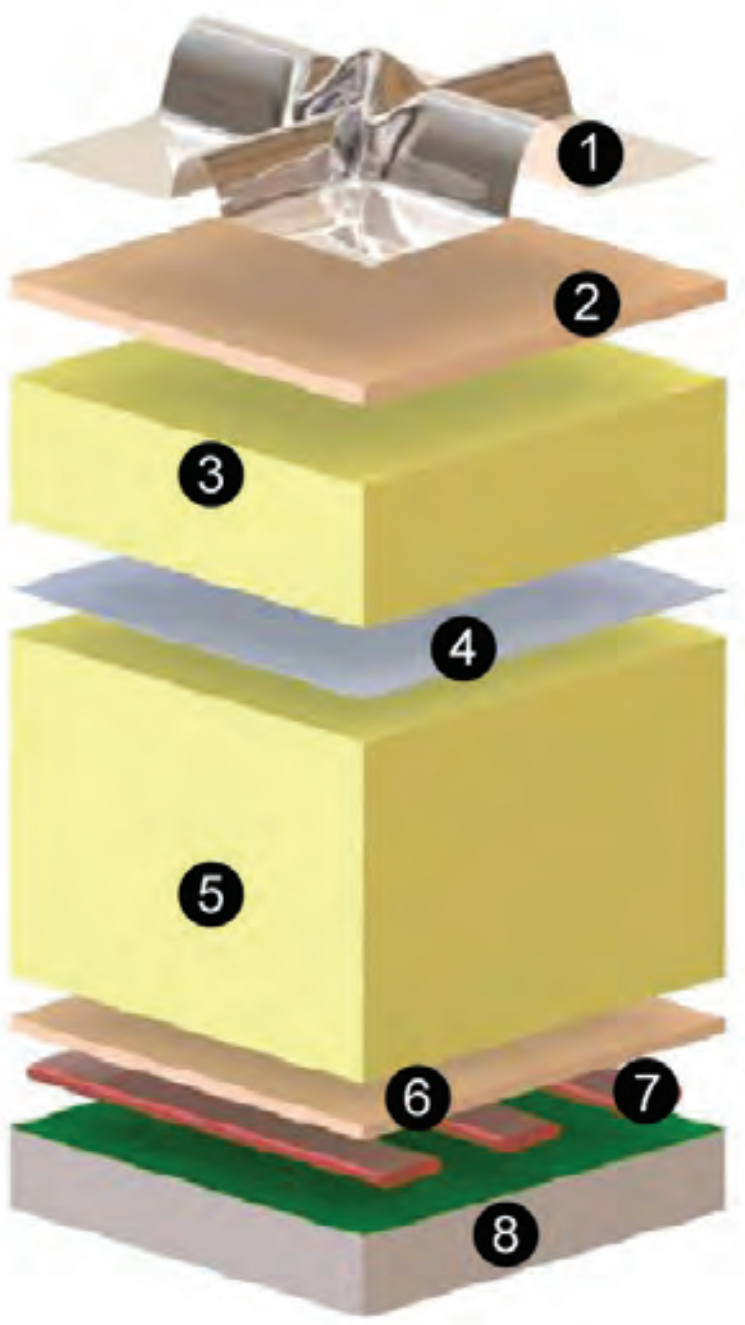

**Figure 3.3.5.7.3-4** Isometric view of one possible construction for the membrane cryostat. The parts, from top to bottom are: 1) stainless steel primary membrane, 2) plywood board, 3) reinforced Polyurethane foam, 4) secondary barrier,5) reinforced polyurethane foam, 6) plywood board, 7) load-bearing mastic, and 8) concrete covered with moisture barrier. [Courtesy LBNE project]

### 3.3.5.8 Detector Options

The ultimate physics goals of the LBNE Science Program cannot be met with a single detector module of either water Cherenkov detector (WCD) or LAr. This leads to the consideration of several options for the configuration of detector modules that can meet the Project's physics goals. Configurations being considered range from two to three water Cherenkov modules with a total fiducial mass of 300 kT, up to three LAr modules with a total mass of around 50 kT, or a hybrid arrangement of modules. The LBNE project is assuming funding would support the construction of two detector modules, each of a size equivalent to the performance of a 100 kT fiducial mass water Cherenkov detector. With this constraint, there are three configurations being studied between the DOE CD-0 and CD-1milestones. Each row of Table 3.3.5.8 shows a possible configuration with either one or two detector types.

The additional cost (~$100 million) of locating a LAr detector at 4850L was not considered worth the physics benefit; however, a shallow location for the LAr detector means an extensive veto system to

| Number of WCD Detectors | WCD Fiducial Mass (kT) | WCD Depth (ft) | Number of LAr Detectors | LAr Fiducial Mass (kT) | WCD LAr Depth (ft) |
|---|---|---|---|---|---|
| 1 | 100 | 4,850 | 1 | 17 | 800 |
| 0 | n/a | n/a | 2 | 17 | 800 |
| 2 | 100 | 4,850 | 0 | n/a | n/a |

**Table 3.3.5.8** The three possible detector location options for LBNE funded from DOE/NSF. A third detector is possible with foreign contribution.

combat background from cosmic rays in proton decay, particularly neutral kaons produced in the adjacent rock that undergo charge exchange in the LAr to produce an isolated $K^+$, mimicking the $p \rightarrow K^+ \bar{\nu}$ decay chain. The LBNE science collaboration is developing reference detector configurations. These configurations include three 100 kT equivalent detector modules and take into account various combinations of technology, and energy thresholds (for water Cherenkov) via photomultiplier coverage (15 to 30%), as well as the possibility of adding gadolinium to the water to enhance neutron capture rates.

Determining what makes the most sensible arrangement for the experiment depends on the results of determining the cost and construction schedule for each technology; it is planned that a down-select be made on the timescale of CD-1.

### 3.3.5.9 Schedule

Advancing the schedule for LBNE and DUSEL will require close coordination between the LBNE project office and the DUSEL Facility in order to carefully choreograph the work necessary for both Projects. In this vein, it is important that technological decisions by the LBNE project be made in close coordination with the DUSEL schedule so that delays and extra costs are minimized.

To this end, the LBNE scientific collaboration has organized a set of working groups to document the physics reach and sensitivity for a complete collection of science goals versus each configuration choice (water Cherenkov, water Cherenkov detector with gadolinium added, LAr, etc.) at the depth believed appropriate for each technology. The science topics include neutrino oscillation physics, proton decay, supernova detection, diffuse neutrinos from relic supernovae, solar neutrinos, ultra-high energy neutrinos, and neutrinos from other astrophysical sources. Using this information, the collaboration, in conjunction with LBNE project management, will make a recommendation on the down-select among the various options, with the expectation of making a configuration choice on the time scale of CD-1.

Clearly it is to the advantage of LBNE to make a technology and depth selection in a timely way so that the progress of both LBNE and DUSEL are not compromised, and the optimum utilization of shared resources can be realized. It is anticipated that the geotechnical work for the large cavities can begin in FY 2011, after the configuration choice is made, including the drilling of boreholes and related laboratory work to assess the rock quality in the vicinity of the proposed location of the large cavities. For siting of the LAr detectors at the 800L, the geotechnical investigations would begin in late FY 2011 or FY 2012.

If water Cherenkov detectors are selected, the excavation of LC-1 begins in FY 2016 and outfitting ends in FY 2018, after which installation of the water vessel, its liner, and the deck can proceed. From the schedule, the long lead item is PMTs, and their procurement begins in 2013, assuming DOE approves CD-3 together with CD-2. Ongoing value engineering studies are considering the possibility of installation of the water vessel in parallel with the cavity excavation. For the two cavities listed in the schedule, commissioning of the experimental equipment for the first would begin in 2021 and the second in 2022.

If a LAr detector is selected, excavation could begin in 2015 after CD-3 approval. The excavation and outfitting duration is about 2.5 years. This would be followed by three years of detector installation with installation of the cryogenic systems, followed by detector installation and filling with LAr. Commissioning of the experimental equipment would begin in 2020.

### 3.3.6 Nuclear Astrophysics

Nuclear astrophysics is concerned with nuclear processes in stars and stellar explosions through charged-particle, neutron, and weak interaction-induced reactions. Critical questions are associated with the origin of elements during the history of our universe, with the sources of neutrino signals from the core of stars and distant supernova explosions, the energy production during stellar evolution and stellar death by explosion, the lifetime of stars and the timescale of stellar explosions. Experimental goals are to identify characteristic new observational signatures associated with stellar processes for neutrino detectors to gamma ray observatories. Experimental nuclear astrophysics is characterized by four major directions: nucleosynthesis processes in stars, which are studied with very low-energy accelerator experiments; explosive nucleosynthesis processes, which require measurements far from stability with radioactive beams; neutron-induced nucleosynthesis in late stellar evolution, which is pursued at reactor and neutron spallation facilities; and, finally, neutrino-induced nucleosynthesis processes, which are still largely confined to theoretical prediction and observation.

The Dakota Ion Accelerators for Nuclear Astrophysics (DIANA) Facility proposed for DUSEL is a next-generation underground nuclear astrophysics accelerator laboratory designed to overcome the experimental limitations of existing state-of-the-art experiments (e.g., the Laboratory for Underground Nuclear Astrophysics [LUNA] at the Gran Sasso Laboratory).[64,65] Once completed, the DIANA Facility will take the leading role in the measurement of critical nuclear-reaction processes at or near stellar-temperature burning conditions. The requirements for the accelerators have been derived from the scientific objects developed by the international astrophysics community.[66] In particular, the described underground accelerator facility will address three fundamental scientific issues in stellar nucleosynthesis: 1) solar neutrino sources and the metallicity of the sun, 2) carbon-based nucleosynthesis, and 3) neutron sources for the production of trans-Fe elements in stars. These are three longstanding, potentially transformational questions of relevance for the understanding of our sun and the chemical evolution of our universe, as outlined in the following section.

#### 3.3.6.1 Nuclear Astrophysics Experiments

Low-energy proton capture and alpha capture reactions have been the focus of intense experimental studies for many decades. However, at the low stellar temperatures associated with these environments, the reaction cross sections are extremely small because of the high Coulomb barrier. This has handicapped all the experimental studies so far and only one of the critical processes, the $^{3}He(^{3}He,2p)^{4}He$ reaction in the pp-chains, has been successfully measured in the solar energy range,[67] in its Gamow window. Nearly all of the “experimental” stellar reaction rates are based on the extrapolation of experimental data, which have been measured at significantly higher energies into the Gamow range.[66] These extrapolations often carry enormous uncertainties, since they require a detailed knowledge of the nuclear structure of the compound nucleus near the particle threshold, as well as a detailed knowledge of the reaction mechanism, different reaction components, and interference effects near the thresholds. These uncertainties often span many orders of magnitude and translate into substantial uncertainties for nucleosynthesis simulations as well as simulations of stellar evolution scenarios.

The experimental difficulties in determining the low-energy cross sections are caused by large background rates associated with cosmic ray-induced reactions, background from natural radioactivity in the laboratory environment, and the beam-induced background on target impurities.[68] An underground location has the advantage that the cosmic ray-induced background is reduced by several orders of magnitude, allowing the measurements to be pushed to far lower energies than now possible. This has been clearly demonstrated at LUNA by the successful studies of critical reactions in the pp-chains[67] and first reaction studies in the CNO cycles.[69]

LUNA is currently the only operating underground facility in the world but there are initiatives for future underground accelerator laboratories in other countries in addition to the DIANA proposal described here. An upgrade of LUNA is being discussed in Italy, since the facility has been extremely successful and has convincingly demonstrated the importance and the advantages of underground accelerator experiments. However, the present LUNA facility is small and limited to the measurement of proton capture reactions below 400 keV with typical proton beam currents between 90 and 400 μA.[65] This is sufficient for capture measurements at low Z target nuclei, but higher beam currents are necessary to extend these measurements into the higher Z range, which is important for stellar burning in massive stars and even in explosive Mg-Al burning in novae. Alpha capture measurements require substantially higher energies than available at LUNA.

The key novel features of the DIANA Facility compared with existing ones will be:

1. The Facility will consist of two accelerators that will cover a wide range of ion beam energies and intensities, with sufficient energy overlap to consistently connect the results to measurements above ground.

2. The Facility beamlines will provide beam to the target stations from both the low and the high energy accelerators. This will allow a particular reaction to be measured with both accelerators in complementary energy ranges with identical target and detector setups. This feature will overcome a major experimental limitation of the currently conducted experiments[65] and will allow DIANA to provide consistent high-precision data over a wide energy range.

3. Additional independent target stations are planned for the 3 MeV accelerator for conducting two experimental campaigns simultaneously or preparing the next experimental campaign. This feature will greatly enhance the ability to carry out the planned science program timely and efficiently, and addresses one of the current limitations at the LUNA Facility, which has only one target station available, since the experimental setups are difficult and time consuming.

4. Both accelerators are designed to be able to incorporate electron cyclotron resonance (ECR) ion sources to increase the beam energy or to vary the accelerated ions (from hydrogen to heavier elements). This unique feature will allow expansion of the scientific goals in the future.

5. The 400 keV low-energy accelerator will be a major technology advance with regard to ion-beam intensity on target in order to address the low count rates close to the Gamow window energies. Advanced target and detector technology will be developed in order to take advantage of its high beam currents.

### 3.3.6.2 Nuclear Astrophysics Candidate Experiments

DIANA, the only candidate experiment, has been proposed by a U.S. collaboration whose goal is to install and operate a deep underground nuclear astrophysics accelerator facility. Led by the University of Notre Dame (UND), it includes the Colorado School of Mines (CSM), Lawrence Berkeley National

Laboratory (LBNL), University of North Carolina at Chapel Hill (UNC), Regis University (RU), Michigan State University (MSU), and West Michigan University (WMU), all having long, distinguished histories in nuclear astrophysics experimentation. They have combined their expertise to develop a state-of-the-art next-generation facility, designed to support a long-term (30+ year) rich and versatile nuclear astrophysics program at DUSEL.

The infrastructure requirements for the DIANA underground accelerator facility at DUSEL are shown in Table 3.3.6.2.

| Requirement | Value/Description | Comment/Justification |
|---|---|---|
| **Layout** | | |
| Depth | 4850L | |
| Footprint | 45 m L x 17 m W | Area usable by experiment |
| Height [m] | 15 | 19 m max usable height at high-energy accelerator dome, see text for details |
| Floor Load [kPa] | 12 | Live load, 3,000 lb concrete requested |
| **Utilities** | | |
| Power [kW] | 1500 | |
| Standby Power [kW] | 100 | |
| Chilled Water [kW] | 1200 | |
| Waste Heat to Air [kW] | 300 | |
| Purified Water [$m^3$] | 0 | |
| Potable Water [lpm] | Nominal use | |
| Compressed Air | Nominal use | |
| Network [Gb/s] | 1 | |
| **Environment** | | |
| Temp. Min [ºC] | 20 | |
| Temp. Max [ºC] | 25 | |
| Humidity Min [%] | 20 | |
| Humidity Max [%] | 30 | |
| Rn Background [$Bq/m^3$] | 30-100 | Experiment to provide Rn scrubbing to achieve |
| **Crane** | | |
| Max. Load [Short Tonne] | 10 | |
| **Occupancy** | | |
| Peak Installation Occupancy [count] | 20 | |
| Installation Duration [months] | 24 | |
| Peak Commissioning Occupancy [count] | 12 | |
| Commissioning Duration [months] | 18 | |
| Peak Operation Occupancy [count] | 6 | |
| Operation Duration [months] | >360 | |

| Requirement | Value/Description | Comment/Justification |
|---|---|---|
| **Cryogens** | | |
| LN Storage | 200 L | |
| LN Consumption | 100 L/day | Access to a $LN_2$ refilling station is requested (ideally at the 4850L) |
| **Major Hazards (Other Than Cryogens)** | | |
| High Voltage | 400 kV<br>60 V<br>5 kV | Low-energy accelerator power supply<br>Magnets power supply<br>Detector power supplies |
| Compressed Gases | H, He, N, Ne, Ar | Small quantities, <1L at STP, injected into ion sources |
| Pressure Vessels | $SF_6$ 4700L at 6 bar | High-energy accelerator tank, $SF_6$ storage tank |
| Asphyxiation | $SF_6$ 4700L at 6 bar | |
| Radioactive Sources | Low activity sealed types | |
| **Assay and Storage** | | |
| Assay Needs | Nominal | Access to Low Counting Facility for material and detector selection |
| Underground Storage | TBD | Low Rn storage area during installation (short term) |

**Table 3.3.6.2** DIANA Facility requirements.

### 3.3.6.2.1 Location and Space Requirements

The specific requirements for DIANA are presented here. In a number of aspects these differ from the baseline design of Laboratory Module 1 (LM-1) described in Volume 5. If the DIANA experiment were selected to be among the first experiments at DUSEL, modifications to LM-1 would be required. The purpose of this and subsequent sections is to present a preliminary version of the DIANA-specific requirements.

The minimum depth required to achieve successful science goals for an underground nuclear astrophysics accelerator facility should be at least similar to the 3100 mwe depth at LNGS,[68] where the LUNA accelerator facility has been operating since 1994.[64.65.67.69] Therefore, the 4850L (about 4300 mwe) at DUSEL will be adequate for the cosmic ray-induced background reduction into detectors. Low-radioactivity concrete (comparable to the radioactivity of the natural rock) in the construction of the walls and floor of the LM (comparable to the natural rock radioactivity) would be beneficial. Air ventilation with low radon content would also be beneficial within the cavity (ideally, at least like the LNGS, maximum 30-100 $Bq/m^3$), together with epoxy sealing to prevent Rn permeation through walls and floor. However, low-radioactivity concrete and low Rn (surface air) ventilation are not part of the current baseline Facility design but are recognized as improved scope options for the Facility.

The DIANA accelerator cavity dimensions shall be $20 \times 50 \times 20$ $m^3$ (W × L × H, S4 proposal) and shall have an underground control room of $8 \times 8 \times 3$ $m^3$ (W × L × H) usable dimensions, placed outside the accelerator cavity, along the DIANA entrance drift, as close as possible to the entrance (Figure 3.3.6.2.2-1). The control room requirements are given in Table 3.3.6.2.1.

The high-energy accelerator tank will require special handling due to its size and weight—3 m diameter, 6.5 m length, 8-10 tonne weight. It will be split into subparts to fit cage size limitations and assembled underground. The possibility to sling the tank under the Yates Cage, and avoid splitting it into parts, will be investigated.

The required height for the high-energy accelerator tank installation and operation is 19 m, and shall be placed between two shielding walls as shown in Figure 3.3.6.2.2-1. The required height for the low and high energy areas (left and right sides of Figure 3.3.6.2.2-1) is compatible with the proposed 15 m hook height of LM-1.

The analyzing magnets will require special handling because of the weight. The DIANA accelerator cavity should have cranes with at least 10 tonne load capacity to accommodate the installation of heavy equipment. The cranes should cover the cavity extension.

| Requirement | Value/Description |
| --- | --- |
| **Layout** | |
| Depth | 4850L, as close as possible to the accelerator cavity egresses (see Figs. 3.3.6.2.2-1 and 3.3.6.2.2-2) |
| Footprint | 8 × 8 $m^2$ (W × L, usable area) |
| Height | 3 m (usable height) |
| Floor Loading | 100 psf live load, 2,000 lb concrete |
| **Utilities** | |
| Power [kW] | 15 |
| Standby Power [kW] | 15 |
| Waste Heat to Air [kW] | 15 |
| Network | At least 1 Gb/s |
| **Environment** | |
| Temperature [ºC] | Standard office environment |
| Humidity [%] | Standard office environment |

**Table 3.3.6.2.1** DIANA underground accelerator control room requirements.

### 3.3.6.2.2 Shielding Requirements: Water Doors and Mazes

Although the deep underground site will reduce the cosmic ray background rate, it must be anticipated that gamma and neutron radiation from decay and reaction processes in the natural underground environment will generate a fairly high background level in the DIANA Facility detector systems.[70] Since the experimental count rate will be extremely low at the energies of astrophysical relevance, the detectors will have to be shielded against the environmental background. This passive shielding will be part of the detector design but also will be advantageous to shield against beam-induced radiation at higher beam energies and is designed to reduce the level of beam-induced radiation below the natural radiation level of the underground environment

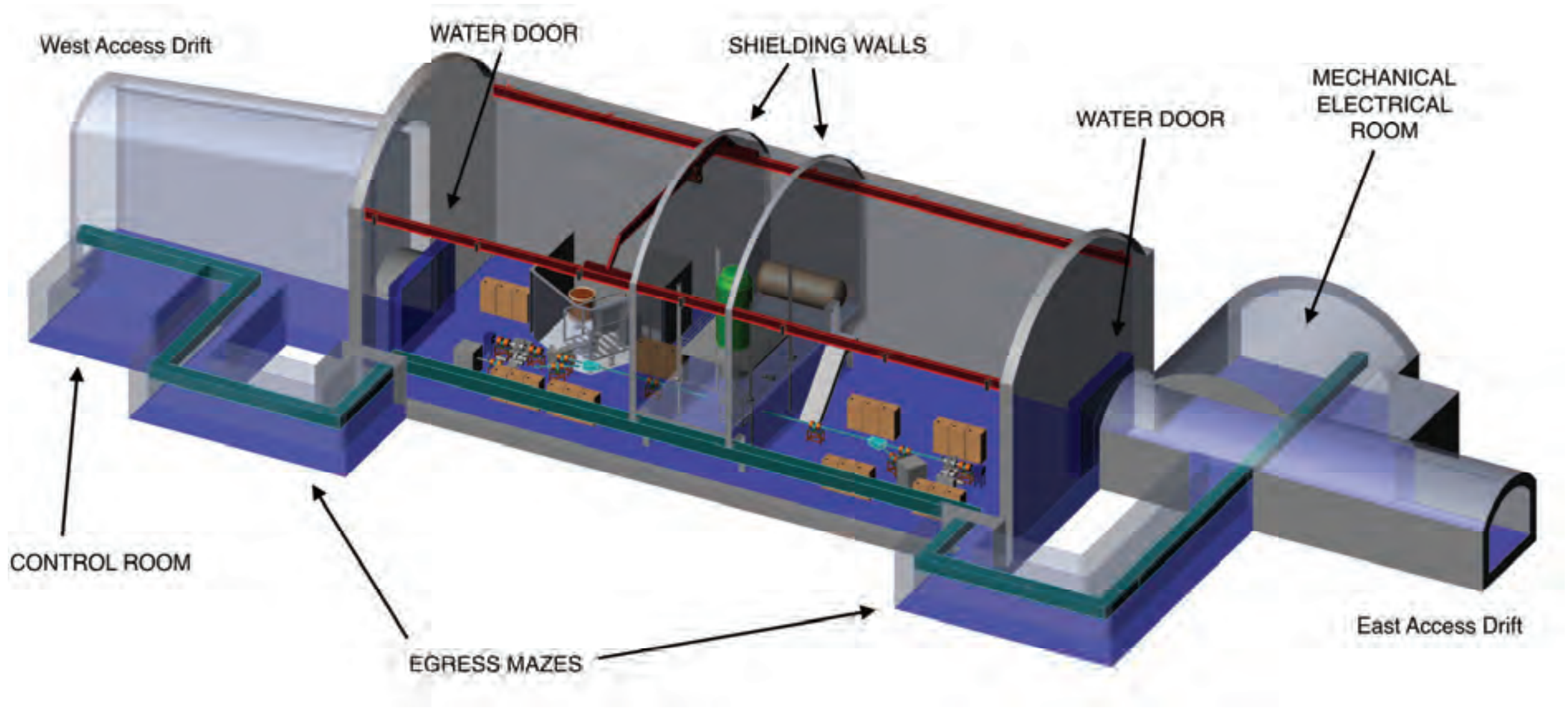

**Figure 3.3.6.2.2-1** Drawing of the required water-shielded doors, and egress mazes that will guarantee that the natural radiation field will be unaffected outside the DIANA Facility cavity. [Courtesy DIANA collaboration]

The DIANA collaboration carried out initial simulations making use of the neutron spectrum produced by the $^{13}C(\alpha,n)^{16}O$, $^{17}O(\alpha,n)^{20}Ne$, $^{22}Ne(\alpha,n)^{25}Mg$, reactions belonging to the DIANA scientific program, assuming high beam intensities (1 mA at 3.0 MeV beam energy $E_{nmax}$ = 5.2 MeV). In addition, the amount of beam-induced radiation that could occur at higher energies if the beam interacts with beam slits or apertures was evaluated. Prompt gamma-ray emission from a beam dump in a high beam-current intensity run can be shielded with relatively simple passive elements, such as a lead shield. DUSEL experimental cavities will be separated by at least 40 m of rock, which effectively shields any beam-induced gamma or neutron fields to negligible levels. However, neutron scattering can occur in the entrance drifts, and needs to be mitigated. For this purpose, the cavity will be designed to include water-shielded doors and specially shaped egresses to effectively reduce any beam-induced radiation outside the DIANA cavity to below the natural radiation levels of the rock walls in the drifts. To verify and optimize the geometry of the shielded doors and egress shape for the DIANA Facility, GEANT4 simulations have been carried out (Figure 3.3.6.2.2-2). During the operation of the DIANA High Energy Accelerator, the water-shielded doors are to stay closed when performing experiments with non-negligible radiation production.

Since the space available at LM-1 is consistent with DIANA space requirements, it has been considered as a possible DIANA location. The proposed experimental layout is shown in Figure 3.3.6.2.2-1. Some modifications to the generic design of LM-1 will be necessary to meet the shielding requirements of the DIANA Facility, if DIANA were selected to be among the initial experiments at DUSEL. The modified plan view of LM-1 with water-shielded doors and the secondary egress mazes is shown in Figure 3.3.6.2.2-2.

In addition, the installation of the water-shielded doors will require routing of the utilities through the secondary egress mazes. Where the utility routing intersects any personnel access drift, the utilities will need to be routed overhead or below grade to prevent possible safety hazard. The nominal cross section of these drifts is $3 \times 3\ m^2$. The roof of these drifts may be crowned at discretion.

In the following, the deviations between the DIANA layout and the generic design of LM-1 are summarized.

1. Egress drifts added on both ends of LM-1 as discussed above and as shown in Figures 3.3.6.2.2-1 and 3.3.6.2.2-2.

2. Western access drift offset 2 m south from the LM-1 center-line as shown in Figure 3.3.6.2.2-2. This is to enable the 8 m wide shielding door to clear the western access drift opening.

3. Mechanical Electrical Room (MER) (Utility Room in Figure 3.3.6.2.2-2) for LM-1 moved eastward by 2 m. In the current design, the western wall of the MER is collinear with the eastern end wall of LM-1. Also in the current design there is a chamfer in the corner between the MER and LM-1 that has been removed. These changes were made so the shielding door can close adequately to provide the intended shielding.

4. In elevation, the 4-m excavation for the entire LM-1 floor is not required for DIANA and has been removed. In other words, both western and eastern access drift floors, the LM-1 floor (with the exception noted in deviation 5 below), and the control room floor are all at the same elevation.

5. In elevation, the area of the LM-1 floor inside the low-energy accelerator shielding room needs to be excavated 1 m deep with respect to the remainder of the LM-1 floor for high-voltage standoff. This represents approximately 64 $m^3$ of excavated material. The location and exact shape of the low-energy accelerator shielding room are still approximate.

6. The height of LM-1 has been reduced from 24 m to 20 m. The reduced 20-m height is also contingent on the requested 1 m local excavation described in deviation 5 above, and a 15 m minimum hook height for the bridge crane.

7. The excavation sequence for the generic LM-1 yields a ramped ceiling for the western access drift. This, however, leaves a very tall opening adjacent to LM-1. The size and cost of the shielding door required to cover this opening would be excessive. An alternate solution is needed to return the opening for the western access drift to the size of the eastern access drift, approximately 6 m wide by 5 m high. One potential solution is to backfill the top of the western access drift opening with concrete, to form a 2 m thick plug.

8. A small local control room ($8 \times 8 \times 3$ $m^3$, see Table 3.3.6.2 and Figures 3.3.6.2.2-1 and 3.3.6.2.2-2) is requested outside the main cavity to house the local control room of the accelerators and data-acquisition systems. Its location is not critical but it does need to be spaced the minimum distance from the egress drift, as shown in Figure 3.3.6.2.2-2.

Reducing the DIANA cavity height (deviation 6) will offset costs associated with the additional excavation for the secondary egress mazes.

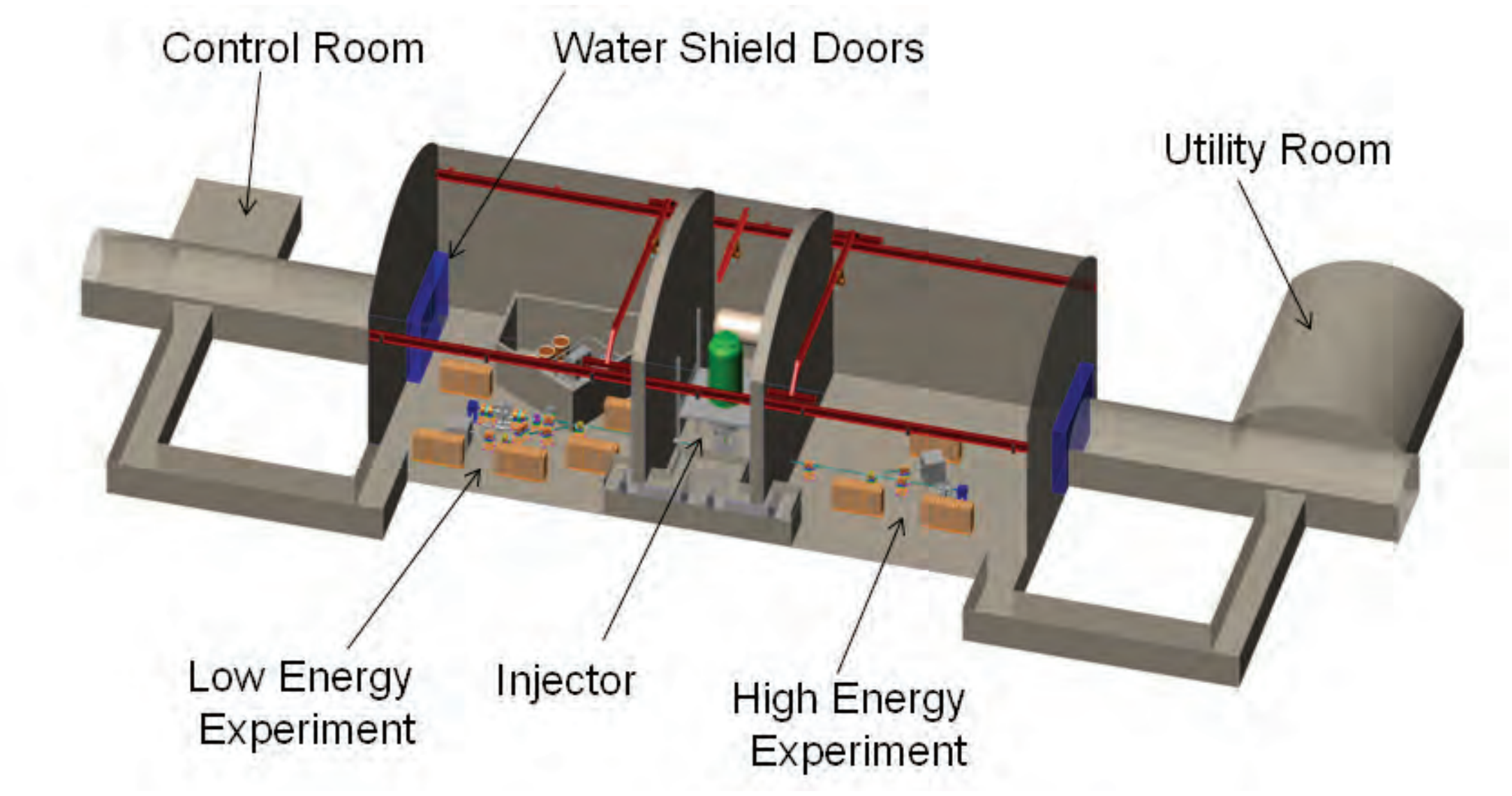

**Figure 3.3.6.2.2-2** Footprint of the required shielded doors, mazes, and control room. [Courtesy DIANA collaboration]

#### 3.3.6.2.3 Schedule and Installation

DIANA will develop Preliminary Design documents for the astrophysics accelerator facility by end of FY 2012. That effort is currently funded by an NSF S4 grant. The ion optics (physics) design together with the conceptual engineering design of the facility will be completed by mid-FY 2011.

The DIANA collaboration is in a unique situation regarding its project maturity. Previously, the low-energy accelerator section had already been designed in fair detail—funded by internal LBNL laboratory-directed research funds before the official DUSEL Project start. In addition, main Project components, e.g., the beamline magnets and the high-energy accelerator, will be procured through commercial companies based on existing technology. Therefore, the DIANA detail engineering effort is minimized and is primarily focused on installation and integration planning.

After completion of the NSF S4 project, DIANA staff would be able to immediately phase into the final engineering planning stage. Based on S4 engineering work, the beamline magnets and the high-energy accelerator procurements can be placed immediately. The low-energy accelerator fabrication could start a few months after completion of the S4 project.

The DIANA collaboration could initially install the low-energy accelerator aboveground while waiting for the underground LM to be completed. This shakedown period could reduce project risks and at the same time allow for the establishment of a productive science program well in advance of first operation at the DUSEL Facility. The location for temporary early deployment of the low-energy accelerator remains to be determined but would not be at the DUSEL site. The high-energy accelerator is planned to be directly installed in the underground LM-1 at 4850L.

### 3.3.7 Biology, Geology, and Engineering (BGE) Experiments

A general overview of the science goals of experiments in biology, geology, and engineering has been presented in Chapter 3.2. In this section, potential BGE experiments based on the approved S4 proposals are described.

### 3.3.7.1 Facility for the Study of Geologic Carbon Sequestration

This proposed Facility is currently the only deep underground laboratory in the world being designed for the controlled study of geologic carbon sequestration. The findings from this experimental Facility will advance carbon-management technology worldwide and help reduce global greenhouse gas emissions.

#### 3.3.7.1.1 Overview of Proposed Research Facility

The proposal is to build an underground experimental Facility to study the vertical flow of $CO_2$ through porous media over realistic length scales that mimic deep sedimentary formations. The Facility is dubbed LUCI, for Laboratory for Underground $CO_2$ Investigations. LUCI is being designed to include three pressure vessels, each with a length of 500 m and a diameter of 1 m. The vessels will be supported within a 3m x 3m vertical shaft and will have an inner column that will be used for housing sensors. The annular space between this column and the outer vessel wall will be filled with brine and sand or other relevant geological material that mimics the strata encountered in sedimentary basins prior to $CO_2$ injection. Thermal and pressure gradients along the length of the columns will mimic real subsurface conditions.

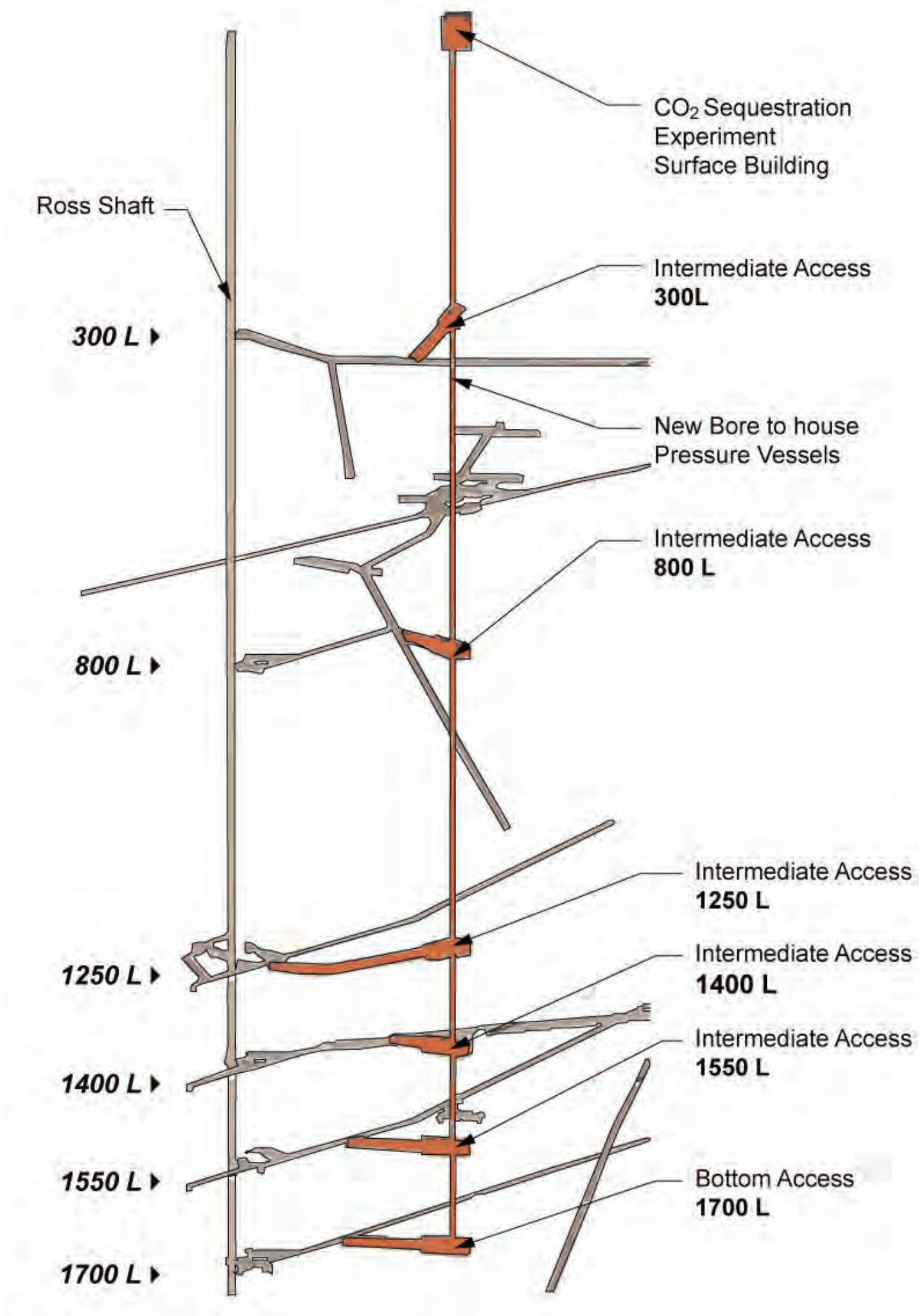

**Figure 3.3.7.1.1** Proposed layout of the LUCI Facility at DUSEL. [DKA]

Within DUSEL, a new vertical borehole will be located ~150 m from the Ross Shaft and will extend from the Surface down to the 1700L (Figure 3.3.7.1.1). Key to the experimental design is the ability to make measurements and sample fluids along the length of the flow columns. The location of the shaft was selected to allow intermediate access to the pressure vessels from the 300L, 800L, 1250L, 1400L, and 1550L. At these levels, new excavations will connect existing drifts with the new shaft. In addition, access to the columns will be possible from an Alimak vertical transporter, an access device that uses vertical track installed on the side of the raise bore to climb and descend within the raise bore.

A complete design could be available by mid-2012. Assuming funding availability for construction, the work on the surface building and site could begin at that time and could be completed in six months. The next step would be the excavations, which would take six months to complete. This could start after the refurbishment of the Ross Shaft. After the excavations, the remaining procurement, fabrication and assembly of the $CO_2$ experimental Facility would take two years.

#### 3.3.7.1.2 Proposed Experimental Investigations

The LUCI Facility will test critical hypotheses needed to understand $CO_2$ vertical flow in the deep subsurface. A goal is to simulate a leak in which $CO_2$ changes from a supercritical fluid to a subcritical gas as it flows up the column. The acceleration in $CO_2$ flow due to increasing buoyancy will be measured, and the extent to which this acceleration is mitigated by Joule-Thomson cooling will be determined. In other experiments involving rock matrices and well cements, $CO_2$-water-rock interactions will be examined and it will be determined whether $CO_2$ will enlarge flow pathways (mineral dissolution) or cause self-sealing (mineral precipitation). Finally, the effects of anaerobic, thermophilic bacteria on $CO_2$ conversion to methane and carbonate will be investigated.

Sensors will monitor governing thermal, physical, and chemical processes. Each vessel will have an inner fluid-filled tube (0.25-m inner diameter) that serves as a proxy well to accommodate a variety of existing well-logging technologies, which provide a testing platform for the development of logging suites specifically supporting $CO_2$ sequestration. For example, a combinable nuclear magnetic resonance tool will be used to discriminate between water- and $CO_2$-filled pores; similar measurements will be conducted using a reservoir saturation tool. Sonic and ultrasonic tools will also be used to image fluids, using differences in acoustic impedance to distinguish liquid from gas phases. These measurements will be used to construct a vertical saturation profile, and to determine how it changes over time as the $CO_2$ plume moves upward. Distributed temperature and pressure sensors will also be deployed outside of the inner tube to provide continuous in situ data. The various data and interpretations from the suite of technologies deployed during the experiments will collectively be used to develop a better understanding of $CO_2$ migration and trapping processes over realistic vertical length scales, to calibrate models that predict the vertical flow of $CO_2$ and brine in porous media, and to compare the spatial resolution and sensitivity of different monitoring tools.

The requirements from DUSEL anticipated for the LUCI Facility are summarized in Table 3.3.7.1.2.

| Requirement | | | |
|---|---|---|---|
| **Layout** | | | |
| Depth | Surface | 300, 800, 1250, 1400, and 1550 Levels | 1700L |
| Footprint [$m^2$] | 30 x 30 | 6 x 20 | 10 x 20 |
| Height [m] | 15 | 5 | 8 |
| Floor Load [kPa] | 3900T (total load) | 98 | 98 |
| Raise Bore | 3m diameter, Surface to 1700L | 3m diameter, Surface to 1700L | 3m diameter, Surface to 1700L |
| **Utilities** | | | |
| Power [kW] | 60 | 60 | 60 |
| Standby Power [kW] | 10 | 10 | 10 |
| Potable Water [lpm] | 40 | 40 | 40 |
| Compressed Air | Nominal use | No | Nominal use |
| Network [Gb/s] | 1 | 1 | 1 |
| **Environment** | | | |
| Temp. Min [ºC] | 20 | 5 | 5 |
| Temp. Max [ºC] | 27 | 40 | 40 |
| Humidity Min [%] | 5 | 0 | 0 |
| Humidity Max [%] | 95 | 100 | 100 |
| **Crane** | | | |
| Max. Load [Ton] | 400 | 0 | 0 |
| **Occupancy** | | | |
| Peak Installation Occupancy [count] | 20 | 5 | 8 |
| Installation Duration [months] | 28 | 8 | 6 |
| Peak Commissioning Occupancy [count] | 10 | 0 | 2 |
| Commissioning Duration [months] | 6 | 2 | 2 |
| Peak Operation Occupancy [count] | 2 | 0 | 0 |
| Operation Duration [months] | 60 | 60 | 60 |
| **Major Hazards (Other Than Cryogens)** | | | |
| $CO_2$ | 600,000 l | 600,000 l | 600,000 l |
| Pressure Vessel | 10MPa | 10MPa | 10MPa |
| **Assay and Storage** | | | |
| Underground Storage | Storage Tanks | 0 | Storage Tanks |

**Table 3.3.7.1.2** Requirements for the LUCI Facility.

### 3.3.7.2 Facility for Monitoring Deformation of Large Underground Rock Masses

Large-scale deployment of fiber-optic sensors appears to be an ideal technology for multispatial and multitemporal measurement of rock-mass response to loading. Fiber-optic monitoring along kilometers of drifts within DUSEL presents a unique opportunity to address questions regarding the mechanical and hydrologic response of rock masses. This effort will result in the world's largest and deepest underground network of fiber-optic strain and temperature sensors, and tiltmeters.

#### 3.3.7.2.1 Overview of Proposed Research Facility

Fiber-optic strain and temperature sensors have been used successfully in civil engineering applications for the structural-health monitoring of bridges, highways, dams, and buildings. They are recognized as a relatively inexpensive, lightweight, versatile, and long-lasting way to monitor structures. Fiber-optic sensors also have great geotechnical potential for monitoring the safety and stability of mines, tunnels, and cavities. This Facility represents the first comprehensive installation of fiber-optic strain and temperature sensors to measure deformation and temperature in a large volume underground. A central component of the effort is to develop underground applications of fiber-optic sensors for scientific and structural health monitoring applications. The main reason for using this emerging fiber-optic technology lies in the cost and efficiency advantages for over-kilometer-length deployments, long-term stability, flexibility of incorporating many types of sensors on a single data-acquisition cable, and a future promise for the methods becoming even better, cheaper, and faster. The fiber-optic network will be supplemented with long-baseline tiltmeters and borehole extensometers.

#### 3.3.7.2.2 Proposed Experimental Investigations

Induced—dewatering, drift and cavity construction, meter-scale loading—and natural—self-weight, Earth tides, seismicity—loading will be monitored at spatial scales ranging from centimeters to hundreds of meters and temporal scales ranging from milliseconds to decades. The deformation-monitoring network consists of installations of fiber Bragg grating (FBG) sensors, Distributed Strain and Temperature (DST) sensing fiber-optic cable, Distributed Temperature Sensing (DTS) fiber-optic cable, and water-level tiltmeters. Sensors will be installed in the drifts of deeper accessible levels, specifically at the 2000L, 4100L, 4850L, 6800L, and 7400L. Each installation of the various sensor sets is designed to work with pre-existing or planned spaces within DUSEL. Sensors will also be installed near planned excavations to record rock deformation before, during, and after construction of large cavities and other LMs to understand how tunnels and large rooms redistribute stress and contribute to straining, activation of sudden failures, and seismicity within the Facility.

The DST fiber will be installed over large areas of drifts within the Laboratory to measure convergence and pillar deformation (Figure 3.3.7.2.2). Continuous fiber will be tensioned with anchors every 0.5 to 2 m along its length. These anchors can be attached to existing rock bolts or to newly installed bolts. The DTS fiber will be placed in winzes down to the maximum depth achievable (currently planned to be 7,700 feet), in shorter-length boreholes, and along drift walls to monitor water inflows and air movement. The tiltmeter array consists of water-level sensors installed within the drifts connected by water and air tubing to measure micrometer-level displacements over a baseline of tens to hundreds of meters.

The combination of different types of sensing techniques will create a network for monitoring strain from the centimeter scale to lengths exceeding 1 km. The tiltmeter arrays will measure deformation over length scales between 30 and 1,000 m. The measurement of rock-mass properties over many spatial and temporal scales requires sensors and instruments that are embedded and stable. The sensing array will take advantage of deformations induced by natural forces such as Earth tides and distant earthquakes, as well as dewatering of the underground facility and construction within the Laboratory. In addition, active experiments on the scale of several meters will be performed using different types of jacks and fluid injection.

Because these fiber-optic sensors have not previously been installed in intact rock, part of the experiment will examine various installation methods for FBG and DST sensors to establish viable mounting techniques that accurately record deformation in the intact rock mass not near-surface deformation of stress-relieved drift wall or artifacts of the installation procedure. In addition to the methods described above, different embedding technologies will be used, including pliable rock strain strips and instrumented rock bolts and cable bolts. To address objectives of this Project, the sensor and tiltmeter arrays will be monitored continuously for changes in strain and temperature. The data from the sensors will be combined with laboratory deformation experiments and finite element modeling to determine the elastic moduli of the rock mass and how they vary over spatial and temporal scales. The deformation-monitoring array will utilize sensors overlapping in their spatial coverage to check for the accuracy and repeatability of these results.

Requirements from DUSEL anticipated for the Rock Deformation Facility are summarized in Table 3.3.7.2.2.

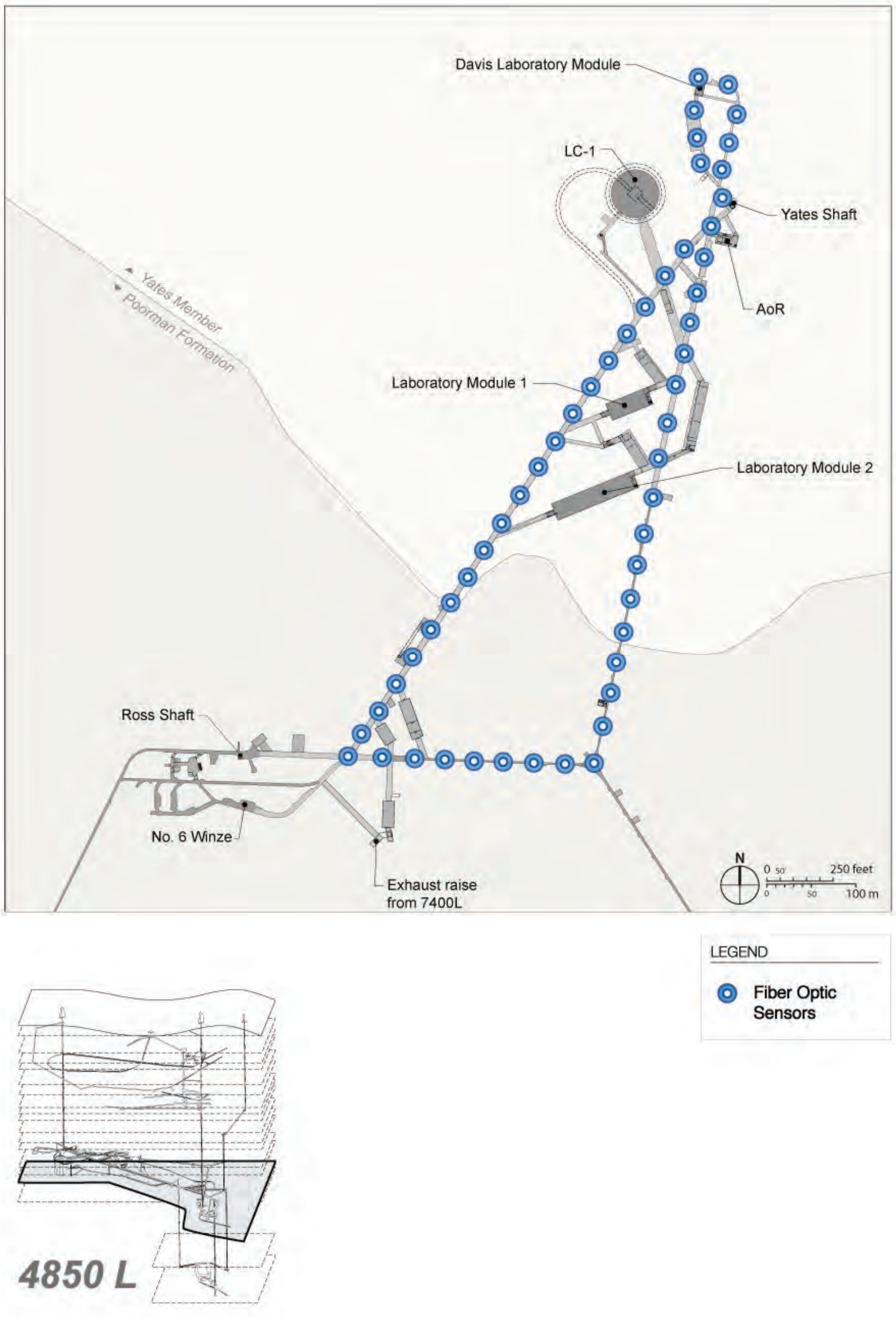

**Figure 3.3.7.2.2** Proposed layout of Distributed Strain and Temperature (DST) fiber-optic sensors (blue dots) within the general area of the LMs, to monitor convergence and deformation of rock. [DKA]

| Requirement | Value/Description |
|---|---|
| **Layout** | |
| Depth | 2000, 4100, 4850, 6800 and 7400 Levels |
| Footprint [m$^2$] | 2.1 x N |
| Height [m] | 3 |
| Floor Load [kPa] | 10 |
| **Utilities** | |
| Power [kW] | 2.5 |
| Standby Power [kW] | 0 |
| Potable Water [lpm] | 16 |
| Compressed Air | Nominal use |
| Network | 1 Gb/s |
| **Environment** | |
| Temp. Min [ºC] | 5 |
| Temp. Max [ºC] | 40 |
| Humidity Min [%] | 0 |
| Humidity Max [%] | 100 |
| **Crane** | |
| Max. Load [T] | 0 |
| **Occupancy** | |
| Peak Installation Occupancy [count] | 10 |
| Installation Duration [months] | 24 |
| Peak Commissioning Occupancy [count] | 2 |
| Commissioning Duration [months] | 6 |
| Peak Operation Occupancy [count] | 0 |
| Operation Duration [months] | 96 |

**Table 3.3.7.2.2** Requirements for the Deformation Monitoring Facility.

### 3.3.7.3 Facility for Studying Coupled Thermal-Hydrological-Mechanical-Chemical-Biological (THMCB) Processes

A large-scale THMCB experimental Facility at depth (4,850 to 7,400 feet) will allow researchers to quantitatively probe the range of coupled THMCB processes taking place at the pore scale, in meter-scale fractures, and within decimeter-scale fluid flow and convection regimes, for time periods of several to tens of years.

### 3.3.7.3.1 Overview of Proposed Research Facility

The purpose of the DUSEL THMCB experimental Facility is to investigate a range of natural and engineered processes by creating a volume of heated rock and fluid that will be instrumented with sensors (mechanical, thermal, hydraulic) and ports for collecting fluid samples (chemical, biological) as a function of space, time, and temperature (Figure 3.3.7.3.1). It is expected that observation/measurement boreholes will be sited to traverse different regions of the heated rock, which are packed-off to isolate a particular fracture or fracture set, into which fluids, gases, or nutrients can be injected to perturb the local THMCB environment. Monitoring ports will be sited along fractures to capture fluids that have been injected elsewhere along the same fracture. In addition to geochemical and isotopic (stable and radiogenic) analyses on sampled fluids, gases and solids, state-of-the-art in situ sampling and monitoring sensors will be employed. The experiments performed at the THMCB Facility at DUSEL would be carried out in a phased approach. The team, with external input and peer review, will refine the necessary initial data, experiments, and modeling that should be performed, prior to starting experiments.

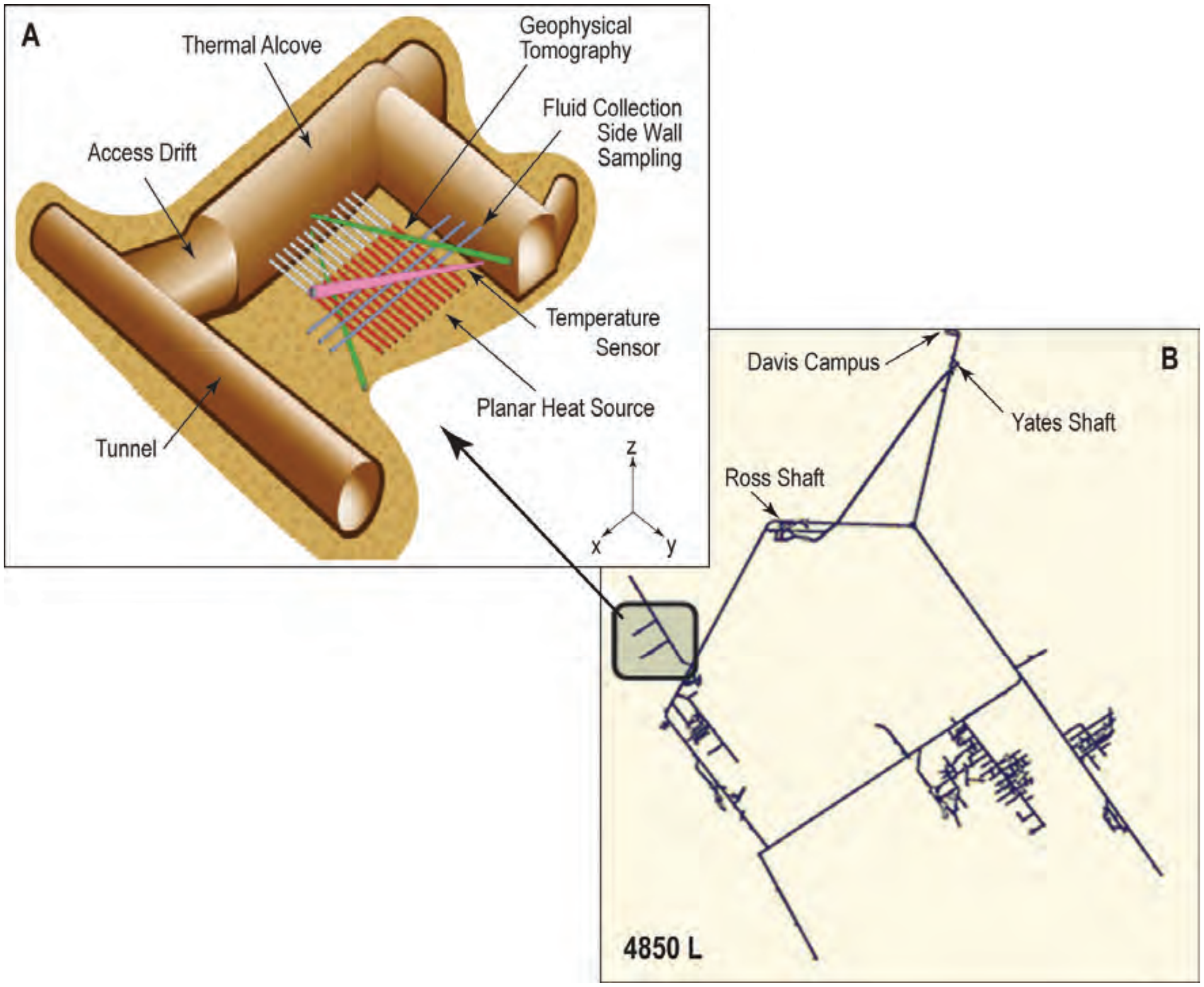

**Figure 3.3.7.3.1** A) conceptual layout of the THMCB Experimental Facility, and B) Location of a potential site for the Facility at the 4850L. [Courtesy THMCB collaboration]

#### 3.3.7.3.2 Proposed Experimental Investigations

Over 4,850 to 7,400 feet depths, lithostatic pressures are significantly greater than those encountered by other experiments that have probed coupled THMCB processes. The metamorphic mineral assemblages making up the rocks at DUSEL are different from the rhyolitic tuffs and granites that have been the host rock of these tests. At DUSEL, rocks are much more anisotropic and are chemically and structurally more heterogeneous. The Fe-rich carbonates and mafic silicate minerals have typically higher dissolution rates than quartz and feldspars in granitic rocks and devitrified tuff. Reaction rates are also highly dependent on reactive surface areas, which in turn are a function of the hierarchical scale of fluid flow, geologic structure, and mineral fabric. Hence, the well-developed metamorphic fabric of the Homestake iron formation and the adjacent lithologies will provide a unique system in which to monitor directional fluid flow and reaction-transport processes under a well-controlled thermal environment. The abundance of ore minerals is an added benefit for the study of the transport, precipitation, and sorption of metals under variable temperature conditions and in fractured rock.

The transport and interaction of fluids, heat, and chemical reactants within a stressed geologic host result in complex feedbacks at a variety of length and time scales. These interactions produce patterns of reaction and mineral redistribution that in turn modify porosity and permeability and are strongly scale dependent. Well-controlled injection/extraction experiments in particular regions of the heated block can be interrogated by in situ probes and sampling, with supporting laboratory experiments, isotopic and (bio)geochemical measurements and reactive transport modeling. This collaboration proposes to address a variety of questions related to these interactions. These include:

- What are the effective reaction rates between minerals and fluids in fractured rock, and how are they controlled by the evolution of the fracture-fluid interface as a function of reaction progress?
- How does the chemistry of fluids and minerals affect the mechanical behavior of fractures, sealing, and permeability evolution under stress?
- At what rates under specific flow and temperature conditions are metals mobilized through water-rock interaction, transported, and concentrated through sorption and/or mineral precipitation (relevant to ore deposition and contaminant transport/immobilization)?
- How do microbiological communities in rocks evolve and migrate in fractured rock undergoing changes in temperature and geochemical environment?
- How does mineralogical and permeability heterogeneity at small scales affect the composition of fluids at a larger scale and how can the effective reaction rates be interpreted from the fluid compositions?

Requirements from DUSEL anticipated for the THMCB Facility are summarized in Table 3.3.7.3.2.

| Requirement | |
|---|---|
| **Layout** | |
| Depth | 4850L |
| Footprint [$m^2$] | 80 x 60 |
| Height [m] | 60 |
| Floor Load [kPa] | 98 |
| **Utilities** | |
| Power [kW] | 300 |
| Standby Power [kW] | 1 |
| Purified Water [$m^3$] | .016 |
| Potable Water [lpm] | 16 |
| Compressed Air | Nominal use |
| Network | 1 Gb/s |
| **Environment** | |
| Temp. Min [ºC] | 20 |
| Temp. Max [ºC] | 30 |
| Humidity Min [%] | 80 |
| Humidity Max [%] | 98 |
| **Crane** | |
| Max. Load [T] | 0 |
| **Occupancy** | |
| Peak Installation Occupancy [count] | 10 |
| Installation Duration [months] | 12 |
| Peak Commissioning Occupancy [count] | 4 |
| Commissioning Duration [months] | 4 |
| Peak Operation Occupancy [count] | 2 |
| Operation Duration [months] | 48 |
| **Major Hazards (Other Than Cryogens)** | |
| Electrical | 300 kW |
| **Assay and Storage** | |
| Underground Storage | Storage Tanks |

**Table 3.3.7.3.2** Requirements for the THMCB Facility.

### 3.3.7.4 Facility for Ecohydrology Studies of Deep Fractured Rocks

DUSEL offers a unique opportunity for integrated geobiological research because of its enormous span of depth and because the host formation comprises a diverse assemblage of minerals and rock. Although groups in other countries are actively researching the subsurface, none has access to a deep, dedicated science facility.

### 3.3.7.4.1 Overview of Proposed Research Facility

The Facility will provide a platform for precise, systematic drilling into the biosphere of a deep continental environment. With kilometer-scale access in three dimensions and a multidecade observational lifetime, it will provide an in situ laboratory for the development of detailed hydrologic and geomechanical conceptual models in complex rock.

The Facility will enable access to groundwater of diverse ages (Figure 3.3.7.4.1). On the south side of the underground workings, water flows from the surface to 1-km depth in less than a year. Water reaching the lower depths of the north side of the underground open areas emanates from rock pores and may be millions of years old. The potential capture footprint extends outward for kilometers from existing underground facility workings and could provide access to about 100 $km^3$ of rock for hydrologic, geomechanical, and microbial biogeographic studies.

The Facility will consist of a series of distributed boreholes drilled/cored sequentially from the surface to ~5-km depth. Drill-site locations have been selected to allow multiple boreholes that will interrogate large volumes of minimally impacted fractured rock in generally north and south directions from the former mine site. Cavities designed to require minimal expansion of existing drifts will accommodate a drill rig and associated supplies. Scheduled activities at each site will be mobilization, drilling/coring, installation of instrumentation, demobilization, and long-term experimentation and monitoring (sporadic visits, months to years). Coring will rely on state-of-the art quality-control procedures for geobiological sampling, including a steam-cleaning station for the drill rods, on-site reverse-osmosis filtration system for the drilling water, a single-pass reverse-flow system for the drilling water, and multiple tracers for potential drilling contamination.

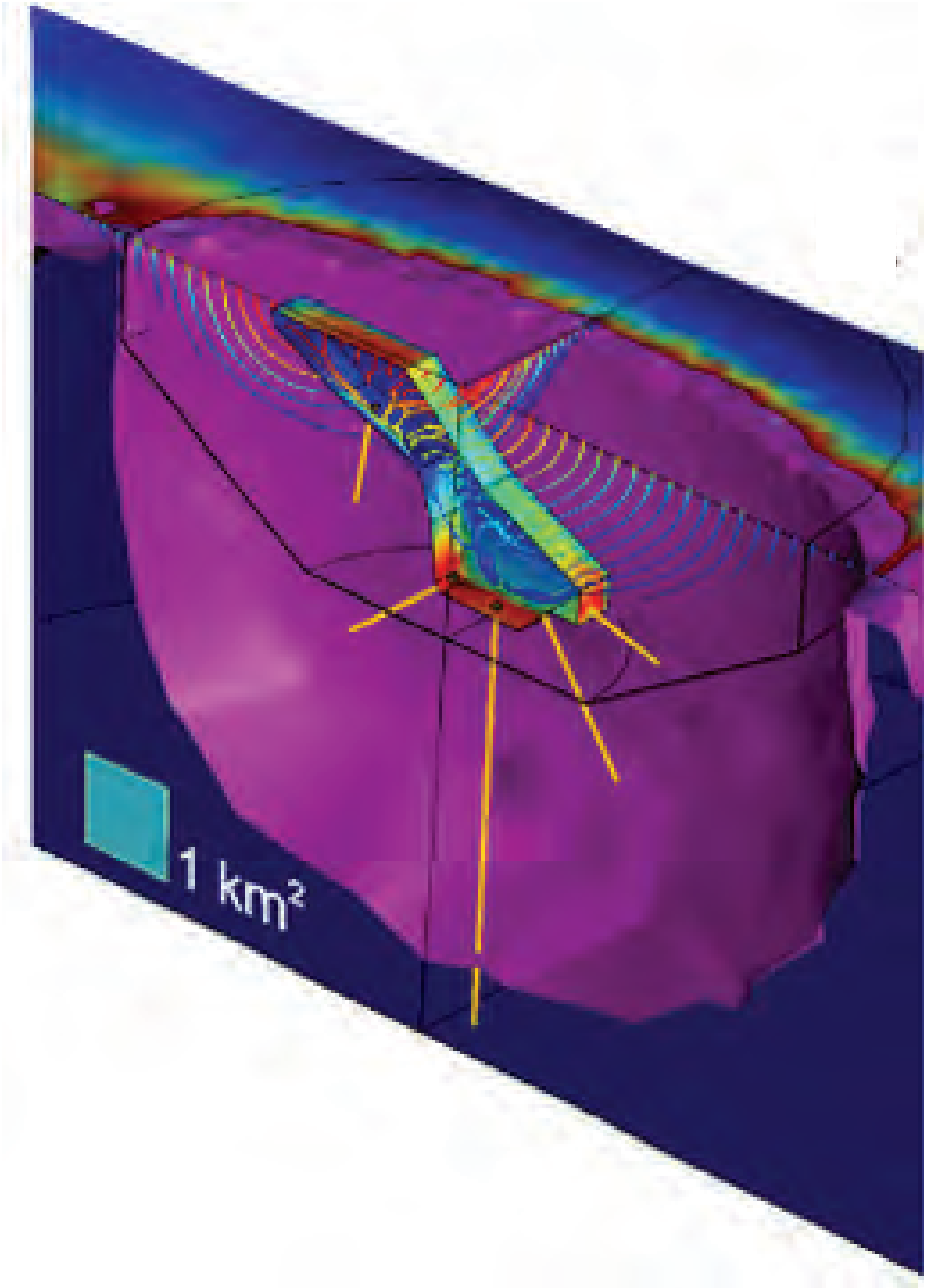

**Figure 3.3.7.4.1** Perspective view of former mine workings (polygonal area at center of diagram) showing simulated particle-flow paths from surface to depth during the excavation period of the former mine to present day. Color shading shows the water flux into the workings. The purple area bounds the region where groundwater has been captured by the underground facility, according to the simulation. Yellow lines are potential cored boreholes for future microbiology and hydrology experiments. [Courtesy Ecohydrology collaboration]

### 3.3.7.4.2 Proposed Experimental Investigations

Large-diameter boreholes will be extended an additional 2-3 km from existing infrastructure at the 7400L, to reach the 121°C isotherm and explore the upper-temperature limit of life. Beyond their use for fluid withdrawal, the boreholes will become experimental stations for conducting in situ transcriptomic and proteomic experiments with mobile underground laboratories, push-pull experiments within the packer-sealed fractures, and cross-borehole experiments—hydraulic, geophysical, and geomechanical—using multilevel packers and induced fluid flow.

The proposed investigations will be guided by the overarching question: What controls the distribution and evolution of subsurface life? The hypothesis being tested is whether these controls are dominated by processes related to geology, geomechanics, and hydrology. The investigation will consist of field studies supported by numerical simulations. The experimental activities will include extending characterization efforts to great depths using deep drilling deployed from the lowest accessible reaches of the Facility. The use of the flooding/dewatering event as a tracer and the hydrologic and mechanical stressor is a theme that cuts across many of the experimental activities.

Requirements from DUSEL anticipated for the Rock Deformation Facility are summarized in Table 3.3.7.4.2

| Requirement | | |
|---|---|---|
| **Layout** | | |
| Depth | 300, 800, 2000, 4100, 4850 and 7400 Levels (exploratory sites) | 4850 and 7400 Levels (observatory) |
| Footprint [$m^2$] | 10 x 10 =100 | 16 x 11 = 176 |
| Height [m] | 5 | 12 |
| Floor Load [kPa] | 98 | 98 |
| **Utilities** | | |
| Power [kW] | 1 | 1 |
| Standby Power [kW] | 0 | 0 |
| Purified Water [$m^3$] | 0 | .016 |
| Potable Water [lpm] | 16 | 16 |
| Compressed Air | Nominal use | Nominal use |
| Network | 1 Gb/s | 1 Gb/s |
| **Environment** | | |
| Temp. Min [ºC] | 5 | 5 |
| Temp. Max [ºC] | 40 | 40 |
| Humidity Min [%] | 0 | 0 |
| Humidity Max [%] | 100 | 100 |
| **Crane** | | |
| Max. Load [T] | 0 | 0 |
| **Occupancy** | | |
| Peak Installation Occupancy [count] | 5 | 5 |
| Installation Duration [months] | 1 | 2 |
| Peak Commissioning Occupancy [count] | 2 | 2 |
| Commissioning Duration [months] | 1 | 1 |
| Peak Operation Occupancy [count] | 0 | 1 |
| Operation Duration [months] | 4 | 36 |

**Table 3.3.7.4.2** Requirements for the Ecohydrology Facility.

### 3.3.7.5 Facility for Studying Cavity Design

The proposed Facility will help transform the fields of rock mechanics and rock engineering by improving existing capabilities in ensuring safety and the satisfactory performance of large cavities and other excavations.

#### 3.3.7.5.1 Overview of Proposed Research Facility

The proposed large underground cavities at DUSEL will need to be fully operational for an extended period of time under demanding conditions pertaining to deformation and safety. This challenges current knowledge of rock-mass behavior. The science vision for the Cavern Design Facility is to determine spatial and temporal characteristics and behavior of rock masses, and to estimate the uncertainty and risk associated with large underground excavations. The results of experiments associated with this Facility will benefit the design, construction, and long-term performance of cavities and other underground structures, thereby contributing to enhanced safety, reduced costs, and completion of the DUSEL Facility.

#### 3.3.7.5.2 Proposed Experimental Investigations

The experiments proposed involve large volumes of rock subjected to complex loading and include monitoring over extended periods of time (Figure 3.3.7.5.2). The experiment will integrate a number of closely related tests associated with the construction and performance of large cavities at the 4850L. They are also relevant to other underground structures at DUSEL and elsewhere.

The experiments include:

- Assessment of the rock-mass characteristics, specifically fracture patterns and fracture behavior (mechanical, hydraulic) as well as petrographic/mineralogic characteristics
- Evaluation of the performance of cavities, mine-by tunnels and large rock pillars performance during construction and operation
- Development of novel construction techniques for faster and safer excavation

Integrating all this will be a risk-analysis procedure in which in situ data will be analyzed using advanced modeling techniques. The risks associated with performance (safety), as well as construction cost and time, will be determined.

Requirements from DUSEL anticipated for the Cavern Design Facility are summarized in Table 3.3.7.5.2.

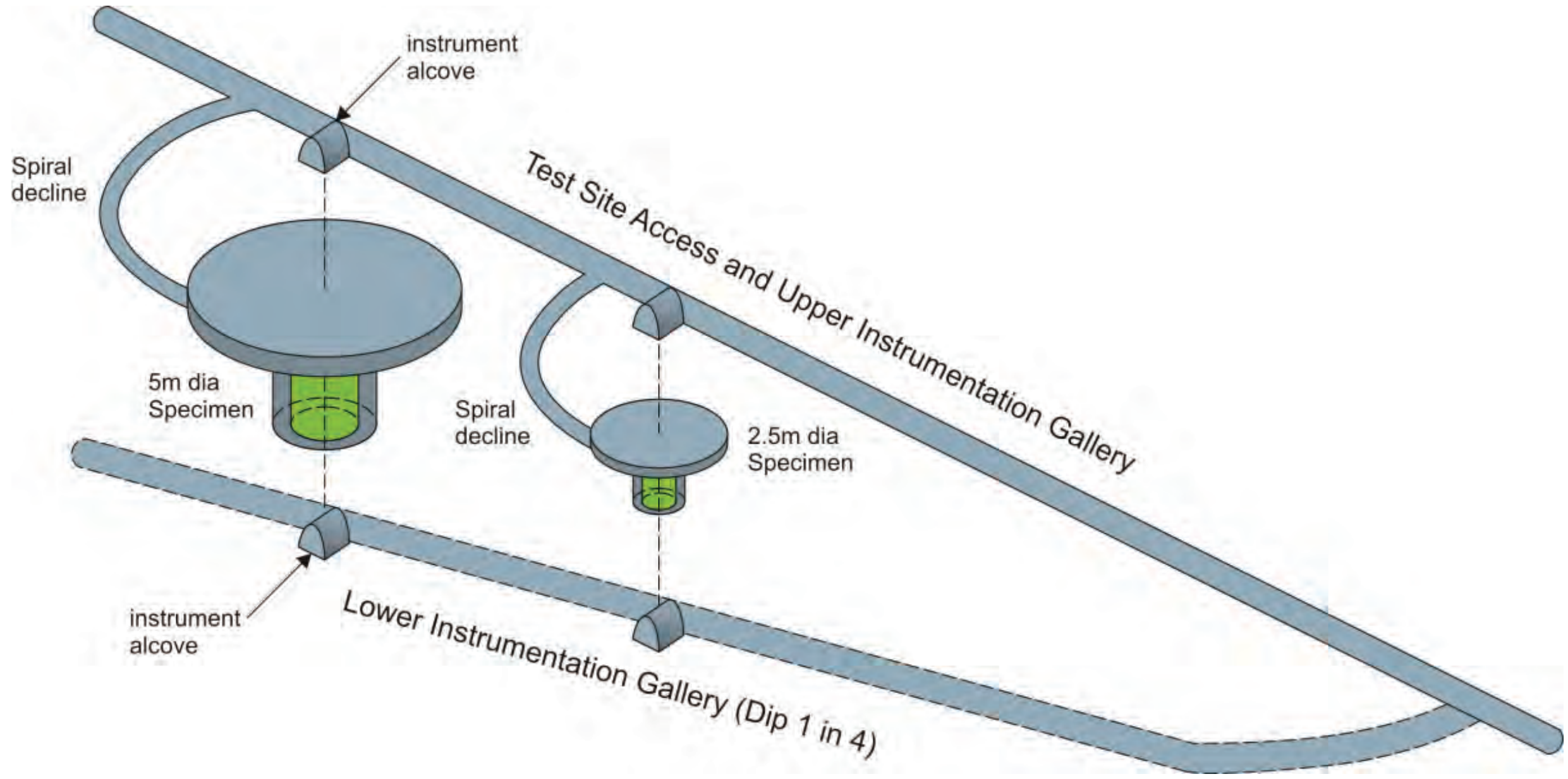

**Figure 3.3.7.5.2** Conceptual diagram of a facility for investigations of complex loading. [Courtesy Cavity Design collaboration]

| Requirement | |
|---|---|
| **Layout** | |
| Depth | 4850L |
| Footprint [$m^2$] | Adjacent to Large Cavity |
| Height [m] | |
| Floor Load [kPa] | |
| **Utilities** | |
| Power [kW] | 250 |
| Standby Power [kW] | 0 |
| Potable Water [lpm] | 16 |
| Compressed Air | Nominal use |
| Network | 1 Gb/s |
| **Environment** | |
| Temp. Min [ºC] | 15 |
| Temp. Max [ºC] | 25 |
| Humidity Min [%] | 0 |
| Humidity Max [%] | 100 |
| **Crane** | |
| Max. Load [T] | 0 |
| **Occupancy** | |
| Peak Installation Occupancy [count] | 5 |
| Installation Duration [months] | |
| Peak Commissioning Occupancy [count] | 2 |
| Commissioning Duration [months] | |
| Peak Operation Occupancy [count] | 0 |
| Operation Duration [months] | |

**Table 3.3.7.5.2** Requirements for the Cavern Design Facility.

## 3.3.7.6 Facility for the Study of Fracture Processes

The fracture processes Facility will address fundamental problems of rock rupture. It will provide access to intact rock, large natural faults, and those created at scales of 1-100 m.

### 3.3.7.6.1 Overview of Proposed Research Facility

At DUSEL, configurations for heating or cooling will be developed to manipulate in situ stresses and create faults. A robust implementation approach will circulate chilled fluid through arrays of subparallel boreholes drilled along vertical planes. This will reduce the horizontal compression normal to the planes of the boreholes while the vertical stress remains unchanged. Borehole arrays have been developed for use at DUSEL that would manipulate stresses at scales from less than 1 m to approximately 10 m. The boreholes in each array would be drilled from common rooms built as stepped cavities to simplify logistics. Instrumentation for monitoring the faulting process will be deployed in additional holes flanking the borehole arrays. The slip patch is expected to span 1 to several meters before it becomes unstable and propagates dynamically along a fault surface. As a result, an experiment to characterize dynamic fault slip may require dimensions larger than 10 m. To accommodate this scale, a patch nucleation experiment has been designed using two thermal panels at different levels along a pre-existing fault (Figure 3.3.7.6.1). A large fault—referred to as the Homestake Fault by the collaboration—has been located in the Facility. There is evidence that the fault is present on multiple levels. The fault is subparallel to the local foliation in the Poorman Formation and extends at least 1.5 km along strike and dip, with a center ~1.5 km deep along the western side of the underground facility. It strikes ~320-340° N, dips ~45-70° NE, and is recognized by a ~0.3-0.5 m thick distinct gouge that contains crushed host rock and black material that appears to be graphite. Although there is no clear evidence for fault displacement, secondary features suggest that it is a normal fault. The size and distinct structure of this fault make it a promising target for in situ experimentation of fault strength, hydrological properties, and slip nucleation processes.

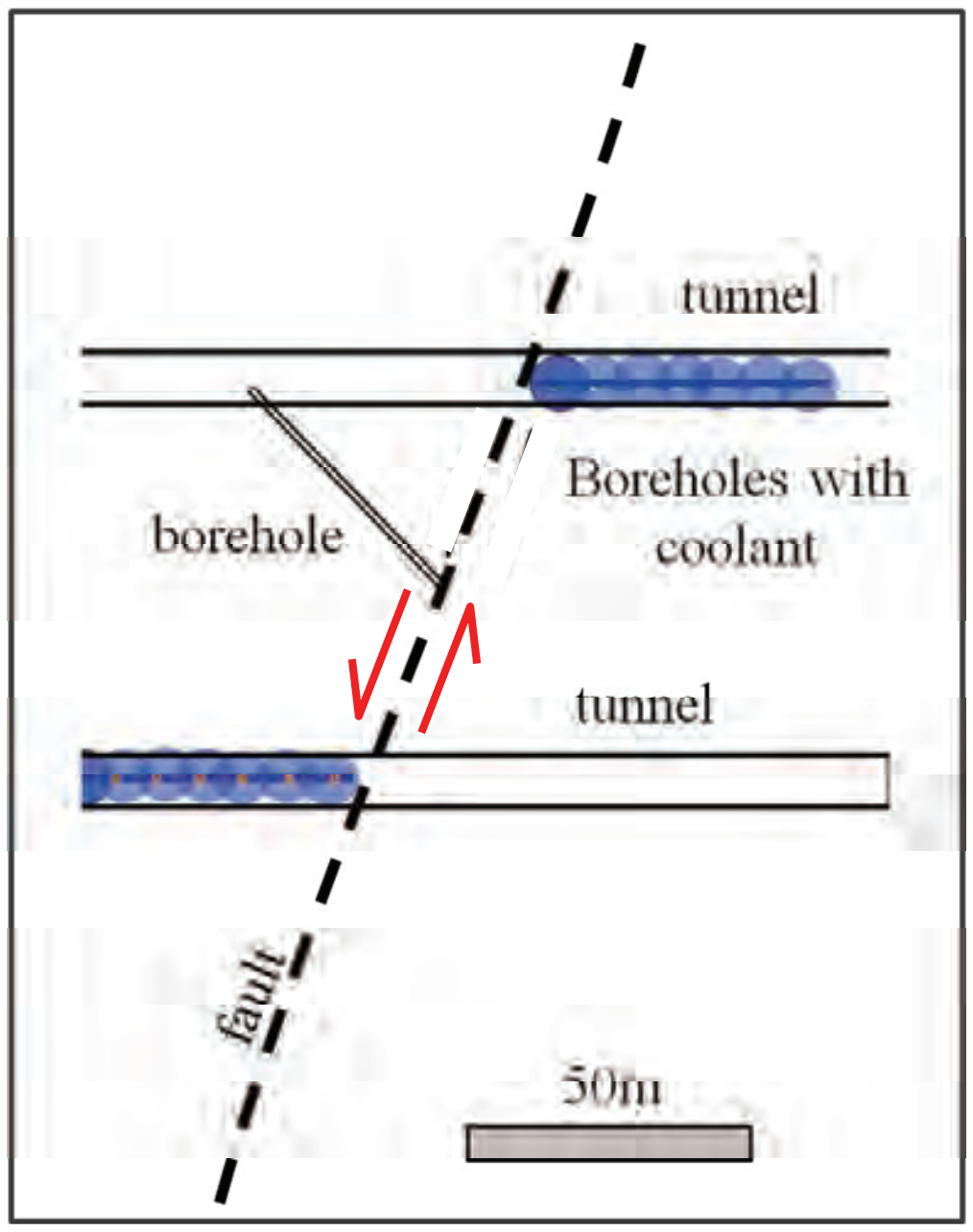

**Figure 3.3.7.6.1** Conceptual Design of a dynamic fault slip experiment. A natural fault is loaded by means of cooling/heating of two thermal panels (arrays of parallel cooling/heating boreholes). Fluid is injected at a given location on the loaded part of the fault to promote slipping. [Courtesy Fracture Processes collaboration]

### 3.3.7.6.2 Proposed Experimental Investigations

Experiments proposed by the Fracture Processes Collaboration are aimed at providing critical data to constrain the extent to which widespread upscaling is valid. This effort will quantify rupture mechanisms in both intact and faulted rock, including the sizes of the smallest frictional slip event, mechanisms of slip triggering and slip nucleation, mechanisms of strength-gain and fault-healing promoted by reactive fluids and other agents, and the role of velocity weakening in the transition between quasistatic and dynamic fault rupture.

The key to experiments associated with this Facility is the creation of carefully controlled faults in crystalline rock. Thermal techniques will be used to locally alter in situ stresses enough to cause faulting in rocks. The general process involves heating to increase compressive stresses, or cooling to reduce them. Because the temperature field can be finely controlled, the stress field can be finely controlled, too. Scaling and numerical analyses show that thermal technique can create differential stresses sufficient to induce faulting within a period of several weeks. Besides being used to study how new faults form in intact rock, the thermal techniques will also be applied in the vicinity of existing faults, to explore the possibility of slips along faults. This is important because most earthquakes are thought to be the result of unstable slip on existing faults. The slip process is likely to be initially localized on a growing patch (or series of patches) that is either weaker than the rest of the fault, or that sustains a locally elevated shear stress. Growth of the slipping patch moderated by the background stress rate may lead to an instability that results in a dynamic shear rupture propagation—this phenomenon is recognized to be related to the triggering of earthquakes.

Requirements from DUSEL anticipated for the Fracture Processes Facility are summarized in Table 3.3.7.6.2.

| **Requirement** | | | | |
|---|---|---|---|---|
| **Layout** | | | | |
| Depth | 2000L | 4100L | 4850 and 6800 Levels | 7400L |
| Footprint [$m^2$] | 36 x 2.1 | 198 x 2.1 | 108 x 2.1 | 18 x 5 |
| Height [m] | 2.5 | 12 | 8 | 12 |
| Floor Load [kPa] | 10 | 10 | 10 | 10 |
| **Utilities** | | | | |
| Power [kW] | 400 | 600 | 400 | 600 |
| Standby Power [kW] | 0 | 0 | 0 | 0 |
| Potable Water [lpm] | 80 | 80 | 80 | 80 |
| Compressed Air | Nominal use | Nominal use | Nominal use | Nominal use |
| Network | 1 Gb/s | 1 Gb/s | 1 Gb/s | 1 Gb/s |
| **Environment** | | | | |
| Temp. Min [ºC] | 0 | 0 | 0 | 0 |
| Temp. Max [ºC] | 30 | 30 | 30 | 30 |
| Humidity Min [%] | 0 | 0 | 0 | 0 |
| Humidity Max [%] | 100 | 100 | 100 | 100 |
| **Crane** | | | | |
| Max. Load [T] | 0 | 0 | 0 | 0 |
| **Occupancy** | | | | |
| Peak Installation Occupancy [count] | 8 | 8 | 8 | 8 |
| Installation Duration [months] | 6 | 6 | 6 | 6 |
| Peak Commissioning Occupancy [count] | 4 | 4 | 4 | 4 |
| Commissioning Duration [months] | 3 | 3 | 3 | 3 |
| Peak Operation Occupancy [count] | 2 | 2 | 2 | 2 |
| Operation Duration [months] | 21 | 31 | 21 | 17 |

**Table 3.3.7.6.2** Requirements for the Fracture Processes Facility.

### 3.3.7.7 Transparent Earth—Observatory for Subsurface Imaging and Sensing

This observatory will be unique in that the underground facility volume is surrounded (sides and bottom) and penetrated by hundreds to thousands of boreholes suitable for instrumentation.

#### 3.3.7.7.1 Overview of Proposed Experimental Facility

The Transparent Earth collaboration proposes installing and operating a permanent and portable geophysical observatory to illuminate the volume of DUSEL. The instrument system will be designed, much like a telescope, to look into particular directions and volumes within the underground facility using different excitation mechanics (e.g., strains, vibrations, electromagnetic field diffusion and propagation, density contrasts).

#### 3.3.7.7.2 Proposed Experimental Investigations

Through this effort, imaging methodologies and procedures will be developed to pursue a variety of fundamental science and engineering objectives. Multiple modalities of geophysical instrumentation within and surrounding the underground facility volume will allow passive and active source measurements of various geo-activities, including rock-mass re-stressing caused by the lowering of the water table, fluid injection and hydraulic fracturing, drilling and excavations during construction of the Laboratory, Earth tide and barometric effects, and daily operations. The deployment of multiple modes of geophysical measurement modalities will provide a large number of constraints for inversions leading to new discoveries. One possibility is the development of new measures for in situ stress, with the possibility of applying these methods to predict rock fracture and pore fluid pressures. Another is the development of new linkages between seismic and electromagnetic Earth science. Further, the proposed large-volume microseismic array will provide the tools needed to study the connection between the rock damage and the seismic waves generated during the geological and engineering processes. This knowledge will be applicable to all geophysics arrays, and provide strong evidence for answering some important questions concerning the energy budget of fracture growth and dynamics, local frictional behavior within a rock mass, seismic scaling laws, and the interpretation of seismic moment tensors.

To complement these activities, the Transparent Earth instrumentation system will perform a wide variety of scientific and engineering experiments. The permanent large-scale seismic array, combined with double-difference tomography, will provide an ongoing measure of facility stability required for occupant safety and the well-being of the experimental Facility. Many of the proposed techniques will be easy to mobilize and operate near new workings, changes in geo-behavior and by new experimental teams.

The nature of extended free-field scattering will be studied. Electromagnetic sensors will be designed to monitor different emission mechanisms and processes at different temporal and volumetric scales. Electrical resistivity, low-frequency electromagnetic, and induced polarization methods will be used to image hydrogeological processes in the underground facility. Portable high-resolution gravity meters will be used to map the underground facility and to evaluate the formation distributions and processes related to lowering water table and surface water changes. The locations of installed and proposed stations are given in Table 3.3.7.7.2-1.

Table 3.3.7.7.2-2 includes requirements anticipated from DUSEL by the Transparent Earth collaboration.

| Level | 2000 (1) | Level | 2000 (2) | Level | 2000 (3) | Level | 2000 (4) |
|---|---|---|---|---|---|---|---|
| Northing | 500 | Northing | -2500 | Northing | -4800 | Northing | -8500 |
| Easting | -2800 | Easting | -500 | Easting | 3500 | Easting | 5500 |

| Level | 4100 (1) | Level | 4100 (2) | Level | 4100 (3) | Level | 4100 (4) |
|---|---|---|---|---|---|---|---|
| Northing | -5000 | Northing | -10400 | Northing | -12000 | Northing | -5500 |
| Easting | -1200 | Easting | 1000 | Easting | 1700 | Easting | 4000 |

| Level | 4100 (5) | Level | 4100 (6) | Level | 4100 (7) | Level | 4550 (1) |
|---|---|---|---|---|---|---|---|
| Northing | -7000 | Northing | -3500 | Northing | -1500 | Northing | -7000 |
| Easting | 5000 | Easting | 4000 | Easting | 5500 | Easting | 1000 |

| Level | 4850 (1) | Level | 4850 (2) | Level | 7400 (1) | Level | 7400 (2) |
|---|---|---|---|---|---|---|---|
| Northing | -13000 | Northing | -8400 | Northing | -10300 | Northing | -9300 |
| Easting | 2150 | Easting | 6000 | Easting | 2500 | Easting | 4000 |

| Level | 7400 (3) | Level | 7400 (4) | Level | 7400 (5) | Level | 8000 (1) |
|---|---|---|---|---|---|---|---|
| Northing | -10500 | Northing | -8500 | Northing | -8800 | Northing | -8000 |
| Easting | 6500 | Easting | 5700 | Easting | 7800 | Easting | 2000 |

**Table 3.3.7.7.2-1** Installed and proposed seismic and E/EM stations.

| Requirement | |
|---|---|
| **Layout** | |
| Depth | 2000, 4100, 4550, 4850 and 7400 Levels |
| Footprint [$m^2$] | 1 x 3 |
| Height [m] | 2.5 |
| Floor Load [kPa] | 20 |
| **Utilities** | |
| Power [kW] | 2.2 |
| Standby Power [kW] | 0 |
| Potable Water [lpm] | 16 |
| Compressed Air | Nominal use |
| Network | 1 Gb/s |
| **Environment** | |
| Temp. Min [ºC] | 10 |
| Temp. Max [ºC] | 25 |
| Humidity Min [%] | 0 |
| Humidity Max [%] | 100 |
| **Crane** | |
| Max. Load [T] | 0 |
| **Occupancy** | |
| Peak Installation Occupancy [count] | 6 |
| Installation Duration [months] | 6 |
| Peak Commissioning Occupancy [count] | 2 |
| Commissioning Duration [months] | 2 |
| Peak Operation Occupancy [count] | 0 |
| Operation Duration [months] | 84 |

**Table 3.3.7.7.2-2** Requirements for Tier-1 Facility for subsurface imaging and sensing.

### 3.3.7.8 Rationale for Access to the 7400L

Five of the seven BGE collaborations (see Table 3.3.1), Ecohydrology, THMCB, Transparent Earth, Deformation Monitoring, and Fracture Processes, have proposed to develop experimental facilities at the 7400L, although many of their proposed activities also involve depths at the 4850L or above. For the Ecohydrology and Fracture Processes collaborations, access to the 7400L provides an enhanced opportunity to extend their experimental activities to deeper environments. For the rest, access is integral to planned activities.

A strong case has been made during and since the DUSEL S1 process that probing the deepest limits of the biosphere constitutes a high priority for subsurface science. An emerging consensus is that access to the deep biosphere is likely to generate major scientific discoveries and transformational science in the coming decade. To this end, the Integrated Ocean Drilling Program (IODP)[71] has funded three drilling legs that will, in the next five years, investigate microbial activity within the sub-seafloor biosphere. Two of these expeditions will probe the hottest and deepest depths where life exists. For the Ecohydrology collaboration, a drilling station at the 7400L will allow partnership with the International Continental Scientific Drilling Program (ICDP)[72] to reach the deepest limits of life.

DUSEL will provide the best opportunity in the United States for studying the deepest regions of the biosphere. It offers the deepest-available platform for initiating drilling. Holes drilled from the 7400L can be extended to great depth, i.e., to the ~121°C isotherm that is hypothesized to be the absolute limit of the biosphere. DUSEL has a geothermal gradient of 20-21°C/km, a value that is typical for continental crust. This contrasts with most environments previously investigated for thermophilic and hyperthermophilic life, (i.e., hot springs at Yellowstone, Kamchatka, and the deep-ocean spreading centers, where the geothermal gradient is ≥100°C/km). DUSEL is thus likely to be more representative of widespread deep-Earth microbial ecosystems. The highest reported temperature for microbial life from deep-sea hydrothermal vents is 122°C and for detectable subsurface microbial activity is ~85°. Given the average annual surface temperature of 7°C and the estimated range in the geothermal gradient, the temperatures correspond to target depths of 5.5 to 5.8 km for 122°C and 3.7 to 3.9 km for 85°C at DUSEL. It is between these two depth estimates that it is expected the transition from biology-dominated geochemical processes to biological processes, or pressure-sensitive life forms, that have yet to be detected, will be seen.

A key question is whether drill-holes originating from the surface can achieve the same scientific goals as those originating deep within an underground facility. While drilling can be done from the surface, it comes at great cost and with significant impacts to the quality of the sampled environment. Estimates of drilling costs from the surface, and from a deep-level platform, based on DOE's Deep Subsurface Science Program and those from mining companies and drilling contractors, show a clear advantage to drilling from within an underground facility (Figure 3.3.7.8). These estimates suggest that drilling from the surface would be 10 times more expensive than drilling from the 7400L. Even with the cost of excavating the cavity and ancillary preparation of the drill site, safety equipment, and a surface laboratory, the difference between surface and deep-level drilling is substantial.

Drilling from the surface requires large-diameter collars for the boreholes, and use of recirculated drilling mud with additives to maintain down-hole pressure above the hydrostatic gradient, and borehole stability. This drilling mud is enriched with microorganisms from the surface structures, holding tank, atmosphere, and shallow formations, where biomass concentration is high. The composition of the microbial community changes with increasing drilling depth, as new formations are encountered, and any indigenous populations are mixed in with the contaminants. In addition, increasing temperature and pressure can select thermophilic bacteria from the contaminating pool of surface organisms. Drilling from depth with smaller-diameter coring tools can reduce the use of contaminating drilling fluids. Further, by initiating drilling at the 7400L, the ambient formation temperature (~ 50°C) acts as an effective geological barrier against surface mesophilic microorganisms that have entered the rock through mining operations or from groundwater penetration. A single one-pass drilling fluid system can be used because the normal infrastructure of the facility enables pumping and disposal of large quantities of drilling water, thereby eliminating contamination of deeper levels of the borehole from shallower levels. Deep drilling from underground, therefore, affords the best opportunity to obtain pristine samples for microbial investigations of the deep subsurface.

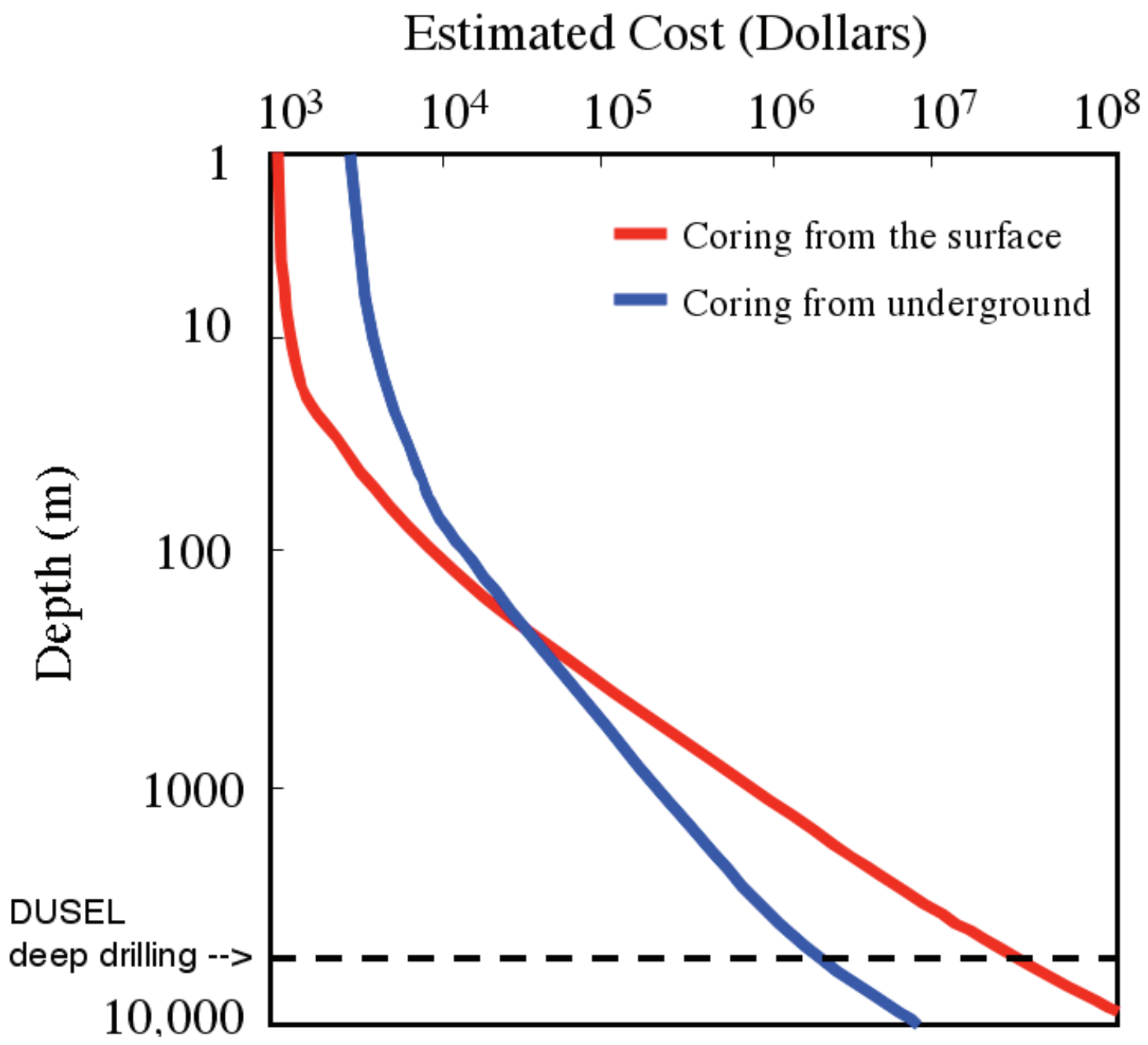

**Figure 3.3.7.8** Comparison of costs for coring from the Surface and from a deep-level platform at the 7400L. [Courtesy Ecohydrology collaboration]

Shortening the length of the borehole by drilling from 2.3-km depth also lowers the cost and increases the feasibility of using multilevel samplers. Further, in situ biogeochemical reaction experiments can be set up utilizing various fluorescent, enriched isotopic, and radiolabeled compounds. The use of such down-hole assemblies to cleanly access deep fracture is essential to achieving the research goals of Ecohydrology. By having a laboratory set up at the drill site on the 7400L and using insulated, high-pressure tubing, the deep borehole water can be brought into the laboratory at ambient temperatures and pressures with shorter transit times than would be the case for Surface operations. This greatly enhances the likelihood of growing hyperthermophiles and barophiles from great depth, and the ability to measure short-lived reactive species and metabolites.

Investigating the nucleation and propagation of dynamic shear rupture (earthquake slip) on a natural fault or plane of weakness in the rock mass is critical to the proposed Fracture Processes experimental facility. With the recent discovery of the Homestake Fault, which likely extends to the 7400L, there is now a unique opportunity to study fault properties at large spatial scales and depths. The size and distinct structure of this fault make it a promising target for in situ experimentation of fault strength, hydrological properties, and slip nucleation processes. Further, dewatering of the underground facility is expected to affect displacements in the fault vicinity. This poroelastic effect provides additional opportunities to characterize the fault better.

Studies of faults are enhanced as the Facility size increases and dimensions of the nucleation patch decrease, as this makes it easier to track both stable and unstable stages of patch development in the same fault slip. The use of the deep DUSEL Facility (7400L) would minimize the size of the nucleation patch, thereby significantly increasing the collaborations' chances of observing the evolution of the dynamic slip patch. Conducting experiments at different proposed depths (4850L and 7400L) would allow comparisons to be made of the phenomena of slip activation, transition to dynamic slip, and the dynamic slip arrest in the shallow crust. Presence of the Homestake Fault through all proposed Facility levels presents an unparalleled opportunity to sample.

Investigations of the THMCB collaboration would benefit from access to the 7400L, as this would help validate various proposed experiments. At this deeper level, the vertical borehole convection model can be tested, with the higher fluid pressures and stresses than those at the 4850L. This is a unique opportunity because of the near-vertical dip of the units. Heaters placed in the deep boreholes will allow for continuous monitoring of deep microbiological activity from ambient temperatures to those approaching or exceeding the current known limits for life, with comparison to deeper borehole observations at those same limits. Further, because there are higher fluid and rock temperatures at these depths, it is likely that the system is closer to chemical equilibrium. This is an excellent starting point for comparing water-rock reactions as a function of initial disequilibrium. In addition, because elevated temperatures and stresses are closer to those expected for typical enhanced geothermal systems, reaction rates and stress effects on permeability will tend to be higher and more representative. More generally, detailed hydrochemical and hydrological studies at the 7400L would help set better boundary conditions on the system at shallower depths (i.e., 4850L) and help evaluate the fluid-flow pathways in the system.

In summary, the Ecohydrology, Coupled Processes, Transparent Earth, Deformation Monitoring, and Fracture Processes collaborations have plans to access the 7400L. For the Ecohydrology group, a drilling station at this depth will greatly facilitate attempts to reach the deepest limits of life. For the Fracture Processes group, the higher stresses at the 7400L minimize the nucleation patch for fault slip and maximize the potential for observing dynamic fault slip. For Coupled Processes, Transparent Earth, and Deformation Monitoring collaborations, access to the 7400L provides an opportunity to extend the volume of investigations, and to validate results from shallower levels.

### 3.3.7.9 Schedule of Activities

From information gathered from the seven S4-funded BGE proposals, the DUSEL Project now has a broad picture of the requirements associated with each of the proposed research facilities. However, as the design of these facilities matures, the requirements will also be better defined, and the Project will need to keep abreast of these evolving requirements.

Two of the seven S4-funded BGE collaborations—Transparent Earth and Deformation Monitoring—are currently evaluating various measurement techniques as part of the Initial Science Program at Sanford Laboratory. They will continue this effort from now until the MREFC-funded construction begins. This will involve interactions with the DUSEL Project for access to the underground and defining utilities such as power and cyberinfrastructure. During this period, the LUCI, Ecohydrology, THMCB, and Fracture Processes collaborations will require periodic access to the underground for surveying and sampling potential locations of their research facilities.

The Transparent Earth, Deformation Monitoring, and Cavern Facility Design collaborations are interested in monitoring rock movement associated with excavations as the DUSEL Final Design is developed. As such, these collaborations will need to participate in the development of the excavation schedules.

There will be increasing interactions over the next three years with each of these seven collaborations as the design for their research facilities matures. These periodic exchanges will include updates on collaboration requirements and the DUSEL design. During this period, it is likely that some of the R&D effort associated with the collaborations will require underground access for testing prototypes. In addition, new proposals for BGE research are likely. These will be evaluated by the DUSEL Project and the DUSEL Program Advisory Committee (PAC).

DUSEL Project engineers and science liaisons will engage in dialogue with the collaborations, understanding their evolving requirements while updating them as the Facility design matures. In addition, the Project will need to provide underground access to the collaborations as they develop the design of their research facilities. While a detailed schedule cannot be realistically outlined at this time, it is clear that these tasks will iteratively proceed through the construction phase of the MREFC-funded process.

### 3.3.8 Low Background Counting

Many important physics results have been obtained in experimental searches for extremely rare events of the kind proposed for DUSEL. These experiments were designed to observe extremely small physics signals, in the presence of huge backgrounds, most due to natural radioactivity. Low-background counting by assaying and selecting materials of sufficiently low radioactivity for detector construction is a necessary process to assure the success of these experiments.

Besides the embedded natural radioactivity, cosmic-ray exposure can lead to in situ production of short- or long-lived isotopes from materials making up the experimental devices. In this case, being able to store the raw material for detector fabrication deep underground as early as possible, and/or to carry out the detector fabrication process underground could potentially improve the experimental sensitivity.

For the success of the next generations of underground scientific programs, it is important to design and construct an underground low-background facility at DUSEL for screening, production, and storage of radiopure materials. The advantages of a deep on-site low-background Facility are many: It could enhance the synergy among the Integrated Suite of Experiments, especially in the area of material assay and radioactivity control, before and during construction; provide spaces for stockpiling ultrapure materials; and make available a platform for developing underground fabrication techniques for ultra-radiopure materials. It could also provide a general-purpose and well-shielded underground laboratory space (with efficient R&D infrastructure) for prototyping new detectors for future experiments. Such a Facility would also satisfy many of the radio-assay needs from other DUSEL biological and engineering experiments.

#### 3.3.8.1 Overview

Radiometric assay of material samples usually has been performed by using $\alpha$, $\beta$, and $\gamma$ screeners in environmentally controlled (e.g., radon-suppressed) and heavily shielded counting stations. With the advance of neutron activation analysis (NAA) techniques and inductively coupled plasma mass spectrometry (ICPMS), the achievable assay sensitivity for certain materials, (e.g., Teflon, copper, etc.) can be lower than those obtained from direct counting alone.

1. **Gamma Counting and Neutron Activation Analysis (NAA).** Low-background $\gamma$-ray spectroscopy using high-purity germanium detectors (HPGe) is a well-developed, mature technology that has served as the prime tool for material selection. Sensitivities down to a few hundred ppt of U and Th are routinely achieved using commercially available large-volume, low-background packaged P- or N-type detectors. The outstanding energy resolution makes them an excellent choice for counting applications where radioisotope identification is important. All current solar neutrino, dark matter, and double-beta decay experiments have been relying heavily on this detection technique. The best sensitivity (~$10^{-12}$g/g U/Th) thus far has been achieved by several of the "GeMPI" class detectors,[73] with special choice of construction materials and an elaborate shielding-enclosure design. By combining $\gamma$ counting and neutron activation analysis, one can increase the assay sensitivity significantly in many cases. Depending on the radioisotope of interest and material composition of the sample, one can place the sample in a flux of neutrons (e.g., inside a reactor) to induce additional characteristic radioactivity that could be gamma counted after the sample has cooled down. Even higher sensitivity can be achieved by chemically separating the unwanted radioisotopes from the sample after neutron activation.

2. **Alpha and Beta Counting.** While direct $\gamma$ counting generally provides superior diagnostic screening information, direct $\alpha$ or $\beta$ counting is sometimes the only means to screen against surface contamination, provide isotope dating, or determine the amount and location of a suitable radiological tracer. A particularly dangerous contamination for a number of experiments is the deposition of radon daughters from the atmosphere, which decay to the long-lived $^{210}$Pb, a low-energy beta emitter, and then to the alpha-emitting $^{210}$Po, with no penetrating radioactivity signature. An ultra-low-background drift chamber (the BetaCage)[74] optimized for detection of <200 keV electrons and alpha particles is under construction as a DUSEL R&D project. An ultrasensitive large-area alpha counter has been developed by the XIA[75] company, the sensitivity of which is only exploited when run underground.

3. **Inductively Coupled Plasma Mass Spectrometry (ICPMS).** The sensitivity of radioactivity assay could be improved significantly by isolating the minute quantity of ions of interest from the entire sample material. One such approach is ICPMS—by utilizing a plasma to ionize the sample, with the resulting ions analyzed by a sensitive mass spectrometer. One advantage of this method is that the original sample mass can be exceedingly small. Depending on the sample preparation process and actual material composition of the sample, ICPMS could achieve a sensitivity range of down to ~$10^{-12}$g/g or better.

4. **Ultra-Sensitive Customized Large-Scale Assay Systems.** The next-generation low-background experiments will need to reach background levels far below what can be screened in even the best HPGe counters. An advanced direct-counting capability with orders of magnitude improvement in sensitivity is therefore needed. Currently, the only type of counting technology that can access such low counting rates are large liquid-scintillator-based detectors (e.g., CTF/Borexino[76] and KamLAND[77]) in large containers. Bulk assay of large amounts of material can be done with a targeted sensitivity at the $10^{-13}$-$10^{-14}$ g/g U/Th level. An advanced direct-counting capability with orders of magnitude improvement in sensitivity can be modeled after these installations. At such sensitivities, it would either have to be housed at the deepest DUSEL level or equipped with a sophisticated shield able to reject background muons, neutrons, and gammas, if run at the 4850L.

### 3.3.8.1.1 Pre-Screening Program for DUSEL Science Support

The proposal for a large scale and multipurpose on-site underground assay Facility, Facility for Assay and Acquisition of Radiopure Materials (FAARM), will be described in detail in the following section (3.3.8.2). However, it is equally important to support early material assay needs from the ISE before the underground laboratory modules would be ready. The research community, led by the AARM (Assay and Acquisition of Radiopure Materials) collaboration, has proposed a strategy to address this situation—by establishing a DUSEL Low Background Counting Facility (DULBCF) Consortium for early screening support. This plan will be based on screeners located both in the Sanford Laboratory Davis Laboratory Module (DLM) counting laboratory and other existing low-background counting sites in the United States and, possibly, facilities abroad.

Besides developing the on-site capability, AARM plans to actively utilize other existing underground sites for early screening support, but integrated under the DUSEL Project umbrella. This includes the building up of a coherent team and staff, establishing a surface main campus at the University of South Dakota (USD), enhancing capabilities at existing low-background counting sites, as well as developing a scheduling tool for multiple-site operations. It is expected there will be close collaboration among the committed sites, such as USD, Soudan Underground Laboratory, and Lawrence Berkeley National Laboratory (LBNL). It is also agreed that all the enhanced off-site equipment will eventually move to FAARM when it is ready.

Figure 3.3.8.1.1 shows a layout of the on-site underground DLM counting laboratory (green shaded area) at the 4850L, next to the Large Underground Xenon (LUX) experiment. This counting laboratory will support AARM's early screening plan. Gamma-counting stations labeled as "CUBED-I" and "Others" will be installed by the Center for Ultralow-Background Experiments at DUSEL (CUBED) collaboration, to support its planned site characterization program. These stations are expected to be in service about six months after the DLM would be accessible. Two additional gamma-counting stations, DUSEL-A and DUSEL-B, each made up of a large (~2.5 kg) HPGe detector, will be added afterward to enhance the overall capability. The laboratory will be partially operated by on-site staff. It is expected that DUSEL-B will have an assay-sensitivity comparable to the GeMPI detectors.

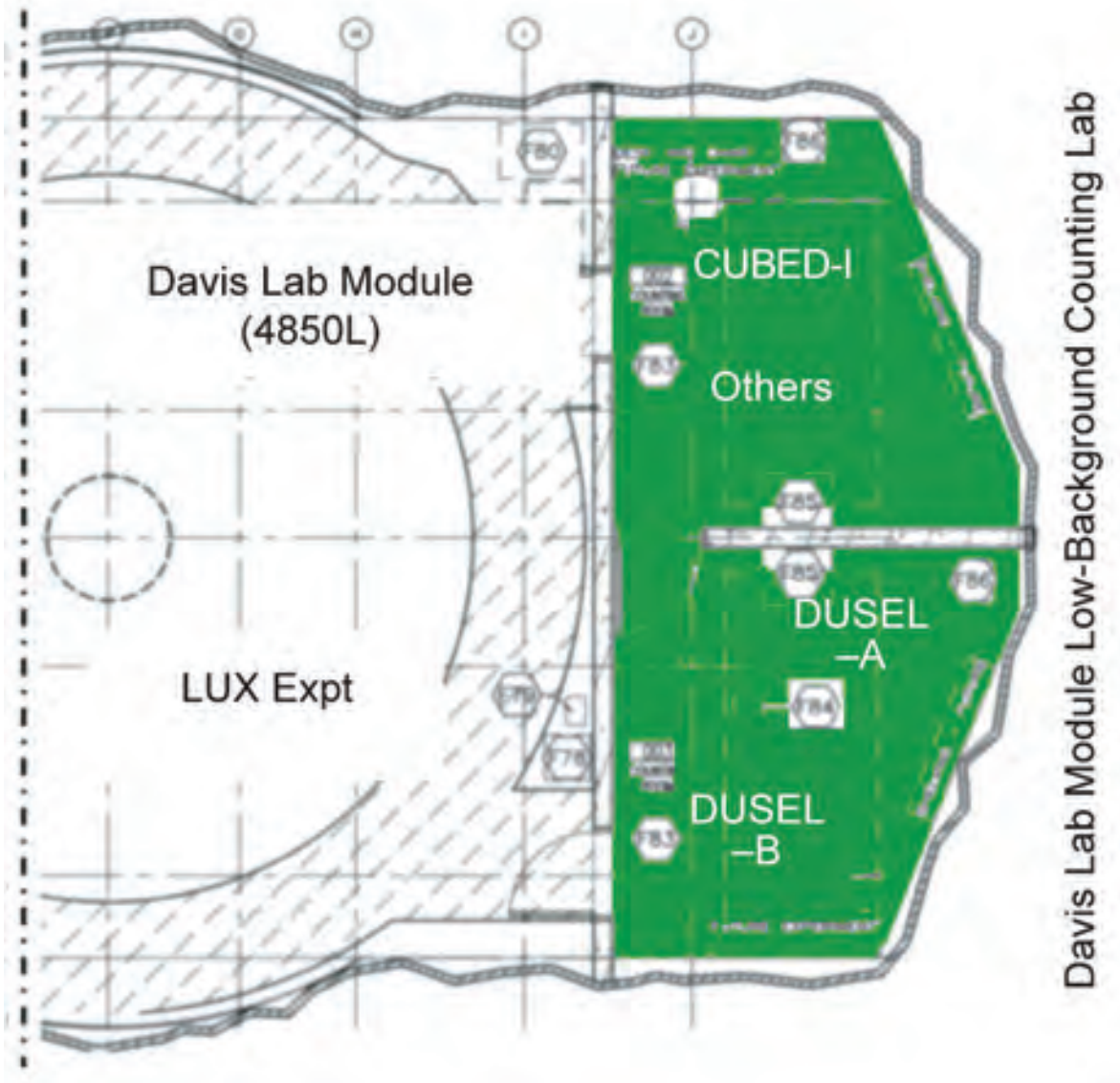

**Figure 3.3.8.1.1** Layout of the DLM counting laboratory at the 4850L. It shares the same LM but is separated from the LUX experiment. [Courtesy SDSTA]

### 3.3.8.2 Low Background Counting Facility—FAARM

A candidate to provide the required capability for DUSEL is FAARM, a proposal from the AARM collaboration. The FAARM proposal seeks resources for the design and implementation of a Facility on the DUSEL site for screening, production, and storage of radiopure materials. Since the first Homestake National Underground Science Laboratory (NUSL) studies seven years ago, such a facility has been identified as a priority. The design of the Facility was explored in the NUSL White Paper[78] and was mapped into the infrastructure matrix[79] during the DUSEL S1 process, with its own technical chapter of the *Deep Science Report*.[80] Its importance was reaffirmed during the November 2007 Town Meeting by the B1 Cross-Cutting Group on Low Background.[81] The clear consensus was that DUSEL must: have world-class facilities capable of providing assay and ultraclean materials support, as well as integration tools to share data; exchange equipment; train personnel; optimize screening throughput (both on site and off site); foster new collaborations in areas of geology, biology, and homeland security; and identify new users in other research fields. There are other on-site and off-site activities that are potentially associated with the FAARM Facility, such as underground storage of ultrapure materials, underground ultrapure material production facilities, clean machine shops and ICPMS, NAA facilities, as well as special fabrication tools such as electrical discharge machines (EDMs) and laser welders.

#### 3.3.8.2.1 The FAARM Proposal

Design of FAARM will optimize economy of scale—combining ultrasensitive and production screening in the same area—and take advantage of common infrastructure (such as purification plants and water-shield engineering at DUSEL). There are four major components in the design (Figure 3.3.8.2.1-1): 1) an active shared water shield for the identification of muons and neutron activation products as well as passive attenuation of external gammas and neutrons; 2) a laboratory space inside the water shield for housing alpha, beta, and gamma screeners; 3) an ultrasensitive immersion-tank system capable of performing large-sample counting; and 4) the provision of a well-shielded space (inside the same water shield) for R&D and prototyping of detectors for future DUSEL experiments.

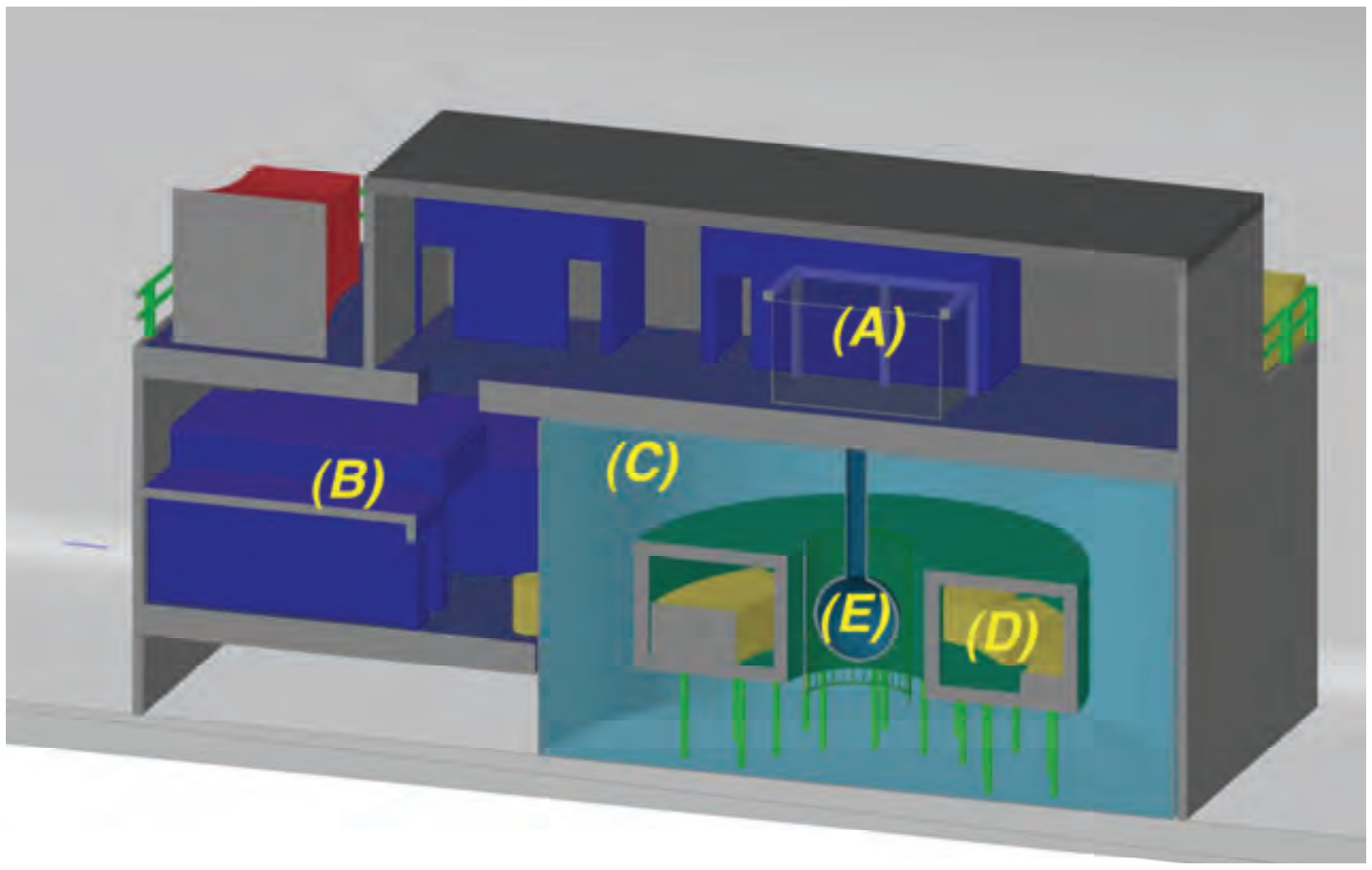

**Figure 3.3.8.2.1-1** A conceptual sketch showing the major components of the FAARM design: (A) clean room with sample preparation, (B) clean machine shop and assembly areas, (C) the water shield, (D) the inner toroidal laboratory, and (E) the ultrasensitive immersion detector system (center). [Courtesy FAARM collaboration]

1. **The Large Water-Shield System.** The purpose of the water shield is to tag muons and attenuate the external background gamma rays and neutrons from the cavity walls (rock and shotcrete), and to provide an inexpensive common shielding for all the moderate-sensitivity and GeMPI-style screeners, while creating ultimate shielding for the innermost ultrasensitive immersion tank system. It can thus share common water-purification infrastructure with other DUSEL experiments that are planning dedicated water shields. Simulations indicate that a minimum water shield thickness of 2.3 m is needed to establish the required sensitivity of the immersion-tank detector. The present shield engineering design is based on a large-diameter stainless steel cylindrical tank filled with processed ultrapure water. The size of the shield is ~15 m diameter and ~8 m height, but the final dimension will depend on the radioactivity level of the surrounding rock and concrete construction materials. The required water is expected to come from the LBNE water system at the 4850L but there will be local processing to improve and maintain the cleanliness of the water. The inner surface of the shield is instrumented with PMTs to create a muon veto and to provide additional rejection of neutron-induced processes in the water and electromagnetic fragments from showers induced by muons in the wall. The option of replacing the water with liquid scintillator or introducing gadolinium or boron to improve rejection efficiency will be explored with dedicated studies over the next two years. If this option is too expensive, the shield will be designed to accommodate this as a later upgrade.

2. **The Inner Screening Laboratory.** Sensitive screeners will be located inside a toroidal inner screening laboratory that forms a tunnel within the water shield (Figure 3.3.8.2.1-1). The exact geometry was optimized and the overall Facility cost cut in half by taking advantage of standard steel tank engineering for the water shield and creating a toroidal screening laboratory that reserves a central shared water space for the most sensitive detector. This saves valuable space, uses the same water system and PMTs, and avoids the cost of an additional containment tank. The walls of the torus laboratory, especially the inner one that faces the immersion-tank system, need to be constructed of radiopure material. Acrylic tunnel design is a staple of large public aquariums and this expertise is being tapped. Several (~ 8 to 10) $\alpha$, $\beta$, and $\gamma$ screeners can be installed inside the torus laboratory. Besides the external water shield, conventional shielding material such as low-background lead or copper could be added to the screener systems if needed.

3. **The Ultra-Sensitive Immersion Tank System.** Designs of such a large-scale screener have been formulated by many groups, and usually consist of a tank made of stainless steel or a cavity lined with a radon-impermeable plastic. It can be filled with liquid scintillator or with doped pure water (~$10^{-14}$ g/g U/Th). It is then accessed by a hermetic top deck and has a nitrogen purge between the liquid surface and deck for handling and insertion of counters and samples into the active volume. To quantify the footprint, cost, placement, and safety issues, the FAARM design (Figure 3.3.8.2.1-2) is modeled after the Borexino Counting Test Facility (CTF)[76] detector, with modifications and improvements (such as low-radioactivity photosensors and new scintillator or water-soluble fluors, etc.) to be designed over the next couple of years. A 2-m diameter transparent nylon vessel filled with liquid scintillator will occupy the central portion of the water shield. To have a more compact detector, low-radioactivity Quartz Photon Intensifying Detector (QUPID) tubes will be immersed in the water shield in a frame surrounding the central scintillator volume.

4. **General R&D Space.** Space will be allocated inside the inner screening laboratory for detector R&D and prototyping for future experiments. This provides a unique shielded underground space for new technologies that otherwise would be unable to afford the type of shielding required to extend their sensitivities and determine their feasibility.

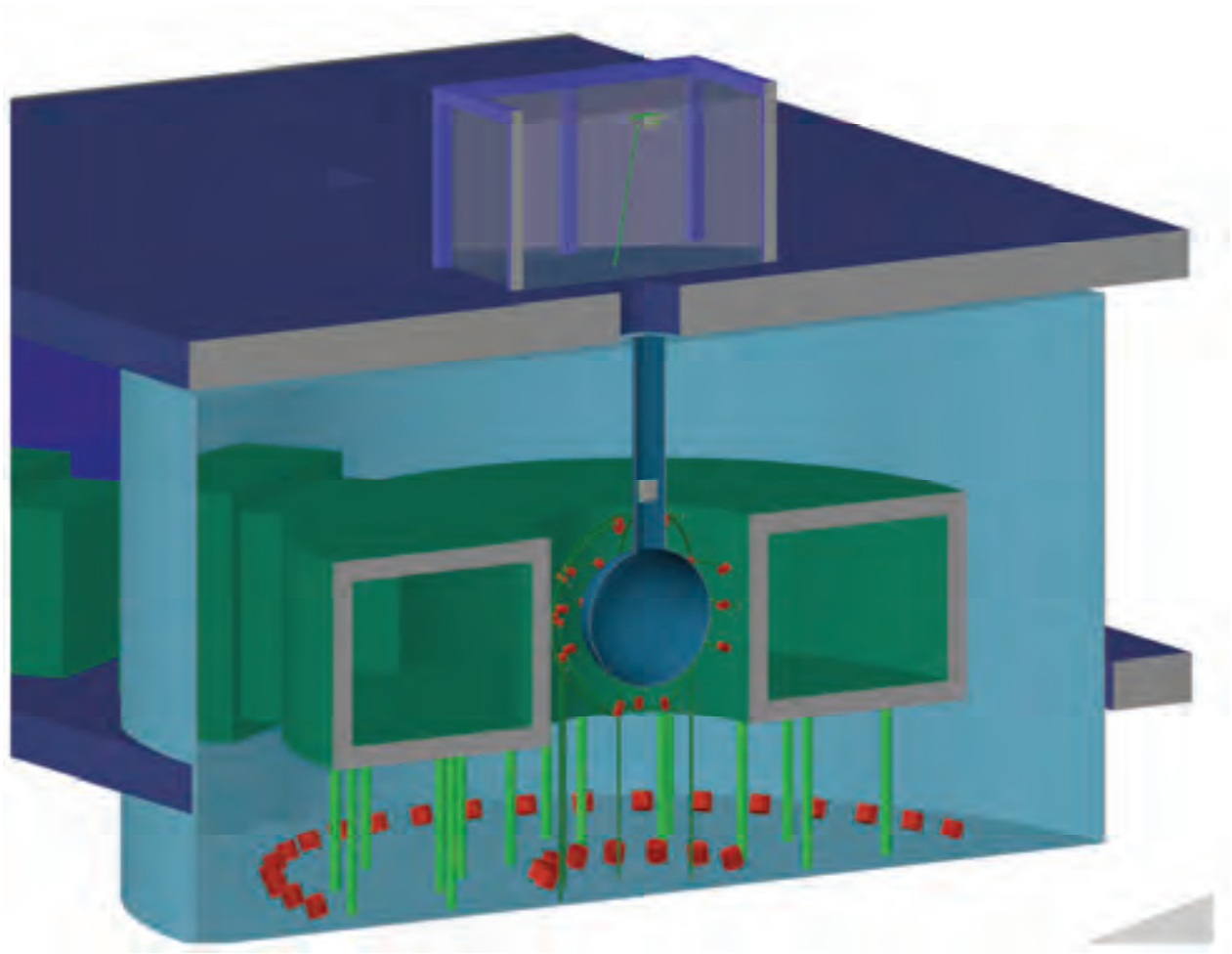

**Figure 3.3.8.2.1-2** The immersion tank system shown with photosensors and upward-looking muon veto photomultiplier tubes. [Courtesy FAARM collaboration]

### 3.3.8.2.1.1 FAARM Location and Space Requirements

The FAARM Facility will be located at the 4850L. The Facility will have two levels, with the main water shield and screening laboratory (including the control room) at the ground level, and a clean room with an assembly area, a small mechanical shop, radon scrubber, local water-processing system, and HVAC support room, etc., situated at the second-floor level (see layout drawings, Figure 3.3.8.2.1.2 in the following section). The general requirements for power, utility, air, etc., are summarized in Table 3.3.8.2.1.1, below.

| Requirement | Value/Description | Comment/Justification |
|---|---|---|
| Depth | 4850L | |
| Footprint | 25 m L x 17 m W | Constrained value |
| Max. Height [m] | 13.25 | Actual value |
| Floor Load [kPa] | 78 | Corresponds to 8-m-high water tank |
| **Utilities** | | |
| Power [kW] | 350 | |
| Standby Power [kW] | 25 | Radon mitigation, emergency lights, some air handling |
| Chilled Water [kW] | 100 | |
| Waste Heat to Air [kW] | 250 | |
| Purified Water [$m^3$] | 1500 | FAARM will further process purified water to 18 Mohm |
| Potable Water [lpm] | Nominal | |
| Compressed Air | Nominal | |
| Network | 1 Gb/s | Nominal |
| **Environment** | | |
| Temp. Min [°C] | 20 | |
| Temp. Max [°C] | 25 | |
| Humidity Min [%] | 20 | |
| Humidity Max [%] | 50 | |
| Rn Background [$Bq/m^3$] | 200 | FAARM radon mitigation to acceptable level |
| **Crane** | | |
| Max. Load [Short Tonne] | 20 | |
| **Occupancy** | | |
| Peak Installation Occupancy [count] | 15 | Installation and commissioning are simultaneous |
| Installation Duration [months] | 30 | |
| Peak Commissioning Occupancy [count] | Same as installation | |
| Commissioning Duration [months] | Same as installation | |
| Peak Calibration Occupancy [count] | 10 | Installation of new screeners, R&D studies |
| Calibration Duration [months] | 0.5 | |
| Calibration Frequency | 4/yr | |
| Peak Operation Occupancy [count] | 6 | 6 Peak/2 Avg |
| Operation Duration [months] | Continuous | |
| **Cryogens** | | |
| LN Storage [L] | 50 | In transportable dewars |
| LN Consumption [L/day] | 4 | Boil-off for pure low radon nitrogen |
| **Major Hazards (Other than Cryogens)** | | |
| Chemistry Lab Operation Waste | Acids, bases, solvents | From sample preparation and processing |
| Water Flood Hazard | 1500 $m^3$ | |
| **Assay and Storage** | | |
| Assay Needs | Assay for others | |
| Underground Storage | FAARM clean storage | Some storage at shallower levels, possibly |

**Table 3.3.8.2.1.1** FAARM requirements.

### 3.3.8.2.1.2 FAARM Layout

A proposed FAARM floor plan is shown in Figure 3.3.8.2.1.2. The overall footprint of the Facility is 25 m x 17 m, in compliance with guidelines for fitting into the DUSEL LMs at the 4850L.

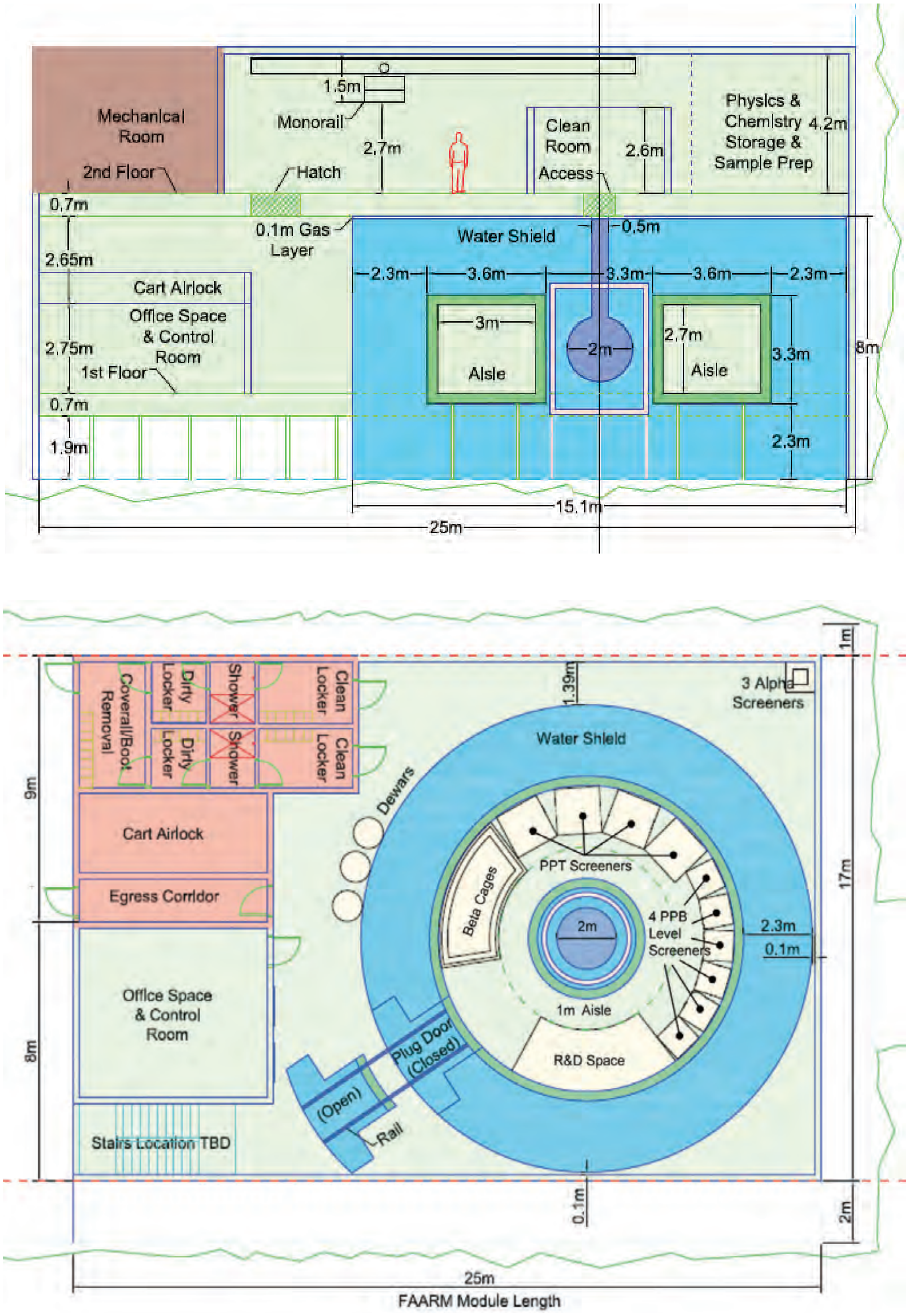

**Figure 3.3.8.2.1.2** Elevation (top) and plan (bottom) views of FAARM. [Courtesy FAARM collaboration]

#### 3.3.8.2.1.3 Major Hazards

There will be a significant volume of scintillation liquid (linear alkylbenzene [LAB]) inside the immersion-tank system. A water-containment plan needs to be in place in case of a major leakage from the main water shield, so as not to affect neighboring experiments in the shared DUSEL LM.

#### 3.3.8.2.1.4 Schedule and Installation

The overall FAARM schedule includes a strategy to establish the DULBCF consortium for early screening using the Sanford Laboratory and multiple sites. The transition from the DLM to full screening at FAARM ensures uninterrupted screening from 2012 onward. The longer-term FAARM schedule has been planned in accordance with DUSEL guidance, with completion of the Conceptual Design expected by 2013 and start of construction in 2018, when the DUSEL 4850L LMs are expected to be available for beneficial occupancy. The DUSEL installation schedule must account for the earliest-possible deployment of FAARM as a service facility for all the experiments. As FAARM is expected to remain in place for the duration of the DUSEL scientific program, its placement within the LM must take this into account.

## 3.3.9 Other Potential Physics Experiments

### 3.3.9.1 Introduction

We provide short descriptions of two examples of potential physics experiments not described in previous sections of this section. Both examples demonstrate unique possibilities at DUSEL so far not available in other underground laboratories. These examples have not been used yet to establish Facility requirements. Further review is needed to determine if their requirements, or those of other experiments that may be proposed, should be included in the process for defining Facility requirements. The DUSEL PAC (see Chapter 3.10) will be consulted over the next few years as part of this process.

### 3.3.9.2 DAEdALUS

The DAEdALUS (Decay At rest Experiment for $\delta_{CP}$ studies At the Laboratory for Underground Science)[82] proposal describes a complementary approach to the LBNE science goal to measure charge and parity (CP) violation in the neutrino sector. The DAEdALUS concept was proposed in early 2010, and was not submitted as part of the S4 process.

The physics of neutrino oscillations has been described in Section 3.2.1.3, *Neutrino Oscillations*. Where LBNE has a fixed source of neutrinos at Fermilab, and near and far detectors between which neutrino oscillations are measured, DAEdALUS proposes to use an ensemble of near (1.5 km), mid (8 km), and far (20 km) cyclotrons on the surface to produce neutrinos from pion decay at rest that are detected in the LBNE water Cherenkov detectors (see Figure 3.3.9.2—three water Cherenkov detectors were assumed in the DAEDaLUS estimates). Timing between the accelerator beam batches allows the experiment to determine the neutrino source among the near, mid, and far cyclotrons.

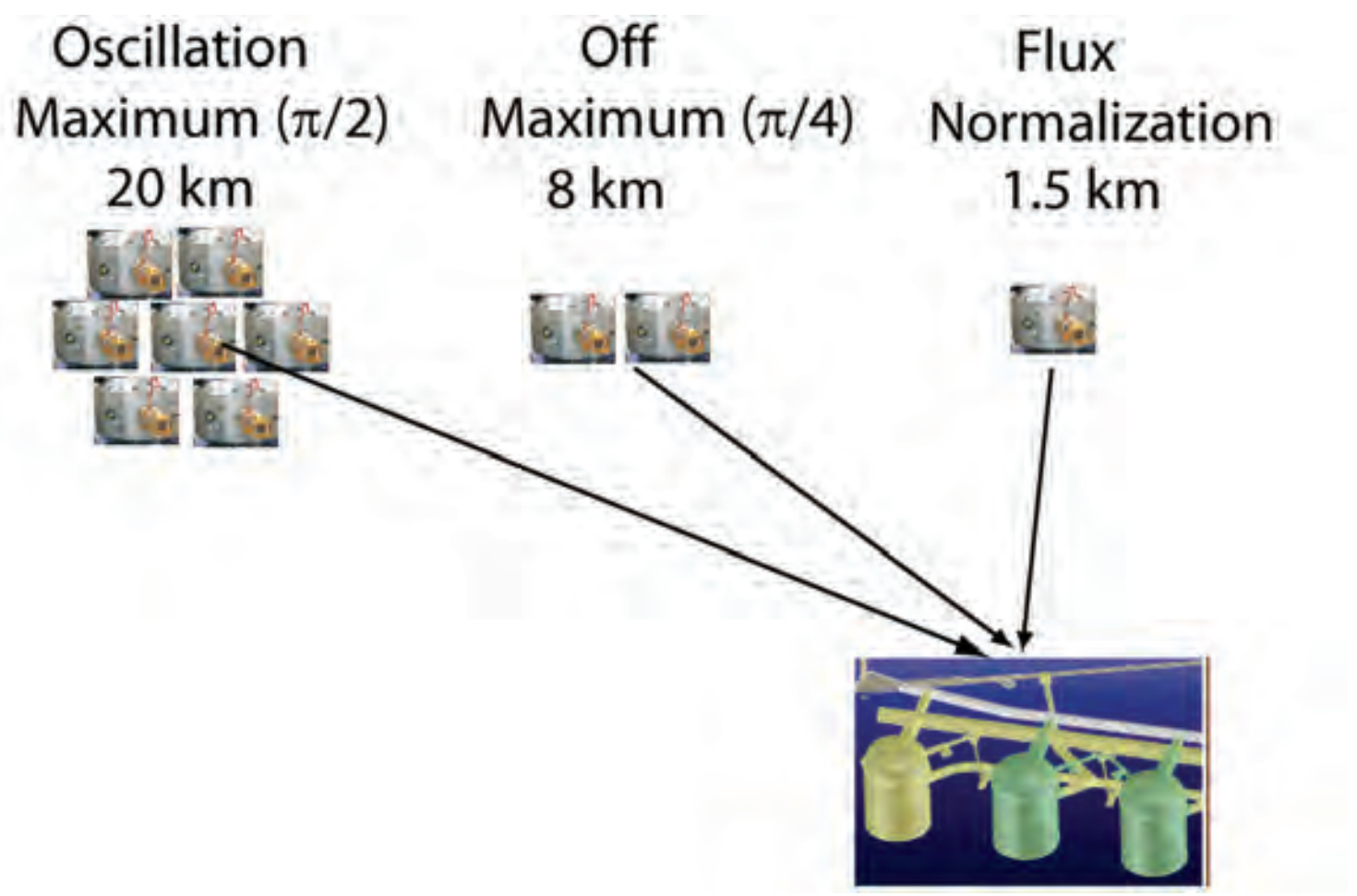

**Figure 3.3.9.2** Schematic diagram showing the relationship of the LBNE water Cherenkov detectors to three complexes of cyclotrons at 1.5 km, 8 km, and 20 km. The oscillation maximum refers to a $\bar{\nu}_\mu$ energy of 40 MeV. The flux is determined from the near accelerator on the surface at 1.5 km from the detector. In this cartoon, the mid and far neutrino flux are increased by adding accelerators, where each cyclotron drawn indicates ~1 MW of beam power. The actual experiment would be phased, and the ultimate design would depend on cost, sensitivity, and choice of cyclotron technology. [Richard Kadel, DUSEL]

A well-defined spectrum of neutrinos from pion decay at rest is produced by the cyclotrons.

Oscillations of $\bar{\nu}_\mu$ to $\bar{\nu}_e$ are detected in the water Cherenkov detector via inverse beta decay: $\bar{\nu}_e + p \rightarrow n + e^+$. Because $\pi^-$ are mostly captured before decay, the fraction of $\bar{\nu}_e$ background in the beam is less than ~ 4 x $10^{-4}$. Since the neutrino source is near the detector, there is no interference between the CP violation and matter effects. Hence, DAEdALUS is sensitive only to CP violation, and not the mass hierarchy. As in the case of electron antineutrinos from supernovae, the products of the inverse beta decay for this experiment are also detected as a double delayed-coincidence signal, with the prompt positron detected through its Cherenkov radiation, and the neutron via delayed capture (~30 μsec) on Gd in the water, releasing a cascade of photons with total energy ~8 MeV of energy of which ~ 4-5 MeV is detected by the PMTs.

The CP measurement sensitivity of DAEdALUS running alone as a function of exposure and a 300 kT water Cherenkov detector can be found in Reference 1. The DAEdALUS results could be combined with LBNE measurements to improve the sensitivity for LBNE measurements of $sin^2 2\theta_{13}$ and $\delta_{CP}$ by statistically combining the results for both neutrino and antineutrino running. Since DAEdALUS uses antineutrinos only, better sensitivity could be realized by running the LBNE with a neutrino beam only. Sensitivity results are also given in Reference 1.

The neutrinos from the DAEdALUS targets are in the same energy range and use the same inverse beta decay detection method (water with gadolinium doping) as diffuse neutrinos from relic supernovae. Whether DAEdALUS operations are compatible with this measurement in LBNE is under investigation.

It is proposed to build the DAEdALUS experiment in three phases. The construction is dependent on current research in high-intensity cyclotrons, currently under way at several institutions.[82]

### 3.3.9.3 NNbar

Recent theoretical and experimental developments[83] in the field of neutrino physics provide reasons for the possible existence of neutron-antineutron transformations or oscillations ( $n \rightarrow \bar{n}$ or NNbar). Discovery of NNbar oscillations would reveal a new force of nature beyond the Standard Model and would shed light on one of the fundamental mysteries of the universe—the origin of matter.

The concept for the NNbar search experiment at DUSEL is illustrated in Figure 3.3.9.3. The source of the neutrons is at or near the surface and it is followed by a cold moderator and a neutron-focusing device. The neutrons then fall inside a 300-m to 1-km long and 4-5-m wide vertical vacuum flight tube (with vacuum better that $10^{-5}$ Pa) where the Earth's magnetic field is cancelled down to ~1 nT to satisfy the quasi-free condition for NNbar transformation. The vertical configuration avoids the detrimental effects of the gravitational force on the subthermal neutron transport. Neutrons transformed in-flight to antineutrons are detected by an annihilation detector.[84] The unique signature of antineutron annihilation makes this experiment background-free and, as such, a single detected event will be a discovery of the $n \rightarrow \bar{n}$ transformation.

The goal of the proposed NNbar experiment at DUSEL is to increase the sensitivity for $n \rightarrow \bar{n}$ transformation by a factor of ~ $10^4$ with respect to the existing intranuclear and free-neutron experimental search limits.[84,85] The same experimental setup at DUSEL[86] with small modifications can be also used for a search for mirror matter (an alternative to supersymmetric WIMP dark matter) by measurement of the neutron flux disappearance.

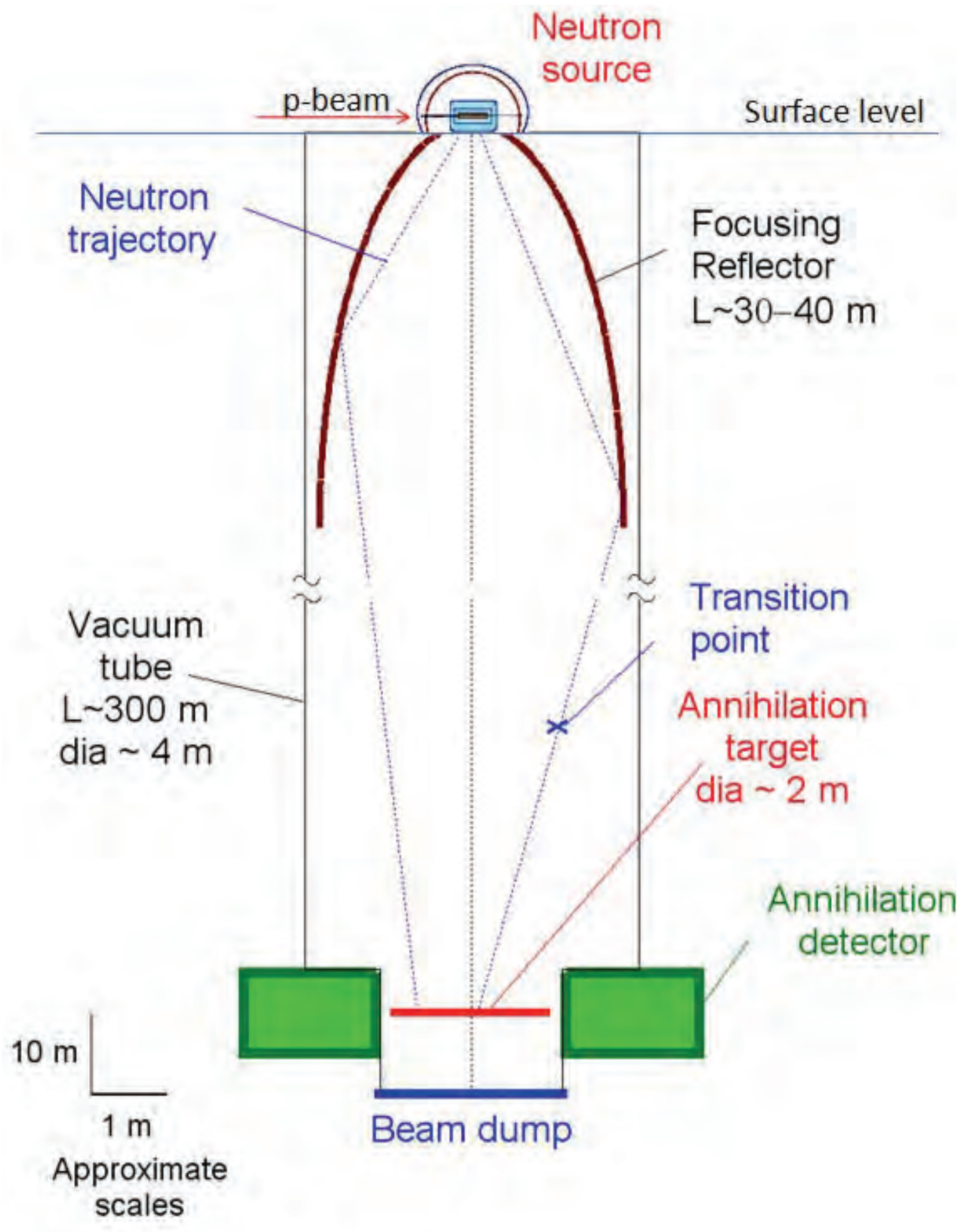

**Figure 3.3.9.3** Schematic view of the NNbar oscillation experiment in a vertical shaft of DUSEL. [Courtesy NNbar collaboration]

The experiment would require an environmentally clean and safe neutron source. Possible neutron sources include recently proposed compact accelerators[82] with a spallation target, arrays of d-d or d-t generators, or commercially available research nuclear reactors. The high neutron flux requires a cooling system at the power level of ~1 MW. Also, a 20-30 kW cryogenic system for the moderator will be required at the surface.

### 3.3.10 Cosmogenic Backgrounds and Depth Requirements for Physics Experiments

The depth requirements for the physics experiments proposing to go into DUSEL are based predominantly on the need to achieve sufficient overburden to reduce cosmogenic backgrounds to acceptable levels. As part of the Preliminary Design process for DUSEL, S4 recipients in physics were asked to respond with specific statements detailing the requested location for their experiment, whether at the 4850L or at the 7400L Campus. Experiments interested in the deployment at the 7400L Campus were asked to submit additional calculations justifying their requests.[87-92] The two experimental areas with the highest sensitivity to cosmogenic backgrounds are neutrinoless double-beta ($0\nu\beta\beta$) decay and dark-matter searches. These experiments drive the need for space at the greatest depths.

A variety of cosmogenic processes can result in backgrounds in physics experiments. These include direct muon interactions, muon capture and decay, muon-induced hadronic and electromagnetic showers, and spallation products. Fast spallation neutrons are particularly troublesome, especially those generated within the rock by muons that never traverse the experimental hall to trigger a veto. The energy spectrum of the fast neutrons extends up into the hundreds-of-MeV range, which makes them highly penetrating, and makes a large number of interaction channels available. For dark-matter searches, the dominant background arises from neutron-nucleus elastic scattering, mimicking the low-energy recoils of Weakly Interacting Massive Particles (WIMP) interactions. In $0\nu\beta\beta$ decay, typically of greater concern are high-energy inelastic neutron interactions, which can result in events in the energy region of interest where the $0\nu\beta\beta$ peak is to be found. Decays of long-lived unstable nuclei formed by muon-induced spallation are also a concern.

To decrease these backgrounds to tolerable levels, the two primary strategies are to go deep enough to reduce the neutron flux, and to moderate the remaining neutrons down to harmless energies. In DUSEL, the fast neutron backgrounds at the 7400L are predicted to be roughly a factor of ~20 lower than at the 4850L. Additional rejection can be obtained from event discrimination and vetoing capabilities, such as the detection of neutron multiple scattering in detectors, the vetoing of electromagnetic and hadronic showers typically accompanying the neutrons, or the detection of the neutrons themselves in active scintillator vetoes. Generally speaking, depth can be traded against a more complex shield design. In their base design, the various S4 experiments have made different choices based on a combination of factors, including perceived uncertainties and risks of a particular shield configuration, design heritage from previous-generation experiments, technology-specific constraints, shield complexity, and costs. Facility-related factors such as access to the two DUSEL levels, timing of the availability, and size of the cavities also played a role.

A critical element in estimating accurately the performance of a particular design is the detailed modeling of the background, which needs to be based on experimental measurements of the neutron yield, energy spectrum, and multiplicity at various depths, and of the correlation with hadronic and electromagnetic cascades. Such data are still very sparse, in particular for the high-energy tail of the neutron distribution that typically dominates the background. Neutrons above 50 MeV not only interact with the detector material but also produce a large number of additional neutrons in the relatively high Z materials needed for support, enclosures, and shields. Because of the rapidly falling neutron-proton cross section, they are unfortunately difficult to moderate. It is not clear yet that the detailed Monte Carlo simulations including GEANT3,[93] Geant4,[94] and FLUKA[95] are sufficiently accurate—the relevant neutron nuclear cross sections are often uncertain—or whether they include all the important physics processes—some may still be unknown. As a result, simulations tend to disagree with the existing data and among each other.[96-100]

The DUSEL low-background community is forcefully engaged in trying to understand in more detail the complex physics processes at play, to use existing data and undertake new measurements to validate Monte Carlos, and to compare simulations among the various groups. A working group, involving in particular the double-beta decay and dark-matter S4 collaborations, is being assembled to build on the work already conducted by the individual collaborations, with the goal of reducing present uncertainties and apparent contradictions, and of quantifying the level of uncertainties. New experimental results from existing setups at various depths as well as from dedicated experiments worldwide should be available within the next two years. Much will also be learned during the pre-DUSEL program, which will result in decreased risk of the proposed experimental designs.

The 1TGe collaboration provided a detailed breakdown of the cosmogenic backgrounds expected for the MAJORANA DEMONSTRATOR prototype experiment that is scheduled to start data-taking in Sanford Laboratory in ~2013. The background is dominated by inelastic ($n$, $n'\gamma$) interactions with the detector, shield, and support materials (Ge, Cu, Pb). The total predicted background, ~1 count in the region of interest per tonne per year (c/ROI/t/y), is at the upper limit of allowed background levels for a tonne-scale $^{76}$Ge 0νββ experiment of similar design to have sufficient sensitivity to cover the phase space corresponding to the inverted neutrino mass hierarchy (see Section 3.2.2). Additional reductions are judged by the 1TGe collaboration to be essential to the experiment's success, especially considering that noncosmogenic backgrounds in the 1TGe detector already contribute ~1 c/ROI/t/y themselves. Alternative shielding designs are under consideration for 1TGe; however, even if prompt fast neutron backgrounds can be moderated, other cosmogenic backgrounds remain problematic. For example, the 1TGe collaboration estimates that the in situ production of $^{77}$Ge at the 4850L is higher than acceptable, given current uncertainties. Until those uncertainties can be addressed or further rejection techniques can be developed, 1TGe is assuming occupancy of the 7400L LM in its design activities.

The EXO collaboration makes a similar argument to that made by 1TGe. They are currently in the midst of commissioning a prototype experiment, EXO-200, at the Waste Isolation Pilot Plant in New Mexico. They plan to investigate cosmogenic backgrounds at the shallower overburden (1585 mwe) provided by this site, but don't expect to have sufficient information to make siting (depth) decisions for DUSEL until the end of the S4 process. In the meantime, given the penetrating nature of the fast neutrons, which they suspect will dominate their cosmogenic backgrounds, the collaboration is reluctant to rely on shielding alone at the 4850L to achieve sufficiently low background, and is assuming occupancy of the 7400L LM in their planning activities.

To reduce neutron-nucleus elastic scattering, which may mimic WIMP-nucleus interactions, the various S4 dark-matter experiments have made different baseline choices based on the designs of previous-generation experiments and the experience gained so far in those configurations. LZD and MAX have chosen to operate at the 4850L with large water shields and/or active liquid scintillator vetoes, and to mitigate the uncertainties in the background by including an ample safety margin in the shield design. GEODM and COUPP are baselining for the 7400L, where the shields can be simpler and much more compact.

The LZD and MAX collaborations intend to use low-mass cryostats and low-Z instrumented water shields. The collaborations' evaluations led to a conclusion that sufficient suppression of cosmogenic activity at the 4850L, with a substantial safety factor, can be achieved using these design features combined with event characterization and fiducial-volume selection. With 4-5 m of water shielding surrounding their inner vessel, the LZD collaboration estimated that fast neutron backgrounds would be more than an order-of-magnitude lower than expected PMT backgrounds before any cuts are applied. Further rejection of these backgrounds is enhanced by the large, monolithic nature of the LZD and MAX liquid-noble time projection chambers. The LZD collaboration estimated that by applying a fiducial volume cut and by rejecting interactions that deposit energy at multiple locations simultaneously within the detector, additional background reduction by over two orders of magnitude can be achieved. The LZD collaboration has also evaluated an option to add a 1-m-thick active liquid-scintillator veto inside the water shield. It is not necessary to reach the baseline background goals, but would provide a significant additional safety margin. The MAX collaboration already has as part of its default option the use of an active, liquid scintillator-based neutron and muon veto.

GEODM does not take this approach, primarily because the shielding requirements for its cryogenic design demand more high-Z material close in to the target material. High-Z material regenerates low-energy neutrons, reducing the effectiveness of instrumented water shields and liquid scintillator vetoes. GEODM estimates that, even with minimal amounts of copper close in to the target and use of a water shield, the raw rate of events due to high-energy neutrons would degrade the experiment's sensitivity by up to an order of magnitude. Rejecting such events via event characterization and detection of high-energy neutrons and their associated secondary particles in an instrumented water shield and possible liquid scintillator veto could provide the safety factor necessary to achieve GEODM's intended sensitivity at the 4850L. However, this would be a significant deviation from the Cryogenic Dark Matter Search design heritage, and therefore would demand a significant testing program, entailing additional costs and risks.

The problem of cosmogenic backgrounds is of course not limited to $0\nu\beta\beta$ decay and dark-matter experiments. It should be kept in mind that for some types of experiments, such as p-p solar neutrino experiments, which were not retained for S4 studies, the issue of cosmogenic spallation products (such as the production of $^{11}C$ in liquid scintillator) would demand a deep location. It is likely that future investigations of yet-unknown physics would have similar requirements.

To summarize, the issue of depth is a multidimensional problem, encompassing the complex and possibly unknown aspects of the physics contributing to the backgrounds, the quantitative uncertainties in the background levels in specific configurations, and the associated scientific risks. Other factors include costs, technological risks, and Facility-related aspects, such as space available at a given depth, access, and timing. It is therefore essential that the Facility design be flexible enough to allow the best experimental strategies to be implemented by providing experimental space at the 7400L, and large-enough halls at the 4850L to be able to accommodate the more complex shields needed at those depths. In parallel, it is important that the experiments further analyze the experience of existing setups, collect additional data, and develop the simulation tools needed to reduce the background uncertainties as much as possible.

## 3.4 Research Activities at the Sanford Underground Laboratory

### 3.4.1 Introduction

Building on the legacy of the Ray Davis chlorine solar-neutrino experiment that began in 1965 at the former Homestake Gold Mine, 19 initial science groups are currently active at Sanford Laboratory. Experiments are being conducted on 12 levels, ranging from the Surface to the 1,480 m (4,850 ft) level to investigate topics in physics, geology, biology, and engineering. Table 3.4.1 shows the list of initial science groups and corresponding levels on which activities are taking place.

Three large-scale physics projects are in process at Sanford Laboratory: the Large Underground Xenon (LUX) dark-matter experiment;[101] the MAJORANA DEMONSTRATOR neutrinoless double-beta decay experiment;[102] and the Center for Ultralow Background Experiments at DUSEL (CUBED),[103] which will be investigating the development of ultrapure crystal-based detectors. However, most of the current projects under way at the Laboratory are smaller in scale and many are associated with the biology, geology, and engineering (BGE) disciplines. Although not specifically funded for characterization studies, much of their work will provide characterization data that will be useful for future projects.

In addition to providing access to existing areas for researchers, SDSTA at Sanford Laboratory is developing new areas for experiments. A former warehouse building on the Surface has been renovated to provide laboratory space for LUX to exercise its deployment procedures and to commission detector systems before moving the experiment underground. The LUX collaboration has had beneficial occupancy in the Surface Facility since December, 2009. Excavation work started in September 2009 and is now complete on the 4850L Davis Campus, which includes the original DLM and the Davis Transition Area. Outfitting is expected to begin in early 2011 and to be completed approximately 10 months later by the end of 2011, followed by beneficial occupancy of the experiments.

The total space at the Davis Campus is roughly 745 $m^2$ (8,000 square feet), with approximately 455 $m^2$ (4,880 square feet) available for scientific activities and equipment. Figure 3.4.1 shows the plan view of the Davis Campus and the two separate areas: the Davis Transition Area (a large portion of which will be dedicated for the MAJORANA DEMONSTRATOR) in a new access drift and the DLM that will be slightly enlarged to accommodate a large water shielding tank for LUX. Prior to the completion of the Davis Campus, a temporary space is being developed near the 4850L Ross Shaft to allow the MAJORANA DEMONSTRATOR project to begin electroforming copper components for its detector.

The SDSTA Science Liaison Department provides logistics support and oversight for the initial science research groups. The Science Liaison Department coordinates closely with SDSTA Operations, as well as the EH&S Department and the DUSEL science integration team.

| Experiment | Description | Levels |
|---|---|---|
| **Physics** | | |
| LUX-350 | Dark matter using Xe | Surface, 4850L |
| MAJORANA DEMONSTRATOR | Neutrinoless double-beta decay using Ge | 4850L |
| CUBED | Crystal growth (Ge; possibly NaI, $CdWO_4$) | TBD |
| Background Characterization (also part of DUSEL Low Background Counting Facility in future) | Muon, neutron, gamma, radon | 800L, 2000L; many levels for Rn; previous incl. Surface, 4550L |
| Vertical Facility (magnetic field background) | N-Nbar, others (e.g., gravity) | Surface, Ross/Yates Shafts |
| **Geology** | | |
| $CO_2$ Sequestration | Environment characterization | 800L, 2000L, 4850L (removed) |
| Deep Underground Gravity Laboratory (DUGL) | Seismic characterization for gravity-wave research | Surface, 300L, 800L, 2000L, 4100L |
| Fiber Sensors | Extensometers, temperature | 4100L |
| Hydrogravity | Dewatering effects on local gravity | Surface |
| Hydrology/Microclimate | Aquifer characterization, groundwater monitoring | Surface, 1250L, 2000L, 2600L, 4850L |
| Petrology, Ore Deposits, Structure-PODS | Core archive and logs, geologic mapping | Surface, 800L |
| Tiltmeter | Rock deformation | 2000L |
| Transparent Earth | Seismic monitoring | 2000L, 4100L |
| **Biology** | | |
| SDSM&T/BHSU | Microbiology | Surface, 300L, 2000L, 4100L, 4850L |
| SDSU | Lignocellulose | 1700L, 2000L |
| Princeton/UTKnoxville | Manifold sampling | 2000L, 4550L, 4850L |
| SDSM&T | Microbiology/Cellulose | 4550L |
| **Engineering** | | |
| Signal Propagation | Electromagnetic transmission | 300L |
| Submersible | Autonomous vehicle navigation, magnetic field background | 1250L |

**Table 3.4.1** Initial science research at the Sanford Underground Laboratory.

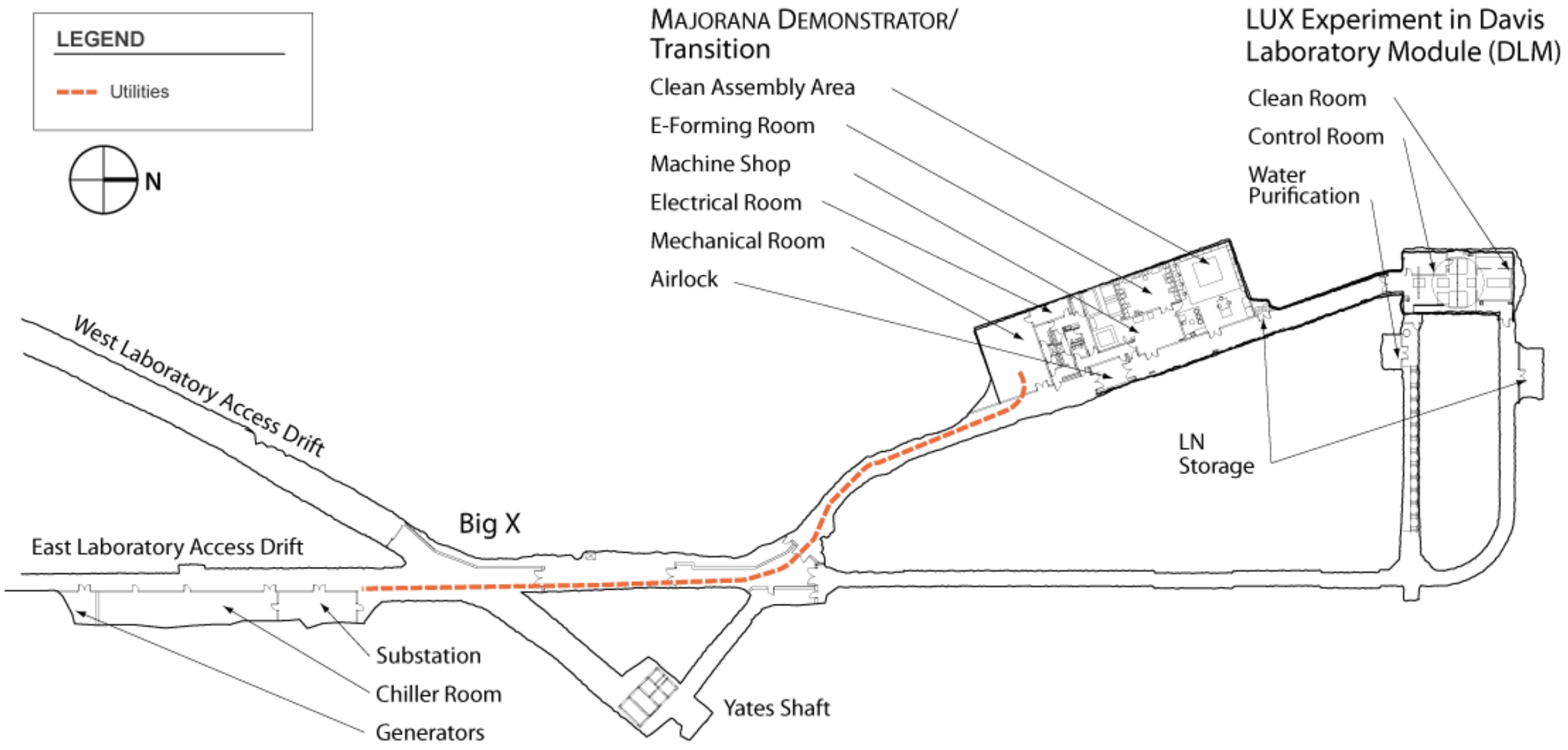

**Figure 3.4.1** Sanford Laboratory 4850L Davis Campus near the Yates Shaft. [J. Willhite, W. Zawada, DUSEL; DKA]

## 3.4.2 Physics Experiments

### 3.4.2.1 LUX and Dark Matter

The LUX collaboration is composed of 13 universities and two U.S. laboratories.[101] The LUX concept is illustrated in Figure 3.4.2.1.

The collaboration is currently deploying the LUX detector at a surface facility at Sanford Laboratory. Detector deployment on the Surface allows full-scale integration and testing of all final detector subsystems before underground deployment. This will allow test commissioning and characterization of the detector in an environment closely emulating the final underground laboratory. This facility will be reconfigured for use by future DUSEL experiments and will become part of the complex of surface buildings in support of the DUSEL scientific program.

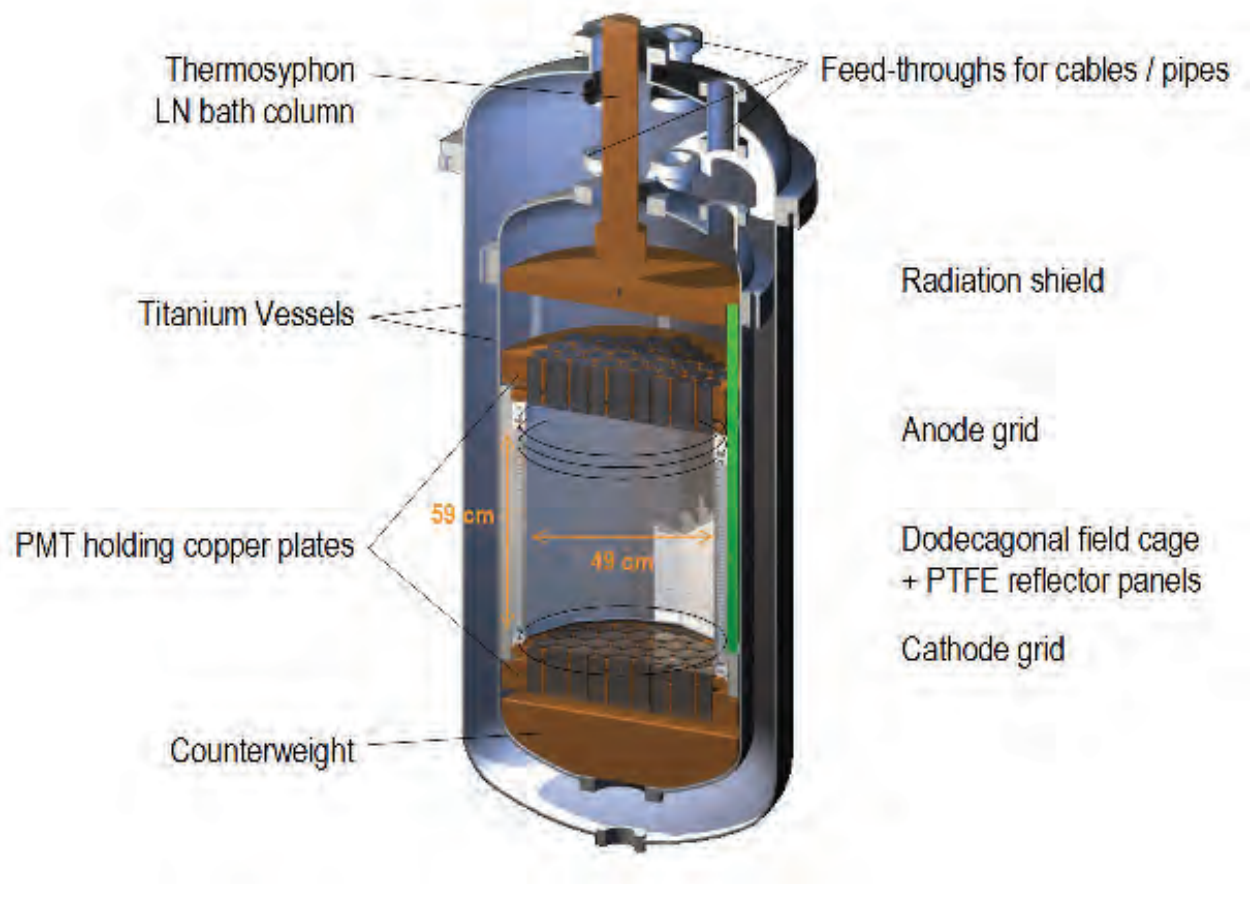

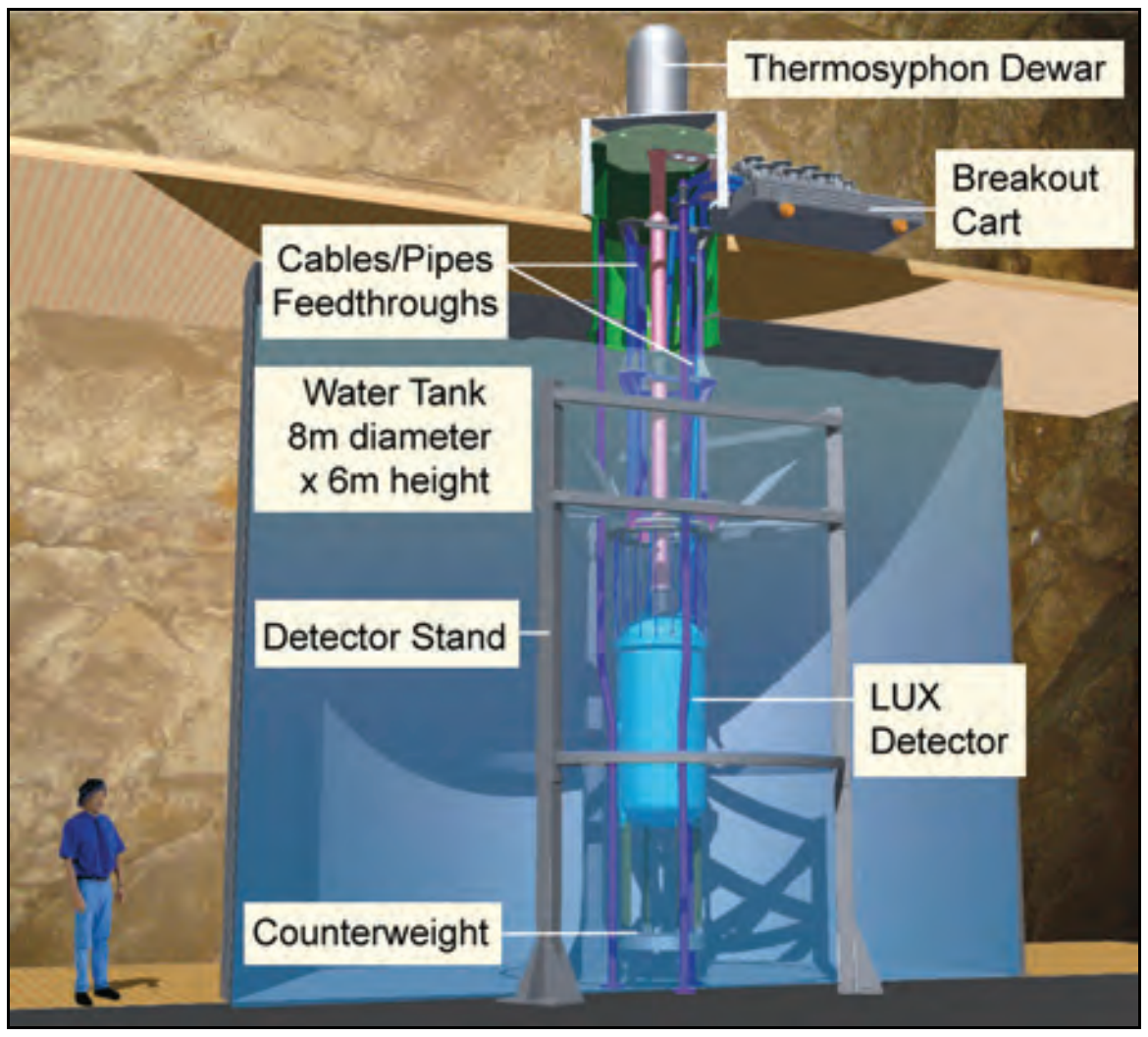

**Figure 3.4.2.1** The LUX detector concept (left) and envisioned in place in the DLM (right). [Courtesy LUX collaboration]

The Surface Facility has been designed to closely emulate both the vertical and horizontal layout of the underground laboratory space (the DLM). The building provides finished laboratory space comprising approximately 190 $m^2$ (2,040 square feet). A vertical shaft descends three stories from the upper floor. The shaft contains a 4-m high/3-m diameter water shield, designed to be a more modest copy of that being placed in the DLM (8-m diameter). The detector will be mounted in the water shield to reduce surface radioactive backgrounds, allowing for detector calibrations as well as providing a direct test of the detector hardware in a water environment. A Class-1000 clean room is maintained for detector assembly and services.

The collaboration gained beneficial occupancy of the Surface Facility in December 2009. Water-tank construction was then completed. Transferring and integration of all LUX subsystems that have been fabricated by the collaboration began in January 2010 and is ongoing. The precision fit of the fully assembled internal structure (that defines drift space, PMT array, and internal plumbing) has been achieved within the titanium cryostats. All mounting and fitting procedures test hardware that will be used in the underground lab, allowing for a characterization of their effectiveness during their final deployment.

The first cryogenic runs of the detector are projected for early 2011, and will be followed by test-filling with liquid xenon. Subsequent runs in early 2011 will include the full 350 kg xenon payload, as well as all 122 PMTs and the full complement of internal hardware and sensors. Data collection and detector-response characterization is expected to begin in the first quarter of 2011, continuing until deployment underground in late 2011. LUX-350 expects to reach a dark-matter cross-section sensitivity of $3\text{x}10^{-46}$ $cm^2$, at a WIMP mass of 60 GeV/$c^2$, which is a factor 100 more sensitive than the best result currently published.

### 3.4.2.2 THE MAJORANA DEMONSTRATOR[102]

In neutrinoless double-beta decay searches, the 1 Tonne Germanium (1TGe) experiment will be the primary endeavor using the $^{76}$Ge isotope. The ultimate experiment will provide unprecedented sensitivity: It is expected to raise the lifetime limit for the $0\nu\beta\beta$ process to the $10^{27}$ year scale or better. The technological complexity of the experiment demands that smaller-scale tests be successfully completed, identifying the best technological approach and experiment design prior to proceeding with the expensive procurement of the full amount of separated isotope. Two coordinated efforts are under way: the MAJORANA DEMONSTRATOR, to be sited at Sanford Laboratory, and GERDA, at Gran Sasso.

The critical path for the DEMONSTRATOR experiment is the production of ultrapure copper for fabrication of the cryostats to hold and cool the germanium crystals. Cosmogenic activation of copper on the surface renders material unacceptably contaminated, for certain critical parts, within a few days to weeks of surface exposure at sea level. Plans are to electroform the most critical copper components in an underground environment, shielded from muons and muon-induced neutrons. In the electroforming process, impurities are left in the baths, providing exceptional purification potential for underground fabrication of the material. However, deposition rates are slow, requiring typically several months to produce a mandrel with sufficient material for machining.

To expedite progress and allow commencement of electroforming activities prior to the completion of the Davis Campus, one of the former Homestake shop areas at the 4850L close to the Ross Station has been designated as the site for a temporary electroforming facility. Preparations for the temporary clean-room area began in December 2009 with the removal of old materials and debris, followed by the installation of new ground support (rock bolts and wire-mesh screen). Shotcrete was applied to the walls and ceiling in August 2010. A modular clean room has been assembled in the renovated space, with utilities and a fire suppression system to be installed by the end of 2010. Rehabilitation work on the Ross Shaft and other underground infrastructure was completed in early September; underground access by experimenters will be available following safety reviews in winter 2010-2011. Operation of the MAJORANA DEMONSTRATOR electroforming baths is expected to begin in February or March 2011.

Once the Davis Campus construction is complete, the DEMONSTRATOR experiment (Figure 3.4.2.2) will be mounted in the Davis Transition Area, the new space designated to house the air-handling and electrical equipment for powering and maintaining the needed cleanliness in the Davis Campus, as well as the dirty-to-clean transition infrastructure for allowing personnel and equipment to move into and out of the Davis Campus. Approximately 280 $m^2$ (3,000 square feet) of space in this Transition Area will be dedicated to the DEMONSTRATOR experiment. Included will be additional electroforming baths, a machine shop, clean (Class 100) assembly hoods, and space for the roughly 2 m x 2 m enclosure of lead and copper to house the two cryostats containing the germanium.

Data-taking with the first modules is anticipated to begin in mid-2012. Full-scale operation will begin in 2014 and is expected to continue for approximately two to five years. In addition to verifying the techniques for achieving necessary background levels, the DEMONSTRATOR is expected to reach a lifetime sensitivity of $10^{26}$ years, sufficient for definitively testing the claimed observation of $0\nu\beta\beta$ decay by Klapdor-Kleingrothaus, et al.

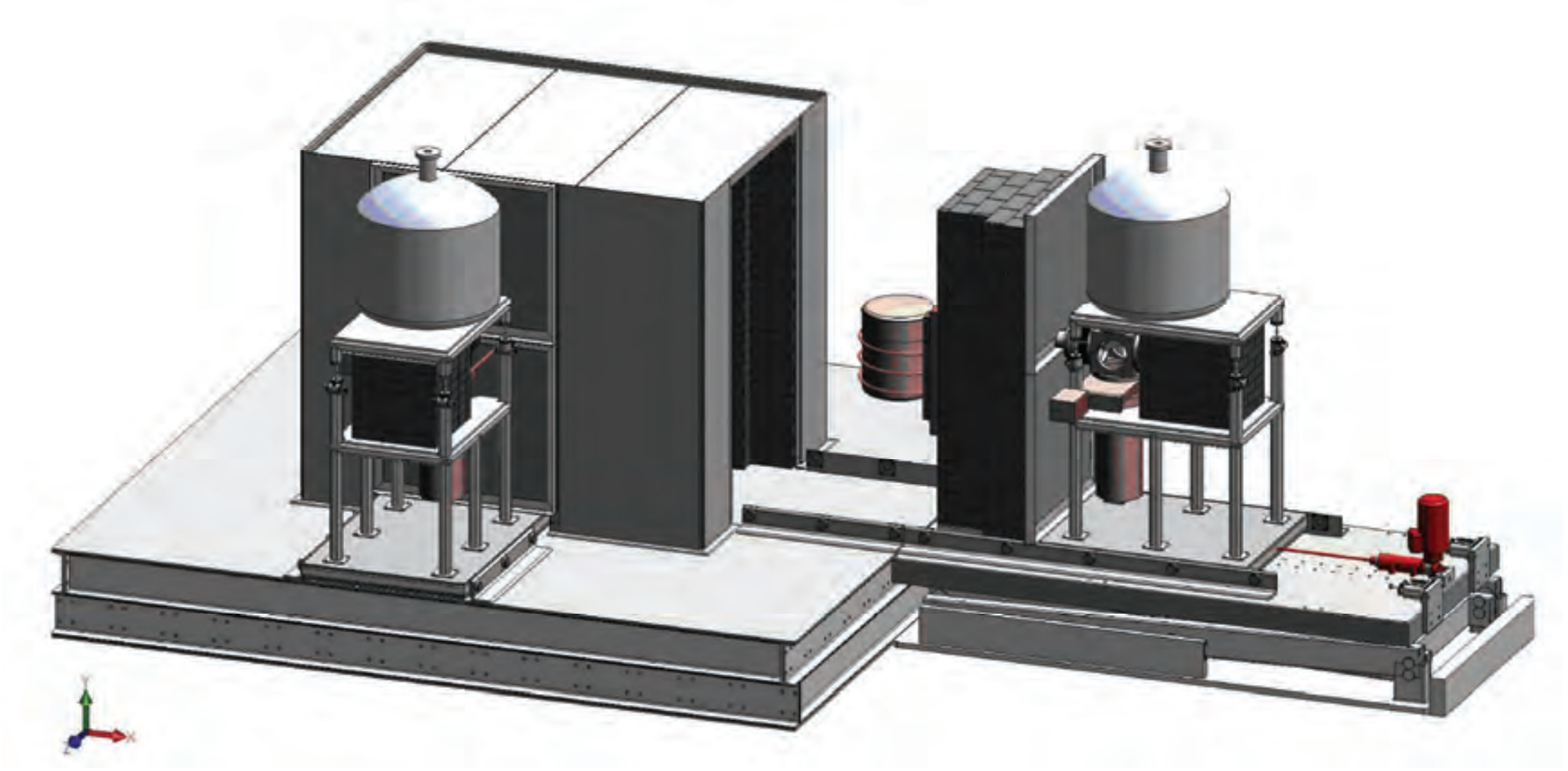

**Figure 3.4.2.2** MAJORANA DEMONSTRATOR apparatus. Two cryostats each containing 20 kg of germanium point-contact detectors mounted in assemblies of ultrapure electroformed copper sit inside a shielded enclosure 2 m on a side of lead and copper. Muon veto counters cover top and bottom surfaces of the shielding. Positive nitrogen flow from the inside of the shield assembly will mitigate radon contamination. [Courtesy MAJORANA collaboration]

### 3.4.2.3 CUBED and Crystal Growth Underground

The Center for Ultralow Background Experiments at DUSEL (CUBED)[103] is supported by the state of South Dakota's 2010 Initiative. One of its initial research focuses is to explore the science and technology for underground material purification and crystal growing. The center, with seven member universities from the state, has been funded by the DOE Experimental Program to Stimulate Competitive Research (EPSCoR) program to develop an underground laboratory to grow detector-grade crystals such as germanium for future experiments at Sanford Underground Laboratory and DUSEL. It has been shown[104] that cosmogenically produced isotopes in germanium crystals manufactured on the surface can create backgrounds for next-generation neutrinoless double-beta decay and dark-matter experiments. To mitigate this problem, it is important for experiments aiming at extremely low background levels (e.g., <~1 count/kg/tonne/year) to have as many detector components as possible (including the detector crystals themselves) manufactured and assembled underground to minimize the total integrated cosmic-ray exposure time. Table 3.4.2.3 shows the estimated rate of cosmogenic radioactivity production in a natural germanium crystal on the surface.[105]

| Natural Germanium (atoms/kg/day) | | | | | Enriched Germanium (atoms/kg/day) | | |
|---|---|---|---|---|---|---|---|
| **Cosmogenic isotopes** | **Lal model** | **Hess model** | **Mei et al.** | **Experiment** | **Lal model** | **Hess model** | **Mei et al.** |
| $^{3}H$ | ~178 | ~210 | 27.7 | - | 113 | 140 | 24 |
| $^{54}Mn$ | 0.93 | 2.7 | 2.7 | 3.3 ± 0.8 | 0.37 | 1.4 | 0.87 |
| $^{60}Co$ | - | - | 2.0 | - | - | - | 1.6 |
| $^{65}Zn$ | 24.6 | 34.4 | 37.1 | 38 ± 6 | 3.12 | 6.4 | 20.0 |
| $^{68}Ge$ | 22.9 | 39.0 | 41.3 | 30 ± 7 | 0.54 | 0.94 | 7.2 |

**Table 3.4.2.3** Calculated and experimental cosmogenic production rates in natural Ge. Calculated production rates in enriched Ge assume 86% $^{76}Ge$ and 14% $^{74}Ge$.[105]

The technology for growing large-volume detector-grade high-purity germanium (HPGe) crystals was developed in the 1970-80s. The growth of HPGe crystals is a very demanding process, requiring the combination of contamination-free equipment, fast-responding diagnostic instrumentation, technological know-how, skilled personnel, and process optimization. Two crystal pullers have been purchased and CUBED researchers are currently developing their skills at surface laboratories based at the University of South Dakota (USD) and the South Dakota School of Mines & Technology (SDSM&T).

The CUBED proposal describes an underground crystal-development laboratory comprising roughly 150 $m^2$ (1,500 square feet). The laboratory is made up of three rooms for mechanical handling and zone refining, crystal growing—equipped with crystal pullers, expected to be developed and finalized during the R&D phase aboveground—and a crystal diagnostics instrumentation and detector-preparation room. The rock overburden required for underground crystal growth could allow for the laboratory to be located at a depth shallower than the 4850L, but there may be advantages to consolidating the initial-science laboratories on that level. Approximately two years of development at surface facilities is planned before underground deployment could begin.

### 3.4.2.4 Other Physics Groups

In addition to the large physics projects described above, several groups are performing fundamental measurements to characterize the Homestake underground environment. One collaboration, interested in developing a vertical facility, collected an initial set of magnetic field measurements in July 2009 that will be useful for the design of such a laboratory.

Physics backgrounds such as fluxes of gamma rays,[106] muons,[107] neutrons (including muon-induced neutrons), and concentrations of radon are being quantified and will be of interest to sensitive neutrino and dark-matter experiments. A group of scientists led by USD, including Regis University, have been working on the radioactive background characterization at different levels (surface, 800L, 2000L, 4550L) since 2008 and they expect to begin measurements on the 4850L starting in early 2011.

Researchers from USD and Black Hills State University (BHSU) are assisting with the radon-monitoring program at Sanford Laboratory that is currently under way. A number of measurements have been taken over different time periods and at various locations; however, the most useful data are obtained from detectors deployed at a given location for several months.[1] Average radon concentrations at the 4850L ventilation supplies are in the range 200-650 $Bq/m^3$—see Figure 3.4.2.4.

Airborne radon daughters from the U and Th chain can be plated out on surfaces of solid components or dissolved in the liquid media during detector construction and could contribute to long-term experimental backgrounds. A well-known case is $^{210}Pb$ from $^{222}Ra$, which has a long enough half-life to be present for the duration of a typical experiment. There are OSHA regulations about permissible radon levels for the work environment as well. The radon level at Homestake has been continuously monitored by SDSTA at Sanford Laboratory at different depths and drift locations underground. Due to the work-in-progress status of the ventilation system, many of the measurements described above are not conclusive. Experiments that are critically sensitive to the airborne radon background could install local scrubbers to reduce the level down to a few tens of $mBq/m^3$ if required. DUSEL is exploring options for introducing small quantities of surface air (~few $Bq/m^3$) by ducting systems such as those employed by the Super-Kamiokande experiment at Kamioka.[108]

The gamma background for the experiments depends in part on the intrinsic U, Th, and K content in the underground rocks surrounding the laboratory. Extensive shielding is required to attenuate these gammas.

A program of radiometric study of rock samples from the former mine operations and core library has been carried out at the Low Background Facility at LBNL with data analyzed and summarized in several reports.[109] One finding was that while most of the bulk rocks are radioactively clean (with U/Th in the sub-ppm scale), there are certain rock intrusions, usually localized, such as rhyolite that are found to have a much higher U/Th content (Table 3.4.2.4).

Due to civil construction needs, a layer of concrete or shotcrete is typically applied to the exposed drift or cavity surfaces underground. These materials also contain U, Th, and K, and depending on the total mass, their contributions to background could be significant. Potential candidate samples of concrete components and aggregates have been assayed (Table 3.4.2.4) and additional measurements are anticipated. These results will be used to guide the material selection and construction planning of the Facility.

[1] Changes in the ventilation system (either due to temperature changes, maintenance, or when ventilation doors are left open for certain activities) can give rise to significant short-term deviations from the baseline.

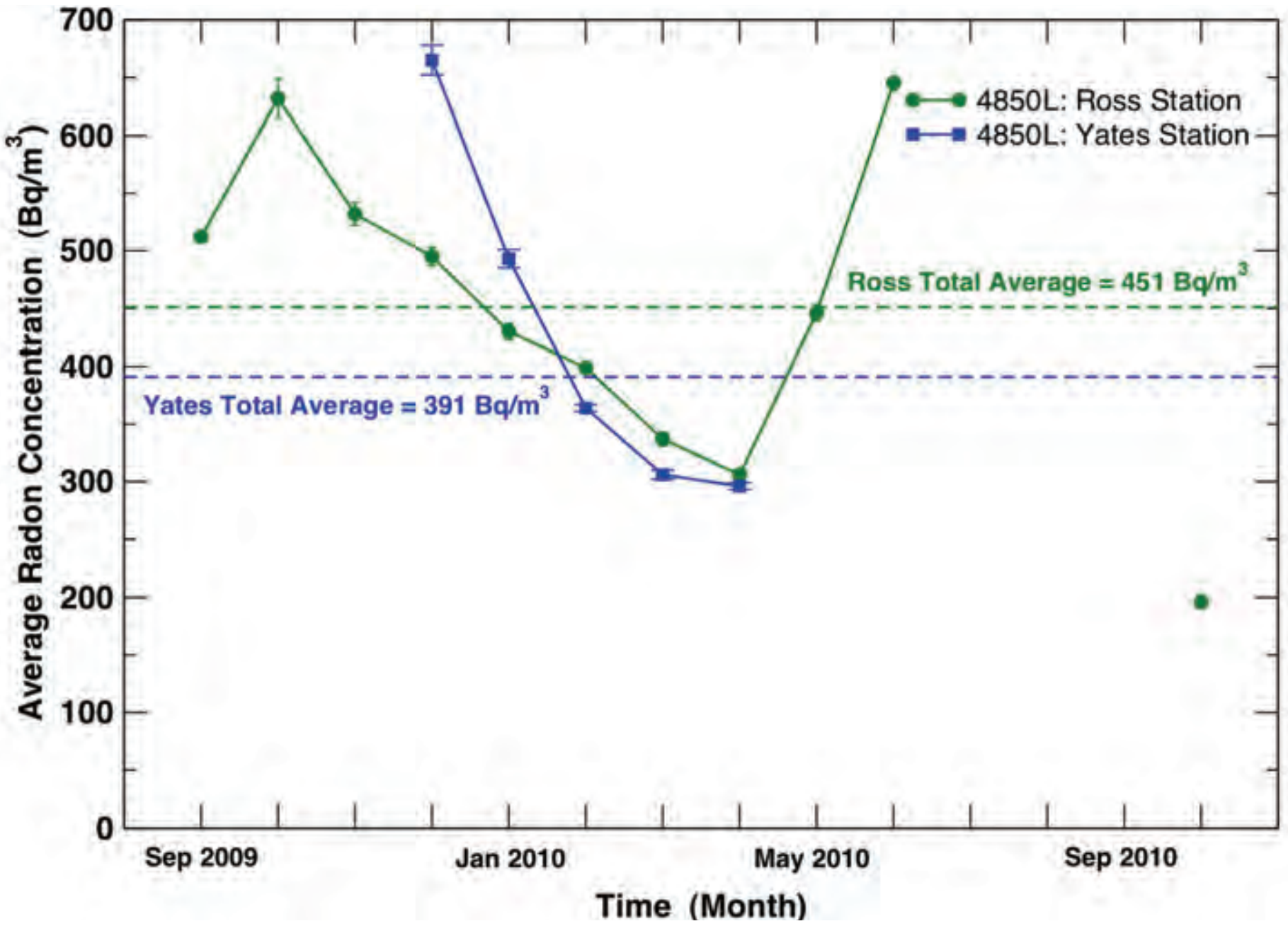

**Figure 3.4.2.4** Sanford Laboratory average monthly radon concentration as a function of time. Measurements are taken at the two main ventilation supplies for the 4850L (data for the Yates Station measurement covers a shorter period of time due to equipment maintenance and construction activities). [Courtesy SDSTA]

| Sample | Uranium (ppm) | Thorium (ppm) | Potassium (%) |
|---|---|---|---|
| **Low-Activity Bulk Rock** | 0.059-0.091 | 0.24-0.30 | 0.18-1.94 |
| **Rhyolite Intrusion** | 4.42-10.9 | 8.76-11.4 | 2.49-7.6 |
| **Typical Local Shotcrete** | 1.62-2.61 | 1.99-3.92 | 0.36-1.28 |

**Table 3.4.2.4** Summary of radiometric results for recent Homestake samples. [Courtesy Low Background Facility, LBNL[109]]

## 3.4.3 Biology, Geology, and Engineering

### 3.4.3.1 Geology

Studies of the geological environment are under way as part of the Early Science program, with several groups engaged in ongoing research. The work by the BGE Early Science community already is yielding results. The $CO_2$ Sequestration group visited the Laboratory in June 2009 to identify potential locations for the proposed geologic carbon sequestration experiment Facility at DUSEL. Part of this effort involved the temporary installation of instruments for environmental monitoring of temperature, pressure, and relative humidity at three different sites (800L, 2000L, and 4850L). This information was then used to identify suitable locations for the $CO_2$ Sequestration Facility along the depth profile.

Another application of geological characterization involves collaborators from the Laser Interferometer Gravitational-wave Observatory (LIGO) project, who are mounting an initiative called the Deep Underground Gravity Laboratory (DUGL) to study seismic noise levels in the frequency range of interest to gravity-wave observations (<10 Hz). This effort may help establish the Homestake environment as a suitable site for the next generation of gravity-wave observatories.[110]

One of the earliest groups to collect data from the Homestake site after re-entry was a team from the South Dakota and Arizona U.S. Geological Survey (USGS) offices. The group uses a very sensitive gravimeter to measure microgravity at particular locations in order to understand the hydrology and groundwater in the Northern Black Hills and more specifically to understand and monitor the Laboratory dewatering process.

Researchers from SDSM&T, in conjunction with SDSTA personnel, are studying water flow along with temperature and humidity dynamics on several levels (surface, 1250L, 2000L, 2600L, 4850L). One aspect of this project involves monitoring the declining water table as the underground facility is dewatered to better understand rock deformation processes. This monitoring is also useful to experiments like the LBNE that are interested in knowing water pressure in areas surrounding the proposed large cavities and LMs.

The Petrology, Ore Deposits, and Structure (PODS) group is combining access to the Homestake drill-core archive and other records with some underground mapping (800L) to pursue topics related to the deposition and mineralization of the gold deposit in Lead. Members of the PODS group continue, as they have in the past, to play a leading role in the stewardship of the drill-core archive.

One of the most synergistic groups that has participated in research at Sanford Laboratory is the Tiltmeter group, comprising personnel from SDSM&T, Fermilab, and the University of Wisconsin at Madison. Combining the hydrology data mentioned above with rock-deformation data measured by tiltmeters will lead to a more complete picture of the dewatering process. The group is also interested in testing different tiltmeter technologies and the most feasible methods to install them. To achieve this, three arrays of two different types of tiltmeters have been installed on the 2000L according to two different methods.

### 3.4.3.2 Biology

Biological sampling of life forms found underground is very active: several new strains have been encountered.

Biologists from BHSU and SDSM&T have collected many samples in the course of their research on metagenomic analysis and bioprospecting of the microbial communities at Homestake following underground facility dewatering.[111] Sample collection sites include surface, 300L, 2000L, 4100L, and 4850L. Interaction with researchers familiar with manifold sampling techniques led to the establishment of the 4100L site, where a borehole packer is being used to pressurize groundwater to sample for micro-organisms through various filters, some of which are used to perform subsequent DNA analysis.

Researchers from South Dakota State University (SDSU) are investigating the enrichment and isolation of lignocellulose-degrading micro-organisms from selected sites at Sanford Laboratory. To date, 13 filamentous fungal cultures have been isolated as well as have 32 unicellular bacteria. Eight of the 32 unicellular bacteria so far examined are able to hydrolyze a form of cellulose at two specific temperatures. Strains that have been identified include three *Bacillus pumilus* strains, two *Bacillus licheniformis* strains, and three *Bacillus subtilis-subtilis* strains. More isolates are being sought.

A workshop was organized by biologists from Princeton University and the University of Tennessee at Knoxville in June/July 2009 to demonstrate manifold sampling techniques that have been used with success in deep South African mines. Researchers from BHSU, SDSM&T, SDSU, University of Tennessee at Knoxville, and Oak Ridge National Laboratory attended the workshop, in which sources of water on the 2000L, 4550L, and 4850L were used to illustrate the sampling methods.

The first publications[112,113] reporting research following the Homestake closure detailed the discovery by an SDSM&T team of a cellulosic-degrading bacterial strain living in soil-like samples collected in May 2008 that were found in the Yates and Ross Shafts as well as in the #6 Winze at the 4550L.[114,115] The results have strong implications for biological conversion of cellulosic agricultural and forestry wastes to commodity chemicals, including sugars.

### 3.4.3.3 Engineering

Several engineering groups from SDSM&T are active at Sanford Laboratory, testing various applications for use in the underground environment.

Data were collected on the 300L by a group from the SDSM&T Electrical and Computer Engineering Department in August 2009 regarding electromagnetic signal propagation in a tunnel environment. The measurements were conducted with multifunctional antennas that the group developed for cryospheric applications at NASA's Goddard Space Flight Center and provided estimates of the tunnel wall material properties up to a depth of a few centimeters.

An SDSM&T group comprising several engineering fields that is developing an autonomous submersible vehicle visited the Laboratory in December 2009 to understand environmental field conditions for a possible future deployment. Two locations were investigated, the 1250L sump near the pump room and the 4850L #6 Winze. Magnetic field measurements were taken to gauge how well the navigation system on the vehicle would work in the underground environment.

Collaborators from the University of Wisconsin-Madison and Montana Tech associated with the GEologic Optical eXtensometer and TiltMeter (GEOX™) DUSEL Project are investigating applications of fiber-optic cable on the 4100L as part of a set of optical extensometer arrays as well as monitoring drift temperature gradient. The full scope of GEOX™ also involves the research with tiltmeters described below, allowing the group to study different deformation and temperature sensors over different length scales. Distributed fiber-optic sensors can extend over kilometers of distance and are highly stable over long times. The objectives of the initial experiments with fiber-optic cable are to develop installation techniques, cross-calibrate them against conventional extensometers, measure rock mass properties at a scale of several meters, and establish its potential for structural health monitoring (SHM) of DUSEL on the scale of the Laboratory volume.

The Transparent Earth collaboration has instrumented three boreholes (2x 2000L, 4100L), each with three accelerometers to allow high-frequency seismic signals to be detected and analyzed when the signal exceeds a defined amplitude level. In addition to the accelerometer instrumentation, each of these sites has an associated tiltmeter. Ultimately, this group intends to deploy a large number of this type of seismic monitoring station throughout the DUSEL Laboratory.

### 3.4.4 Early Science and DUSEL Construction

The DUSEL MREFC-funded Project is anticipated to begin in early 2014, and underground construction will start later in 2014. Rehabilitation activities in the Ross Shaft and elsewhere will take place prior to 2014. The MAJORANA DEMONSTRATOR anticipates continued operation through 2017. A Generation-2 dark-matter experiment located in the 4850L Davis LM could be implemented after completion of the LUX experiment and possibly operate into 2017. The CUBED collaboration expects to begin development in surface facilities in the next two years and then, if successful, move operations into an underground area where it may operate for some years. Their activity is also closely connected with the production of Ge for potential future DUSEL experiments. The BGE activities described above are expected to continue, at least in part, well into the period when DUSEL construction begins. New activities in BGE may also arise in the next few years. The DUSEL construction schedule will be devised to preserve the operation of the early science experiments consistent with meeting overall Project critical milestones. Some interruptions (months) in the operation of LUX, MAJORANA DEMONSTRATOR, and CUBED may be required to switch electrical, water, and IT infrastructure during the DUSEL construction period. Every reasonable effort will be made to minimize the disruptions, and these periods will be included in the DUSEL Project schedule.

## 3.5 Integrated Suite of Experiments

Potential experiments for the DUSEL Facility have been described in Chapter 3.3. The ongoing and planned scientific programs at Sanford Laboratory were presented in Chapter 3.4. The initial experimental program at the completed DUSEL Facility remains to be determined. We anticipate that the dark-matter and $0\nu\beta\beta$ experiments under construction and soon to be installed at Sanford Laboratory will continue operation into the period of the construction of the DUSEL Facility for as long as justifiable by their scientific output. In addition, BGE experiments already under way or anticipated at Sanford Laboratory will continue. It is likely that additional investigations, or scientific directions, prior to the completion of the DUSEL Facility will be proposed. Those will be reviewed by the DUSEL PAC (see Chapter 3.10) and the integrated management of Sanford Laboratory and the DUSEL MREFC-funded Project. All future experiments proposed to be housed in the DUSEL Facility will be similarly reviewed. The proposed means for selection and management of the experimental program are summarized in Chapter 3.10.

The design of the DUSEL Facility is driven essentially by the needs of the experimental program. Although this program has not yet been defined in detail—indeed, it will take some years to establish a specific set of first experiments—the preliminary requirements needed for the design of the civil construction aspects of the Facility are being, indeed must be, established now. The requirements for the civil construction of the Facility are guided by our expectations and goals for a generic Integrated Suite of Experiments (ISE). The process for obtaining these requirements and a summary of the requirements are given below in Chapters 3.6-3.8. In the present chapter, we provide an overview of the key elements of the generic ISE as it pertains to obtaining the Facility requirements for underground civil construction. Science-driven requirements for the projected first experiments at DUSEL that guide the definition of the generic ISE were described in Chapter 3.3.

### 3.5.1 Long Baseline Neutrinos

A long-baseline neutrino experiment (LBNE) will be included in the ISE. The selection among the options for implementation of this experiment (Section 3.3.5) will ultimately determine the Facility requirements. The current baseline design for the DUSEL Facility describes in detail the requirements and implementation of one large cavity and associated infrastructure for a water Cherenkov detector. Options—the addition of an additional cavity for a water Cherenkov detector, a cavity of larger size than the baseline, and facilities and infrastructure for a LAr detector (or detectors)—have been described, and requirements for the options gathered but at a less-detailed level at this time. A selection among the possible options is anticipated to be made by mid-2011 prior to the start of the Final Design of the Facility, currently anticipated to begin in early 2012.

### 3.5.2 Proton Decay

The ability to search for proton decay will be an integral part of the LBNE detectors and thus will be included in the ISE. Separate detectors aimed solely at proton decay searches are not currently included in the ISE. Specific requirements that enable the proton-decay search will be included in the Facility design.

### 3.5.3 Detection of Other Neutrinos

The detection of solar neutrinos, neutrinos from supernovae, and other astronomical sources (see Section 3.3.5.9) are possible goals for the large detectors for long-baseline neutrinos and proton decay and thus part of the ISE. Separate detectors aimed solely at detection of these neutrinos are not currently included in the ISE. Specific requirements that enable detection will be considered in the Facility design.

### 3.5.4 Dark Matter Experiments

At least one Generation Three (G3) (see Section 3.3.3) dark-matter experiment will be included in the ISE, chosen from those to be proposed.

### 3.5.5 $0\nu\beta\beta$ Decay

At least one $0\nu\beta\beta$ experiment (see Section 3.3.4) will be included in the ISE, chosen from those proposed. Designs are being evaluated at the 7400L that include a variety of shielding configurations (Cu/Pb, LAr, water) so that the Facility can be designed while retaining the flexibility to accommodate different designs.

### 3.5.6 Nuclear Astrophysics

The Dakota Ion Accelerators for Nuclear Astrophysics (DIANA) (Section 3.3.6) is currently the only proposal for this area of science. The final Facility design at the 4850L will allow aspects of DIANA to be implemented if it is selected to be one of the first DUSEL experiments.

### 3.5.7 Low Background Counting and Materials Assay

Basic aspects of infrastructure for low-background counting and materials assay will be included in the design of the DUSEL Facility in support of the experimental program. More advanced R&D and a Facility in this area are represented currently by the Facility for Assay and Acquisition of Radiopure Materials (FAARM) proposal (Section 3.3.8). The Facility design at the 4850L will allow aspects of FAARM to be implemented if it is selected to be one of the first DUSEL experiments.

### 3.5.8 Biology, Geology, and Engineering

BGE experiments, as currently proposed, and the related Facility requirements are described in Section 3.3.7. The Facility design will accommodate a subset of the proposed experiments (or experiments proposed in future). The number and type of experiments will be determined later, based on reviews of scientific merit, funds available, and appropriateness to the Facility (see also Chapter 3.10).

### 3.5.9 Support Activities and Staging Areas

The underground Facility will provide areas for machining, low-activity materials fabrication, and other support activities for the experimental program. An initial scope for such areas will be part of the baseline Facility design.

Staging areas for assembly and installation will be included in the ISE requirements. An initial scope for such areas will be part of the baseline Facility design.

### 3.5.10 Research and Development

Areas for R&D of new concepts for physics and BGE experiments are an important aspect of the experimental program. An initial scope for such areas, both above- and below-ground, will be part of the Facility design.

## 3.6 Integrated Suite of Experiments (ISE) Requirements Process

Integral to development of a Preliminary Design for the Homestake DUSEL Facility is an understanding of the infrastructure required to implement the ISE. DUSEL science and engineering staff have determined requirements of potential experiments via workshops, meetings with experiment collaborations, and through a Web-based system in which experimenters specify the necessary infrastructure required to implement their proposed experiments.

The development of a Preliminary Design for the DUSEL Facility has proceeded in parallel with the development of proposals for experiments to be incorporated into the ISE. However, the development of the Facility design has proceeded on a different timeline than the experiments. In particular, the project schedule requires that the Facility Preliminary Design Report (PDR) be complete before the final ISE is selected and, in fact, before experiment designs are fully matured. The strategy for accomplishing this has been based on developing a detailed understanding of proposed experiments and to develop Facility designs to accommodate a generic ISE, as described in Chapter 3.5.

### 3.6.1 Science Liaison Organization and Activities

The DUSEL Science Liaison Group consists of scientific specialists corresponding to the various disciplines represented by the proposed experiments. Engineers experienced with science integration are assigned to work with the scientists—see Appendix 3.A

Several ISE workshops have been held, providing opportunities for the science liaison staff to interact with potential experimenters. These included two workshops organized by the DUSEL Experiment Development and Coordination (DEDC) Committee, held April 21-26, 2008, and September 30-October 3, 2009. The workshops were organized around work groups focusing on areas of experimental interest. These meetings provided the opportunity for direct discussion with potential experimenters about their facility infrastructure requirements.

A Web-based database system for experimenters' entry of infrastructure requirements was created following the April 2008 ISE workshop. This system was upgraded and linked to the DUSEL document management system in December 2009. The database, along with direct interactions between experiments and the science liaison staff, have provided the basis for the experiment-specific requirements tables contained in Sections 3.3.3 through 3.3.8.

Much of the information supplied to DUSEL from the experiments has been accomplished through a formal schedule of deliverables, outlined in Table 3.6.1-2. Following the last of the three formal deliverables, one-on-one detailed meetings will be held with each experiment. The meetings' purposes will be to review and refine the information supplied through the formal deliverables, including cost, schedule, and potential funding.

| Date | Deliverables |
|---|---|
| 12/18/09 | Team roles and responsibilities, project contacts<br>Science and project objectives, phases, and evolution |
| 1/29/10 | Location of proposed experiment<br>Experiment layout drawing, showing all major subsystems<br>Requirements and hazard identification, via entries to the database |
| 3/19/10 | Cost estimate, Conceptual Design level or better<br>Schedule, Conceptual Design level or better. To include major milestones through design, assembly, and installation.<br>Depth document, justification for those requesting installation at the Deep-Level Campus |

**Table 3.6.1** Experiment collaboration scheduled deliverables.

### 3.6.2 Requirements Flow-Down Process

The requirements process for the Preliminary Design phase involves transforming the DUSEL scientific program goals into an ISE. From the ISE comes the high-level system functional and operational requirements; these inform the DUSEL civil Facility design. The input into this process consists of a multitude of individual experiment needs and the output is a clear set of requirements needed to build the Laboratory facilities. This process, which assures that the DUSEL design is linked to documented experiment needs, is shown in Figure 3.6.2-1. ISE requirements that are directly flowed into the detailed low-level design requirements are documented in Appendix 9.F, Integrated Suite of Experiments Interface Requirements Document (IRD).

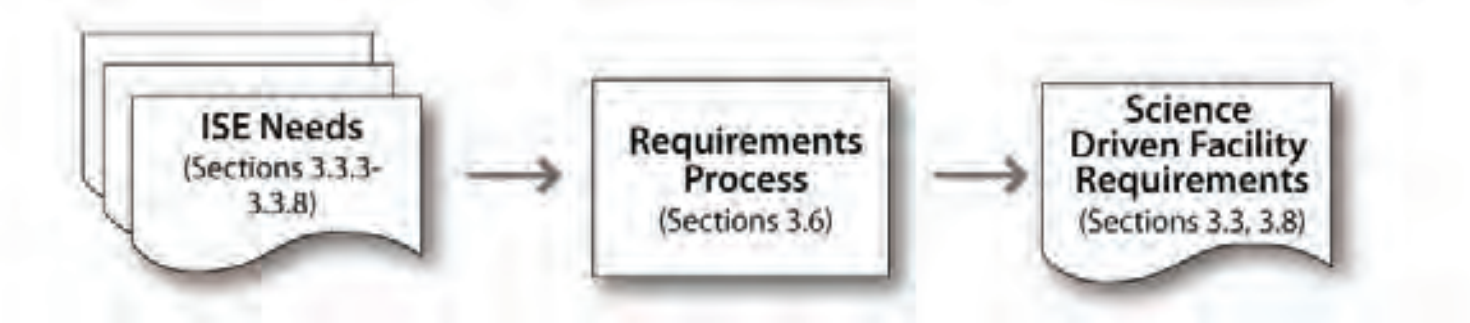

**Figure 3.6.2-1** Requirements process inputs and outputs. [W. Kalinowski, DUSEL]

This volume of the PDR discusses specific experiment proposals and related requirements in Sections 3.3.3 through 3.3.8. Generic requirements were developed for dark-matter and neutrinoless double-beta decay experiments, as there are multiple proposals within these two categories; these are discussed in Chapter 3.7. Finally, in Chapter 3.8, science-driven requirements are associated with laboratory locations and facilities. The current section describes the high-level process employed by the team.

The ISE consists of experiments that physically reside in different areas in DUSEL. Physics experiments will typically be housed within dedicated laboratory spaces that are part of the Mid-Level Laboratory (MLL) and Deep-Level Laboratory (DLL) campuses. The biology, geology, and engineering (BGE) experiments will typically reside in drifts or other areas outside the main campuses; these areas are designated as other levels and ramps (OLR). Table 3.6.2 lists the physics experiments that were directly considered in the determination of the requirements of the large cavity (LC) and laboratory modules (LMs). Figures 3.6.2-2 and 3.6.2-3 illustrate the requirements flow-down for the LMs. Table 3.8.3-1 summarizes proposed experiments by level. Table 3.8.3-2 summarizes the available access in linear feet for each level, and the total power available to support the proposed experiments by level.

All spaces associated with the LBNE will be dedicated solely to this experiment. Requirements associated with this experiment are discussed in Section 3.3.5. Chapters 3.6, 3.7, and 3.8 will discuss requirements and processes associated with LMs and OLR, which are shared spaces.

The following logic has been used to associate experiment requirements with the two MLL LMs (LM-1 and LM-2) and the single DLL LM (LMD-1):

- LM-1 is 50 m long and will accommodate either the proposed DIANA nuclear astrophysics experiment or a single large dark-matter, double-beta decay, or low-background counting experiment. DIANA is unsuitable to be close to other experiments, so it would be the only occupant. For other experiments, approximately half the space would be available for R&D and prototyping activities. Requirements will be based on the most challenging individual requirement among the experiments. Since DIANA has the highest power requirement, it will be used to set the power and other associated requirements, such as chilled water. All other requirements are driven by large cryogenic dark-matter or double-beta decay experiments.
- LM-2 is 100 m long and is assumed to be compatible with three physics experiments of any variety other than DIANA. Requirements will be based on the most challenging combination of three.
- Although four proposed experiments have designated the DLL as their preferred or required location, LM-1 and LM-2 will be compatible with installation of these experiments as well. Full justification for installation for experiments at the deep level will be based on experiments that are currently under way or that are planned to start in the next few years. We therefore do not want to preclude the possibility of installing one of these at the MLL.
- LMD-1 is 75 m long and is assumed to be compatible with installation of two experiments. Only requirements associated with the four experiments asking for DLL installation will be used to set requirements for the LMD-1. The most challenging combination of two will be used to set requirements where they are compatible within the constraints associated with the baseline design.
- Experiment envelope size is based on constraints associated with the baseline LM designs. Note that current layouts for several of the experiments proposed for installation in LMD-1 are not compatible with these envelope definitions.
- Requirements for OLR are organized by level. This is discussed in detail in Section 3.8.4.

Figures 3.6.2-2 and 3.6.2-3 illustrate the flow of experiment requirements into Laboratory Facility requirements. Note that all the generic neutrinoless double-beta decay requirements are the same as those for LM-1 and LM-2, with the sole exception of size constraints. Generic dark-matter constraints are different, as there are no proposed cryogenic dark-matter experiments for LMD-1.

| Facility | Experiment Name | Experiment Type |
|---|---|---|
| Large Cavity | LBNE | Long Baseline Neutrino |
| | | Proton Decay |
| | | Other Neutrinos |
| LM-1 | EXO | Neutrinoless Double Beta Decay |
| | 1TGe | |
| | LZD | Dark Matter |
| | MAX | |
| | GEODM | |
| | COUPP | |
| | DIANA | Nuclear Astrophysics |
| | FAARM | Low Background Counting |
| LM-2 | EXO | Neutrinoless Double Beta Decay |
| | 1TGe | |
| | LZD | Dark Matter |
| | MAX | |
| | GEODM | |
| | COUPP | |
| | FAARM | Low Background Counting |
| LMD-1 | EXO | Neutrinoless Double Beta Decay |
| | 1TGe | |
| | GEODM | Dark Matter |
| | COUPP | |

**Table 3.6.2** Experiments considered in the derivation of the laboratory module and large cavity Facility requirements.

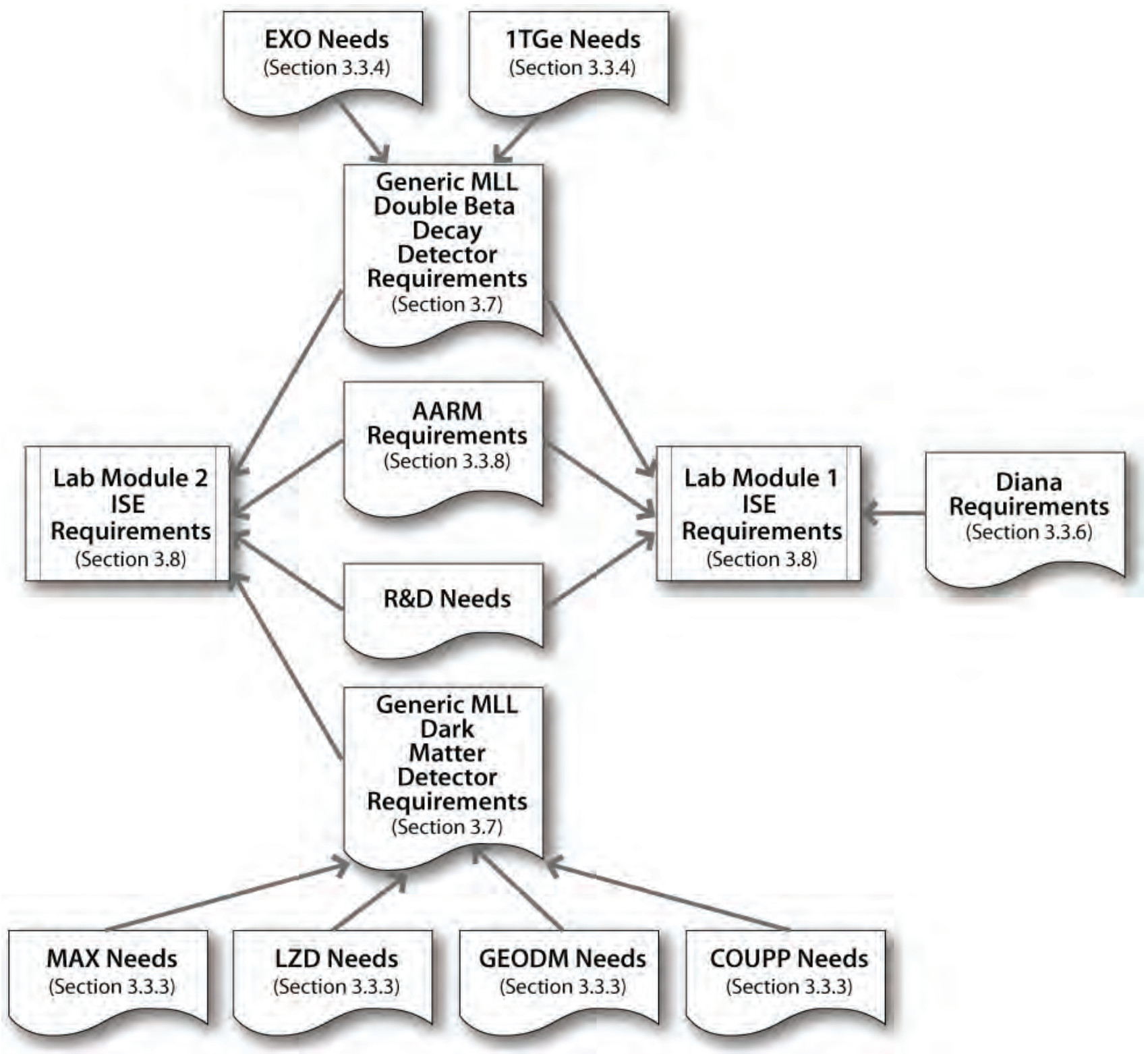

**Figure 3.6.2-2** Flow of Integrated Suite of Experiments needs into LM-1 and LM-2 requirements. [W. Kalinowski, DUSEL]

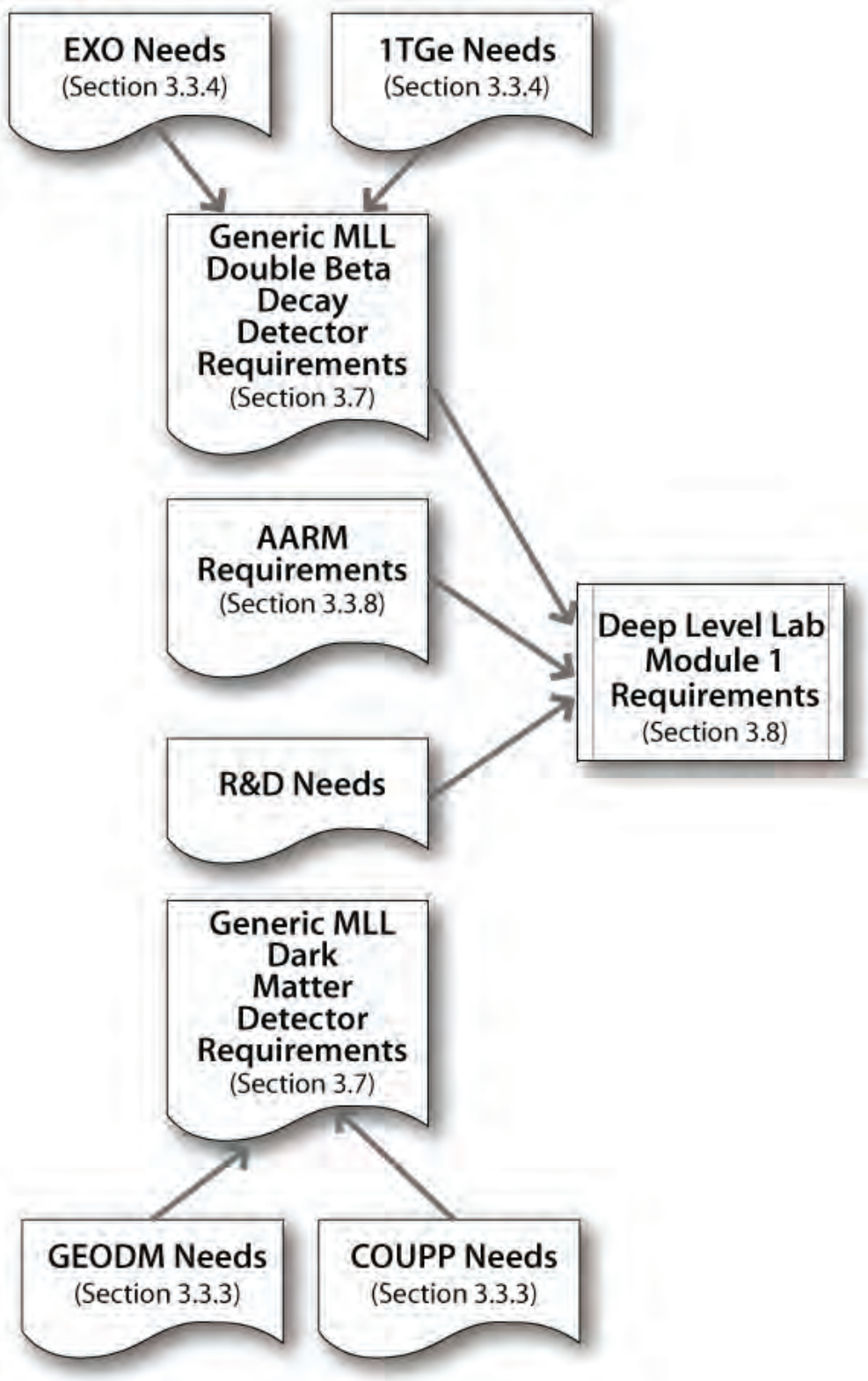

**Figure 3.6.2-3** Flow of Integrated Suite of Experiments needs into LMD-1 requirements. [W. Kalinowski, DUSEL]

### 3.6.3 Physics Campus Layouts

Figure 3.6.3-1 shows a typical experiment installation in LM-1. This figure shows the proposed LZD dark-matter experiment installed. The allowance for a single-experiment installation is 25 m long. This will leave half of the 50-m-long space available for prototypes and general R&D space.

Figure 3.6.3-2 shows typical installation of three experiments in LM-2. The figure shows FAARM, the proposed low-background assay facility; and MAX, the proposed dark-matter experiment comprising two cryogenic detectors, one liquid xenon and the other LAr. The MAX installation shows a shared clean room than can access either detector. This design feature could be used for other combinations of cryogenic detectors installed in the same LM.

Figure 3.6.3-3 shows a typical installation of two detectors in LMD-1. The figure shows the proposed Germanium Observatory for Dark Matter (GEODM) experiment, and the Cu/Pb shield configuration for 1TGe. Note that installation of these experiments with layouts as shown would be problematic, as there is no space left for lay-down and assembly. The maximum allowance for an experiment installation would be 25 m to allow for lay-down and assembly space. While the GEODM layout fits within this constraint, the 1TGe layout is approximately 45 m long.

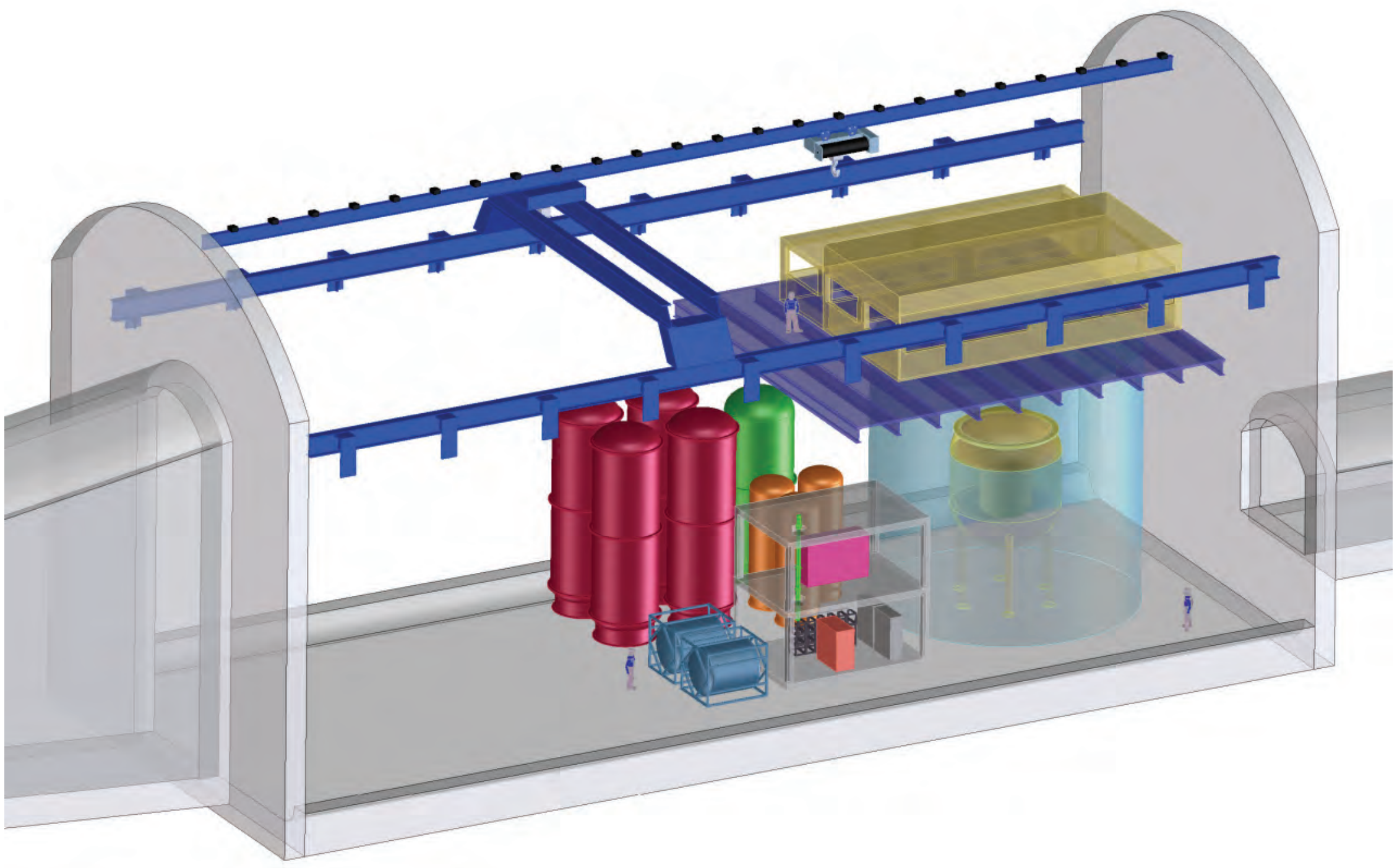

**Figure 3.6.3-1** Typical experiment installation in LM-1. [Dave Plate, DUSEL]

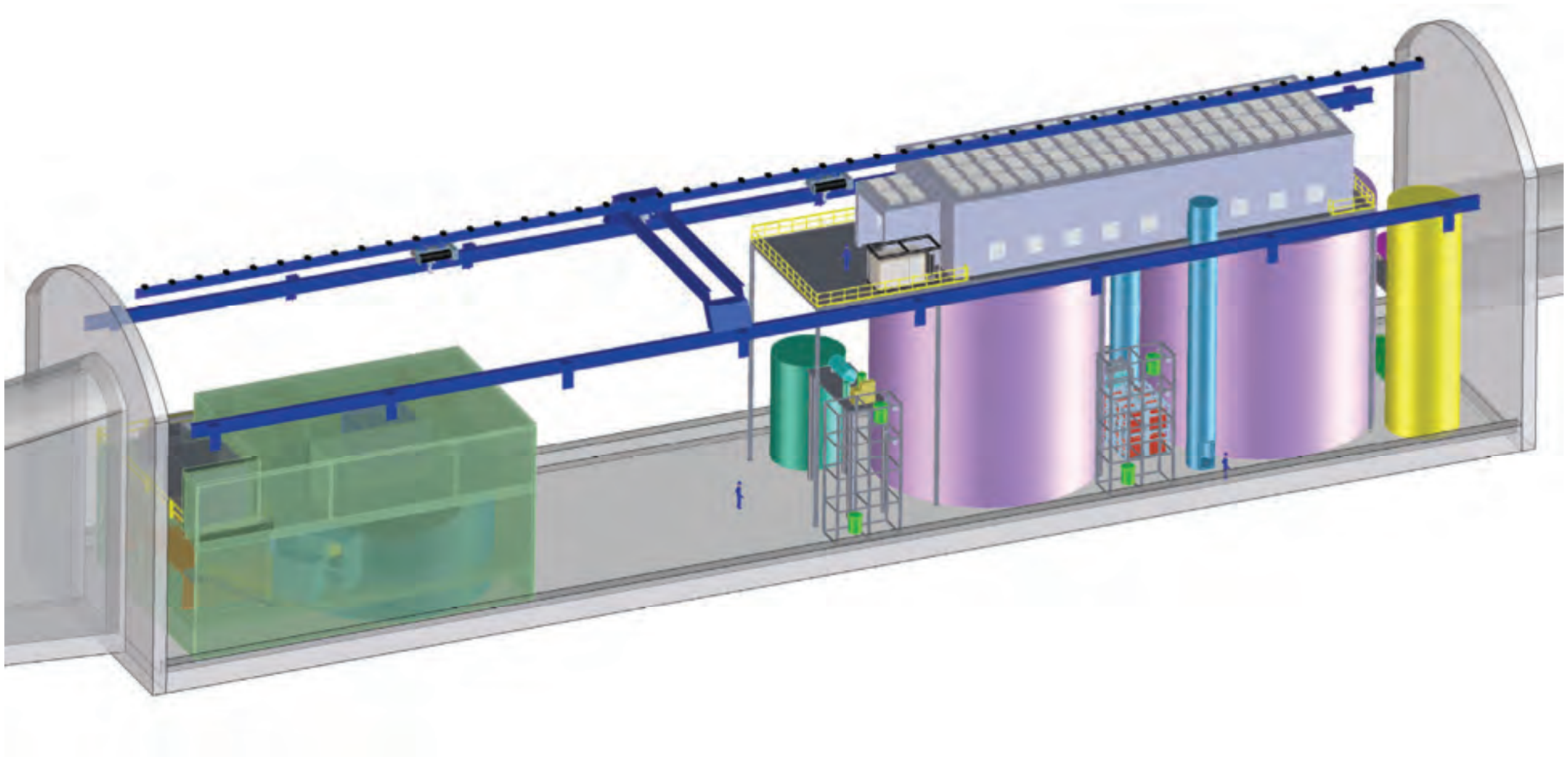

**Figure 3.6.3-2** Three experiment installations in LM-2. [Dave Plate, DUSEL]

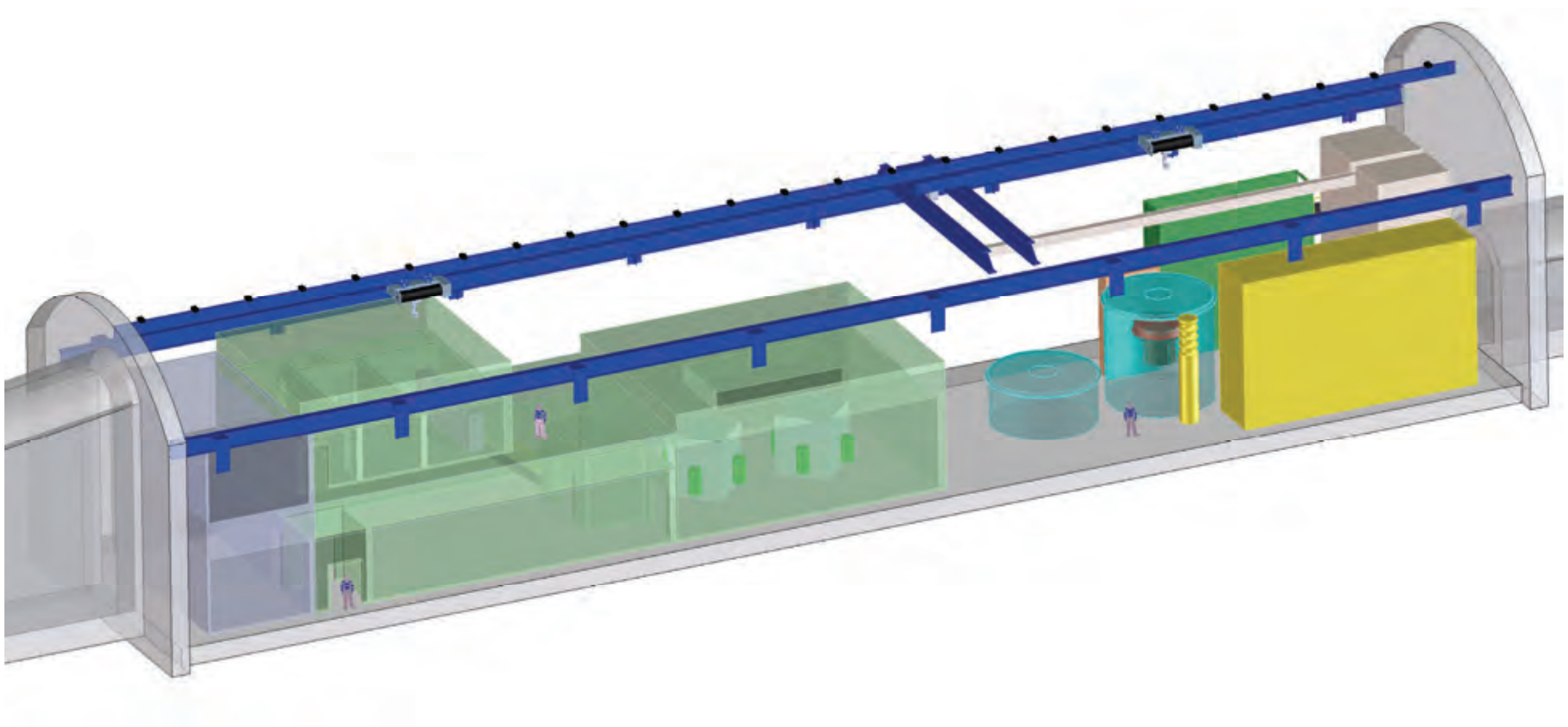

**Figure 3.6.3-3** Two experiment installations in LMD-1.[Dave Plate, DUSEL]

## 3.7 Generic Physics Requirements

The requirements of specific proposed experiments, as outlined in Sections 3.3.3 through 3.3.8, form the basis for generic experiment requirements, in the cases where multiple proposals exist for a given scientific category. The three generic categories—for which requirements are outlined in Tables 3.7-1, 3.7-2, and 3.7-3—are dark matter at MLL, dark matter at DLL, and neutrinoless double-beta ($0\nu\beta\beta$) decay at DLL. A separate table for generic ($o\nu\beta\beta$) decay for MLL is not included; this would differ from that for DLL only in the constrained size of experiments. For cases of a single proposal within a category, the specific experiment requirements were used. Single-proposal physics categories include nuclear astrophysics and low-background counting; for these, refer to the appropriate section above. Requirements for the LBNE project are outlined in Section 3.3.5. Requirements for BGE experiments are summarized in Section 3.3.7.

A generic requirement, for example for the dark-matter category, is, in general, the most demanding requirement among the specific requirements of the group of proposed experiments in cases where that requirement is consistent with the Facility baseline design. In some cases, specific experiment requirements are not consistent with the baseline design; in these instances, the generic requirement is based on constraints imposed by the baseline Facility design. These inconsistencies are addressed as options within Trade Studies discussed in Section 3.8.6.

| Requirement | Value/Description | Comment/Justification |
|---|---|---|
| **Layout** | | |
| Depth | 4850L | |
| Footprint | 25 m L x 17 m W | Constrained value |
| Height [m] | 19 | Constrained value |
| Floor Load [kPa] | 200 | Corresponds to 20-m-high water tank |
| **Utilities** | | |
| Power [kW] | 300 | LZD + 33% |
| Standby Power [kW][1] | 70 | GEODM |
| Chilled Water [kW] | 160 | GEODM |
| Waste Heat to Air [kW] | 240 | LZD, 60 kW chilled water |
| Purified Water [$m^3$] | 4000 | 19 m h, 17 m dia. water shield |
| Potable Water [lpm] | Nominal use | |
| Compressed Air | Nominal use | |
| Network | 10 Gb/s | Nominal |
| **Environment** | | |
| Temp. Min [ºC] | 20 | Nominal, may be tighter in clean rooms |
| Temp. Max [ºC] | 25 | |
| Humidity Min [%] | 20 | Nominal, may be tighter in clean rooms |
| Humidity Max [%] | 50 | |
| Rn Background [$Bq/m^3$] | Meets OSHA and other applicable codes | |
| **Crane** | | |
| Max. Load [Short Tonne] | 20 | Nominal bridge crane |

| Requirement | Value/Description | Comment/Justification |
|---|---|---|
| **Occupancy** | | |
| Peak Installation Occupancy [count] | 24 | MAX single detector |
| Installation Duration [months] | 24 | GEODM |
| Peak Commissioning Occupancy [count] | 20 | LZD |
| Commissioning Duration [months] | 12 | |
| Peak Operation Occupancy [count] | 10 | |
| Operation Duration [months] | 60 | |
| **Cryogens** | | |
| LN Storage | 3,000 L | LZD |
| LN Consumption | 500 L/day | |
| LXe Storage | 6,700 L | 20 T LZD detector |
| LXe Consumption | NA | |
| LAr Storage | 21,000 L | 20 T Ar MAX + 10 T storage |
| LAr Consumption | NA | |
| **Major Hazards (Other Than Cryogens)** | | |
| Liquid Scintillator | 100 T | LZD, MAX single detector, type TBD |
| Water Flood Hazard | 4000 $m^3$ | |
| **Assay and Storage** | | |
| Assay Needs | Nominal | |
| Underground Storage | 100 $m^2$ | Depleted Ar for MAX |

**Table 3.7-1** Generic dark-matter requirements for MLL. Experiments: COUPP, GEODM, LZD, and MAX.

[1] Standby power is meant to provide systematic shutdown or continuous operation, as appropriate, for critical equipment that will suffer damage in a power outage. According to NFPA 520, Sections 6.6, 6.7, 6.8, the transition from the instant of failure of the normal power source to an alternative power source shall not exceed 60 seconds.

| Requirement | Value/Description | Comment/Justification |
|---|---|---|
| **Layout** | | |
| Depth | 7400L | |
| Footprint | 25 m L x 12 m W | Constrained value |
| Height [m] | 11 | Constrained value |
| Floor Load [kPa] | 100 | Corresponds to 10-m-high water tank |
| Clean Room | Class 1000<br>12 m x 25 m x 10 m | GEODM |
| Clean Room | Class 100<br>10 m x 15 m x 10 m | GEODM |
| **Utilities** | | |
| Power [kW] | 240 | GEODM + 33% |
| Standby Power [kW] | 70 | GEODM |
| Chilled Water [kW] | 160 | GEODM |

| Requirement | Value/Description | Comment/Justification |
|---|---|---|
| Waste Heat to Air [kW] | 80 | GEODM |
| Purified Water [$m^3$] | 1700 | COUPP |
| Potable Water [lpm] | Nominal use | |
| Compressed Air | Nominal use | |
| Network | 10 Gb/s | Nominal |
| **Environment** | | |
| Temp. Min [ºC] | 20 | Nominal, may be tighter in clean rooms |
| Temp. Max [ºC] | 25 | |
| Humidity Min [%] | 20 | Nominal, may be tighter in clean rooms |
| Humidity Max [%] | 50 | |
| Rn Background [$Bq/m^3$] | Meets OSHA and other applicable codes | |
| **Crane** | | |
| Max. Load [Short Tonne] | 20 | Nominal bridge crane |
| **Occupancy** | | |
| Peak Installation Occupancy [count] | 24 | GEODM |
| Installation Duration [months] | 24 | GEODM |
| Peak Commissioning Occupancy [count] | 5 | GEODM |
| Commissioning Duration [months] | 9 | |
| Peak Operation Occupancy [count] | 10 | COUPP |
| Operation Duration [months] | 120 | GEODM |
| **Cryogens** | | |
| LN Storage | 200 L | GEODM |
| LN Consumption | 100 L/day | COUPP cover gas option |
| LHe Storage | 200 L | GEODM |
| LHe Consumption | 2 L/day | |
| **Major Hazards (Other Than Cryogens)** | | |
| CF3I | 1600 kg | 500 kg/module, nontoxic, ODH |
| Water Flood Hazard | 1700 $m^3$ | |
| **Assay and Storage** | | |
| Assay Needs | Nominal | |
| Underground Storage | NA | |

**Table 3.7-2** Generic dark-matter requirements for DLL. Experiments: COUPP, and GEODM.

| Requirement | Value/Description | Comment/Justification |
|---|---|---|
| **Layout** | | |
| Depth | 7400L | |
| Footprint | 25m L x 12m W | Constrained value; current layouts exceed this |
| Height [m] | 11 | Constrained value; current layouts exceed this |
| Floor Load [kPa] | 1670 | EXO lead shield option |
| **Utilities** | | |
| Power [kW] | 400 | EXO + 33% |
| Standby Power [kW] | 60 | EXO |
| Chilled Water [kW] | 250 | EXO |
| Waste Heat to Air [kW] | 150 | EXO |
| Purified Water [$m^3$] | 4000 | 19 m h, 17 m dia. water shield |
| Potable Water [lpm] | Nominal use | |
| Compressed Air | Nominal use | 1TGe will use air pads to move Pb shielding |
| Network | 10 Gb/s | Nominal |
| **Environment** | | |
| Temp. Min [ºC] | 20 | Nominal, may be tighter in clean rooms |
| Temp. Max [ºC] | 25 | |
| Humidity Min [%] | 20 | Nominal, may be tighter in clean rooms |
| Humidity Max [%] | 50 | |
| Rn Background [$Bq/m^3$] | Meets OSHA and other applicable codes | |
| **Crane** | | |
| Max. Load [Short Tonne] | 40 | Nominal monorail crane |
| **Occupancy** | | |
| Peak Installation Occupancy [count] | 20 | Generic experiment estimate |
| Installation Duration [months] | 24 | |
| Peak Commissioning Occupancy [count] | 10 | |
| Commissioning Duration [months] | 12 | |
| Peak Operation Occupancy [count] | 10 | |
| Operation Duration [months] | 60 | |
| **Cryogens** | | |
| LXe Storage | 3,400 L | 10T EXO |
| LXe Consumption | NA | |
| LAr Storage | 21,000 L | 1TGe GERDA style. 30 T used in GERDA |
| LAr Consumption | NA | |
| LN Storage | 10,000 L | 1TGe. Assume 5 days storage—tied to consumption rate |

| Requirement | Value/Description | Comment/Justification |
|---|---|---|
| LN Consumption | 2,000 L/day | Based on 2 L/day per 1000 detectors |
| **Major Hazards (Other Than Cryogens)** | | |
| HFE Refrigerant | | EXO—nontoxic |
| Water Flood Hazard | 4000 $m^3$ | |
| **Assay and Storage** | | |
| Assay Needs | Nominal | |
| Underground Storage | 5 m x 6 m x 3 m | 1TGe |

**Table 3.7-3** Generic neutrinoless double-beta decay requirements for DLL. Experiments: 1TGe and EXO.

## 3.8 Science-Driven Facility Infrastructure Requirements

To translate experiment requirements into Facility requirements, the various experiment categories must be associated with particular locations. This section describes requirements for the LMs, other 4850L support, other levels and ramps (OLR), and surface requirements. Trade studies applicable to science requirements are discussed in the final subsection. The process through which these values were determined is described in Chapter 3.6.

### 3.8.1 Laboratory Module Science Requirements

Figure 3.8.1-1 shows a cross section for LM-1 and LM-2, with the permissible experiment envelope indicated by the crosshatched area. The 19 m maximum height is limited by bridge crane clearance. The maximum 17 m width allows for personnel egress and a utility corridor. The allowed length for a single experiment is 25 m to accommodate lay-down and assembly space. Figure 3.8.1-2 shows the LMD-1, with the same constraints.

Tables 3.8.1-1, 3.8.1-2, and 3.8.1-3 list science requirements for LM-1, LM-2, and LMD-1, respectively. Utility requirements are for direct experiment support only; requirements for lighting, ventilation, cranes, and other infrastructure are not included.

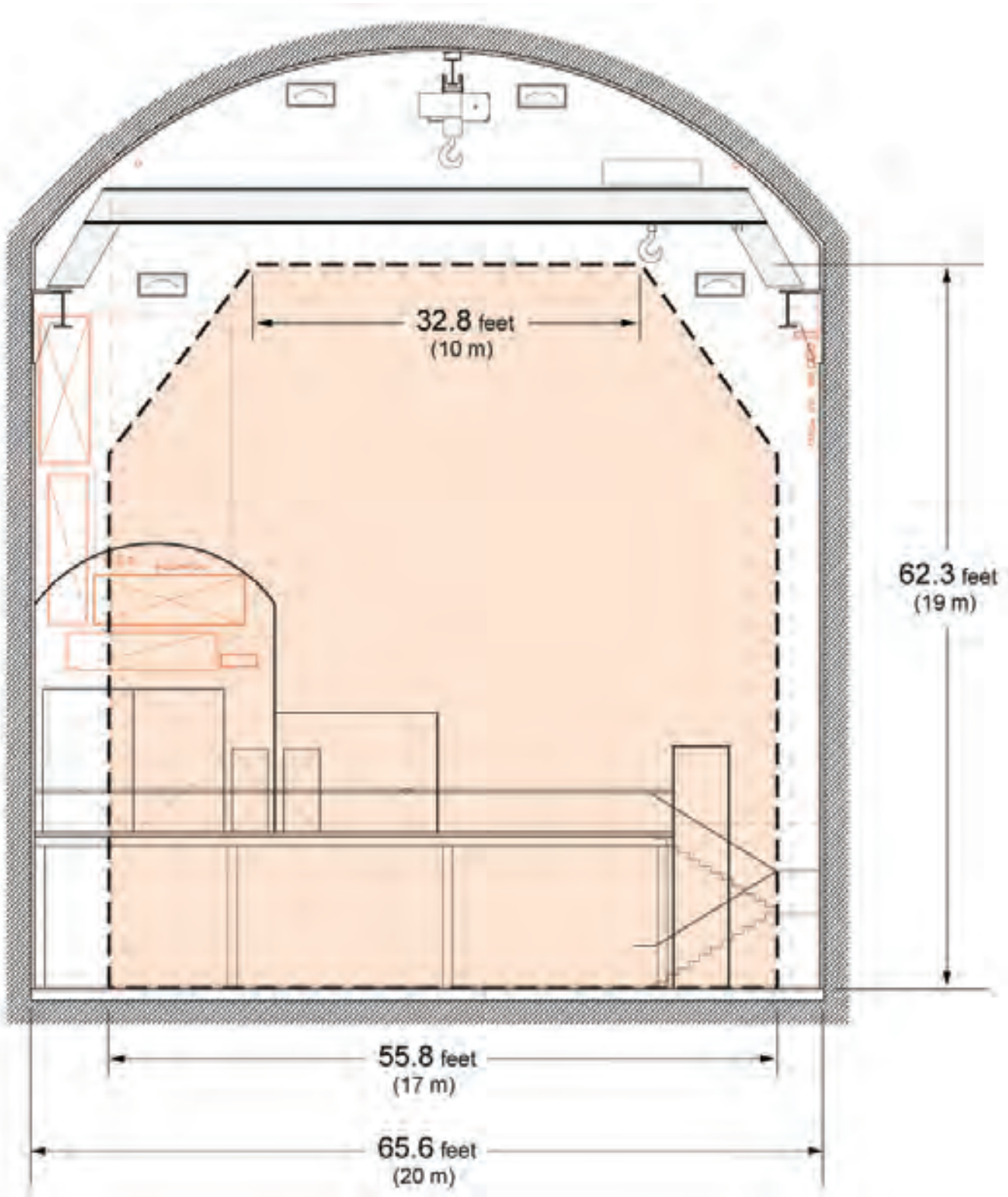

**Figure 3.8.1-1** MLL LM cross section and experiment envelope. [DKA]

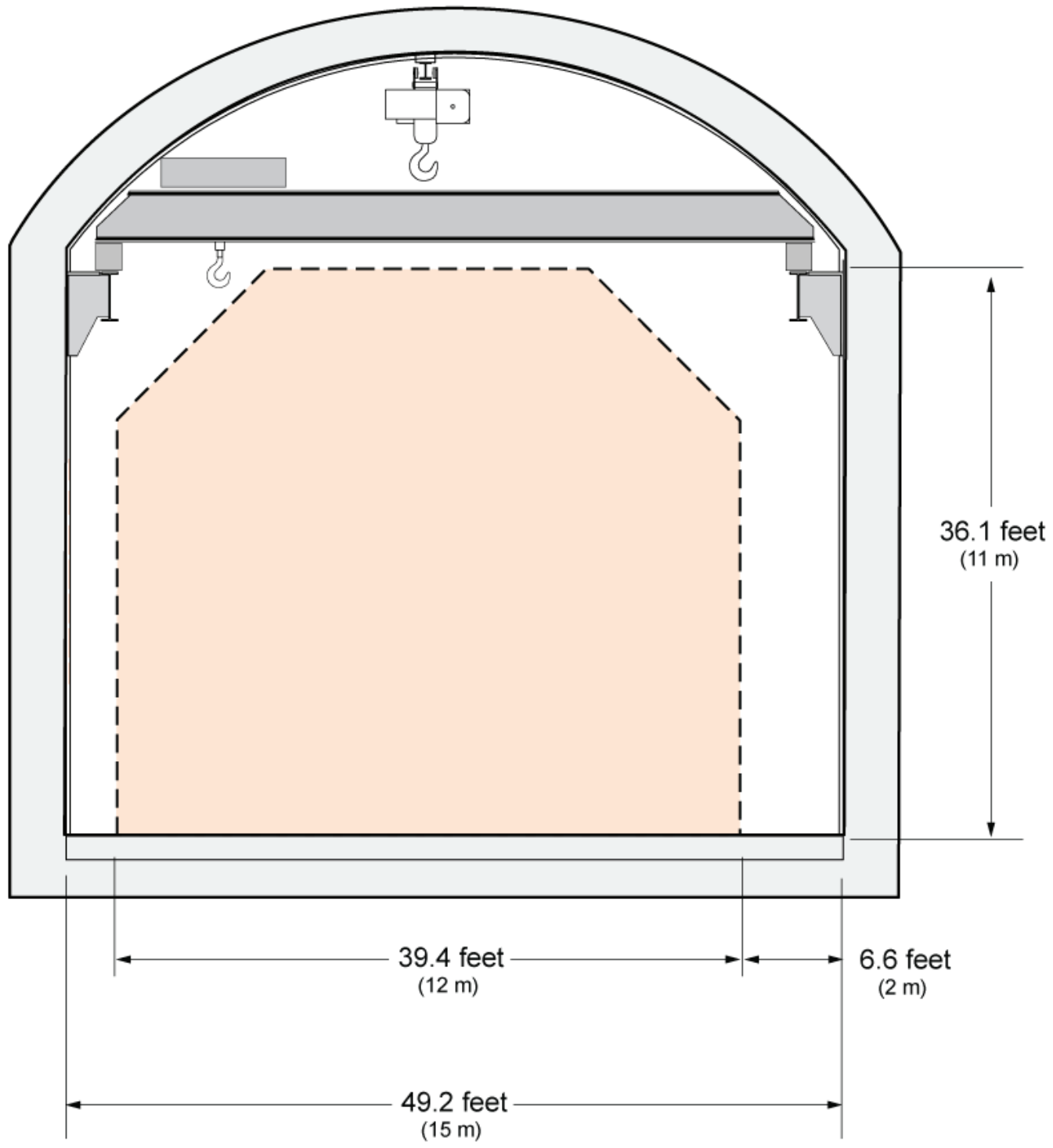

**Figure 3.8.1-2** DLL LM cross section and experiment envelope. [DKA]

| Requirement | Value/Description | Comment/Justification |
|---|---|---|
| **Layout** | | |
| Footprint | 50 m L x 20 m W | Nominal size |
| Height [m] | 24 | Nominal size, crowned roof |
| Floor Load [kPa] | 1670 | EXO lead shield case |
| **Utilities** | | |
| Power [kW] | 2000 | DIANA 1.5 MW load + 33% margin |
| Standby Power [kW] | 100 | DIANA detectors and cryo pumps |
| Chilled Water [kW] | 1800 | |
| Waste Heat to Air [kW] | 1400 | |
| Purified Water [$m^3$] | 4100 | Assume water tank 19 m high, 17 m dia.; inner volume 9 m high, 5 m dia. |
| Potable Water [lpm] | Nominal use | |
| Compressed Air | Nominal use | |
| Network | 10 Gb/s | |
| **Environment** | | |
| Temp. Min [ºC] | 20 | Generic lab environment temp. values |
| Temp. Max [ºC] | 25 | |
| Humidity Min [%] | 20 | DIANA |
| Humidity Max [%] | 45 | Nominal expected value; DIANA max. 30% |
| Rn Background [$Bq/m^3$] | Meets OSHA and other applicable codes | |
| **Crane** | | |
| Max. Load [Short Tonne] | 40 | EXO lead shield case |
| **Occupancy** | | |
| Peak Installation Occupancy [count] | 25 | |
| Installation Duration [months] | 24 | |
| Peak Commissioning Occupancy [count] | 20 | |
| Commissioning Duration [months] | 6 | |
| Peak Operation Occupancy [count] | 9 | |
| Operation Duration [months] | 60 | Assume typical 5-year experiment. Laboratory Module lifetime is 30 years. |
| **Cryogens—Target will be either LXe or LAr** | | |
| LXe Storage | 20 T | LZD |
| LXe Consumption | NA | |
| LAr Storage | 30 T | MAX |
| LAr Consumption | NA | |
| LN Storage | 500 L | |
| LN Consumption | 50 L/day | |

| Requirement | Value/Description | Comment/Justification |
|---|---|---|
| **Major Hazards (Other Than Cryogens)** | | |
| Liquid Scintillator | 100 T | LZD, MAX |
| High Voltage | 400 kV, 100 mA | DIANA |

**Table 3.8.1-1** Science requirements for LM-1.

| Requirement | Value/Description | Comment/Justification |
|---|---|---|
| **Layout** | | |
| Footprint | 100 m L x 20 m W | Nominal size |
| Height [m] | 24 | Nominal size, crowned roof |
| Floor Load [kPa] | 1670 | EXO lead shield case |
| **Utilities** | | |
| Power [kW] | 1100 | EXO + GEODM + FAARM + 33% |
| Standby Power [kW] | 160 | MAX (2 detectors) + EXO |
| Chilled Water [kW] | 840 | EXO + GEODM + FAARM + 50% |
| Waste Heat to Air [kW] | 820 | |
| Purified Water [$m^3$] | 10000 | Assume 2 water tanks 19 m high, 17 m dia. with inner volume 9 m high, 5 m dia. + FAARM |
| Potable Water [lpm] | Nominal use | |
| Compressed Air | Nominal use | |
| Network | 10 Gb/s | |
| **Environment** | | |
| Temp. Min [ºC] | 20 | Generic lab environment—maybe tighter control within clean rooms |
| Temp. Max [ºC] | 25 | |
| Humidity Min [%] | 20 | Generic lab environment—maybe tighter control within clean rooms |
| Humidity Max [%] | 50 | |
| Rn Background [Bq/$m^3$] | Meets OSHA and other applicable codes | |
| **Crane** | | |
| Max. Load [Short Tonne] | 40 | EXO lead shield - nominal monorail capacity |
| **Occupancy** | | |
| Peak Installation Occupancy [count] | 40 | Assume installations staggered by 1 year, 20 per experiment—two overlap |
| Installation Duration [months] | 48 | |
| Peak Commissioning Occupancy [count] | 20 | Assume commissioning does not overlap |
| Commissioning Duration [months] | 6 | |
| Peak Operation Occupancy [count] | 18 | Assume 6 per experiment |

| Requirement | Value/Description | Comment/Justification |
|---|---|---|
| Operation Duration [months] | 60 | Assume typical 5-year experiment. Laboratory Module lifetime is 30 years. |
| **Cryogens** | | |
| LXe Storage | 20 T | LZD |
| LXe Consumption | NA | |
| LAr Storage | 30 T | MAX |
| LAr Consumption | NA | |
| LN Storage | 1200 L | Assume 2 Cryogen detectors (500L each) + FAARM (200L) |
| LN Consumption | 150L/day | Assume 50L/day per experiment |
| **Major Hazards (Other Than Cryogens)** | | |
| Liquid Scintillator | 200T | Two DM cryogen detectors (100T each), type TBD |

**Table 3.8.1-2** Science requirements for LM-2.

| Requirement | Value/Description | Comment/Justification |
|---|---|---|
| **Layout** | | |
| Footprint | 75m L x 15m W | Nominal size |
| Height [m] | 15 | Nominal size, crowned roof |
| Floor Load [kPa] | 1670 | EXO lead shield case |
| **Utilities** | | |
| Power [kW] | 650 | EXO + GEODM + 33% |
| Standby Power [kW] | 100 | EXO + 1TGe |
| Chilled Water [kW] | 650 | EXO + GEODM + 50% |
| Waste Heat to Air [kW] | 420 | |
| Purified Water [$m^3$] | 0 | Purified water must be generated by the experiment. No purified water supply is provided to LMD-1. |
| Potable Water [lpm] | Nominal use | |
| Compressed Air | Nominal use | |
| Network | 10 Gb/s | |
| **Environment** | | |
| Temp. Min [ºC] | 20 | Generic lab environment—maybe tighter control within clean rooms |
| Temp. Max [ºC] | 25 | |
| Humidity Min [%] | 20 | Generic lab environment—maybe tighter control within clean rooms |
| Humidity Max [%] | 50 | |
| Rn Background [Bq/$m^3$] | Meets OSHA and other applicable codes | |
| **Crane** | | |

| Requirement | Value/Description | Comment/Justification |
|---|---|---|
| Max. Load [Short Ton] | 40 | EXO lead shield—nominal monorail capacity |
| **Occupancy** | | |
| Peak Installation Occupancy [count] | 40 | Assume installations staggered by 1 year, 20 per experiment—two overlap |
| Installation Duration [months] | 36 | |
| Peak Commissioning Occupancy [count] | 20 | Assume commissioning does not overlap |
| Commissioning Duration [months] | 6 | |
| Peak Operation Occupancy [count] | 12 | Assume 6 per experiment |
| Operation Duration [months] | 60 | Assume typical 5-year experiment. Laboratory module lifetime is 30 years. |
| **Cryogens** | | |
| LXe Storage | 10T | EXO |
| LXe Consumption | NA | |
| LAr Storage | 30T | 1T Ge, Ar option |
| LAr Consumption | NA | |
| LN Storage | 10000L | 1TGe. Assume 5-day storage—tied to consumption rate |
| LN Consumption | 2000L/day | Based on 2L/day per 1000 detectors—probably too high |
| LHe Storage | ~200L | GEODM |
| LHe Consumption | ~10L/day | |

**Table 3.8.1-3** Science requirements for LMD-1.

### 3.8.2 Other MLL Campus Support Requirements

#### Davis Campus

The Davis Campus has been developed for the LUX dark-matter experiment and the MAJORANA DEMONSTRATOR neutrinoless double-beta decay experiment as part of the Initial Science Program as described in Chapter 3.4. The LUX detector is a 350-kg LXe detector with a water shield; the DLM and water shield have been designed to house a 3-T LXe detector. After the LUX experiment is complete, this space could be used for moderate-size cryogenic detectors (e.g., Generation-2 dark matter) or R&D space.

The Davis Transition Area houses the MAJORANA DEMONSTRATOR as well as the Cu electroforming and clean machine shop space that support this experiment. The support room is sized to house an expanded electroforming facility that will be capable of supporting the 1TGe experiment as well as other experiments that need low-background Cu. This expansion will require relocation of the clean machine shop space elsewhere.

#### Shops

The MLL Campus will have a clean machine shop as well as a general machine shop. The area for each will be approximately 60 $m^2$.

### 3.8.3 Other Levels and Ramps (OLR) to Support Biology, Geology, and Engineering (BGE) Experiments

Other Levels and Ramps (OLR) is a general term intended to represent the infrastructure at locations other than those directly related to the physics experiments at the 4850 and 7400 levels. exist: 1) those associated solely with the science requirements and 2) those that are necessary for the operation and maintenance of the Facility. OLR that are necessary for operations will also be available for scientific use as needed and as accessible. Therefore, a substantial number of drifts and ramps will be available to contain DUSEL infrastructure and a variety of BGE experiments. Safe access will be provided, but these areas will not be maintained at the same level as those more frequently occupied. The discussion below includes figures showing locations of proposed experiments and the associated requirements. In OLR spaces, unless noted otherwise, DUSEL will be responsible for maintaining ground support and providing power, data network, and ventilation. In most areas, ventilation will be continuous flow-through. In some cases, ventilation will be provided on an as-needed basis using localized fans.

Table 3.8.3-1 summarizes the proposed BGE locations by level. Table 3.8.3-2 summarizes the total accessible area, in linear feet, and the total estimated power required to power the full set of installations summarized in Table 3.8.3-1, with the exception of $CO_2$ Sequestration, which will have power fed from its surface facility. The total power for each level includes only the power to support the proposed BGE experiments on that level; it does not include power for general infrastructure support, or the power for physics experiments within the developed MLL and DLL Campuses. The funding allocation associated with providing power to the OLR has been capped, based upon a cost estimate for providing the power listed in Table 3.8.3-2.

Figures 3.8.3-1 to 3.8.3-8 show level diagrams for the 300L, 800L, 2000L, 4100L, 4550L, 4850L, 6800L, and 7400L. The areas open to experiment are indicated. The site number indicators, as designated in Table 3.8.3-1, will be used throughout the level diagrams to indicate experiments.

| Level | Site | Experiment |
|---|---|---|
| 300 | 1 | $CO_2$ Sequestration |
| | 2 | Ecohydrology |
| 800 | 3 | $CO_2$ Sequestration |
| | 1, 4 | Ecohydrology |
| 2000 | 1, 3, 5 | Ecohydrology |
| | 3 | Fractured Processes |
| | 3 | GEOX™ (Distributed around 3) |
| | 1, 2, 5 | Transparent Earth (Broadband Seismic Array) |
| | 1, 2, 5 | Transparent Earth (Earth Passive EEM1) |
| | 1d, 4d | Transparent Earth (Earth Electrical Array, Distributed around 1d and 4d) |
| 4100 | 2, 4, 6 | EcoHydrology |
| | 2, 5 | GEOX™ (Distributed around 2 and 5) |
| | 1, 3, 4, 6 | Transparent Earth(Broadband Seismic Array) |
| | 1, 3, 4, 6 | Transparent Earth (Earth Passive EEM1) |
| | 3 | Transparent Earth (Earth Passive EEM2) |
| | 1-3, 4d | Transparent Earth (Earth Electrical Array, Between 1 and 3, around 4d) |
| | 1, 3, 6 | Transparent Earth (Active Seismic Monitoring |
| 4550 | 1 | Transparent Earth (Broadband Seismic Array) |
| | 1 | Transparent Earth (Earth Passive EEM1) |
| | 1 | Transparent Earth (Active Seismic Monitoring) |
| | 1 | Transparent Earth (USGS Calibration Site) |
| 4850 | 1 | Cavity Design |
| | 1 | Cavity Monitoring |
| | 3 | Coupled Processes |
| | 2, 6 | Ecohydrology |
| | 3 | Fractured Processes |
| | Distributed | GEOX™ (Distributed around main triangle) |
| | 5, 6 | Transparent Earth (Broadband Seismic Array) |
| | 5, 6 | Transparent Earth (Earth Passive EEM1) |
| | 5, 6 | Transparent Earth (Active Seismic Monitoring) |
| | 4 | Transparent Earth (Earth Passive EEM2) |
| | 4 | Transparent Earth (Active Seismic Stress) |
| | 6d | Transparent Earth (Earth Electrical Array, Distributed around 6d) |
| 6800 | 1 | GEOX™ (Distributed around 1) |
| 7400 | 2 | Ecohydrology |
| | 1 | Transparent Earth (Broadband Seismic Array) |
| | 1 | Transparent Earth (Earth Passive EEM1) |
| | 1 | Transparent Earth (HPPP, MicroGravity, SQUID) |

**Table 3.8.3-1** Proposed BGE locations by level.

| Level | Total Linear Feet | Total Power [kW] |
|---|---:|---:|
| 300 | 1,740 | 23 |
| 800 | 8,010 | 23 |
| 2000 | 15,810 | 481 |
| 4100 | 16,940 | 742 |
| 4550 | 3,770 | 2 |
| 4850 | 11,390 | 978 |
| 6800 | 1,800 | 425 |
| 7400 | 880 | 628 |

**Table 3.8.3-2** Summary of access and power by level.

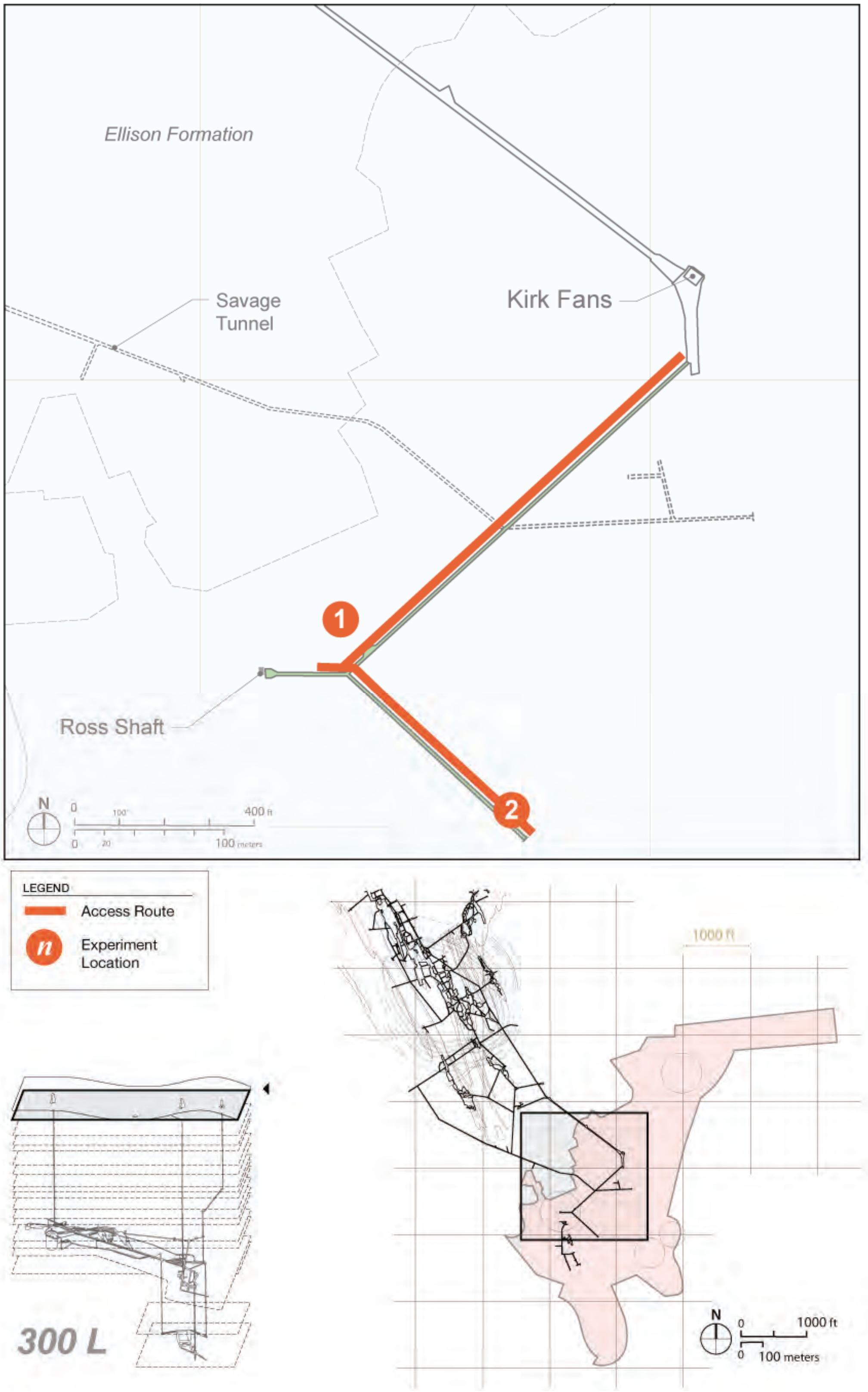

**Figure 3.8.3-1** Level map for the 300L. A number of initial science groups have deployed instruments near site #1, including the DUGL group, a physics and geoscience group interested in identifying free surface vibrations as part of an evaluation for an advanced LIGO installation. [DKA]

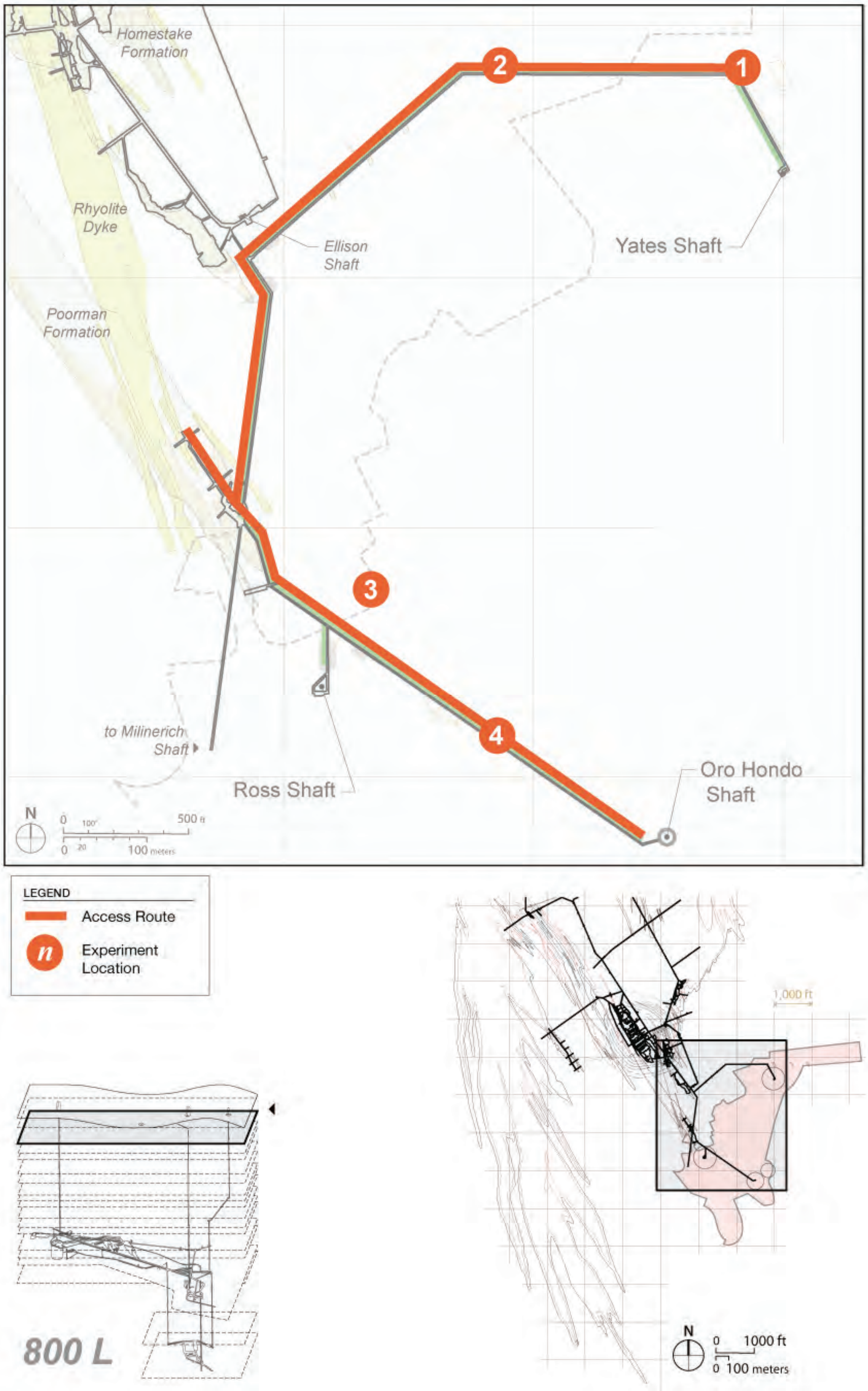

**Figure 3.8.3-2** Level map for the 800L. A number of initial science groups have deployed instruments near sites #3 and #4, including the physics background characterization group and the DUGL group. Shielding materials for the MAJORANA DEMONSTRATOR are currently being stored ~100 m to the west of site #3. [DKA]

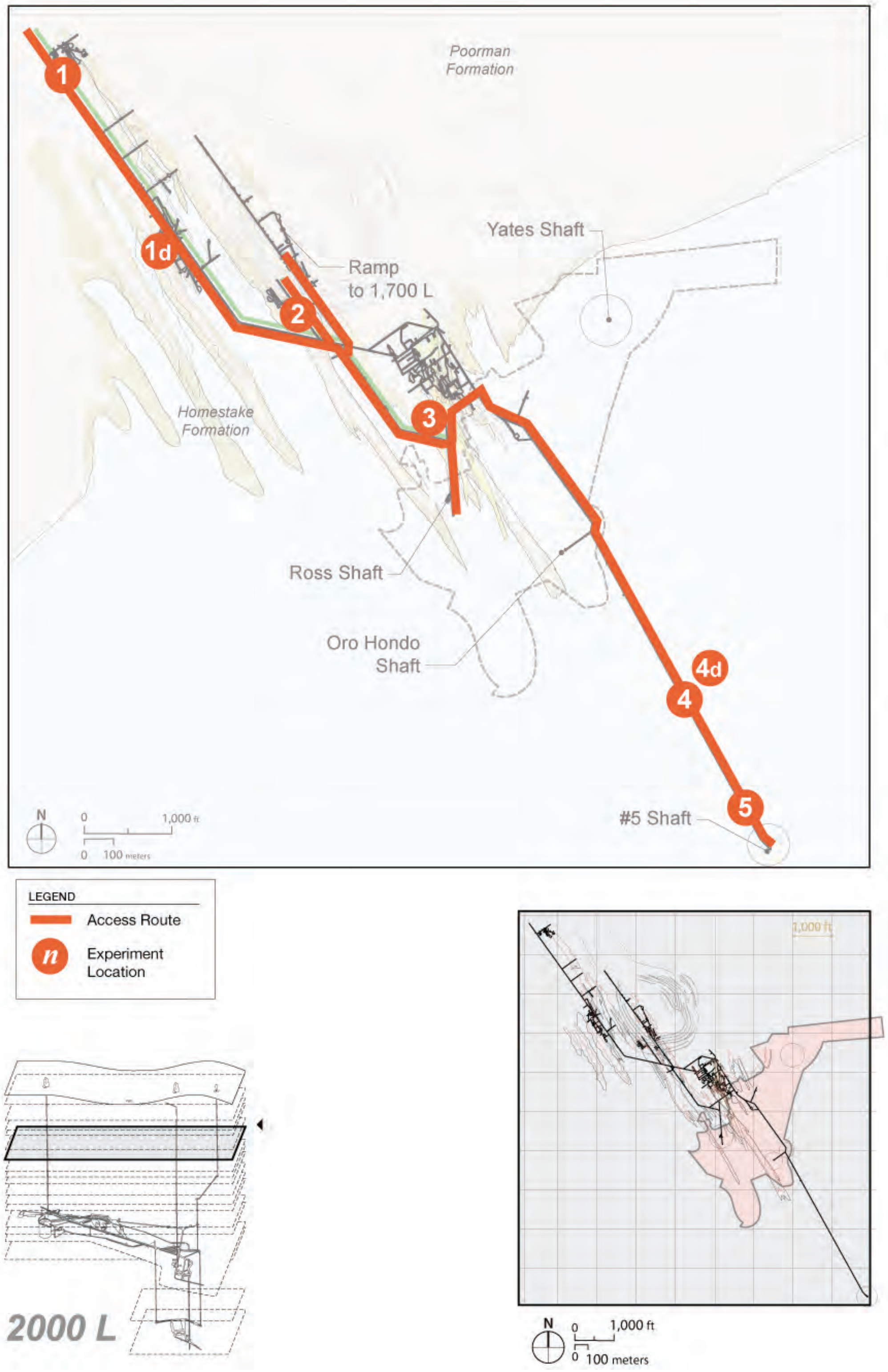

**Figure 3.8.3-3** Level map for the 2000L. This provides access along the NW-SE long axis of the Homestake infrastructure. The numbers refer to potential sites for BGE instrumentation. [DKA]

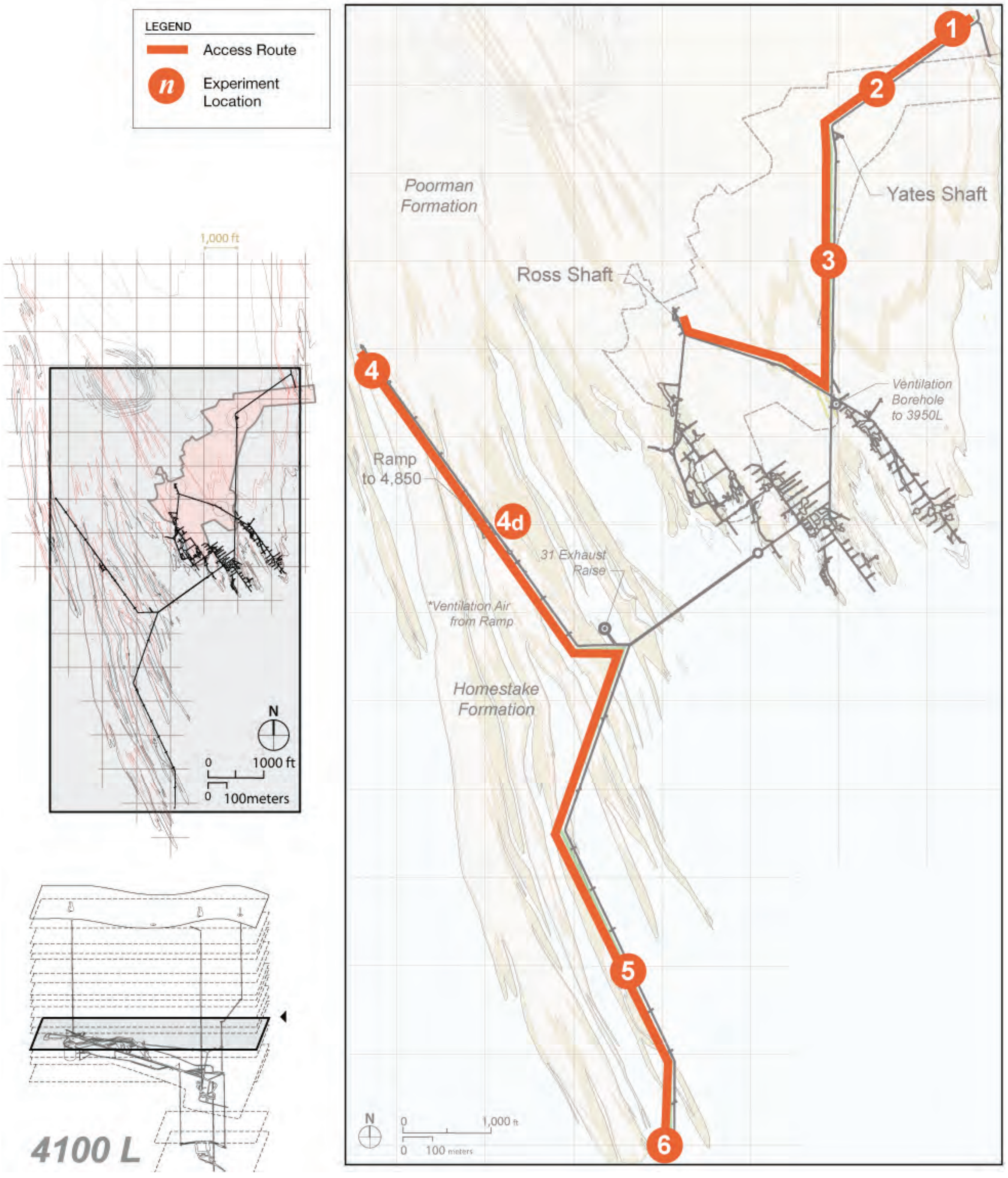

**Figure 3.8.3-4** Level map for the 4100L. This level provides access in NE-SW orientation across the Homestake infrastructure. Collaborations most interested in this level include Ecohydrology, Fracture Processes, Transparent Earth. [DKA]

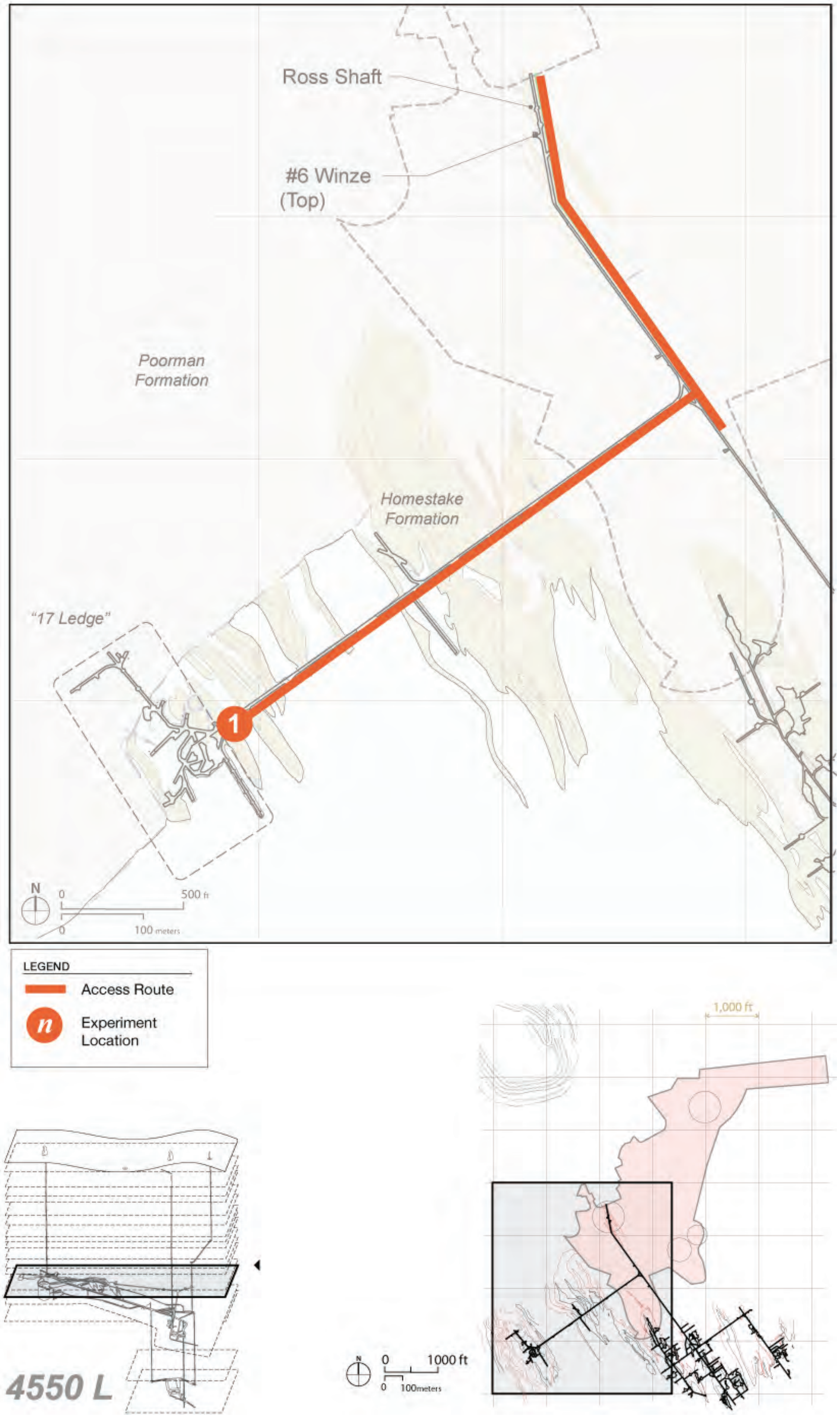

**Figure 3.8.3-5** Level map for the 4550L. Much of this level will be necessary for operations maintenance because the hoistroom for the #6 Winze is located at this level as well as a ramp system. [DKA]

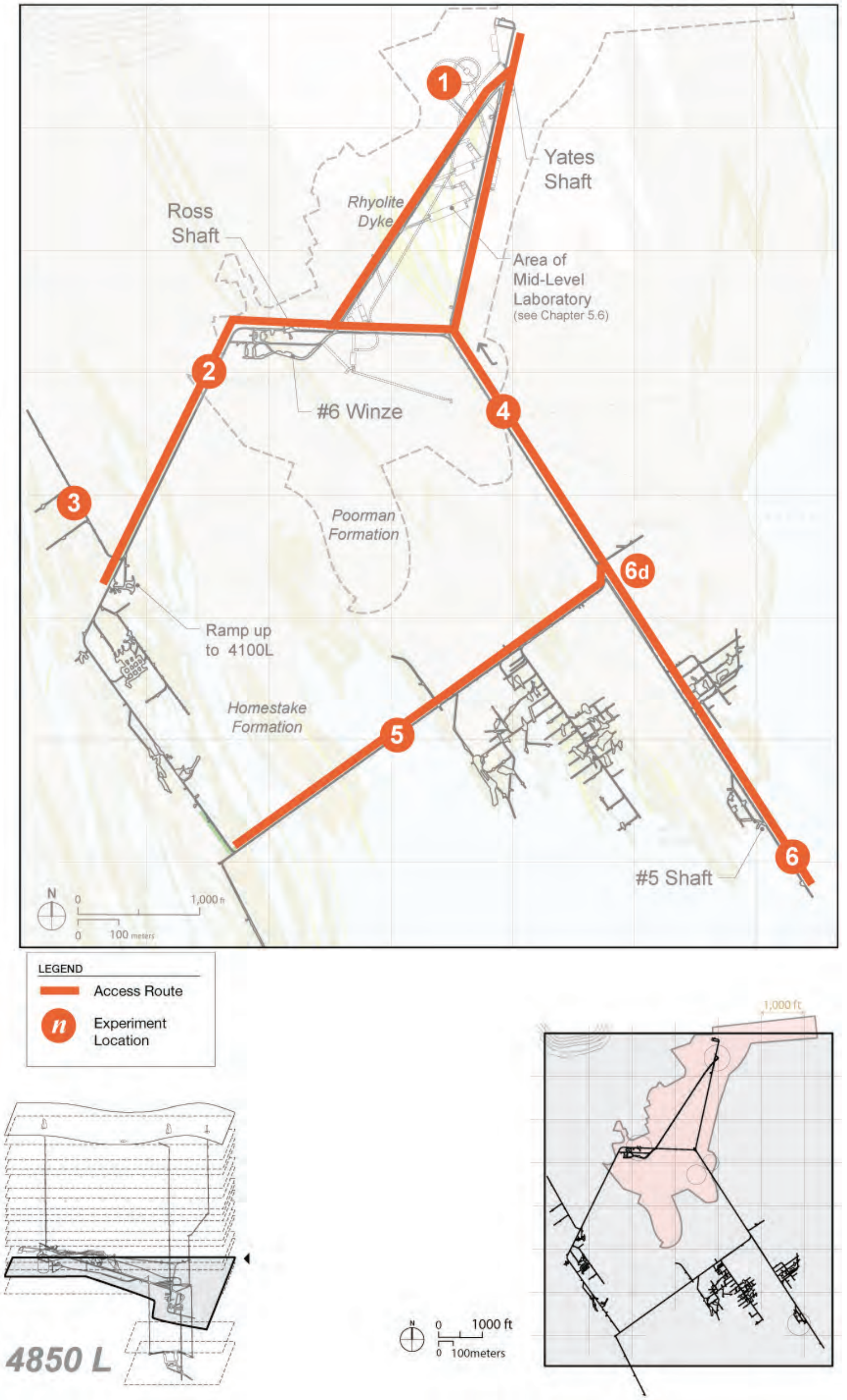

**Figure 3.8.3-6** Level map for the 4850L This is one of the main campus levels. In addition to the physics area near the Yates Shaft, the Fracture Processes and Coupled Processes collaborations have plans for a laboratory located at #3 in the diagram. Transparent Earth and the GEOX$^{TM}$ collaborations would also have interests at other numbered locations on this level. [DKA]

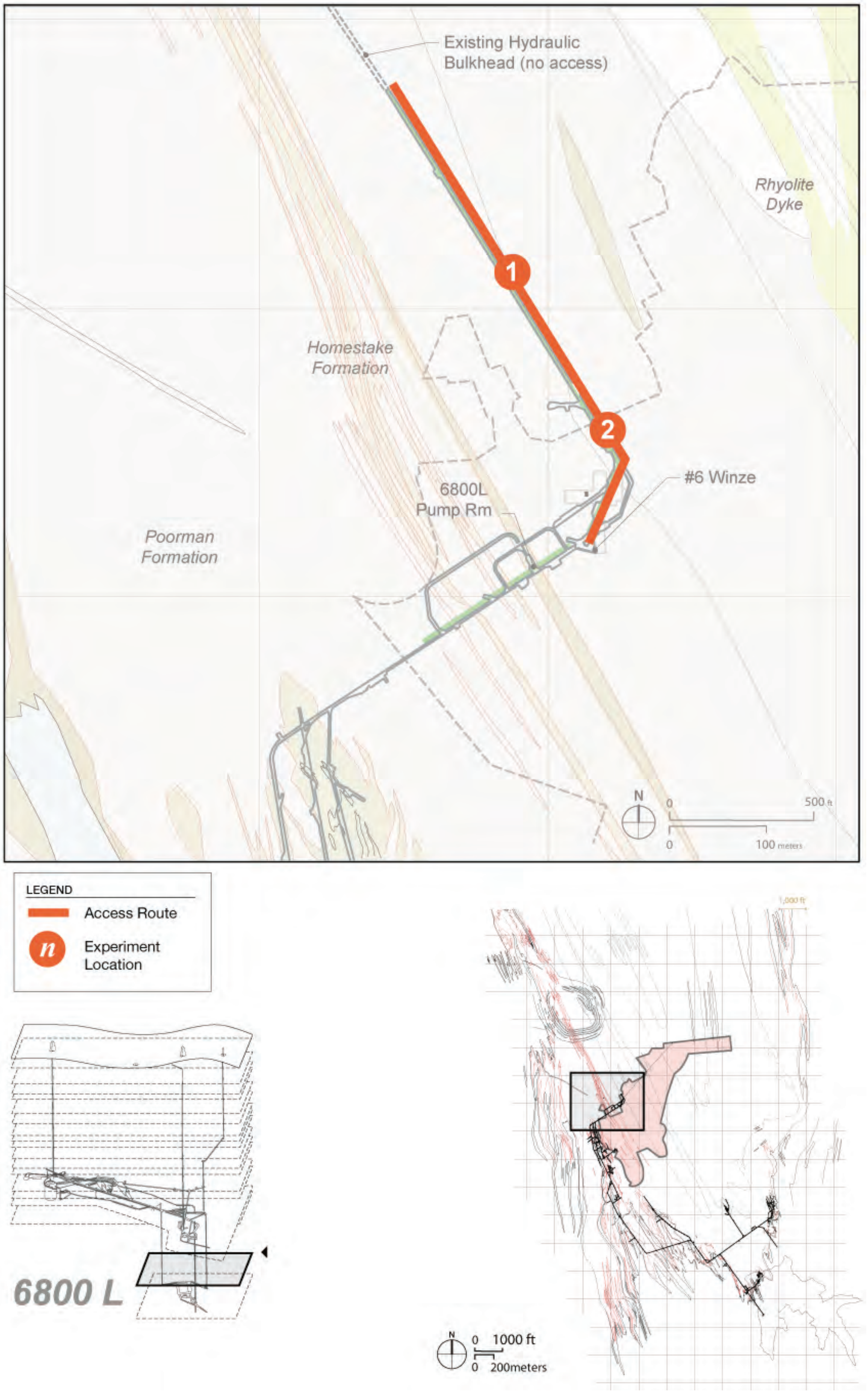

**Figure 3.8.3-7** Level map for the 6800L.This is primarily an operations level that requires access to the North Drift Plug for monitoring purposes. The GEOX$^{TM}$ collaboration as well as Transparent Earth would have an interest in using this level as well. [DKA]

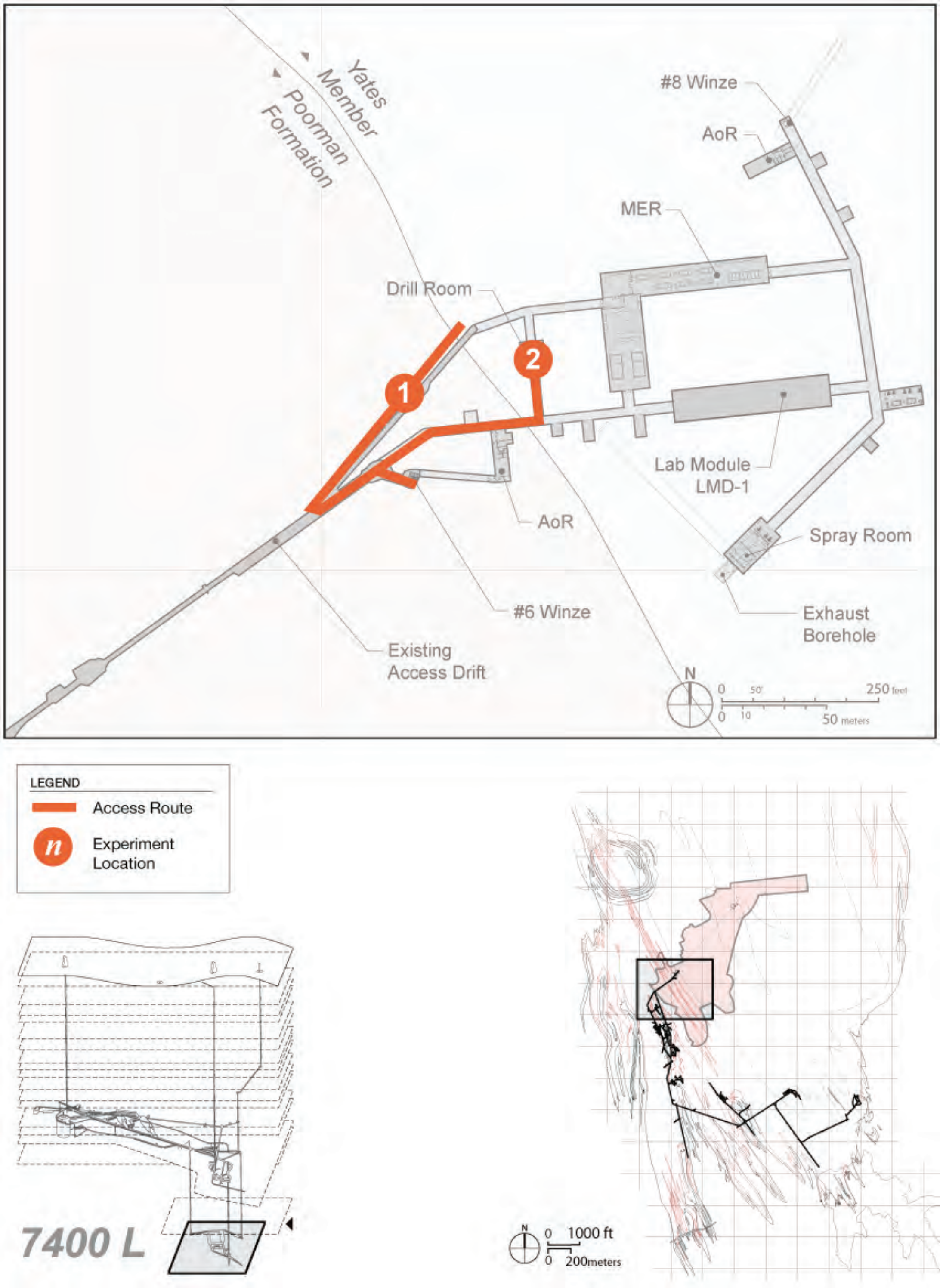

**Figure 3.8.3-8** Level map for the 7400L. Although some of this level is shown as part of Other Levels and Ramps, the main activity is the 7400L campus. Fracture Processes and Couple Processes also have an interest in using this level as part of their deep studies of faulting and related processes. [DKA]

### 3.8.4 Surface Requirements

Surface facilities to support science will include experiment-specific as well as shared spaces. Experiment-specific spaces will generally include a control room and a limited number of dedicated offices. Shared spaces will include additional offices and meeting rooms, electronic and mechanical shops, shipping and receiving, storage, laboratories, and assembly space.

Two high-bay surface assembly areas will be provided for assembly of critical elements before transporting underground. The assembly spaces have approximately the same floor space as will be allocated in an LM. The high bay will be 10 m high. Refer to Volume 5, Chapter 2 of this document for more details on the Surface Facility design.

### 3.8.5 Trade Studies Related to Science and Facility Requirements

A number of Trade Studies have been initiated to consider optional Laboratory configurations relative to the baseline design. The objective is to assess the additional cost or savings of a particular option and to assess its impact on the potential for fulfilling the Laboratory's science mission. The Trade Studies are described below. Many of the studies will continue beyond the completion of the Preliminary Design during the period leading up to the start of the Final Design. Results will be presented here for those that are complete; status and the plan for resolution will be described for the others. The cost differences presented in this section include direct and indirect costs, and Management Reserve (in FY 2010 dollars). Cost differences compared to the baseline are given to the nearest $0.1 million.

For Sections 3.8.5.1 and 3.8.5.2, a simple parametric scaling of the excavation-related costs was used to estimate the cost difference compared with the baseline. Excavation-related costs include excavation of LMs and related access drifts, ground support, and shotcrete and concrete floors. Calculated excavation cost variations in length and height assume linear scaling with volume; this reasonably represents the cost of rock removal. An increase of width or span will result in increased rock stress that is nonlinear with width. This will require additional ground control, and therefore we expect that cost will scale with width at an order higher than linear. The Project has assumed a quadratic scaling for these parametric studies; this is assumed to be conservative. Cost differences related to ventilation and other utilities have been included.

Direct costs include estimates for labor and materials. A 60% factor is assumed for indirect costs; 40% Management Reserve is assumed for LM-1 and LM-2 at the Mid-Level Campus; 50% Management Reserve is assumed for LMD-1 at the Deep-Level Campus. The Project recognizes that these estimates are highly preliminary. More detailed estimates of costs will be needed if the baseline LM dimensions are changed.

#### 3.8.5.1 LM-1 Size and Configuration for DIANA

If the nuclear astrophysics accelerator proposal, DIANA, is part of the ISE, it will be installed as the sole experiment in LM-1. Neutron backgrounds are one of the noise sources for experiments proposed for DUSEL; a reduction in neutron backgrounds provides the primary motivation for deploying deep underground. Since DIANA is a potential neutron source, it is important that it is sufficiently shielded so that neutron exposure for other experiments is not affected by the presence of DIANA. This will require some specialization in the form of an egress maze at both entries. The required height will be reduced to 20 m rather than 24 m, and will not require the floor to be below the MLL grade. As this experiment, if approved, will likely operate through the duration of the Laboratory life, another experiment likely will not replace it in the future. It therefore makes sense to consider the associated increased costs or savings associated with a DIANA-specific cavity design.

A conceptual plan view of DIANA with the desired modifications is presented in Section 3.3.6. How these modifications might look in the context of the overall 4850L layout is shown in Figure 3.3.6.2.2-1. The overall cost of an LM designed specifically for DIANA is expected to be less than the cost of the baseline LM-1 configuration by about $1.7 million, as summarized in Table 3.8.5.1.

| Change | Cost Difference($M) | Comments |
|---|---:|---|
| Decrease height 24 m → 20 m | -0.55 | Linear scaling of volume @ $138/m$^3$ |
| Remove slope in entry drift | -0.26 | Linear scaling of volume @ $400/m$^3$ |
| Add East entrance maze | +0.15 | Linear scaling of volume @ $400/m$^3$ |
| Add West entrance maze | +0.16 | Linear scaling of volume @ $400/m$^3$ |
| Ventilation change | -0.27 | Linear scaling of volume @ 45/m$^3$ |
| Subtotal | -0.77 | Direct costs |
| Total | -1.7 | Direct + indirect costs + contingency |

**Table 3.8.5.1** Trade Studies as described in the text for DIANA in LM-1.

### 3.8.5.2 LM-2 Size and Configuration

The baseline cross sections for LM-2 are shown in Figure 3.8.1-1. This Trade Study looked at the cost difference for 20% variations in the height, width, or length of LM-2. The results are summarized in Tables 3.8.5.2-1 to 3.8.5.2-4. The cost of large excavations is believed to be particularly sensitive to the maximum span of the excavation.

| Change | Cost Difference ($M) | Comments |
|---|---:|---|
| Increase excavation volume | 1.10 | Linear scaling of volume @ $138/m$^3$ |
| Addition to entry drift | 0.43 | Linear scaling of volume @ $400/m$^3$ |
| Ventilation change | 0.54 | Linear scaling of volume @ $67/m$^3$ |
| Subtotal | 2.07 | Direct costs |
| Total | 4.6 | Direct + indirect costs + contingency |

**Table 3.8.5.2-1** Option 1: Increase height from 24 m to 28 m.

| Change | Cost Difference ($M) | Comments |
|---|---:|---|
| Increase excavation volume | -1.10 | Linear scaling of volume @ $138/m$^3$ |
| Addition to entry drift | -0.43 | Linear scaling of volume @ $400/m$^3$ |
| Ventilation change | -0.54 | Linear scaling of volume @ $67/m$^3$ |
| Subtotal | -2.07 | Direct costs |
| Total | -4.6 | Direct + indirect costs + contingency |

**Table 3.8.5.2-2** Option 2: Decrease height from 24 m to 20 m.

| Change | Cost Difference ($M) | Comments |
|---|---|---|
| Increased width | 2.65 | Quadratic scaling of width |
| Ventilation change | 0.52 | Linear scaling of volume @ $67/m$^3$ |
| Increased struct./archit. | 0.27 | Linear scaling of area @ $667/m$^2$ |
| Increased misc. electrical | 0.45 | Linear scaling of area @ $1120/m$^2$ |
| Increased misc. plumbing | 0.11 | Linear scaling of area @ $285/m$^2$ |
| Increased crane width | 0.13 | Scaling of width @ $32,800/m |
| Subtotal | 4.13 | Direct costs |
| Total | 9.3 | Direct + indirect costs + contingency |

**Table 3.8.5.2-3** Option 3: Increase width from 20 m to 24 m.

| Change | Cost Difference ($M) | Comments |
|---|---|---|
| Increased volume | 1.21 | Linear scaling of volume @ $138/m$^3$ |
| Ventilation change | 0.52 | Linear scaling of volume @ $67/m$^3$ |
| Increased struct./archit. | 0.27 | Linear scaling of area @ $667/m$^2$ |
| Increased misc. electrical | 0.45 | Linear scaling of area @ $1,120/m$^2$ |
| Increased misc. plumbing | 0.11 | Linear scaling of area @ $285/m$^2$ |
| Increased bridge crane length | 0.22 | Scaling of length @ $11,200/m |
| Increase monorail length | 0.11 | Scaling of length @ $5600/m |
| Subtotal | 2.89 | Direct costs |
| Total | 6.5 | Direct + indirect costs + contingency |

**Table 3.8.5.2-4** Option 4: Increase length from 100 m to 120 m.

### 3.8.5.3 LMD-1 Size and Configuration

The baseline cross sections for LMD-1 are shown in Figure 3.8.1-2. This Trade Study looked at 20% variations in height, width, or length of LMD-1. The initial design and cost estimates for excavations at the 7400L are at a Conceptual Design level, without the benefit of in-depth site and geotechnical investigations. The results of the Trade Study are summarized in Tables 3.8.5.3-1 to 3.8.5.3-3.

| Change | Cost Difference ($M) | Comments |
|---|---|---|
| Increase excavation volume | 1.60 | Linear scaling of volume @ $475/m^3$ |
| Ventilation change | 0.52 | Linear scaling of volume @ $153/m^3$ |
| Subtotal | 2.12 | Direct costs |
| Total | 5.6 | Direct + indirect costs + contingency |

**Table 3.8.5.3-1** Option 1: Increase height from 15 m to 18 m.

| Change | Cost Difference ($M) | Comments |
|---|---|---|
| Increased width | 3.15 | Quadratic scaling of width |
| Ventilation change | 0.39 | Linear scaling of volume @ $153/m^3$ |
| Increased struct./archit. | 0.17 | Linear scaling of area @ $774/m^2$ |
| Increased misc. electrical | 0.32 | Linear scaling of area @ $1421/m^2$ |
| Increased misc. plumbing | 0.08 | Linear scaling of area @ $372/m^2$ |
| Increased crane width | 0.11 | Scaling of width @ $36100/m |
| Subtotal | 4.23 | Direct costs |
| Total | 11.08 | Direct + indirect costs + contingency |

**Table 3.8.5.3-2** Option 2: Increase width from 15 m to 18 m.

| Change | Cost Difference ($M) | Comments |
|---|---|---|
| Increased volume | 1.43 | Linear scaling of volume @ $475/m^3$ |
| Ventilation change | 0.46 | Linear scaling of volume @ $153/m^3$ |
| Increased struct./archit. | 0.17 | Linear scaling of area @ $774/m^2$ |
| Increased misc. electrical | 0.32 | Linear scaling of area @ $1421/m^2$ |
| Increased misc. plumbing | 0.08 | Linear scaling of area @ $372/m^2$ |
| Increased bridge crane length | 0.20 | Scaling of length @ $13100/m |
| Increase monorail length | 0.10 | Scaling of length @ $6560/m |
| Subtotal | 2.76 | Direct costs |
| Total | 7.23 | Direct + indirect costs + contingency |

**Table 3.8.5.3-3** Option 3: Increase length from 75 m to 90 m.

## 3.9 Systems Engineering for the Integrated Suite of Experiments

Systems Engineering related to the ISE is focused on the following areas:

- Experimental needs solicitation and documentation
- Development and maintenance of a top-level ISE requirement set per LM or OLR area
- Interface control and management between the civil facilities and the experiments
- Value engineering and Trade Studies
- Change control and management

More detailed information about Systems Engineering on DUSEL may be found in Volume 9, *Systems Engineering*.

### 3.9.1 Interface and Requirements Management

During the DUSEL Preliminary Design phase, Systems Engineering developed and released the Integrated Suite of Experiments Interface Requirements Document (ISE-IRD), which can be found in Appendix 9.F. The ISE-IRD is the repository for requirements between the science program and the Facility design. The document also captures constraints placed by the Facility design on experiment hardware. The requirements presented in this PDR volume for the LMs, LBNE water Cherenkov detector, and the BGE experiments on the OLR are captured in the ISE-IRD. The ISE-IRD is subject to the DUSEL Configuration Control process: It requires the signatures of the Deputy Project Director, the Science Project Manager, and Facility Project Manager before obtaining Change Board approval. Changes and updates will be made to the ISE-IRD as appropriate to support the continued maturation of both Facility and science program designs.

The ISE-IRD is a key part of the overall DUSEL requirement structure that is described in more detail in Volume 9, *Systems Engineering*. The requirements in the ISE-IRD are stored in the DOORS requirements management database and linked to lower-level Facility requirements. This linking process systematically ensures that every requirement contained in the ISE-IRD has a lower-level requirement “child” that fulfills the science need. Noncompliances between the requirements set and the DUSEL design can then be systematically identified, tracked, and managed. The DUSEL requirements management system ensures that all stakeholders are involved in the creation of requirements, allows integration among different levels of the organization, and formalizes the connection between the science needs and the Facility design.

During the period after Preliminary Design and during Final Design phase, Systems Engineering will continue to be engaged in requirements definition and maturation. As changes occur in the program, Systems Engineering thoroughly assesses impacts to the various stakeholders by examining the linkages between requirements, thereby helping ensure an informed project decision occurs. As experiments are selected for DUSEL, Systems Engineering will lead the development of Interface Control Documents between the Facility and the experiment. These documents will be in greater detail than the Interface Requirements Document and ensure a smooth integration of the experiment into the Facility.

### 3.9.2 Value Engineering, Trade Studies

Systems Engineering supports Value Engineering and Trade Studies for the DUSEL Facility, many of which involve issues related to capability available for the resident science investigations. Systems Engineering participates in the assessment of the priority of trades, the Trade Study approach, and the overall approval and impact assessment to the Facility design baseline as it develops.

### 3.9.3 Configuration Control

Configuration Control is a vital aspect of any complex project because it provides a clear baseline by offering official documentation and an official change process. DUSEL Systems Engineering developed the DUSEL Configuration Control Plan and is, along with project management, responsible for its execution. Systems Engineering will ensure that the proper procedures are followed and the agreements are adequately reviewed by and disseminated to the proper parties. The ISE-IRD is the key configured item related to the science program that is under configuration control.

## 3.10 Organization and Management of the Research Program

This chapter provides a short summary of plans for organization and management of the research program. We first outline the roles of the U.S. funding agencies (NSF and DOE) for elements of the DUSEL research program. We then summarize the proposed roles and responsibilities for experiments by the management of the DUSEL MREFC-funded Project, including the evolution from the current organization for the operation of Sanford Laboratory and the DUSEL design team. We then describe the general considerations expected to pertain to the organization and management of experiments.

### 3.10.1 Agency Roles

NSF and DOE have formed a DUSEL Joint Oversight Group (JOG) to coordinate and oversee DUSEL program elements. The agencies are developing their roles in the different elements of the potential DUSEL facility and scientific program. Nonfederal funding of experiments is anticipated, and may include funds from private individuals, industrial partners, and the state of South Dakota. In addition, participation from funding agencies outside the United States is expected, consistent with a world-leading scientific program at DUSEL.

The steward of a program element will provide the majority of the funding for that element and also will accept the risk inherent in funding the program element. Additional scientific topics may arise during the course of defining the DUSEL experimental program. The appropriate agency roles will be defined as required.

### 3.10.2 DUSEL Management of the Experimental Program

The management of the DUSEL MREFC-funded Project and related Operations is described in detail in Volume 7, *Project Execution Plan*, of this PDR. We summarize here the salient features that pertain specifically to the experimental program.

**Program Advisory Committee**

A Program Advisory Committee (PAC) with representation from all scientific disciplines that form the DUSEL scientific program advises the Vice Chancellor for Research at the University of California, Berkeley. This committee will review all experiments proposed to be located at the DUSEL site.[2]

A scientific program is in existence at Sanford Laboratory and is expected to continue during the initial phase of construction of the DUSEL Facility (see Chapter 3.4, *Research Activities at the Sanford Underground Laboratory*). In early 2006, the Homestake collaboration convened a PAC to review the 80 Letters of Intent (LOIs) that were submitted following a solicitation to the underground science community. Several experiments suitable for initial deployment in a so-called Early Implementation Program (EIP) were identified, including the Large Underground Xenon (LUX) experiment, the MAJORANA DEMONSTRATOR, and several geophysics and biology initiatives. By early 2008, the dewatering and reopening of the underground facility had progressed to the point that it became possible to consider the development of a deployment timetable for an early science program. The Sanford PAC was reconvened to recommend a prioritized list of experiments that could be realistically installed within the reopened areas of the facility deemed safe and suitable. LUX and the MAJORANA DEMONSTRATOR were identified as top priorities from the seven physics initiatives reviewed. SDSTA was encouraged to

2 Close coordination with the Fermilab PAC is planned for consideration of the LBNE Project and other experiments reviewed by the Fermilab PAC to minimize duplication of reviews.

support several of the smaller-scale biology and geophysics efforts. The Sanford PAC is no longer active and all review of existing or planned scientific programs at Sanford Laboratory and at the DUSEL site will be performed by the DUSEL PAC.

### Science Liaison and Integration

The Scientific Liaison Department at Sanford Laboratory is currently responsible to the Director of Sanford Laboratory for oversight of the early (pre-DUSEL) experimental program at the Laboratory. The Liaison Department is headed by an experienced physicist and includes technical-support personnel. The Liaison Department calls on other resources—EH&S, Operations, and Engineering—at Sanford Laboratory to support the experimental program. The DUSEL science integration team (described in Chapter 3.6, *Integrated Suite of Experiments (ISE) Requirements Process*) is primarily responsible for the interface to the experimental community and for the scientific input to requirements that determine the future DUSEL Facility design. The team also has an ad hoc role in aspects of engineering and EH&S review of experiments at Sanford Laboratory and Memoranda of Understanding (MOUs) between these experiments and Sanford Laboratory. The organization of the support of experiments at Sanford Laboratory and later at the DUSEL Facility is in a transition phase. A unified organization, combining all functions, is expected to be in place by mid- to late 2011 under the auspices of the DUSEL LLC (see Volume 7).

The costs of the experimental program and its support are enumerated in two Work Breakdown Structure (WBS) elements. DUS.SCI.EXP includes the MREFC funds devoted to experiments. This element also includes experiment design and R&D activities and support for research time of scientific members of the DUSEL staff. The costs of personnel and related materials and supplies in support of the integration of experimental program are included under the WBS element DUS.SCI.SUP. The costs are further described in Volume 2*, Cost, Schedule, and Staffing*, and additional description is provided in Volume 10, *Operations Plans*.

The unified DUSEL science integration team will be responsible for supporting the integration of experiments into the Facility during the design, construction, and installation phases of these experiments. The team will then become responsible for common aspects of operational and maintenance support of experiments as they complete commissioning and begin operations. The team will concurrently support R&D at the DUSEL Facility and integration of new experiments as they are conceived, proposed, accepted, built, installed, and operated. It should be noted that the primary support of experiments, in all phases, is the responsibility of the experimental teams and the associated funding sources. The science integration team is and will be responsible for the interface of experiments to the Facility, for establishing and maintaining a safe working environment, and for coordination and implementation of common aspects to support multiple experiments. In the sections below, we describe briefly the major elements of the team.

Some operational support of experiments, including consumables, is currently provided by the SDSTA at Sanford Laboratory. It is currently planned that this support from SDSTA will continue through FY 2012. A complete transition to federal support (NSF and DOE) by FY 2013 is anticipated. The phasing and detailed planning for this transition remains to be developed with the federal agencies and SDSTA. These non-labor costs for support of experiments are included under the WBS element DUS.SCI.EXP.

### Organization of Experiments

The DUSEL Facility will be open to all proposals from any scientific discipline. Proposals will be subject to the review process described briefly below. It is anticipated that proposals for experiments at DUSEL will undergo a phased review process by the DUSEL management and the PAC in a manner very similar to the review of experiments at accelerator laboratories or other user facilities. The DUSEL management team, in consultation with the PAC and the agencies, will develop more detailed guidelines for experimenters and the associated review process by fall 2011. Many of the BGE experiments do not require long lead times for R&D. They could be ready for deployment whenever the facility is ready to host them.

The financial and technical scope of the experiments at the DUSEL Facility will vary greatly. The number of scientists participating in a given experiment will also vary from single-investigator-driven science to international collaborations of hundreds of physicists and students. Despite this large variation, all experiments will share some common features; other features of the organization and management will be adapted to the scope, funding, and nature of a particular experiment or experimental program at DUSEL.

#### General Requirements for Experiments

All future experiments at DUSEL will share the following features:

- Review by DUSEL management regarding the suitability of the proposed experiment for the DUSEL Facility
- Review and approval by the appropriate elements of the DUSEL organization of all relevant EH&S issues
- Review and recommendations by the DUSEL PAC
- A signed MOU between the designated representative(s) of an experiment and DUSEL management
- A General Services Agreement (GSA), to be amended and signed yearly by the experimental representative(s) and DUSEL management, that specifies services and items to be provided by DUSEL to the experiment and the responsibilities of the experiment group
- Addenda or other documents related to the MOU as required to meet fiduciary or other requirements
- A well-identified organizational structure and defined points of contact with clearly defined responsibilities on the part of the experiment team
- Implementation of experiments done according to codes and standards established by DUSEL

The organization of a particular experiment or R&D activity will depend on the nature of the experiment, its financial scope (including the participation of non-U.S. parties), and the relevant agency steward. We note that Memoranda of Understanding and General Services Agreements (annual agreements specifying the support from SDSTA) are now in place between the LUX and MAJORANA DEMONSTRATOR experiments and SDSTA. The process for establishing these agreements, and the documents themselves, provide experience and templates for future experiments at DUSEL.

#### Major Experiments

Major experiments[3] will have dedicated management in the form of a spokesperson (or co-spokespersons), a project manager, a project office adequately staffed to provide engineering and project-management support, and an organizational structure commensurate with the scope and complexity of the experiment. The responsibility to form and staff the management team for a major experiment rests with the experiment. DUSEL management reserves the right to review the management structure and management team of major experiments, including the necessity to concur in the selection of project manager. The roles and responsibilities of the experiment management team and the DUSEL organization will be documented in a corresponding MOU.

---

[3] We assume here that a major experiment is one with a project cost of $5 million or more.

The DUSEL science integration team involved directly in aspects of the experimental program is anticipated to be small. Its primary focus will be on the experiments' interfaces with the DUSEL Facility, including the implementation of common infrastructure shared among experiments. Such interfaces will be documented for each experiment in an Interface Control Document (ICD) that will provide the basis for all interfaces between a given experiment and the DUSEL Facility. The mechanism for review, approval, oversight, and reporting for each major experiment will be developed on a case-by-case basis by the experiment management, DUSEL management, and the funding agencies (primarily the steward agency). In some cases, substantial in-kind or financial contributions from outside the United States may be needed to complete a proposed experiment. DUSEL management may choose to form an international finance board for a given experiment or group of experiments as needed and in close collaboration with the management of the respective experiments.

#### Experiments at User Facilities

To implement the scientific program, DUSEL may include long-term experimental facilities that will have a changing set of users. Examples of this include an accelerator facility for nuclear astrophysics, an advanced low-background counting facility, or some aspects of potential experiments in BGE. In such a case, DUSEL management may form an advisory body, in addition to the PAC, to provide advice and recommendations and to provide periodic reports to the PAC.

#### Other Experiments

Experiments or R&D efforts not falling under the "major" classification will be subject to the general conditions and requirements described earlier. The organization of these efforts will be tailored to the circumstances of the individual proposed experiment by DUSEL management, the proponents, and the steward agency.

#### Center for Underground Science and Engineering

The MREFC-funded Project consists of both the construction of the Facility and construction of the experiments that will be hosted in the facility. The coordination between the experiments and the Facility is of great importance. The science liaison and support structure will be critical to maintain that coordination. Scientists, including post-doctoral fellows, affiliated with DUSEL will provide the liaison and support functions. Many of these scientists will also be expected to participate in analysis of data and research as experiments begin to acquire data.

To this accomplish these goals, a Center for Underground Science and Engineering will be established that will provide a framework for collaboration among the principal institutions (U.C. Berkeley and SDSM&T) responsible for DUSEL. Funding for the centers is expected to be derived, in part, from the DUSEL operational budget.

### 3.10.3 User Community

The user community for DUSEL is substantial and estimated to include 750-1,000 interested scientists and engineers. The DUSEL Experiment Development and Coordination (DEDC) group was formed in January 2008 to address the development of the initial experiments to be included in the DUSEL Facility. Through workshops and working groups, the DEDC provided input and assistance to the DUSEL Project staff to develop the requirements for the DUSEL Facility. The mission of the DEDC was completed in early 2010 and a new user body was formed, the DUSEL Research Association (DuRA). This Association comprises the full cadre of members of the DUSEL scientific community with representation by a smaller Executive Committee to liaise with the Facility and with the funding agencies. The Executive Committee is elected by the general user community and is representative of the much broader scientific community intending to propose experiments at DUSEL.

### 3.10.4 Experimental Program Costs and Funding

The overall scope of the initial experimental program at the DUSEL Facility will, in part, be determined by funds available from U.S. and non-U.S. funding agencies, research institutions, and other sources. The DUSEL Facility design is based on housing and supporting an initial generic experimental program—the Integrated Suite of Experiments (ISE) (see Chapter 3.5, *Integrated Suite of Experiments*) when set. The following is a summary of the proposed MREFC funding requested to implement a world-leading program of scientific research. The results of experiments in the next few years will inevitably influence the nature of the future experiments at DUSEL. In addition, funding for the ISE must come from a number of sources outside the MREFC funding to realize the proposed program.

NSF funding for experiment design and R&D is currently in place, primarily through the funding of S4 proposals. The S4 funding is expected to be exhausted by the end of FY 2012. However, the design of potential experiments, and the related R&D, will not be completed. NSF funding to complete Preliminary and Final Design of the first DUSEL experiments, and the related R&D, must continue smoothly after S4 funding ceases. We propose that funding outside the MREFC funding for completion of experiment design and R&D be in place from FY 2013-FY 2016 for the first experiments at the DUSEL Facility. These design and R&D costs are under WBS element DUS.SCI.EXP.

The proposed MREFC funding for experiments is based in part on preliminary cost estimates provided by the NSF S4 awardees (see Section 3.3.1). The S4 awardees were requested to provide cost information in the form of FTEs for different types of personnel (scientists, engineers, technicians, etc.) and procurement costs. They were also requested to provide a very preliminary schedule. The DUSEL Project controls team used this information to formulate a rough cost estimate. It must be emphasized that these cost estimates are at a very early stage. The DUSEL team applied a uniform costing model for personnel to all estimates. This model used composite rates for personnel based on a 50:50 mixture of university and national laboratory labor. The costs of scientific personnel were not included in these estimates for physics experiments. Given the very preliminary nature of these estimates, we believe a risk-based contingency of 50% is appropriate at this time. MREFC-funded costs are under WBS element DUS.SCI.EXP.

The cost range for specific experiment types is given in Table 3.10.4-1. The costs shown include 50% contingency and are in $M (FY 2010). The costs shown do not include R&D or Final Design activities. Installation costs are included, but not operations. The cost range for $0\nu\beta\beta$ experiments includes a very

rough estimate of scope contingency as well. A large fraction (30-40%) of the cost of the $0\nu\beta\beta$ experiments is projected to be in separated isotopes. A reduction in scope (target mass) thus translates directly into a substantial reduction in cost. We have attempted to include such a reduction (factor of two from the maximum mass proposed) but recognize well that the cost of $0\nu\beta\beta$ experiments can only be set for a given design with a specified scientific reach. The cost range of the nuclear astrophysics facility is based solely on the DIANA proposal. The upper range is as proposed and the lower range assumes scope contingency or staging is possible, an assumption made by the DUSEL Project team. The cost range of providing low-background counting is significant. The upper range is based on the full FAARM proposal. The lower range reflects the possibility of implementing only a more limited capability at the DUSEL site. The cost range of BGE experiments is very uncertain. The range given in Table 3.10.4-1 is based on the S4 BGE proposals but the upper range is not a sum of the cost estimates of these current proposals. The range reflects the Project's judgment regarding a plausible range of proposals for initial BGE experiments.

| Experiment Type | Cost Range |
|---|---|
| Dark-matter experiments (per experiment) | 80-100 |
| $0\nu\beta\beta$ experiments (per experiment) | 220 - 300 |
| Nuclear astrophysics facility | 30 -45 |
| Advanced low background | 2-15 |
| Biology, geology, and engineering experiments | 60-180 |

**Table 3.10.4-1** Cost range of potential initial DUSEL experiments, as explained in the text. Cost range is in FY 2010 million dollars.

The total MREFC funding proposed for experiments is $300 million. Of this, a fixed amount of $125 million is allocated to the Long Baseline Neutrino Experiment (LBNE) project, which will provide all other funds required, including contingency, for the LBNE far-detectors. A total of $175 million is allocated to other experiments—see Table 3.10.4-2.

The ultimate allocation to specific experiments within the $175 million envelope will be determined by scientific merit, interactions with the steward agencies, and non-U.S. agencies, and will include thorough review by the DUSEL PAC.

However, to advance planning of the future DUSEL experimental program, we present here a description of a model for experiment types within the $175 million envelope. We emphasize this is a preliminary model intended to initiate further development of a robust experimental program and is consistent with meeting the key scientific goals of the DUSEL MREFC proposal.

MREFC funds, including contingency, would be allocated with high priority such that one or more Generation-3 (G3) dark-matter experiments would be realized as part of the initial DUSEL experimental program. Additional financial support from DOE and non-U.S. sources is needed to create a robust program of multiple G3 dark-matter experiments. MREFC funds may be allocated to allow funding to complete a G2 dark-matter experiment to be located in the Davis Laboratory Module, if a compelling scientific case is made and if such an experiment has negligible impact on the cost and schedule of the DUSEL Facility. A decision to solicit proposals for G2 experiments in the Davis Campus will be made by the end of 2011, in consultation with the funding agencies and after review by the DUSEL PAC. Initial funding of such a G2 experiment for the Davis Campus must come from non-MREFC sources. The earliest selection of such an experiment could be made by mid-2012.

A fixed amount of MREFC funding (no contingency, similar to the LBNE case) would be allocated to $0\nu\beta\beta$ experimental activities, and most of the funding (and all of the contingency) would be provided by DOE Office of Nuclear Physics (ONP) and other sources.

An allocation of MREFC funding for a nuclear astrophysics facility may be made and is contingent upon review by the DUSEL PAC of scientific merit and availability of funding. The total MREFC funding of this aspect of the scientific program can only be determined after thorough review and a better understanding of potential funding from DOE and non-U.S. sources. A decision to proceed with an underground accelerator to accomplish this scientific program must be made relatively early. Modifications to the current generic design of the candidate Laboratory Module (LM-1 at the 4850L) are required to accommodate an underground accelerator. Thus a decision to design LM-1 specifically for an underground accelerator must be made by mid-2012.

MREFC funding for low-background counting, material assay, and related activities is planned to allow for the minimum support that is required at the DUSEL site that cannot be provided by other facilities. MREFC funding may allow for an advanced facility to meet future needs, including the option of a phased approach to such a facility.

Funding for BGE experiments would allow a science program in these areas to start early in the MREFC-funded period and would provide the basis for continued activities that are likely to be also supported by funds from multiple sources outside the funding.

<table>
<tr><th>Experiment Type</th><th>MREFC Funds Proposed</th></tr>
<tr><td>Long-baseline neutrino, proton decay</td><td>125</td></tr>
<tr><td>Dark-matter experiments</td><td rowspan="5">175</td></tr>
<tr><td>$0\nu\beta\beta$ experiments</td></tr>
<tr><td>Nuclear astrophysics</td></tr>
<tr><td>Low background counting</td></tr>
<tr><td>Biology, geology, and engineering</td></tr>
<tr><td>TOTAL</td><td>300</td></tr>
</table>

**Table 3.10.4-2** Proposed MREFC funding of experiments. Proposed funding in FY 2010 million dollars.

### 3.10.5 Schedule of the Initial Experimental Program

This section provides a high-level overview of the schedule for experiments at the DUSEL site through the end of the MREFC-funded period. The need to begin Final Design of the DUSEL Facility in early 2012 and the plan to begin construction of the Facility (MREFC-funded Project start) in January 2014 are critical near-term schedule constraints. A more detailed description of the schedule is given in Volume 2, *Cost, Schedule, and Staffing*.

#### Long Baseline Neutrino and Proton Decay

The schedule for the LBNE project is under active development and depends in detail on the final choice of technologies for the LBNE far detectors. A Critical Decision-1 (CD-1) review is anticipated in the fourth quarter of 2011, when a more developed schedule will be available. A CD-2 review is currently anticipated in 2013. The CD-3 milestones are under development and CD-3a (for long-lead-time items) and CD-3b milestones may be developed. Final CD-3 approval is anticipated for the last quarter of 2014. Underground construction of a large cavity could begin in FY 2015 and be ready for experiment installation to begin by 2019, assuming one or more water Cherenkov detectors are selected by the LBNE project. Experimental installation would follow. A schedule for the liquid argon (LAr) option for LBNE and its impact on the DUSEL Facility schedule is under development.

**Dark Matter Experiments**

The U.S. dark-matter community has proposed a road map leading from the current generation of dark-matter experiments (G1) through implementation of G3 dark-matter experiments at DUSEL (see Section 3.3.3, *Dark Matter Experiments*). The U.S. community proposes to make a selection among the various experimental techniques such that G3 experiments can be fully realized at DUSEL by 2018-2020. They propose a choice of technique be made about 3½ years before the beneficial occupancy of the relevant LM for housing an experiment. It is recognized that this is an aggressive schedule and may not take full advantage of the information obtained from sustained operation of G2 dark-matter experiments. For dark-matter experiments at the 4850L, this implies a technology choice in 2014. We have assumed about four years between selection of a G3 dark-matter experiment (by mid-2014) and when the relevant 4850L LM is ready to begin experiment installation (2018). The duration of assembly and installation for these experiments is not yet well understood but is estimated to be up to two years. This would lead to a start in data-taking by 2019-2020. The LM at the 7400L is expected to be ready for experiment installation in 2019. Assembly and installation at the 7400L will be more difficult than at the 4850L (for an experiment of similar complexity) and a two-year or longer duration is likely to be needed, leading to data taking at the 7400L by 2021-2022.

**Neutrinoless Double Beta Decay Experiments**

The $0\nu\beta\beta$ experiments, as represented by the 1TGe and EXO examples, anticipate starting construction (first procurements) no earlier than the latter half of 2015. This approximate date is driven primarily by the need to obtain data from ongoing experiments—the MAJORANA DEMONSTRATOR, GERDA, and EXO200—and the time needed for Preliminary and Final Design. Both the 1TGe and EXO experiments want to be located at the 7400L. The time for production of the enriched isotopes needed for these experiments is significant, as is the fabrication of the remainder of the experimental apparatus. These example $0\nu\beta\beta$ experiments desire beneficial occupancy of the LM in early 2018, which is not consistent with the current planning for the 7400L (experiment installation to begin by the second quarter of 2019). However, planning for these experiments is at an early stage and constraints from availability of funding or technical choices have not yet been fully evaluated. The assembly and installation durations of these experiments are long (see Section 3.3.4, *Neutrinoless Double-Beta Decay Experiments*)—potentially three years, roughly, after beneficial occupancy. Depending on the type of experiment, staged operation may be possible, but full operation of multitonne-scale detectors would begin three to four years after beneficial occupancy of the 7400L LM, that is in approximately 2022.

**Nuclear Astrophysics Facility**

The design of an accelerator facility for nuclear astrophysics experiments, as represented by DIANA, is relatively well advanced. We have already noted in a previous section that a choice to proceed with an underground accelerator needs to be made by mid-2012 if LM-1 at the 4850L is to be a "custom" design for an underground accelerator. Final Design of the accelerator complex could begin in late 2012, and some aspects of the accelerator could be built and operated on the surface (not at the DUSEL site), prior to the availability of the underground LM, if funds were available and if scientifically justified. The schedule for deployment of the accelerator underground is, of course, tied to the beneficial occupancy of LM-1 at the 4850L (2018). The duration of installation and the phasing of commissioning of the accelerators are not yet well understood, but operations could likely begin in 2019.

### Low-Background Counting and Material Assay

Low-background counting and material-assay capabilities are needed during the design, R&D, and early construction phases of the initial dark-matter, $0\nu\beta\beta$, and LBNE projects. This early capability would be provided partly by capabilities at Sanford Laboratory located in the Davis LM and by other facilities located elsewhere in and outside of the United States. The large FAARM Facility requires occupancy of LM-2 (or LM-1 if a nuclear astrophysics facility is not selected to be among the initial experiments). Implementation of FAARM after beneficial occupancy is expected to take one to two years, and thus would be available by late 2019 or 2020. It would therefore benefit future experiments and R&D on pushing the limits of low-background counting. In our planning, we assume a choice to proceed with FAARM or similar facility would be taken by about mid-2015.

### Biology, Geology, and Engineering Experiments

Elements of a number of BGE experiments are already under way in a modest way at Sanford Laboratory (see Chapter 3.4). The implementation of BGE experiments could begin in a phased way very early in the MREFC-funded era, as continuations of ongoing work in 2014 and then as new experiments in 2015-2016. Some potential BGE experiments are of sufficiently short duration that they could begin early in the MREFC-funded era and be completed well before completion of the Project. The critical issue for some experiments would be access to the 4850L and later to the 7400L; phasing implementation of these experiments with construction will be an important constraint. Some proposed BGE experiments are of significant scope and require new excavation and related work beyond that included in the DUSEL baseline design. These experiments would be reviewed by the DUSEL PAC in 2012-2013, such that the appropriate selections can be made, potentially leading to timely start of such experiments in 2014-2015.

### Long-Term DUSEL Schedule

The DUSEL Facility, once completed, would have an operational lifetime of more than 30 years and would serve multiple generations of experiments. The descriptions in this Volume have emphasized the characteristics and Facility requirements for the initial DUSEL experiments. However, the Facility design is sufficiently flexible, and upgradeable, to accommodate future generations of experiments in dark matter, neutrinoless double-beta decay, neutrino oscillations, and possibly proton decay. An accelerator dedicated to nuclear astrophysics is in itself a facility with decades of anticipated experiments. Experimental studies in underground biology and geosciences are also anticipated to be relevant over many decades.

## Volume 3 References

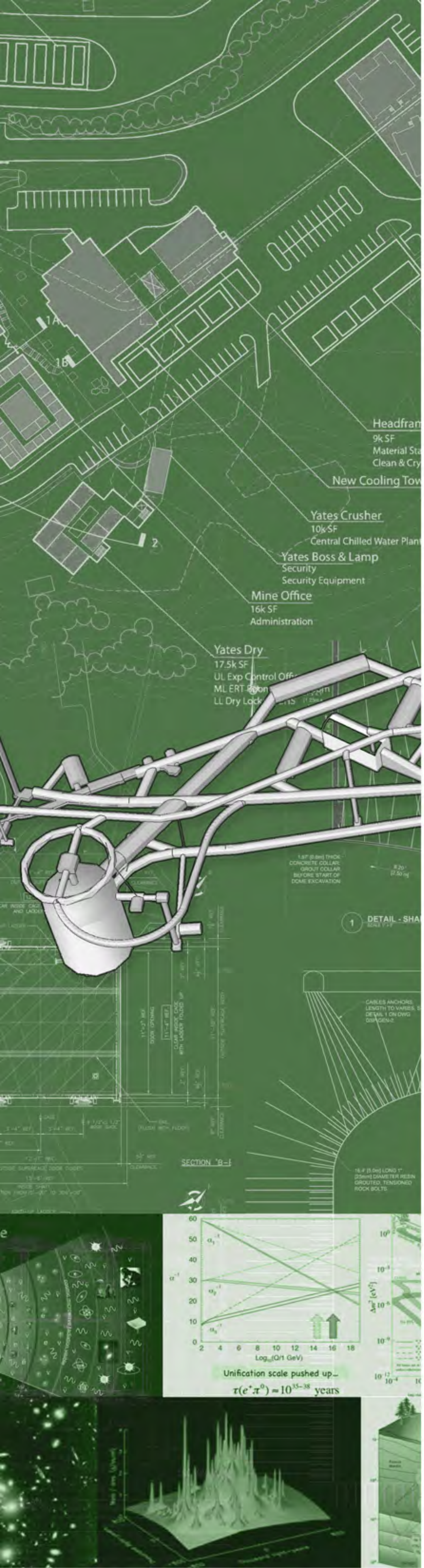

# Preliminary Design Report

May 2011

# Volume 4:

# Education and Public Outreach

DUSEL
Deep Underground Science and Engineering Laboratory

This page intentionally left blank

# Education and Public Outreach

## Volume 4

### 4.1 Sanford Center for Science Education Overview

The Sanford Center for Science Education (SCSE) will be the facility to house the education and public outreach arm of DUSEL. Precipitated by a generous pledge from philanthropist T. Denny Sanford and consistent with the commitment of the National Science Foundation (NSF) to education and public engagement, the SCSE is envisioned to exceed expectations for education and public outreach at a national scientific research facility. The SCSE will support innovative programs to expand educational opportunities for a broad range of audiences, from the general public to school groups to educators to scientists.

With the creation of the SCSE comes a special opportunity to integrate education and public outreach into the DUSEL Project. DUSEL's research frontiers are captivating, and its geographic and cultural setting provides an attractive environment to structure programs—drawing upon the scenic beauty of the Black Hills, the area's geology and ecology, its history, and the culture of its native people. The SCSE will feature engaging activities and interactive exhibits that build understanding and spark imagination.

The distance of DUSEL at Homestake from urban centers, the relative inaccessibility of the underground environment, and the desire to have wide-ranging national and international impact necessitate development of especially compelling and robust off-site and virtual program elements. Key to DUSEL's education and public outreach vision is the weaving together of all three programmatic facets: on-site, off-site, and online education. Select audiences will access the 4850L directly, and all will have access to the underground through remote video, data transmission, and computer simulation. Students will be able to design their own inquiries, under the mentorship of a scientist or engineer either on site or thousands of miles away, using data from DUSEL experiments.

Due to the historical importance of the Black Hills area to the regional American Indian cultures and also because of the low representation of American Indians in scientific and engineering disciplines, American Indians represent an especially important audience. Other historically underrepresented groups in the region, especially within the physical sciences and engineering, include women, girls, and rural Americans. DUSEL remains deeply committed to serving all of these audiences.

With leadership from its Education Governing Board and guidance from its Education Advisory Committee and Cultural Advisory Committee (Appendices 4.A, 4.B, and 4.C), the Project has developed and adopted the following mission, vision, and goals. Programs and organizational functions have been defined. All reflect scientific and broader community input gathered through numerous workshops and planning sessions over the past nine years. Together, the mission, vision, and preliminary program roster have guided the design of the proposed educational facility.

#### 4.1.1 Mission, Vision, and Goals

The mission of DUSEL's Education and Public Outreach Program is to draw upon DUSEL's science and engineering program, its human resources, its unique Facility, and its setting within the Black Hills to develop and provide rich, inquiry-driven learning experiences that engage and connect diverse audiences of students, educators, scientists, engineers, and the general public.

DUSEL's education and public outreach program objectives and goals are to deepen understanding of the nature and value of science and its application through engineering and, in particular, to inspire and prepare future scientists, engineers, science educators, and a skilled technical workforce. More detailed specification of these goals, objectives, and core values appears in the *Statement of SCSE Mission, Vision, Goals, Core Values, and Functions* (Appendix 4.D).

DUSEL's education and public outreach program vision is to be a world leader, a resource, and a catalyst for the improvement of science education, serving as a model educational enterprise within a scientific research facility that will draw upon and generate knowledge about the teaching and learning of science and engineering.

### 4.1.2 Functions of the Sanford Center for Science Education

To realize the mission and vision of the DUSEL education and public outreach program, innovative programs and exhibits will be developed, and a facility will be constructed on the Yates Campus. The functions performed by the education and public outreach organization will include:

- Operation of a visitor center open to the general public and to organized school groups
- Development and implementation of programs and exhibits on site, off site, and online
- Development and dissemination of resource materials for teachers, students, scientists, and the general public
- Development and maintenance of a sophisticated digital presence
- Staffing of a science liaison office to engage and support the DUSEL science groups in science education
- Staffing of a cultural outreach office to develop and nurture educational partnerships
- Arranging and conducting on-site and off-site classroom experiences and workshops
- Arranging and conducting educational tours of surface and underground facilities
- Fundraising, grant writing, and other development functions

Further details are presented in Chapter 4.3.

### 4.1.3 Facility Components

Requirements for the education and public outreach facility and its public, instructional, and administrative spaces are listed in Table 4.1.3 and further described as part of the architectural design in Chapter 4.2. Additional detail also appears in Chapter 5.2, *Surface Facility and Infrastructure*.

| Function | Facility / Program Requirements |
|---|---|
| **Education and Public Outreach Programs - On Site** | 7,500-sq.-ft. exhibit space<br>3,000-sq.-ft. outdoor exhibit space (seasonal)<br>(2) 1,000-sq.-ft. classrooms<br>100-seat theater (1,000 sf)<br>Visitor services (cloakroom, restrooms, food service)<br>Resource library for SCSE staff, scientists, and visiting educators<br>Access to scientific data and facility monitoring data from underground<br>Interaction space (for scientists, public, students, and educators)<br>Safe and convenient access and parking for private automobiles and buses |
| **Education and Public Outreach Programs - Off Site** | Videoconference and other distance-learning capability<br>Materials storage<br>Access to scientific data and facility-monitoring data from underground<br>Access to larger theater/auditorium (local or regional) |
| **Administrative and Support Functions** | Offices and other personnel support spaces (conference room, copy room, restrooms)<br>Exhibit and materials storage and support space<br>Information technology (IT) and media support |

**Table 4.1.3** SCSE Facility and program requirements.

### 4.1.4 Institutional Planning

Education and public outreach plans have been informed by extensive engagement of the DUSEL scientific community, at town meetings, conferences, and workshops dating back to 2001. Representatives of local and regional educational communities, tribal communities, national laboratories, and other members of the national science education community have contributed—as part of the DUSEL town meetings, conferences, and workshops and through focus groups, site visits, and smaller meetings devoted specifically to development of the education and public outreach program. Institutional planning has also been informed by an initial market assessment and analysis commissioned to establish industry benchmarks and attendance projections (Appendix 4.E).

## 4.2 Baseline Facility Preliminary Design

The baseline SCSE Facility design is described in detail in Volume 5, *Facility Preliminary Design*, and the corresponding estimates for Facility Construction are discussed in Volume 2, *Cost, Schedule, and Staffing*.

### 4.2.1 SCSE Surface Facility Design Principles

The Project's architectural and engineering consultant for the design of the DUSEL Surface Facility, HDR CUH2A, has developed concepts for new and repurposed buildings on the Yates Surface Campus. The SCSE is one of two new surface buildings proposed for construction. New construction presents an opportunity to provide the site with both a new *modern public entrance* and a new image, integrating the site's cultural and historical context with the excitement of new frontiers in science. The SCSE's role as a center for science education and public outreach, both regionally and globally, makes its presence and prominence on the site especially important.

The building will be designed to have a low impact on the land, taking a sustainable approach to its materials and systems wherever possible. Though some contrast is desirable, materials and aesthetic expression of the SCSE will be complementary to the industrial design that is common among all existing buildings on the site.

A collection of architectural design goals has been developed by the Project with HDR to help guide the development of the SCSE. The goals align with the overall Project goals and the SCSE mission. They also help to engage public support across a wide array of stakeholders. These design goals are to:

- Respect the historical context of the site through preserving the vertical impact of the Yates Headframe, respecting the scale of the "Homestake Mine Office" (Administration Building), and relating to the Homestake Mine in terms of site context (choice of material, etc.)
- Reflect the cultural heritage of the site through an understanding of the SCSE's place in the Black Hills, respecting the land through sustainable site design, and enhancing the views
- Become a campus bridge through its position as the front door and image of DUSEL, connecting the different levels of the laboratory, and providing a welcoming place for the public as well as scientists and staff to gather and interact
- Invoke a sense of the underground, whether with the building itself or the elements it contains, providing the public with the experience of being underground and a sense of the science being conducted underground
- Provide a "wow" factor, generating high levels of interest among students, families, educators, scientists, and members of the general public
- Maintain flexibility in its design of spaces for multiple uses and in the possibility of future expansion
- Provide a comfortable learning environment in a center that demonstrates sustainable construction, and which integrates history, culture, and science, all within the design of the building itself
- Meet the functional and relational requirements of the education and public outreach program and its staff

### 4.2.2 Sustainability and Energy Efficiency

The Project has determined that a Leadership in Energy & Environmental Design (LEED) Silver certification will be targeted for new construction projects, including the SCSE, while continuing to evaluate standards for repurposed buildings. The Project seeks to follow Sustainable Sites Initiative (SSI) program guidance in developing outdoor public areas, including the outdoor elements of the SCSE. Every effort will be made to manage it as a sustainable site, considering storm-drainage systems, water management, convenient access to public transportation as available, orientation of the Facility for solar gain and natural lighting, and landscape design for environmental quality. See Section 5.2.3 for discussions on the LEED system and the SSI program as they apply to DUSEL.

### 4.2.3 Cultural Considerations

The Project seeks to understand and reflect the rich history and cultures associated with its location in the Black Hills and to be a valued member of the regional community. The Black Hills are sacred to many American Indians. The Black Hills have also provided a livelihood for generations of miners. These are important components of the SCSE cultural context.

Through a wide array of programs and partnerships, the Project through the SCSE will endeavor to increase appreciation for and understanding of science and engineering among groups historically underrepresented within these disciplines. Because of DUSEL's location, a special opportunity exists to engage and collaborate with members of American Indian communities across the region. Learning from, serving, and supporting American Indian community members represents one of three "essential elements" of the education and public outreach program (see *SCSE Essential Elements Document*, Appendix 4.F). The effort to make the SCSE inclusive, inviting, and beneficial for American Indian community members will lead to greater inclusion and enhanced educational opportunities for everyone, including other underrepresented groups.

In March 2010, the Project's Cultural Advisory Committee, the education and outreach planning team, and HDR organized a set of workshops to obtain input on principles of site design for the DUSEL Surface Campus. The first meeting was attended by leaders of American Indian communities, and the second by community leaders in the northern Black Hills. Notes from these two meetings are in the *Community Input on Facility Design Principles*, Appendix 4.G.

As a follow-up, in June 2010, DUSEL and the SDSTA organized two public open house forums in Lead and Rapid City, South Dakota. At these sessions, aspects of the Project design were presented to regional community members in an informal setting through posters and conversations, connecting community members with project staff representing the areas of operations; engineering; science; environment, health, and safety; education; and cultural outreach.

In September 2010, project team members met individually with members of the Cultural Advisory Committee and American Indian leaders in the region to identify key American Indian elders and community and education leaders to contribute to education and public outreach facilities and program planning. Building relationships with these individuals and the tribal communities they represent is a continuing top priority for the education and outreach planning team.

### 4.2.4 Space Specifications

Through a series of meetings and in concert with educational program plans, the following Preliminary Design programming requirements have been developed for the SCSE Facility. These initial prescriptions will be revisited as part of more comprehensive strategic business planning for the education and public outreach program, to be conducted over the next five years that will include further market research, content development, and overall operational analysis. These planning efforts for the SCSE Facility and the DUSEL Education and Public Outreach Programs are funded through a complementary NSF award (PHY-0970160). Specifications for public, educational, and administrative and support spaces are shown in detail in the following listings. Positioning of these various spaces within the SCSE Conceptual Design is shown in Figure 4.2.4-1. The SCSE location on the Yates Campus is shown in Figure 4.2.4-2.

#### Public Space Programming

- **Exhibit space.** 7,500 net sq. ft.—to be supplemented with outdoor space in the summer
- **Theater.** 1,000 sq. ft.—a 100-seat theater between the lobby and the exhibit space. The theater will be used for orienting educational audiences, for multimedia presentations, and for scientific colloquia.
- **Café.** 900 sq. ft.—a coffee/sandwich shop for the public, DUSEL staff, and users
- **Main Building lobby.** 1,800 sq. ft.—a gathering place for families and groups

#### Educational Space Programming

- **Classroom A.** Wet lab, 1,000 sq. ft.—a teaching laboratory to be used primarily by students and teachers
- **Classroom B.** Distance learning classroom/media lab, 1,000 sq. ft.—a state-of-the-art classroom with distance learning and multimedia capabilities to be used by students and teachers in on-site and off-site workshop settings
- **Educator Resource Center and Library.** 700 sq. ft.—a physical space and resource center to support educators, scientists, and science education researchers

#### Administrative and Support Space Programming

- **Exhibit storage and support.** 1,000 sq. ft.—space for preparation of exhibits and storage of supplies
- **IT media support/control room.** 400 sq. ft.—a support room for the distance learning classroom/media lab as well as the interface to monitoring programs
- **Five administrative offices.** 600 sq. ft.—administrative offices for five permanent staff
- **10-15 cubicle workspaces.** 750 sq. ft.—for additional permanent, guest, and seasonal staff
- **Administrative meeting and support.** 250 sq. ft.—a conference room and gathering area for the staff
- **Copy/coffee room.** 150 sq. ft.—administrative office support

The total baseline scope space planned for the SCSE Facility is summarized in Table 4.2.4. Scope options considered by HDR in their Final Conceptual Design report are noted in Section 4.2.4.1.

| Type of Programming | Gross Sq. Ft. | Net Sq. Ft. |
|---|---|---|
| Public | 17,235 | 11,200 |
| Educational | 4,455 | 2,700 |
| Support | 5,070 | 3,150 |
| Total | 26,760 | 17,050 |

**Table 4.2.4** Program summary for SCSE Conceptual Design. "Gross square feet" is the overall footprint area; "net square feet" is usable interior space.

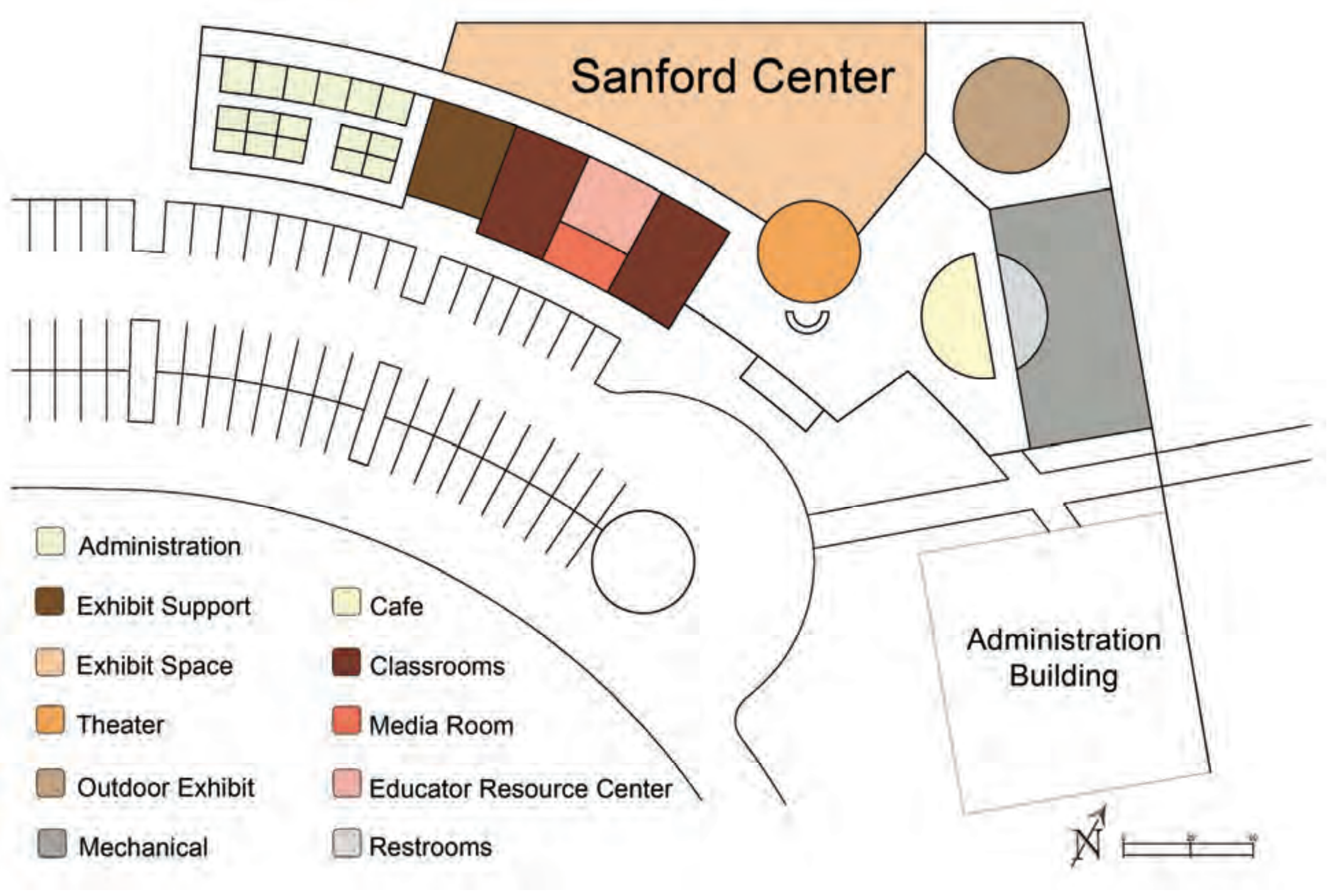

**Figure 4.2.4-1** SCSE Facility floor plan. [Adapted from HDR]

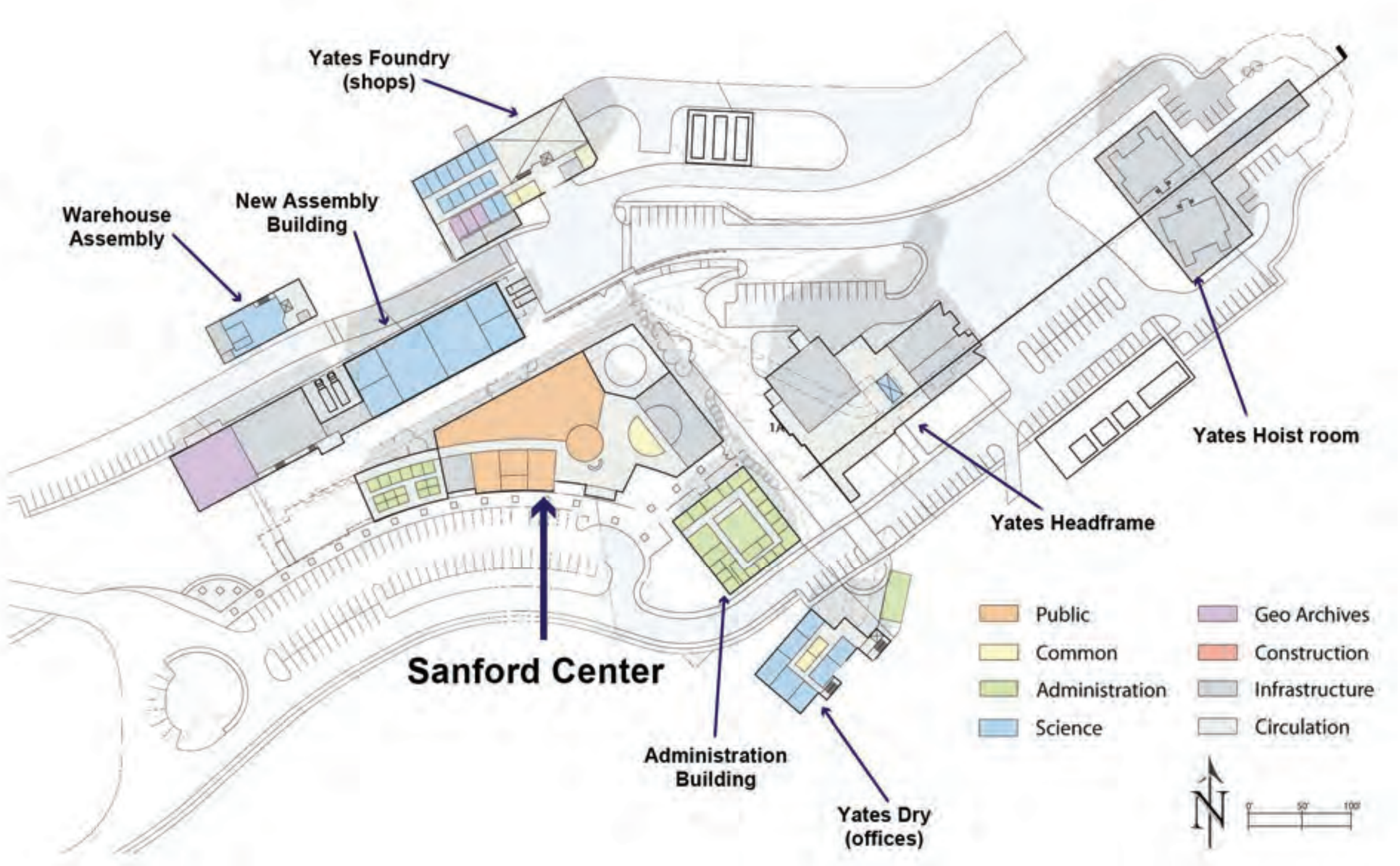

**Figure 4.2.4-2** Buildings, including the SCSE, in the Yates Campus. [Adapted from HDR]

### 4.2.4.1 Scope Options

Scope options defined and prioritized for the SCSE Facility include:

- **Lecture hall.** 3,300 sq. ft.—a 275-seat lecture hall would be valuable for serving large public audiences as well as for scientific conferences, complementing the 100-seat theater. The lecture hall would add capability to host scientific meetings, conferences, and larger public events. Possible locations include the west side of the SCSE shown in Figure 4.2.4-1 or the west side of the New Assembly Building shown in Figure 4.2.4-2.
- **New entrance improvements.** Access to the Yates Campus along Ellison Road and overflow parking. Described further in Section 5.2.5.3.
- **Additional exhibit space.** Extending the interior space to 10,000 net sq. ft.
- **Materials Center.** 1,000 sq. ft.—support and storage space for educational materials

### 4.2.5 Construction Schedule

The Surface Campus has been developed to a conceptual level with a target opening for the SCSE during the summer of 2018. Further Surface Campus design work is planned for 2011. A refined construction schedule will be generated as part of that design package.

### 4.2.6 Underground Access

Based on input from a wide array of key stakeholders and project reviewers, a program requirement has been established for DUSEL to provide access to the science deployments at the 4850L for a limited number of educators, students, and the general public. Partially in support of this requirement, the decision was made to locate the SCSE at the Yates Campus, in close proximity to the Yates Headframe. These requirements are laid out in the *Underground Experience as Education Requirement*, Appendix 4.H.

The Project is planning safe access to the underground for groups of 15 to 20 escorted visitors at any one time. The Project has also been working with science groups to consider visitor access when designing their laboratory spaces. The minimum age, number of escorts, and regularity of tours are yet to be established. An orientation/gathering space near the Yates Shaft at the 4850L is required and has been included within the Preliminary Design.

## 4.3 Baseline Education Program

The education and public outreach program development will be funded outside the DUSEL Major Research Equipment and Facilities Construction (MREFC) award, which can only be applied to construction activities. Nonetheless, the physical requirements of the SCSE Facility are driven by the envisioned program. Full-scale development of the baseline education program, including continued definition, prioritization, prototyping, evaluation, and refinement, is under way, primarily with outside funding, and is scheduled to continue through Final Design and Construction phases. The development of a robust suite of education programs by the opening of the SCSE in 2018 is funded by a complementary NSF award (PHY-0970160) for the period 2010-2015.

### 4.3.1 SCSE Content Areas

The research agenda described in Volume 3, *Science and Engineering Research Program*—and the major science and engineering topics addressed in that agenda—provide compelling content for the SCSE. The four main content areas, each with a complementary roster of fundamental questions that drive the research, include the Underground Universe (physics/astrophysics), Dark Life (microbiology), the Restless Earth (geosciences), and Ground Truth (engineering).[1] The unique Black Hills ecosystem and the importance of preserving it suggest that the interplay of chemical and environmental studies can also provide rich educational opportunities.

Facilitated by the firm of David Heil & Associates, Inc. (DHA), initial content development efforts for the SCSE have drawn upon local, regional, and national experts in science, education, and public outreach. DHA's *Content Development Report*, included as Appendix 4.I, is a preliminary look at topics, audiences, and delivery mechanisms for the SCSE. The SCSE has tremendous potential to convey the nature of scientific and engineering processes and how a community of scientists and engineers engages in these processes. These more general themes correspond well with increasing global interest in Science, Technology, Engineering, and Mathematics (STEM)-related educational goals and the development of workforce skills for the 21st century.

### 4.3.2 Programs by Audience Sector

The Project is focusing on four key audience sectors: educators, students, scientists and engineers, and the general public. Each will benefit from tailored content development and delivery based on industry norms and specific audience interests, needs, and expectations identified through additional targeted market research, which is currently under way. The Project is also considering special strategies to engage and support American Indian audiences throughout the region. Tables 4.3.2-1 through 4.3.2-5 offer sample programs by audience sector, categorized by level of engagement.

**Educators.** DUSEL's Education and Public Outreach Program will provide STEM-appropriate content, rich professional development, and research experiences for K-12 teachers; for higher education faculty, including tribal college and technical institute educators; and those who work outside formal education settings in places such as science museums. The content for the educator community parallels that associated with the students they serve. The education offerings will be well aligned to local, state, and national education standards.

| Level of Engagement | Impact | Delivery |
|---|---|---|
| **Casual** | National/International | Web sites, webcasts, virtual tours, and simulations |
| **Deeper** | Regional/National | On-site and off-site (regional and national professional meetings) workshops and graduate-level coursework combined with materials to take back to the classroom; on-site workshops and courses may include opportunity for educators to go underground |
| **Committed** | National | Research experiences, underground experiences |

**Table 4.3.2-1** Strategies for educator engagement.

**Students.** DUSEL's Education and Public Outreach Program will reach and serve students from kindergarten through graduate school, helping to inspire and cultivate the next generation of scientists, engineers, and science educators. From the student audiences will develop future scientists and engineers who solve problems and fuel the economies of South Dakota, the nation, and the world. Broad student engagement also serves to build appreciation and understanding of science and engineering across the citizenry.

| Level of Engagement | Impact | Delivery |
|---|---|---|
| **Casual** | Regional/National/International | School group visits, perhaps overnight, to the SCSE, including hands-on science activities and use of exhibits; other programs include virtual tours and Internet resources |
| **Deeper** | Regional/National | On-site summer workshops and field camps for secondary school students and undergraduates, including underground experience; weekend science events |
| **Committed** | Regional/National | Research internships for undergraduates and high school students |

**Table 4.3.2-2** Strategies for student engagement.

**Scientists and Engineers.** DUSEL users, scientists, and engineers will serve DUSEL's Education and Public Outreach Program by helping to educate other audiences, but they, themselves, are a distinct and important audience to be served. DUSEL's Education and Public Outreach Program will help foster a sense of community throughout the DUSEL research collaborations, providing cultural and intellectual enrichment across disciplines. DUSEL's Education and Public Outreach Program will offer guidance, especially to graduate students and postdoctoral fellows, on how to communicate science effectively to school and public audiences and will share educational research and best practices. Scientists and engineers will develop funding proposals of their own that will benefit by partnering with DUSEL's Education and Public Outreach Program for broader impact and outreach. DUSEL's Education and Public Outreach Program can facilitate outreach opportunities for the science groups and ensure that their educational messages and materials are appropriate for their intended audiences and are designed, implemented, and evaluated to ensure efficacy and fidelity in the field.

| Level of Engagement | Impact | Delivery |
|---|---|---|
| **Casual** | National/International | Web site |
| **Deeper** | National/International | Sophisticated presentation materials/videos about DUSEL and its science; consultation and interfacing about broader impact possibilities |
| **Committed** | National/International | Mentorship/networking programs for postdocs and graduate students stationed on site; collaborative work on specific programs, guidance on best practices in education and evaluation of program impact |

**Table 4.3.2-3** Strategies for scientist and engineer engagement.

**General Public.** The general public represents DUSEL's largest audience segment. Local residents and tourists will come to visit the SCSE and experience DUSEL science firsthand. Outreach programs will extend DUSEL's Education and Public Outreach Program's impact regionally and, by way of the Internet, globally. DUSEL's Education and Public Outreach Program has an opportunity to introduce tens to hundreds of thousands of adults, families, and seniors each year to compelling science and engineering through memorable on- and off-site experiences. DUSEL's Education and Public Outreach Program will serve as a unique added destination amid an already well-established tourist region, becoming a centerpiece for science learning. While high throughput is often the focus when tourist audiences are considered, in the case of the SCSE and the DUSEL Education and Public Outreach Programs, success will be measured in terms of educational impact.

| Level of Engagement | Impact | Delivery |
|---|---|---|
| **Casual** | Regional/National | On-site interpretation, off-site interpretation at partner institutions, Web sites, virtual tours and simulations, webcasts |
| **Deeper** | Regional/National | Multi-day workshops for adults, including authentic science experience, marketed nationally (e.g., Elderhostel model), perhaps in partnership with other Black Hills institutions, and possible underground experience |
| **Committed** | National | Extended workshops for adults, including authentic science experience and underground experience |

**Table 4.3.2-4** Strategies for public engagement.

### Preliminary Strategies for Reaching Previously Underserved Audiences

The Project is deeply committed to partnering with and serving the needs of communities and audiences historically underrepresented within science and engineering disciplines. This will be especially important for rural and American Indian students and families throughout the region. The Content Development Report—informed by regional science education experts and tribal members—identifies several strategies for building relationships with American Indian communities and for continued engagement of leadership from these communities moving forward. Building trust and relationships among leaders of these communities requires significant investment of time and effort. Intensive involvement of the Project's Cultural Liaison and guidance from the Cultural Advisory Committee are yielding strong benefits. Concerted effort is under way and increasing.

| Level of Engagement | Impact | Delivery |
|---|---|---|
| **Casual** | Regional/National | Off-site engagement of elementary-middle school students, teachers, and parents in rural communities, including reservations; presentations and presence at regional and national meetings related to American Indian education, women in science, and other areas of equity |
| **Deeper** | State/Regional | Multi-day workshops and laboratory experiences for American Indian high school students, conducted in partnership with existing state and federal initiatives; weekend workshops on site for students from reservations across the region |
| **Committed** | National | Research internships |

**Table 4.3.2-5** Strategies to engage and support underserved audiences.

### 4.3.3 Interactive Exhibits

Educational activities on the exhibit floor will be highly engaging and interactive. A few simple prototypes have already been developed and deployed through an annual science festival at Sanford Laboratory that attracts more than 600 visitors on one Saturday each July. World-class scientists, master teachers, and summer interns staff a variety of stations, inviting visitors of all ages to explore sub-surface life forms at microscope stations, to detect the presence and to gain understanding of cosmic radiation using a diffusion cloud chamber, and to develop scientific models of particle interactions using computer simulations (to name just a few). More sophisticated and polished successor activities and exhibits will be developed and new ideas will continue to flow from the creativity of the laboratory and educational community.

The exhibit floor may include a few museum pieces, scientific artifacts, and storyboards around its perimeter, but its heart will be a vibrant instructional environment where learners test and build scientific ideas for themselves, engineer solutions to laboratory-related challenges, and engage with leaders across the scientific, engineering, and education domains.

### 4.3.4 Underground Educational Experience

As DUSEL is a national underground science and engineering research facility, it is imperative that select audiences and laboratory guests have access to underground research activity. One of the most compelling components of the laboratory educationally and what will set it apart from all other human endeavors is the depth and magnitude of the enterprise. Much of this can be conveyed remotely and virtually, and the vast majority of DUSEL's Education and Public Outreach Program audience members will only experience the underground from a distance, but it is expected that a few thousand visitors per year will physically visit experimental activity underground at the Mid-Level Laboratory Campus at 4850 feet below ground.

Part of the educational experience in visiting the underground will involve safety training and the use of appropriate safety equipment. Topics of educational interest include the measures required for safe access and extended occupancy, seeing and engaging with scientists within the underground environment, and perhaps collecting one's own data, such as cosmic radiation flux at depth. Operationally, it will be important that time spent underground by education audiences be kept brief and that visits be structured and coordinated to minimize interference with the science and to maximize safety.

An underground experience at shallow to intermediate depth is also possible, though not proposed for funding within the MREFC-funded construction project. A shallow to intermediate depth experience could accommodate many more visitors per year. Such an area could be developed for exclusive or predominantly educational use.

### 4.3.5 Digital / Virtual Presence

Underground sensors and webcams will provide continuous access to underground environments and control rooms. Students using data from DUSEL's experiments, from among the many relevant disciplines, from thousands of miles away or right on site will be able to design their own inquiries under the mentorship of a scientist or engineer who himself or herself may be on site or thousands of miles away.

A key component of the SCSE digital presence is an effort titled Virtual DUSEL (vDUSEL). vDUSEL is being developed for the SCSE through a partnership with Dakota State University (DSU). The vDUSEL mission is to share the wonders of the deep underground science being conducted at DUSEL with those online and to excite and engage them in the active learning of science. vDUSEL will have several core roles. First, it will act as a virtual science center to connect online visitors to the physical SCSE. It is important that a strong relationship exist between vDUSEL and DUSEL's Education and Public Outreach Program, positioning the virtual environment as a catalyst to excite distant audiences to visit the SCSE in person to see and learn firsthand. Second, vDUSEL will contribute immersive environments within the interpretive center for visitors who cannot directly experience the underground and will provide follow-up activities when they return home. These roles will be integrated into the Preliminary Design of the exhibit space. Third, vDUSEL will deliver innovative online educational content in support of DUSEL's programs. Through computer simulation, visitors will be able to explore the entire laboratory, travel inside detectors, watch simulated particle trajectories, experiment with particle interactions, and change experimental conditions. Planning funds are in place to integrate vDUSEL into DUSEL's overall Education and Public Outreach Program.

### 4.3.6 Relationship of Baseline Educational Program to Facility Requirements

**On Site.** The SCSE facility components listed in Section 4.2.4 will enable and host an exciting education and outreach program. It is expected that the exhibit area will be used primarily for the general public in the tourist season and for classroom visits and local residents during the remainder of the year. Research internships for both students and teachers will be organized in partnership with the science collaborations and will draw primarily upon space in the scientific areas of the laboratory. The SCSE will provide space for seminars, classes, and social interactions, utilizing the theater, classrooms, and lobby area.

**Off Site and Virtual.** Support of these programs within the SCSE facility primarily requires storage, work space, and office space, as well as a server room. Storage space allocated within the current facility design, in combination with cold storage elsewhere on the DUSEL site and nearby, is considered adequate.

## 4.4 Operations and Preliminary Business Plan

A full business plan for DUSEL's Education and Public Outreach Program, now under development, is scheduled for completion in 2012. What follows are highlights from a preliminary business plan completed in 2009, informed by an initial market analysis and a project development plan, both commissioned in the fall of 2009 and completed in the spring of 2010. All three planning documents have been reviewed by project management and accepted by the Education Governing Board. The organizational concepts and business elements described herein will be refined through targeted market research that is currently under way and in the development of the full business plan.

Below is a preliminary look at each of the following components: governance/organizational structure, attendance projections, operations plan, staffing, budget, and timeline. The institutional profile of the SCSE includes

- Funding for facility construction as described in Volumes 2 and 5 with a contribution from T. Denny Sanford
- Projected annual attendance of 64,500
- Paid staffing of ~17 FTE (funded through multiple sources)
- Starting endowment of $15 million, held by the SDSTA, also donated by Mr. Sanford
- Target opening date in 2018

### 4.4.1 Governance and Organizational Structure

Figure 4.4.1 shows how the Education and Public Outreach department is envisioned to fit within the overall DUSEL organization. Key to this structure is that the SCSE will be a governed and funded entity within the laboratory, but also retain the ability to raise funds from private sources. The Education and Outreach Director will be an employee of DUSEL and report to the DUSEL Directorate. There will be an affiliated but independent Foundation with its own Board of Governors. The Education and Outreach Director will apprise the Board of Governors of SCSE activities and submit budgetary requests. This model will facilitate the blending of public and private funding streams—for construction, for basic operations, and for ongoing development and implementation of programs and exhibits.

The Education Advisory Committee (EAC) will provide guidance to both the DUSEL Central Project Directorate and the SCSE Board of Governors. The EAC is already established and has guided the Project since the spring of 2008. The EAC's charter and current composition, including local, regional, national, and international experts in science and education, are in Appendix 4.B.

A second key advisory body is DUSEL's Cultural Advisory Committee (CAC). This committee was established to build the Project's cultural understanding, foster development of relationships and partnerships throughout the region, and guide the establishment of a diverse project workforce. The CAC will continue advising the DUSEL Central Project Directorate and helping to shape education and outreach efforts through the rest of Facility design and beyond.

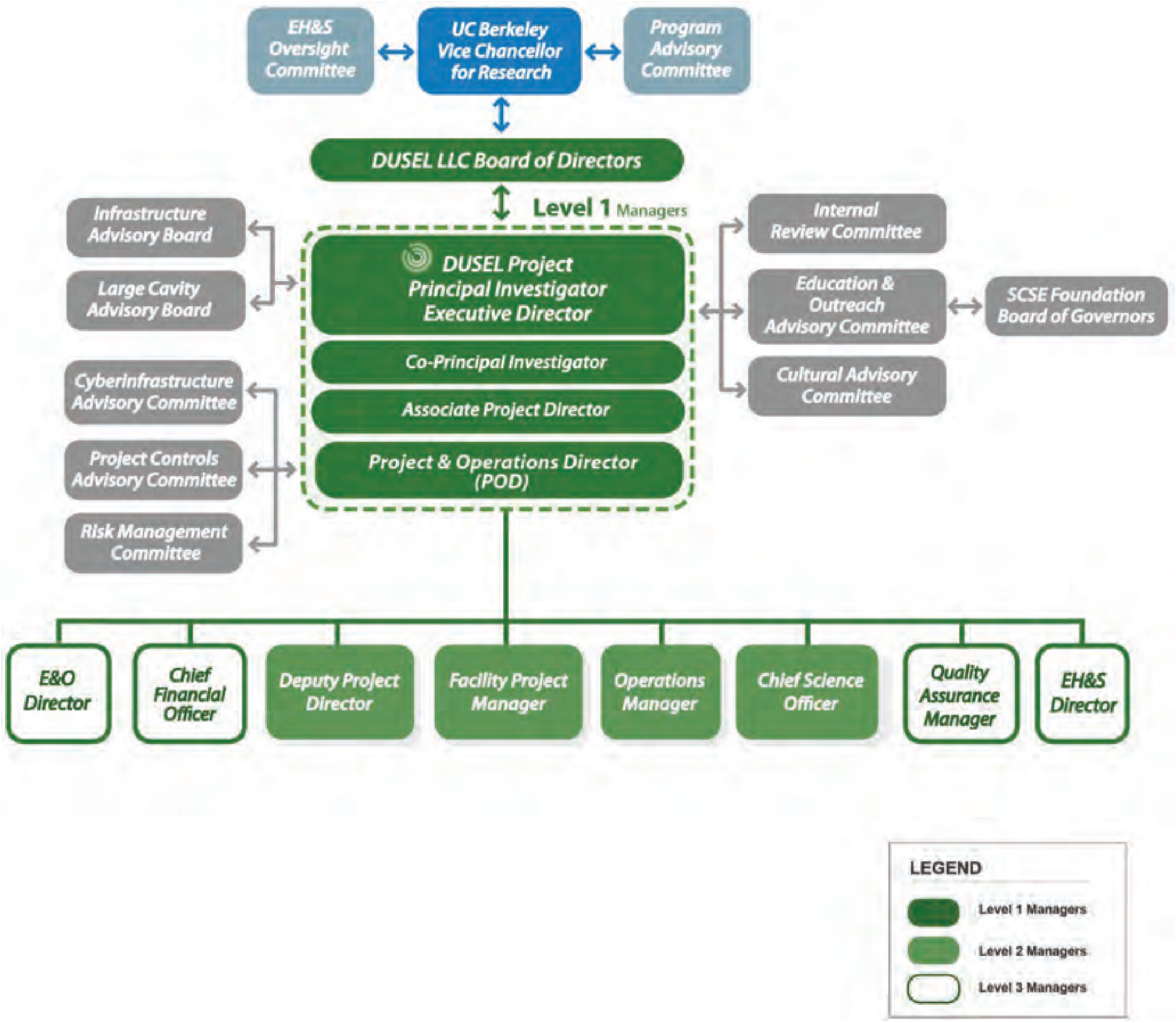

**Figure 4.4.1** DUSEL LLC organizational structure with emphasis on education and outreach. [DKA]

An Education Governing Board (EGB) has been established in advance of the opening of the SCSE to govern the combined fiscal and human resources and the diverse interests of a wide array of stakeholders, including the DUSEL Project, the SDSTA, the state of South Dakota, other partner organizations, and major donors. The charter and composition of the EGB are in Appendix 4.A. Once the SCSE has been formally established, the EGB's functions will be subsumed by DUSEL's governance structure, the DUSEL Central Project Directorate, and the SCSE Foundation Board of Governors.

### 4.4.2 Market Analysis and Attendance

In the fall of 2009, the EGB commissioned a preliminary market survey for the SCSE that was prepared by DHA (funded through Black Hills State University). The resulting *Market Assessment & Analysis Report*—included in Appendix 4.E—provides attendance projections and industry analysis for the SCSE. The projections and analysis are grounded in an understanding of the broader context for the SCSE, including:

- Current plans for the facility and programming
- Local, regional, and state demographics
- National and international trends in science, technology, engineering, and mathematics education

### 4.4.2.1 Projected Attendance

Annual attendance projections for the SCSE, shown in Table 4.4.2.1, were produced by DHA using a regression model developed to account for exhibit-space area, measures of regional demographics, population density, and tourism rates. An input estimate of 10,000 square feet of exhibit space was used for the purpose of the model. While the Conceptual Design of the SCSE Facility includes 7,500 square feet of indoor exhibit space, 3,000 square feet of outdoor exhibit space are envisioned to supplement in summer, which will be the time of highest demand.

| Conservative | Projected | Possible |
|---|---|---|
| 41,500 | 64,500 | 96,750 |

**Table 4.4.2.1** Annual attendance estimates for the SCSE.

The “Conservative,” “Projected,” and “Possible” attendance estimates offer different scenarios for visitation patterns for the SCSE. While the “Conservative” estimate may be achievable with limited marketing and programming, the “Possible” estimate would likely require extensive marketing efforts and strategically designed programming.

### 4.4.2.2 Seasonal Trends

Table 4.4.2.2 provides monthly attendance estimates. These were developed using five years of data for six cultural attractions in the Black Hills, Badlands, and Lakes region of South Dakota. These rates reflect a strong seasonal effect in attendance, with the months of May through September accounting for 85% of the total annual attendance and with 46% of total annual attendance expected to be concentrated in June and July.

| Month | Rate | Conservative (41,500 annual) | Projected (64,500 annual) | Possible (96,750 annual) |
|---|---|---|---|---|
| **January** | 1% | 359 | 558 | 837 |
| **February** | 1% | 569 | 884 | 1,326 |
| **March** | 2% | 902 | 1,402 | 2,103 |
| **April** | 3% | 1,178 | 1,831 | 2,746 |
| **May** | 11% | 4,568 | 7,099 | 10,649 |
| **June** | 23% | 9,558 | 14,856 | 22,284 |
| **July** | 23% | 9,382 | 14,582 | 21,873 |
| **August** | 15% | 6,326 | 9,832 | 14,749 |
| **September** | 14% | 5,604 | 8,710 | 13,065 |
| **October** | 5% | 2,065 | 3,210 | 4,815 |
| **November** | 1% | 493 | 766 | 1,149 |
| **December** | 1% | 496 | 771 | 1,156 |

**Table 4.4.2.2** Seasonal attendance estimates for the SCSE.

The seasonal trend in visitor attendance has important implications for both program and facility planning. The SCSE will need to shift the focus of programming efforts throughout the year to accommodate the variations in on-site attendance. For example, programmatic efforts may shift between outreach/off-site efforts during the winter months and on-site services during the summer months, a pattern that will have important ramifications for SCSE staffing. Seasonal attendance projections have been factored into an industry standard "design day" approach[2] to consider facility square footage. Using the "Possible" projections, the SCSE Facility would need to accommodate up to 1,000+ visitors on a given day at peak times, while anticipating 45 visitors on a Saturday during the winter season. This wide variation highlights the need for multipurpose spaces within the Facility that can accommodate shifts in visitor volume, programming, and service needs throughout the year.

### 4.4.2.3 Visitor Segmentation

The three major visitor segments that SCSE will serve on site include school groups, local and regional audiences, and tourist audiences. The geographic location of the SCSE is important in determining both the potential size and proportion of these visitor segments. The majority of school groups will travel no more than two hours for a one-day field-trip; local and regional audiences making a day trip to the SCSE are most likely to reside within one to two hours of the center, with a greater proportion of repeat visitors coming from within a one-hour drive. The SCSE's location within western South Dakota makes proximity to major tourist attractions, such as Mount Rushmore, an important factor in determining tourist visitation.

Most informal science facilities rely on school groups to provide the bulk of the visitor numbers during the academic year; on average, 20% of total attendance numbers are due to this audience sector. DHA has used that figure in determining the attendance numbers of Table 4.4.2.1. However, 20% of the projected attendance of 64,500 equates to a penetration rate of 36% and 49% for school-age youth within a 100- and 50-mile radius, respectively, for all grades. Reaching these high percentages of student populations in the region will require specialized programming and marketing. These numbers will be further refined throughout 2011 and into 2012, informed by further target market engagement of local and regional schools.

Local and regional demographics play an important role in determining market penetration for science centers and museums. Despite ongoing efforts among museums to engage underserved audiences, the majority of museum visitors tend to be well-educated and affluent groups with discretionary income to spend on educational experiences. Although DUSEL's Education and Public Outreach Program is determined to reach underserved audiences through effective programming and outreach efforts, for the purposes of the initial market analysis, DHA considered education and income level to indicate a conservative value of 2% penetration within the local/regional populace in a 100-mile radius, corresponding to 7% of the projected attendance figure.

Tourists are expected to make up the remaining 73% of visitor numbers, for a total annual attendance of 29,000-66,000. The midpoint of this range, 47,000, is 105% of the current annual attendance at the Homestake Visitor Center in Lead, South Dakota. To increase this number, the SCSE will need to employ vibrant marketing strategies and partnerships to draw tourists to Lead. Partnerships with regional institutions such as the Journey Museum in Rapid City, South Dakota; Mount Rushmore National Memorial; and Crazy Horse Memorial are likely. Results from the 2007 Mount Rushmore Visitor Survey, which indicated that 48% of visitors to Mount Rushmore visited Deadwood (10 minutes from the Homestake site) on their trip to the Black Hills area, suggest that local partnerships with Deadwood-based institutions, such as the Adams Museum, are of comparable importance to partnerships with attractions having more national draw.

### 4.4.3 Operations Plan

Tentative plans envision the SCSE to be open seven days per week in summer and six days per week the rest of the year, with hours to vary seasonally. Students or groups attending overnight camps will be supervised by SCSE staff and designated guardians at all times, and hours of operation will be extended for these activities. Further discussion on the operations of the SCSE can be found in Volume 10, *Operations Plans*.

SCSE Operations will be staffed with full-time and part-time employees. Many positions that serve tourist audiences may be staffed with volunteers. Association of Science and Technology Centers (ASTC) statistics show that a science center of the envisioned size generally needs 15-20 full-time staff for operation and upward of 75 volunteers.

An SCSE Operations Manager will be hired to manage the day-to-day activities and interface with the DUSEL Operations department in the areas of purchasing and logistics, facility services, visitor services, information technology and exhibit animation, and accounting and finance.

Elements considered for operation of the SCSE:

- Science Visitor Center Exhibit Hall
- Outdoor exhibits/nature trail/grounds
- Theater (and possible lecture hall)
- Computer animation room and equipment
- Classrooms
- Ticketing
- Café
- Parking area/structure or shelter for queuing
- Trolley/transport from remote parking areas
- Maintenance space and equipment
- Office space and equipment
- Rest areas/restrooms
- Storage
- Orientation space and protective equipment storage for surface and underground tours

#### 4.4.3.1 Staffing Plan

A preliminary staffing plan for the Education and Public Outreach department is shown in Figure 4.4.3.1. It is expected that with DUSEL-specific R&RA funding, investment by the SCSE Foundation, and SCSE-earned income, approximately 17 full-time employees will be supported as of facility opening, anticipated in 2018. Other staff will be added as external funding is obtained for specific programs and/or exhibit development. The staffing plan will be refined as part of the full business planning exercise in 2011-2012. Further discussion on staffing levels for the Project are described in Volume 2, *Cost, Schedule, and Staffing*, and Volume 10, *Operations Plans*.

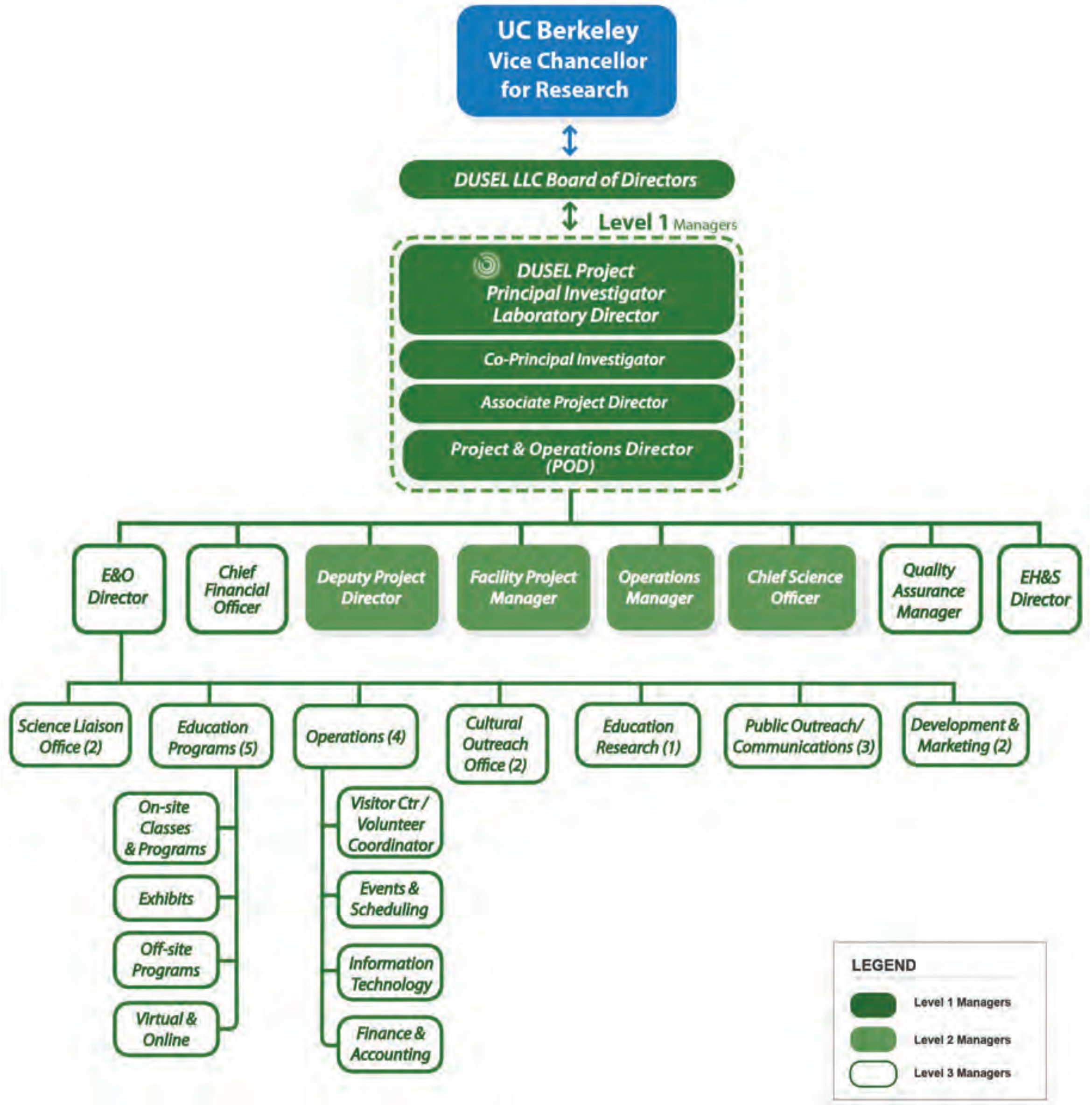

**Figure 4.4.3.1** DUSEL LLC organization chart showing preliminary E&O staffing, across all funding sources. Numbers in parentheses indicate FTE within each function. Education and Public Outreach staff will complement and draw heavily on centralized DUSEL staff performing similar functions. [DKA]

**Science Liaison Office.** The Science Liaison Office will be staffed by trained scientists tasked with building and maintaining close relationships between the DUSEL Education and Public Outreach Programs and the scientific collaborations working at DUSEL. The science liaison staff members will work closely with the User Support Office, described in Volume 10, and support the scientific collaborations within the education domain and will recruit from the scientific collaborations to support SCSE programs. This office will also support summer internships and research experiences for educators.

**Cultural Outreach Office.** The Cultural Outreach Office will pay special attention to building partnerships and providing educational services to audiences historically underrepresented in science and engineering disciplines. This office will perform laboratory-wide functions in addition to serving as a key component of DUSEL’s Education and Public Outreach department.

**Education Programs.** This office requires the largest staff support. Personnel will have expertise across the full science education spectrum, including program development, implementation, evaluation, exhibit design, computer simulations, K-12 curriculum, teacher training, and university-level faculty development. The base education program staff will be supplemented through funding from outside grants. Coordination with the vDUSEL effort, under development through collaboration with Dakota State University (DSU), will reside within this office, which will also work closely with the Science Liaison, Cultural Outreach, Public Outreach and Communications offices.

**Public Outreach and Communications.** Public Outreach and Communications will engage and educate the general public through the coordination of public lectures and programs, DUSEL's Web site, and scientific press releases. This office will perform DUSEL-wide functions in addition to serving as a key component of DUSEL's Education and Public Outreach Program team. The staff described in this area are budgeted through other DUSEL departments as described in Volumes 2 and 10.

**Education Research.** One Ph.D.-level science education researcher will support the Education Programs Office in designing impact studies, analyzing impact data, interfacing with external evaluators, staying current, and publishing within the science education literature.

**Operations.** This department will interface with overall DUSEL Operations, especially in the areas of Business Services including finance and administration; Environment, Health, and Safety; Information Technology, and Facility Operations services. The SCSE Operations Office will also handle recruitment, training, and management of volunteers and the coordination of special events.

**Development and Marketing.** This fundraising arm of the DUSEL's education and public outreach efforts at the SCSE will provide grant-writing support, donor cultivation, and membership services. All staff and activities within this department will be supported exclusively with nonfederal funds.

#### 4.4.3.2 Operations Budget

Budgeting analysis and resulting figures will be refined as the full business plan is developed during the Final Design phase and as discussions with funding agencies and representatives of T. Denny Sanford progress. The preliminary operations budget estimates for MREFC-funded items are included in Volume 2, *Cost, Schedule, and Staffing*, and corresponding staffing descriptions are presented in Volume 10. Additional funding will be required to support the full operations of the SCSE beyond the MREFC and R&RA budget levels described in Volume 2. Detailed budget information will be included in full business plan.

#### 4.4.3.3 Fundraising

In the process of being marketed to various constituents, the SCSE will be branded with the unique features of its mission, vision, and place within DUSEL.

The SCSE will have a dedicated staff for public relations and marketing. In addition, the SCSE will have a dedicated development officer to work with private donors and foundations. The SCSE Foundation will provide funding for these personnel.

### 4.4.4 Program Budget and Timeline: Final Design to Opening Day (FY 2012-FY 2018)

The budget estimate to develop, construct, and outfit the SCSE and to conduct pre-operations for the period beginning in fall 2011 through a projected opening day in 2018 is presented in Volume 2, *Cost, Schedule, and Staffing*. The majority of required funding is committed or already in hand, thanks in large part to the generous pledge of $20 million from philanthropist T. Denny Sanford. Exhibit design and fabrication will require a separate funding source to complete.

From FY 2012 through FY 2018, the SCSE requires development within the areas of Facility, Program, and Institutional Support. A high-level development schedule appears in Table 4.4.4. The Facility design will continue as part of the overall DUSEL design effort. Program and Institutional development will be accomplished with complementary funding that was recently awarded by NSF to support those specific domains. All the while, the Project will be building education and outreach capacity, prototyping programs and exhibits, and pursuing additional funding.

As of FY 2011, the Project has dedicated staffing for education and public outreach plus collaborative support from many other staff members across the Project. From FY 2011 through FY 2018, dedicated staffing levels need to ramp up from the current level of four to 17 to support an operating SCSE and associated programs. This ramp-up in staffing, as included in the project budget is described in Volume 2.

| Design / Construction / Institutional Development | Target Date / Span |
|---|---|
| **Target Market Research** | 2010-2011 |
| **Completion of Preliminary Facility Design** | 2011 |
| **Full Business Plan** | 2012 |
| **Conceptual Design of Exhibit Space** | 2012 |
| **Preliminary Exhibit Design** | 2013 |
| **Final Facility Design** | 2012-2013 |
| **Final Exhibit Design** | 2013-2016 |
| **Facility Construction** | 2017-2018 |
| **Exhibit Fabrication and Installation** | 2017-2018 |
| **SCSE Facility Opening Day** | 2018 |

**Table 4.4.4** High-level timeline for SCSE development.

## 4.5 Summary and Conclusions

DUSEL's Education and Public Outreach Program has a clear mission and vision coupled with great promise to deliver. Strong partnerships and advisory bodies are in place and a robust planning process has been laid out, is funded, and is under way. A compelling roster of educational programs is under development, with prototypes being implemented, evaluated, and refined using outside fiscal resources. The Conceptual Design of a 27,000-square-foot SCSE Facility will support the envisioned roster of programs well on opening day and will allow for future program growth as additional resources are secured.

The overall program budget for design, pre-operations, construction, and outfitting of the SCSE is appropriate for the magnitude of the DUSEL enterprise, and reflects significant leveraging of DUSEL funding (R&RA and MREFC Awards) with outside resources. Approximately 85% of the overall program budget that is required to get to opening day in 2018 is in-hand or identified, with only 15% remaining to be obtained externally.

A full business plan is under development with outside resources that will ensure sustainable operation of the SCSE. As with the preconstruction and construction budget, the envisioned operations budget includes a balance of funding sources, with over 50% due to come from non-NSF sources, including interest generated from the SCSE endowment already secured.

## Volume 4 References

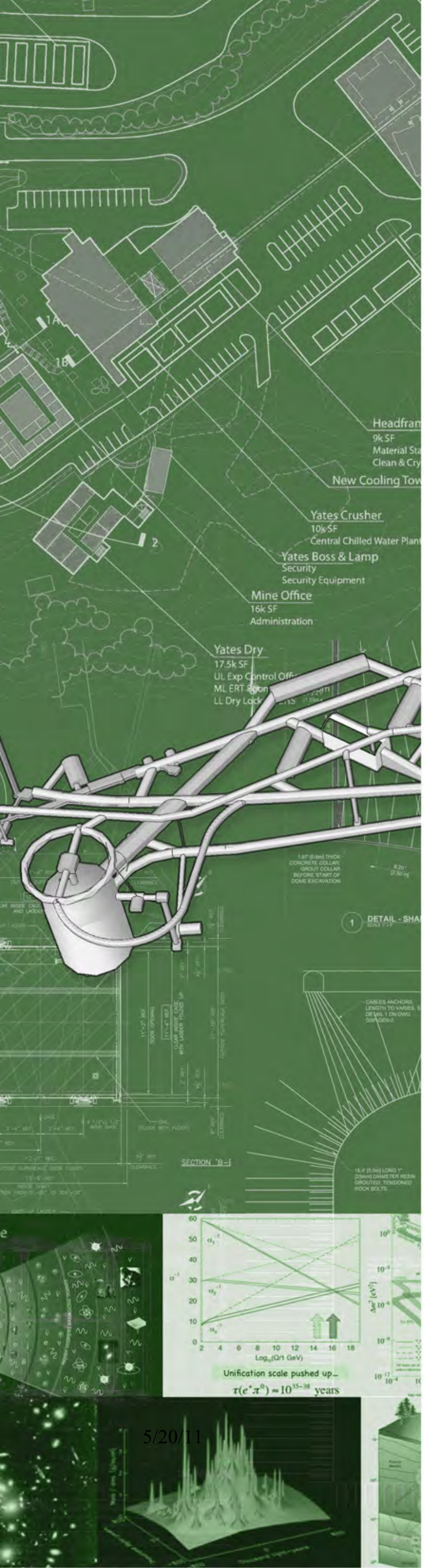

# Preliminary Design Report

May 2011

## Volume 5:
## Facility Preliminary Design

DUSEL
Deep Underground
Science and
Engineering Laboratory

This page intentionally left blank

# Facility Preliminary Design

## Volume 5

### 5.1 Facility Design Overview

This Chapter outlines the DUSEL Facility Preliminary Design and the current site conditions as assessed during the Preliminary Design phase. The approach used to develop the Preliminary Design, the facility requirements and interfaces, and applicable codes and standards that drive the Preliminary Design presented herein are included to provide a context for the design. The Facility design described in this Preliminary Design Report (PDR) is responsive to the scientific goals established by the Project and to the allocated construction funding targets provided by the National Science Foundation (NSF), which represents a maturity level of 30% of "construction ready" documents. Finally, the Volume outlines DUSEL plans for the Final Design, Bidding, and Construction phases, including construction sequencing and acquisition plans. An overview of the Facility design contract scopes and their relationship to the overall Facility design is provided in Appendix 5.A, which should be downloaded, when possible, and used for reference and orientation while reading Volume 5. This overview is intended to provide a cross reference between the design contract deliverables and the major systems that constitute the design.

#### 5.1.1 Design Context

The primary goal of the DUSEL Facility development effort is to provide a facility where science and engineering research are the sole facility uses. This program develops both the facilities for scientific use as well as the support facilities required for science research and daily laboratory operations and maintenance activities. The Facility will accommodate science and engineering research programs from multiple disciplines in both surface facilities and underground facilities for selected experiments, many of which require shielding from surface and cosmic sources with up to 6,000 meters water equivalent (mwe) depth. The *Science and Engineering Research Program* and its requirements are discussed in Volume 3. It presents the science goals, collaborations and techniques, facility requirements, and experiment development plans and discusses the relationship between the facility and the science collaborations. The DUSEL Facility plans also address requirements in support of the *Education and Public Outreach* program discussed in Volume 4.

The DUSEL Facility design delivers a facility and supporting infrastructure plan that will enable expansion to address future science experiments and engineering requirements. At the Facility 60% Preliminary Design development milestone in July 2010, construction cost targets required the review of and in some cases the removal of growth capacity within design subsystems. Some of these items have been prioritized as scope options and are presented in Chapter 5.10. The design presented in this chapter does support the DUSEL science program and facility requirements baseline established during the PDR process. Furthermore, the DUSEL Facility supports DUSEL requirements for cost-effective facility operations and maintenance for a nominal life expectancy of at least 30 years of operations.

While this Volume focuses primarily on facility construction that will be funded under the DUSEL Major Research Equipment and Facility Construction (MREFC) account, the facility designs also include aspects that will be implemented under operations and maintenance activities and funded with NSF

Research and Related Activities (R&RA) funding. These R&RA-funded items are focused on rehabilitation maintenance items to provide safe access to the Facility and to reduce or mitigate risk. Throughout Volume 5 where activities will be performed using R&RA funding, those items are specifically noted, as they are not a part of the MREFC-funded construction baseline. The implementation of these R&RA-funded items is also addressed in Volume 10, *Operations Plans*. Although implemented through operations and maintenance activities, these elements of the design were developed alongside the MREFC-funded elements to ensure that their implementation was cohesive and resulted in a well-integrated DUSEL Facility—regardless of the funding source.

#### 5.1.1.1 Requirements Basis for DUSEL Facility Development

The DUSEL Facility design has included detailed input gained through close coordination with all components of the DUSEL Project and the DUSEL science collaborations. The facility requirements and interface requirements developed during the Preliminary Design phase are the result of detailed discussions with representatives of the scientific program (Volume 3), laboratory operational planning (Volume 10), education and public outreach (Volume 4), and environment, health, and safety (EH&S) (Volume 6) and defined through processes put in place and documented by systems engineering (Volume 9) and project controls (Volumes 7 and 8). The detailed discussion on the requirements generation process with links to the requirements documents is located in Volume 9, *Systems Engineering*.

The facility requirements were provided to the architecture and engineering teams under contract with DUSEL for the Preliminary Design development to ensure the Project's requirements were formally addressed in the resulting Preliminary Designs. A formal requirements compliance matrix was completed by each design contractor and provided to the DUSEL Project to ensure each designer addressed the requirements in the design. These compliance matrices are included in Volume 9, *Systems Engineering*. The DUSEL design was developed using a requirements set that represents the classes of experiments planned for DUSEL.

#### 5.1.1.2 Funding Overview for DUSEL Facility

The budget for the DUSEL MREFC-funded Facility construction is approximately $575 million and is discussed in Volume 2, *Cost, Schedule, and Staffing*.

The Education and Public Outreach Program center, with facility construction partially funded through MREFC funds, is also funded through a generous gift from philanthropist Mr. T. Denny Sanford to support the development and operations of the Sanford Center for Science Education (SCSE). The SCSE is addressed in the Surface Facility design presented in Chapter 5.2.

To meet the overall Facility requirements, the MREFC-funded construction scope includes the development of two underground laboratory modules (LMs) at the 4850L Mid-Level Laboratory (MLL); one LM at the 7400L Deep-Level Laboratory (DLL); underground areas to support biology, engineering, and geology research located outside of the MLL and DLL Campuses; as well as infrastructure systems and Surface Facility to support the underground research initiatives.

As part of the DUSEL Project, but separate from the MREFC-funded construction activities, there exist operations and maintenance-related facility rehabilitation projects to ensure safe access to the DUSEL Facility and reduce Project risk. The Facility design includes the technical details to guide many of the operations activities that are outlined in Volume 10, *Operations Plans*. The design descriptions provided in Volume 5 outline both MREFC- and non-MREFC-funded design elements. Those items addressed outside the MREFC funding profile are specifically noted throughout as having funding provided outside of the MREFC scope through one of the following sources: NSF R&RA, private funding such as support from T. Denny Sanford, and funding from individual science collaborations such as DOE's Long Baseline Neutrino Experiment (LBNE) project.

#### 5.1.1.3 Site Ownership

The South Dakota Science and Technology Authority (SDSTA) owns the site planned for DUSEL construction, the current Sanford Laboratory at Homestake. Barrick Gold Corporation agreed to donate the former Homestake Gold Mine to the state of South Dakota in 2001 for use as an underground scientific laboratory. The SDSTA was created by an act of the South Dakota Legislature in 2004 to accept the property from Homestake, and to own, manage, and develop the site to make it available for scientific research. The SDSTA is a body corporate and politic, meaning part corporation and part governmental entity. As a corporation, it is governed by a Board of Directors (appointed by the Governor) and has its own bylaws. For some purposes (such as taxation) it is considered a governmental entity.

Closely related to facility planning and management, the DUSEL Project is obligated by the provisions of the Property Donation Agreement (PDA), entered into in May 2006 between and among the SDSTA, Barrick, and the state of South Dakota. The PDA provides primarily for the indemnification of Barrick with respect to existing and future liabilities that might result from ownership of the site by the SDSTA. It also requires the SDSTA, and thus DUSEL through its relationship to the SDSTA, to secure and maintain adequate funding and management and technical capabilities to ensure a safe and sustainable operation at the Homestake site, including provisions for proper levels and types of insurance for both operations and construction activities at the Homestake site.

#### 5.1.1.4 Site Context

The SDSTA currently operates and maintains Sanford Laboratory at the Homestake site in Lead, South Dakota. The Sanford Laboratory property comprises 186 acres on the surface and 7,700 acres underground. The Sanford Laboratory Surface Campus includes approximately 253,000 gross square feet (gsf) of existing structures. Using a combination of private funds through T. Denny Sanford, South Dakota Legislature-appropriated funding, and a federal HUD Grant, the SDSTA has made significant progress in stabilizing and rehabilitating the Sanford Laboratory facility to provide for safe access and prepare the site for DUSEL construction. These efforts have included dewatering of the underground facility and mitigating and reducing risks independent of the DUSEL efforts and funding.

The following figures provide a context for the Sanford Laboratory site. Figure 5.1.1.4-1 illustrates Sanford Laboratory's location within the region as a part of the northern Black Hills of South Dakota. Figure 5.1.1.4-2 outlines the Sanford Laboratory site in relationship to the city of Lead, South Dakota, and points out various significant features of Lead: Sanford Laboratory and surrounding property that still remains under the ownership of Barrick. Finally, Figures 5.1.1.4-3 and 5.1.1.4-4 provide perspectives of the Sanford Laboratory Complex from a surface and aerial view of the property and its surroundings. These views illustrate the varied topography found throughout the site.

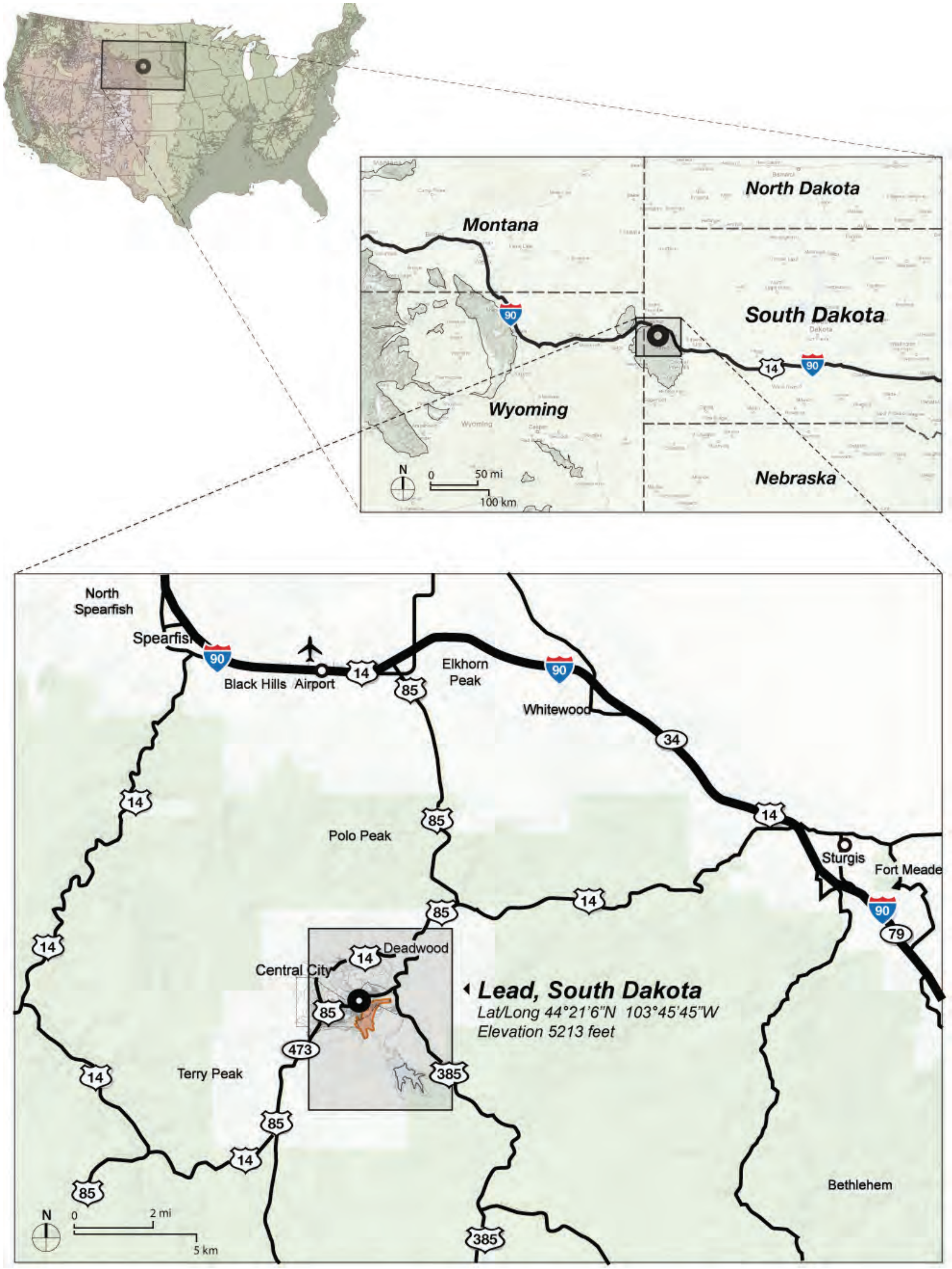

**Figure 5.1.1.4-1** Regional context showing the city of Lead, South Dakota. [DKA]

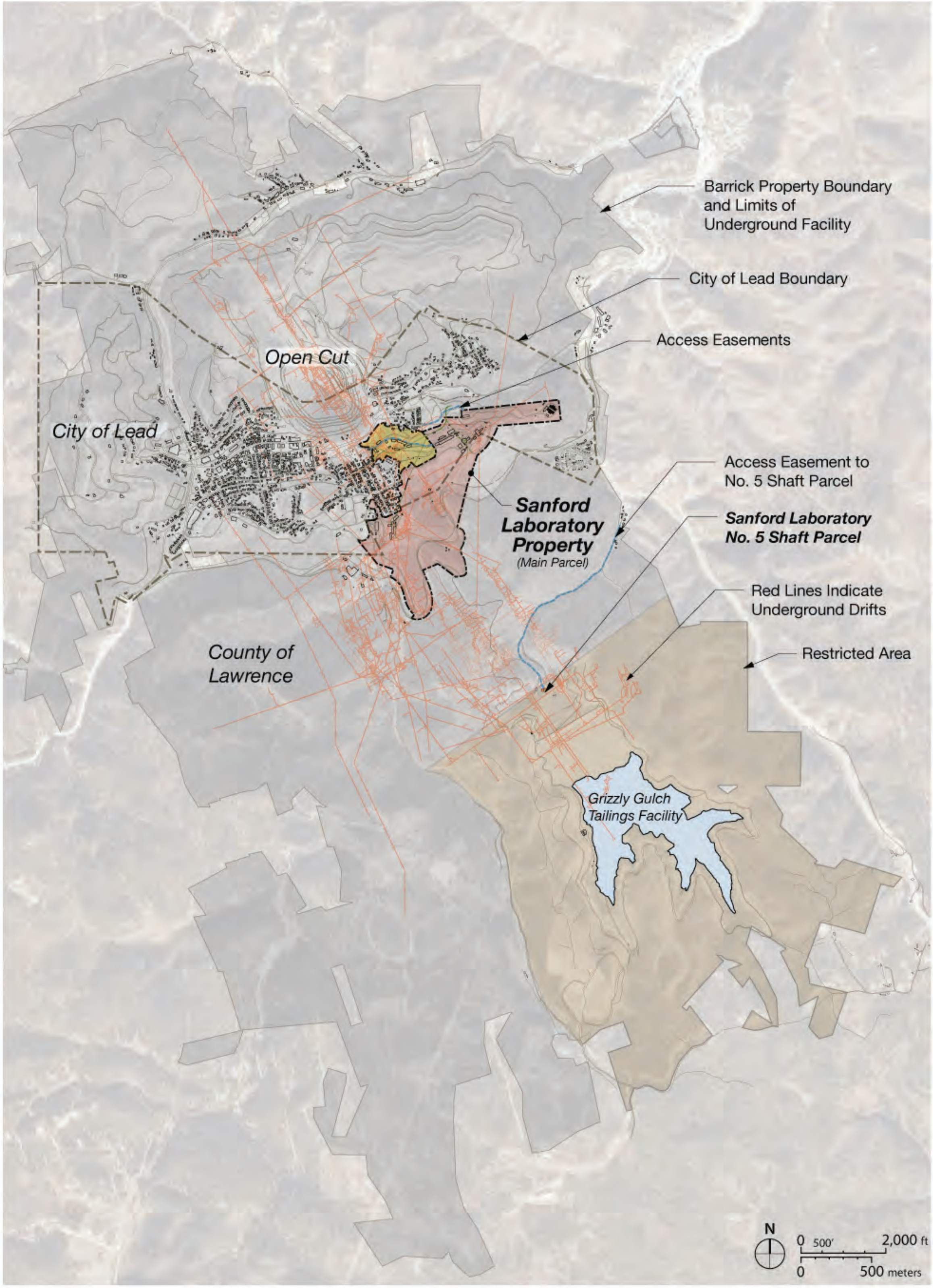

**Figure 5.1.1.4-2** DUSEL Complex shown in the context of the city of Lead, South Dakota, and the property remaining under ownership of Barrick. Area shown in yellow is a potential future expansion of the SDSTA property. [DKA]

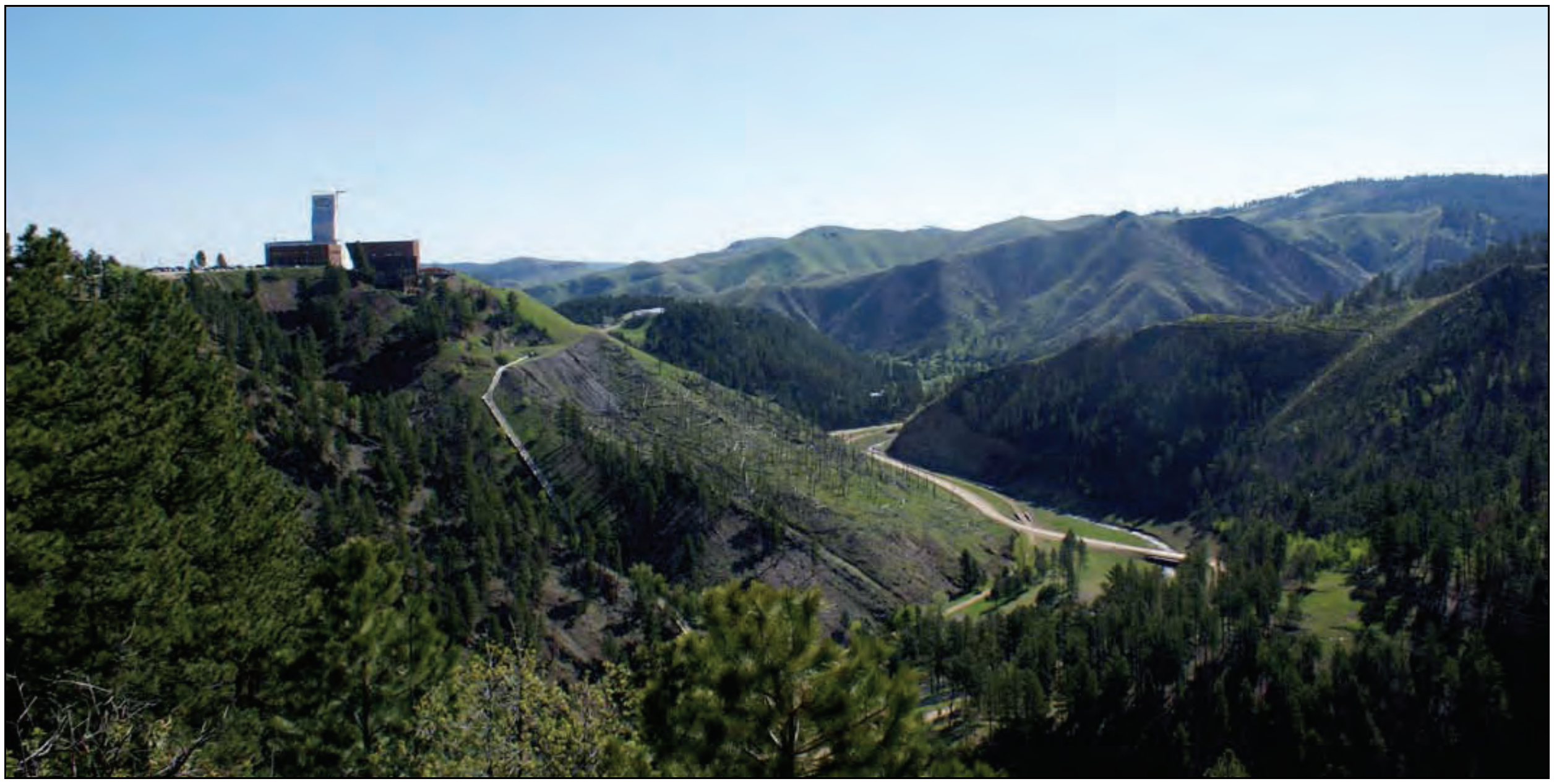

**Figure 5.1.1.4-3** DUSEL Yates Campus shown on left and Kirk Canyon to right. [DKA]

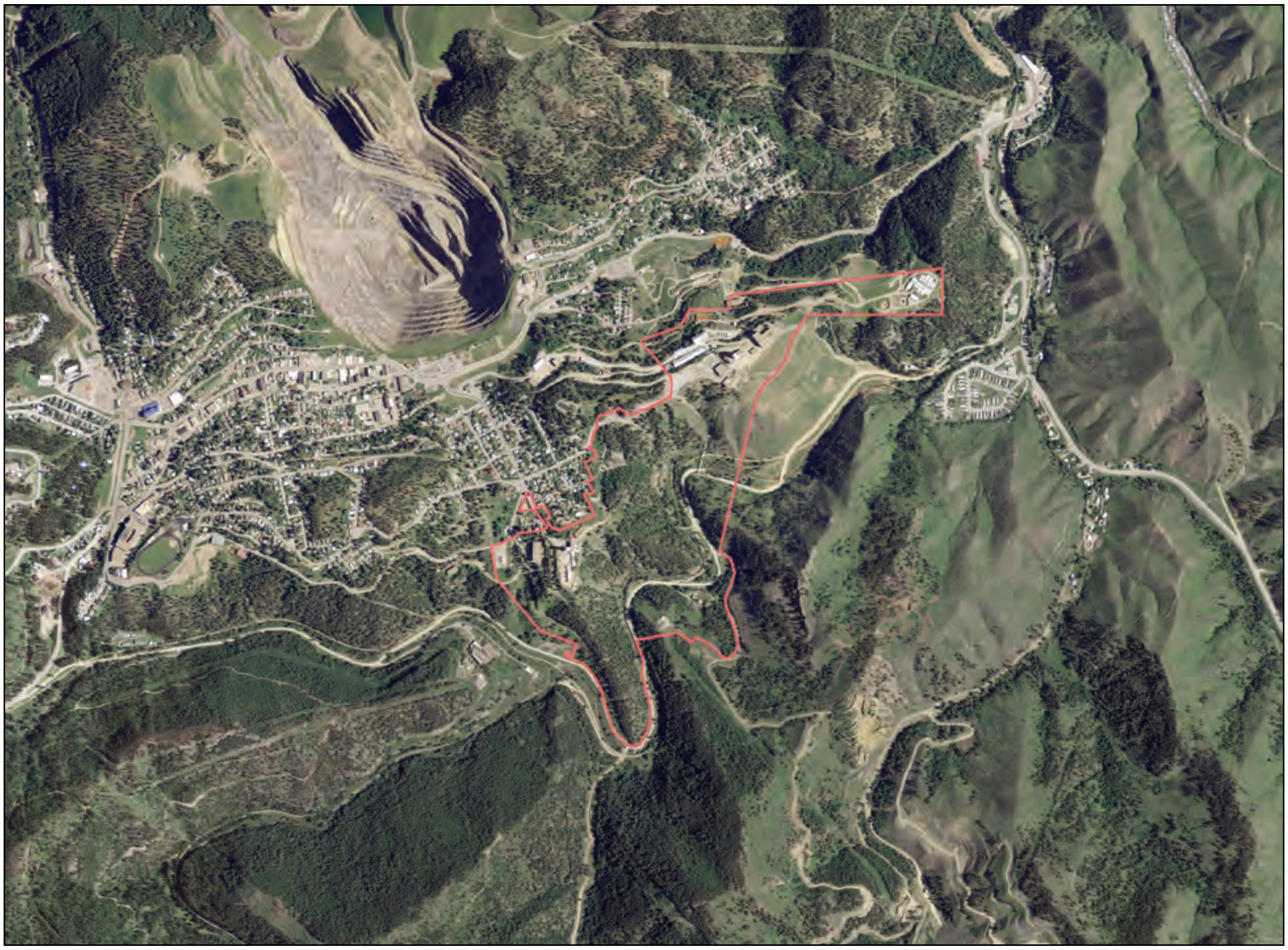

**Figure 5.1.1.4-4** Aerial view of Sanford Laboratory (boundary in red) and the adjacent city of Lead. [DKA]

### 5.1.2 Major Facility Scope Elements

The DUSEL Facility consists of surface and underground campuses and supporting infrastructure at the Homestake site and is illustrated in Figure 5.1.2. The Ross Surface Campus will be used primarily for construction and operations support, while the Yates Surface Campus will support science and administrative activities, education and public outreach functions, and the Waste Water Treatment Plant (WWTP). Also located on the surface is the Ellison Campus, discussed in Chapter 5.2, which is not currently owned by the SDSTA and is not part of the MREFC scope, but may provide important site access in the future. Surface infrastructure elements—including power, water supply, and communications—interface with the underground infrastructure to support underground science facility operations.

Three major underground science laboratory regions are located within the DUSEL Facility design. The MLL Campus is located at the 4850L and includes two laboratory modules (LMs) for science use. Also on the 4850L but not included in the MLL scope is one large cavity designed to support a 100 kT water Cherenkov detector (WCD) for the LBNE. This large cavity (LC-1) is included in the Facility design presented in this PDR, but the scope is wholly within the LBNE project's responsibilities. NSF, through the MREFC budget, will make a financial contribution to the LBNE development but the facility is not included in the facility scope of the DUSEL MREFC budget. The DLL Campus is located at the 7400L and includes one LM and a drill room to support ecohydrology experiment activities. The Other Levels and Ramps (OLR) include portions of various underground levels located within the laboratory footprint that will host a wide range of biology, geology, and engineering experiments. Certain underground infrastructure elements such as power substations and pumping installations for underground dewatering maintenance are located on the same levels as the OLR.

The underground infrastructure supports the underground laboratory facility operations. The DUSEL underground infrastructure includes the shafts, winzes, and hoists to provide access to and egress from the underground facility. It also includes systems such as fire and life safety, ventilation, water inflow management and dewatering, electrical power, cyberinfrastructure and communications, transportation, waste rock handling, and plumbing. These systems provide the lifeline between the underground laboratory space and the Surface Facility. Figure 5.1.2.1 illustrates the Ross and Yates Shafts, which provide access to the underground, and the Oro Hondo fan facility, which provides an exhaust point for underground exhaust ventilation.

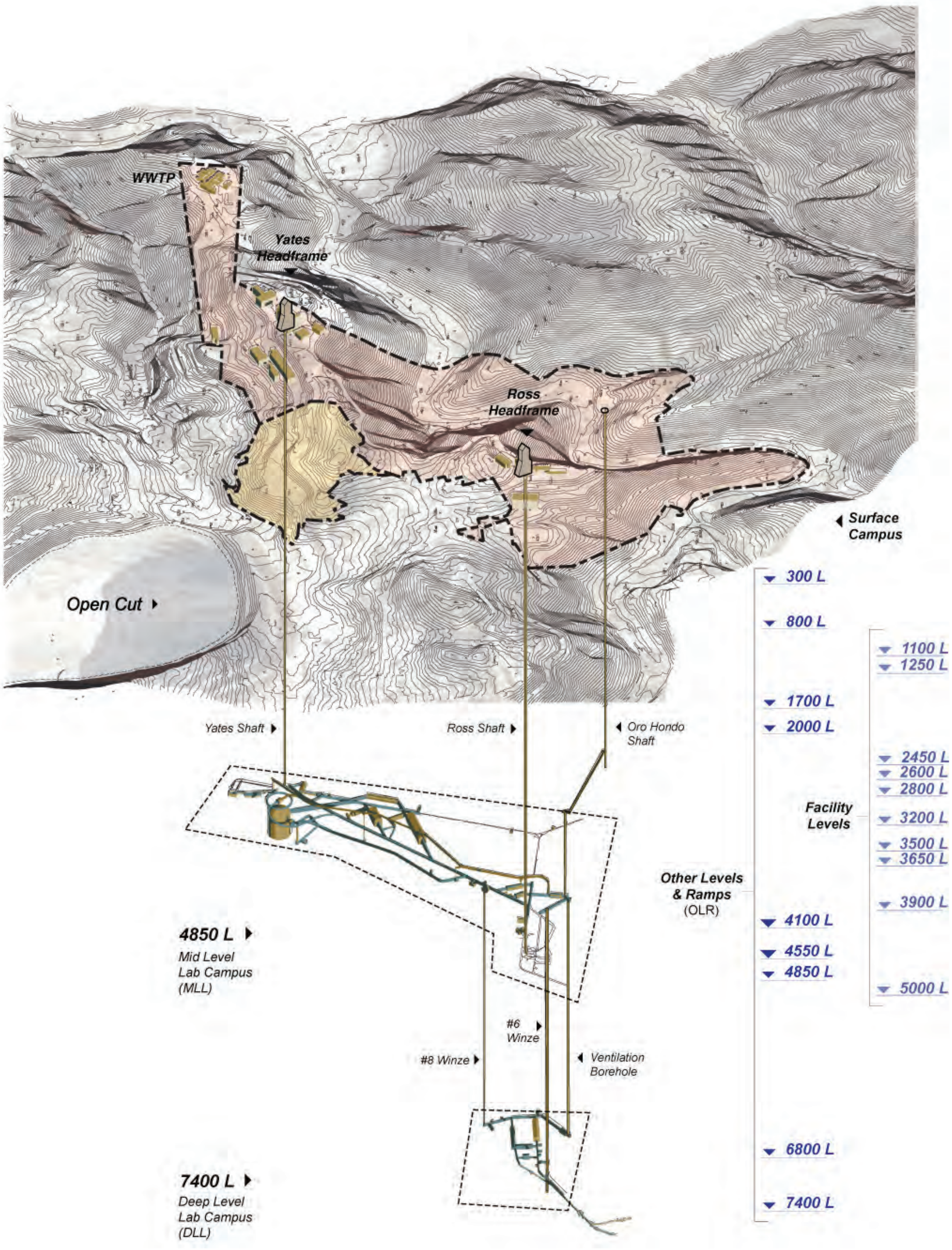

**Figure 5.1.2** Major elements of the DUSEL Complex. [DKA]

### 5.1.2.1 Surface Campus

The DUSEL Surface Facility has two distinct campuses, Ross and Yates, which provide a natural separation of surface functions. The Yates Campus will be developed as the primary campus for science, administration, and education and outreach activities. The Ross Campus will be developed with a primary mission of supporting construction, operations, and maintenance activities.

The surface development includes a mix of adaptive reuse of many of the existing surface buildings and infrastructure donated by Barrick along with the addition of two new buildings on the Yates Campus—the Sanford Center for Science Education (SCSE) to support the education and public outreach program and a new science experiment Assembly Building to support experiment surface assembly and fit checks. The existing surface structures total approximately 253,000 gsf on the Yates and Ross Campuses. For DUSEL construction, 205,000 gsf of structures will be reused. Through operations activities under R&RA funding, 48,000 gsf of structures will be removed over time. There is 42,000 total gsf of new facility construction planned for the Assembly Building and the SCSE. The planned Surface Facility configuration is shown in Figure 5.1.2.1 and outlines plans for new and reused structures. Deferred maintenance is required on the reused surface buildings and will be performed through operations activities with R&RA funding.

The Ellison Campus, as depicted in Figure 5.1.2.1, was not included in the original PDA between the Barrick Gold Corporation and the SDSTA. The Ellison Campus holds potential for future use in support of DUSEL to include additional level ground for building sites, limited reuse of existing buildings, additional parking, and most importantly, potential for a new access road to the Yates Campus from the city of Lead for employee and visitor access that provides site access with less grade and residential traffic than the current access route. In 2010, the SDSTA secured an agreement with Barrick for an option to purchase the Ellison Campus in the future. The design of the new access road was included in Surface Facility designs as a scope option.

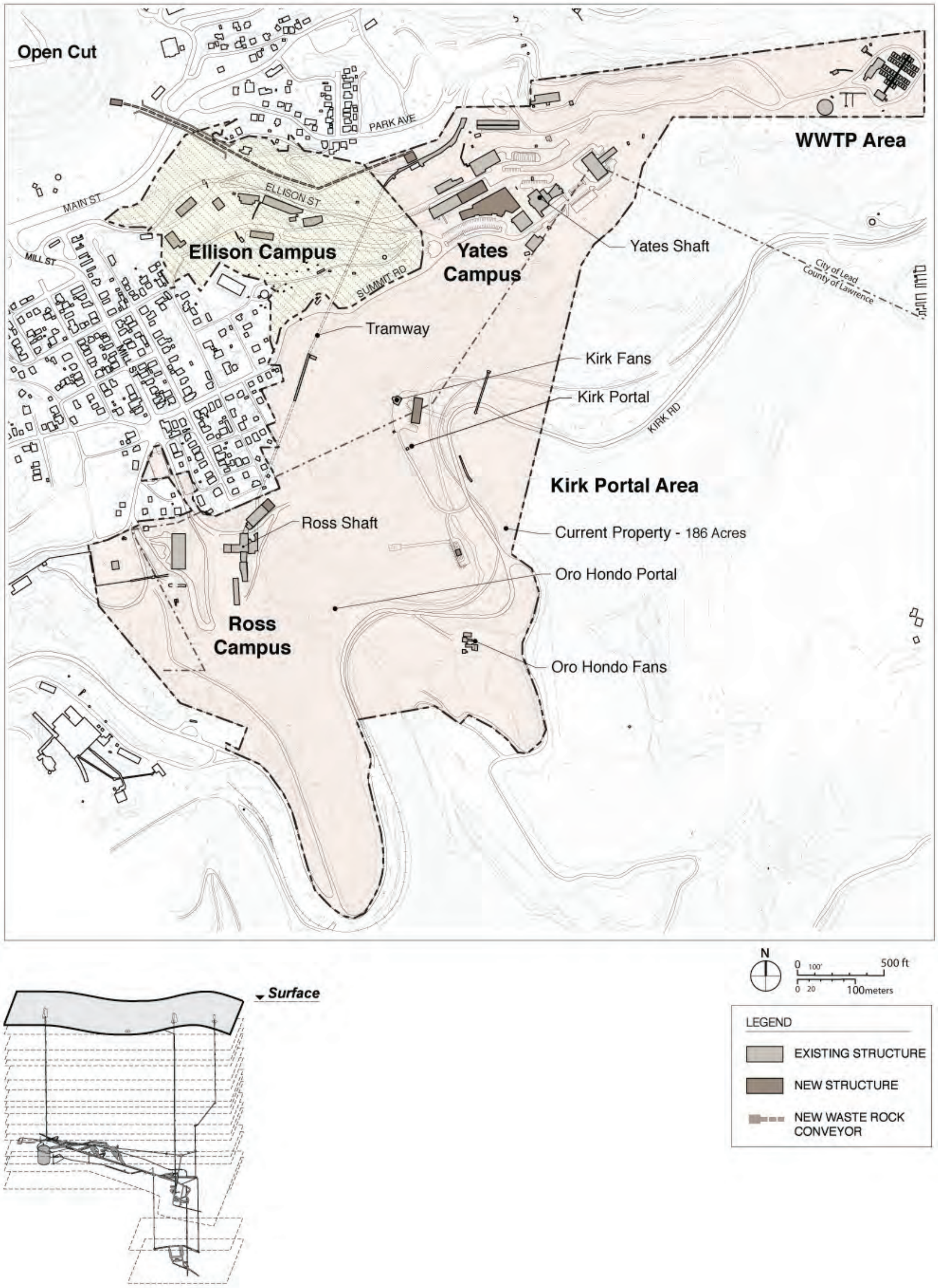

**Figure 5.1.2.1** Major surface elements of the DUSEL Surface Campus. [DKA]

### 5.1.2.2 4850L Mid-Level Laboratory (MLL) Campus

The largest underground campus at DUSEL and the primary focus during Preliminary Design is the MLL Campus located at the 4850L. The MLL Campus includes Laboratory Module 1 (LM-1) and Laboratory Module 2 (LM-2), and the Davis Laboratory Module (DLM). The Large Cavity 1 (LC-1) also resides on the 4850L but is intentionally not included in the MLL. The LBNE project funds the LC-1 construction separately from the MREFC budget. Figure 5.1.2.2-1 provides a plan view of the MLL Campus and Figure 5.1.2.2-2 an isometric perspective to show its relationship to surrounding support facilities, including an exhaust ventilation drift located at the 4700L and a decline ramp to the 5060L to facilitate excavation and access to the bottom of LC-1. The mechanical and electrical rooms (MERs) located immediately adjacent to the LMs are included in the scope of the MLL, whereas the Ross and Yates Shaft MERs on the 4850L are included in the underground infrastructure scope.

The DLM will be developed in advance of DUSEL construction through the SDSTA and funded through the generous gift from Mr. T. Denny Sanford in support of early science activities. DUSEL construction efforts will provide interfaces to the DLM to connect it to the DUSEL installed infrastructure.

LM-1 and LM-2 are designed to house a generic suite of experiments and therefore have similar sectional sizes and configuration. LM-1 will hold two small or one larger experiment and LM-2 up to three experiments nominally. Table 5.1.2.2 shows the dimensions of the planned elements of the MLL Campus plus the LC-1 to show the main components of the 4850L.

| Experiment Space | Width ft (m) | Height ft (m) | Length ft (m) | Floor Area $ft^2$ ($m^2$) | Finished Volume $yd^3$ ($m^3$) |
|---|---|---|---|---|---|
| LM-1 | 66 (20) | 79 (24) | 164 (50) | 10,764 (1,000) | 29,422 (22,495) |
| LM-2 | 66 (20) | 79 (24) | 328 (100) | 21,528 (2,000) | 58,845 (44,990) |
| DLM | 30 (9) | 50 (15) | 60 (18) | 1,800 (167) | 3,333 (2,548) |
| LC-1 | --- | 272 (83) | 180 (55) diameter | 25,575 (2,376) | 243,210 (185,947) |

**Table 5.1.2.2** MLL Campus science space dimensions.

Several ancillary spaces support MLL Campus functions with electrical and mechanical equipment, maintenance shops, storage rooms, communication distribution rooms, Areas of Refuge (AoRs), chiller and chiller spray rooms, sumps, and excavation mucking bays. Both LMs will have their own utility room that will accommodate their mechanical and electrical equipment. There is also space allocated for general purpose and clean machine shop space for science. The design of the MLL is discussed in Chapter 5.6. The design of LC-1 is addressed in Chapter 5.7. For reference, the scope of the LC-1 is captured in a separate Level-3 Work Breakdown Structure (WBS) element DUS.FAC.LGC.

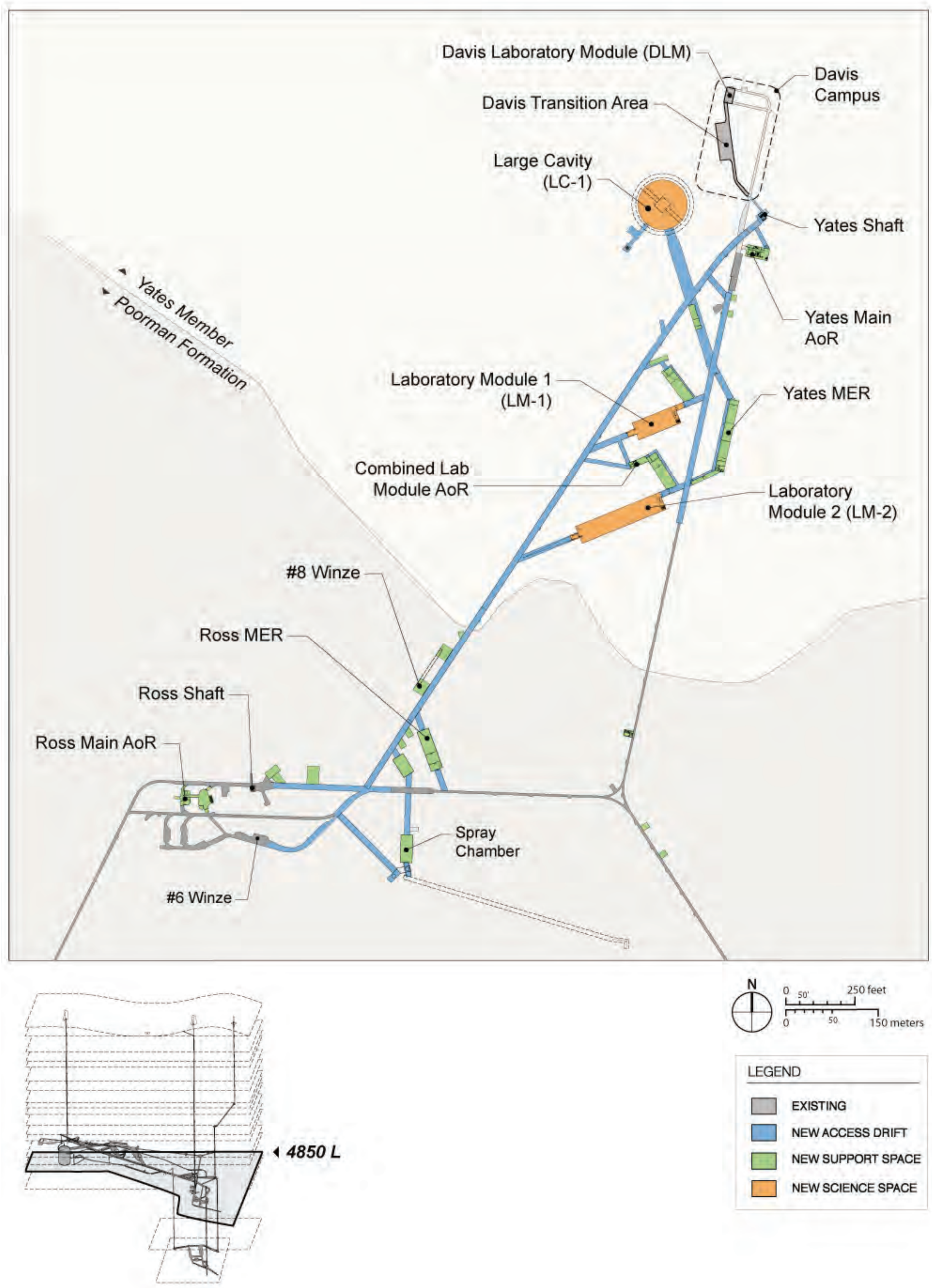

**Figure 5.1.2.2-1** Plan view of the MLL Campus. [DKA]

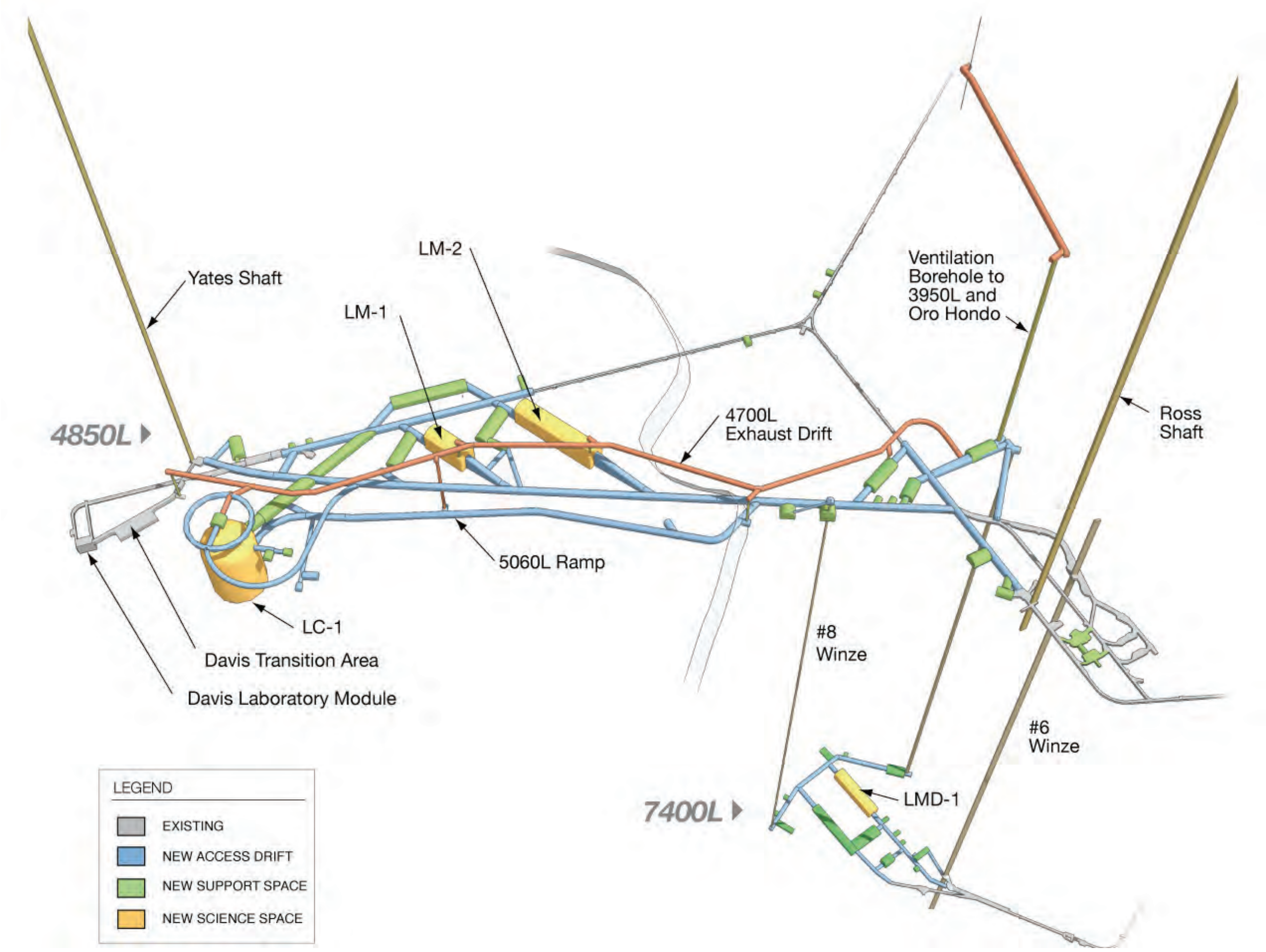

**Figure 5.1.2.2-2** Three-dimensional view of MLL and DLL Campuses. [DKA]

### 5.1.2.3 7400L Deep-Level Laboratory (DLL) Campus

The DLL Campus includes one Laboratory Module (LMD-1) to host physics experiments and also includes a drill room to support ecohydrology science research. The drill room falls within the scope of the DUSEL OLR. Nearly all of the development on the DLL Campus will represent new excavations. Given the facility has not yet been dewatered to the 7400L, the DLL excavation design represents extrapolation of geologic information from the 4850L. Figures 5.1.2.3-1 and 5.1.2.3-2 provide an overview of the DLL and its relationship to surrounding facilities.

The DLL Campus design, similar to the MLL, was developed using requirements that represent the classes of experiments planned for DUSEL. Initial planning and assignment of specific potential Integrated Suites of Experiments (ISEs) has been performed by the DUSEL Project to support test fits and requirements proofing. The results of these analyses are outlined in Volume 3, Sc*ience and Engineering Research Program*. Table 5.1.2.3 provides overall space dimensions for the DLL related science spaces.

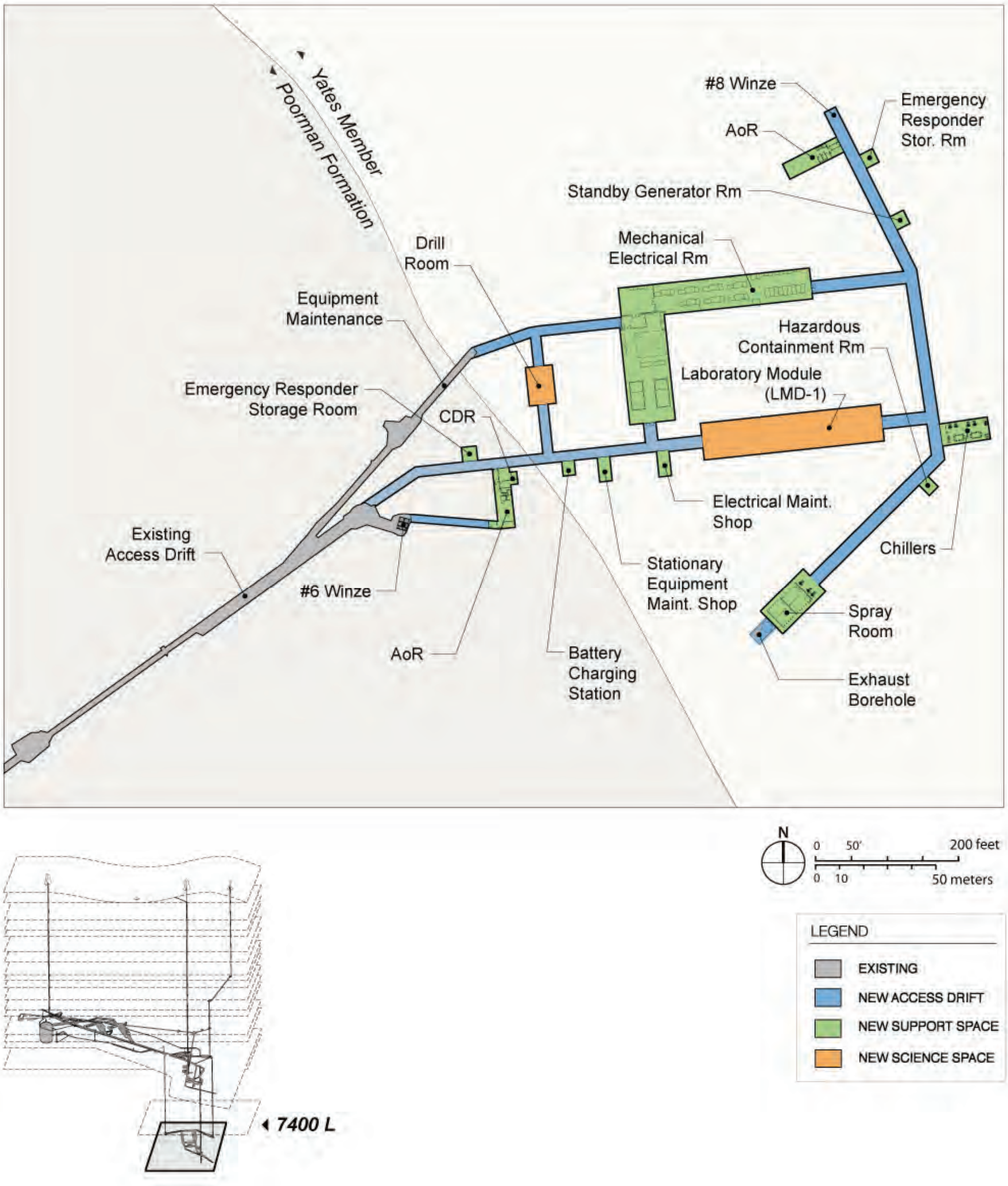

**Figure 5.1.2.3-1** Plan view of the DLL Campus. [DKA]

| Experiment Space | Width ft (m) | Height ft (m) | Length ft (m) | Floor Area ft$^2$ (m$^2$) | Finished Volume yd$^3$ (m$^3$) |
|---|---|---|---|---|---|
| LMD-1 | 49 (15) | 49 (15) | 246 (75) | 12,103 (1,125) | 19,470 (14,898) |
| Drill Room | 36 (11) | 36 (11) | 53 (16) | 1,894 (176) | 2,274 (1,740) |

**Table 5.1.2.3** DLL Campus science space dimensions.

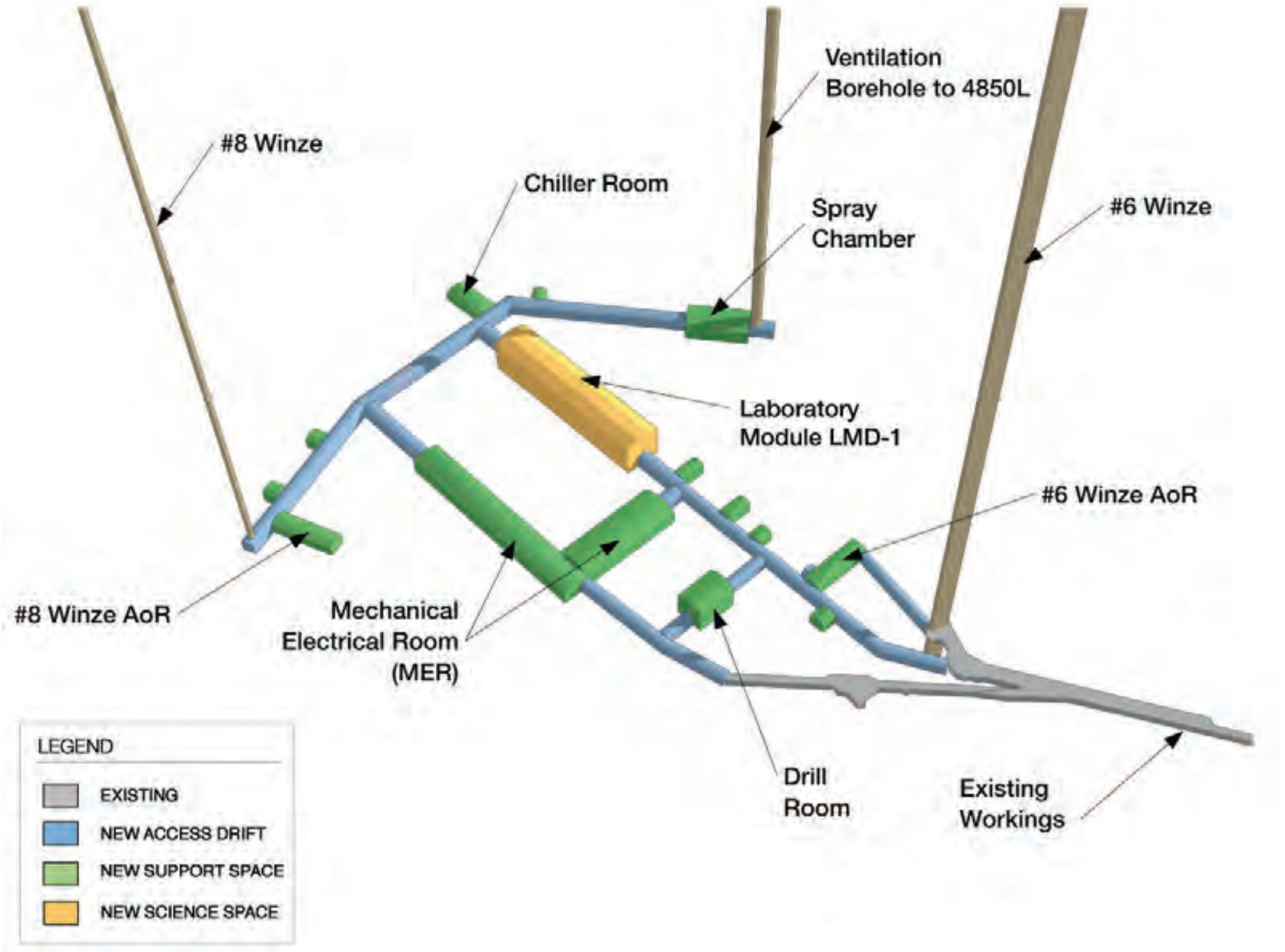

**Figure 5.1.2.3-2** Three-dimensional view of DLL Campus. [DKA]

### 5.1.2.4 Other Levels and Ramps (OLR)

Although DUSEL physics experiments will be located primarily at the newly constructed MLL and DLL Campuses, DUSEL biology, geology, and engineering experiments in the DUSEL underground space will be located mainly in areas previously excavated during mining operations. These areas are referred to as Other Levels and Ramps (OLR) and will support both facility operations and science use. The OLR provide approximately 19 miles (30 km) of laboratory space distributed on selected levels within the underground facility. Through R&RA funding as part of DUSEL operations activities, the OLR drifts will be improved through installation of ground control systems, ventilation, storm-water management controls, and utilities such as power, water, and cyberinfrastructure to provide safe access and support for science operations. Facility infrastructure, including dewatering pump stations, electrical substations, and water pressure reducing stations to support facility operations, may be located on the same level as the OLR experiments, but these are included in the underground infrastructure scope.

Excavated levels exist every 100 vertical feet for the first 1,100 feet, and every 150 feet from 1,100 to 8,150 feet. The OLR spaces that will be available to science experiments in addition to the MLL and DLL are: 300L, 800L, 1700L, 2000L, 4100L, 4550L, 4850L, 6800L, and 7400L. Levels that will be used to support the facility systems and operations are: 1100L, 1250L, 2400L, 2600L, 2800L, 3200L, 3500L, 3650L, 3900L, 5000L, and 6800L. These facility-related OLR may be used for science in the future, but infrastructure to support science is currently not planned. Figure 5.1.2.4 outlines the collective OLR scope planned for science and Facility uses.

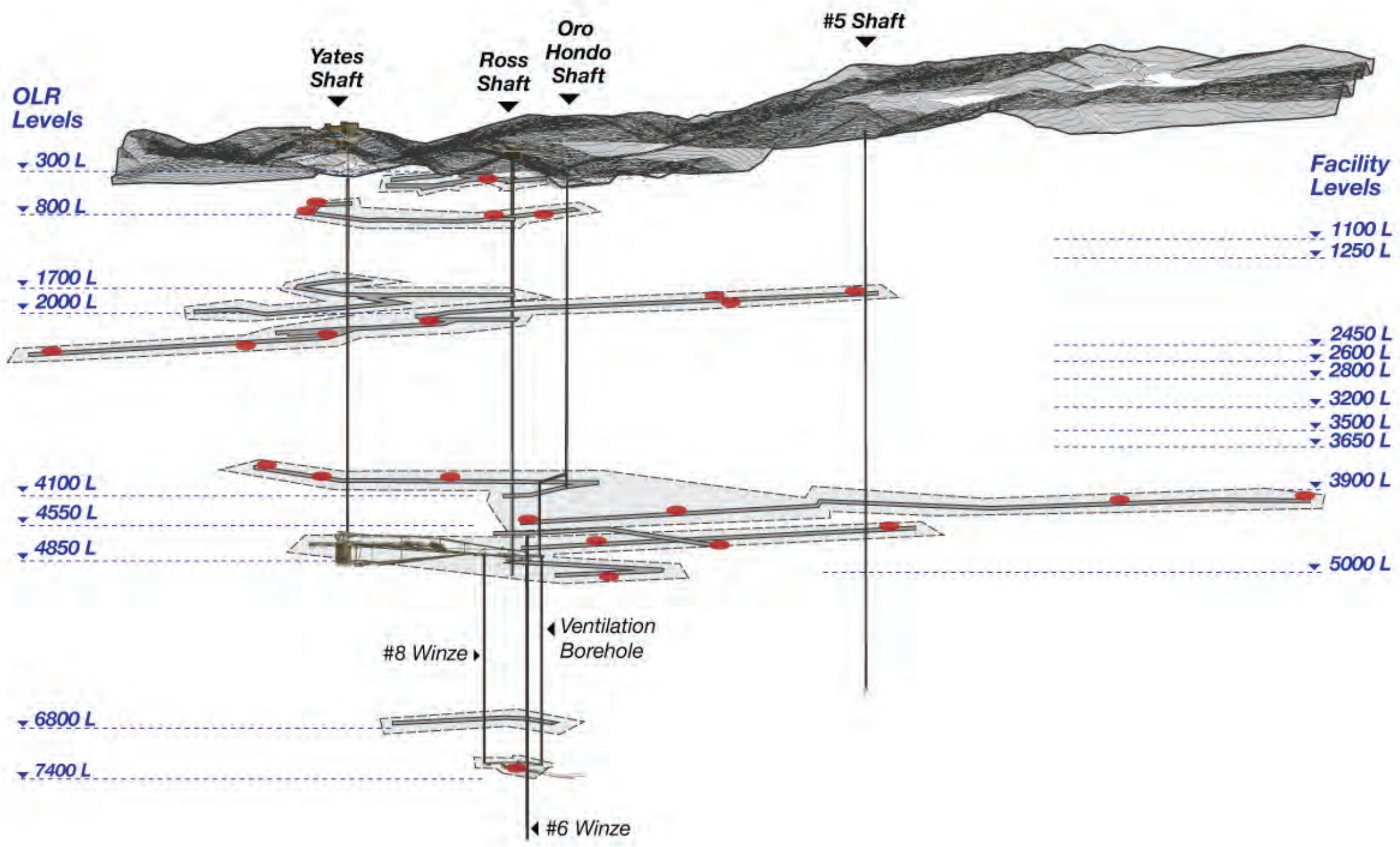

**Figure 5.1.2.4** OLR planned for science activities and facility support with approximate science locations noted in red. [DKA]

### 5.1.2.5 Underground Infrastructure (UGI)

The previous sections outlined the major elements of the surface and underground laboratory campuses necessary to support science and facility operations. The Underground Infrastructure (UGI) supports these campuses and links the underground laboratory space and facility infrastructure to the surface.

UGI includes the shafts and hoisting systems to provide access and egress to the major underground spaces, including the Ross and Yates Shaft hoists, headframes, and shaft furnishings to convey materials and personnel from the Surface Campus to the MLL Campus on the 4850L and OLR between. UGI also includes winzes to traverse from the MLL Campus on 4850L to the DLL Campus on the 7400L and OLR between. These winzes to the DLL Campus include the existing #6 Winze and a new #8 Winze. The rehabilitation of these hoists and shaft systems to provide safe access to the underground will be sponsored in part through R&RA funding. The R&RA-funded work includes the replacement of shaft furnishings and replacement of hoist electrical and mechanical components.

In addition to hoisting and shaft systems, the UGI includes a number of other elements:

**UGI MREFC-Funded Elements**

- Life safety systems and areas of refuge
- Maintenance shops, utility rooms, storage and containment areas
- Drifts and ramps required for both primary and secondary access and egress
- Cyberinfrastructure, controls, and monitoring systems
- Air quality and ventilation systems, including a new ventilation borehole to the DLL Campus

- Electrical power distribution systems
- Water inflow management systems
- Chilled-water systems
- Plumbing systems
- UGI R&RA Operations elements
- Waste handling systems
- Dewatering systems
- Material handling and personnel transportation systems

### 5.1.3 Facility Work Breakdown Structure (WBS)

The WBS for the DUSEL Project is thoroughly described in Volume 2, *Cost, Schedule, and Staffing*, of this PDR. The WBS, as part of the configuration management process, is a controlled document with all changes requiring Configuration Control Board (CCB) approval. The development of the DUSEL Facility is a Level-2 element of the WBS (DUS.FAC). The DUSEL WBS Dictionary provides a description of each WBS item. The Facility portion of the WBS is divided into seven levels and further subdivided to address the specific areas within each scope. Figure 5.1.3 shows the top-level WBS elements for the DUSEL Facility.

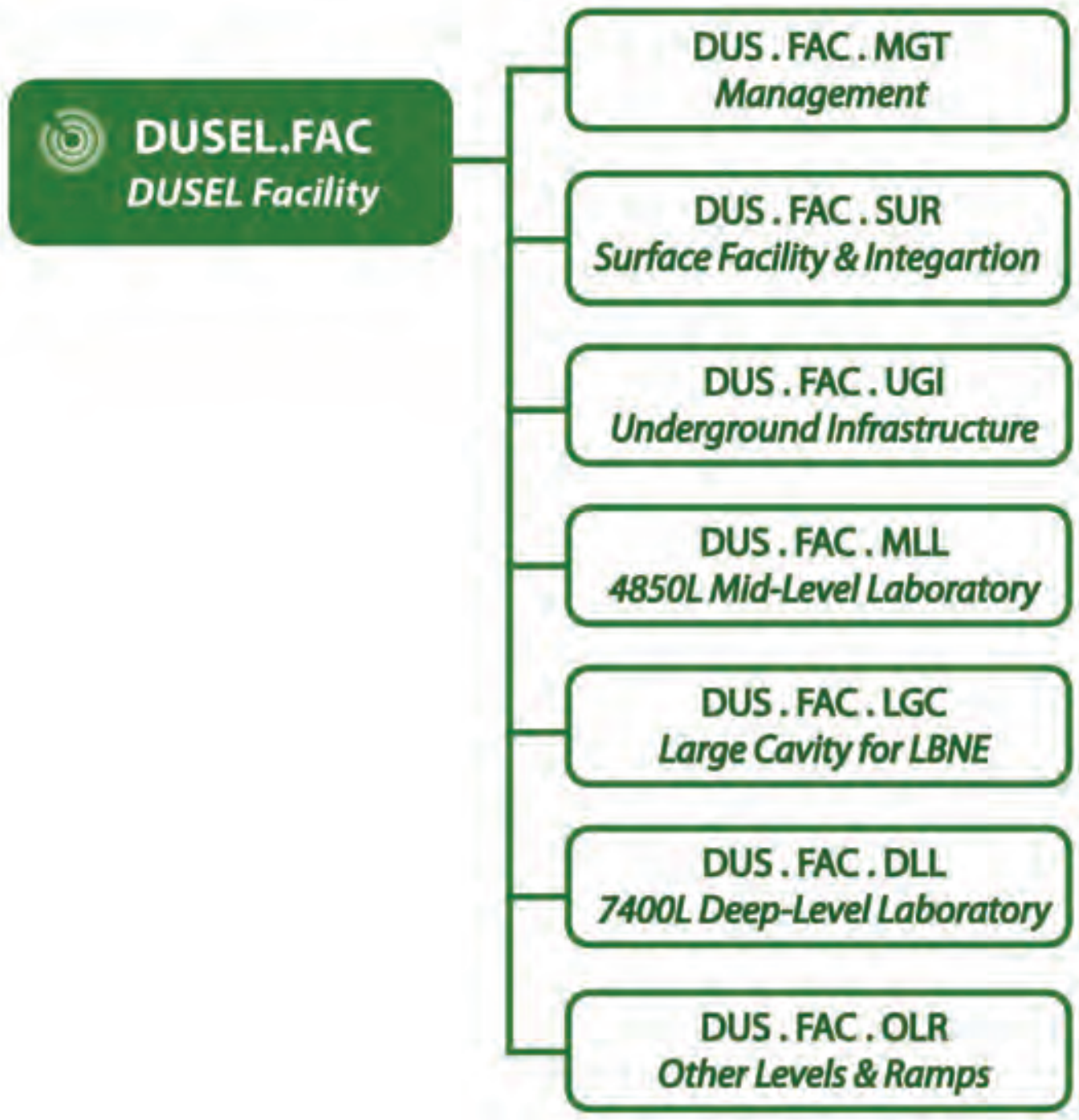

**Figure 5.1.3** DUSEL Facility Work Breakdown Structure. [DKA]

Table 5.1.3 lists each of the Level-3 scope sections discussed in this Volume 5. As there are many interfaces and crosscutting systems within the Facility, this table is intended to outline where discussions on these key elements can be found within this Volume.

| Facility WBS Element | Element Description | PDR Discussion Location |
|---|---|---|
| **DUS.FAC.MGT** | DUSEL Facility Management | Chapter 5.1 |
| **DUS.FAC.SUR** | Surface Facility and Infrastructure | Chapter 5.2 |
| **DUS.FAC.UGI** | Underground Infrastructure | Chapter 5.4 |
| **DUS.FAC.MLL** | Mid-Level Laboratory | Chapter 5.6 |
| **DUS.FAC.LGC** | Large Cavity for LBNE | Chapter 5.7 |
| **DUS.FAC.DLL** | Deep-Level Laboratory | Chapter 5.8 |
| **DUS.FAC.OLR** | Other Levels and Ramps | Chapter 5.9 |

**Table 5.1.3** DUSEL Facility major WBS elements and location of key discussions within Volume 5.

## 5.1.4 Facility Interfaces

### 5.1.4.1 Surface Campus Interfaces

#### 5.1.4.1.1 Interfaces with Underground Systems and Infrastructure

Numerous interfaces exist between the surface and underground scopes of work, including water (potable, purified, and industrial), electricity, communications, cyberinfrastructure, ventilation air, shaft air heating in the winter months, and underground facility dewatering. The physical interface points between the surface and the underground are the shaft collars at the Ross and Yates Shafts.

#### 5.1.4.1.2 Interfaces with Integrated Suite of Experiments (ISE)

Specific areas of the Surface Campuses are developed for dedicated ISE use, while other areas are shared with science collaborations and DUSEL Operations. The Yates Dry will have portions of the building dedicated to experiment control rooms and shared meeting rooms for the collaborations. The Yates Dry will house a server room, as part of the cyberinfrastructure backbone, where data collected by the experiments underground can be received on the surface, stored, and processed on the surface. The Yates Dry will also have a locker room/dry facility for scientist use, both short-term and long-term.

The Foundry will be renovated to house laboratory and support modules for the ISE as well as shared areas including hazardous material storage, electronic and machine shop, a meeting room, break room, and a campus-wide shipping and receiving facility. The science community will also interact with staff and the general public through the SCSE, where classrooms and meeting rooms will be available.

### 5.1.4.2 Mid Level Laboratory (MLL) Campus Interfaces

#### 5.1.4.2.1 Interfaces with Underground Infrastructure

The interface of the MLL with the UGI design at the 4850L is very important because the backbone utility services and infrastructure from the surface to the 4850L are provided through this scope of work. Thus, the designs of both UGI and the Underground Laboratory (UGL) design scopes require close interaction and coordination.

**Shafts and Hoisting.** The Yates Shaft Service Hoist payload and cage dimensions are designed to accommodate conveyance of science equipment, materials, and personnel. The Yates and Ross Shaft services compartments are designed to accommodate the utility installations required to support laboratory operation. The conveying of science equipment and materials to the DLL will be through the #6 Winze, while the #8 Winze is designed for secondary egress only.

**Life Safety Systems and Areas of Refuge.** Areas of Refuge (AoR) have been designed at the Yates, Ross, and #6 Winze stations to accommodate the anticipated occupancy during laboratory operations. A compressed-air system is included in the design to provide emergency breathable air in the AoRs in the event of an emergency. The ventilation system has been sized to accommodate evacuation of various hazardous environmental events. The fire water supply system includes provisions for a conventional sprinkler system, monitor nozzles, fire hoses, and a water mist system for areas with sensitive electrical equipment in all laboratory spaces. Rescue equipment will be included in the facility to facilitate evacuation of laboratory personnel.

**Maintenance Shops, Utility Rooms, Storage, and Containment Areas.** The design of all facilities associated with maintenance shops, utility rooms, storage, and containment areas has considered the needs of anticipated laboratory users.

**Drifts and Ramps Required for Access, Egress, and Ventilation.** Sizes of drifts and ramps required for access, egress, and ventilation were determined through expected load dimensions of science equipment and utilities required to support laboratory operation. The turning radii of the drift accesses were designed to accommodate science equipment dimensions.

**Cyberinfrastructure (CI) Controls and Facility Management System (FMS)**. The MLL requires high-speed data connectivity and redundancy, and also requires that DUSEL fire control and safety systems monitor the underground systems and remote operation of underground equipment from the surface. The design of the CI backbone distributed through the shaft and access drifts considered the anticipated requirements for data communications and transfer by the experiments on the MLL Campus. Consolidated monitoring of all facility management systems and life safety systems is located in the Ross and Yates Campuses and 4850L and 7400L satellite control rooms.

**Material Handling Systems**. The selection of the type of material handling equipment took into consideration the type of science equipment and material that need to be transported. This includes consideration of possible cryostats and other large containers. Unit-load automated guided vehicles (AGVs) will be employed at the Facility for movement of material and equipment from the shafts to the end user locations. Consideration has been given to accommodate the handling of bulky and unusually shaped objects.

**Air Quality and Ventilation Systems.** The quantity of fresh air provided to the Facility was determined by experiment requirements. Exhaust air from the laboratories will be directed to a separate exhaust drift above the laboratory spaces, providing a means to exhaust hazardous gases from laboratory spaces.

**Electrical Power Distribution Systems.** The estimated electrical loads for each experimental space have been included in the total facility load determination and have been factored into the design for upgrades of the substations. Separate feeders from the surface substation will be provided for each LM and the Large Cavity to enable a minimum of interference between facility loads and laboratory loads. Standby power generation is included in the design to ensure power is provided to critical laboratory systems and life safety systems. Emergency power is provided to critical life safety systems through the use of uninterruptable power source (UPS).

**Chilled Water Systems.** The chilled-water systems are designed to accommodate all expected heat loads from science experiments and electrical systems. At the 4850L, the chilled-water system is designed with four chiller units, of which three are required to be operational at any given time to enable continued experiment operation if one unit is down. At the 7400L, the system is designed with two chiller units, of which only one is expected to be required for operation.

##### 5.1.4.2.2 Interfaces with Excavation Design

The MLL interface with the excavation design is a crucial component of developing the correct usable space for all the infrastructure and scientific needs. Nearly all of the MLL design incorporates new excavations. In addition to the space provided for the MLL, the excavation design includes the concrete floor and shotcrete applied to walls and ceilings. Another interface is the design of the rock anchors that support the cranes in the LMs.

##### 5.1.4.2.3 Interfaces with the Integrated Suite of Experiments (ISE)

The design of the MLL is driven by the needs of the ISE program. The ISE program is not fully developed; thus, the MLL is guided by the requirements of the generic ISE. The process for obtaining these requirements and a summary of the requirements are given in Volume 3, *Science and Engineering Research Program*.

##### 5.1.4.2.4 Interfaces with Other Levels and Ramps (OLR)

OLR utility services on the 4850L will be provided to the OLR from the main Mechanical Electrical Room (MER).

#### 5.1.4.3 Large Cavity for LBNE Interfaces

LBNE interfaces to the remainder of the Facility are very similar to the MLL interfaces delimited in the previous section. Only differences from the MLL interfaces will be highlighted in the following discussion.

##### 5.1.4.3.1 Interfaces with the Surface Facility

The LBNE surface purified water plant will be located in the existing Yates Motor Generator Room. The existing Yates motor generators will be removed and replaced with modern, smaller electrical equipment that will fit in the hoistroom.

##### 5.1.4.3.2 Interfaces with Underground Infrastructure (UGI)

**Life Safety Systems and Areas of Refuge.** Areas of Refuge (AoRs) are supplied at both the 4850L and the 5060L. Secondary egress from the LC-1 is provided via an exit stairway from the 4850L to the 5060L. This stairway also provides secondary egress from the 5060L.

### 5.1.4.4 Deep-Level Laboratory (DLL) Campus Interfaces

The DLL interfaces to the remainder of the Facility are very similar to the MLL interfaces delimited in the previous section. Only differences from the MLL interfaces will be highlighted in the following discussion.

#### 5.1.4.4.1 Interfaces with Other Levels and Ramps (OLR)

The primary interface with the OLR is the drill room on the 7400L. This drill room is located in the DLL Campus and is included in the OLR scope. Utility services will be provided to the OLR from the main MER in each campus, as shown in the UGI design. Many other UGI services provided to the 7400L will be shared between LMD-1 and the drill room, including the use of the conveyance, water, ventilation, communications, and AoRs.

### 5.1.4.5 Other Levels and Ramps (OLR) Interfaces

The design of the OLR must account for interfaces during design and construction with the excavation activity; the main campus levels at 4850L and 7400L, the surface, and cyberinfrastructure.

#### 5.1.4.5.1 Interfaces with Excavation Design

Despite an emphasis on maximizing the use of existing excavations for OLR experiments, the spaces provided are not adequate to house some experiments; therefore, additional excavation will be required at some experiment sites.

#### 5.1.4.5.2 Interfaces with Main Campuses at 4850L and 7400L

The OLR scope also includes areas at the 4850L and 7400L. Utility services will be provided to the OLR from the main MER in each campus, and a coordination plan will be developed to provide cage access to these levels for all occupants. The AoRs provided at these levels are sized to provide safety to experiments defined within the OLR footprint. These experiments must not impact physics experiments on these levels that may be sensitive to vibration, noise, and radiation.

#### 5.1.4.5.3 Interfaces with Surface Facility

Interfaces with the design of the Surface Facility are primarily related to occupancy, utilities, and logistics. Parking, staging areas, material deliveries, and limited office space will be required to interface stakeholders of the OLR with other parties.

#### 5.1.4.5.4 Interfaces with Underground Infrastructure

The OLR design has a key interface with the UGI because the OLR provide ancillary spaces such as electric rooms and pipe pressure-reducing station, egress, and support of life safety systems.

**Shafts and Hoists.** The OLR are accessed through either the Ross or Yates Shafts. OLR accessible from both shafts and thus having two means of egress include: 800L, 1700L, 4100L, 4850L, 2000L and 4550L via the 17 Ledge Ramp, and the 4850L.

**Drifts and Ramps Required for Access, Egress and Ventilation.** The OLR will utilize existing excavations for access, egress, and ventilation. Refurbishment of existing excavations will be required and will be performed as part of the DUSEL Facility operations outside of the DUSEL MREFC funding.

**Cyberinfrastructure Controls and Monitoring Systems.** Data transfer and communications will be needed in the OLR to collect and archive experimental data. A network of fiber-optic backbone cables will be distributed through the Ross and Yates Shafts to support data collection and communications for OLR experiments. Cables will be terminated in junction boxes at the shaft stations, from which fiber will be distributed to the various experiments.

**Material Handling Systems.** The existing material handling systems, consisting of track equipment such as locomotives and rail cars, will continue to be used. These will be provided for and maintained by DUSEL Operations & Maintenance.

**Air Quality and Ventilation Systems**. The design of the facility ventilation system includes OLR requirements of 10,000-30,000 cfm.

**Electrical Power Distribution Systems.** The design of the facility includes substations at the 300L, 800L, 1700L, 2000L, 4100L, and 4850L that are sized to provide service for anticipated OLR experiment needs.

**Life Safety Systems.** The design for Facility life safety systems will include an alarm system at the various shaft stations that will be activated in the event of an emergency. Communications will be tied into the fiber-optic backbone also provided to the shaft stations. It is expected that a Leaky Feeder system will also be provided to the various shaft stations. This system will be particularly useful to provide communications to personnel travelling within OLR areas not located at a specific experiment. Stench gas will be used as a means of communicating an emergency event that requires Facility evacuation to personnel working outside of the main laboratory campuses.

### 5.1.4.6 Underground Infrastructure (UGI) Interfaces

#### 5.1.4.6.1 Interfaces with Excavation Design

Infrastructure required to support excavation activities will be provided to the MLL and DLL Campuses. The following summarizes key infrastructure requirements during excavation activities. Interfaces discussed in previous sections have not been restated herein.

**Ventilation.** Prior to the commencement of major excavation activities, it will be necessary to excavate a new 14 ft (4.3 m) diameter borehole from the 4850L to the 3950L to provide exhaust air for excavation. The drift from this new borehole to the Oro Hondo Shaft on the 3950L will be enlarged as well. Fresh air for excavation will be provided to the campuses via the Ross and Yates Shafts to the 4850L and via the #6 Winze to the 7400L. Exhaust ventilation from the 4850L will be via a new 4850L to 3950L borehole and Oro Hondo Shaft. Exhaust from the 7400L will be via a new 8 ft (2.4 m) diameter borehole that will connect to the 4850L at the 4850L to 3950L borehole.

**Waste Rock Handling.** Prior to the commencement of any excavation activities, it will be necessary to complete the rehabilitation of the Facility waste handling system. The capacity of this system will be equivalent to what was in place during mining. There are a number of components to the Facility waste rock handling system, including refurbishing the Ross Shaft hoisting system, the Ross Shaft crushers, and the tramway; procuring track haulage equipment; and installing a surface conveyor to the Open Cut from the tramway dump. In addition, the existing production hoisting system will be refurbished in the #6 Winze and a new loading pocket will be constructed at the 7500L.

**Electrical Power.** The power requirements to support excavation will be significantly less than the power requirements for laboratory operations. Much of the excavation equipment will be diesel powered. Electrical equipment such as drill jumbos, pumps, air compressors, lights, and shop equipment will be powered by temporary electric infrastructure provided to the MLL Campus during construction. Temporary power to the 7400L Campus will be provided through the #6 Winze from a substation on the 4850L.

**Dewatering.** Wastewater generated from excavation will not exceed the existing dewatering system capacity. Because the solids content of the water is a concern, settling sumps will be required to clarify the water prior to discharge into the Facility dewatering system via the #6 Winze and Ross Shaft.

**Industrial Water Supply.** Industrial water will be necessary during excavation for dust control and drill operation. The Ross Shaft is equipped with an existing industrial water supply line large enough to meet the requirements of the excavation activities. Excavation crews will install temporary water pipelines from the Ross Shaft to the advancing excavation headings.

### 5.1.5 Design Development Resources and Teaming

The design development has been pursued through a combination of a focused DUSEL Project Facility design and construction team, five outsourced architecture and engineering services contracts, a construction management services contract, and an architecture and integration services contract. This collective DUSEL Facility design development team of DUSEL Project staff and contractors has the responsibility to develop and integrate the four design scopes into a single comprehensive facility design, including construction drawings, specifications, cost and schedule estimates, and a risk-based contingency analyses. The outsourced design-related contracts and the prime contractor include:

- Geotechnical Engineering Services—RESPEC
- Surface Facility and Infrastructure Design—HDR CUH2A
- Underground Laboratories Design—Arup USA
- Underground Infrastructure Design—Arup USA
- Underground Excavation Design—Golder Associates
- Construction Management Services—McCarthy Kiewit DUSEL, a Joint Venture
- Architecture and Integration—Oppenheim Lewis, Inc. with Dangermond Keane Architecture

The alignment of Facility scope to the design contracts is shown in Figure 5.1.5.

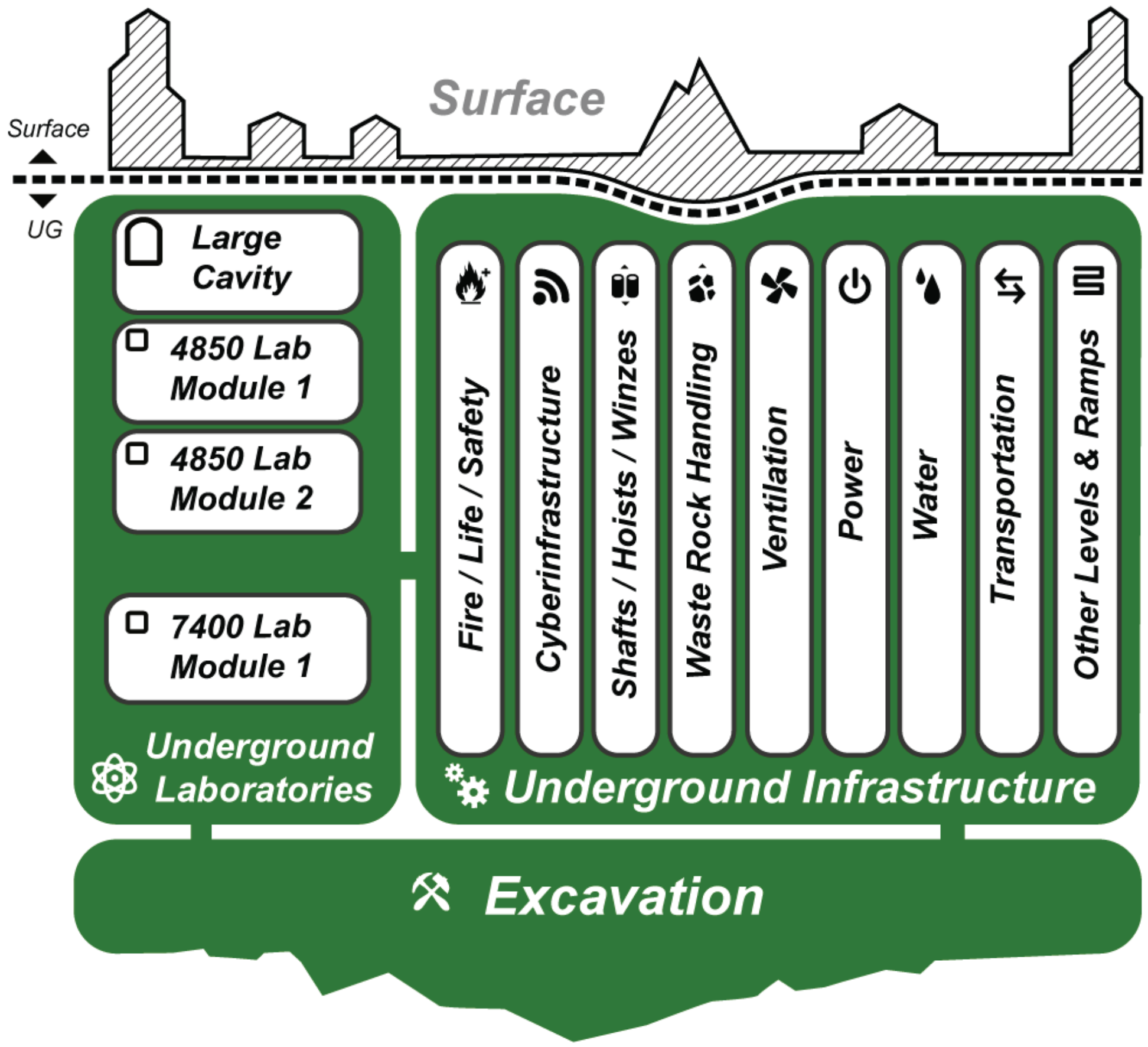

**Figure 5.1.5** DUSEL Facility major design scope distribution. [DKA]

### 5.1.5.1 Organizational Overview of the Facility Project Team

The collective Facility Project Team of DUSEL Project staff and contractors has the responsibility to develop and integrate the four design scopes of work (Underground Infrastructure Design, Underground Laboratory Design, Excavation Design [including Geotechnical Studies], and Surface Infrastructure Design) into this single consolidated Preliminary Design Report. The facility team staff is primarily based in Lead, South Dakota, at Sanford Laboratory.

The DUSEL Facility team interfaces with every other aspect of the DUSEL Project Team, including senior management and project management, Project Controls, Systems Engineering, Science, Operations, and current SDSTA at Sanford Laboratory Operations. Methods of integration of the design team—Project staff and outsourced contractors—are described in sections below. The Facility Project Team is a substantial portion of DUSEL staff, budget, and efforts. Management of these components is a key to the success of the DUSEL Project.

### 5.1.5.2 DUSEL Project Staff

The DUSEL Project staff supporting facility design development is a cross-functional team of skill sets, including architecture; project management; and engineering disciplines such as mining, civil, mechanical, and electrical. This DUSEL Facility team represents a mix of South Dakota School of Mines and Technology (SDSM&T), University of California at Berkeley (UC Berkeley), Lawrence Berkeley

National Laboratory (LBNL), SDSTA staff, and outsourced architecture and project management consultants. The DUSEL Facility team provides the project management, architecture, and engineering oversight and leadership for facility design development, including direct oversight of the execution of outsourced design contracts through daily interaction, oversight, and monitoring of contractor activities. This staff provides the foundation for continued design advancement through the planned Transition and Final Design phases for the Project on into Construction. Figure 5.1.5.2 outlines the current Facility team composition and structure. Although geographically dispersed, throughout the Preliminary Design period the team uses a combination of telephonic, video, and face-to-face meetings with team members and outsourced contractors to integrate the Facility requirements and resulting designs. Monthly interface meetings with DUSEL Facility, design and construction management contractors, Science, and Systems Engineering staff were key to integrating designs and interfaces and supporting cost and schedule reconciliation activities as well as Value Engineering (VE) efforts.

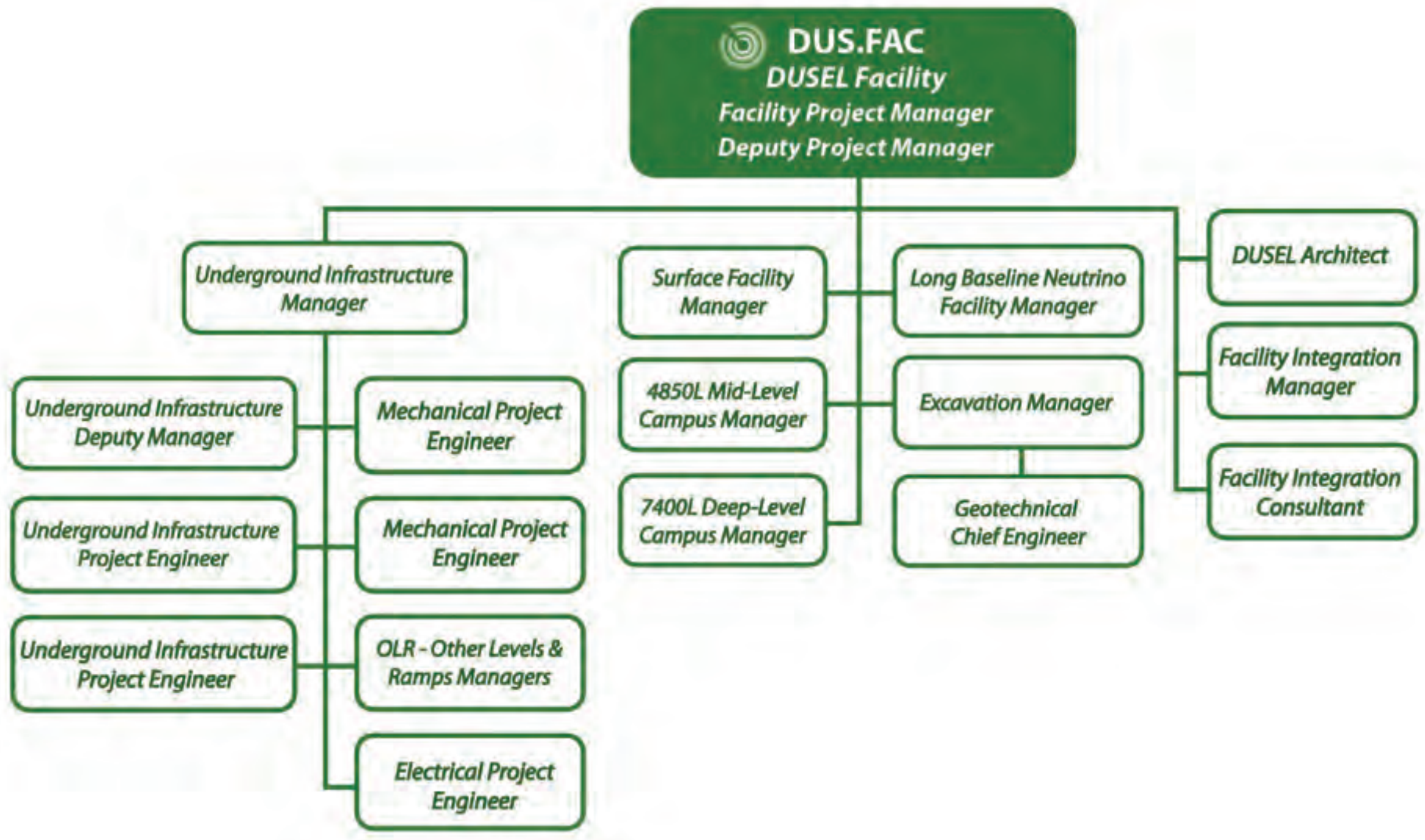

**Figure 5.1.5.2** Current DUSEL Facility team organization. [DKA]

### 5.1.5.3 Design Services Contracting Approach

The procurement process for each of the major design contracts followed a similar model to maintain consistency in the Project's approach to selecting and procuring contracts for major outsourced services.

For each scope of work, a DUSEL staff technical representative helped define the specific requirements of the scope of work. An openly competed Request for Proposal (RFP) process was used to select design teams for the design and engineering of the individual scopes for DUSEL at Homestake. The areas required for design of DUSEL Facility were divided into five scopes of work: Geotechnical Engineering Services, Surface Assessment and Design, Excavation Design, Underground Laboratory Design, and Infrastructure Assessment and Design.

The individual RFPs were developed and issued using standard contracting procedures developed for the DUSEL Project. These procedures are outlined in detail in the contract documentation binders that were prepared for each of the major contracts. The contracts were openly competed with a budget established by the DUSEL management team. Given the funding and schedule constraints, the scopes of the RFPs were developed to prioritize the needs to achieve the PDR milestones and to provide sufficient information to inform parallel design scopes of work.

DUSEL received proposals in response to each of the RFPs and competing firms were interviewed and selected to continue into contract negotiations. The selection committees for each contract competition identified the bidding contractors that best met the selection criteria and were best suited to provide services for each scope of work. The committees then made selection recommendations to DUSEL management prior to proceeding with contract negotiations.

A complete DUSEL Contract Management Plan (CMP) has been established to provide an overall framework for the management of the major DUSEL contracts. The CMP outlines roles and responsibilities of the central personnel involved with contract administration and management. It provides the structure and processes used in the management of DUSEL contracts and describes the communications required for successful execution of the contracts. The purpose of the CMP is to ensure successful control and management of DUSEL contracts so that deliverables outlined in each contract are on schedule, within allocated budgets, and of excellent quality. The DUSEL Project relies heavily on successful interaction with its contractors. The CMP seeks to define a framework that promotes successful contract management and oversight to ensure that DUSEL-awarded contracts provide the best value in support of the Project's requirements.

DUSEL contracts support the CMP and provide further instructions on the execution of contract-related responsibilities and processes. The CMP is updated when appropriate to ensure that the material included reflects the current DUSEL Project's major contracts and current contract management and oversight approach. A copy of the CMP is available for reference and is included in Appendix 9.D.

#### 5.1.5.4 Outsourced Design Services Contracts

As described in Section 5.1.5, the assessment and design of the facility and infrastructure for the DUSEL Facility at Sanford Laboratory was divided into five scopes that are integrated through the efforts of DUSEL Project staff and Construction Manager (CM).

The five outsourced design teams each report to a DUSEL technical representative (also known as a contracting officer's technical representative [COTR]). The DUSEL Project staff meets regularly with the designers and CM to discuss interfaces among the scopes of work. Each of the contracts with the design teams and CM was scoped to allow for a continued contracting relationship beyond Preliminary Design for Final Design and Construction Administration. The contractors must have authorization to proceed into the next project phases based upon performance and availability of funding. This contracting structure ensures continuity of the design team through the full Project while providing DUSEL an option to make team changes if required. The following describes the scope of each design contract and major deliverables.

**Geotechnical Engineering Services.** To support the early design efforts, the first outsourced service contract was put in place in January 2009 with RESPEC as the prime consultant, with support from Golder Associates, Lachel Felice & Associates, and Connors Drilling. This early contract was established to perform geotechnical investigations with the limited funds available early in the PDR phase, and to provide initial geotechnical data, testing, and analysis to inform the bidding, selection, and design requirements development for the four design and engineering scopes described below. This contract was completed during the Preliminary Design phase and has closed. Future geotechnical excavation investigations will be performed through the Excavation Design contract. The primary items within this initial geotechnical scope of work include:

- Geotechnical investigation and analysis of the 300L and 4850L
- Geotechnical mapping, core drilling, testing, analysis, and modeling to support laboratory module and large cavity placement decisions and design
- Development of a geotechnical data packing to support the excavation design contract

**Surface Facility and Infrastructure Design.** This contract was awarded in May 2009 to HDR CUH2A with a team of subconsultants. HDR's subconsultants include Architecture Incorporated (architecture and design), Wyss Associates (site planning and landscape architecture), MacDonald & Mack Architects (historic preservation and adaptive reuse), Albertson Engineering Inc. (structural engineering), West Plains Engineering Inc. (mechanical and electrical engineering), American Engineering Testing Inc. (geotechnical, environmental, materials, and forensics), and Parametrix (cost consulting).

The primary items within the Surface Facility Infrastructure and Design scope of work include:

- Assessment of Surface Facility, buildings, and infrastructure
- Assessment of existing surface site conditions and environmental hazards
- Providing cost estimates for reuse/rehabilitation of assessed items
- Providing design of Surface Campus, utilities, and infrastructure to support underground experimental facility and other required Surface Facility
- Full architectural and engineering services for the design of the DUSEL Surface Campus
- Development of full cost estimates and risk analyses for items within scope of work

**Underground Laboratory Design.** This contract was awarded in February 2010 to Arup USA, with subconsultants who specialize in laboratory planning and research facility design. Arup's subconsultants include Davis Brody Bond Architects (architectural design), Research Facilities Design (laboratory planning), SRK Consulting (mining, water, and environmental consultants), and TSP Inc. (engineering, architecture, and construction consulting). The primary items under this scope of work include:

- Design and analysis necessary for construction of underground laboratory facility and build-out for installation of scientific experimental equipment
- Design of two LMs and one large cavity on the 4850L and one LM at the 7400L to be ready for the experiment build-out by individual scientific collaborators
- Development of full cost estimates and risk analyses for items within scope of work

**Underground Infrastructure Design.** In March 2009, the Underground Infrastructure Design contract was awarded to Arup USA Inc., with subconsultants including SRK Consulting (mining, water, and environmental consultants), and G.L. Tiley and Associates (architectural and systems design).

The primary items under this scope of work include:

- Assessment of the underground infrastructure, including hoists, shafts, dewatering pumps, ventilation systems, duct systems, and all utilities

- Design of underground infrastructure to support two LMs and one large cavity on the 4850L and one LM on the 7400L
- Development of the Ross and Yates Shafts, finishing designs to maintain continued safe underground access
- Development of full cost estimates and risk analyses for items within scope of work

**Excavation Design.** The underground Excavation Design contract was awarded in October 2009 to Golder Associates, with specialized subconsultants working under their contract. Golder's team of subconsultants includes Lachel Felice & Associates (design of structural linings in underground facilities and geotechnical characterization), Atkinson Construction (tunneling and mining contractor), and RESPEC (geotechnical investigations).

The primary items under this scope of work include:

- Geotechnical site investigations and testing to support the development of the design, including but not limited to geotechnical mapping, core drilling, and testing analysis and modeling to support the design of underground laboratory and underground infrastructure
- Design of excavations for two LMs and one large cavity on the 4850L and one LM on the 7400L, including rock stability analysis and description of ground support for all openings. All associated drifts and ancillary spaces on both levels are also included.
- Design of liners, floors, and surface finishes to prepare the sites for construction and installation of research equipment and instrumentation
- Development of full cost estimates and risk analyses for items within scope of work

**Construction Manager.** A Construction Manager (CM) firm, a joint venture of McCarthy and Kiewit, was retained in April 2010 to provide construction management expertise to the Project during Preliminary and Final Design. During the design phases, the CM is responsible for providing the following major items within the scope of work:

- Development of independent cost and schedule estimates, reconciled with the estimates provided by each design contractor
- Constructability review of all design documents
- Development of a construction acquisition plan
- Design of integration support to the DUSEL Project staff to integrate cost, schedule, and risk from the major design contracts

The CM contract is structured to continue through construction without requiring rebidding or reselection. During the construction phase, the CM will likely be retained "at risk" for the majority of the DUSEL construction scope. "At risk" refers to a construction delivery method where the CM would hold all of the construction subcontracts and would be wholly responsible for the subcontractors' and the CM's performance. DUSEL would have a single construction contract with the CM in an at-risk arrangement. Certain activities may be more appropriate for an agency relationship, where DUSEL would hold subcontracts, and will be defined through the development of construction acquisition plans.

The CM is working during the design phases to develop acquisition plans and construction sequencing plans. These plans are outlined in Chapter 5.10. The construction sequencing schedule and acquisition plan will look at the procurement of long-lead items as well as strategies to centralize the purchase commodities and materials to reduce overhead across the subcontractors for the Project.

The early involvement of the CM in the preconstruction phases provides the CM early insight into the DUSEL requirements and design and allows them to provide feedback to the Project on the design and its constructability, which will reduce future change orders due to ambiguities in the design.

#### 5.1.5.5 Other Outsourced Services

Other outsourced service contracts were used in the development of the Preliminary Design, primarily supporting the DUSEL Facility Project Team in architecture, design contract procurement, and engineering support. Specifically, Oppenheim Lewis Inc. (OLI) of San Francisco, California, provided leadership to the design and construction management contract procurement process, design integration, cost estimation, and VE consulting. OLI was supported by subconsultant Dangermond Keane Architecture (DKA) of Portland, Oregon. DKA provided support to the development of the facility design contract scopes of work, developed initial site programming and requirements, and developed the DUSEL architecture long-range development plan along with DUSEL Facility Operations concepts. Finally, consulting contracts were used for geotechnical engineering, excavation design, and risk management and risk financing support. This type of support is planned to continue forward into the Final Design phase and Construction, as the DUSEL Project regards these outsourced services support arrangements to be more cost and technically effective than relying solely on in-house support staff.

#### 5.1.5.6 DUSEL Facility Integrated Product Team Structure

The overall DUSEL Facility team is organized around cross-functional work teams called Integrated Product Teams (IPTs). These teams are centered primarily on the individual design scopes and construction management, with an overall integration IPT to oversee and manage the scope, cost, schedule, and technical integration of the overall facility design efforts. Each IPT is directly responsible for its scope, schedule, budget, and deliverables for its assigned WBS elements. In conjunction with design activities, these teams meet weekly and include members of the DUSEL Facility team, outsourced architecture and engineering contractors, and science and systems engineering. Formal risk, action items, Trade Studies, engineering issues, and VE database structures that are available to the entire team were used during the Preliminary Design phase and will continue into the Transition and Final Design phases. Figure 5.1.5.6 outlines the DUSEL Facility IPT structure.

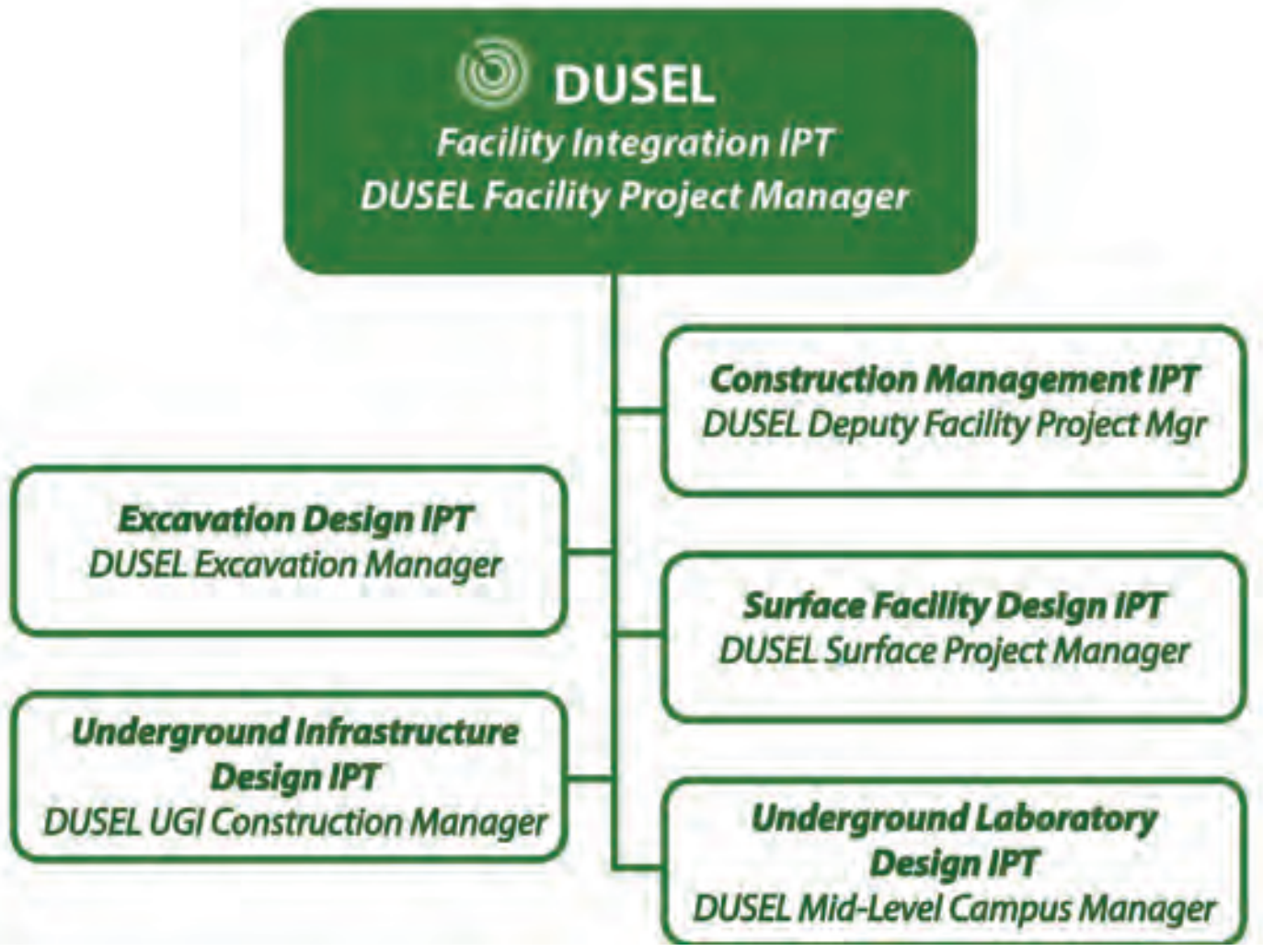

**Figure 5.1.5.6** DUSEL Facility design development Integrated Product Teams. [DKA]

### 5.1.6 Design Development Approach

In fall 2008 and spring of 2009, DKA developed an initial Facility Master Plan and Architectural Program that laid out broad concepts for campus access and an order-of-magnitude estimate of surface space needs. The programming effort was based on interviews with likely science teams, visits and examinations of other underground laboratories, and interviews with DUSEL and SDSTA staff and scientists. The preliminary Master Plan defined the broad functional and space requirements for the Facility. This document also formed a scope basis for securing consultants to perform detailed assessments of the site and to prepare the PDR design.

Concurrent with the development of the PDR development and Master Plan, a *Long Range Development Plan (LRDP)* (Appendix 5.B) was begun. The LRDP outlines the broad current and future development program and envelope for the surface and underground portions of DUSEL. It also defines the development vision and broad design concepts for the Facility.

The 7400L design was limited to a Conceptual level of detail as well, since the level was not accessible during the Preliminary Design phase. The surface work included an assessment of most existing buildings and utilities, prioritized according to likely candidates for reuse and/or importance in the function of underground activities. The surface design included an expansion of the Surface Master Plan, based on detailed assessment and subsequent design work. The needs and requirements of the scientific community are discussed further in Volume 3, *Science and Engineering Research Program*, of this PDR.

#### 5.1.6.1 Site Assessment

The design consultants were hired to perform an in-depth site assessment of the existing facility to prepare the PDR design. These assessment activities were focused on understanding underground geotechnical characteristics and rock quality, the state of the underground infrastructure, and the condition of the existing Surface Facility and infrastructure.

In January 2009, DUSEL contracted RESPEC to perform geotechnical assessments of the 300L and the 4850L. Geologic mapping and laser scanning were performed on these two levels to develop an improved profile of the underground rock characteristics to support design. In early 2009, the Project determined that a planned campus at the 300L was not required by science. Experimenters were surveyed and they suggested that the Project invest in the 4850L and 7400L Campuses instead. Further geotechnical investigations were then targeted only at the 4850L with a successful core drilling campaign to collect rock samples for in situ and laboratory testing and analysis to further characterize the 4850L MLL Campus for the siting of LM-1, LM-2, and LC-1. The investigation results are reported in Chapter 5.3.

In March 2009, a contract for UGI site assessment and design was awarded to Arup USA to investigate the various elements of the UGI and document their condition to support UGI Preliminary Designs. The assessment results are outlined in Chapter 5.4 and include reviews of systems such as shafts and hoisting, ventilation, hydrology, dewatering, health and safety, electrical, cyberinfrastructure, and wastewater treatment.

Finally, an assessment and design contract was awarded in May 2009 to a team led by HDR CUH2A to assess the existing surface buildings and site infrastructure for the Ross, Yates, and portions of the Ellison Campuses. Due to funding limitations, the surface site assessment was divided into three phases, of which two have been completed to support the Preliminary Design report. The third phase will be performed in advance of Final Design. The surface assessment results are described in Chapter 5.2.

### 5.1.6.2 Basis of Estimate and Facility Scoping

In June of 2009, the Arup UGI team was tasked with completing a Basis of Estimate (BOE) study to develop a detailed construction estimate and schedule for development of the DUSEL underground laboratory facility. The BOE scope included three LMs on the 4850L, one large cavity on the 4850L and one LM on the 7400L. In addition, provisions for future expansion to a total of three large cavities on the 4850L and three LMs on the 7400L were included. The resulting cost estimate was not within MREFC funding targets. The BOE scope was reduced to fit the MREFC funding profile. The scope and estimates were updated twice after the initial report was completed starting in late 2009 and completing in February 2010 to form the “BOE Option B” scope, which became the MREFC scope outlined in this PDR.

A similar study to develop a construction scope for the DUSEL Surface Campuses was led by DKA, working in close coordination with HDR CUH2A, the Surface Facility design contractor. The resulting scope is the Surface Facility scope outlined in Chapter 5.2.

### 5.1.6.3 Preliminary Design Development

As the underground and surface BOE scoping studies came to a conclusion in late 2009, the design development efforts accelerated for all scopes, with plans to complete the Preliminary Designs in late 2010. The four main design contracts were focused on the development of the following design deliverables:

1. Basis of Design (BOD)
2. 30% Preliminary Design
3. 60% Preliminary Design
4. 90% Preliminary Design
5. 100% Preliminary Design

For each design stage, the deliverables included an updated BOD, design drawings and specifications, a detailed construction cost estimate using a bottom-up estimate approach, a construction schedule including construction sequencing recommendations, and a risk analysis. At each milestone, the consultant teams delivered these designs for review and approval by the DUSEL Project team. Design reviews included presentations by designers as well as in-depth analysis of BOD documentation and drawings. Iterations to the designs were performed according to feedback by DUSEL design team leads.

In advance of the 60% design delivery, the McCarthy Kiewit construction management team was added to develop independent cost and schedule estimates at 60%, 90%, and 100% Preliminary Design and the 7400L Conceptual Design Report (CDR) delivery to compare directly with the designer estimates during formal estimate reconciliation and VE sessions with the DUSEL Project staff and design firms. The preconstruction services obtained from McCarthy Kiewit included constructability analyses at 90% Preliminary Design and also the development of a construction acquisition plan. The first cost and schedule reconciliation and VE discussions were held for the 60% Preliminary Design deliverable.

### 5.1.6.4 Value Engineering

VE has been important during the Preliminary Design phase to manage the design and requirements within the Project budget. At each of the major estimating milestones during Preliminary Design, a specific and thorough VE process was implemented and is documented in the Volume 9, *Systems*

*Engineering*, and in the DUSEL Value Engineering Management Plan (VEMP) included in Appendix 9.AB. VE is the process in which alternative approaches to accomplish the design requirements are generated and evaluated for cost and constructability. These alternatives provide value to the Project—most often by decreasing the estimated construction costs to fit with budgetary constraints but also by introducing better ways to satisfy the facility requirements.

At 60% Preliminary Design, the construction cost after reconciliation exceeded the established Design to Cost budget established through the BOE process in late 2009. VE items were incorporated into the design and reduced the estimated Project cost to realign the facility designs with NSF MREFC cost targets. DUSEL issued Design to Cost targets to the design firms and CM to provide direction on cost targets expected to be achieved with the Preliminary Design. Along with reconciliation, the VE process was repeated at the 90% and 100% Preliminary Design and 7400L CDR delivery stages. In conjunction with the cost reconciliation activities, site rehabilitation scope items more closely aligned with operations activities using R&RA funding were identified. These included work to provide safe access to the existing facility and to reduce overall risk.

All segments of the DUSEL Project were involved in the review and approval of VE. As VE items were ready for implementation into the design, they were reviewed and approved by the DUSEL CCB. Once VE was approved, formal VE Directives were issued, and the list of approved items is included in Appendix 9.AC. These VE Directives were provided in writing to the designers and CM to outline VE guidance for incorporation of VE into the next design stage. The approved Preliminary Design VE items are outlined in Volume 9, *Systems Engineering*.

### 5.1.6.5 Design Trade Studies

A series of design Trade Studies were performed to evaluate various design options during the Preliminary Design phase. These are described and located in Volume 3, *Science and Engineering Research Program*, and Volume 9, *Systems Engineering*.

### 5.1.6.6 Constructability Review

A formal process to review the design for constructability was implemented during the Preliminary Design phase. McCarthy Kiewit reviewed the design at each reconciliation point and provided constructability feedback to the designers. A formal constructability analysis was documented at 90% and 100% Preliminary Design and is included in the corresponding McCarthy Kiewit Deliverable, included as Appendix 2.B. The constructability review included the following elements: drawings and specifications, construction budget and schedule including sequencing, and constructability.

### 5.1.6.7 Design Integration

A key early decision was to divide the DUSEL Project's engineering and design scope into the contract scopes defined in Section 5.1.5 and to openly compete each scope to ensure the DUSEL Project had the best qualified team for each specialized portion. Once the decision was made to contract with multiple firms for the design and engineering services, the Project Team recognized the critical need to define and manage scope interfaces. An initial scope interface document was developed before contracting with the four design teams, with agreement from each of the scope-specific technical representatives. This scope clarification and delineation document was provided with the descriptions of each of the individual scopes of work during the bidding and selection process.

When all design contractors were identified and under contract, an initial two-day Scope Coordination Workshop was held in October 2009 with all DUSEL technical representatives and key members of each of the four design scopes to establish a working relationship and coordination among Project Team members. A substantial portion of the workshop focused on identifying and discussing areas where the design teams anticipated gaps or overlaps among the design scopes. Key areas of responsibility on the highest-priority interfaces were assigned to individual designers to advance critical items to meet the Project schedule milestones. Starting in April 2010, as the CM was added to the Facility Team and throughout the remainder of the Preliminary Design phase, monthly interface discussions were held to resolve open issues and coordinate design changes at the design documents matured. The resulting discussions, questions, and process greatly strengthened the quality of the design product.

In addition to the face-to-face workshops and weekly coordination calls, the Project established several databases available to Project team members—both outsourced contracts and Project staff—through DocuShare to track action items, Trade Studies, VE items, and engineering suggested items.

#### 5.1.6.8 Interface with LBNE

The LBNE project and the LBNE facility at DUSEL play a major role in the overall DUSEL Facility development. Throughout the Preliminary Design period, the LBNE project office from Fermi National Accelerator Laboratory (Fermilab) and other major LBNE collaboration members such as Brookhaven National Laboratory (BNL) have participated directly in the development of facility and interface requirements for DUSEL. LBNE participation included direct attendance at weekly DUSEL teleconferences with the design firms, design reviews, and cost reconciliation meetings. The LBNE project was included throughout the Preliminary Design. LBNE will sponsor two direct staff positions within the DUSEL Facility team—a Facility Deputy Project Manager role and a Level-3 LBNE project Engineering role. During Preliminary Design, LBNE funded two specific tasks with the DUSEL design contractors through the DUSEL Project. The first task was a study by Golder Associates to analyze the feasibility of alternative shapes and sizes for large cavities on the 4850L. The second was a study by HDR to refine requirements and design concepts for accommodation of LBNE Surface Campus facility needs.

The DUSEL Project has assigned a specific Level-3 WBS element to the LBNE Large Cavity, DUS.FAC.LGC. This WBS element captures all the costs directly associated with the Large Cavity. There are infrastructure costs that are shared across the Project that are apportioned to the LBNE. At 60%, 90%, and 100% of the Preliminary Design phase, a detailed apportionment of the LBNE-related costs was developed in collaboration with the LBNE project, including specific algorithms for the apportionment of shared infrastructure. The apportionment results are documented and managed by the DUSEL Project Controls team in association with the overall Project cost estimation process.

### 5.1.7 Codes and Standards

DUSEL has special challenges when it comes to codes and standards. Nearly unprecedented in the United States as a facility type, a deep underground science laboratory does not fit neatly into conventional code classifications. The typical lines of authority and the expertise of code officials require special application to address the DUSEL code needs. Furthermore, the DUSEL property is currently half in the city of Lead, South Dakota, and half in Lawrence County. Steps are being taken to clarify the lines of authority and to identify which code sections apply to different parts of the Facility.

#### 5.1.7.1 Authority Having Jurisdiction (AHJ)

Currently, the city of Lead is the Authority Having Jurisdiction (AHJ) for DUSEL at Homestake. The DUSEL Project and its contractor design teams have had a series of meetings with the city of Lead to review codes and permitting requirements. The city of Lead's adopted codes will form the basis for design, and Lead will be the technical review authority. It is anticipated that the city of Lead will utilize additional code review consulting support to address the technical challenges of DUSEL. Likewise, the Project will not rely solely on the city of Lead and its consultants for technical expertise. DUSEL will form a series of technical committees that will determine best strategies for code application to different parts of the Facility, recommend standards that should be applied in addition to regulations, and perform peer reviews on selected design elements.

The Sanford Laboratory site project lies within a National Historic District, and the DUSEL Project is subject to review by the South Dakota State Historic Preservation Office (SD SHPO) for compliance with The National Historic Preservation Act of 1966 as Amended.

Environmental permitting for water and sewer main and service lines shall be designed and installed in accordance with the rules and regulations of the South Dakota Department of Environment and Natural Resources (SD DENR).

#### 5.1.7.2 Codes and Standards Approach

Due to the unique nature of the Facility and the different regulatory agencies involved, there will be requirements that conflict with one another, are not able to be met, or are not appropriate to the Facility. In addition, there will be areas that require interpretation as to how different codes will or will not apply.

A model list of additional codes and standards has been developed by DUSEL Environment, Health, and Safety (EH&S). This list will form a starting point for analysis of DUSEL Facility designs, and it is expected that design teams will use this list as a starting point for their own code analyses; however, it is not meant to be exhaustive, and design teams are required to recommend changes or additions to this list. The current design teams preparing the PDR for the DUSEL MREFC-funded Construction have performed a detailed analysis of life safety and building codes as a basis for their work.

#### 5.1.7.3 Building Codes and Standards

For DUSEL, the city of Lead has adopted the 2009 edition of the International Code Council (ICC) codes, including the International Building Code (IBC), the International Fire Code, the International Electrical Code, and the International Plumbing Code. The IBC suite of codes is clearly applicable to the Surface Facility, but has limited usefulness for an underground facility. Lead has adopted the National Fire Protection Association (NFPA) Section 520 as the applicable code for underground construction to supplement IBC-2009.

The fire-safety systems for this facility are being designed to comply with the applicable NFPA fire-safety codes and standards, which are referenced by the IBC, including:

- NFPA 13 2007, Standard for the Installation of Sprinkler Systems
- NFPA 14 2007, Standard for the Installation of Standpipe and Hose Systems
- NFPA 70 2008, National Electrical Code
- NFPA 72 2007, National Fire Alarm Code

- NFPA 80 2007, Standard for Fire Doors and Other Opening Protectives
- NFPA 110 2005, Standard for Emergency and Standby Power Systems

Additionally, the following NFPA codes and standards provide requirements and recommendations generally accepted in the industry as good practice for specialized use and storage of hazardous liquids, including cryogens, and fire-safety requirements for underground buildings and laboratory hazards:

- NFPA 520 2010, Standard on Subterranean Spaces
- NFPA 45 2004, Standard on Fire Protection for Laboratories Using Chemicals
- NFPA 55 2010, Compressed Gases and Cryogenic Fluids Code
- NFPA 30 2008, Flammable and Combustible Liquids Code

Where NFPA codes and standards reference NFPA 101, Life Safety Code, or NFPA 5000, Building Construction and Safety Code, as the base building code, i.e., for egress, it is proposed that either NFPA 520 or the IBC take precedence.

#### 5.1.7.3.1 Occupational Health and Safety

As the DUSEL Facility will be an occupied underground facility and not an operating mine, the Occupation Safety and Health Administration's (OSHA) regulations 29 CFR 1926 and 29 CFR 1910 are considered the most appropriate standards for health and safety, particularly for surface activities. The Mine Safety and Health Administration's 30 CFR standards will be employed as a best practice for underground activities when the OSHA standards do not sufficiently address a given hazard. Additional details concerning occupational health and safety can be found in Volume 6, *Integrated Environment, Health, and Safety Management*.

#### 5.1.7.4 National Register of Historic Places and SD SHPO

Nearly all of the former Homestake Gold Mine is listed in the National Register of Historic Places (NRHP). The DUSEL portion of the District includes all of the Ellison Campus and most of the Ross and Yates Campuses. The nomination was prepared in July 1974 and the district was formally listed in December 1974. A 1998/2000 amendment to the nomination expands the District further and extends the period of significance from 1920 to 1948. This extension allows for the inclusion of the Ross and Yates Campuses, which were developed after 1920, and for the change from stamping crushers to rotary crushers. Although small portions of the site are outside the boundaries of the District, they will be considered historic. The somewhat arbitrary existing boundaries follow paved streets and extensions of those streets; during the review process with the SHPO, it is highly likely that the portions of the site outside the current boundaries will be evaluated and found to have historic significance.

National Register designation mandates that any activities receiving federal funds or licensing that have an effect on the property go through the Section 106 review process of the National Historic Preservation Act. The net result of this review is that any work on the historic property—buildings, site, or equipment—must comply with the rules and regulations of the SD SHPO. The SD SHPO regulations require compliance with The National Historic Preservation Act of 1966 as Amended, which generally means DUSEL must meet The Secretary of the Interior's Standards for the Treatment of Historic Properties and, more specifically, the Standards of Rehabilitation.

#### 5.1.7.5 Accessibility—The Architectural Barriers Act (ABA) and Americans with Disabilities Act of 1990 (ADA)

As a federally funded Project, DUSEL shall comply with accessibility standards under the Architectural Barriers Act (ABA), which applies to facilities designed, built, altered, or leased with federal funds. The ABA Standards, adopted by the United States (U.S.) General Services Administration (GSA) and effective on May 8, 2006, were developed by the U.S. Access Board. Also, since the Laboratory property is owned by the state of South Dakota, the DUSEL Project must meet the requirements of the Americans with Disabilities Act of 1990 (ADA) Title II, which covers state-owned government facilities (Title 28, Code of Federal Regulations, Part 35). DUSEL will need to comply with the ADA Standards, found in the Appendix to the ADA (Appendix A of Title 28, Code of Federal Regulations, Part 36). The ADA design standards are contained in the 2010 Americans with Disabilities Act Accessibility Guidelines for Buildings and Facilities (ADAAG), which were developed by the U.S. Access Board, and accepted by the Department of Justice.

The current legally enforceable version of the ADA is the 2010 edition. All facilities constructed after March 15, 2012, which includes the DUSEL Facility, are required to comply with the 2010 version. Facilities completed before that date are considered “existing facilities” and must be brought up to ADA standards when significantly altered. There is a limited exception for registered historic facilities, when the State Historic Preservation Office (SHPO) agrees that modifications to meet ADA Standards would destroy the historic value of the Facility. In such cases, alternative accessibility means may be employed. DUSEL has adopted a formal Facility ADA Policy, addressing these issues, and is included as Appendix 5.C (DUSEL Facility ADA Policy). It is understood that certain portions of the underground laboratory facility will be difficult to make fully compliant with ADA regulations. The DUSEL Facility ADA Policy does address this issue and the approach to managing these exceptions.

### 5.1.8 Budget, Schedule, Risk Overview

#### 5.1.8.1 Facility Construction Budget

The DUSEL Project budget is discussed in Volume 2, *Cost, Schedule, and Staffing.* The specific details of the budget are excluded from this section.

#### 5.1.8.2 Construction Cost Estimation Process

The detailed plans for the cost estimating process for the Project budget are discussed in Volume 2. This section is intended to address the cost estimating during the facility design process, including the independent cost estimate that is provided by the CM. This process was developed in partnership with the DUSEL Project Controls and supports their process for developing a full DUSEL Project budget.

Project Controls provided the designers and CM detailed instructions and baseline formats to use in developing the facility cost estimates. The Construction Specifications Institute MasterFormat (CSI) 2004 has been selected for DUSEL and all estimates confirm to this standard.

Starting at 30% Preliminary Design for the surface scope and at 60% Preliminary Design for the underground design scopes, each design team prepared a full construction cost estimate and a parallel, independent cost estimate was generated by the CM. The detail of the two estimates for each scope of work was reconciled. The reconciliation process included a detailed item-by-item review with both estimating teams and a third party facilitator, OLI.

Through the Preliminary Design phase of the Project, the estimates for each scope have included a recommended contingency from the designer that corresponds with the risks and assumed quantity of unknown information. Contingency levels are presented in Volume 2, *Cost, Schedule, and Staffing*. These contingencies are recommendations and the DUSEL Project has factored the designer's recommendations into the development of the contingency values used for the construction related work.

#### 5.1.8.3 Construction Schedule Overview

The complete construction schedule is provided in Volume 2. In addition, a summary of the construction sequencing is outlined in Chapter 5.10. Based upon an assumption that construction will begin in February 2014, initial activities will include the completion of waste rock handling systems and internal ventilation systems to support excavation and infrastructure construction. In mid-2014, the excavations of tunnels, the LMs at the 4850L, and the large cavity to support the LBNE will begin as part of underground construction activities along with improvements to the underground infrastructure. Excavation activities on the 4850L include LM-1, LM-2, and LC-1.

In parallel with excavation activities in 2014-2016, the rehabilitation of the Yates Shaft will include the installation of a Supercage conveyance to provide for movement of construction and large experiment equipment to the 4850L.

In May 2018, facility outfitting of the two LMs will complete and early access for science will begin for experiment installation. In January 2019, early access for experiment installation for LBNE in LC-1 will begin. On the DLL, excavation activities will occur in 2016 to 2018 with facility outfitting complete in April 2019, allowing early science access and commencement of experiment equipment installation.

The Surface Facility construction activities are scheduled to occur across this entire construction duration. The surface infrastructure upgrades and construction that are required to support underground construction will take place in 2014-2015 and the construction of the new facilities, including the SCSE, are currently scheduled to occur in 2017-2018. The renovations of existing facilities and the site finish are scheduled to occur at the end of the Construction phase, approximately 2019-2020.

#### 5.1.8.4 Risk Management

In close coordination with the DUSEL Systems Engineering Team, the Facility Team provides primary input to the risk-management process, including supporting risk board discussions, identification of risks, evaluation of probability and impact, and outlining mitigation plans for facility-related risks. This includes directly contributing to the development of the Risk Register and developing analyses to assess risks for their impact on cost and schedule performance. The overall risk-management process is outlined in the DUSEL Risk Management Plan, Appendix 9.C and discussed in Volume 9, *Systems Engineering*. Furthermore, each facility design contractor performed formal risk assessments throughout the development of Preliminary Designs. These assessments resulted in Risk Registry entries that were captured and managed in the DUSEL Risk Registry and provided contingency assessments with associated confidence values for a given level of cost and schedule contingency for each design scope. Again, these assessments provided by the design contractors were incorporated as appropriate into the overall DUSEL Risk Registry, management approach, and development of the baseline cost and schedule estimates for construction.

### 5.1.9 Quality Assurance/Quality Control

During the Preliminary Design phase, a Quality Assurance Surveillance Plan (QASP) was developed to provide a systematic method to evaluate and monitor the performance for the stated contract, including what will be monitored, surveillance methods, and how monitoring results are documented. Each contractor is expected to have internal quality assurance plans as well. The DUSEL QASP is centered on ensuring that deliverables from contractors are of an acceptable level of quality before they are accepted by the Project. The QASP does not detail how the contractor accomplishes the work. Rather, the QASP is created with the premise that the contractor is responsible for management and quality control actions to meet the terms of the contract. It is DUSEL’s responsibility to be objective, fair, and consistent in evaluating performance.

#### 5.1.9.1 Performance Standards

Performance standards define desired services. DUSEL performs surveillance to determine whether the contractor meets or does not meet these standards. Surveillance includes ensuring that scope requirements as defined under the WBS and the Project requirements have been fully addressed. It also means ensuring that the design is at the designated level of maturity, with an achievable work plan up to and completing the design phase. This includes cost/schedule estimates, meeting design-to-cost parameters, and risk analysis—discovering whether all risks have been identified and risk-based contingency recommendations provided.

#### 5.1.9.2 Method of Surveillance

The DUSEL technical representatives use the surveillance methods listed below in the administration of this QASP. The DUSEL Project’s approach to review and evaluation of the deliverables and thus the contractor’s performance is based on:

- An understanding of the performance-based nature of the contract and the expectations for all major deliverables
- Knowledge of the contractor’s performance baseline in response to the contract requirements
- Awareness of the type and level of associated risks and hazards
- Insight into the technical and management approaches to mitigating programmatic risks and controlling hazards
- Constructability, cost effectiveness, and interface with other major plan components

Evaluation and oversight efforts identify areas with indications of poor or suspect contractor performance as early as possible to ensure that the contractor is addressing these areas before a performance issue results in deliverables being delivered late or with poor quality. In general, the DUSEL Project’s intent is to allow the contractor to perform to or exceed the contract requirements and to hold the contractor accountable for providing deliverables that are responsive to the Project’s requirements within the established cost and schedule baseline.

### 5.1.10 Management of the DUSEL Facility

The operation of the Facility during and after construction has been a key consideration during the design process—this is true for operations from both the facility perspective as well as the usability of the laboratory by the scientific community. The life-cycle costs were evaluated during the design process, including consideration of VE to optimize the design that will be operated by DUSEL staff and experimenters for years to come. A newly established Limited Liability Corporation (LLC) will construct and operate the DUSEL laboratory. The DUSEL operations and maintenance activities are more fully described in Volume 10, *Operations Plans*, including details on the execution of R&RA funded activities that are related to the DUSEL Facility.

## 5.2 Surface Facility and Infrastructure

This chapter outlines the DUSEL Preliminary Design for the Surface Facility. The majority of the existing surface facilities have been assessed through two phases of site assessment to inform the design process. The Surface design has been completed to 30% of the Preliminary Design and reflects the known existing conditions and incorporates the science needs documented in the requirements baseline. The following chapter addresses improvements for the surface infrastructure, including the adaptive reuse of most existing buildings and removal of a small portion of the existing structures. The chapter addresses the construction of new facilities and replacement of surface infrastructure to support both surface and underground operations. The Surface Facility design described herein conforms to the allocated construction funding targets provided by the NSF and represents a maturity level of 10% of "construction ready" documents.

### 5.2.0 Overview

The DUSEL Surface Complex is composed of three main campuses. The Surface Facility and Infrastructure scope of work covered in this chapter includes all facility and infrastructure systems on the surface complex; all utilities and facilities underground are covered in other sections of this Volume. The overall Surface Complex covers approximately 186 acres. The three Surface Campuses are the Ross Campus, Yates Campus, and Ellison Campus, which are supported as well by ancillary and support infrastructure and facilities. A diagram of the full complex is included in Figure 5.2.1, below.

This section provides an overview of the Surface Complex planning process, including:

- Surface Facility design strategies
- A review of existing facilities and site assessments
- A proposed site plan for DUSEL
- A review of sustainability strategies to be employed by DUSEL
- A summary of Surface Facility and infrastructure design to date
- Surface Facility program requirements
- A list of scope contingencies
- A summary of deliverables to complete the Preliminary and Final Designs for the DUSEL Surface Campus

### 5.2.1 Surface Complex Overview and Planning Summary

To accomplish the DUSEL mission, the Surface Facility of the former Homestake Gold Mine will be transformed to provide access to the underground for science, facility operations, and maintenance, and to develop the necessary surface support infrastructure for science, operations, and construction of the Facility.

To accomplish this plan, the DUSEL Complex is divided into distinct campuses to separate the science and administration functions from the Operations and Construction functions. This separation is created in a natural fashion by using the two campuses at Yates and Ross. The Yates Campus will be developed as the primary campus for science, administration, and education portions of the DUSEL program. The Ross Campus will be developed as the operations, maintenance, and construction center for the DUSEL Complex. A diagram of the current configuration of the Surface Facility is shown in Figure 5.2.1.

DUSEL will also include an educational program, which will be housed in a new education building, the Sanford Center for Science Education (SCSE). Here, the science program and DUSEL research will be presented to the public through an education and public outreach program designed to enhance the understanding of and appreciation for scientific research within the education community and among the general public.

Although the Ellison Campus was part of the initial assessment completed during Preliminary Design, there is currently no work scheduled for the Ellison Campus. Future work, should funding become available, would include demolition of most of the existing structures and design of new facilities to support the DUSEL program. No further designs have been developed to date for this campus; additional discussion is included in Section 5.2.6.1.

Of the 253,000 gross square feet (gsf) of existing surface structures on the Yates and Ross Campuses, 205,000 gsf will be reused and 48,000 gsf of structures will be removed; there will be 42,000 gsf of new facility construction.

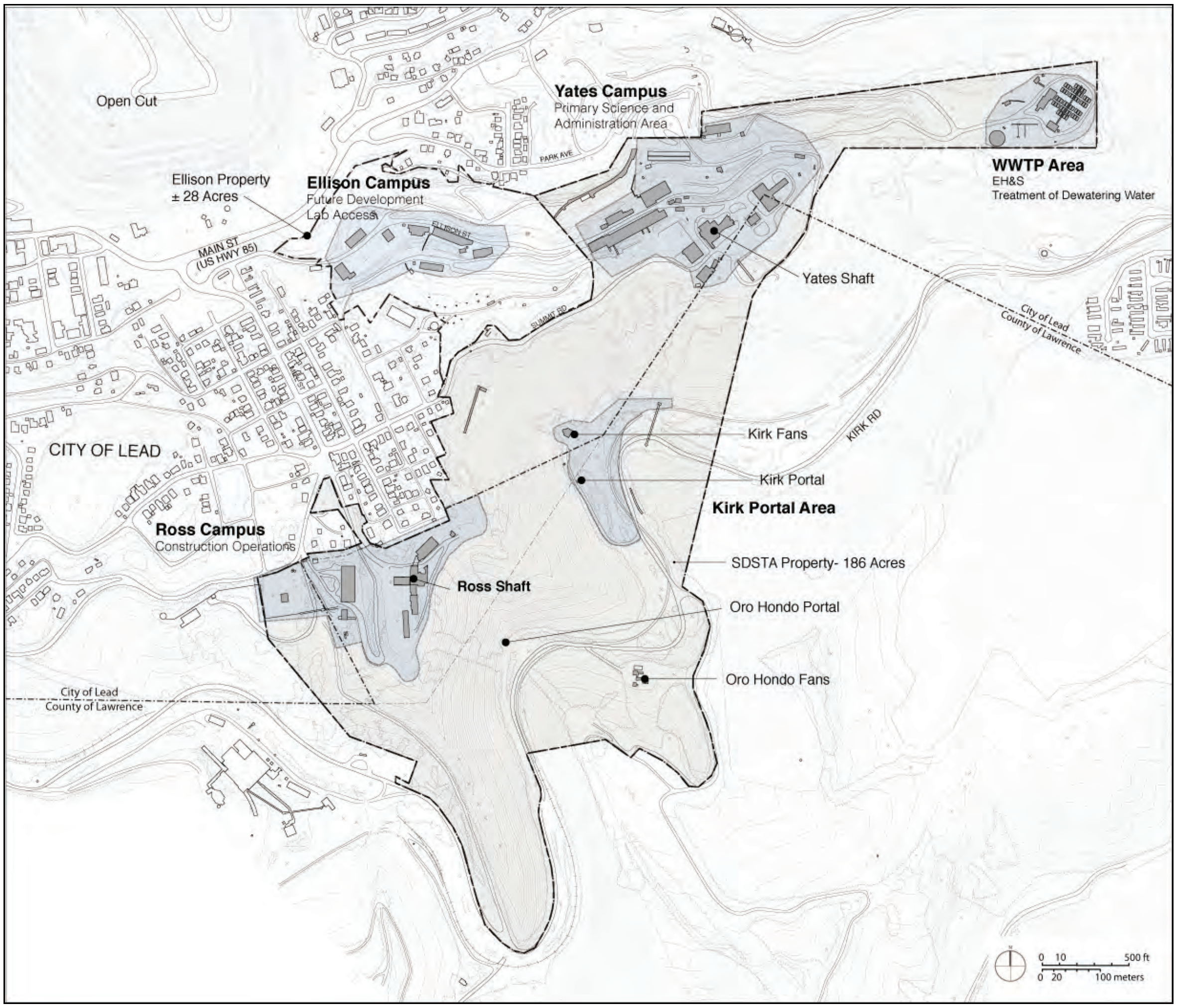

**Figure 5.2.1** DUSEL Campus overview. [DKA]

The DUSEL Project Team evaluated the desired Project scope and programs with the target budget for the Surface Facility and site. The team also evaluated the full DUSEL Complex for reuse opportunities, new construction, and demolition, and reviewed the full scope of the desired programmatic requirements. Based upon those evaluations, the team determined that the full program could not be accommodated within the allocated budget for surface facilities. Therefore, a specific set of alternates was developed, and these scope options could be developed in the future, should funds become available. A detailed discussion on these items can be found in Section 5.2.8, *Scope Options, Scope Contingencies, and Value Engineering*.

**Project Goals and Objectives—Surface Complex**

Goals and objectives established for the Surface Campuses include:

- Provide a surface complex to support a dedicated multidisciplinary laboratory for underground science research.
- Provide a surface complex to support a multiple level underground science campus.
- Develop an overall Master Plan that creates an expandable and sustainable Facility while embracing the rich historic aspects of the region.
- Provide a facility, the SCSE, to house the Education and Outreach Program, which will meet the requirements of the DUSEL education and public outreach program.
- Develop a facility that embraces and enhances the many diversified cultural aspects of the region.
- Develop a facility that will benefit the local community and region.

To assist prioritizing the desired program elements and decisions made during the Preliminary Design, Project values were established by the Surface Project Team to guide the decision-making process as the site is designed and ultimately developed. As design decisions are made for the Surface Campus, these Project values are considered as part of the list of decision-making process:

1. Cost/budget
2. Science/foundation
3. Education/outreach
4. Risk mitigation
5. Schedule for underground support infrastructure
6. Identity/image/historic context
7. Cultural
8. Adaptability
9. Sustainability

### 5.2.1.1 Surface Facility and Infrastructure Requirements

The primary codes governing development of the Surface Campus will be the 2009 International Building Code, International Residential Building Code, International Fire Code, International Electrical Code, and the International Plumbing Code. Further discussion on the applicable codes, standards, and the Authority Having Jurisdiction (AHJ) is included in Volume 6, *Integrated Environment, Health, and Safety Management*.

A draft list of applicable codes and standards, prepared by the DUSEL Environment, Health, and Safety (EH&S) Department, was addressed in the design process and are included in Appendix 6.D, *EH&S*

*Standards for DUSEL/Sanford Laboratory*. These codes and standards are also addressed in the consultant's report included in Appendix 5.D, *Final Report, Phase 1 (30%) Preliminary Design Report (PDR) Surface Facilities and Campus Infrastructure*.

### Surface Facility Requirements

The discussion below is a summary of the key requirements for the Surface Facility. The process of developing these requirements by Systems Engineering is discussed in Volume 9, *Systems Engineering*, of this PDR. The full set of requirements for the Surface Facility and infrastructure systems can be found in Volume 9. The Surface Facility will include the following general departments and organizations:

**Administration.** The Administration Building will house the executive administrative offices, Operations, Engineering, and EH&S staff. The building will have offices, workstations, meeting rooms, and support spaces for the administration of the DUSEL Project. In the current scope of work, the Administration Building will remain as is, with individual upgrades and reconfigurations undertaken on an as-needed basis through funding from operations outside of the Major Research Equipment and Facilities Construction (MREFC) budget.

**Experiment Assembly.** The experiment assembly facility will be a flexible, high-bay preassembly space used to replicate the underground laboratory space and test experiments prior to delivery and installation to the underground laboratory modules (LMs). The assembly facility will be approximately 11,000 gsf and will include a high-bay experiment assembly area of 6,000 gsf, equipment area of 2,000 gsf, and a flexible staging area of 2,000 gsf.

**Dedicated Experiment Facilities.** Dedicated experiment facilities on the Surface Campus will consist of individual clusters of space for approximately seven underground experiments. Each cluster will be fitted out for the specific needs and makeup of each experiment and may consist of experiment control rooms, offices, and laboratory technician workstations. These spaces are designed to be flexible to accommodate various or changing experiments.

**Shared Experiment Facilities and Infrastructure.** In addition to the assembly areas, shared experiment facilities consisting of spaces shared by all experiments will be provided. These include flexible research and support LMs, repair shops, staging, and storage areas. Basic procedural and laboratory bench space will be provided with a small number of fume hoods to serve immediate laboratory needs.

**Geological Archive.** The regional Geological Archive is a cold-storage facility to house the enormous collection of core samples from the Homestake Mining Company. In addition to the existing collection, the archive will house core taken during construction of the DUSEL Facility and will consolidate regional core samples currently stored off site into a comprehensive regional archive, in cooperation with the United States Geological Survey (USGS). In addition to approximately 4,000 sf of cold storage to house 43,000 core samples, an office and records space, layout areas, and a geological analytical laboratory will be provided.

**Construction Operations.** Construction offices, staging areas, and storage will be provided to support the ongoing development of the underground lab. These requirements are currently being developed with McCarthy Kiewit, Construction Manager for DUSEL, and are being considered across all scopes of work.

**Infrastructure to Support Campus Buildings and Facilities.** Approximately 60,000 sf of existing facilities and infrastructure will be upgraded and maintained to support the ongoing operations of and future development for the DUSEL Complex, including hoist and headframe buildings, the Waste Water

Treatment Plant (WWTP), electrical substations and power distribution systems (including normal, standby, and backup power supplies), ventilation for underground facilities, ramps and tunnels, and waste rock handling systems on the surface (see Section 5.4.3.9 for details on waste rock handling).

As an example, the electrical power systems are designed to supply power for normal daily operations (supporting both facility and science power requirements) as well as provide for backup power requirements:

- The normal power requirements for the Surface Campuses and the WWTP will be 5.0MW.
- Support for underground access, such as hoist operations for personnel, equipment, and waste rock removal, ventilation, and compressed air, will be 12.5 MW.
- Underground facility infrastructure systems will require 11.3 MW, and
- Experiments in the underground LMs are anticipated to require 14.5 MW.
- Backup power, produced by on-site surface generators, will provide:
    - 10.0 MW that will be used for hoisting, ventilation fans, compressed air, and WWTP operations for surface support activities, and
    - 6.6 MW to support underground facilities.
    - Backup generators will be diesel powered; storage tanks will be designed for each generator for a 96-hour operation period. Currently, no on-site storage is available.
    - Throughout the Preliminary Design process, the DUSEL staff, the design teams, McCarthy Kiewit, and the science collaboration representatives thoroughly reviewed and evaluated the backup power requirements. The original Basis of Design (BOD) documents proposed 45 MW of backup power, which was reduced, through the design process, to the amounts shown above.

Other infrastructure systems include cyberinfrastructure and communication systems to support science and daily operations, water (potable, industrial, and purified for experiment use), sewer, air flow for ventilation and cooling, and facility dewatering. The requirements for the complete compliment of infrastructure systems can be found within the respective subsections of this Volume; however, the current condition of and Preliminary Design for these critical infrastructure systems are primarily discussed in Chapter 5.4, *Underground Infrastructure Design*.

**Education and Outreach.** The SCSE facility development is covered in detail in Section 5.2.6.3.4 and in Volume 4, *Education and Public Outreach*, of this PDR. The discussion in this Chapter 5.2 focuses solely on the areas of the program that have a facility design impact.

#### Sustainability and Energy Efficiency

The DUSEL Surface Campus will be developed using the U.S. Green Building Council's (USGBC's) Leadership in Energy & Environment Design (LEED) ranking and evaluation system for the two new buildings and the American Society of Landscape Architects (ASLA) and others' Sustainable Sites Initiative (SSI) program for development of the site, landscaping, water management, construction, operations, and maintenance. A more detailed discussion of the surface sustainability plan can be found in Section 5.2.3, *Sustainable Design,* below.

#### Facility Site and Infrastructure Program Requirements

As the DUSEL Project progresses into Final Design, the following design activities will be realized:

- The Yates Campus design will maximize parking area, improve traffic flow—both vehicular and pedestrian—and minimize vehicular/pedestrian conflict locations.
- The site storm-water management plan will be updated for both the Yates and Ross Campuses to comply with State of South Dakota regulations and to augment the SSI.
- New water and sanitary sewer mains and services will be designed on the Yates Campus to meet the demands of the new and renovated facilities.
- A new electrical substation will be designed for the Yates Campus to meet demands not only of the surface but also the underground campus. A new distribution line will be designed to support the underground as well as the surface needs.
- All Yates Campus driving surfaces will be designed for new hard-surface paving to accommodate increased traffic and to control dust.
- The Yates Campus pedestrian and parking areas will be designed with proper lighting for safety and security and to maximize the use of the available area.
- The entire DUSEL Complex will be networked through a prime server facility located in the Yates Dry with a backup facility on the Ross Campus providing state-of-the-art data and communication connectivity. A detailed review of the cyberinfrastructure systems can be found in Chapter 5.5, *Cyberinfrastructure Systems Design*, of this Volume of the PDR.

### 5.2.2 Surface Facility Design Strategy

#### 5.2.2.1 Interfaces with Underground Systems and Infrastructure

Numerous interfaces exist between the surface and underground scopes of work, including water (potable, purified, and industrial), electricity, communications, cyberinfrastructure, ventilation air, shaft heating in the winter months, and laboratory dewatering. These scope interfaces are coordinated and discussed in Chapter 5.1, *Facility Design Overview*, of this PDR. The physical interface point between the surface and the underground will be the shaft collars at the Ross and Yates Shafts. Surface design and construction activities will provide the utilities to the shaft collar, where they will be taken underground as part of the underground infrastructure work.

#### 5.2.2.2 Interfaces with Integrated Suite of Experiments (ISE)

Specific areas of the Surface Campuses will be developed strictly for dedicated and specific ISE use, while other areas will be shared with other science collaborations and DUSEL staff.

The Yates Dry (Figure 5.2.6.3.2-2) will have portions of the building dedicated to experiment control rooms and shared meeting rooms for the collaborations. The Yates Dry will house a server room, as part of the cyberinfrastructure backbone, where data collected by the experiments underground can be received on the surface. The Yates Dry will also have a locker room/dry facility for scientist use, both short term and long term. Long-term use will be on an as-available basis. The Yates Bosses' Office area will be converted to a safety equipment and safety training area, where scientists will be trained and can check out personal protective equipment (PPE) prior to going underground. See Appendix 5.D (Page 4.58 of *Final Report, Phase 1 [30%] Preliminary Design Report [PDR] Surface Facilities and Campus Infrastructure*, prepared by HDR CUH2A), for a conceptual layout of the Yates Dry.

The Foundry (Figure 5.2.6.3.2-3) will be renovated to house laboratory and support modules for the ISE as well as shared areas that will include hazardous material storage, electronic and machine shop, a meeting room, break room, and campus-wide shipping and receiving facility. A small hand tool inventory will be maintained by DUSEL for science use via a check-out system. See Appendix 5.D (Page 4.59 of *Final Report, Phase 1 [30%] Preliminary Design Report [PDR] Surface Facilities and Campus Infrastructure*), for a conceptual layout of the Foundry.

The science community will also interact with staff and the general public through the SCSE, where classrooms and meeting rooms will be available.

### 5.2.3 Sustainable Design

During the Preliminary Design process for the Surface Facility, the Project held a series of community meetings in March 2010 with various constituencies of the Project neighbors and stakeholders to seek feedback from the community and to understand what is most important to the communities relative to site design. The Project invited attendees to two meetings and included participants from the American Indian communities and neighboring businesses, schools, and residential areas. Sustainability was indicated as a high priority through the planning process at both meetings and it was requested that the Project address sustainability and maximize existing resources as much as possible.

Two programs have been selected to be the core of the DUSEL sustainability program: LEED and the newly established SSI.

LEED is a third-party certification program and nationally accepted benchmark for the design, construction, and operation of high-performance green buildings. LEED is a voluntary, consensus-based, market-driven building rating system based on existing proven technology. It evaluates environmental performance from a whole-building perspective over a building’s life cycle, providing a definitive standard for what constitutes a green building. The version referred to as LEED v3 is the version for which the DUSEL Project will be registered.

Although LEED certification is not a requirement of NSF funding, the Project recognizes that sustainable design has value beyond the USGBC certification. Sustainable design seeks to reduce, or preferably eliminate, a negative impact on the environment through planning, design, and operations. Decisions are made throughout the design and planning process regarding materials, energy efficiency, reuse of existing materials, materials that can be reclaimed and recycled after initial use, use of rapidly renewable and responsibly harvested materials, reduced visual and physical impact on the site, and consideration for creating environments that are not harmful to building occupants.

Designing sustainable research and science laboratories is a unique challenge. LEED requirements are typically based on reducing the impact on the environment through reducing a building’s energy use. Laboratories and research facilities are very energy demanding and therefore must be considered by a different set of criteria for sustainable design that includes increased efficiency, reduction of hazardous materials, increase in the use of products with higher recycled content, and improvement of the environment for building occupants. These standards, difficult to achieve in laboratories built on the Earth’s surface, are compounded in complexity for underground laboratories. The Project has therefore stated that all design decisions will be made stressing sustainability criteria, but a LEED rating will not be set as a goal for the underground facilities.

SSI is a pilot program to promote the development of sustainable sites. The initial project selection/pilot project portion of the program runs through 2012. The DUSEL Project was fortunate to be selected as one of the pilot projects for this program. Selection as an SSI Pilot Project will put DUSEL at the forefront in identifying sustainable design principles for multibuilding campuses and postindustrial facilities. The program seeks the development of sustainable sites through a series of design criteria detailed below.

The SSI program is organized through the American Society of Landscape Architects, the Lady Bird Johnson Wildflower Center, and the United States Botanic Garden to promote the development of sustainable sites. The program seeks the development of sustainable sites through the following criteria:

- **Selection**—through a process of site location that preserves existing resources and repairs damaged systems
- **Predesign Assessment and Planning**—by planning for sustainability at the onset of the Project
- **Site Design—Water**—by establishing a design that protects and restores processes and systems associated with the site's hydrology
- **Site Design—Soil and Vegetation**—by establishing a design that protects and restores processes and systems associated with the site's soil and vegetation
- **Site Design—Materials Selection**—by establishing a design that reuses/recycles existing materials and supports sustainable production practices
- **Site Design—Human Health and Well-Being**—by establishing a design that builds a strong community and sense of stewardship
- Development of design that minimizes the effects of **Construction**-related activities
- Development of design that incorporates a long-term sustainable **Maintenance and Operations** plan
- And finally, development of a plan that incorporates **Monitoring and Innovation** planning and techniques for exceptional performance and improvement of the body of knowledge for long-term sustainability

With this understanding of sustainable design, the Project has placed a priority on selections and decisions throughout the design process to reuse and repurpose buildings and materials where feasible, and to make all design decisions with an eye toward sustainability and understanding on the impact to life-cycle costs.

#### 5.2.3.1 LEED Overview

Early in the Preliminary Design process, a workshop was held to familiarize DUSEL staff with the LEED program, determine viability of a sustainable design, and set preliminary goals for the Project. It was determined through this workshop that attaining at least LEED Silver was realistic for the new building construction. This goal was limited to new construction, as there were too many unknowns at that time to consider a LEED goal for renovated structures. As the Project proceeds to Final Design, the consideration of a LEED goal for renovated buildings will be revisited.

On completion of an early review of the Project, a goal of LEED Silver was established as a minimum goal for the two new buildings on the Surface Campuses: the SCSE and the Science Assembly Building. As the Project's Construction Manager, McCarthy Kiewit, has completed its initial review, it appears LEED Gold may be attainable without adding undue cost to the Project. As DUSEL progresses into Final

Design, the design team and the DUSEL surface team will determine specific categories that will be used to establish the LEED goal to be attained. At the present time, DUSEL will strive for LEED Silver at a minimum, with serious consideration being given, during the completion of Preliminary and Final Design, to upgrading this goal to LEED Gold.

For DUSEL, the steps in the LEED process are:

- Build a feasibility matrix of strengths, weaknesses, opportunities, and threats (SWOT) to establish formal goals for the Project.
- Determine which version of LEED fits DUSEL: Multiple Building and Campus Application Guide, or New Construction.
- Formally decide which buildings are to be certified (SCSE and Science Assembly are currently recommended).
- Register buildings online with the USGBC.
- Assign tasks and responsibilities early to ensure inclusion in early design efforts.
- Establish a goal for the first design review with LEED; early submittals allow for changes in Final Design without undue costs.
- Construction review, second review, will establish how well the contractor is fulfilling obligations to the Project.
- The Certification Process will be finalized upon completion of building construction.

A detailed review of the LEED program and potential credit points can be found in Appendix 5.D (Appendix 2, *LEED Workshop Document* of the *Final Report, Phase 1 [30%] Preliminary Design Report [PDR] Surface Facilities and Campus Infrastructure*).

#### 5.2.3.2 Sustainable Sites Initiative Overview

SSI is dedicated to fostering a transformation in land development and management practices that will bring the importance of ecosystem services to the forefront. For the purpose of SSI, land practices are defined as sustainable if they enable natural and built systems to work together to "meet the needs of the present without compromising the ability of future generations to meet their own needs," as stated in the Sustainable Sites Program manual. To this end, the Guiding Principles detailed below not only inform the work of SSI but also inform all aspects of sustainable site development:

- **Do no harm.** Make no changes to the site that will degrade the surrounding environment. Promote projects on sites where previous disturbance or development presents an opportunity to regenerate ecosystem services through a sustainable design.
- **Precautionary principle.** Be cautious in making decisions that could create risk to human and environmental health. Some actions can cause irreversible damage. Examine a full range of alternatives, including no action, and be open to contributions from all affected parties.
- **Design with nature and culture.** Create and implement designs that are responsive to economic, environment, and cultural conditions with respect to the local, regional, and global context.
- **Use a decision-making hierarchy of preservation, conservation, and regulation.** Maximize and mimic the benefits of ecosystem services by preserving existing environmental features, conserving resources in a sustainable manner, and regenerating lost or damaged ecosystem services.

- **Provide regenerative systems as intergenerational equality.** Provide future generations with a sustainable environment supported by regenerative systems and endowed with regenerative resources.
- **Support a living process.** Continuously re-evaluate assumptions and values and adapt to demographic and environmental changes.
- **Use a systems thinking approach.** Understand and value the relationships in an ecosystem and use an approach that reflects and sustains ecosystem services; re-establish the integral and essential relationship between natural processes and human activity.
- **Use a collaborative and ethical approach.** Encourage direct and open communication among colleagues, clients, manufacturers, and users to link long-term sustainability with ethical responsibility.
- **Maintain integrity in leadership and research.** Implement transparent and participatory leadership, develop research with technical rigor, and communicate new findings in a clear, consistent, and timely manner.
- **Foster environmental stewardship.** In all aspects of land development and management, foster an ethic of environmental stewardship—an understanding that responsible management of healthy ecosystems improves the quality of life for present and future generations.

A complete description of the prerequisites and anticipated achievable credits for SSI can be found in Appendix 5.D (Appendix 3 of HDR's *Sustainable Site* of the *Final Report, Phase 1 [30%] Preliminary Design Report [PDR] Surface Facilities and Campus Infrastructure*).

#### 5.2.3.3 Construction Material Considerations during Design

During the design effort, consideration is being given to the sustainability of various construction materials throughout the Project. Some considerations include:

- Embodied energy and embodied greenhouse gases. The production of cement for use in concrete accounts for approximately 5% of global greenhouse gas emissions. Reduction of the amount of cement in DUSEL's concrete mixes through the use of high-volume fly ash will be considered as a means of reducing the Project's carbon footprint.
- Use of locally harvested and manufactured materials and products will be considered.
- Use of rapidly renewable materials will be considered.
- Recycled and reclaimed products will be given consideration. Reuse of materials from buildings and structures to be demolished on site will be considered. For example, some of the old structures have high-quality heavy timber-framed structural members that can be reused in new construction. Old concrete slabs and structures may be pulverized on site and used as base course for new slabs and roadways.
- High-performance materials and assemblies will be used for new building envelopes to reduce energy consumption.
- Wood products will be selected from sustainably harvested sources.
- Low-VOC and formaldehyde-free products will be chosen for building interiors.

#### 5.2.3.4 Water and Energy Use

As part of the SSI, close attention will be paid to DUSEL surface water use for landscaping. Where possible, water will be reclaimed and reused to reduce the amount of potable water consumed for irrigation. Plant species with reduced irrigation demands will be chosen.

DUSEL will consume enormous amounts of energy in the laboratory facilities and in support of the science. Reducing the use of energy in surface buildings will be achieved through high-performance building envelopes in new construction, aggressive retrofitting of existing building envelopes, high-performance mechanical systems, and integrated energy management systems, etc. Belowground, mechanical systems will be right-sized and chosen for efficiency.

Energy being consumed will be recaptured where possible. In-line generators may make use of head pressure when bringing water down the 7,400 vertical feet to DUSEL's lowest level. Water being pumped from the lowest levels of the Facility, warmed by the natural ambient temperatures of the rock, may be used to melt snow in outdoor walkways. Other considerations for reclaimed use of water and energy are being considered across the site, particularly where they have a minimal to low impact on construction cost and reduce future operating costs.

### 5.2.4 Existing Facilities and Site Assessment

Site and facility assessments were performed during Preliminary Design by HDR CUH2A to evaluate the condition of existing facilities and structures on the Ellison, Yates, and Ross Campuses. The assessments reviewed the condition of buildings proposed for continuing present use, new use, or potential demolition. Building assessments were performed in the categories of architectural, structural, mechanical/electrical/plumbing (MEP), civil, environmental, and historic. Site assessments looked at the categories that included civil, landscape, environmental, and historic. Facility-wide utilities such as electrical, steam distribution lines, water, and sewer systems were also assessed. The assessment evaluation was completed in three phases. The detailed reports are included in the appendices of this PDR as noted and are titled:

- *Phase I Report, Site Assessment for Surface Facilities and Campus Infrastructure to Support Laboratory Construction and Operations* (Appendix 5.E)
- *Phase II Site and Surface Facility Assessment Project Report* (Appendix 5.F)
- *Phase II Roof Framing Assessment* (Appendix 5.G)

The sites and facility assessments outlined above and performed during Preliminary Design were completed in the three phases listed above and include a review of the following:

- Buildings proposed for reuse were evaluated for preliminary architectural and full structural, environmental, and historic assessments.
- Buildings proposed for demolition were evaluated for preliminary historic assessments.
- Preliminary MEP assessments were performed on the Ross Substation, #5 Shaft fan, Oro Hondo fan, Oro Hondo substation, and general site utilities for the Ross, Yates, and Ellison Campuses.
- The Waste Water Treatment Plant received preliminary architectural and structural assessments and a full MEP assessment.
- Preliminary civil assessments of the Kirk Portal site and Kirk to Ross access road were also completed.

A final phase of assessment work, scheduled for 2011, will complete environmental, historic, architectural, structural, and MEP evaluations for the balance of the structures and facilities on the site.

#### 5.2.4.1 Building Assessment Results

Results of the building assessment work, as detailed in the three reports referenced above, show that the buildings on the Ross and Yates Campuses were architecturally and structurally generally suitable for reuse or continued use with some upgrades or modifications. However, with the exception of the Drill and Bit Shop and Old Compressor building, the structures on the Ellison Campus were deemed not suitable for reuse, see Figure 5.2.6.1.

#### 5.2.4.2 Site Civil Assessment

Results of the civil assessment found in the *Phase I Report, Site Assessment for Surface Facilities and Campus Infrastructure to Support Laboratory Construction and Operations* (Appendix 5.E) and *Phase II Site and Facility Assessment, Project Report* (Appendix 5.F) showed the following results:

- Water and sewer utilities on both the Ross and Yates Campuses need replacement.
- Roadway and parking lot surfaces need replacement and regrading. Drainage ways and steep slopes need maintenance.
- Retaining walls and transportation structures are in useable condition, with some maintenance, except for two failing retaining walls, one at lower Yates Campus and one on the Ellison Campus.
- Retaining walls and transportation structures need maintenance in the form of drainage improvements and minor repairs to section loss due to rust and erosion.
- Existing fencing and guardrails are a very inconsistent pattern of chain link, wood, and steel; much of the fencing is deteriorating or collapsed.
- Abandoned equipment/scrap-metal piles around the sites represent traffic and health hazards.
- Pedestrian and traffic separation is poorly defined.
- Existing traffic signs are faded and do not meet *Manual of Uniform Traffic Control Devices* (MUTCD) standards.

The Civil Site Assessment recommendations can be found in Appendix 5.E (Section 4, Page 4(1) of the *Phase I Report, Site Assessment for Surface Facilities and Campus Infrastructure to Support Laboratory Construction and Operations*); and Appendix 5.F (Section 2, Page (2.1) – 39 of the *Phase II Site and Facility Assessment Project Report*).

As the Preliminary Design is advanced to 100%, these recommendations will be considered in the development of the Final Design for the Surface Campus.

#### 5.2.4.3 Landscape Assessment

The landscape assessment, found in Appendix 5.E (*Phase I Report, Site Assessment for Surface Facilities and Campus Infrastructure to Support Laboratory Construction and Operations)*; and Appendix 5.F (*Phase II Site and Surface Facility Assessment Project Report)* noted many of the same items as the site civil assessment: drainage issues, erosion concerns, abandoned equipment, and scrap metal. Soil conditions were noted as well as rock escarpments and soil stability concerns.

#### 5.2.4.4 Site Electrical Assessment

The site electrical assessment, detailed in Appendix 5.E (*Phase I Report, Site Assessment for Surface Facilities and Campus Infrastructure to Support Laboratory Construction and Operations)*; and Appendix 5.F (*Phase II Site and Surface Facility Assessment Project Report)* found the electrical distribution condition on all three campuses to range from fair to excellent, depending on the age of the equipment. The Ross Campus recommendations generally consisted of upgrades to increase reliability. The Yates Campus recommendations call for a new substation to replace the old abandoned East Substation. For the Ellison Campus to be suitable for DUSEL programs, it will require a complete rebuild of the substation and distribution systems, as the electrical system has been dismantled.

#### 5.2.4.5 Site MEP Assessment

The site MEP assessment, found in Appendix 5.E (*Phase I Report, Site Assessment for Surface Facilities and Campus Infrastructure to Support Laboratory Construction and Operations)*; and Appendix 5.F (*Phase II Site and Facility Assessment Project Report)* evaluated the natural gas and steam distribution systems. Natural gas is provided to the site at three locations and appears to have the capacity required to meet surface needs as they are currently understood. However, the natural gas supply is an interruptible supply (non-firm) and thus cannot be guaranteed. Either an upgrade to Montana-Dakota Utilities (MDU, local natural gas supplier) supply lines (outside the scope of this Project) or an alternate fuel/heating source will be needed to meet the surface needs. The steam boiler systems have been dismantled and should not be reused. The existing components represent placeholders for routing for new distribution if steam is re-employed.

The site telecommunications service currently is provided by Knology Inc., Rapid City, South Dakota, and a fiber-optic data connection is from the South Dakota Research, Education and Economic Development (REED) Network (see Chapter 5.5, *Cyberinfrastructure Systems Design*, for details on these service providers). Both services are quite new and have historically been very reliable. The site distribution system is a mix of copper and fiber, copper being quite old and fiber very new. The Ross and Yates Campus' recommendations are to increase reliability as the campuses are developed. The Ellison Campus will require a complete build, as no such services exist on the campus.

#### 5.2.4.6 Environmental Assessment

The environmental assessment, found in Appendix 5.F (*Phase II Site and Surface Facility Assessment Project Report)* looked for contamination from lead-based paint (LBP); polychlorinated biphenyls (PCBs) contained in electrical equipment, lubrication oils, and hydraulics; asbestos-containing building materials (ACBMs); heavy metals; the historic presence of petroleum hydrocarbons and chlorinated solvents; molds; historic uncontrolled discharges of domestic sewage; industrial wastewater; and storm-water runoff. Environmental results showed some LBPs in various locations across both the Ross and Yates Campuses. No PCB concentrations above EPA regulatory standards were encountered, and no heavy metals above EPA regulatory standards were found.

#### 5.2.4.7 Historic Assessment

The former Homestake Gold Mine site is a major component of the Lead Historic District. Most of the DUSEL Complex is within the historic district; thus, work on the DUSEL site must conform to the National Historic Preservation Act of 1966, as Amended. These standards recognize that historic buildings and sites must change with time if they are to meet contemporary needs but that alterations to meet these needs can be done in a manner that is sensitive to the historic property. Figure 5.2.4.7-1 is a historic photograph showing the former Homestake Mining Company milling operation and components of the Yates Campus. Figure 5.2.4.7-2 shows the boundaries of the Lead historic district.

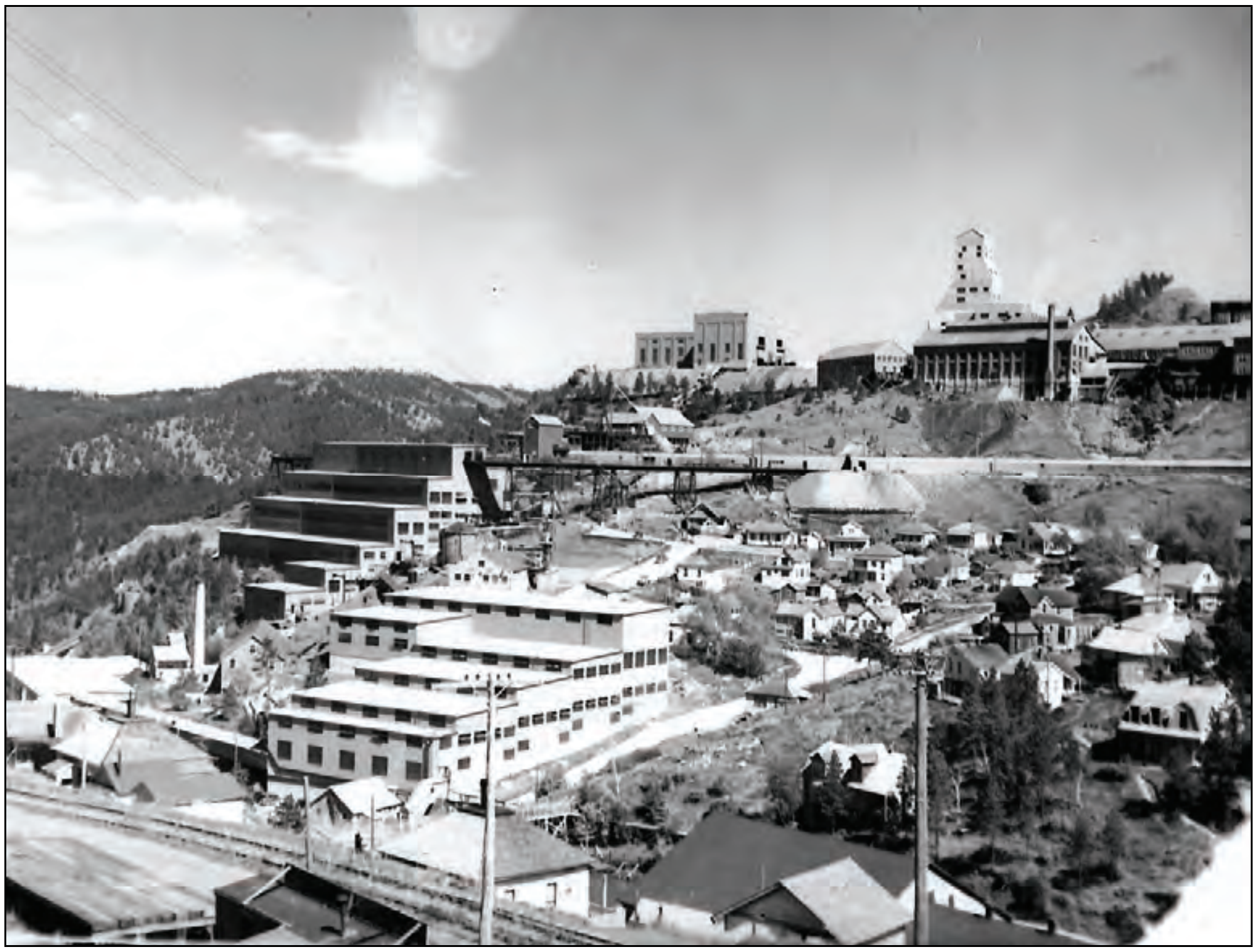

**Figure 5.2.4.7-1** Historic photo of milling operation, Yates Headframe, Hoist, and Foundry. [Courtesy HARCC]

The historic assessment consisted of the full assessment of 10 transcendent and eight support buildings. Transcendent buildings have the most significant historic value and represent an operation that was unique or limited to the site. Support buildings represented a function or activity that, although performed on the site, could have been done off site. Of the 10 transcendent buildings, nine were deemed to have significant historic value while one held only moderate historic value. Seven of the support buildings held moderate historic value, while the eighth has only limited historic value. Sixteen other buildings received a preliminary historic assessment. Two were deemed to have moderate historic value, 13 held limited historic value, and the last was deemed to be of limited historic value.

To assist the DUSEL Project in understanding the historic requirements for the Project, a meeting was held with the South Dakota State Historic Preservation Office (SD SHPO) in June 2010. The DUSEL team provided a Project overview for the SD SHPO staff and took a site tour so the SHPO staff could develop an understanding of the Project. The SD SHPO staff members were pleased, for the most part, with the direction the design team was taking for the Project. They provided recommendations for documentation and preservation options that will need to be addressed during Final Design to meet mitigation requirements for facilities that will ultimately be removed.

It should be noted that the historic assessment prepared for this portion of the overall site assessment is not the formal historic assessment that will be required to comply with the Environmental Impact

Statement (EIS) for the Project. The EIS is being completed by NSF and has been contracted to the Argonne National Laboratory.

The Project risks relating to the historic preservation requirements have been noted and will be evaluated fully as part of the 100% Preliminary Design for the Surface Facility.

The entire historic assessment process and results can be viewed in Appendix 5.E (*Phase I Report, Site Assessment for Surface Facilities and Campus Infrastructure to Support Laboratory Construction and Operations)*, and Appendix 5.F (*Phase II Site and Surface Facility Assessment Project Report)*.

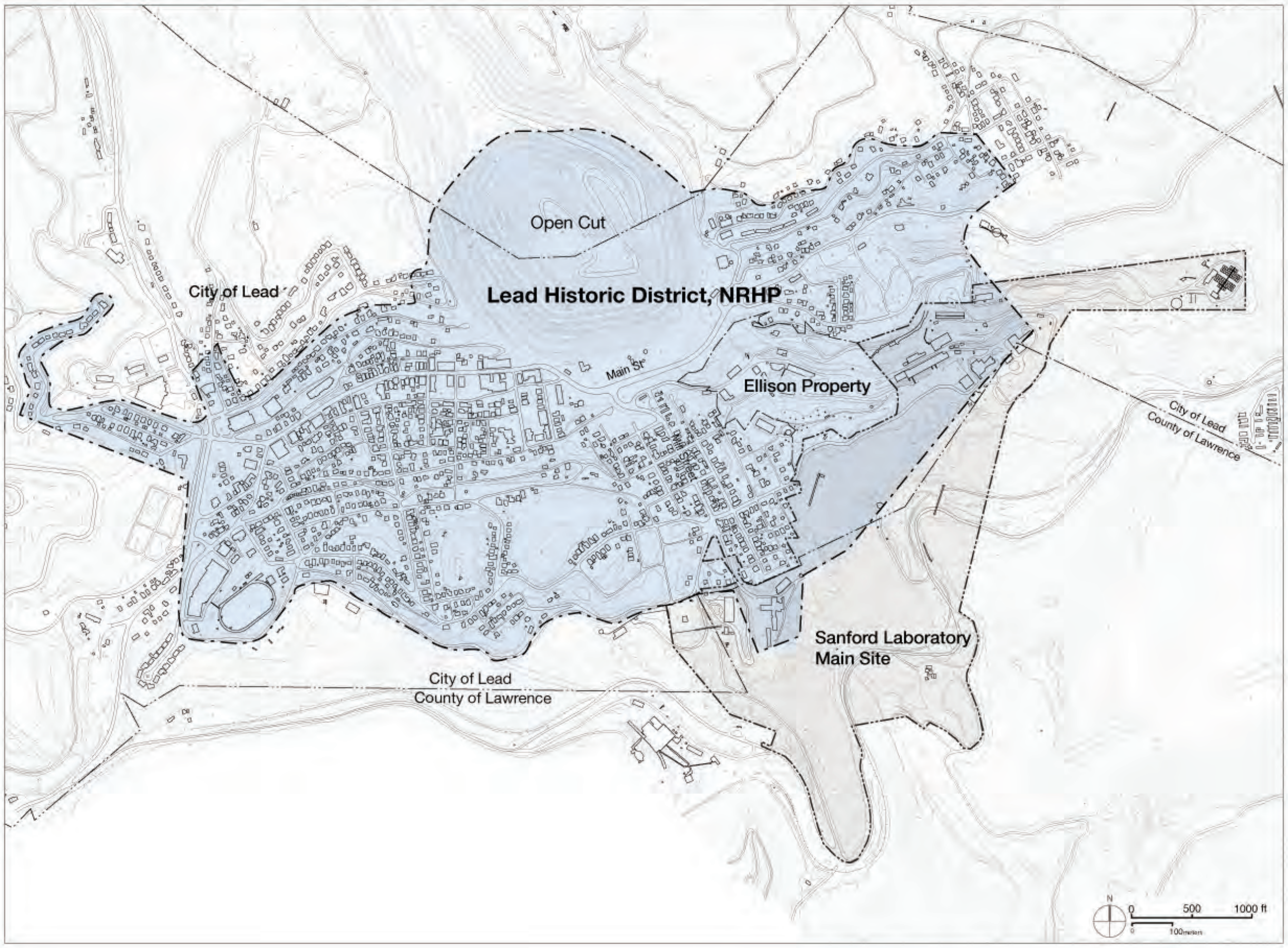

**Figure 5.2.4.7-2** Map of Lead Historic District. [DKA]

## 5.2.5 Site Planning and Campus Master Plan

Site planning efforts have been documented in two parts. First, the *DUSEL Long Range Development Plan (LRDP)* (Appendix 5.B), prepared by Dangermond Keane Architecture (DKA), has been developed, which defines the broad concepts and strategies for current and future DUSEL surface site development. Second, a specific site surface Master Plan has been developed for the MREFC-funded Project, as part of HDR's 30% Preliminary Design effort. See Appendix 5.D (*Final Report, Phase 1 [30%] Preliminary Design Report [PDR] Surface Facilities and Campus Infrastructure*).

#### 5.2.5.1 Ross Campus Access, Traffic, and Parking

Primary construction access to the Ross Campus will be up Mill Street (Figure 5.2.5.1). An alternate (back) route skirting the south side of the ridge upon which DUSEL sits may be used as a winter access when snow makes Mill Street unusable. This route consists of a series of city streets, from South Mill Street to Houston Street, to Pavilion Street and West Summit Street, arriving at Highway 14A and 85. This alternate route will not be the preferred access for construction operations.

Staff and construction crew parking will be in two parking lots on DUSEL property that sit immediately outside the Ross Access Gate. One lot is accessed from Mill Street the other from South Mill Street.

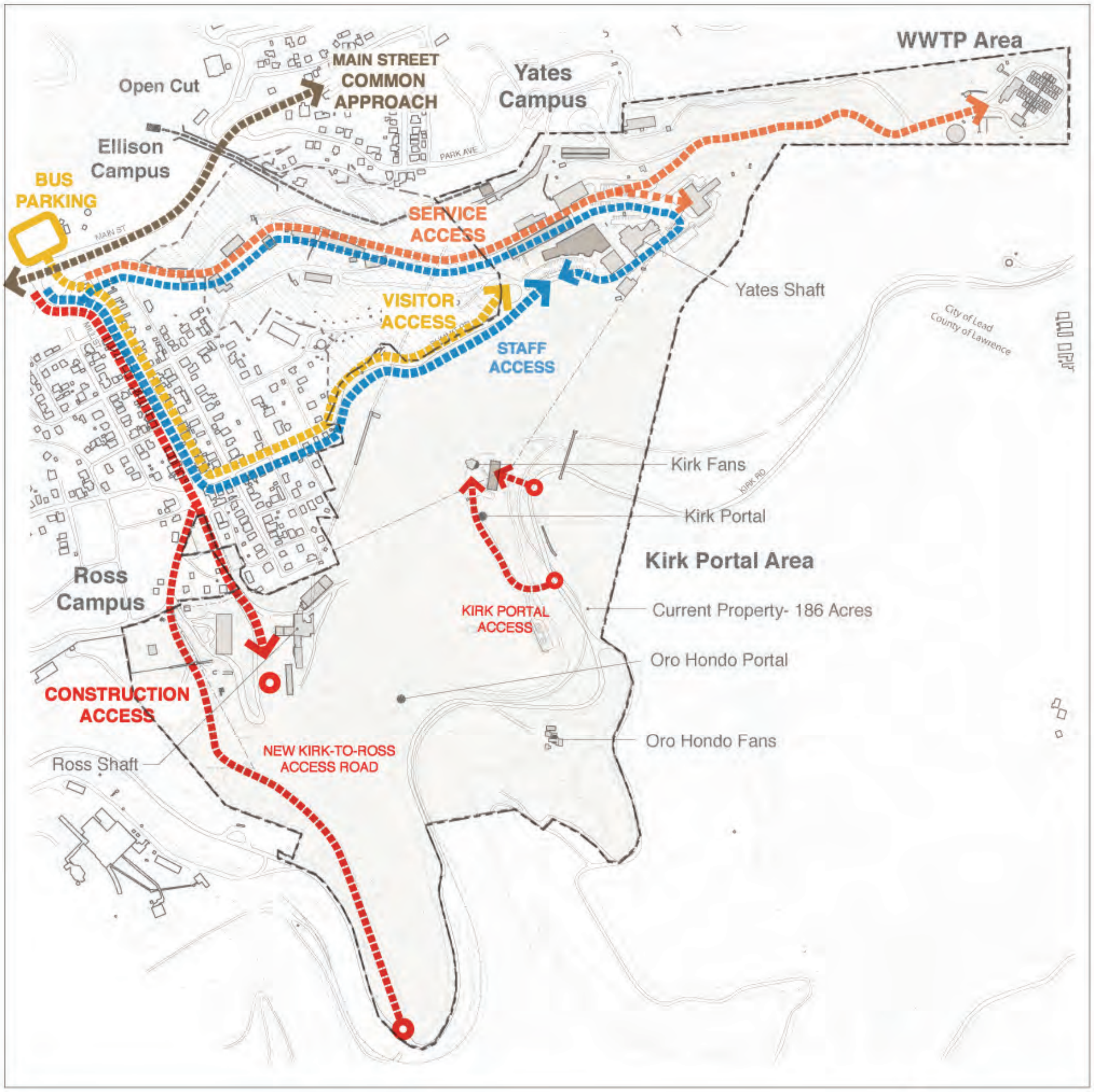

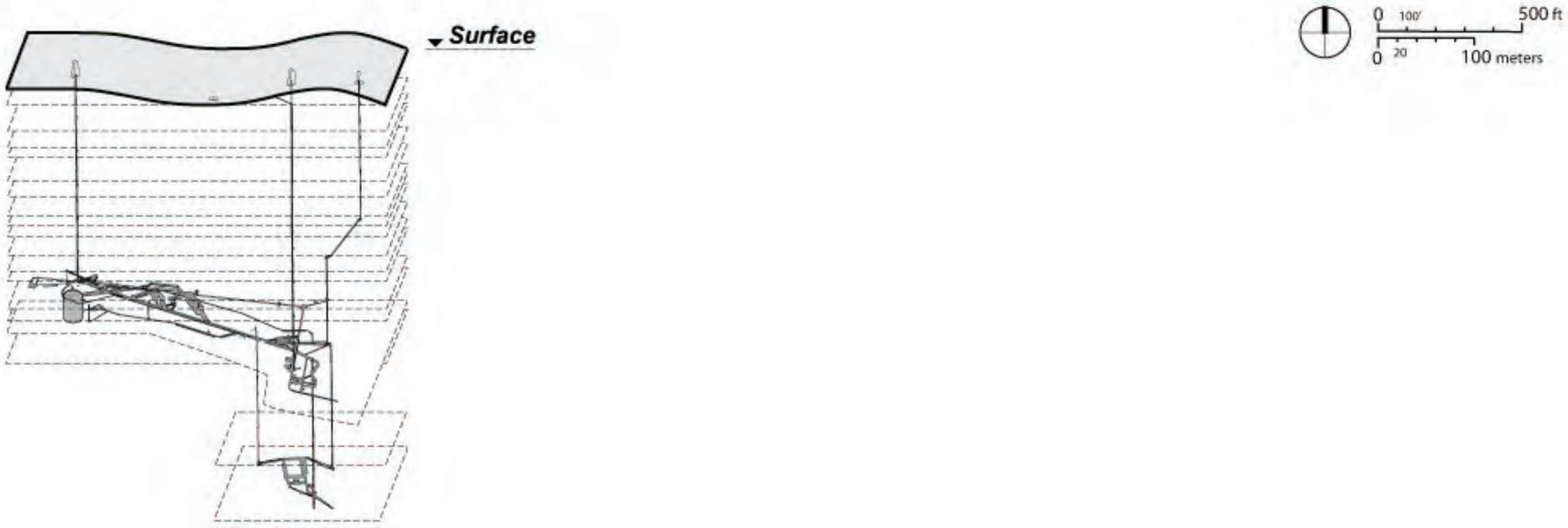

**Figure 5.2.5.1** Future development access and Circulation Plan. [DKA]

### 5.2.5.2 Yates Campus Access, Traffic, and Parking

Initially, the visitor access to the Yates Campus will be up Mill Street to East Summit Street, which leads directly into the Yates parking lot. In the future —a deferred scope option—a new public access will be created along the Ellison Road. This new access will follow the present Ellison Road to the vicinity of the Ellison Boiler Building, where it will proceed uphill to the west end of the Yates parking lot (see Figure 5.2.5.2-1). This new route will eliminate the traffic flow through the residential areas along Mill Street and East Summit Street and will be a much flatter grade than the present Mill Street access, making for safer winter driving. Staff, service, and construction access will be up the Ellison Road to the lower Yates areas.

Parking at the Yates Campus will be at a premium and the design for parking is shown in Figure 5.2.5.2-2. Initial studies for the SCSE indicate 175-200 spaces are needed, including employee spaces. Thus, parking areas will be established in all reasonable areas across the Yates Campus. Primary parking will be in the main Yates parking lot adjacent to the SCSE. Smaller parking areas will be developed in middle and lower Yates Campus areas, where feasible. These will include, for example, spaces for security personnel off the lower level of the Yates Dry (proposed location of the Security offices); spaces for warehouse personnel near the Foundry (proposed location for Receiving); spaces for hoist operators and shaft technicians in the middle Yates yard around the Hoist and Headframe; spaces for scientists near the LUX Surface Facility (Old Homestake Warehouse) and around the new Science Assembly Building.

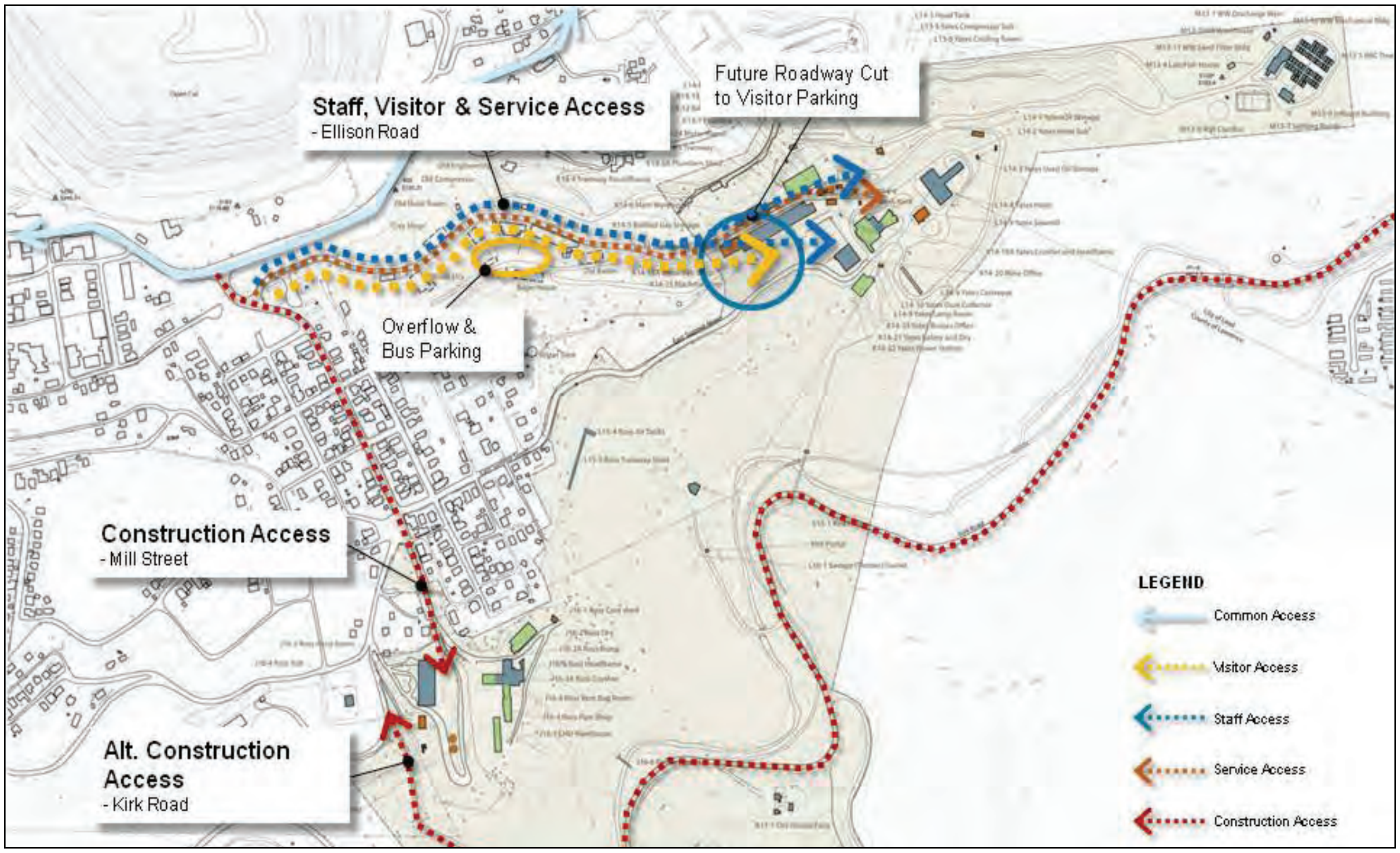

**Figure 5.2.5.2-1** Future access (deferred scope) to the Yates Campus. [DKA]

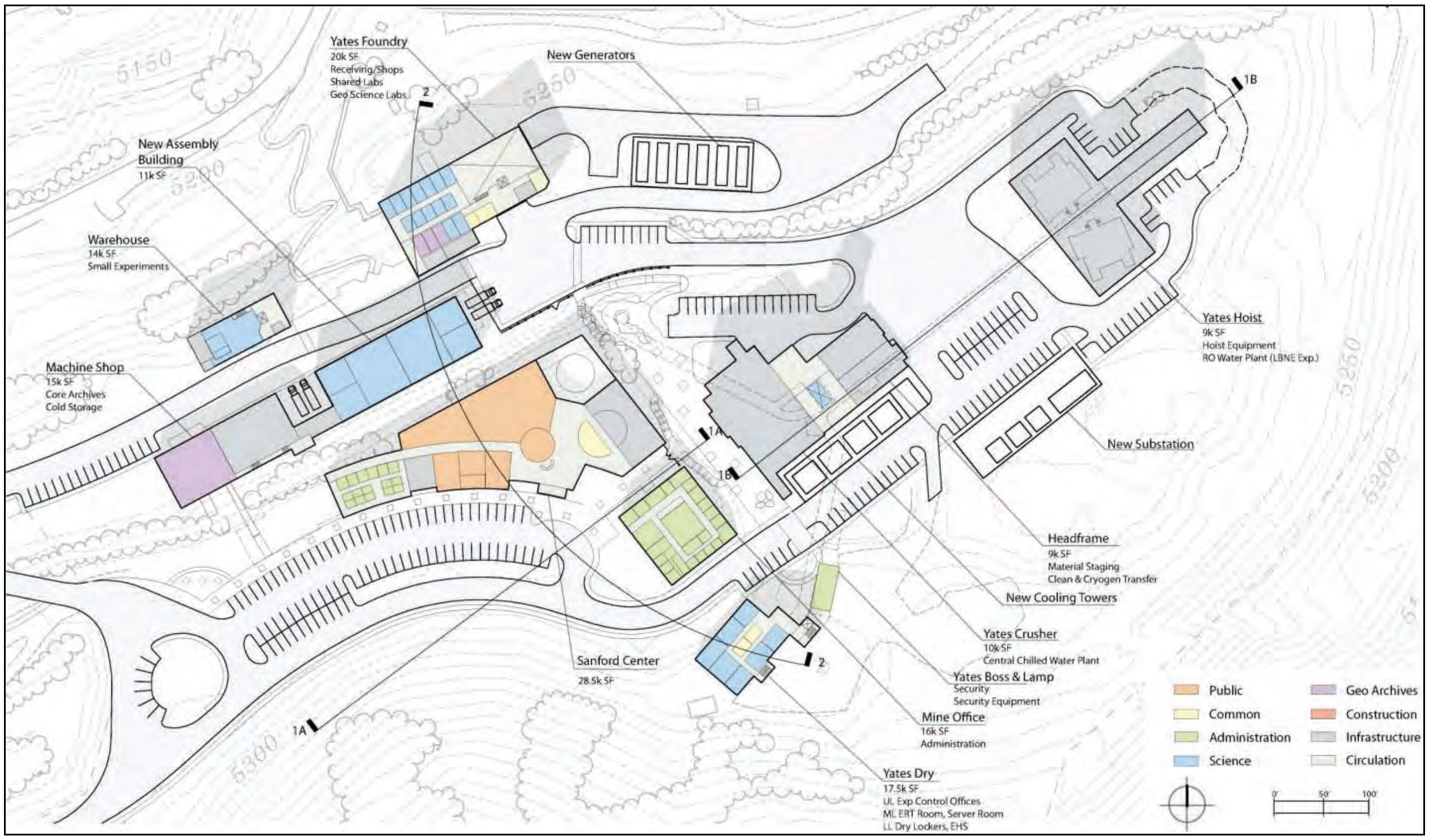

**Figure 5.2.5.2-2** Yates Campus Plan showing distributed parking areas. [HDR]

## 5.2.6 Surface Facility and Infrastructure Design

### 5.2.6.1 Ellison Campus

No work will be performed on the Ellison Campus for the initial development of the DUSEL Campus. During the Phase I Assessment process it was determined, with the exception of the Old Compressor and the Drill & Bit buildings, all buildings and infrastructure on the Ellison Campus had deteriorated to the point that reuse is not feasible, see Appendix 5.E (*Phase I Report, Site Assessment for Surface Facilities and Campus Infrastructure to Support Laboratory Construction and Operations*). Several scenarios for the future development of the site have been proposed, but will be developed at some future date when and if funding is available. Figure 5.2.6.1 shows the future building development plan for the Ellison Campus, including demolition of most of the existing structures, if funding becomes available and the site is developed in support of the DUSEL Project. A prime use for the Ellison Campus would be overflow parking to meet the overall Complex parking demands, which cannot be met on the Yates Campus. The two useable buildings could be used to address scope options needs (see Section 5.2.8), such as visiting scientist offices or a Commons Building, Figure 5.2.5.2-1, above, shows potential Visitor and Overflow parking as well.

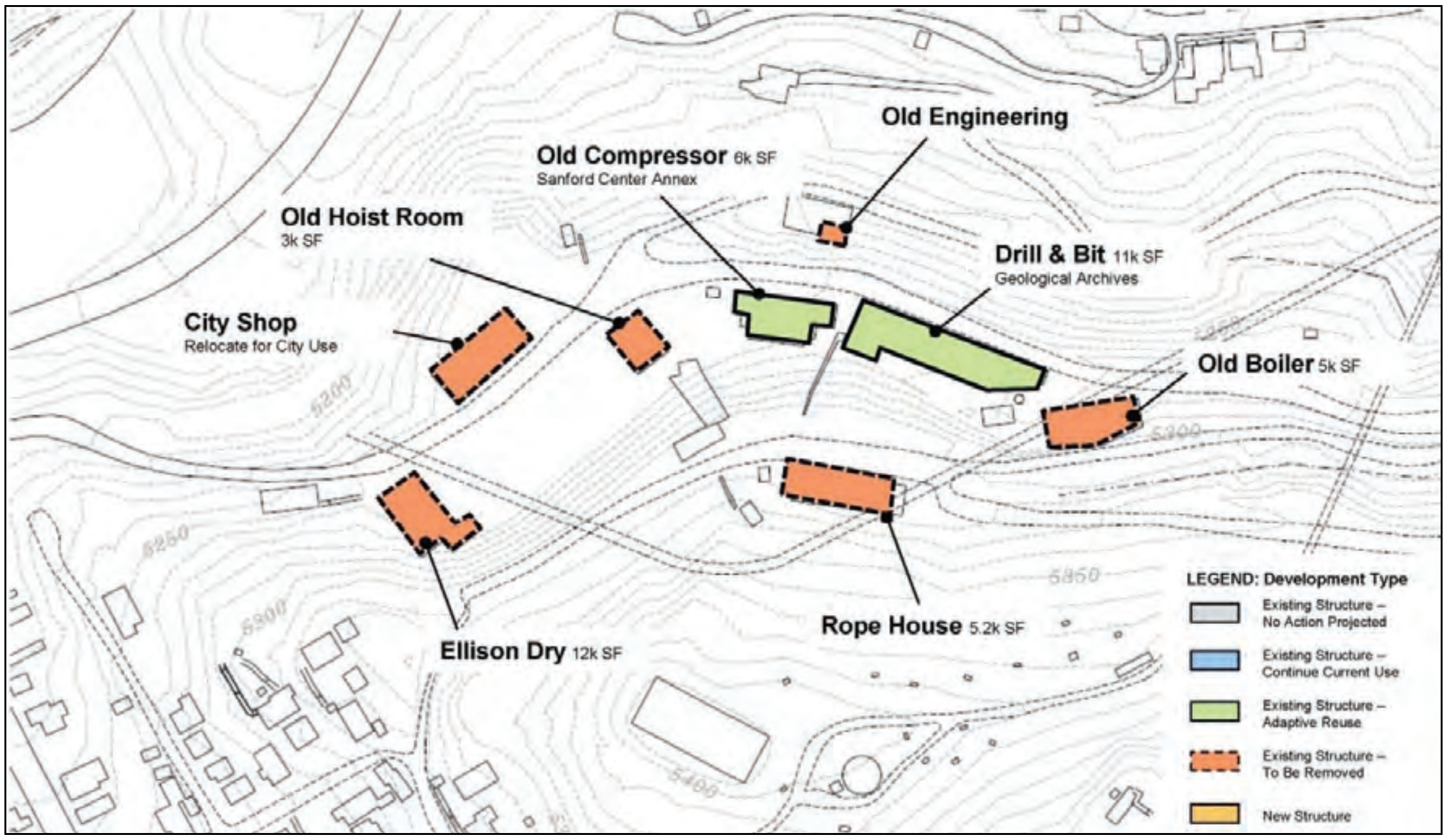

**Figure 5.2.6.1** Future development for the Ellison Campus. [HDR]

## 5.2.6.2 Ross Campus

The Ross Campus has been programmed to become the construction and operations access for the underground facility. Thus, work and improvements at the Ross Campus will be limited to items necessary to support the construction efforts. Figure 5.2.6.2 shows the Ross Campus building program plan.

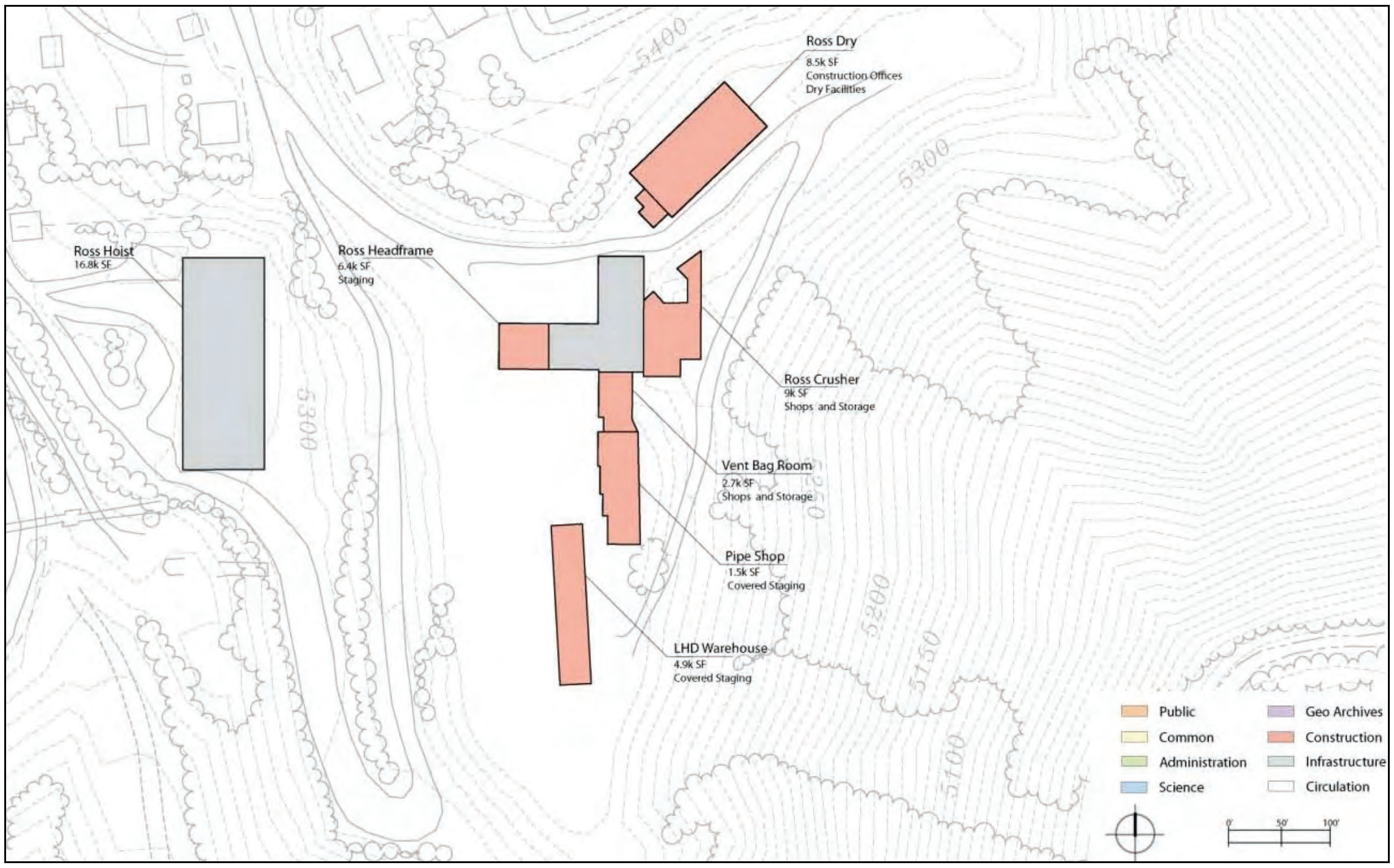

**Figure 5.2.6.2** Ross Campus building program plan. [HDR]

##### 5.2.6.2.1 Ross Buildings, Removal

Buildings and facilities scheduled for removal on the Ross Campus are the Ross Boiler, Ross Core Shack, and the Ross Oil Tanks/Storage facility. During the early stages of Project development, DUSEL staff determined that because of building condition, location, previous use, or size, these structures have no future use for DUSEL.

##### 5.2.6.2.2 Ross Buildings, Continued Use

Buildings on the Ross Campus will be upgraded to provide weather-tight, secure facilities. Upgrades will include roofing replacement, door and window improvements, repairs to brick and mortar, and roof flashing. The Ross Crusher and Hoist Building will need structural modifications to the roof truss members, as shown in the *Phase II Roof Framing Assessment* (Appendix 5.G). These structural modifications are necessary to comply with building code roof loadings.

Buildings on the Ross Campus to be upgraded are the Ross Dry, Crusher and Headframe, Hoist Building, Pipe Shop, and Vent Bag Room.

The Ross Dry will contain dry facilities and office areas, while the remainder of the Ross buildings will retain their present uses.

Ancillary structures in poor condition or not programmed for use will be removed.

##### 5.2.6.2.3 Ross Infrastructure

Infrastructure improvements on the Ross Campus will be limited to site grading and drainage improvements around the existing buildings. The gravel parking and access areas will be regraveled and graded to keep runoff away from the buildings. A concrete drainage pan will be installed on the uphill side of the Ross Dry to control hillside runoff after excess native material has been removed. Although the Phase I Assessment report, (Appendix 5.E) recommends the replacement of the site utilities, no improvements to the site water and sewer system are planned. The recommended water and sanitary sewer improvements were removed during the Value Engineering process. Drainage areas around retaining walls and buildings will be cleaned and regraded to provide proper control of storm-water runoff. Upgrades to the site electrical, communications, and cyberinfrastructure will also be needed but on a much smaller scale than other infrastructure items (see Chapter 5.5, *Cyberinfrastructure Systems Design*).

#### 5.2.6.3 Yates Campus

The Yates Campus will become the science and administration center for the DUSEL Complex. Several existing buildings will be modified for science and facility administration and support uses. Two new buildings will be constructed: the SCSE and the Science Assembly Building. Buildings not programmed for use or to be found in unusable condition will be removed. Figure 5.2.6.3 shows the Yates Campus building development plan.

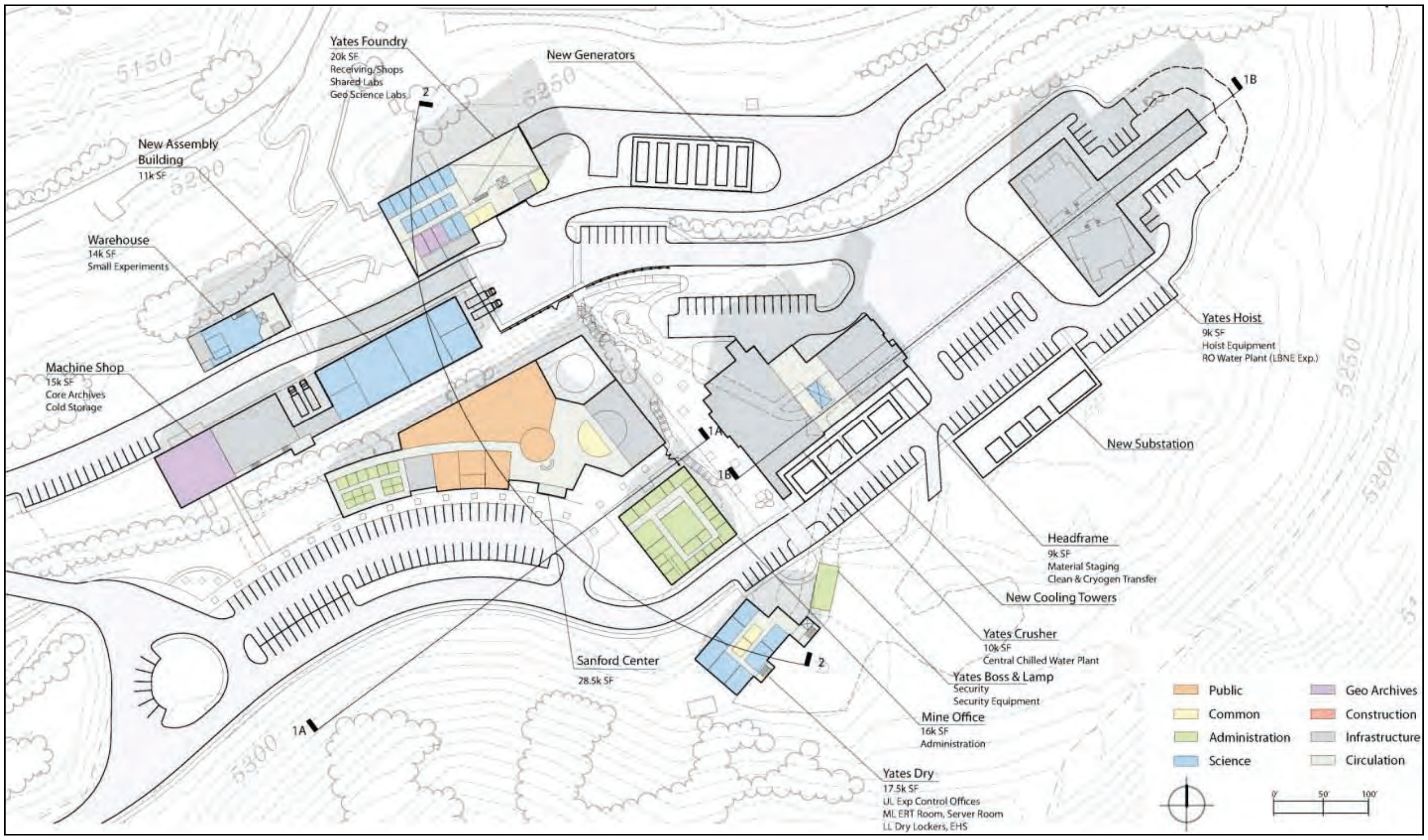

**Figure 5.2.6.3** Yates Campus building development plan. [HDR]

### 5.2.6.3.1 Yates Buildings, Removal

Buildings and facilities scheduled for removal on the Yates Campus are Bottled Gas Storage, Wash Rack, Metalizing/Paint Shop, Paint/Old Bit Shop, Battery/Slusher/Shovel Repair shop, Sand Blast booth, Iron House, the easterly portion of the Machine/Metal Fabrication Shop, Yates Sawmill, Yates Oil Storage, Yates Used Oil Storage, and Yates Dust Collector. During the early stages of the Project development, DUSEL staff determined that because of the buildings' condition, location, previous use, or size, they have no future use for DUSEL.

### 5.2.6.3.2 Yates Buildings, Continued and Repurposed Uses

The Yates Hoist and Headframe will be retained and upgraded to continue their present uses.

The Yates Crusher building will be reprogrammed for other future uses. Presently, no DUSEL program is scheduled for this building; however, the Long Baseline Neutrino Experiment (LBNE) or others will most likely have programs for which this location will be highly desirable.

The Yates Dry will be reconfigured to house experiment control rooms, meeting rooms, emergency response rooms, engineering control rooms, computer server rooms, EH&S offices, men's and women's dry facilities, and building support facilities. The reuse plan for the Yates Dry is shown in Figure 5.2.6.3.2-2.

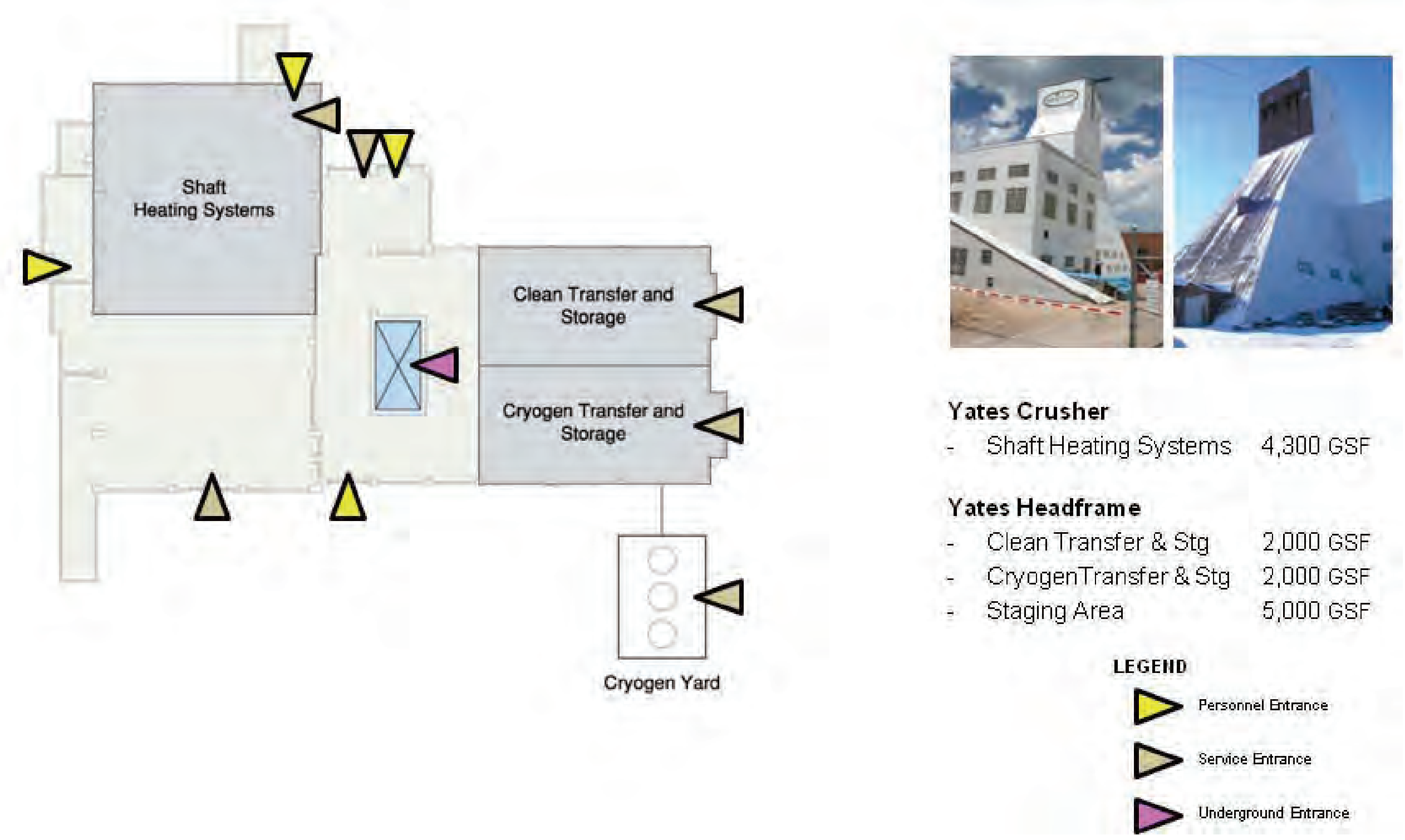

**Figure 5.2.6.3.2-1** Yates Crusher and Headframe Reuse Plan. [HDR]

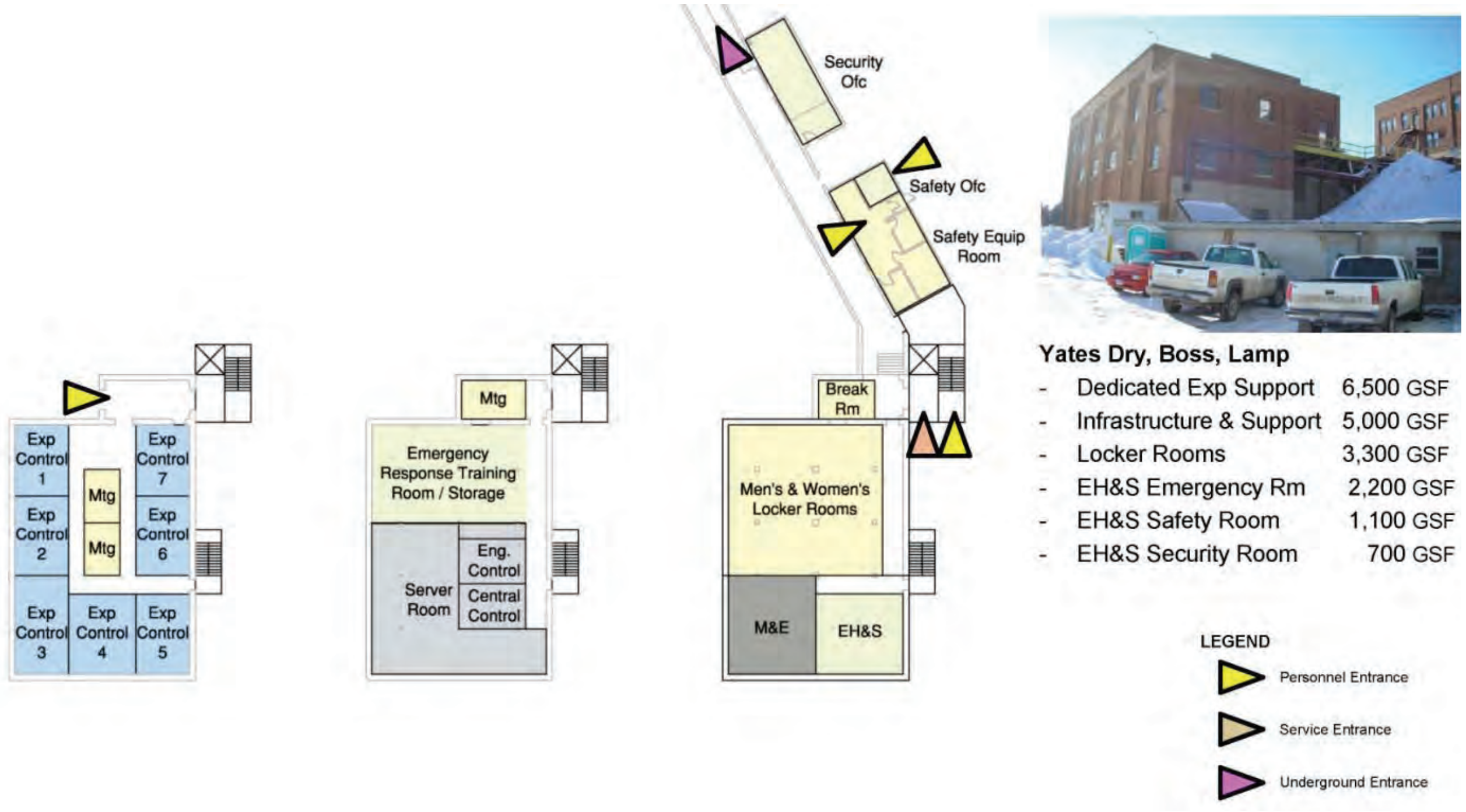

**Figure 5.2.6.3.2-2** Yates Dry Reuse Plan. [HDR]

The Foundry building will be renovated to house laboratories, hazmat offices, meeting rooms, a break room, a common-use electronics shop, a common-use machine shop, a shipping and receiving warehouse and offices, a maintenance shop, and building support. The Foundry Building reuse plan is shown in Figure 5.2.6.3.2-3.

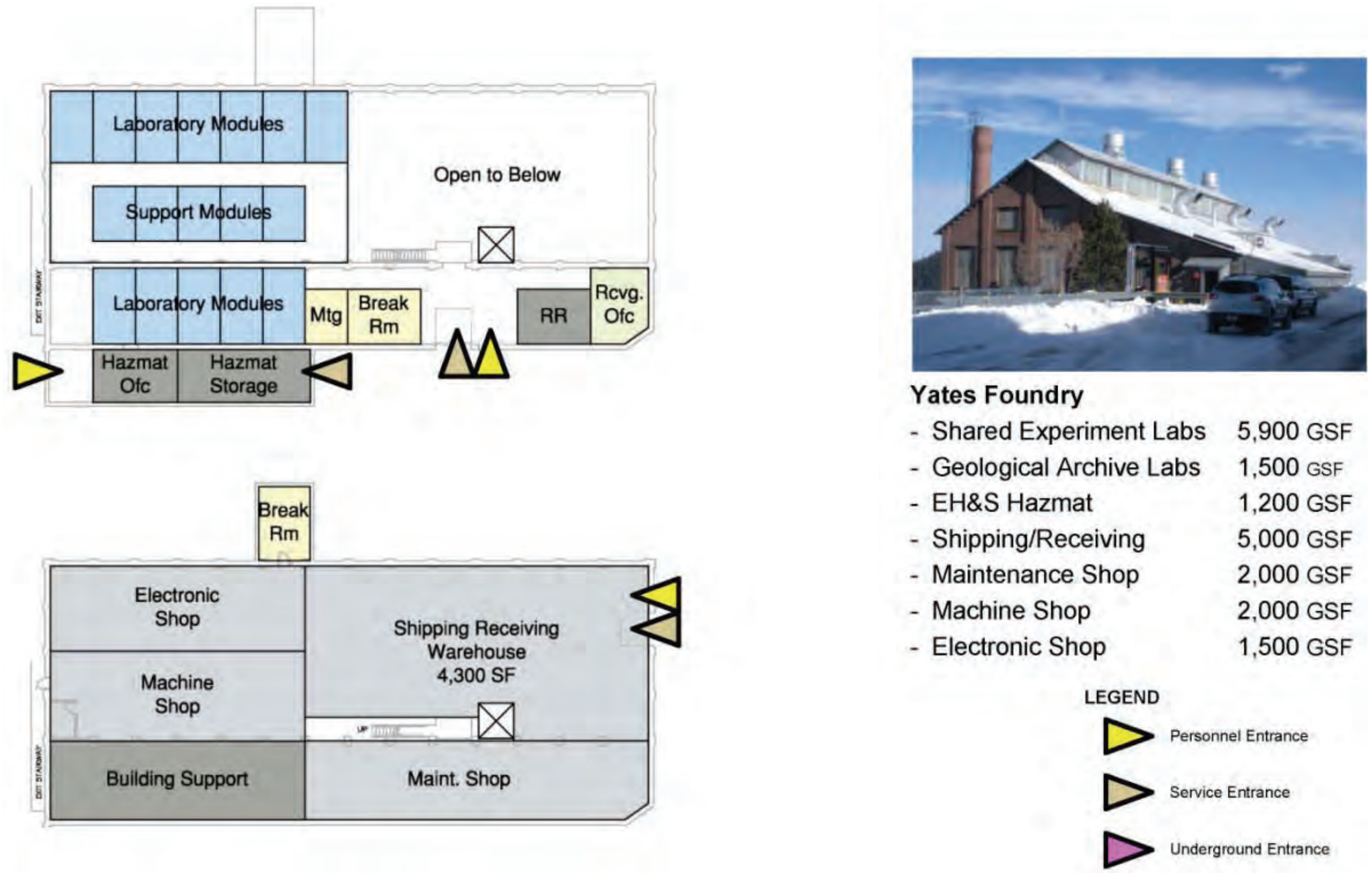

**Figure 5.2.6.3.2-3** Foundry Reuse Plan. [HDR]

The Hoist Generator portion of the Yates Hoist building is scheduled to be the home for the Reverse Osmosis (RO) Water Plant for the LBNE. This RO plant will also be used to supply purified water to other underground experiments that may need it. The design and construction of the RO Plant will be funded by LBNE. The Yates Hoist building reuse plan is shown in Figure 5.2.6.3.2-4.

A portion of the Machine/Fabrication Shop will be removed for the construction of the new Science Assembly Building. The remaining portion of the Machine/Fabrication Shop will be reprogrammed to house the Core Archive and Core Study area (also referred to as the Geological Archive Labs), see Figure 5.2.6.3.5.

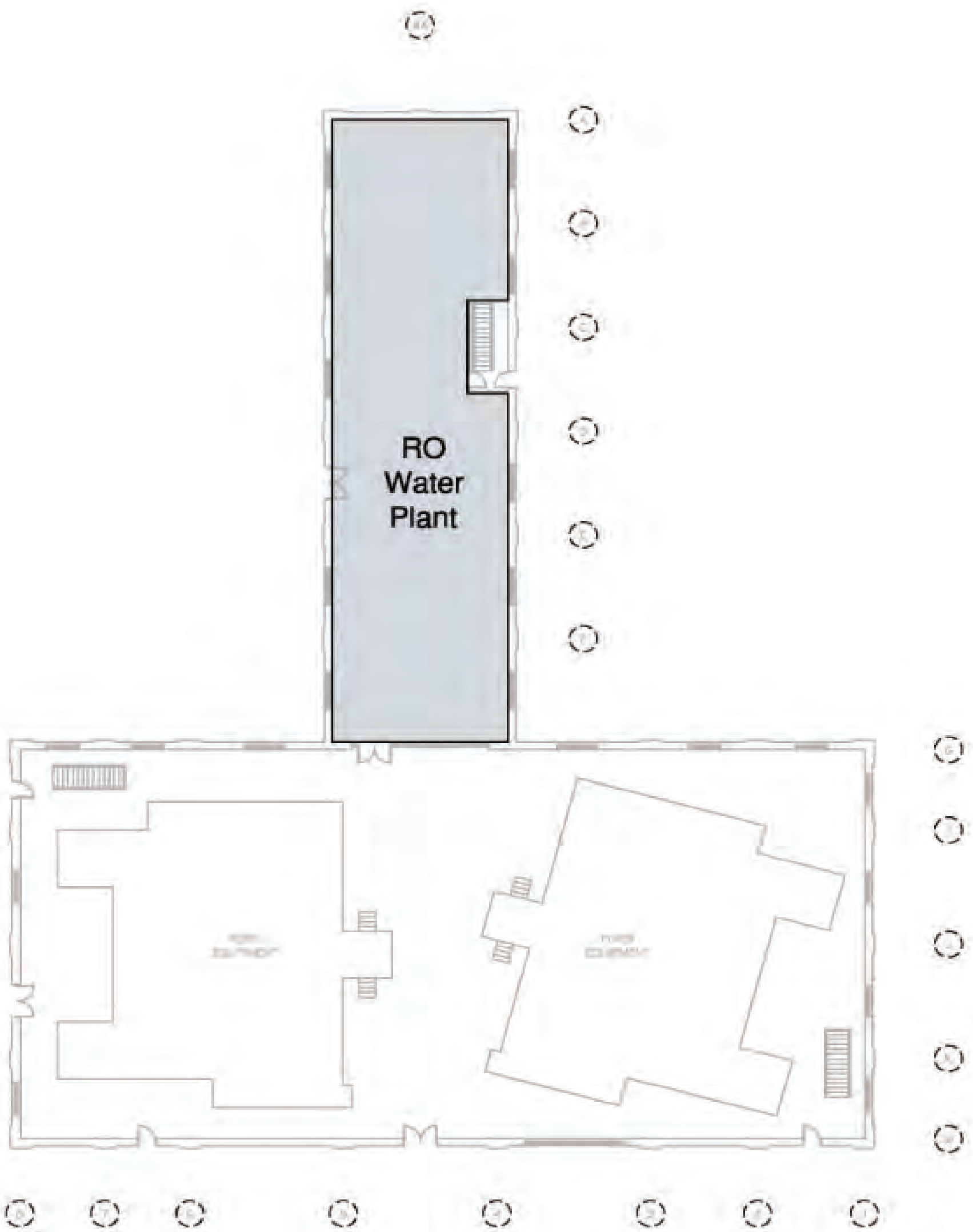

**Figure 5.2.6.3.2-4** Yates Hoist Building Reuse Plan. [HDR]

### 5.2.6.3.3 Yates Infrastructure

Improvements to the Yates Campus infrastructure will include new and resized water and sewer mains and services, upgrades to the surface communications, and cyberinfrastructure utilities (Chapter 5.5, *Cyberinfrastructure Systems Design*). Parking areas and roadways will be reconfigured and re-graded to improve drainage, and will be resurfaced with hard surface pavement, concrete, or asphalt. Site fencing, guardrails, and retaining walls will be replaced or repaired as needed. The storm water management plan will establish improvements, policies, and procedures to be incorporated for overall site storm water management. These improvements will be coordinated with the Sustainable Sites Initiative (SSI) Pilot Program referred to in Section 5.2.3, *Sustainable Design*.

### 5.2.6.3.4 Sanford Center for Science Education

The goal of the Sanford Center for Science Education (SCSE) will be to house a world-class education and outreach facility and represents a unique opportunity for DUSEL to establish an identity separate from the Homestake Mine, as well as to create a new campus center to represent the modern and progressive scientific research taking place deep underground.

During the Preliminary Design process, it was determined the that the SCSE was best suited to be in a new building and that, given its importance to the DUSEL Complex, it should be located on the Yates Campus close to the Yates Headframe, to provide access to the underground. Furthermore, because modifications to the Administration Building are currently deferred due to funding limitations, the SCSE will take on the additional role as the new front door to the DUSEL campus and will be located in physical proximity to the Administration Building, the current entry point for visitors to the site.

The SCSE will incorporate facilities and programs that feature the unique science and engineering of DUSEL, with a goal of 7,500 sf exhibit space; 2,000 sf classrooms; a 100-seat 1,000 sf theater; 3,000 sf of outdoor exhibit space and Education and Administrative support space, creating a total facility with a total gross area of 26,760 sf. See Volume 4, *Education and Public Outreach*, Chapter 4.2, *Baseline Facility Preliminary Design*, for the complete space allocation. The SCSE will reflect the cultural and historic heritage of the Black Hills in the overall building design and exhibit and education programs.

Special programs developed for the SCSE will include a limited underground experience and an on-site interpretation in the proximity of the Yates Shaft, and an off-site interpretation in the form of exhibits, distance-learning capabilities, material storage, and a teacher resource center. See Chapter 4.3, *Baseline Education Program*, for the complete SCSE programming plan.

Programmatic goals for the SCSE that affect the facility design and planning are described in the *Final Report, Phase 1(30%), Preliminary Design Report, Surface Facility and Campus Infrastructure* (Appendix 5.D).

The SCSE will support the science activities with a lecture hall, classrooms, interaction space, and cafe. See Figure 5.2.6.3.4 for a description of the SCSE program as understood at 30% Preliminary Design.

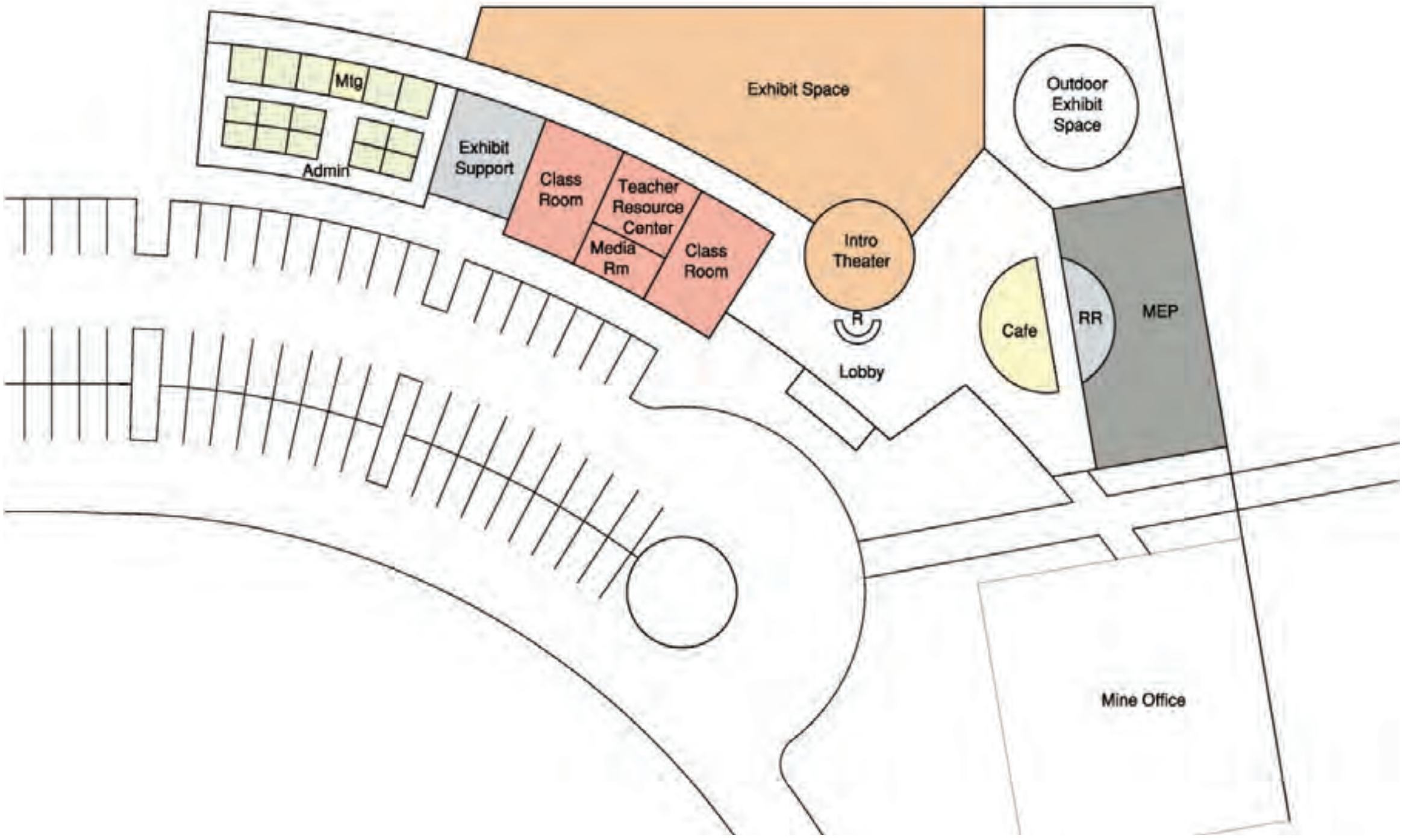

**Figure 5.2.6.3.4** Program for the SCSE. [HDR]

#### 5.2.6.3.5 Science Assembly Building

A new Science Assembly Building will be constructed on the present location of the Machine/Metal Fabrication Shop (a portion of the current structure will be demolished to make room for the new structure). This will be a high-bay assembly area suitable for erection of underground experiments to test-fit and possibly test-run prior to the experiments' move to their underground locations. It will be designed and constructed to blend in with the present exterior architecture of the lower Yates area, and to conform to historic preservation requirements. Figure 5.2.6.3.5 shows the development plan for the Science Assembly Building.

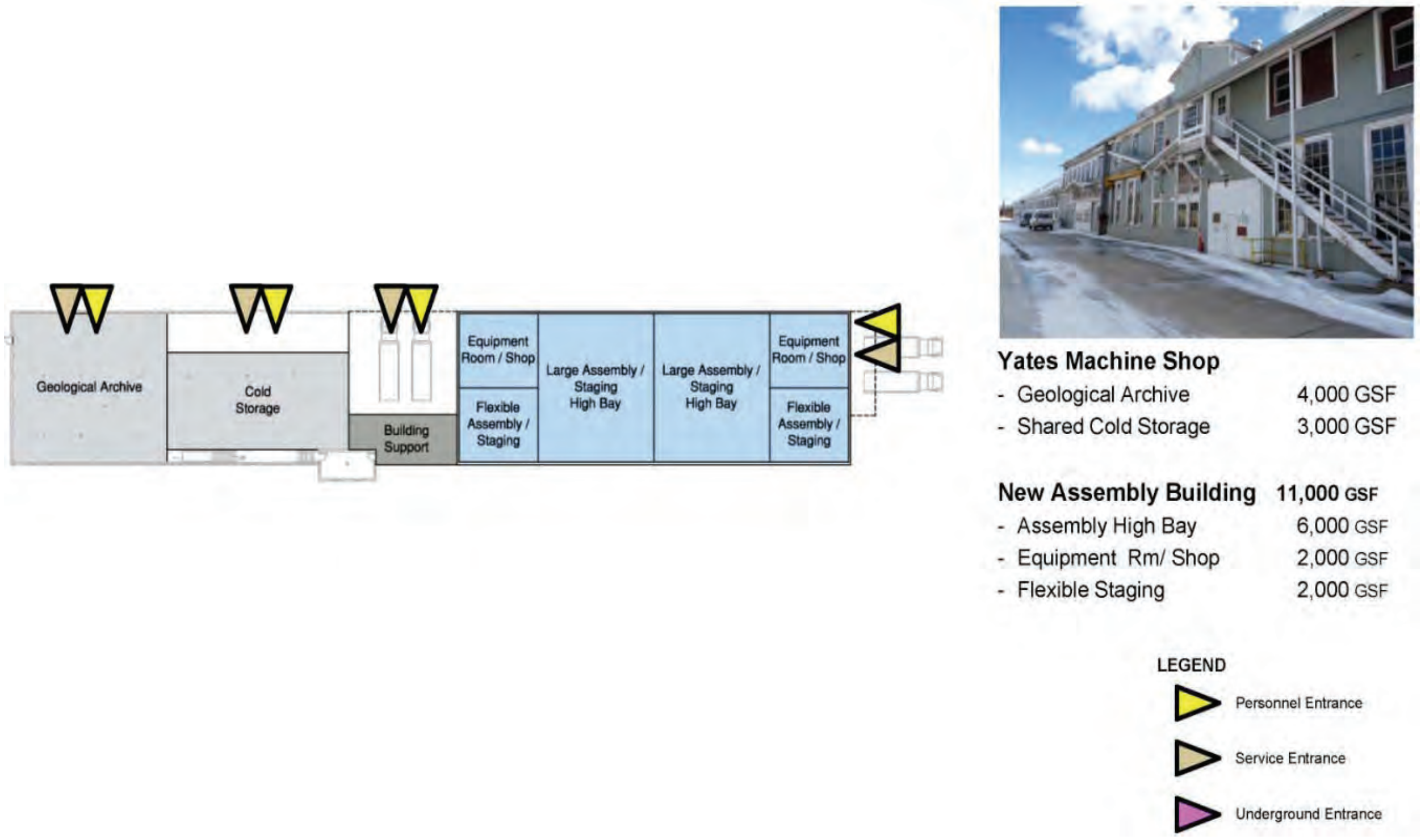

**Figure 5.2.6.3.5** Science Assembly Building Concept. [HDR]

### 5.2.6.4 Other Site Facilities and Infrastructure

The DUSEL Complex includes a number of other infrastructure systems and facilities that support the development and access of the Yates, Ross, and Ellison Campuses described above.

#### 5.2.6.4.1 Waste Water Treatment Plant

The Homestake Waste Water Treatment Plant (WWTP) was originally constructed to treat the effluent water from the gold extraction process and the natural inflow water removed from the mine. With the closing of the mining process, the plant was modified by SDSTA to treat the water being removed from the potential laboratory levels. The water presently being removed from the underground levels has acquired a high iron content from its extended contact with the native rock; thus, treatment changes were required. It is anticipated the iron content in the removed water will drop once the facility is dewatered, but the treatment plant will still be needed, as continued dewatering will be required to remove the natural inflow into the facility.

The DUSEL WWTP was evaluated for present and future treatment capacities based on plant operating data and field observations; see Appendix 5.F, *Phase II Site and Surface Facility Assessment Project Report.* Results of this evaluation show the facility is currently performing effectively and meeting water quality treatment goals as established by the current (in effect) Discharge Permit issued and administered through the South Dakota Department of Environment and Natural Resources (SD DENR). Due to decreased loadings on the plant, some portions are overdesigned, and energy savings could be realized by altering current operating practices. Short-term and long-term maintenance/upgrade items have been identified that will prolong the life and increase the capacity of the facility.

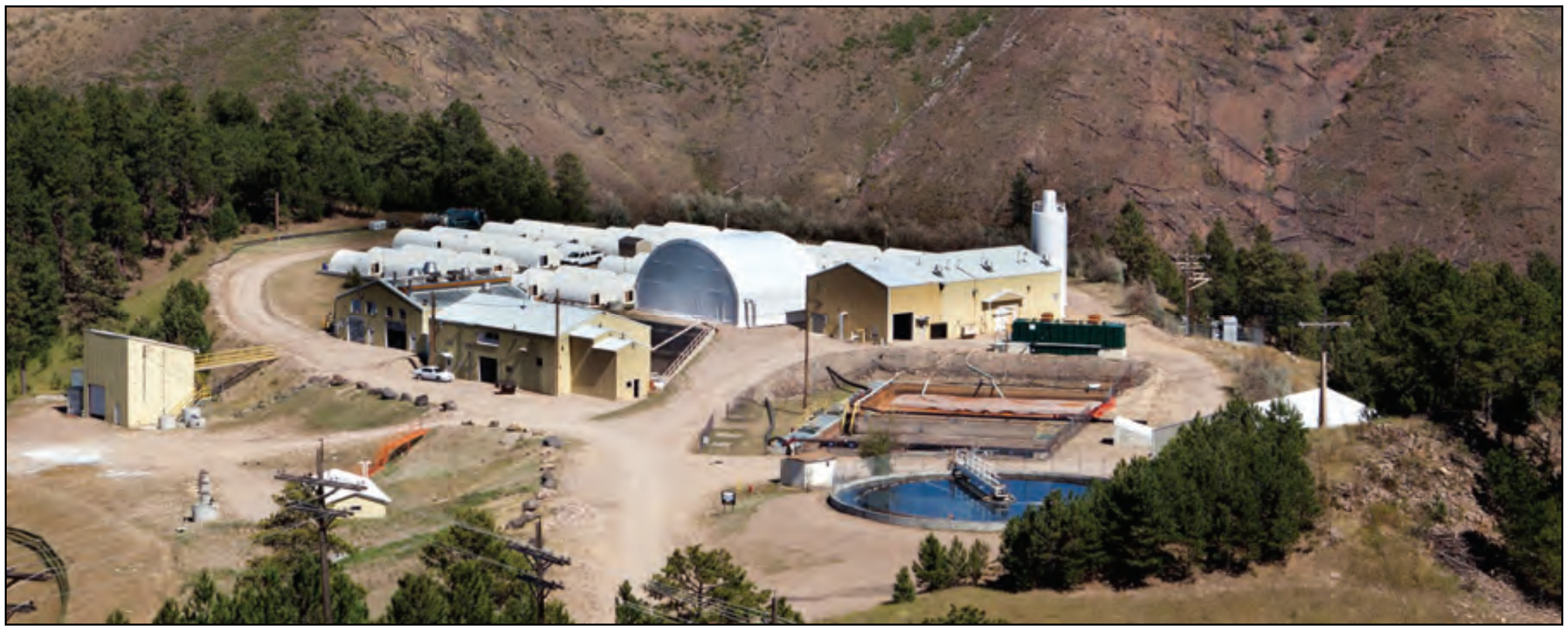

**Figure 5.2.6.4.1-1** Waste Water Treatment Plant. [John Scheetz, DUSEL]

Short-term improvements include:

- Repair of the Grizzly Gulch (Decant) pipeline (a project currently being undertaken by the SDSTA Operations and Maintenance)
- A new backwash, residual flocculation tank, or purchase of the existing rented tank at a negotiated price (SDSTA is presently renting a tank for the backwash portion of the treatment process. This would be a long-term financial improvement, not a process improvement.)
- Aeration for the Mill Reservoir mine water (the Mill Reservoir is currently used as a settling pond in the treatment process of the mine discharge water), see Figure 5.2.6.4.1-2
- Rehabilitation of the existing Parshall Flume located within WWTP
- Conversion of temporary infrastructure such as piping and wiring to permanent
- Diversion of Primary Filter Backwash back to the WWTP
- Reduction in the number of rotating biological contactor (RBC) units in use daily
- Upgrade of the sludge dewatering equipment

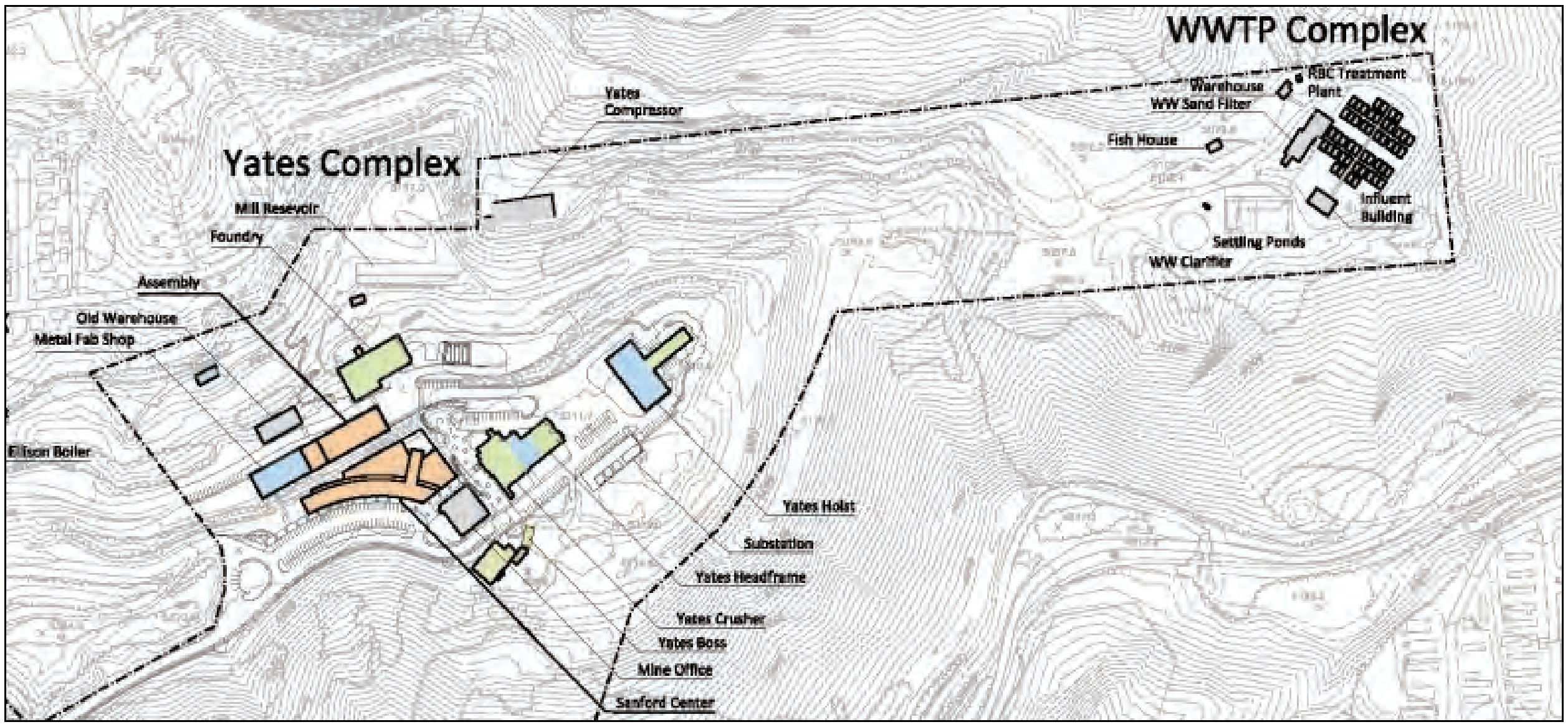

**Figure 5.2.6.4.1-2** WWTP and Mill Reservoir location on Yates Campus. [HDR]

Long-term improvements depend largely on any modifications to the existing Discharge Permit that might be imposed by the SD DENR. Potential modifications to the existing permit may involve more restrictive limits on total dissolved solids (TDS)/specific conductivity, temperature, and nutrients. Treatment options investigated to reduce these discharge parameters were:

**TDS/specific conductivity**

- Softening of mine water to precipitate $Mg(OH)_2$ and $CaSO_4$
- Membrane treatment/evaporation of a mine water side stream
- Permitting an alternative discharge location to a larger stream temperature
- A cooling tower to reduce discharge water temperature
- Aeration to increase evaporation and reduce discharge water temperature nutrients
- Denitrification to reduce total nitrogen concentrations

See Appendix 5.F (Section 2, Pages (2.2)-6 thru 21 of *Phase II Site and Surface Facility Assessment Project Report, April 2, 2010*) for WWTP Assessment results and recommended upgrades.

There are no plans at present for major alterations to the WWTP. The potential for a change in the present Discharge Permit and its impact on the plant does represent an element of risk; however, the SD DENR would allow a reasonable time period in order to come into compliance with a new permit.

### 5.2.6.4.2 Oro Hondo and #5 Shaft

The Oro Hondo (Figure 5.2.6.4.2-1) and the #5 Shaft (Figure 5.2.6.4.2-2) fans were initially proposed to be upgraded to provide ventilation to the underground. It was determined through the Preliminary Design process that it is not economically viable to plan for continued use of the #5 Shaft. The Oro Hondo and #5 Shafts are discussed in greater detail in Sections 5.4.2.3 and 5.4.3.3.

The Surface scope of work for the fans will consist of upgrades to the buildings housing the fans and access to the fan site. These evaluations will be completed as part of the Phase III Site Assessment to be undertaken in 2011.

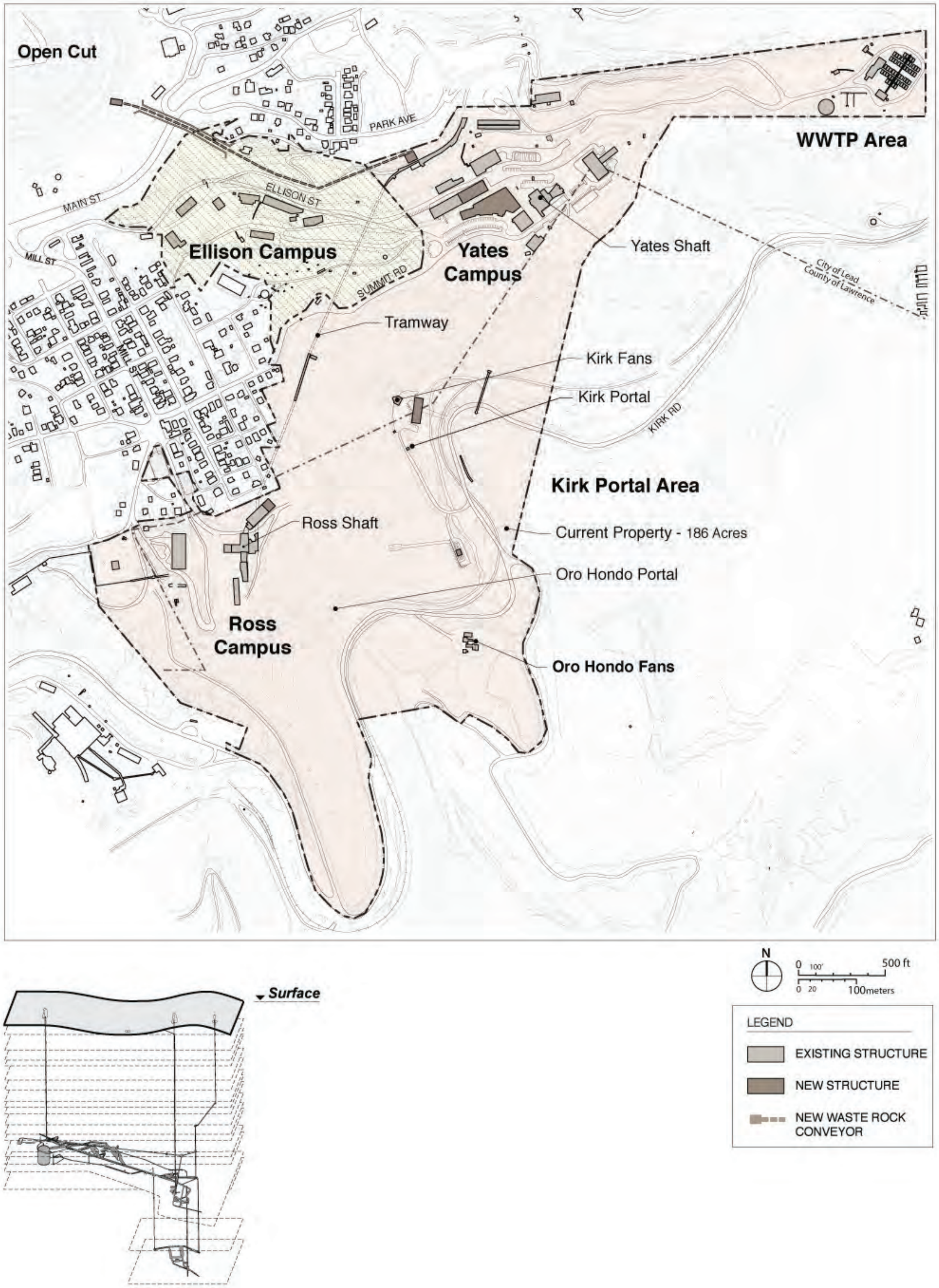

**Figure 5.2.6.4.2-1** DUSEL Surface Campus showing Oro Hondo fan location. [DKA]

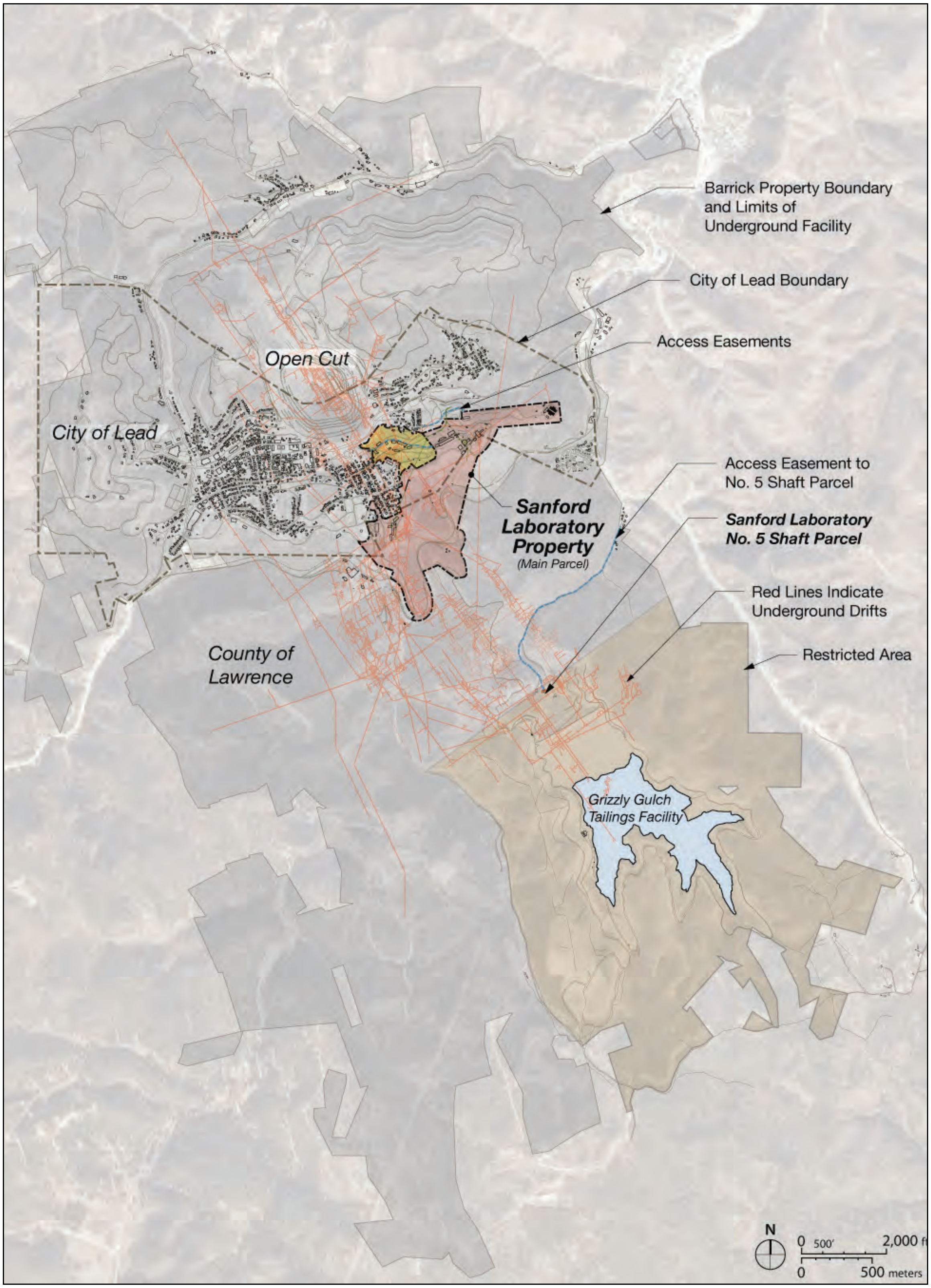

**Figure 5.2.6.4.2-2** #5 Shaft fan location. [DKA]

#### 5.2.6.4.3 Kirk Portal

The Kirk Portal was initially proposed as a potential access to the underground to provide contractor access for materials and supplies, in the same fashion used by Homestake. The Surface Assessment evaluated the access to the Portal. Road upgrades recommended included regrading to correct drainage concerns, installation of a more effective storm-water control system, and rip-rap installation along the Whitewood Creek bank.

Again through the Preliminary Design process, it was determined that the use of the Kirk Portal would not be an appropriate option for the initial construction efforts for DUSEL. The Portal site has limited storage area for either employee parking or for material and equipment storage prior to their move to the underground. Thus, improvements to this area will not be part of the MREFC-funded program for DUSEL.

Although the Kirk Portal site will not be included in the DUSEL MREFC-funded design initially, LBNE is preparing a Conceptual Design for a liquid argon (LAr) detector, which could use the site. The site could provide a location for support facilities and a drive-in access for a LAr detector to be placed on the 800L.

#### 5.2.6.4.4 Tramway Facility

The Tramway Facility will retain the same functions for DUSEL that it held for the Homestake Mining Company. It will be a utility corridor between the Ross and Yates Campuses as well as a rock haulage route. Present utilities found within the tramway include potable water, city sewer, communications line, fiber-optic lines, old compressed-air lines, and the decant water pipeline from Grizzly Gulch.

It will also be used as the rock haulage route for excavated material removed from the underground. The assessment of the tramway will be completed as part of the Phase III Site Assessment to be undertaken in the spring and summer of 2011.

#### 5.2.6.4.5 Kirk to Ross Access Road

Homestake constructed an access road from the Kirk Road along Whitewood Creek up the hillside to the Ross Campus. The road's sole purpose was construction access to the Ross Campus. A preliminary assessment of this roadway, referred to as the Kirk to Ross Access Road, was undertaken as a portion of the Phase II Site Assessment, see Appendix 5.F (*Phase II Site and Surface Facility Assessment Project Report*). DUSEL's interest in the road is also for construction access to relieve construction-related traffic on Mill Street. On completion of the preliminary assessment, it was determined the re-commissioning of this road would be an unwarranted expense to the Project and thus it was dropped from further consideration.

### 5.2.7 Vehicle and Mobile Equipment Program Requirements

The vehicles and mobile equipment requirements on the Surface Campuses are designed to support the ongoing operations and maintenance of the facilities. Some of these requirements are coordinated with the underground infrastructure and transport systems.

#### 5.2.7.1 Program Requirements

Vehicles and mobile equipment will be needed for day-to-day maintenance and operations of the Facility, including emergency operations and science support.

#### 5.2.7.2 Existing Vehicles and Mobile Equipment to Be Repurposed

The present inventory of vehicles and mobile equipment on the DUSEL Complex is owned by the SDSTA or leased from the State of South Dakota. SDSTA-owned vehicles consist of seven 1-ton rated or less pickups or trucks with utility beds, two vans (one cargo, one equipment), and one eight-passenger suburban. Four additional pickups are leased through the state of South Dakota motor vehicle pool.

SDSTA currently owns mobile equipment consisting of one Kubota utility vehicle (two-person all-terrain vehicle with utility box); and two John Deere wheel loaders with bucket accessories, forks, blades, etc.

Assorted SDSTA-owned miscellaneous equipment necessary for day-to-day maintenance is also on site. This equipment consists of lawnmowers, snow removal equipment (blades, sanders, etc.), sprayers, welders, air compressors, winches, hand tools, etc.

Other equipment such as man lifts, cranes, and other specialized equipment has been rented short term or leased long term, on an as-needed basis.

#### 5.2.7.3 Vehicles and Mobile Equipment to Be Acquired

Material and equipment movement underground will be accomplished with the use of the automated guided vehicle (AGV) system and is discussed in detail  in Section 5.4.3.7.3, *Material Handling and Personnel Transport Preliminary Design* This system will consist of electric battery powered, wire-guided engines or locomotives with customized wire-guided carts that are pushed or pulled by the engines. The wire guides will be embedded in the concrete floor during construction. Carts will travel in the cages with their payloads and be reconnected to engines on each campus level. DUSEL will provide customized shipping containers, suitable for science use, to move scientific materials and equipment to the underground. These containers will be sized to fit on the Yates Cage. Underground approved all-terrain vehicles (ATVs) will be used by maintenance and emergency personnel.

On the surface, an AGV or other motorized equipment will be used to load and unload container carts from the cage. Additional material handling equipment, forklifts, and loaders are required to transport the volume of materials anticipated at the two surface warehouses and the Science Assembly Building. The surface use carts must be capable of transporting the containers from the Science Assembly Building and other science-use areas on the Yates Campus to the Yates Headframe loading area.

As existing equipment ages and useful life extension becomes uneconomical, buy-or-lease decisions will be made for future procurements as needs arise.

### 5.2.8 Scope Options, Scope Contingencies, and Value Engineering

Through the development of the 30% Preliminary Design for the surface, several programmatic elements were identified as either scope alternates or deferred scope items because of budgetary constraints. Several items also were studied as facility alternates that were ultimately set aside, also because of budgetary limitations. These items were carried through the 30% Surface Preliminary Design as Scope Options or Deferred Scope. The complete list of the deferred alternatives can be found in Appendix 5.D (*Final Report, Phase 1 [30%] Preliminary Design Report [PDR] Surface Facilities and Campus Infrastructure*). In summary, the key programmatic elements deferred include, in no order of priority:

- A 275-seat lecture hall for science use as well as SCSE
- Visiting scientist office space, providing flexible workspace for up to 10 visiting scientists
- The Commons building, consisting of additional lecture space, meeting and breakout rooms, and a full food-service cafeteria
- User Lodge—extended-stay rooms for visiting researchers
- New Ellison entrance improvements—overflow parking, and access road to the Yates Campus
- Expanded space for SCSE exhibitions, teacher resource center, and dormitories

It is anticipated that these deferments will leave the DUSEL Campus as a whole with a severe shortage of office and meeting space. Until such time as these elements can be funded, use of alternate sites within the City of Lead may be needed to supplement program needs.

A new Ellison Hill access will provide a safer winter access to the site and will remove the Yates Campus vehicle traffic from the present residential areas along Mill and East Summit Streets.

A list and preliminary high-level cost estimate of a subset of these scope options can be found in Chapter 5.10, Final Design and Construction Acquisition Plan.

Value Engineering/Cost Reconciliation workshops were held at each deliverable milestone within the design process. This effort was undertaken to meet the MREFC budget for the surface improvements. The results of these workshops generated Value Engineering items, Cost Reduction Items, and logical budget move items, which were included within the 30% Surface Preliminary Design. They included:

- Moving building maintenance items and building demolition activities to Operations (Budget Move)
- Reducing the Science Assembly Building sidewall height (Cost Reduction)
- Eliminating any upgrades to the Yates Headframe to Machine Shop ramp (Cost Reduction)
- Not removing the Ross Warehouse, demolition cost savings (Cost Reduction)
- Requiring a secondary egress stairway for the Yates Dry, making this an exterior stairway rather the interior (Value Engineering)
- Sizing Surface Power systems to meet underground loads, reducing redundancy and emergency power loads (Value Engineering)
- Combining Site Emergency Power Generator enclosures into a single building rather than a separate enclosure for each generator (Value Engineering)
- Eliminating the Backup/Remote Command Center at the Ross Campus (Cost Reduction)

### 5.2.9 Summary of Surface Deliverables for Final Design

Deliverables required to complete Final Design consist of a wide range of items, from completion of Preliminary Design items to Final Design documents. The items outstanding include:

- Phase III Site and Facility Assessment Report (This will complete assessment work.)

- A complete Site Survey (will establish property boundaries and provide topography data necessary to complete Final Design work)
- Completion of Preliminary Design Report (Work complete to date on Surface Preliminary Design is at 30% level.)
- LBNE Report on Surface Facility needs prepared by HDR
- Final Design Elements:
  1. Final Design Report
  2. Cost Estimate with Basis of Estimate
  3. Construction Schedules adapted to Funding profile
  4. Plans, specifications, renderings, and other supporting design documents for two new buildings (SCSE and Science Assembly Building)
  5. Plans and specifications for specialized systems (shaft heating system, surface HVAC/mechanical)
  6. Plans, specifications, renderings, and other supporting design documents for renovated buildings on Yates Campus (Dry Foundry, Headframe, Hoist Building, Bosses' Offices, Lamp Room, Yates Ramp)
  7. Plans and specifications for surface infrastructure upgrades (water, sewer, roadway and parking lot surfacing, IT/data, security, etc.)
  8. Plans and Specifications for Science Support Facilities (LBNE RO Plant, etc.)

## 5.3 Geotechnical Site Investigations and Analysis

The field and laboratory data collected to date, and the initial geotechnical analyses performed, indicate the rock conditions are adequate for DUSEL construction and no adverse geological features are present that could not be mitigated (Section 5.3.3.1). Specific conclusions extracted from the preliminary geological and geotechnical assessment reports summarize rock conditions, data reliability, and the constructability of the proposed excavations for Large Cavity (LC-1), Laboratory Module 1 (LM-1), and Laboratory Module 2 (LM-2), (Section 5.3.3.2). The site investigations produced design input used in numerical modeling (Sections 5.3.3.3 and 5.3.3.4), and design of ground support and excavation sequence (Section 5.3.4). The following section reviews the investigation and analysis results and outlines the resulting excavation Preliminary Design planned for the DUSEL underground campuses.

### 5.3.1 Introduction and Summary of Results

The purpose of the geotechnical investigations was to provide information for the excavation and stabilization of the laboratory modules (LMs) and the large cavities supporting the Long Baseline Neutrino Experiment (LBNE). Characterization of the rock mass (Sections 5.3.2 and 5.3.3) was accomplished through a program of mapping existing drifts and rooms in the vicinity of the new laboratory excavations, drilling and geotechnical logging, and laboratory measurements of the rock properties. The recovery of more than 4,500 feet of core and mapping were performed to determine if discontinuities in the rock exist that would cause difficulties in the construction and maintenance of either the LMs or the large cavities. In general, the proposed locations of the excavations do not appear to be complicated by geologic structures that cause undue difficulties for construction. This information, along with measurement of in situ stresses, allowed initial numerical modeling (Section 5.3.3.3) of the stresses associated with the anticipated excavations. The numerical modeling was then used to design the ground support and rock bolting system that will ensure that the large cavity, in particular, remains stable. The excavation design, which includes methods of excavation and sequence of excavation, is described in Section 5.3.4, followed by the means by which the excavations will be monitored to ensure their long-term stability in the following sections. The overall analysis of the work indicates that the rock in the proposed location is of good to excellent quality for the purposes of the Project, that preliminary numerical modeling shows large cavities of the size envisioned can be constructed, and that a workable excavation design has been developed.

The geotechnical investigations were initiated in January 2009 and headed by RESPEC Inc., with Golder Associates and Lachel Felice & Associates as the main subcontractors. The initial scope was modified to include the addition of a water Cherenkov detector (WCD) of the 100 kT class from initially a design of LMs only. The scope was further modified, resulting in the requirement for the potential for up to two WCDs into the DUSEL Preliminary Design Report (PDR) effort. In mid-2010, the PDR scope was narrowed to one WCD.

In mid-2009, an initial geotechnical program was executed, first on the 300L, then on the 4850L of the former Homestake Mine in Lead, South Dakota. This program included site mapping, reconnaissance-level geotechnical drilling and core logging, in situ stress measurements, optical and acoustic televiewer logging, numerical modeling, laboratory testing, initial surveying, and generation of a three dimensional (3-D) Geological and Geotechnical Model. Additional tasks added in 2010 included characterization of ground vibrations from blasting associated with the Davis Campus excavation activities, groundwater monitoring, and data conversion to Vulcan software format. Over the course of the geotechnical contract, the raw data and preliminary geotechnical assessments and geo-model updates were transferred to DUSEL as a Vulcan database[1,2,3,4,5] and to the DUSEL Excavation Designer (Golder Associates). A *Geotechnical Engineering Summary Report* (Appendix 5.H) was completed in March 2010, which recommended additional drilling and mapping to address data gaps and reduce uncertainty in the LC-1, LM-1, and LM-2 locations for the Final Design phase.

As part of the design process, the DUSEL Project engaged two advisory boards to provide expert review of the geotechnical investigation and excavation design efforts, described in Section 5.3.9. The Geotechnical Advisory Committee (GAC) is an internal committee that focuses primarily on geotechnical investigation and analysis. The Large Cavity Advisory Board (LCAB) is an internal high-level board that focuses on geotechnical investigations and excavation design of the LC-1 in support of the LBNE. The Geotechnical Engineering Services GES contract was reviewed by the GAC and the LCAB and contained the following elements:

- The mapping program included 4,400 ft (1,340 m) of drifts mapped in detail and 2,600 ft (793 m) of drifts and large openings (Davis Campus that included Davis Laboratory, Transition Area, and associated connecting drifts).
- The drilling program included 5,399 ft (1,646 m) of HQ (4-inch) diamond drilling, which incorporated continuous logging, continuous core orientation, detailed geotechnical and geological logging, full depth continuous televiewer imaging, and initial groundwater monitoring.
- The in situ stress measurement program included stress measurements in three locations; two sites in amphibolite and one site in rhyolite for the total of eight measurements (six in amphibolite and two in rhyolite).
- The laboratory testing program included uniaxial compressive strength tests (80 samples that incorporated elastic constants and failure criteria), indirect tensile strength tests (40 samples), triaxial compressive strength tests (63 samples), and direct shear strength of discontinuities (36 samples).

The geotechnical site investigations area on the 4850L, showing boreholes, in situ measurement stations, and planned cavities within the triangle of drifts between the Ross and Yates Shafts, is presented in Figure 5.3.1-1.

The goal of the geotechnical program was to collect data of adequate quantity and quality required to design the DUSEL Facility in good rock to fulfill science needs and long-term stability. To fulfill that goal, the following geotechnical design criteria (based on rock mechanics principles, constructability, and safety) were established: a) rock quality ratings[5] indicate good rock (Rock Mass Rating [RMR>60] and NGI- Q [Q>10]); b) no adverse geological and structural features present (not present on a large scale or present on a small scale but could be mitigated; some of the features of concern included rock burst potential, shear zones, large open joints, weak rock, low Rock Quality Designation [RQD] on a large scale, weak contact zones, extreme rhyolite, and high stress zones); c) predictable rock properties; and d) predictable state of stress in the rock mass. Addressing the above issues and other challenging project constraints and requirements, the following approach to the completion of the Preliminary Design was undertaken: a) accelerated schedule; b) concurrent site investigations, data collection, interpretation, and design; c) building and updating the 3-D model of geology; d) incremental drilling, coring, and concurrent geotechnical assessment based on a decision tree; e) flexibility to quickly adjust geotechnical investigations to gather data that can best support design; f) rapid transfer of data to the designer; g) feedback from the designer and the advisory groups; and h) maintaining data quality.

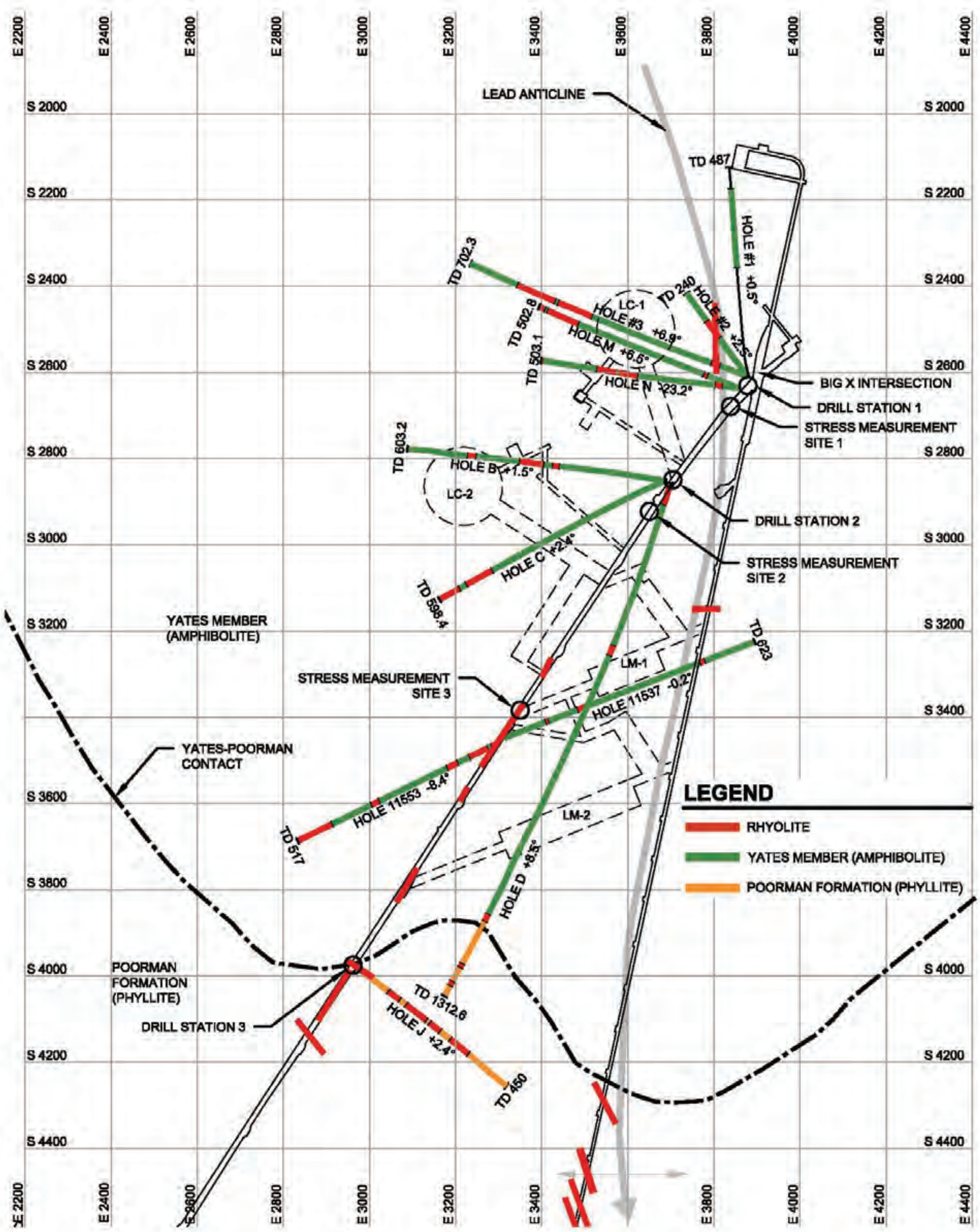

**Figure 5.3.1-1** General geologic map at the 4850L and location of drill holes. [Golder Associates]

### Conclusions

The field and laboratory data and the geotechnical analyses, assessment, and modeling performed revealed that:

a) Overall, the rock is of good quality (RMR>60 and Q>10).
b) No adverse geological or structural features are found that could not be mitigated.
c) Rock mass properties vary but are predictable.
d) In situ state of stress is favorable.

Based on this preliminary assessment, the suitability of cavity placement is as follows:

a) The present LC-1 location is adequate within one diameter (55 m).
b) The present location of LM-1 and LM-2 is adequate within the Yates Member portion of the triangle of drifts between the Yates and Ross Shafts.
c) The geotechnical database assembled to date is adequate for the Preliminary Design.
d) Further site-specific geotechnical investigations (including mapping of new excavations, additional drilling and coring, additional laboratory, and in situ tests, and ground monitoring) are needed for the Final Design.
e) The LC-2 and LC-3, if implemented, would require additional drilling and site investigations at an exploratory level. General location of main laboratory facilities at the 4850L and related geology contacts are presented in Figure 5.3.1-2.

The geotechnical site investigations relate to the DLL at 7400L in the following ways:

- Known data from the 4850L to be extrapolated to the 7400L.
- Unknown geotechnical parameters to be assumed and inferred.
- A site-specific investigation program to be implemented at the 7400L when access is available or an early drilling program from the 4850L is deemed practical and cost efficient.

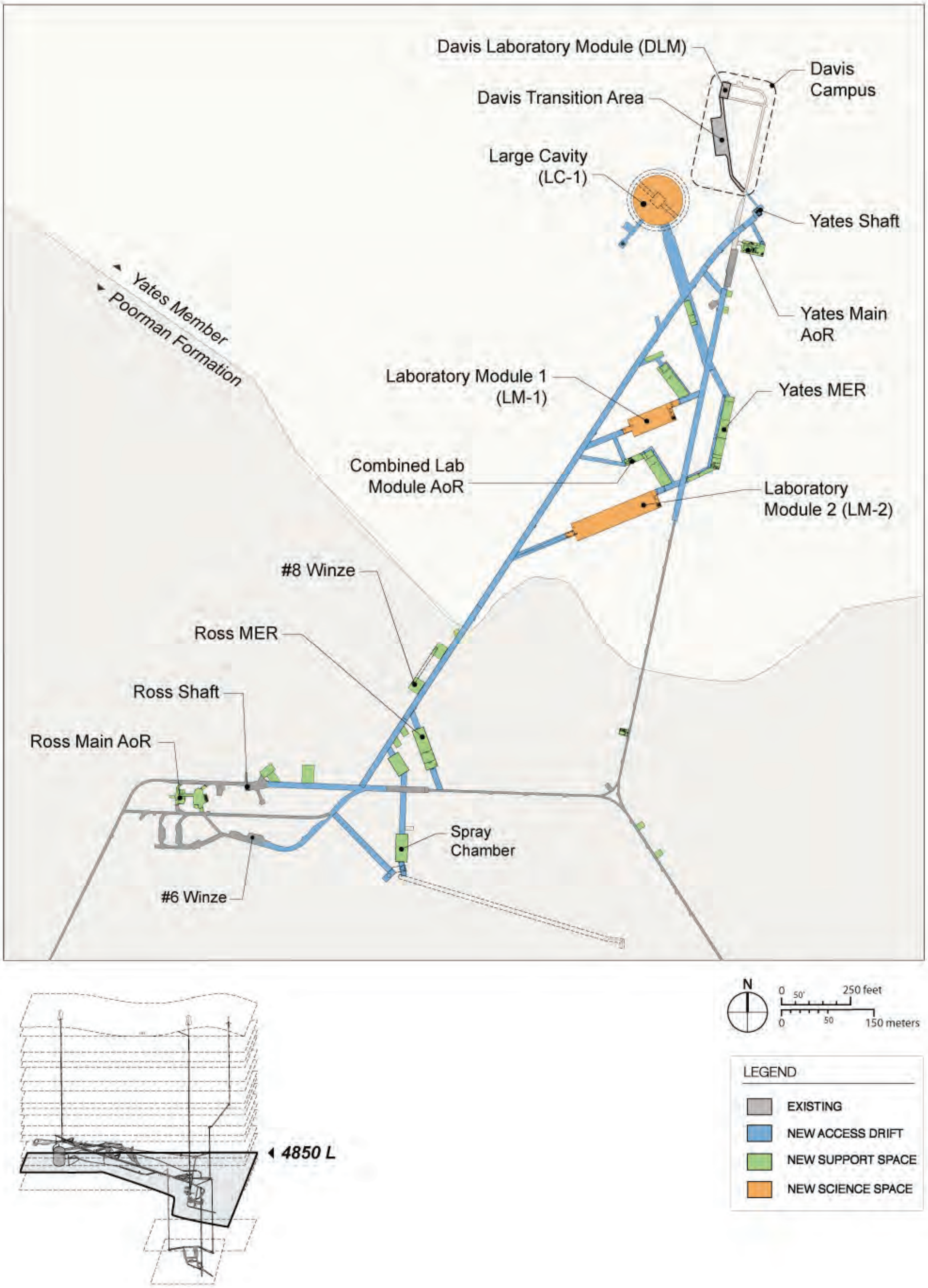

**Figure 5.3.1-2** General location of Mid-Level Laboratory Campus. [DKA]

## 5.3.2 Preliminary Site Investigations and Geotechnical Studies

### 5.3.2.1 Overview of the Site Geology

The geology of the DUSEL site has been studied during the 125 years of operations at the Homestake Mine and more recently as part of the studies for the DUSEL Project. The three major units encountered in the DUSEL underground are, from youngest to oldest: the Ellison Formation, Homestake Formation, and Poorman Formation. These rock units consist of interbedded schists, metasediments, and amphibolite schists. The Yates Member (Unit) is the lowest stratigraphic unit of the Poorman Formation. Exposed on the 4850L (in the triangle of drifts between the Yates and Ross Shafts) are the Yates Unit and the Poorman Formation, as well as Tertiary Rhyolite Dikes. Exposed on the 300L are the Ellison and the Northwestern Formations. The overall geology at DUSEL is a well-defined stratigraphic sequence of schists and phyllites, as shown in Figure 5.3.2.1. These rocks are of high strength and low deformability except when influenced by fracturing, folding, and (dike) intrusions. Alignment of mica (biotite) can produce a moderately developed metamorphic fabric causing planes of weakness, and faults can contain graphite at deeper levels. Therefore, the rock properties can be anisotropic, which affects both strength and deformability.

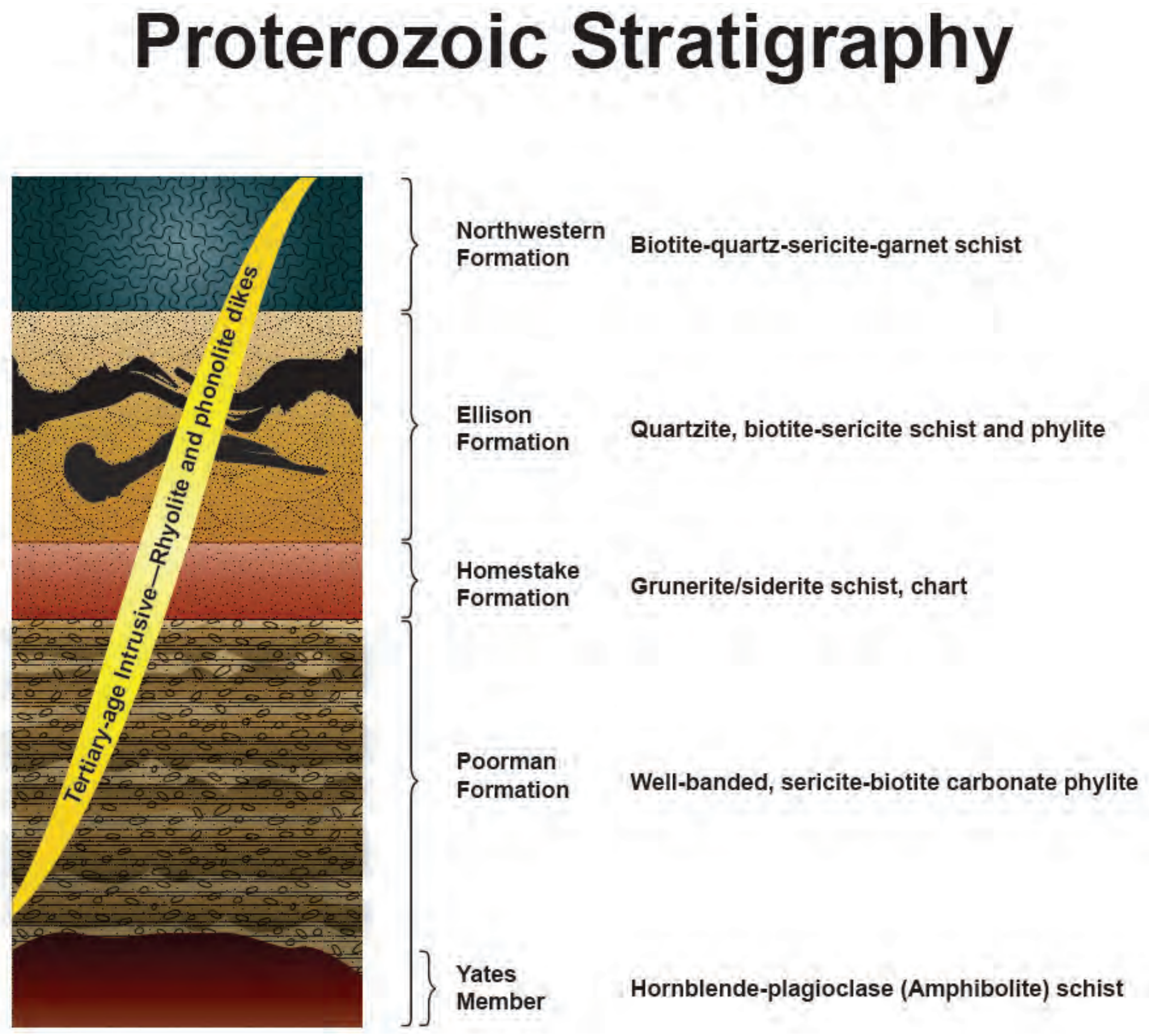

**Figure 5.3.2.1** Stratigraphy of Homestake rocks. The yellow crosscutting feature represents the young Tertiary-age dikes that intrude the Proterozoic Precambrian metasedimentary sequence. [Zbigniew Hladysz, DUSEL]

The large-scale structural geology of the DUSEL site contains a series of interacting synclines and anticlines. This pattern of multiple, overlapping folding and deformation events occurs at all scales such that local folding can be seen in the rock fabric at the scale of inches, feet, and tens of feet. Fold deformations result in complex yet systematic variations in rock mineralogy and structure. Folding creates multiple repeats of the broad-scale formational geology. As a result, most areas close to the ore zones (known in the mine as ledges) show multiple anticlines and synclines. Such structures occur right across the mine width at any given level. At the 4850L, the formations are apparently domed up over the main amphibolite body forming the core of the Yates Member, at least near the proposed cavity locations.

Multiple faults were identified within the Yates Member. Apparent vertical offset is typically small, on the order of inches or feet. It is not possible to rule out the occurrence of such faults at any given location. Five large-scale faults have been identified that transect the Homestake Mine site. Faulting is clearly evident on drift walls at the 4100L, along with evidence of multiple periods of metamorphism and structural dislocation. These faults occasionally exhibit graphitic inclusions, which may amplify their effect on structural stability. Given the observed fault frequency in the 4100L drift (spaced by 50 feet to 200 feet), it is possible that significant undetected faults could exist at proposed LC locations. However, the extent of this faulting, and the significance for siting and construction of the LC, does not appear to be troublesome based on the core logging and borehole televiewing information obtained from the drilling program. It will not be possible to rule out the occurrence of such faults at any given location until the information obtained from these locations is integrated and model projections are completed.

Distribution and orientation of joints and discontinuities in the rock mass are of concern both because of their effect on rock mass geomechanical properties (strength, deformability, anisotropy) and because kinematic stability issues may exist affecting the DUSEL cavities and their access drifts. More detailed discussion of the geo-structural model is provided in Section 5.3.3.

### 5.3.2.2 Drift Mapping

**Mapping of the 300L**

Site mapping activities at the 300L included mapping of the Kirk Drift, the Oro Hondo Drift, portions of the Ross Shaft, as well as the available surface outcrops at the portal. Data[6,7] collected included: geology, hydrogeology (water seepage), weak/shear zones, rock alterations and rock structure along the walls and roofs of drifts, and drift infrastructure (pipes, rock bolts, utilities). In total, approximately 1,400 feet of drift were mapped both manually and by laser scanner. In early 2010, site investigations were terminated, and available funding was focused on the 4850L investigation effort.

**Mapping of the 4850L**

The objective of the mapping program was to characterize the rock mass exposed in the existing drifts on the 4850L (Mid-Level Campus). To accomplish this objective, site mapping at the 4850L focused on detailed evaluation of the existing excavations, geological structure (discontinuities, foliation planes, faults, shear zones, inflows, etc.), hydrogeology (water seepage), and rock alteration. This mapping activity focused on the primary area of interest, which is the triangle of drifts between the Ross and Yates Shafts up to the Yates and Poorman contact in both drifts (see the geology map in Figure 5.3.1-1). The data collection included more than 4,300 feet of detailed mapping of the walls and backs of drifts in the Yates Formation part, including the Yates/Poorman contact zone (the upper half of the triangle). Due to funding constraints, no mapping was performed in the Poorman part (the bottom part of the triangle) at that time. Mapping of the Poorman will be included in the additional site investigations for the Final Design phase. The mapping area at the 4850L is shown in Figure 5.3.2.2.

The data collected from the mapping efforts were analyzed to delineate and identify dominant faults, joint sets, and fractures within the rock mass.[5] Additionally, the data collection included the rock mass characterization parameters needed to estimate RQD and calculate Q and RMR. The rock discontinuity data (strike and dip data only) were evaluated with stereographic plots to determine representative dip directions and dip angles of the predominant discontinuity sets. Multiple analyses were conducted, including separating bedding (foliations), joints, faults, and veins, to see how the orientations varied. The full dataset was then broken down and evaluated as sections, or domains. This iterative process resulted in identification of a total of 11 structural domains at the 4850L. Water inflow for this project area was estimated as minor to no inflow. Only three of the 11 structural domains were found to have any seeps and they were all found within the transition zone between the Yates Member and Poorman Formation.

Mapped discontinuity data were analyzed to determine overall discontinuity orientations and dominant structural domains, and a rock mass analysis was conducted to determine RMR and Q values. Based on the entire dataset, two dominant bedding/foliation orientations were identified (B/F/J-3 and B/F/J-4), along with two other joint sets and fault orientations, as shown in Table 5.3.2.2-1. The data were also broken down and evaluated based on type of discontinuity (foliation, fault, joint, vein) to evaluate the discontinuity orientations by feature. The entire dataset was broken down and evaluated along structural domains based on the results of the stereonet and discontinuity analyses as well as the observed structure and overall geology. In all, 11 domains in the study area were identified: the Davis Complex is a single structural domain and the Ventilation Drift and Exhaust Drift were each divided into five structural domains, as shown in Table 5.3.2.2-2.

The rock mass characterization analysis was conducted using the collected dataset to calculate Q and RMR values for each of the 11 structural domains at the 4850L. The domains were characterized by changes in overall rock mass quality (bulk-estimated RQD and Geologic Strength Index (GSI)), continuity of foliation planes, and lithology. Each of these domains was evaluated separately to determine the overall rock mass characteristics.

The Q system is a commonly used rock engineering tool for the empirical evaluation of ground support requirements in hard rock tunnels. The following parameters related to tunnel excavation stability and ground support are used to determine Q: RQD, Joint set number (Jn), Joint roughness number (Jr), Joint alteration (Ja), Joint water reduction factor (Jw), Stress Reduction Factor (SRF), and Excavation Span and Excavation Support Ratio (ESR). The results of the Q analysis along the drifts mapped on the 4850L show that the calculated values of Q range from 0.5 to 3.4 (reflecting an SRF of 5) and from 1.0 to 16.5 (reflecting an SRF of 1). The Q varied within the domains.

The RMR system was used to evaluate rock mass quality. The parameters involved in calculating RMR include the strength of the intact rock, RQD, discontinuity spacing, condition of discontinuities, groundwater, and an adjustment for discontinuity orientation relative to the excavation. The results of the RMR analysis, as shown in Table 5.3.2.2-2, indicate that the calculated values of RMR range from 51 to 77 along the drifts and varied within the sections.

GSI is a method of estimating the effect of geological conditions on the reduction in rock mass strength based on structure and surface conditions of the rock mass. GSI values were estimated regularly along the drifts while mapping, and the average value was determined for each domain. The overall range of GSI values is from 35 to 95, with an average GSI of 62 for the total dataset.

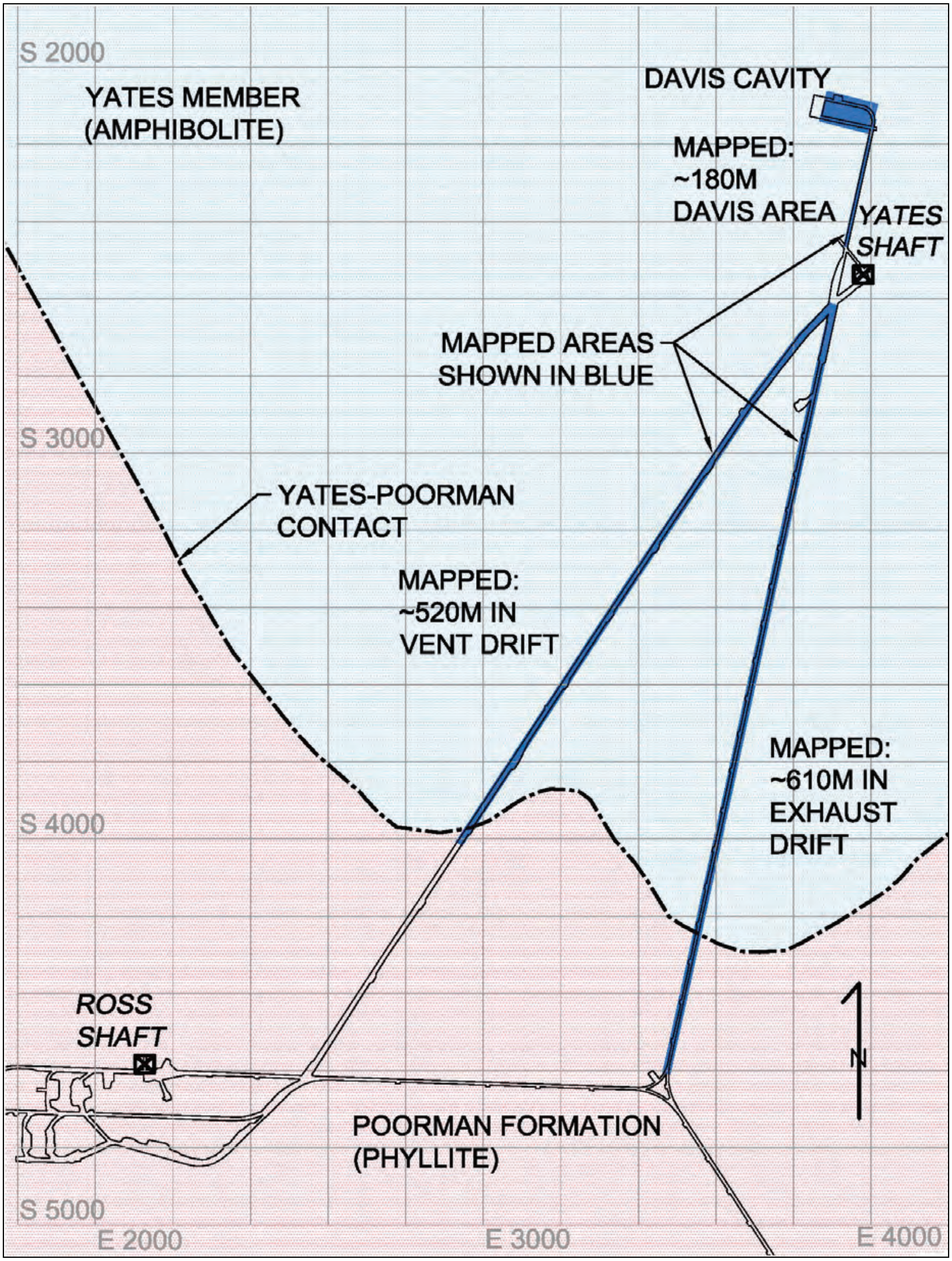

**Figure 5.3.2.2** Mapped area on the 4850L shown in blue. Units consist of the Poorman Formation and the Yates Member. The rhyolite intrusions are not shown in this figure. [Golder Associates]

| I.D. [a] | Total Dataset (1,495 Points) | Joints (1,149 Points) | Faults (194 Points) | Bedding (132 Points) | Veins (17 Points) |
|---|---|---|---|---|---|
| J/F-1 | 63/003 | 63/001 | 53/016 | — | — |
| J-2 | 88/265 | 88/265 | — | — | — |
| B/F/J-3 | 41/109 | 36/115[b] | 40/113 | 46/103 | — |
| B/F/J-4 | 43/200 | 40/197[b] | 38/183 | 51/215 | — |
| B-5 | — | — | — | 31/155 | — |
| V-1 | — | — | — | — | 89/061 |
| V-2 | — | — | — | — | 66/227 |

[a] Explanation: J = Joint set, F = Fault with visible offset, B = Bedding/Foliation planes, V = Veins. Sequence of name is based on dominant features within (i.e., J/F has both joints and faults but is dominated by joint data). Orientation is given as Dip/Dip Direction (i.e., 63/003).

[b] Minor data point clustering relative to other sets for this feature type.

**Table 5.3.2.2-1** Overall representative discontinuity orientation by feature type. [RESPEC]

| Domain | Phase | Stations | Rock Type | Avg. RQD | GSI | RMR (No Deductions) | RMR (With Deductions) Best Case (–5) | RMR (With Deductions) Worst Case (–12) | Q (SRF=5) | Q (SRF=1) |
|---|---|---|---|---|---|---|---|---|---|---|
| 1A | 1 | 600' | Davis Complex Amphibolite | 85 | 70 | 77 | 72 | 65 | 1.4 | 7 |
| 2A | 2 | 0+00–4+50 | Amphibolite and Rhyolite | 74 | 64 | 75 | 70 | 63 | 3.3 | 16.5 |
| 2B | 2 | 4+50–9+22 | Amphibolite | 76 | 65 | 75 | 70 | 63 | 3.4 | 17 |
| 2C | 2 | 9+22–14+00 | Amphibolite and Rhyolite | 70 | 65 | 65 | 60 | 53 | 0.9 | 4.5 |
| 2D | 2 | 14+00–16+85 | Amphibolite and Rhyolite | 68 | 63 | 58 | 53 | 46 | 0.9 | 4.5 |
| 2E | 2 | >16+85 | (biotite) Schist/Phyllite | 35 | 50 | 51 | 46 | 39 | 0.2 | 1 |
| 3A | 3 | 0+00–3+75 | (biotite rich) Amphibolite | 74 | 64 | 67 | 62 | 55 | 3.2 | 16 |
| 3B | 3 | 3+75–5+85 | (biotite-rich) Amphibolite and Rhyolite | 77 | 66 | 77 | 72 | 65 | 0.8 | 4 |
| 3C | 3 | 5+85–15+50 | (biotite-rich) Amphibolite | 83 | 70 | 74 | 69 | 62 | 1 | 5 |
| 3D | 3 | 15+50–16+05 | Amphibolite Schist | 79 | 64 | 61 | 56 | 49 | 2 | 10 |
| 3E | 3 | 16+05–18+60 | Schist/Phyllite and Rhyolite | 40 | 43 | 54 | 49 | 42 | 0.5 | 2.5 |

**Table 5.3.2.2-2** Rock Quality Designation, Rock Mass Rating, Geologic Strength Index, and Q Results. [RESPEC]

### 5.3.2.3 Drilling and Core Logging

The geotechnical drilling program on the 4850L included the following tasks:

- Coring of two 6-inch-diameter boreholes (BHs) to collect samples for laboratory testing
- Drilling of seven HQ-3 diameter BHs aimed at suggested locations of the LMs and LCs

- Limited geotechnical surveys of two BHs (BH1 and BH2), previously advanced by Sanford Laboratory into the area of the planned Transition Cavity in the Davis Campus
- Core orientation for the entire drilled footage
- Geotechnical core logging, including description and orientation of discontinuities
- Point-load testing of rock samples at approximately 10-foot intervals
- Core photographs
- Optical and acoustic televiewer (OTV and ATV) logging for each BH
- Preliminary observations of groundwater flow and shut-in groundwater pressures in BHs

Before the initiation of the DUSEL studies (Appendices 5.H, 5.I), no drifts or BHs penetrated the rock mass at the locations where DUSEL large cavities are being considered. However, the regional scale geology was thought to be sufficiently consistent that the rock at the 4850L in that location was expected to consist of Yates Unit schists and phyllites similar to those encountered in adjacent 4850L drifts, as shown in the historic Homestake data illustrated in Figure 5.3.2.3-1. This was confirmed by the field data obtained in the drilling program. The lithologies in the majority of BHs consist mainly of amphibolite of the Yates Unit with rhyolite intrusions in the form of dikes. Additionally, BHs advanced in the southern portion of the project area encountered phyllite units of the Poorman Formation. However, given the complexity of the folding and deformation within the formations, the local mineralogy and rock quality is expected to vary significantly.

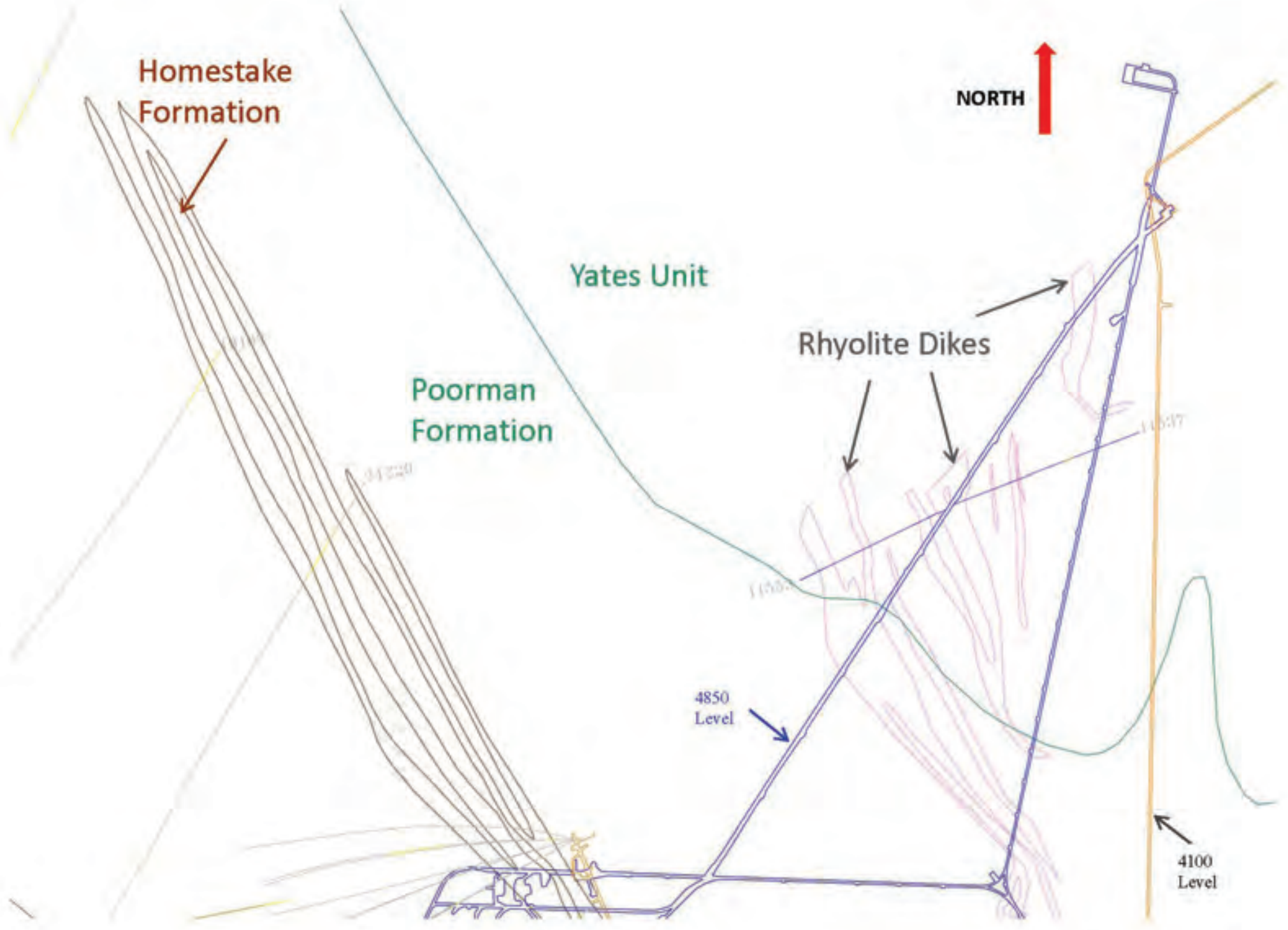

**Figure 5.3.2.3-1** Homestake historical data. [Golder Associates]

Connors Drilling of Montrose, Colorado, was retained to perform all drilling and auxiliary work for the project. For underground drilling at the 4850L, Connors used a 20HH Underground Electric/Hydraulic Core Drill rig capable of drilling holes in full range of vertical and horizontal directions. All drilling was done with an HQ-3 (triple-tube) wireline system. This HQ system drills 3.782-inch-diameter holes and produces 2.4-inch-diameter core. The drilling progress ranged from 30 to 140 feet per day, averaging 70 feet per day, with the slowest drilling occurring in the rhyolite zones.

The drilling program[8] entailed completion of seven BHs drilled from three drill stations located along the Ventilation Drift on the 4850L. Two additional BHs were drilled into the area of the planned Davis Campus excavation at the request and expense of the South Dakota Science and Technology Authority (SDSTA). The locations of the drill stations and orientations of the drill holes are shown in Figure 5.3.1-1, and information on the BHs and their alignment is given in Table 5.3.2.3-1. The total footage drilled was 5,400 feet.

Surveys of BH direction were performed after approximately every 50 feet of drilling using a Reflex ACT digital survey instrument. At each survey point, measurements of BH azimuth, inclination, gravity roll angle, magnetic field strength, and temperature were recorded. The BH surveys were instrumental in determining the orientation of discontinuities in the rock mass. After core retrieval, measurements for core recovery and RQD were collected. The field geologists then documented major encountered lithologies, their contacts, and observed discontinuities. Each discontinuity was examined and marked on the core as having natural or mechanical origin. Potential natural discontinuities (i.e., contacts, veins, joints, and foliations) were then marked with a depth measurement on the core. Discontinuity logging included measurements of the orientation of discontinuities relative to the axis of the BH, which can be corrected into their true 3-D orientation using the BH survey information. Point-load testing was conducted on core samples to estimate the uniaxial compressive strength (UCS) of intact rock and to aid in rock strength classification. High-resolution core photographs were taken and saved in both jpeg and raw image (CR2) formats.

| Borehole | Drill Station | Target | Total Depth (ft) | Average Azimuth[a] | Average Dip[b] | Drilling Start Date | Drilling End Date |
|---|---|---|---|---|---|---|---|
| BH1 | 1 | Davis Campus | 487 | 354.1 | 0.5 | 8/22/2009 | 8/26/2009 |
| BH2 | 1 | Davis Campus | 240 | 323.2 | 2.5 | 8/26/2009 | 8/27/2009 |
| BH3 | 1 | Davis Campus/LC-1 | 702.3 | 291.7 | 6.9 | 8/28/2009 | 9/5/2009 |
| BHM | 1 | LC-1 | 502.8 | 292.1 | 6.5 | 9/11/2009 | 9/14/2009 |
| BHN | 1 | LC-1 | 503.1 | 277.9 | –23.2 | 9/7/2009 | 9/9/2009 |
| BHB | 2 | LC-2 | 603.2 | 277.0 | 1.6 | 11/14/2009 | 11/18/2009 |
| BHC | 2 | LC-3 | 598.4 | 242.3 | 2.5 | 10/21/2009 | 10/28/2009 |
| BHD | 2 | LM-1 and LM-2 | 1,312.6 | 203.8 | 8.6 | 10/2/2009 | 10/19/2009 |
| BHJ | 3 | LM-3 | 450 | 128.4 | 2.4 | 11/20/2009 | 11/22/2009 |
| | | **Total Footage** | **5,399.4** | | | | |

[a] Relative to true north. Includes 8.8 degree east correction for magnetic declination.

[b] Degrees from horizontal; positive is up, negative is down.

**Table 5.3.2.3-1** Summary of geotechnical borings at the 4850L. [RESPEC]

### Televiewer Logging

The main purpose of televiewer logging (in-hole imaging), shown in Figure 5.3.2.3-2, was to augment the geotechnical logging and core orientation data, thus providing a check on the accuracy and completeness of geotechnical information. Acoustic and optical televiewer techniques were used, depending on the presence or absence of water in the BH. Both techniques produce a continuous, oriented, 360-degree image of the BH wall. The televiewer also records bearing and inclination so that the azimuths of the detected discontinuities are measured. The locations and orientations of discontinuities logged manually and those recorded by televiewer are in good agreement, which provides assurance regarding the reliability of the geotechnical data generated during the investigation.

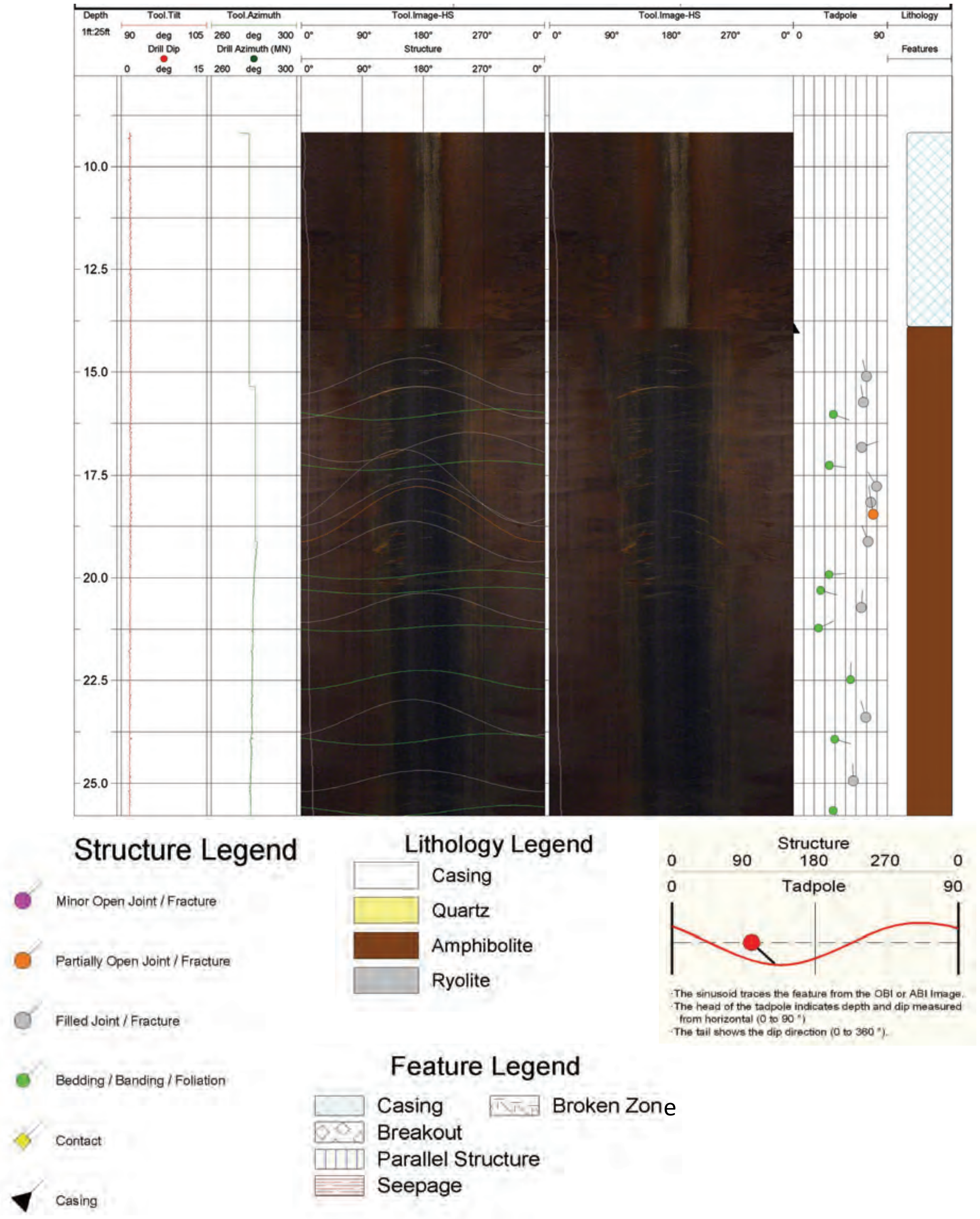

**Figure 5.3.2.3-2** Televiewer log and interpretation example from BH 3. [Golder Associates]

### Data Reduction

The compilation and reduction of geotechnical data[8] recorded during the field investigation further enhanced information on geotechnical conditions expected in the vicinity of the proposed excavations. The field data include summaries of lithologies encountered in each of the BHs advanced in 2009. The encountered lithologies in the majority of BHs consisted mainly of amphibolite with rhyolite intrusions, consistent with expected conditions in the Yates Unit. Additionally, BHs advanced in the southern portion of the project area encountered phyllite units of the Poorman Formation. Central to the effective characterization of the rock mass was an understanding of the nature of the natural discontinuities. This rock mass characteristic was defined in the collected core by the percentage of recovered core and calculation of RQD, calculations of the Q factor, and calculation of RMR for each of the core runs. The average Q and RMR values obtained for each of the BHs are summarized by BH lithology in Table 5.3.2.3-2. A large amount of logs, photographs, images, tables, and computed data exists for further evaluation. In general, core recovery was very good, with percentages well above 90% for the majority of runs in all BHs. The findings from the geotechnical drilling program were intended to provide a basis for Preliminary Design of excavations at the 4850L and for elaboration and implementation of an advanced geotechnical investigation program contemplated to generate the data needed for Final Design of excavations planned at the 4850L.

| Borehole | Q | | | RMR | | |
|---|---|---|---|---|---|---|
| | Amphibolite | Rhyolite | Phyllite | Amphibolite | Rhyolite | Phyllite |
| 3 | 28 | 23 | - | 84 | 79 | - |
| M | 44 | 27 | - | 83 | 81 | - |
| N | 24 | 68 | - | 80 | 86 | - |
| B | 24 | 40 | - | 79 | 83 | - |
| C | 61 | 18 | - | 81 | 72 | - |
| D | 45 | 28 | 15 | 82 | 82 | 81 |
| J | - | 47 | 35 | - | 80 | 81 |

**Table 5.3.2.3-2** Summary of Rock Mass Rating and Q Values for core from the geotechnical borings at the 4850L. [Golder Associates]

### Water Flow

Nine geotechnical BHs were drilled at the 4850L. All BHs were drilled through 10-foot- to 13-foot-long standpipes (HQ-diameter casings), grouted into the formation, and equipped with shut-off valves to control groundwater if encountered during drilling. The grouted-in casings were tested to withstand a groundwater pressure of 13,800 kPa (2,000 psi) without leaks. If leaks around the casings were noted, the casings were regrouted and retested until competent seals were attained and documented. The majority of the BHs encountered small quantities of groundwater. Observed flow rates range from 0.42 liters per minute (L/min) in BH3, measured on September 21, 2009 (about two weeks after completion of drilling), to 0.02 L/min observed in BHC on November 14, 2009. Detailed information on the flow rates can be found in the drilling report. Flow rates have declined significantly in BHs that initially showed the greatest discharges. The reduction in flow rates over time is expected to continue. Based on examination of core from all geotechnical BHs, the rock matrix does not appear to have any significant hydraulic conductivity or porosity. Therefore, the water is assumed to be present only in interconnected discontinuities. The discontinuities appear tight, and drilled formations have a limited ability to yield water. The groundwater monitoring is an ongoing task[9,10] and will continue as long as the BHs are available and accessible.

### 5.3.2.4 In situ Stress Measurements

As part of the geotechnical site investigations, a total of 10 overcoring stress tests,[11] utilizing hollow inclusion cells (HI-cell), were conducted, of which eight tests were successful. The three locations for the stress-measurement holes with lengths up to 8 meters (26 feet) were selected along the 4850L Ventilation Drift, as shown in the drilling map above. Six of the measurements were conducted in the amphibolite of the Yates Unit, and two of the measurements were conducted in the rhyolite. The in situ stress measurements address the lack of data from these rock types and at these depths and this location compared with the historical data available from Homestake database. The overcore samples are tested in biaxial cells to derive elastic properties, which are used for the stress calculation. This configuration tested variability within the amphibolite itself along with variability to a different rock (the rhyolite). Because of its foliations and calcite veining, the amphibolite is complex both in terms of structure and lithology. Water and cell temperatures were monitored during the stress tests to ensure that thermal effects did not affect the quality of the tests. Maximum observed temperature deviations of 4°C were not significant enough to affect the test results.

The solution for homogenous, isotropic rock was used to calculate the stress from the overcored strains. Two out of 10 measurements did not produce results, leaving two rhyolite and six amphibolite measurements. As with any overcoring method, the stress calculation depends on the characterization of the elastic properties of the rock, which are usually determined for the specific overcore sample using a biaxial cell. The average Young's modulus and Poisson's ratio values determined from the biaxial cell in the field and the comparable values obtained from the laboratory testing program and calculated stress tensor components and their orientations are presented in Tables 5.3.2.4-1, 5.3.2.4-2, and 5.3.2.4-3, correspondingly.

| Lithology | Biaxial Cell | | Laboratory Values | | | |
|---|---|---|---|---|---|---|
| | E (MPa) | $\nu$ | E (Mean) (MPa) | E (S.D.) (MPa) | $\nu$ (Mean) | $\nu$ (S.D.) |
| Amphibolite | 85.9 | 0.29 | 89.0 | 15.0 | 0.23 | 0.04 |
| Rhyolite | 61.5 | 0.36 | 70.0 | 28.0 | 0.21 | 0.06 |

**Table 5.3.2.4-1** Elastic constants. [RESPEC]

| Lithology | $\sigma_{HMax}$ (MPa) | Degrees from North | $\sigma_{HMin}$ (MPa) | $\sigma_{v}$ (MPa) | $\sigma_{oct}$ (MPa) | $\tau_{oct}$ (MPa) |
|---|---|---|---|---|---|---|
| Amphibolite | 46.5 | 27.3 | 36.1 | 39.5 | 40.7 | 9.1 |
| Rhyolite | 37.2 | 140.9 | 30.1 | 58.2 | 41.2 | 13.5 |

**Table 5.3.2.4-2** Stress components. [RESPEC]

| Test No. | Rock Type | Major ($\sigma 1$) | | | Intermediate ($\sigma 2$) | | | Minor ($\sigma 3$) | | |
|---|---|---|---|---|---|---|---|---|---|---|
| | | Magnitude (Mpa) | Bearing (°) | Dip (°) | Magnitude (Mpa) | Bearing (°) | Dip (°) | Magnitude (Mpa) | Bearing (°) | Dip (°) |
| SM-02 | Amphibolite | 43.6 | 15 | 18 | 35.2 | 256 | 56 | 30.8 | 115 | 28 |
| SM-03 | Amphibolite | 76.8 | 209 | 48 | 48.7 | 310 | 10 | 41.5 | 49 | 40 |
| SM-04 | Amphibolite | 59.7 | 224 | 31 | 47.9 | 44 | 59 | 40.3 | 134 | 0 |
| SM-05 | Amphibolite | 43.6 | 185 | 21 | 27.0 | 275 | 1 | 19.0 | 8 | 69 |
| SM-06 | Amphibolite | 61.1 | 22 | 9 | 38.7 | 113 | 66 | 31.7 | 316 | 22 |
| SM-07 | Amphibolite | 34.2 | 133 | 51 | 27.8 | 359 | 30 | 24.9 | 255 | 23 |
| SM-08 | Rhyolite | 60.8 | 258 | 75 | 38.4 | 145 | 6 | 26.9 | 54 | 14 |
| SM-09 | Rhyolite | 59.5 | 175 | 75 | 34.8 | 308 | 10 | 30.7 | 40 | 11 |

**Table 5.3.2.4-3** Measured in situ principal stress magnitudes and orientations. [Golder Associates]

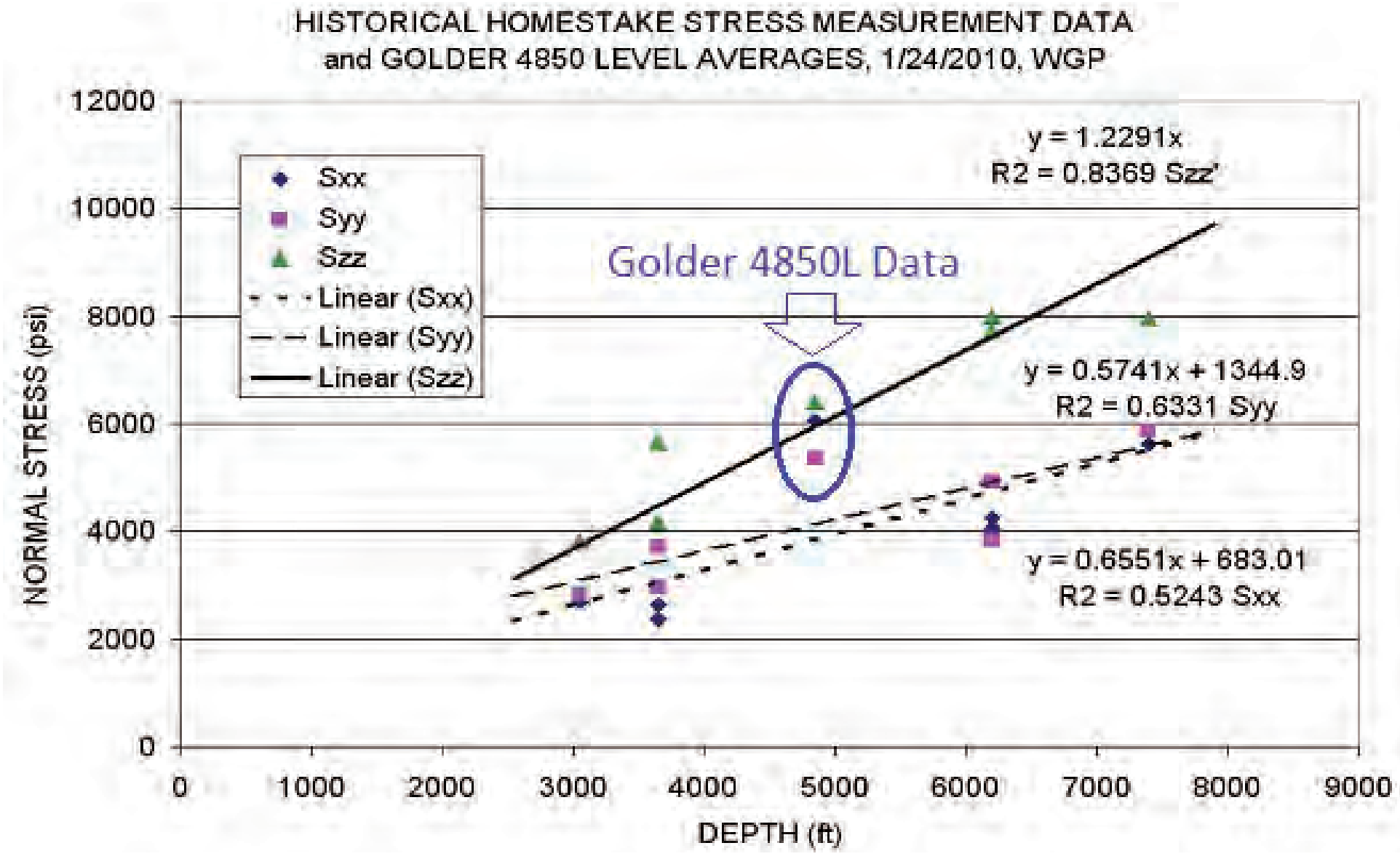

**Figure 5.3.2.4-1** Comparison of in situ stress measurements. [William Pariseau]

On average, the standard deviation for principal stresses and stress components was about 11 MPa for the amphibolite and 1 MPa for the rhyolite. The large variability in the amphibolite could be explained by natural rock variability or inconsistencies in methodology. Review of the tests did not reveal any indication of problems, and therefore it was concluded that the results are correct and accurately represent the in situ state of stress. The rhyolite measurements were more consistent, but only two tests were conducted. The same procedure was used for both the rhyolite and amphibolite tests; therefore, support is provided for the conclusion that variability in the results is a function of the rock and not the stress measurement methodology. In addition, review of the statistical quality and quantity of strain measurements indicated they were appropriate and sufficient. The variability seen in the laboratory testing results supports the rock variability hypothesis. Although when looking at the individual tests and the resulting variability one may suspect that the spread is too large, there are three stress data features that indicate their usefulness in support of the design: a) the maximum horizontal stress component has a northeast trend overall, even though for the rhyolite it is northwest; b) the maximum principal stress is subvertical; and c) the new data are congruent with the historical stress data (see Figure 5.3.2.4-1). Graphical representation of the principal stress magnitudes and plots of vectors are shown Figures 5.3.2.4-2, 5.3.2.4-3, and 5.3.2.4-4.

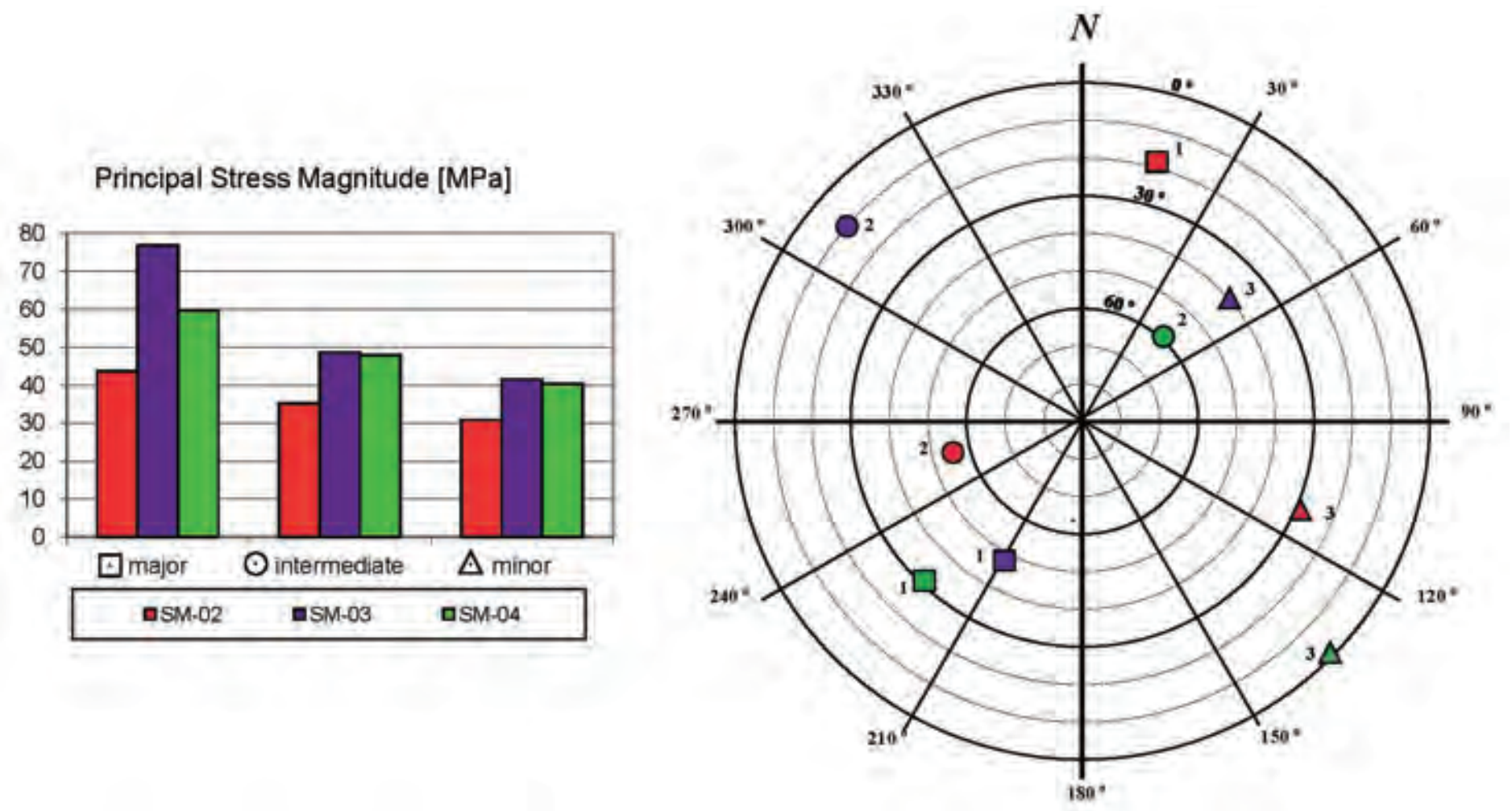

**Figure 5.3.2.4-2** Principal stresses measured at Station SMS I. [Golder Associates]

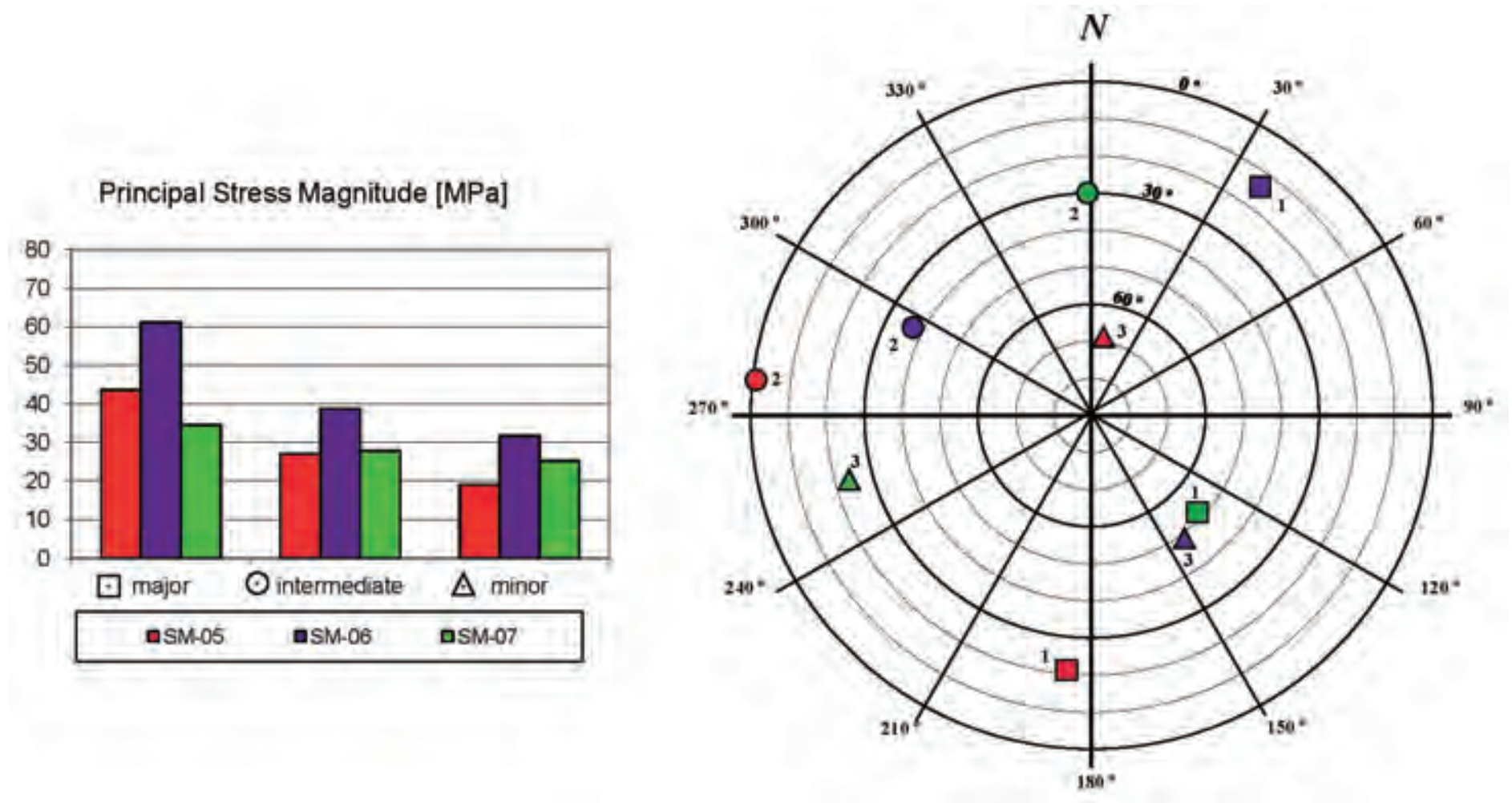

**Figure 5.3.2.4-3** Principal stresses measured at Station SMS II. [Golder Associates]

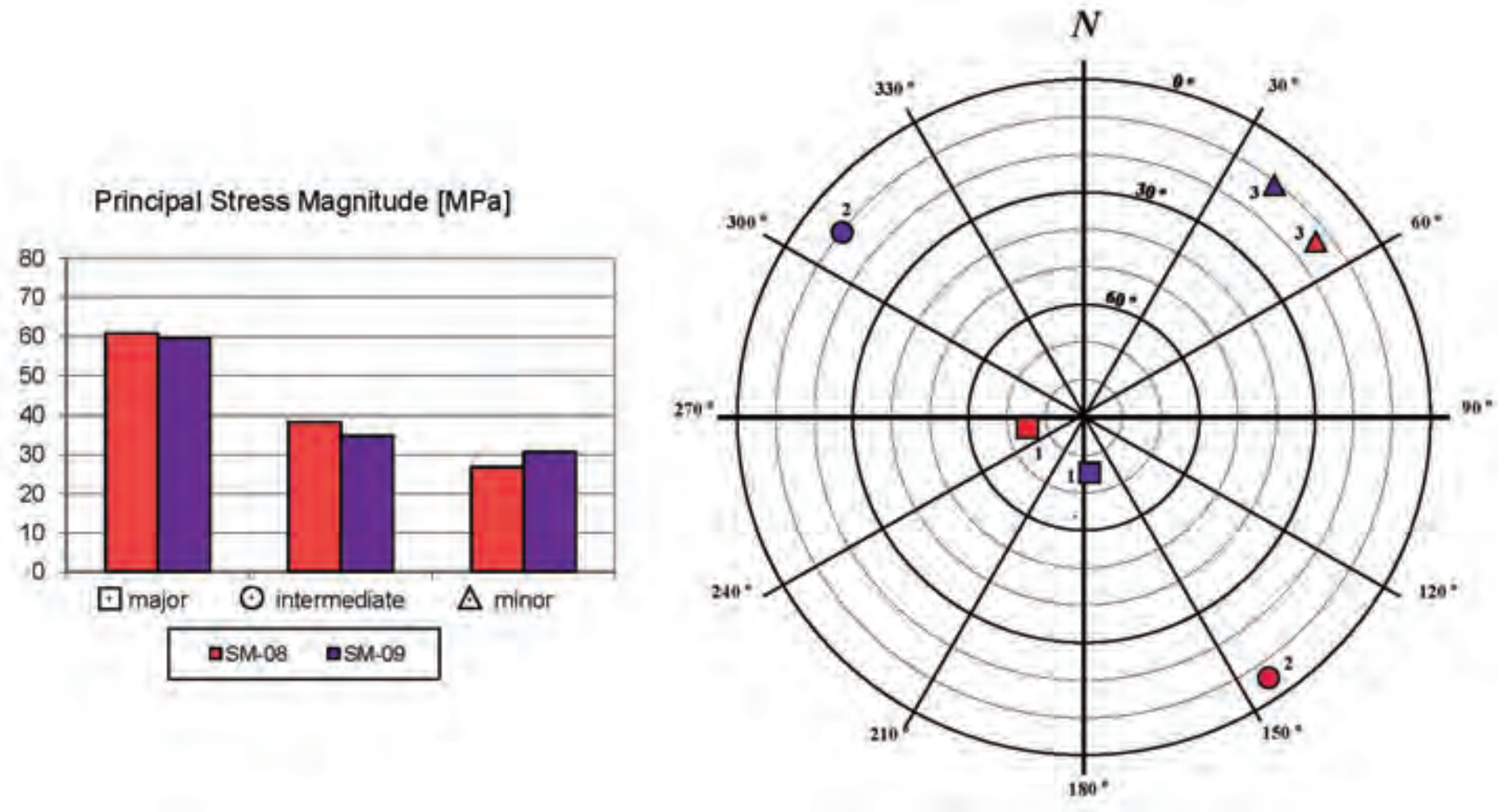

**Figure 5.3.2.4-4** Principal stresses measured at Stations SMS III. [Golder Associates]

The quality of in situ stress data can be addressed by comparing measured results with calculated vertical stresses and other stress indicators. Comparison of the vertical stress to the lithostatic stress shows that the average measured vertical stress of 44.2 MPa (average of amphibolite and rhyolite) is very close to the lithostatic value of 40.4 MPa (at the 4850L assuming rock density of 2,800 kilograms per cubic meter). The laboratory-measured densities for amphibolite 100, amphibolite 392, and rhyolites are 2,950, 2,900, and 2,550 kg/m3, respectively. The average calculated vertical stress for the rhyolite, 58.2 MPa, is higher than the lithostatic value. According to Pariseau, the vertical stress component for the 4850L is 41.8 MPa. The average horizontal stresses in the amphibolite are higher than Pariseau's (1985) gradient values by 1.4 and 1.8 times for the maximum horizontal and the minimum horizontal stresses, respectively. The average of the maximum horizontal stresses appears to be similar to the lithostatic stress, while the minimum horizontal values are 80% of lithostatic. Even with the high variability, average values of the tests are reasonable compared to lithostatic stresses. While the average stresses appear reasonable, the significant variability adds complexity to the state of stress. The numerical modeling in Section 5.3.3.3, however, shows that the rock is sufficiently robust to not be materially affected by this variability.

#### 5.3.2.5 Laboratory Tests

Laboratory testing[12,13,14,15] included uniaxial compressive strength tests with axial, lateral, and shear deformation measurement; indirect tensile (Brazilian) strength tests; triaxial compression tests; direct shear strength tests on representative discontinuities; rock density; and digital photograph recording of test specimens before and after testing. Material properties of interest include strength criterion (friction angle and cohesion) for intact rock and joints and representative values of strength and elastic properties (perhaps anisotropic) for intact rock.

In planning the laboratory testing, the decision was made to drill 15-centimeter (6-inch) diameter core (see Figure 5.3.2.5-1) so that laboratory test specimens (nominally 5-centimeter [2-inch] diameter) could be subcored in different directions (e.g., three orthogonal directions, other orientations, and oriented with the sample foliation). Five sets of core were delivered to the RESPEC laboratory. All of the core had a nominal diameter of about 15 centimeters (6 inches) and arrived in pieces ranging in size from rubble to about 0.6 meter (2 feet) long. Each set of core was recovered from a different BH; BHs were known as DS-1, 100-1, 100-2, 394, and rhyolite. BH locations are shown in Section 5.3.1 with a layout of the drilling program completed on the 4850L. The first BH (DS-1) was a product of the drilling program, and the other four BHs resulted from in situ stress testing. All laboratory testing was completed using samples from three BHs—100-2, 394, and rhyolite—which are identified as the stress measurement sites 1, 2, and 3 in Figure 5.3.1-1. The overall testing effort comprised 49 Brazilian tests, 54 uniaxial stress tests, 29 triaxial tests, and 18 direct shear tests.

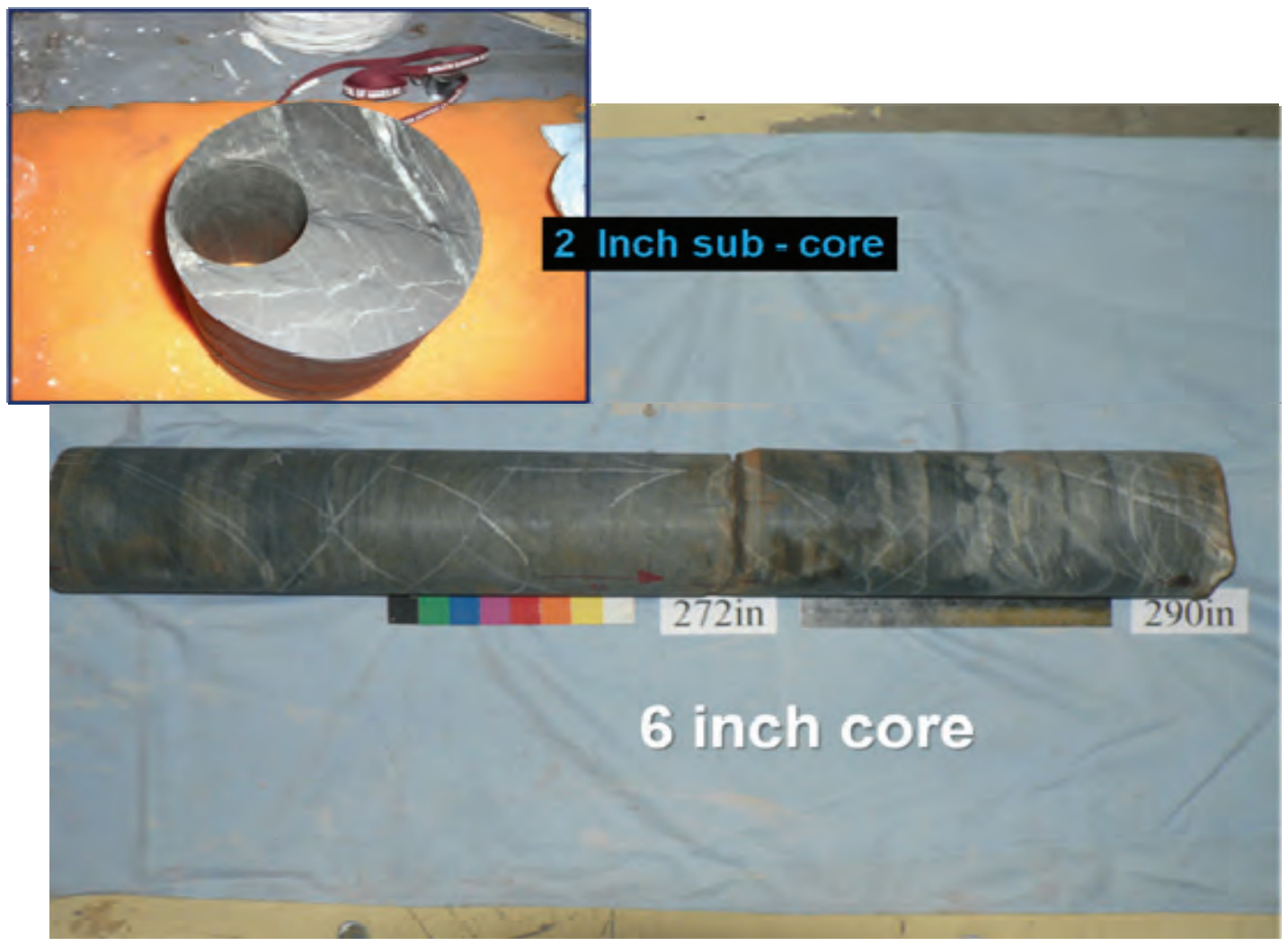

**Figure 5.3.2.5-1** Six-inch core preparation for testing. [Golder Associates]

**Results**

The results from the laboratory tests can be categorized as physical properties, strength properties, and deformational properties (elastic moduli). The first of those results is represented by a bulk density that was determined by measuring and weighing the specimens before they were tested. The density measurements can be summarized in a plot of the density values as a function of their depth along the nearly horizontal BH as shown in Figure 5.3.2.5-2. The density measurements in the plot indicate that the physical properties of the rock will vary along the BH length but do not show any pattern indicating that the density is a function of location along the BH. The density measurements were averaged for each BH and the results are given in Table 5.3.2.5-1.

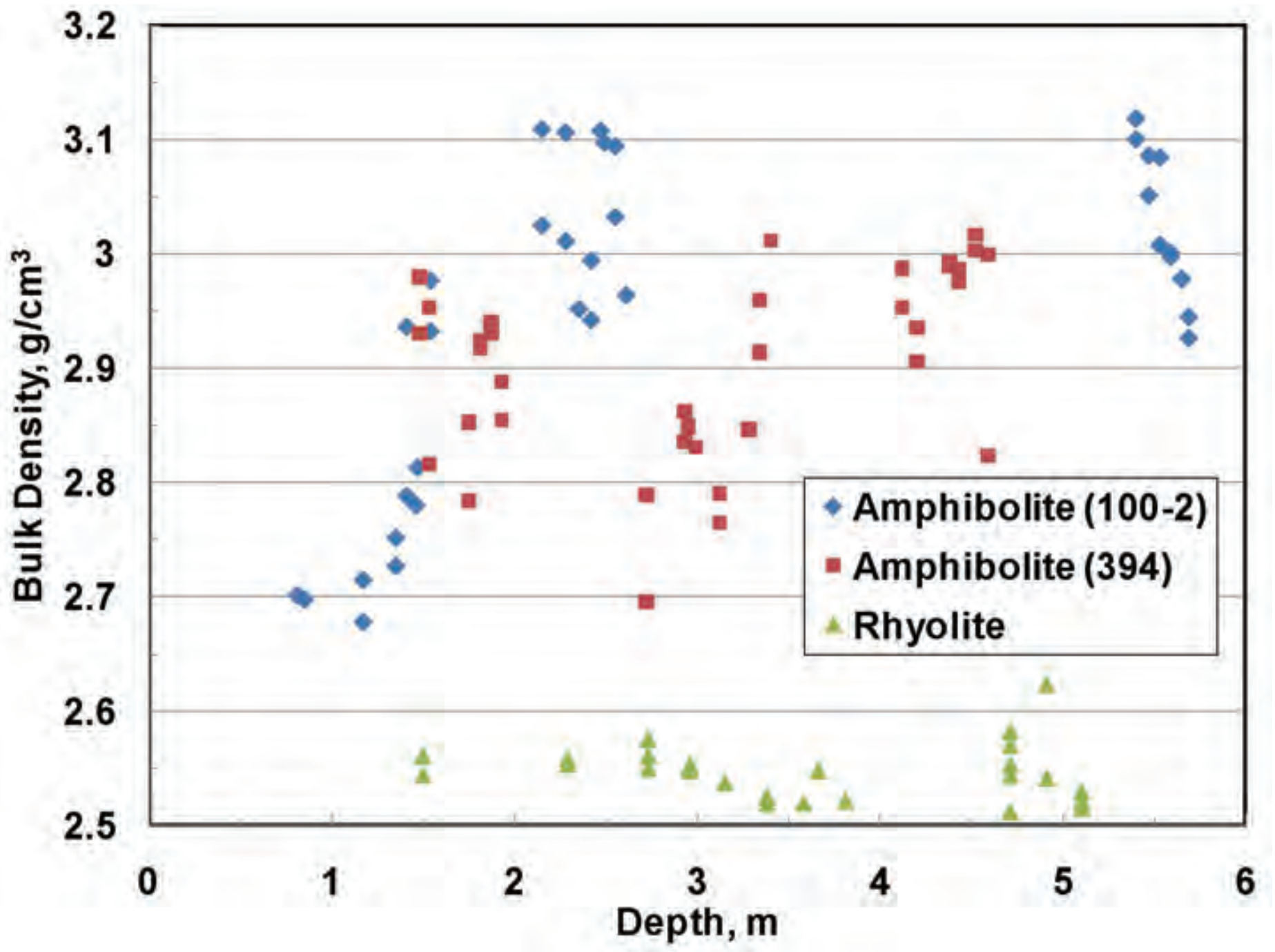

**Figure 5.3.2.5-2** Density measurements for three BHs at the 4850L. [RESPEC]

| Test Location | Number of Specimens | Average Density (g/cm3) | Standard Deviation (g/cm3) |
|---|---|---|---|
| Amphibolite 100-2 | 36 | 2.95 | 0.14 |
| Amphibolite 394 | 36 | 2.90 | 0.08 |
| Rhyolite | 31 | 2.55 | 0.02 |

**Table 5.3.2.5-1** Summary of density measurements. [RESPEC]

The elastic properties and the tensile and compressive strength properties for amphibolite and rhyolite are presented in Tables 5.3.2.5-2 and 5.3.2.5-3, correspondingly. The elastic properties are defined as Young's modulus, Poisson's ratio ν and shear modulus, *G*; and they were all determined in the uniaxial stress tests based on strain gage measurements. The amphibolite properties represent the average of the two amphibolite BHs, 100-2 and 394.

| Rock Type[a] | E (GPa) | | | | ν | | | | G (GPa) | | | |
|---|---|---|---|---|---|---|---|---|---|---|---|---|
| | Mean | S.D. | Low | High | Mean | S.D. | Low | High | Mean | S.D. | Low | High |
| Amphibolite | 89 | 15 | 69 | 103 | 0.23 | 0.04 | 0.14 | 0.25 | 30 | 12 | 10 | 41 |
| Rhyolite | 70 | 28 | 50 | 90 | 0.21 | 0.06 | 0.17 | 0.25 | 31 | 9 | 24 | 37 |

[a] 36 amphibolite tests and 18 rhyolite tests for a total of 54 tests.

**Table 5.3.2.5-2** Comparison of amphibolite and rhyolite elastic properties. [RESPEC]

| Rock Type | Uniaxial Compressive Strength (MPa) | | | | | Tensile Strength (MPa) | | | | |
|---|---|---|---|---|---|---|---|---|---|---|
| | No. Tests | Mean | S.D. | Low | High | No. Tests | Mean | S.D. | Low | High |
| Amphibolite | 36 | 115 | 52 | 33 | 216 | 36 | 14 | 7 | 0.2 | 35 |
| Rhyolite | 18 | 111 | 55 | 28 | 223 | 13 | 10 | 4 | 5 | 20 |

**Table 5.3.2.5-3** Comparison of amphibolite and rhyolite strength. [RESPEC]

The direct shear strength results were obtained for joints, which were only observed to be present in the amphibolites. The rhyolite did not exhibit any visually apparent joint sets that could be selected for testing. Because the direct shear tests were performed over a range of normal stresses, a Mohr-Coulomb strength model could be fitted to the data to represent the strength of the amphibolites. This was done for both the peak strength of the joints and the subsequent residual strength values that were observed after failure. The Mohr-Coulomb model has two parameters: the cohesion and the angle of internal friction. The parameter values determined by fitting to the data are given in Table 5.3.2.5-4.

| Rock Type | Number of Tests | Type | Cohesion (MPa) | Friction Angle (°) |
|---|---|---|---|---|
| Amphibolite | 18 | Peak | 12.0 | 47 |
| Amphibolite | 18 | Residual | 3.6 | 42 |

**Table 5.3.2.5-4** Comparison of peak and residual Mohr-Coulomb strength parameters for joints in amphibolite. [RESPEC]

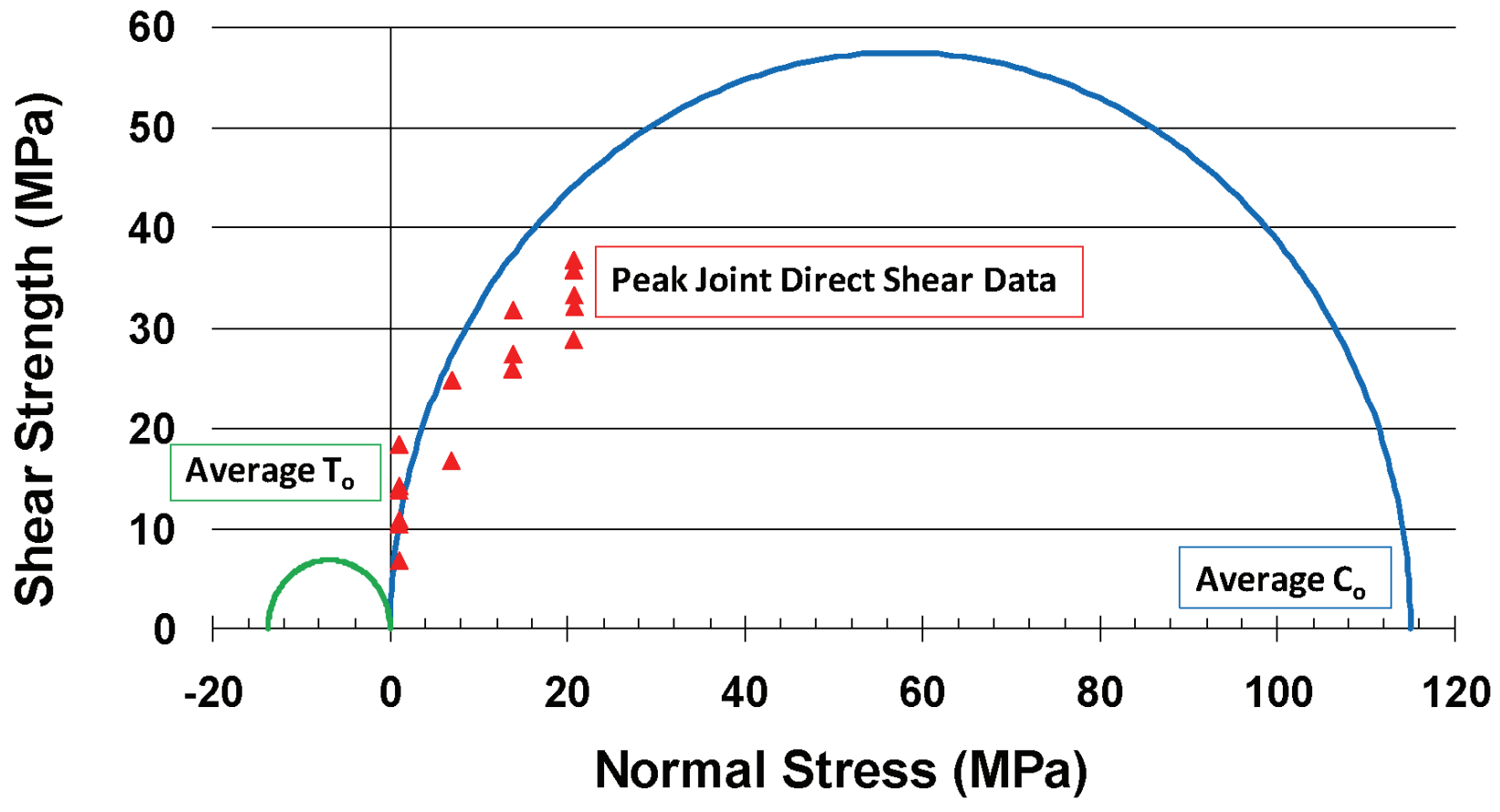

**Figure 5.3.2.5-3** Mohr-Coulomb plot for the direct shear data. [W. Pariseau]

The direct shear data for peak joint failure in the amphibolites is compared to average uniaxial compressive strength ($C_0$) and average tensile strength ($T_0$) for the amphibolites, as shown on a Mohr-Coulomb plot in Figure 5.3.2.5-3. The uniaxial compressive strength ($C_0$) and average tensile strength ($T_0$) can be related to the joint cohesion and angle of internal friction for a linear Mohr-Coulomb envelope. This suggests that the peak joint strengths are very high and in the range of 70 to 75% of the intact amphibolite. The limited direct shear testing program completed to date will be continued in the Final Design phase on more representative rock samples from site-specific areas to fully evaluate a more representative dataset and to determine what if any implications to design may exist.

The triaxial tests results were used to calculate the elastic properties and failure criterion (cohesion and friction angle) for both amphibolite and rhyolite. Tables 5.3.2.5-5 and 5.3.2.5-6 summarize these parameters.

| Rock Type | No. of Tests | E (GPa) | | | | ν | | | |
|---|---|---|---|---|---|---|---|---|---|
| | | Mean | S.D. | Low | High | Mean | S.D. | Low | High |
| Amphibolite | 25 | 83 | 15 | 48 | 109 | 0.27 | 0.04 | 0.23 | 0.45 |
| Rhyolite | 4 | 56 | 4 | 51 | 60 | 0.19 | 0.02 | 0.17 | 0.20 |

**Table 5.3.2.5-5** Summary of deformation properties for amphibolite and rhyolite. [RESPEC]

| Rock Type | Cohesion (MPa) | Angle of Internal Friction (°) | Tension Cut-Off (MPa) |
|---|---|---|---|
| Amphibolite | 37 | 24 | 14 |
| Rhyolite | 38 | 20 | 10 |

**Table 5.3.2.5-6** Summary of Mohr-Coulomb strength parameters for amphibolite and rhyolite. [RESPEC]

To create a representative dataset for estimating the cohesion and the friction angle in the Mohr-Coulomb strength model, the triaxial data from this testing program were combined with the uniaxial strength data for the amphibolite and rhyolite specimens. The resulting combined datasets and least squares fits for failure envelopes are shown in Figures 5.3.2.5-4 and 5.3.2.5-5.

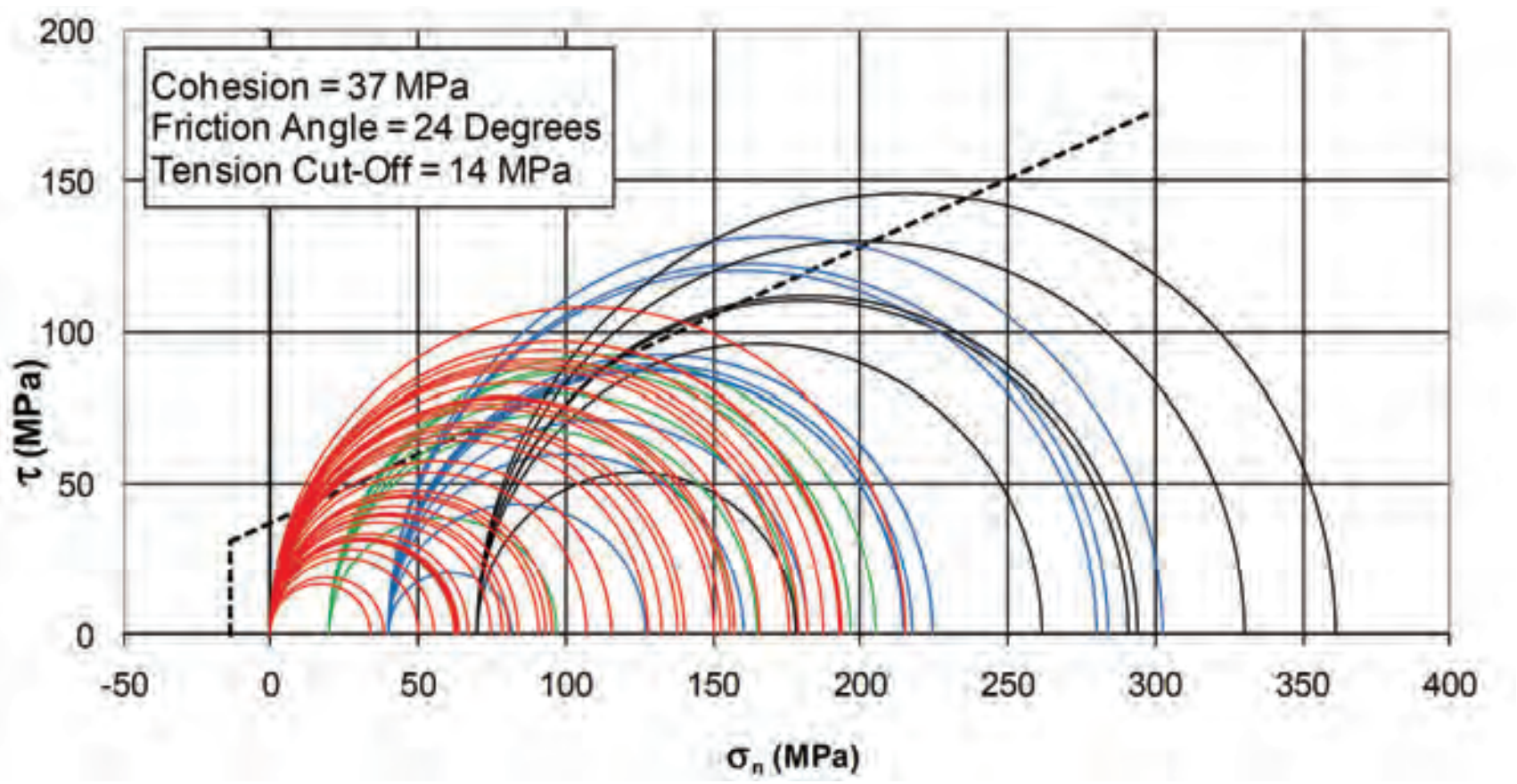

**Figure 5.3.2.5-4** Mohr-Coulomb failure criterion for amphibolite. [W. Pariseau]

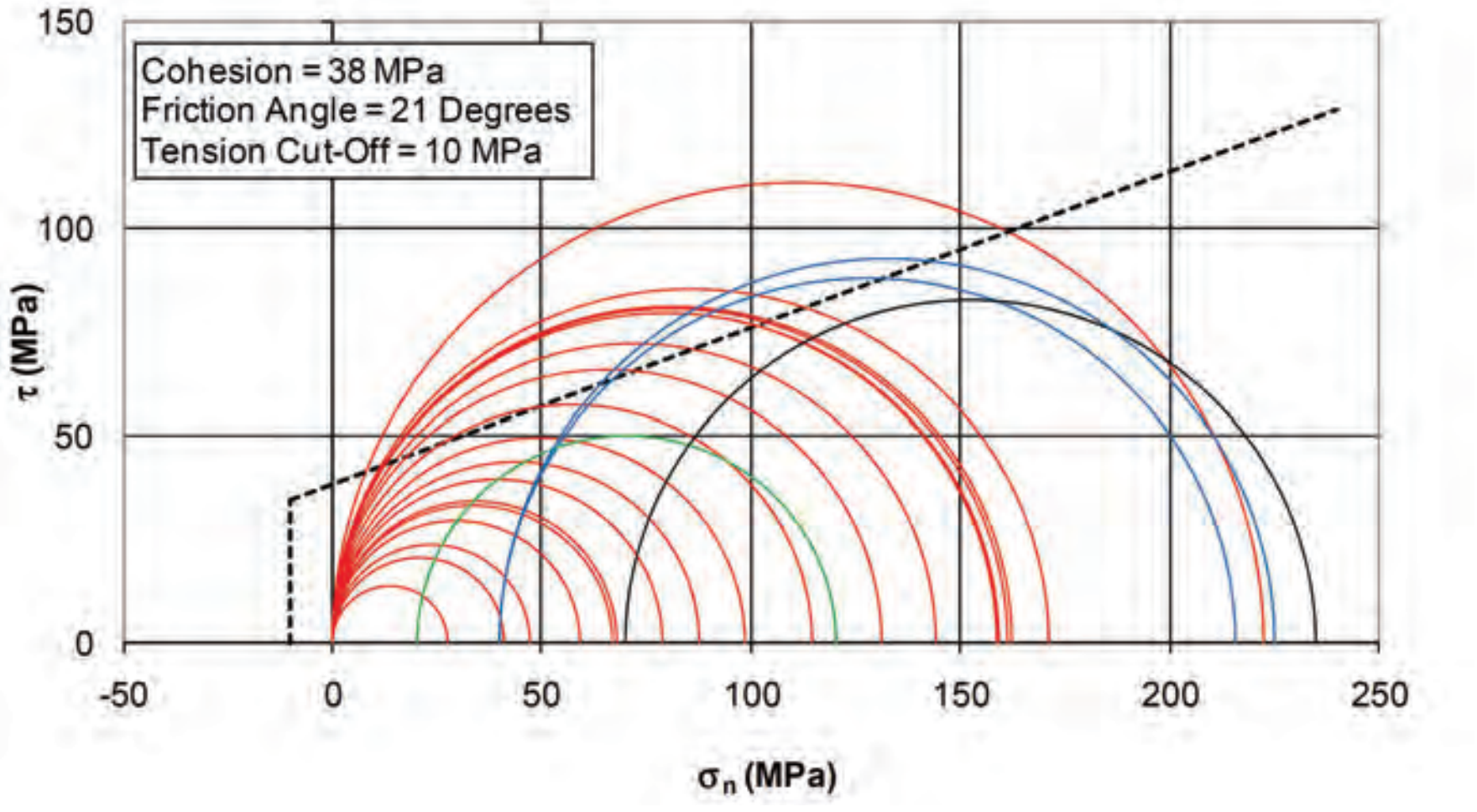

**Figure 5.3.2.5-5** Mohr-Coloumb failure criterion for rhyolite. [W. Pariseau]

An assessment of the anisotropy in material properties was made, and it was determined that any anisotropy that may exist in the amphibolites in triaxial compression is masked by the overall variability of the rock. Significance of anisotropy to the Large Cavity design requires further evaluation be conducted during Final Design.

In addition to the laboratory tests performed by RESPEC, a University of Utah (UU) student team led by Dr. William Pariseau carried out independent laboratory tests16 as part of a class laboratory and senior design project. The student team donated their time and effort into DUSEL Project. The results greatly enhanced the project geotechnical database. The somewhat higher strength of the UU samples may be attributed to a better quality rock core taken from the Hole N while the RESPEC samples were taken from the 6-inch cores in an area disturbed by the ventilation drift. The combined results (UU and RESPEC) are presented in Figures 5.3.2.5-6 and 5.3.2.5-7 and Table 5.3.2.5-7. Generally, the response was quite linear to the elastic limit with failure occurring suddenly and violently, often with disintegration of the entire test specimen. The differences in failure are associated with differences in fitting procedures as seen in the figures.

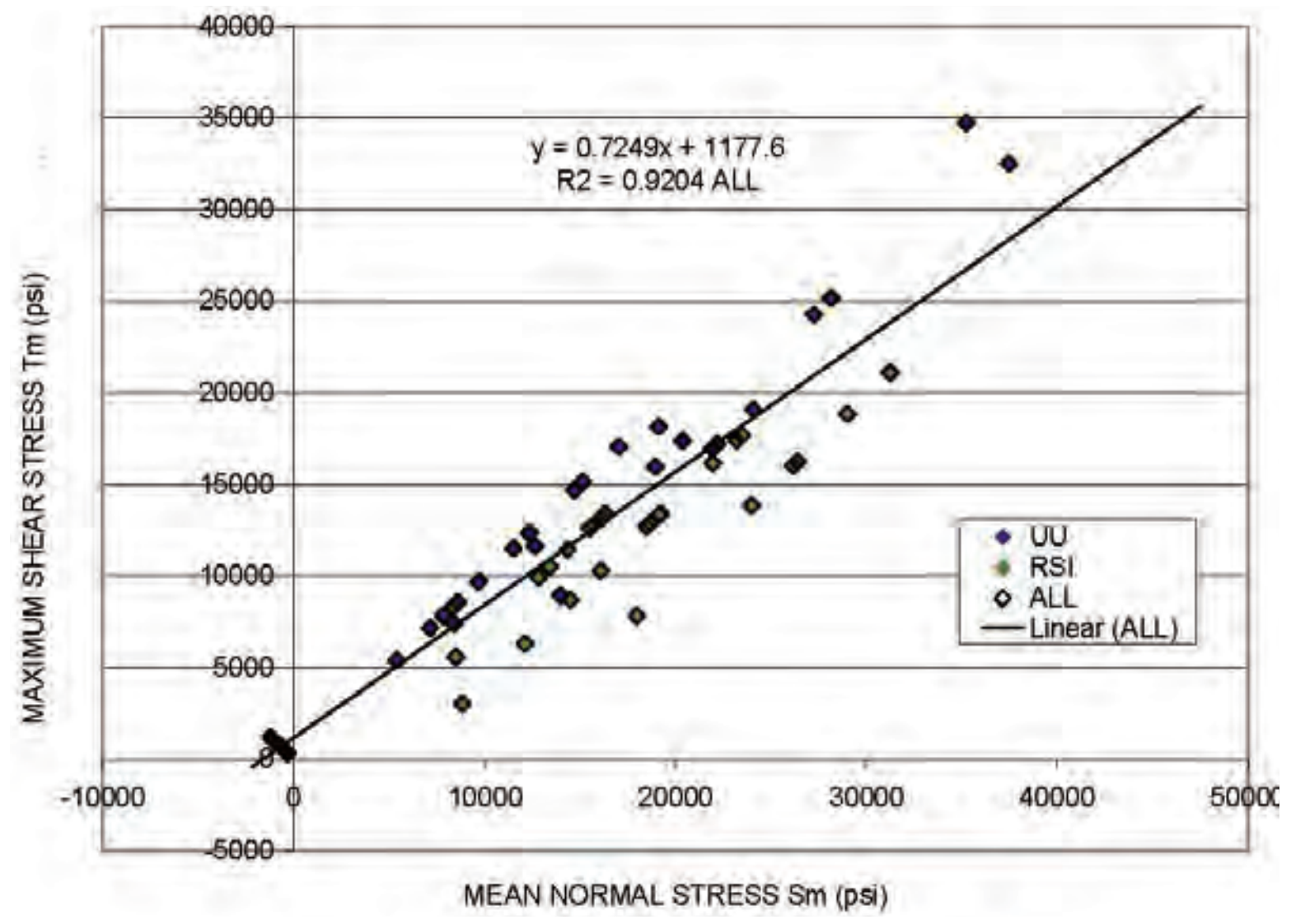

**Figure 5.3.2.5-6** Failure criterion for amphibolite. [W. Pariseau].

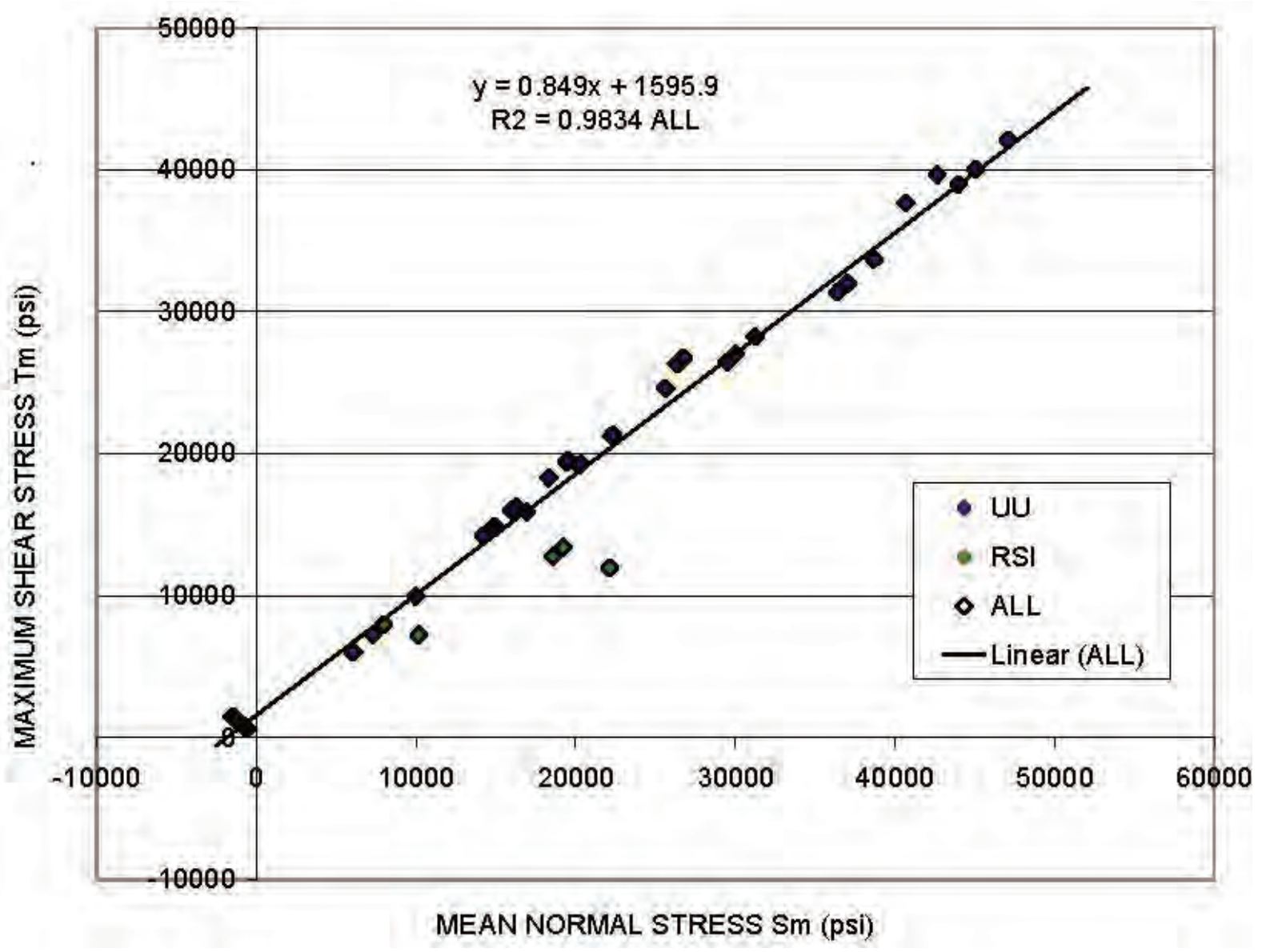

**Figure 5.3.2.5-7** Failure criterion for rhyolite. [W. Pariseau]

| Institution | Rock | *G*, GPa | $T_0$, MPa | $C_0$, MPa | *C*, MPa | $\phi_0$ |
|---|---|---|---|---|---|---|
| RESPEC | Amphibolite | 89 | 11 | 115 | 38 | 24 |
| University of Utah | Amphibolite | 79 | 11 | 151 | 12 | 46 |
| RESPEC | Rhyolite | 70 | 10 | 111 | 38 | 20 |
| University of Utah | Rhyolite | 60 | 13 | 213 | 21 | 58 |

**Table 5.3.2.5-7** Combined laboratory strength results.

## 5.3.3 Geotechnical Modeling and Analysis

### 5.3.3.1 Initial Compilation and Reduction of Geotechnical Data

Compilations, reductions of geotechnical data recorded during the field investigation, and preliminary baseline analyses were performed to evaluate rock conditions in the vicinity of the proposed excavations. The first step in the reduction of this data was to digitize the logs produced for each BH during the drilling program.[8]

The lithologies in the majority of BHs consisted mainly of amphibolite of the Yates Member with rhyolite intrusions. Additionally, BHs advanced in the southern portion of the project area encountered phyllite units of the Poorman Formation. Rock structure data, including location and orientation of discontinuities, were recorded and presented using stereographic projection as a series of stereonets for each drill hole and each geologic domain. An example in Figure 5.3.3.1-1 shows stereonets for joints, veins, foliation, and all discontinuities for BH3.

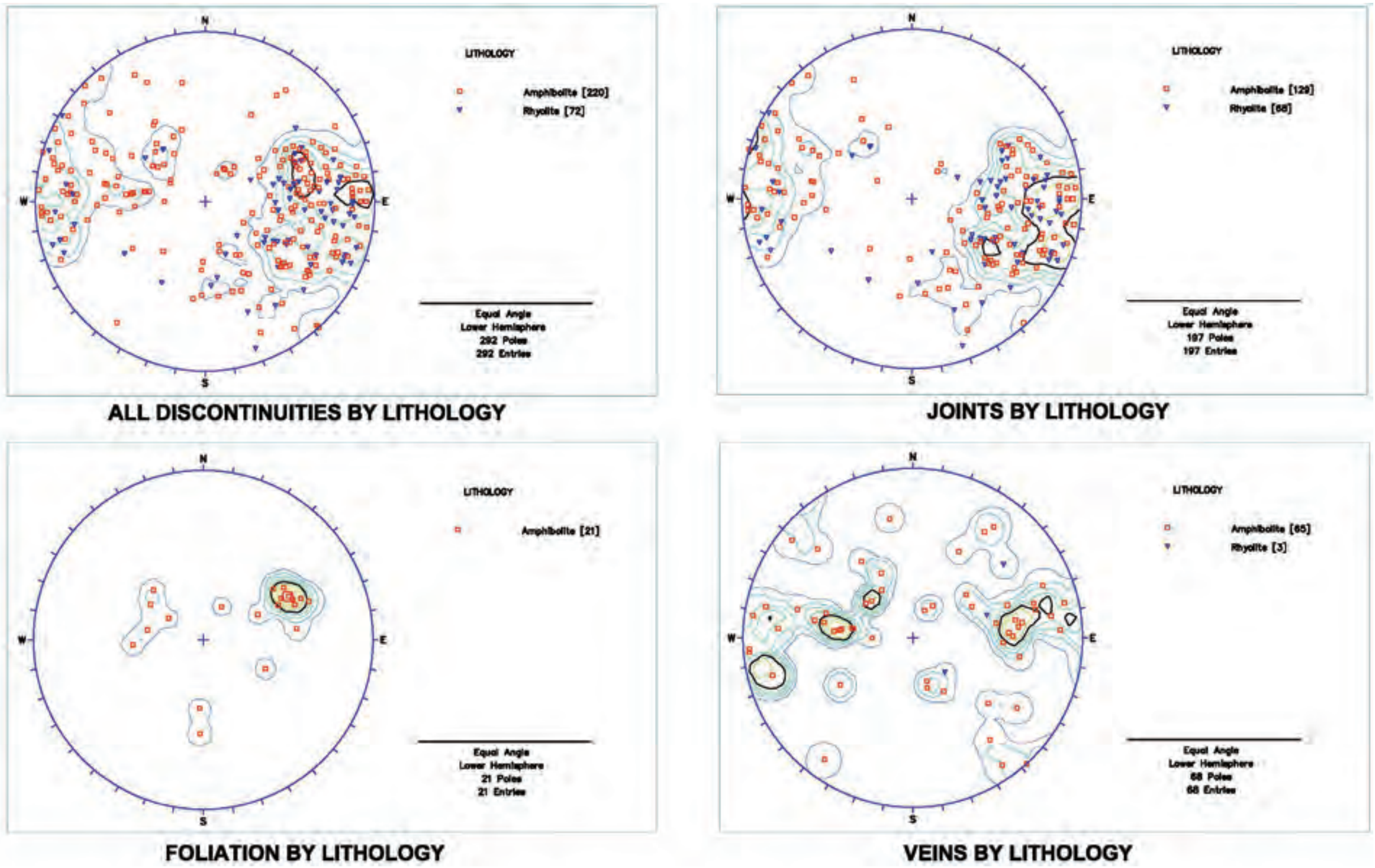

**Figure 5.3.3.1-1** Stereonet plots of corrected discontinuity data. [RESPEC]

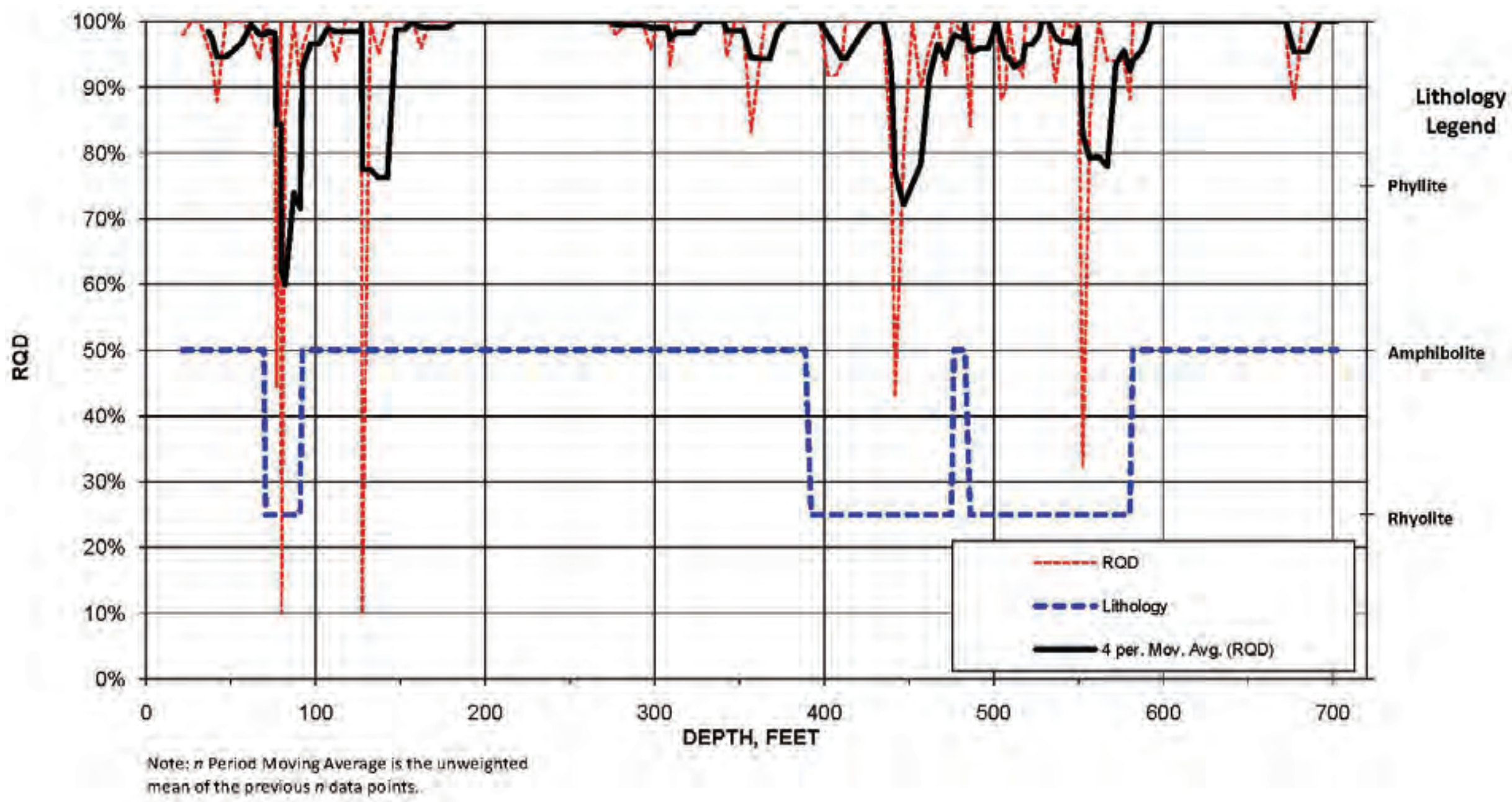

**Figure 5.3.3.1-2** Plot of RQD and lithology vs. depth. [Golder Associates]

In general, it was found that the data recorded during the field core logging correlated well with the data collected by the televiewer BH logging. Also, the BH orientation data collected during televiewer logging validated the BH survey data collected during the drilling program.

Crucial to the effective characterization of the studied rock mass is an understanding of the nature of the natural discontinuities. This rock mass characteristic was primarily defined in the collected core by the percentage of recovered core and calculation of the Rock Quality Designation (RQD) for each advanced run. Core recovery is the length of core recovered for each run expressed as a percentage of the total length of the run. Core recoveries were generally very good, with percentages well above 90% for the majority of runs in all BHs. The RQD value for each run of core was calculated as the percentage of the total run length comprised of intact core pieces greater than 4 inches (0.1 m) in length. An example in Figure 5.3.3.1-2 shows the RQD variation with depth for BH3. Rock Mass Rating (RMR) and Q values (Q) (see Section 5.3.2.2 for full description of RMR and Q), were also calculated for each run of core. Both the RMR and Q are standard geotechnical assessment tools used worldwide in tunnel engineering.

Examples of the Q and RMR distribution along BH3, shown in Figures 5.3.3.1-3 and 5.3.3.1-4, demonstrate good quality of the encountered rocks.

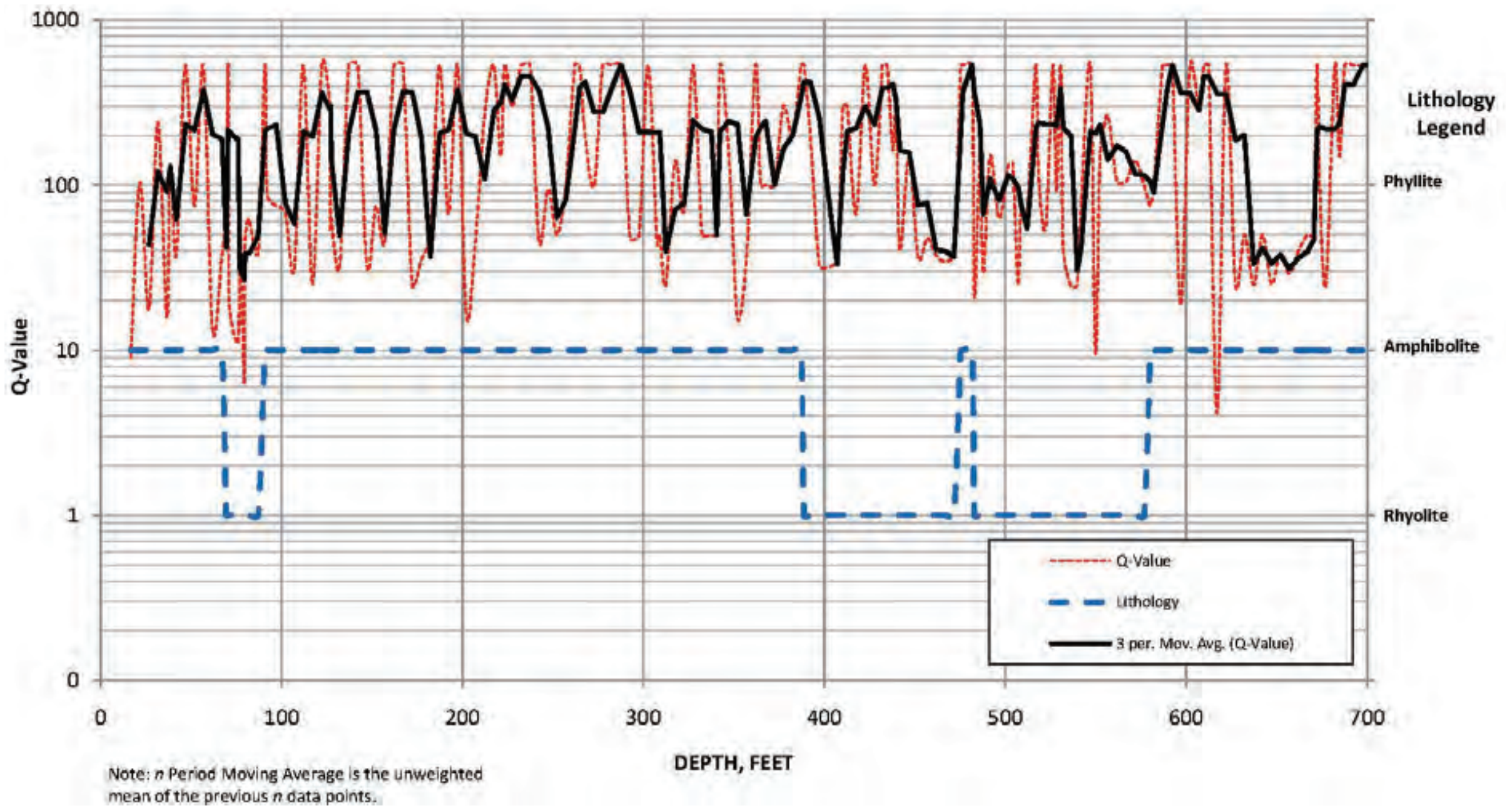

**Figure 5.3.3.1-3** Plot of Q value and lithology vs. depth. [Golder Associates]

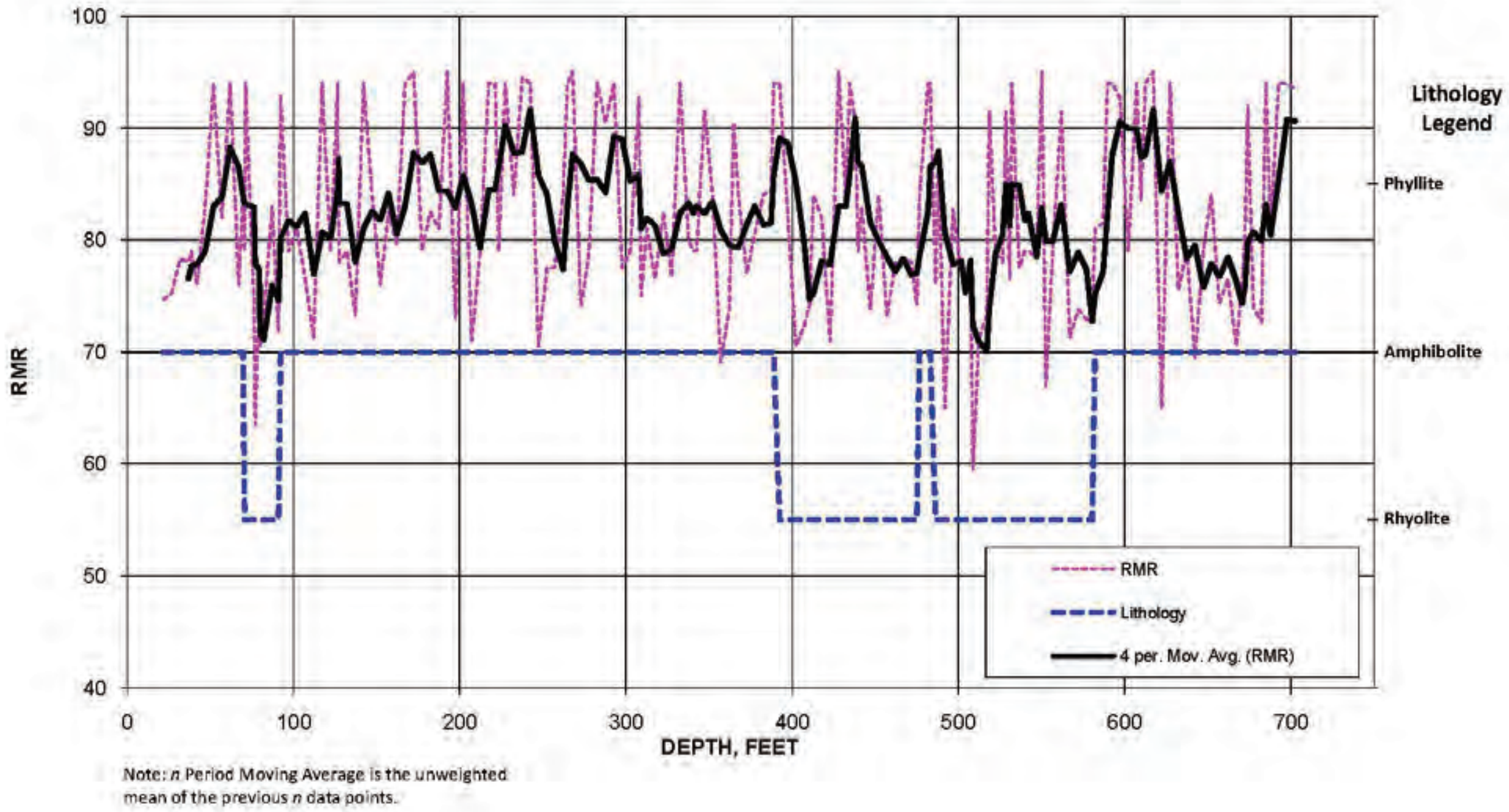

**Figure 5.3.3.1-4** Plot of RMR and lithology vs. depth. [Golder Associates]

### Field Data Conclusions

The field and laboratory data and the initial geotechnical analysis indicate:

a) Overall, the rock is of good quality (RMR>60 and Q>10).
b) No adverse geological or structural features are present that could not be mitigated.
c) Rock mass properties vary but are predictable.
d) In situ state of stress is favorable.

Based on this initial assessment, the suitability of the planned excavations placement is as follows:

a) The present LC-1 location is adequate within one diameter (55 m).
b) The present location of LM-1 and LM-2 is adequate within the Yates Member portion of the triangle.
c) The geotechnical database assembled so far is adequate for the Preliminary Design.
d) Further site-specific geotechnical investigations (including mapping of new excavations, additional drilling and coring, additional laboratory and in situ tests, and ground monitoring) are needed for the Final Design, as discussed in Section 5.3.5.
e) The LC-2, if implemented, would require additional drilling and site investigations at an exploratory level.

#### 5.3.3.2 Preliminary Geological and Geotechnical Assessment

It was recognized early in the project that the data collected during site investigations needed to be coalesced to provide a cohesive understanding of structural geological conditions that may exist across the 4850L. Furthermore, it was understood that the emphasis of these analyses should be on the geology in the vicinity of the proposed location for the LC-1 and the locations of the LM-1 and LM-2. While all proposed excavations are of significant size, LC-1 is the largest, and because of its size and depth is the most challenging. The LC-1 excavation is currently planned as a 180 ft (55 m) diameter, 282 ft (86 m) high upright cylindrical cavity with a domed crown. The largest-span excavation in the world (the Gjørvik Mountain Hall ice rink arena excavation in Norway) is slightly wider than the proposed LC-1 span (200 ft [61 m] versus 180 ft [55 m]). Several powerhouse cavities have been constructed with heights of 165 to 197 ft (50-60 m), but again, the height of the LC-1 and its depth are without precedent.

The 4850L Campus has been evaluated,(Appendix 5.I) on a reconnaissance—level, geologically and geotechnically from three main perspectives: a) geologic interpretation and lithologic model development, b) geologic interpretation of discontinuities, and c) geotechnical characterization of the rock mass. The 3-D model of geology has been continually updated as new information becomes available. Important aspects of these efforts, first, included the examination of the lithological fabric of the rock mass to determine the rhyolite intrusion history and formulate an appropriate geologic interpretation, stressing principally the geometry of the rhyolite bodies and foliation in the amphibolite host rock in the vicinity of the LC-1 area.

Secondly, the litho-structural fabric data gathered from the completed geotechnical drill holes and drift mapping were plotted and analyzed with respect to fracture characteristics (orientation, surface condition, and fracture density). The purpose of this analysis was the identification of structural features of engineering significance inherent in the rock units themselves. Furthermore, the available geotechnical and geomechanical data were analyzed to identify any possible structures (faults) that may be present with the subject area of the 4850L.

Based on the orientations of veins, joints, and the foliation fabrics presented in the stereonets developed from the oriented core, the 4850L was divided into three structural domains: a) a northwest segment—Domain I, b) a southern segment principally encompassing the phyllites of the Poorman Formation—Domain II, and c) a southeast structural segment consisting dominantly of Yates Member amphibolites—Domain III. Characteristics of the three structural domains defined above, and shown in Figures 5.3.3.2-1, 5.3.3.2-2, and 5.3.3.2-3, are as follows:

**Domain I:** comprises mostly the northwest area of the 4850L west of the Lead Anticline axis and encompasses the complete area proposed for the large cavities planned west of the existing Ventilation Drift. Structural fabric orientations within this domain are dominated by two major discontinuity sets within the amphibolite and one discontinuity set within the rhyolite dikes. The conjugate sets were observed within the amphibolite for joints, foliation, and veins. Within Domain I, an additional subdivision, Domain Ia, was made based on observations of deviations in the orientations of the conjugate pairs of joints and veins in the amphibolites (observed in data from BHC).

**Domain II:** comprises approximately a third of the southern 4850L Campus in the area of the Poorman Formation between the vent and exhaust drifts. Structural fabric orientations in this domain are dominated by two joint sets within both the phyllite and rhyolite units. The dominant joint set is observed in both phyllite and rhyolite units. The observed vein orientations are somewhat random in both phyllite and rhyolite units, and the foliation shows evidence of folding.

**Domain III**: comprises the eastern part of the 4850L Campus, principally encompassing the area of BHD lying between (and probably including) the vent and exhaust drifts. Domain III extends chiefly within the Yates Member amphibolite just west of the crest of the Yates Anticline. Foliation is slight. Joints and veins form two strongly developed sets.

The structural fabric illustrated by the stereonets and the inferred domain boundaries also shown in Figures 5.3.3.2-1, 5.3.3.2-2, and 5.3.3.2-3 suggest that some form of structural dislocation occurs between the zones east and west of the vent drift. This observed dislocation is possibly the result of regional disposition of the Poorman-Yates contact in the vicinity of the vent drift, suggesting possible faults. Further analysis of the RQD data, fracture frequency (FF), and inferred peak friction angle (PFA) calculated from the Q System parameters (joint-roughness coefficient and joint-alteration values), revealed that there are specific zones of low RQD, high FF, and low PFA, indicating a possible presence of faults, as shown in Figures 5.3.3.2-4 and 5.3.3.2-5.

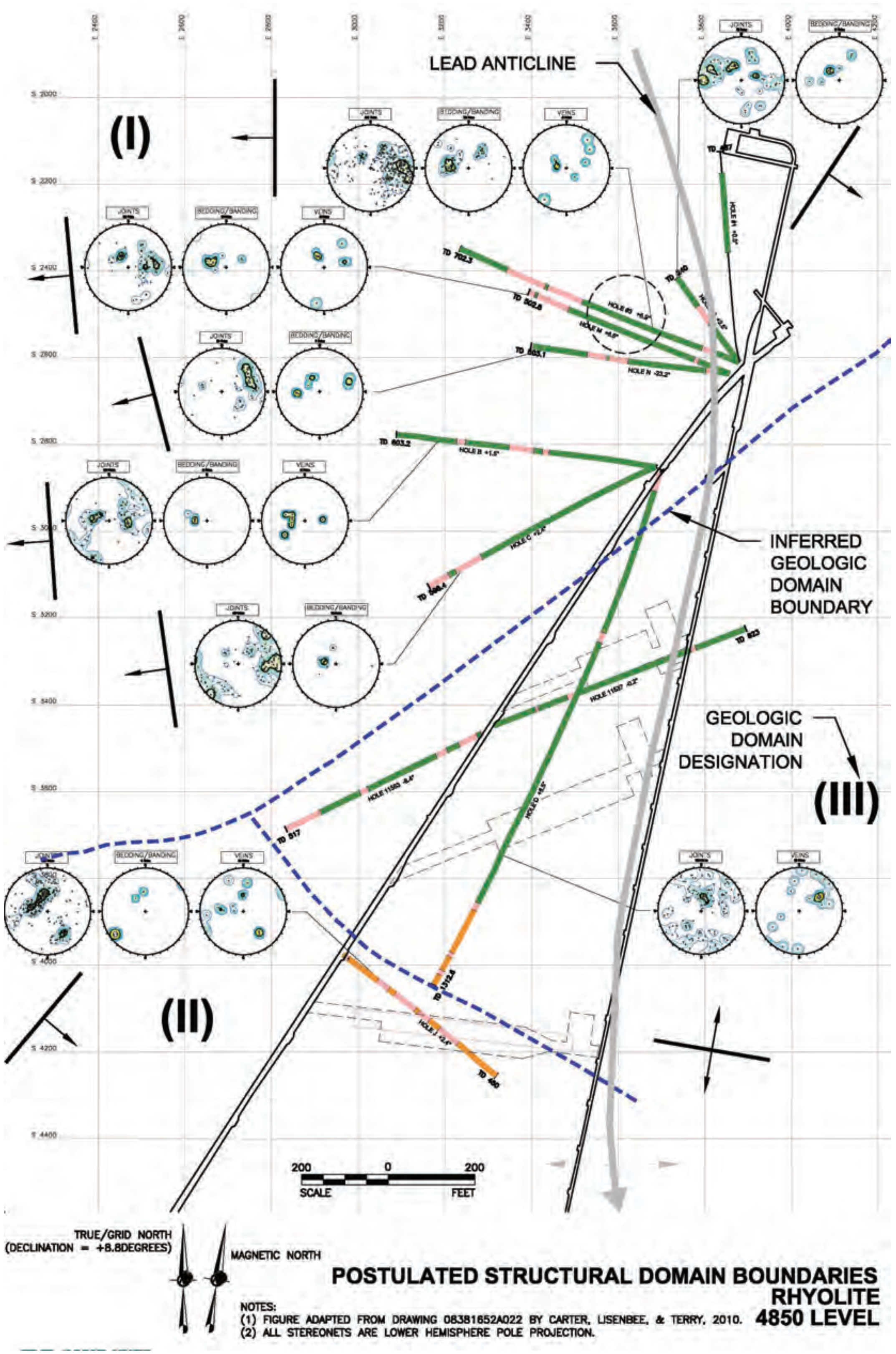

**Figure 5.3.3.2-1** Postulated structural domain boundaries, amphibolite, 4850L. [Golder Associates]

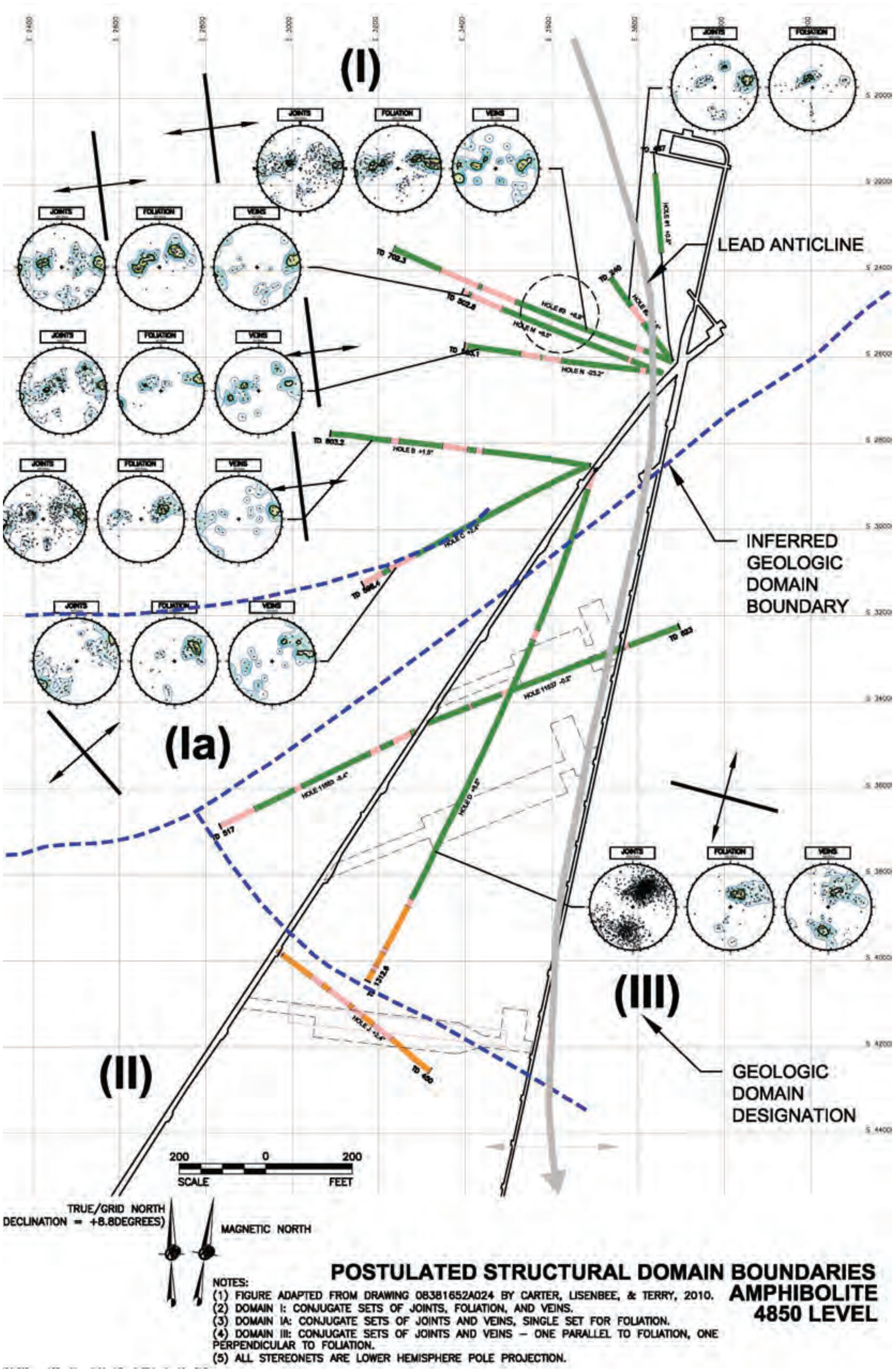

**Figure 5.3.3.2-2** Postulated structural domain boundaries, rhyolite, 4850L. [Golder Associates]

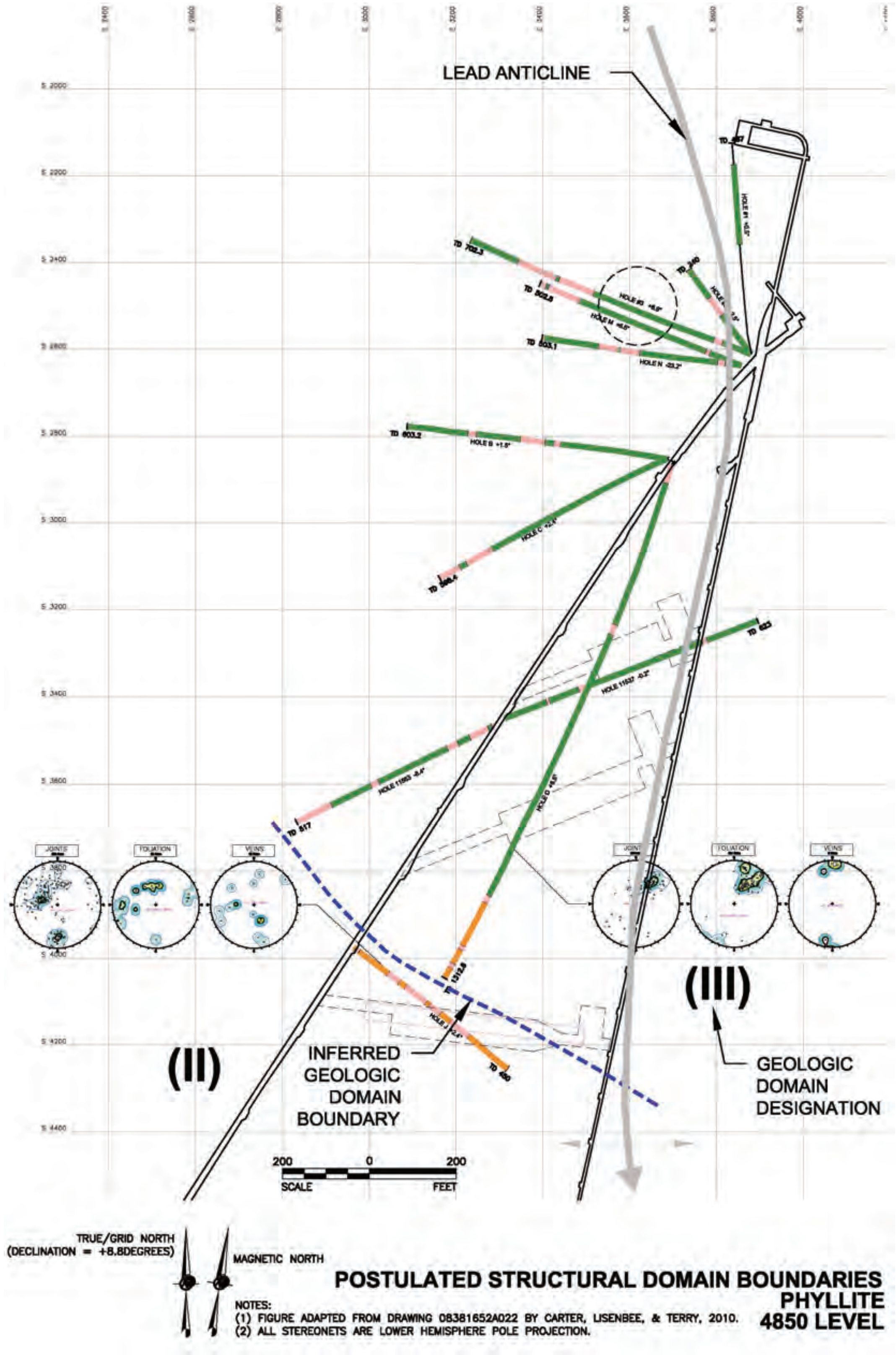

**Figure 5.3.3.2-3** Postulated structural domain boundaries, phyllite, 4850L. [Golder Associates]

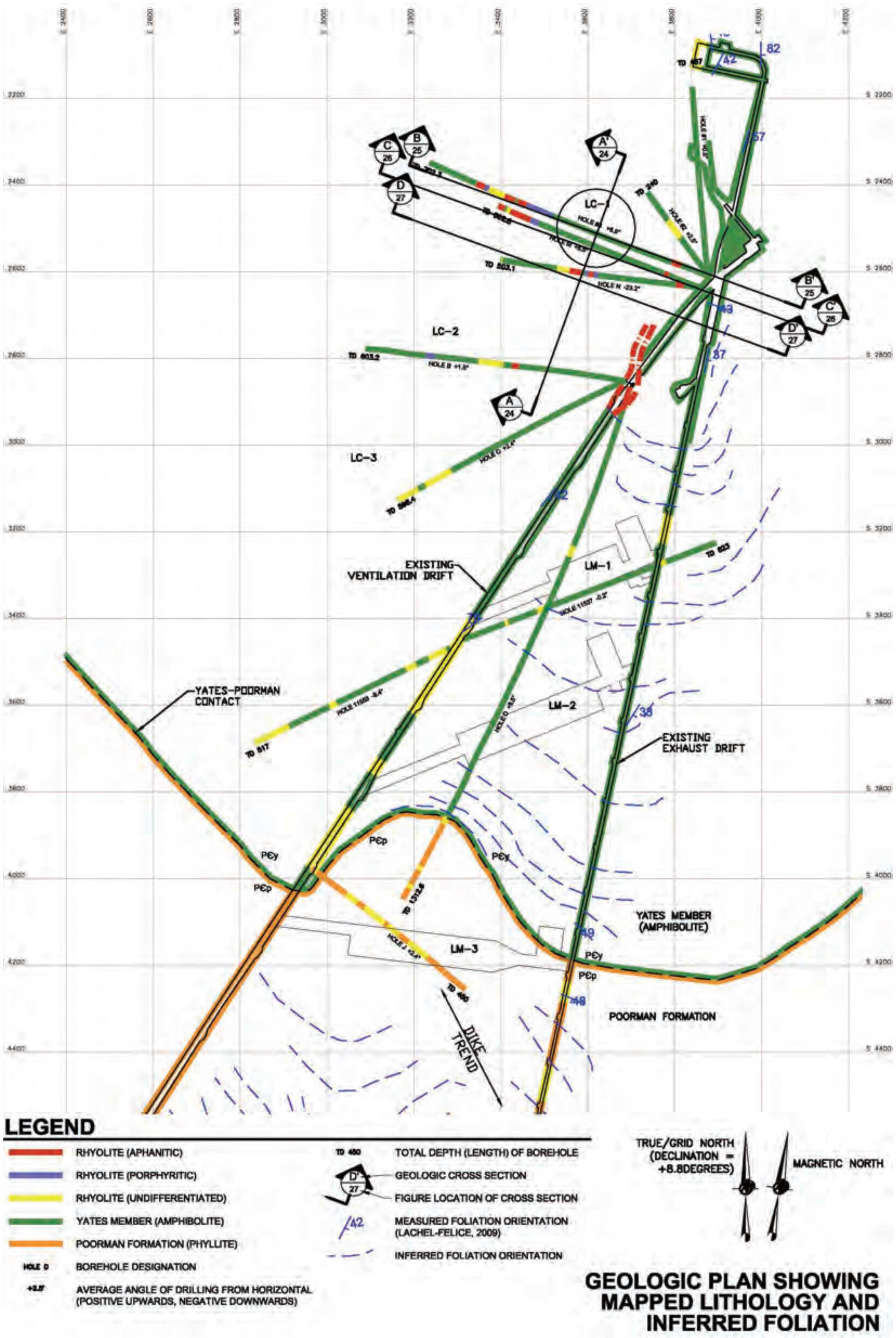

**Figure 5.3.3.2-4** Geological plan showing mapped lithology and inferred foliation. [Golder Associates]

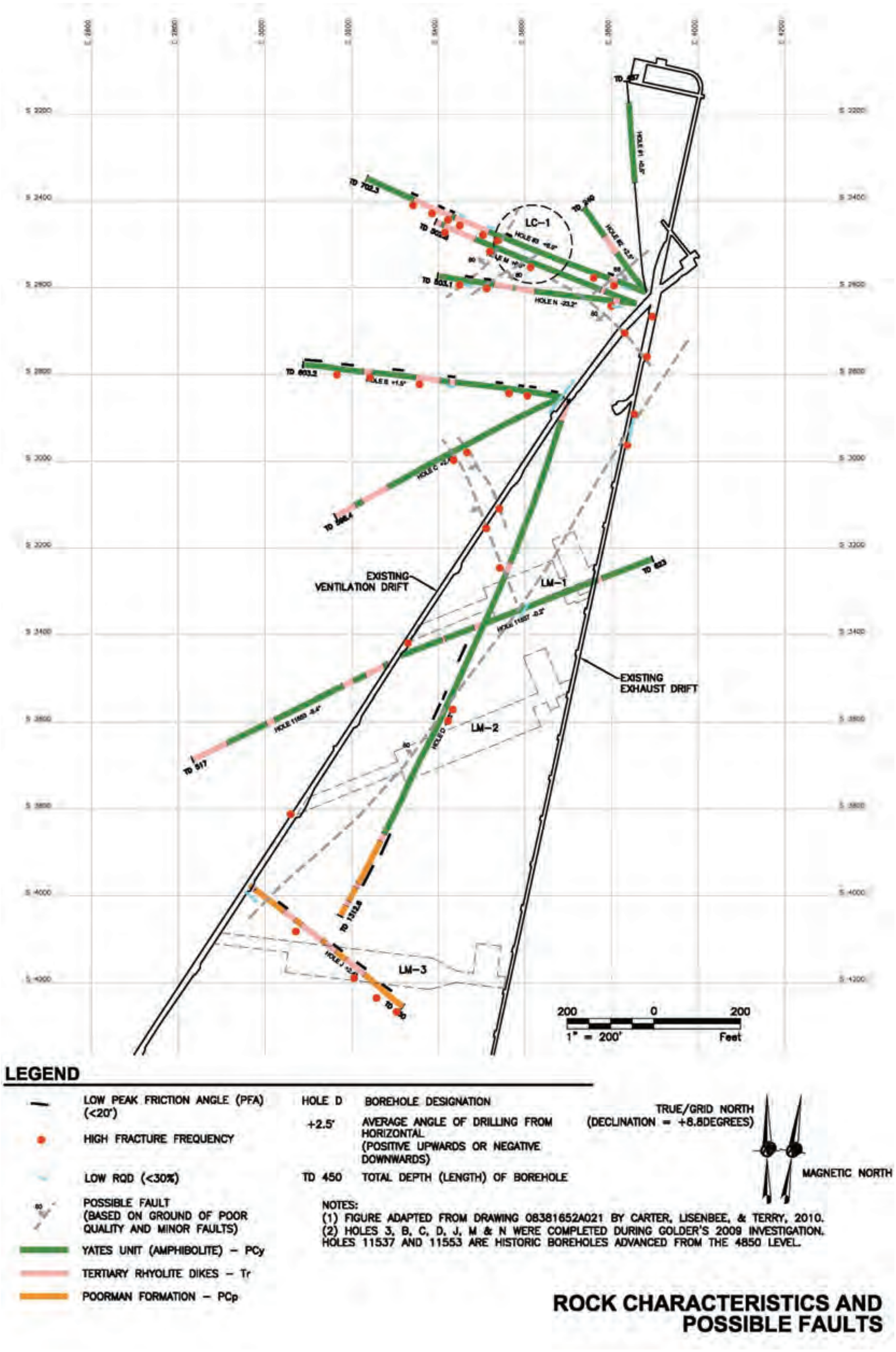

**Figure 5.3.3.2-5** Rock characteristics and possible faults. [Golder Associates]

### Geology of the LC-1 Area

As is clear from the data, two rock units are expected to be present in the LC-1 area—the Precambrian Yates Member amphibolite and the NNE-trending Tertiary rhyolite dike swarm. The structural characteristics of the Yates amphibolite as encountered in the drifts and seen in the BH core are likely due to the proximity of the mapped units to the Lead Anticline axis. In the vicinity of LC-1, foliation displays dip directions ranging from southerly (proximal to the Lead Anticline axis) to easterly/westerly (with increasing distance away from Lead Anticline). Dip angles in this area are observed to range from approximately 10°-30° with the dip angles generally increasing with increased east or west distance from the anticline axis. The Davis Laboratory Module and the Transition Cavity lie in the east limb of the anticline where foliation dips are approximately 30°E. The axis of the Lead Anticline lies between these cavities and the proposed LC-1 site. Two Tertiary rhyolite bodies are present in the LC-1 area, the large LC-1 dike in the cavity area and a smaller dike to the east near to the vent drift. The LC-1 dike is approximately 100 ft (30 m) thick; the second dike about 15 ft (4.5 m) thick, as shown in Figure 5.3.3.2-6. The LC-1 body is composite in nature and contains multiple intrusions of older porphyritic (quartz/feldspar) rhyolite and younger flow-foliated, aphanitic rhyolite as well as minor inclusions of the amphibolite country rock. Based on best-fit correlation of the contact information, it is considered that both dikes trend N25°E and dip to the east at about 40°, crosscutting the foliation within the amphibolite. These orientations will be confirmed through additional drilling as soon as it becomes feasible.

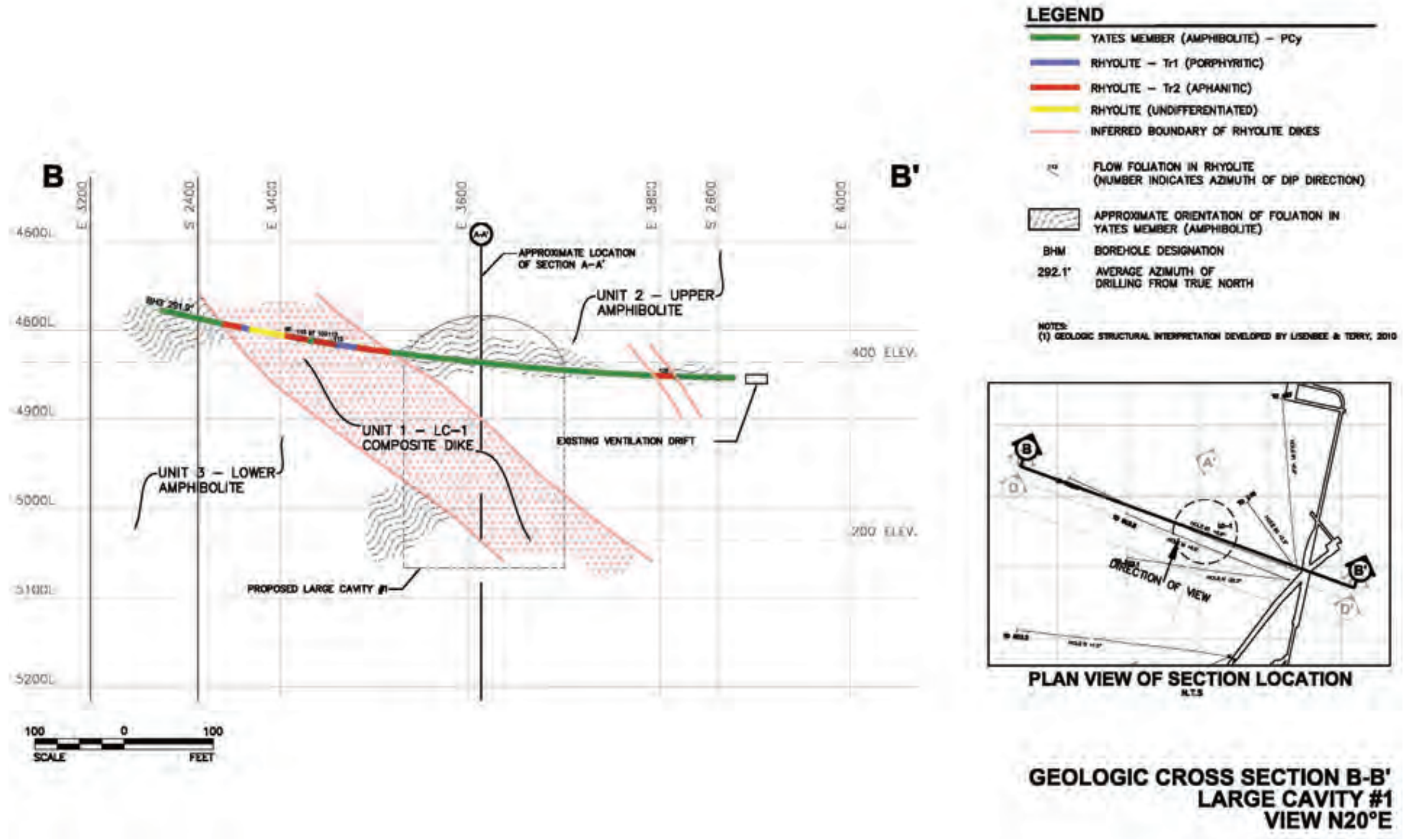

**Figure 5.3.3.2-6** Geological cross section B-B' LC-1—View N20ºE. [Golder Associates]

Several zones of core disking were observed[8] in the core taken from the various holes drilled into the LC-1 cavity area. Examination of the disking in several of the holes through the LC-1 cavity suggests that the phenomenon is restricted to only the higher modulus rocks, of which three types were identified: a) Tertiary rhyolites, b) some rare Tertiary igneous breccia zones (which appear to be completely recrystallized/annealed and thus are integral with the adjacent parent rock mass), and c) Precambrian quartz veins. The zones observed did not exceed 2 ft (0.6 m) in width and mostly consisted of only two to four pieces of disked core.

In addition to ubiquitous joints and veins, the collected geotechnical information indicates the potential presence of several faults. However, it must be noted that the faults that were mapped in the drifts have displacements of just a few inches. Most are annealed and probably not significant to rock quality. Some unannealed faults, however, were identifiable in the core. The margin of the eastern rhyolite dike in BH3 is cut by one such a fault. This structural feature strikes N47°E and dips 66°E. Recent observations of a steep structure in the Transition Cavity suggest that at least one such feature might have an extent of several hundred feet. Additionally, based on the projection of areas of lower rock quality observed in the BHs, one fault may cross close to or through the LC-1 cavity, trending northwest/southeast, and two others may extend into the cavity area with a southwest/northeast trend. There are other potential faults interpreted from the extensive datasets regarding RQD, PFA, and FF derived from the BH data and drift maps. In the vicinity of the LC-1, combinations of these features suggest several possible faults (as shown in Figure 5.3.3.2-7), one of which may be projected to the northwest along the margin of the proposed cavity site, and two which trend northeast into the site. Limited control suggests a steep dip (approximately 80°) for all of these postulated faults.

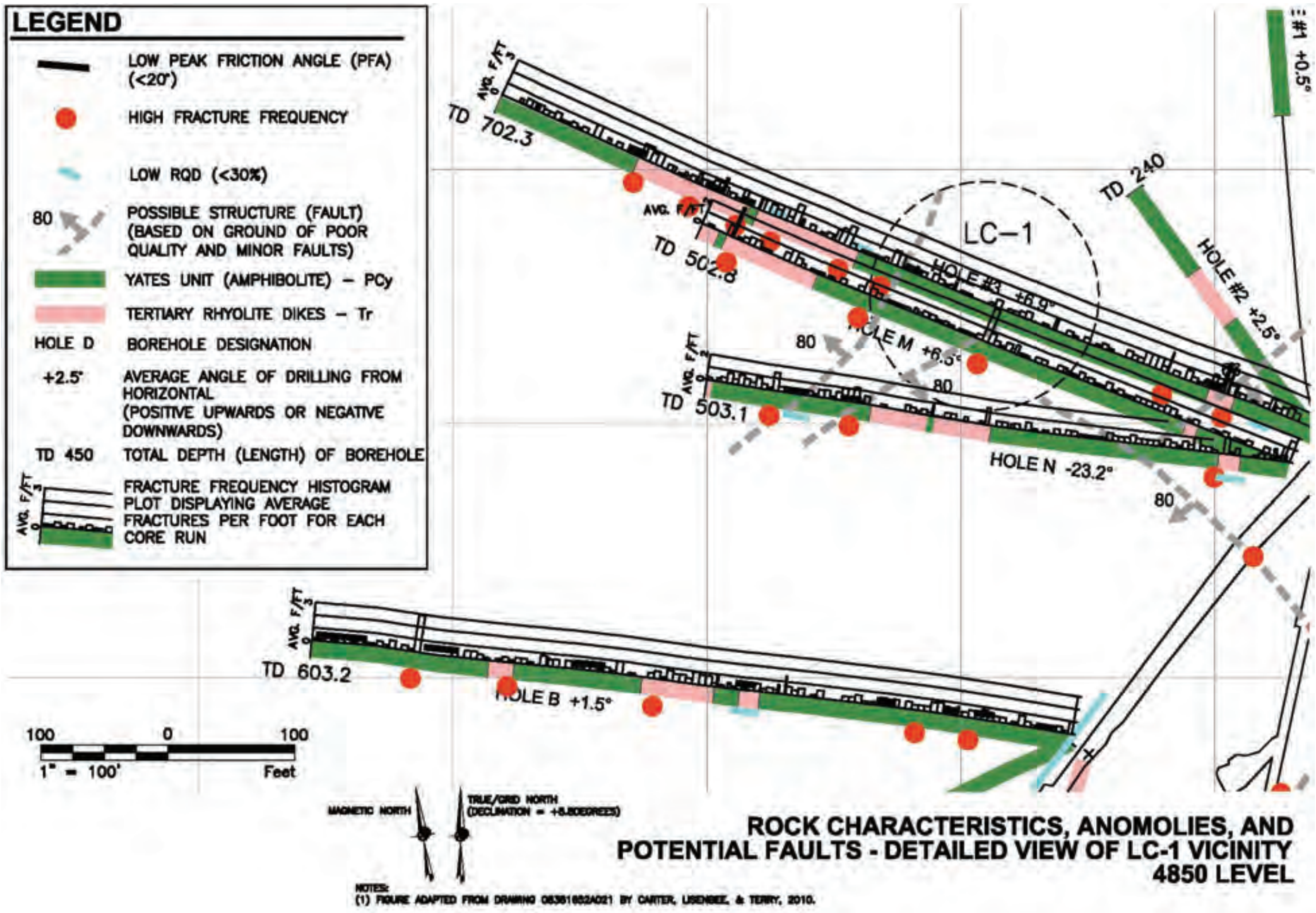

**Figure 5.3.3.2-7** Rock characteristics, anomalies, and potential faults—detailed view of LC-1 vicinity: 4850L. [Golder Associates]

In summary, the correlations of geological and geotechnical data suggest that the immediate vicinity of the LC-1 can be subdivided into three rock mass units (see Figure 5.3.3.2-8):

**Unit 1:** A central, relatively more fractured, rhyolitic section comprising interdigitating flow foliated rhyolitic dike intrusions of at least two injection ages, with local fingers of the parent amphibolite host rock also present. The contacts of the rhyolites generally appear to be annealed; the younger of the two intrusions is more strongly fractured, particularly on the margins with the parent rock mass, and one faulted dike contact was observed in drill core.

**Unit 2** and **Unit 3**, respectively: Includes an upper and a lower amphibolite section, which, based on preliminary assessment of the geotechnical data, may in fact be slightly different, as it appears that lower RQDs and higher FF counts characterize the upper unit, suggesting it may be of inferior rock mass quality with respect to the lower unit, but this needs checking in detail with subsequent follow-up work in the Final Design phase.

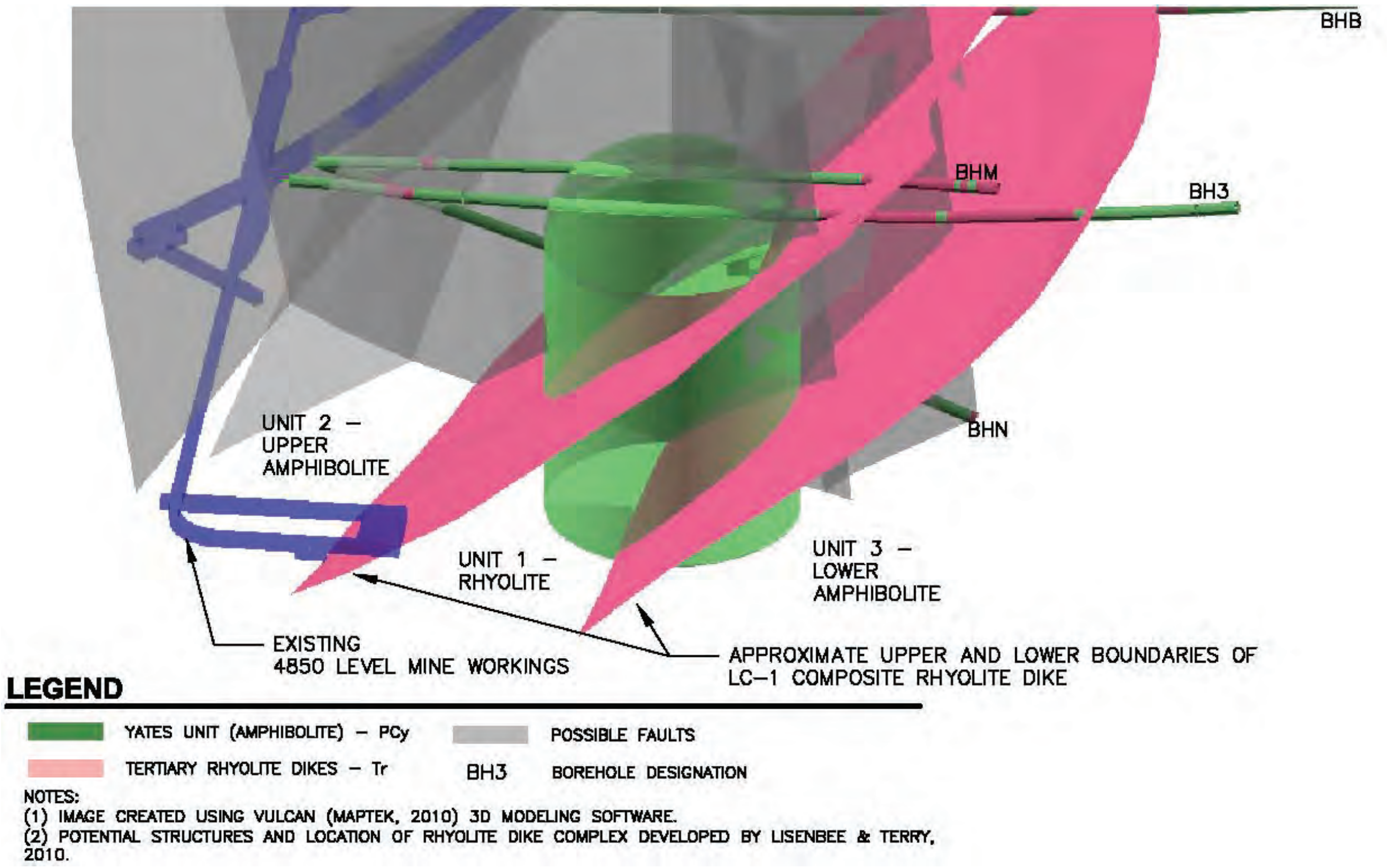

**Figure 5.3.3.2-8** Isometric view of geological structures in the vicinity of LC-1. [Golder Associates]

### Geotechnical Assessment of LC-1 Area

Based on the developed geostructural model (Appendix 5.J) and performed geotechnical assessments and analyses, rock mass quality for Unit 1—the main rhyolite dike swarm—is expected to be good to very good. However, within any hydrothermally altered or fault margins to the dykes, rock mass quality locally might be expected to be only in the poor to fair range. Rock mass quality for Unit 2—the Upper Amphibolite Unit—based on current information, appears inferior to the quality of Unit 3—the Lower Amphibolite Unit. These observations have engineering significance not just for the crown, but also for the walls and the invert of the cavity. The decreased quality of Unit 2 points to the need for greater attention to wedge and block failure evaluation for the crown and upper parts of the cavity. The relatively high rock mass quality for Unit 3 and the inference also of likely annealed contacts with Unit 1—the rhyolite dike swarm zone—(particularly if the basal margin is all comprised of older [Tr1] rhyolite), points to the need to address potential stress concentration, envelope geometry, and floor-related heave effects for the lower zones of the cavity. A summary of rock mass classifications, recalculated probabilistically for the LC-1 areas, is presented in Table 5.3.3.2. Although the present estimates of rock mass characterization are based on the reconnaissance-level investigation, the interpretations presented by Golder do appear to exhibit convergence, suggesting that the crown, sidewalls, and invert conditions of the main LC-1 cavity could be quite different. In addition, it can also be noted that toward the best quality end of the rock mass scale, the upper amphibolite and the lower amphibolite and the rhyolite are almost indistinguishable one from the other, which is also evident from observations in the drifts. Despite these observations of rock mass properties, the collection and evaluation of additional information on the three currently indentified zones will be required for a comprehensive rock mass characterization. Additional verification of the models will be accomplished through the mapping of the current Davis Campus on the 4850L and driving an exploratory drift to the LC-1 area.

| Rock Mass Unit | Mean RMR | Mean Q | Estimated GSI Range | | |
|---|---|---|---|---|---|
| | | | Min | Max | Mean |
| Unit 2 – Upper Amphibolite | 67 | 11.1 | 47 | 80 | 67 |
| Unit 1 – Competent Rhyolite | 90 | 53.3 | 59 | 89 | 82 |
| Unit 1 – Extreme Rhyolite | 41 | 0.83 | 16 | 63 | 42 |
| Unit 3 – Lower Amphibolite | 78 | 28.3 | 59 | 79 | 76 |

**Table 5.3.3.2** Probabilistic ratings of rock mass classifications. [Golder Associates]

The generated geologic and geotechnical data; the development of lithological, structural, and geotechnical models that have refined understanding of the existing conditions on the 4850L; and the analyses and assessments performed revealed that nothing within the current state of our understanding of the existing conditions precludes the subject rock mass as a suitable host medium for the proposed excavations. Furthermore, the data collected during the preliminary investigation have provided sufficiently complete information for the development of a Preliminary Design for LC-1, LM-1, and LM-2. It is, however, understood that verification of the lithology, geological structure, and properties of various rock mass units is needed to continue during Final Design. Verification of the current model through additional drilling and extended site investigations will allow refinement of the design, reduction of the risk, and minimization of the cost of potential overdesign.

### Specific Conclusions Regarding Geology and Geotechnical Rock Conditions of LC-1, LM-1, and LM-2 Area

The generation of the lithologic, structural, and geotechnical models from the review of current geologic and geotechnical data has demonstrated that no factors were identified that would disqualify the proposed location of LC-1 and LMs from consideration for large excavations. A number of important facts related to the constructability of the proposed large excavations on the 4850L were identified. These are as follows:

1. Marked petrological and geomechanical differences exist between the rocks of the Poorman Formation (phyllite and schists), the Yates Member (primarily amphibolites), and the two different types of rhyolite. There is clear evidence suggesting that structural fabrics differ significantly across the 4850L. The main areas of differing fabrics are exemplified through the various domains.

2. Two Tertiary rhyolite dikes are present in the LC-1 area. A large complex dike approximately 100 ft (30 m) thick, striking at approximately N25°E and dipping at approximately 40° to the east, is present in the LC-1 cavity area. Rhyolites in this unit were observed to be relatively more fractured than the other rock units encountered. The second, smaller dike, approximately 15 ft (4.5 m) thick, exists just to the west of the existing ventilation drift. At this stage, the presence of rhyolite above the crown and/or below the invert of LC-1 cannot be ruled out.
3. Two dominant sets of joints are observable within the studied amphibolite units in the vicinity of the LC-1 cavity. One strikes southeast and dips steeply to the southwest and the other strikes ENE and dips moderately-to-steeply to the SSE. Only one southeast-striking set appears to be present in rhyolite units within the same area.
4. Observations of the geotechnical properties of the upper and lower amphibolite units, Units 2 and 3, respectively, suggest that the upper, Unit 2, may be of lower rock mass quality than the lower, Unit 3. The observed orientation of foliation fabric in the vicinity of the crown and invert of the proposed LC-1 cavity may require close attention in the design, and need further investigation. Specifically, it is likely that the observed SSE-trending fabric will be of greatest significance to the design and construction of the proposed cavities. The recent ground control issues in the new Transition Cavity of the Davis Campus are evidence of this potential issue. Significant evidence exists to suggest the intersection of several possible faults in the rock units of the 4850L (Section 5.3.3.2).
5. Measured in situ stress conditions are highly variable within the amphibolite units on a small scale. This variability is likely the result of the natural heterogeneity in the rock. The magnitudes of the measured principal stresses, however, are in strong agreement with previously measured in situ stresses and the subsequently developed lithostatic models. Some core disking was observed in core collected during the recent geotechnical investigation. These observations appear to be limited to higher modulus rocks identified as the Tertiary rhyolites, some Tertiary igneous breccia zones, and Precambrian quartz veins. These zones were limited in their extents and are not believed to represent significant problems to the constructability of the proposed excavations.
6. Resolution of much of the uncertainty in the current data can be achieved through additional investigations during Final Design. This additional work should include: a) additional mapping and laser scanning of existing drifts and new excavations with emphasis on measurements of joint density and persistence; b) extended, site-specific drilling (areas of LC-1, LM-1, and LM-2), and in situ and laboratory testing of rock mass properties, including shear strength of joints; and c) driving an exploratory drift into the area of LC-1, its mapping and ground monitoring.

#### 5.3.3.3 Initial Numerical Modeling

The initial numerical modeling task[17] was initiated at the early stage of the geotechnical engineering project when very little specific information (i.e., geological structure, excavation designs, material properties, and locations) was available for directed numerical analyses. It has, however, fulfilled a preliminary goal of evaluating the geometries, orientations, ground control needs, and proximity of the new deep excavations and access drifts. Even though design calculations could not be performed early on, numerical modeling was used to evaluate the relative significance of parameters that are of importance to the design calculations, such as material properties, material anisotropy, cavity shapes, and spacing between large cavities. To investigate these various parameters, the numerical modeling effort focused on gathering existing information (rock properties and initial in situ conditions) available at that time, analytical solutions for different opening shapes in isotropic and anisotropic rock, 2-D numerical analyses, and 3-D numerical analyses. As new information became available from the geotechnical site investigations, numerical modeling was continued by the excavation designer (Golder) as part of the Preliminary Design effort.

The initial numerical modeling investigations were conducted in parallel with the tasks designed to develop the site-specific conditions and properties. Therefore, a set of conditions and properties based on available information was developed for conducting the analyses. In most cases, a baseline quantity was determined, and then a range of values encompassing the baseline value was included in the analyses. A literature review of past work was conducted to determine appropriate reasonable ranges in the input conditions required to perform stability calculations for the large cavities and LMs at the 4850L Campus. These input conditions included: 1) in situ stress conditions, 2) elastic deformational properties, 3) strength properties, and 4) joint properties. Various cavity geometries and the distance between cavities (cavity spacing) were also investigated. Information was gathered, qualified, and referenced regarding existing magnitudes and ranges in these input conditions. All of the numerical analyses conducted in this study were elastic. However, the states of stress computed from the elastic numerical analyses were evaluated by comparing the computed states of stress in the model to a proper strength criterion.

**Analytical Solutions**

Analytical solutions are quick, efficient tools for providing an estimate on the impact that a change in problem input (e.g., geometry, boundary conditions, and/or material properties) might have on the calculated results. In the case of a large, isolated cavity in an initially stressed medium at Homestake, these solutions provide stress concentration measures to estimate increases in the stress magnitudes above the pre-excavation in situ stress level. Therefore, existing analytical solutions were modified to calculate the stress concentration factors for combinations of opening shapes, isotropic and anisotropic material properties, and far-field stress boundary conditions. The analytical solutions were compared with the 2-D and 3-D programs to verify that the programs computed anisotropic elastic solutions correctly.

The total of 27 numerical modeling combinations of opening shape, material properties, and far-field boundary conditions were analyzed and compared to the analytical solutions, demonstrating that the numerical analysis programs compared very well with the analytical solutions.

**Two-Dimensional Analyses**

Candidate shapes identified for the large cavities include: domed right-cylinder, horseshoe-shaped horizontal prism, circular binocular, and triaxial ellipsoid. The domed right-cylinder and horseshoe-shaped horizontal prism were reasonably represented using axisymmetric and 2-D finite-element software, respectively. These two shapes were evaluated in the 2-D analyses portion of this study under a multitude of possible input conditions. The primary output of the 2-D finite-element analyses was the assessment of the structural stability of an isolated cavity. Ranges in the in situ stress state, material deformational properties, intact rock strength, and joint strength conditions were investigated in a parametric manner to help identify the impact of each input parameter on the stability of the cavities. Simulation results were evaluated and presented in terms of strength ratios (SRs) using intact strength properties. Results of the parametric study were used to identify the most-likely, the most-favorable, and the least-favorable set of input conditions for the overall stability of the underground cavities for the range of input parameters evaluated. Examples of 2-D plane strain and axisymmetric analyses are shown in Figures 5.3.3.3-1 and 5.3.3.3-2, correspondingly.

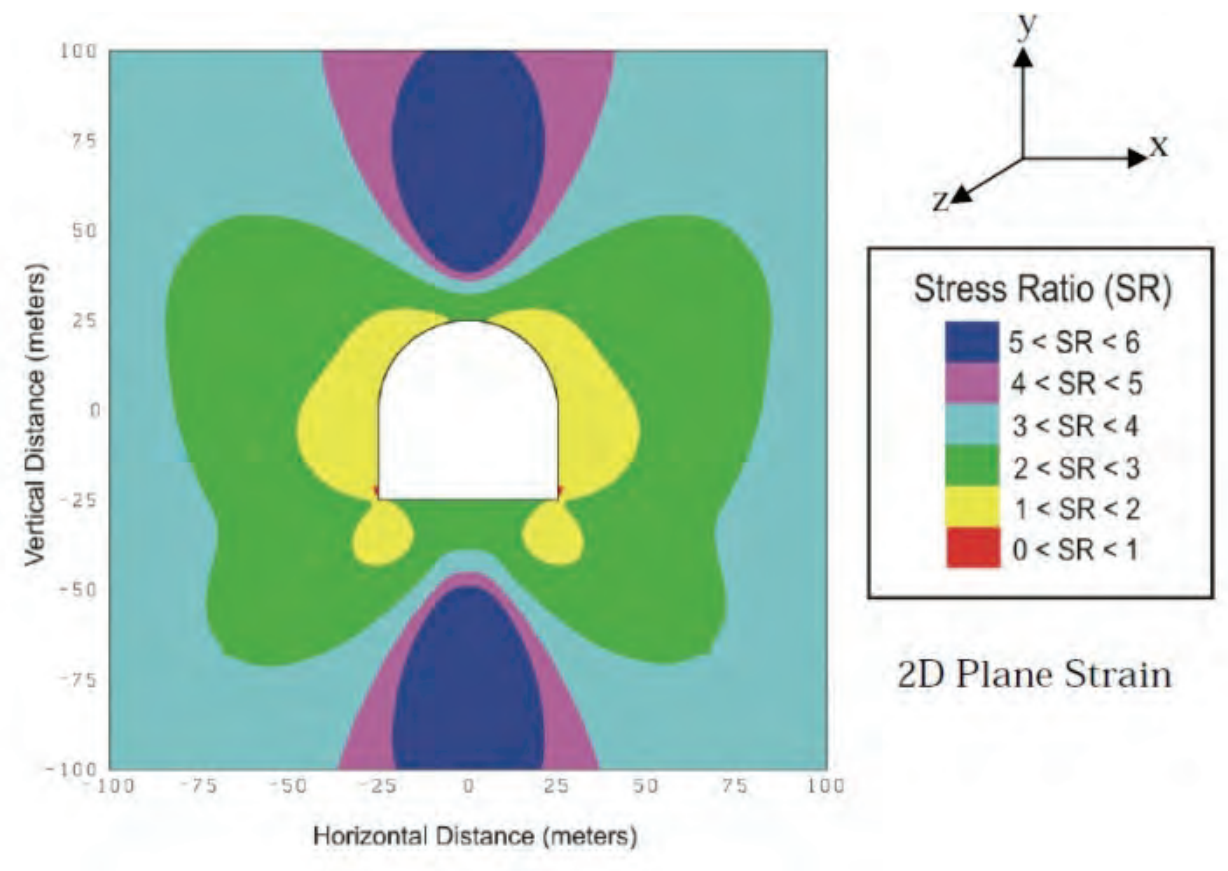

**Figure 5.3.3.3-1** 2-D plane strain analysis. [RESPEC]

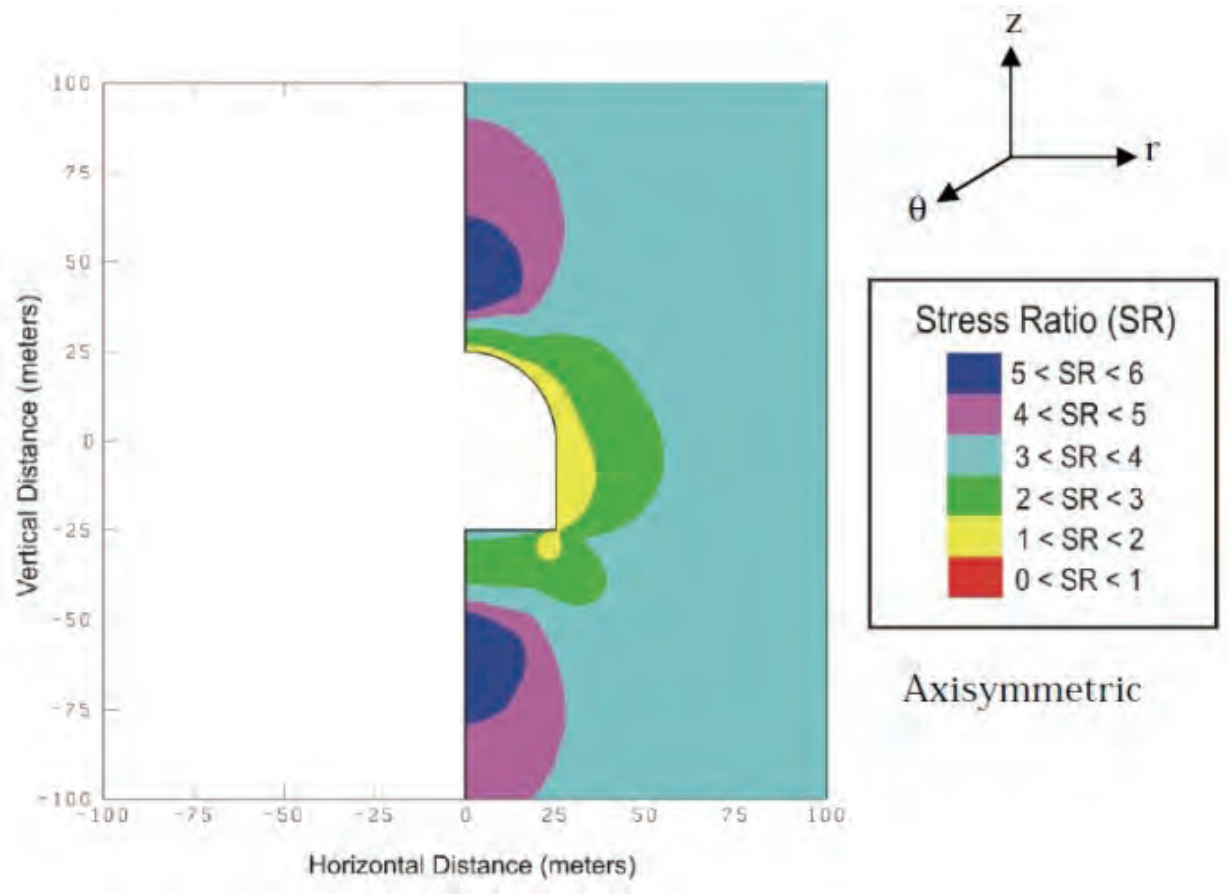

**Figure 5.3.3.3-2** 2-D axisymmetric analysis. [RESPEC]

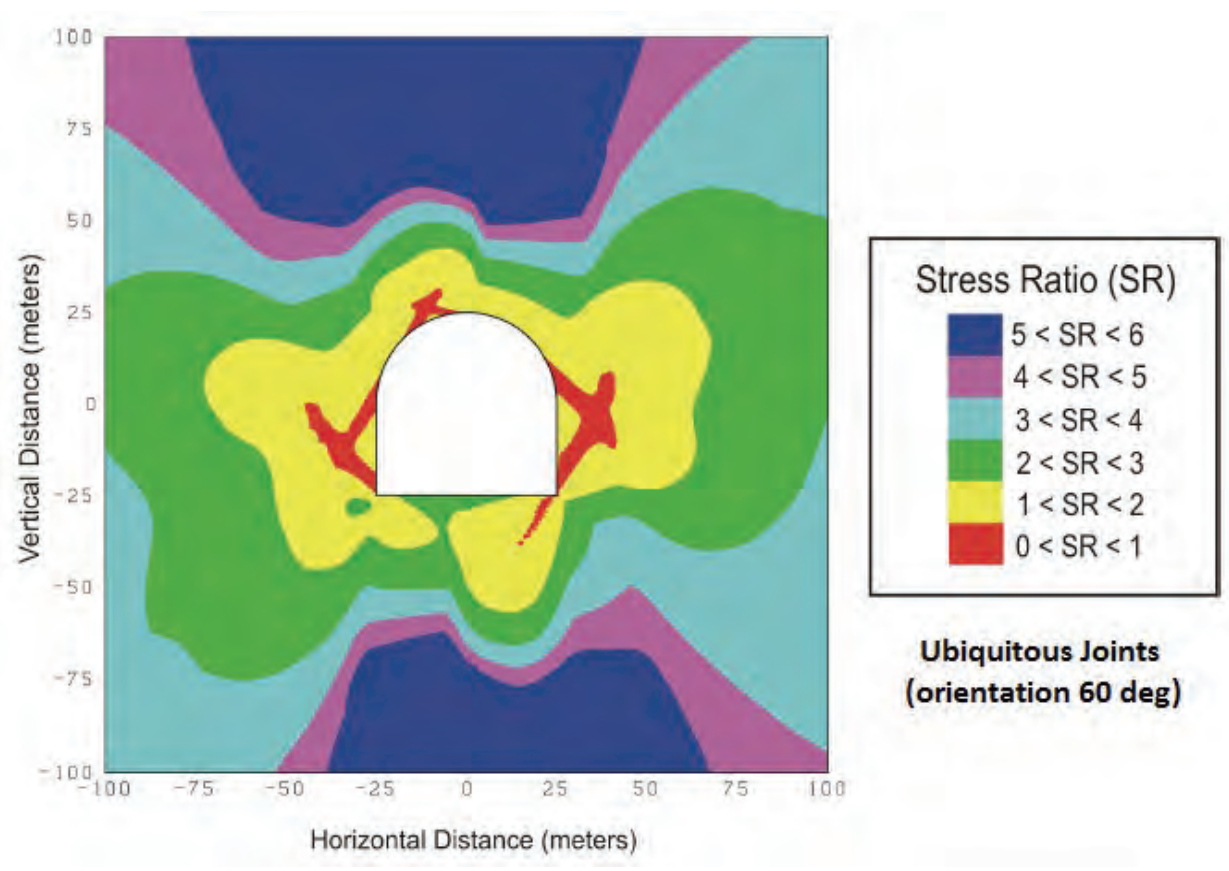

**Figure 5.3.3.3-3** Ubiquitous joints (orientation 60°). [RESPEC]

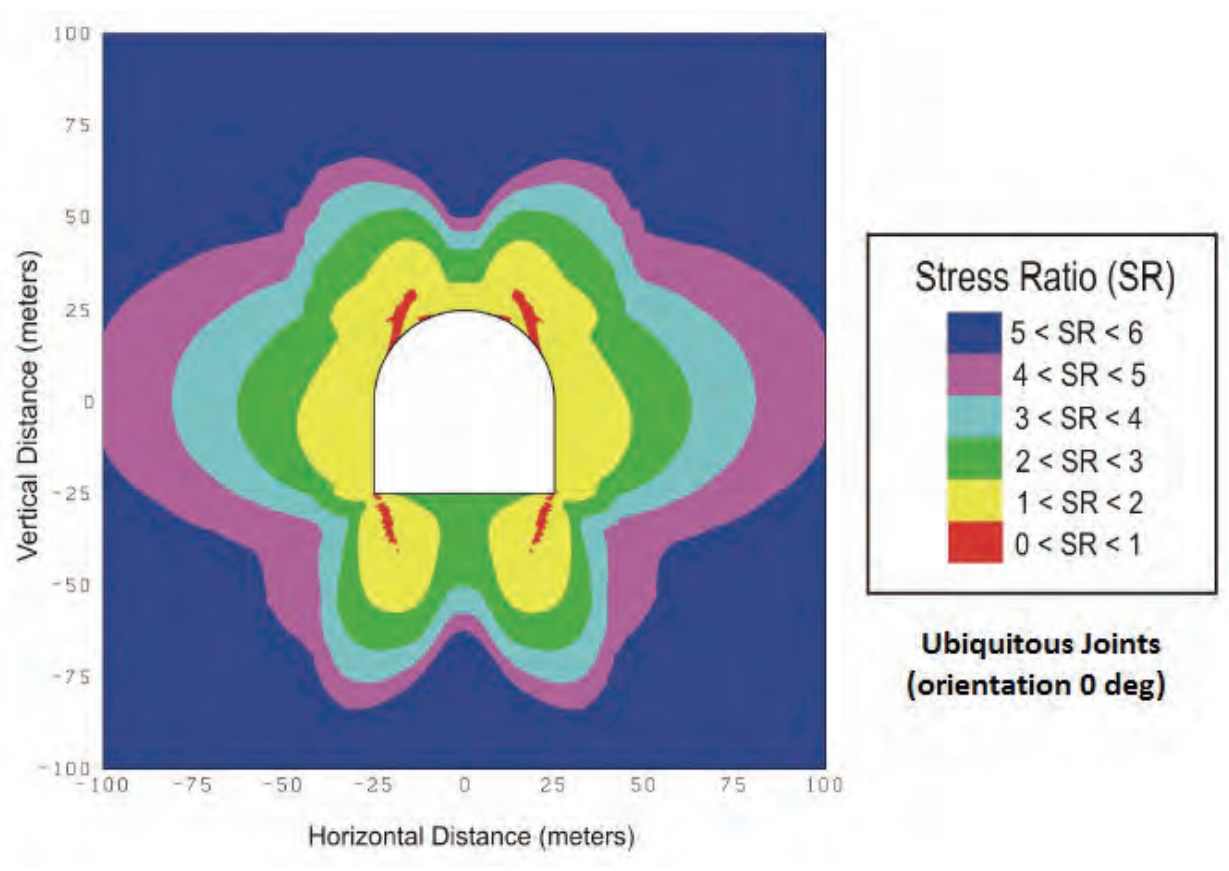

**Figure 5.3.3.3-4** Ubiquitous joints (orientation 0°). [RESPEC]

A total of 64 elastic numerical analyses using SPECTROM-32 finite elements software[17] were performed based on 24 primary models and 40 supplemental models. These models were used to examine the stability of an isolated cavity to variations in magnitude and orientation of in situ stress, elastic deformational properties, intact rock strength, and joint strength. In each model, the area (or equivalently, the volume) of the disturbed zone (DZ) (i.e., locations where SR < 1 surrounding the cavity) was calculated. The extent of the DZ was used to compare the results of the various analysis parameters over their input ranges. Two examples presented in Figures 5.3.3.3-3 and 5.3.3.3-4 illustrate finite element modeling of rock mass with ubiquitous joints.

The 2-D calculation effort used the input conditions over a selected range in their values to identify which of these input conditions or parameters were important in assessing the structural stability of an isolated DUSEL cavity. Stability of the host rock was assumed to occur when the calculated stress ratio (ratio of strength to stress) was greater than unity. The overall stability of the cavity was quantified by comparing the volume of the host rock that had a stress ratio less than unity to the volume of the cavity.

The 2-D modeling results show that:[17]

1. Over the expected ranges in conditions encountered at Homestake, the effects of intact rock strength (magnitude and orientation) and joint strength (magnitude and orientation) on the 2-D stability of an isolated cavity are of highest importance.
2. Over the expected ranges in conditions encountered at Homestake, the effects of in situ stress, strength orthotropy, and cavity shape on the 2-D stability of an isolated cavity are of medium importance.
3. Over the expected ranges in conditions encountered at Homestake, the effects of elasticity and in situ stress orthotropy on the 2-D stability of an isolated DUSEL cavity are of low importance.
4. The most-favorable cavity stability environment is defined in terms of: a) isotropic in situ stress state, b) isotropic elastic properties, c) isotropic intact rock strength, d) domed right-cylinder cavity geometry, and e) no joints.
5. The least-favorable cavity stability environment is defined in terms of: a) orthotropic in situ stress state, b) orthotropic elastic properties, c) orthotropic strength properties, d) horseshoe-shaped horizontal prismatic cavity geometry, and e) reduced joint strength at a stratification angle of $60^{\circ}$.
6. The structural stability of an isolated cavity should not be compromised under isotropic (most favorable) site conditions.
7. The structural stability of an isolated unsupported cavity may be compromised under orthotropic (least favorable) site conditions. Obviously, these conclusions and their relative importance are only valid for the ranges of the parameter values considered in the 2-D analyses.

**Three-Dimensional Analyses**

Three-dimensional analyses using the finite difference method (FDM) (Appendix 5.I) were conducted to enable the evaluation of cavity shapes not amenable to 2-D analyses and to investigate 3-D aspects of the large cavities. Cavity shapes evaluated in the 3-D calculations include: domed right-cylinder, triaxial ellipsoid, and horseshoe-shaped horizontal prism. The results of the 2-D calculations for the most-favorable (isotropic) and least favorable (orthotropic) environments were used to focus the 3-D calculations to further investigate the least-favorable environments. Therefore, the 3-D calculations of cavities examined stability for expected variations in a) cavity shape, b) orthotropic strength, c) excavation sequence, and (d) cavity spacing. Stress ratios were calculated for the host rock surrounding the cavities against potential rock damage using the 3-D orthotropic strength model and associated strength properties.

Results were expressed in terms of the volume of rock for which the stress ratio (SR) is less than unity and exhibits the potential for damage. The extent of the damage zone (DZ) was used to compare the results of the various analysis parameters over their input ranges and to determine the relative importance of the parameters on cavity stability. Examples of modeling results for the isotropic and orthotropic models are presented in Figures 5.3.3.3-5 and 5.3.3.3-6.

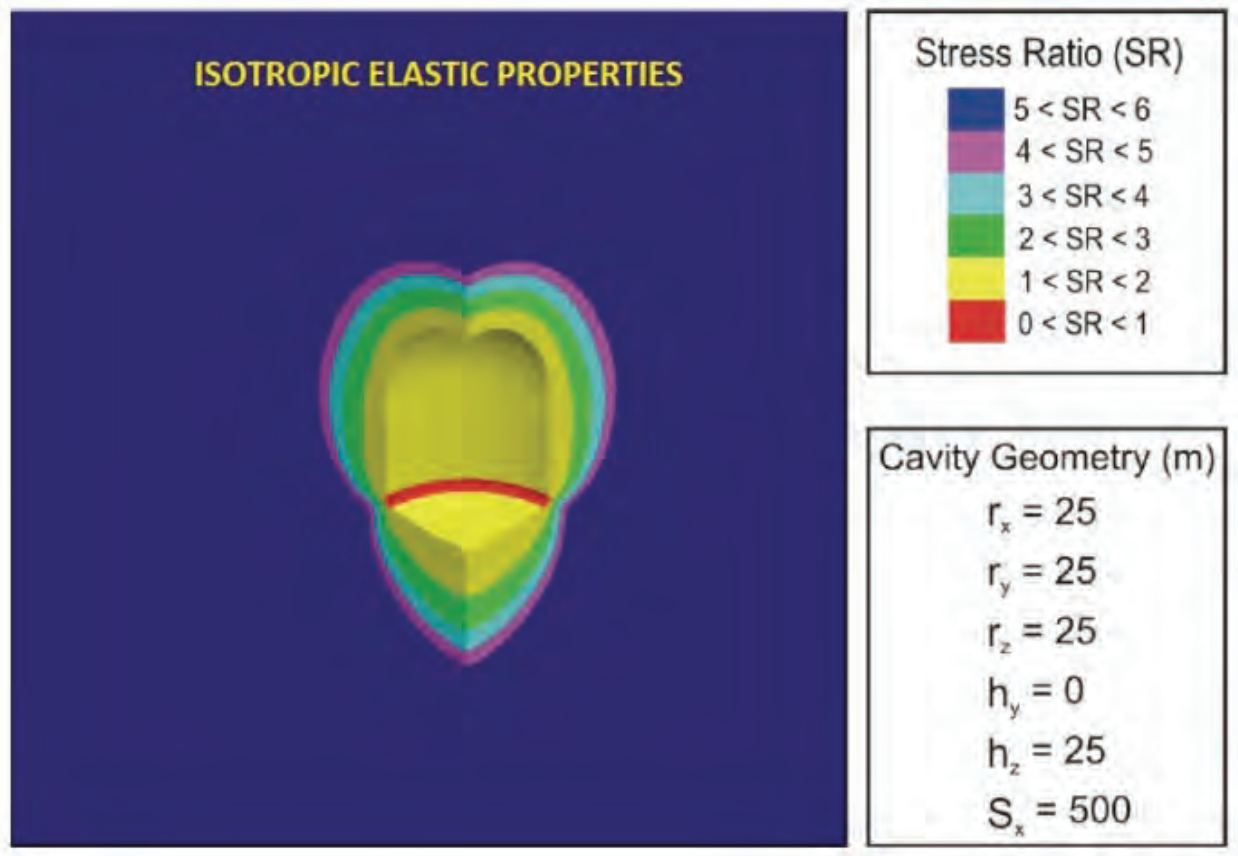

**Figure 5.3.3.3-5** Isotropic elastic properties. [RESPEC]

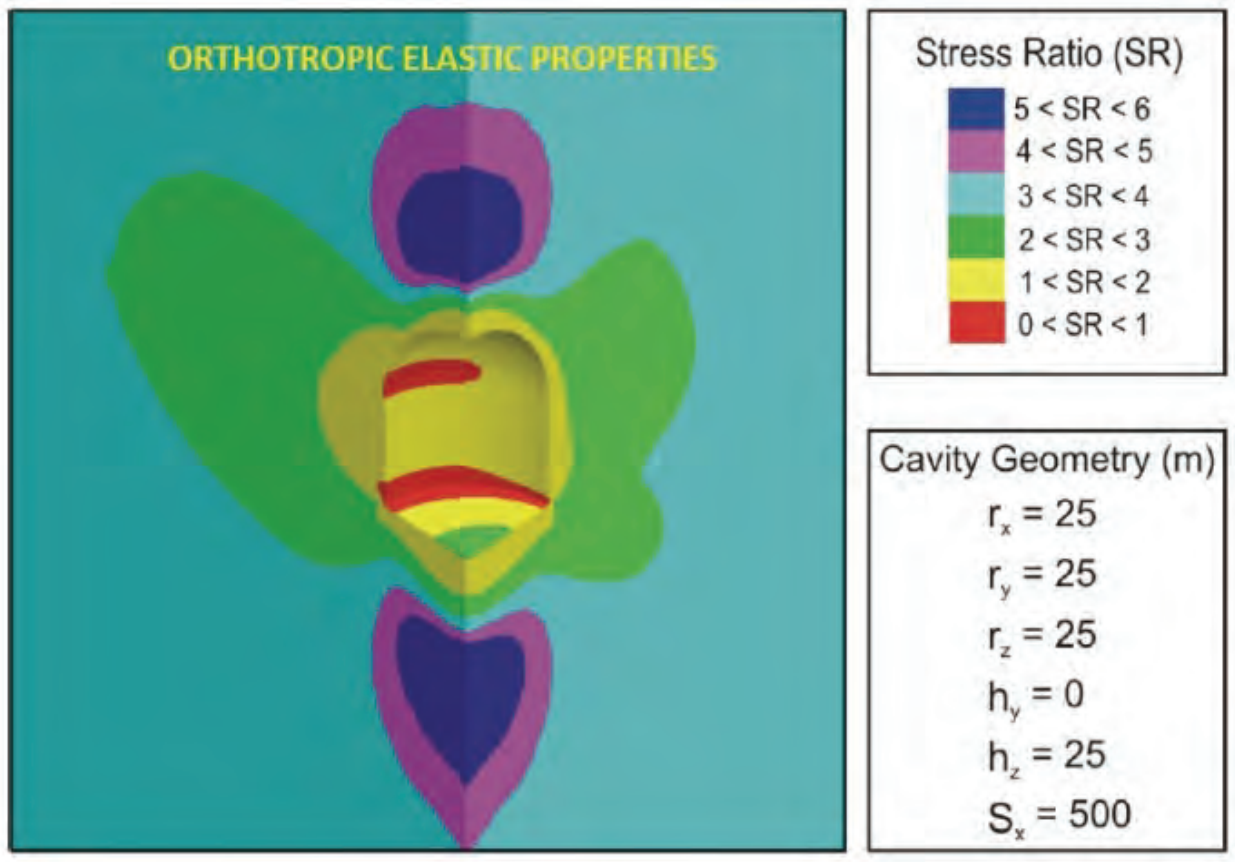

**Figure 5.3.3.3-6** Orthotropic elastic properties. [RESPEC]

For all cavity shapes investigated and the ranges in parameters investigated, the 3-D modeling results showed that:

1. Of the three cavity shapes considered (i.e., domed right cylinder, triaxial ellipsoid, and horseshoe-shaped horizontal prism), the most stable configuration is the domed right-cylinder cavity and the least stable is the horseshoe-shaped horizontal prism.
2. The effect of joint strength on the overall 3-D stability of the cavities is of high importance.
3. The effect of orthotropic intact rock strength on the overall 3-D stability of the cavities is of medium importance.
4. The effect of excavation sequence on the overall 3-D stability of the cavities is minimal.
5. A minimum cavity separation distance of two cavity diameters should be sufficient to limit structural interaction of multiple cavities. Naturally, these conclusions and their relative importance are only valid for the ranges of the parameter values considered in the 3-D analyses.

### 5.3.3.4 Numerical Modeling of Alternative Shapes of Large Cavities

The initial numerical modeling performed by RESPEC was continued by Golder within the design scope, and incorporated evaluation of the stability of alternative sizes and shapes, including cylindrical as well as mailbox cavity geometries.[18] The study also included estimates of incremental cost changes associated with the increase in cavity radius and estimated incremental cost changes per meter of length in the mailbox cavity design. Six variants of large cavity sizes and shapes (Table 5.3.3.4) were evaluated by conducting the following analyses: 2-D finite element method (FEM), 2-D distinct element method (DEM), and 3-D finite difference method (FDM). The FEM analyses comprised both plane strain and axisymmetric models. The six variants combined two shapes (cylindrical cavities and mailbox-shaped cavities) with three different sizes, corresponding to fiducial volumes of 100, 150, and 300 kT. Five of the models respected the currently accepted cavity heights, based on the limiting pressures for the photomultiplier tubes (PMTs); one cylindrical model (300 kT) investigated the possibility of building a deeper cavity.

| Case | Fiducial Volume (kT) | Shape[a] | Excavation Volume ($yd^3$) | Length (ft) | Width or Diameter[b] (ft) | Wall Height[c] (ft) | Dome Height (ft) |
|---|---|---|---|---|---|---|---|
| 1 | 100 | LC | 238,239 | - | 180.4 | 210.0 | 62.3 |
| 2 | 150 | LC | 348,401 | - | 215.9 | 210.0 | 70.5 |
| 3 | 300 | LC | 689,673 | - | 285.4 | 225.4 | 98.4 |
| 4 | 300 | LC | 593,913 | - | 215.9 | 391.1 | 70.5 |
| 5 | 150 | MB | 349,887 | 378.9 | 105.0 | 210.0 | 35.0 |
| 6 | 300 | MB | 669,281 | 547.9 | 134.5 | 210.0 | 44.8 |

[a] LC denotes cavity with the cylindrical base; MB stands for mailbox shaped cavity.

[b] Denotes cavity diameter for LC analyses and cavity width for MB analyses.

[c] Measured from cavity bottom to spring-line (base of the dome).

**Table 5.3.3.4** Case scenarios and cavity geometries. [Golder Associates]

The models include a 98 ft (30 m) thick rhyolite dike swarm, in the more prevalent amphibolites of the 4850L, dipping at approximately 35° and transecting the cavity span, which represents the current understanding of the geological conditions at the location of LC-1 on the 4850L. Figure 5.3.3.4-1 presents the geology for FEM software.[18] Figure 5.3.3.4-2 shows rock structure modeled with DEM software.

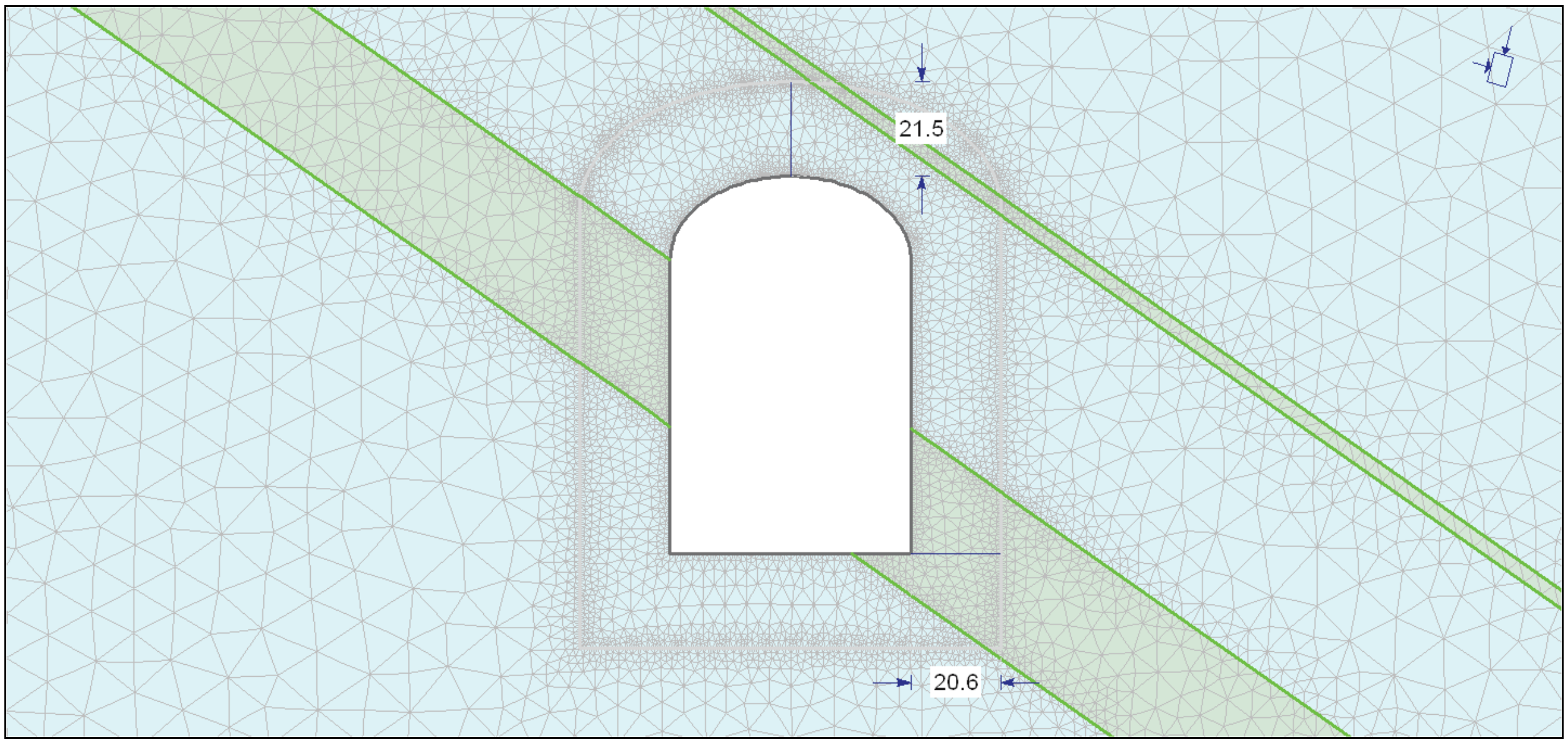

**Figure 5.3.3.4-1** Typical mesh pattern for FEM analyses showing rhyolite dikes. [Golder Associates]

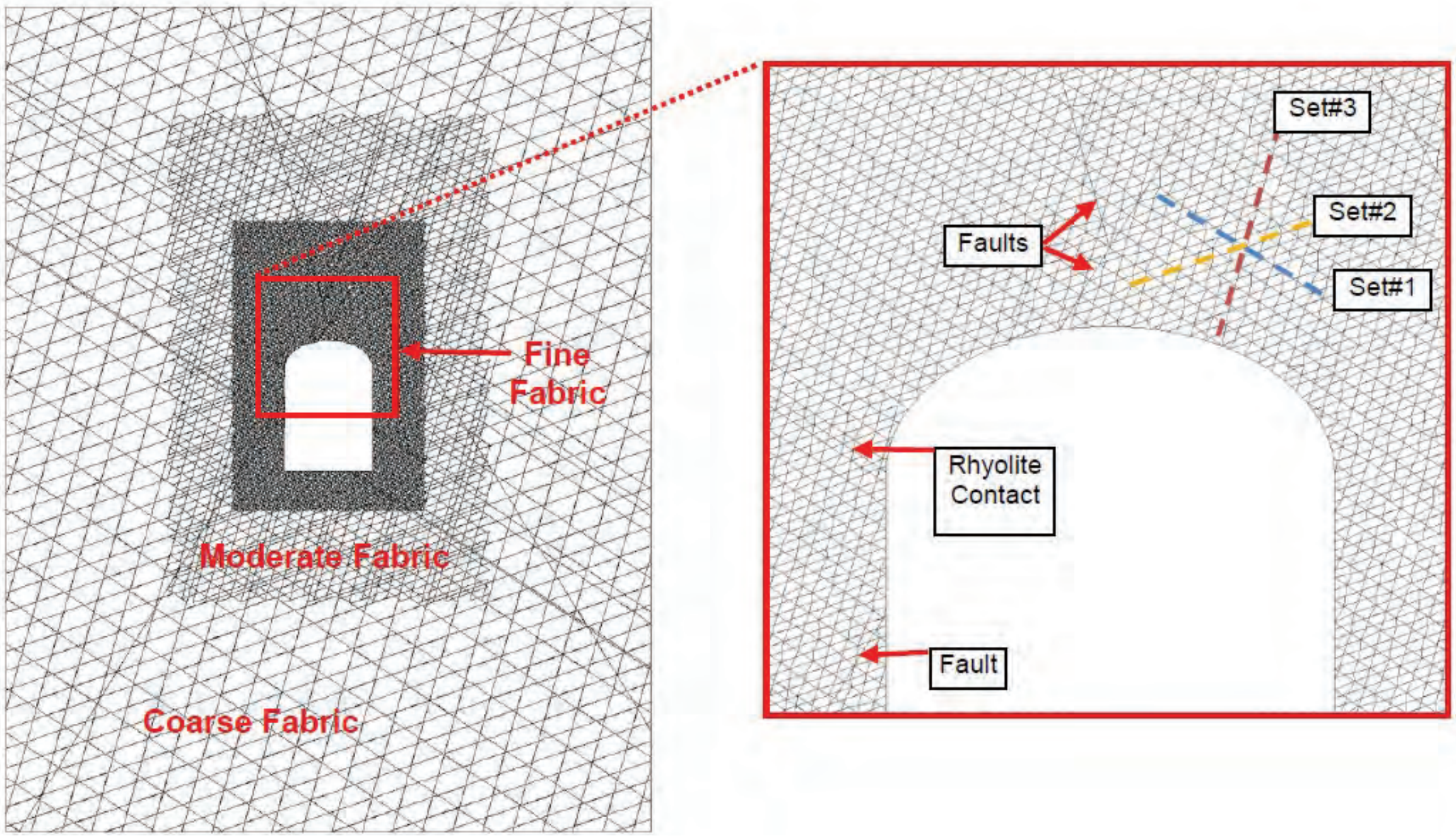

**Figure 5.3.3.4-2** Structural fabric used in DEM models. [Golder Associates]

The displacements in the crown and in the walls of the LC-1 are shown in Figures 5.3.3.4-3 and 5.3.3.4-4, correspondingly.

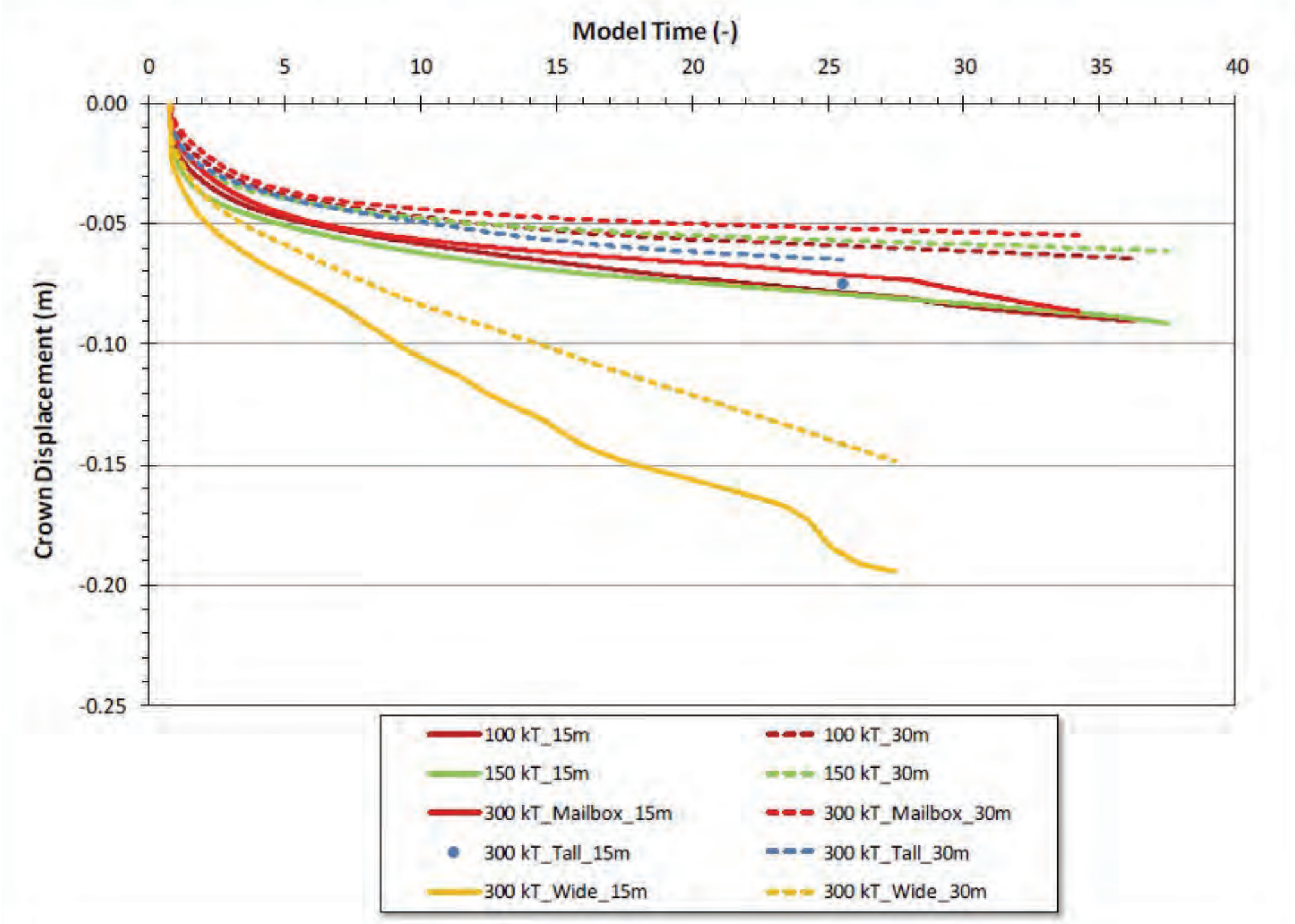

**Figure 5.3.3.4-3** Crown displacements. [Golder Associates]

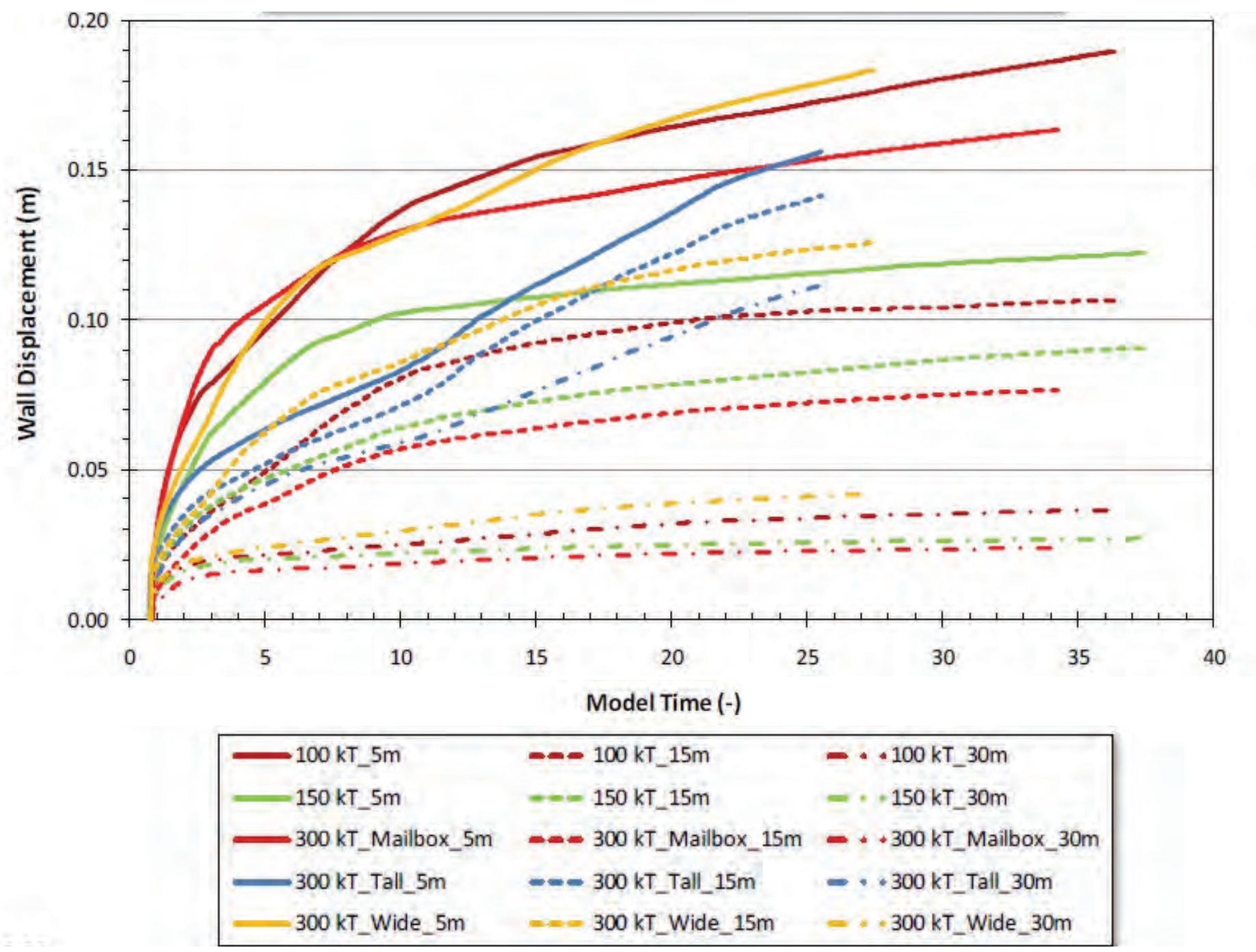

**Figure 5.3.3.4-4** Wall displacements. [Golder Associates]

The results of the analyses suggest several conclusions regarding cavity shapes and sizes.

1. Domed, upright, cylindrical cavities are feasible up to spans of about 217 ft (66 m). However, the depth of support starts to reach lengths of 82 ft (25 m) or more in the roof and upward of 98 ft (30 m) in the walls.
2. For domed, upright, cylindrical cavities with spans larger than 217 ft (66 m), the depth of disturbance in the roof becomes quite large (20 m) and control of the roof is an issue. These spans are not recommended.
3. Domed, mailbox-shaped cavities are feasible up to spans of 131 ft (40 m). The depth of support starts to reach lengths of 82 ft (25 m) or more in the roof.
4. Mailbox-shaped cavities with vertical walls up to 197 ft (60 m) tall are feasible. However, control of the flat-sided walls becomes critical and support will reach depths well in excess of 98 ft (30 m).
5. Control of the side walls becomes more difficult as the height increases and the length of the cavity exceeds twice the height.

Cost analyses for the different cavity alternatives were conducted by considering direct costs for excavation and support installation. Costs were developed based on the up-to-date information developed for the LC-1 design and current industry standards. These costs were provided for comparative purposes only (i.e., they should be refined after selecting a specific engineering alternative).

Costs per unit volume of excavation were also estimated for each alternative for a more direct comparison. Cost analyses indicate the following:

1. Cavities with large area-volume ratios (ratios between the supported surface area and the volume of excavation) are likely to exhibit higher cost-per-unit volume.

2. Consequently, mailbox cavities are likely to be more expensive than cylindrical cavities, assuming the same excavation volume.
3. For cavities with fiducial volumes ranging from 100 kT to 300 kT, the relationship between the total direct cost-per-unit volume and the area-volume ratio is approximately linear.

### 5.3.4 Design for Ground Support and Stabilization

Rock support design for the major excavations, ancillary rooms, and drifts is based on the site characterization and rock mass assessment presented in Sections 5.3.1, 5.3.2, and 5.3.3, and refined in the final excavation design report (Appendix 5.I). The following methods were used to design rock support for the excavations:

- Empirical method—the Tunneling Quality Index (NGI-Q) (Appendix 5.I)
- Kinematic analyses based on the current layout of the major excavations and drifts and all structural data available to date
- Numerical analyses to estimate the depth of rock mass damage

**Empirical Methods**

The LC-1, LM-1, and LM-2 excavations in particular present a great challenge in engineering, designing ground support, and stabilization with their unprecedented spans and depth of location. The support recommendations in Grimstad and Barton's Tunneling Index Quality chart (Appendix 5.I) are based on experience from smaller openings. However, the method is believed by the design team, the Geotechnical Advisory Committee (GAC), and the Large Cavity Advisory Board (LCAB) to be appropriately relevant and was therefore considered acceptable to estimate initial support requirements. Refinement of the design of support spacing and capacity was based on kinematic analyses and numerical analyses as discussed in detail in the Golder Associates Preliminary Design Report (Appendix 5.I).

**Kinematic Analyses**

Estimates of the continuity of the predominant joint sets were used to establish realistic wedge sizes that could exert load on the supports. Wedge (kinematic) analyses were then performed to confirm the length and spacing of the cables and further refine the layout. Although the level of stress at the 4850L is favorable, temperature effects and long-term stability require that the potential for large wedges be addressed. Due to the interpreted existence of subvertical faults in the cavity area, as reported in the *Geological and Geotechnical Assessment Report* (Appendix 5.J), joint continuities of up to 66 feet (20 m) were considered.

**Numerical Analyses**

Numerical analyses included the 2-D axisymmetric finite element method, the 2-D distinct element method, and the 3-D finite difference method (Appendix 5.I).[17] Rock properties and input stress field used were based on the investigation performed to date. Examples of the distinct element modeling of the geology and structural fabric for the LC-1 and LMs are shown in Figures 5.3.4-1, 5.3.4-2, 5.3.4-3, and 5.3.4-4, correspondingly.

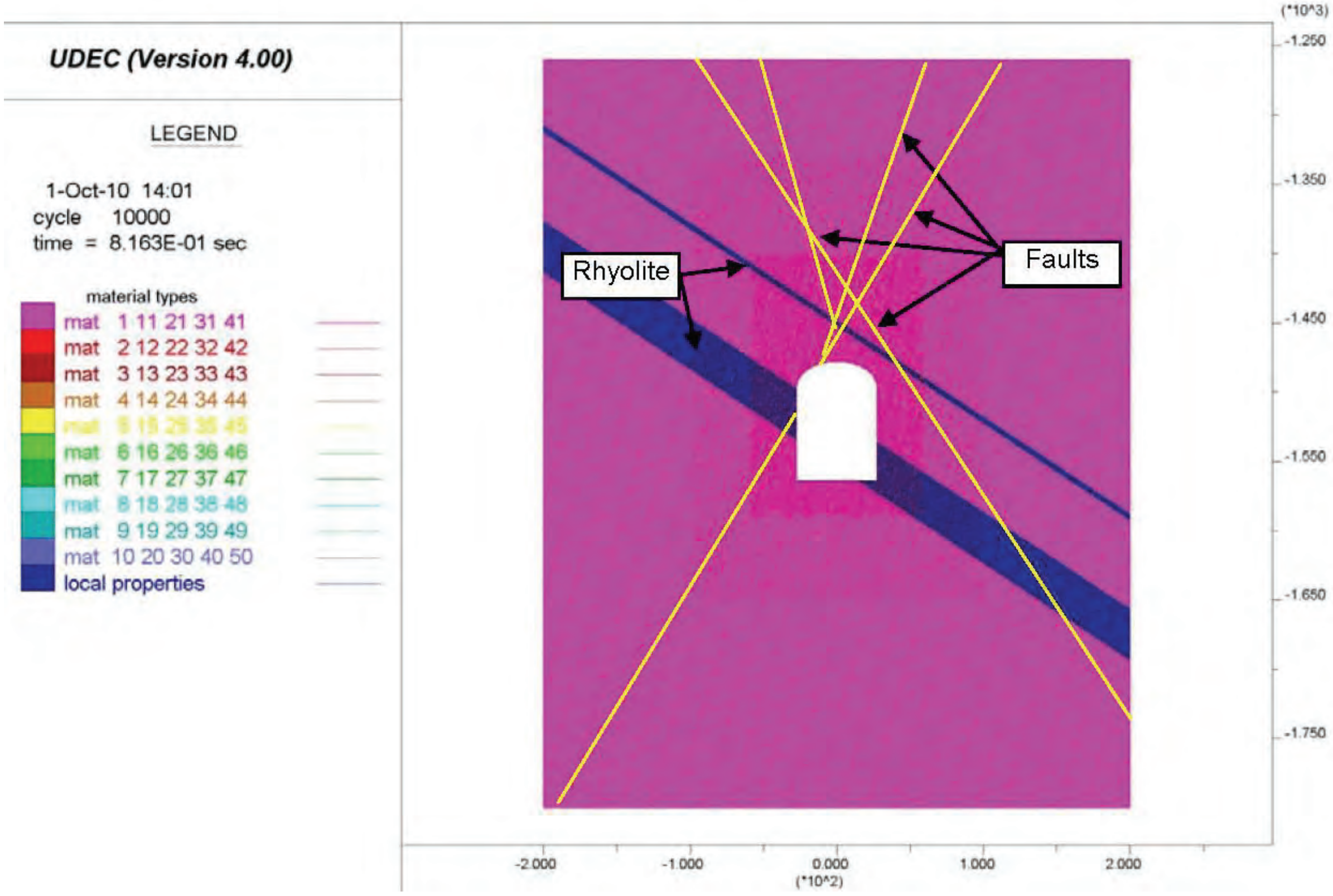

**Figure 5.3.4-1** Distinct element modeling of the geology of the LC-1. [Golder Associates]

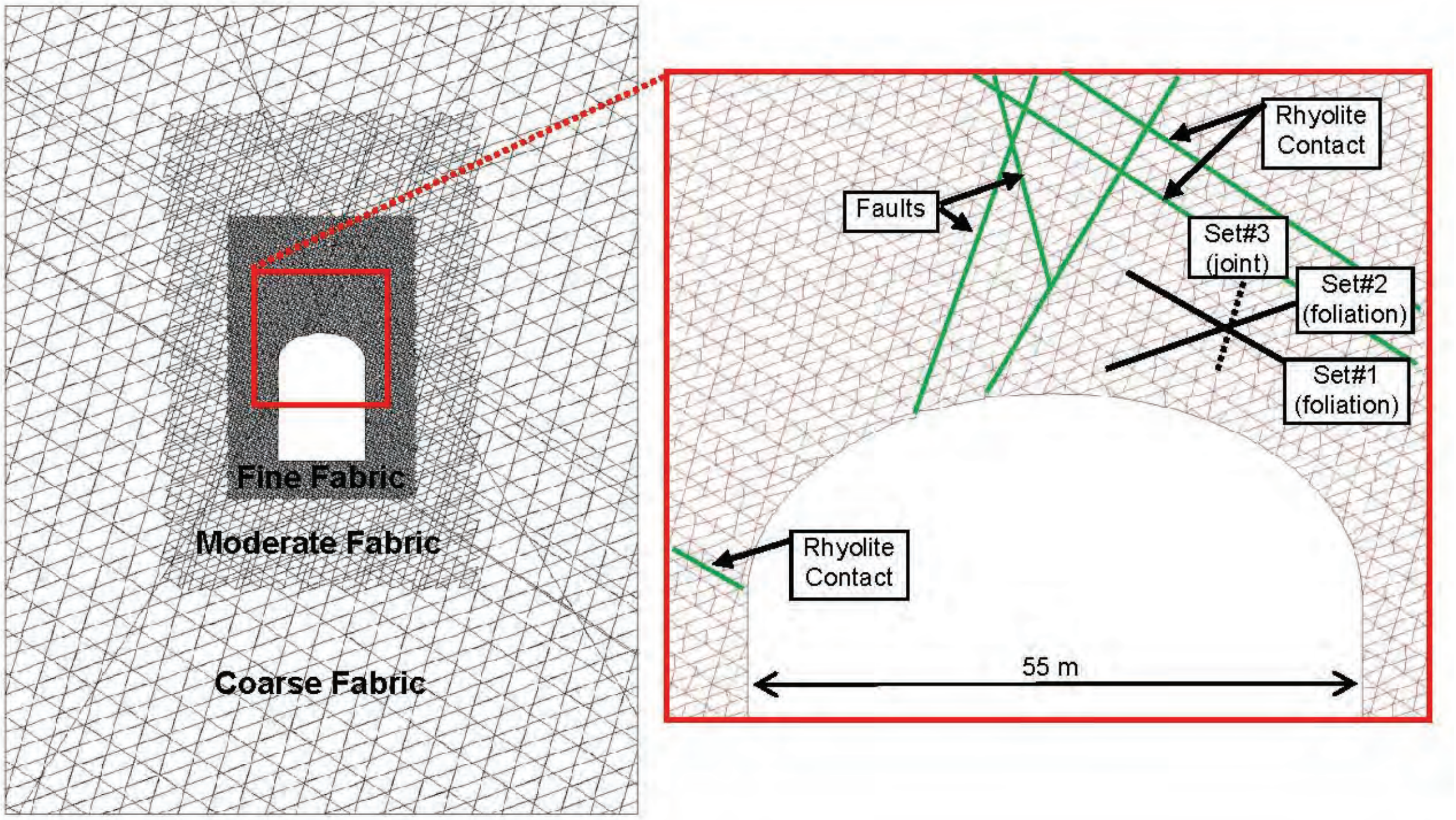

**Figure 5.3.4-2** Distinct element modeling of the structural fabric of the LC-1. [Golder Associates]

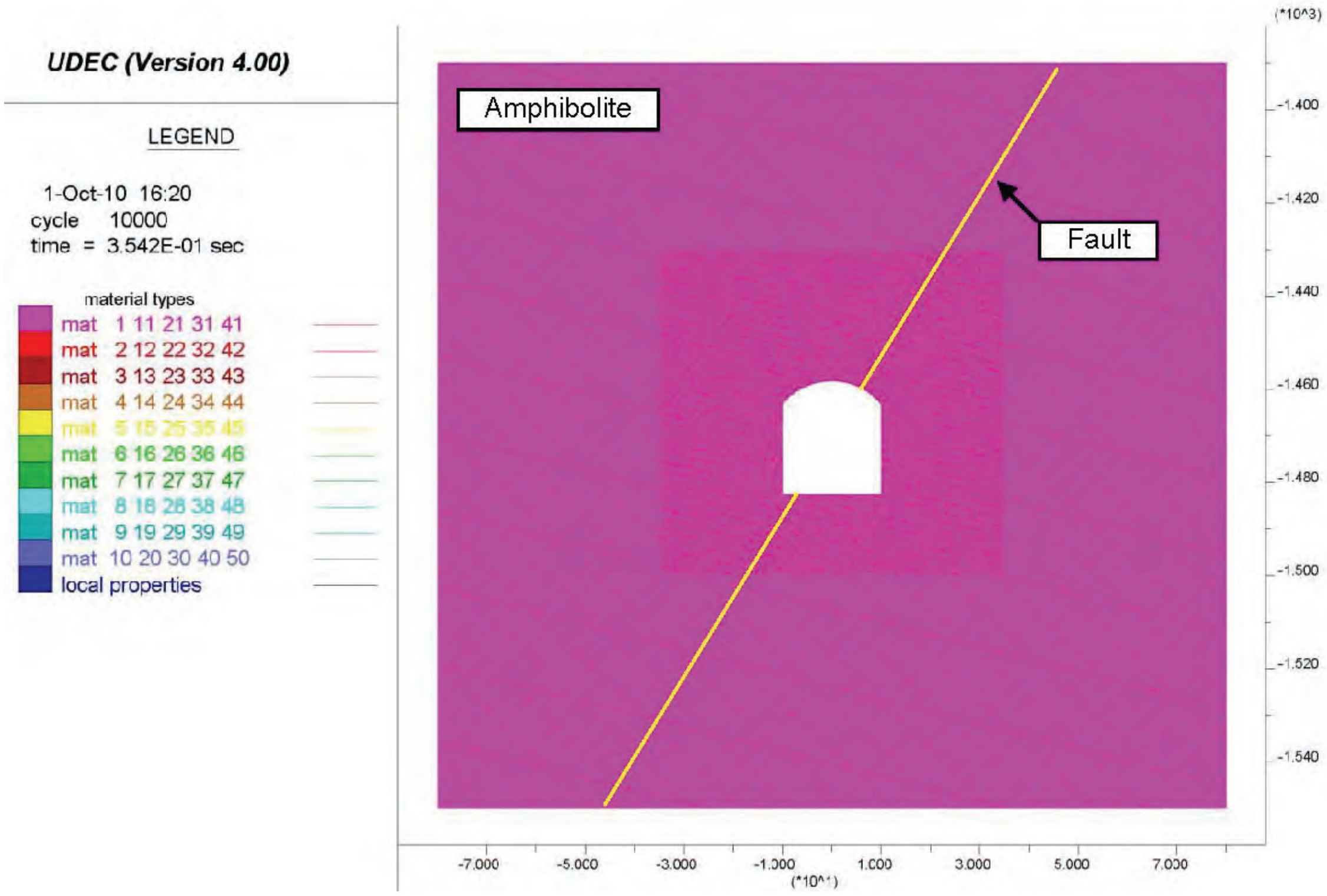

**Figure 5.3.4-3** Distinct element modeling of the geology of the LMs. [Golder Associates]

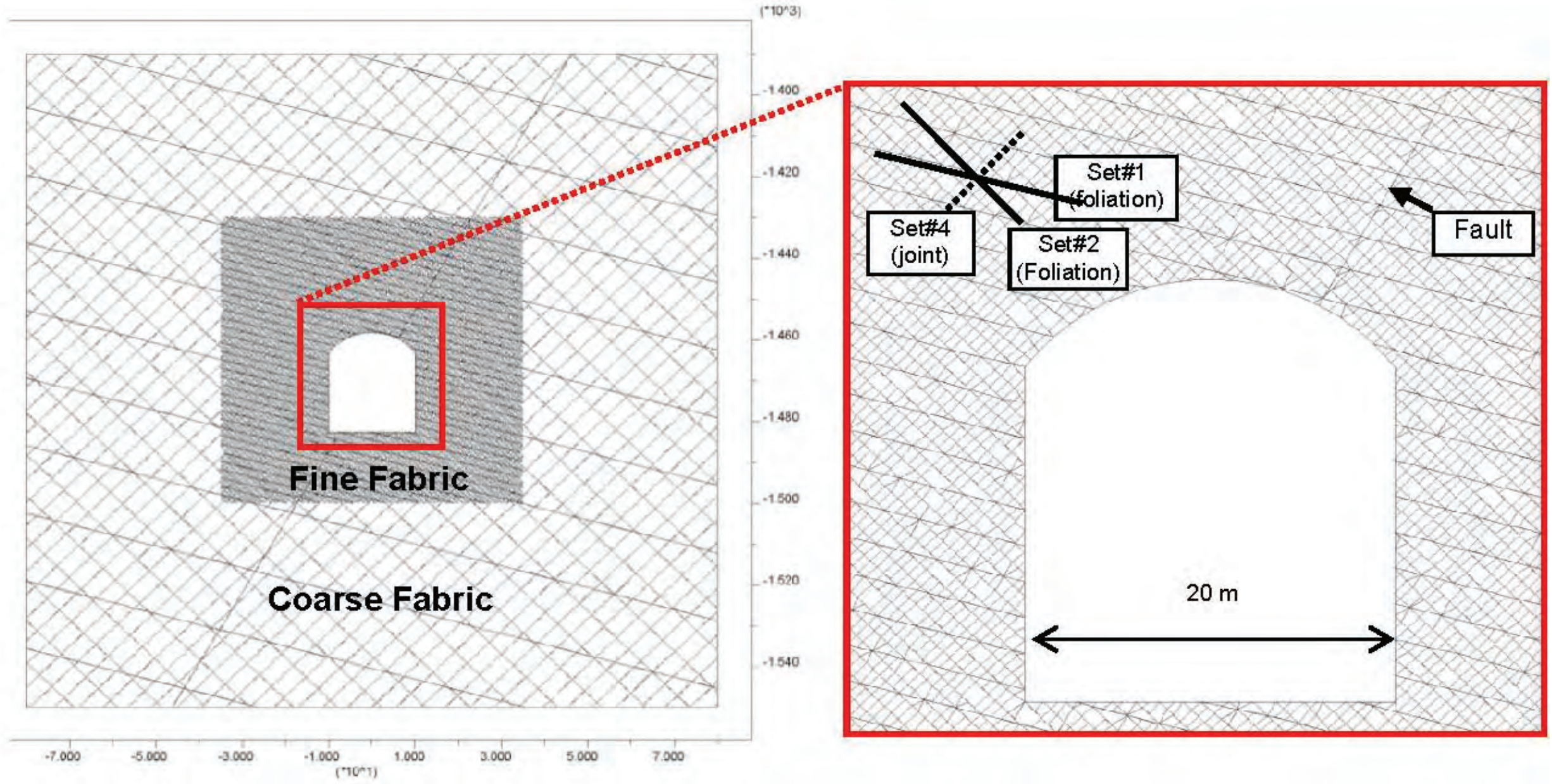

**Figure 5.3.4-4** Distinct element modeling of the structural fabric of the LMs. [Golder Associates]

### Ground Support Modeling of LC-1

Wedge analyses were used to confirm the length and spacing of the cables and further refine the layout. Mesh and shotcrete are not required for the stability of the walls; however, they are included in the support drawings for safety reasons (control of small, loose rock and protection of workers).

Two-dimensional models took into consideration the 98.4 feet (30 m) thick rhyolite dike swarm dipping at approximately 35˚, transecting the cavity span, as well as the smaller rhyolite dike at depth into the crown. The models also assumed a few persistent fault-type structures as shown on the preliminary geological model (Appendix 5.J).

Various ground support scenarios were modeled for the LC-1 cavity geometry, including:

- **Model 1.** Full excavation, unsupported
- **Model 2.** Full excavation, 65.6 feet (20 m) fully bonded cables and 16.4 feet (5 m) rock bolts
- **Model 3.** Full excavation, 65.6 feet (20 m) debonded cables and 16.4 feet (5 m) rock bolts
- **Model 4.** Sequential excavation, 65.6 feet (20 m) debonded cables

The success of a support system design relies not only on providing the appropriate capacity to carry the loads imposed by the excavation, but also on matching the stiffness of the support system to the stiffness of the rock mass, i.e., providing compatibility of deformation. The unsupported models were instrumental in identifying failure mechanisms and the excavation potential depth of influence so that the adequate type of support could be selected. Generally, the bonded cables tend to control displacements better than the unbonded cables. However, the loads on the bonded cables tend to be concentrated at the locations where they cross the joints; therefore, the displacements over such small distances result in large strains on the cables and locally overstress them. The debonded cables fare much better in that regard because the total deformation resulting from the opening joints is distributed over the full debonded sections of the cable. The maximum loads on the debonded cable model were considerably less than the bonded cable model, in some cases as low as 50%. In addition, the resulting deformations from the debonded cable model are only marginally larger than the bonded cable model.

The other significant insight from the analyses resulted from the comparison of the cable loads between the full excavation model (excavation performed in one step) and the staged excavation. The more critical stage in the excavation is not its final geometry; it occurs in the excavation of the top of the dome (second lift), when the cavity geometry induces the highest stresses in the excavation sequence. The consequence of having to transition through such geometry is increased loads in the roof cables and some reduction in parts of the side walls.

The LC-1 Plane Strain models considered two rhyolite bands—a thick band that intersects the cavity, and a thin band slightly overlying the cavity—both dipping at 35° from horizontal within a domain of amphibolites. The axisymmetric analysis considered only a horizontal, thick rhyolite band.

The effectiveness of the support systems was evaluated by comparing the supported and unsupported model output with respect to:

- Depths of yielded elements
- Intersecting shear bands in zones of yielded elements
- Total displacements in the vicinity of the excavations
- Yielding of support elements

In the interpretation of results, support was deemed effective if it prevented the formation of intersecting shear bands (regions of localized, high shear strains), which initiate from excavations and extend into the rock mass. Such intersecting shear bands in zones of yielded materials indicate development of failure mechanisms that make it kinematically feasible for material to collapse into excavations.

Interpretation of FEM modeling results indicates that the support system defined above significantly reduces total displacements in the near-excavation rock mass and prevents the formation of intersecting shear bands in the yielded zones. Only a few bolts in the system yielded; the others had loads well below the bolt anchor capacity. All the prestressed cables experienced loads below 80 tons, i.e., loads lower than 80% of anchor capacity.

Three-dimensional models of the LC-1 included two scenarios. The first model considered the amphibolite as a rock mass continuum with Generalized Hoek-Brown (Appendix 5.J) properties and the rhyolite also as a continuum, but with Brittle Hoek-Brown properties; the second model considered the amphibolite to have a set of ubiquitous joints. This was done to assess the impact of this flat joint set in a true 3-D setting and to corroborate the 2-D analyses and their assessment of the support performance. No support was applied in the 3-D models.

The geology in the LC-1 area shows the presence of a rhyolite band of about 98.4 feet (30 m) to 114.8 feet (35 m) thick intersecting the cavity diagonally from the springline of the dome to the bottom. This rhyolite band was incorporated into the model mesh and material properties.

The continuum model shows limited yielding around the cavity, which is in good agreement with the FEM method axisymmetric results. However, the displacements on the walls of the cavity are quite high, approximately 11.8 inch (30 cm), but only on the skin of the cavity. This is an indication of spalling, as the displacements subside to <0.8 inch (2 cm) past the first element on the boundary.

The ubiquitous-joint model shows that both the crown and the floor of the excavation can have large yield zones, mainly due to delamination of the foliation. This can lead to large wedge-shaped zones approximately 49.2 feet (15 m deep), requiring deep support. This is in good agreement with the findings from the distinct element models. However, it should be noted that, in both approaches, the joint sets, discrete or ubiquitous, are continuous and therefore these results are quite conservative. The springlines of the cavities and the edges of the floor are the highest stressed areas of the cavities.

The 2-D analyses can be much more detailed than the 3-D analyses, especially when it comes to the support interaction, because of the ability to develop very fine meshes. The good agreement between the 3-D models and the 2-D models gives the necessary confidence in the support design, which was based on the 2-D models.

For the base case of LC-1, the following support system was selected:

- 66 feet (20 m) long cables at 8.2 feet (2.5 m) x 8.2 feet (2.5m) spacing in the dome and walls. The cables have an anchor capacity of 100 tonnes, which requires 4 strand anchors in a 5 inch (125mm) hole sheathed and stressed with faceplates and clamps. They are installed from within the cavity.
- 16.4 feet (5 m) long rock bolts of 1 inch (25 mm) diameter at 49.2 inch (1.25 m) x 49.2 inch (1.25 m) spacing in the walls and dome. The bolts are grouted and tensioned with two speeds of resin, and have dome faceplates with hemispherical nuts. A wire mesh is applied to excavation faces before rock bolting, and then 4 inch (100 mm) thick shotcrete applied.

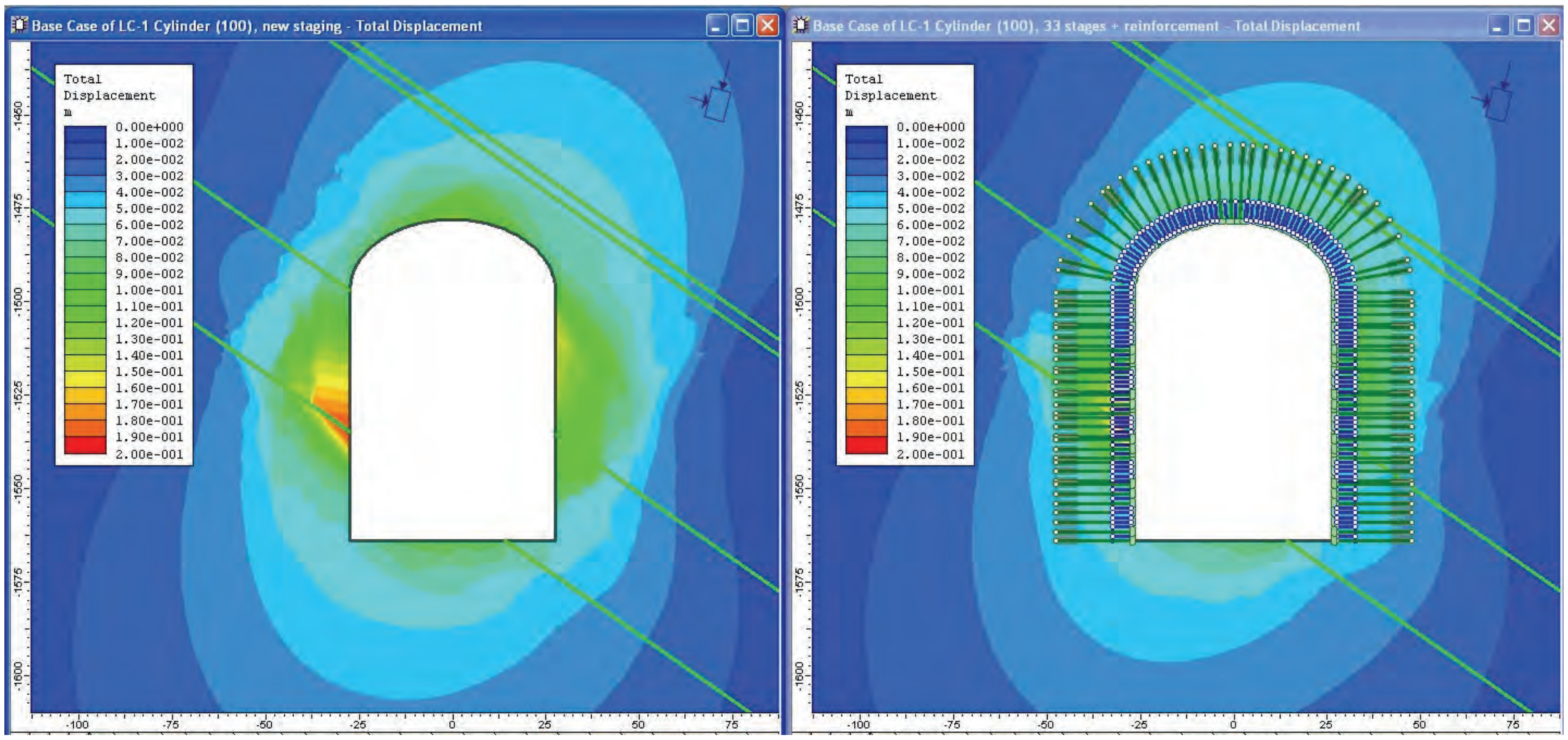

**Figure 5.3.4-5** FEM modeling of ground supports for the LC-1. Left: illustration modeling without ground support. Right: illustration modeling with ground support. [Golder Associates]

An example of finite element modeling of ground supports is illustrated in Figure 5.3.4-5.

**Ground Support Modeling of Laboratory Modules 1 and 2 (LM-1 and LM-2)**
Initial support analyses for the roof and walls of LM-1 and LM-2 were empirically performed according to the charts proposed by Grimstad and Barton (Appendix 5.I).

Wedge analyses were used to confirm the length and spacing of the cables and further refine the layout. Mesh and shotcrete are not required for the stability of the walls; however, they are included in the support drawings for safety reasons (control of small, loose rock and protection of workers).

The rock mass qualities for the LM-1 and LM-2 area are similar to the LC-1 area, and as such, the same rock mass properties and the same software codes as those for LC-1 were used.

The planned LM excavations have the 2-D same-section geometry; therefore, only one model section was assessed.

The rock mass characterization work for the LM area identified four main discontinuity trends. This fabric was incorporated into the distinct element models by calculating apparent dips and joint spacing with the assumed section geometry. An inferred fault is depicted in the preliminary geological model; therefore, a persistent fault feature dipping at 60˚ was assumed to crosscut the cavity in order to assess the potential implications of such a feature.

Ground support details were similar to those utilized in the LC-1 models. The following models were developed for assessing the LM excavations:

- **Model 1.** Single stage excavation, unsupported
- **Model 2.** Single stage excavation, 33 ft (10 m) bonded cables spaced at 8.2 ft (2.5 m) and 8.2 ft (2.5 m) reinforcement dowels spaced at 4.1 ft (1.25 m). The cables were pretensioned to 40% of the assumed maximum cable capacity (2 strand cable = 500 kN).

The results generally show substantial relaxation in the crown, following the main trend of foliation and jointing. With the use of ground support, the crown was controlled, and the degree of yielding in the walls and crown was substantially reduced. Based on the factored cable loads, more appropriate cable spacing would be 8.2 feet (2.5 m) in the crown and 9.8 feet (3 m) in the walls. Also, increasing the cable length to 33 feet (10 m) in both the crown and walls is warranted in order to anchor the cables in undisturbed ground.

Only plane strain analyses were performed for the LMs, due to their geometry.

For the LMs, the selected support system comprised the following:

- 33 ft (10 m) long cables at 8.2 ft x 8.2 ft (2.5 m x2.5 m) spacing in the crown and walls
- The cables have an anchor capacity of 50 tons, which requires 2 strand anchors in a 4 in (100 mm) hole sheathed and stressed with face plates and clamps. They are installed from within the cavity.
- 16 ft (5 m) long rock bolts of 1 in (25-mm) diameter at 4.1 ft x 4.1 ft (1.25 m x 1.25 m) spacing in the walls and dome. The bolts are grouted and tensioned with two speeds of resin, and have dome faceplates with hemispherical nuts. A wire mesh is applied to excavation faces before rock bolting, and then 4 in (100 mm) of shotcrete without fibers applied. The wire mesh and shotcrete components of the support system were not numerically simulated.

Since the geology at the proposed locations of the LMs has not been defined to the same extent as the LC-1 area, two simplified geological conditions were considered in the analysis of LM-1 and LM-2. The first case assumed that the cavities would be excavated in a domain consisting only of amphibolites with conventional Generalized Hoek-Brown parameters (Appendix 5.I). In the second case, a domain consisting entirely of rhyolite rock mass with brittle parameters was assumed. These two cases are sufficient for evaluating the performance of the proposed support system at this stage of excavation design.

The effectiveness of the support systems for LM-1 and LM-2 was evaluated using the same methodology as that employed for the LC-1 modeling. The model results clearly indicate the adequacy of the support system for the LMs. In addition to preventing intersecting shear bands and reducing total displacements, the system also significantly reduces the zone of yielding. Similar to the performance of the LC-1 support, only a few bolts in the LM support system yield. As well, all cable loads are below 80% of the cable anchor capacity.

The wire mesh and shotcrete components of the support system were not simulated by numerical modeling.

An example of finite element modeling of ground supports is illustrated in Figure 5.3.4-6.

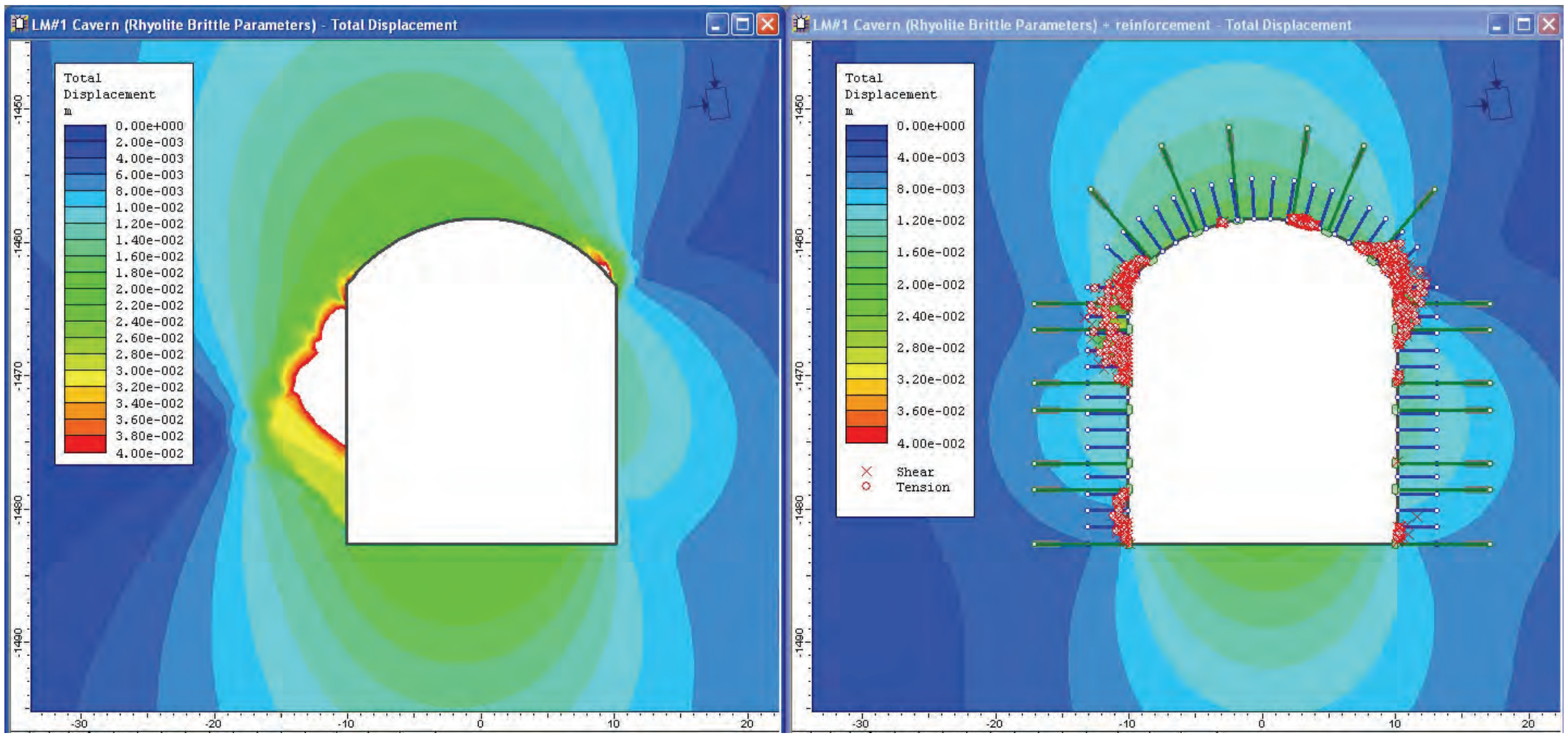

**Figure 5.3.4-6** FEM modeling of ground supports for the LMs. Left: illustration modeling without ground support. Right: illustration modeling with ground support. [Golder Associates]

### Ground Support Modeling of Access Drifts, Boreholes, and Ancillary Rooms

Typical rock support for the access drifts to LC-1, LM-1, and LM-2, and campus access drifts as well as the smaller auxiliary drifts and ancillary spaces is based on empirical methods and kinematic analyses according to the charts proposed by Grimstad and Barton (Appendix 5.I).

The access drifts are relatively long and traverse different rock conditions, which will require slight modification to ground support as excavation proceeds.

The typical rock support for drifts comprises 7.9 feet (2.4 m) long #8 rock bolts spaced at 4 feet (1.25 m) x 4 feet (1.25 m) and wire mesh in the crown of the drift installed at the heading, with shotcrete applied after the excavation has advanced. Kinematic analyses and insight gained from the stress analyses of the larger LMs have shown this support to be adequate.

The rock support analysis for larger ancillary excavations is also based on empirical methods and kinematic analyses according to the charts proposed by Grimstad and Barton (Appendix 5.I) and FEM analysis. The selected support system for ancillary excavations the selected support system comprised the following:

- 16-ft (5-m) long rock bolts of 0.8-in (20-mm) diameter at 4.1 ft x 4.1 ft (1.25 m x 1.25 m) spacing in the walls and dome. The bolts are grouted and tensioned with two speeds of resin, and have dome faceplates with hemispherical nuts. A wire mesh is applied to excavation faces before rock bolting, and then 4 in (100 mm) of shotcrete without fibers applied.

The rock bolts were modeled in 2-D as fully bonded bolts with faceplates. They were assigned small residual capacities (≈10% of tensile capacity).

The vertical raise bores should be stable without support as the maximum excavation induced stresses should not exceed approximately 64 MPa. Should control of small-dimension loose wallrock material be required, shotcrete can be applied.

**Ground Support Modeling Campus-Wide**

The modeling presented in the previous sections addresses the support requirements and the prediction of the rock mass behavior for each individual major opening of the 4850L. To address issues of potential for interference and influence from one opening on the others, concerns with the size of rock pillars left between rooms, and drift-drift or room-drift intersections with less favorable geometries, a campus-wide model was developed with a boundary element code. This is an elastic analysis that allows identification of zones of overstress resulting from either openings that are too close to each other or zones with poor geometry.

The analysis results generally indicate there is no interference between the large excavations (i.e., LC-1, LM-1, LM-2, Transition Area, and the Davis Laboratory). Adequate distance was provided in the layout to avoid problems arising from the proximity of these openings.

The pillars between the LMs and the ancillary spaces seem to be of adequate size to ensure that they will stay in their elastic state, except for a few localized zones in the corners formed by the intersections of the ancillary spaces and the drift. Design modifications—chamfered corners—have been incorporated to alleviate these overstress zones. The level of stress can be safely handled and special attention to these design elements will be integrated during detailed Final Design for intersections. At the completion of Preliminary Design, the layout of the 4850L Campus does not present any difficulties that would require relocation of any of the facilities. An example of campus-wide modeling is illustrated in Figure 5.3.4-7.

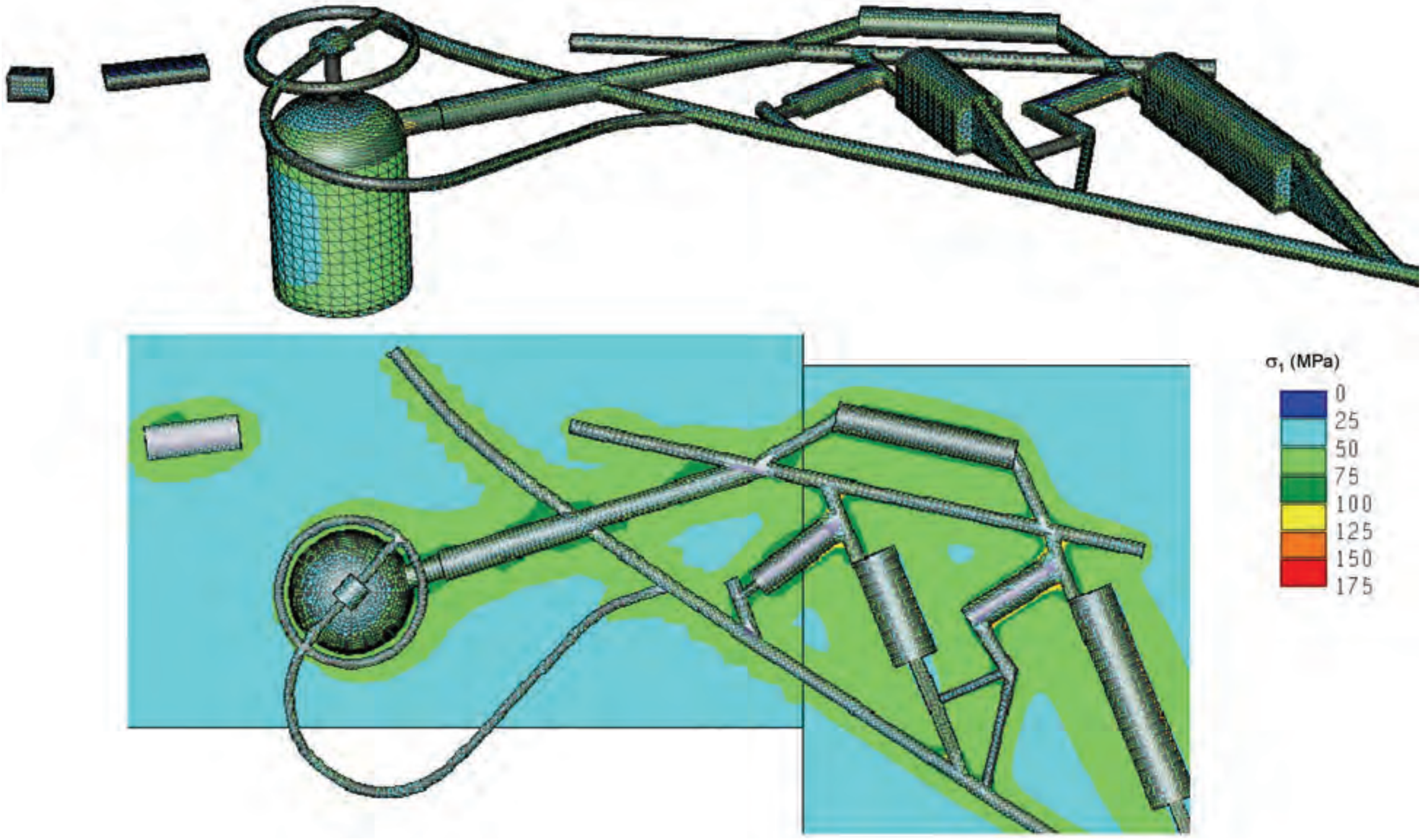

**Figure 5.3.4-7** Campus-wide modeling with FEM. [Golder Associates]

## 5.3.5 General MLL Excavation Sequence

Excavation of the horizontal and inclined elements of the 4850L Campus will be performed using modern drill-and-blast methods. Vertical elements such as the ventilation and emergency-egress raises will be excavated using raise-bore techniques. High-quality, smooth-wall civil construction methods and resulting quality of excavation are expected.

The excavation sequence and schedule at this level of completion were developed under the following assumptions:

- The enlargement of the Yates Shaft Station will be done as part of the Yates Shaft refurbishment contract.
- The excavations of the West Laboratory access drift and East Laboratory access drift near the Yates Shaft should be scheduled to minimize disruption experiments in the Davis Campus area.
- The spiral access drift that goes to the construction access area above LC-1 will have been completed as part of the final-design geotechnical investigation, prior to the start of excavation.
- Ross Shaft rehabilitation will have been completed prior to the start of excavation of the 4850L Campus.
- All excavation materials and equipment for DUSEL excavations will be accessed through the Ross Shaft.
- Waste-rock handling systems will be established and available for use at the start of the excavations for the 4850L Campus.
- All muck removal will be through the Ross Shaft. Maximum capacity of the haulage at the Ross Shaft will be 3,300 tons/day.

The overall construction schedule is expected to take 48 months from beginning of mobilization to end of demobilization.

In general, excavation will start from the Ross Shaft and progress along the West Laboratory access drift toward LM-1 and the Yates Shaft, then south along the East Laboratory access drift. The Areas of Refuge (AoRs) and utility rooms will be excavated as the headings of the excavations arrive at each location.

The proposed development sequence is organized in the following steps:

Initial excavation items:

- Cap and powder magazines
- Drifts and rock breaker for muck handling through the Ross Shaft, construction equipment garage, and AoR near the Ross Shaft
- Drifts at the 4850L and 3950L to provide access for the raise bore to the Oro Hondo Shaft
- Slash of the existing drift on the 3950L for ventilation exhaust with emphasis on excavating the Big X intersection in combination with drifting to bottom of new ventilation raise on 4850L
- Raise bore to 3950L for ventilation exhaust
- Rock support and short connecting drift for the AoR near the Ross Shaft

Start of main campus excavations:

- Start slash of West Laboratory access drift
- Start excavation of overhead ventilation drift at south end

- Ross Shaft mechanical-electrical room and sectionalizing switches
- Start laboratory module excavations:
  - Initial heading through LM-2 to East Laboratory access drift to establish ventilation circuit
  - Continue slash of West Laboratory access drift; construct small openings along drift as they are encountered
  - 4850L-5060L access ramp and drift
  - LM-1 and associated utility room

- LC-1 utility excavations:
  - Continue slash of West Laboratory access drift into Big X area
  - Yates Shaft AoR and access drift
  - Big X intersection
  - LC-1 access and utility drifts
  - Complete LM-1 and LM-2
  - Complete 5060L access drift into bottom of LC-1

- East Laboratory access drift and LC-1:
  - Raise bore for LC-1 muck removal
  - Start East Laboratory access drift from north end at Big X
  - Yates Shaft mechanical/electrical room
  - LC-1 crown excavation
  - Ventilation access drift on 5060L
  - 5060L AoR

- Complete LC-1 and related excavations:
  - LC-1 cylinder excavation
  - Calibration room and drifts

**Large Cavity Excavation Design**

The excavation for LC-1 will have a finished inside diameter of about 180 feet (55 m) and height of approximately 275 feet (84 m), as shown below in Figure 5.3.5-1 and Table 5.3.5. It will be an upright domed right-cylindrical cavity with an ellipsoidal crown. Depending on the type and thickness of the final interior waterproof containment vessel lining selected by LBNE during the CD2 process, the final fiducial volume of the LC-1 may change. The containment vessel liner is being designed by the LBNE and, as currently planned, will be a separate installation process from the excavation. The highest point of the arched roof will be about 62 feet (19 m) above the springline (the inflection point of an excavation from the sidewall into the dome), which is at the sill (the surveyed floor of the level) elevation of the 4850L.

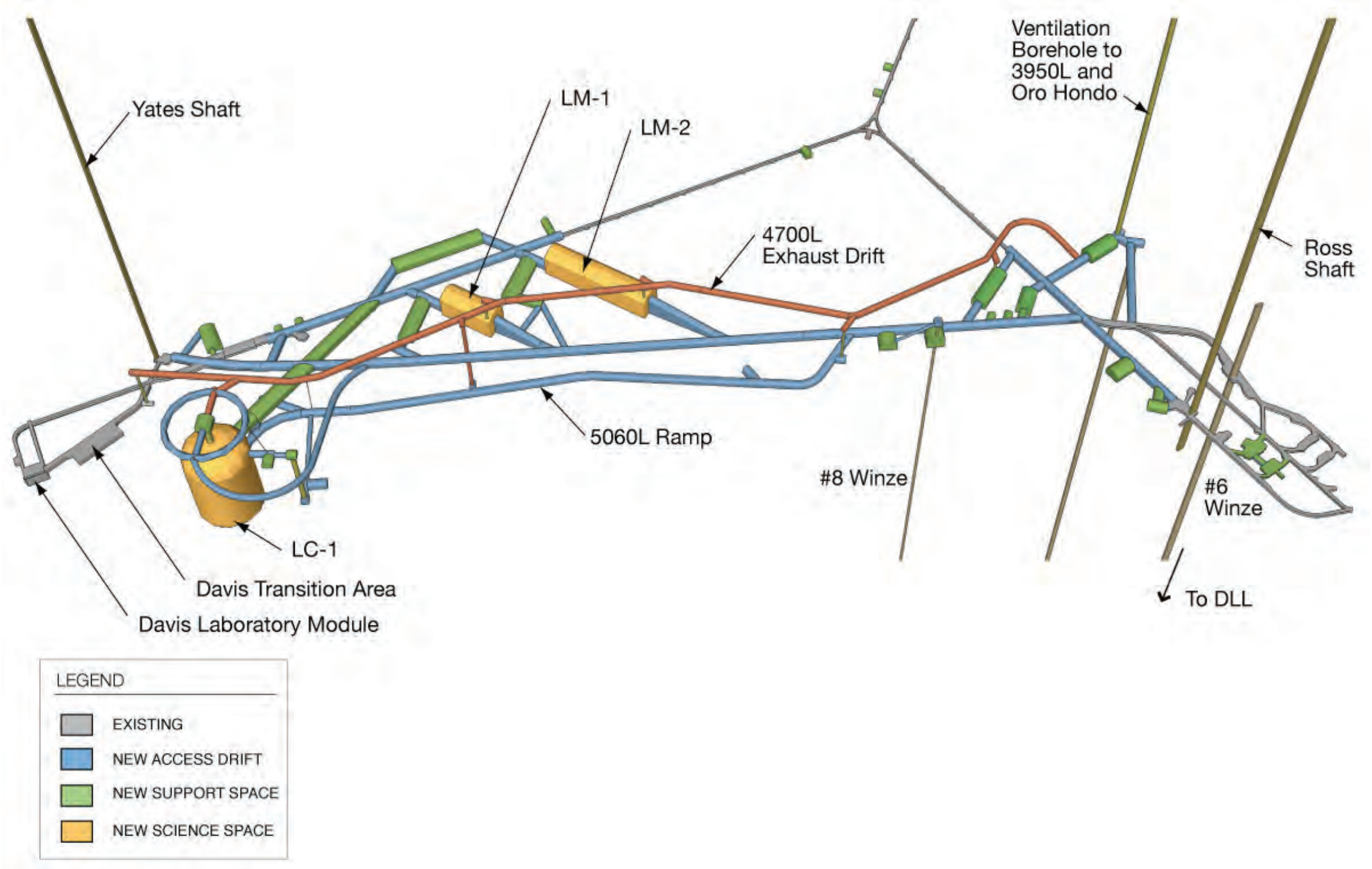

**Figure 5.3.5-1** Isometric view of the MLL Campus. [DKA]

| Experiment Space | Width (m) | Height (m) | Length (m) | Floor Area ($m^2$) | Finished Volume ($m^3$) |
|---|---|---|---|---|---|
| LM-1 | 20 | 24 | 50 | 1000 | 22,495 |
| LM-2 | 20 | 24 | 100 | 2000 | 44,990 |
| LC-1 | -- | 83 | 55 (diameter) | 2376 | 185,947 |

**Table 5.3.5** Space available to experiments.

Excavation access for LC-1 will be via the 4700L LC-1 access drift, which connects to the 4850L West Laboratory access drift. Access to the bottom of the cavity will be necessary for the removal of excavated rock material from the LC-1. This will be provided by a construction access drift that will ramp down from near the Ross Shaft to the 5060L (Appendix 5.I).

Excavation will initiate through a borehole (BH) access drilled from 66 feet (20 m) above the apex of the crown. The excavation will progress with widening of the BH access into the dimensions of the dome. Excavation will progress from this BH access downward and will connect with an incline drift from the 4850L sill to provide larger equipment access and secondary egress. As excavation proceeds, installation of prescribed ground support will be applied utilizing tensioned grouted cable bolts, bolts, wire mesh, and shotcrete as shown in Figure 5.3.5-2.

A native water-collection membrane lining of geosynthetic composite strip drains will be installed vertically against the perimeter of the cavity. The strips will have width of 6 inches (15 cm) and will be spaced 5 to 7 feet apart (about 2 m). The strips will act as vertical drains. The strips will be installed beneath a shotcrete layer, will extend for the entire height of the LC-1, and will discharge into the drain underneath the footing slab. The drain collecting the groundwater seepage will be directed into a sump on the 5060L.

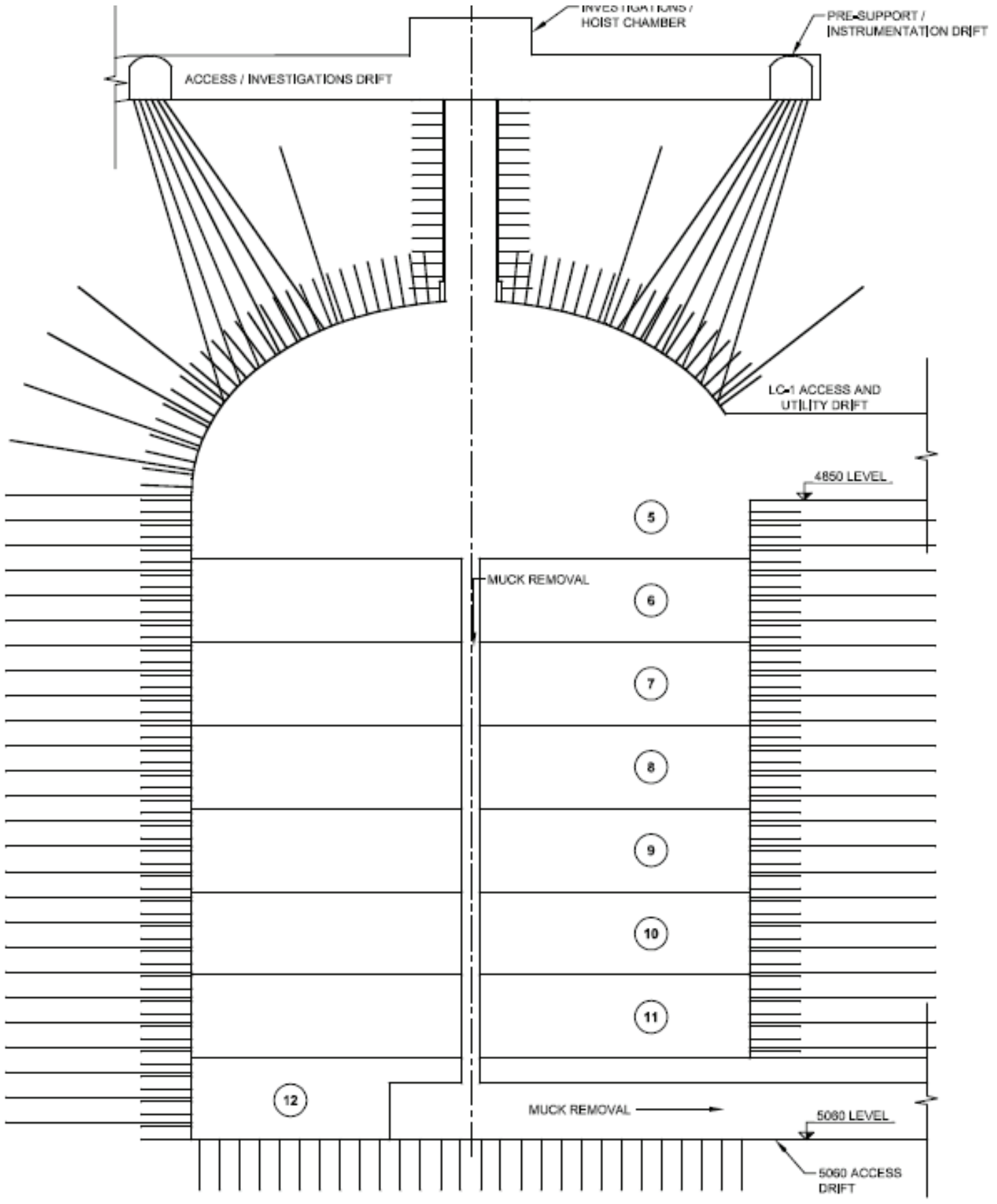

**Figure 5.3.5-2** Excavations sequence and ground supports of LC-1 (Stage 5 to completion). [Golder Associates]

In summary, the excavation sequence for the LC-1 as detailed in the Golder Associates Final Preliminary Design Report (Appendix 5.I) is described as follows:

- **Pre-Final Design.** Excavation of the spiral investigations and vent drift 66 feet (20 m) above the crown of LC-1will be done during the investigations phase of the project, prior to Final Design. A Final Design expanded geotechnical site investigation core drilling program will be completed from this drift, created during the excavation phase, and will support the Final Design.
- **Stage 1.** Excavate 5060L access drift into the center at the base of LC-1. Drill the raise bore between the hoist chamber in the vent drift and the 5060L. Excavate the circular ventilation drift access and connect to the 4700L central ventilation drift. Slash the top 85 feet (26 m) of the raise bore from the vent drift into the dome of LC-1 and line the 66 feet (20 m) of the slashed shaft with concrete above the dome.
- **Stage 2.** Using access through the vent drift above LC-1, excavate the top level of the dome of LC-1 and install rock bolts and cable anchors from the circular ventilation drift access and from within the dome excavation.

- **Stage 3.** Using access through the vent drift, excavate the second level of the dome and install support. Excavate an access ramp from the 4850L at 15% grade into the third level of the LC-1 excavation.
- **Stage 4.** Using access from the 4850L, excavate and support the fourth excavation level of the dome.
- **Stage 5 to completion.** Excavate benches in the tank section of LC-1 down to the 5060L, installing rock support from each bench as the excavation proceeds. Access would be from the 4850L and from the vent drift.

The LCAB committee met twice during Preliminary Design and Dr. Evert Hoek, LCAB chair, communicated on an as-needed basis in the latter stages of Preliminary Design, to provide design modifications and recommendations to the initial geotechnical evaluation and excavation Preliminary Design of the LC-1. LCAB's comments regarding the Preliminary Design include the following recommendations;

1. **Obtain better definition of the rhyolite intrusive geometry through additional drill definition.** The proposed 2011 geotechnical site investigation would conduct close-spaced drilling to provide detailed geologic and geotechnical definition of the LC-1 excavation site. This proposed program is designated as a critical path item to progress the design during Final Design.
2. **Acquire representative joint persistence data.** The joint persistence data is actively being obtained from the ongoing Davis Campus laser scanning project. The data will allow for improved descriptions of existing fracture networks, improved 2-D and 3-D modeling of cavity stability, and refinement of ground-support designs. It was felt the current assumption-based modeling was too conservative, leading to overdesign of the ground support requirements.
3. **Refine excavation approach and sequence design.** Excavation options during Preliminary Design include consideration for a top-down excavation approach in combination with an internal-to-dome ramp access from the 4850L. Excavation design is constrained by the requirement to minimize wallrock damage while excavating in the safest and most economical manner. Further excavation design evaluation will be conducted during Final Design.
4. **Refine ground support design and installation technique.** Refined modeling, as stated in Item 3 above, may alter the current design requirements for ground support as well as installation sequence. The Preliminary Design includes the options to pre-drill ground support cable bolt holes from an overhead drift and/or allow for pre-installation of cable bolts. This option is designed to be done in combination with installation of cable bolts from inside the dome excavation. Due to the length, weight, and rigidity of the 20-meter-long cable bolts, further evaluation will be required during Final Design of the preferred logistics and installation techniques.

#### Laboratory Modules 1 and 2 (LM-1 and LM-2) Excavation Design

The excavation sequence of LM-1 and LM-2 are similar (Appendix 5.I). The access drift to the crown of the cavity will initially ramp up from the West Laboratory access drift to provide access for excavation and support of the crown of the LM. Details of the construction drifts and benching sequence will depend on the contractor's selected means, methods, and equipment. In brief, the proposed excavation sequence is as follows:

- **Stage 1.** Excavate and support the crown of the access drift and the central crown drift in LM-1. Install temporary glass fiber rock bolt support above the crown of the access drift where it is within the LM.
- **Stage 2.** Excavate and support the side drifts in the crown to widen the excavation to the full 65.6 feet (20 m).
- **Stage 3.** Slash the central portion of the crown so the full width of the LM is excavated and supported. Install crane supports; install cranes.
- **Stage 4.** Excavate the first bench below the crown drifts in the LM and in the access tunnel. Install support in the walls.
- **Stage 5.** Excavate the remainder of the LM and the access tunnel by benching to final invert level. Install wall support as the benching proceeds.

The excavation line for the walls and crown is 0.3 feet (0.1 m) outside of the clearance envelope to provide clearance for the rock bolt heads and shotcrete and for convergence during construction. The excavation clearance required for the LM is 65.6 feet (20 m) wide, 78.7 feet (24 m) high, and LM-1 is 164 feet (50 m) long while LM-2 is 328 feet (100 m) long. The LM cavities will be 62.3 feet (19 m) to the springline. The floor will be 13.1 feet (4 m) below the sill elevation of the 4850L access drifts. The excavation line for the walls will be 10 centimeters (cm) outside of the clearance envelope to provide clearance for the rock bolt heads and shotcrete. The crown shape is a semicircular arch.

A 20-ton-capacity bridge crane and a 40-ton-capacity monorail crane will be required by the science experiments. The rails for the bridge crane will be supported by corbels secured to the walls of the cavity with rock bolts and the rail for the monorail crane will be supported from ceiling anchor points with rock bolts. The rock bolt specifications and design will be incorporated into the draft final PDR after further discussions with the groups responsible for designing the crane itself.

The texture of the LM walls is designed with "as-smooth-as-possible" shotcrete application (smoothness criteria will be defined by experiment requirements during Final Design). Floors are standard 5,000 psi-rated smooth concrete with perimeter drains on one wall for collection of LM discharge waters into a central sump and pump system.

Primary personnel and nominal equipment access is provided from the East Drift via the East Laboratory access drift, which also allows access to the laboratory's utility room. Personnel and large-dimension equipment access is also provided from the West Drift via entry through the West Laboratory access drift ramp.

#### Access Drifts, Boreholes, and Ancillary Rooms Excavation Design

Excavation of the horizontal and inclined elements of the 4850L Campus will be performed using modern drill-and-blast methods. Excavation will proceed with standard drift configuration drill-blast patterns with emphasis on high-quality, smooth-wall civil engineering blast pattern design and explosive loading/ignition procedures. Removal of waste material will utilize load-haul-dump (LHD) excavators. High-quality, smooth-wall civil construction methods and resulting quality of excavation are expected. Typical excavation equipment will consist of single, twin, or triple-boom drill jumbos, LHDs, rock bolt drills, and various support equipment.

Vertical elements such as the ventilation and emergency-egress raises will be excavated using raise-bore techniques.

**Drainage from Drifts and Laboratory Modules**

Where seepage or damp areas are observed during the excavations of the drifts and LMs, drain holes with slotted PVC pipes and geocomposite strip drains will be installed prior to shotcreting. The seepage will then be directed to the floor drain at the sides of the drifts or LMs. The seepage flow will be collected at sumps, which will be positioned along the drifts and LMs. From there, water will be pumped into the overall Campus water management system. The design of the sumps-and-drains drainage system will be finalized during the detailed design phase of the project, when a topographic survey of the invert elevations along the existing drifts in the 4850L is available.

### 5.3.6 Requirements for Additional Future Geotechnical Investigations and Analysis

The data and information developed to date appear adequate for preliminary-level excavation designs. Verification of the lithology and geological structure with emphasis on foliation, discontinuities—including their mechanical properties—and the presence of rhyolite dikes is needed to progress to Final Design. Verification of the current geological and geotechnical model will allow refinement of the design, risk reduction, and minimization of the cost of potential overdesign.

Golder has proposed[19] a comprehensive investigation to develop the data needed for Final Design. This investigation will include the following three components:

**Component 1.** L C-1 Confirmatory Drilling—The scope of this component is based on Golder's evaluation of existing data and recommendations. It includes the drilling of four BHs into the rock mass at the proposed location of LC-1 along with associated logging and testing. On completion of drilling and logging, instrumentation allowing long-term measurement of rock response to the excavation of the proposed exploratory drift will be installed in selected BHs. The data on rock response to excavation will be of paramount importance to final design of LC-1.

**Component 2.** Final Design Investigations—The second component includes excavation of an exploratory drift to the area of the crown of LC-1, followed by mapping, drilling, and logging of 12 BHs, televiewer imaging, geophysical (BH tomography), and stress measurements within the rock mass in the vicinity of the proposed location of LC-1. This component also includes short-term and long-term monitoring instrumentation in the exploratory drift. The second component also includes the advancement of eight BHs to investigate the rock mass in the vicinity of the proposed locations of LM-1 and LM-2 and three BHs within the area considered for excavation of the large Yates Shaft mechanical/electrical utilities room east of the East Laboratory access drift. This component will also include all associated core logging, drift mapping, and in situ and laboratory testing. This drilling and testing program is recommended to develop the data needed for Final Design of the excavations for the LMs and mechanical/electrical room.

**Component 3.** Additional Required Work—The third component includes additional mapping and scanning of the openings on the 4850L, a comprehensive survey of the 4850L, and surveying down the Ross and Yates Shafts to establish vertical control on the 4850L. This component will also include continued blast-induced vibration monitoring of ongoing construction activities on the 4850L, as well as those associated with the construction of the proposed LC-1 exploratory drift. Component 3 also includes one vertical BH to be advanced along the proposed alignment of the vertical ventilation raise between the 4850L and 3950L.

## 5.3.7 Underground Monitoring

### 5.3.7.1 General requirements

Geotechnical and structural monitoring have become standard practice in both the mining and tunneling industries and are now a key element of any large project's risk management approach. In underground construction projects, these essential components of risk management translate usually into performance monitoring with the use of geotechnical and structural instrumentation. The instrumentation and techniques used in tunneling are more comprehensive, integrated, and sophisticated than those used for mine monitoring. The major difference is the excavation's long-term stability requirement for civil engineering construction vs. the temporary nature of mining excavations. Despite the differences in planning, implementation, complexity, and extent, the principles and instrumentation in both engineering settings are much the same. Since most of the DUSEL's future excavations will be placed at great depth (4,850 to 7,400 feet below surface)—much deeper than relatively shallow civil tunnels—geotechnical monitoring practice will be based on experience available from the tunneling and mining industries.

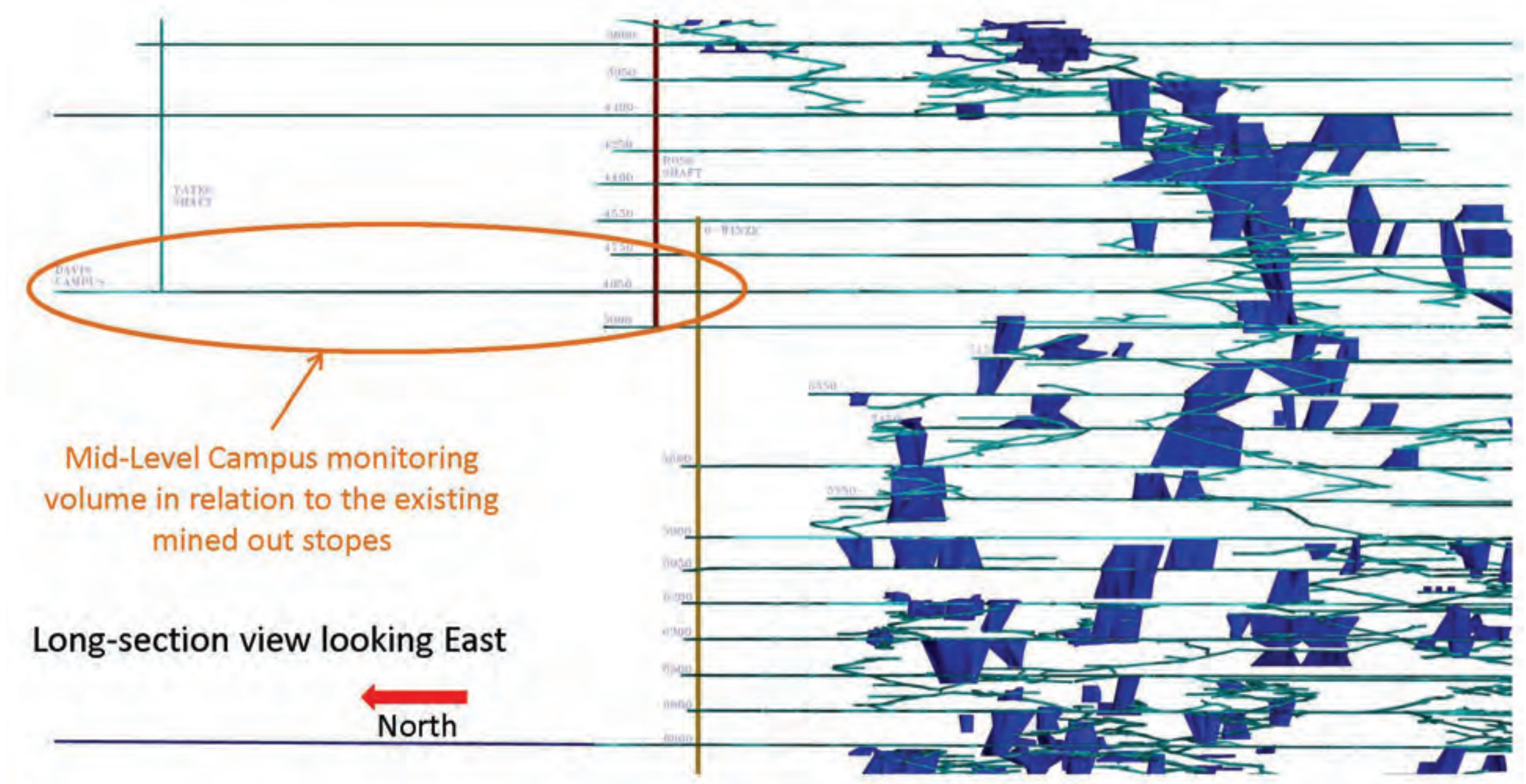

**Figure 5.3.7.1-1** DUSEL Mid-Level Campus monitoring area in relation to the existing mined-out Homestake stopes. [Golder Associates]

It will also be important to implement a well-defined monitoring program as early as possible to be able to capture reference ground characteristics before any rock movement may take place. This initial information will become a valuable component of the Geotechnical Baseline Report (GBR), which will be used for management of an excavation contractor's risks. Recent data available from large underground construction projects reveal that the geotechnical performance monitoring return on investment is several times its cost by helping to manage risk.

Although most of the ground monitoring instruments will be installed before DUSEL construction takes place, the ground monitoring plan will be implemented in the following three stages, each having distinctly different purposes to fulfill different project objectives:

1. **Before construction**, the purpose will be to collect and record the initial values of geotechnical parameters required for design or to initiate a geotechnical project, and/or to develop the GBR. Checking the validity of the assumptions and addressing data gaps will help address the initial risks.
2. **During construction**, the goal will be to confirm the validity of the design of the excavations and to ensure safety during construction and operation of already completed facilities. An early warning of excessive ground deformations or excessive support loads will help manage risks.
3. **After construction**, the primary objective will be to monitor and verify the overall behavior of the excavation, including rock/support interactions during operation, i.e., excavation performance.

With current technology, direct assessment of rock response can only be performed through the measurement of displacements either as the absolute displacements—or the relative displacements—of a number of points on the boundaries of the excavation, or within the rock mass. Even pressure, or stresses and forces, are measured indirectly through displacements, strains, or deformations.

### 5.3.7.2 Ground Movement Monitoring of the Davis Campus

Ground movement monitoring already implemented at the Davis Campus (excavations associated with the LUX/MAJORANA early experiments) included excavation monitoring with extensometers and monitoring of ground vibrations from blasting.

### 5.3.7.3 Excavation Monitoring with Extensometers

Monitoring of displacements in the rock surrounding an excavation is normally done by one of the following methods:

1. **Convergence measurements.** Convergence multipoint stations or simple tape extensometers to measure displacement/closure of the perimeter of an excavation
2. **BH extensometers (single or multiple point).** To measure displacement in the rock mass surround an underground excavation. Usually, BH extensometers are placed (anchored or grouted and recessed) in drill holes perpendicular to the back/roof or walls of an excavation.
3. **LIDAR (Light Detection and Ranging)/laser scanning and surveying.** More sophisticated techniques for the measurement of the movement of the perimeter of an excavation. The perimeter scanning could be combined with the geotechnical scanning of an exposed rock structure, and scanning of the excavation blasting rounds for blasthole control, overbreak management, and smooth blasting (controlled blasting), shotcrete thickness, and infrastructure inventory.

The rock movement data, after appropriate data reduction process and data analysis, will be used for the following purposes:

1. General performance of an excavation
2. Stability analysis
3. Calibrations of predictions and back-analysis for geotechnical modeling and calibration of numerical models

4. Verification of the design output (scaling of displacements and strengths—laboratory values vs. in situ values)
5. Early warning system—safety
6. Risk management

Although the specifications described above will have immediate applications for the Davis Campus assessment, the data to be collected will be used as important baseline information for DUSEL's Final Design and its construction.

#### 5.3.7.4 Davis Campus Excavation Monitoring

The original Davis Campus excavation monitoring plan was developed by CNA Consulting Engineers (the designer of the new LUX/MAJORANA excavations under contract with the SDSTA). The CNA plan did not include any rock mass monitoring parameters other than excavation convergence and rock displacements. Following the CNA recommendations, in April 2010, SDSTA personnel installed three extensometers in the back of the Davis Cavity and four in the back of the Davis Transition Area. It appears the extensometers were installed too late within the excavation schedule, and the major ground movements (due to excavating of the new LUX/MAJORANA excavations) had already occurred. No extensometer readings, other than the setup readings, have been collected to date. It was decided that it will be extremely useful, however, to continue this program and use the data for DUSEL's geotechnical baseline database for overall excavation performance purposes and risk management. Excavation convergence measurements, although recommended by CNA, have not been implemented by the SDSTA. In April 2010, the current extensometer monitoring activities in the Davis Campus area were incorporated into the DUSEL geotechnical program as one of the synergistic DUSEL/SDSTA activities.

The objectives of the monitoring program for the Davis Campus area were as follows:

1. Continue to monitor rock movement around the Davis Campus area utilizing extensometers already installed by the SDSTA, and install additional instruments as needed.
2. Perform LIDAR/Laser scanning of the Davis Campus for geotechnical assessment, shotcrete thickness measurements, and excavation convergence baseline.
3. Monitor ground vibrations due to blasting during ongoing excavation activities (the Transition Cavity and the Big X).

A diagram showing components of the ground monitoring plan for the Davis Campus is shown in Figure 5.3.7.4-1. Figure 5.3.7.4-2 shows the location of the extensometers in the Davis Cavity, Davis Transition Area, and the Big X area (intersection of the ventilation and the exhaust drifts).

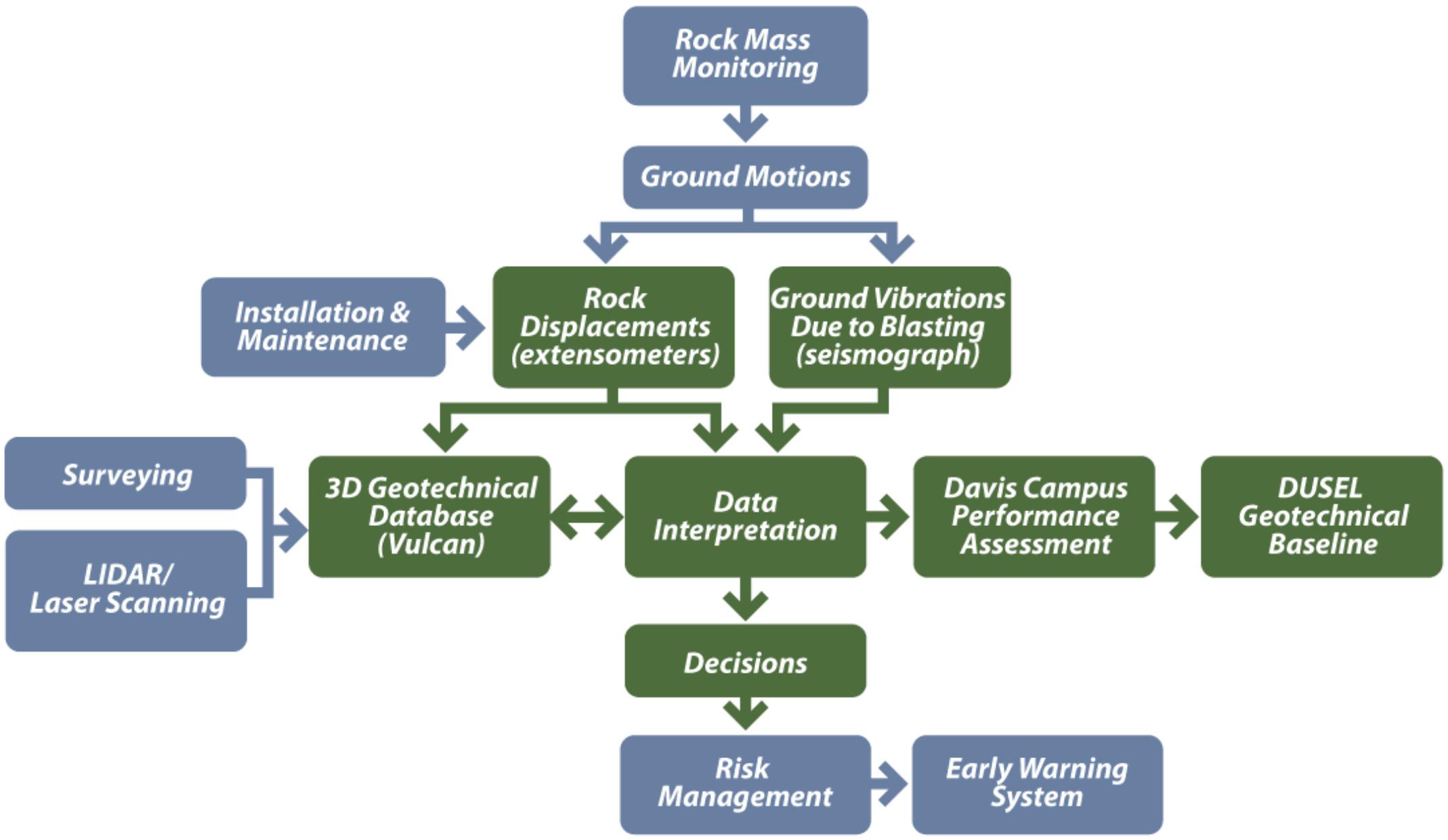

**Figure 5.3.7.4-1** Ground-monitoring plan for the Davis Campus. [Z. Hladysz; DKA]

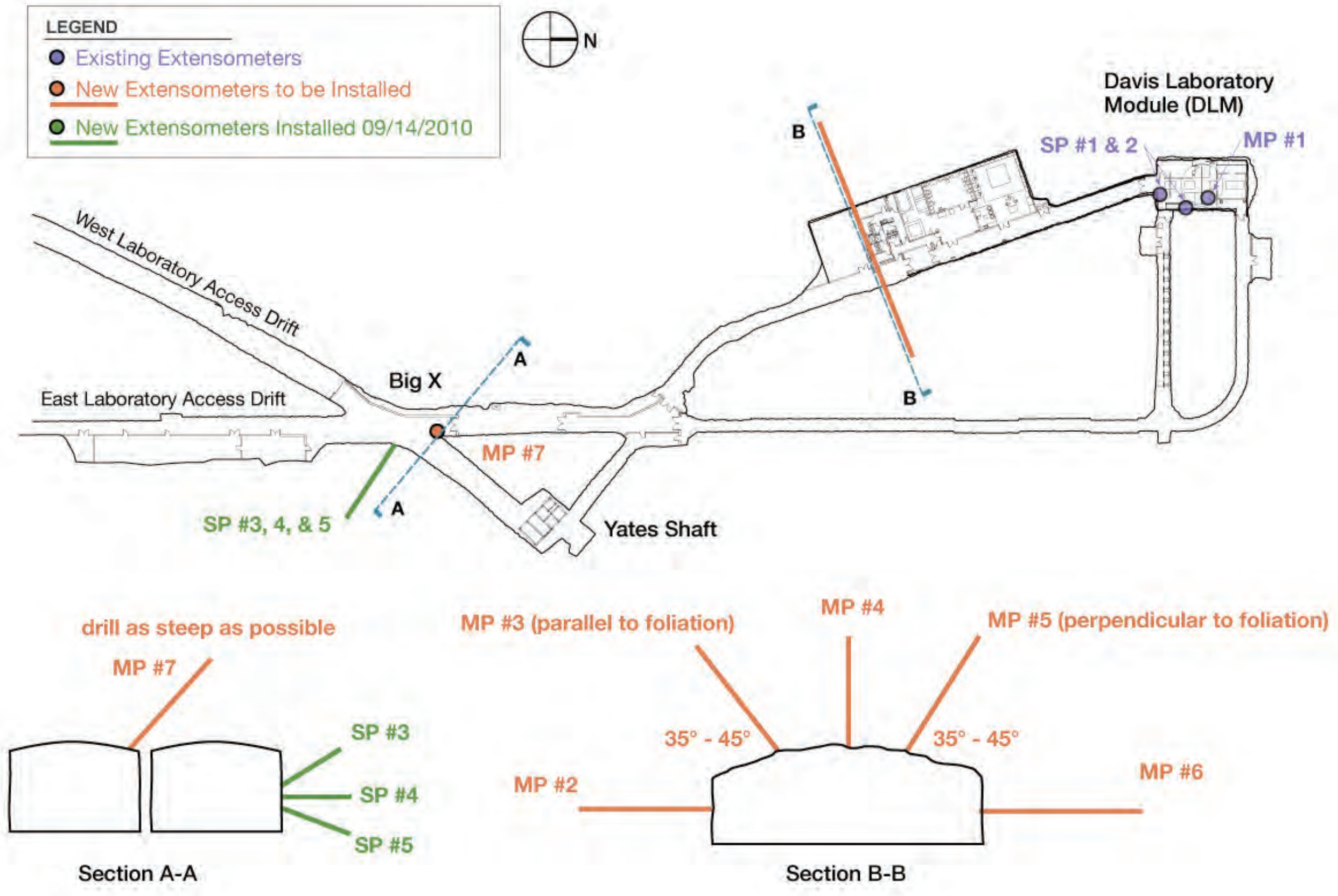

**Figure 5.3.7.4-2** Location of the extensometers at the Davis Campus.[20] [DKA]

### 5.3.7.5 Monitoring of Ground Vibrations from Blasting

Current schedules call for occupation of some of the laboratories during construction of the remaining excavations. When explosives are used for rock breakage, some part of the explosive energy is dissipated as ground vibration and air overpressure. Typically, these factors are dependent on a number of variables such as rock type, type of explosives used and charge per delay, timing configuration, and distance from the point of blasting to point of observation. Therefore, the level of vibration produced by the excavation activities may, in some instances, be of concern for the experiments operating concurrently. In order to quantify the effect of the blasting as a function of weight of explosives and distance, a program of blast monitoring was undertaken using blasts conducted as part of the excavation of new areas on the 4850L during 2010.

Analysis of the blast monitoring program is under way but the scope of the investigations was as follows:

- Blast vibration monitoring, analysis, and interpretation, including rock microseismic signature
- Parameters and variables monitored and measured: peak particle velocity (PPV) and blast frequencies
- Collecting pertinent information regarding blast parameters, such as explosive charges, blasting pattern, and delays, concurrently with the monitoring activities
- Deliverables will include rock constants, vibration prediction, and input needed for effective and safe blast design.

### 5.3.7.6 Monitoring of DUSEL Large Excavations

Monitoring of the large excavations (LC-1, LM-1, and LM-2) is an integral part of the design. Instrumentation should be installed as early as possible for the large excavations. In the case of LC-1, some instrumentation should be installed before excavation begins, to establish baselines and capitalize on as complete a ground response history as is feasible. Instrument reading frequencies should be adapted to the different phases of the project and into the operational period of the Facility. Information acquired during the investigation and early construction will help calibrate the models used for assessment of the stability of the cavities, refine support system designs, and allow for more accurate projection of future ground movements and deformations. Information acquired prior to and through the construction period will be used to monitor and document the construction progress in the context of varying geotechnical conditions and help to manage the associated risks. The installation of the instrumentation should ensure its durability in the presence of excavation activities. This is of utmost importance, especially for LC-1, because it will undergo frequent thermal and pressure fluctuations during its operational life. Long-term monitoring is intended to verify that the excavations are performing as intended and will aid in selection of proper rectification measures should they be needed. Typical instrumentation systems for underground excavations consist of:

- Single and multiple point extensometers
- Tape extensometers and convergence measurements (pins)
- Load cells and pressure cells
- Smart cables
- Inclinometers/tiltmeters
- Piezometers
- Thermistors

- Seismographs and accelerometers

Other types of monitoring, which do not involve installed instrumentation and would be used only during construction and/or at long time intervals during operation of the Facility, include:

- Scanners (laser/LIDAR)
- Surveying instruments
- Surveying benchmarks and optical survey points

The monitoring program should be treated in a holistic manner; therefore, it is best used when different instruments are installed at the same location, usually in arrays to complement one another and to corroborate behavior. It is common in tunneling to establish an interval at which arrays of instrumentation are installed at the same station. A similar approach has been adopted here for the large excavations.

### 5.3.7.7 Groundwater Monitoring

The nine geotechnical BHs that were drilled at the 4850L during the site investigations (see Section 5.3.2.3) have been monitored continuously. The monitoring program was designed to provide data in support of DUSEL activities related to LC-1 and LMs. Eight of the nine holes intercepted the groundwater system(s) and produced varying amounts of flow.

Groundwater monitoring[9,10] has consisted of two primary elements: 1) measurement of flow rates and water quality, and 2) pressure buildup testing and analysis. Data for Element 1 was generated from the ending date of BH completion (August-November 2009) and was continued until the BHs were shut in on June 2, 2010. Data from Element 2 were all generated since the BHs were shut in on August 21, 2010.

### 5.3.7.8 Seismic and Microseismic Monitoring

**Seismic Hazard Analysis**

Golder has analyzed (Appendix 5.I) existing information on historic earthquakes, faults, and current estimates of seismic hazard at the DUSEL site.

The DUSEL site is located within the western part of the North American craton, away from known areas of historic large earthquakes and active tectonic deformation with Quaternary-active faults and folds. Although the DUSEL site is located in a tectonically uplifted area, the regional geologic history of the initiation, growth, and uplift of the Black Hills indicates that the area has probably remained tectonically stable over at least the Quaternary Era (about 2 million years) and probably for much longer. The recent tectonic geologic history, historic seismicity, and seismic hazard mapping are consistent with a relatively low seismic hazard at the DUSEL site. Historic earthquake activity within about 188 miles (300 km) of the DUSEL site is low. The five records of felt earthquakes in Lead, South Dakota, since 1928 indicate that only infrequent, low-intensity earthquake shaking has been experienced. From the excavation standpoint, the need for a campus-wide seismic monitoring (earthquakes only) is not necessary.

### 5.3.7.9 Microseismic Monitoring

Microseismic monitoring is commonly used in mining due to the ever-changing nature of the geometry of the mine. Changes in geometry bring about changes in the stress field around the excavations and result in measurable seismic responses. This is used both as an indicator of where and how the stresses are being redistributed to mitigate risk, as well as a system for early warning of potential instabilities. It is also used to control (smooth blasting technique) and refine (overbreak control) blasting patterns.

The current preliminary monitoring plan, focused at this point on the ground monitoring plan of large excavations, will be refined during Final Design effort. This will include design of a network of geophones to be installed in strategic locations, particularly in the vicinity of large excavations, and large and essential intersections and service rooms. The number, specific type, campus-wide locations, and vertical locations on selected levels will be integrated with the infrastructure and science needs. Regardless of a specific pattern and network size, the geophones will be installed prior to excavation of the new DUSEL infrastructure, particularly LCs and LMs. The geophones installed in this manner will serve the purpose of monitoring the construction process, measuring the impact of large excavations on other excavations, future responses to filling and emptying of the LCs, including the cooling effects, and providing a baseline configuration for long-term monitoring and detection of any indicator of excavation instability.

### 5.3.8 Excavation Air Blast Modeling and Mitigation

Atmospheric over pressurization caused by uncontrolled air blast management during excavation process can cause significant damage to existing facilities. Changes in geometry as excavations progress bring about changes in the air pressure response around the excavations and may result in measurable overpressure responses. Modeling is commonly used in excavation design due to the ever-changing nature of the geometry of the excavation and the need to eliminate the occurrence of air blast damage to existing facilities. The modeling will define operational procedures, air door locations, and bulkhead construction to minimize or eliminate the propagation of uncontrolled air blasts. This will include planning with an emphasis on execution of good blast round design, as well as drilling and loading quality control (correct hole spacing/burdening, stemming, timing, and adjusting for geologic conditions). DUSEL will conduct modeling and evaluation of mitigation measures during the Final Design phase, once configuration of the facilities and the excavation sequence are fully defined.

### 5.3.9 Advisory Committees and Advisory Boards

DUSEL currently engages three advisory groups to assist with quality assurance, review, guidance, and recommendations of project geotechnical data collection and analysis, geotechnical and numerical modeling, excavation design and sequencing, and evaluation of infrastructure design. These groups include the Large Cavity Advisory Board (LCAB), Infrastructure Advisory Board (IAB), and Geotechnical Advisory Committee (GAC).

#### Large Cavity Advisory Board and Infrastructure Advisory Board

The LCAB and IAB convene concurrently to provide independent review, recommendations, guidance, and assistance to the DUSEL Project Team regarding geotechnical data collection and analysis, geotechnical and numerical modeling, excavation and ground support design, and evaluation of infrastructure design and sequencing. In particular, the LCAB provides review and recommendations on geotechnical assessment programs, including core, core logs, geological modeling, geotechnical assessment programs, in situ tests, laboratory test results, geotechnical interpretation, geotechnical modeling and interpretations, design of excavation, and ground support. The Board also reviews the proposed geotechnical assessment programs for Final Design and advises the Project Team on the adequacy of the proposed future Mid-Level Laboratory (MLL) and Deep-Level Laboratory (DLL) geotechnical programs, including obtaining addition drill hole core coverage, in situ tests, and laboratory tests to achieve Preliminary Design and, to the degree possible, the Final Design for LM-1, LM-2, LC-1, and ancillary support excavations, including the placement of LMs and the LC and their proximity and placement to other planned excavations.

The LCAB evaluates proposed excavation designs for LM-1, LM-2, LC-1, and ancillary support excavations, including excavation techniques, strategies, cavity monitoring approaches, ground support, and rock stabilization. The Board also advises the Project on any updates to the placement of a potential LC-2. These evaluations include the recent project excavations, ground support, and progress in enlarging the Davis Cavity complex.

**Geotechnical Advisory Committee**

The GAC provided independent review and recommendations, guidance, and assistance to the DUSEL Project team on geotechnical issues. Specifically, these reviews include current geotechnical design features and geotechnical design criteria, and the providing of recommendations on what is needed for the design of the DUSEL Facility to fulfill science needs and long-term stability requirements of excavations. Recommendations to DUSEL assist the geotechnical team in maintaining objectivity with contractors and consultant interactions. Evaluation of the current geotechnical database is conducted and recommendations submitted to DUSEL regarding data gaps and data needed for the completion of the design. The GAC made recommendations on priorities for the current and the future geotechnical plans to develop credible, safe, and cost-effective technical design, and identify and recommend potential geotechnical research opportunities with facility design and construction.

The GAC addressed the specific tasks and issues regarding appropriateness of the MLL (4850L) geotechnical design criteria; establishment of reliable numerical models and inclusion of rock mass discontinuities; geotechnical data gaps and design gaps; site-specific geotechnical investigations needed for Final Design; geological-geotechnical model updates and interpretation of data; long-term geotechnical basis ground monitoring, excavation performance, and further needs; and recommendations regarding the Geotechnical Baseline Report (GBR) for construction and future work (LC-2, LM-3, Long Baseline Neutrino Experiment [LBNE] exploration, 7400L, and other levels). The DUSEL MLL geotechnical work has advanced from an initial exploratory type of geotechnical investigation to design-oriented and more Final Design site-specific investigations and the GAC evolved in its advisory role as well. At the conclusion of Preliminary Design, the responsibilities of the GAC were assumed by the LCAB and the IAB as described above.

## 5.4 Underground Infrastructure Design

This chapter provides an overview of the DUSEL underground facility infrastructure, beginning with a description of the infrastructure scope and the activities required to support construction. Following this description, each infrastructure system is discussed, including a description of current conditions, design requirements, and the Preliminary Design. The information provided is intended to give the reader an overview of the design. Additional details on the design of each scope, as well as how the infrastructure designs interface with each other, can be found in the reference material noted throughout the section. Interfaces between the infrastructure scope and other scope elements of the DUSEL Project are discussed in Chapter 5.1, *Facility Design Overview*.

### 5.4.1 Underground Facility Infrastructure Summary and Overview

In January 2010, a team led by Arup USA with subcontracts to SRK Consulting and G.L Tiley & Associates commenced with the Preliminary Design for Underground Infrastructure (UGI). The design conforms to the scope that was identified in the final iteration of the *Basis of Estimate* (BOE) process as Option B (Appendix 5.K) and is captured in the *Underground Infrastructure Basis of Design (BOD) Report* (Appendix 5.L) as discussed in Chapter 5.1, *Facility Design Overview*. The scope of this work includes all underground infrastructure to support laboratory construction and operation for laboratories on the 4850L. Design for infrastructure to support laboratories on the 7400L was performed separately and completed as a Conceptual Design Report as described in Chapter 5.8, *Deep-Level Laboratory Design at the 7400L (DLL)*.

The baseline scope for underground infrastructure to support laboratory construction on both main campus levels includes the following components (section reference noted parenthetically):

- Life safety systems and Areas of Refuge (Section 5.4.3.1)
- Maintenance shops, utility rooms, storage and containment areas (Section 5.4.3.5)
- Drifts and ramps required for access, egress, and ventilation (Section 5.4.3.6)
- Material handling systems (Section 5.4.3.7)
- Air quality and ventilation systems (Section 5.4.3.8)
- Waste handling systems (Section 5.4.3.9)
- Electrical power distribution systems (Section 5.4.3.10)
- Dewatering systems (Section 5.4.3.11)
- Water inflow management systems (Section 5.4.3.12)
- Chilled water systems (Section 5.4.3.13)
- Plumbing systems (Section 5.4.3.14)
- Cyberinfrastructure controls and monitoring systems (Chapter 5.5)

In addition to these components, infrastructure specific to providing access for each major underground campus is described below.

**4850L**

- Yates Shaft Hoist, Headframe, and shaft infrastructure upgrades (Section 5.4.3.2)
- Ross Shaft Headframe upgrades (Section 5.4.3.2)
  - Ross Shaft Hoist and steel furnishings upgrades are planned to be performed prior to the Major Research Equipment and Facilities Construction (MREFC) funded Project

**7400L**

- #6 Winze upgrades and refurbishment (Section 5.4.3.3)
- #8 Winze (new borehole and hoist for egress) (Section 5.4.3.3)
- New ventilation borehole (Section 5.4.3.8)

Note that while Other Levels and Ramps (OLR) experiment utilities are included in the Work Breakdown Structure (WBS) as part of UGI, discussions for these are contained in Chapter 5.9, *Design and Infrastructure for Other Levels and Ramps*. Designs for OLR ground support, excavation, and utilities were not included in the BOD scope.

**Infrastructure Advisory Board**

An Infrastructure Advisory Board (IAB) was formed during Preliminary Design to provide a third-party review of the infrastructure design. The IAB reviews the plans for near-term inspections, rehabilitation, and deferred-maintenance programs. These reviews include the infrastructure Preliminary Design reports and plans; adequacy of the plans for shaft rehabilitation and accomplishing the deferred maintenance; adequacy of the current designs for major infrastructure, including access, ventilation, fire and life safety; proposed use of the Yates Shaft prior to the start of construction; and adequacy of plans to provide dual egress to the underground during shaft infrastructure rehabilitation and upgrade.

## 5.4.2 Facility Infrastructure Required to Support Construction

Substantial infrastructure is required to enable the construction of the DUSEL Project. Much of this exists from former mining operations. This section describes both the existing infrastructure to be reused or refurbished, and new infrastructure that will be installed to allow for construction activities.

Laboratory construction will occur in three stages: excavation, infrastructure installation, and experiment installation. The infrastructure requirements for each of these stages are considerably different, but in general, excavation requirements exceed other infrastructure requirements and can be used as the base requirements for all construction needs.

### 5.4.2.1 Electrical Infrastructure during Construction

The amount of power supplied to the Facility, and the currently installed surface substation capacity, are sufficient to provide the Facility needs from the present time and into construction.

Limited power is available underground for excavation and construction during the early phases of the Project. Power for early-phase construction activities will be derived from the Ross Shaft dewatering system, which has an excess capacity of approximately 700 KW. The dewatering system will also supply power for early science at the Davis Campus. It is possible that large storm events could result in the interruption of construction activities to provide sufficient power to operate the dewatering system at full capacity (two pumps per station as opposed to one per station during normal operation) while still maintaining power for early science.

The power requirements range from 2.5-3.0 MW during peak excavation and construction phases. Peak demand is dependent on the construction schedule, with peak loads occurring during the Large Cavity (LC-1) excavation. The Yates Shaft rehabilitation will be complete before peak power is required and supplemental power will be provided through the Yates Shaft to a temporary substation on the Yates side of the 4850L Campus. Electrical equipment for excavation and construction include battery charging stations, lighting, pumps, portable ventilation fans, raise-bore machines, and jumbo drills. Compressors and rock haulage equipment will be diesel powered.

The surface Ross Substation has adequate switchgear (2400 volt) and capacity required for the upgrades of the Ross Hoists and waste rock handling system rebuild. Additional equipment will be installed to provide backup power to the Ross Service Cage Hoist motor and provide for future critical loads underground. The additions will be performed using Research and Related Activities (R&RA) funds to complete the Ross Shaft and Hoist refurbishment prior to the commencement of DUSEL underground construction.

Construction of the new Yates Substation on the surface is an important part of DUSEL construction. Coordination with other surface work is essential, as all existing, refurbished, and new facilities on the Yates Campus will be powered from this substation. Excavation, trenching, and duct bank installation will create congestion in the Yates Headframe and Shaft areas. Power for the Yates Hoist motors and drives upgrades will come from the new Yates Substation. Backup generators for the hoists will be installed in conjunction with new substation construction.

Currently, power to the Davis Campus is provided through a cable in the 4850L West Drift between the Ross and Yates Shafts. This cable will be removed during initial excavation activities. A new cable will be installed from the Ross Shaft across the 4100L to a new borehole connected directly to the campus. At the completion of the Project, the Davis Campus will be powered from the new Yates mechanical electrical room (MER). No underground services in the Yates Shaft can be installed until the shaft refurbishment is completed and the conveyances are operational.

### 5.4.2.2 Mechanical Infrastructure during Construction

As the mechanical infrastructure needs during construction are different from those during operations, temporary accommodations will be planned to support construction activities, including water supply, heat for the shafts, power, and compressed air. Some items, such as compressed air, will be provided and maintained by the construction contractor, but this section describes services provided for construction use by the SDSTA.

Drilling, blasting, and rock handling underground requires approximately 5,000 gallons per day of industrial water to cool drill bits and suppress dust. Drinking water will be supplied by the contractors. Sanitary facilities will include chemical toilets or a similar technology that does not require water. Water is currently supplied through piping in both the Ross and Yates Shafts. A 6 in (152 mm) industrial water supply follows the Ross Shaft from the surface to the 4850L. This pipe has a 1000 gpm capacity for construction, and will remain in use after the construction is complete to provide additional capacity for fire control and other Facility uses. A 2 in (51 mm) potable and 4 in (102 mm) industrial water pipe were installed by the SDSTA in the Yates Shaft to support early science. While the pipes installed by the SDSTA may be useful during early construction, the early rehabilitation of the Yates Shaft will limit their usefulness for anything other than shaft rehabilitation. As with the power, a new water service crossing the 4100L from the Ross to a new borehole near the Yates will maintain fire protection for the Davis

Campus during construction. Note that industrial water is untreated water directly from the supply to the city of Lead (primarily surface streams). Potable water is derived from this supply, but the city filters it and adds fluorine and chlorine to inhibit bacterial growth.

The existing dewatering system capacity is adequate to remove all construction water from the underground as well as maintain the pool level.

As discussed in Section 5.4.3.2, the Ross Shaft will be in full operation when construction begins, including skips and crushers. The Yates Shaft will only be able to provide secondary egress during rehabilitation activities; therefore, Ross will be the primary conveyance for both materials and personnel during construction.

Heating the shaft during cold weather is required to protect the shaft infrastructure. As evidenced by the damage in the #5 Shaft (Section 5.4.3.4), ice can form in the shaft during cold weather and destroy critical shaft infrastructure. Therefore, the heating system is not installed to provide human comfort; it is to prevent ice formation in the intake shafts (Ross and Yates) that can damage the infrastructure. The requirements for this heating system are detailed in HDR's *Surface Facilities and Campus Infrastructure* report (Appendix 5.D).

Ventilation requirements to support construction and to support operations are calculated in different ways due to the different activities taking place. During construction, the primary drivers of air volume requirements are the amount of diesel equipment in operation and number of people working. Arup describes details on the design for ventilation during each phase of construction in Section 12.3 of its *UGI Basis of Design Report* (Appendix 5.L).

### 5.4.2.3 4850L Construction Requirements

Infrastructure requirements will change during the phases of construction. The initial task will be to make the waste rock handling system operational and ensure that waste rock can be processed and transported to the Open Cut; a large portion of this work will be completed prior to the beginning of the MREFC-funded Project. As the remainder of the waste handling system is completed, work underground will be limited to Area of Refuge (AoR) development and mobilization activities, with the primary infrastructure requirement focused on scheduling cage availability in the Ross Shaft.

As worker safety is a primary requirement, plans for temporary AoRs, scheduling and sequencing of construction activities to ensure safe Project development, and all work conditions were evaluated during Preliminary Design and will continue in Final Design. At the start of Construction and parallel to making the waste rock handling system operational, an AoR near the Ross Shaft will be developed in an existing excavation, and will later be used as a permanent AoR. A secondary egress route will be provided in the Yates Shaft even during Yates Shaft rehabilitation, as described in Section 5.4.3.2.3.

Once the waste rock handling system is operational, excavation will begin. Several phases of excavation were modeled for ventilation and are discussed in Section 12.3 of Arup's *UGI Basis of Design Report* (Appendix 5.L), Conveyance schedule priority will transition from excavation mobilization to infrastructure installation as spaces are finished and available for utility installation.

Once the Yates Shaft is rehabilitated, it will provide a second means of access for delivery of materials and people to the underground. As both shafts are outfitted with new plumbing and electrical, these services will become available for construction use as well. The large cross section of the Yates

Supercage allows for larger material deliveries to support construction such as mobile equipment, chillers, air handling units, and transformers.

It is anticipated that the 4850L Campus will be available approximately one year before the LC-1 excavation is complete. This will allow experiment installation in the lab modules (LMs) concurrent with LC-1 excavation. No temporary services are planned for experiments, so their installation and operation will depend on the schedule of infrastructure installation. Careful planning to schedule cage time to support science installation concurrent with construction activities will be required.

#### 5.4.2.4 7400L Construction Requirements

The #6 Winze will be rehabilitated by the time the 4850L Campus LMs are constructed. This will include installation of electrical and plumbing services, as well as a temporary ventilation duct and a cooling system. Development of the 7400L will begin after the LM excavations are complete at the 4850L. The Ross skipping system will transport rock from both the LC-1 and the 7400L excavations during 7400L development.

As on the 4850L, the 7400L will be developed in sequenced phases. Since the 7400L footprint is much smaller than the 4850L, there will be limited staging space at this level. It may be necessary to complete excavation, and then complete infrastructure installation, followed by experiment installation. Secondary egress routes will not be available for significant periods of construction, so controlling the total number of people working on this level and installing temporary AoRs will be critical. The contractor will need to provide these temporary AoRs until permanent facilities are developed.

Development of the 7400L is currently planned in two phases (see Chapter 5.8, *Deep-Level Laboratory Design at the 7400L [DLL]*). The first phase will provide a ventilation path, secondary egress, AoR, and drill room. During this phase, a temporary power installation will service the drill room and water for drilling will also be installed. In phase two, the permanent MER will be added along with the lab module (LMD-1). All temporary services will be transferred to the permanent system prior to experiment installation.

### 5.4.3 Preliminary Design

Each infrastructure system included in this discussion is included in the Preliminary Design; Preliminary Design is defined as completing approximately 30% of the total design process. This phase develops the complete flow of each system, defines interfaces between systems, and develops preliminary layouts of each design. Design details, such as precise plumbing diagrams, electrical connection points, and ventilation door design, are necessarily limited at this stage and will be fully developed and detailed during Final Design. Due to the complexity of these systems, it is critical to understand the requirements, interfaces, and overall layouts of the infrastructure systems. The discussion below summarizes the work completed during Preliminary Design on all infrastructure systems. More detailed information on each item can be found in the individual design reports referenced at the end of this section and included in the Appendix to Volume 5 of this Preliminary Design Report.

Much of the existing infrastructure installed throughout the operation of the Homestake Mining Company (HMC) will be utilized as part of this Project. Through the Underground Infrastructure design contract, the Arup/SRK/Tiley team completed assessments of the current conditions of the underground facility infrastructure. The current condition subsections of each portion of the infrastructure summarize the findings of these reports, supplemented by information gathered by SDSTA employees, discussions with former HMC employees, and the DUSEL Design Team during Preliminary Design. Flooded conditions following mine closure had a significant impact on the underground spaces, with the highest water level reached in August 2008 at 4,529 feet below the Ross Shaft collar (top of shaft).

Part of the Preliminary Design process was to establish a list of requirements from both the scientific community and the facility. These requirements provide guidance for what is needed to perform experiments and maintain a functioning facility. This section describes the key requirements for each area of the infrastructure design. For a more detailed discussion on the requirements development process see the Systems Engineering discussion in Volume 9.

### 5.4.3.1 Life Safety

Life safety is a value critical to the design of every major project. As an underground facility, a primary focus for life safety in the DUSEL Project revolves around events that could impact the ability to safely escape or, if escape is not immediately possible, isolate people from events underground. Fires underground are of particular concern, as the ventilation system will both deliver oxygen to a fire, and convey smoke throughout the Facility if not properly controlled. This section describes the design controls included to provide safety during this type of event, as well as other safety controls for normal operations. Volume 6 describes the environmental health and safety program in greater detail and Section 5.1.7.2, *Codes and Standards Approach,* describes the codes used as a basis for Preliminary Design.

#### 5.4.3.1.1 Current Life Safety Systems

The HMC operated under the jurisdiction of the Mine Safety and Health Administration (MSHA). Upon re-entry of the underground, most of the Homestake mine rescue team was reassembled, complete with emergency response equipment. As the DUSEL Facility is not an operating mine, the Occupational Safety and Health Administration (OSHA) is the regulatory agency with jurisdiction, but practically speaking, the emergency response philosophy remains similar to an underground hardrock mine.

Currently, construction personnel are required to complete site-specific safety training. Site-specific safety training addresses actions required in case of an underground emergency and the use of personal protective equipment (PPE). The full set of safety training requirements can be found in the DUSEL Environment, Health, and Safety (EH&S) Manual that is publically available on DocuShare and also included in Appendix 6.B.

Notification of an underground emergency will use the facility-wide ventilation system. Ethyl mercaptan, more commonly known as stench gas, is dumped into the ventilation system at the surface and quickly disperses underground. This is a commonly used and effective method of notification, with the exception of identified areas where the ventilation system does not reach, such as the tramway and 300L where ventilation is directly from the surface. Documented action plans and the tag in/out board manage the locations of people working underground.

There are currently no AoRs underground and evacuation is the understood response to an underground emergency. There are two forms of access to the 4850L—the primary access is the Ross Shaft and the secondary egress is the Yates Shaft until the Ross Shaft rehabilitation begins in advance of construction. At that time, the primary and secondary access/egress paths will be reversed. Primary and secondary access/egress varies on OLR levels. The current number of people underground is administratively limited by the speed and capacity of the hoists.

#### 5.4.3.1.2 Life Safety Requirements

The safety of personnel both constructing and operating the Facility is the highest-level requirement and highest priority in the Facility design. Specific guidelines and codes help define these requirements, combined with the best practices and lessons learned from counterpart facilities to design and provide a safe work environment. Section 3.2 of Arup's *UGI Basis of Design Report* (Appendix 5.L) discusses the challenges presented by the unique nature of this Project. The key drivers for defining life-safety requirements focus on providing safe egress from the underground, or safe refuge until safe egress is possible. Controlling an emergency event is an important component of the life-safety strategy and is accomplished by controlling airflow and sprinkler systems, and providing communication infrastructure throughout the Facility to maximize response time for occupants.

The number of occupants at any given time for each laboratory level will be limited and monitored to ensure that AoRs are sufficient and egress routes are not stressed. Figure 5.4.3.1.2 shows the design occupancy limits for the laboratory. Occupancy limits were developed through discussions with design contractors, potential experiment representatives, and facility operations personnel. Each group provided an estimated maximum expected occupancy at any given time, and these occupancies were then detailed in a chart to determine maximum occupancy by year. This was reconciled with the expected completion dates for AoRs to determine maximum capacity requirements. Temporary AoRs will be required for contractors during the Construction Phase to accommodate increased occupancy rates.

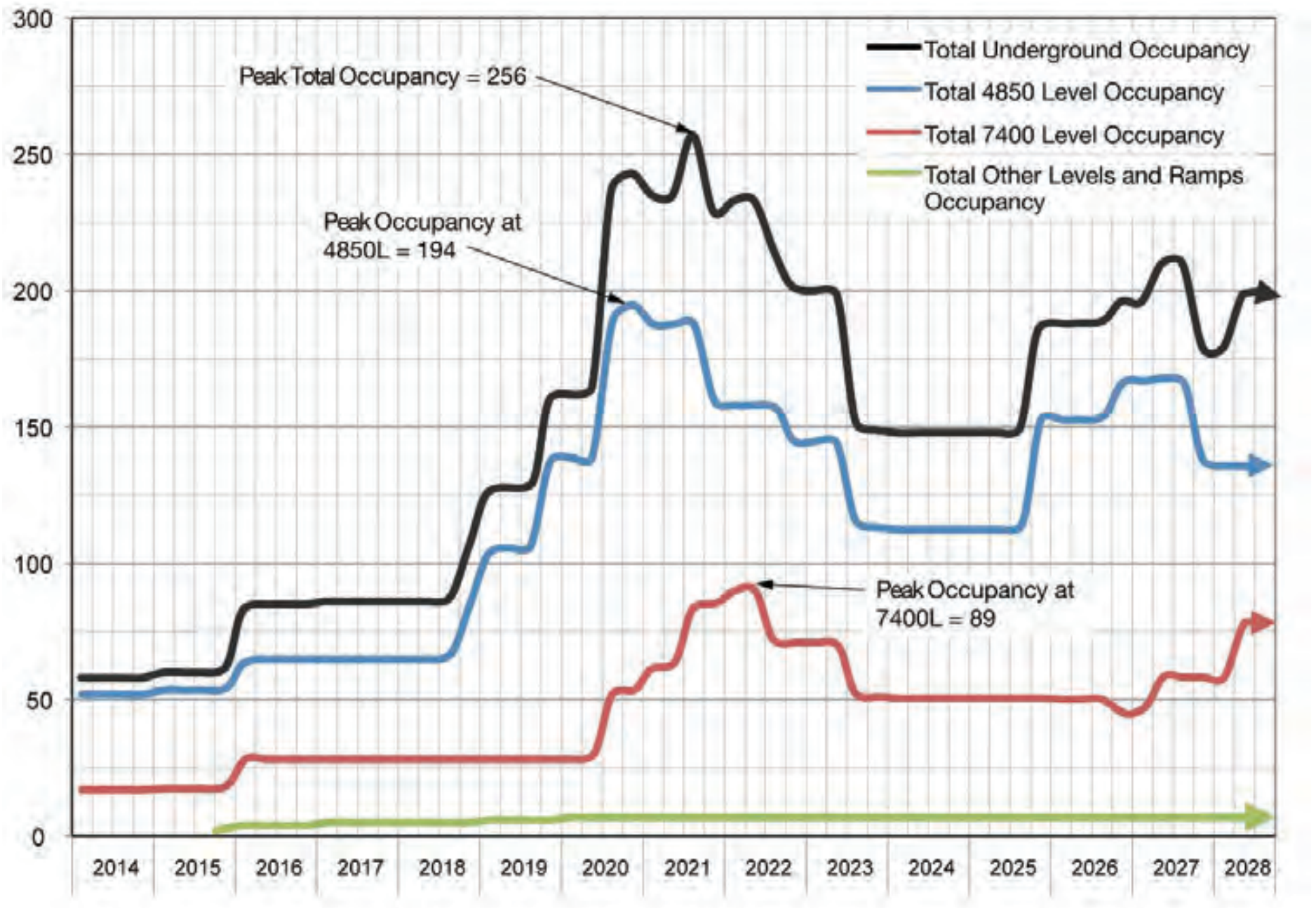

**Figure 5.4.3.1.2** Occupancy design chart. [DKA]

**Authority Having Jurisdiction (AHJ)**

Identifying the AHJ is critical to ensure that the design meets all building and regulatory requirements. Section 5.1.7.2, *Codes and Standards Approach*, and Volume 6, *Integrated Environment, Health, and*

*Safety Management*, further discuss the multiple agencies with applicable codes and requirements. A third-party analysis of the design was commissioned with Hughes Associates, Inc. to help ensure that the design appropriately addresses the applicable codes and standards. The report from this review is discussed briefly in Chapter 6.7, *Independent External EH&S Reviews*, and is included in Appendix 6.E, *Third Party Review of the Fire and Life Safety (FLS) Design for DUSEL*. As most of the property is located within the city of Lead, the city's building department will be the local AHJ, but the Facility will also conform to the other applicable codes and requirements for safe operation of an occupied underground facility.

#### 5.4.3.1.3 Life Safety Preliminary Design

Inherent to an underground facility are a number of fire safety-related challenges that are addressed in the design of the DUSEL Facility. Typical building codes do not adequately address the specific needs and hazards present in this type of facility. The design process requires careful examination of all code systems and best practices to determine the full complement of design elements required and ensure the safety of both the Facility and its occupants.

A number of hazards associated with the operation of an underground laboratory must be considered, and mitigated when possible, in the design process:

- Fire hazards due to materials in the Facility—including insulation and other flammable materials, chemical use, and the presence of flammable/combustible liquids required for experiment operations
- Accidental release of gases that are used for scientific experiments, and which displace oxygen
- Explosion hazards from pressurized vessels, explosive chemicals, and cryogen expansion
- Release of toxic and/or other irritant chemicals

Combustible materials must be limited and restricted in terms of use, type, and quantity allowed by code, good practice, and management and operational procedures. Items that pose risk to the Facility and its occupants, such as cryogens and other hazardous materials, will be tracked through a Radio Frequency Identification (RFID) system and by DUSEL staff to verify the quantity of these materials present at all times underground.

Compartmentalization through cross drift and fire separation doors provides a means to isolate an event that may compromise underground occupant safety. These fire doors will have a minimum two-hour fire rating. Placement of the doors is shown in Figure 5.4.3.1.3. These doors can be operated both manually and remotely from the central command and control center on the surface (the command and control center is discussed in Chapter 5.5, Cyberinfrastructure Systems Design).

Automatic fire suppression systems will be provided throughout all areas of the main campuses, including LMs, LC-1, AoRs, drifts, shafts and winzes, and MERs, in accordance with the IBC, NFPA 45, NFPA 55, and NFPA 13. Specialized science equipment installed by collaborations may require dedicated suppression systems, which will be coordinated with the DUSEL Facility during experiment design, but provided by and at the expense of the collaboration. Standpipes, also known as hose stations, will also be provided throughout the Facility, including OLR and experimental areas to provide coverage for all occupied areas.

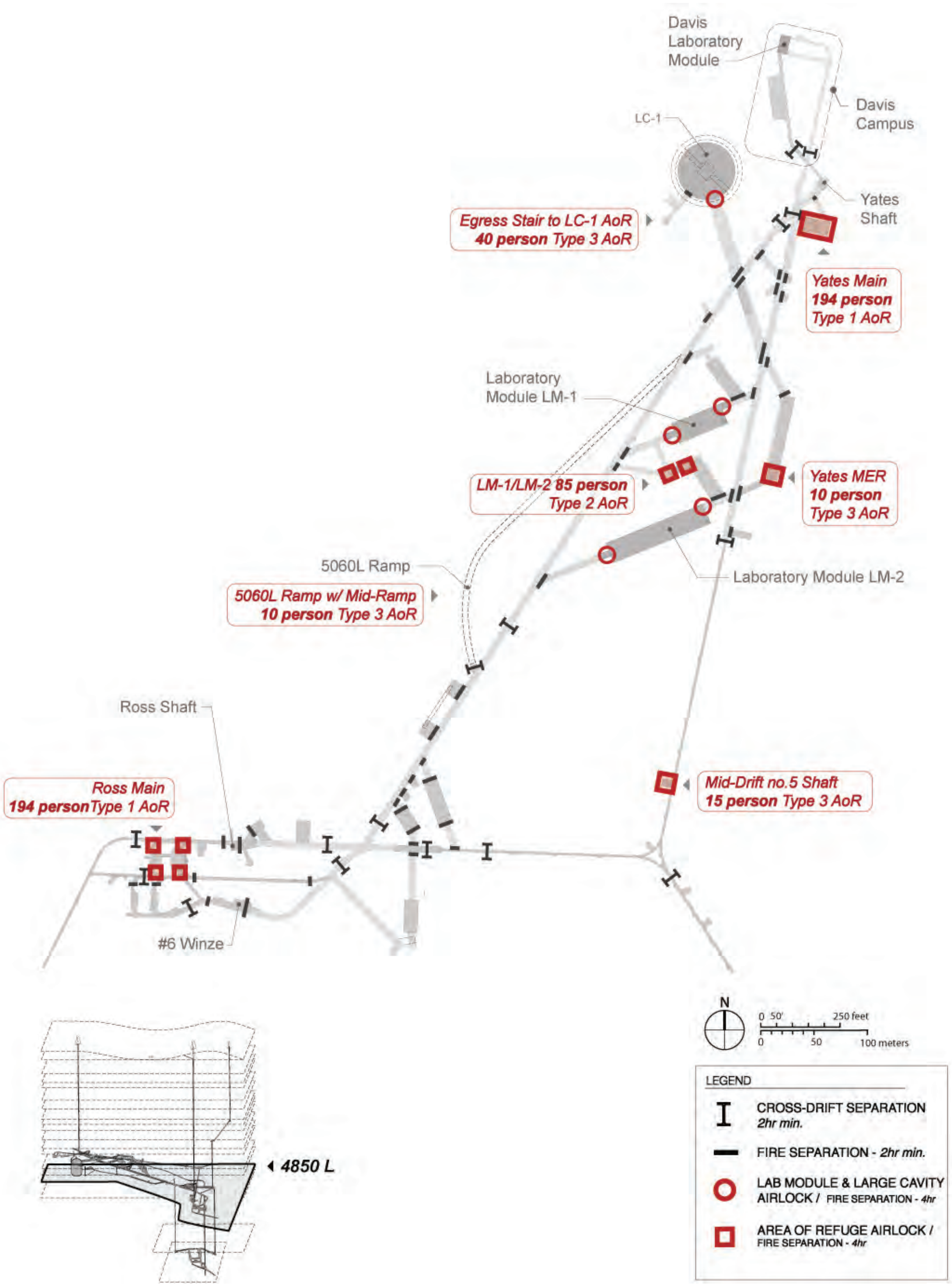

**Figure 5.4.3.1.3** AoRs and compartmentalization. [DKA]

**Areas of Refuge (AoRs)**

The Fire/Life Safety (FLS) evacuation strategy for the Facility is an operational and training precedent. Emergency response scenarios will vary depending on whether occupants are instructed to evacuate the Facility or find refuge in an AoR. Mass notification through a public address system on the main campus levels will provide clear direction. The OLR notification and evacuation procedures include the use of odorants in the ventilation system described in the current conditions in Section 5.4.3.1.1, and radio communications to notify laboratory personnel of an emergency event. All personnel going to OLR will need authorization and adequate safety training, and must carry requisite PPE at all times.

Travel distances dictate the frequency of AoRs in the main laboratory campuses and there are three classifications for AoRs. The descriptions below are excerpted from the Arup *UGI Basis of Design Report* (Appendix 5.L).

- **Type 1: Shaft AoR -** located at the Yates Shaft and the Ross Shaft/No. 6 Winze.

  Yates and Ross/No 6 Winze are each sized to accommodate 194 occupants, based upon the predicted peak occupancy load at the 4850L.

  Both Type 1 AoRs provide an AoR for occupants and a Satellite Command and Control Center for emergency response. Type 1 AoRs are accessed from the main drifts and have a secondary connection directly to the hoist lobbies, since two means of egress are required per code (i.e. the IBC requires two means of egress when occupant loads exceed 49 persons) and recommended by good life-safety practice. In addition, the secondary egress allows occupants to leave the AoR via an air lock, pass through a protected intermediate drift, and reach the hoist without the risk of traversing a potentially hazardous area.

  Each Type 1 AoR will have dedicated electrical, emergency electrical and mechanical service rooms and plumbing connections/water service for toilet rooms (including accessible toilets).
- **Type 2: Common Area AoR -** A Type 2 AoR is located between LM-1 and LM-2 and will be accessed from the West drift and from both LMs. The layout allows occupants to access the AoR from their work area without passing through the main access drifts, as well as exit from the AoRs into the West drift without having to pass back through the LM where the hazard may be. All access points to the AoR will have air locks. The LM AoR is sized to accommodate 85 occupants, which includes 66 Science Occupants (20 in LM-1 and 46 in LM-2) and the remaining for Underground Operations occupants.
- **Type 3: Remote AoR -** provided along drifts, remote areas where travel distances to other AoRs are extended, the Yates main ancillary MER spaces, and adjacent to the emergency egress stair/borehole at the 5060L.

  Type 3 AoRs are exclusively for emergency use, and may also serve as secondary staging areas for emergency response to events remote from the shafts. Type 3 AoRs will have a single means of entry/egress, provided with an airlock. The AoRs located mid drift will be sized to accommodate a transient population that may be within the drift including maintenance staff i.e. 10 -15 persons, plus storage and systems support areas.

### 5.4.3.2 Ross and Yates Shafts

The Ross and Yates Shafts provide the only access from the surface to the underground, and are therefore critical to the function of the Facility. Both shafts provide service from the surface to the 4850L, though not every intermediate level is serviced from both shafts. The shafts also provide a path for all utilities from the surface to the underground.

#### 5.4.3.2.1 Current Conditions of the Ross and Yates Shafts and Hoists

The Ross and Yates Shafts were both installed in the 1930s and have operated since installation. These shafts, along with their furnishings, hoists, and cages, were well maintained during mining operations, but have experienced some deterioration as described in this section.

**Ross Shaft**

**Shaft Furnishings**

The Ross Shaft is rectangular in shape—14 ft 0 in (4.27 m) by 19 ft 3 in (5.87 m), measured to the outside of the set steel. The shaft collar is at elevation 5,354.88 ft (1,632.17 m) and the 5000L is the bottom at elevation 277.70 ft (84.64 m) above sea level. Service is provided to 28 levels and three skip loading pockets. The shaft is divided into seven compartments: cage, counterweight, north skip, south skip, pipe, utility, and ladder way. Sets are made from various lengths of 6-inch structural steel wide-flange beams positioned to maintain compartment spaces in the horizontal plane. Sets are vertically spaced on 6 ft (1.83 m) centers throughout most of the shaft. Spacing at stations is 7 ft (2.13 m) and there are short correction sets to make up for uneven elevations between stations. Sets are connected to one another using studdles and suspended from bearing beams located at station sets (typically) spaced vertically each 100 ft (30.5 m) to 150 ft (45.7 m). The bearing beams carry the weight of the set steel plus any dynamic loads produced by the conveyances in the shaft. Sets are secured to the rock wall using wooden wedges and blocking. Two wood guides are positioned in each compartment carrying a conveyance. These keep the conveyance travelling within tight tolerances in the compartment. Blocking holds the position of the set steel stationary to maintain guide alignment. Figure 5.4.3.2.1-1 is a plan view of a typical shaft set in the Ross Shaft.

The shaft was in operation until the mine closed in 2003. Deterioration through corrosion and wear on the shaft steel, including studdles, sets, and bearing beams, is evident. Detailed site investigations were conducted by Arup through its subcontractor, Tiley. The results of their investigations are included in Section 3.4 of the Arup *Preliminary Infrastructure Assessment Report* (Appendix 5.M). Based on their visual assessment, the findings indicate that as much as 50% of the steel furnishings will need to be replaced to enable full operation of the shaft to be restored. Tables 5.4.3.2.1-1 and 5.4.3.2.1-2, respectively, represent a guide to Tiley’s initial assessment and a summary assessment of the shaft furnishings.

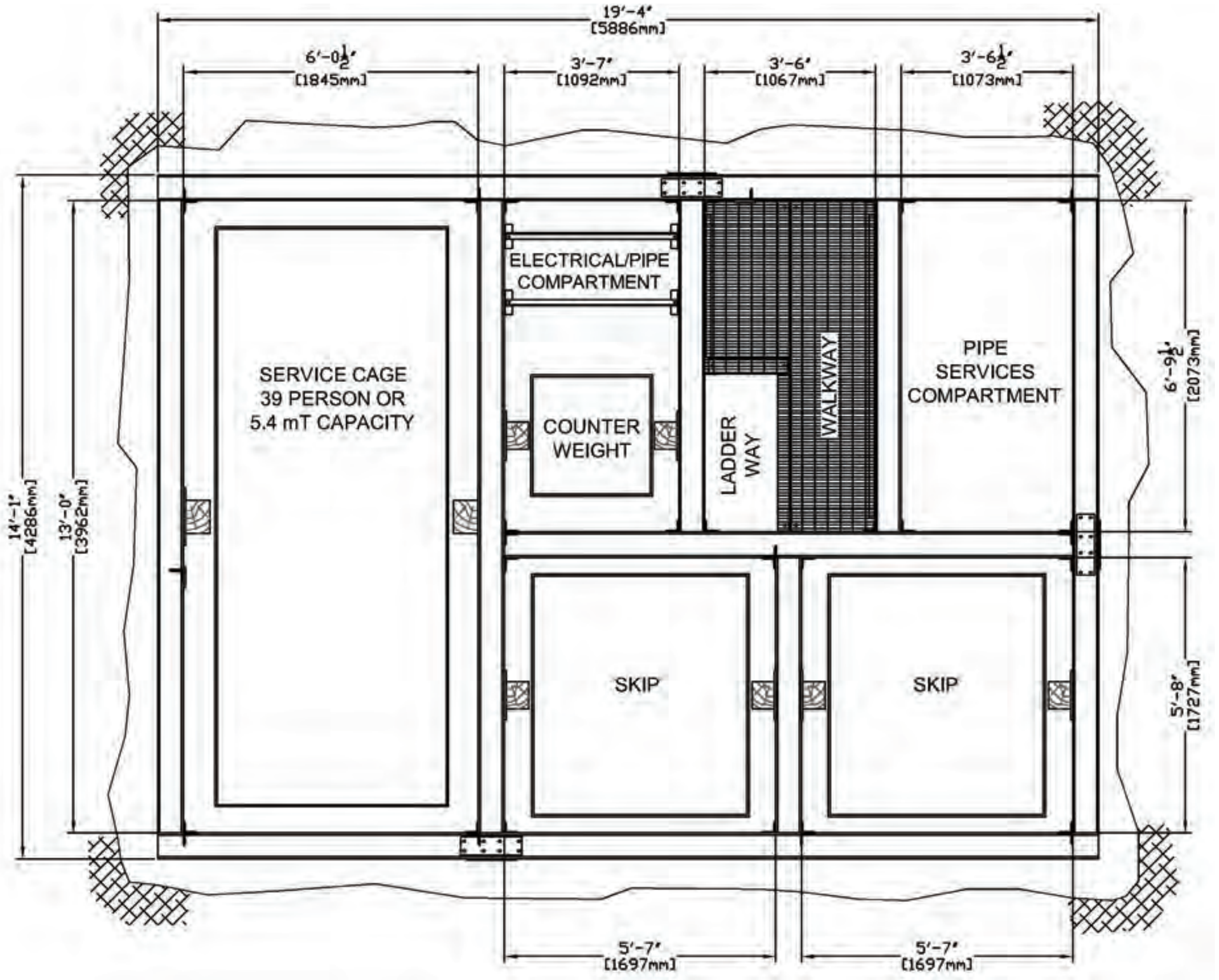

**Figure 5.4.3.2.1-1** Ross Shaft, typical shaft set. [SRK]

| Risk Category | Risk Description | Recommended Action Required |
|---|---|---|
| **R1** | ACCEPTABLE CONDITION - no action required for 2 years unless conditions change | No action required at this time |
| **R2** | FAIR CONDITION - Primary structural element is sound but may have minor section loss, cracking, decay or water damage | Review conditions at next scheduled shaft inspection |
| **R3** | SERIOUS CONDITION - Loss of section, deterioration, spalling, decay, infestation or scour have seriously affected primary structural components. Local failures are possible | Fix within 6 months |
| **R4** | IMMINENT FAILURE CONDITION - Major deterioration or section loss present in critical structural components, or obvious vertical or horizontal movement affecting structure stability. Hoisting duty loading should be discontinued until corrective action taken | Fix Immediately |
| **R5** | FAILED CONDITION - Out of service | Fix Immediately |

**Table 5.4.3.2.1-1** Risk assessment rating. [G.L. Tiley and Associates]

| | Set Steel | Cage guides | CWT. Guides | North Skip Guides | South Skip Guides |
|---|---|---|---|---|---|
| | % | % | % | % | % |
| R1 | 6 | 36 | 37 | 1 | 3 |
| R2 | 18 | 57 | 48 | 82 | 63 |
| R3 | 50 | 6 | 13 | 14 | 33 |
| R4 | 25 | 1 | 2 | 3 | 1 |
| R5 | 1 | 0 | 0 | 0 | 0 |

**Table 5.4.3.2.1-2** Ross Shaft furnishings visual assessment. [G.L. Tiley and Associates]

HMC initiated a shaft rehabilitation program in 1989 that included the replacement of studdles as well as various pieces of set steel. The rehabilitation efforts were continued until the mine shut down in 2003, and were restarted by Dynatec during SDSTA re-entry program in 2006. Approximately 7% of the total shaft steel has been replaced by either Dynatec or the SDSTA operations staff. The primary goal of this program was to establish safe access for dewatering.

### Ross Hoisting System—Overview

The production and service hoists at the Ross Shaft are located on the surface in a dedicated hoistroom west of the shaft. The service hoist operates the service cage and the production hoist operates the production skips. Both the production and service hoists are bicylindrical, conical-shaped double drums. Figure 5.4.3.2.1-2 is a photograph of these two hoists. As part of the Arup Site Investigation contract, Tiley conducted detailed inspections of both the electrical and mechanical components of each hoist. Their findings are included in Section 2.1 of the Arup *Preliminary Infrastructure Assessment Report* (Appendix 5.M).

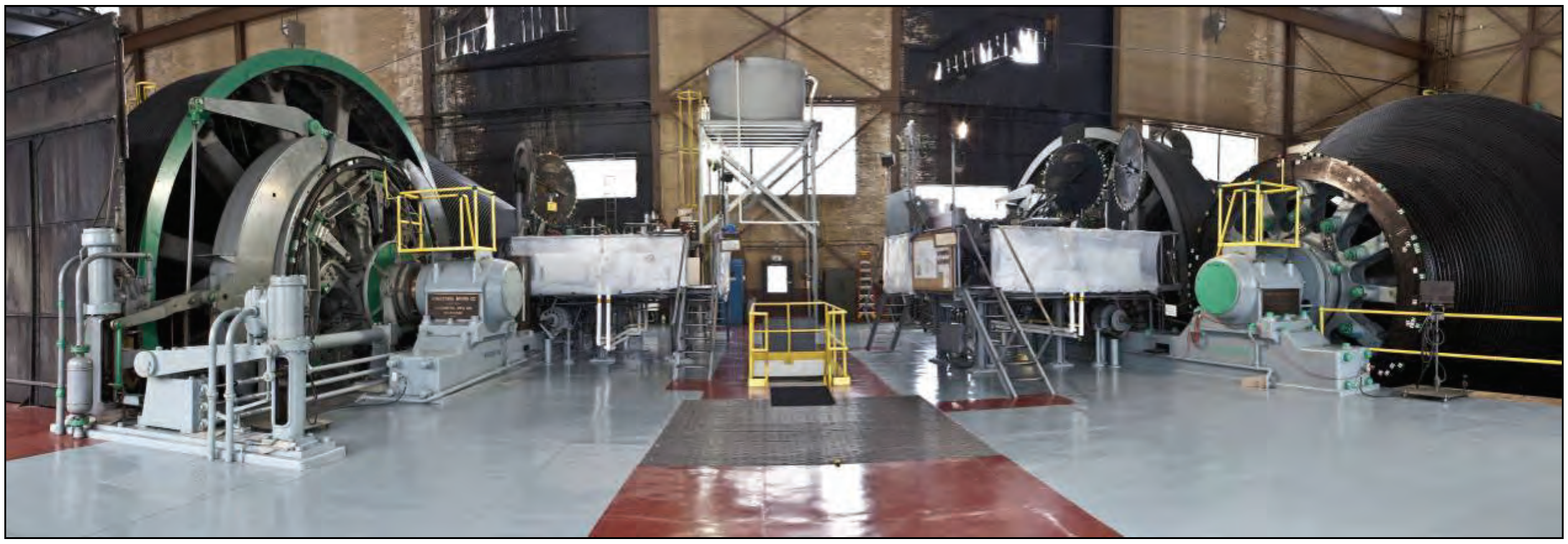

**Figure 5.4.3.2.1-2** Ross Hoists. [Matt Kapust, DUSEL]

### Ross Hoisting System—Electrical

Megger tests were performed by Tiley on both the production and service hoist motors and generators. In general, the overall condition of the Ross Hoist motors and generators was determined to range between good and poor. Armature windings were found to be in the worst condition, primarily due to significant buildups of carbon and dirt. Field windings were found to be in good condition. Tables 5.4.3.2.1-3 and 5.4.3.2.1-4 summarize the conditions of the Ross service and production hoist motors based on the Megger Tests:

| | Motor | Motor | Gen | DC Exciter | AC Induct Motor |
|---|---|---|---|---|---|
| **Armature Winding (MΩ)** | **Result** | 0.10 | 0.10 | 0.44 | 5.40 |
| | **Min. Criteria** | 1.00 | 1.00 | 1.00 | 2.00 |
| **Field Winding (MΩ)** | **Result** | 277.0 | 0.24 | >550 | >2200 |
| | **Min. Criteria** | 1.00 | 1.00 | 1.00 | 1.00 |
| **Winding Condition** | | Poor | Poor | Poor | Poor |
| **Brush Clearance** | | Poor | Good | Good | Good |
| **Commutator Condition** | | Good | Poor | Good | Good |

**Table 5.4.3.2.1-3** Ross service hoist test results summary. [G.L. Tiley and Associates]

| Motor | | Motor 1 | Motor 2 | Gen 1 | Gen 2 | DC Exciter | AC Induct Motor |
|---|---|---|---|---|---|---|---|
| Armature Winding (MΩ) | Result | 0.04 | 0.70 | 0.10 | 0.02 | 0.92 | 5.30 |
| | Min. Criteria | 1.00 | 1.00 | 1.00 | 1.00 | 1.00 | 2.00 |
| Field Winding (MΩ) | Result | 526.00 | 276.00 | 0.07 | 0.10 | 33.10 | 644.00 |
| | Min. Criteria | 1.00 | 1.00 | 1.00 | 1.00 | 1.00 | 1.00 |
| Winding Condition | | Poor | Good | Poor | Poor | Poor | Poor |
| Brush Clearance | | Poor | Good | Poor | Poor | Good | Good |
| Commutator Condition | | Good | Good | Poor | Good | Good | Good |

**Table 5.4.3.2.1-4** Ross production hoist test results summary. [G.L. Tiley and Associates]

Based upon the motor evaluations, the SDSTA Operations team has undertaken a comprehensive motor maintenance program, including motor cleaning with dry ice. Further details regarding the electrical conditions of the Ross Shaft Hoists is included in Sections 2.1.1.2 and 2.1.2.2 of the Arup *Preliminary Site Assessment Report* (Appendix 5.M).

### Ross Hoisting System—Mechanical

Dynamic brake tests by Tiley show that the emergency braking systems are adequate for reduced speeds and loads, but not acceptable for the normal loads or the higher rate of speed required during construction and laboratory operation. The main drum bearings are a Babbitt type and wear has tightened the side clearances. Inspections revealed that both production drums have several cracks in the shells. These have been documented and are being monitored for further damage. The drum tension rods' torques were tested and adjusted to the manufacturers torque specifications. No new or crack propagation has been seen since this work was performed. Further details regarding the mechanical conditions of the Ross Shaft Hoists are included in Sections 2.1.1.1 and 2.1.2.1 of the Arup *Preliminary Site Assessment Report* (Appendix 5.M).

### Ross Hoisting System—Headframe

The general condition of the Ross Headframe steel was determined to be in an acceptable to good condition with no reported major structural deficiencies. Some structural strengthening will be included in the future design to meet MSHA Standard 57.19035. This will include modifications to allow installation of a detaching hook mechanism, eliminating the need to provide structural strengthening to resist the entire load of the hoist in the event that the cage is pulled to the top of the headframe.

### Yates Shaft

### Shaft Furnishings

The Yates Shaft is rectangular in shape—15 ft 0 in (4.572 m) by 27 ft 8 in (8.433 m) measured to the outside of the set timbers. There are two cage compartments and two skip compartments as shown in Figure 5.4.3.2.1-3. In addition to the cage and skip compartments, there are two other compartments in which shaft services are located. The shaft collar is at 5,310.00 ft (1,618.49 m) elevation and the 4850L is the bottom level at elevation 376.46 ft (114.75 m) above sea level. Service is provided to 18 levels plus two skip-loading pockets. Sets are made up of various length and size timbers located to maintain compartment spaces. Sets are vertically spaced on 6 ft (1.83 m) centers throughout most of the shaft, with stations and correction sets as the exception. Timber bearing beams are spaced intermittently throughout the shaft and carry the weight of the timber sets from above through 10 timber posts positioned around the perimeter of the shaft at each set. Wood guides are also positioned in each compartment, carrying a conveyance to keep the conveyance travelling within tight tolerances in the compartment. Wood blocking holds the position of the sets stationary to maintain guide alignment.

The Yates Shaft is timbered except for a fully concrete-lined portion from the collar to the 300L. Recent repairs include full set replacement from the concrete portion to the 800L and additional set repair below this level where deemed critical. A summary of the overall conditions of the shaft sets as determined by Tiley is presented in Table 5.4.3.2.1-5.

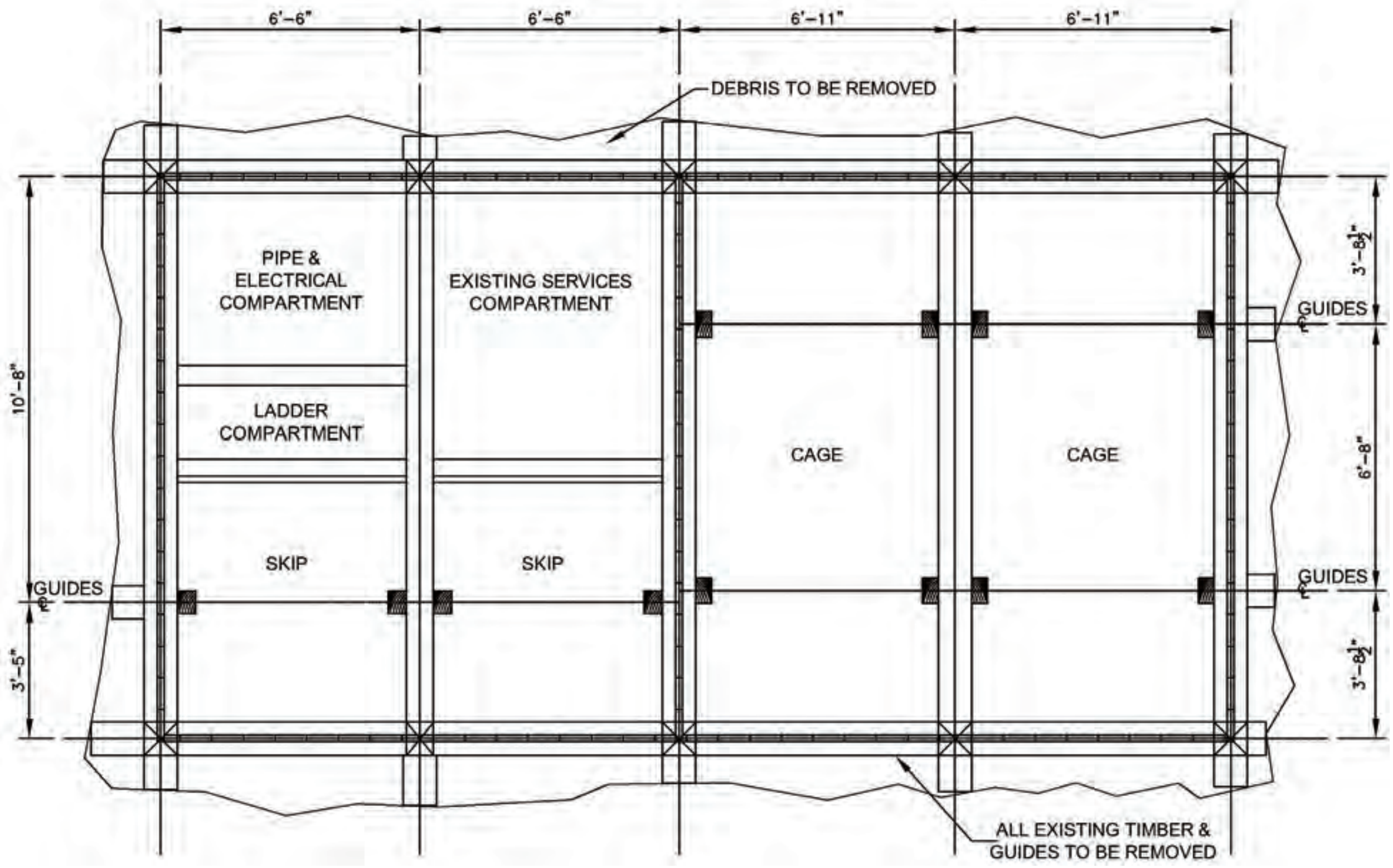

**Figure 5.4.3.2.1-3** Plan view of typical shaft set in the Yates Shaft. [Adapted from SRK]

| | Set Wall Plates & Dividers | North Cage Guides | South Cage Guides | Posts | North Skip Guides | South Skip Guides |
|---|---|---|---|---|---|---|
| | % | % | % | % | % | % |
| **R1** | 5 | 13 | 9 | 31 | 0 | 1 |
| **R2** | 57 | 43 | 55 | 65 | 67 | 68 |
| **R3** | 33 | 38 | 34 | 2 | 29 | 28 |
| **R4** | 4 | 6 | 2 | 0 | 3 | 3 |
| **R5** | 1 | 0 | 0 | 0 | 1 | 0 |

**Table 5.4.3.2.1-5** Yates Shaft set visual inspection results. [G.L Tiley and Associates]

Finite Element Analysis (FEA) modeling by Tiley (Appendix 5.M) showed that a dogging load produced by the cage would require vertical joint reinforcement, guide connection modifications, and additional new bearing beam installations. A dogging load occurs when emergency stop devices, called dogs, dig into the guides to stop the cage if the wire rope loses tension. The east and west wall plates are divided into two pieces, making the removal of a timber divider to make room for the Supercage structurally unsecure. Based on these factors, the support system in the Yates will only be used until it can be replaced as part of the MREFC-funded Project.

### Yates Hoisting System—Hoists

Similar to the Ross Shaft, there is both a production and service hoist at the Yates Shaft. The configuration of the hoists for the Yates Shaft is nearly identical to that of the Ross, with the only difference that the rope size for the production and service hoist are the same at the Yates. The Yates Shaft hoists are located on the surface in a dedicated hoistroom east of the shaft. As part of the Arup Site Investigation contract, Tiley conducted detailed inspections of both the electrical and mechanical components of each hoist. Their findings are included in Section 2.2 of the Arup *Preliminary Infrastructure Assessment Report* (Appendix 5.M).

### Yates Hoisting System—Electrical

Megger tests were conducted by Tiley on both the production and service hoist motors and generators. In general, the overall condition of the Yates Hoist motors and generators was determined to range between good and poor. Both the armature and field windings for the service hoist were found to be in poor condition. Only the armature windings for the production hoist were found to be in poor condition. In all cases, the primary cause for poor results was significant buildup of carbon and dirt. Table 5.4.3.2.1-6 summarizes the conditions of the Yates Service Hoist motors:

| | Motor | Motor 1 | Motor 2 | Gen 1 | Gen 2 | DC Exciter | AC Induct Motor |
|---|---|---|---|---|---|---|---|
| Armature Winding (MΩ) | Result | 0.00 | 0.10 | 0.10 | 0.30 | 76.00 | 1.60 |
| | Min. Criteria | 1.00 | 1.00 | 1.00 | 1.00 | 1.00 | 2.00 |
| Field Winding (MΩ) | Result | 11.40 | 0.00 | 0.31 | 0.17 | >550 | 135.00 |
| | Min. Criteria | 1.00 | 1.00 | 1.00 | 1.00 | 1.00 | 1.00 |
| Winding Condition | | Poor | Poor | Poor | Poor | Good | Poor |
| Brush Clearance | | Poor | Poor | Good | Good | Good | Good |
| Commutator Condition | | Good | Good | Good | Good | Poor | Poor |

**Table 5.4.3.2.1-6** Yates Service Hoist test results summary. [G.L. Tiley and Associates]

| | Motor | Motor 1 | Motor 2 | Gen 1 | Gen 2 | DC Exciter | AC Induct Motor |
|---|---|---|---|---|---|---|---|
| Armature Winding (MΩ) | Result | 0.10 | 0.10 | 0.20 | 0.70 | 536.00 | 1.60 |
| | Min. Criteria | 1.00 | 1.00 | 1.00 | 1.00 | 1.00 | 2.00 |
| Field Winding (MΩ) | Result | 124.00 | >550 | 10&20 | 17&70 | >550 | 5.30 |
| | Min. Criteria | 1.00 | 1.00 | 1.00 | 1.00 | 1.00 | 1.00 |
| Winding Condition | | Poor | Poor | Poor | Poor | Poor | Poor |
| Brush Clearance | | Poor | Poor | Good | Good | Good | Good |
| Commutator Condition | | Good | Good | Good | Good | Poor | Poor |

**Table 5.4.3.2.1-7** Yates Production Hoist test results summary. [G.L. Tiley and Associates]

Similar to the Ross Shaft Hoists, the SDSTA Operations team has undertaken a comprehensive motor maintenance program including motor cleaning with dry ice. Further details regarding the electrical conditions of the Yates Shaft Hoists is included in Section 2.2.1.2 and Section 2.2.2.2 of the Arup *Preliminary Site Assessment Report* (Appendix 5.M).

**Yates Hoisting System—Mechanical**

Inspections by Tiley revealed that both the Yates Production and Yates Service Hoists have no cracks in the drum shells. The drum tension rods were found to be slightly below the original equipment manufacturer torque specifications and the operations group from the SDSTA has subsequently adjusted the torque to the correct values. A number of bearing clearance measurements indicate wear requiring maintenance. Braking tests performed indicate all brakes operating according to specification.

The Yates Service Hoist is planned to be used for the Supercage in the future, while the Production Hoist will be removed and can provide spare parts for both the Ross and Yates.

Further details regarding the condition of the Yates Hoists' mechanical condition can be found in Section 2.2.2.1 and Section 2.2.1.1 of the Arup *Preliminary Site Assessment Report* (Appendix 5.M).

**Yates Hoisting System—Headframe**

As with the Ross, the Yates Headframe is in acceptable to good condition, with no records of structural deficiencies. Also similar to the Ross, the structure will be modified during the MREFC-funded Project to allow for the installation of a detaching hook mechanism.

#### 5.4.3.2.2 Ross and Yates Shafts and Hoists Requirements

The Ross and the Yates Shaft hoisting facilities will be the two means to convey both materials and people in and out of the underground facilities. The capacity of the two hoists, developed using science requirements, has partially defined the size of access drifts at the 4850L and has guided the design of equipment to be transported on these levels.

The size requirement for the Yates Shaft is constrained by the cross-sectional area of the existing shaft, as described in Section 5.4.3.2.1. This shaft will not be used for waste rock removal, which allows the design of the shaft to be reconfigured to combine multiple compartments and provide one much larger conveyance (Supercage) as well as a second conveyance, both of which support the scientific program.

The capacity of the main hoist for the Supercage is limited by the capacity of the existing system, and the primary limiting factor is the wire rope strength. Cargo capacity is further limited by the weight of the rope and conveyance, discussed further in the next section.

Design for future auxiliary personnel conveyance in the Yates Shaft is included in the design documentation but is not included in the facility baseline cost estimate. This conveyance is designed for personnel only, and therefore the requirements are defined by the cross-sectional area of the cage. This establishes the number of people that can fit within the cage, which in turn establishes the weight capacity needed. All requirements for this conveyance are code based.

The Yates Shaft will be the primary means of entrance and egress for the scientific community. As such, the design will consider primarily safety and then the functionality for transporting sensitive scientific equipment. The transport of the scientific equipment requires a very well-aligned shaft to reduce cross-shaft acceleration. This rehabilitation effort will be completed with R&RA funding.

The Ross Shaft will not be significantly modified from the existing configuration. The requirements for this shaft are safety, performance, and code driven and defined by the existing configuration. This shaft will be used for construction and routine facility maintenance, OLR access, and secondary egress path for the finished underground campuses. This designation minimizes the requirements for both aesthetics and cross-shaft acceleration. This design and rehabilitation effort is to be completed with R&RA funding.

The key specifications of each of these shafts and hoists can be seen in Table 5.4.3.2.2.

| Specifications | | Yates Shaft | | Ross Shaft | |
|---|---|---|---|---|---|
| | | Service Hoist Conical DD | Auxiliary Hoist Blair | Service Hoist Conical DD | Production Hoist Conical DD |
| Production required | tons/day | N/A | N/A | N/A | 3,300 |
| Payload | Mass (tons) | 20 | 5 | 6 | 11 |
| | Personnel | 90 | 7 | 60 | N/A |
| Rope mass | Mass (lbs) | 36,000 | 10,403 | 21,785 | 31,415 |
| Conveyance mass | Mass (lbs) | 30,000 | 8,000 | 9,000 | 16,500 |
| Number of ropes | | 1 | 2 | 1 | 1 |
| Rope size | inch dia. | 1.875 | 0.750 | 1.625 | 1.875 |
| Rope strength | lbs | 435,000 | 57,600 | 258,000 | 348,000 |
| Rope FoS | | 4.2 | 4.1 | 6.0 | 4.98 |
| Cage inside dimensions per deck | No. of Decks | 1 | 1 | 2 | N/A |
| | height (ft) | 11'-6" | 8' | 7'-0" | N/A |
| | width (ft) | 11'-4" | 3'-6 3/4" | 4'-8" | N/A |
| | Length (ft) | 12' | 5'-8½" | 12'-4½" | N/A |
| Guides | | wood | steel | wood | wood |
| Shaft | Length (ft) | 5,000 | 5,000 | 5,000 | 5,000 |
| | From | surface | surface | surface | surface |
| | To | 4850L | 4850L | 5000L | 5000L |
| Set spacing (ft) | | 20 | 20 | 6 | 6 |
| Hoisting speed | ft/min | 1,500 | 1,600 | 1,500 | 1,650 |
| Hoist Power Rated | HP | 2,500 | 1,250 | 1,500 | 2,400 |
| Motor Speed | rpm | 360 | 78 | 340 | 375 |
| Skip cycle time (one-way) | mins | N/A | N/A | N/A | 3.60 |
| Cage travel time (one-way) | mins | 4.50 | 4.00 | 4.58 | N/A |
| Cage load/unload time | mins | 5.00 | 1.50 | 7.00 | N/A |
| Cage total time (one-way) | mins | 9.50 | 5.50 | 11.58 | N/A |
| Availibility (after all planned maintenance) | hours/day | 18 | 18 | 18 | 18 |
| Production capacity | tons/day | N/A | N/A | N/A | 3,300 |
| Slinging capacity at 150fpm | tons | 16 | | 4 | |

**Table 5.4.3.2.2** Ross and Yates Shaft specifications (auxiliary hoist not included in MREFC-funded scope). [G.L Tiley and Associates]

### 5.4.3.2.3 Ross and Yates Shafts and Hoists Preliminary Design

Inspections and tests in the Ross Shaft discussed above indicate that shaft repairs are necessary to maintain the hoist capacities required to support both construction and facility operation. The rehabilitation program will repair the Ross Shaft for safe and adequate operation through the Design Development and Construction phases. Rehabilitation work is anticipated to have a useful life of approximately 10 years. This work, with the exception of the headframe upgrades and new utility installations, will be completed by the SDSTA using R&RA funding prior to the DUSEL MREFC-funded Project construction, and therefore a detailed discussion on the rehabilitation work is not included in this report. The Yates Shaft will also be funded under R&RA, but the work will be done within the Project schedule using the contract management group. The work for the Yates Shaft is described in detail in this section.

**Ross Shaft**

The Ross Headframe repair is included in the MREFC funding profile. Repairs are planned to begin early in 2014 and will include adding reinforcement to the sheave decks. The Ross Cage will be equipped with a detachable hook arrangement to substantially decrease the loads on the deck and back legs in the event of a runaway cage. The headframe will also be equipped with a catchment system to keep the conveyance from falling if the rope were ever to detach. Plans include provisions for hoist emergency backup power in the event of power line loss.

**Yates Shaft**

The timber in the Yates Shaft, even if substantial repairs to the current conditions were made, presents a fire risk and has high maintenance requirements. The re-equip options studied during Preliminary Design included a completely concrete-lined shaft compared with installing new steel sets attached to concrete rings spaced on 20 ft (6.1 m) intervals vertically with shotcrete applied between rings. Although providing another degree of reduced maintenance, the fully concrete-lined shaft was not chosen due to cost. The proposed ring arrangement accommodates the new Supercage compartment and will provide significantly less downtime than timber for shaft maintenance. A multiple-deck work stage is planned, with each deck manned as required to perform specific work steps. These steps include removal of timber and loose rock, rock support, concrete ring placement, set steel, pipe, and guide installation. The work stage then would be moved down and the process repeated. There are areas in the shaft where the walls are too far from the shaft cross section to allow for a simple concrete ring installation. In these areas, a configuration involving the installation of a bearing beam to support the set steel will be employed. Only two short areas of the shaft require this type of steel assembly.

Figure 5.4.3.2.3-1 shows the original Yates Shaft timbered layout. Figure 5.4.3.2.3-2 shows the new arrangement with the larger Supercage compartment with a counterweight. Also shown is the design for an auxiliary cage compartment located at the south skip location. This auxiliary cage has been included in the design, but is not part of the baseline estimate and will only be installed if budget contingency can support the cost. A divider has been taken out and the depth has been extended to make room for a slightly larger cage compartment. The auxiliary cage would operate as a single conveyance and would not have a counterweight. The remaining compartments will be utilized for ventilation air, electrical, and pipeline utilities. The conveyances are to be equipped with detachable hooks so that the structural steel reinforcing of the headframe can be minimized. Roughly 94,000 pounds of steel must be added to make the headframe structurally sound.

Figure 5.4.3.2.3-3 shows the new hoist configuration at the Yates Shaft, which will have a cage with counterweight plus an auxiliary cage. The hoist requirements for the auxiliary cage are significantly lower than provided by the existing hoists. The repair cost plus the operating costs do not justify the expense of rebuilding an existing hoist. Therefore, a new auxiliary hoist consisting of a two-rope single drum Blair hoist is included in the design, but not in the cost estimate. The location will be in the existing Yates Hoist building sitting in the location of the underwind drum of the production hoist. That drum must be dismantled and a new foundation installed in the pit. Substantial doweling and pinning into the existing floor and rock will keep the hoist in place. The existing Yates Service Hoist will be utilized for the new Supercage. Upgrades to this hoist include a new braking system utilizing a high pressure hydraulic release system with spring set. Systems of this nature are able to take up brake clearance at a fast rate, then slowing the brake application to decelerate the hoist at a rate that is safe regardless of location in the shaft. The existing friction clutches will either be modified or replaced with a positive engagement jaw type clutch. The pinion shafts will also be set up with brakes to allow personnel to ride while one drum is unclutched, a feature that is valuable while shaft rehabilitating. The drum main bearings will either be sent out for new Babbitt or scraped to obtain proper operating clearances. The new hoist can be seen

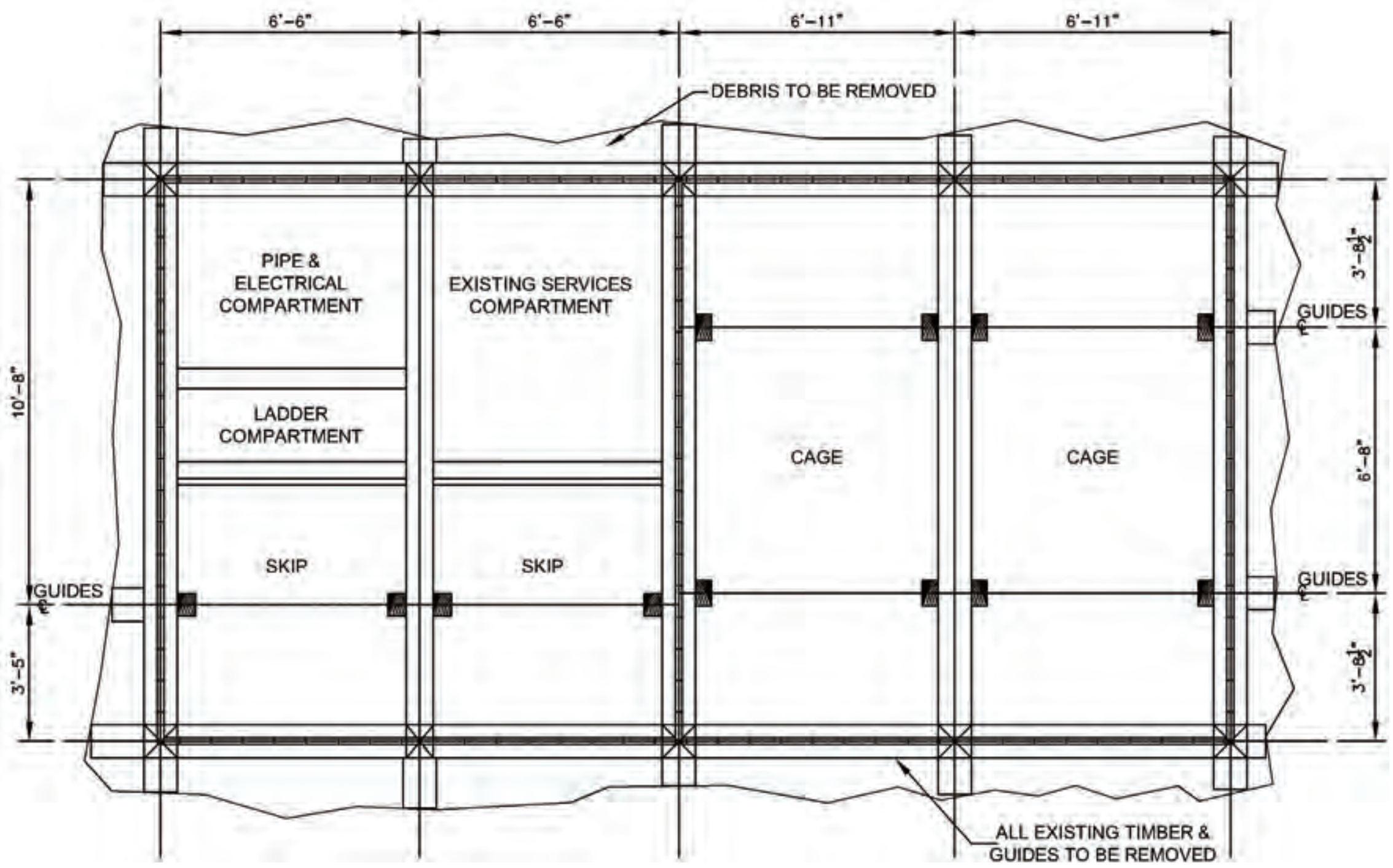

**Figure 5.4.3.2.3-1** Existing Yates Shaft layout. [Adapted from SRK]

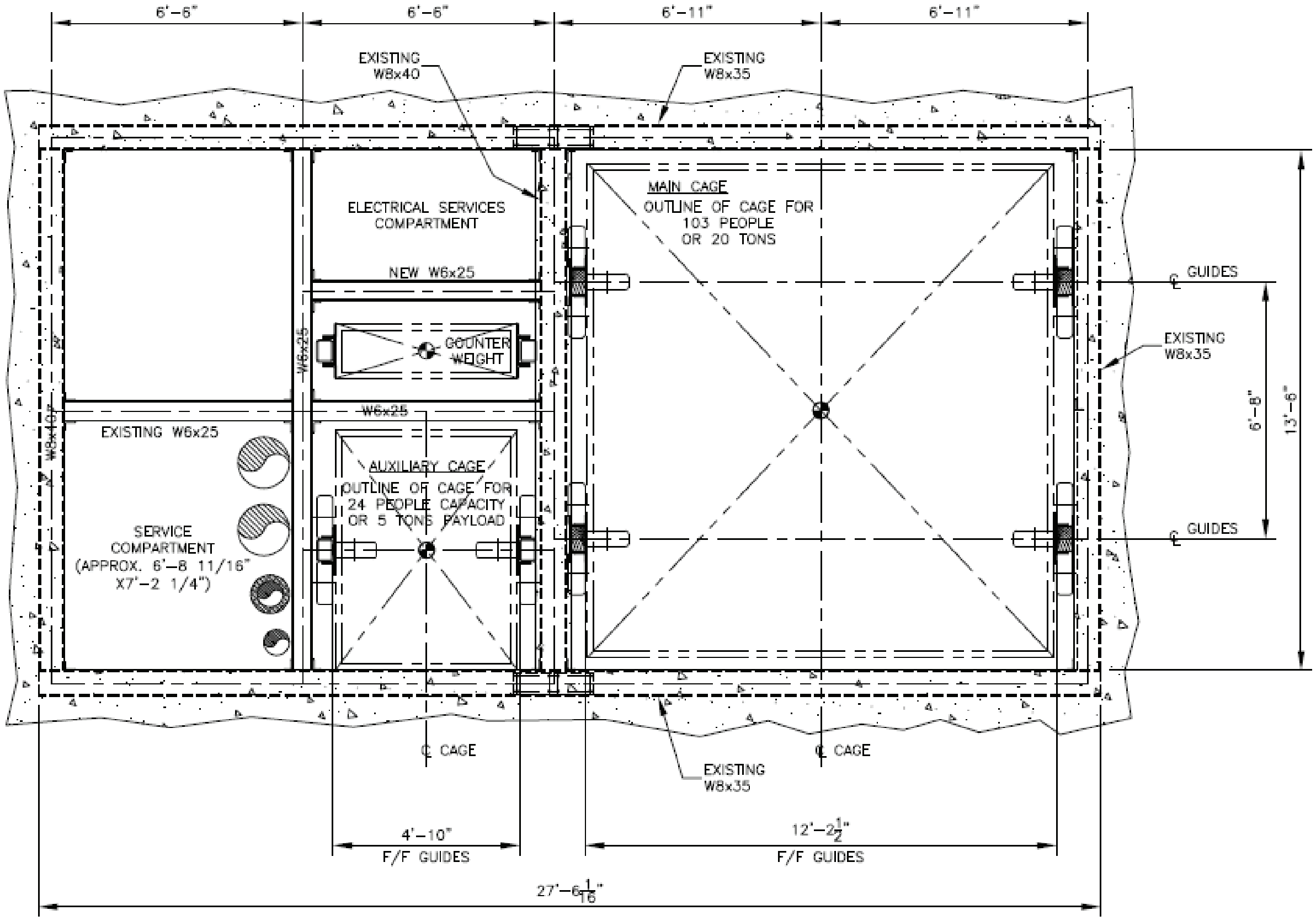

**Figure 5.4.3.2.3-2** Preliminary Yates Shaft design layout. [G.L Tiley and Associates]

toward the upper portion of the drawing. New AC drives and motors will be installed. Plans include provisions for hoist emergency backup power in the event of power line loss.

The current control system uses a motor-generator to vary the Yates Hoist speed and Lilly controllers to prevent overspeed, overwind, and underwind protection. A 43.75-ton flywheel isolates the motor from rapid changes in load, and also provides a brief supply of standby power, allowing the hoist to make at least one full trip from the bottom to top in the event of power loss. This system will be replaced with a variable frequency drive (VFD), modern speed and limit sensors, and a diesel generator for standby power. This hoist, along with the two hoists for the Ross Shaft and the underground winze hoists, will be capable of remote control.

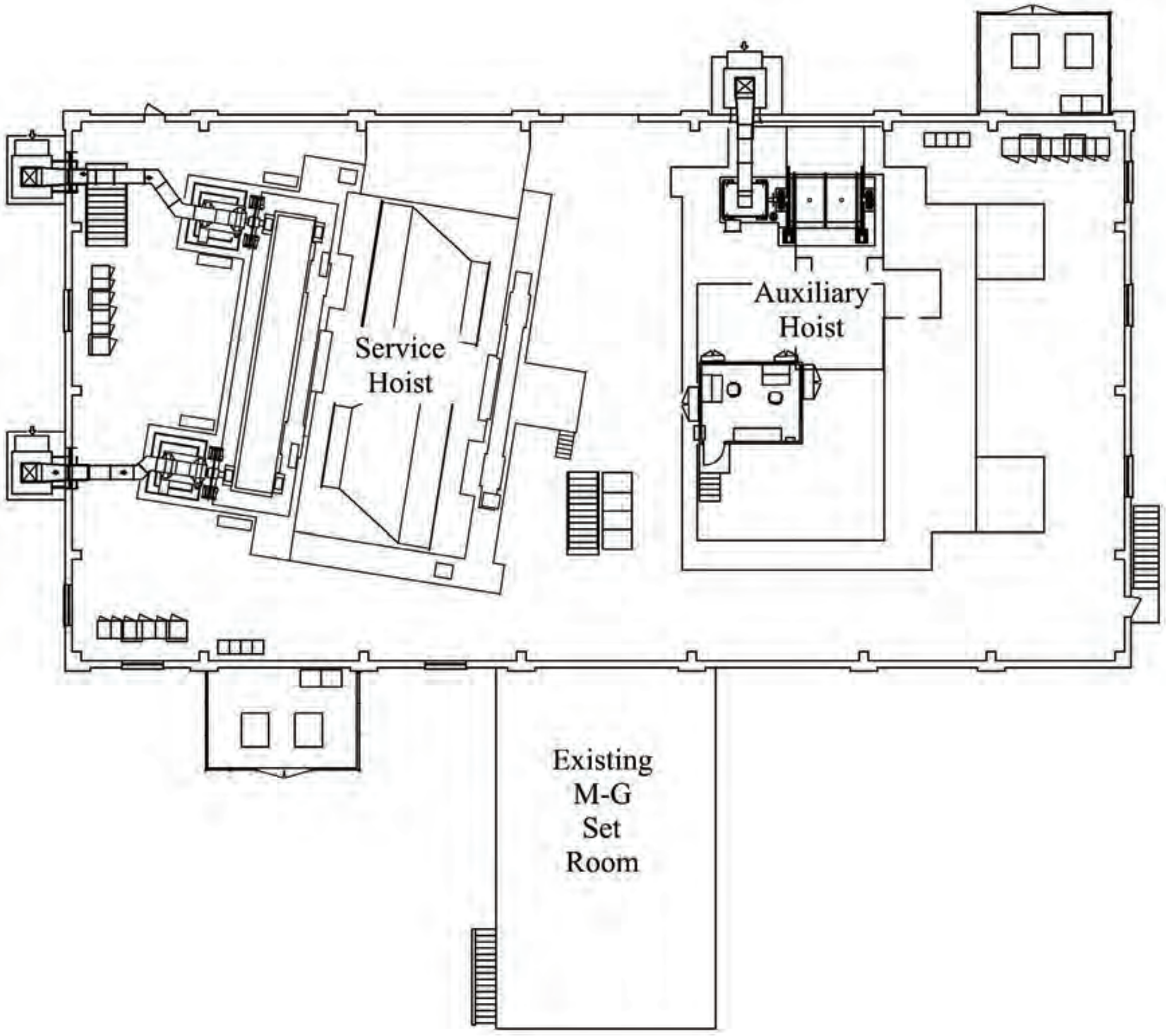

**Figure 5.4.3.2.3-3** Preliminary Yates Hoist layout. [SRK]

### 5.4.3.3 #6 and #8 Winzes

The #6 and #8 Winzes provide access from the 4850L to the 7400L. The #6 Winze also provides access to other intermediate levels, but the #8 Winze will not connect to any additional levels. The #8 Winze is designed for emergency use only. As with the Ross and Yates Shafts, these winzes also provide a path for all utilities from the 4850L to the 7400L.

#### 5.4.3.3.1 Current Condition of the #6 and 8 Winzes and Hoists

The #6 Winze is an existing winze that was used and maintained until mine closure. With the hoistroom at the 4550L, the entire winze and all its mechanical components were submerged prior to dewatering, and much still remains below the water level. The #8 Winze will be an entirely new installation. Since access is not currently available below the 5000L, the design of these two winzes has only been completed to a Conceptual stage, providing less detail than described in previous sections.

**#6 Winze Shaft Infrastructure**

The #6 Winze is rectangular in shape—14 ft 1⅛ in (4.296 m) by 17 ft 1½ in (5.220 m), measured to the outside of the set steel. The shaft collar, which is also the hoistroom, is on the 4550L at 678.50 ft (206.81 m) elevation and the 8000L is the bottom level, at elevation 2729.57 ft (831.97 m) below sea level. Service is provided to 16 levels and two skip loading pockets. The shaft is divided into seven compartments: the cage, counterweight, east skip, west skip, ladder way, electrical, and pipe. Sets are made up of various length and size structural steel wide flange beams located to maintain compartment spaces. Sets are vertically spaced on 8 ft (2.44 m) centers throughout most of the shaft, with correction sets for stations as the exception. Sets are connected to one another using studdles suspended from bearing beams located vertically each 100 ft (30.5 m) to 150 ft (45.7 m). The bearing beams carry the weight of the set steel plus any dynamic loads produced by the conveyances in the shaft. Sets are secured to the rock wall using drill pins holding hardened cement bag blocking. Two wood guides are positioned in each compartment carrying a conveyance. These are used to keep the conveyance travelling within tight tolerances in the compartment. Blocking holds the position of the set steel stationary to maintain guide alignment. Figure 5.4.3.3.1-1 is a plan view of a typical shaft set in the #6 Winze.

The overall condition of the #6 Winze shaft furnishings is unknown at this time. This shaft was completely submerged and Figure 5.4.3.3.1-2 shows the condition of the #6 Winze station on the 4850L.

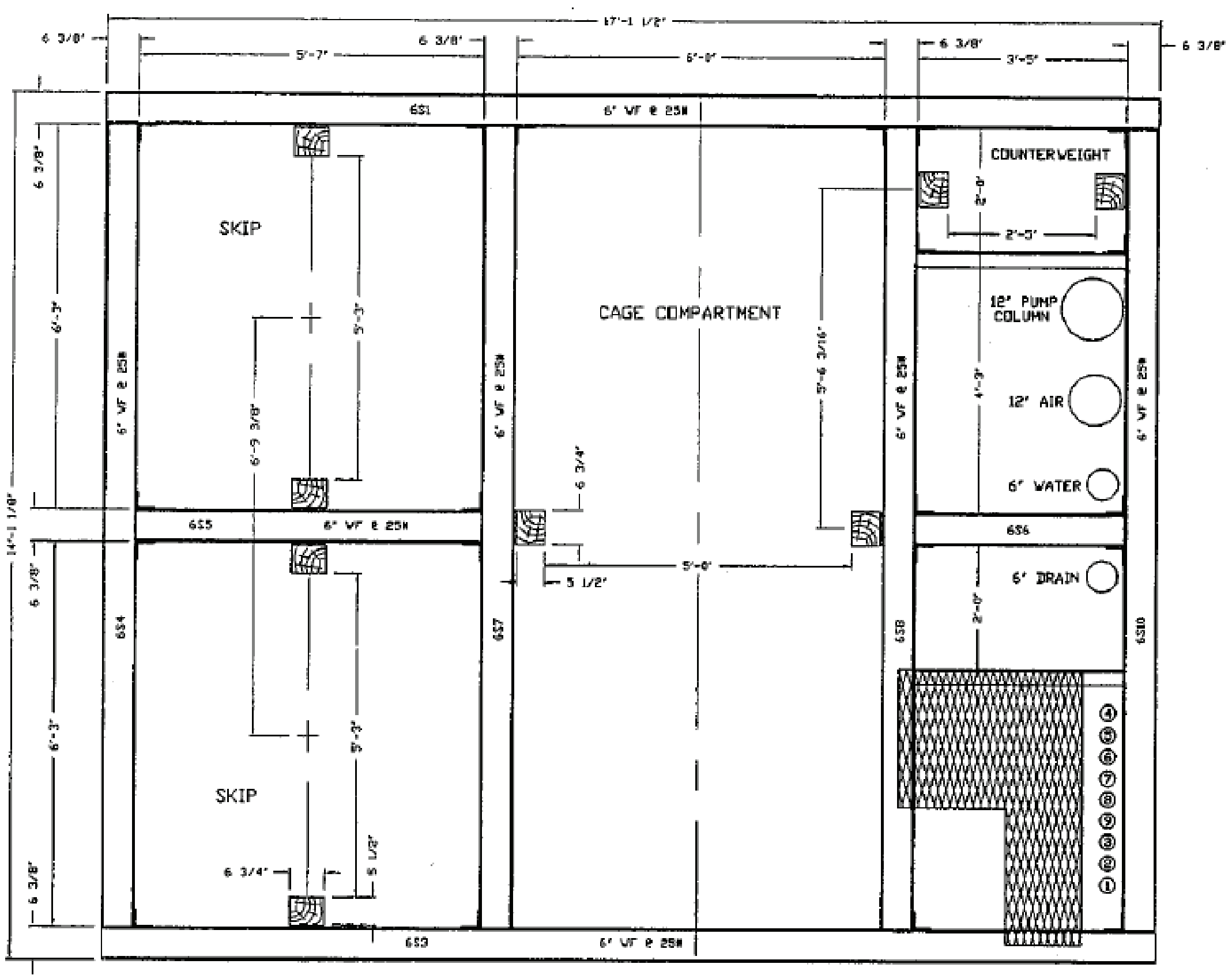

**Figure 5.4.3.3.1-1** Plan view of typical shaft set in the #6 Winze. [HMC]

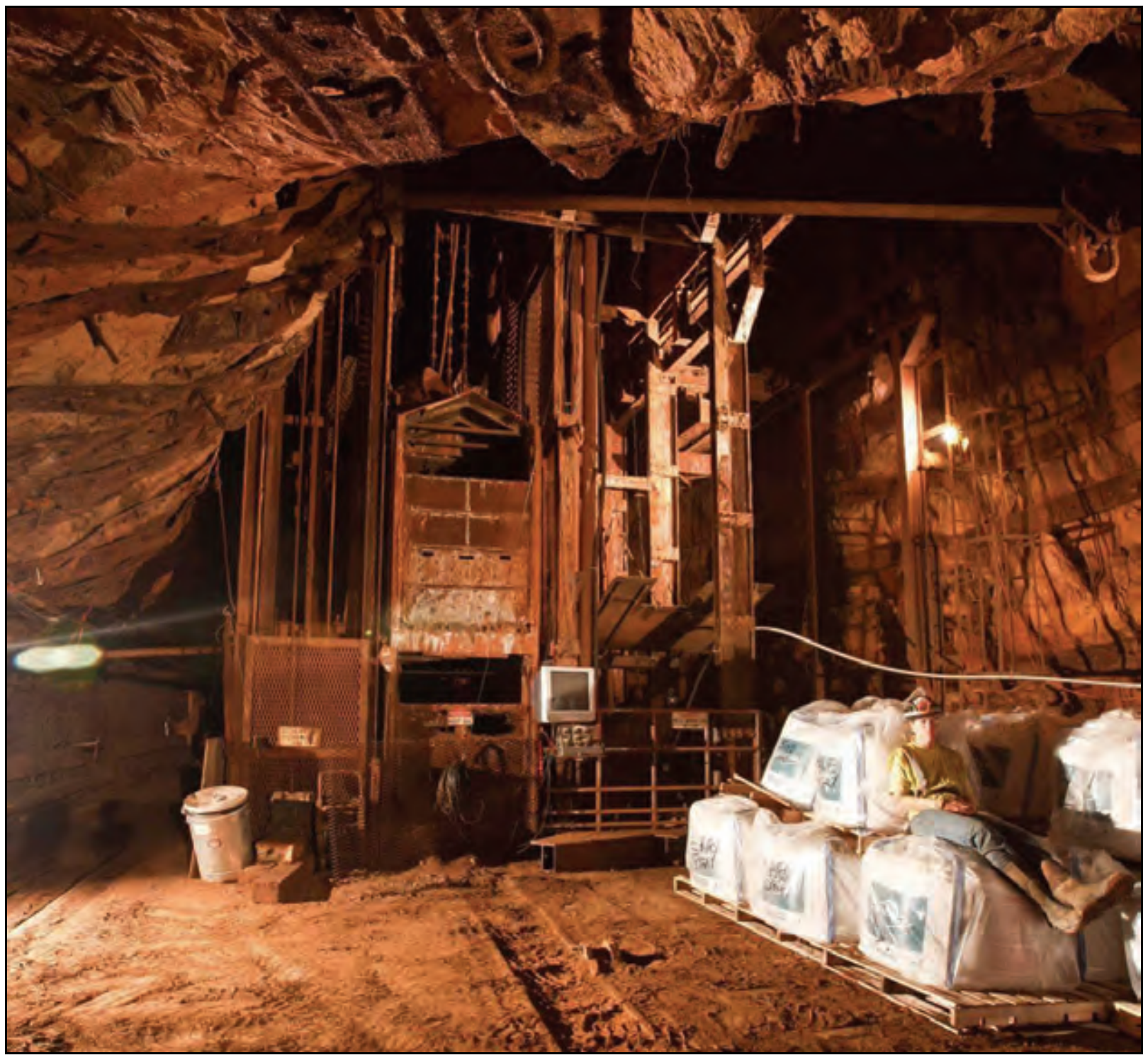

**Figure 5.4.3.3.1-2** #6 Winze station on 4850L. [SDSTA]

**#6 Winze Hoists**

The hoists for the #6 Winze are multi-rope friction winders manufactured by ASEA in Sweden and were installed in 1973. Drum diameters are 7 ft 7 in (2.311 m) for the service hoist and 6 ft 9 in (2.057 m) for the production hoist. The ropes run on polyurethane treads clamped to the drum with aluminum spacers. Drives for both are thyristor-controlled DC motors powering the hoists through gear reducers. The production hoist had a history of rope problems. The diameter-to-length ratio is excessive for these hoists, so very close attention to rope lengths and drum-groove diameters is required. Production rope lives were a matter of a few months until a compacted strand rope was used, increasing this life to two years. Rope length must still be checked daily and groove diameters weekly to obtain the two-year life. Service hoist ropes lasted six to eight years with the same maintenance. The braking systems consisted of a disk brake with hydraulic pressure release and spring set. The system had multiple methods of fluid dump in the event of a valve failure assuring brake set. Full-speed emergency tests were conducted on a yearly basis to check deceleration speeds and the hoists were set with minimal effort.

The dewatering process was not initiated in time to keep the hoists located at the 4550L from being submerged. Consequently, all the electrical systems are not serviceable and were removed by the SDSTA. The hoists along with their corresponding motors are still in place. Hoist ropes are still attached to the drums, and conveyances are attached to the ropes; however, there is evidence that some ropes have completely corroded and disconnected from conveyances.

#### 5.4.3.3.2 #6 and #8 Winzes and Hoists Requirements

The dimensions and capacity of the #6 and #8 Winzes define the cross sections of drifts and capacity of material handling equipment in response to science access requirements.

The #6 Winze will provide primary access and egress to the DLL and other levels and ramps below the 5000L. This winze also contains the skipping system that will be used to remove waste rock during excavation. As the demand for waste rock removal will be considerably less than was required for Homestake mine operations, no significant modifications are planned for this winze or hoist. The requirements for both the conveyance and the hoist are defined by the existing design. The sizes of the existing utility compartments limit the number and size of pipes and conduits that can be routed to these levels.

Designating the #8 Winze for emergency use eliminates some requirements, such as dogging mechanisms, from the design which reduces overall cost while ensuring safe egress capability. The size and capacity of this winze, conveyance, and hoist have been designed to meet the requirements to evacuate the total anticipated occupancy of the 7400L. Space is also provided in this winze for redundant emergency power. Specifications for these two winzes can be seen in Table 5.4.3.3.2.

| Specifications | | No. 6 Winze | | | | No. 8 Winze |
|---|---|---|---|---|---|---|
| | | Service Hoist | | Production Hoist | | Emergency Egress |
| | | Koepe | | Koepe | | |
| | | Headrope | Tailrope | Headrope | Tailrope | Single drum |
| Production required | tons/day | N/A | | 1,000 | | N/A |
| Payload | Mass (tons) | 6 | | 6 | | 1 |
| | Personnel | 60 | | n/a | | 6 |
| Rope mass | Mass (lbs) | 22,785 | 22,878 | 30,222 | 30,222 | 2066 |
| Conveyance mass | Mass (lbs) | 9,000 | | 10,000 | | 2500 |
| Number of ropes | | 3 | 3 | 6 | 3 | 1 |
| Rope size | inch dia. | 1.000 | 1.188 | 0.750 | 1.250 | 0.625 |
| Rope strength | lbs | 123,200 | 113,938 | 60,000 | 122,000 | 40200 |
| Rope FoS | | 8.44 | 10.72 | 7.17 | 12.11 | 6.6 |
| Cage inside dimensions per deck | No. of Decks | 2 | | N/A | | 1 |
| | height (ft) | 7'-0" | | N/A | | 7 |
| | width (ft) | 4'-9" | | N/A | | 5 |
| | Length (ft) | 12'-4" | | N/A | | varies |
| Guides | | wood | | steel | | rope |
| Shaft | Length (ft) | 2,950 | | 2,950 | | 2600 |
| | From | 4550L | | 4550L | | 4850 |
| | To | 7500L | | 7500L | | 7400 |
| Set spacing (ft) | | 8 | | 8 | | n/a |
| Hoisting speed | ft/min | 1,500 | | 1,500 | | 1000 |
| Hoist Power Rated | HP | 600 | | 600 | | 200 |
| Motor Speed | rpm | 950 | | 500 | | 900 |
| Skip cycle time (one-way) | mins | N/A | | 2.87 | | N/A |
| Cage travel time (one-way) | mins | 2.53 | | N/A | | 3.53 |
| Cage load/unload time | mins | 7.00 | | N/A | | 1 |
| Cage total time (one-way) | mins | 9.53 | | N/A | | 4.53 |
| Availability (after all planned maintenance) | hours/day | 18 | | 18 | | 20 |
| Production capacity | tons/day | N/A | | 1,000 | | N/A |
| Slinging capacity at 150fpm | tons | 6 | | | | |

**Table 5.4.3.3.2** #6 and #8 Winze specifications. [G.L. Tiley and Associates]

### 5.4.3.3.3 #6 and #8 Winzes and Hoists Preliminary Design

#### #6 Winze

As previously mentioned, the condition of the #6 Winze is mostly unknown due to flooding as of this report; however, because the winze was in good condition prior to shutdown, the current assumption is that only 20% of the steel sets will need to be replaced, but all of the services will need to be replaced. The shaft rehabilitation will replace steel in kind as needed and will not modify the existing design. As the Facility is dewatered below the 5000L, options will be reviewed to enable the completion of site investigations.

The SDSTA has removed all of the electrical drive cabinets, leaving only the motors and main hoist equipment. During construction, both drums and gear cases will be removed and sent to be rebuilt as required to original equipment specifications. New AC drives and hydraulic braking systems will be purchased and installed. New drum shells will be installed in order to wind rope in multiple layers to assist in the shaft rebuild. After the rehabilitation, new head and tail ropes will be installed to match the location for the new skip loading system to be located on the 7500L. Rock from the 7400L excavations must be skipped through the #6 Winze to the Ross Shaft and out through the waste rock handling system. Plans include provisions for hoist emergency backup power in the event of power line loss.

#### #8 Winze

The #8 Winze, consisting of an 8 ft (2.44 m) diameter borehole, will provide an emergency egress route from the 7400L Campus. The excavation of this winze will utilize a raise boring machine that will be positioned on the 4850L and will drill a directionally drilled pilot hole down to the 7400L. There, an 8 ft (2.44 m) diameter cutter head will be attached to the drill string and the winze will be back reamed to the diameter required. A work platform will be installed and the borehole will be rock supported and lined with shotcrete. The new winze will carry several utility backup systems for the 7400L including electrical, ventilation air, and water. The cage will be rope guided, and since it is only used for emergency it will not require the emergency dogging system. The cage capacity is estimated at six persons. The winze requires approximately 15 ft (4.6 m) of extra travel below the 7400L sill. To access that level, a ramp will be driven down and will be used for shaft cleanup, water pump sump, and an area to set the anchors for the guide ropes.

The #8 Winze Hoist configuration is a single drum single rope located on the 4850L. A new hoistroom, rope raise, and sheave deck areas will require proper excavations. Figure 5.4.3.3.3-1 shows the general arrangement of the #8 Winze Hoistrooms, while Figure 5.4.3.3.3-2 shows a typical shaft plan view of the #8 Winze.

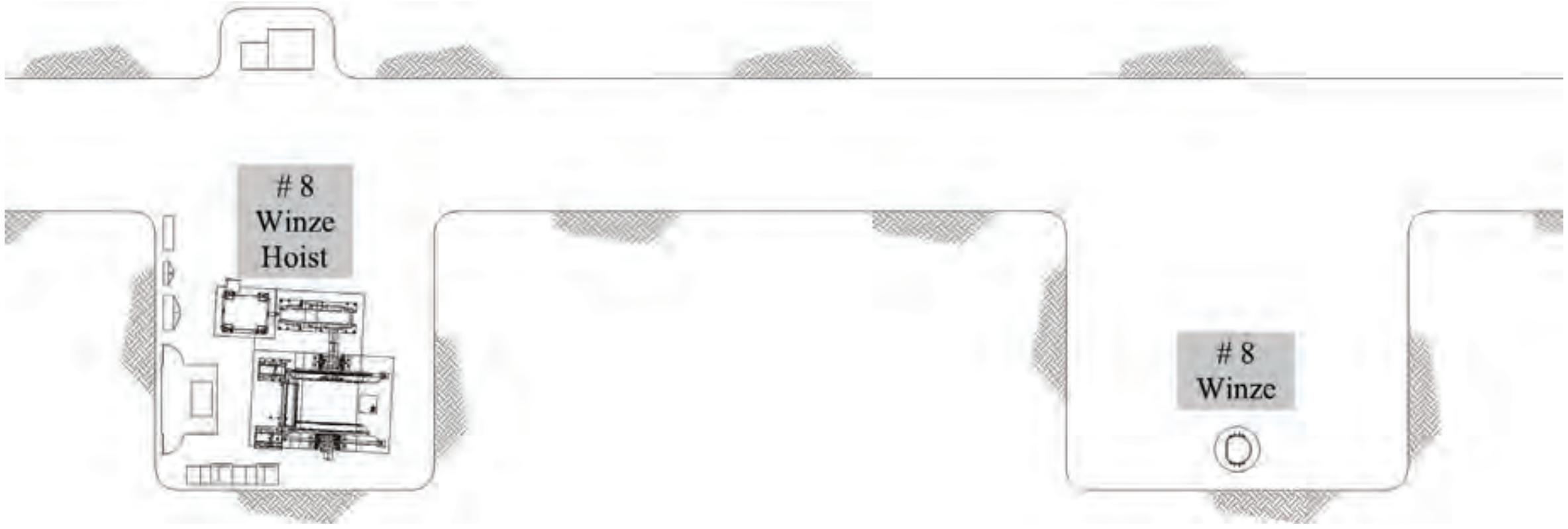

**Figure 5.4.3.3.3-1** #8 Winze Hoistroom general arrangement. [SRK]

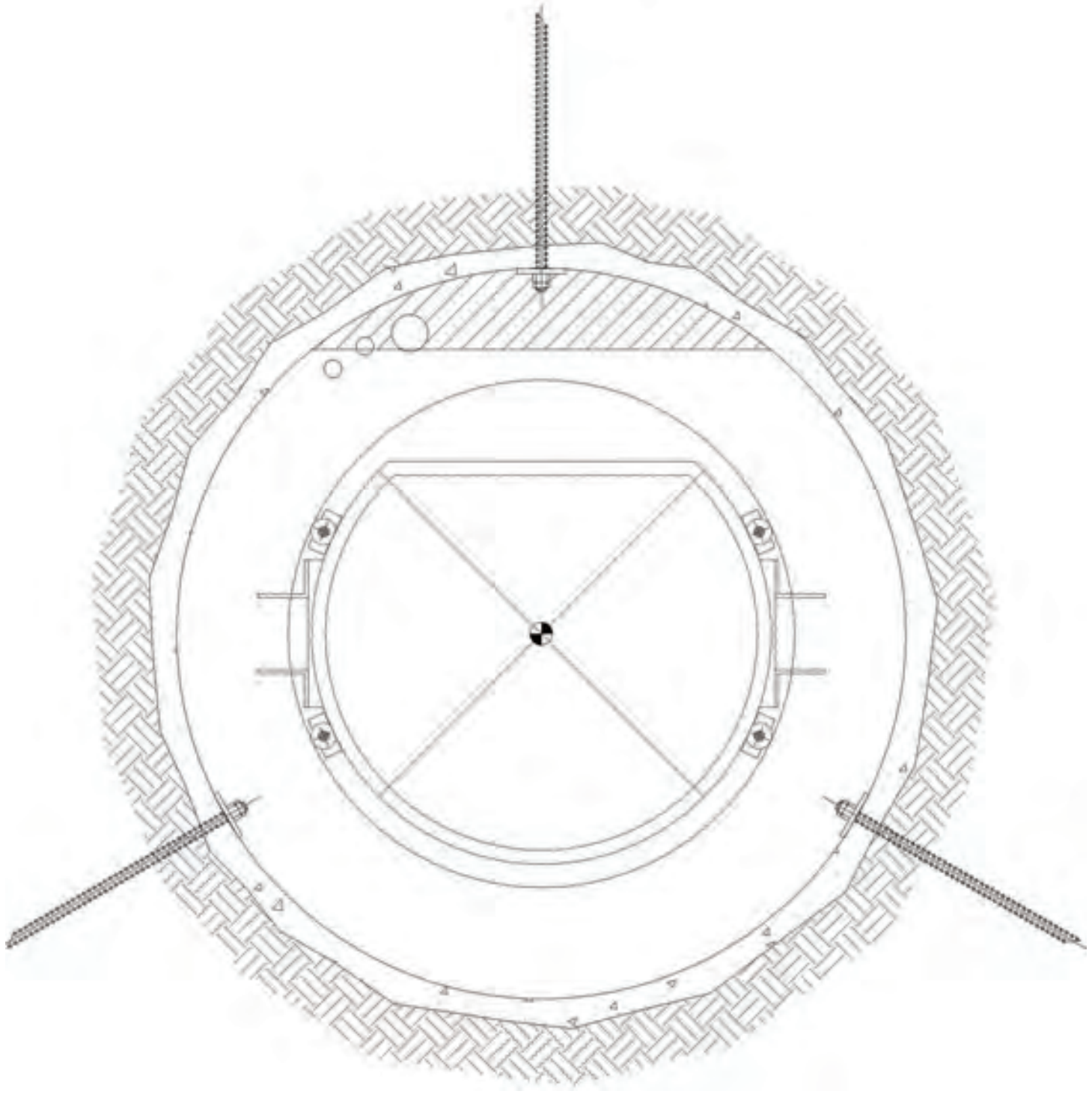

**Figure 5.4.3.3.3-2** Typical shaft plan #8 Winze. [SRK]

### 5.4.3.4 #5 and Oro Hondo Shaft

Two primary exhaust ventilation shafts are currently in use for underground ventilation: the #5 Shaft and the Oro Hondo Shaft. The #5 Shaft was originally installed as an access shaft with a hoist system for personnel access to the underground, and operated as an air intake, as is common practice for shafts with conveyances. The Oro Hondo Shaft was designed as an exhaust shaft when first constructed.

#### 5.4.3.4.1 Current Condition of the #5 and Oro Hondo Shafts

**#5 Shaft**

The #5 Shaft is circular in shape and 16 ft (4.88 m) in diameter. The collar is located approximately 5,135 ft (1,565 m) in elevation, ending underground on the 5600L, and is concrete lined for the first 300 ft (91.4 m). This shaft was originally outfitted with shaft steel, guides, and a cage for transporting personnel and materials. The #5 Shaft was serviced by a double drum hoist, which has been out of service for several years, using the shaft for intake ventilation only.

To better understand existing conditions in the #5 Shaft, SDSTA Operations crews conducted a camera survey in November 2009. The survey was limited; a partial blockage was identified approximately 500 ft (152 m) from the collar. During Homestake operations, problems occurred with ice buildup in the shaft. During winter months, cold, fresh air would downcast into the shaft and cause the freezing of water inflows near the collar. It is believed that ice released during thawing conditions contributed to the failure of shaft furnishings. Pictures from the camera survey reveal the lack of shaft furnishings and the pile of debris approximately 500 ft (152 m) from the collar. Figure 5.4.3.4.1-1 shows the pile of debris. To mitigate the risk, plans are in place to proceed without the #5 Shaft prior to the MREFC-funded Construction, and the MREFC-funded baseline ventilation plan does not include the use of the #5 Shaft (see Section 5.4.3.8.3).

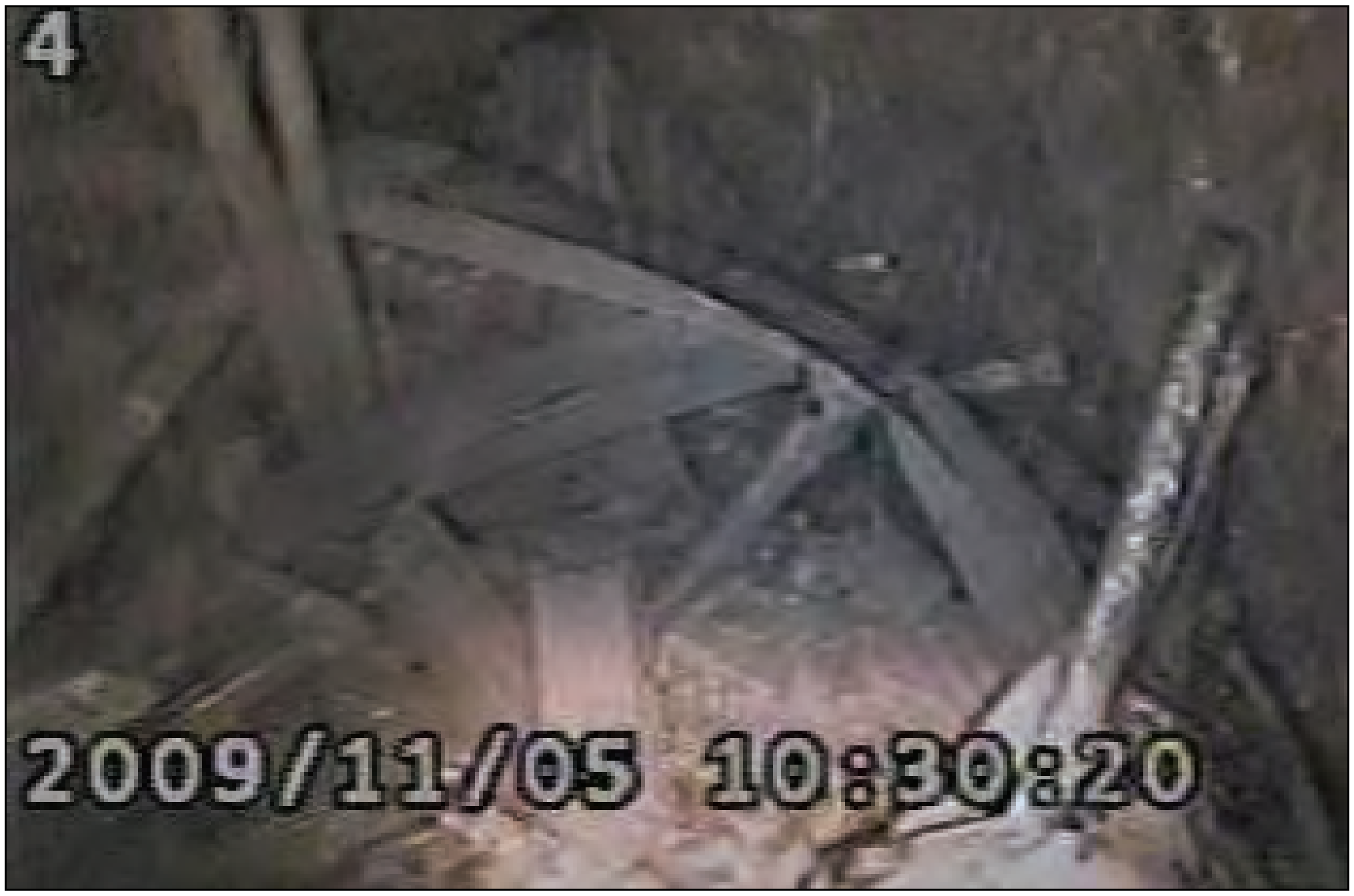

**Figure 5.4.3.4.1-1** Blockage in the #5 Shaft 500 ft (152 m) below the collar. [SDSTA]

**Oro Hondo Shaft**

The collar of the Oro Hondo Shaft is approximately 4,940 ft (1505.7 m) in elevation and approximately 7 ft (2.13 m) wide by 14 ft (4.27 m) long, rectangular in shape, though the shapes and dimensions vary along its length. As this is a ventilation shaft only, there are no shaft furnishings located within the shaft. The top 135 ft (41.15 m) of the shaft is concrete lined. Restrictions in air flow of the rectangular shaft resulted in slashing the shaft from the 2300L to the 3950L. This portion of the shaft does not contain any ground support and rock spalling required Homestake to muck rock out the bottom of the shaft from the 4100L every one to two years. The shaft has not been mucked out since the closure of the mine, and the rock has accumulated to approximately 50 ft (15.24 m) below the 3500L.

The Oro Hondo Shaft has been filmed twice. The first survey by the SDSTA in November 2009 proved the shaft was clear; however, the spinning of the camera could not be controlled and ground conditions were difficult to evaluate. In the summer of 2010, Zapata Engineering performed a laser/camera survey of the Oro Hondo Shaft and the results are referenced in Laser Mapping and Video Imaging of the Oro Hondo Airshaft. The laser scan identified several areas that have experienced significant deterioration. Figure 5.4.3.4.1-2 represents the areas where rock has collapsed, causing large voids in the shaft walls, and provides recommendations for additional ground support, though this is not included in the MREFC-funded Preliminary Design cost estimate. Deterioration of the shaft walls does not negatively affect the capacity to exhaust air from the underground.

A 3,000 hp (2237 kW) 1986 American Davison centrifugal fan was used by HMC to exhaust air from the underground through the Oro Hondo Shaft. This fan has been refurbished by the SDSTA and is currently the main exhaust ventilation fan. New 350 hp (261 kW) ventilation fans were installed at the collar of both the Oro Hondo and #5 Shafts in the fall of 2009, and it is anticipated that these fans will continue to be serviceable with appropriate maintenance for both DUSEL construction and operations, though the #5 Shaft is not included in the ventilation design for operations.

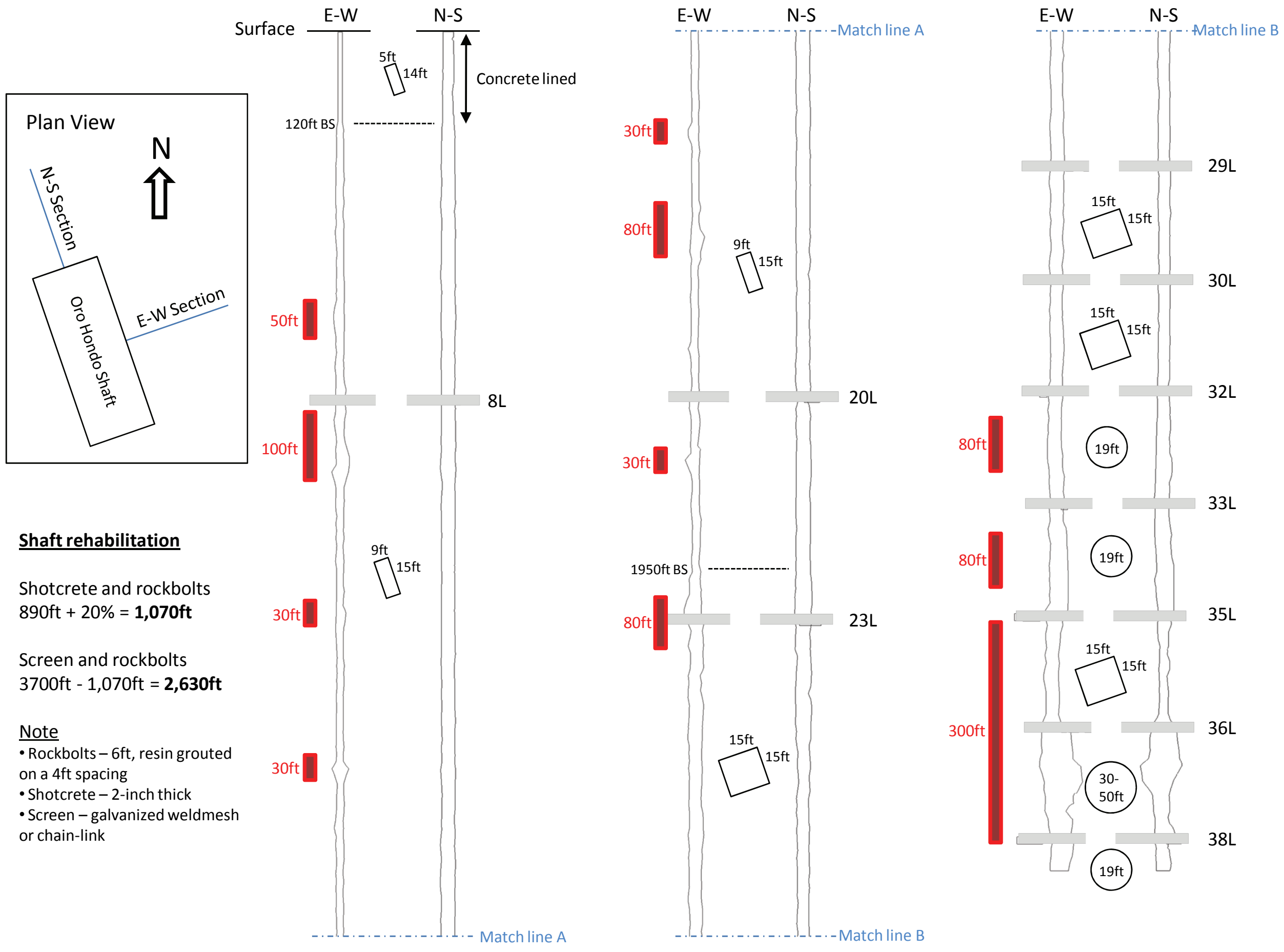

**Figure 5.4.3.4.1-2** Oro Hondo Shaft survey results. [SRK]

### 5.4.3.4.2 #5 and Oro Hondo Shaft Requirements

The #5 Shaft was planned to provide a ventilation exhaust path for the underground facilities. The significant deterioration of this shaft described above led the Project team to eliminate this shaft from the design. As a result, the Oro Hondo Shaft will accommodate all underground ventilation air requirements.

The ventilation requirements described in Section 5.4.3.8.2 establish the total volume of air required to pass through the Oro Hondo Shaft. If this shaft were designed as a new excavation, the cross-sectional area would be defined to limit the velocity in the shaft and reduce the pressure required for the fan. Since it is not a new excavation and will not be modified, the existing cross-sectional area combined with the exhaust volume requirements define the fan requirements. The existing Oro Hondo Fan is capable of meeting these requirements.

### 5.4.3.4.3 #5 Shaft and Oro Hondo Shaft Preliminary Design

#### #5 Shaft

Early in the design process, the #5 Shaft was considered for use as a pathway for exhaust ventilation as well as emergency egress during construction. As discussed above, the deteriorating condition of the #5 Shaft and the corresponding cost of rehabilitation prevented it from being included in the baseline design.

#### Oro Hondo Shaft

Historically, the Oro Hondo Shaft was utilized as the primary conduit for exhaust ventilation. The deteriorating ground conditions within the shaft present a risk to its long-term operation. Three options were considered for the future use of the Oro Hondo Shaft as the primary pathway for exhaust ventilation at DUSEL.

#### Option 1—Refurbish the Oro Hondo Shaft

This option considers the installation of conventional ground support to prevent further collapse and deterioration, including rock bolting, screening, and shotcrete. Rehabilitation activities could not commence in this area without first rerouting the existing overhead power lines within the shaft area. A temporary headframe could then be installed and a single work deck stage would be lowered by winches mounted on the surface. A single drum single rope hoist would operate the sinking bucket, which would serve as the access to the work deck and appropriate ground support could then be installed for long-term stability.

#### Option 2—Excavate a New Exhaust Raise

It is estimated that a new exhaust raise could be excavated for approximately the same cost as refurbishment described in Option 1. An important consideration relating to this option is waste rock removal. Depending on the method of excavation, waste rock would either need to be hoisted to the surface in a shaft sinking method or conveyed through the network of waste rock dumps to the 4850L loading pocket. Waste rock generated from sinking a new shaft would require a new designated surface dump location, which would pose environmental concerns. One advantage of a new raise is that it would intersect multiple desired levels underground and alter the ventilation system to reduce unnecessary flow paths and therefore reduce power requirements for the fan.

#### Option 3—Routine Maintenance

The third option would be to continue to operate as was done historically by regularly removing muck as the shaft deteriorates. This option carries the most risk and highest cost for ongoing operations. Rock that has fallen can be accessed at the bottom of the shaft (4100L). The drift used to access the bottom of the shaft is narrow and requires the use of a track mucker. Waste rock storage on this level is limited and requires rehabilitation. Because this area currently houses several science experiments, science traffic would be stopped for a period of time every one to two years to allow for removal of the fallen rock resulting in interruptions to the science collaboration access.

Since the shaft is currently functioning in accordance to ventilation needs, Option 1 or 2 will not be pursued as part of the baseline scope of the Project, but will be considered as options for future capital operations expenditure.

The exhaust ventilation fan on the Oro Hondo Shaft is monitored at the Ross Headframe and the Waste Water Treatment Plant (WWTP). Operators will receive an alarm if the fans should surge or shut down. Carbon monoxide (CO) sensors have been installed in the Oro Hondo Exhaust Fan.

### 5.4.3.5 Maintenance Shops, AoRs, Common Rooms, Storage and Containment Rooms

The DUSEL Project will require many ancillary spaces for maintenance, AoRs, storage, and various utilities outside of the main science spaces. This section describes the design for these spaces.

#### 5.4.3.5.1 Current Condition of the Maintenance Shops, AoRs, Common Rooms, Storage and Containment Rooms

On the 4850L, only two existing spaces are intended for reuse as ancillary space. These two spaces, located near the Ross Shaft, were historically used as a drill repair shop, supervisor's office, and electric shop. By adding a connection between them, they will become the Ross AoR for DUSEL. One section of this space was rehabilitated by the SDSTA with a shotcrete lining in 2010 for use as an electroforming laboratory to support early science experiments. A third excavation used for electrical installations during Homestake operations near the Yates Shaft in the East Laboratory Access Drift has not yet been designated for a specific use, but will likely be used for storage. This space housed electrical equipment during the SDSTA development of the Davis Campus. Several existing excavations near the Yates Shaft are being used to redevelop the Davis Campus, including the Davis Laboratory Module (DLM) itself (the large cavity where the original Ray Davis experiment was housed). These spaces will be upgraded and made functional in advance of DUSEL construction outside of the MREFC-funded Project. Figure 5.4.3.5.3-1 shows the areas discussed in this section.

At the 7400L, an existing drift between the #6 Winze and the new DLL Campus location will be used as a maintenance shop. Figure 5.4.3.5.3-3 shows the 7400L, including this drift. Until this level is accessible, the condition of this drift will remain unknown. It is expected to require both additional ground support and repair of existing ground support. All other ancillary spaces will be new excavations.

#### 5.4.3.5.2 Maintenance Shops, AoRs, Common Rooms, Storage and Containment Rooms Requirements

The need for and use of ancillary spaces will be considerably different during Construction and Operations phases of the Project. During Construction, the requirements include spaces to transfer excavated material efficiently, to store and assemble equipment, to install temporary equipment (compressors, transformers, etc.), and for people, including both temporary AoR and break areas. Once the Facility is a fully operational laboratory, the use of these spaces will change to include permanent mechanical, electrical, and plumbing installations, permanent AoRs, areas for equipment maintenance and assembly, fabrication and assembly of laboratory equipment, and areas to store small quantities of materials needed to support experiments. All of these areas will have specific life/safety requirements, including but not limited to fire protection, containment berms, and fire doors. Requirements for noise and lighting controls in these areas vary by space and can be found in detail in Arup's *UGI Basis of Design Report* (Appendix 5.L, Chapter 7).

#### 5.4.3.5.3 Maintenance Shops, AoRs, Common Rooms, Storage and Containment Rooms Preliminary Design

For clarity, the Preliminary Design of ancillary spaces has been separated into three areas and is discussed below: OLR, 4850L, and 7400L.

**OLR**

Very limited ancillary space is anticipated for OLR. Experiment specific excavations are discussed in Chapter 5.9, *Design and Infrastructure for Other Levels and Ramps (OLR)*. Spaces will be modified or created at the 300L, 800L, 1700L, 2000L, 2600L, 3500L, 4100L, 4550L, and 6800L to support mechanical and electrical installations. These spaces will have minimal finishes, but will include fire protection and fire doors to prevent smoke from travelling down the Yates Shaft, Ross Shaft, or #6 Winze to the main campuses.

**4850L (including 5060L)**

A variety of ancillary spaces exist on the 4850L as the main campus for the laboratory. Figure 5.4.3.5.3-1 provides an overview of the 4850L with many of the ancillary spaces highlighted. Six AoRs will be built on the 4850L and 5060L to provide safe refuge for personnel in an emergency for up to 96 hours (see Section 5.4.3.1.3). Two MERs provide power for loads that are not equipment specific and three are dedicated to electrical distribution equipment.

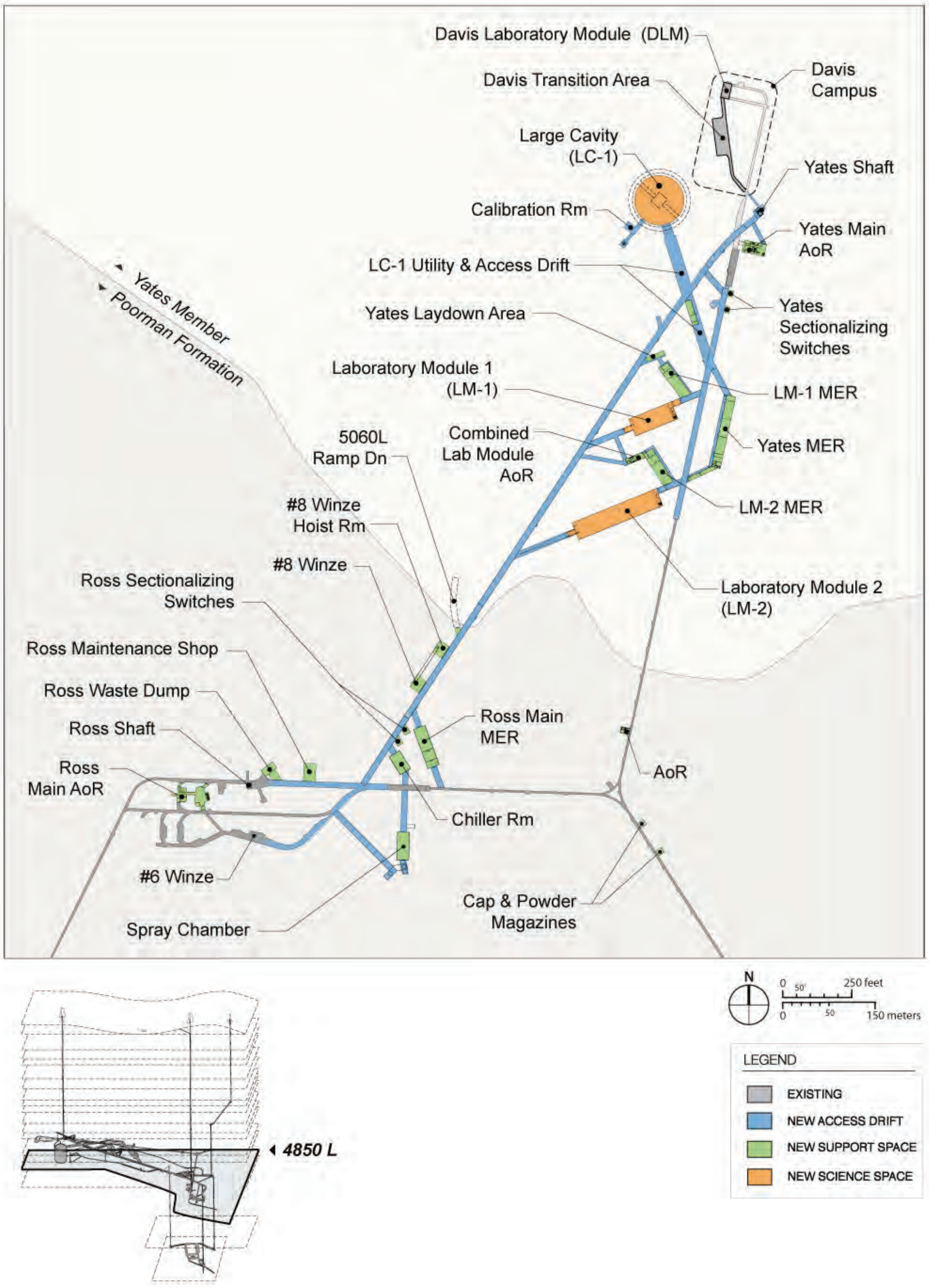

**Figure 5.4.3.5.3-1** 4850L ancillary spaces. [DKA]

Contained within the same excavated space as the Yates MER and its access drifts are rooms for charging mobile equipment batteries, mobile equipment maintenance, electric shop, and a stationary equipment shop (See Figure 5.4.3.5.3-2). These rooms provide space for Facility workers to maintain the equipment and services on the level. Each LM and the LC-1 also have MER spaces housing their electrical switchgear, air compressors (LMs only), and air-handling systems. Large spaces are required for installation of the chilled-water system near the Ross Shaft. Laydown spaces near each shaft are included in the design. The space near the Ross Shaft will be important during the Construction Phase to allow staging and assembling equipment out of the material hauling path. Two excavations will be added outside the main campus for powder and cap magazines to support construction.

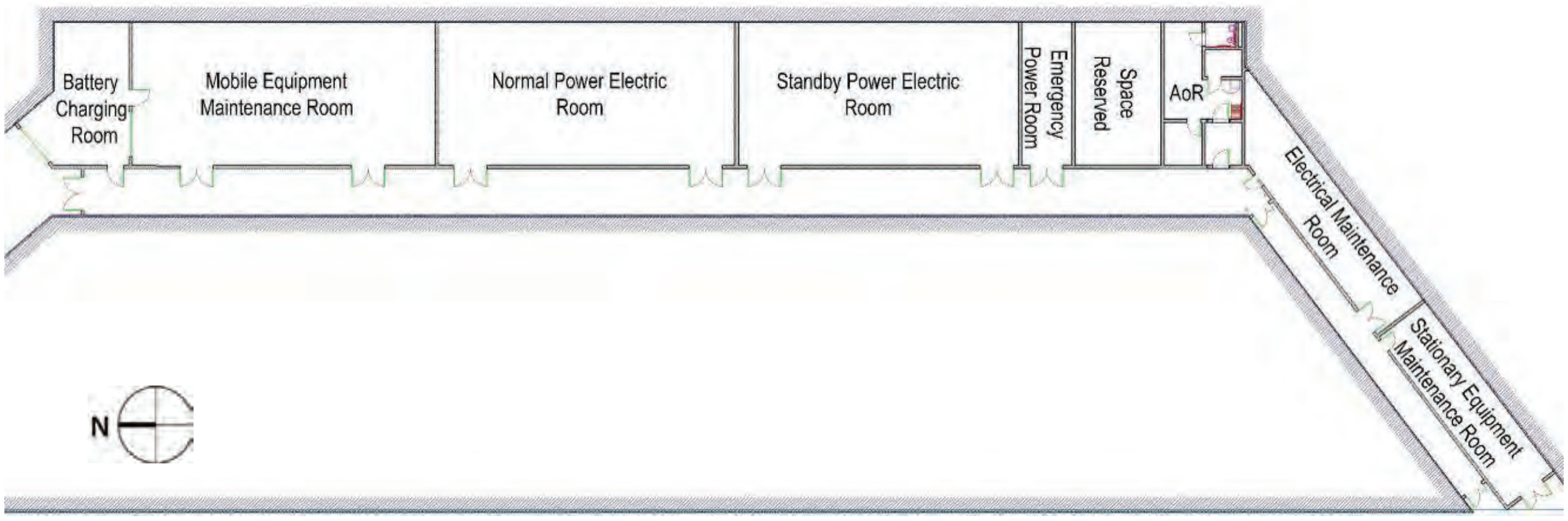

**Figure 5.4.3.5.3-2** Yates MER space. [Arup]

**7400L**

The 7400L has similar requirements to the 4850L and therefore has similar installations. Figure 5.4.3.5.3-3 shows an overview of the 7400L with the ancillary spaces highlighted. With a much smaller footprint, only two AoRs are included at this level, one near each winze. Small spaces will be built outside these AoRs to store emergency response equipment. A maintenance shop, mentioned in the current conditions, will utilize existing space to provide an area for facility maintenance. A single MER space in an L configuration (Figure 5.4.3.5.3-4) provides space for all electrical and mechanical needs to support the LM and facility. This includes electrical switchgear, the air-handling system, and an air compressor. Unlike the shared space at the 4850L, the battery-charging, stationary equipment maintenance shop, and electrical maintenance shops each have separate excavations. Standby power to this level is provided through two redundant backup generators located in another excavation. A hazardous-materials storage room will be included in the exhaust ventilation path near the chiller and spray chamber rooms for the chilled-water system.

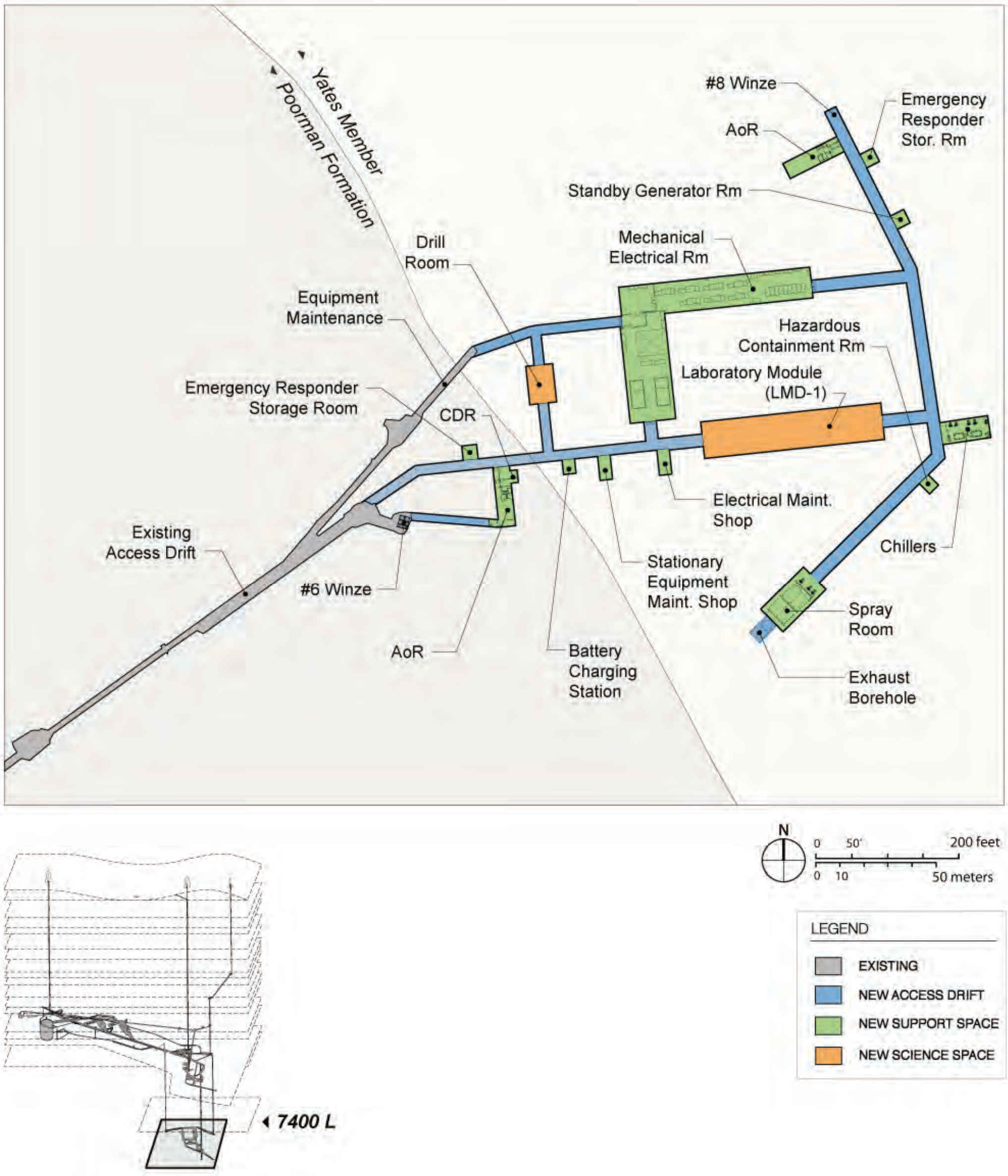

**Figure 5.4.3.5.3-3** 7400L ancillary spaces. [DKA]

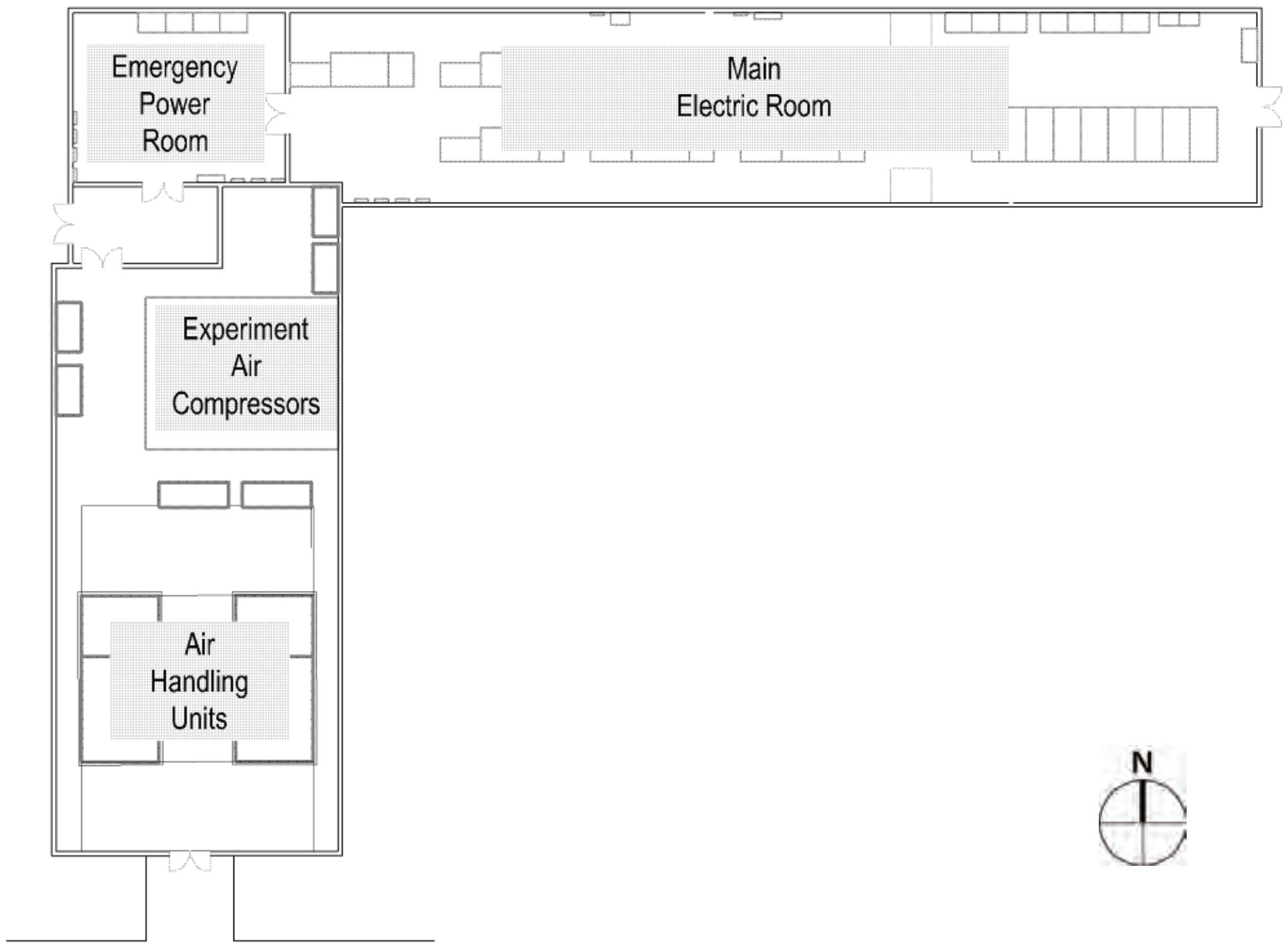

**Figure 5.4.3.5.3-4** 7400L Plant Room. [Arup]

### 5.4.3.6 Drifts and Ramps Required for Access, Egress, and Ventilation

Many excavated areas outside of the main campus areas on the 4850L and 7400L are required to provide access to facilities, related infrastructure, OLR experiments, or provide flow paths for water or ventilation. In many cases, these areas make use of existing excavations created by HMC.

#### 5.4.3.6.1 Current Condition of the Drifts, Ramps, and Raises for Access, Egress, and Ventilation

The underground facility consists of over 300 miles of existing excavations that were created over the 125 years of HMC operations. The condition of the excavations varies widely throughout the Facility. As of the writing of this report, only areas above the 5000L have been assessed, due to the flooded condition of the Facility below this level.

Existing excavations that are of interest for DUSEL include the following:

1. Lab campuses at the 4850L and 7400L
2. Access and egress, ventilation pathways, and utility rooms to support the construction and operation of LMs at these levels
3. OLR used specifically for Earth-science related experiments

The first two items are discussed in this section, while the third is discussed in Chapter 5.9, *Design and Infrastructure for Other Levels and Ramps*.

SRK Consultants, a subcontractor to Arup, performed geotechnical site investigations in excavations required for access, egress, and ventilation. Details regarding their findings are included in Section 4 of the Arup *Preliminary Site Assessment Report* (Appendix 5.M). In general, their findings indicate rock conditions in areas required to support laboratory operation vary depending on location, but all will require installation of new ground support to provide long-term access. The type and quantity of ground support required depends on the specific ground conditions in a given area, the size of the opening, the particular geologic features, and the end use of the space.

#### 5.4.3.6.2 Drifts, Ramps, and Raises for Access, Egress, and Ventilation Requirements

Drifts and ramps provide access for people, equipment, and utilities throughout the underground campus. The size of these excavations is dictated by the maximum size of construction and science equipment passing through them (limited by the shaft cages), code-required clearance, and utility space needs. The finishes (walls, ceiling, and floors) in each area are designed specific to the type of use. In areas with frequent science use, smooth concrete floors provide a level path for material handling, and shotcrete is placed on walls and ceilings for ground support and to enhance light and provide aesthetic improvements. In less-traveled areas that are primarily for Facility needs, the floors may be gravel and there may be bare rock walls with rock bolts and welded wire mesh for ground support.

#### 5.4.3.6.3 Drifts, Ramps, and Raises for Access, Egress, and Ventilation Preliminary Design

The primary paths from the shafts to the laboratory spaces are designed to be large enough to allow passage of the largest single item possible to pass from the primary hoisting system for the level. On the 4850L, this is defined by the Supercage, while at the 7400L it is limited by #6 Winze cage. An imaginary box 52 in (1.3 mm) wide x 83 in (2108 mm) tall on either side of this large item allows a safe egress path in an emergency. Floors in these drifts will be smooth concrete, and walls (ribs) and ceiling (back) will have rough shotcrete finishes. An example cross section is shown in Figure 5.4.3.6.3.

Secondary paths not intended for science use, such as the ramp from 4850L to the 5060L, have less strict requirements. In these cases, concrete floors may be provided for ease of construction, but are not required for operational use. In a similar fashion, ground control for the walls and ceiling may be less aesthetically pleasing, using bolting and screening without shotcrete. Areas not intended for occupancy at all, such as the 4700L and 3950L ventilation paths and the exploratory drift for the Large Cavity, will have very limited finishes with the primary purpose of maintaining an open cross section. These areas will only be accessed by qualified and authorized workers, and will be fully inspected during passage through the areas. Excavations near the shafts for services down the shafts will have a variety of finishes, depending on the ground conditions in each area, criticality of services, and interfaces with OLR.

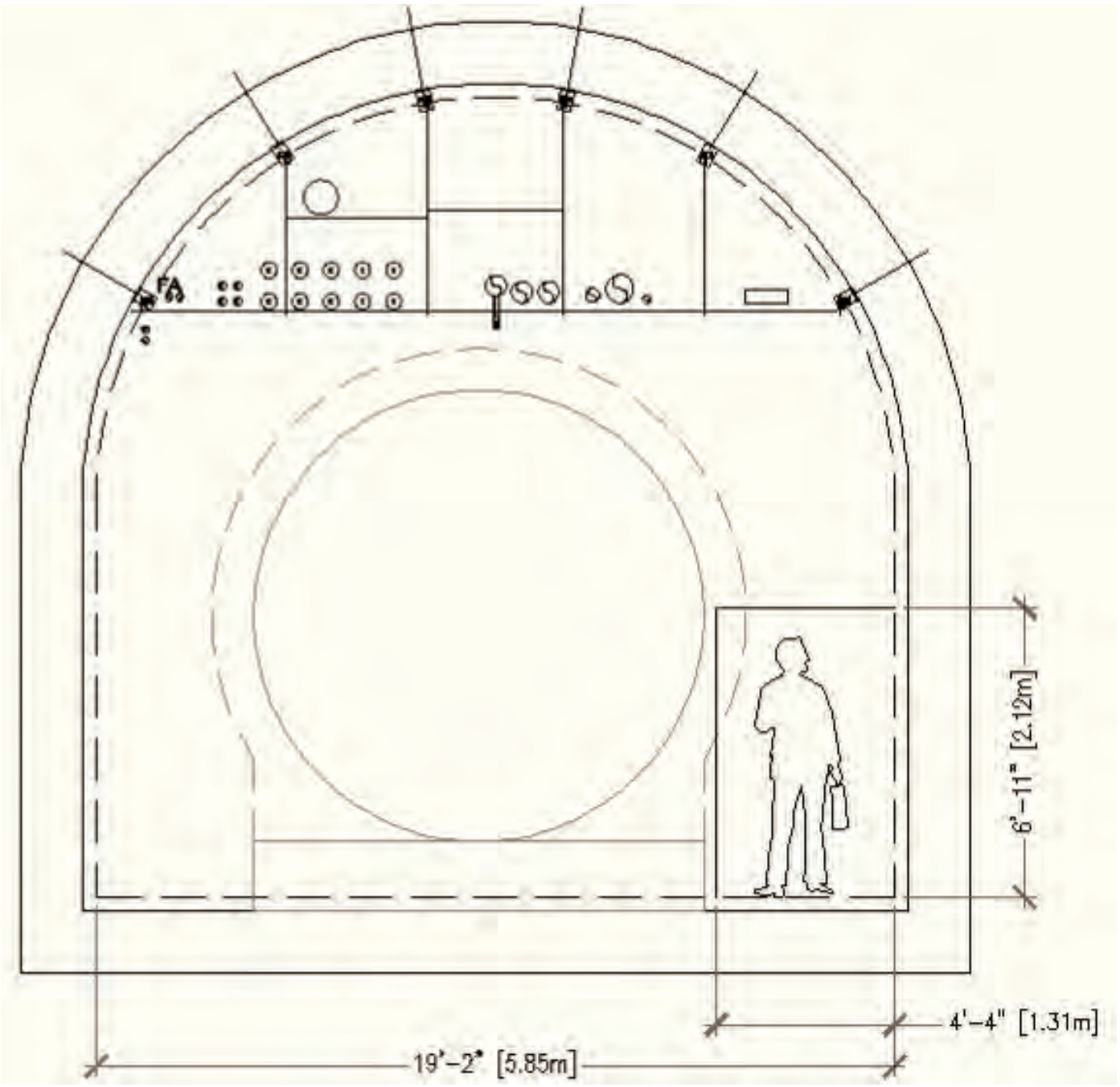

**Figure 5.4.3.6.3** Example of a cross section in a main access drift at the 4850L. [Arup]

### 5.4.3.7 Material Handling and Personnel Transportation

Large volumes and quantities of materials will be managed in the underground spaces for the DUSEL Project. The infrastructure design included a preliminary investigation to determine the best method for managing these materials while maintaining safe access for personnel travelling in the same area and ensure sensitive scientific equipment is not damaged in transport. With a footprint covering several miles, personnel transportation will also be needed for convenience, maintenance, and emergency response.

#### 5.4.3.7.1 Current Conditions of the Material Handling and Personnel Transportation Systems

The HMC transported all supplies and materials underground using track equipment, primarily via the Ross Shaft. The conveyances in both the Ross and Yates Shafts are equipped with rail so loads can be easily transferred onto the conveyance and from the conveyance to the underground rail system. Upon Homestake closure, most locomotives and corresponding flat cars were sold; some of the rolling stock (track equipment) was left underground and a portion of this equipment was submerged underwater. As areas were dewatered, equipment in salvageable condition was recovered and refurbished. The rail in occupied areas of the Facility has been restored by the SDSTA for locomotive traffic, allowing materials and supplies to be delivered to working areas. Rolling stock that was recovered and refurbished includes a number of rock cars, flat cars, timber trucks, and man cars. Two 1½-ton electric locomotives were purchased by the SDSTA. A skid steer completes the existing underground fleet of material handling equipment.

Two ATVs and several rail based man cars are the only dedicated personnel-transport vehicles on site. One of the ATVs is a dedicated mine rescue vehicle and is kept on the surface for emergency situations. The other is used underground for operation and maintenance needs. Man cars conveyed by locomotives are primarily used to transport working crews from the shafts to the working areas and back.

#### 5.4.3.7.2 Material Handling and Personnel Transport Requirements

Many of the key requirements for the material handling systems are defined by the capacities of the hoisting systems as described in Sections 5.4.3.2 and 5.4.3.3. In addition to the internal dimensions of the cage, each hoist can suspend a load beneath the cage, delivering a longer load than would otherwise be possible. The limits of this load have been defined as a 4.9 ft (1.5 m) x 4.9 ft (1.5 m) x 42 ft (13 m) box for the Yates Shaft. As discussed in 5.4.3.2, the material handling system at each main campus level must be sized to accommodate transport of anything that can physically fit down the shafts and winzes. Once the material handling system is loaded at the shafts, it must prevent inadvertent contact with utilities or surfaces, and navigate the delivery routes. An example of the routes at the 4850L is shown in Figure 5.4.3.7.2.

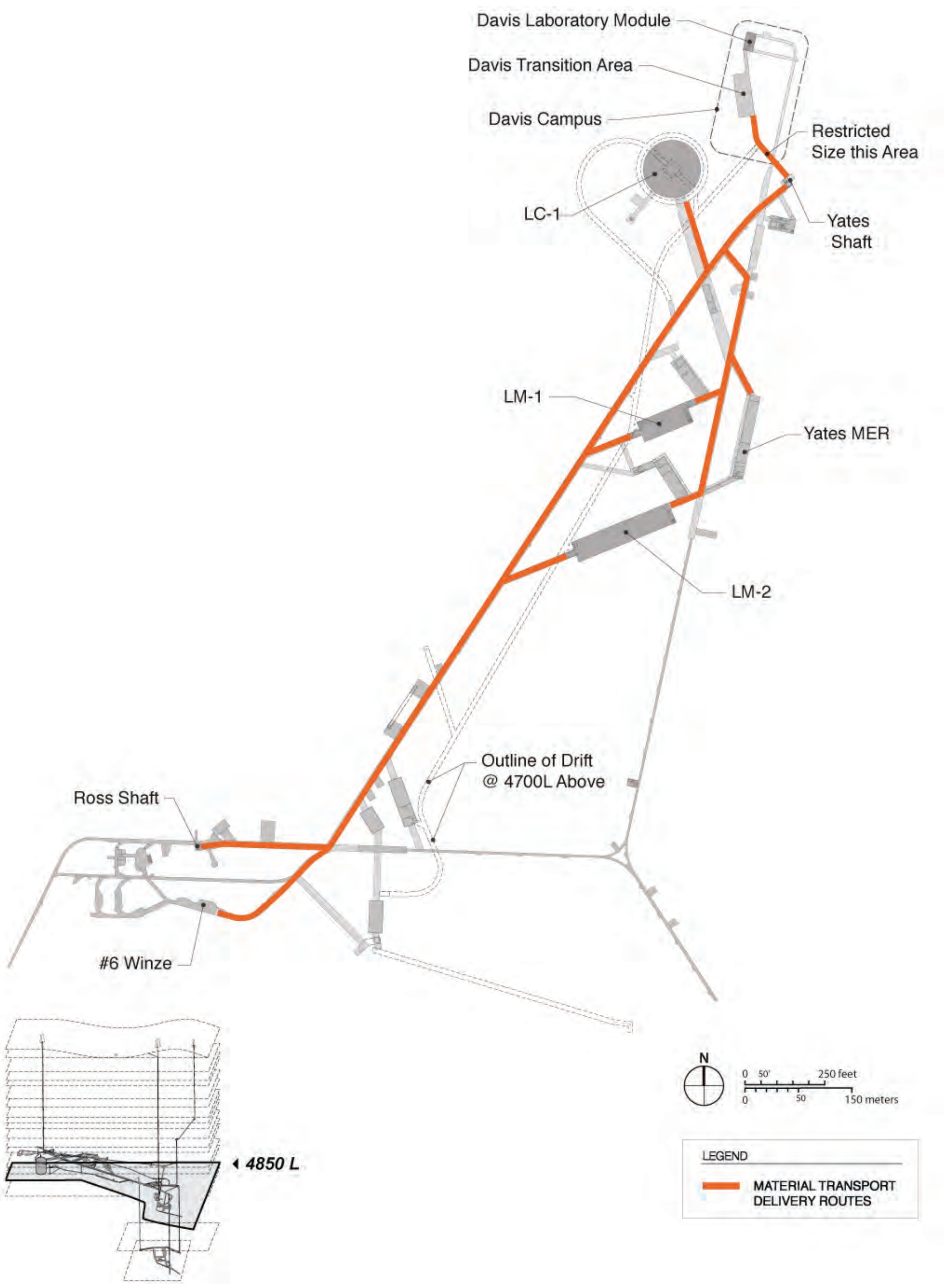

**Figure 5.4.3.7.2** Material transport delivery routes at 4850L. [DKA]

Transportation for people in the underground campuses is not expected to be a significant requirement. A study referenced in Arup's *UGI Basis of Design Report* (Appendix 5.L), Section 8.2, explains that given the distances between the shafts and the LMs, most people would choose to walk rather than use a form of transportation. Under this assumption, requirements for transportation systems will be primarily driven by the facility needs for parts and tools, transport of larger equipment, and emergency response needs. A form of transportation may be desirable for travel from the Yates to the #6 Winze to access the 7400L.

### 5.4.3.7.3 Material Handling and Personnel Transport Preliminary Design

Two design options considered for material handling included track and trackless systems. Trade Study #377 (Appendix 9.Y) captured the comparison of these options. Historically, HMC operations used a track-guided system to guide cars in the underground spaces. The proposed trackless system includes a guidance system embedded in concrete to prevent inadvertent contact with walls or utilities. This system uses vehicles known as automated guided vehicles (AGVs). Excavation and construction crews will remove the existing track on the 4850L and 7400L once their work has commenced, so either option requires an entirely new installation. The cost differential between the systems is substantial and the design recommendation moving forward is the AGV system on the 4850L and 7400L Campuses. The existing rail systems will remain in places for all OLR.

Materials and supplies outside of the main laboratory campuses and in remote locations will be serviced via the rail, much as it is currently. AGVs are mobile robots that sense floor-based markers (wires, tags, magnets) or can be controlled by radio frequency communication. The Unit-Load (see Figure 5.4.3.7.3-1) AGV is designed with on-board load-carrying accommodation; the load is directly controlled and therefore eliminates the need for a tow cart.

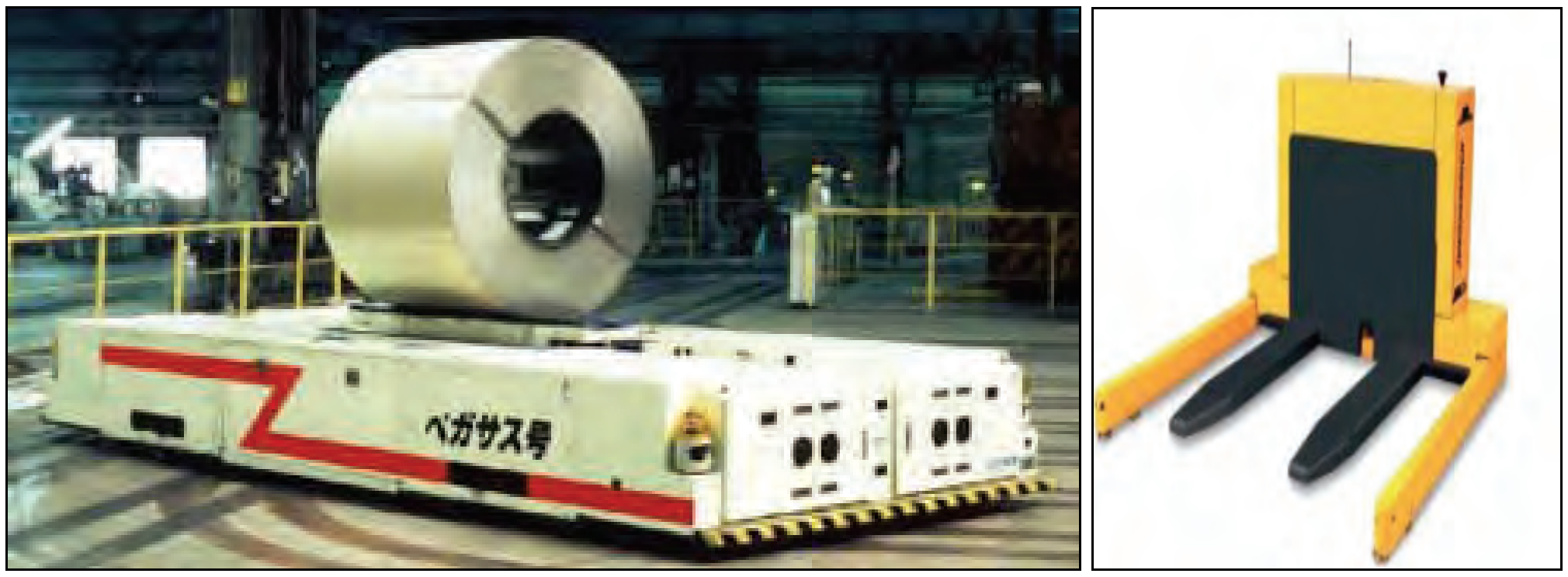

**Figure 5.4.3.7.3-1** Unit-Load AGV examples.

The Ross Shaft will be the dedicated artery for supplies and personnel during the construction and excavation phases of the Project. Once the Yates Shaft has been rehabilitated, it will be the dedicated science conveyance. Materials and supplies will be delivered to the Yates Warehouse and stored for inventory or transported directly to the Yates Headframe for distribution underground. To maintain flexibility, a custom pallet designed to the parameters of the Yates Cage will be used in conjunction with the Unit-Load AGV. Equipment and supplies will be limited to the dimensions of the cage and capacity of the hoist. Loads can be hung beneath the cage to increase the length possible for delivery to as much as 42 ft, but these loads must be delivered at a slower rate, and therefore this capability is of limited use. Figures 5.4.3.7.3-2 and 5.4.3.7.3-3 show some examples of loads that may be transported in the Yates Shaft.

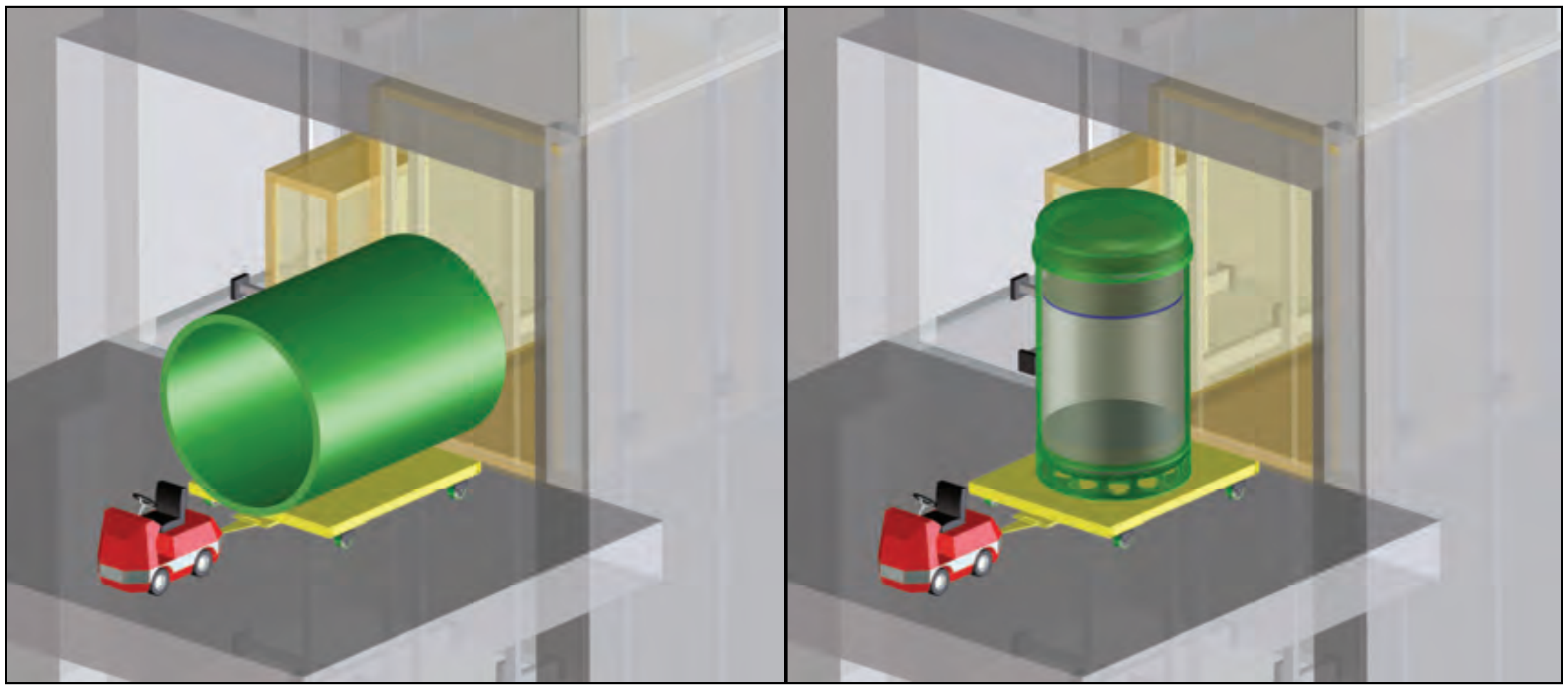

**Figure 5.4.3.7.3-2** Examples of large loads in Yates Supercage. [David Plate, DUSEL]

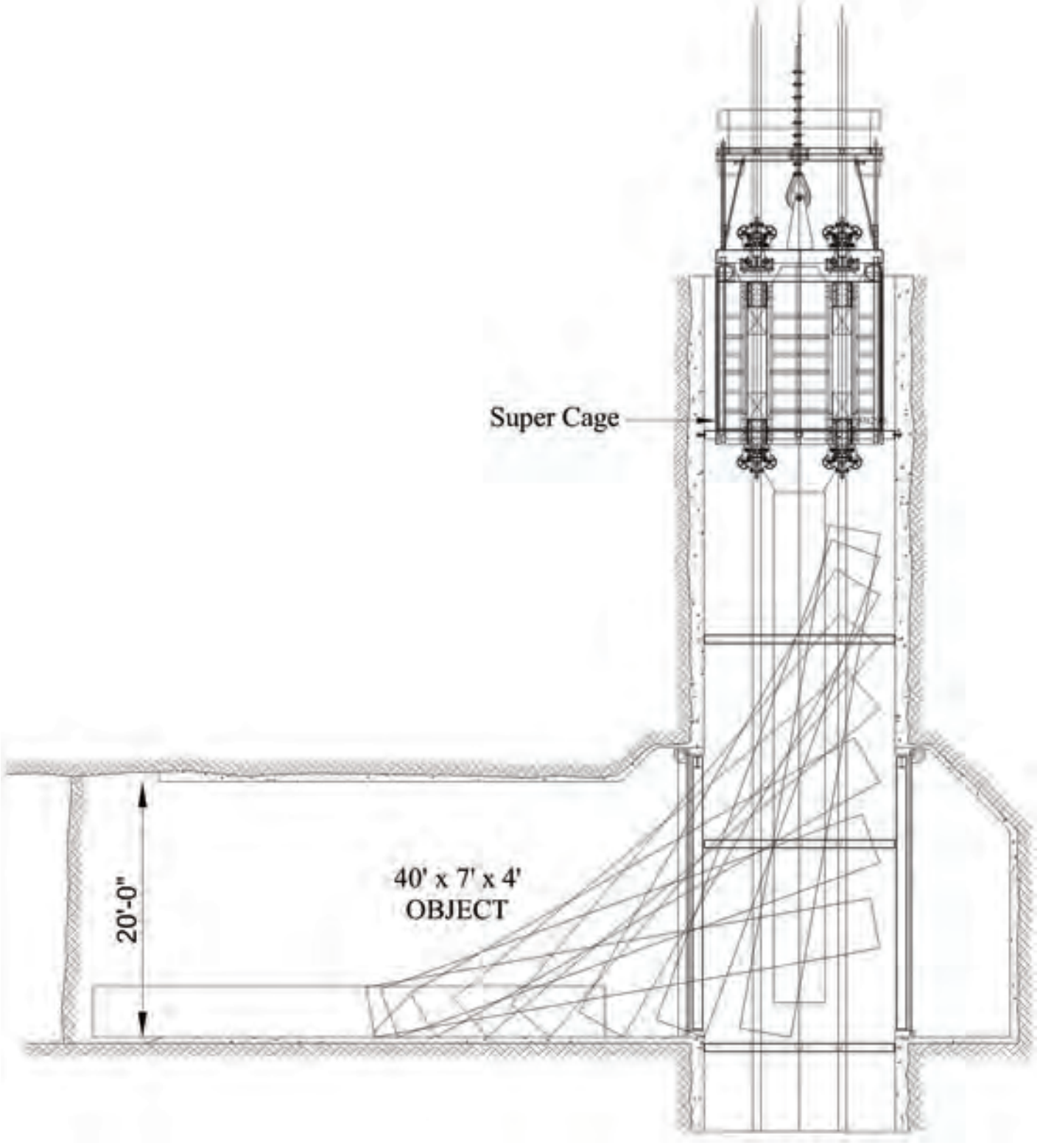

**Figure 5.4.3.7.3-3** Example of a long load at the 4850L Yates Shaft station. [Arup]

DUSEL operations and management will have to ensure the size of equipment and materials is compatible with the conveyances on the levels the materials will pass through. Scheduling deliveries and training personnel are both key to the success of the material handling program. DUSEL operations and maintenance personnel will be responsible for the operation of the material handling equipment.

The design of the AGV system is not fully developed at this stage of the design process. Various options exist for guidance systems, including embedding tags or wires in the concrete floors, sensors on the vehicle itself, a slot in the floor, or radio guidance. This technology is improving as artificial intelligence systems are developed, and it is reasonably anticipated that any system purchased for use at the end of construction will advance from what is currently available.

As previously discussed, the needs for personnel transportation are expected to be very limited, though may be desired for travel from the Yates Shaft to the #6 Winze. At the Preliminary Design stage, Arup included examples of equipment that could be used for personnel movement (Section 8.2 of Arup's *UGI Basis of Design Report* [Appendix 5.L]). A small fleet of personnel vehicles similar to those used in airports have been used as a basis for estimation purposes in the Preliminary Design.

### 5.4.3.8 Air Quality and Ventilation

The quality and quantity of breathable air in occupied spaces underground is critical for the safety and comfort of the occupants. Contaminants from equipment, tools, and people must be removed to the surface and replaced with fresh air to maintain a habitable environment. This section focuses on how the air is directed through the underground. More discussion on filtering the air for laboratory spaces in included in Section 5.4.3.13, *Chilled Water Systems*, and Chapters 5.6 and 5.8, *Mid-Level Laboratory Design at the 4850L (MLL)* and *Deep-Level Laboratory Design at the 7400L (DLL)*, respectively.

#### 5.4.3.8.1 Current Conditions of Air Quality and Ventilation

The current ventilation system for the Facility utilizes a similar overall design as was used during HMC operations. In general, fresh air is provided to the Facility through both the Yates and the Ross Shafts and exhaust air exits the Facility through the Oro Hondo and #5 Shafts. Figure 5.4.3.8.1-1 shows the ventilation flow path in place.

A key consideration in the development of this plan is to limit air entering the footprint through the old Open Cut mine workings. Open Cut mine workings can pose a threat for spontaneous combustion due to timber in the old stopes and pyrrhotite (Appendix 5.N) in the sand backfilled stopes. Figure 5.4.3.8.1-2 shows an SDSTA-installed Kennedy Stopping used to isolate the Open Cut workings from the ventilation footprint. This stopping is constructed of galvanized steel and has a door to allow passage and controlled airflow when travelling beyond the door.

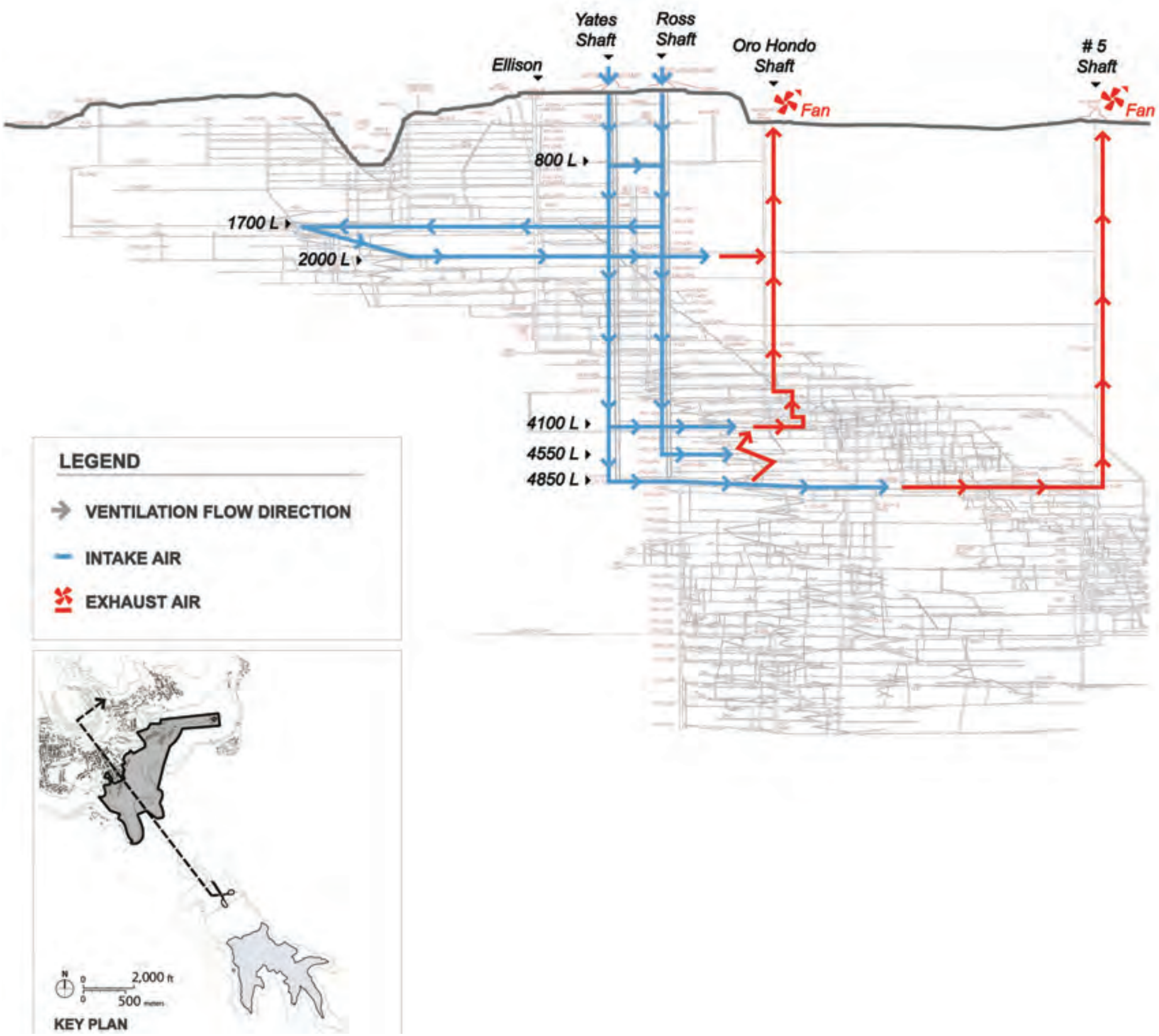

**Figure 5.4.3.8.1-1** Existing ventilation flow path. [DKA]

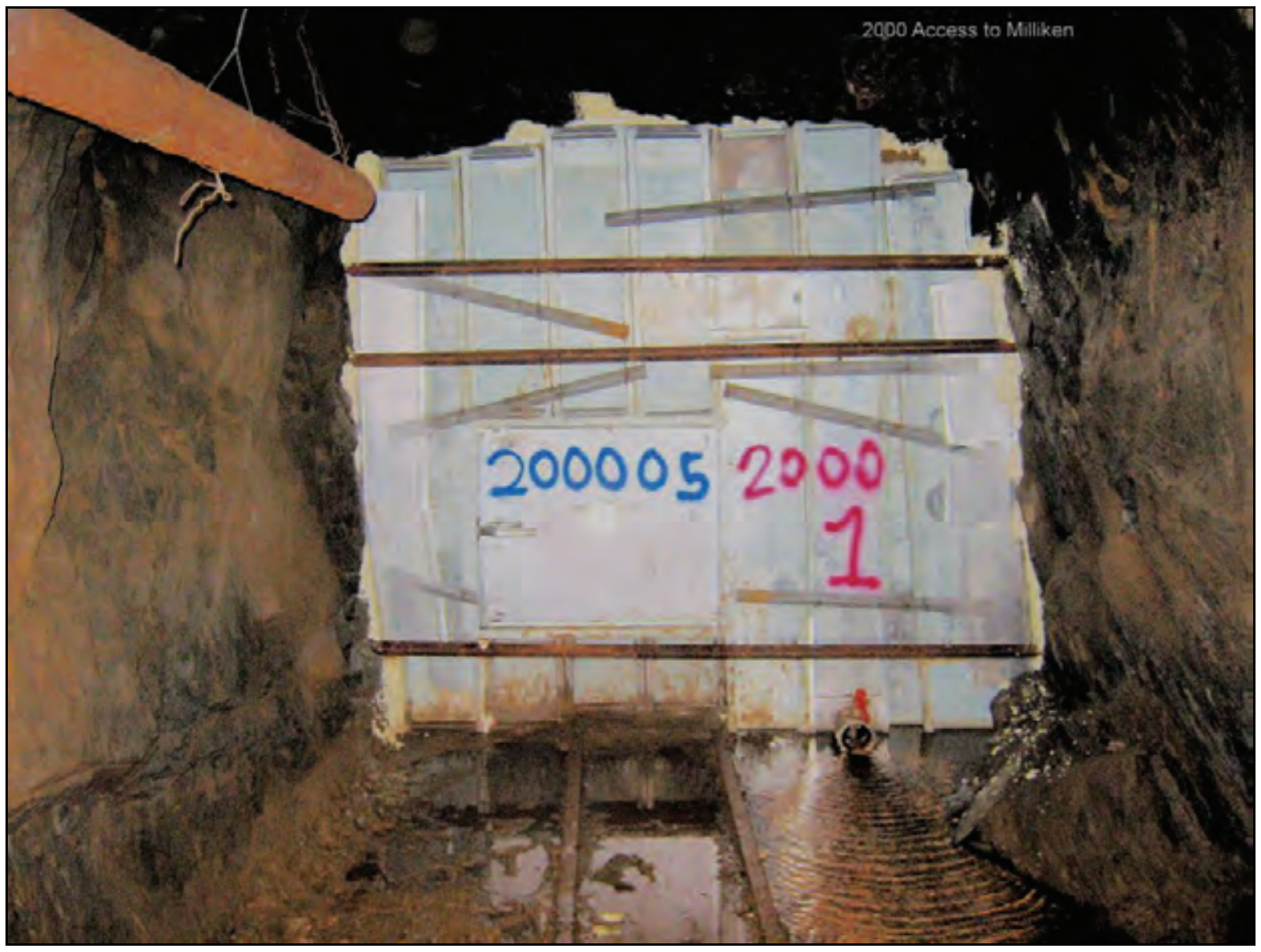

**Figure 5.4.3.8.1-2** Kennedy Stopping on the 2000L. [SDSTA]

The #5 Shaft will be used in the short term during re-entry and portions of the initial DUSEL construction. As discussed in Section 5.4.3.4, the condition of the #5 Shaft results in a partial restriction in ventilation flow. There is a risk that another failure could lead to a complete blockage at any time, but there are some advantages to using the shaft in its current condition, as it adds flexibility to the system. To mitigate the risk of complete blockage, plans are in place to proceed without the #5 Shaft prior to MREFC-funded Construction if it does fail, and the ventilation plan (baseline MREFC-funded design) does not include the use of the #5 Shaft.

Primary exhaust air is drawn through the Oro Hondo Shaft utilizing the existing American-Davidson 3,000 hp (2,237 kW) centrifugal fan. This fan is currently being operated at approximately 50% of design capacity, drawing approximately 225,000 cfm to sustain the Facility through re-entry and portions of DUSEL construction. A Spendrup 350 hp (261 kW) vane-axial fan with a variable frequency drive (VFD) has been added by the SDSTA to provide reversibility for emergency events. This fan can also be used as a backup to the centrifugal fan. Figure 5.4.3.8.1-3 shows a picture of the current fan configuration at the Oro Hondo Shaft collar. The system is flexible enough to adapt to changing requirements and allows for redundancy, with two fans at the Oro Hondo Shaft and one at the #5 Shaft. In conjunction with the main fans, there are a number of auxiliary fans in place underground for localized ventilation.

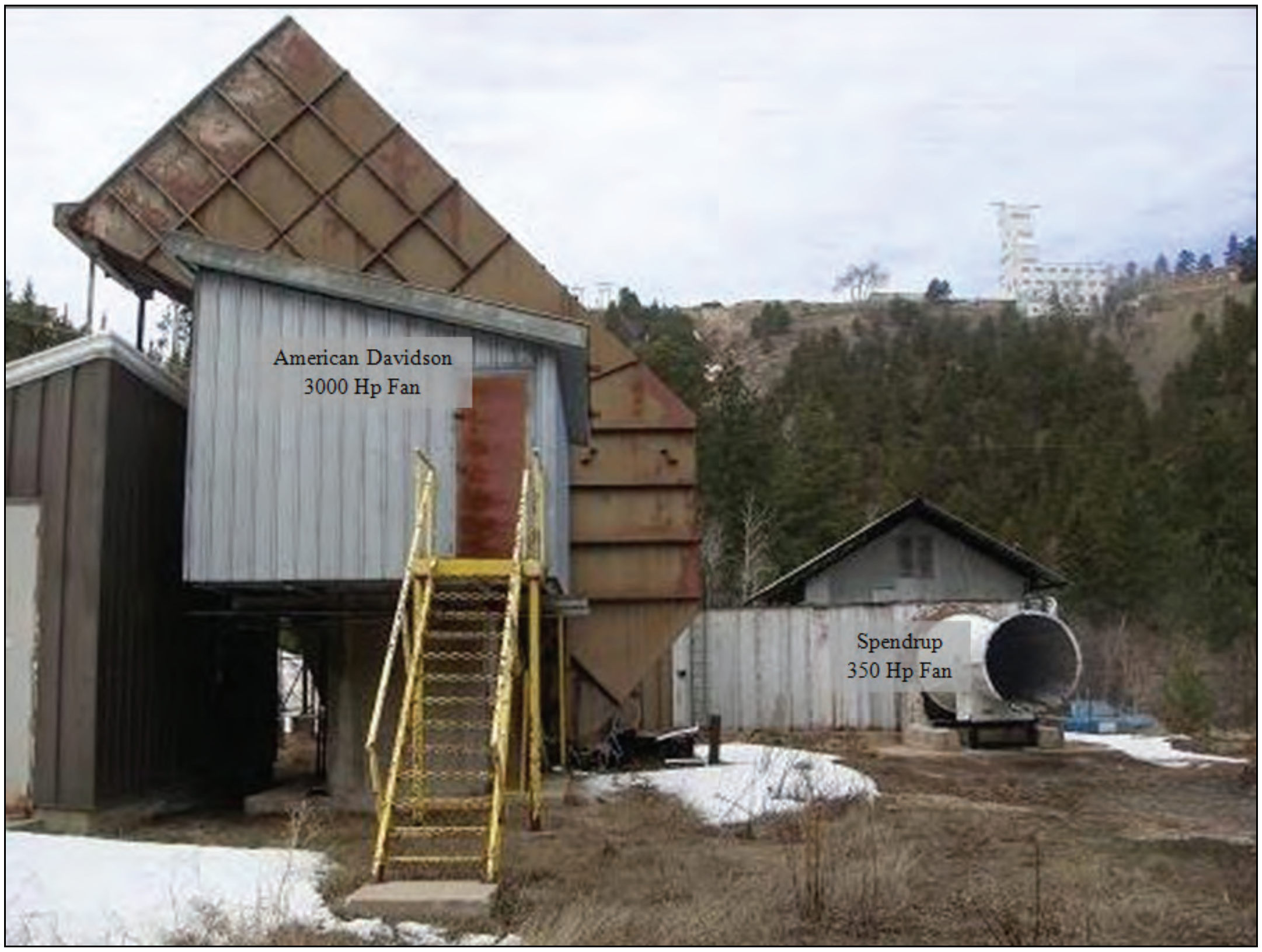

**Figure 5.4.3.8.1-3** Oro Hondo Shaft current configuration. [SDSTA]

#### 5.4.3.8.2 Air Quality and Ventilation Requirements

Air quality and ventilation requirements are very closely related. Providing temperate, clean air in areas both inside and outside experimental spaces requires sufficient volume of ventilation air to allow filtering, humidity control, and heat removal. Key drivers of these requirements for operation include occupancy, fresh-air exchanges per hours required for experimental spaces, and chilled-water system heat removal capacity. The ability to quickly clear smoke and/or hazardous gases is also included in the ventilation design requirements. Air velocity in ducts and drifts must be controlled to provide physical comfort in drifts and reduce noise in ducts. Air velocity in unoccupied spaces must also be controlled to reduce pressure loss, and therefore power requirements for the ventilation fans.

#### 5.4.3.8.3 Air Quality and Ventilation Preliminary Design

The design of the ventilation system for DUSEL will accommodate multiple stages of DUSEL construction and operation and expectations that early science experiments will continue to operate during DUSEL construction. A fully operating DUSEL Facility must have a robust ventilation system that provides adequate air volumes for emergencies such as smoke removal from a fire, or cryogen removal, as well as climate-controlled air through air conditioning and dehumidification. All work required for ventilation controls outside of the main campuses on the 4850L and 7400L will be completed using R&RA funds, leaving only those items on the main campuses within the MREFC budget.

The air quality and ventilation systems will have seven phases of construction. These phases are described in detail in Section 12 of Arup's *UGI Basis of Design Report* (Appendix 5.L). Figure 5.4.3.8.3-1 shows Phase 7 of the ventilation plan once laboratory facilities are in operation.

Design of ventilation system phases will support the varied demands of early science and laboratory construction. During construction, contaminants will have to be removed from the drifts, ramps, raises, etc. Some of these contaminants include dust, diesel particulate matter, blast fumes, and heat load. Numerous auxiliary ventilation systems will be used to provide localized air during the construction process. Air doors will be strategically located throughout the Facility to minimize leakage and provide safe separation of fresh and contaminated air.

The long-term ventilation concept will utilize an overhead drift at the 4700L to remove routine exhaust as well as smoke and other contaminants in the event of an emergency. The DLL will be provided with fresh air from the #6 and #8 Winzes and exhaust through a new borehole to the 4850L to 3950L borehole (Oro Hondo system). This system provides for two separate fresh-air sources on both the 4850L and 7400L.

Control of the air flows in each ventilated space is achieved through use of doors, dampers, and regulators. Strict control of the flow settings will be maintained by the Facility to ensure proper ventilation. An example of an air door can be seen in Figure 5.4.3.8.3-2. The ability of the door to swing in opposite directions on each side allows it to be opened with ventilation pressures. Once the Facility has been established and is in operation, these systems will be set in an optimum semi-permanent pattern (open, closed, or closed with a damper to allow some flow). Control of these systems is needed for emergency events such as a fire or cryogen release to isolate the impacted areas and provide a fresh air path for egress. The entire ventilation path can be reversed if needed for this type of emergency. The ventilation system is connected to the standby power system to ensure operation even with loss of power.

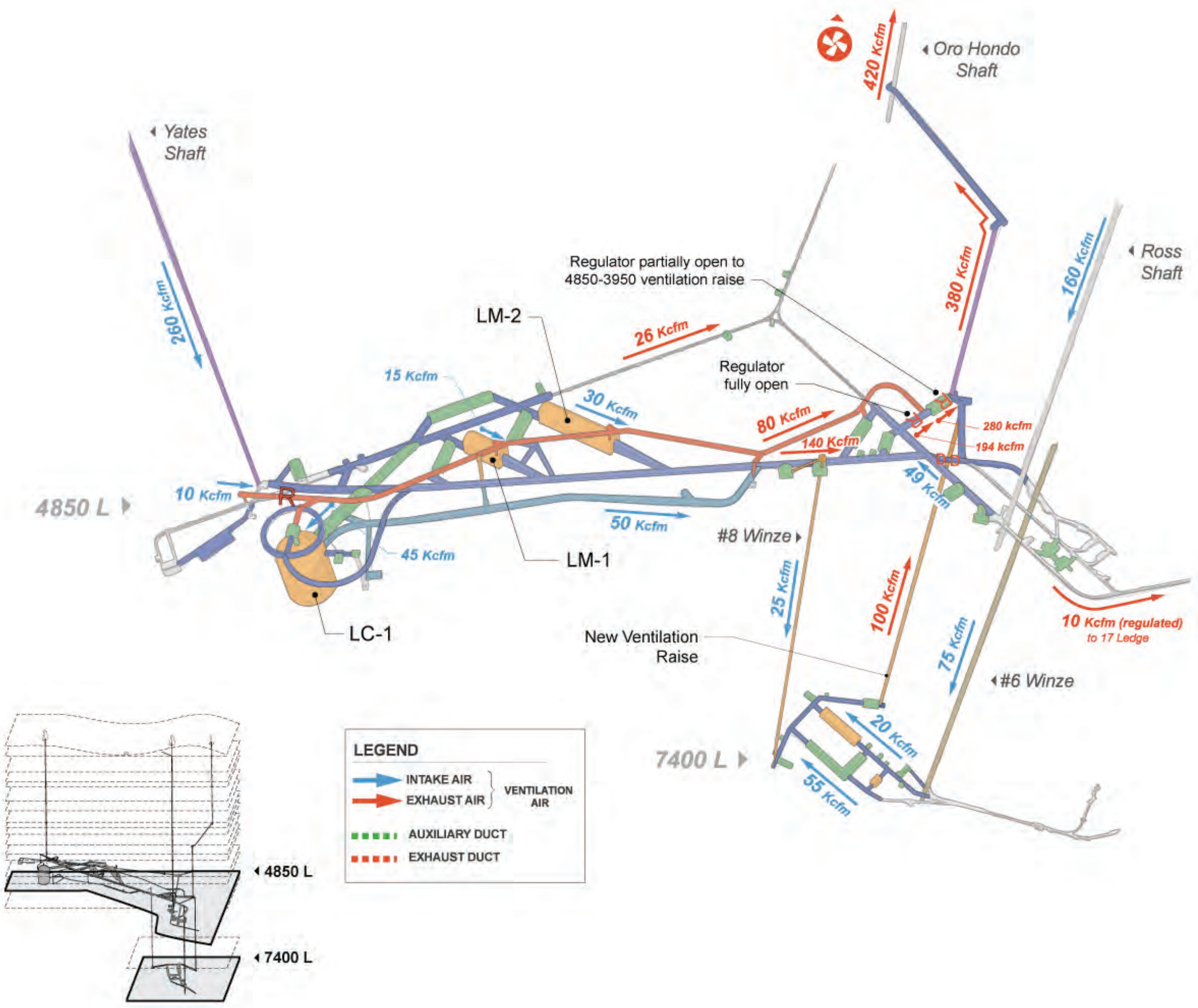

**Figure 5.4.3.8.3-1** Phase 7 of the ventilation plan. [DKA]

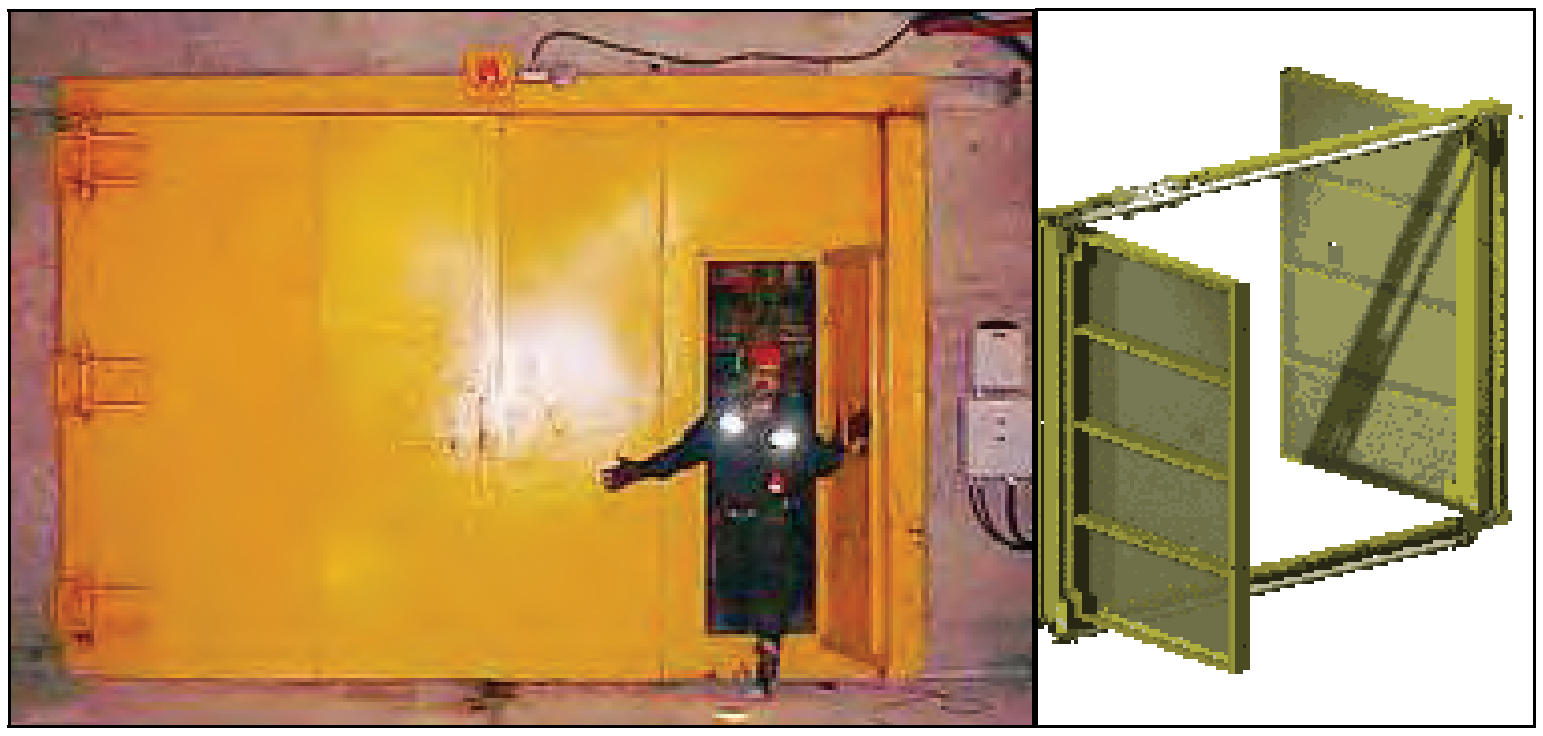

**Figure 5.4.3.8.3-2** Example of an air door and how it operates. [Minedoor]

## 5.4.3.9 Waste Rock Handling and Disposal

Excavations to create underground laboratory spaces will generate large volumes of waste rock material that must be removed to the surface. Unlike all other components of UGI, the waste rock handling system will only be necessary during construction. Despite the limited duration of use, the system is still designed for a long life to provide the ability to support future expansion.

##### 5.4.3.9.1 Current Conditions of the Waste Rock Handling and Disposal

HMC managed rock, or ore, from three separate locations. Underground material was brought to the surface using skips in both the Ross and Yates Shafts. At the headframe of each shaft, the material was crushed to a nominal ¾ in, passed through ore bins, and was transported via underground rail to the mill system. The underground rail passed through a level called the tramway at approximately 125 ft (38 m) below the collar of the Yates Shaft. The third supply of ore was the Open Cut, where material was transported with haul trucks to a surface crushing system. A pipe conveyor (the longest in the world when it was constructed in 1987) delivered the material overland to the mill system.

During construction, DUSEL will remove the excavated waste rock material from the underground for disposal, with no intention of further processing. The Yates Shaft will primarily provide science access and will be rehabilitated during a significant portion of construction. The Ross Shaft will be the means of removing of material from the underground during DUSEL construction.

The Ross skipping system allows material to be transported at a rate of 3,300 tons per 18-hour day, allowing six hours of downtime for maintenance, breaks, shift changes, etc. The loading pocket at the 5000L for the 4850L will be cleaned of any accumulated sand during the skip pocket rehabilitation prior to MREFC-funded Construction. The loading system uses a chute with two gates to meter a full load into the skips. This chute was underwater and will require rehabilitation, including cleaning, new liners, and new gate cylinders. At the top of the headframe, a scroll opens the bottom of the skip and allows material to fall into a small bin. Both the scroll and the bin appear to be in serviceable condition, but the design and cost estimate include minor repairs to the system. The skips are expected to require replacement twice during construction, based on equipment life that was experienced during historic production. At the discharge of the skip dump bin is an existing chain feeder, using large chains to hold the material in the bin. The chains are in continuous loops around a drive system to control the flow from the bin to the primary gyratory crusher. Below this crusher is a short belt conveyor feeding a secondary cone (a.k.a. standard) crusher. A belt magnet is suspended above the conveyor to remove tramp metal and protect the remainder of the system. The standard crusher feeds onto another conveyor, transporting material to a vibrating screen. This screen will be removed to allow material to pass directly into the existing 5,000-ton fine-ore bin (FOB). The system includes a large dust collector with ducts to each transfer point.

While both crushers are in serviceable condition, the budget estimate assumes complete rebuilds of these crushers, which will ensure reliable operation during construction. The electrical service equipment is outdated and will be replaced. The belt conveyors will have all belting and most idlers replaced. The dust collector will have all bags replaced and the ducting will be reworked to optimize effectiveness.

The base of the FOB is at the tramway level (~125 ft [38 m] below the Ross Shaft collar). Six discharge gates allow for continuous loading of rail cars. These gates will be replaced, as they corroded beyond repair. The existing cars are not large enough to meet the cycle times required for construction, but the axles and wheels can be reused with new bodies. New locomotives will be purchased by the Project using R&RA funds. At the point where the tramway exits the underground, the existing steel-sided building is in disrepair and will be replaced. All other equipment associated with this material handling system, including the original pipe conveyor, has been removed from the site.

Waste rock removal from the 7400L will utilize the skipping system that is in place in the #6 Winze. This winze is underwater after mine closure, so rehabilitation is anticipated. The loading pocket formerly used is located at the 8100L and is not practical for future use with most development occurring at the 7400L. A new skip pocket will be installed at approximately the 7500L. Waste rock skipped up the #6 Winze will be dumped into the existing waste rock bin that feeds the Ross Shaft loading pocket. From there, the existing passage allows for a direct transfer of material from the skip dump for the #6 Winze to the Ross skip system. This waste pass system between the 4600L and 5000L is expected to be in serviceable condition and will be tied into the new loading pocket at 7500L. Figure 5.4.3.9.1 depicts how the waste pass system works.

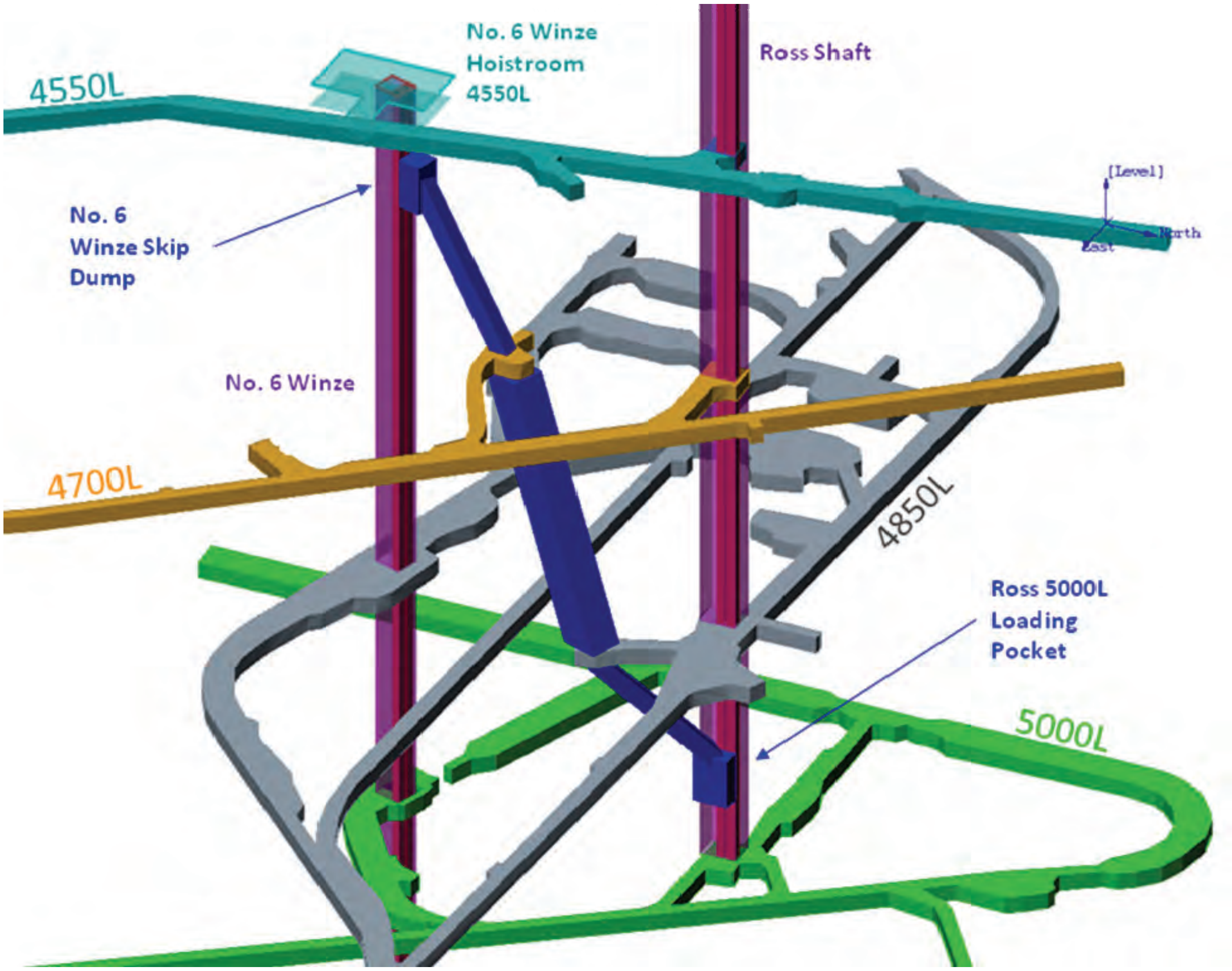

**Figure 5.4.3.9.1** Depiction of existing waste pass system. [SRK]

### 5.4.3.9.2 Geochemical Analysis of Waste Rock and Acid Rock Drainage (ARD)

Very limited geochemical characterization data from Homestake operational period is available for the proposed excavation rock (Yates Formation), as it is not spatially associated with the gold-bearing Homestake Formation. Consequently, new geochemical characterization of the waste rock is important to determine the rock’s interaction with the environment. Specifically, the Project needs to understand whether the rock is chemically reactive in the presence of water and oxygen, and if so, how would this affect its ultimate disposition? Geochimica Inc. performed an analysis captured in a technical memorandum to the SDSTA (Appendix 5.N).

Geology, mineralogy, and representative sampling are critical to establish the basis for a geochemical interpretation of waste rock. Detailed geology was performed on the waste rock by Geochimica and the SDSTA geologist. Detailed mapping of the East and West Main Access drifts was performed along with the careful examination of drill core created for the geotechnical evaluation of the waste rock. The geologic units of these areas and core project into the anticipated waste rock excavation zones. A total of 33 samples were collected and are representative and proportionate to the identified lithologies. Ten of those samples were examined in detail by optical and X-ray diffraction analyses to confirm the nature of the sulfide and carbonate minerals, as well as to confirm lithologic descriptions. The samples were sent for analysis to ACZ Laboratories, which is well known for its QC/QA and expertise in this type of work.

Five geochemical tests were performed on each sample:

1. Trace element chemistry
2. Static acid-base accounting (modified Sobek methodology—this is a standard North American acid-base accounting method)
3. Determination of carbonate neutralization potential by analysis of total inorganic carbon
4. Direct measurement of acid-base potential by the net acid generation (NAG) test
5. Analysis of dissolved metals and anions in effluents from the NAG test to evaluate potential mobility of trace elements if sulfides were permitted to oxidize quantitatively

The tests were designed to:

a) Determine if portions or all of the waste rock likely would generate acidic water when in contact with storm water and oxygen.
b) Determine if there is sufficient acid-neutralization capacity in minerals in the rocks to neutralize potential acid generated.
c) Determine total metals from the rocks
d) Determine potential leachable metals (under oxidized conditions) from the rock.

The results of the geochemical study indicate that approximately 30% of the rock, primarily sulfide-rich, rhyolitic intrusives, will potentially generate acid drainage. Only 5-10% of this rock is somewhat likely to produce significant acidity, the other 20-25% is expected to generate only very low acidity. The remaining 70% of the total rock expected to be mined is not acid forming and contains significant carbonate able to neutralize potential acid formation.

The trace metal analysis for the 33 samples shows concentrations (including selenium and arsenic) are at or below concentrations that exist in the rocks of the Open Cut. This favorable result indicates that trace metal loading to the mine water will be no more than what currently exists.

The NAG procedure confirms the standard North American acid-base accounting results and provides further evidence that trace metals from the waste rock are unlikely to impact mine water quality and potential permit compliance. The waste rock samples that were leach-tested were crushed (< 75µm) and oxidized by solution of hydrogen peroxide to maximize their acid-forming potential and surface-area for leaching of metals and metalloids. The leachate from this extraction was analyzed for trace metals and again confirmed that trace metal loading to mine water will not be not be significant and impact permit compliance, given that the disposal will be within the Open Cut.

The work performed by Geochimica validates what is observed in practice by other rock units found in the mine area. Sulfide minerals (primarily pyrrhotite and pyrite) oxidize, creating soluble sulfate, iron, and small amounts of acid. The acid is neutralized by carbonates leaving the sulfate in solution with any other metals that do not precipitate with the increase of pH to 7-8 s.u. Calcium, potassium, and sulfate typically are soluble in the pH-neutral drainage water and are primarily responsible for creating high total dissolved solids (TDS) levels.

The TDS levels and the presence of metals in this drainage make this water nondischargeable without treatment under National Pollutant Discharge Elimination System (NPDES) rules for new permits. Consequently, this water will need to be captured and treated. The Open Cut is an ideal place to capture this water. Compared with impacts from the fractured rock of the Open Cut and between the Open Cut and the water level in the underground facilities, the incremental mass loading from the small amount of new rock from the underground excavation will have minimal impact on water quality in the groundwater. Further, all Open Cut water drains to the underground facility and is captured through an existing pump and treatment system discussed in Section 5.4.5.10. The treated water discharges through an existing NPDES permit that has no limits for TDS.

#### 5.4.3.9.3 Waste Rock Handling and Disposal Requirements

Approximately 1.65 million tons of waste rock will be excavated to create space for the facility construction at both the 4850L and 7400L. A detailed summary of each excavation space volume is included as part of Golder Associates Inc.'s *PDR Final Report, DUSEL Excavation and Design Services Contract* (Table 4.1) (Appendix 5.I). The waste rock from the excavation will be relocated to the Open Cut via an overland conveyor, similar to one used during HMC operations as described in the next section, and the design team has been mindful of the impact this activity may have on the local community. The design will accommodate more stringent noise and dust requirements than other portions of the Project may require. In an effort to limit public exposure to this process, all material will be transported through residential areas only during a 10-hour daytime period, which requires a higher design capacity than a 24-hour operation would allow. A limit of 45 dBA at the property boundary has been established to further minimize the public impact. Extreme weather conditions experienced in Lead, South Dakota, must also be considered in the development of design requirements.

#### 5.4.3.9.4 Waste Rock Handling and Disposal Preliminary Design

Multiple options were considered through the design process for disposal of waste rock material using different disposal sites, transportation methods, and transportation paths. Based on the quantity of material to be removed, the Open Cut was selected as the most viable location for final disposal. This location has an open permit for this purpose, eliminating this risk from the Project. In addition to a low environmental impact, the Open Cut also has the lowest impact on the community, as it is isolated and already perceived as being part of a mine site.

Once the final disposal site was identified, two primary methods for transportation were considered: trucking and conveying. A detailed cost analysis of reasonable routes for these options was performed by SRK and Arup with the result showing a lower overall cost for the conveying option. This option also reduces the community impact that would be experienced with as many as 130 round trips per day using trucks. This analysis was included as part of the 60% Basis of Design Report[21] from Arup (cost estimate, Appendix L). Further details comparing other aspects of these options are captured in Trade Study #379 (Appendix 9.0).

A final consideration to define the design was to determine the most appropriate conveying technique. As previously mentioned, HMC operations used a pipe conveyor to transport material from the Open Cut to the mill. This conveyor was removed after HMC operations ceased. This type of conveyor has unique attributes that make it ideal for the application. It uses belting similar to common conveyors in industry, but rolls this belting into a tube shape overlapped at the top of the tube (Figure 5.4.3.9.4-2). This creates a seal to prevent dust and spills and allows the conveyor to turn both horizontally and vertically to more closely follow the hilly terrain. The turning capability eliminates the need for transfer points between straight-line conveyors, limiting dust, maintenance, and noise.

#### Material Sizing

A description of the existing waste rock handling equipment to be reused was discussed in the current conditions section above. Because the excavated material only needs to be disposed of, and not processed, the ultimate size of the material is dictated by the limitations of the transportation equipment. The most limiting equipment for this Project is the pipe conveyor. In this case, a nominal 4 in (102 mm) size has been defined. This is small enough to maintain a relatively small conveyor, but large enough that crushing energy requirements are low. The primary crusher has the capability of producing 4 in (102 mm) minus material, but the secondary crushing process will produce a more consistent sizing and reduce potential problems with the pipe conveyor.

Dust control for the crushing system will be managed through modifications to ductwork in the existing dust-collection system. A pulse-jet dust collector will capture dust from the crushers and conveying transfer points, and reintroduce the dust into the fine ore bin. The fan for this dust collector is located on the opposite side of the headframe building from the air intake, preventing dust infiltration to the underground if the filters fail. In addition, walls and doors are in place between the crushing system and the shaft to prevent any dust not captured by the collection system from being introduced into the underground air intake. These walls and doors also isolate the noise generated by the crushing system from the common travel paths to the shaft conveyance.

#### Transport to Open Cut

Once the material passes through the existing crushing system(s) and fine ore bin, it will be loaded into a set of 20 new rail cars designed to carry 10 tons per car. Two new 15-ton battery-operated locomotives will pull these cars through the tramway a distance of over 2,300 ft (700 m). Spare batteries will be kept charged to ensure that the locomotives can operate a full 10-hour shift every day. The tramway rail will be realigned with new ballast. Some ground support upgrades are also included to ensure stability for the Project duration. Where the tramway comes out of the ground, the existing building will be rebuilt for the length required to allow the train to pull completely past the dumping station and empty the cars. An existing maintenance room will also be rehabilitated in this building.

At the point where the tramway daylights, a new dumping station will be excavated to allow two cars to empty at the same time, as shown in Figure 5.4.3.9.4-1. A surge bin ensures consistent feed onto the pipe conveyor. A typical pipe conveyor can be seen in Figure 5.4.3.9.4-2. Experience described by former HMC employees as well as more recent owners of pipe conveyors has shown that inconsistent material flow is the most frequent cause of pipe conveyor issues. The bin will be steel sided inside a large concrete retaining area. This 300-ton bin will be designed for mass flow, with the entire bottom live—i.e., a belt feeder will seal the entire bottom length and width of the bin. A short belt will transfer the material to the pipe conveyor, where material is transported approximately 1,700 ft (520 m) to the edge of the Open Cut. The steep walls of the Open Cut allow material to cascade to the bottom with limited fall height while still providing sufficient space to store the entire volume required (Figure 5.4.3.9.4-3). The entire route of the waste rock handling system from the Ross Shaft to the Open Cut is shown in Figure 5.4.3.9.4-4. A preliminary analysis of the geotechnical stability of the Open Cut rim at the discharge location determined that the site is suitably stable for this purpose. More information about this system and geotechnical analysis can be found in Section 15.10 of the Arup *UGI Basis of Design Report* (Appendix 5.L).

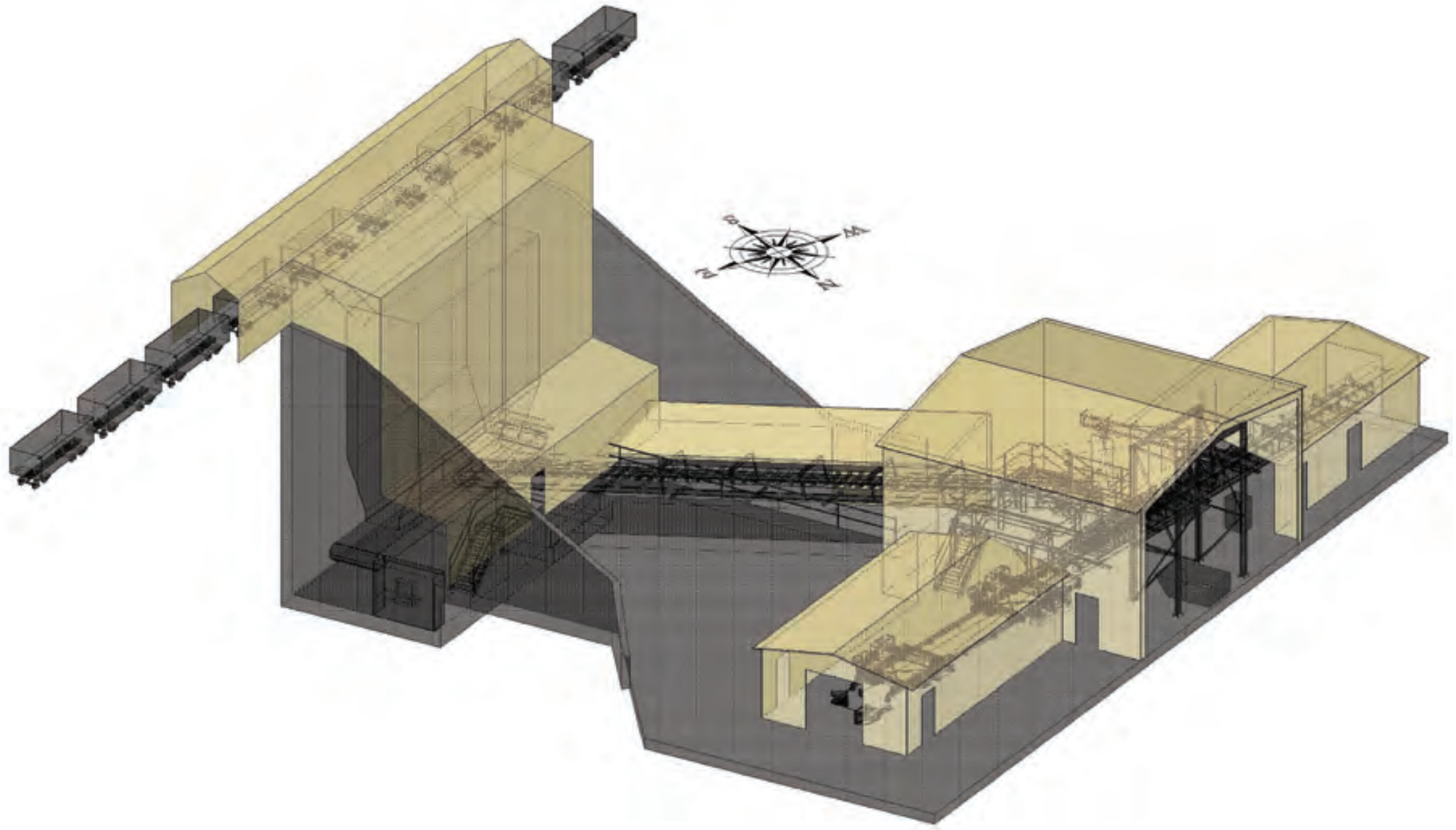

**Figure 5.4.3.9.4-1** Car unloading station. [SRK]

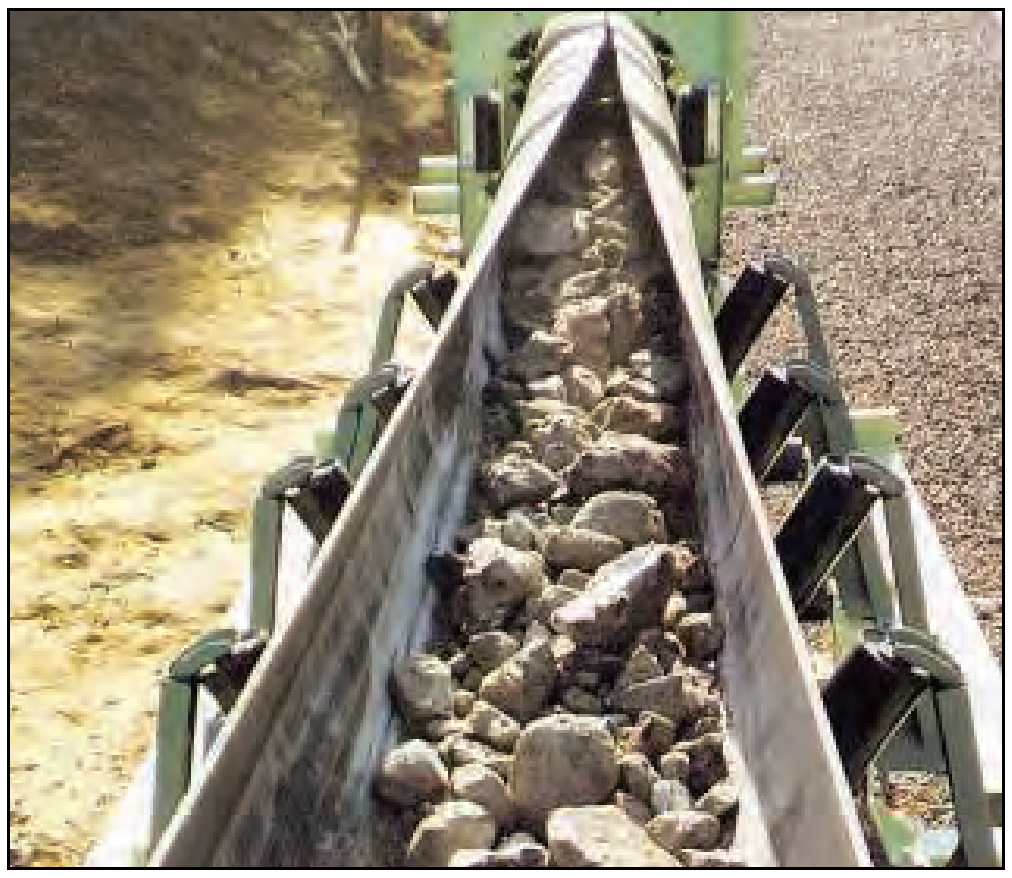

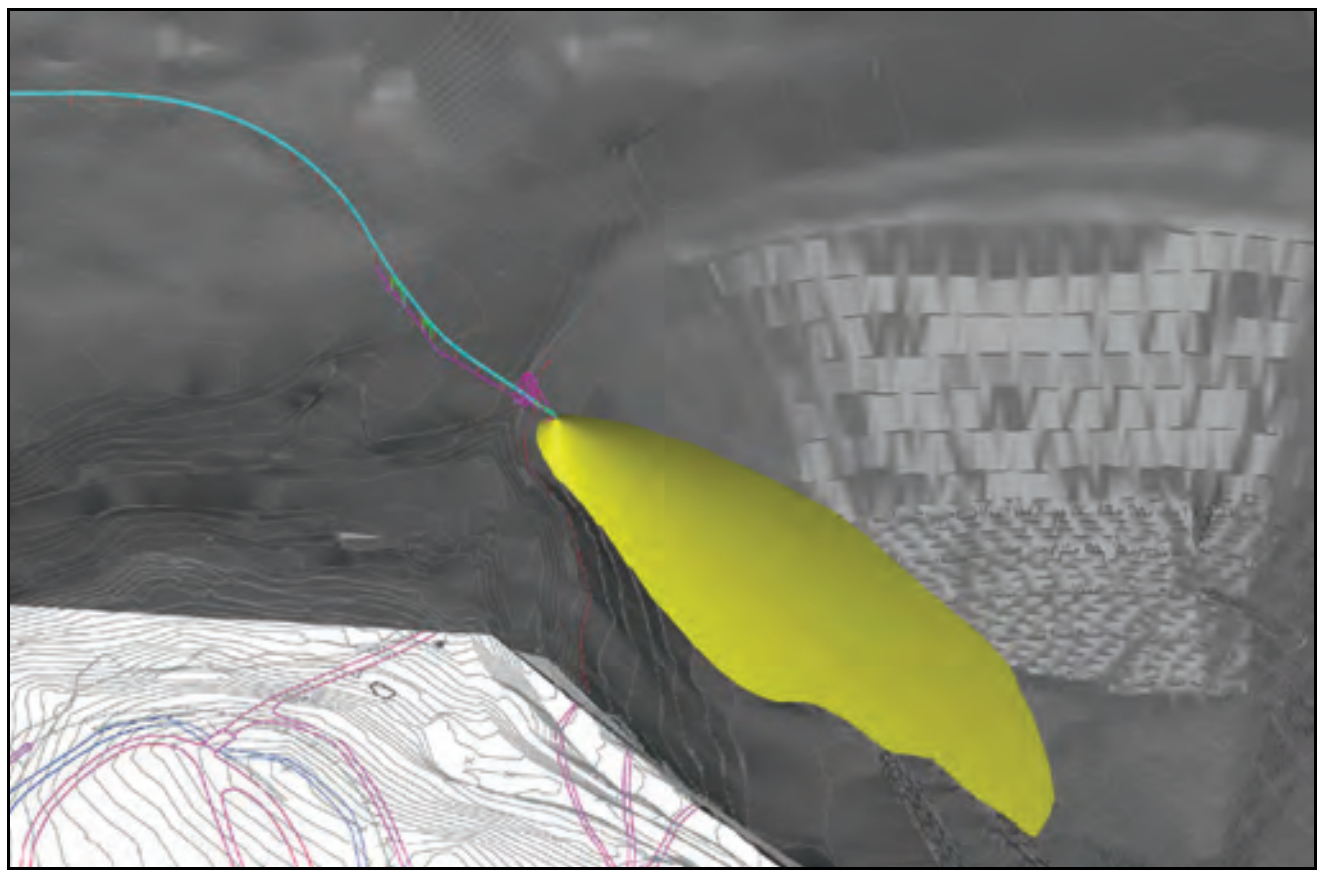

**Figures 5.4.3.9.4-2** Typical pipe conveyor. [SRK] **5.4.3.9.4-3** Excavated volume in Open Cut. [SRK]

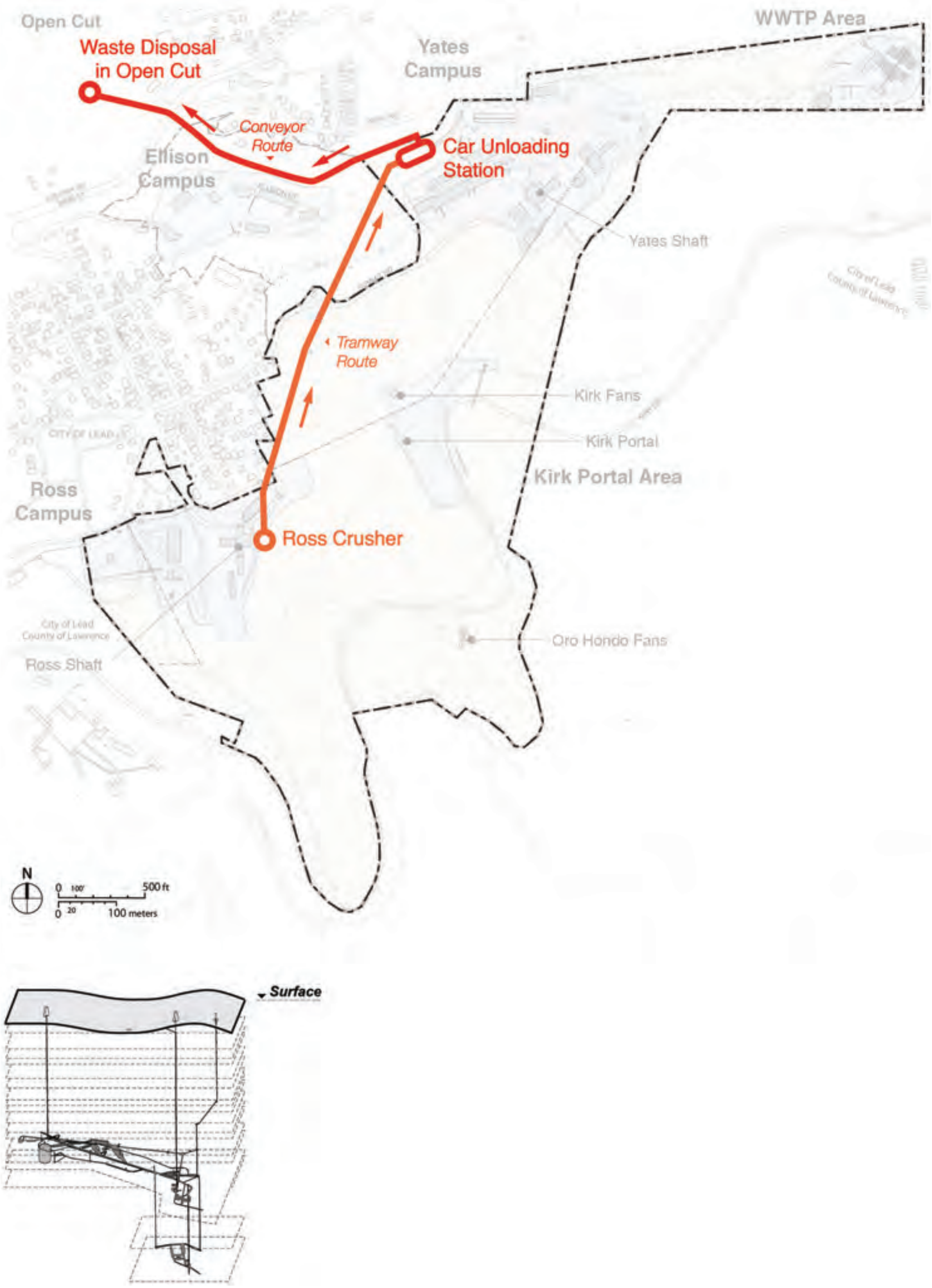

**Figure 5.4.3.9.4-4** Waste Rock Handling System route. [DKA]

### Controls

Several controls are included in the waste rock handling system design to protect both the equipment and the community. The existing belt magnet previously described provides a first defense against belt damage due to rock bolts, loader bucket teeth, etc. Prior to the pipe conveyor rolling into the pipe configuration, an additional magnet followed by a metal detector will catch both ferrous and nonferrous metals and shut down the system before damage is done. A scale on this belt protects against over- or underloading the conveyor, preventing issues experienced with similar conveyors. Standard safety controls, including pull cords, drift switches, zero-speed switches, and guarding provide further protection for both the equipment and operators. A full building enclosure around the car dump, surge bin, and pipe conveyor feeding point will contain noise and spills, should they occur. The entire length of the pipe conveyor will be enclosed and fencing will be provided to eliminate public access. Figure 5.4.3.9.4-5 shows a depiction of what the conveyor may look like as it passes over Main Street in Lead and into the Open Cut. A combination of dust collection and suppression will ensure that all environmental standards are met or exceeded. The Facility Management System (FMS) will create interlocks to limit the potential for human error. More discussion on the FMS can be found in Chapter 5.5, *Cyberinfrastructure Systems Design*.

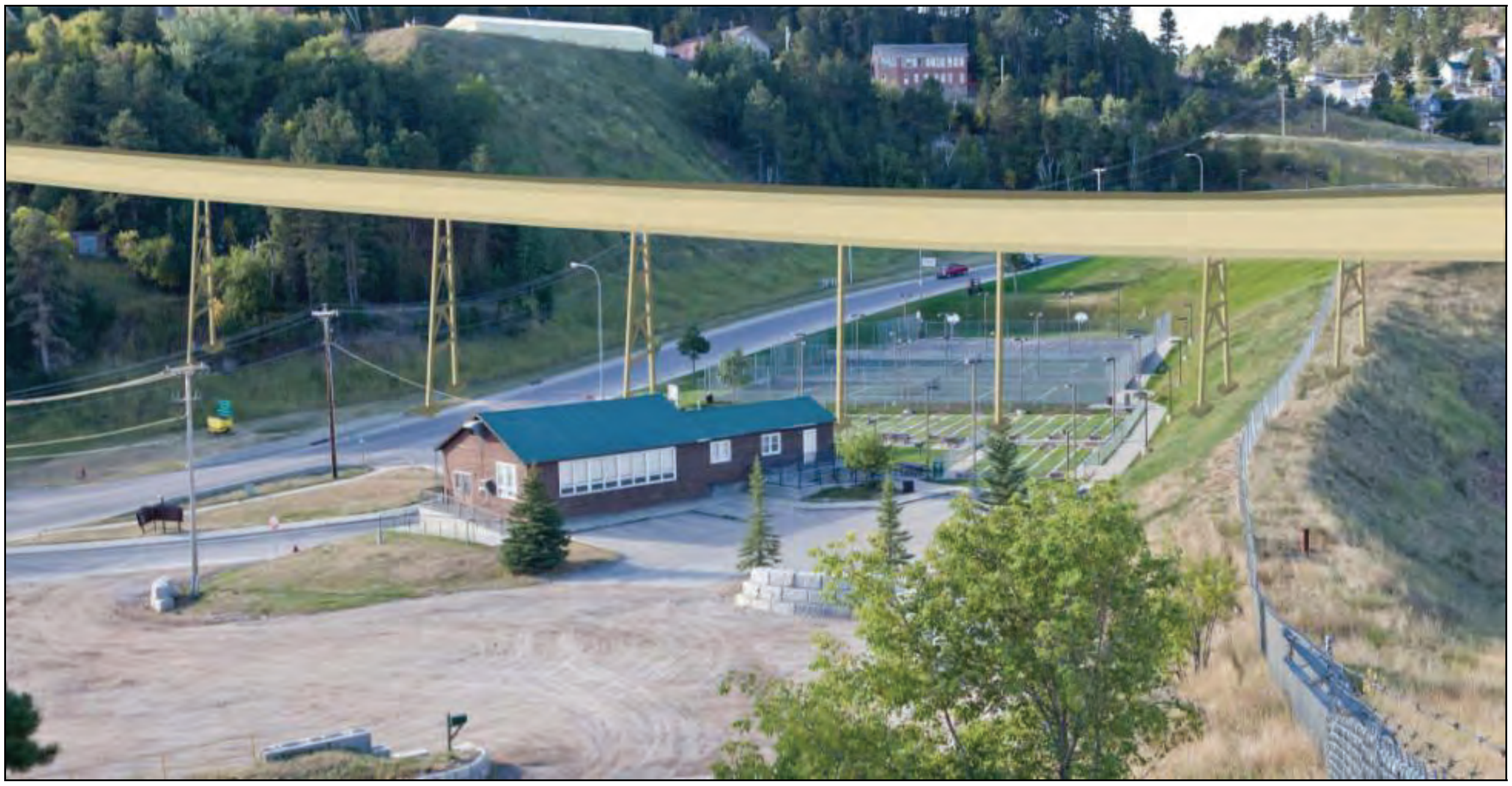

**Figure 5.4.3.9.4-5** Depiction of what the pipe conveyor will look like to the community. [SRK]

**Schedule**

Installation of the waste rock handling system is a critical item in the schedule for underground development. What little space is available underground for material disposal will be filled by early excavation activities occurring outside of MREFC funds. To facilitate the development of the MREFC-funded Project, many of the activities related to this system have been transferred to R&RA budgets. All refurbishment activities, including the crushers, conveyors, and tramway, will be performed by the SDSTA Operations staff with support from manufacturers' representatives. Purchase of the locomotives and rail cars will be done with assistance from the Construction Management contractor, through R&RA funding. All civil work will be completed in 2013 to ensure that weather does not prevent this work in early 2014, when MREFC funds are expected to be available.

### 5.4.3.10 Electrical Power Distribution

Electrical power distribution is a key infrastructure component throughout the DUSEL Facility. Due to the physical separation and functional differences between the underground spaces, each LM, LC, AoR, and ancillary space will be treated as an individual facility as if the campus were located on the surface.

Each major area will contain its own electrical distribution system and services and will function independently of the other facilities.

#### 5.4.3.10.1 Current Condition of the Electrical Power Distribution

As the Project site was historically an operating gold mine, robust electrical infrastructure was in place to operate energy-intensive systems to crush, grind, and process gold ore. Upgrades to the system were completed throughout HMC operations, providing the Project with a solid foundation from which to build a new infrastructure to support laboratory development. The discussion below focuses on the infrastructure that will be used and/or modified as part of the Project.

**Distribution**

The East substation, located near the Foundry Building, originally supplied power to the Surface Facility, the Yates Hoists, headframe and compressors, the WWTP, and underground power via the Yates Shaft. Due to chronic problems with buried 69 kilovolt (kV) feeders to the East substation, it was decommissioned as a primary substation and is used for 12kV distribution only. The 69kV transformers were removed from this substation and used to upgrade the Ross and Oro Hondo substations. The East substation is now fed at 12kV from the Oro Hondo substation and distributes power to the Surface Facility, Yates Hoists, and the WWTP. The Yates Compressors are no longer in use and have been disconnected. Two existing 12kV feeders extend from the East substation down the Yates Shaft and were used by HMC. The cables have been tested, and verified to be in good condition between the surface and the 4100L, but are not useable below the 4100L. The feeders are not serving any loads at this time and will be removed during the Yates Shaft rehabilitation.

The Oro Hondo substation has 20 megavolt-amps (MVA) of capacity and supplies power to the Oro Hondo ventilation fans, #5 Shaft equipment, the Grizzly Gulch Dam area, and the old East substation. Recent upgrades to this substation include the addition of a 10 MVA, 69/12kV transformer that was moved from the East substation, a new switchgear building with 12kV switchgear, and a new battery and charging system for the 120 VDC substation control system.

The Ross substation is the most robust distribution system at the Facility. A 20 MVA, 69/12kV transformer and 12kV switchgear were moved from the East to the Ross substation and are now used to provide 12kV power to the underground facilities via the Ross Shaft. An existing 10 MVA transformer at the Ross substation supplies 2400 volt power to surface electrical systems at the Ross Campus. Recent upgrades to this substation include a new battery and charging system for the 120 VDC substation control system.

**Systems and Equipment**

Many of the systems described throughout this chapter include descriptions of existing electrical equipment and their conditions. Examples include ventilation fans (Section 5.4.3.8.1), hoist motors (Sections 5.4.3.2.1 and 5.4.3.3.1), and the waste rock handling system (Section 5.4.3.9.1). Refer to these sections for more detailed description of conditions of specific electrical equipment.

Most of the electrical systems that service the surface facilities are in usable condition. However, local codes require that when additions or modifications to an existing building or structure are performed, the associated electrical systems must be upgraded to meet current state and local codes. This is of little consequence, as most existing structures that will be refurbished will require electrical upgrades to meet increased energy demands.

The existing electrical distribution infrastructure at the Yates Campus is not adequate to meet facility requirements. The power requirements for the Yates Hoists, the underground LMs, and to some extent the new surface buildings such as the SCSE, exceed the capacity and expandability of the current electrical systems. The Yates rock-crushing system will be demolished and the current 2400 volt Yates substation and Yates Hoist substation will be decommissioned. A new Yates substation will be constructed to meet the emerging needs of DUSEL surface and underground infrastructures as well as to continue existing services to the Yates Dry and Administration building. The new substation will be located south of the Yates Headframe. A new overhead 69kV line will be required to feed the substation and will extend across Kirk Canyon from the Oro Hondo substation.

As water levels rose after mine closure, the electrical switchgear, motors, and cabling became immersed, resulting in deterioration of the equipment. Upon reopening of the mine and gaining access to the 4250L, it was determined that most of the electrical infrastructure was unusable. A pair of medium-voltage, 250 MCM cables, extending from the Oro Hondo substation, through the Kirk Portal, and down the Ross Shaft to the 4250L, were found in salvageable condition and used to re-establish dewatering pumps at the 1250L, 2450L, and 3650L. Currently, the cables feeding the dewatering system have been cut at the 300L and spliced to new cables extending down from the Ross substation. Cable #1 supplies the 1250L, 2450L, and 3650L pump stations; Cable #2 supplies underground power substations at the 4550L and 4850L and a dewatering pump at the 5000L.

A few existing cables in the upper levels of the Facility are being used to provide limited power to shaft stations and early science experiments. Even though these cables have passed insulation and ground tests, they have experienced some degradation; and therefore, long-term use of the cables is not planned.

#### Redundancy

The WWTP is the only system at this Facility that currently has a backup generator system.

The Yates and Ross Hoists are powered by AC/DC MG sets. Backup power to the hoists is achieved through energy stored in flywheels driven by the MG sets. Under typical conditions, this arrangement will provide approximately 20 minutes of reserve power to the hoists. A 2400 volt connection between the Ross substation and the Yates Hoist switchgear has been installed so that if power from the Oro Hondo substation is lost for an extended period, the Yates Hoist can be back-fed from the Ross substation and the hoist operation can be maintained.

Cables extending from the Oro Hondo substation through the Kirk Portal to the Ross Shaft are no longer in use and can be used as 12kV backup feeders to the underground power and dewatering distribution system in the Ross Shaft if necessary.

#### Reliability

Investigations into power system reliability were conducted during Preliminary Design. Written records were not maintained for outage events at Homestake prior to 2001; only two known significant issues occurred during the 1990s. One issue involved a series of power outages due to lightning strikes on the 69kV overhead distribution lines. This problem was corrected by installing static lines above the power conductors to direct lightning strike currents to ground. A second significant event involved a six-hour power outage that affected the entire Black Hills region in the late 1990s. This outage was due to a problem at Black Hills Powers WYGEN power plant located near Gillette, Wyoming.

In e-mail correspondence from Jim Keck of Black Hills Corporation dated June 10, 2010, titled *DUSEL Design - Outage Probability/Security Plan* the following information was provided:

- Our electronic supervisory control and data acquisition (SCADA) records for the Kirk Sub show no breaker operations or bus outages since 2001. The records do not go back any further.
- Our OMS (Outage Management System) records and tabulates outage data. Since 2003, we have had three significant widespread outage events that stand out—they were all snowstorms/blizzards:
  - April 18-19, 2006
  - May 1-2, 2008
  - November 5-6, 2008
- Homestake / Sanford Laboratory did not lose power during any of those events.
- Other significant events that took place during the past 10 years where Homestake / Sanford Laboratory did not lose power include:
  - Grizzly Gulch fire—2002
  - Ultra Light plane flies into transmission lines near Spearfish—2007

None of these events resulted in loss of power to the Sanford Laboratory facility.

Since the reopening of the underground facility, the primary points of failure in the electrical distribution system have been underground medium-voltage cable terminations in the Ross Shaft. The SDSTA has worked diligently to resolve issues with cable terminations and has made significant improvements in the integrity of the cabling and the power distribution systems in general.

#### 5.4.3.10.2 Electrical Power Distribution Requirements

Fire and life safety are major driving forces behind the electrical distribution design. The system will be appropriately sized to accommodate both the support infrastructure and the science experiments' energy demands. The sequence and transition from construction to operations, and related energy demands, are a factor in the distribution design. For example, the experiments will not be outfitting or operating during the first half of the excavation and construction period, and the Ross Skips and waste rock handling system will not be in operation once excavation is complete. Other factors that affect the distribution system design include:

- Energy availability from supplier
- Power availability to laboratories during electrical maintenance
- Space limitations for electrical transformers and switchgear
- Space limitations in shafts for distribution cables
- System reliability and redundancy
- Power quality and the affect of individual loads on the entire system
- Operating costs
- Energy efficiency
- Capacity for future growth

**Code Requirements**

The existing electrical infrastructure was designed to meet the requirements of the National Electric Code (NEC), the National Electric Safety Code (NESC), and MSHA regulations. The transition to a science user facility means that new codes, standards (primarily OSHA), and practices must be employed to meet the regulatory requirements applicable to an occupied underground facility. Redefining the purpose of the underground facility as a laboratory does not change the unique conditions, constraints, and challenges of developing infrastructure systems at depth.

As discussed in Section 5.1.7.2, *Codes and Standards Approach*, NFPA 520 is the section of the code that specifically addresses underground facilities. One of the challenges in determining code requirements is determining where code conflicts exist and which one has priority. For example, parts of NFPA 520 conflict with other NFPA, International Building Code (IBC) codes, and OSHA regulations that are intended primarily for the Surface Facility. The DUSEL Project Team, in cooperation with DUSEL EH&S and the design contractors, have identified the applicable codes and intend to meet or exceed applicable codes and standards for the DUSEL electrical distribution system while maintaining an efficient and cost-effective design. The NEC, NESC, and NFPA 520 are the predominant codes used in developing the electrical system Preliminary Design.

### 5.4.3.10.3 Electrical Power Distribution Preliminary Design

The design for electrical power distribution includes several subsystems both on the surface and underground. These subsystems will take advantage of existing infrastructure described in Section 5.4.2.9 as much as possible. Figure 5.4.3.10.3-1 shows an overall view of the electrical infrastructure system. Each subsystem is explained in detail below.

**Incoming Power**

Sanford Laboratory is fed from the Black Hills Power (BHP) Kirk switch, located in Kirk Canyon. The Kirk switch is fed from BHP's Yellow Creek substation, located approximately 2 miles south of Sanford Laboratory (See Figure 5.4.3.10.3-2), part of BHP's main 230kV transmission line providing power to the Black Hills area. The Kirk switch is configured to provide 32 MVA to the DUSEL Facility. Upgrades to the Kirk switch by BHP will include replacing the existing oil circuit breakers with modern SF6 gas-filled circuit breakers to ensure availability of spare parts and decrease maintenance costs. Sanford Laboratory may require additional capacity from BHP, but this will not be fully defined until the Final Design phase of the DUSEL Project.

Power enters the DUSEL Facility at 69kV on overhead power lines at two points: the Ross substation located near the Ross Campus on Houston Street and the Oro Hondo substation located in Kirk Canyon. The on-campus substations reduce the 69kV to 12,470 volts (12kV) for distribution to the Facility's internal infrastructure. A new substation will be located south of the Yates Headframe to service the Yates Campus. The Yates substation will be supplied at 69kV as an extension of the overhead line feeding the Oro Hondo substation.

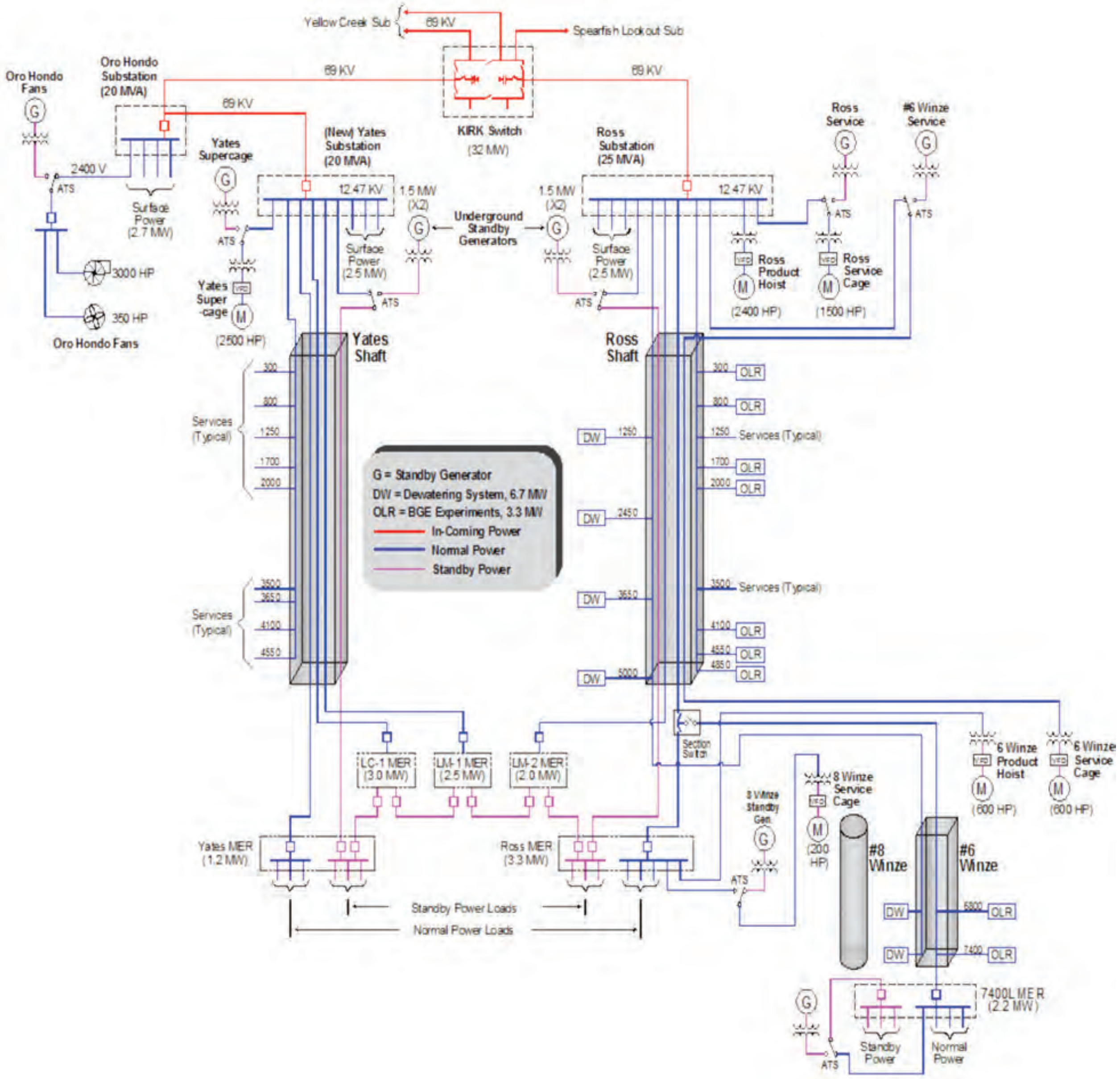

**Figure 5.4.3.10.3-1** Electrical power distribution system. [Paul Bauer, DUSEL]

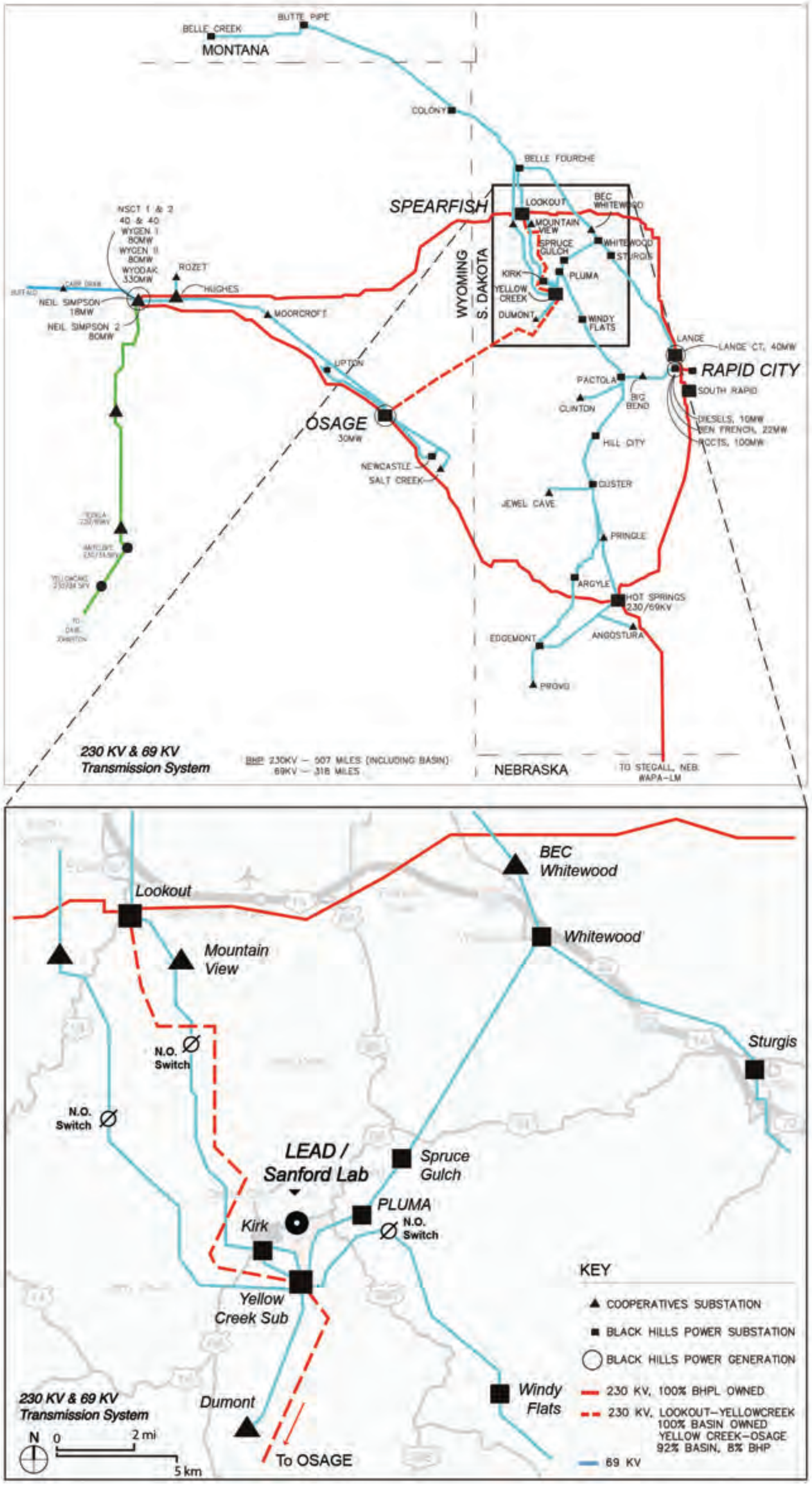

**Figure 5.4.3.10.3-2** Incoming power map. [DKA, with data from BHP]

## Surface Distribution

The Ross substation capacity will be upgraded to meet the demands of the underground DUSEL Facility on the Ross side, plus surface loads at the Ross Campus. The Ross substation distributes power at 2400 volts to the Ross Hoistroom (service and production hoists), the Ross Headframe substation (waste rock handling system), the Ross Boiler Room, and the Whitewood Creek substation (powers the Kirk Fans) and the Ellison Campus. To provide backup power to the Ross Hoists and critical underground loads, 12kV standby generators will be installed at the Ross substation.

The Oro Hondo substation provides power at 2400 volts to the 350 hp and 3,000 hp Oro Hondo fans. This substation also supplies power at 12kV to feed the #5 Shaft fans, hoistroom and heaters, and the Grizzly Gulch Dam monitoring wells and decant barge pumps.

The new substation at the Yates Campus replaces the existing Yates Hoist substation, Yates Headframe substation, and East substations. This new substation will be built with sufficient capacity to meet the demands of the underground DUSEL Facility on the Yates side, plus all surface loads at the Yates complex. The new Yates substation will distribute power at 12kV to the existing foundry, compressor, and wastewater treatment plant substations. The SCSE and all other new or existing/refurbished buildings on the Yates Campus will be powered from the new Yates substation at 2400 volts. Figure 5.4.3.10.3-3 shows the surface distribution plan.

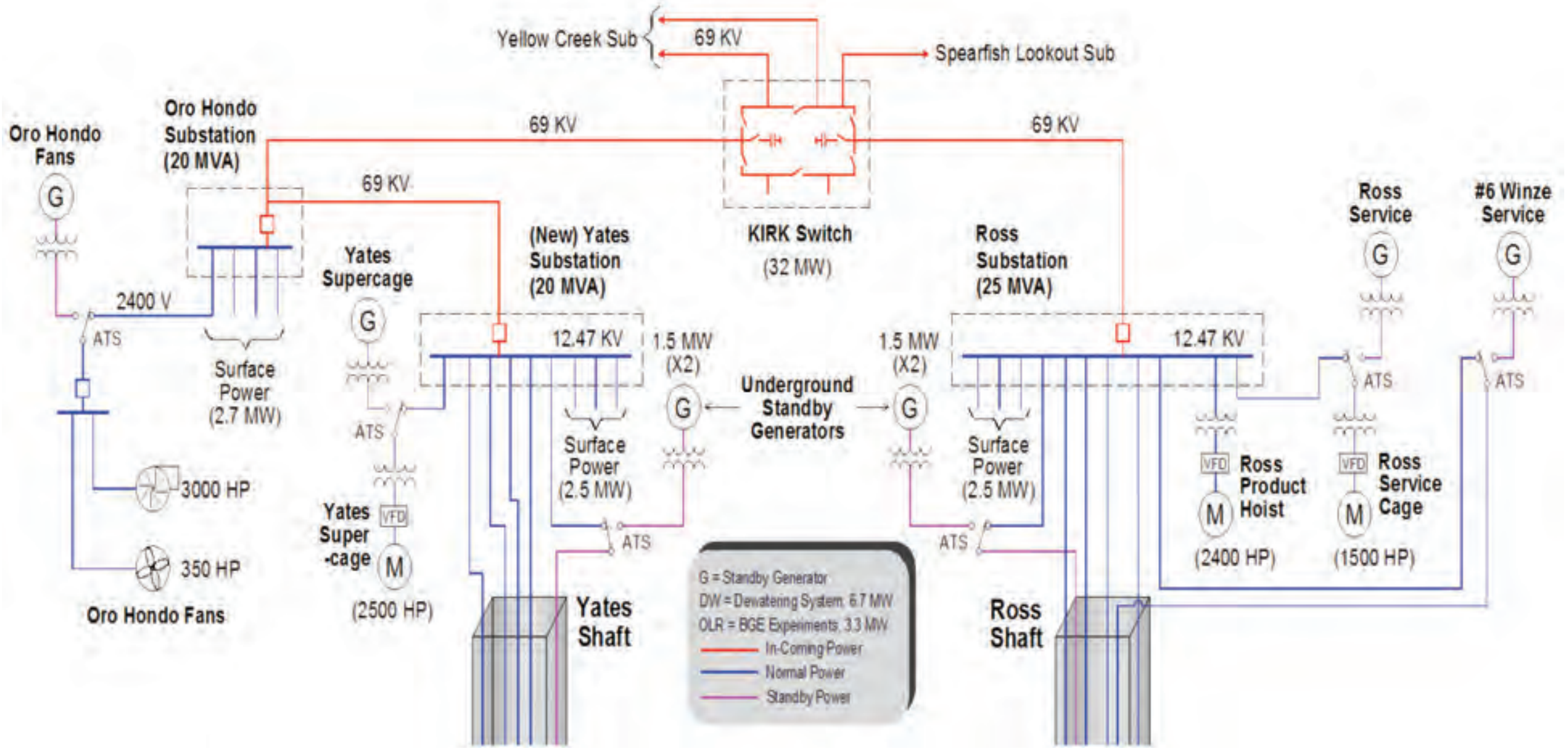

**Figure 5.4.3.10.3-3** Surface electrical distribution. [Paul Bauer, DUSEL]

### Underground Distribution

The underground electrical system is composed of normal power distribution, which is the primary distribution system that services the bulk of the electrical load, and standby power distribution, a generator backup system that supplies power only to essential equipment if normal power is lost. The standby distribution systems is designed to provide redundancy of backup power for underground support facilities, three science facilities at the 4850L Campus, and a science facility and drilling station at the 7400L Campus.

### Normal Power

Each underground facility will have its own MER where the electrical equipment is located. Each MER will have a dedicated feeder cable that originates at either the Ross or Yates surface substation.

The primary reason for dedicated feeders is to allow power distribution switchgear to be located at the surface substations. This reduces the MER space requirements for electrical equipment, resulting in significant cost savings.

Additional benefits for dedicated feeders include:

1. **Power Quality**. Electrical noise, harmonics, and transient voltage spikes created in one facility or system will have less affect on the others.
2. **Reliability.** A cable failure on one dedicated feeder will only affect its associated system, not the entire underground campus.
3. **Stability.** The electrical load on each dedicated feeder is small compared with the total underground load, resulting in lower current flows per cable, less cable heating, less electromagnetic interference (EMI), and reduced voltage drop.

There are some disadvantages of the dedicated-feeder approach. An increased number of feeders requires more room in shaft utility areas; even though the dedicated feeders are smaller, the total weight of the feeders increases, and increases the shaft load. An increased number of cables requires more splices, which takes more space at shaft stations every 600-900 ft and creates additional potential failure points.

Each dedicated feeder delivers 3-phase power at 12kV using mine-rated armored cables. All power cables are rated for low-smoke, flame-retardant, and zero-halogen, and include a separate grounding conductor in addition to the 3-phase conductors. Cables in shafts are specially designed to include three steel messenger cables for vertical hanging support.

To maintain system integrity, all MERs are fitted with transient voltage surge suppression (TVSS), ground fault circuit interrupters (GFCI), power factor correction capacitor banks, and provisions for an isolated ground system for sensitive electronics and communication equipment.

The Preliminary Design includes individual feeders from both shafts to provide redundancy of normal power for experiments and facility. Through the Value Engineering process, the redundancy for normal power has been eliminated from the cost estimates and this will be updated in the design documents during Final Design. The power distribution is now divided between the two shafts so that if an outage event occurs, it will only impact power in a portion of the underground campus. This presents significant savings for the Project without affecting standby power redundancy, assuring power availability for both personnel safety and critical experimental needs. All figures in this PDR reflect the changes from the Preliminary Design Value Engineering process. Note that the drawings in Arup's supplemental documents (Appendix 5.O) do not reflect these changes and therefore the impact of the Value Engineering can be seen by comparing the PDR to the design documents. Restoration of redundancy, if deemed necessary, would require additional feeders in the shafts and on the levels, as well as space for transfer switches in the MER rooms. The current excavation plan would support this, but additional savings may be realized by reducing excavated space during final design.

#### Standby Power

*NFPA 520 Standard on Subterranean Spaces* refers to backup power systems as "alternative power supplies." The standard further defines the meaning of alternative power supplies to be "standby" power and “emergency” power. There are important distinctions between the two types of systems. Table 5.4.3.10.3-1 summarizes these differences.

| Standby Power | Emergency Power |
|---|---|
| Transition from normal to standby power cannot exceed 60 seconds. | Transition from normal to emergency power cannot exceed 10 seconds. |
| Standby systems must be capable of providing power for a minimum of 4 hours. | Emergency systems must be capable of providing power for a minimum of 90 minutes. |
| Equipment classified as Standby loads:<br>• Electric-driven fire pumps<br>• Mechanical air-handling systems for AoRs and exit passageways<br>• Smoke-control systems for AoRs and exit passageways<br>• Standby lighting for AoRs<br>• Standby lighting for smoke-control MERs<br>• Two-way communication systems | Equipment classified as Emergency loads:<br>• Fire detection and alarm systems<br>• Exit sign illumination<br>• Emergency lighting<br>• Fire command center lighting |

**Table 5.4.3.10.3-1** Summary of alternate power supply definitions. [NFPA 520]

The NEC (also NFPA 70) similarly categorizes emergency systems, legally required standby systems and optional standby systems. Emergency systems are intended to supply power to equipment “essential for safety to human life” in the event of normal power loss. Legally required standby systems are intended to supply nonemergency-classed loads that “could create hazards to rescue or fire-fighting operations.” Optional standby systems are intended for powering loads where “damage to the product or process” could result from a loss of power. The NEC allows the local Authority Having Jurisdiction (AHJ) to classify equipment and systems into one of these categories.

For the DUSEL Project, loads for emergency power are limited to lighting and communications. Meeting the 10-second response time requires large uninterruptible power source (UPS) systems for these loads. Experiment-specific UPS systems are provided by the experiments. Emergency power follows the code requirement of 90 minutes operation, while standby power has been designed to operate for 96 hours rather than the 4 hours dictated. This provides safety for underground personnel if evacuation is not immediately possible. Standby loads include hoists, AoRs, ventilation fans, controls for emergency response (closing ventilation doors, sprinkler systems, etc.), breathing air compressor, experimental loads for critical assets (primarily cryogen cooling), and critical dewatering at the 5060L due to anticipated leakage of the water Cherenkov detector (WCD).

The distribution system for underground standby power uses redundant generators located near the Ross and Yates surface substations, each capable of powering the entire standby and emergency loads underground. Switches at the 4850L MERs allow a series-looped connection of these systems such that a failure of one standby power feeder will not disable the system, providing single-fault tolerance. Figure 5.4.3.10.3-4 shows the system of generators and where they are located on the surface.

Emergency power is provided by battery-backed UPSs connected to the standby power system. The UPS allows critical systems to ride-through the switching and generator starting sequences without interruption of service. The UPS also provides a second level of redundancy for 90-minute duration. Figure 5.4.3.10.3-5 illustrates the relationship between standby and emergency power distribution.

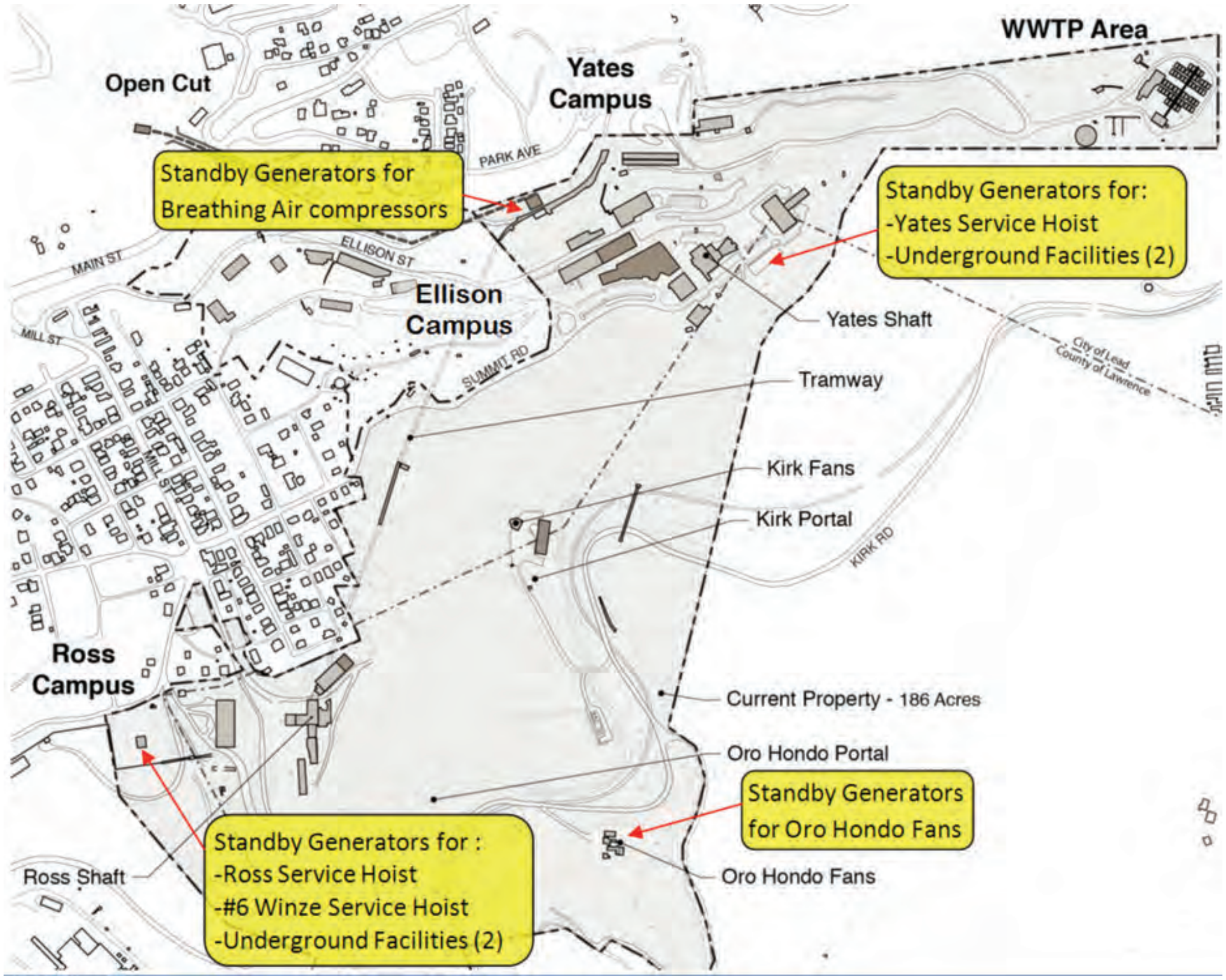

**Figure 5.4.3.10.3-4** Standby generator map. [Paul Bauer, DUSEL]

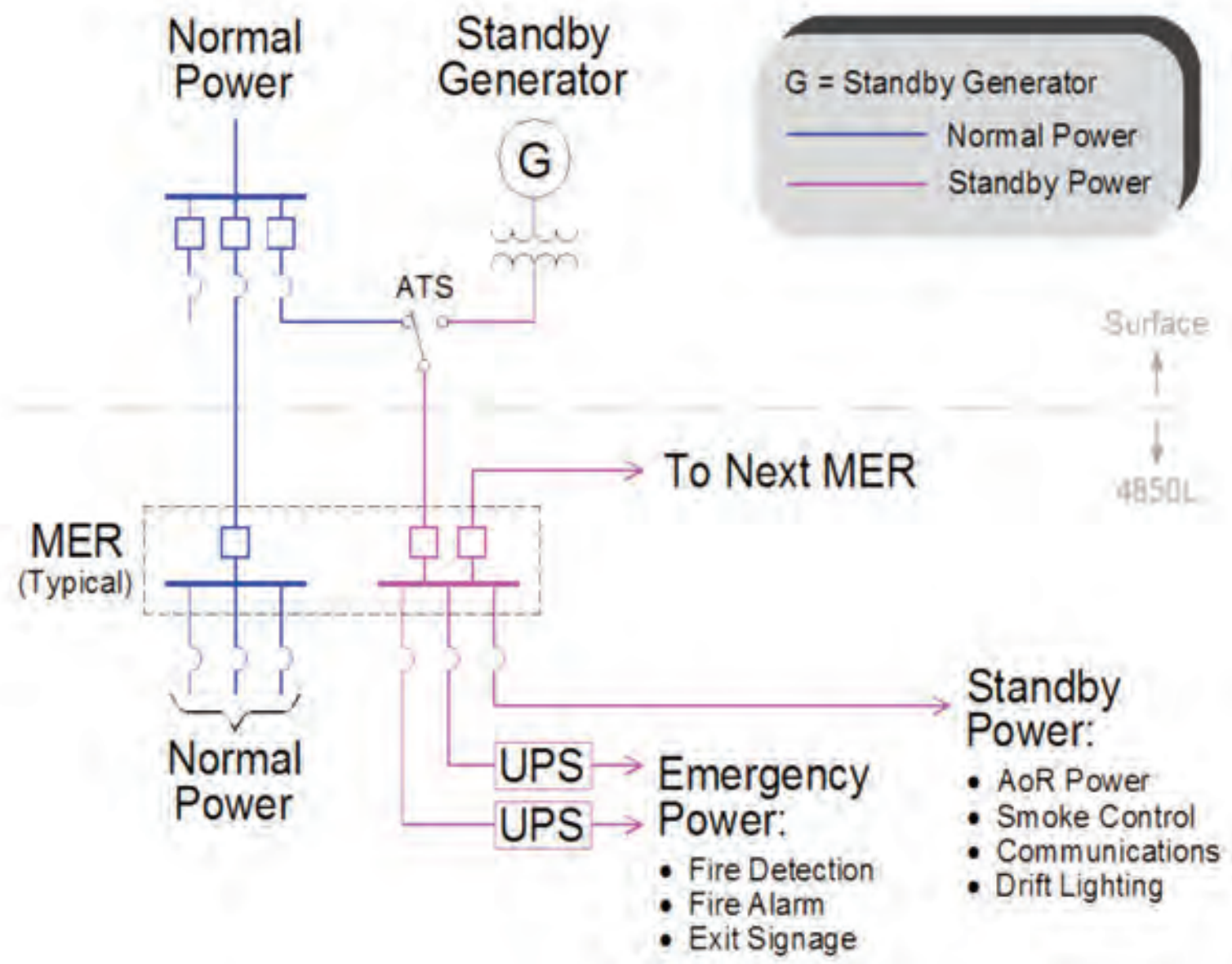

**Figure 5.4.3.10.3-5** Standby/emergency power distribution. [Paul Bauer, DUSEL]

Stand-alone standby generators are required for each hoist based on the 2009 IBC code, Sections 3003, 3007, and 3008, and also because of the loading profiles of these systems. Another stand-alone generator will be provided for the Oro Hondo Fan due to its isolation from the Ross and Yates Campuses. An additional stand-alone standby generator will be provided for the breathing air compressors located at the Yates Campus compressor substation.

#### Power Distribution for Underground Systems

The underground electrical infrastructure is divided into separate facilities/systems:

- #6 Winze Hoists
- #8 Winze Hoist
- Dewatering System
- Large Cavity (LC-1)
- Lab Module 1 (LM-1)
- Lab Module 2 (LM-2)
- Other Levels & Ramps (OLR) System
- 4850L Support System
- 7400L Campus (LMD-1 and all Facility power)

The Large Cavity, LM, and Facility MERs all provide electrical services to similar types of equipment. Typical equipment that is common to all underground MERs includes:

- Access drift lights and receptacles
- Air-handling units
- Communication enclosures
- MER utilities
- Sump pumps
- Supply and exhaust fans

Equipment and systems that are uniquely powered from specific MERs are outlined in the system descriptions later in this section.

**#6 Winze.** The #6 Winze Service and Production Hoists for 7400L access will be fed by a single 12kV power cable from the Ross surface substation. The major power requirement for the #6 Winze is two 600 hp electric motors controlled by variable frequency drives (VFD). Transformers are required to step down the voltage from 12kV to 690 volts and provide isolation for the VFDs.

**#8 Winze.** The #8 Winze for secondary egress from the 7400L will be fed from the 4850L Ross MER via a dedicated isolation transformer. This hoist uses a 200 hp motor and VFD and the standby power for the hoist will be provided by a stand-alone generator also located on the 4850L.

**Dewatering System.** Power distribution for the dewatering system comprises five underground substations at the 1250L, 2450L, 3500L, 4850L, and (future) 6800L. The substations receive power from the Ross surface substation at 12kV and supply power to 12 lift pumps at 4160 volts. The total horsepower of the pumps is 8,900 hp (6.5 MVA) and each pump is supplied with a soft-start controller to limit inrush current during starting.

**Large Cavity.** The WCD receives power at 12kV on a dedicated feeder from the Yates surface substation. The LC-1 MER houses the switchgear, transformers, and panel boards required for facility support. In addition to the common equipment listed previously, the following equipment is served by LC-1 MER:

- 480 volt connections for experiment-specific power
- 5060L sump pumps
- Cavity dome lighting and utility power
- Monorail crane system

Power for the detector equipment is derived from a 480 volt transformer located in the LC-1 MER. Transformers and panel boards for lower voltage levels to feed experiment-specific equipment, electronics, and instrumentation are the responsibility of the science collaborations for both design and fit-out. Experiment-specific equipment will be located within the cavity dome area or calibration drift, as there is no space allocated in the MER for this equipment.

**Lab Module 1.** LM-1also receives power on dedicated 12kV feeders from the Yates surface substation. The LM-1 MER houses the switchgear, transformers, and panel boards required for LM-1 support systems. Equipment unique to the LM-1 MER includes:

- 20-ton bridge cranes and 40-ton monorail crane
- 480 volt connections for experiment specific power
- Air compressor (for experiments)
- LM lighting and utility power

Power for laboratory equipment is derived from a 480 volt transformer located in the LM-1 MER. Transformers and panel boards for lower voltage levels to feed experiment-specific equipment, electronics, and instrumentation are the responsibility of the collaboration for both design and fit-out. Experiment-specific equipment must be located within the LM space, as there is no space allocated in the MER for this equipment. LMs will be fitted with LED light fixtures to reduce the effect of EMI noise on sensitive instruments.

**Lab Module 2.** The configuration of the electrical system for LM-2 is the same as that for LM-1 with the exception that more electrical equipment is required to support the larger module size and expected number of installed experiments. Power for LM-2 is received on a dedicated 12kV feeder from the Ross surface substation.

**Ross and Yates MERs (substations).** The 4850L Ross and Yates MERs are powered from their respective surface substations. Redundancy to these substations, including an interconnecting tie between them, was eliminated during the Value Engineering process for Preliminary Design. Figure 5.4.3.10.3-6 shows the distribution system on the 4850L including the LMs, LC-1, Ross, and Yates MERs.

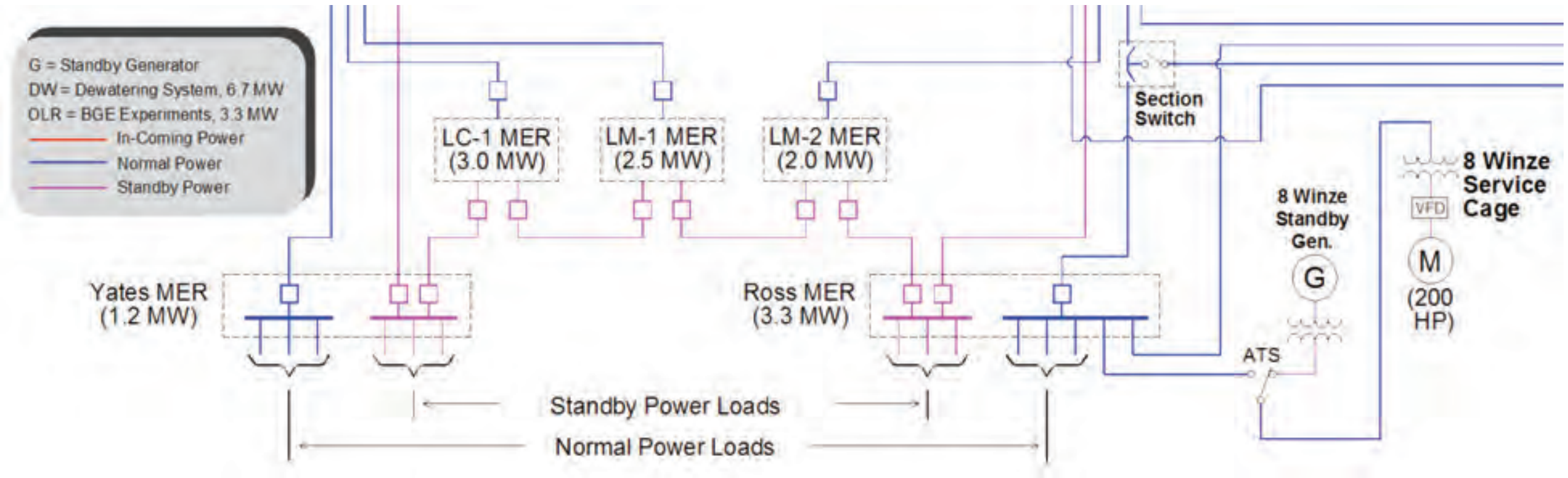

**Figure 5.4.3.10.3-6** 4850L power distribution. [Paul Bauer, DUSEL]

**Other Levels and Ramps.** Power for OLR is provided through two substations in each shaft at the 1700L and 4100L (Figure 5.4.3.10.3-7). These substations step the power down from 12kV to 4160 volt for distribution to other levels with experiments. Lighting and convenience power is provided through further step-down transformers on OLR as required. Experiment locations will be provided with transformers delivering both 480 volt and 120 volt power. This system design is a result of the power requirements and the long distances between the substations and the final points of use at the experiment locations. Using 4160 volt service for the majority of the distance allows for smaller cables while maintaining a maximum 5% voltage drop.

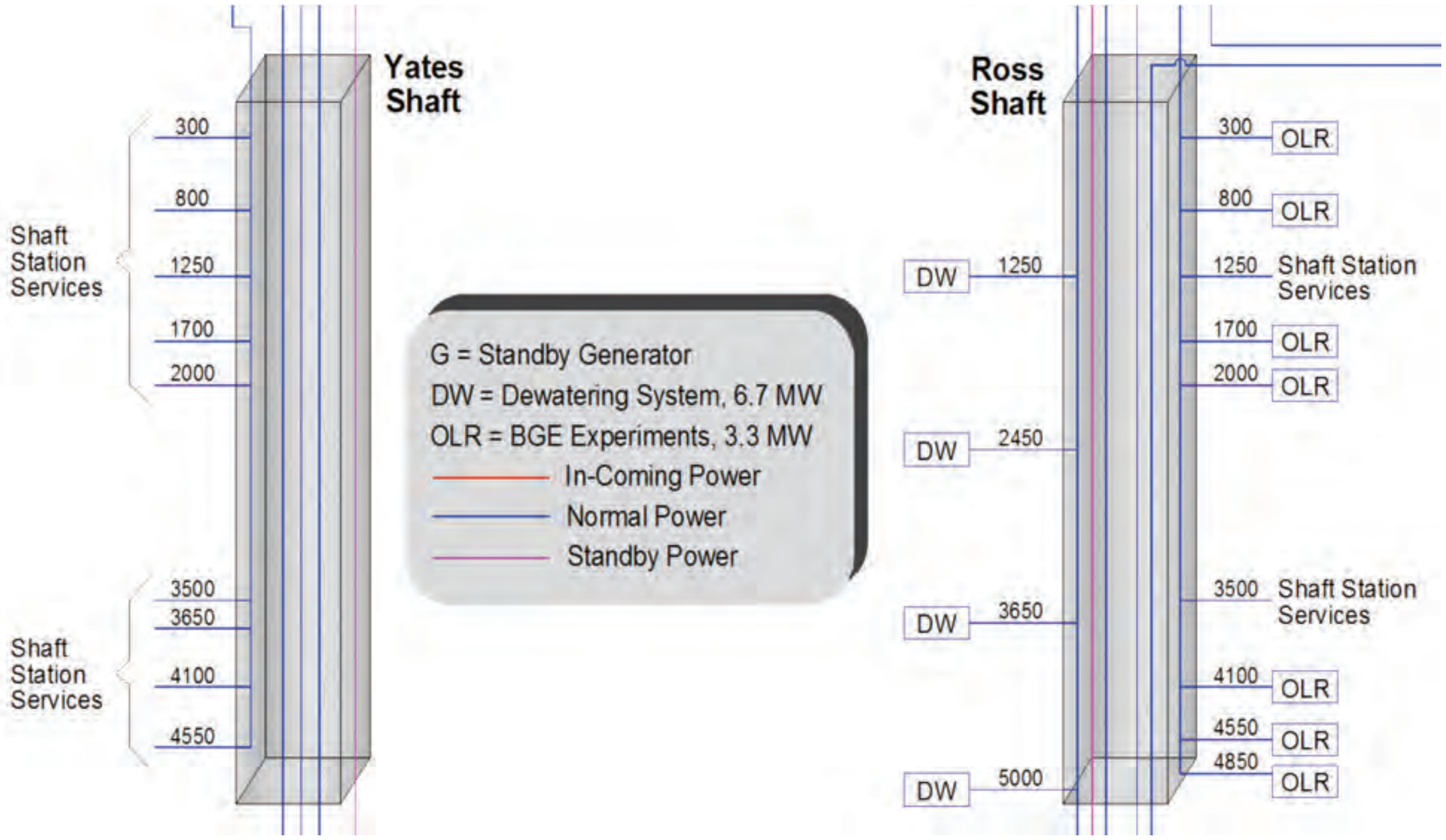

**Figure 5.4.3.10.3-7** OLR and shaft power distribution. [Paul Bauer, DUSEL]

**4850L Support Systems.** Support systems are divided into two groups: those systems required for safe occupancy, and systems necessary to manage and maintain the Facility. The support systems include drift lighting, communication and control systems, AoRs for life safety, plumbing and pumping, ventilation and climate control, maintenance shops, and battery-charging stations for facility management. Two-hour rated fire separation between normal power and power used for life-safety systems is a code requirement, as outlined previously in this section. Separation of conductors is accomplished by running two separate feeders from the surface to each MER, one for normal and one for standby power, where each feeder takes an alternate path down opposing shafts. Each MER is divided into two rooms separated by 2-hour rated fire walls, one for normal power and one for standby power to house the distribution switchgear for each system. Equipment is connected to the switchgear depending on whether it is a normal facility load or an emergency/life/safety load.

**7400L Campus.** The design philosophy of feeder separation used on the 4850L is not used in the design of the 7400L electrical system design. All power for this level is provided via a single system, including power needed for deep drilling to support an OLR experiment, facility infrastructure, and experiments (LM). Complete redundancy is provided with both normal and standby power via the #6 Winze and #8 Winze. As with the 4850L, redundant normal power was eliminated through Value Engineering during Preliminary Design to reduce Project cost. Figure 5.4.3.10.3-8 shows the distribution system for the 7400L.

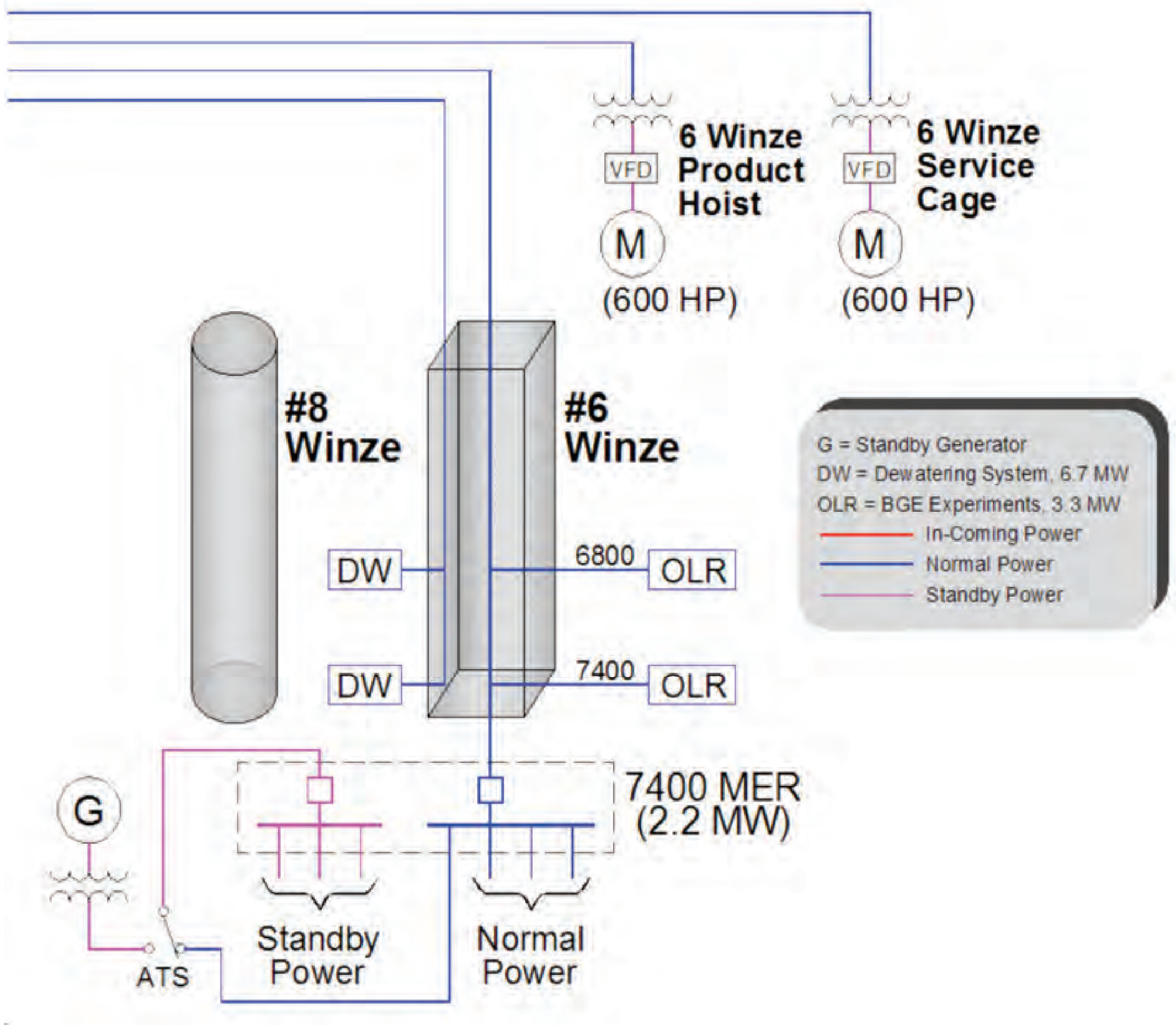

**Figure 5.4.3.10.3-8** 7400L power distribution. [Paul Bauer, DUSEL]

### Energy Demand

Some systems have well-defined power requirements and are not expected to change as the design develops. The systems with well-defined loads include the dewatering system, hoists, ventilation fans, waste rock handling, and the WWTP. Energy demand for the rest of the Facility is based on assumptions for systems such as the chiller plants, LM support (lighting and air-handling), AoRs, OLR, and the loads that are directly related to the experiments as provided by the science collaboration. Table 5.4.3.10.3-2 provides the power estimates of connected loads that were used to establish the power requirements for the entire Facility at 100% Preliminary Design.

| WBS | Source* | Description | Normal Power (kW) | Standby Power (kW) |
|---|---|---|---|---|
| **Surface** | SY | Air Compressors | 2653 | 3980 |
| | OS | Oro Hondo Vent Fans + Grizzly Gulch | 4000 | 3750 |
| | SR | Ross Service & Production Hoists | 3170 | 1830 |
| | SR/SY | Shops & Buildings | 2549 | 460 |
| | SR | Waste Rock Handling | 1602 | |
| | SY | Waste Water Treatment | 1500 | 580 |
| | SY | LC-1 Water Plant | 737 | |
| | SY | Yates Supercage | 2000 | 3000 |
| **UGI** | SR | Dewatering Pumps | 6528 | |
| | SR | #6 Winze Service & Production Hoists | 980 | 735 |
| | R | #8 Winze Service Hoists | 150 | 225 |
| | R/Y | 4850L Facilities Support Power | 390 | 90 |
| | L2/R/Y | 4850L Areas of Refuge (AOR) | 567 | 567 |
| | R | 4850L Chiller Plant | 2325 | |
| | LC | 5060L Areas of Refuge (AOR) | 79 | 79 |
| | SR | 7400L Facilities Support Power | 264 | 114 |
| | DL | 7400L Areas of Refuge (AOR) | 61 | 61 |
| | DL | 7400L Chiller Plant | 349 | |
| **UGL** | SY | $CO_2$ Sequestration | 420 | |
| | SR | Other Levels & Ramps (Earth Science) | 3302 | |
| | Y | 4850L Early Science & Support | 1511 | 160 |
| | SY | 4850L LC-1 Experiment + Support | 2988 | 110 |
| | SY | 4850L LM-1 Experiments + Support | 2539 | 100 |
| | SR | 4850L LM-2 Experiments + Support | 2009 | 160 |
| | DL | 7400L LMD Experiment + Support | 1048 | 130 |
| | DL | 7400L Drill Room | 258 | |
| | | **TOTAL CONNECTED LOAD =** | **44.0 MW** | **16.1 MW** |

* DL-7400L MER, LC-LGC MER, R-4850L Ross MER, Y-4850L Yates MER
OS-Oro Hondo Substation, SR-Surface Ross Substation, SY-Surface Yates Substation

**Table 5.4.3.10.3-2** DUSEL Facility total connected load.

The values given in Table 5.4.10.3-2 represent connected loads. At the Preliminary Design phase, the initial design loads are based on the total connected load for the highest demand combination of experiments that might be included in the Final Design configuration. Other factors will be considered during Final Design when estimating the actual power requirements of the Facility. Some factors that will be considered during Final Design include:

- Using diversity and demand factors, a peak load of 60-70% of the connected load is expected. (26-32 MW)
- Breathing air compressors will never run except during maintenance/testing and in emergencies. (2.7 MW)
- The waste rock handling, LC-1 surface water plant, and drilling for BGE experiments will not operate after commissioning. (2-3 MW)
- Equipment operation, energy management strategies, and equipment with intermittent duty cycles (hoists, sumps, pumps, etc.) will all contribute to reducing the effective continuous energy demand of the Facility.

### 5.4.3.11 Dewatering and Pumping

As demonstrated by the conditions found on re-entry, continual dewatering of the Facility is necessary to manage natural water inflow to the underground. Pumping water from as far as 8,000 feet below the surface requires a specialized pumping system with high energy demand. Ensuring reliability of the pumping system provides security for both occupants and equipment in the planned underground spaces.

#### 5.4.3.11.1 Current Condition of the Dewatering and Pumping Systems

**Current Operations**

The SDSTA is currently dewatering the Facility using the same system from the 5000L upward that was utilized during HMC operations. This Ross Shaft system is operational from the 5000L to the surface and provides for a maximum capacity of 2,300 gpm. At the conclusion of Preliminary Design, the Facility below the 5200L is flooded. A temporary deep-well pumping system is installed in the #6 Winze and provides for dewatering of the Facility to the 7700L. The capacity of this deep-well system averages 1,500 gpm.

Access to the Deep-Level Laboratory (DLL) campus is dependent on the completion of the dewatering to the 7700L. Given that the development of the DLL is a priority and access is required to conduct site investigations for excavation and infrastructure design, efforts are being made to accelerate the rate of dewatering. In particular, the installation of a parallel deep water pumping system in the #6 Winze is being investigated that would enable the pumping rate from deeper levels of the Facility to match the Ross Shaft system. In addition to providing for an increased dewatering rate from the pool, a parallel system will provide redundancy to ensure uninterrupted dewatering. The WWTP capacity is designed to manage this volume.

**Existing System Conditions**

To manage the constant inflow of water into the underground areas, six major underground dewatering pump stations are in place. Four of these, located on the 1250L, 2450L, 3650L, and 5000L, are associated with the Ross Shaft. Two pumps are located within each station and all four Ross Shaft stations utilize the same model of pump. The remaining two stations are located on the 6800L and 8000L associated with the

#6 Winze. As the 8000L is not part of the Preliminary Design scope, the design of the pump stations at this level is not included in this PDR.

Two pumps per station are required to achieve pumping rates of at least 1,500 gpm to support dewatering efforts. HMC began pump upgrades to the four Ross Shaft-based pump stations beginning in the mid-1970s. The new pumps were Ingersoll Rand 6HMTA-3 stage units operating at 3600 rpm. The 6HMTA pumps operated at around the 81% efficiency range, which is quite good for this style of pump. However, due to the age of these pumps and associated high maintenance costs, it has been decided to replace all the existing HMTA pumps with new equivalents.

Upgrades to the Ingersoll Rand 6x11 DAD 4 stage #6 Winze pumps on the 6800L occurred in 1989. Homestake did not perform excavations during this upgrade to allow an adequate distance between the sump and pump, and consequently the inlet piping design is not correct. The distance from the elbow to the pump inlet is too short to obtain reliable extended service. The first-stage impellers were subject to excessive wear due to cavitation early in their life. These are a newer style of pump that has the diffuser cast into the casing. The two pumps located on the 6800L have been submerged for several years and are still submerged, which makes predicting their condition somewhat difficult. There is one complete spare on hand and one stainless case with various parts.

All five pump stations require sumps to draw water for the pump suction as well as to discharge from the previous level. Sumps located on the 3650L and 6800L are equipped with settling areas where solids can drop out. Clear water decants into a clean-water sump for the pump suction. A thorough inspection and cleaning will be required for all sumps and is included in the Preliminary Design.

The dewatering pipelines in the Ross Shaft are 12 in (305 mm) diameter by .375 in (9.5 mm) wall thickness grade A-106 seamless. HMC replaced a section between the 2450L and 1250L in the early 1990s and began the installation of a new column between the 5000L and 3650L just prior to closure. This section was finished in August 2010 by the SDSTA and is now in operation. The latest two new sections of pipe are joined with grooved couplings and at each length supported by shaft set steel. During HMC operations, removal of the old pipeline was not prioritized, so there are two to three sections of 12-in (305-mm) pipe installed in the shaft in some locations. These old sections are being removed by the SDSTA to allow installation of any new column.

#### 5.4.3.11.2 Dewatering and Pumping Requirements

Due to the native groundwater conditions, Facility dewatering will be a requirement as long as DUSEL is in operation. It is expected that the Facility will be dewatered to the 7700L before the start of the MREFC-funded phase. Pumping requirements beyond the Facility dewatering consider groundwater infiltration as well as wastewater from construction, excavation, and operation. Infiltrating groundwater enters the Facility at an annual average of approximately 750 gpm based on historic measurements by Homestake. The majority of this is collected in the 3650L and 5000L pump stations. High precipitation events can lead to a significantly increased infiltration rate. The system capacity of 2,300 gpm, running two pumps in parallel, is adequate to maintain a dewatered facility.

A reliable pumping and piping system is required for development of the underground infrastructure. Repairs to the pipe column, a component of the dewatering system, require shaft time to maintain. Conversely, the pumps can be changed or repaired while the shaft remains in full operation in support of Construction or Operations. Pump repairs and/or changes are to be expected over the full life of the Project.

#### 5.4.3.11.3 Dewatering and Pumping Preliminary Design

Several options were considered for an upgraded pumping system. These options considered existing pipeline refurbishment vs. new, pump consolidation, type of pumps, and operating cost efficiency.

The investigation revealed that a system similar to what is currently installed is sufficient to provide the reliability required for laboratory development. The pump system will continue to utilize the five existing pump stations and all work will be performed by SDSTA staff using R&RA funds. A typical pump station at the 3650L is shown in Figure 5.4.3.11.3. Upgrades to the four Ross Pump Stations include new pumps, valves, and piping. The station on the 6800L must be excavated to allow a streamlined suction pipe system. Since the IR6x11DAD pumps are more modern than other pumps in place, they are planned to be refurbished (not replaced).

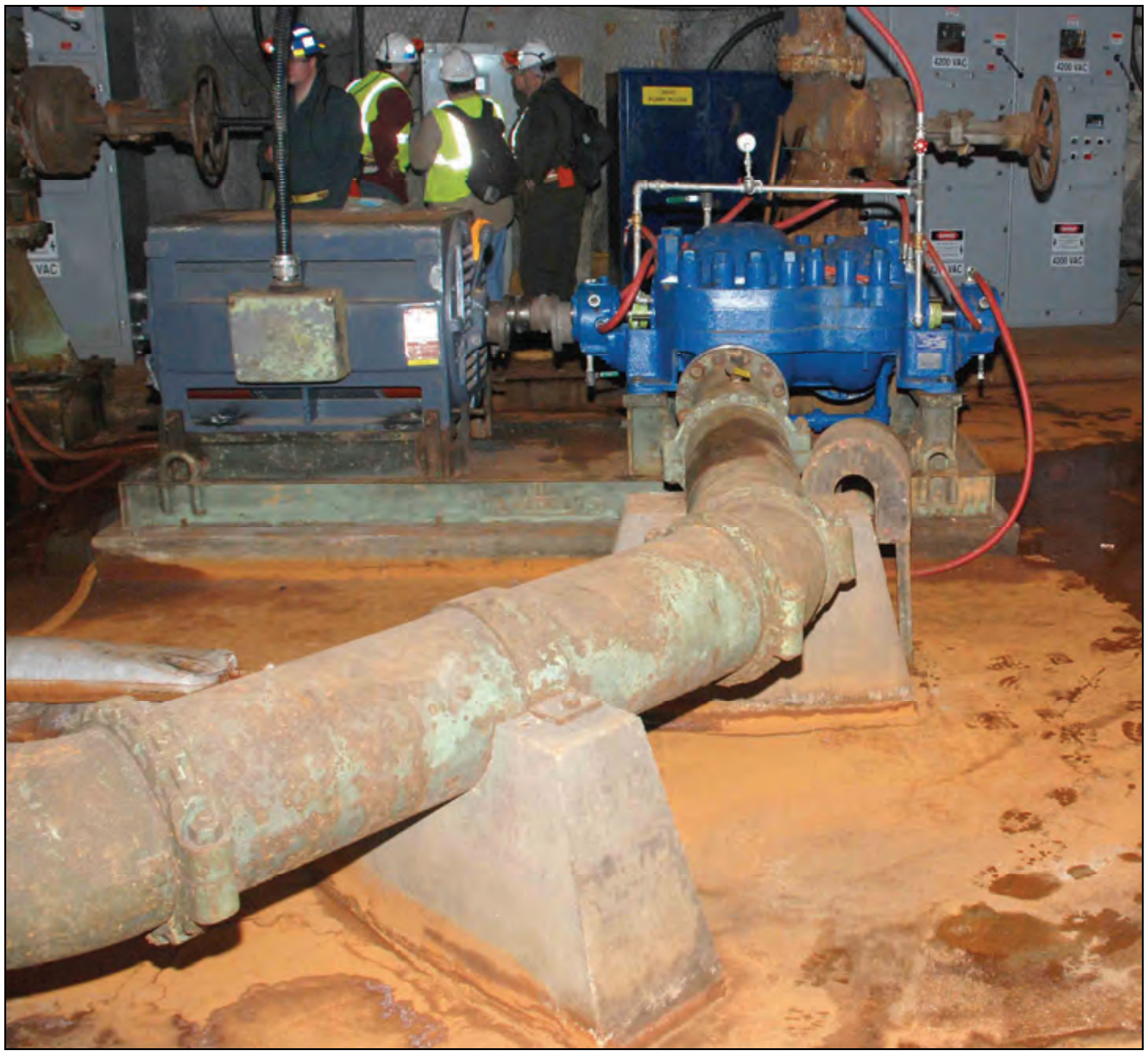

**Figure 5.4.3.11.3** Pump station at the 3650L. [SDSTA]

In the Ross Shaft, two of the four sections (5000L to the 3650L and 2450L to the 1250L) of pipe column have been replaced recently by the SDSTA. One other section requires replacement between the 3650L and 2450L. While rehabilitating the shaft, the SDSTA will remove all existing pipelines and replace this section of pipe. The column from the 1250L to the surface was tested in the early 1990s by HMC and at that time revealed very little corrosion loss. Consequently, it was determined that replacement of this section is not required.

All existing sumps will be cleaned. A new settling system was installed by the SDSTA on the 3800L to handle the waste from the 3650L settling sumps. Other levels will have to be pumped and the slime moved either by rail car or piped to an old stope, depending on the level.

During the excavation process, water is normally required for drilling and dust suppression. The contractor must set up settling sumps for solids collection before the decant water is allowed to pass into the groundwater discharge system. Similarly, during operation, laboratory personnel will be required to ensure that no environmentally unacceptable contaminants are allowed to reach the dewatering system

**Electrical and Controls**

When the SDSTA rebuilt the current dewatering system, new robust 4160 volt soft starters and switchgear were purchased and installed. This equipment is in use today and is in good condition and acceptable for use during construction and ongoing operation. Additional or new 12kV wiring to the transformers will be required. No emergency generator backup is planned for these stations.

Pump operation is handled through a Programmable Logic Controller (PLC) system with sump level controls through float switches; remote monitoring and control is typical. Additionally installed devices during the upgrade will consist of flow meters, temperature, vibration, and pressure transmitters. These devices will help monitor the status of the pump system.

## 5.4.3.12 Water Inflow Management

To complement the dewatering system discussed in the previous section, a design has been developed to direct the normal inflow paths of naturally occurring groundwater away from occupied spaces, capturing it where possible in the dewatering system before it reaches the bottom pool, 7,700 feet below the surface. This design provides less water flow in occupied spaces during normal operation and protection from large inflows during storms.

### 5.4.3.12.1 Existing Water Inflow Management

Groundwater inflows enter the Facility during normal weather at an average rate of approximately 750 gpm. Large storm events can increase this quantity to an amount in excess of the current system's dewatering capacity. The majority of the water from rainfall events enters the Facility through the Open Cut. Section 14 of the *UGI Basis of Design Report* (Appendix 5.L) summarizes over 100 years of area rainfall event data to support the necessity for water inflow controls. Figure 5.4.3.12.1-1 shows an aerial view of the Open Cut to the south. Note the Ross and Yates Shafts, along with the Grizzly Gulch Tailings Facility to the south. Figure 5.4.3.12.1-2 shows the direct and run-on water inflow zones into the Facility via the Open Cut. The arrow in this figure depicts the direction of the photo in Figure 5.4.3.12.1-1 for reference. Historic mine workings extend from the bottom of the Open Cut, dipping at a steep angle downward toward Grizzly Gulch. These old mine workings are composed of multiple raises, ramps, shafts, winzes, and sand-filled stopes, which act as conduits during rainfall events. Unless directly addressed, the hazard this poses during larger rainfall events includes saturating old mine workings, potentially resulting in water to build up behind old walls that are not designed to hold a hydraulic load. The goal of water inflow management systems for DUSEL is to directly address these risks and to actively control water inflows to protect the underground facility, property—including valuable science equipment and, most importantly, people.

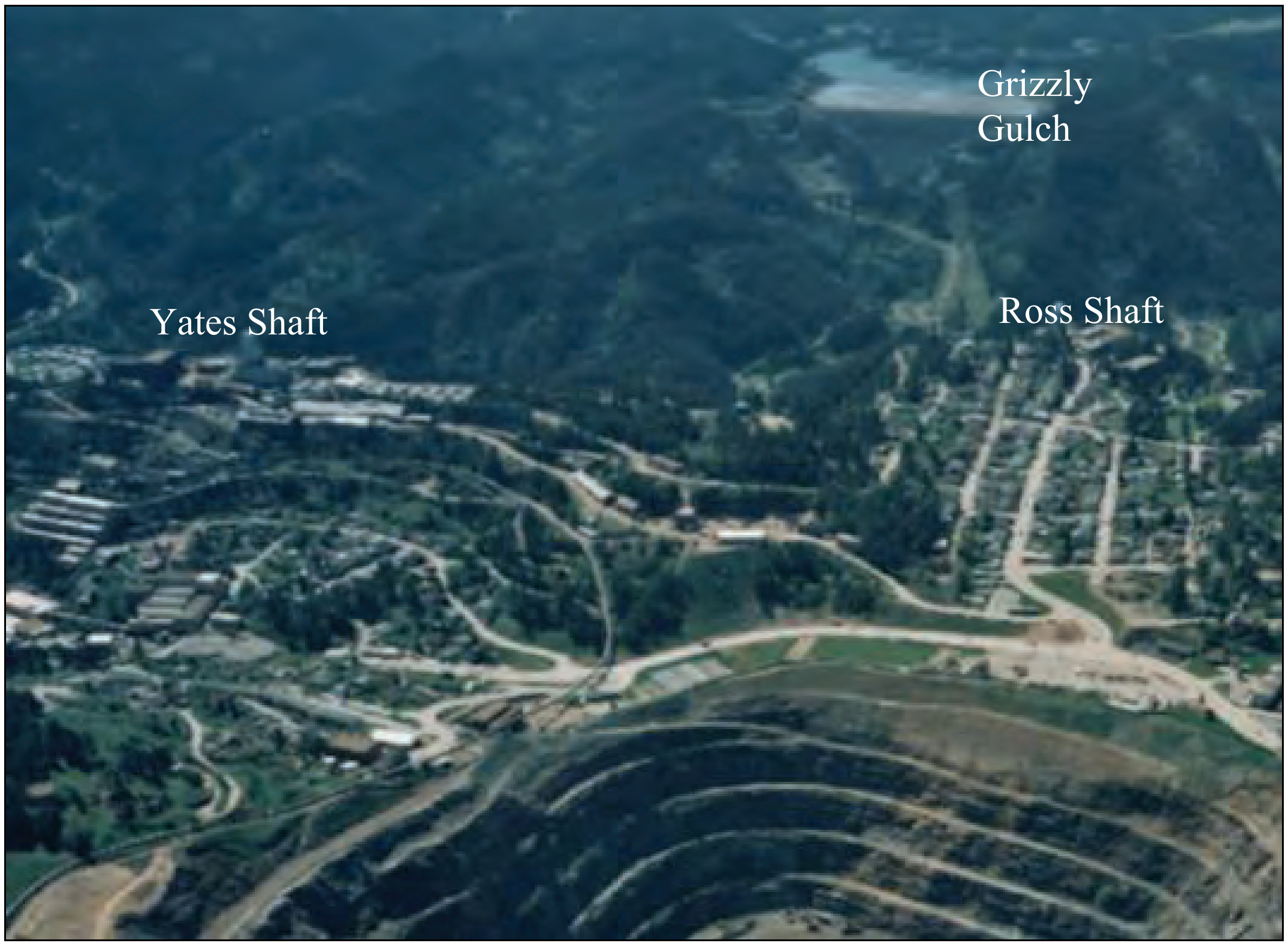

**Figure 5.4.3.12.1-1** Aerial view of Open Cut workings looking to the south.

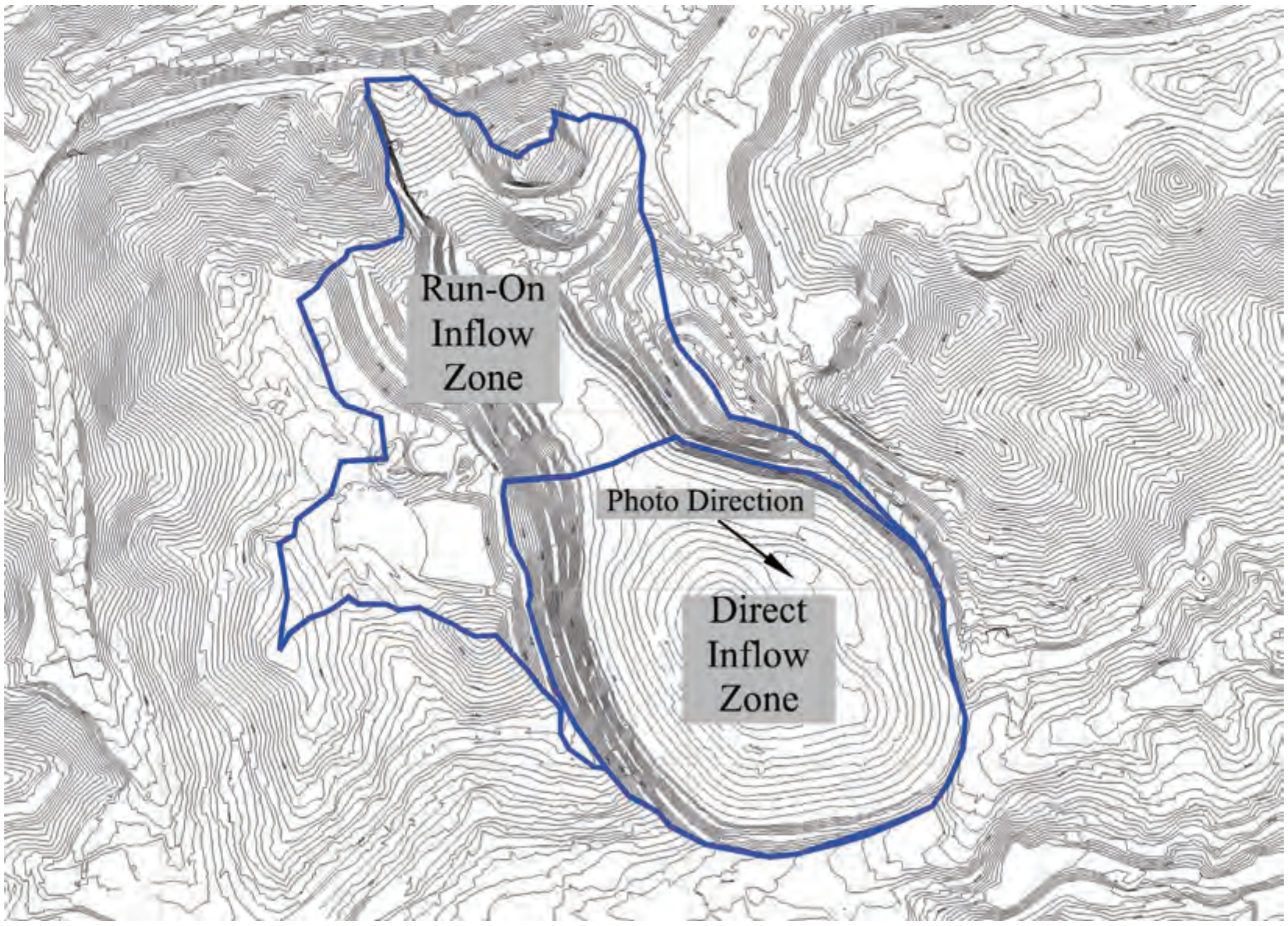

**Figure 5.4.3.12.1-2** Plan view of the Surface Water Catch Basin showing run-on flow path to the Facility through the Open Cut. [SRK]

During HMC operations, certain water-inflow controls were in place to provide mitigation from large rainfall events. Some of these mitigation measures and their purposes are listed below:

- **Sumps.** Provide surge capacity to store baseflow water and limited stormwater to slow its flow to the bottom pool. This water is metered out over time at a slower rate than it enters the underground facility to reduce the inflow through old workings (water was also utilized in HMC operations and WWTP cooling). Homestake utilized a 1.9 million gallon sump and pumping system on the 1100L. There are no plans to use this area for pumping in the future because a large portion of inflows bypass the sump. It will be utilized, however, for limited surge capacity and metered into the future inflow control system.
- **Drainholes.** Provide pathways for localized water inflows to areas where the water can be controlled (i.e. pump stations, sumps, etc.)
- **Hydraulic Bulkheads.** Walls constructed to support a head of water. An example of this would be the historic 6800L north drift plug.
- **Water Diversion Walls.** Walls constructed to less than the height of the drift that will protect an area from a limited rush of water and divert the water to a separate flow path. They are not designed to hold back a head of water beyond the height of the wall.

Even though limited controls were in place during HMC operations, historically the mine was evacuated during very large rain events until the surge slowed. This is a risk until the new designed controls are in place.

Since re-entry by the SDSTA, a campaign of installing new water controls was executed for safe access to begin the infrastructure assessement and mitigation process. These water-control measures included installing more than 60 water diversion walls to protect areas for personnel access. This water inflow-control system is considered a temporary measure until Final Designs have been developed and a long-term system has been constructed. The current controls are designed and installed with the understanding that the Facility, until DUSEL designed inflow controls are installed, may still need to be evacuated for a period of time during a large storm event. During re-entry inspection, numerous examples of the need for water-inflow controls were noted. Some of these are listed below:

- **Sand inundation on the 3650L after storm events in 2008.** A rain event flooded areas not available for inspection to cause a release of sand onto the 3650L. This sand eventually flowed down the Ross Shaft to the 4850L and 5000L. Figure 5.4.3.12.1-3 shows a diversion wall on the 3650L that held back a portion of the sand on the level and provided ample warning time to allow for evacuation.
- **Evidence of water behind stope walls in Ross Pillar ramp system.** Water was observed flowing from numerous Ross Pillar ramp walls and surrounding country rock just below the 3650L. These walls have become dry since the initial inspections.
- **Inspection of the bottom of Millikin Winze.** Inspection of the bottom of the Millikin Winze on the 3500L showed evidence of ponded water that built up above the sill brow and released a muck pile of slough rock already stored in place. Water, rock, and finer solids flowed over a distance of 800 ft to damage the existing Homestake booster fans on the 3500L.
- **Seepage from existing walls or surrounding rock.** Figure 5.4.3.12.1-4 shows an example of an existing mine wall seeping water. Numerous examples of seepage have been identified.

Other examples of water inflow risk potential have been identified, further confirming the need for long-term controls to protect future laboratory infrastructure. Until future diversion controls are in place, a rigorous inspection and monitoring process will be implemented. Pumping, pool level, and discharge sump level status is presently monitored, providing an indication of underground water inflow conditions.

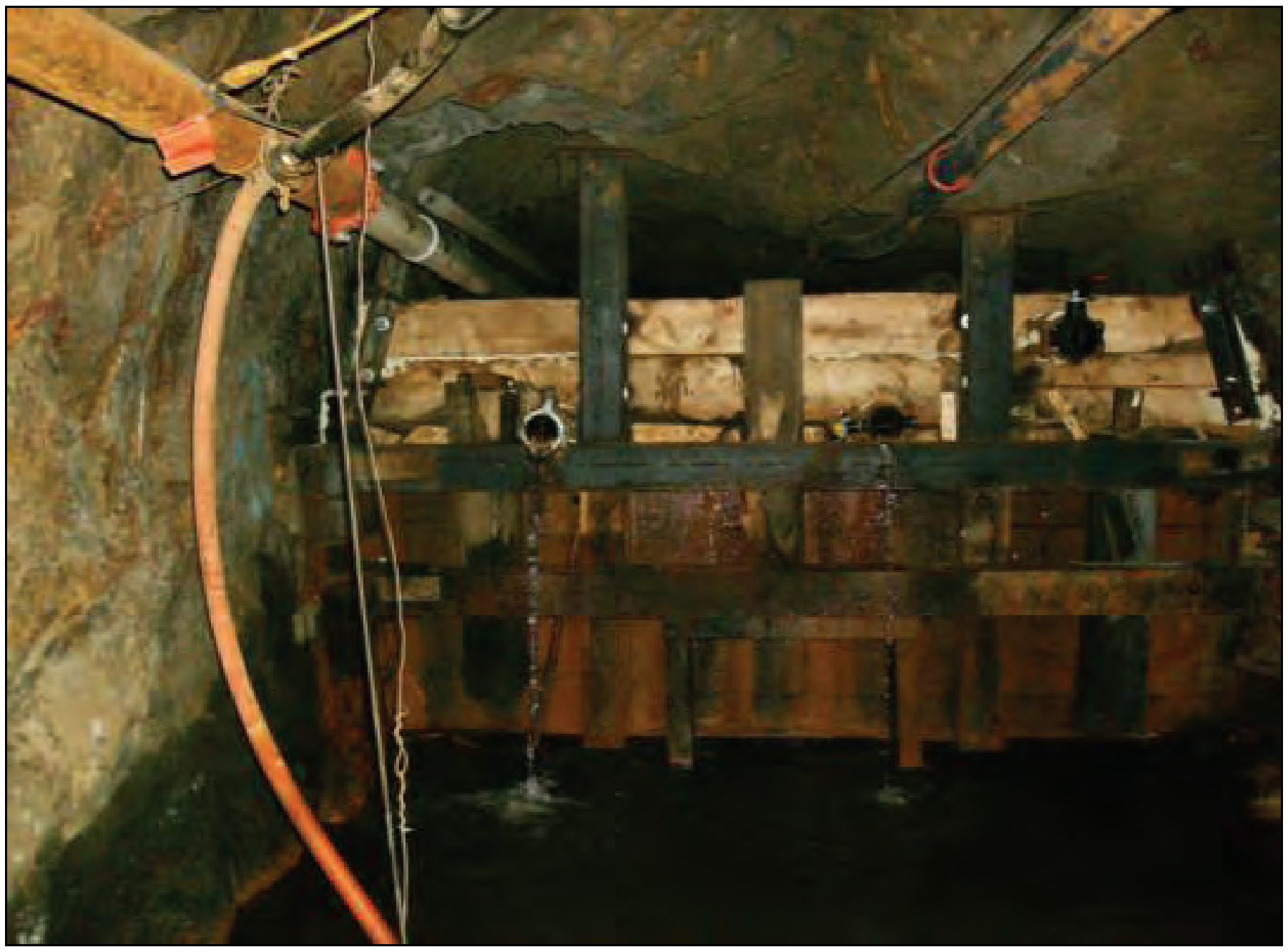

**Figure 5.4.3.12.1-3** SDSTA water diversion wall on the 3650L. [SDSTA]

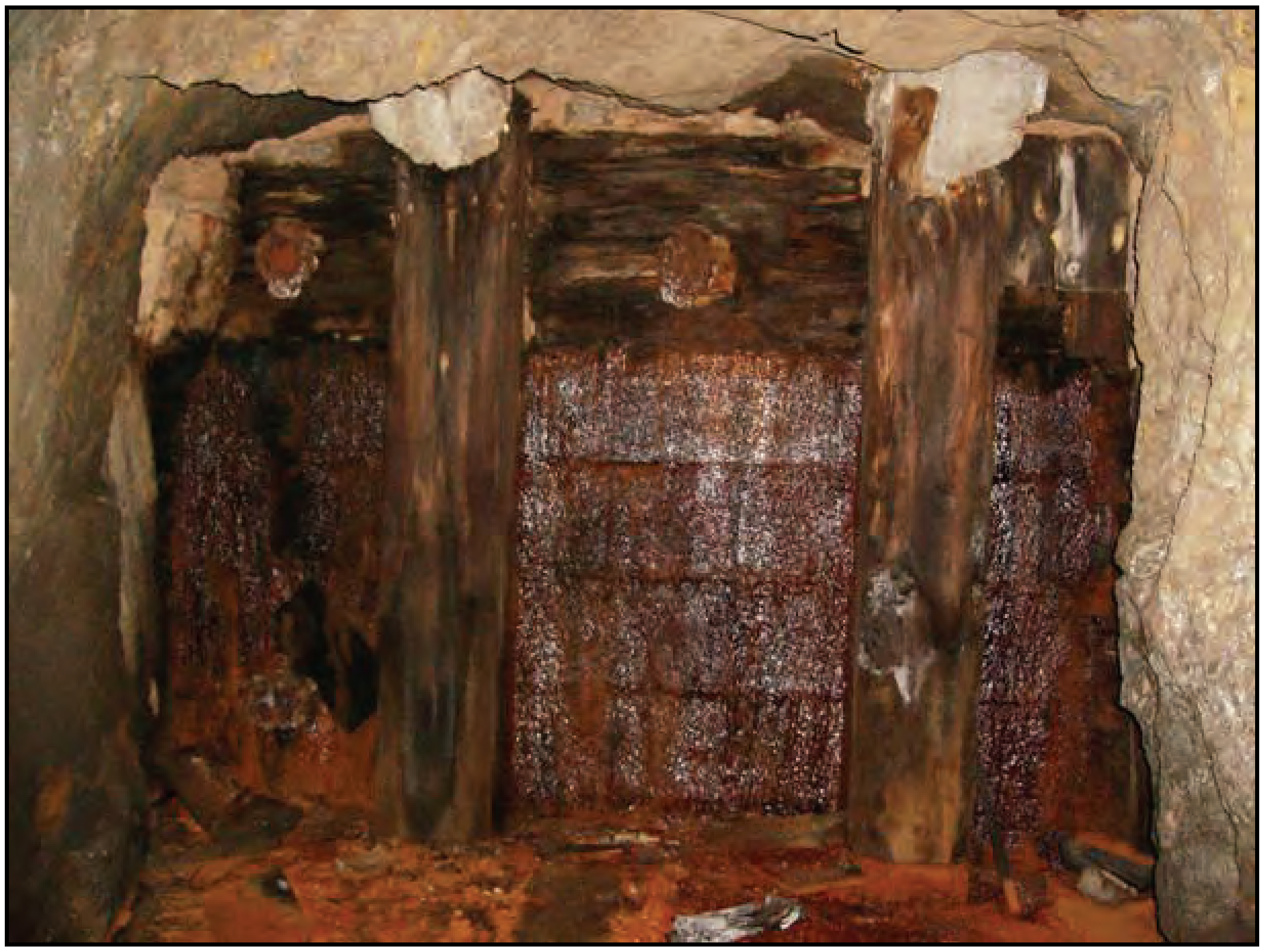

**Figure 5.4.3.12.1-4** Leaking former Homestake mine stope wall on the 2000L. [SDSTA]

### 5.4.3.12.2 Water Inflow Management Requirements

Base flow and storm-water inflow data have been compiled and analyzed to determine the amount and rate of water that needs to be controlled. The key requirement allows for a safe pathway of large rain events to the bottom pool (7700L), protecting all infrastructure and occupied spaces. Long- and short-term monitoring of water inflows will be a necessity, as groundwater conditions can change at any time in an underground environment.

### 5.4.3.12.3 Water Inflow Management Preliminary Design

The primary design strategy for water inflow control is to allow water a free flow path without ponding, except at the lowest levels of the Facility, and strategic control locations during storm events. Historic inflow data has proved that water will enter the DUSEL Facility from time to time due to storm events at flow rates higher than what is possible to store and pump out at the same rate. With this in mind, a cascade system to the pool with strategic surge capacity sumps is necessary. This will allow for water to be directed to the lowest levels of the Facility utilizing mostly existing infrastructure. New surge capacity sumps on the 3500L and 3950L will slow down water inflows and keep the majority of water flowing from the Open Cut from reaching the bottom pool. Figure 5.4.3.12.3-1 shows a cross section of the water inflow control system. Section 14 of the Arup *UGI Basis of Design Report* (Appendix 5.L) shows this system level by level and describes in more detail the types of controls that need to be installed.

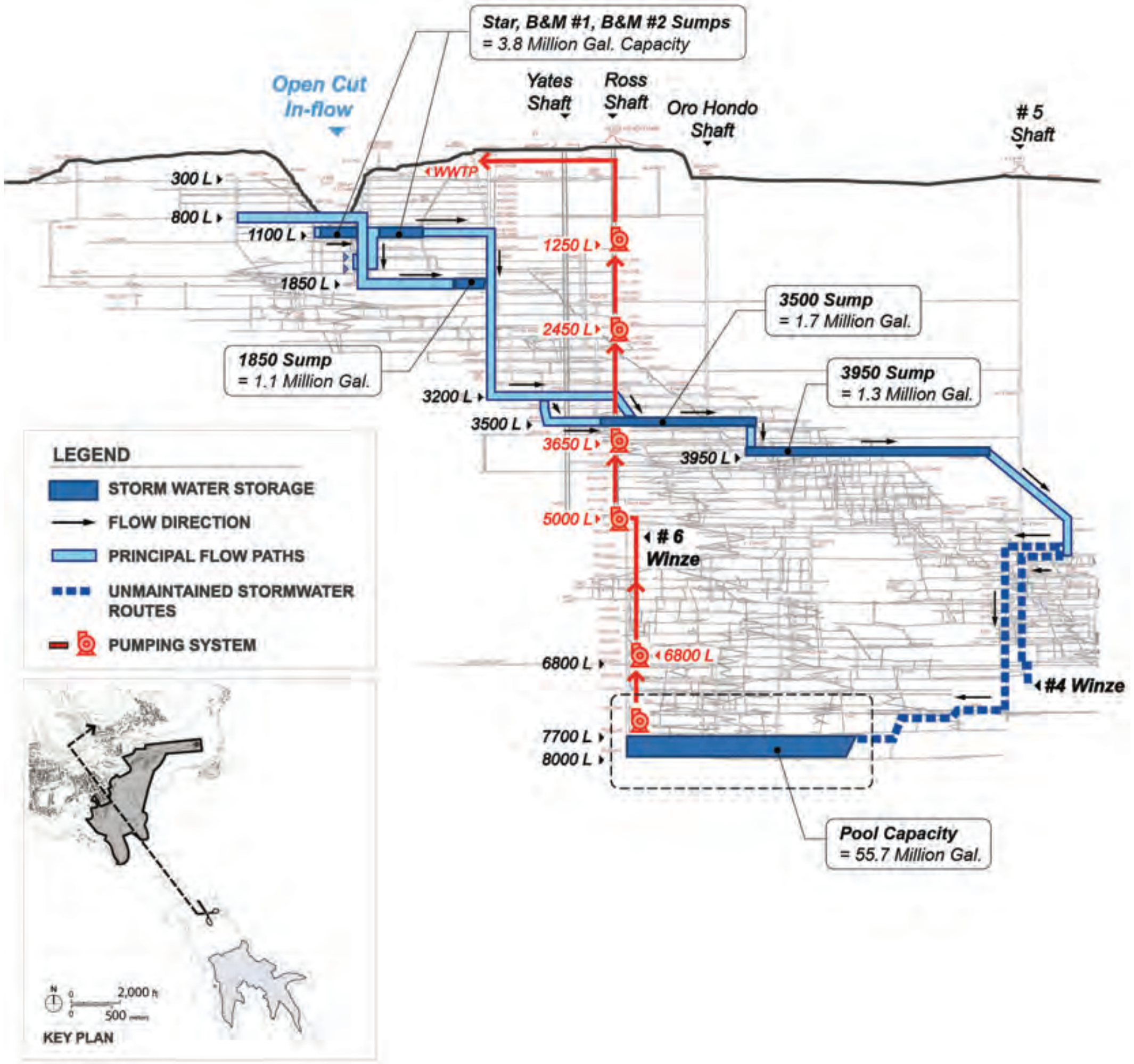

**Figure 5.4.3.12.3-1** Water inflow management plan flow path. [DKA]

SDSTA re-entry diversion walls will be replaced during DUSEL construction with a more robust concrete design. The 3200L is the most critical level in the system where significant water inflows from the upper levels of the Facility flow uncontrolled and are a risk to the Ross Shaft. At the bottom of the Ellison Shaft, a bulkhead containing a large-diameter pipe will be installed that will carry predominant water inflows past fractured zones of rock to a borehole leading to the 3500L sump. Figure 5.4.3.12.3-2 shows the design of this system.

Water in the 3500L sump will be kept at a low volume to provide storm event surge capacity. Base flows and most rain events will be metered into the 3650L pumping station, where it can be pumped to the surface. When large rain events overwhelm the surge capacity of the 3500L sump (1.7 million gallons), water will overflow through a raise to the 3950L. From there, the 3950L will hold 1.3 million gallons of water until it overflows to an existing Alimak Raise to the south that will deliver water to the bottom pool well out of the way of any planned infrastructure. A total of 55.7 million gallons of empty sump will need to be kept below the 7400L to accommodate surge capacity and ensure flooding does not occur on this level. To provide this surge capacity, the Facility will need to be dewatered to the 7700L.

Monitoring systems will be installed on diversion walls and sumps to monitor if a surge of water intercepts a wall. Sump levels will be used to interlock pumping systems.

Future efforts during Final Design will refine the plan to monitor inflows during rain events. This will allow for increased understanding of accessible inflow observation sites and continuation of water inflow data refinement. All work required to control the inflow of water will be performed using SDSTA staff under R&RA funds.

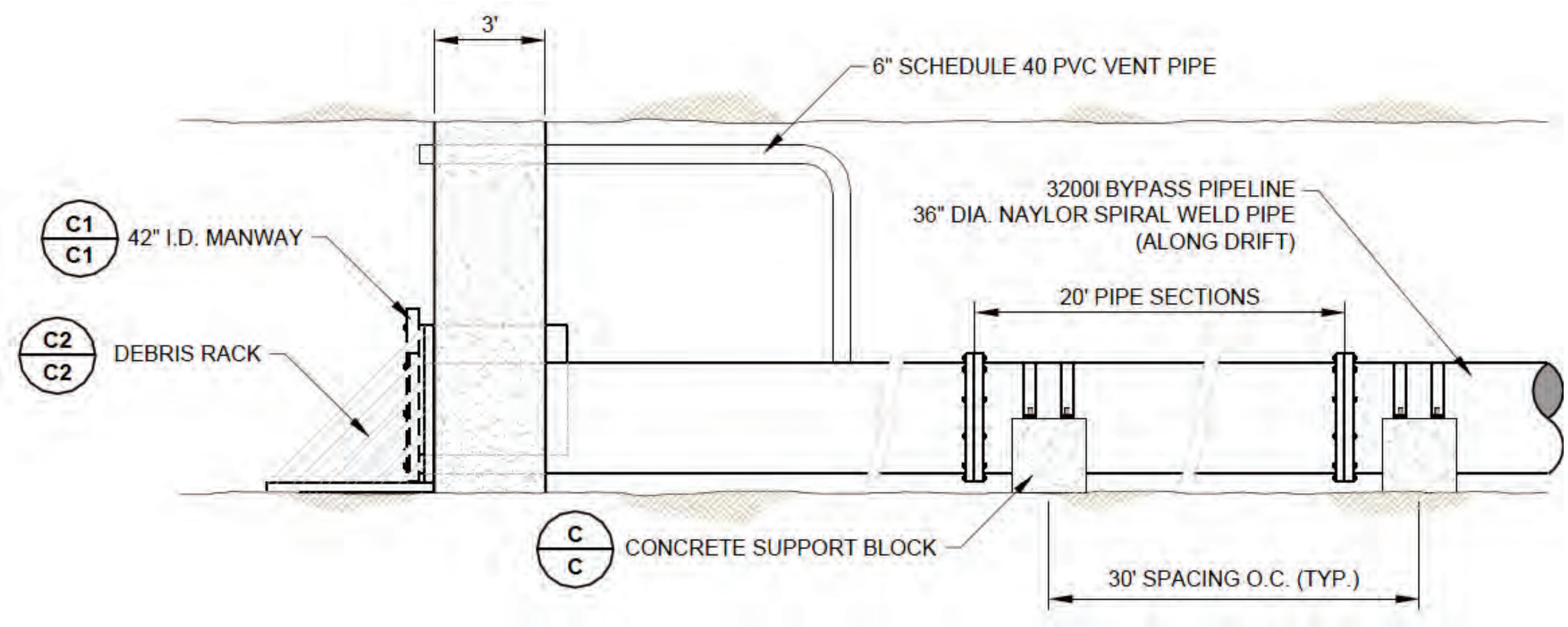

**Figure 5.4.3.12.3-2** Proposed 36-inch (914 mm) pipe and bulkhead at the 3200L. [SRK]

## 5.4.3.13 Chilled Water System

Providing a climate-controlled environment at the 4850L and 7400L will require a heat removal system to remove heat generated by the rock, people, and equipment on these levels. A common solution for widespread campuses is the use of centrally cooled chilled-water distributed where cooling is needed to capture heat. This heat is then rejected into the end of the exhaust air stream to be removed to the surface. The section below describes the current conditions and Preliminary Design of the chilled water system.

### 5.4.3.13.1 Current Condition of the Chilled Water Systems

Despite the difference between a production facility and a science laboratory, the historic facility can be used to help define the design of DUSEL cooling systems. None of the chilled water equipment used by HMC is available for future use, as the systems previously installed were damaged by water. The environmental conditions of the underground site can be inferred from the data gathered during Homestake production. These conditions are expected to change somewhat due to change in ventilation schemes, lining of the shafts, and excavations, but these changes should improve the environment (less particulate matter, lower temperatures, and lower humidity) and make the chilled-water systems more efficient.

### 5.4.3.13.2 Chilled Water Requirements

There are three primary factors that define the required cooling capacity of the chilled-water systems. First, there are natural heat loads due to surface weather conditions and virgin rock temperature (VRT). The surface weather condition impacts are minor in comparison with other heat loads, as auto-compression and evaporative cooling from the surface to the 4850L Campus and DLL allow the ventilation air to reach more stable conditions. Historic HMC records show temperature variations of less than 10°F from the cold winter to hot summer days. The VRT elevates at deeper levels, as can be seen at the 7400L, where the VRT is 125 °F (51.6 °C).

The second primary heat load is from mobile equipment, primarily during excavation, when a large fleet of diesel-powered equipment will be in use. After construction, the heat load from diesel-powered equipment will become less significant. As the chilled-water system will be installed after the excavations are complete, the design will not account for the heat loads generated by the larger mobile equipment, but will consider the smaller fleet required to support operations.

The third and the most predominant heat load is from the electrical equipment. With an underground facility, there is no free cooling as there would be on the surface through the use of the atmosphere as an infinite heat sink. Any energy used underground will ultimately become heat that must be removed from the facility through either air or water. With as much as 27 MW of connected load underground, it is very important to define peak and standard operating loads, along with diversity factors. The connected loads are identified in Appendix G of Arup's *UGI Basis of Design Repor*t (Appendix 5.L), but diversity and actual loads will develop once the scientific program and requirements are further developed.

A requirement that is less dependent on electrical load, but still a major factor in the design, is the temperature of the chilled water. The WCD is currently planned to be cooled to 13 °C (55 °F) using the chilled water, requiring that the chilled water be kept below this temperature with enough differential to provide adequate cooling capacity. The lower the chilled-water temperature, the less efficient (and more costly) the heat exchanger system will be. With a standard design for chilled water at 42 °F, this specific requirement will be easily met.

#### 5.4.3.13.3 Chilled Water Preliminary Design

Two options were considered during Preliminary Design for the chilled-water system. The first option, which was included in the design through the Arup's 60% Basis of Design Report,[21] was to place chillers and cooling towers on the surface in the Yates Campus. In this option, chilled water would pass down the Yates Shaft through a series of pelton wheels (energy recovery turbines) to the 4850L. At this level, the piping would split a portion to the 7400L through more pelton wheels. Very large pumps would be required to return this water to the surface at a rate of 5,000 gpm—over 3 times the rating of the current dewatering system.

The second option, and the option that is included in the 100% Preliminary Design configuration, is to place the entire chilled-water system underground. Chillers and spray chambers will be installed in excavations on the 4850L and 7400L.

Both options have benefits. In the first option, it is possible to tap off of the pipe columns to service OLR. As spot coolers could be used for cooling OLR if needed, and no experiments have expressed a desire for this, this was not a major consideration. Having the chilled-water system on the surface allows for free (atmospheric) cooling during colder weather, which is a significant portion of the year in Lead, South Dakota. The cost savings from this benefit is more than offset by the high costs involved with pumping the large volume of chilled water over 1.5 miles (2.4 km) vertically. The capacity of the underground chilled-water system is limited by the ventilation volume, since all heat must ultimately be removed using this air. Ventilation volumes as designed for the Project are sufficient to support the heat loads from the underground chilled-water system, but would (and could) be increased for future expansion.

Cost estimates for both options were performed by DUSEL staff and verified by both ARUP and McCarthy Kiewit. This analysis is captured in Trade Study #380 (Appendix 9.V). The comparison showed a significant savings for both initial installation and ongoing operational expense for the underground installation.

The design of the chilled-water system considers using an n+1 philosophy. This philosophy takes the total load and divides it between a number of chillers (n), then adds one chiller (+1) of the same size. At the 4850L, four chillers are designed with each delivering 625 tons of cooling, or 33% of the estimated maximum load. At the 7400L, two 625-ton chillers are provided, either of which is capable of supporting the entire required cooling load. This allows one chiller to be down at any time for maintenance without loss of capacity. The same philosophy will be applied to the heat rejection system, which functions much like a large version of any air conditioning system. Heat removal will be provided through the use of spray chambers, where hot condenser water will be pumped to nozzles spraying into the ventilation air at the end of the occupied ventilation circuit. More nozzles will be provided than required to provide redundancy and extra capacity to support maintenance. A portion of the water is evaporated and carried to the surface. The remainder is cooled and returned to the chiller system to continue through the cooling circuit. The evaporated water is continuously replenished with industrial water from the common supply provided through the plumbing system.

Chilled water from the chillers is fed into a storage sump for distribution throughout the level. Redundant pumps provide the mechanism for circulating this water through the heat exchangers and back to the chiller. All connection points are placed in parallel to provide approximately the same temperature of water regardless of distance from the chillers. The piping is insulated to reduce heat gain during transportation.

Each LM and the LC-1 is provided with an air-handling unit (AHU) that includes a series of filters and an induced draft fan that pulls air through a heat exchanger supplied with chilled water. Only a portion of the air is fresh ventilation air, with a larger portion recirculated within the cavity. The recirculation allows the heat to be carried out of the facility in the chilled-water system, which is designed to return with a 16 °F temperature rise. Less-complex systems are designed for electrical and mechanical rooms that use a fan blowing through a heat exchanger to cool the areas. Taps at each experimental area will be connected directly to equipment by the experiments to remove heat from equipment without having the heat transfer first through the air.

The chilled-water system will be controlled by the Facility Management System (FMS). Controls at each AHU and fan coil unit will vary flow rates through the individual heat exchangers to control ambient temperatures. No standby cooling is included in the design. Any experiment-specific cooling required during power outages will be the responsibility of the collaborations.

### 5.4.3.14 Plumbing

#### 5.4.3.14.1 Current Plumbing Systems

As an operating mine, a robust plumbing system existed in the underground spaces to provide water and compressed air to support mining operations. Two additional water lines were installed by the SDSTA to support development of the Davis Campus for early science. Most of this plumbing has been or will be removed during shaft rehabilitation efforts. The only exceptions are the dewatering system as described in Section 5.4.3.11 above, and a 6-inch industrial water line in the Ross Shaft. Industrial water will be required during construction and future laboratory operation, so this existing line will be repaired where needed and used as the permanent supply. To address the pressures created by travelling through nearly a mile of vertical shaft, pelton wheels are installed at the 2450L and 5000L. These wheels provide a pressure break and capture the energy, generating electricity. The pelton wheel systems require rehabilitation, which will be completed by the SDSTA using R&RA funds.

Service to the DUSEL site is supplied from the city of Lead, with an 8-inch and 12-inch industrial water line and a 6-inch and 4-inch potable water line at the Yates and Ross Campuses, respectively. The water source comes from mountain streams west of Lead, with two separate water sources diverted into pipes feeding the city. This system was originally installed by HMC and has provided a reliable water source since installation. The city can provide up to 1,600 gpm of industrial water year-round without modification to their existing infrastructure. This flow rate can be temporarily augmented for fire protection by lowering the city's 1.6 million gallon industrial water storage tank, which is maintained at or above 50% full at all times. Potable water will be restricted to less than 1,000 gpm without modifications to their system. No detailed studies have been performed to determine maximum flow availability with modifications, but a conservative estimate is that these values could be doubled with additional pumps. No agreement is in place currently for any set value of water availability.

Managing water after it was used in equipment during HMC operations was done using natural drainage augmented by infrequent sumps or boreholes to levels below. This allowed the water to eventually reach the dewatering system for removal from the mine.

#### 5.4.3.14.2 Plumbing Requirements

Plumbing systems are required for compressed air and water. Two compressed-air systems are included as part of the UGI scope. A small compressed-air system is included in the mobile equipment maintenance shop to support small tools with requirements defined by the tools expected for use. Additional compressed-air systems are provided for each LM, included as part of the Underground Laboratory (UGL) scope. The second compressed-air system is required to provide breathing air for AoRs. The requirements for this system are derived from NFPA 520, which dictates 20 cfm/person. To ensure adequate air supply to all AoRs, no diversity is applied to this number, resulting in a volume defined by the total capacity of all AoRs. In addition to volume requirements, the air is cleaned and dried to breathing air standards.

Water supplied to the Facility can be divided into three primary categories:

- Industrial water comes from the city of Lead's water supply, which originates from a system of mountain streams and creeks. This water is not treated in any way.
- Potable water is taken from the industrial water stream but treated by the city of Lead. This treatment uses sand filtration and adds fluoride and chlorine.
- Purified water is provided by the Long Baseline Neutrino Experiment (LBNE). As the current design of this experiment requires 100 kT of purified water, a reverse osmosis system provided by LBNE will remove all impurities from industrial water supply and use stainless steel pipe to deliver this water to the 4850L. At the 4850L, a connection will be provided as part of the MREFC-funded Project to allow other experiments access to this water.

Each of these water systems has different specific requirements, but only general requirements are included in this document. For both industrial and potable water, the primary volume requirement is defined for fire protection since these two types provide redundancy. A total of 3,250 gpm is required from each water type: 1,000 gpm for sprinklers, 250 gpm for a water hose connection, and 2,000 gpm for water monitors in the LMs. These monitors are remotely operated nozzles that allow fire fighting from outside of the affected spaces. Using the 1,650 gpm difference between 1,600 gpm supply and 3,250 gpm fire fighting requirement, and assuming the worst case of a 50% full city water supply tank, eight hours of flow can be provided at this rate before the tank is empty and flow drops to the 1,600 gpm base. Normal-use industrial water will be for toilet flushing, experiment use, and make-up for the chilled-water system. The chilled-water system make-up water will be the highest demand, with volumes estimated around 200-300 gpm. Another 600 gpm of industrial water will be used as LC-1 is filled with purified water from the LBNE system, but this will be a temporary requirement. Ongoing make-up water for LC-1 is expected to require less than 100 gpm.

For the purified-water system, all requirements are defined by LBNE outside of this scope of work. A 3-inch stainless steel pipe is connected to the LBNE supply to match the pipe size included in their design.

Water removal from the underground is primarily managed with the dewatering system described in Section 5.4.3.11, but additional systems are required on the 4850L and 7400L to capture all water used in the course of normal operation and direct it through pipes to the dewatering system. Flow requirements for this drainage system are also primarily derived from fire protection needs. To ensure environmental contaminants are not introduced into the dewatering system, experimental space sumps will be required to be tested prior to discharge into the main drainage system. If contaminants are found, the experiment will be required to treat the water, or the water will be manually removed via tanks for proper disposal at the expense of the collaboration.

#### 5.4.3.14.3 Plumbing Preliminary Design

The design of plumbing systems includes plumbing for compressed air and water, both into and out of the underground facilities. The water coming out of the underground is discussed in Section 5.4.3.11, *Dewatering and Pumping*. This section discusses the remaining plumbing systems. Figure 5.4.3.14.3 shows an overview of all water systems.

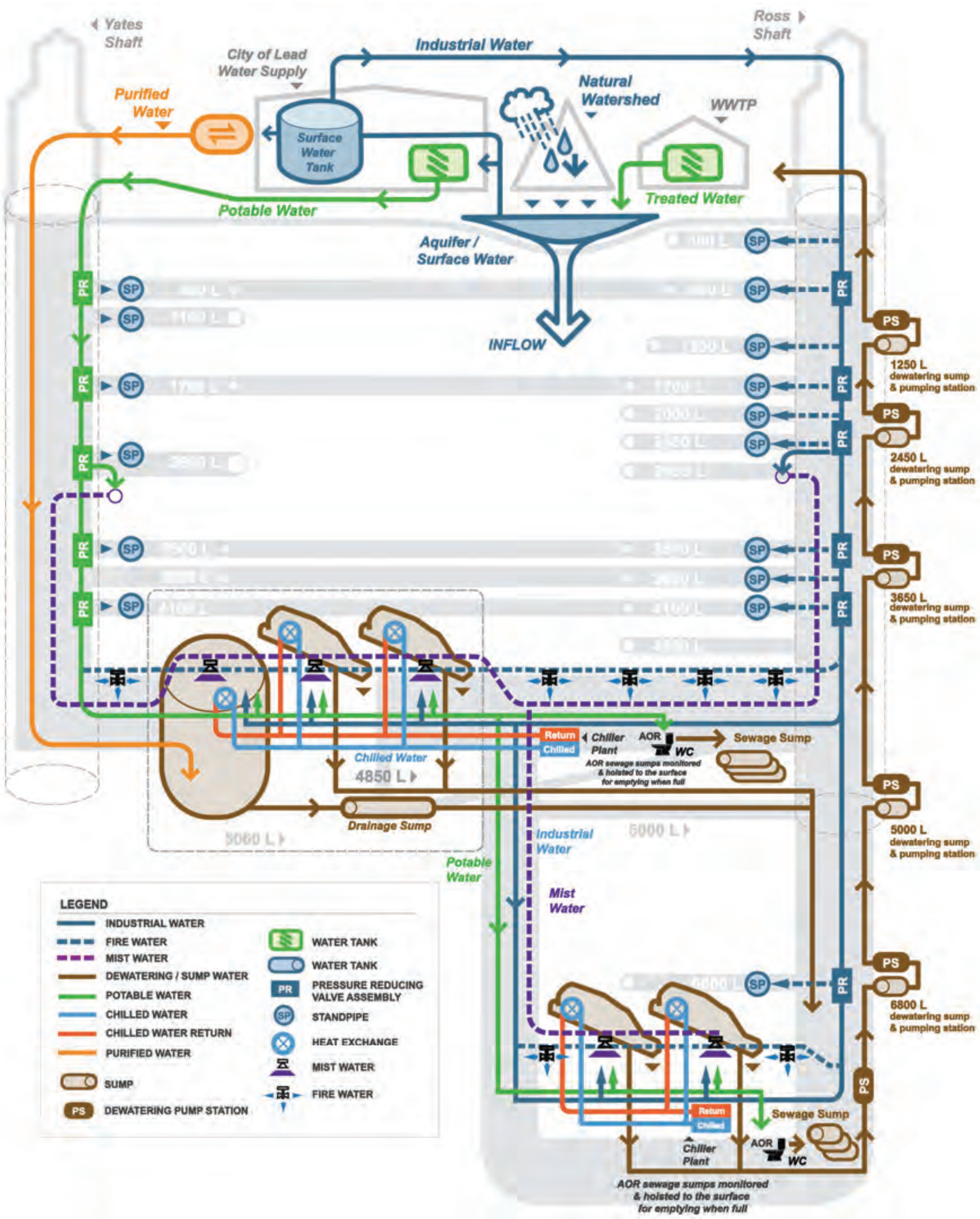

**Figure 5.4.3.14.3** Water system overview. [DKA]

### Compressed Air

The compressed-air system included for use in the equipment maintenance shop has not been designed in detail for this report, but will be a simple design typical to any maintenance facility with a compressor, filter, dryer, and receiver tank tied to a small pipe distribution system. For the LMs, as part of the UGL scope, a similar system will be included as part of the Final Design, but will include two compressors for each LM for redundancy.

For breathing air, a requirement of 15,980 cfm has been communicated to the Surface Design Team to design the compressors. The UGI scope takes the air using a 10-inch pickled steel pipe from the shaft collar to the 4850L in the Yates Shaft and forms a loop around the 4850L. Two branches from this loop supply the 7400L through the #6 Winze. Redundancy is provided at each AoR with bottled oxygen and either $CO_2$ scrubbers or soda lime curtains, depending on the size of the AoR. These consumable items (oxygen and scrubbers/curtains) will be provided using R&RA funding. Each AoR connects to the compressed-air system and oxygen through the AHU in the AoRs. Both compressed air and bottled oxygen use regulators to control the volume and pressure of air delivered to the occupied spaces of the AoR.

**Industrial Water**

Industrial water will use an existing pipe in the Ross Shaft as described above. A branch will be added at each level intended for future science or facility use to provide fire protection at these levels. Pressure-reducing valves will be installed at these levels and standpipes to allow hose connections. Any water required for experiments at each level can be taken from these lines, but the cost for pipe will be within the science collaborations budget. The pelton wheels mentioned previously will provide the primary means of pressure reduction, but new pressure-reducing valves will be added to allow these wheels to be bypassed if needed.

At the 2600L, a branch line will be installed to provide high pressure to the 4850L for use in a water mist system. The water mist system requires fine filtration to prevent nozzle plugging and will only be used for equipment protection in areas requiring this form of protection. For the MREFC-funded Project, the only systems using water mist are the communications data rooms.

At the 4100L, another branch in the main water line separates water to be used for sprinklers, standpipes, and water monitors from water intended for experiment and facility use. This separation ensures adequate volume will be available for fire control independent from other uses. From this point, the industrial water, now separated into three pipes, continues to the 4850L. At the 4850L, industrial water is provided to each experimental area, where it is capped for future connections. Industrial water provides make-up water for the chilled-water system and water for flushing toilets in AoRs. A line continues to the 7400L in the #6 Winze, splitting at the 6800L into fire protection water and general service water. At the 7400L, this water again serves the chilled-water system and AoRs.

Also at the 4100L, a branch will be added prior to MREFC-funded Construction to provide fire protection water to the Davis Campus via a new borehole between the 4100L and 4840L. This is discussed in Section 5.4.2.2.

**Potable Water**

Potable water will be provided through a new 8-inch water line in the Yates Shaft. This service was designed into the Yates Shaft to provide redundancy for fire protection so a major event in either shaft would not cut off water to the levels below. The concept for the potable water line is similar to that of the industrial water, splitting at each level intended for future science or facility use to provide fire protection. Pressure control for this water will be provided through a series of pressure-reducing valves as opposed to pelton wheels used for the industrial-water system. Pelton wheels are typically used in high volume applications, where the energy recovered justifies the capital and maintenance cost for them. Except when used for fire control, potable water will not have a high demand, and will only be used for drinking. The water mist division for potable water is at the 2450L as opposed to the 2600L for industrial water, but the fire control water is split on the same level (4100L) as the industrial water. At the 4850L, the water is also distributed similar to the industrial water line, but will not be used for the chilled-water system. In the AoRs, potable water is used as the primary supply of drinking water, while a supply of bottled water sufficient for the AoR capacity for 96 hours provides secondary drinking water.

Potable water continues to the 7400L through the #8 Winze through two lines. One of these lines is intended for use as water mist and the other for fire control and general use. The #8 Winze will not include any connections to existing levels, so only one pressure-reducing valve will be included at the bottom of the winze.

#### Purified Water

As previously discussed, the purified-water system is primarily designed by LBNE. The only portion included in the MREFC-funded Project is a 3-inch stainless steel pipe on the 4850L from the LBNE system to the LMs.

### 5.4.4 Conclusions

As outlined in this section, during the Preliminary Design phase the DUSEL Project has completed a comprehensive set of site investigations to inform the Preliminary Design presented in this section. While in some cases the required rehabilitation and upgrades are significant, the conditions are understood and the resulting design addresses the current conditions in the context of the future laboratory requirements to provide a viable facility for the long term. The Preliminary Design is responsive to the needs outlined in the facility requirements to support operation of a safe laboratory that supports the planned DUSEL science agenda as outlined in Volume 3, *Science and Engineering Research Program*.

Additional information describing the sequence of design and construction for the infrastructure and other facility components can be found in Chapter 5.10, *Final Design and Construction Acquisition Plan*. Chapter 5.10 also describes scope options that will be considered if additional funding is identified to support additional design.

## 5.5 Cyberinfrastructure Systems Design

The Cyberinfrastructure (CI), Information Technology (IT), and data systems at DUSEL will constitute an expansion of the developing Sanford Laboratory network, which itself will continue to evolve over the coming several years. The Sanford Laboratory network continues to address immediate needs, whereas DUSEL will address future, expanded laboratory requirements. In the lifetime of DUSEL, including design, construction, and operations, many new services will be provided, and the CI team, with Systems Engineering Integration Team (SEIT) support, will track science and enterprise requirements and modify the technology road map accordingly. Throughout the Final Design phase, CI technologies will be regularly examined to ensure the Facility design incorporates current technology with a defined upgrade path as CI technology evolves. Options will be kept open as late as possible, especially in the case where competing emerging technologies exist.

### 5.5.1 Cyberinfrastructure Systems Summary and Overview

CI is the term coined by the National Science Foundation (NSF) to describe the high-performance systems, networks, and services designed specifically to meet the needs of modern scientific research. In providing a robust, flexible, and extensible CI architecture in support of science and laboratory operations requirements, DUSEL will need to leverage cutting-edge information technologies and be flexible enough to incorporate yet-to-be-developed technologies into the design. The overall DUSEL CI design goals address three main components: support for science experiments, including high-bandwidth science data transmission, computing and storage, and experiment monitoring and control; support for laboratory business functions; and support for laboratory facility operations, including monitoring, command, control, and management of facility systems, which include life safety and emergency response support systems. Throughout the CI design process, several key principals have guided CI design efforts: flexibility and scalability to adjust for future requirements and technology evolution, use of open standards by default to avoid proprietary solutions, adopting the best of what already exists at the Sanford Laboratory and best practices from other science institutions and commercial industry, and using a risk-based management approach in the design of DUSEL CI systems. To guide CI design, the Project is addressing NSF data preservation requirements and plans to complete a Data Management Plan prior to Final Design.

Given the complexities of deep underground laboratory operations using advanced experiment technologies, the DUSEL CI architecture is designed to meet a demanding set of requirements that exceed standard facility IT systems. The design of the DUSEL CI systems is based on requirements provided by scientists and engineers during the Preliminary Design phase that are outlined in Volume 3, *Science and Engineering Research Program*. For example, one of the primary CI functions is the transmission of science experiment data from the underground laboratories to the surface, where experiment-provided CI storage and computing systems are located to provide longer-term data archival and science computing cluster capabilities. In support of experiment operations, scientists require high data bandwidth transmission capabilities, accessible worldwide, to share science data with large, diverse collaborations in an open form to facilitate scientific discovery. These large data transmissions exceed what is generally available from current firewall and network security device technologies. Advanced network security approaches using firewall-less adaptive monitoring and control are required to securely manage these high-bandwidth data flows. Conversely, these same collaborations require highly secure data connections accessible to a very limited user group for monitoring and control of experimental equipment to guarantee safe experiment operations. Both requirements push the limits of CI security technologies and practices in two extremes.

In the case of business and facility operations domains served by DUSEL CI systems, data must be stored and backed up in off-site locations not only to prevent information loss in the case of a hardware failure, but also to make data accessible at a moment's notice to support operations continuity and disaster recovery plans—including running communications and operations from a remote command and control center geographically separated from the primary DUSEL command and control center.

In general, the major requirements for CI support include needs for:

- 10 Gbps network bandwidth providing continuous connectivity to the Internet and Internet2—network connections between the underground laboratory space and science collaboration control rooms and offices located on the surface—both on and off the DUSEL site
- Robust network services and security that accommodate high-bandwidth data transfers using advanced firewall-less technologies employed at other major science laboratories
- Distance communications support such as video teleconferencing and voice data communications
- Infrastructure to support education and public outreach activities, including emerging visualization technologies

Off-site connectivity, allowing high-speed transfer of data from DUSEL to collaborators' home institutions around the world, is also a critical item for consideration in the design of the CI architecture.

DUSEL CI planning, implementation, and operations efforts are reviewed by an independent Cyberinfrastructure Advisory Committee (CIAC) with current membership listed in Section 5.5.8. The CIAC, along with the DUSEL staff who are responsible for maintaining the infrastructure during installation and operation of DUSEL, work closely with the DUSEL science liaisons and Facility design team to ensure both the facility and scientific objectives are met.

In order to advance the CI and Monitoring and Controls subsystem designs, a System Integration Plan (SIP) was developed during the Preliminary Design phase by Arup USA. The SIP development effort included guidance from a SIP Steering Committee comprising representatives of the DUSEL and SDSTA staffs, including members of the Facility, EH&S, and Science teams. The SIP includes the analysis of existing systems and recommendations for further development of the CI and Monitoring and Controls subsystems during the Final Design phase. The complete SIP report is included in Appendix 5.P.

### 5.5.2 Assessment and Condition of Current Cyberinfrastructure Systems

In support of early science, laboratory administration, and current operations, Sanford Laboratory deployed early networking and communications capabilities for both the surface and underground facilities funded outside of the MREFC budget. Sanford Laboratory will continue to develop these early network activities to meet immediate needs, whereas the DUSEL CI designs will be deployed in conjunction with the facility construction to support DUSEL Facility and experiment operations. Figure 5.5.2-1 shows the existing high-level surface network topology. For comparison, a similar diagram showing the high-level DUSEL surface network topology is shown in Figure 5.5.4.1-1.

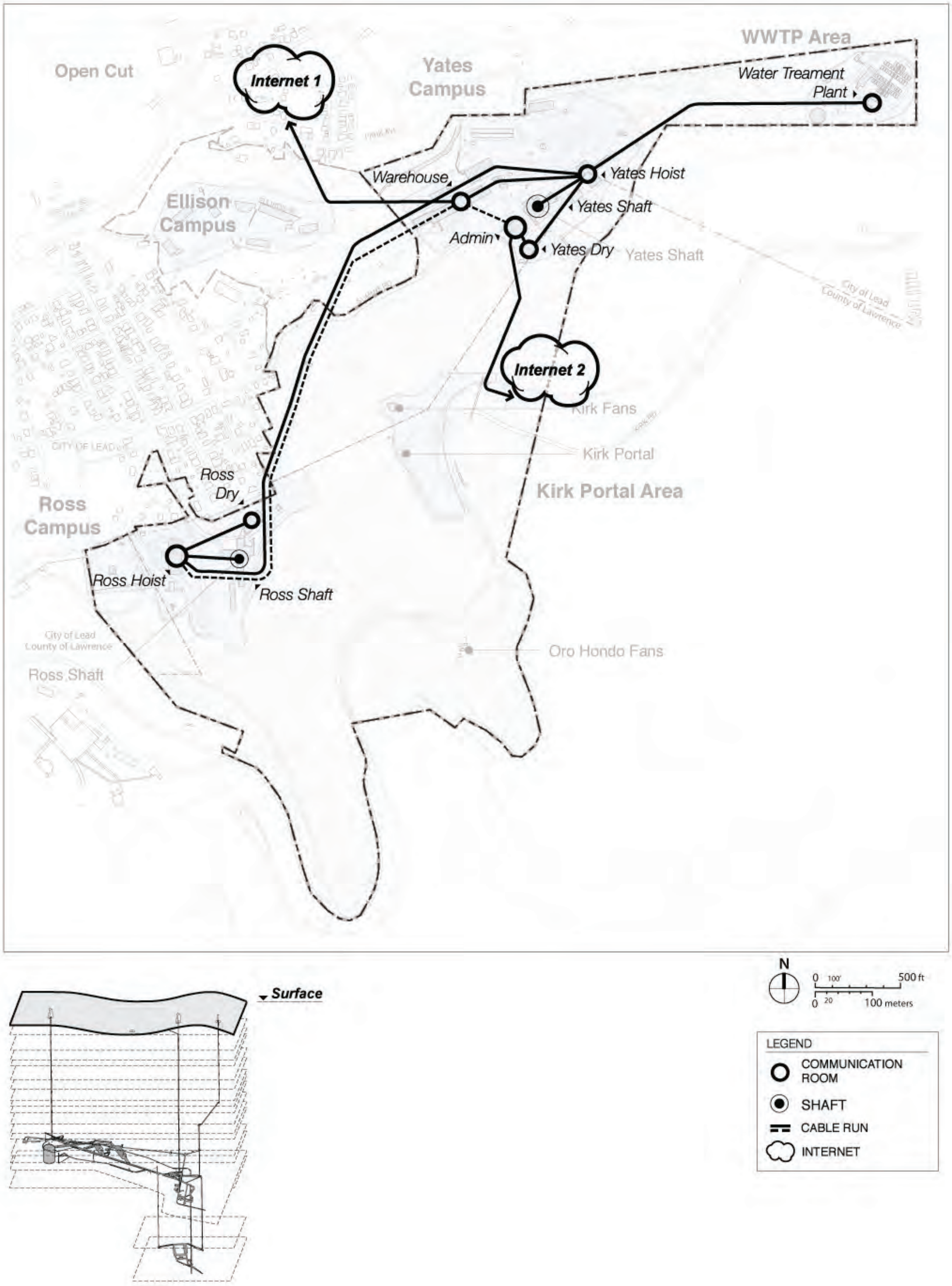

**Figure 5.5.2-1** Current Sanford Laboratory surface network topology. [DKA]

### Networking and Computing

The current network infrastructure at Sanford Laboratory includes a central IT room at the Yates Administration Building, housing file servers, storage systems, network patch panels, switches, and connections to the Internet and Internet2.

The existing surface fiber-optic network exists in a star topology with the Yates Hoistroom as the focal point. From the Yates Hoistroom, separate fiber trunks extend to the Waste Water Treatment Plant (WWTP), the Warehouse (also called the Sanford Laboratory Surface Laboratory), the Ross Hoistroom, the Administration Building, and the Yates Headframe. From the Ross Hoistroom, the fiber is extended to the Ross Headframe and the Ross Dry Building. The Yates Dry Building is fed from the Administration Building. Secondary Ethernet connections exist throughout the Facility using copper media and/or wireless access points.

Twelve and twenty-four strand single-mode fiber-optic cables currently are installed in the Ross Shaft and are used in the Programmable Logic Controller (PLC) control network for monitoring and control of the dewatering pump system, video cameras to monitor dewatering pumps and shaft stations, Global Positioning System (GPS)-based timing signals used by early science experiments, connectivity for Internet access and video broadcasting, and a Voice Over Internet Protocol (VoIP) telephone system.

### Controls, Monitoring, and Security

Sanford Laboratory has real-time, PLC-based control systems that monitor the underground, WWTP, and limited other surface based systems. These other systems include underground air quality, such as air flows and carbon monoxide (CO) levels, exhaust fan status, control, and fault status; dewatering pump system status, including pressures, voltage and current levels, temperatures, and water reservoir levels; and electrical distribution system status and performance. Current control and monitoring equipment use General Electric (GE) 9030 PLCs and GE Proficy iFix Human/Machine Interface (HMI) software. HMI computers provide audible alarms to operators and allow them to visually monitor the Facility on process-flow-based graphic screens, start and stop pumps, and trend and store system data. Figure 5.5.2-2 shows a typical existing HMI screen from the WWTP.

In addition to the current HMIs, GE touch screen QuickPanels are located in the pump rooms and hoist operator stations. The QuickPanels have reduced graphic capabilities and less functionality than the iFix HMI, but still allow operators and technical staff to monitor and control equipment from remote locations. Figure 5.5.2-3 shows a typical QuickPanel and enclosure at a hoist operator workstation.

The Sanford Laboratory security systems include gated entrances (Figure 5.5.2-4 shows the Ross Campus entrance gate), proximity badge access control locks on gates and building entrances, and a video surveillance system. Badges are created and managed by Sanford Laboratory staff. Electronic asset tracking is not currently used at the Facility but is planned for DUSEL.

The video surveillance system can be monitored remotely over existing data networks. Figure 5.5.2-5 is a sample screen shot of the surveillance camera monitoring system.

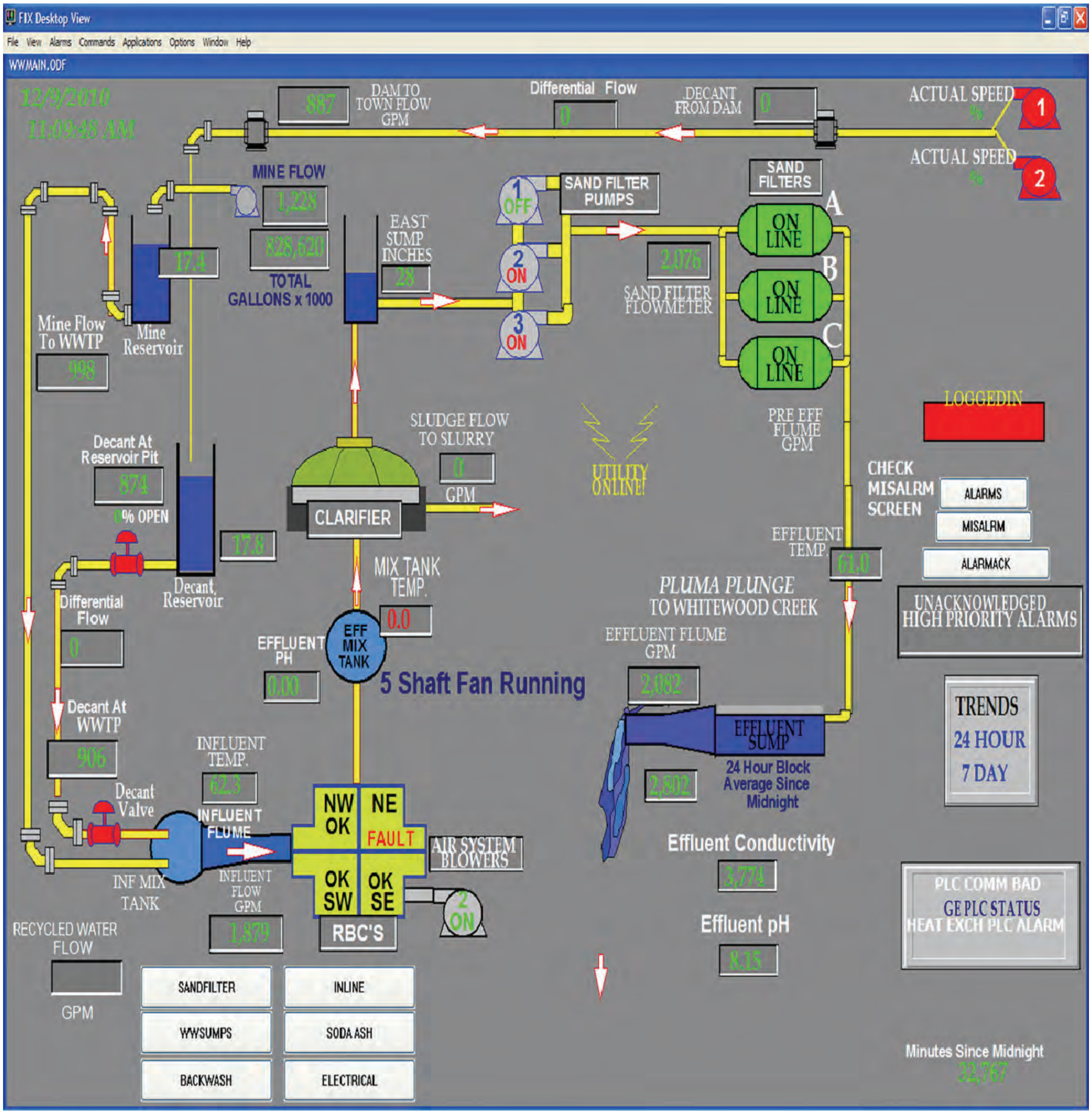

**Figure 5.5.2-2** Screenshot of an existing Waste Water Treatment Plant (WWTP) HMI screen.

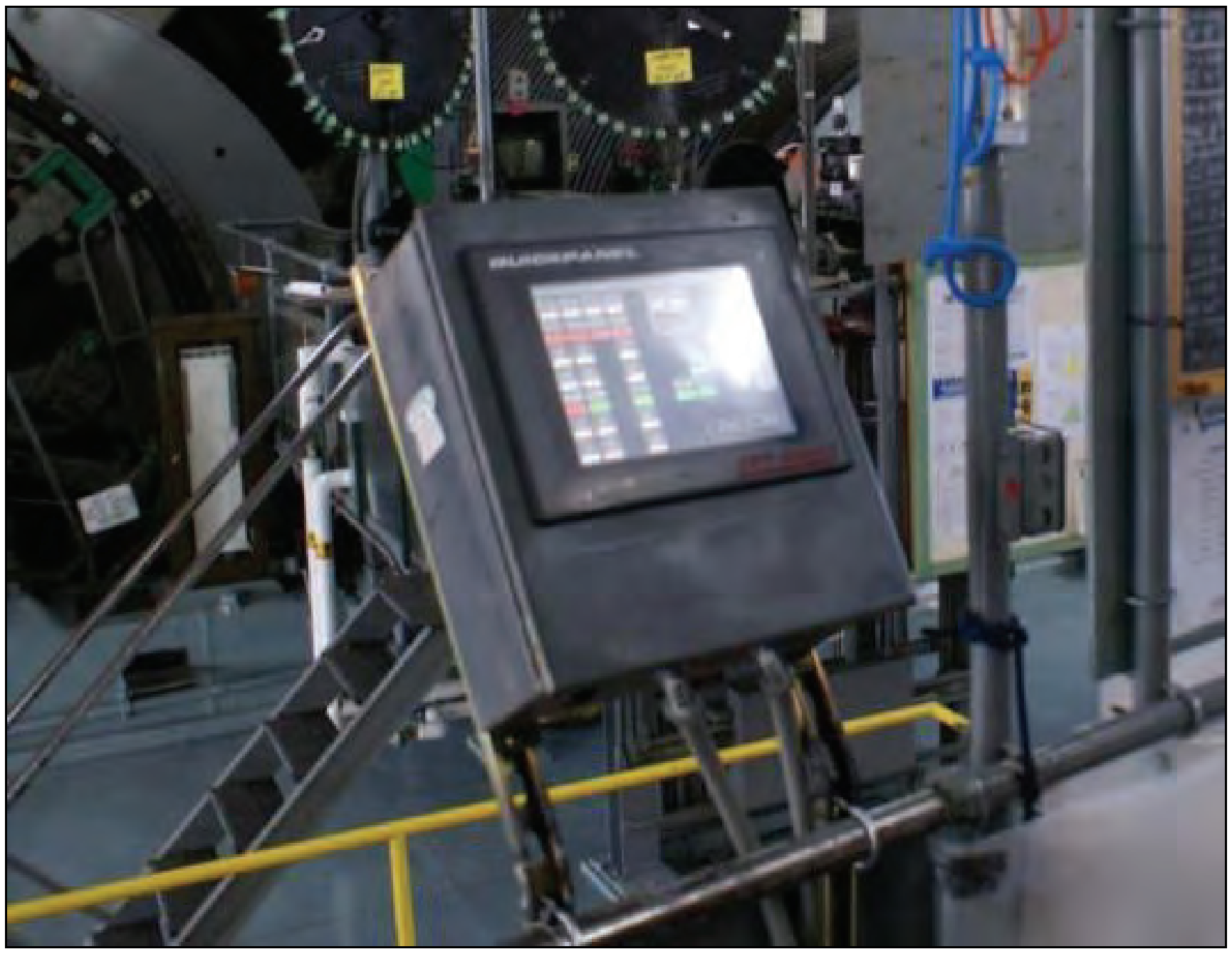

**Figure 5.5.2-3** Existing QuickPanel operator Interface at a hoist operator work station.

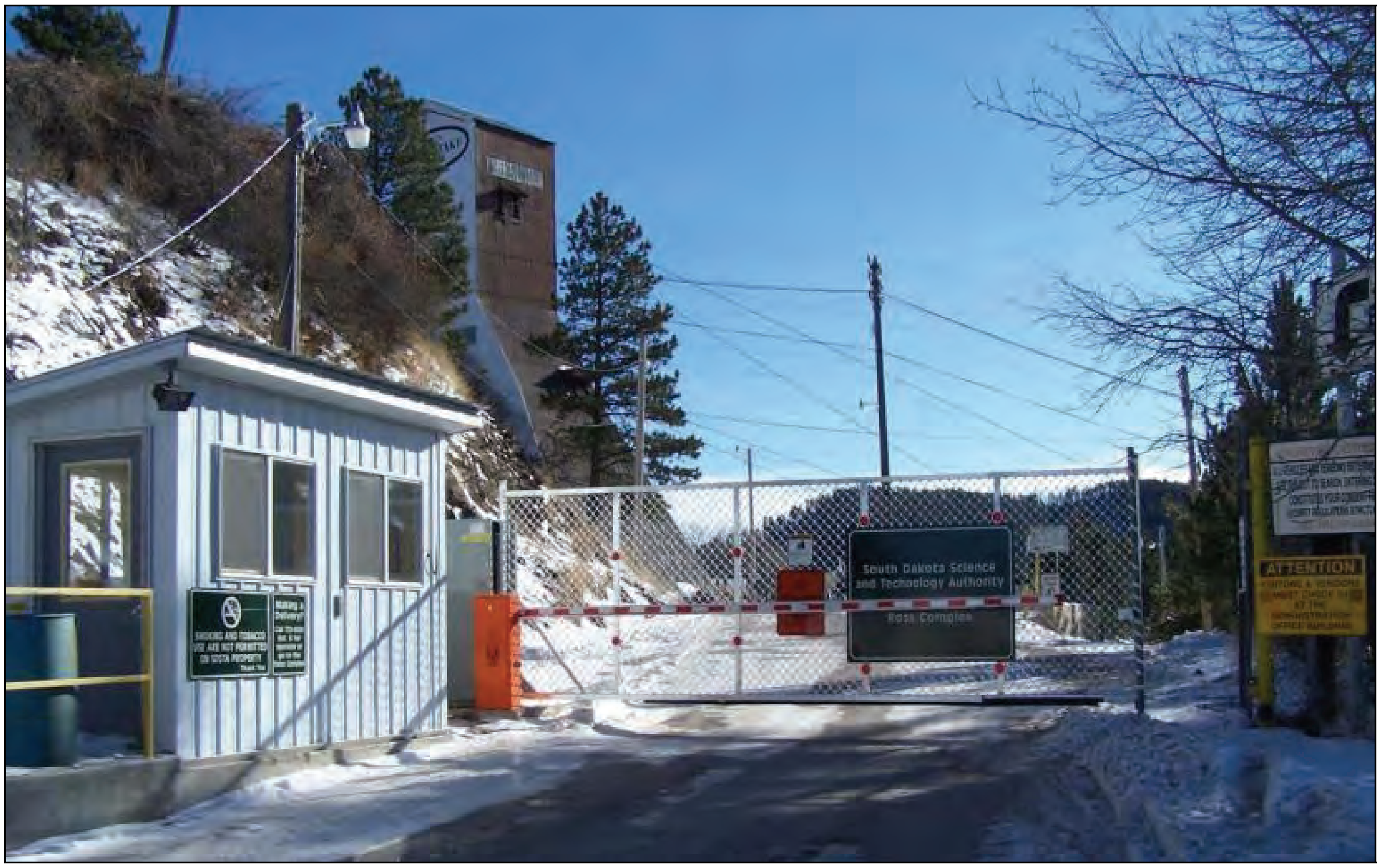

**Figure 5.5.2-4** Ross Complex entrance gate.

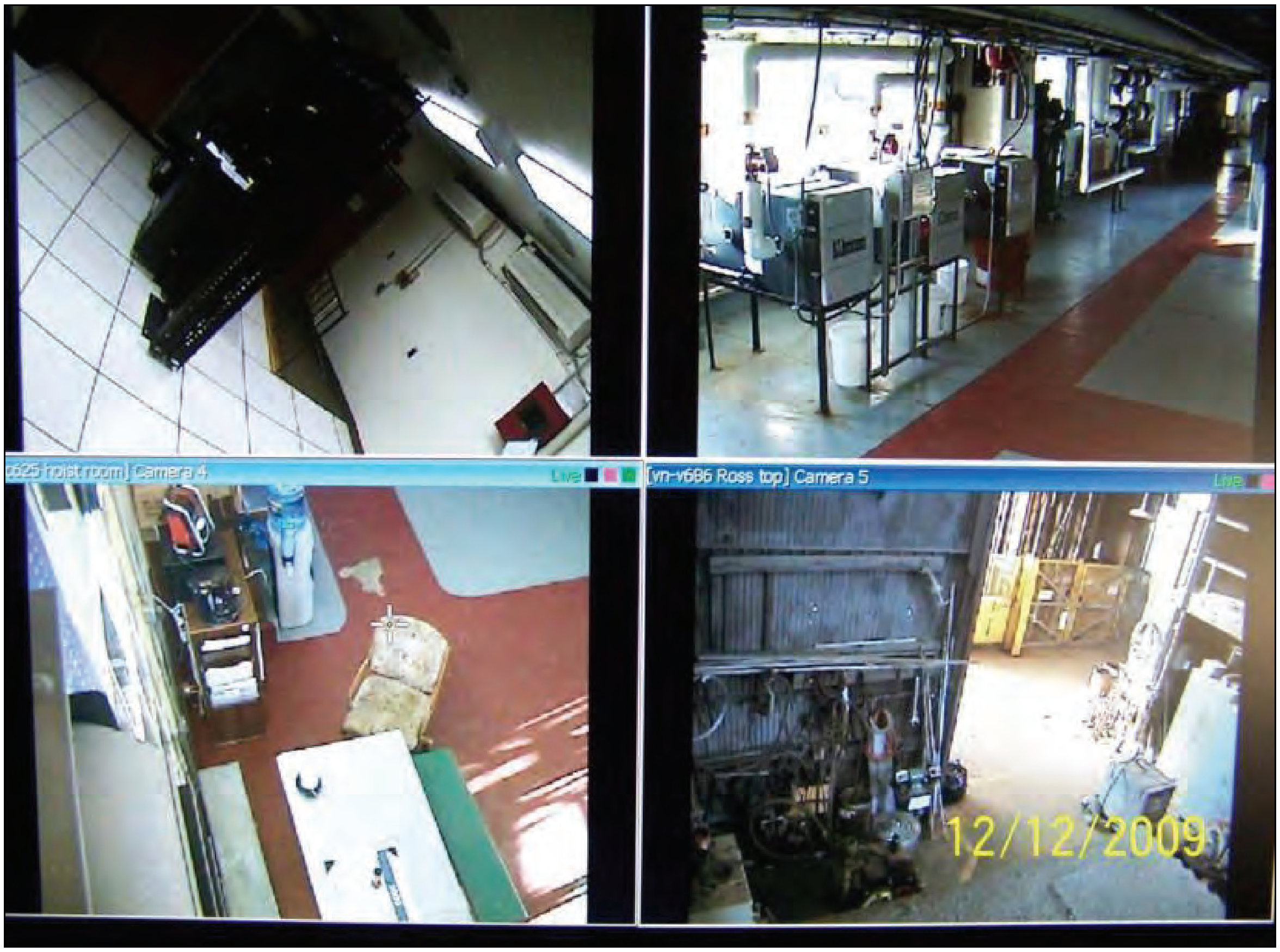

**Figure 5.5.2-5** Surveillance camera monitor.

**Voice Communications**

The current communication systems available at Sanford Laboratory include systems on the surface and underground. The surface systems include Toshiba Strata digital phone systems, two-way Motorola radio system, and commercially provided cellular phone systems provided by Alltel Wireless—transitioning to AT&T in 2011.

Underground communications systems include a FEMCO Leaky Feeder telephone system, providing communications between underground and the surface; a separate 50-pair copper phone system serving as a backup to the FEMCO system; a VoIP-based system utilizing the existing Sanford Laboratory fiber-optic network; and a two-way Motorola radio system.

The FEMCO Leaky Feeder system is heavily used and, given its age, is scheduled for replacement during DUSEL construction. The 50-pair copper phone system is functional and will remain in place near term but is not included in the future DUSEL communications plan. The existing surface phone system is in good working order and will be utilized for DUSEL. Both the surface and underground radio systems are in good working order. The Sanford Laboratory is currently deploying a VoIP system—primarily in the underground science locations. Recent fiber-optic cabling and telephone installations are in good condition but the overall topology will be reconfigured for DUSEL and new cabling installed to support expanded underground science spaces.

### 5.5.3 Key Cyberinfrastructure Facility Requirements

The following is an outline of the major requirements driving the CI systems in support of DUSEL facility operations, business functions, and science experiment.

**Networks and Computing.** DUSEL CI systems shall provide a high-performance, reliable network backbone consisting of routers and switches spanning the surface and underground facilities that are capable of supporting the IEEE 802.3ae (10 Gbps bandwidth standard) and extensible to the IEEE 802.3ba (40/100 Gbps standard) without replacing the fiber cable plant. These network systems must support wide area connectivity to both commodity Internet (Internet1) and Research and Education networking (Internet2), including connections to DOE’s Energy Sciences Network (ESnet). Wireless communication is required on the Surface Campus and on major underground campuses. Additionally, the network shall include redundancy and provide for single fault tolerance to maintain laboratory operations.

Computing resources will be provided to support facility and business operations laboratory functions as briefly described in Volume 10, *Operations Plans*. Computing resources in support of DUSEL science will also be required. The baseline approach for science computing support during Preliminary Design is that the experiment community will provide its own computing resources and utilize the core DUSEL network for data communications. The DUSEL Facility will provide conditioned computing space in the Yates Dry Building and will investigate the need for a computing cluster, enterprise storage, and data backup capabilities in preparation for the Final Design phase. Currently, due to Project funding constraints the Facility design does not include computing or storage capabilities for science but will work to develop support for such services on a reimbursable basis in the future.

**Monitoring, Control, and Security.** DUSEL CI systems will provide a monitoring and control system that meets a wide range of requirements, including a central command and control center to support Facility operations—both surface and underground—through a robust facility management system. Information shall be easily accessible to facility operations staff, site security, IT staff, and laboratory management staff in real time. This command and control center will house the Facility Management System (FMS), communications and security capabilities, and will provide robust graphical interfaces. Capabilities for remote monitoring and management to support emergency response activities and interfaces with first responders shall be addressed by the CI systems.

Monitoring and control systems will support operation of a safe and functional Facility, including air quality, dewatering, electrical, fire detection and suppression, hoists, water inflow, and ventilation systems. Other activities include tracking of personnel and high value and safety critical assets, data trending, and alarming. Monitoring and controls data will be made available to operations, maintenance, science experimenters, education and outreach personnel, and support systems, as appropriate.

**Voice Communications**. DUSEL CI systems will provide voice communications systems both on surface and underground to support laboratory operations, including technologies such as VoIP telephone system utilizing the core backbone, two-way radio systems, cell phone service (surface only), and FEMCO telephone system in support of underground operations.

## 5.5.4 Cyberinfrastructure Preliminary Design

### 5.5.4.1 Networks

The following discussion outlines the facility designs for both wired networks, including local and wide area networks (WANs), and wireless networks on the surface and underground.

The DUSEL CI network backbone integrates administrative networks, research networks, wireless networks, and publically accessible networks, while maintaining security within the infrastructure. The network will support high-speed data throughput and integrity and will provide services to multiple physical locations, including surface buildings and underground laboratories. The planned high-level surface network topology is shown in Figure 5.5.4.1-1. The design provides redundancy and resiliency to withstand hardware failures and accidental fiber cuts. In most cases, secondary diverse paths for network connectivity, including WAN connectivity, will exist across the site.

The DUSEL network design is built around Virtual Local Area Networks (VLANs) and individual experiments will occupy dedicated VLANs or sets of VLANs, which will be isolated from each other in a way that balances risk mitigation, operational efficiency, and scientific productivity. Typically, a VLAN number will map directly to an IPv4 and IPv6 subnet. The technology road map discussed in Section 5.5.5 describes the implementation of a Multiple Protocol Label Switching (MPLS) network at DUSEL. In this case, VLAN numbers will be replaced with MPLS tags, but the concept of logically separating the network remains the same. All networks will be designed to support both IPv4 and IPv6. Sanford Laboratory currently has a block of IPv4 addresses from the American Registry of Internet Numbers (ARIN). The production network will make use of private IP addresses and devices that do not need a public address but will use a private address and access the Internet through Network Address Translation (NAT).

All areas of the DUSEL Complex will have access to core services. These are the minimum set of services that a visiting user would expect to find including: World Wide Web (WWW), e-mail, Domain Name System (DNS, specifically secure DNS-DNS Sec), the Dynamic Host Configuration Protocol (DHCP) and the Network Time Protocol (NTP). All services will be dual stack, and support both IPv4 and IPv6.

**Wired Networks—Internal to the DUSEL Facility**

The core routers and switches in the DUSEL backbone network will support 1 Gbps and 10 Gbps connections. Edge switches that connect to end-hosts will provide 10/100/1000 Mbps connections. These switches will also have the ability to provide 10 Gbps connections for end-hosts (i.e., workstations or servers) to support high-bandwidth science requirements. The underground laboratory modules (LMs) will be provisioned with multiple 10 Gbps connections and Other Levels and Ramps (OLR) with multiple 1 Gbps connections. Although the exact number of connections will be refined during Final Design, growth capacity in fiber network and physical connections has been provided in the current design to address the emerging requirements as the experiment requirements evolve prior to construction. The planned DUSEL network implements a ring topology to provide redundancy to guard against a potential fiber cut or scheduled system maintenance.

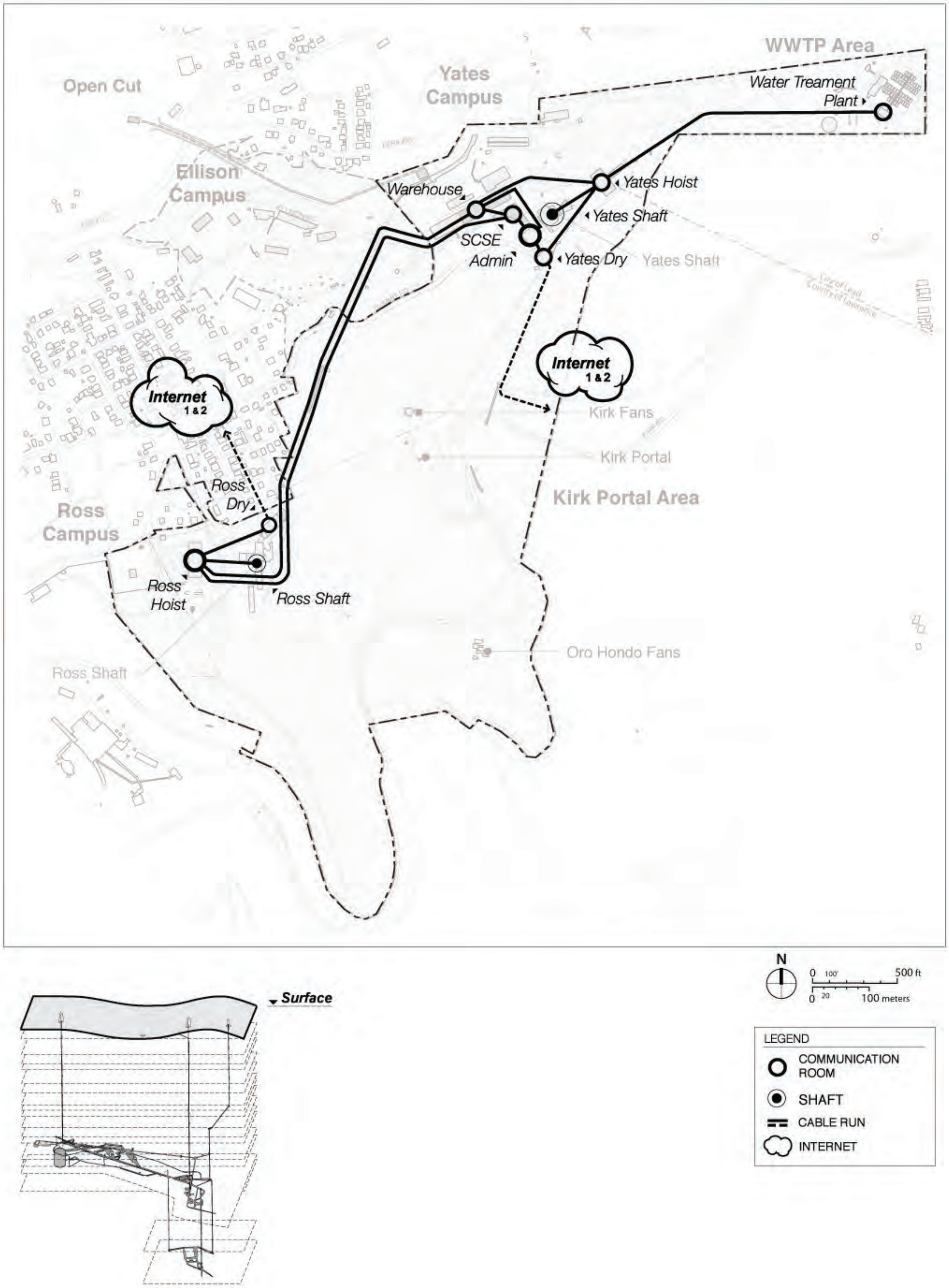

**Figure 5.5.4.1-1** Future DUSEL surface network topology. [DKA]

The CI network backbone, as diagramed in Figure 5.5.4.1-2, serves both underground and surface facilities over distances that exceed the capability of current copper media; therefore, single-mode fiber-optic cables are planned for the core network backbone because single-mode fiber can support distances up to 10,000 meters. The technology road map in Section 5.5.5 discusses support for future 40/100 Gbps connections. Currently, science requirements indicate such interfaces are not a necessary, given the increased expense. To minimize signal degradation, the length of fiber cable runs is maximized to minimize cable splices required. Fusion splicing is planned during installation to minimize signal losses at splice connections, and cables will be looped at shaft stations to provide vertical stress-relief on the cable's fiber strands. The fiber count for individual cable runs is outlined in Figure 5.5.4.1-2 as well.

Redundancy is built in to the fiber-optic backbone by providing multiple cables to communication rooms throughout the site. The objective is to run separate backbone cable runs between the main communication rooms (MCRs) on the surface, the communications distribution rooms (CDRs) located near the shafts on the 4850L, and then to communications enclosures (CEs) within each LM providing separate, diverse pathways to create a ring topology for redundancy. Communications distribution enclosures (CDEs) are used for network distribution at the OLR. Where two isolated pathways for cable routing are not available, alternate approaches will be used to reduce the risk of connectivity interruptions using multiple separated cables along a common pathway.

The primary focal point of the network backbone is the MCR located in the Yates Dry building. This is where fiber-optic cable terminations, core switches/routers, servers, and radio and telephone equipment will be located. The command and control center (CCC) and the science collaboration offices/control room will be located in the same building, allowing reliable and secure connectivity to the MCR and IT infrastructure. The Yates Administration Building IT room will provide additional space for network equipment. A backup MCR will be located at the Ross Dry building, where secondary fiber-optic cable terminations, core switches/routers, servers, and radio and telephone equipment will be located.

In addition to the backbone ring topology between the surface and 4850L Campus, Figure 5.5.4.1-2 shows the fiber-optic cabling serving OLR outside of the Mid-Level Laboratory (MLL) Campus. These levels are accessible from both the Yates and Ross Shafts and will be connected through 48-strand fiber-optic home runs between each OLR area and its respective MCR on the surface.

The fiber connections to OLR are required to support: voice communications, monitoring and control of pump stations, electrical substations, groundwater inflow, ventilation and air quality instrumentation, biology, geology, and engineering (BGE) experiments distributed throughout the Facility at remote locations as discussed in Chapter 5.9, *Design and Infrastructure for Other Levels and Ramps (OLR)* and Volume 3, *Science and Engineering Research Program*. These locations require network connectivity for remote access, Internet/Intranet access, time synchronization signals, data transfer, and data storage.

After installation, if the DUSEL network begins to suffer congestion, extra capacity can be added. Additional fiber and link speed can be upgraded by replacing electronics or optics in the network switches. For example, a gigabit network interface card (NIC) can be upgraded to a 10 Gbps NIC. The single-mode fiber-optic cabling planned for DUSEL is capable of supporting both 10 Gbps and the new 40/100 Gbps standards. These types of technologies will continue to be evaluated throughout the design period so that technology choices can be made as close to construction start as possible to extend the planned life of CI systems.

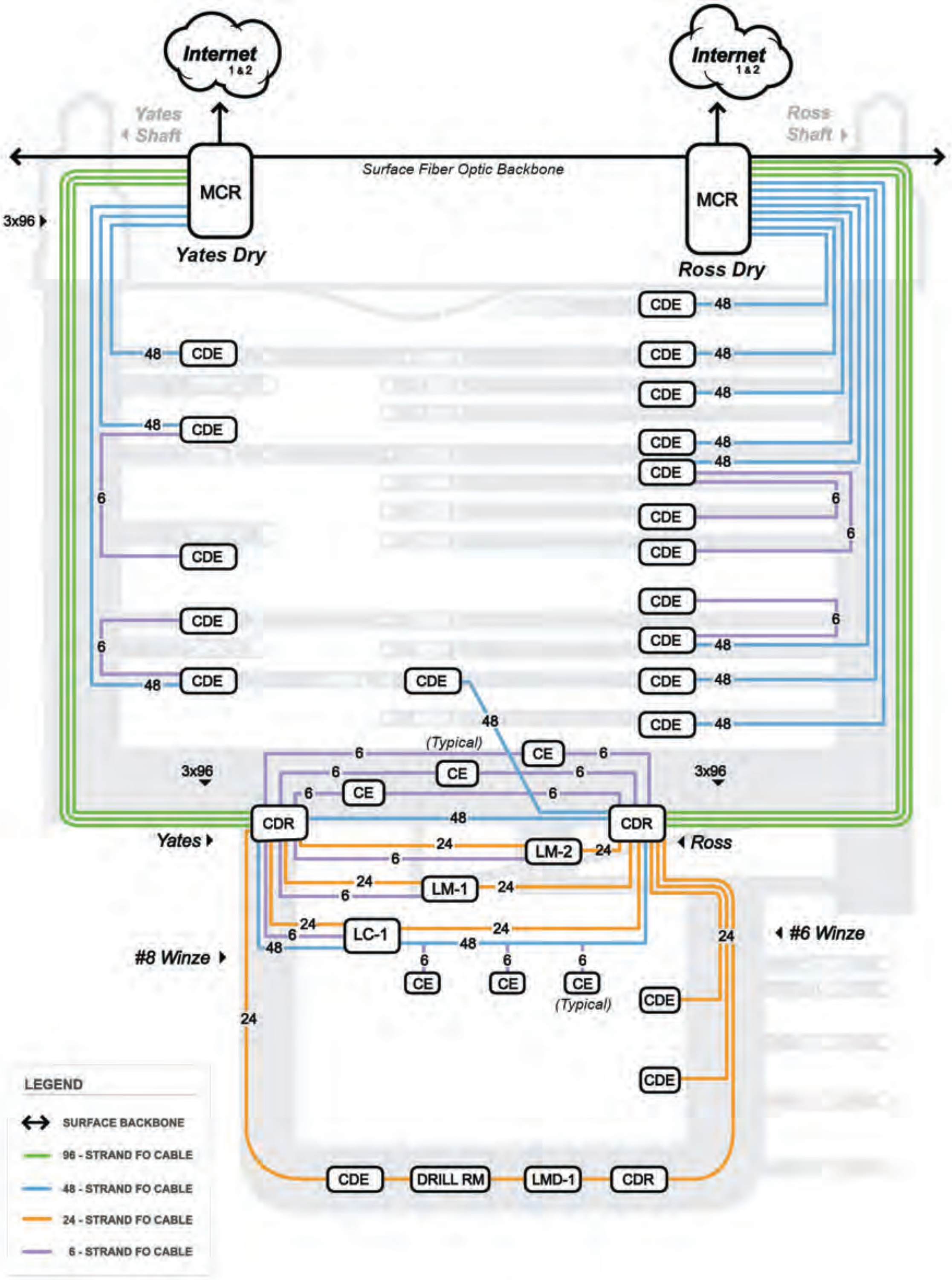

**Figure 5.5.4.1-2** Surface to underground fiber-backbone topology. [DKA]

**Wired Networks—Wide Area Connections and Interfaces to DUSEL**

Wide Area Network (WAN) connectivity beyond the physical site is a key component of the CI network. Currently, Sanford Laboratory has commodity (also known as Internet1) connectivity via Knology's IP service and research and education (R&E) connectivity (often referred to as Internet2[1]) via the State of South Dakota's Research Education and Economic Development (REED) network. The Knology link is 100 Mbps currently and can be upgraded to meet DUSEL needs. The REED link supports up to 5-10 Gbps links currently. REED connects to the Great Plains Network (GPN), which is the regional connector to the Internet2 backbone network and provides connectivity to the Department of Energy's Energy Sciences Network (ESnet). Internet2 provides DUSEL with connectivity to the international research community.

Both the Knology and REED connections to the Sanford Laboratory network occur at the Yates Campus. The REED connection operates on fiber leased from South Dakota Network (SDN) Communications. In addition, Knology provides a cable modem service to the Ross Campus. If the Knology connection is disrupted, commodity traffic can traverse the REED connection. If the REED connection is disrupted, the Knology connection does provide a backup network, albeit at a lower performance level until REED is restored. Additional commodity and R&E network connectivity for DUSEL is being investigated.

Knology is able to provide a network connection to Rapid City that would provide direct connection for DUSEL to REED through a second location. The connection would take a completely separate path from the existing connection at the Yates Campus. Currently, REED has a primary (multiple 10 Gbps) connection across South Dakota via Pierre, the state capital, to Sioux Falls and a 1 Gbps backup connection between the same endpoints routed across the southern part of the state. Out-of-state connectivity, including the uplink to GPN, is from Sioux Falls. REED is working to create an additional link from Sioux Falls to North Dakota. DUSEL will work with REED to ensure diverse multiple paths out of the state are also available, possibly going west to Denver and connecting to the Front Range GigaPop (FRGP) regional network.

**Wireless Networks**

DUSEL will provide ubiquitous wireless on the surface. The usual 802.11 protocol family will be supported, including 802.11n. The 802.11n specification support speeds up to 300 Mbps. Wireless will also be provided in some underground areas such as major campuses at 4850L and 7400L, the Areas of Refuge (AoRs), and other localized areas as required by experiments. Wireless access may need to be controlled due to potential interference with experiments and this will be evaluated during Final Design to ensure the Facility design does not interfere with science operations. The technology road map in Section 5.5.5 discusses the use of emerging wireless technologies for DUSEL to provide higher bandwidth access and longer geographic reach.

### 5.5.4.2 Monitoring and Controls

The DUSEL CI systems include monitoring and control capabilities that are essential to the safe, reliable, and efficient operational control of the Facility. Monitoring predominantly addresses the operation and safety of the Facility, but it also supports science, engineering, and education. Science monitoring requirements are discussed in Volume 3, *Science and Engineering Research Program*. The monitoring and control system will integrate with the communications infrastructure systems to support the open exchange of information between systems. DUSEL provides the communications backbone and is responsible for facility climate and Fire-Life-Safety (FLS) systems monitoring. The facility and FLS data will be available to the experiments. Experiment operations data will also be available to DUSEL.

[1] The term Internet2 refers to the Internet2 network itself as well as other national and regional research and education backbone networks such as the Energy Sciences Network (ESnet) and the Northern Tier Network (NTN).

DUSEL will work with each experiment on hazard analysis, but each experiment is responsible for its own operation. Conceptually, DUSEL would only intervene with science intentionally in the following ways: 1) mandatory evacuation during life-threatening emergency events, (2) dropping power to an LM if a fire event is occurring within that LM, (3) evaluating and controlling equipment and materials that are being taken into the underground facility. A power outage does not necessarily constitute a need to evacuate. Standby power is provided for lighting and critical laboratory equipment/systems for a minimum of 96 hours. It is assumed that senior laboratory managers will always be present and will be trained to make decisions regarding laboratory operation, retreat to AoRs, or evacuation during emergency events.

The components of the monitoring system include monitoring of baseline conditions, performance assessment monitoring, and compliance monitoring.

**Baseline Conditions.** Baseline monitoring is necessary to establish initial conditions prior to the start of excavation and construction. These measurements provide reference data to evaluate the impact of excavation and construction to inform the Facility and experiment design. Baseline measurements will be stored and compared to real-time measurements to detect structural and environmental changes in the Facility and determine whether changes are due to natural causes or anthropogenic (human occupancy) activities. Examples of monitoring activities include geotechnical stability through rock movement, drift and native rock temperatures, humidity, location and discharge rates of water, water table locations, radon, oxygen and carbon monoxide gas levels. The Sanford Laboratory is collecting early baseline data in support of current operations and DUSEL design efforts. Current measurements include rock movement and stability, water flows and levels, ventilation, and air quality.

**Infrastructure Monitoring and Control.** The Facility infrastructure system, described primarily in Chapter 5.4, *Underground Infrastructure Design*, provides input to the Facility Management System (FMS) and output signals that can control these systems. The control system, for example, provides output signals that can start and stop motors, open and close valves, set the position of dampers, and adjust the speed of pumps and fans to maintain Facility environmental conditions, including performance trending and visual display of current and historical information along with generation of reports.

**Performance Assessment Monitoring.** Performance assessment is an analysis that identifies the features, processes, or events that affect the DUSEL Facility (e.g., flooding, fires, cryogen release, earthquakes, structural failure, etc.) and the probability of such events occurring. It allows DUSEL staff to examine the impact of events on the performance of the Facility. While this type of assessment is typically based on theoretical and statistical risk analysis, the monitoring and control system will support ongoing Facility performance assessment that measures the operations and readiness of hazard detection, fire suppression systems, emergency equipment, egress routes, AoRs, and life-support systems, and also provides notification of changes to laboratory conditions.

**Compliance Monitoring.** Compliance monitoring is a continuous process of obtaining information to determine if air and water pollution controls are operating correctly and that the laboratory is being constructed and operated within the permitted limits established by local, state, and federal regulatory agencies. The DUSEL monitoring and control systems will track discharge and emission points at the WWTP, ventilation systems, and Waste Rock Handling Systems for permit compliance.

An analysis was performed to document DUSEL monitoring requirements. Figure 5.5.4.2-1 illustrates the commonalities and differences in the monitoring needs for each functional group identified during the requirements generation process.

The DUSEL Facility will provide the necessary infrastructure and instrumentation to support the common monitoring requirements of the Facility along with an FMS that serves as the central framework for all Facility monitoring, control, and communications systems. DUSEL operations will maintain and operate the FMS and work with related groups to implement additional monitoring needs as they are defined.

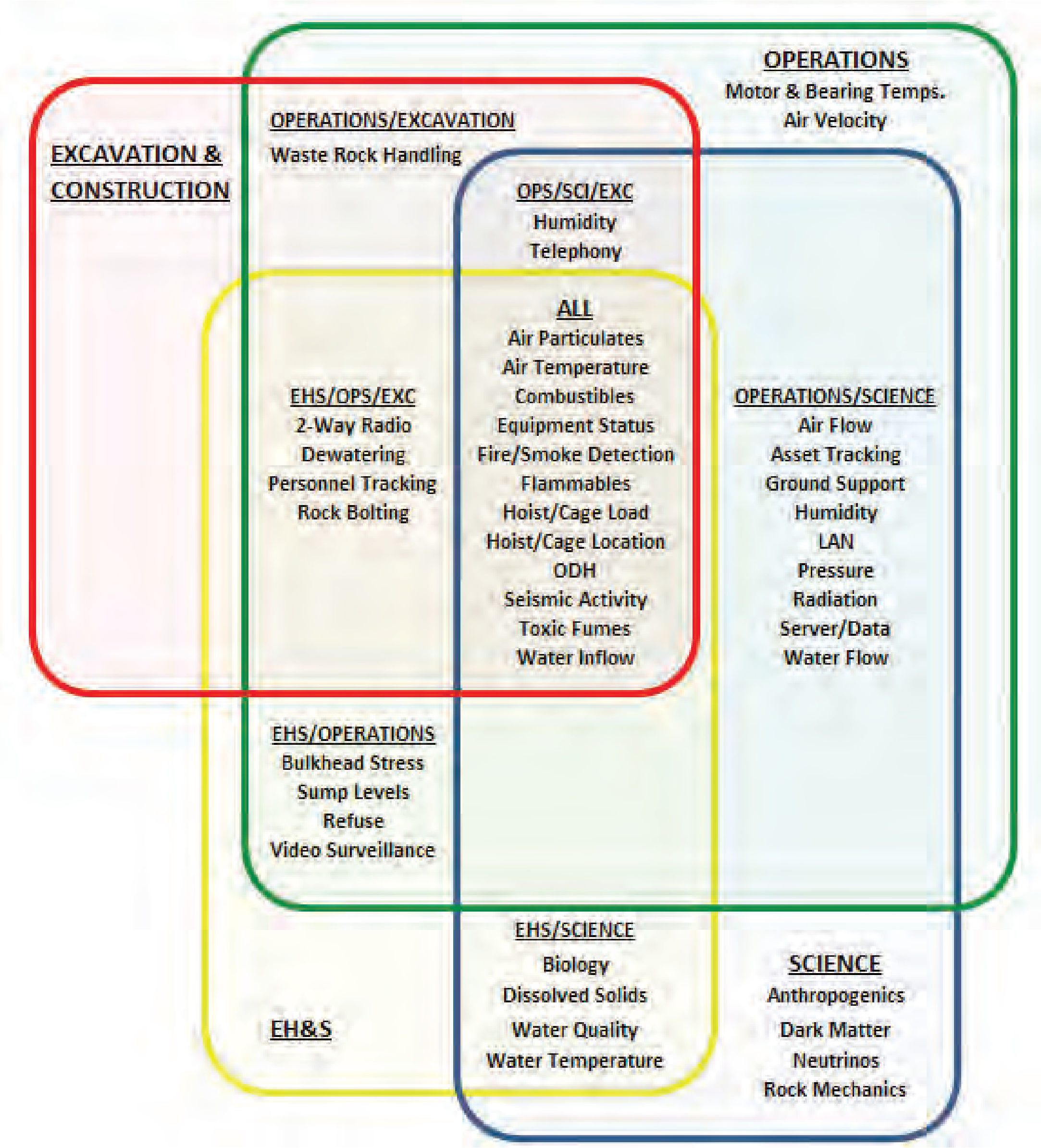

**Figure 5.5.4.2-1** Monitoring overview and requirements.

### Facility Management System (FMS)

The DUSEL FMS is a PC-based HMI system in a centrally located command and control center (CCC). DUSEL operators will use the CCC to monitor and control equipment and systems throughout the surface, underground, and support facilities that make up the DUSEL Complex. The FMS receives signals and data from monitoring and control equipment, displays information graphically, processes data, generates alarms if predetermined thresholds are crossed, and pushes data to a central database maintained by DUSEL that is backed up and maintained off site as well to support disaster recovery operations.

The DUSEL FMS will be a standards-based system that utilizes common interfaces to ease integration and create a single, unified control system. Selected equipment vendors will supply equipment that complies with specifications and protocols defined by DUSEL FMS standards. The FMS will provide automated control functions that direct the operator’s attention to changes or events, reducing the opportunity for human error. This flexible, scalable, and modular system allows for system modifications and future growth, and reduces single points of failure to ensure reliable Facility operations.

DUSEL monitoring and control systems include a variety of instrument, control, and communication components. Equipment will be industrial-quality, durable, and reliable to withstand the environment in the underground campuses. The science experiments will have similar control and data-collection systems and will use the same servers and databases managed by the DUSEL CI team. It is also assumed that the BGE experiments, located outside of the main campuses in OLR, will use personal computers or data logger technology to collect, store, and transfer experimental data to the DUSEL servers/databases.

Figure 5.5.4.2-2 illustrates the FMS architecture and components of the system.

### Network Performance Monitoring

The performance of the network backbone is central to an effective DUSEL CI system. Backbone performance must be measured to understand system status and to facilitate maintenance and future expansion activities. Network performance is measured in two ways, both of which will be leveraged by DUSEL CI monitoring: active measurements where test traffic is injected onto the network, and passive monitoring where real traffic is observed.

Devices for active measurement will be located with routers and switches, except for equipment located in CEs due to space restrictions. Each monitoring machine, typically a standard PC, will gather latency and packet loss data. Data gathered from the individual machines will be centrally collated and analyzed to provide early warning of performance issues. The development of a characteristic signature may help diagnose an issue before users become aware of it.

Passive monitoring will be gathered centrally using industry-standard tools, including the Multi Router Traffic Grapher (MRTG). Figure 5.5.4.2-3 shows a typical MRTG graph detailing incoming traffic (green) and outgoing traffic (blue) in bits per second over a period of 33 hours.

### Other Monitoring Systems

Additional systems to maintain the safety, security, and control of the DUSEL Facility are required and will be independent of the FMS and Information and Communications Technology (ICT) systems, but will interface as required. All systems will be available to the control room operators at the CCC to provide a complete and unified control system. Following are brief descriptions of these additional monitoring systems.

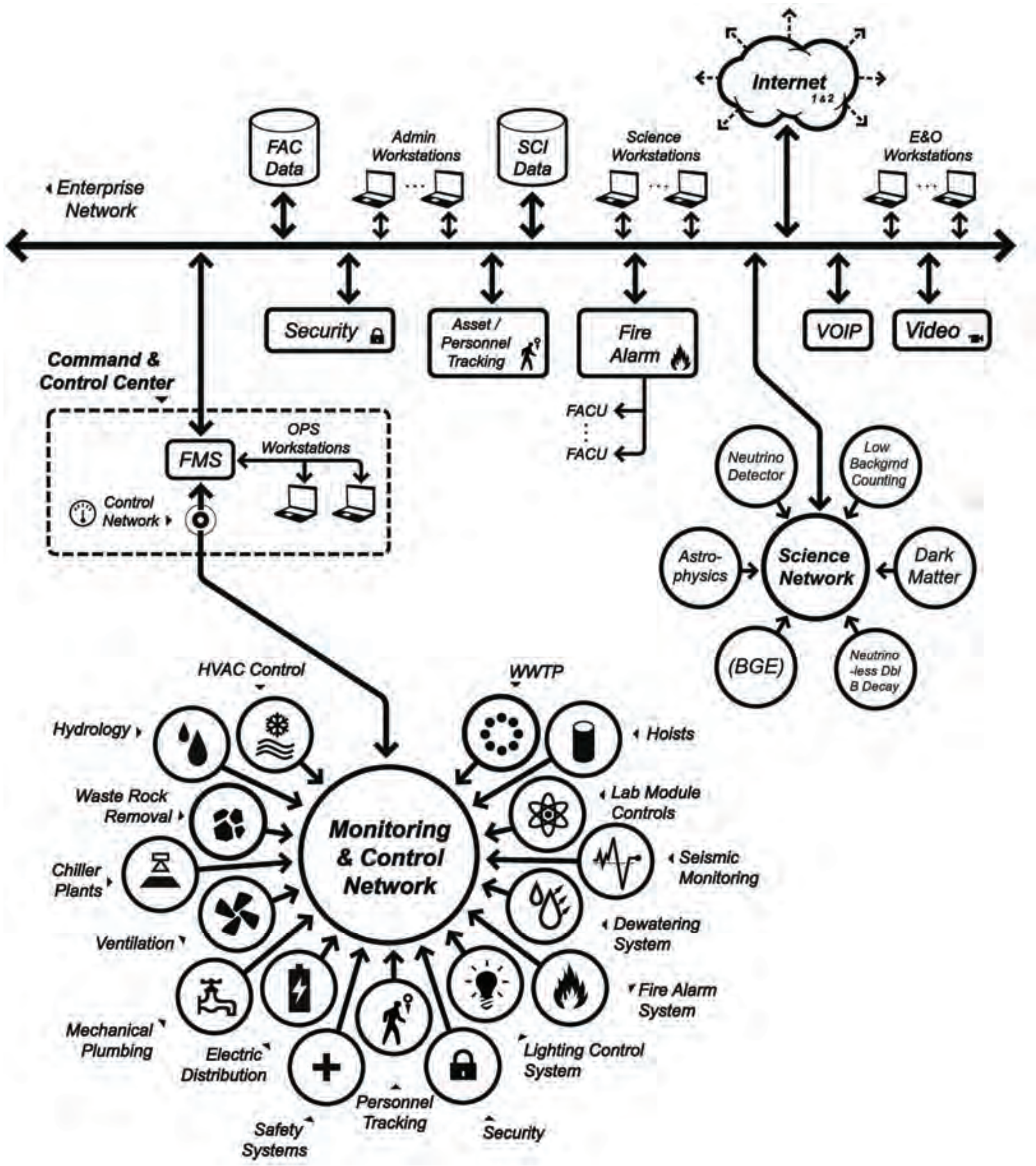

**Figure 5.5.4.2-2** Facility Management System (FMS) architecture. [DKA]

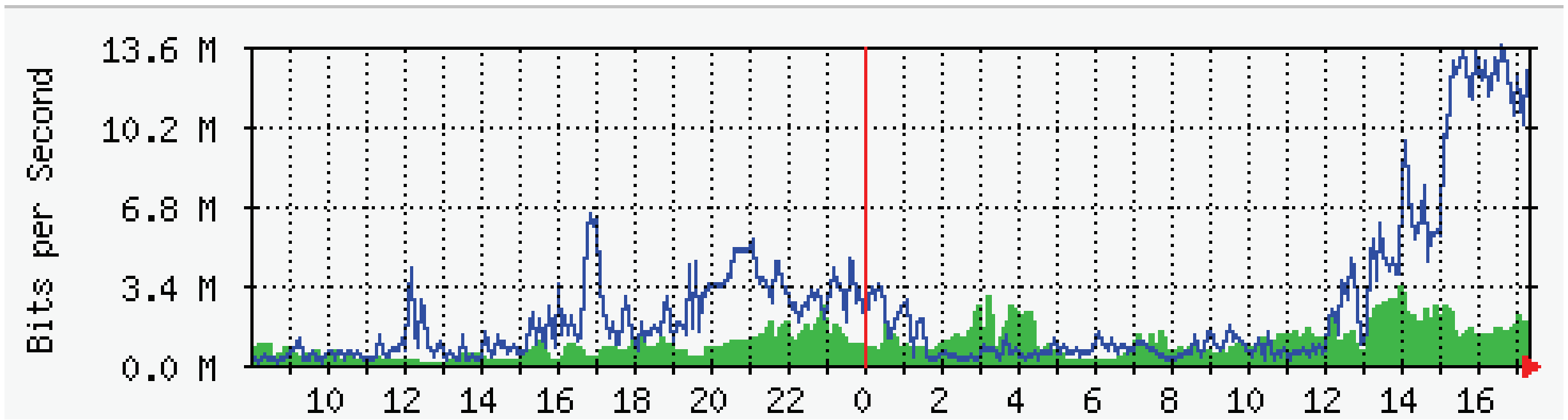

**Figure 5.5.4.2-3** A Multi Router Traffic Grapher (MRTG) graph monitoring network performance.

**Fire Alarm System (FAS).** The FAS is a stand-alone system that will meet or exceed the requirements of the NFPA and IBC codes. Interfaces will be provided to connect to and exchange data with the FMS. The FAS consists of a graphical user interface located at the CCC that connects to fire alarm control units (FACUs) distributed throughout the DUSEL surface and underground facility over a dedicated fiber-optic network. The underground fiber cable is installed in a loop so that damage to the cable anywhere in the loop will not affect the operation of the system (single-fault tolerance). FACUs will be located at every shaft station in both the Yates and Ross Shafts and #6 Winze. Every device on the FAS network is addressable so the exact location of an event, or the identification of a failed device, will be reported to the system.

Fire detection devices such as alarm pull stations, smoke and heat detectors, oxygen and CO sensors, and water spray flow switches will connect to the FACUs. The FACUs will be capable of initiating fire suppression devices and controlling doors, dampers, and fans where appropriate to isolate potential hazards. The requirements for related infrastructure components of the FAS are also discussed in Chapter 5.4, *Underground Infrastructure Design*. Each FACU has an annunciation panel that will display information about the system status and location of an emergency event. Speakers, strobe lights, and stench gas in the ventilation system when appropriate will be used to assist the evacuation of underground facilities. The speakers will be capable of providing instructions to Facility occupants unique to the specific emergency situation, directing occupants to either seek refuge in a designated AoR or evacuate the Facility.

**Electronic Security System (ESS)**. The DUSEL ESS has three components to form a complete site-wide security system: Access Control and Alarm Monitoring System (ACAMS), Asset/Personnel Tracking System (APTS), and a Video Surveillance System (VSS).

The ACAMS provides an electronic access control system that uses proximity card identification badges to control entry through gates and doors across the DUSEL surface and underground campuses. IP addressable card readers and electronic locks will be networked together and connected to an ACAMS server located at the CCC. Control room operators and security personnel will monitor and receive alerts from the ACAMS.

The APTS tracks the location of high-value assets including science and Facility personnel (wearing identification badges) using radio frequency identification (RFID) tags. APTS provides control center operators and security personnel with immediate identification and location of the RFID tags and thus the asset or person of interest.

The VSS uses IP addressable digital video cameras, the local area network, and the fiber-optic backbone to create a facility-wide video monitoring and surveillance system. The system includes both fixed and pan-tilt-zoom cameras, a network video management system and recorders to provide control room operators and security personnel with site-wide video surveillance capabilities.

### Command and Control Center

A CCC will be located at the Yates Dry building next to the MCR and below the science collaboration offices/control room. Figure 5.5.4.2-4 shows a Preliminary Concept design for the CCC.

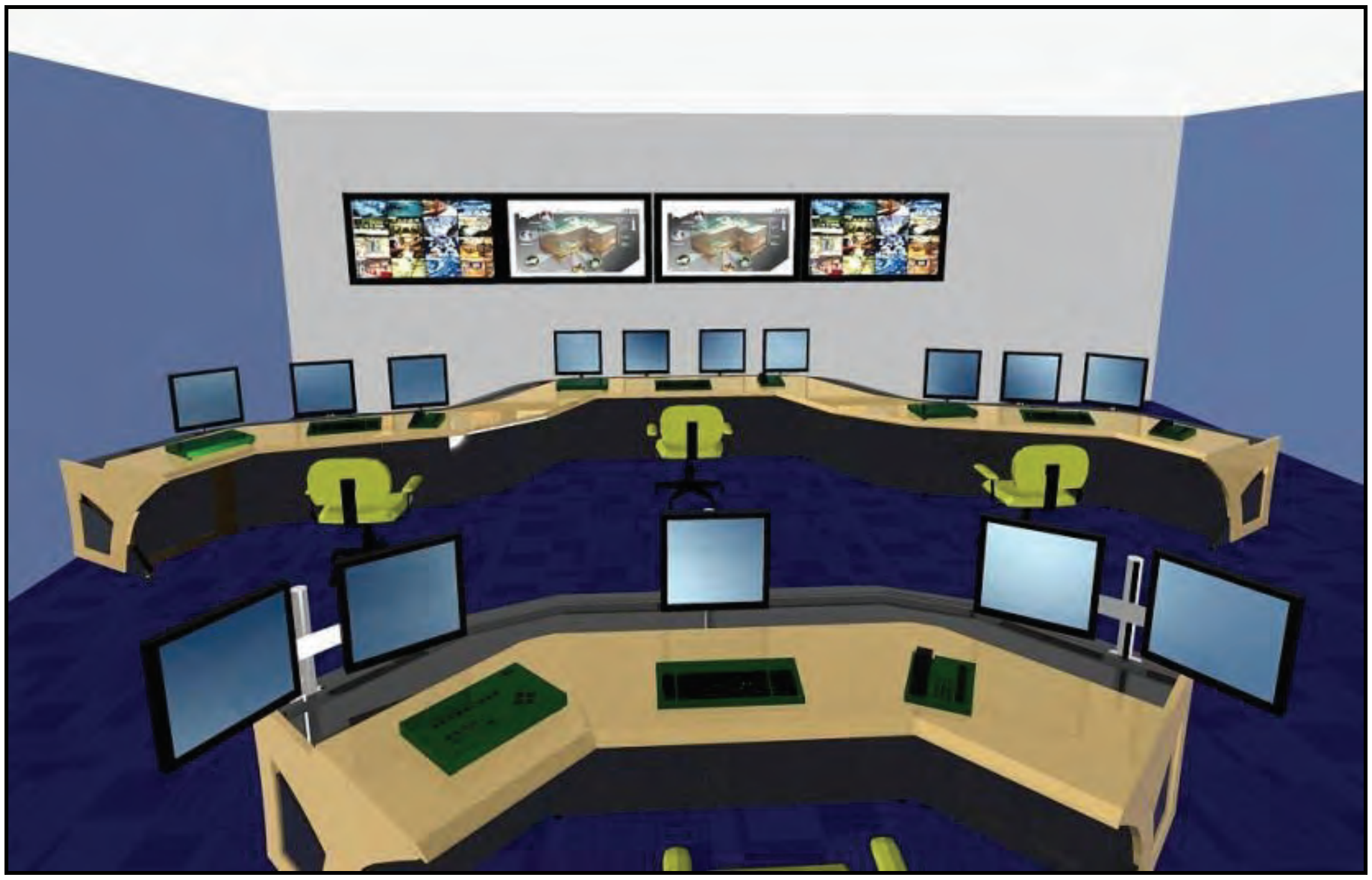

**Figure 5.5.4.2-4** Conceptual drawing of the Command and Control Center. [ARUP]

The CCC provides a central location to manage, monitor, and control all DUSEL activities both surface and underground. The CCC houses the FMS, FAS, ESS (ACAMS, APTS, VSS), telephone system, two-way radio communications system, and the CI management system. The CCC will assist the operations staff in coordinating and scheduling maintenance and operations activities as well as troubleshooting and analyzing system faults and equipment failures. The CCC will convert to an emergency command center for first responders, rescue teams, fire fighters, and evacuation procedures in the case of an emergency. Satellite CCCs are located underground at both the Ross and Yates 4850L AoRs. These satellite CCCs are equipped to fully communicate and operate the FMS and FAS as the main surface CCC. They provide first responders and rescue teams with an underground command center to assemble, obtain current information, evaluate conditions, and communicate between the other command centers and the AoRs.

### 5.5.4.3 Facility-wide Interfaces Related to Cyberinfrastructure Systems

Typically, the physical interface with the CI system will be a network switch port facing the non-DUSEL equipment, such as the equipment in the experiment's communications room (CR) or CE; however, this is only an administrative boundary. Equipment that requires public IP addresses, whether owned by the experiments or by DUSEL, will be configured with an address block for that portion of the DUSEL network. These are administered by DUSEL to ensure the smooth network operations. However, administration of individual addresses may be shared between DUSEL and an appropriate local administrator. Services such as the Dynamic Host Configuration Protocol (DHCP) and the Domain Name System (DNS) are provided by DUSEL but administration may also be shared with experiments.

End-hosts often have built-in copper network interface cards (NICs). Edge switches have copper interfaces and also have the ability to provide a limited number of fiber connections. Copper connections will be used in the underground laboratories where distances permit and fiber connections will be used for longer distances. Research equipment may come with single-mode fiber, multimode fiber, or copper

NICs. Within the industry standards, DUSEL will be flexible in order to ensure the necessary connectivity is achieved. Non-standard interfaces will not be supported.

Experimental equipment, including computing systems for experiment control and data processing and storage, is the sole responsibility of the experiment. Communications equipment in the laboratories provided by the Facility and connected to the enterprise network, such as telephones, will be maintained by DUSEL. Equipment outside the experimental areas and connected to the science network will be maintained by the responsible collaboration; DUSEL will provide space, power, and cooling for experiments to house and maintain their own equipment. Experiment collaborations will have remote access both on and off site via the network to their experiments for data access, monitoring, and control.

#### 5.5.4.3.1 Surface Facility

The primary location for IT equipment and servers is the Yates Dry building. Additional space in the Yates Administration Building will also be utilized. DUSEL will host experimental equipment, including servers and storage devices. DUSEL will provide space, power, and cooling for experiments to house and maintain their own equipment.

#### 5.5.4.3.2 Underground Facilities

The two MCRs at the Yates Dry and Ross Dry are the physical interface points between the backbone cables for the underground facilities and the Surface Campus. The communications distribution rooms (CDRs), located near the two shafts on the main laboratory campus levels, are the interface points between underground infrastructure and the underground laboratory designs and will be coordinated between the scopes. The LMs and large cavity CRs and CEs are the interface points between the underground laboratory scope and the experimental installation and will be coordinated. The physical interface between the CI and the experiment's equipment is the port on the switch in the CR or CE facing the experimental equipment.

### 5.5.5 Cyberinfrastructure Technology Road Map

Future requirements and evolving CI technologies need to be considered early so that plans to incorporate them into DUSEL can be formed and supported. Because of the unknown future requirements, the technology road map focuses on tracking, implementing, and incorporating new technology into the Facility design.

**Network Advances.** Current network requirements (10 Gbps links) will be upgraded to 40/100 Gbps links when necessary, and upgraded again as speeds move toward terabits per second (Tbps). DUSEL will be prepared to offer dynamic network provisioning with the use of the ESnet On-demand Secure Circuits and Advance Reservation System (OSCARS) software and the Internet2 Inter-Operability Network (ION) service. The feasibility of using multiple technologies will be assessed and weighed against the requirements of the science community. Also, technologies such as Multi Protocol Label Switching (MPLS) will be used to provide capabilities beyond simple layer 2 Virtual LANs (VLANs) to provide separation of logical networks on a single physical network.

DUSEL will assess the possibility of adopting Wimax or long-term evolution (LTE) technology for wireless network connectivity, which provides a wider reach and higher speeds. DUSEL will track technology advances in domains applicable to the DUSEL Facility and those of interest to the DUSEL science community. Modern off-the-shelf servers are expected to meet the requirements at an affordable cost, including the use of cloud computing services. Identity management is central to the DUSEL technology road map and will be used to facilitate access to DUSEL resources, especially for visiting scientists. DUSEL will leverage InCommon[22] and Eduroam[23] services to authenticate users through their home institution.

**Green Computing.** Green computing policies aim to ensure that the setup and operations of IT systems result in a reduced carbon footprint. DUSEL programs and goals require substantial use of IT systems and as a result, significant power consumption. By aligning the CI strategy with green computing policies, DUSEL can realize significant benefits, including operational cost savings. Several approaches will be implemented, and some are already in use, to reduce the environmental impact of DUSEL CI systems. These approaches include server virtualization and clustering, green equipment disposal practices, replacing paper-based systems with online systems, and reducing staff travel through use of high-definition video conferencing.

### 5.5.6 Cyberinfrastructure Advisory Committee

The DUSEL Cyberinfrastructure Advisory Committee (CIAC) advises the Project on CI and communications requirements, designs, construction approaches, operations issues, and Lessons Learned through the design, implementation, and operation of CI capabilities. The DUSEL CI Chief Engineer serves as the lead day-to-day project interface to the CIAC and reports CIAC status to the Project Director. The CIAC has provided input to develop the DUSEL architectural principles, and outlined the service model. Members of the CIAC have also provided technical expertise in the design of the LAN architecture, network technologies, and server virtualization.

## 5.6 Mid-Level Laboratory Design at the 4850L (MLL)

This chapter presents an overview of the Preliminary Design of the facility laboratory outfitting specific to the two laboratory modules (LMs), LM ancillary spaces, and also incorporating Davis Campus utility infrastructure at the 4850L. The design includes providing the infrastructure between the main facility infrastructure and the experiments, along with providing a finished core structure ready for experimental outfitting. Included in this section is an overview of the Mid-Level Laboratory (MLL) design, beginning with the scope, design requirements, design strategy, and existing conditions. Following this, the core utility systems provided to the LMs are discussed. The descriptions in this section are intended to give the reader an overview of the design. Additional details on the design of each scope, as well as how the infrastructure design interfaces with other scopes, can be found in the reference material noted throughout the section.

### 5.6.1 Overview and Planning Summary

The largest underground campus at DUSEL is located 4,850 feet below ground. A major part of the 4850L Campus is the MLL, which will include two laboratory modules (LM-1 and LM-2) and the Davis Campus. The Davis Campus is anticipated to be constructed by the SDSTA prior to the Major Research Equipment and Facilities Construction (MREFC) funded Project and is discussed further in Sections 5.6.7 and 5.4.2.1. The Large Cavity (LC-1) for the Long Baseline Neutrino Experiment (LBNE) is also located on the 4850L Campus and is discussed in Chapter 5.7, *Large Cavity for the Long Baseline Neutrino Experiment*. Although LC-1 is located on the 4850L, the LC-1 scope is not part of the MLL design.

As discussed in Section 5.1.5, *Design Development Resources and Teaming*, through an open solicitation process, DUSEL selected Arup USA for the underground laboratory design scope of work and Golder Associates for the excavation design scope of work. Both scopes were centered on development of Preliminary Designs for the development of DUSEL's MLL Campus.

The scope of the MLL Preliminary Design Report (PDR) includes the following major components:

- Two laboratory modules (LMs), LM-1and LM-2
- Ancillary spaces required to support the MLL operations, including dedicated MLL Area of Refuge (AoR)
- Mechanical electrical rooms (MERs); and
- Utility room elements dedicated to the two LMs

The schematic diagram in Figure 5.6.1-1 shows the relative placement of the 4850L components with primary laboratory access to this level from the surface via the Yates Shaft, and secondary egress and construction access via the Ross Shaft. The shaft infrastructure, including the #6 Winze used to access the 7400L Campus, are discussed in detail in Chapter 5.4, *Underground Infrastructure Design*.

LM-1 and LM-2 are designed to house a generic suite of physics experiments and therefore have similar sectional sizes and configuration. A decision was made by DUSEL management, informed with input from the DUSEL science liaisons, to use the same design configuration for both LMs to provide more flexibility for initial experiments as well as for the life of the DUSEL Facility. Generally, LM-1 will hold one (isolated) experiment while LM-2 will hold three experiments, which are discussed in Volume 3, *Science and Engineering Research Program*.

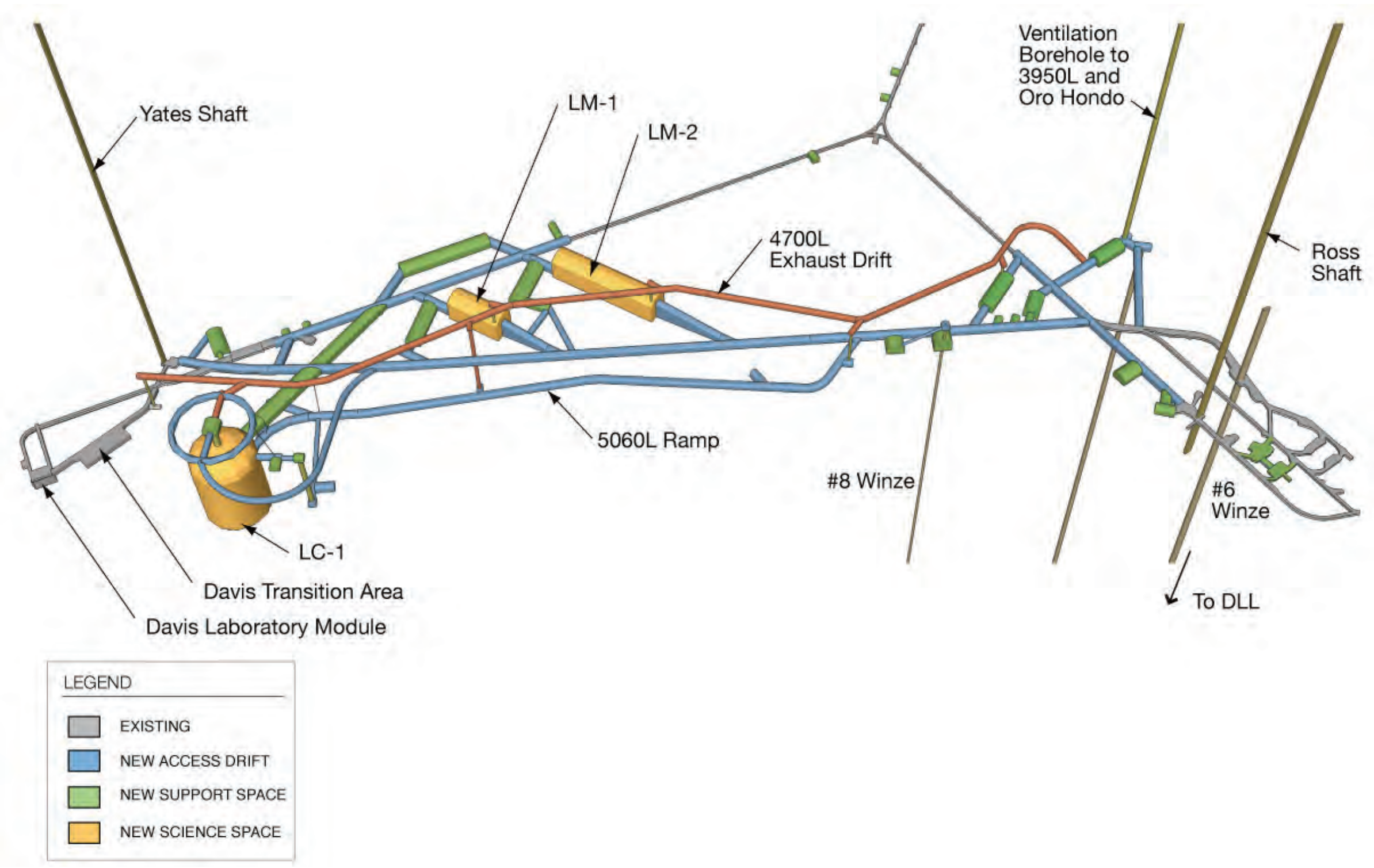

**Figure 5.6.1-1** Schematic diagram of the 4850L Campus. [DKA]

This section covers the preliminary Arup MLL Underground Laboratory (UGL) design for the LM-1 and LM-2 at the 4850L (Appendix 5.Q and Appendix 5.R). The Preliminary MLL UGL design includes the facility outfitting and dedicated infrastructure to support the MLL. The *Arup Underground Infrastructure (UGI*) design (Appendix 5.L), discussed in Chapter 5.4, includes the facility-provided backbone utilities and infrastructure supporting UGL. The UGL interface with UGI and the other scopes of work are discussed in Chapter 5.1, *Facility Design Overview*. Figure 5.6.1-2 represents which areas are included in the UGL and the UGI scopes of work and the associated demarcation between the two design scopes in relationship to the LMs.

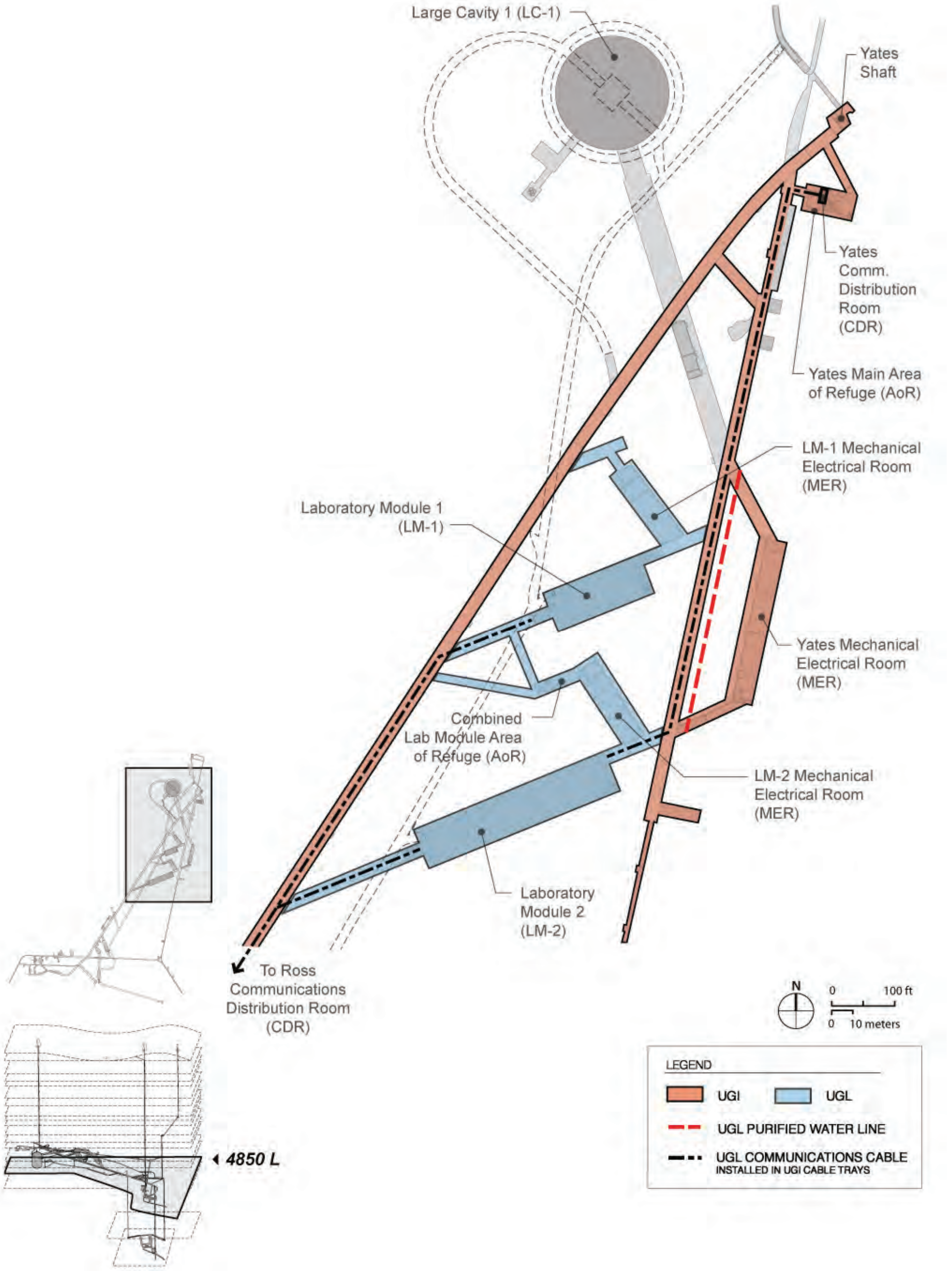

**Figure 5.6.1-2** 4850L area representation of UGL and UGI scopes of work pertaining to the LMs. The UGI scope of work includes the facility-provided backbone utilities and infrastructure supporting UGL. [DKA]

## 5.6.2 Facility and Infrastructure Requirements for the 4850L

Detailed requirements for the MLL have been developed, reviewed, and approved. A thorough discussion of the requirements structure and the development process can be found in Volume 9, *Systems Engineering*. The requirements listed in the PDR summarize the formal requirements included as Appendix 9.I to the PDR. The detailed method for extracting user requirements from the science collaborations and turning them into a cohesive set of requirements for the LMs is described in Chapter 3.6, *Integrated Suite of Experiments (ISE) Requirements Process*. The key driving requirements from science that emerged from this process are listed in Chapters 3.6, 3.7, and 3.8. The functionality and performance mandated by this set of requirements form the basis for the DUSEL Facility design as presented throughout Volume 5, *Facility Preliminary Design*. All key driving requirements of the design have been either met or otherwise addressed by this report.

The ISE has defined the requirements for the MLL, and these requirements have defined the MLL design. The ISE program is not fully developed; thus, the MLL design is guided by the requirements of the generic ISE and will be refined during Final Design as ISE requirements mature.

The design of LM-1will accommodate either the proposed Dakota Ion Accelerators for Nuclear Astrophysics (DIANA) experiment or a single large dark-matter, neutrinoless double-beta decay, or low-background counting experiment. DIANA is unsuitable for close proximity to other experiments, so it would be the only occupant, although the installation of DIANA would require the addition of "shielding mazes" at both entries. This is because DIANA is a potential neutron source, and it is important that it is sufficiently shielded so that other experiments are not affected by its operation. If DIANA is chosen and the requirements are further developed to customize LM-1 to best suit DIANA, the LM-1 design would be modified to be not as tall and without a recessed floor. If different experiments are chosen, approximately half the space would be available for R&D and prototyping activities. Requirements are based on the most challenging individual requirement among the candidate experiments. Since DIANA has the highest power requirement, it is used to set the power and other associated requirements, such as chilled water. All other requirements are driven by large cryogenic dark-matter or neutrinoless double-beta decay experiments.

The design of LM-2 is assumed to be compatible with three physics experiments of any variety other than DIANA. Requirements are based on the most challenging combination of the three possible experiments.

The dimensions of the laboratory areas are included in Table 5.6.2. The LC-1, which is not part of the MLL design, is shown for size comparison to the other laboratory spaces on the 4850L.

Figures 5.6.2-1 and 5.6.2-2 represent examples of typical experimental installations in LM-1 and LM-2. It is anticipated that the open space in each figure will be used for laydown and assembly of the experiments. However, once experiments are assembled, it could potentially be used as R&D space.

| Experiment Space | Width ft (m) | Height ft (m) | Length ft (m) | Floor Area $ft^2$ ($m^2$) | Finished Volume $yd^3$ ($m^3$) |
|---|---|---|---|---|---|
| LM-1 | 66 (20) | 79 (24) | 164 (50) | 10,764 (1,000) | 29,422 (22,495) |
| LM-2 | 66 (20) | 79 (24) | 328 (100) | 21,528 (2,000) | 58,845 (44,990) |
| Davis Lab Module (DLM) | 30 (9) | 50(15) | 60 (18) | 1,800 (167) | 3,333 (2,548) |
| Davis Transition Area (DTA) | 50 (15) | 17.5 (5) | 140 (43) | 7,000 (650) | 4,537 (3,469) |
| LC-1 | - | 272 (83) | 180 (55) diameter | 25,575 (2,376) | 243,210 (185,947) |

**Table 5.6.2** 4850L laboratory space dimensions.

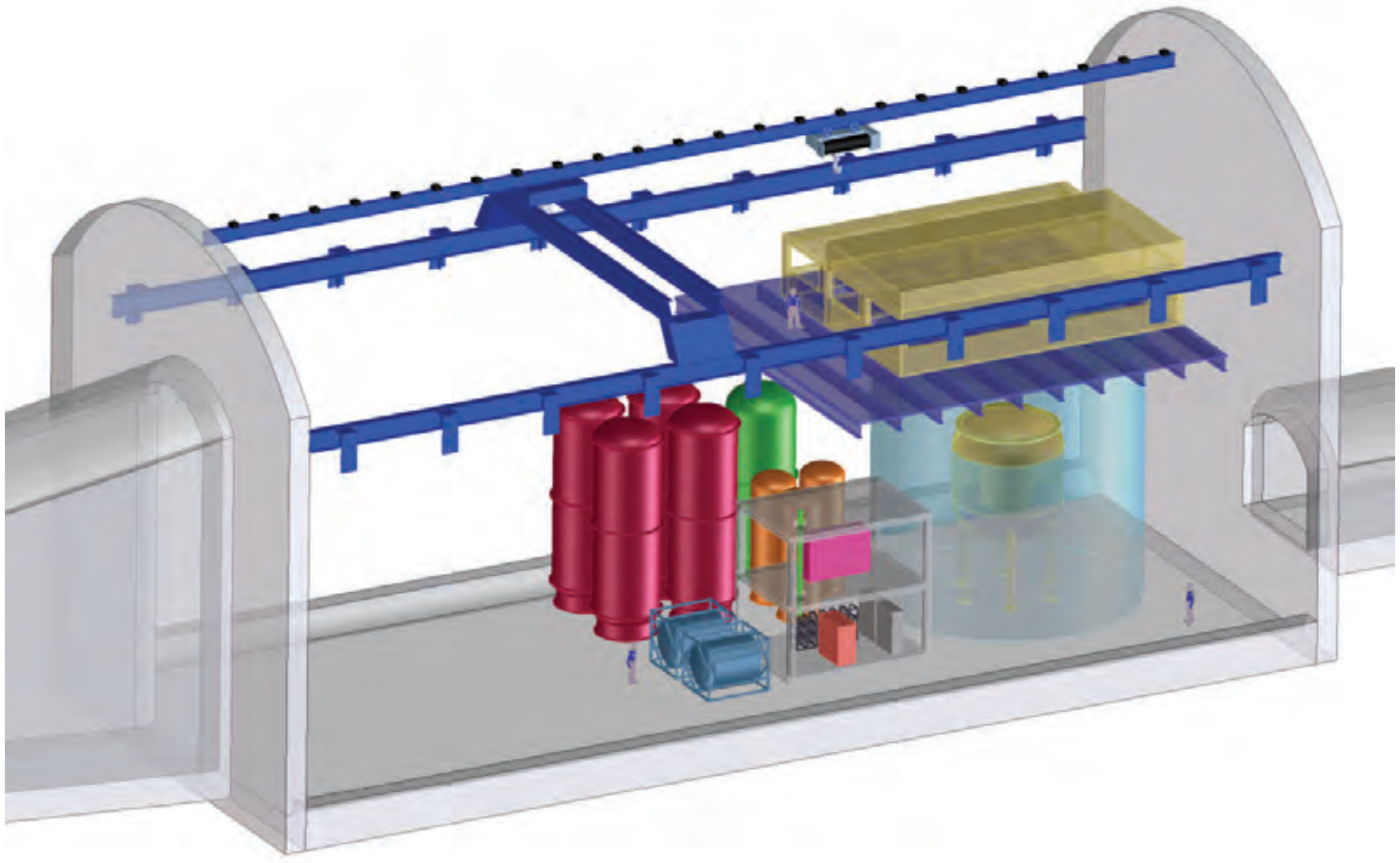

**Figure 5.6.2-1** Example dark-matter experiment in LM-1. [Dave Plate, DUSEL]

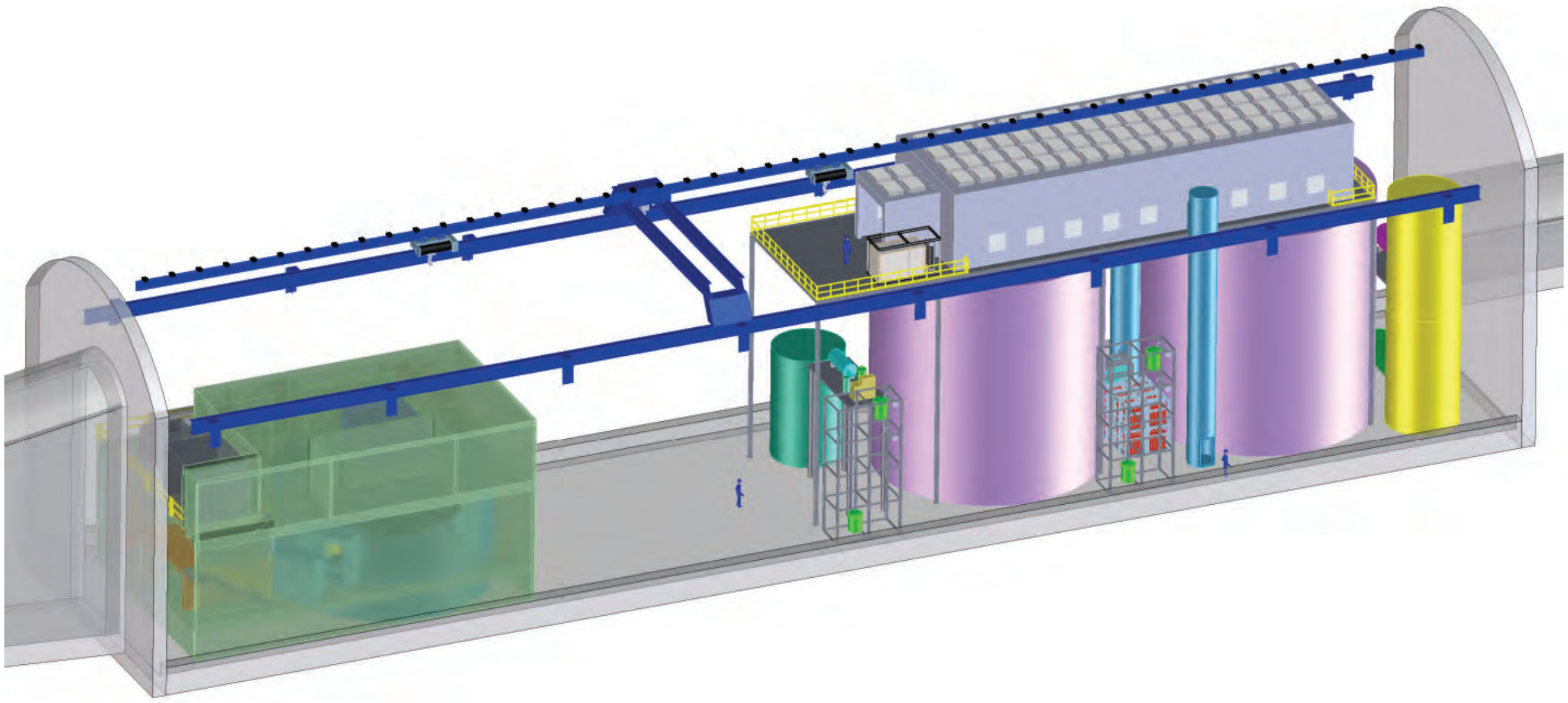

**Figure 5.6.2-2** Three examples of experiments in LM-2 representing a low–background assay facility and two dark-matter experiments. [Dave Plate, DUSEL]

Figure 5.6.2-3 shows a cross section for LM-2 depicting the permissible experiment envelope (shaded). LM-1 and LM-2 have similar cross-sectional views. The 62.3 ft (19 m) maximum height is limited by bridge crane clearance. The maximum 55.8 ft (17 m) width allows for personnel egress and a utility corridor on either side. Also shown is the monorail crane that is centered over the experimental envelope.

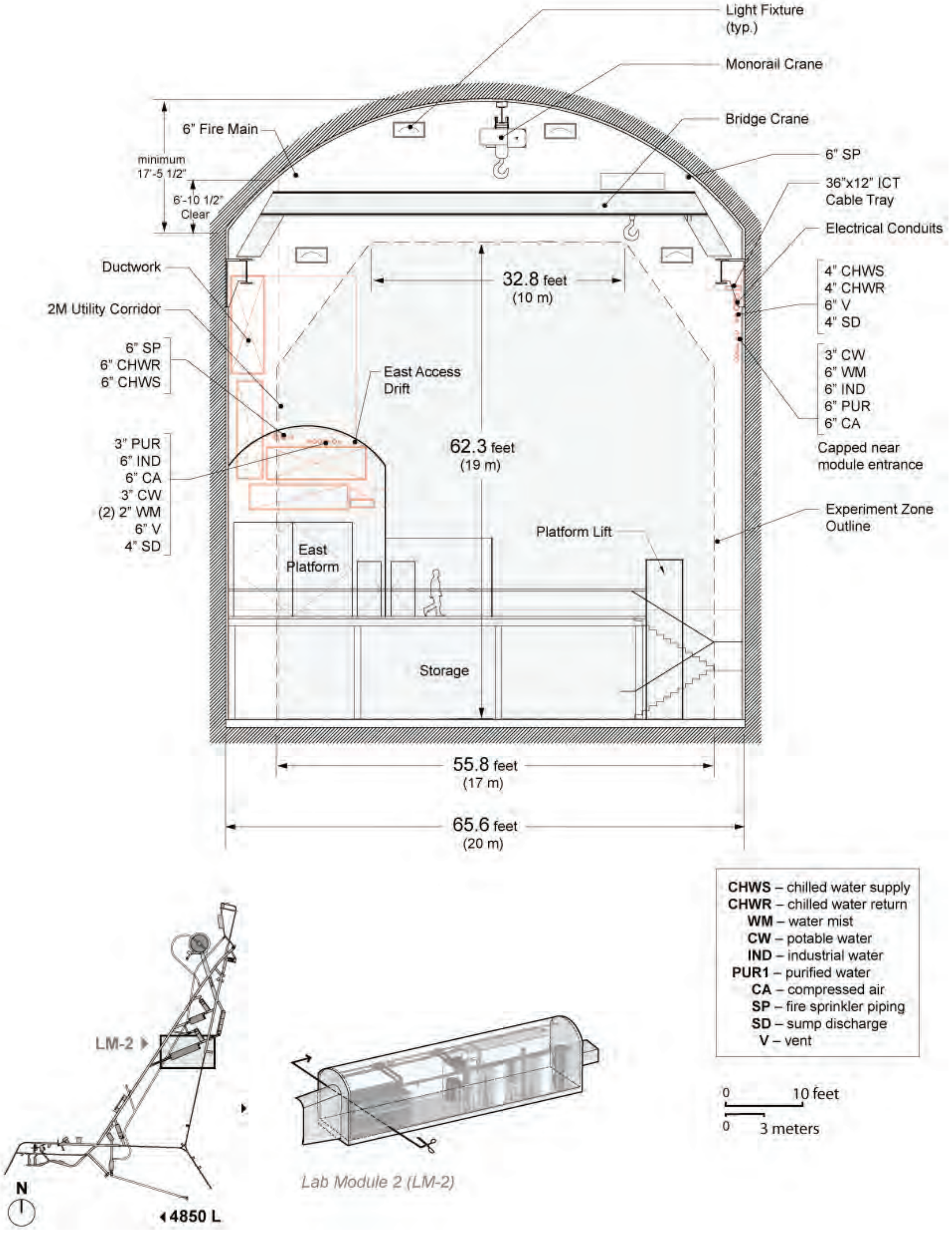

**Figure 5.6.2-3** LM-2 East entrance cross-sectional view with experiment envelope and utilities. LM-1 and LM-2 cross-sectional views are similar. [DKA]

The design also includes concrete floors and shotcrete applied to walls and ceilings. The floor of the LMs will be 13.1 ft (4 m) below the sill elevation of the 4850L access drifts. This is for containment in case of catastrophic leak of water, liquid scintillator, or cryogen.

The following are the key experiment requirements for LM-1 and LM-2:

- Space—as discussed above
- Power
    - Normal (LM-1: 2000 kW; LM-2: 1100 kW)
    - Standby (LM-1: 100 kW, LM-2; 160 kW)
- Chilled water (LM-1: 1800 kW; LM-2: 840 kW)
- 20-ton bridge crane
- 40-ton monorail crane
- Purified water (LM-1: 1.1 M gallons; LM-2: 2.6 M gallons)
- Network communications/IT (10 Gbps)
- Fresh air supply at one air exchange per hour (LM-1: 13,200 cfm, 22,500$m^3$/h; LM-2: 26,600 cfm, 45,200$m^3$/h)
- Exhaust ventilation for cryogen release and smoke (100,000 cfm, 170,000$m^3$/h)
- Fire sprinkler / Water mist protection
- Potable and industrial water
- Environmental / humidity control (LM-1 and LM-2: 68 to 77 °F/20 to 25 °C / LM-1 RH: 20% to 45%; LM-2 RH: 20% to 50%)

### 5.6.3 Underground Laboratory Design at the 4850L—Design Strategy

The design of the MLL LMs will share as many attributes as possible with the Deep-Level Laboratory (DLL) Conceptual Design and the LC-1 Preliminary Design. The designs of the LMs at the MLL and DLL are very similar except in their dimensions. Also, all the LMs share the design concept of dedicated MERs for each LM. Because the MLL, DLL, and LC-1 share many of the same requirements and services, the design interfaces are critical. Future maintenance and serviceability will benefit from the similar design strategy, as well as from the same backbone utility services provided through the UGI design.

The excavation design contractor has performed preliminary modeling of excavation utilizing projection of MLL rock characterization, rock support, and rock removal methods. The MLL excavation design, including the design for ground support and stabilization and also the geotechnical analysis, is discussed in Chapter 5.3 and Golder Associates' Preliminary Design *Final Report* (Appendix 5.I).

The DUSEL UGL design, through MREFC funding, will provide the following utilities to the MLL via connections to the UGI facility-provided systems: power, potable water, industrial water, fire-sprinkler piping, water-mist piping, chilled water, fresh air, conditioned air, floor drainage, and data acquisition equipment.

### 5.6.4 Existing Facilities and Conditions Assessment

The Homestake dewatering pumps were turned off in June 2003 and the mine began to flood from the 8000L. The water reached 4,529 ft beneath the shaft collar at the surface in August 2008 before pumping resumed. The 4850L had been flooded for approximately two years by that time and was dewatered in May 2009. Since that time, the 4850L has been dry and has been continually occupied. Rehabilitation has taken place in preparation for future science development.

Excavations for the two physics early science experiments, Large Underground Xenon (LUX) and the MAJORANA DEMONSTRATOR, near to and in the existing Davis Laboratory Module (DLM) have been completed. Currently, preparations for the infrastructure outfitting for the LUX and MAJORANA DEMONSTRATOR experiments are currently in progress.

Geotechnical drilling and surveys in support of Preliminary Design have been completed, and additional assessments are planned for 2011 to support the Final Design phase. The existing overall geotechnical conditions of the 4850L are very good in regard to ground conditions and proposed supporting excavations. The LMs, LC-1, and most of the ancillary spaces will be located in new excavations. The entire infrastructure provided as part of the UGL design is new.

There are no existing hazardous materials identified on the 4850L. Prior to closure of Homestake Gold Mine, Homestake removed all hazardous materials.

### 5.6.5 Preliminary Design for Core Utility Systems

This section comprises a summary-level discussion of the key core utility systems included in the Preliminary Design of the MLL. Figure 5.6.5 represents the LM-2 longitudinal section, with the ramp access shown on the left, platform access on the right; the smaller section on the right is the LM-2 MER. Figure 5.6.2-3 represents the cross-sectional view of the LM-2 general utilities entering LM-2 along with the cranes. The general sectional views for LM-1 and LM-2 are similar. Because the utility system design is similar between LM-1 and LM-2, only LM-2 layouts will be represented in the following sections.

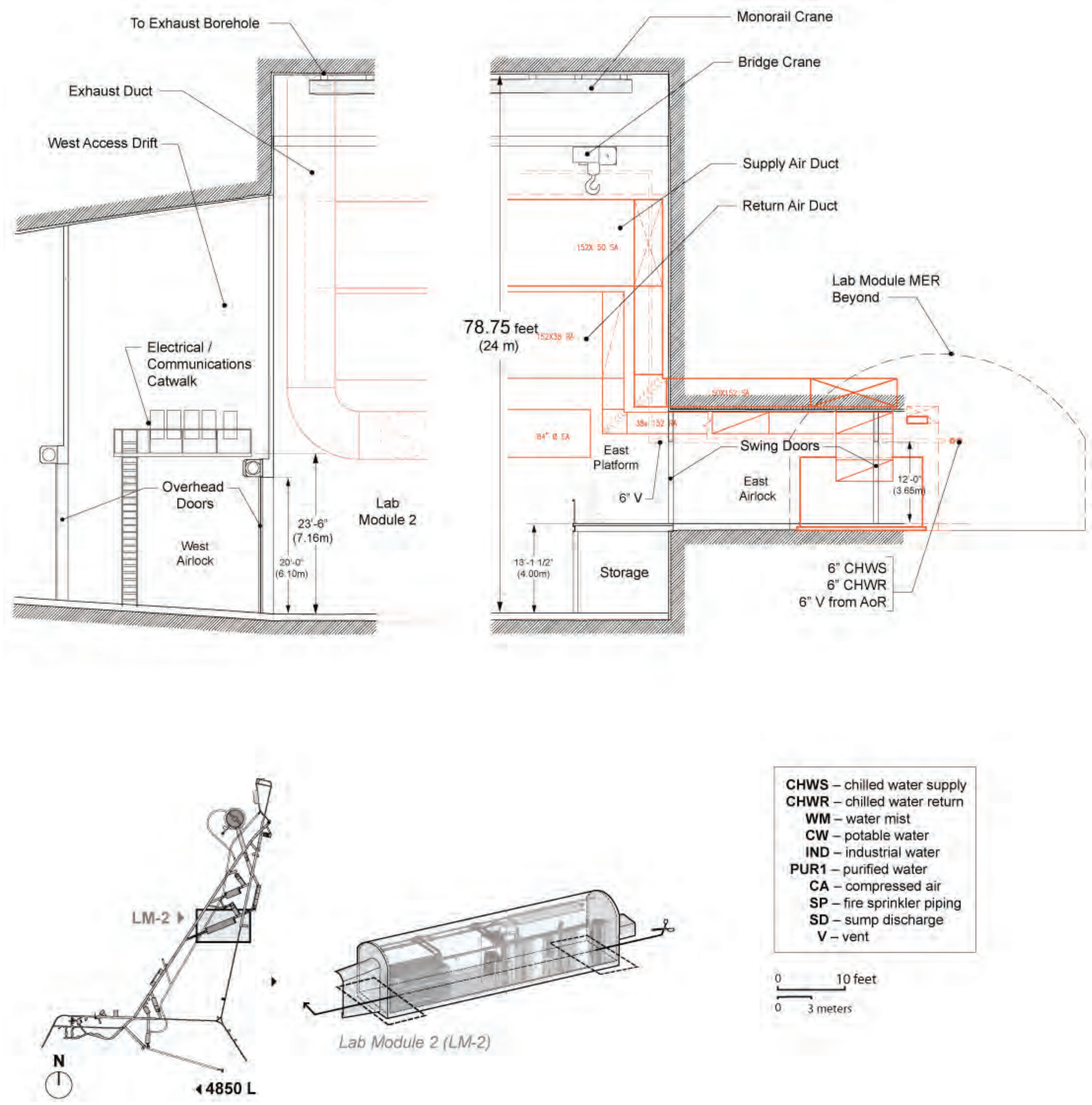

**Figure 5.6.5** LM-2 longitudinal section and LM-2 MER cross section depicting the HVAC and exhaust ducting. [DKA]

### 5.6.5.1 Ventilation, Chilled Water, and Air Conditioning Systems

The DUSEL Facility ventilation system described in Section 5.4.3.8.3, *Air Quality and Ventilation Preliminary Design*, will be used to supply fresh air and remove exhaust at the 4850L. The ventilation path on the 4850L will enter in the Yates Shaft. The exhaust path from the 4850L will be via a new borehole to the 3950L and then exit the existing Oro Hondo exhaust shaft to the surface. A dedicated new overhead drift (4700L) exhaust system will be provided for normal and hazardous exhaust from the LMs and Davis Chamber.

The LMs' ventilation will be provided from air-handling units (AHUs) using air drawn directly from the access drifts. This air will be filtered to Minimum Efficiency Reporting Values (MERV) 15 standards. The HVAC system design, provided as part of the UGL facility-level systems, will include air-handling equipment and main ductwork located inside the LMs. The make-up air supply rate is designed to be one air change per hour and the remainder of the air required for cooling will be provided by recirculating air within the LMs. Filtration equipment for areas requiring air cleanliness above MERV 15 will be provided as part of the fit-out by the experiment collaborations.

The exhaust system will be provided for emergency exhaust events such as a cryogen release or fire within the LMs. The cryogen exhaust inlet will be placed near the floor to capture the heavier cryogens, with a second inlet near the top of the LM to capture smoke from a fire. This system is not intended to handle the effects of a catastrophic event such as a tank rupture. Depending on the situation, dampers will control the flow rates from the LM at low and high levels. Exhaust in the unaffected LM would be shut off in an emergency event, resulting in the exhaust system of the affected LM being capable of 100,000 cfm (170,000 $m^3$/h). This extraction rate is based on airflow requirements for smoke management. Based on experience gained from the Gran Sasso National Laboratory, the emergency exhaust system will need to be sized to accommodate a cryogenic material release rate of 5,900 cfm (10,000 $m^3$/h), which is much lower than what is required for smoke management. The design assumptions for cryogen release rates will need to be confirmed after the science requirements are further refined.

Specialty equipment rooms, such as telecom rooms, will be supplied by dedicated computer room air-conditioning (CRAC).The LM AHUs will include supply and return fans controlled by variable frequency drives, chilled-water cooling coils, mixing dampers, pre-filters, and final filters to provide filtration to MERV 15.

Trade Study #380, *Central Utility Plant* (Appendix 9.V), determined that the most economical method for cooling the 4850L facility was through the use of an underground chilled-water system. Based on this Trade Study, a chilled-water system and spray chamber are planned to be installed near the ventilation borehole on the 4850L to the 3950L as part of the UGI design. The heat will be exhausted from the ventilation borehole.

This system will remove heat loads from surrounding or ambient heat loads, mobile equipment, personnel, experiments, and electrical equipment. A circulation system with insulated piping will allow the chilled water to capture heat loads either through AHUs or by directly cooling equipment.

Further processing of the chilled water for experiment cold storage and other specialized laboratory requirements, if needed, will be completed by the experiment collaborations.

Figure 5.6.5.1 represents a depiction of the general mechanical layout of LM-2, LM-2 MER, and combined LM AoR. Also shown are the electrical rooms located in the LM-2 MER.

Additional information on the ventilation, chilled water, or air conditioning systems can be found in Arup USA, *Preliminary Design Report DUSEL Underground Laboratory Design, UGL Basis of Design Report, 100% PDR REV1*(Appendix 5.Q), Page 20-23, December 2010.

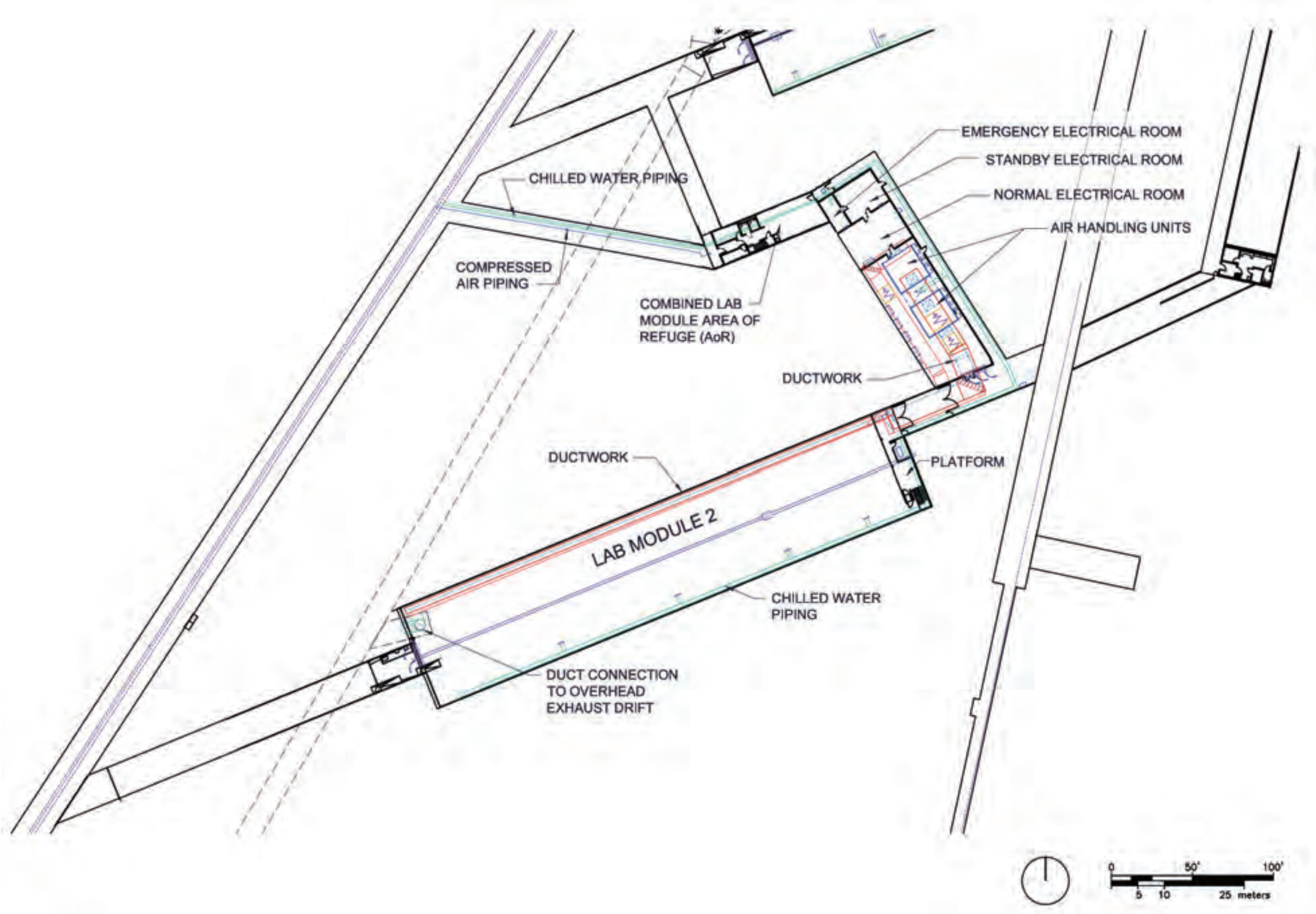

**Figure 5.6.5.1** Layout showing LM-2, LM-2 MER, and combined LM AoR. The layout includes a general depiction of the mechanical utilities for LM-2 along with the electrical rooms in the MER. [DKA]

### 5.6.5.2 Facility Management System

The heating, ventilation, and air conditioning system, along with other facility systems, will be controlled and monitored via a Direct Digital Control System through the Facility Management System (FMS). The system architecture is part of the UGI scope and is further described in Chapter 5.5, *Cyberinfrastructure Systems Design*. Laboratory airflow, pressurization and temperature, and miscellaneous systems will be monitored and controlled to maintain a safe and comfortable working environment for laboratory personnel.

Additional information pertaining to the Facility Management System can be found in Arup USA, *DUSEL Underground Laboratory Design, UGL Basis of Design Report, 100% PDR REV1* (Appendix 5.Q), Pages 23-24, December 2010.

### 5.6.5.3 Electrical

The power for the LMs as discussed in the Arup. *DUSEL Underground Infrastructure Design, UGI Basis of Design Report*, *100% PDR REV1*(Appendix 5.L), Pages 34-38, December 2010 includes a dedicated, redundant 12kV feeder from the Ross and Yates surface substations. The redundant 12kV feeder was removed through the Value Engineering process (UGI #55; Appendix 9.AC) as a cost reduction. The 100% PDR cost and schedule includes a dedicated single 12kV feeder originating at the surface and routed down the Yates Shaft for LM-1 and down the Ross Shaft for LM-2.

LM-1 and LM-2 will each have a dedicated MER (Figure 5.6.1-2) that accommodates medium-voltage gear, step-down transformers, and panel boards. According to the NFPA 110, Section 7.2, the normal power equipment will have its own dedicated room separate from standby and emergency power equipment for increased reliability. The 12kV standby power for the LMs will originate from an automatic transfer switch located at the surface level and supplied by surface-level generators. Redundant 12kV standby power feeders will be routed down the Yates and Ross Shafts and will be connected in a loop system at the 4850L to each LM and the Ross and Yates substation.

Figures 5.6.5.1 depicts the location and types of the electrical rooms in the LM-2 MER. Once at the 4850L, 12kV power will be stepped down to 480 volt at the LM electrical rooms. Standby power will provide egress lighting and power to support laboratory equipment critical to maintaining experiment integrity or prevent release of hazardous materials. Fire, life, and safety system equipment will be fed by emergency uninterruptable power sources (UPS) that are connected to the standby power system.

Transformers and panel boards to feed experiment-specific equipment, electronics, and instrumentation are the responsibility of the experiment design/fit-out. The LMs will be fitted with LED light fixtures to reduce the effect of EMI noise on sensitive instruments. The other dedicated LM spaces will be fitted with fluorescent lighting. The LMs will also be outfitted with electrical receptacles for general use and fire alarm devices.

Electrical service will be provided to the following equipment specific to the LMs:

- Bridge and monorail cranes
- 480 volt connections for experiment-specific power
- Air compressor (for experiments)
- LM lighting and utility power
- Lights and receptacles
- AHUs
- Communication rooms/enclosures
- MER utilities
- Sump pumps
- Supply and exhaust fans

The following systems will be provided with standby power including, but not limited to:

- Mechanical air handling systems and smoke control systems for all refuge areas
- Standby lighting required for refuge areas and smoke control mechanical equipment rooms
- Two-way communication

In addition, LM standby power is provided for loads where “damage to the product or process” could result from a loss of power, e.g., cryogen pumps and orderly shutdown of science-related equipment.

Emergency power definition is based on NFPA 520 and is limited to the following systems:

- Fire detection systems
- Fire alarm systems
- Exit sign illumination

- Emergency lighting
- Fire alarm systems

Additional information pertaining to the LM-1 and LM-2 power can be found in Arup USA, *DUSEL Underground Laboratory Design, UGL Basis of Design Report, 100% PDR REV1* (Appendix 5.Q)*,* Pages 24-27, December 2010.

### 5.6.5.4 Plumbing

Plumbing services to the LM-1and LM-2, LM combined AoR, and LM MERs will tie in to the plumbing services provided by the UGI design with the exception of the purified water system, which is provided by LBNE.

These plumbing services include:

- Potable water
- Sump pump discharge
- Purified water
- Industrial water
- Sanitary vent
- Safety equipment
- Fire sprinkler piping
- Water mist

Both potable and industrial water are supplied by the city of Lead, South Dakota. Potable water will be used for domestic consumption while industrial water will be available for construction, mechanical, and experiment applications. Further discussions of the anticipated volumes that will be available are discussed in Section 5.4.3.14.1.

The purified water will be provided by the LBNE water purification plant on the surface. The LBNE will provide the piping down the Yates Shaft. At this point, the LM purified water piping will tie into the LBNE piping and extend to the LMs.

The potable water, industrial water, purified water, and main water mist lines will have isolation valves and will be capped and available for connection during experiment installation.

The combined LM AoR will have a vented sump with a manually operated sewage ejector. The sewage ejector will be emptied by the DUSEL Facility maintenance staff into a portable container after a signal from the ejector control panel to the Facility Management System (FMS) indicates that the sump is full. No provisions are made for sewage collection in the LMs. Restroom usage for LM occupants will be at the combined LM AoR, as shown on Figure 5.6.1-2.

The floor will be sloped at 1% to the northwest corner in each LM. At the northwest corner of each module there will be an open sump with two submersible sump pumps to collect spills, condensate, and groundwater inflow. The sump pumps will automatically discharge water when inflow reaches a certain height. These sumps will not be used to collect hazardous drainage, and will be provided with a manual shutdown to prevent spreading of the hazard in the event of a hazardous release in either of the modules.

LM sumps will discharge to the drift drainage system at 4850L.The LM MERs will also have sloped floors and sumps similar in design and function as the LMs.

Instrument-grade compressed air will be supplied to each LM from an air compressor in the LM MER and capped at the entrance to LM.

Figure 5.6.5.4 represents a general layout of the plumbing systems that are planned for the LM-2.

Additional information pertaining to the LM-1 and LM-2 plumbing can be found in Arup USA, *DUSEL Underground Laboratory Design, UGL Basis of Design Report,100% PDR REV1* (Appendix 5.Q), Pages 29-31, December 2010.

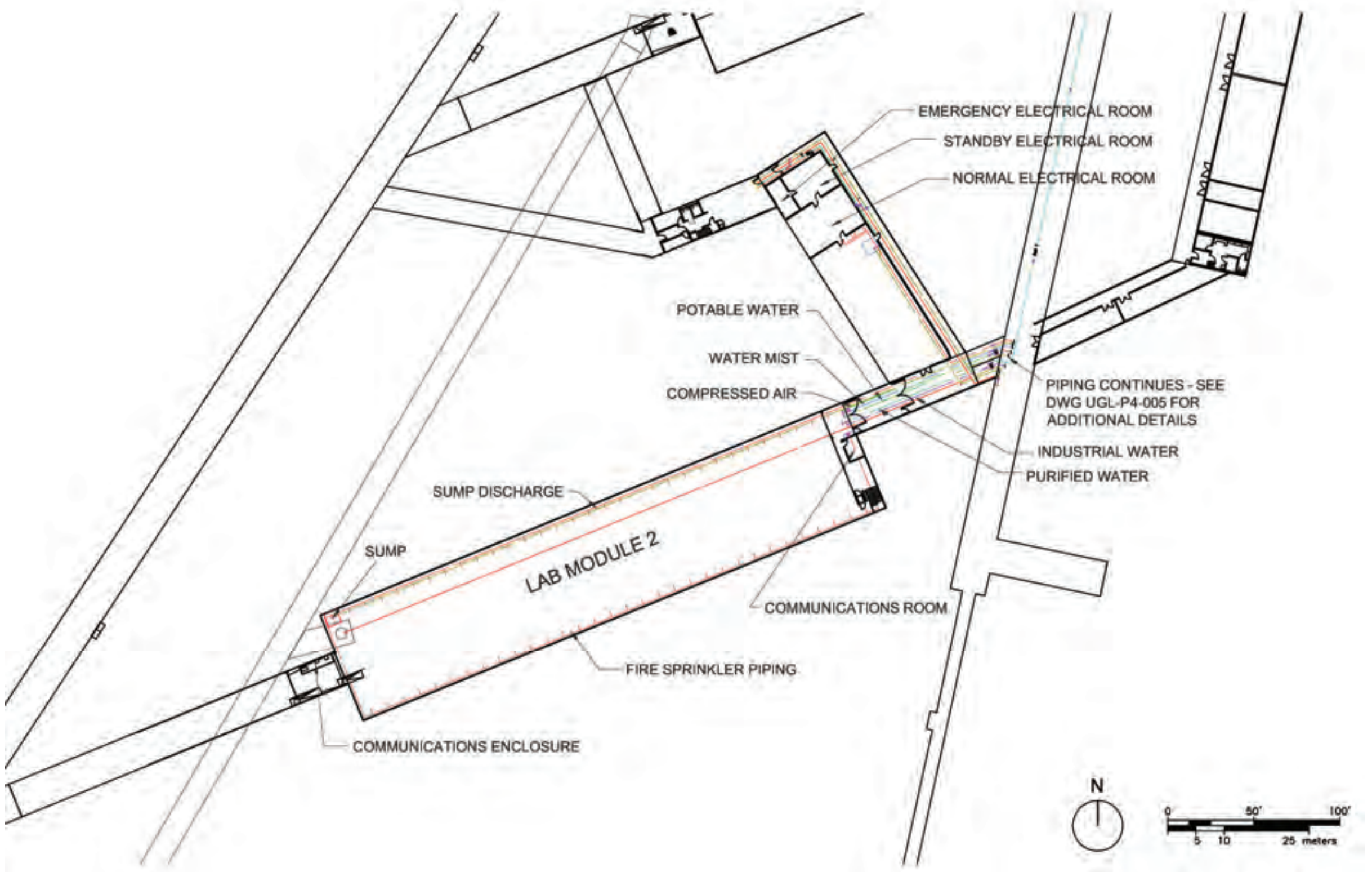

**Figure 5.6.5.4** Plan view depicting the general plumbing layout for LM-2. Also shown are the communications rooms/enclosure in LM-2. [DKA]

### 5.6.5.5 Fire Protection

The fire protection system will consist of a combined standpipe and automatic wet sprinkler systems. The standpipe will have hose valve outlets located throughout LMs, LM MERs, and at the entrance to the combined LM AoR. All points throughout the facility will be within 200 ft (61 m) of a hose station.

There will be separate fire sprinkler zones for each of the following; LMs, LM MERs, and combined LM AoR. Sprinkler systems in each zone include automatic sprinklers, control valves, drain/test valves, water-flow switches, and tamper switches.

The LMs will be provided with connections to a water mist fire suppression system. A valved and capped connection will be provided in each LM, near the entrance. Water mist fire suppression system will be provided in the communications rooms as part of the design. Other than the communications rooms, water mist fire suppression systems in other areas are not part of this Project and will be provided by the science collaborations during experiment installation as needed.

The UGI plumbing design includes the capability to provide supplemental fire suppression by means of remote-controlled water monitor nozzles. These could be connected to the fire sprinkler mains and could operate in conjunction with the sprinkler system. Installation of the remote monitor nozzles is not part of this Project but could be done by others as part of experiment installation.

Each LM will have basic notification devices installed to alert laboratory occupants in the event of a fire. Notification devices will consist of speakers and strobe lights. Manual pull stations will be provided at each LM egress. A phone will be installed at each LM that will connect directly to the main fire alarm panels on the surface.

An air sampling system will be installed as an early detection of a fire condition within the LMs. The air sampling system will be connected into the fire alarm system. In addition, oxygen depletion, carbon dioxide, and nitrous oxide detectors will also be installed integral with the air sampling system. Activity of the LM fire alarm systems shall communicate with the Satellite Command and Control Centers at the 4850L Ross and Yates AoRs as well as the surface Command and Control Center.

Primary emergency egress from the LMs would be the Yates Shaft, with the Ross Shaft being secondary. If the shafts or drifts were inaccessible, AoRs would provide a protected environment for occupants during an emergency event such as a fire or cryogen leak. The main AoRs are located at each of the exit points from the 4850L at both Ross and Yates Shafts. Refer to Section 5.4.3.1.3, *Life Safety Preliminary Design*, for detailed description of these AoRs. In general, the AoRs are sized for anticipated personnel on the 4850L. Services provided include water, restroom, air source, and communications.

The combined LM AoR is located between LM-1 and LM-2 and will be accessed from the West Laboratory Access Drift and from both LMs. The layout allows occupants to access the AoR from their work areas without passing through the drift, as well as exit from the AoRs into the drift without having to pass back through the LM where the hazard may be. All access points to the AoR will have air locks. The combined LM AoR is sized to accommodate 85 occupants.

Additional information concerning the fire strategy, including evacuation and fire protection for the LMs, can be found in Arup USA, *DUSEL Underground Laboratory Design, UGL Basis of Design Report*, *100% PDR REV1* (Appendix 5.Q), Pages 8-18, 26, and 32, December 2010.

### 5.6.5.6 Information and Communications Technology (ICT)

The ICT systems include:

- Structured cabling system (SCS)

The SCS consists of the ICT infrastructure for the underground LMs, and will include:

- ICT spaces
- Backbone distribution
- Horizontal distribution

The SCS design will be capable of and include:

- Support of various systems in the underground environment, such as wired and wireless Local Area Network (LAN), Voice Communications, Radio Communications, Electronic Security, Monitoring, Audiovisual, etc.
- Support for connectivity to networks beyond the LAN such as the Internet, Internet2, and Wide Area Networks (WANs)
- Support for business and research requirements of laboratory staff and other users of the facility
- Support for expansion and upgrades to address future requirements and emerging technologies
- Capability to support 10 Gbps Ethernet in the backbone and 1 Gbps Ethernet to the desktop/equipment outlet

The SCS design will provide an allowance for 25%-50% additional spare capacity in rooms and pathways for future growth and be based on uniform cable distribution with a dual star topology in the LMs.

Redundant backbones will be installed to the 4850L as part of the UGI design, one in the Yates Shaft and the other in the Ross Shaft. This will provide redundant connectivity for the LMs. The two backbones will originate at the surface level and end at 4850L in communications distribution rooms (CDRs) located near the Ross and Yates Shafts. The CDRs are the points of interface between the UGI and the UGL. The backbones will extend from the two CDRs to the communications spaces serving the LMs. A communications room (CR) will be provided in each LM. There will also be communications enclosures (CEs) in addition to the CR to ensure compliance with standard requirements for the maximum horizontal cable length. CEs will also be provided in the combined LM AoR and MERs.

Horizontal distribution will be provided in combined LM AoRs, LM access drifts, and other ancillary/support spaces required to support connectivity for telephones, FMS devices, electronic security system (ESS) devices, monitoring devices, and other technology systems. The horizontal distribution in the LMs for support of the experiments is not included in the DUSEL design and will be provided as part of the experimental installation with non-MREFC funding.

DUSEL will provide network service based on a Transmission Control Protocol/Internet Protocol (TCP/IP) standard to a demarcation point established at network switches located in the communications rooms and communications enclosures. The experiments in the LMs will interface with the network by plugging their devices into the switch to access the overall facility network and utilize WANs over the Internet and Internet2. As outlined in Chapter 5.5, *Cyberinfrastructure Systems Design*, the facility side of the network demarcation will be capable of supporting 10 Gbps in the backbone and 1 Gbps to individual devices and 10 Gpbs to devices on an as-needed basis. The science side of the demarcation will be provided by LBNE and is not included in the facility design scope.

### 5.6.6 Scope Options

Due to the ISE and other science collaborations' request for LM-1 and LM-2 to accommodate a variety of experiments, DUSEL completed two Trade Studies to analyze the cost, modifying the size of the excavations and supporting infrastructure. These Trade Studies are referred to as Trade Study #409, *LM-1 Modified for DIANA* (Appendix 9.Z) and Trade Study #368, *LM-2 Size and Configuration Options*, (Appendix 9.AA). Further information on these Trade Studies is found Section 3.8.5.

LM expansion options were not included in this PDR due to funding constraints. However, there are locations suitable to site a future third LM near the planned LMs. Additional geotechnical site investigations would be required to confirm this.

Value Engineering Item UGL #23 (Appendix 9.AC) removed a second finish shotcrete coat over the top of the base layer of shotcrete in the LMs. The second coat may be desired to provide for a cleaner and easier-to-maintain environment. The estimated value of the shotcrete finish coat in LM-1 and LM-2 was $2 million.

Additional Value Engineering items were included in the UGL design. Those include:

- Elimination of vacuum pumps, (UGL #6) for the LMs, resulting in $100,000 savings. It was anticipated that the vacuum system could be branched off to equipment or bench valves and used for laboratory processes requiring vacuum. It was determined that it was not a requirement of the collaborations (Appendix 9.AC).
- Reduction in number of sumps (UGL #22) in each LM, resulting in $275,000 savings. The original sump design in the LMs included multiple sumps with floor drains; the number of sumps was reduced to one sump per LM and a 1% slope was included in lieu of floor drains. The sumps are intended for nonhazardous spills, condensate, and groundwater (Appendix 9.AC).
- Removal of vibration controls (UGL #17) for the AHUs and pumps, resulting in $300,000 savings. It was determined that this was also not a requirement of the collaborations (Appendix 9.AC).

### 5.6.7 Davis Campus

#### Assessment of Existing Conditions at Start of the MREFC-funded Project

The Davis Campus consists of two laboratory spaces and miscellaneous supporting spaces for the LUX and MAJORANA DEMONSTRATOR experiments. It is anticipated that these two experiments will be in operation at the start of the MREFC-funded Project. The Davis Transition Area (DTA) is a newly excavated structure completed by Sanford Laboratory crews that will feature a mechanical room, electrical room, toilet facilities, and liquid nitrogen alcove. Also included in the DTA is the MAJORANA DEMONSTRATOR with space for the machine shop, clean e-forming room, and control room. Personnel and supplies will enter the DTA through a passageway with an air lock midway through the adjacent drift. Progressing north through this passageway, the connection drift is the upper access of the Davis Laboratory Module (DLM).) The Davis Campus including the DTA, DLM, and supporting spaces are represented in Figure 5.6.7-1. The Davis Cavity was excavated by Homestake Mining Company in the1960s for a science experiment. Sanford Laboratory excavation crews enlarged the height of the existing Davis Cavity by 8 feet to provide space for a tank that is 25 feet wide by 20 feet tall. Decking will surround the tank and the cavity can be accessed from the top or bottom. The enlarged Davis Cavity is now known as the DLM. This laboratory space includes a control room, laboratory, clean room, and counting room. A photograph inside of the DLM, just after shotcrete application, is shown in Figure 5.6.7-2.

Support spaces for the Davis Campus include a liquid nitrogen storage area, water treatment cut-out, and the chiller/electrical room. Most of these facilities utilize spaces were existing excavations outfitted for the needs of the campus with the exception of the chiller/electrical room is a new excavation.

Additional information concerning the Davis Campus including the LUX and MAJORANA DEMONSTRATOR experiments can be found in Sections 3.4.1, 3.4.2.1, and 3.4.2.2, respectively.

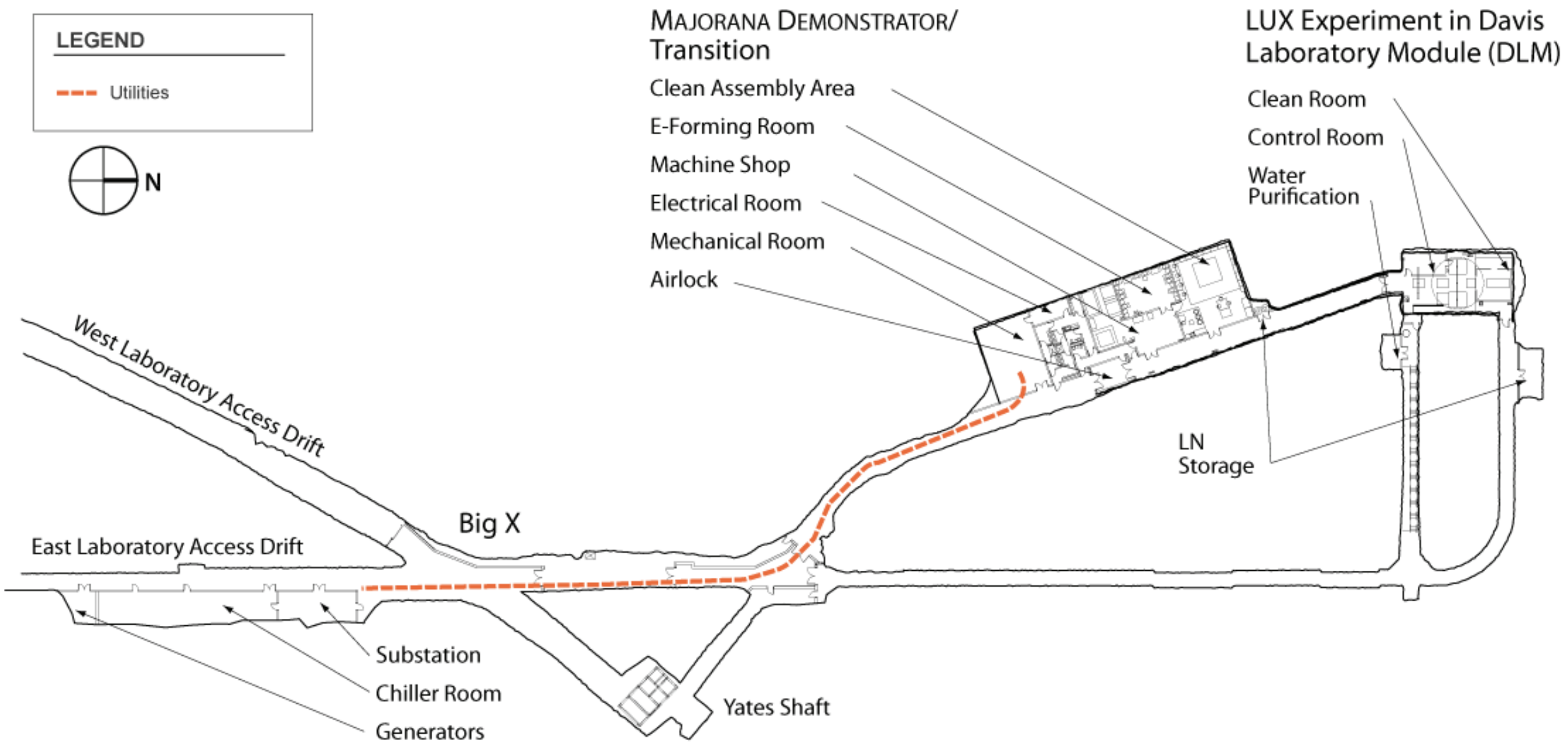

**Figure 5.6.7-1** Davis Campus layout, including the DTA, DLM, and supporting spaces. [J. Willhite, W. Zawada, DUSEL; DKA]

**Requirements for Lab Alterations and Upgrades to Transition from Early Science to DUSEL**
Electrical power and communications cable for the Davis Campus is routed through the Ross Shaft and cable that is run through the West Main Access Drift to the Campus. DUSEL construction activities will interrupt the power sources and communications for the Davis Campus. Options are being considered to minimize delays to operations of the campus during DUSEL construction. These options are discussed in Chapter 5.10, *Final Design and Construction Acquisition Plan*.

DUSEL construction activities include routing the Davis Campus power functionalities through the planned Yates MER located off the East Laboratory Access Drift. The plan is to have shared utilities between the Davis Campus and DUSEL Facility.

Industrial water, potable water, and fire water will be supplied to the Davis Campus via the Ross and Yates Shafts. Water lines located within the Yates Shaft will have to be rerouted through the Ross Shaft during the rehabilitation of the Yates, further described in Section 5.4.2.2. On completion of the DUSEL chiller facility, chilled-water lines for the Davis Campus will be rerouted to these chillers and the Davis Campus chillers can be removed.

Ventilation for the Davis Campus is provided through the Yates Shaft. Intake and exhaust air is ducted through the Campus and discharged into the underground exhaust system. Six fire doors are located throughout the Campus and will help isolate specific areas in case of an emergency. The Davis Campus Cryogen Safety Committee analyzed the Oxygen Deficiency Hazard (ODH) for various cryogenic releases for the LUX and MAJORANA DEMONSTRATOR experiments. The results of this analysis have been incorporated in the Davis Campus exhaust ventilation design.

DUSEL construction activities include an overhead exhaust drift at the 4700L. Once this excavation has completed, the LUX and MAJORANA DEMONSTRATOR experiments will experience a minimum amount of disruption until the ventilation ducting can be rerouted through the new overhead exhaust drift.

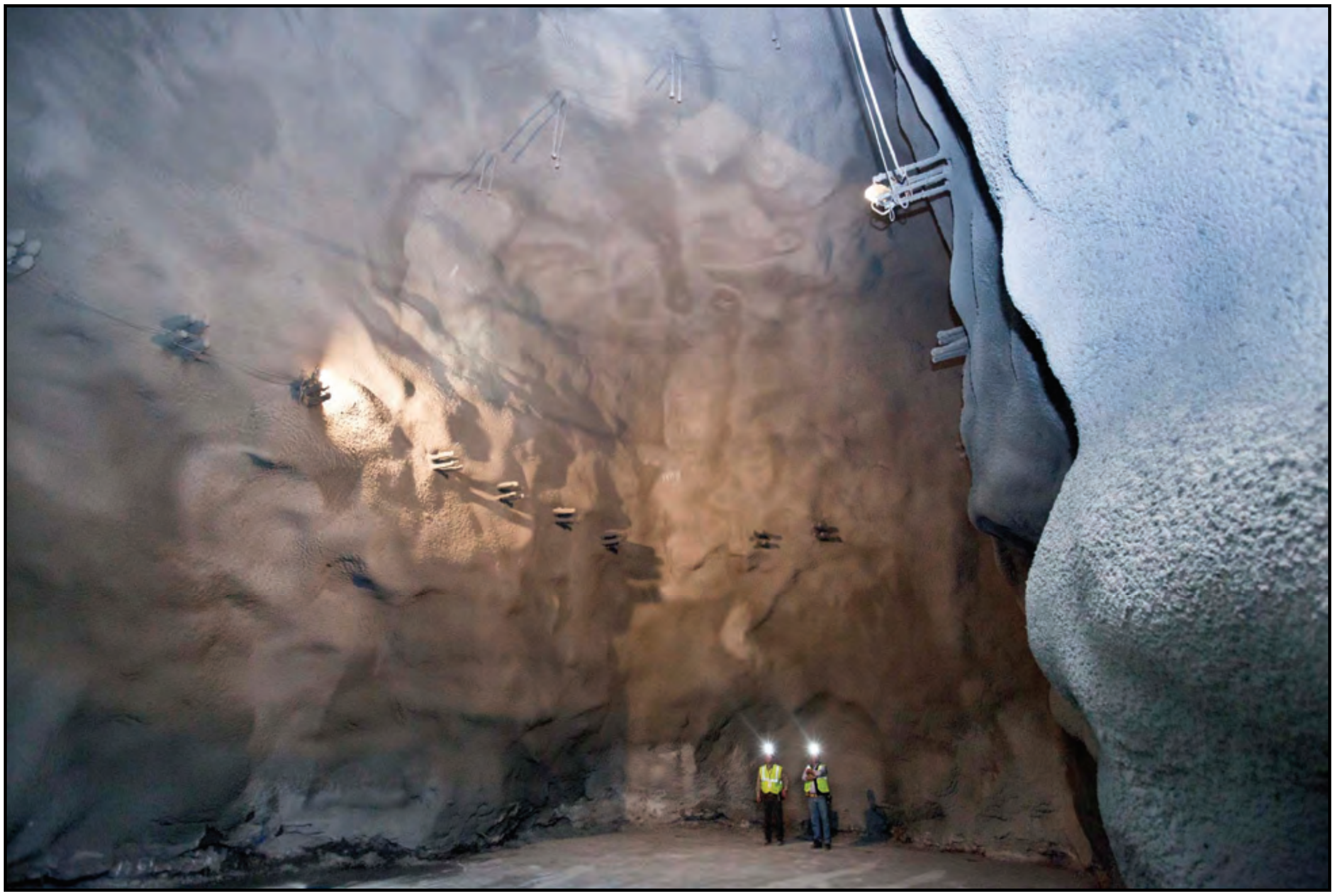

**Figure 5.6.7-2** Inside the DLM excavation facing south after shotcrete application; DLM dimensions are: length 60 ft (18 m) x width 30 ft (9 m) x height 50 ft (15 m). [Bill Harlan, DUSEL]

### 5.6.8 Conclusion

This chapter presented an overview of the Preliminary Design of the DUSEL Facility laboratory outfitting specific to the two LMs and LM ancillary spaces; it also incorporated Davis Campus utility infrastructure at the 4850L. The Preliminary Design addresses the current laboratory requirements and provides a solid foundation on which to optimize and build the Final Design. The Preliminary Design is responsive to the needs outlined in the facility requirements to support operation of a safe laboratory that supports the planned DUSEL science agenda as outlined in Volume 3, *Science and Engineering Research Program*.

Additional information describing the underground facility infrastructure design and construction for the infrastructure and other facility components can be found in Chapter 5.4, *Underground Infrastructure Design.*

## 5.7 Large Cavity for the Long Baseline Neutrino Experiment

This chapter will present an overview of the Preliminary Design of the facility laboratory outfitting specific to the Large Cavity (LC-1) and associated ancillary spaces at the 4850L. The design includes the infrastructure between the main facility infrastructure and the experiment, along with providing a core structure ready for experimental outfitting. Included in this section is an overview of the LC-1 design, beginning with the scope, design requirements, design strategy, and existing conditions. Following this, the core utility systems provided to the LC-1 as well as scope options and scope contingency are discussed. The descriptions in this section are intended to give the reader an overview of the design. Additional details on each design, as well as how the infrastructure design interfaces with other design scopes, can be found in the reference material noted throughout the section.

### 5.7.1 Overview and Planning Summary for the Large Cavity

This section covers the Underground Laboratory (UGL) design for the LC-1 at the 4850L, Mid-Level Campus as shown in Figure 5.7.1-1. The UGL design includes the LC-1 dedicated infrastructure required to support the Long Baseline Neutrino Experiment (LBNE). The Underground Infrastructure (UGI) design, discussed in Chapter 5.4, includes the Facility-provided backbone utilities and infrastructure. The UGL interface with UGI and the other scopes of work are discussed in Chapter 5.1, *Facility Design Overview.*

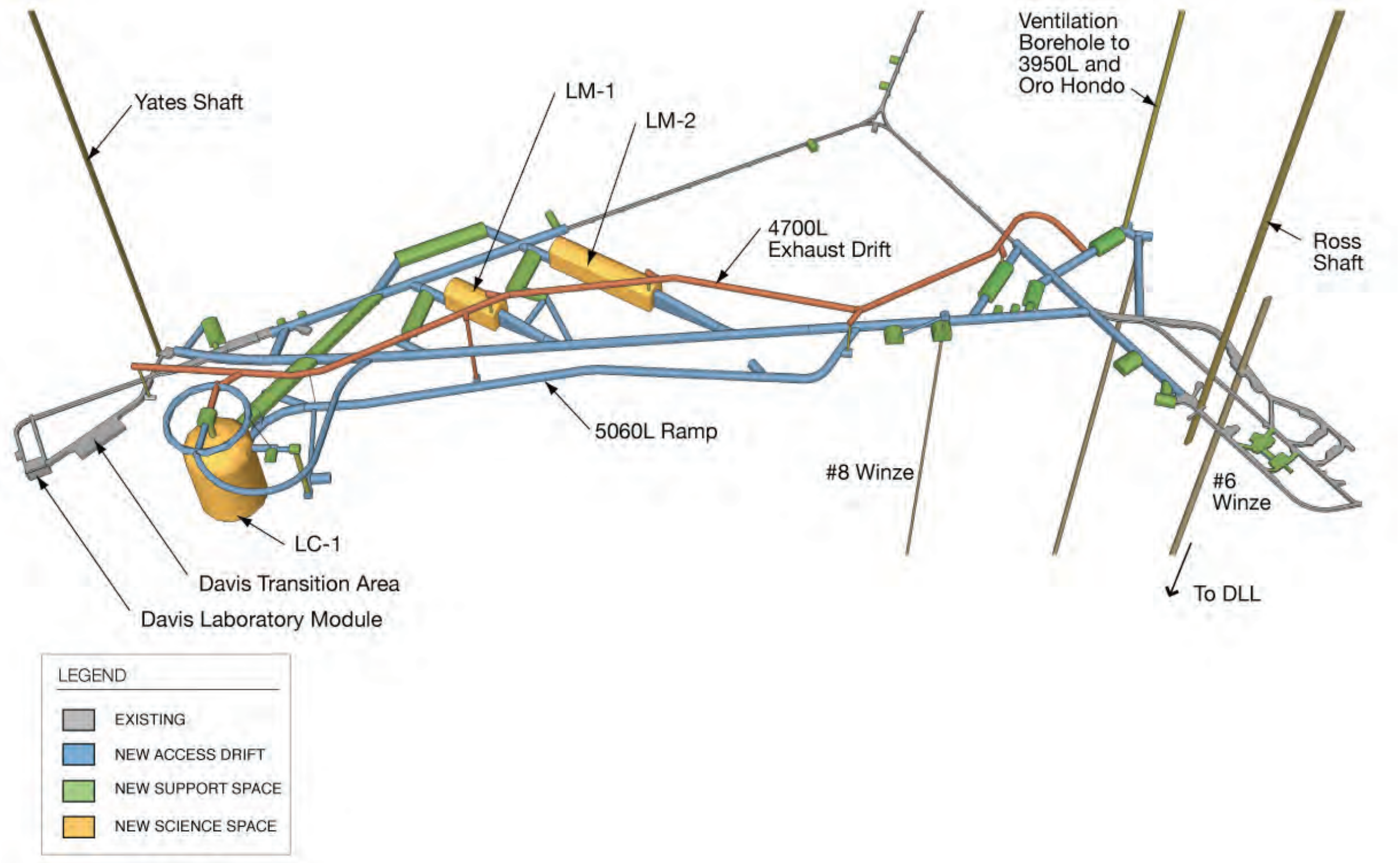

**Figure 5.7.1-1** DUSEL 4850L Mid-Level Campus layout depicting the Davis Campus, LC-1, LM-1, and LM-2, along with Yates and Ross Shafts and the #6 Winze. [DKA]

Figure 5.7.1-2 represents which areas are included in the UGL and the UGI scopes of work and the associated demarcation between the two design scopes in relationship to the LC-1. Further, all the outfitting inside the LC-1 and the calibration drift, along with all experimental outfitting, and the purified water system, are the responsibility of the LBNE.

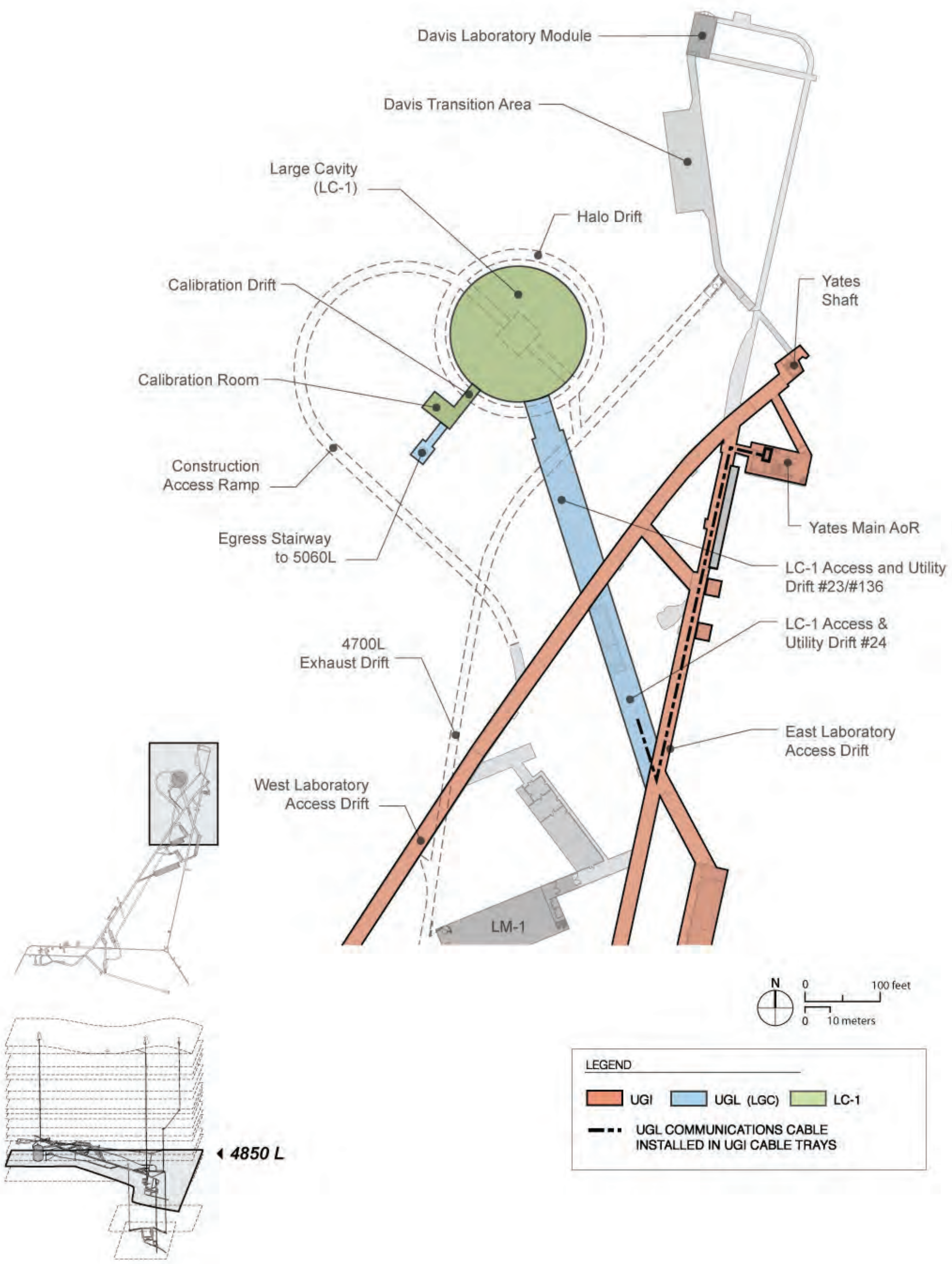

**Figure 5.7.1-2** 4850L area representation of UGL and UGI scopes of work pertaining to the LC-1. The UGI scope of work includes the facility-provided backbone utilities and infrastructure supporting UGL. [DKA]

The water Cherenkov detector (WCD) configuration facility at the DUSEL site is funded primarily through the Department of Energy (DOE), supplemented by National Science Foundation (NSF) S-4 funding under the direction of the University of California at Davis. Both DOE and NSF are providing funding for the construction of the LC-1. NSF is providing construction funding of the LC-1 through the experiment portion of the Major Research Equipment and Facilities Construction (MREFC)-funded Project at a fixed amount. DOE will lead the funding of the construction costs and steward the design and construction. The design of the LBNE is managed by Fermi National Accelerator Laboratory (Fermilab). Although the LBNE facility design is represented in the UGL design scope, the construction of the WCD facility is outside of the DUSEL Facility portion of the NSF MREFC funding package. A portion of the DUSEL infrastructure is common to both DUSEL and LBNE facilities. The costs of this shared facility infrastructure have been apportioned between DUSEL and the LBNE. These costs are reflected in Volume 2, *Cost, Schedule, and Staffing*. The DUSEL portion of this shared infrastructure is funded through the MREFC budget.

The LC-1 represents one of the more technically challenging elements of DUSEL's Preliminary Design. Consequently, the DUSEL Project initiated geotechnical and initial design elements in advance of the DOE's issuance of CD-0 for the LBNE project and the establishment of the LBNE project. The LBNE project is well integrated into DUSEL Project, with the DOE project now leading the design and engineering aspects.

To support the LBNE, the WCD is being designed within DUSEL using the mature technology of Cherenkov radiation in water to detect neutrino interactions. For the LBNE, those neutrinos will originate from an accelerator at Fermilab in Batavia, Illinois. The development of and rationale for the experiment is discussed in Volume 3, *Science and Engineering Research Program*, and information provided here is for background to support the discussion on Facility development. The detector will contain a 100 kT fiducial mass of highly purified water, beneficial both for its external radiation shielding attributes and transmission of optical photons. Photomultiplier tubes (PMTs) connected to a data-acquisition system will detect Cherenkov light emitted by neutrino interactions within the water mass. Figure 5.7.1-3 shows a sectional view of the proposed LC-1 excavation and the WCD described above.

The separation of responsibility between LBNE and DUSEL for the LC-1 excavation and the vessel is defined by the "neat line" of the cavity. The "neat line" is defined as a virtual surface within which no part of rock wall or ground support may intrude, as shown in Figure 5.7.1-3. The configuration of the region outside this virtual surface is the responsibility of the DUSEL excavation design. The configuration of the region inside this virtual surface is set aside for installation of the experimental equipment and is the responsibility of the LBNE collaboration.

The limits of size for the detector are principally determined by cavity constructability, clarity of the water, and maximum hydrostatic pressure that may be applied to submersed PMTs. Space occupied by vessel wall, liner, and PMTs reduces the fiducial volume of the detector below the volume of the excavation. Space must also be allocated outside the detector for installation of a support deck that will function as a light shield to the detector while also facilitating assembly, maintenance, and operations and providing a storage location for data systems and cables attached to PMTs. Access and Utility Drifts #23/136 and #24, as shown in Figure 5.7.1-2, will accommodate the anticipated spaces and the systems that will be required to support a functioning WCD. Drifts #23/136 include the following LBNE-provided systems: clean room, calibration equipment storage room, control room, gas blanket generation room, and the gadolinium removal system. Drift #24 will house the LBNE water purification system and the Facility-provided electrical room.

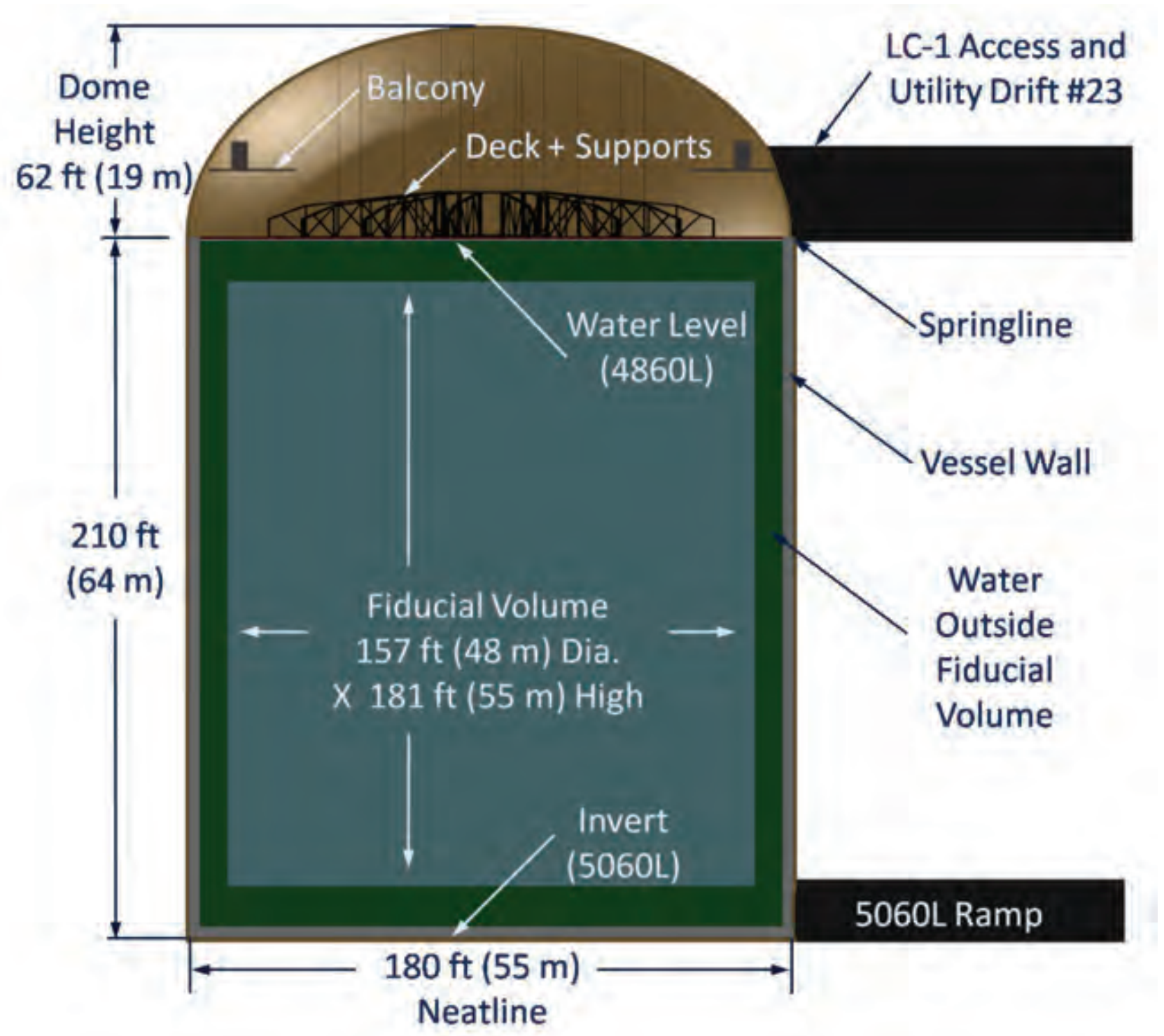

**Figure 5.7.1-3** Sectional view of the Large Cavity excavation and water Cherenkov detector. [Dave Taylor, DUSEL]

At DUSEL, a depth of approximately 4,850 feet from the surface to the top of the detector meets physics requirements for shielding based on analysis documented in DUSEL Depth Document accepted by the LBNE in 2008.[24] The experiment has required interfaces for electrical, heating and cooling, water supply and disposal, and cyberinfrastructure, in addition to protection for life/safety events such as fire or smoke. To assemble an experiment of this scale underground will also require planning of assembly sequencing, surface and underground laydown space, hoist use, and a commissioning plan.

The baseline Facility-level UGL design at the 4850L includes:

- Two standard Laboratory Modules (LMs)
- One large cavity for LBNE
- Necessary ancillary spaces to support the campus-level operations, including Areas of Refuge (AoRs), storage spaces, mechanical electrical rooms (MERs), and utility room elements dedicated to the LMs and LC-1.

The 5060L ramp will initially be used for waste rock removal from excavation for the LC-1. While a second LC is not included in the baseline, the 5060L ramp could allow for its excavation using the same rock removal path (see Section 5.7.7, *Scope Options and Scope Contingencies*).

## 5.7.2 Facility and Infrastructure Requirements for the Large Cavity (LC-1)

Detailed requirements for the MLL have been developed, reviewed, and approved. A thorough discussion of the requirement structure and the development process can be found in Volume 9, *Systems Engineering*. The actual requirements listed in the Preliminary Design Report (PDR) summarize the formal requirement sets included as an appendix to the PDR (Appendix 9.F). The key driving requirements from LBNE are listed in Section 3.3.5.7.3. The functionality and performance mandated by this set of requirements form the basis for the DUSEL Facility design as presented throughout Volume 5, *Facility Preliminary Design*. All key driving requirements of the design have been either met or otherwise addressed specifically by this design report.

Key Facility and infrastructure design requirements include:

- Space—for LC-1 and associated LC-1 ancillary spaces
- Electrical power—normal (2248 kW) and standby (110 kW)
- Potable and industrial water
- Chilled water—1411 kW
- Exhaust ventilation (100,000 cfm, 170,000 $m^3$/h)
- Fresh air supply at one air exchange per hour (18,000 cfm, 30,600$m^3$/h)
- Environment and humidity control (64 to 77 °F/18 to 25 °C; RH 30% to 50%)
- Network IT connections (10 Gbps)
- Fire protection systems including fire sprinkler, standpipes, and water mist
- Dewatering systems including both the native and vessel discharge water systems at the 5060L, described below

### 5.7.3 Facility Large Cavity Design Strategy

Initial requirements were created by the LBNE collaboration to define room sizes, electrical power, ventilation, and cooling water needed to support a WCD at the 4850L. DUSEL Science Liaison physicists/engineers and DUSEL Facility engineers have worked closely with LBNE to refine these requirements and more specifically identify interfaces. These were delivered to the design contractors to support the Preliminary Designs. Weekly Project meetings and monthly interface meetings with LBNE and DUSEL representatives and design contractors were used to resolve requirements and design issues.

A surface water purification plant is required to fill the WCD, and the LBNE will provide all design, funding, and equipment necessary for this system. The equipment includes reverse osmosis, filtration, uranium/thorium removal, UV, degasification, and chillers, which will all be installed in the existing Yates Motor Generator Room adjacent to the Yates Hoistroom. Purified water piping from the surface water purification water plant to the WCD at the 4850L is within the scope of the LBNE and is not included in the UGL design. Pressure-reduction equipment and pumps to empty the WCD will be installed by the LBNE at levels along the Yates Shaft.

Golder Associates, the excavation design contractor, has performed extensive modeling of excavation, rock support, and rock removal methods. LBNE funded Golder Associates to evaluate the optimal WCD geometry for the 4850L,[25] which would provide a stable and maximum fiducial volume capacity.

The study concluded that the upright dome cylinder geometry was the optimal stable design and could potentially be excavated to dimensions that would allow for a possible 200 kT fiducial volume, if hydrostatic pressures on the detectors were not a limiting issue. However, issues of constructability do become more critical with designs beyond the existing 100 kT volume. Additional studies would be required to definitively state the design requirements to reach a 200 kT capacity.

The geotechnical analysis is discussed in Chapter 5.3.

Utility requirements to support the science experiments have been communicated to the UGL and UGI design contractor to size adequate support equipment. Based on layouts created for that support equipment, the UGL/UGI design team determined the required size for excavations to accommodate the support equipment.

The DUSEL UGL design, through MREFC funding, will provide the following utilities to the WCD at the 4850L via connections to the UGI facility-provided systems:

1. **Utilities.** Electricity, potable water, industrial water, fire-sprinkler piping, water-mist piping, chilled water, fresh air, conditioned air, and data-acquisition equipment will be supplied to the LC-1 by the UGL scope of work. Additionally, the LBNE has requested space at the LC-1 entrance to position an experiment-furnished monorail crane to facilitate installation of the experiment equipment including vessel, liner, PMTs, and deck. Since much of the equipment located in the LC-1 cannot be installed until after the experiment installs the deck, supporting utilities provided by DUSEL will be terminated at the LC-1 entrance until experimental fit-out may be completed.
2. **Sumps and dewatering.** Due to the quantity of water contained in the WCD, planning for vessel failure, drainage, and/or leaks has been a significant part of the process. Sumps that collect vessel-leakage water and native water behind the vessel wall will be installed and operated at the 5060L. As baselined, the purified water in the detector has no impurities, and disposal into the facility dewatering system is acceptable. The process of mixing purified water with the facility groundwater will reduce the percentage of impurities in the groundwater. The sump and associated plumbing will be designed to accommodate either filtering the water prior to mixing or total isolation until it reaches the surface. This isolation may become necessary if an additive such as gadolinium is added to the water for experiment enhancement.

### 5.7.4 Facility Existing Conditions at Start of Construction

The proposed location of the LC-1 has no excavations that intersect it although drilling and coring have investigated the area at a preliminary level. The current geologic assessment and conditions are discussed in Chapter 5.3, *Geotechnical Site Investigations and Analysis*. The existing conditions of infrastructure to support this space are discussed in Chapter 5.4, *Underground Infrastructure (UGI) Design*. To support the WCD, the existing Yates Motor Generator Room will be repurposed to house the WCD purified water plant. The existing Yates Motor Generators will be removed and replaced with modern, smaller electrical equipment that will fit in the Yates Hoistroom., as described in Section 5.4.3.2.3.

No existing hazardous materials have been identified within the 4850L Campus proposed for the WCD site.

### 5.7.5 Excavation Design

A geotechnical investigation was conducted on the 4850L to support the early development of the underground design work scopes.

The geotechnical investigation provided information to support the Preliminary Design of the LC-1.These investigations did not reveal any features that would disqualify the proposed location for the excavation of LC-1. Further discussion of the geotechnical site investigations, ground support design, and excavation design are found in Chapter 5.3, *Geotechnical Site Investigations and Analysis,* and also in Golder Associates, *Preliminary Design Final Report* (Appendix 5.I).

The excavation for LC-1 will have a finished inside diameter of 180 feet (55 m) and height of 272 feet (83 m), as shown above in Figure 5.7.1-3. It will be an upright cylindrical cavity with an ellipsoidal crown. The containment vessel liner is being designed by the LBNE and will be a separate installation process from the excavation.

Excavation access for LC-1will be via the Construction Access Ramp and LC-1 Access and Utility Drift #23/#136, which connects to the West Laboratory Access Drift. Access to the bottom of the cavity will be necessary for the removal of excavated rock material from the LC-1. This will be provided by a 5060L ramp that will ramp down from near the Ross Shaft to the 5060L. The Halo Drift will allow access for the installation of cable bolts in the crown of the LC-1.

A native water-collection membrane lining of geosynthetic composite strip drains will be installed vertically against the perimeter of the cavity. The strips will have width of 6 inches (15 cm) and will be intermittently spaced. The strips will act as vertical drains. The strips will be installed beneath a shotcrete layer, will extend for the entire height of the LC-1, and will discharge into the drain underneath the concrete invert. The drain collecting the groundwater seepage will be directed into a sump on the 5060L.

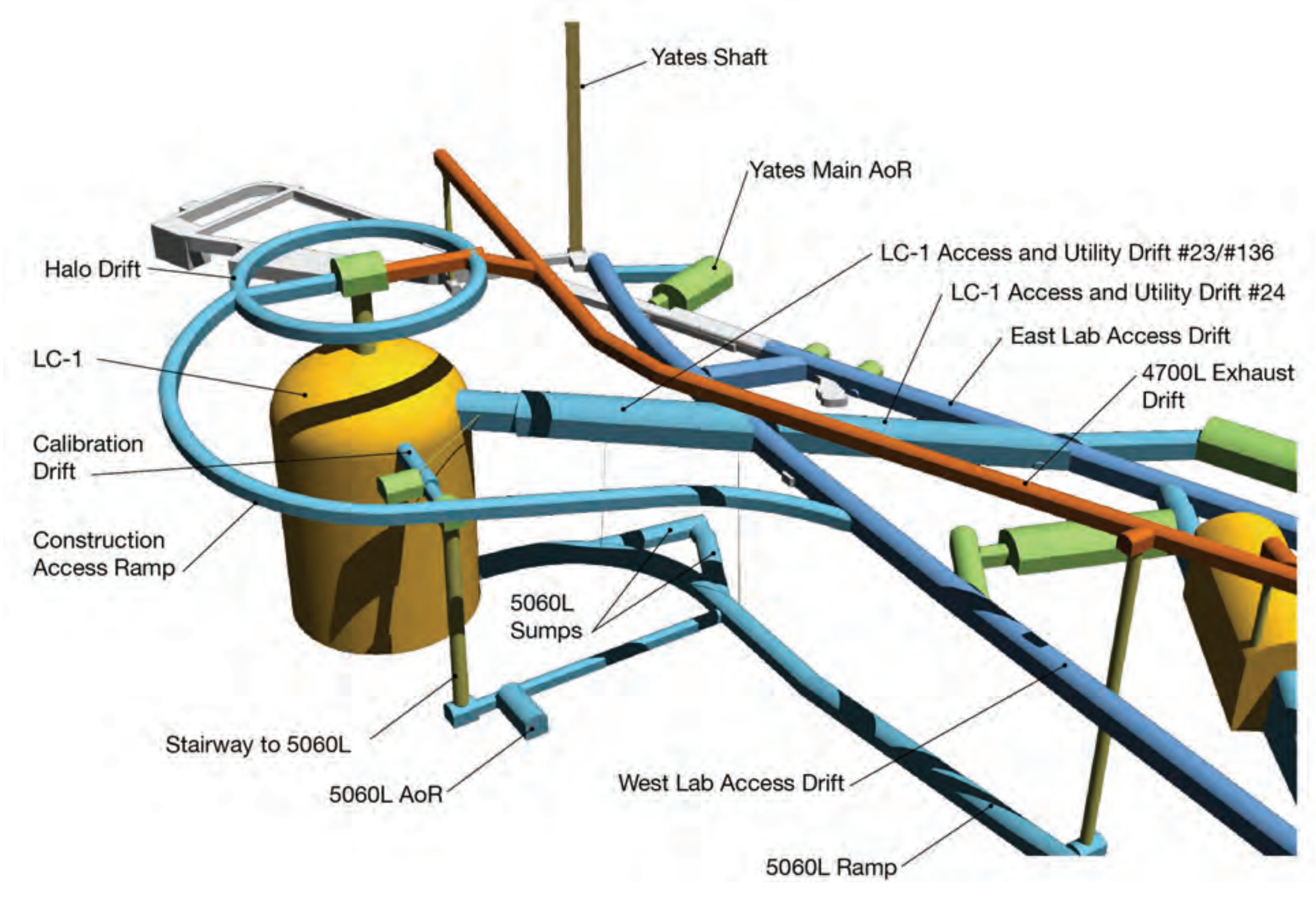

**Figure 5.7.5** LC-1 and LC-1 ancillary spaces. [DKA]

### 5.7.6 Preliminary Design of Core Utility Systems

The DUSEL Facility will provide infrastructure to the LC-1 as described below. Facility infrastructure inside the LC-1 dome and calibration drift is included in the DUSEL Preliminary Design, although the cost is not included in the MREFC budget; the design and costs for these areas are the responsibility of the LBNE. Because the infrastructure inside the dome cannot be installed until after the deck inside the dome is installed, as shown in Figure 5.7.6 (LBNE responsibility), the work inside the dome will need to be completed by an LBNE-funded contractor. The demarcation between the DUSEL Facility and the LBNE for all utility services entering the dome will be at the entrance to the LC-1 intersection of LC-1 Access and Utility Drift #23/#136 and the LC-1 dome.

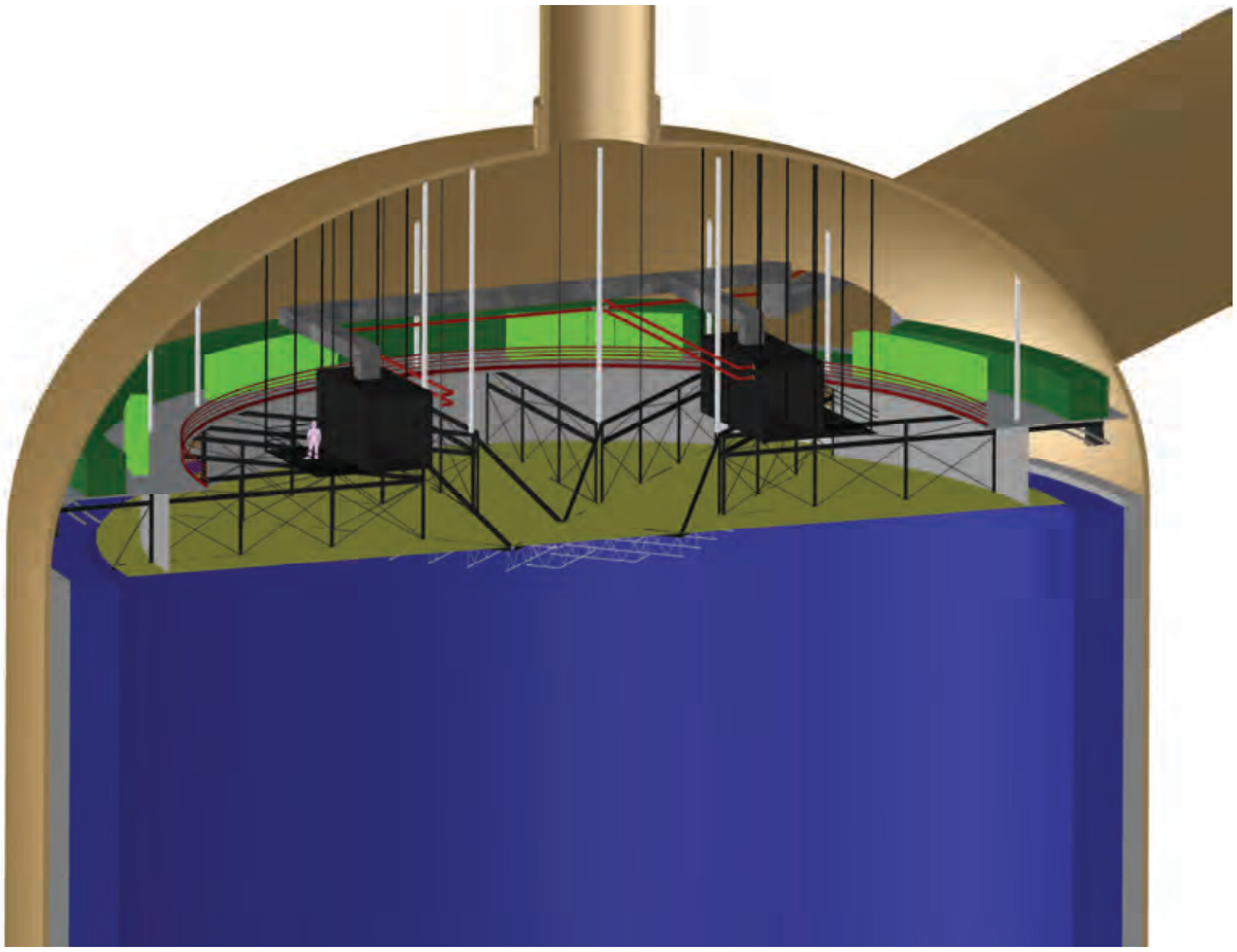

**Figure 5.7.6** WCD deck configuration. [Jeff Dolph, LBNE; Dave Taylor, DUSEL]

### 5.7.6.1 Ventilation, Chilled Water, and Air Conditioning Systems

The DUSEL Facility ventilation system described in Section 5.4.3.8.3, *Air Quality and Ventilation Preliminary Design*, will be used to supply outside air and exhaust at the 4850L. Fresh air will be provided by air-handling units (AHUs) using the air directly from the access drifts. This air will be filtered to Minimum Efficiency Reporting Value (MERV) 15 standards. The HVAC system design provided as part of the facility-level systems for the LC-1 will be limited to air-handling equipment and main ductwork located inside the dome of the LC-1. All mechanical, electrical, and plumbing (MEP) distribution within the LC-1 will be part of the experiment design scope by the LBNE.

The make-up air supply rate is designed to be one air change per hour and the remainder of the air required for cooling will be provided by recirculating air within LC-1. A dedicated exhaust system in the top of the LC-1 dome will be provided for hazardous exhaust with an extraction rate of 100,000 cfm (170,000 $m^3$/h). This exhaust will be diluted when connected into the ventilation drift at the 4700L. The smoke extraction system that serves the LMs and the LC is a shared system. It is assumed that only one emergency incident occurs within the LMs or LC at any one time.

The LC-1 space will be supplied with connections to the chilled-water system to provide direct cooling as required. Chilled water will also be supplied to the AHU cooling coils for air conditioning in the LC-1 and ancillary spaces.

AHUs for the LC-1 will include supply and return fans controlled by variable frequency drives, chilled-water cooling coils, mixing dampers, pre-filters, and final filters. The HVAC system will be placed in the dome of the LC-1 above the 4850L. This reduces the space that the equipment would occupy in the utility room and makes more space available for LBNE use.

Due to requirements specific to each area, separate air systems are provided for the LC-1 and ancillary spaces. Specialty equipment rooms, such as telecom rooms, will be supplied by dedicated fan coil units. The refuge area on the 5060L will be cooled with an air-cooled direct expansion (DX) unit. The air handler and DX systems for this refuge area will also be serviced with emergency power.

Figure 5.7.6.1 represents a general layout of the mechanical systems that are planned for the LC-1.

Additional information on the ventilation, chilled-water, and air conditioning systems can be found in Arup USA, *Preliminary Design Report DUSEL Underground Laboratory Design, UGL Basis of Design Report, 100% PDR REV1* (Appendix 5.Q), Page 20-23, December 2010.

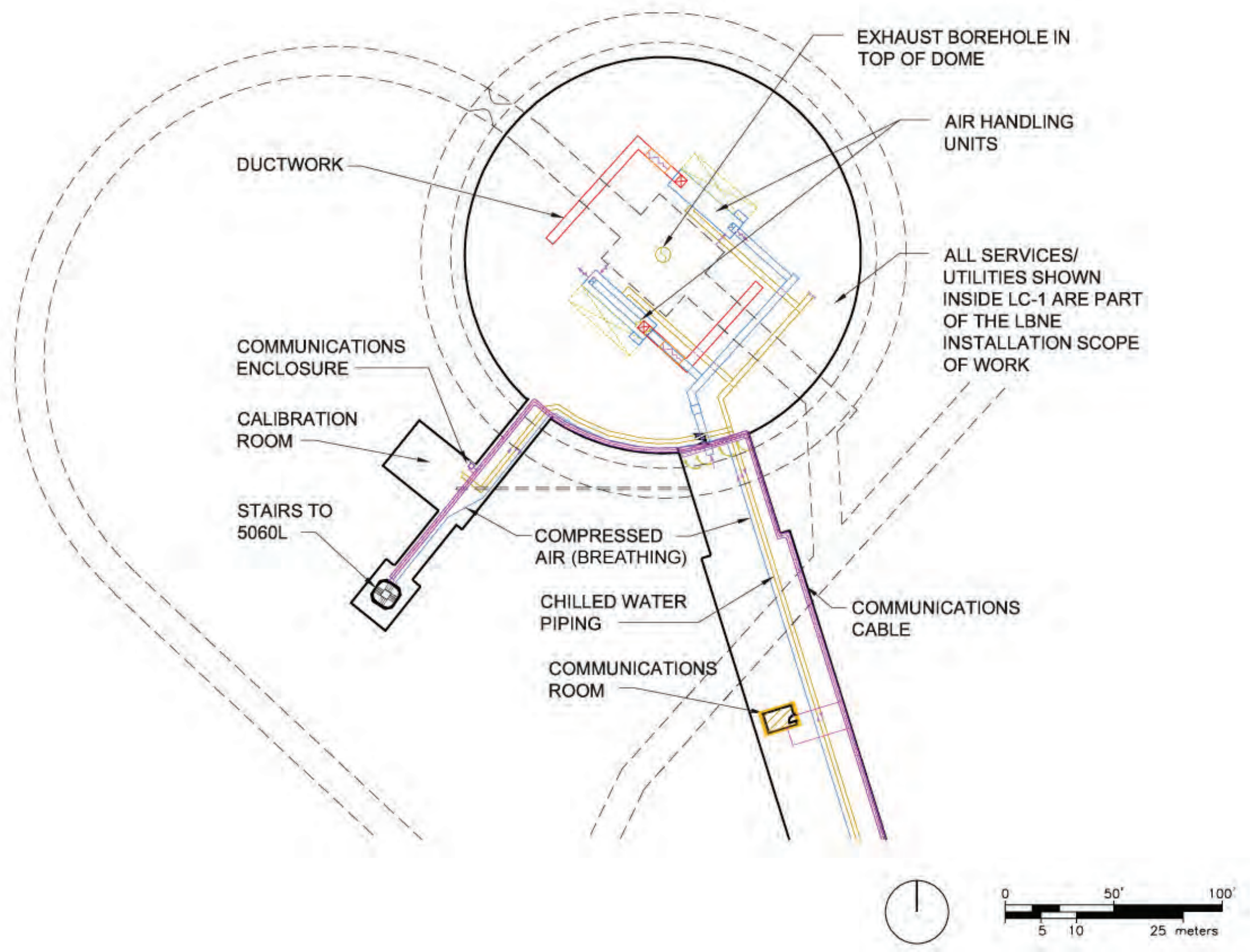

**Figure 5.7.6.1** Plan view of the mechanical and communication systems planned for the LC-1. [DKA]

### 5.7.6.2 Facility Management System

The heating, ventilation, and air conditioning system, along with other facility systems, will be controlled and monitored via a Direct Digital Control System through the Facility Management System. The system architecture is part of the UGI scope and is further described in Chapter 5.5, *Cyberinfrastructure Systems Design*. Laboratory airflow, pressurization, temperature, and miscellaneous systems will be monitored and controlled to maintain a safe and comfortable working environment for laboratory personnel.

Additional information pertaining to the Facility Management System can be found in Arup USA, *DUSEL Underground Laboratory Design, UGL Basis of Design Report, 100% PDR REV 1* (Appendix 5.Q), Pages 23-24, December 2010.

### 5.7.6.3 Electrical

The power for the LC-1, as discussed in the Arup USA Inc, *DUSEL, Underground Infrastructure Design, UGI Basis of Design Report, 100% PDR REV1* (Appendix 5.L), Pages 34-37, December 2010, includes a dedicated, redundant 12 kV feeder from the Ross and Yates surface substations. The redundant 12 kV feeder was removed through the Value Engineering process (UGI #55, Appendix 9.AC) as a cost reduction. The 100% PDR cost and schedule include a dedicated single 12 kV feeder originating at the surface and routed down only the Yates Shaft.

The LC-1 will have a dedicated electrical room located in LC-1 Access and Utility Drift #24, shown in Figure 5.7.1-2**,** that accommodates medium-voltage switchgear, step-down transformers, and panel boards. The normal LC-1 power will have a dedicated room separate from standby and emergency power equipment to increase system reliability. The LC-1 electrical room will contain 15kV switchgear with an automatic switchover circuit breaker.

Electrical service will be provided to the following equipment specific to the LC-1:

- Access drift lights and receptacles
- Air-handling units
- Communication enclosures
- Mechanical Equipment Room (MER) utilities
- Sump pumps
- Supply and exhaust fans
- 480 V connections for experiment-specific power
- 5060L sump pumps
- Cavity dome lighting and utility power
- Monorail crane system

The LBNE will install the necessary panels and wiring for all of the associated support spaces for the LC-1. All wiring will be low-smoke/zero halogen. Special cables required for the PMTs will not meet this requirement, so special fire-control measures are required for these cables and will be the responsibility of the LBNE.

All switchgear, switchboards, and panel boards will have integral transient voltage surge suppression (TVSS). All main 480V circuit breakers will have ground fault protection function.

The following systems will be provided with standby power including, but not limited to:

- Mechanical air-handling systems and smoke-control systems for all refuge areas
- Standby lighting required for refuge areas and smoke-control mechanical equipment rooms
- Two-way communication

In addition, LM standby power is provided for loads where "damage to the product or process" could result from a loss of power, e.g., sump pumps and orderly shutdown of science-related equipment.

Emergency power definition is based on NFPA 520 is limited to the following systems:

- Fire detection systems
- Fire alarm systems
- Exit sign illumination
- Emergency lighting
- Optional standby systems intended for powering loads where "damage to the product or process" could result from a loss of power, e.g., sump pumps and proper shut down of science-related equipment.

Additional information pertaining to the LC-1 power can be found in Arup USA, Inc., *DUSEL, Underground Laboratory Design, UGL Basis of Design Report, 100% PDR REV 1* (Appendix 5.Q), Pages 24-28, December 2010.

#### 5.7.6.4 Plumbing

Plumbing services to the LC-1, LC-1 Areas of Refuge (AoRs), and MERs will tie in to the facility plumbing services provided by the UGI design.

These services do not extend into the LC-1 and will be capped at the entrance to the LC-1 (Figure 5.7.6.4). These services and include:

- Potable water
- Industrial water
- Fire sprinkler piping
- Water mist
- Sump pump discharge
- Drainage of water within the drifts, LC-1, and utility rooms

Purified water will be supplied and piped to the LC-1 from the LBNE surface water purification plant. This will be an LBNE responsibility for funding, design, and construction.

Drainage will be routed to the facility drainage collection systems. The drainage will either be collected by DUSEL Operations or discharged into the facility dewatering systems and pumped to the surface water treatment plant as appropriate. Grey and black water will be collected by DUSEL Operations and transported to the surface for treatment. Groundwater and vessel leakage water (free of gadolinium or other additives) may be discharged into the facility dewatering system. Experimental wastewater that requires special waste treatment will be the responsibility of the LBNE.

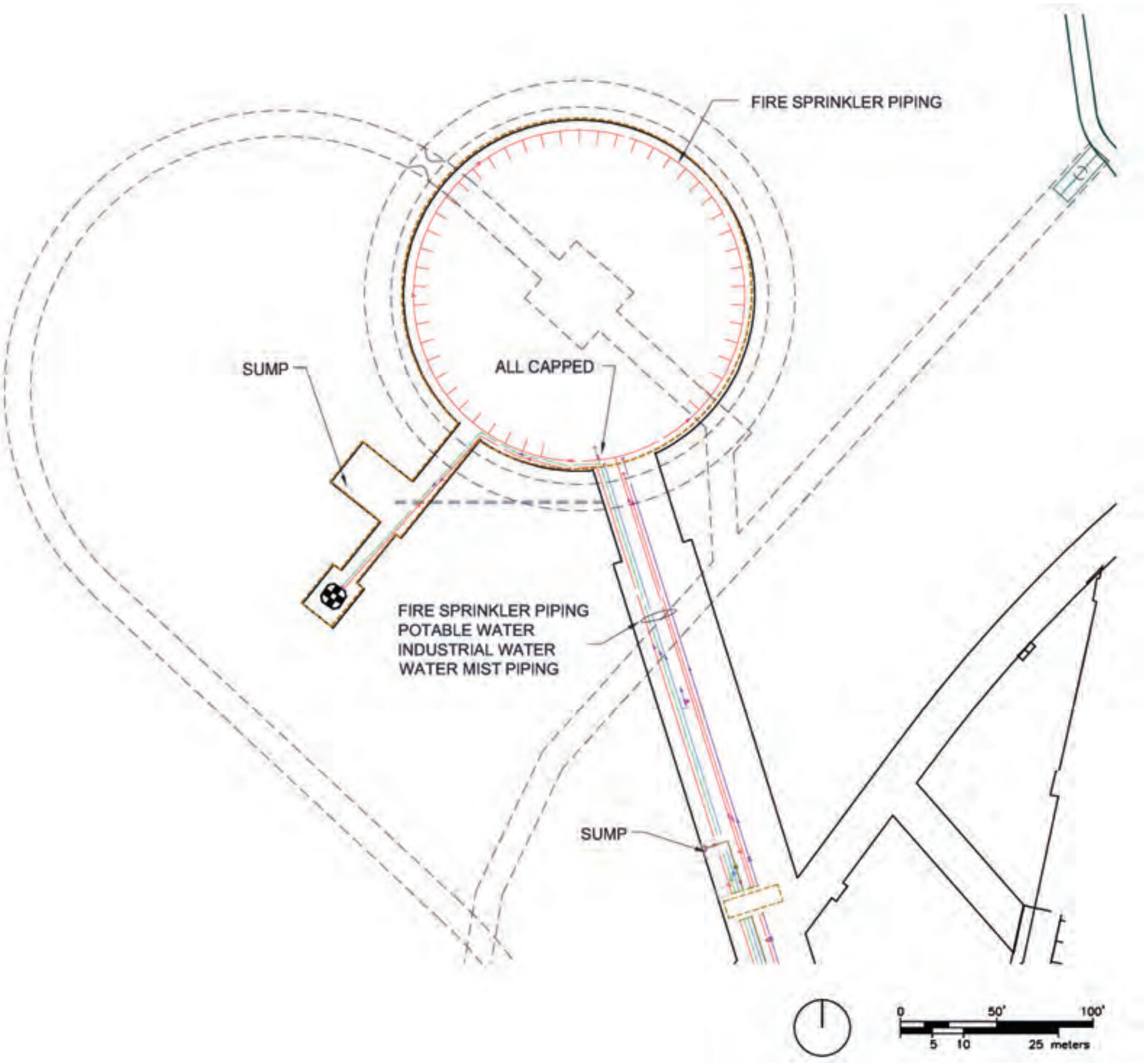

**Figure 5.7.6.4** Plan view of the plumbing and fire sprinkler piping for LC-1. [DKA]

Conditioning the water to fill or maintain the LBNE vessel is not included in the DUSEL Project. All piping and equipment required for water treatment, including chemicals, drainage, and ventilation, are also not part of the DUSEL Project but are the responsibility of the LBNE.

At the 5060L, there will be four sumps: a native water collection sump used to collect water that accumulates between the vessel wall and the surrounding membrane; a second sump to collect any water that leaks from the detector, isolating it for sampling and filtering as needed; a third sump to be used as a settling sump collecting water from these other two sumps as well as level drainage; and this sump overflows into the fourth sump, the facility operations sump, which then ties to the main facility dewatering system in the Ross Shaft. Additional information pertaining to the LC-1 plumbing can be found in Arup USA, Inc., *DUSEL, Underground Laboratory Design, UGL Basis of Design Report, 100% PDR REV1 (*Appendix 5.Q), Pages 29-32, December 2010.

### 5.7.6.5 Fire Protection

The fire protection system will consist of a combined standpipe and automatic wet sprinkler systems. The standpipe will have 2½-inch hose valve outlets for those trained in handling heavy hose streams. The hose valve outlets will be located throughout LC-1, MERs, and at the entrance to each AoR. All points throughout the underground facility are designed to be within 200 ft of a hose station.

The LC-1 will be provided with connections to a water mist fire suppression system. A valved and capped connection will also be provided in the LC-1 access drift and at the entrance to the LC-1. Water mist fire suppression will be provided in the communications rooms and in the LC-1 access drift. Other than the communications rooms, water mist fire suppression systems in other areas are not part of this Project and will be designed and built (if required) by the collaborations during experiment installation.

A Class III standpipe system incorporating hose cabinets and hose racks for use by occupants and emergency responders will be provided in the LC-1 access drift. The UGI design includes the capability to provide supplemental fire suppression by means of remote-controlled water monitor nozzles at a later date. These will be connected to the fire sprinkler mains within each space and will operate in conjunction with the sprinkler system. Installation of the remote monitor nozzles is not part of this Project but shall be done by LBNE as part of experiment installation. Refer to the fire/life safety section of this report, Section 5.4.3.1, for more information.

If required, additional fire suppression in the LC-1 to meet specific experiment needs will be the responsibility of the LBNE collaboration and will be installed during experiment installation.

The LC-1 access and utility drifts will have notification devices installed to notify laboratory occupants of a fire or other facility-wide hazard. Notification devices will consist of speakers and strobe lights. Manual pull stations will be provided at each egress from the LC-1. A phone will be installed at the LC-1 to connect directly to the main fire alarm panels on the surface. All notifications, initiation, and signaling line circuits for the LC-1 will terminate at a dedicated fire alarm panel installed with the emergency electrical room dedicated to the LC-1.

An air sampling system will be installed in the LC-1 drifts, and connected to the fire alarm system, to aide in early detection of fire conditions. No fire alarm devices will be installed in the LC-1 as a part of the DUSEL UGL scope of work. LBNE will be responsible for installing fire alarm annunciation and notification devices that are compatible with the facility-wide fire alarm system.

Primary egress from the LC-1 will follow the path of the West Main Access Drift to the Yates Shaft, where one of the two main 4850L 194-person AoRs is located.

Secondary egress from the LC-1 is provided via a staircase (represented in Figure 5.7.6.5) from the 4850L to the 5060L. There is a 40-person AoR provided at the 5060L near the base of the staircase. The secondary egress path proceeds up the ramp leading to the 4850L near the Ross Shaft. There is also a 10-person AoR located midway up the 5060L ramp. Once on the 4850L, occupants can proceed either toward the Ross Shaft or Yates Shaft for egress to the surface.

Additional information concerning the Fire Strategy, including evacuation and fire protection for the LC-1, can be found in Arup USA, Inc., *DUSEL, Underground Laboratory Design, UGL Basis of Design Report, 100% PDR REV1* (Appendix 5.Q), Pages 8-19 and 32-33, December 2010.

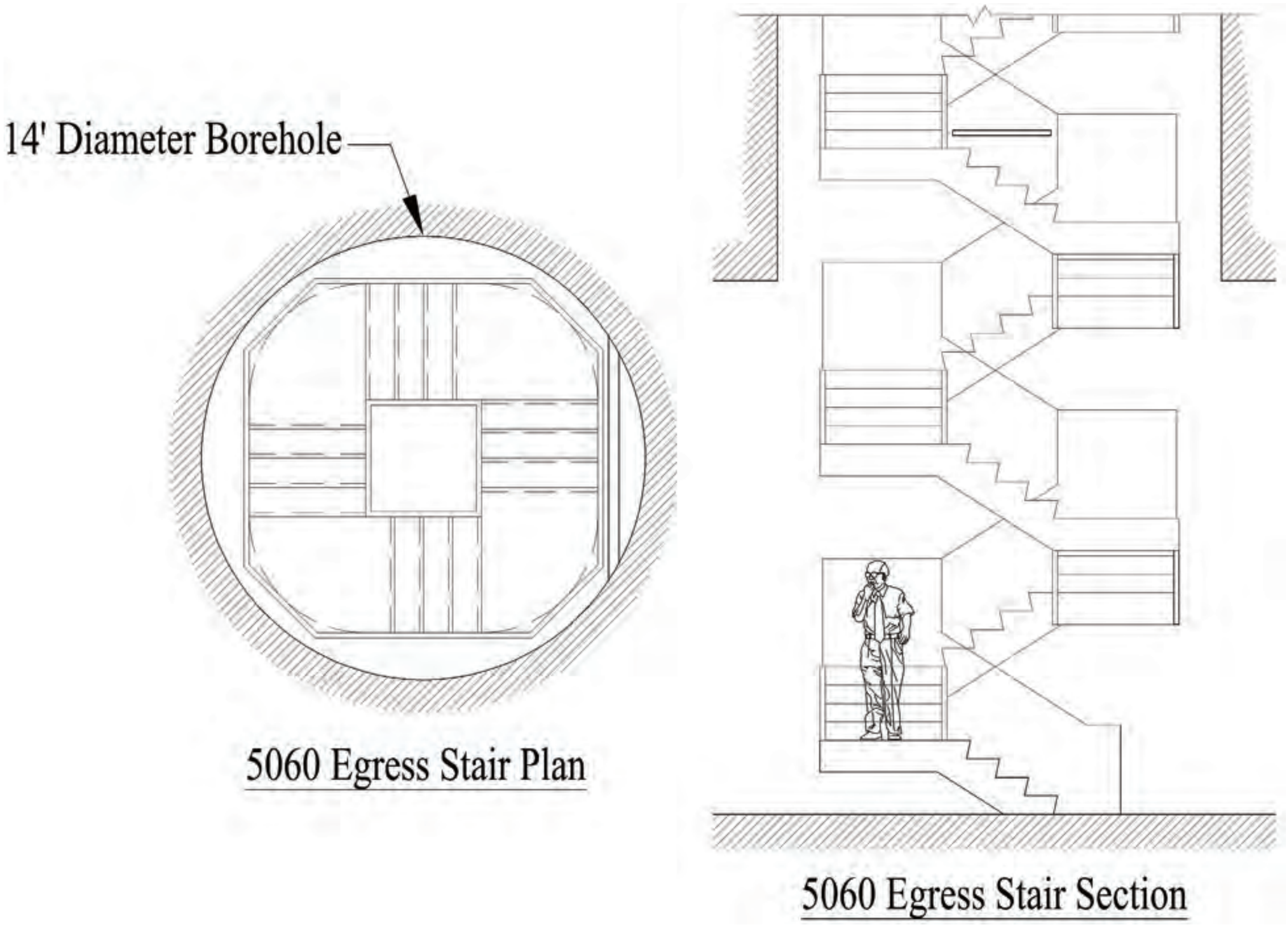

**Figure 5.7.6.5** 4850L to 5060L secondary egress stairway. [Arup]

### 5.7.6.6 Information and Communications Technology (ICT)

The ICT systems supporting LC-1 will consist of the following:

**Structured Cabling System (SCS*)***
The SCS consists of the ICT infrastructure for the LC-1, and will include:

- ICT spaces
- Backbone distribution
- Horizontal distribution

The SCS design will be capable of and will:

- Support various systems in the underground environment, such as wired and wireless LAN, Voice Communications, Radio Communications, Electronic Security, Monitoring, Audio Visual, etc.
- Support connectivity to networks beyond the LAN such as the Internet, Internet2, and Wide Area Networks (WANs)
- Support the business and research requirements of the laboratory staff and other users of the facility

- Support for expansion and upgrades to address future requirements and emerging technologies
- The SCS will have the capability to support 10 Gbps Ethernet in the backbone and 1 Gbps Ethernet to the desktop/equipment outlet

The SCS design will provide an allowance for 25%-50% additional spare capacity in rooms and pathways for future growth and will be based on uniform cable distribution with a dual star topology within the LC-1.

The SCS will provide connectivity to LC-1 and associated access and utility drifts.

Redundant backbones will be installed to the 4850L as part of the UGI design, one in the Yates Shaft and the other in the Ross Shaft. This will provide redundant connectivity for the LC-1. The two backbones will originate at the surface level and end at 4850L in communications distribution rooms (CDRs) located near the Ross and Yates Shafts. The CDRs are the points of interface between the UGI and the UGL. The backbones will extend from the two CDRs to the communications spaces serving the LC-1. A communications room (CR) will be provided in LC-1 access drift. There will also be communications enclosures (CEs) in addition to the CR to ensure compliance with standard requirements for the maximum station (horizontal) cabling length. CEs will also be provided in the AoRs, MEP equipment rooms, and other ancillary/support spaces that cannot be served by the CRs/CEs located in the LC-1. Figure 5.7.6.1 depicts the CRs and enclosures serving LC-1 on the 4850L.

Horizontal (station) distribution will be provided in AoRs, LC-1 access drifts, and other ancillary/support spaces required to support connectivity for telephones, FMS devices, electronic security system (ESS) devices, monitoring devices, and other technology systems. The horizontal distribution in the laboratory facilities for support of the experiments is not included in the facilities portion of the MREFC and will be provided by the experiments.

The DUSEL Facility will provide network service based on a Transmission Control Protocol/Internet Protocol (TCP/IP) standard to a demarcation point established at network switches located in the CRs and CEs. The LBNE will interface to the network by plugging their devices into the switch to access the overall facility network and utilize WANs over the Internet and Internet2. As outlined in Chapter 5.5, *Cyberinfrastructure Systems Design*, the facility side of the network demarcation will be capable of supporting 10 Gbps in the backbone and 1 Gbps to individual devices and 10 Gpbs to devices on an as-needed-basis. The science side of the demarcation will be provided by the LBNE and is not included in the facility design scope.

### 5.7.7 Scope Options and Scope Contingency

In preparation for the LBNE CD-1 review, the LBNE project is evaluating various configuration options for the far site detector at DUSEL. Three configurations have been established for evaluation with the intent to finalize the selection to a single configuration by April 2011.

**Option 1.** One 100 kT WCD on the 4850L and one liquid argon detector on the 800L

**Option 2.** Two 100 kT WCDs on the 4850L and no liquid argon detector

**Option 3.** No WCD and two liquid argon detectors on the 800L

Various fiducial volumes of both detectors are also being considered. Infrastructure to support three 100 kT WCDs was addressed in the 30% and 60% UGL and UGI Preliminary Design reports from Arup.

This was removed between 60% and 90% of Preliminary Design to reflect only what is required to support the experimental spaces included in the scope of work. The impact of each option on the overall DUSEL Facility design is understood and would include the following systems:

1. **Electrical.** A dedicated feeder is included in the design for the single WCD. Removal or addition of a WCD would eliminate or duplicate this feeder. The surface design would also be modified to incorporate the change in load. Liquid argon detector options would be supplied directly from the Oro Hondo Substation.
2. **Water.**
    a. Potable and industrial water supplies to the 4850L are adequate to support any configuration chosen. For liquid argon options, new supplies would be required from the Ross Shaft and design of the water column would be modified to accommodate the additional capacity required.
    b. Purified water is specific to the WCD, but supplied to LM-1 and LM-2 because it is available. If no WCD is installed, purified water for other experiments would be provided by the experiments.
    c. The dewatering system would be impacted in each option based on the requirements to remove leakage water from the WCD.
3. **Ventilation.** The total ventilation requirements for the 4850L assume a single WCD. Changes to this would impact the total flow requirements. In addition, the overhead ventilation drift at the 4700L would follow a different path in any of the three options. The liquid argon options would provide completely isolated ventilation systems with no additional impact on the 4850L facility design.
4. **Excavation.** The excavation requirements for the single WCD are well defined and can be extrapolated for any option. Liquid argon options would be completed separately from the main facility. Modifications to the waste handling system would be required with any option including liquid argon.
5. **Schedule.** Each option considered would require a full schedule analysis to determine the impacts. Items such as conveyance schedule, blasting schedules, and safe access through muck transport paths would have significant changes based on the options considered. Liquid argon options may impact waste handling schedules if done concurrently with development of the 4850L or 7400L.
6. **Shafts.** Shaft design impacts would be limited to the pipe column for purified water and access requirements for either the WCD or liquid argon options.

### 5.7.8 Conclusion

This section presented an overview of the Preliminary Design of the Facility laboratory outfitting specific to the Large Cavity and associated ancillary spaces at the 4850L. The Preliminary Design addresses the current experiment laboratory requirements and provides a solid foundation on which to optimize and build the Final Design. The Preliminary Design is responsive to the needs outlined in the facility requirements to support operation of a safe laboratory that supports the planned DUSEL science agenda as outlined in Volume 3, *Science and Engineering Research Program*.

Additional information describing the underground facility infrastructure design and construction for the infrastructure and other facility components can be found in Chapter 5.4, *Underground Infrastructure Design.*

## 5.8 Deep-Level Laboratory Design at the 7400L (DLL)

This chapter provides an overview of the Conceptual Design of the Facility outfitting for the Laboratory Module (LM) and ancillary spaces at the 7400L. The design includes providing the infrastructure between the main Facility infrastructure and the experiments, along with providing a finished core structure ready for experimental outfitting. Included in this chapter is an overview of the DLL design, beginning with the scope, design requirements, design strategy, and existing conditions. Following this, the core utility systems provided to the LMs are discussed. The descriptions in this chapter are intended to give the reader an overview of the design. Additional details on each design scope, as well as how the infrastructure design interfaces with each design scope, can be found in the reference material noted throughout the chapter.

### 5.8.1 Overview and Planning Summary

This chapter will cover the Deep-Level Laboratory (DLL) Underground Laboratory (UGL) design for Lab Module 1 (LMD-1) at the 7400L. The DLL UGL design includes the facility outfitting and dedicated infrastructure to support the LMD-1. The Arup Underground Infrastructure (UGI) design (Appendix 5.L), discussed in Chapter 5.4, includes the Facility-provided backbone utilities and infrastructure supporting UGL. The UGL interface with UGI and the other scopes of work are discussed in Chapter 5.1. The DLL proposed location is in a currently undeveloped area north of and accessible from the #6 Winze. The 7400L design scope shown in Figures 5.8.1-1 and 5.8.1-2 includes LMD-1 and the drill room. Also included for spatial reference are the mechanical and electrical plant rooms (perpendicular and parallel to LMD-1), existing #6 Winze, new #8 Winze, and the new vent raise to the 4850L.

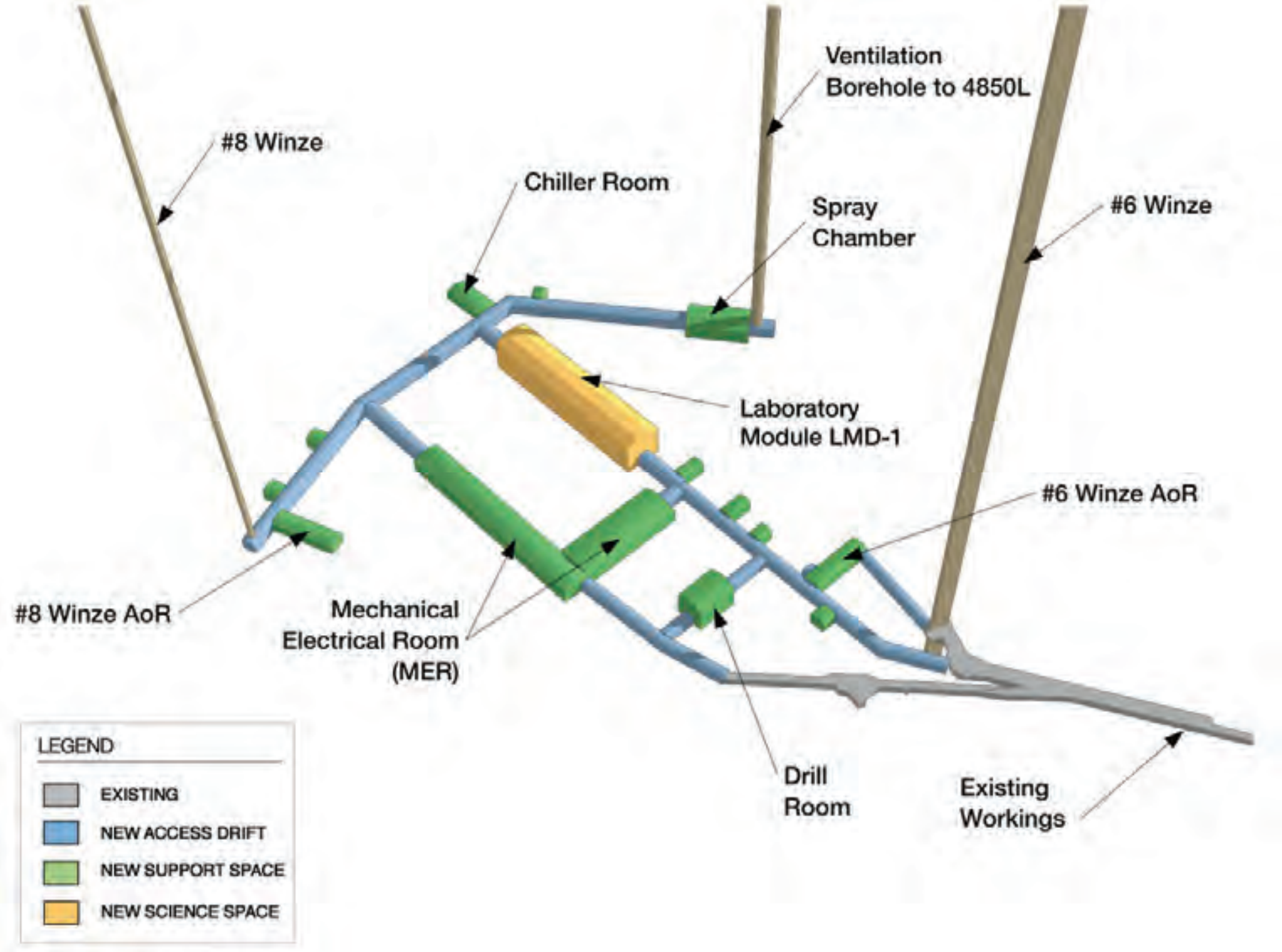

**Figure 5.8.1-1** Conceptual Design 7400L layout. The new #8 Winze will provide secondary egress from the 7400L and the new ventilation borehole will provide exhaust ventilation from the 7400L. [DKA]

Figure 5.8.1-2 represents areas are included in the UGL and the UGI scopes of work and the associated demarcation between the two design scopes in relationship to the LMD-1.

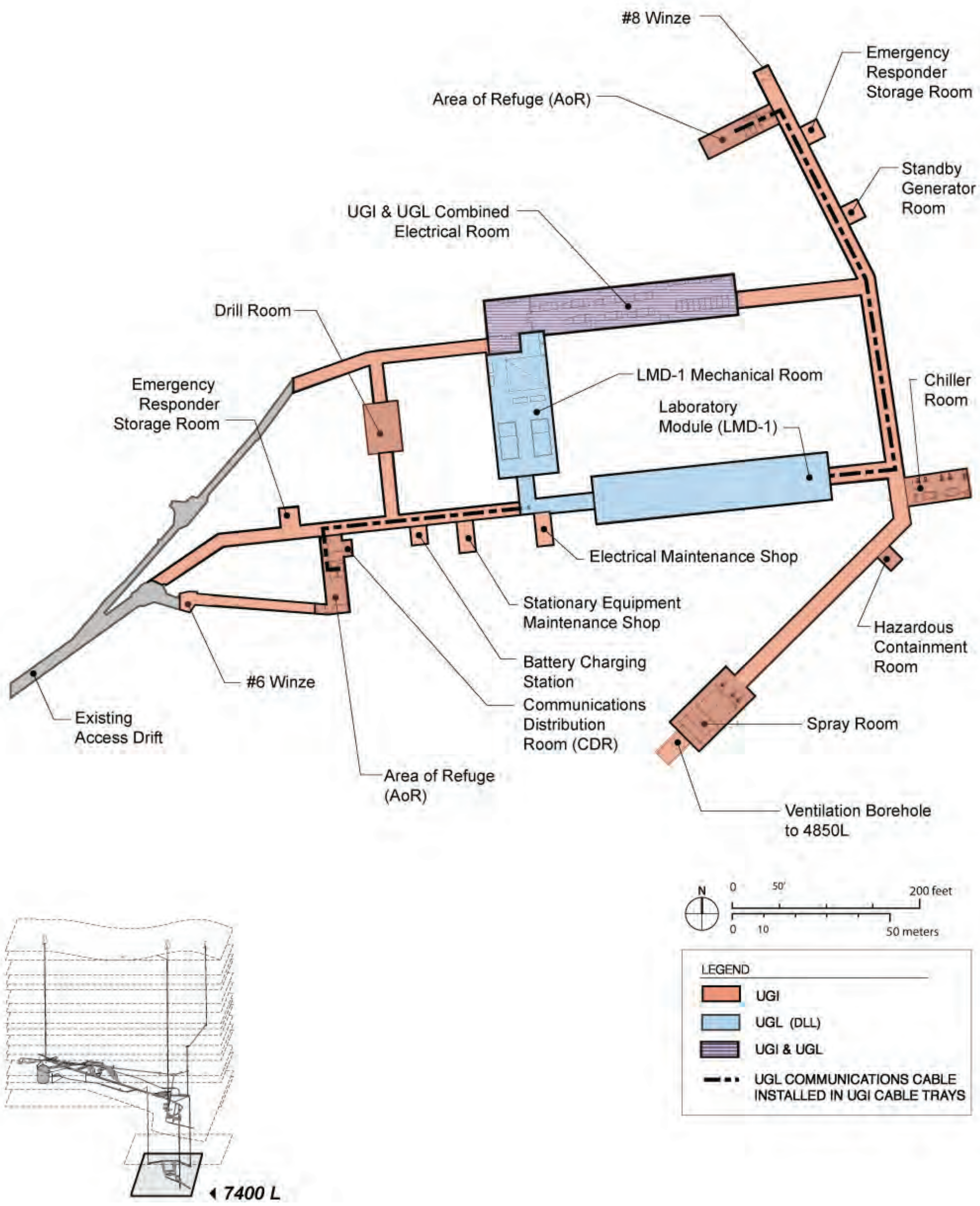

**Figure 5.8.1-2** 7400L area representation of UGL and UGI scopes of work pertaining to the LMD-1. The UGI scope of work includes the Facility-provided backbone utilities and infrastructure supporting UGL. [DKA]

As documented in Trade Study #385 (Appendix 9.M), the plans for the 7400L to 4850L secondary egress ramp and the #31 Exhaust Shaft were eliminated. These were replaced with the addition of a secondary egress winze (#8 Winze) and a new separate exhaust raise, which are included in the UGI design.

The 7400L has been underwater since 2003, and as of September 30, 2010, the water level in the DUSEL Facility was approximately 5,150 feet below ground level. Relatively limited design information is currently available and no geotechnical site investigations have been completed in the area of planned developments due to the lack of access to the 7400L, although mining was conducted on that level to south in different geological units. The DUSEL Project and South Dakota Science and Technology Authority (SDSTA) have implemented dewatering system improvements to accelerate water removal because the DLL design progression is directly related to gaining access to the 7400L. The current expectation is that the underground laboratory will be dewatered to the 7400L by mid-2013.

Thus, the design for the 7400L Campus is at a Conceptual level based on current 7400L information available. The excavation design at the 7400L relies on extrapolation of geotechnical information obtained for work on the 4850L and on core logs from Homestake Mining Company. Verifying the extrapolations will require confirming a number of assumptions that have been made once the 7400L becomes available for re-entry. Once the 7400L is dewatered and rehabilitated to a point where re-entry is safe and feasible, extensive geotechnical surveys will be completed.

### 5.8.2 Experiment Requirements for the LMD-1

Detailed requirements for the LMD-1 have been developed, reviewed, and approved. A thorough discussion of the requirement structure and the development process can be found in Volume 9, *Systems Engineering*. The requirements listed in the PDR summarize the formal requirement sets included as Appendix 9.F. The detailed method for extracting user requirements from the science collaborations and turning them into a cohesive set of LM requirements is described in Chapter 3.6, *Integrated Suite of Experiments (ISE) Requirements Process*. The key driving requirements from science that emerged from this process are listed in Chapters 3.6, 3.7, and 3.8.The functionality and performance mandated by this set of requirements form the basis for the DUSEL Facility design as presented throughout Volume 5, *Facility Preliminary Design*. All key driving requirements of the design have been either met or otherwise addressed specifically by this PDR.

The ISE has defined the requirements for the DLL, and these requirements have driven the DLL design. The ISE program is not fully developed; thus, the DLL is guided by the requirements of the generic ISE and will be refined in Preliminary and Final Design as ISE requirements mature.

The design of LMD-1 is configured to house two experiments, a combination of solid-state dark-matter and/or neutrinoless double-beta decay experiments. Alternatively, one of the dark-matter experiments may be a bubble chamber detector. Currently, four experiments have requested LMD-1 installation space, and these experiment requests were used to establish the initial set of requirements. The four experiments include Neutrinoless Double Beta Decay (EXO and 1TGe) and Dark Matter (GEODM and COUPP). LMD-1 will have many of the same requirements and features as the 4850L Mid-Level Laboratory (MLL) Modules. Figure 5.8.2-1 provides an example test-fit of two neutrinoless double-beta experiments occupying LMD-1.

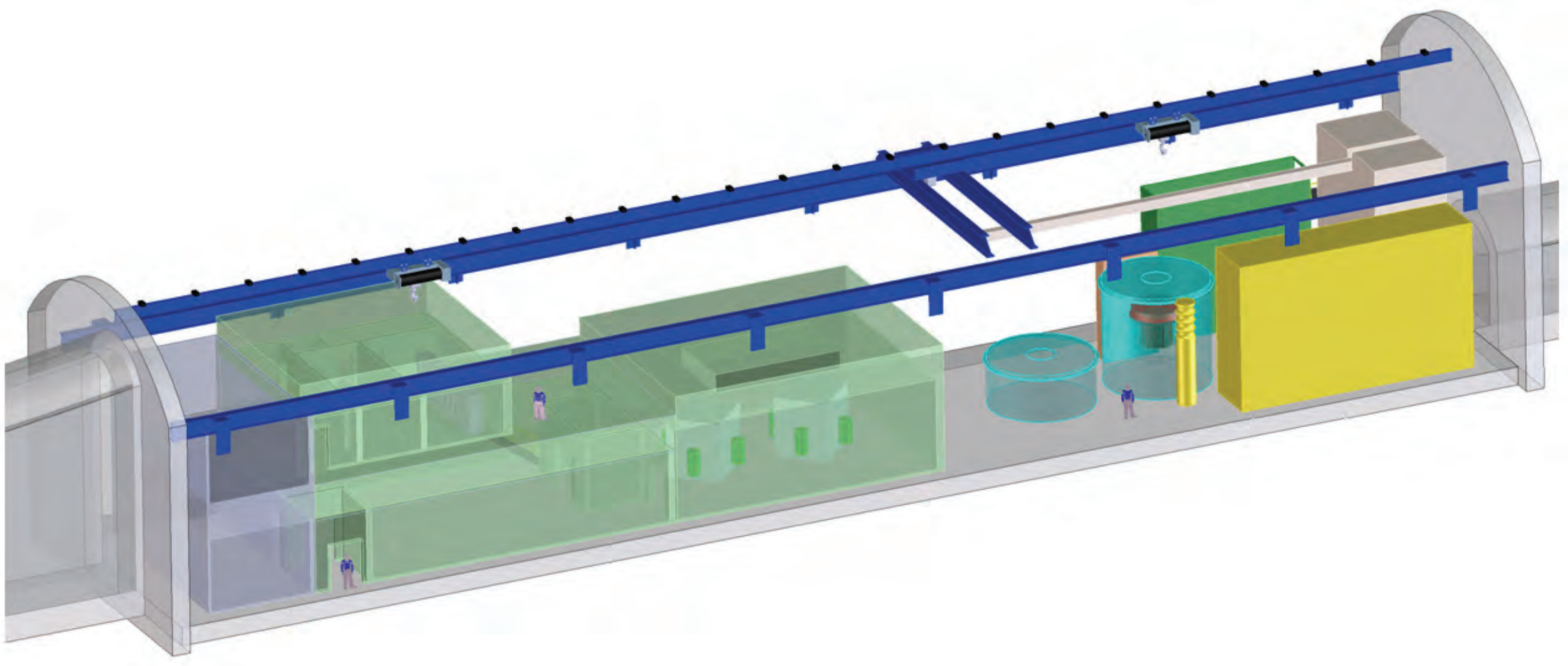

**Figure 5.8.2-1** Two typical example experiments in LMD-1. [Dave Plate, DUSEL]

The planned LMD-1 has a smaller section than LM-1 and LM-2 on the 4850L—49.2 ft (15 m) wide and 49.2 ft (15 m) high. The length of LMD-1 is 246.1 feet (75 m). Figure 5.8.2-2 shows a cross-sectional view of the LMD-1 depicting both the bridge and monorail cranes along with the shaded envelope designated for experiment use.

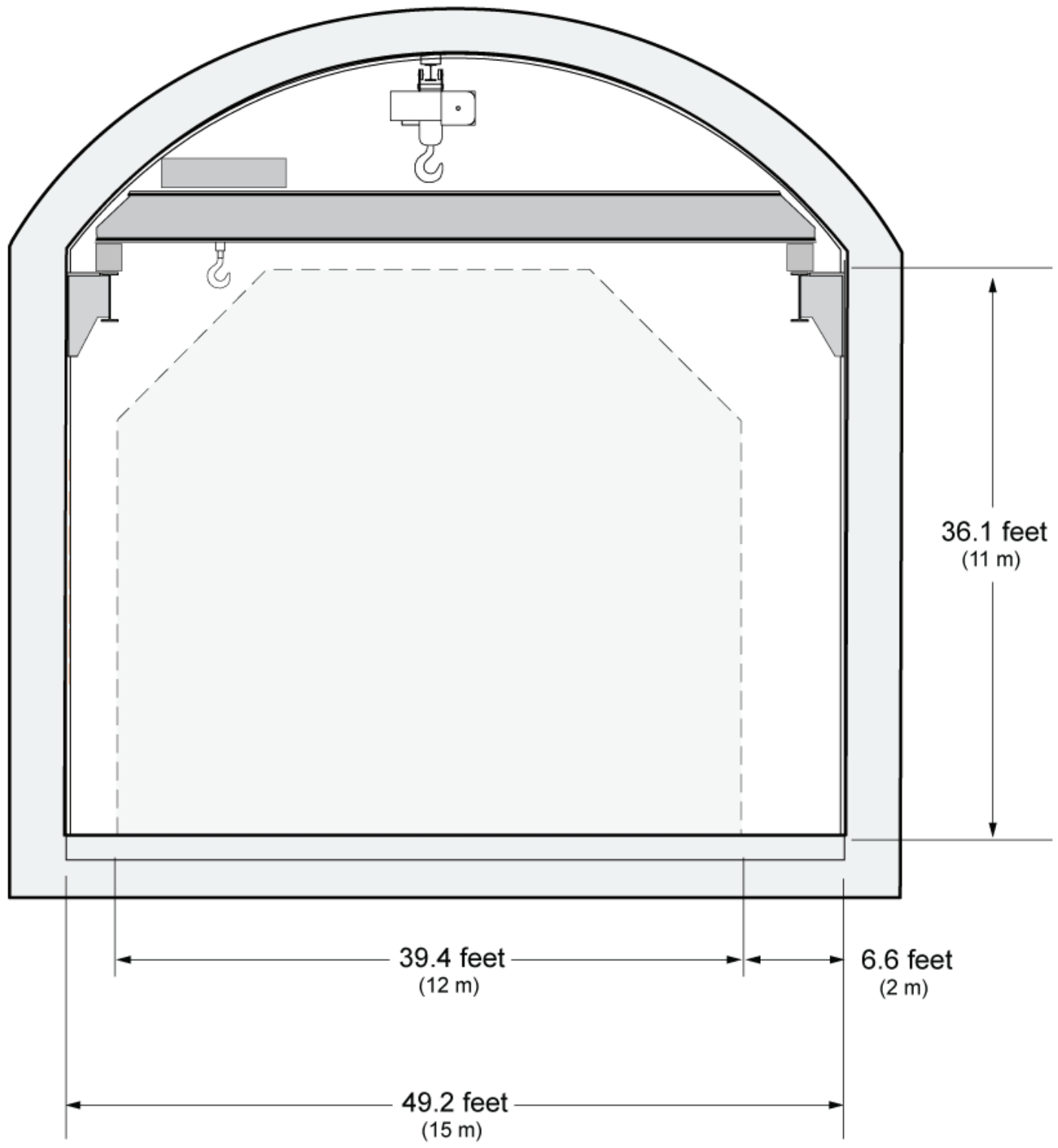

**Figure 5.8.2-2** LMD-1 cross section and experiment envelope. [DKA]

The following are the key experiment requirements driving the LMD-1 design:

- Space – as discussed above
- Power—normal (650 kW) and standby (100 kW)
- Chilled Water (650 kW)
- 20-Ton Bridge Crane
- 40-Ton Monorail Crane
- Network Communications/IT (10 Gb/s)
- Exhaust Ventilation for cryogen release and smoke (100,000 cfm, 170,000 $m^3$/h)
- Fresh air supply at one air exchange per hour (10,000 cfm, 17,000 $m^3$/h)
- Fire Sprinkler/Water Mist Protection
- Potable and Industrial Water
- Environmental/Humidity Control (68 to 77 °F/20 to 25 °C, RH 20 to 50%)

Please refer to Chapter 5.4, *Underground Infrastructure (UGI) Design*, for a detailed description of facility infrastructure supporting the construction of LMD-1.

### 5.8.3 Underground Laboratory at the 7400L Design Strategy

The design of the DLL has incorporated many of the same aspects from the MLL design. The MLL design interface is critical because the MLL and DLL share many of the same requirements and services. Maintainability and serviceability are improved by ensuring that the designs and interfaces are common for both the MLL and DLL laboratory spaces. The MLL and DLL also share much of the same backbone utility services and infrastructure provided through the UGI design. The main difference, beyond laboratory sizing, between the LMs in the MLL and DLL is the floor elevation of the LMD-1 is at the same elevation as the 7400L main access. This is different than MLL LM-1 floor elevations, which are recessed 13.1 ft (4 m) lower than the main access drifts designed to contain water or cryogen release. As the science requirements mature, the Final Design will likely include a similar recessed floor as in the MLL LMs to better accommodate a wider range of experiments.

Arup's *UGL 7400L, Conceptual Design Report* (Appendix 5.S), Page 1, describes the scope split into two phases of construction. Phase 1 consists of a bare-bones scope that allows for drill room operations only. Phase 2 consists of the fully expanded scope to include the UGL and associated ancillary spaces in addition to Deep Drilling. The bare-bones option, as outlined in Phase 1, of establishing the drill room and then the LMD-1 at a later date, is no longer being considered. These two distinct phases have been combined into a single excavation and outfitting approach for the 7400L. However, the excavation sequence does still follow a phased approach to establish the ventilation and the secondary egress route as soon as possible for safety and efficient construction.

Driven by the dewatering schedule, the DLL occupancy schedule lags that of the MLL. Several critical interfaces, such as the interface with UGI and the corresponding backbone utilities, are continually monitored and fostered to ensure the successful development and eventual buildout of the DLL. Overall management of the design process and core integration is discussed further in Chapter 5.1, *Facility Design Overview*.

Nearly all the DLL design incorporates new excavations. In addition to the space provided for the DLL, the excavation design includes the concrete floors and shotcrete applied to walls and ceilings. The excavation design contractor has performed preliminary geotechnical modeling of excavation utilizing extrapolation of the MLL rock characterization. The Conceptual Design has considered rock support and rock removal methods to address the additional stresses expected at the 7400L. The geotechnical analysis is discussed in Chapter 5.3.

The drill room, located in the DLL Campus, is intended to be used for drilling to levels below that where any biological life has ever been identified. The drill room is included in the Other Levels and Ramps (OLR) scope. Utility services will be provided to the OLR from the main electrical room, as shown in Arup's *UGI 7400L, Conceptual Design Report* (Appendix 5.T). Many other UGI services provided to the 7400L will be shared between the LMD-1 and the drill room, including shaft access, water, ventilation, communications, and Areas of Refuge (AoRs). In addition to the drill room, there are other experiments planned for the DLL campus. These include biology, geology, and engineering (BGE) experiments to be located in the drill room and existing drifts. The DLL requires high-speed data connectivity and redundancy for experimental needs and also requires that DUSEL fire control and safety systems monitor the underground systems and provide basic control for remote operation of underground equipment from the surface.

The DUSEL UGL design will provide the following utilities to LMD-1 via connections to the UGI facility provided systems: ventilation, chilled water, air conditioning, Information and Communications Technology (ICT), electrical, plumbing, and fire protection.

### 5.8.4 Excavation Design

The DLL Campus is anticipated to lie mainly within the Yates Member of the Poorman Formation. The Yates Member is overlain by phyllite of the Poorman Formation, which is exposed adjacent to and south of the #6 Winze. The Poorman Formation may be intruded by Tertiary-age rhyolite dikes in this area, similar to what is found on the 4850L although, perhaps, to a lesser extent. The detailed discussion on the known geology of the 4850L (from which the 7400L is extrapolated) is included in Chapter 5.3, *Geotechnical Site Investigations and Analysis*.

#### Excavation Summary

The initial site characterization and design were based on:

- Extrapolation of data from geotechnical/geological investigations at the 4850L
- Historic data, mostly acquired from the original Homestake Mine
- Professional judgment resulting from past experience with similar projects

While the information included within Parts A and B of the Golder Associates *7400L Conceptual Design Report* (Appendix 5.U) is deemed suitable for the current level of design, further investigation is required prior to advancing the design to higher levels of refinement.

Geology of the DUSEL site is complex in terms of stratigraphy, tectonics, and intrusive alterations. At the scale of the 7400L, four main lithologic zones plus intrusives have been inferred to be present. These zones include:

- Zone I: Yates Unit Amphibolites
- Zone II: Poorman Formation Phyllites
- Zone III: Homestake and Ellison Formations

Of the above units, the Zone I Yates Amphibolites are considered most suitable for excavation, based on this extrapolation of geotechnical data from the 4850L.

Although the quality of rocks that will host the campus is expected to be good, the magnitude of in situ stress at the 7400L is expected to be 45% higher than the corresponding stress measured on the 4850L based upon extrapolation of stresses from the 4850L. This elevated in situ stress suggests that significantly different behavior can be expected between geologically comparable rock masses at the 4850L and 7400L.

Considering the relatively high in situ stress and lack of location-specific data, the decision was made to utilize the Yates Member as the primary location for the DLL. Two layouts and orientations of excavations for the DLL are proposed in the Golder Associates *7400L Conceptual Design Report* (Appendix 5.U), Pages ES-1 and ES-2.

- The Baseline Layout, which is not considered optimal from a geotechnical standpoint due to the oblique angle between the long axis of the major excavations and inferred orientation of local geological structures
- An Alternative Layout, with the orientation of the long axes of the major excavations perpendicular to the inferred local geologic structures, improving the stability of openings. Although this layout is recommended and may be chosen after additional geotechnical studies are performed, the Baseline Layout was chosen in this report for scheduling and costing purposes to best support integration with other design scopes.

The inventory of excavations at the DLL shows the finished floor area of all excavations is 81,333 square feet (7,560 $m^2$). The nominal excavation volume is 71,459 cubic yards (54,676 $m^3$). The total volume of waste rock generated will be 119,116 cubic yards (91,140 $m^3$), taking into consideration potential assumed 10% overbreak and a 150% swell factor.

The excavation development of the 7400L is planned in two phases:

- Excavation Phase 1 includes the major ventilation elements, secondary egress drift, the drill room, and AoRs.
- Excavation Phase 2 includes the remainder of the 7400L Campus, including enlargement of the Laboratory Module (LMD-1) and associated plant room to their final dimensions.

The geologic and geotechnical data suggest that the excavation of the DLL will be complicated by high stress conditions causing ductile deformation in the weaker rocks and spalling or strain bursting in the stronger rocks. This is not unusual for excavations at great depth but does require specific techniques be employed to produce stable excavations. The Homestake Mining Company's (HMC) records document rock bursts and microseismic events at depths similar to that of 7400L Campus. High in situ stress dictates the need for de-stress blasting techniques to be applied as discussed in Golder Associates *7400L Conceptual Design Report* (Appendix 5.U), Pages 17-19, in the more competent zones and yielding support in the more deformation-prone zones.

Excavation for Phase 1 development of the DLL is further subdivided into three sections based on ventilation available for construction equipment. The excavation of Phase 2 structures is not expected to be limited by ventilation. Figures 5.8.4-1 and 5.8.4-2 present the phased excavation plan for the 7400L Campus and 4850L support facilities, including the Phase 1 subdivisions.

Development of the 4850L facility supply ventilation and secondary egress for the DLL is planned in one phase, concurrent with the MLL and the Large Cavity excavation.

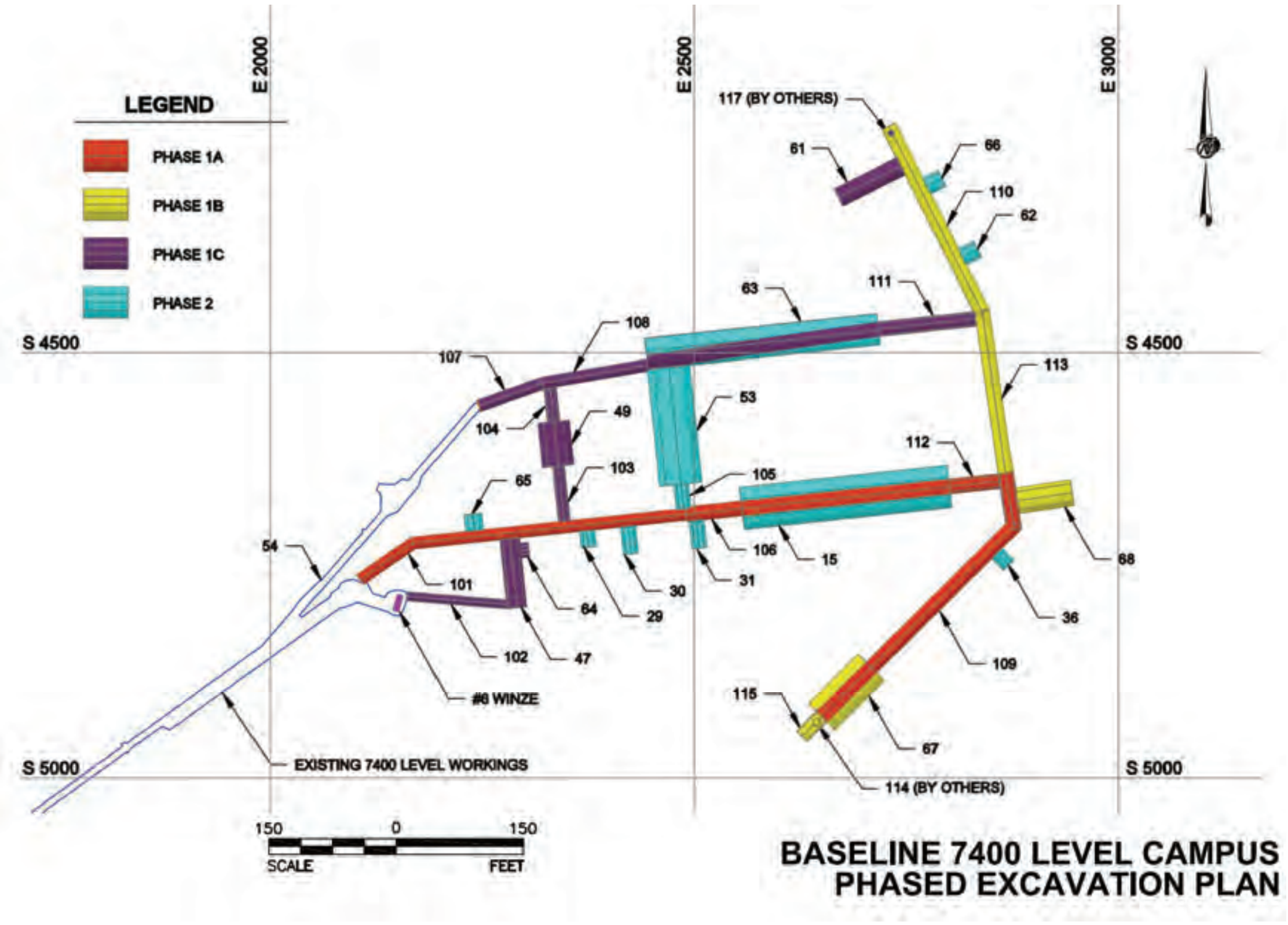

**Figure 5.8.4-1** Phased excavation plan at the 7400L. [Golder Associates]

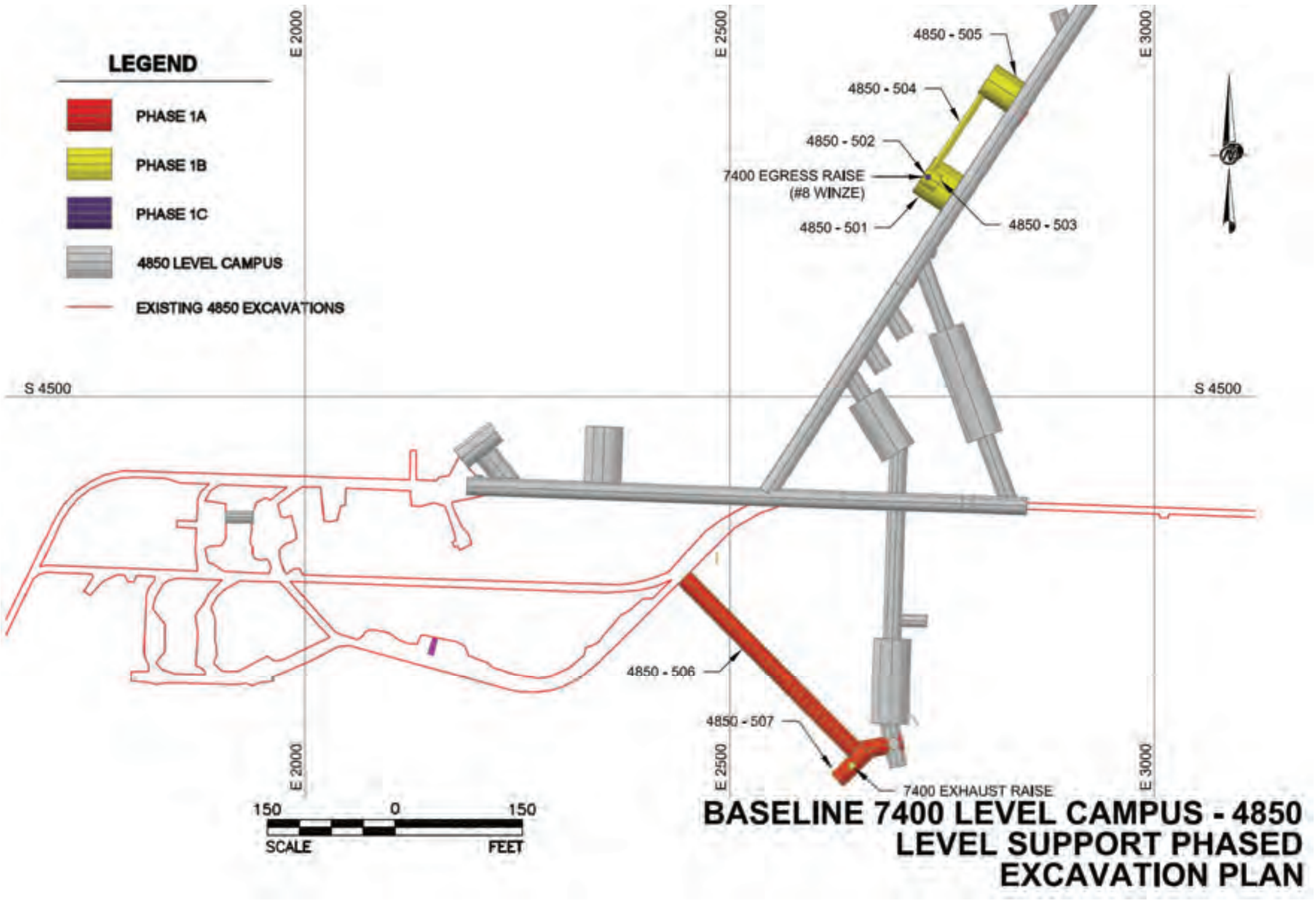

**Figure 5.8.4-2** Phased excavation plan at the 4850L to support the 7400L. [Golder Associates]

### 5.8.5 Conceptual Design for Core Utility Systems

This section comprises a summary-level discussion of the key core utility systems included in the Conceptual Design for the LMD-1. Figure 5.8.5 represents a general layout of the core utility systems planned the LMD-1 that are further described in this section.

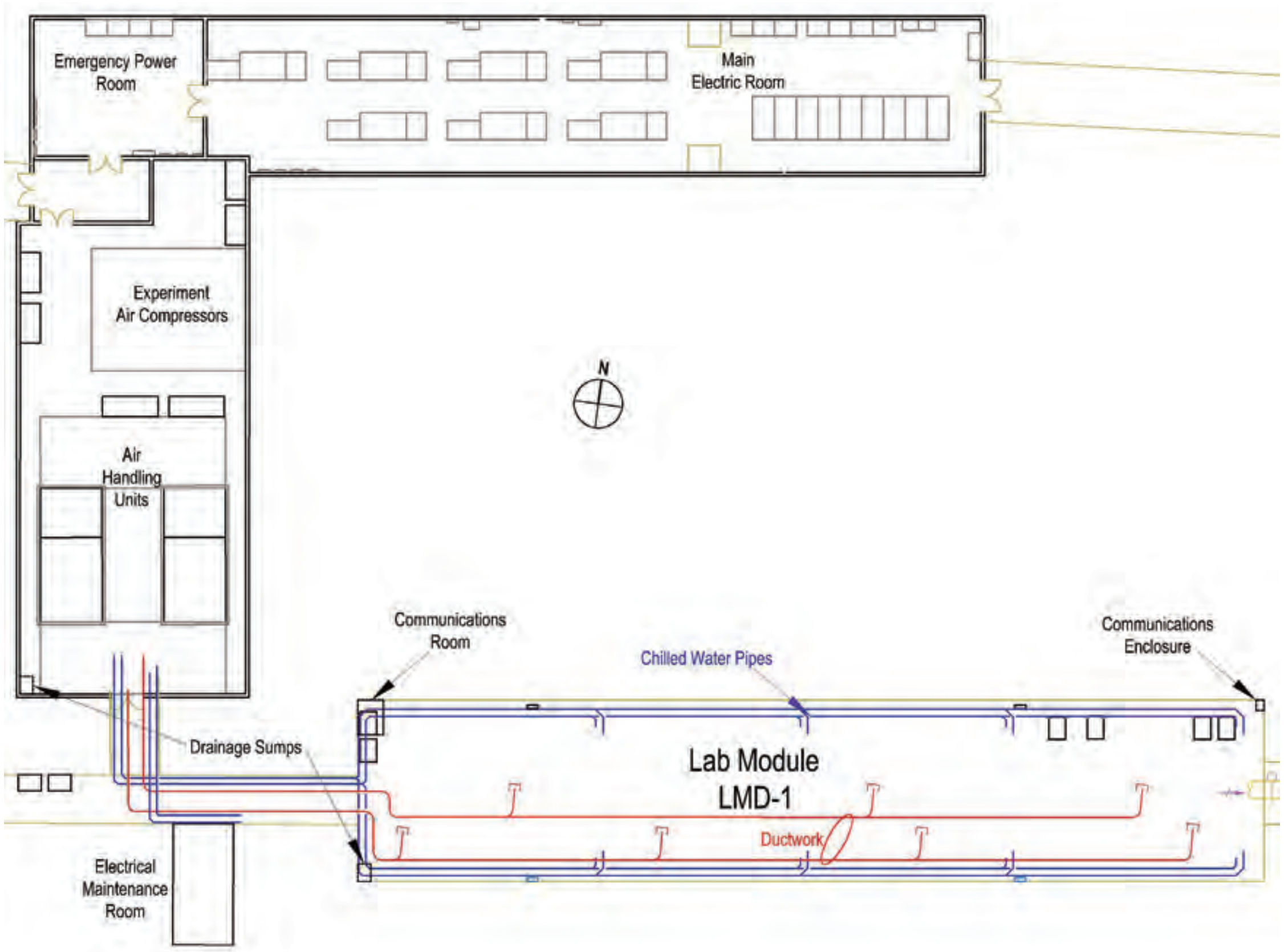

**Figure 5.8.5** 7400L LMD-1 depicting the core utility systems along with mechanical and electrical rooms. [Joshua Willhite, DUSEL]

#### 5.8.5.1 Ventilation, Chilled Water, and Air Conditioning Systems

The DUSEL Facility ventilation system described in Section 5.4.3.8.3, *Air Quality and Ventilation Preliminary Design*, will be used to supply fresh air and exhaust at the 7400L. The primary ventilation source of air for the 7400L will be via the Ross Shaft to the 4850L, then through the #6 Winze to the 7400L. The air will then exit out a new borehole from the 7400L to the 4850L main facility exhaust system. The #8 Winze will also provide fresh air to the level to ensure egress through a ventilation intake.

LMD-1 ventilation will be provided through air-handling units (AHUs) drawing supply air directly from the access drifts. This air will be filtered to minimum efficiency reporting values (MERV) 15 standards. The HVAC system design, provided as part of the UGL facility-level systems for the LMD-1, will include air-handling equipment and main ductwork located inside LMD-1. The make-up air supply rate is designed to be one air change per hour and the remainder of the air required for cooling will be provided by recirculating air within LMD-1. Filtration equipment for areas requiring air cleanliness above MERV 15 will be provided as part of the fit-out by the experiment collaborations.

Dedicated exhaust ducting from LMD-1 will be provided for normal and hazardous exhaust in the experiment areas for hazards such as cryogens or smoke.

Ventilation air to AoRs will also be provided from the UGI main facility ventilation system under normal conditions. Also included in the UGI design are two redundant sources of breathable air—compressed air and bottled oxygen—which are provided in the event that the ventilation system is contaminated or inoperable. Motorized dampers will be provided to control the flow of ventilation. The mechanical and electrical room (MER) will be positioned to allow direct access to the supply air drifts and exhausted into the exhaust drifts.

Specialty equipment rooms, such as the communications distribution room (CDR), will be outfitted with dedicated fan coil units for additional cooling.

The AHUs will include supply and return fans controlled by variable frequency drives, chilled-water cooling coils, mixing dampers, pre-filters, and final filters to provide filtration.

Trade Study #380 (Appendix 9.V), evaluated during design by the DUSEL Project Team, determined that the most economical method for cooling the 7400L facility was through the use of an underground chilled-water system. Based on this Trade Study, a new chilled-water system and spray chamber are planned for installation near the ventilation borehole on the 7400L as part of the UGI *Conceptual Design Report* (Appendix 5.T).

Additional information concerning the 7400L ventilation, chilled-water system, and air conditioning can be found in Arup USA, Inc., *DUSEL—Deep Underground Science and Engineering Laboratory, UGL 7400L Conceptual Design Report, FINAL—REV2* (Appendix 5.S), Pages 17-20, November 2010.

This system will remove heat loads from surrounding or ambient heat loads, mobile equipment, personnel, experiments, and electrical equipment. A circulation system with insulated piping will allow the chilled water to capture heat loads either through AHUs or by directly cooling equipment.

Further processing of the chilled water for experiment cold storage and other specialized laboratory requirements, if needed, will be completed by the experiment collaborations. These experiment-provided systems would be located in the LMD-1 mechanical and electrical room (MER).

### 5.8.5.2 Information and Communications Technology (ICT)

The ICT systems support the DLL in meeting various requirements, including voice and data communications for DLL operations and science data communications. ICT systems provide the communications backbone for plant monitoring and control as well. The following outlines the ICT system components.

**Structured Cabling System (SCS)**

The SCS consists of the ICT infrastructure for the underground LM, and will include:

- ICT spaces
- Backbone distribution
- Horizontal distribution

The SCS design will be capable of and include:

- Support of various systems in the underground environment, such as wired and wireless Local Area Network (LAN), Voice Communications, Radio Communications, Electronic Security, Monitoring, Audiovisual, etc.
- Support for connectivity to networks beyond the LAN such as the Internet, Internet2, and Wide Area Networks (WANs)
- Support for the business and research requirements of the laboratory staff and other users of the facility
- Support for opening day and emerging technologies
- Capability to support 10 Gigabit Ethernet in the backbone and 1 Gigabit Ethernet to the desktop/equipment outlet

Redundant backbones will be installed from 4850L as part of the UGI design (Appendix 5.T), one in the #6 Winze and the other in the #8 Winze for egress. This will provide redundant connectivity for laboratory facilities. Redundant backbone extension cables will be provided from the 7400L CDR to LMD-1. Horizontal distribution in LMD-1 will be provided as part of the experiment installation.

Additional information concerning the 7400L IT/Communications can be found in Arup USA Inc., *DUSEL—Deep Underground Science and Engineering Laboratory, UGL 7400L, Conceptual Design Report, FINAL*—REV2 (Appendix 5.S), Pages 24-28, November 2010.

#### 5.8.5.3 Electrical

The power for the LMD-1—as shown in the Arup USA Inc., *DUSEL—Deep Underground Science and Engineering Laboratory, UGI 7400L, Conceptual Design Report, FINAL—REV2* (Appendix 5.T), Pages 28-33, November 2010—includes a dedicated, redundant 12 kV feeder from the Ross and Yates surface substations. The redundant 12 kV feeder from the surface to the 4850L was removed through the Value Engineering process (UGI VE#55; Appendix 9.AC) as a cost reduction. The 100% Preliminary Design cost and schedule include a dedicated single 12 kV feeder originating at the surface and routed down only the Yates Shaft and the #6 Winze. The design of the electrical feeder from the 4850L to the 7400L will include a single system, including that needed for deep drilling (OLR experiment), facility infrastructure, and experiments. Complete redundancy is provided for standby power by standby generators located on the 7400L. As with the 4850L, redundant normal power to the 7400L was eliminated through the Value Engineering process (UGI VE#175; Appendix 9.AC) to reduce project cost.

Once at the 7400L, 12 kV power will be transformed to 480 V at the electrical room, shown in Figure 5.8.1-2, and distributed to the LMD-1 and the drill room. Standby power will provide egress lighting and power to support laboratory equipment critical to maintaining experiment integrity or prevent release of hazardous materials. Fire, life, and safety system equipment will be fed by emergency uninterruptable power sources (UPSs) that are connected to the standby power system.

The LMD-1 will have its own dedicated electrical switchgear room within the Main Electrical Plant Room to accommodate medium-voltage gear, step-down transformers, and panel boards.

The LMD-1 will also be outfitted with electrical receptacles for general use, LED lighting, and fire alarm devices.

Electrical service will be provided to the following equipment specific to the LMD-1:

- 20-ton bridge crane and 40-ton monorail crane
- 480 V connections for experiment-specific power
- Air compressor (for experiments)
- LMD-1 lighting and utility power
- Lights and receptacles
- Air handling units
- Communication enclosures
- MER utilities
- Sump pumps
- Supply and exhaust fans

The following systems will be provided with standby power including, but not limited to:

- Mechanical air-handling systems and smoke-control systems for all refuge areas
- Standby lighting required for refuge areas and smoke-control mechanical equipment rooms
- Two-way communication

In addition, LM standby power is provided for loads where "damage to the product or process" could result from a loss of power, e.g., cryogen pumps and orderly shutdown of science-related equipment.

Emergency power definition is based on NFPA 520 and is limited to the following systems:

- Fire detection systems
- Fire alarm systems
- Exit sign illumination
- Emergency lighting

Additional information concerning the 7400L electrical system can be found in Arup USA Inc., *DUSEL—Deep Underground Science and Engineering Laboratory, UGL 7400L, Conceptual Design Report, FINAL—REV2* (Appendix 5.S), Pages 20-22, November 2010.

#### 5.8.5.4 Plumbing

The LMD-1 plumbing systems will provide potable water, industrial water, fire sprinkler piping, water mist, and sump pump discharge. Both potable and industrial water are supplied by the city of Lead, South Dakota. Potable water will be used for domestic consumption while industrial water will be available for construction, mechanical, and experiment applications.

Both the potable and industrial water lines will have isolation valves, will be capped, and will be available for connection during experiment installation.

Instrument-grade compressed air will be supplied to the LMD-1 from an air compressor in the LMD-1 MER.

The LMD-1 floor will be sloped at 1% to one corner, where the sump pump will be located.

Additionally, there will be an 18–inch-wide, 6–inch-deep trench running along the wall of the module. This trench is in place to prevent hazards from spreading out of the module. The LM and MER will have open sumps used to collect water, leaks, and spills that occur in the rooms. The sump from LMD-1 will discharge into a sump pit in the MER. The discharge line from MER sump will discharge, after it is tested, into the main facility drainage system.

Additional information concerning the 7400L plumbing can be found in Arup USA Inc., *DUSEL—Deep Underground Science and Engineering Laboratory, UGL 7400L, Conceptual Design Report, FINAL—REV2* (Appendix 5.S), Page 22-23, November 2010.

#### 5.8.5.5 Fire Protection

Fire protection will be provided to the LMD-1 by facility fire mains, a system that consists of a combined standpipe and automatic wet sprinkler systems. The standpipe system will have hose outlets at each entrance to the LMD-1 and MER.

There will be separate fire sprinkler zones for the LMD-1 and MER. Sprinkler systems in each zone include automatic sprinklers, control valves, drain / test valves, water flow switches, and tamper switches.

In addition, the LMD-1 will be provided with a capped connection to the water mist system, to be fully installed with future experiment fit-out. The water mist system will only provide fire suppression for the LMD-1 communication room as part of the DUSEL MREFC-funded Construction. The water mist system was selected for two primary reasons: 1) the water mist system decreases potential damage to equipment in the case of false alarms or noncatastrophic events; and 2) because it is not a chemical fire suppressant, it is considered to be a more environmentally conscious choice. Supplemental fire suppression capability will be provided to the LM by means of remote water monitor nozzles. The installation of the remote monitor nozzles will be done as part of experimental installation.

Refuge areas provide a protected environment for occupants during an emergency event, such as a fire or cryogen leak. The Areas of Refuge (AoRs) are located at each of the exit points from the 7400L, the #6 Winze and the #8 Winze. Refer to Section 5.4.3.1.3, *Life Safety Preliminary Design*, for a detailed description of these AoRs. In general, the AoRs are sized for a total anticipated occupancy of 85 on the 7400L. Services provided include water, restroom, air source, and communications. The DLL AoRs are patterned similar to those that will be found at the MLL.

### 5.8.6 Laboratory Expansion Options

While the DLL is at a Conceptual level and includes only a single LMD-1, expansion options for additional LMD-1s are being considered, as shown in Figure 5.8.6 and discussed in Golder Associates, *7400L Conceptual Design Report, Part B—Conceptual Design of Excavations, Contract D10-04, Engineering and Design Services for Excavation—DUSEL* (Appendix 5.U), Section 3.2, Page 15, October 1, 2010. As the design progresses, consideration will be given to the impacts caused by materials and personnel movements on ongoing experiments if additional LMs were constructed. Consideration will also be given to egress routes during construction of the additional LM in the DLL.

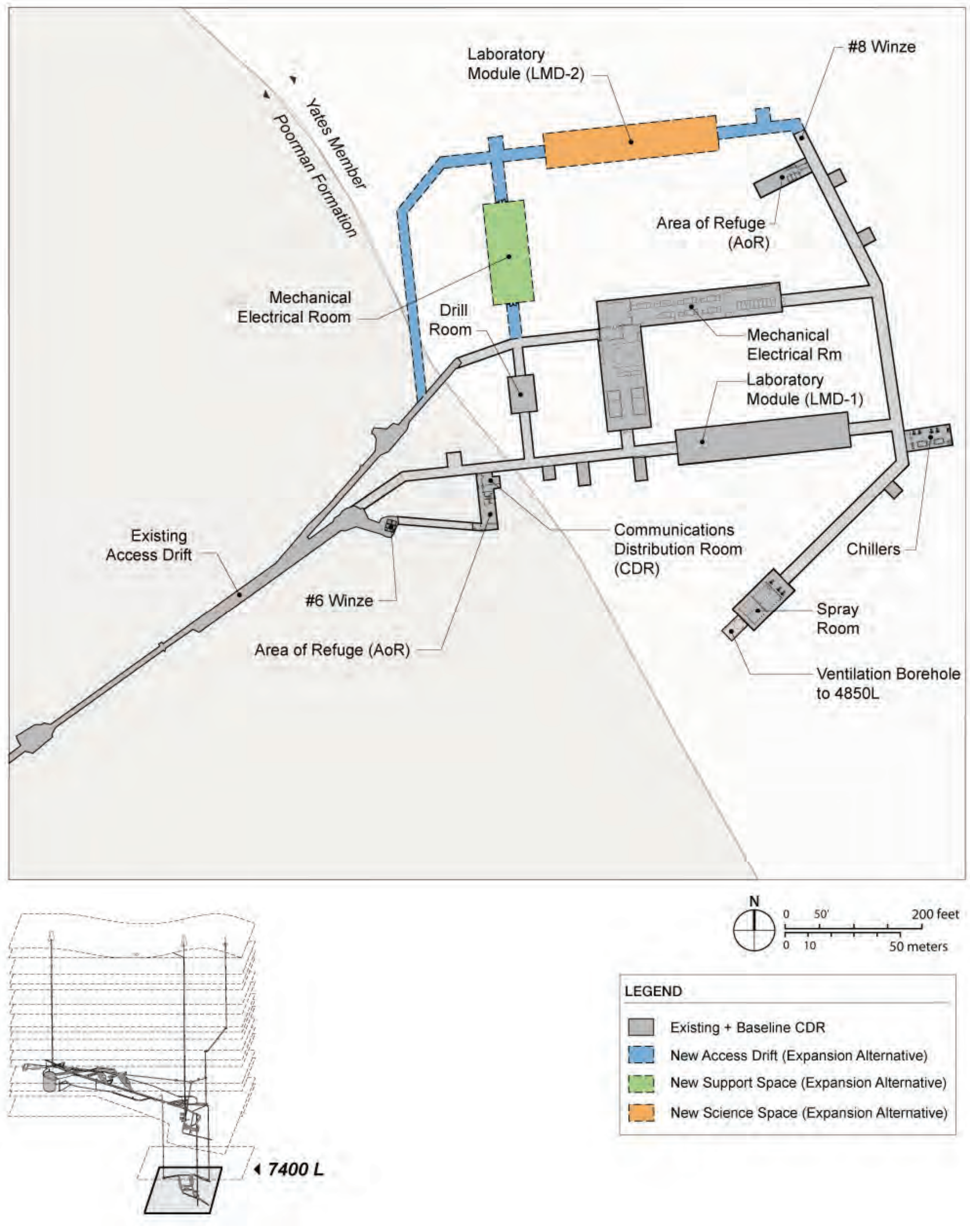

**Figure 5.8.6** 7400L alternative campus expansion. [DKA]

Although this area is currently inaccessible due to being underwater, preliminary evaluation of the geological and geotechnical setting indicates expansion of additional LMs may be permitted. Due to the science collaborations' request for the LMD-1 to be larger to accommodate a variety of experiments, DUSEL completed Trade Study #408 (Appendix 9.R) to analyze the additional cost of increasing the size of the excavations and supporting infrastructure. Section 3.8.5 provides further discussion on Trade Study information completed on the LMD-1. The permissible LM size will be determined after accessing the 7400L and completing the geotechnical site investigations.

### 5.8.7 Conclusion

This section presented an overview of the Conceptual Design of the facility laboratory outfitting specific to the 7400 LMD-1 and ancillary spaces. The Conceptual Design addresses the current laboratory requirements and provides a solid foundation on which to optimize and build the Preliminary and Final Designs. The Conceptual Design is responsive to the needs outlined in the facility requirements to support operation of a safe laboratory that supports the planned DUSEL science agenda as outlined in Volume 3, *Science and Engineering Research Program*.

Additional information describing the underground facility infrastructure design and construction for the infrastructure and other facility components can be found in Chapter 5.4, *Underground Infrastructure Design.*

# 5.9 Design and Infrastructure for Other Levels and Ramps (OLR)

## 5.9.1 Overview and Planning Summary for OLR

The excavations created during the lifetime of the Homestake Gold Mine present an opportunity to explore hundreds of miles of underground space with the existing underground levels to support science discovery. Within Homestake, excavated underground levels exist every 100 vertical feet for the first 1,100 feet (335 m), and every 150 feet (45.7 m) from 1,100 (335 m) to 8,150 feet (2.48 km). Horizontally, the workings extend approximately 16,400 feet (5 km) in a north-south direction and 9,850 feet (3 km) in an east-west direction although, in general, the workings move to the south as they are deeper. The mining operation had no need to maintain some of these levels once the gold ore had been removed, so their future use will require varying degrees of rehabilitation to ensure safe access. Two primary science collaborations have interest in exploring these levels outside of the main laboratory campuses. Biologists are interested in studying life underground, and therefore can benefit from the most expansive footprint possible. Geologists would like to study the rock itself and would benefit from a variety of levels and geological formations in which to study their properties and changes throughout the underground space. Additionally, an experiment has been proposed to study $CO_2$ sequestration and has primary interest in underground access along a vertical line near the Ross Shaft that extends from the surface to the 1700L. A detailed discussion on the scientific activities can be found in Volume 3, *Science and Engineering Research Program*.

The DUSEL Conceptual Design Report (CDR) generated in 2007 contained a preliminary list of levels with scientific interest. A series of assessments by the Project since the CDR release has reduced the desired scope to a footprint large enough to meet science needs, but small enough to be economically feasible for the Project. Providing safe access is critical to opening these spaces for science use by drift rehabilitation, proper ventilation consistent with the overall laboratory ventilation scheme, and storm-water drainage diversion to prevent damage to the experiments. Finally, levels to support facility infrastructure operations (i.e. levels with dewatering pumps, electrical substations, and water pressure reducing stations) were aligned with science levels where feasible to consolidate the needs of operations and science and reduce overall costs while meeting science and operations requirements. These criteria were used to define the OLR footprint at the levels shown in Table 5.9.1, showing associated total linear footage.

As noted in Table 5.9.1, some early science experiments will be in place on levels above the 4850L before the start of DUSEL construction. The early experiments require services such as power, communications, and transportation. Hazard mitigation controls—such as ground support, infrastructure cleanup, and floor leveling/stabilization—are also needed prior to construction of experiments.

Existing ground support installed within the OLR footprint will be mostly in the form of spot bolting in the walls and ceilings of levels used to support science. This method will be continued for infrequently visited OLR areas. Routine inspections will be conducted throughout the footprint to identify and mitigate any hazards that may develop over time. DUSEL Facility infrastructure technicians will assist in the installation of ground support in these areas. Areas that are visited more frequently will be pattern bolted to designs suitable for the localized region.

| Level or Ramp | Acces-sible Linear Footage | UGI Facility Linear Footage | OLR Linear Footage | OLR Linear Footage (Escape/Venti-lation Only) | Ground Support | Ventilation (Flow Through) | Ventilation (Auxiliary) | Power (110 V) & Data Network | Maintained for Science | Maintained for Operations |
|---|---|---|---|---|---|---|---|---|---|---|
| 300 | 1,740 | 0 | 1,740 | 0 | X | X | | X | X | |
| 800 | 8,010 | 0 | 8,010 | 0 | X | X | | X | X | |
| 1700 | 9,190 | 0 | 0 | 9,190 | X | X | | | X | |
| 2000 | 15,180 | 0 | 15,180 | 0 | X | X | X | X | X | |
| 4100 | 16,940 | 2,180 | 14,760 | 0 | X | X | X | X | X | X |
| 4550 | 3,770 | 3,770 | 0 | 0 | X | X | | X | | X |
| 4850 | 11,390 | 5,360 | 6,030 | 0 | X | X | | X | X | X |
| 6800 | 4,850 | 4,850 | 0 | 0 | X | X | | X | X | X |
| 7400 | 4,900 | 4,900 | 0 | 0 | X | X | | X | X | X |
| Ramp, 1700 to 2000 | 2,190 | 0 | 0 | 2,190 | X | X | | | X | |
| Ramp, 4100 to 4850 | 4,990 | 0 | 4,990 | 0 | X | X | | X | X | X |
| | | | | | | | | | | |
| **Total Linear Footage** | 83,780 | 21,060 | 51,340 | 11,380 | | | | | | |
| **Total Linear Footage Available for Science =** | | | | | | 51,340 | | | | |
| **Total Linear Footage Available for Science if Shared Facility Space is Available =** | | | | | | 72,400 | | | | |

**Table 5.9.1** OLR levels showing linear footage and use details.

### 5.9.2 Facility and Infrastructure Requirements for OLR

Physicists, biologists, and geologists are not the only groups motivated to access levels outside the main campuses at the 4850L and 7400L. Providing services this far below grade presents unique engineering challenges and is, therefore, of interest to a variety of engineering disciplines. Ensuring safety at all levels adds challenges to the access and infrastructure requirements. Shaft rehabilitation and maintenance also require access to levels outside the main laboratory footprint to support operations.

#### 5.9.2.1 Services

Installing electrical cables to provide power and communications to both the OLR and the main campuses can only be done by adding splice points and substations at strategic locations in the shafts; the cables selected for this project are limited to 800-900 vertical feet in any single run. A 12 kV service line supplies power to the main substations for the OLR, and main substations at the 1700L and 4100L provide power for the OLR at 4160 volts (V). This voltage allows smaller wire sizes for long runs to reduce voltage loss (<5% loss). Transformers convert this to 480 V and then to 240/120 V near the experiment location. Only one science installation on the 4850L has requested a 480 V supply, but it is possible to make this available in other areas depending on experimental requirements. The current understanding is that most experiments plan to use the power for computing needs and lighting. Distribution from the 1700L substation provides power to the 300L, 800L, 1700L, and 2000L. The 4100L substation provides power to the 4100L and 4550L. Section 5.4.3.10 describes this system in detail.

Communications infrastructure is provided at all levels where future access is anticipated, as discussed in Chapter 5.5, *Cyberinfrastructure Systems Design*. Fire alarms will also be installed where practical, with an emphasis on areas with timber support that carry the potential for spontaneous combustion.

Providing water at depth needs to consider a balance between required operating pressures and increased pipe cost. Industrial water is available in the Ross Shaft and potable water is available at the Yates Shaft. The current design pressure for potable and industrial water is less than 1,000 feet of head (430 psi). Every level intended for future use will be provided with a connection from each water line passing the level. These connections will have pressure-reducing valves to provide 120 psi at the connection. The primary intention of this connection is for fire protection, but it will be available for science needs as well.

### 5.9.2.2 Safety

Several hazards exist in an underground facility that must be addressed throughout the design and construction of the vertical footprint, including issues such as ventilation, fire safety, rapid water inflows, and ground control. The design must include consideration for emergency egress and/or Areas of Refuge where applicable. Ventilation air should be isolated to only service the occupied areas. Allowing uncontrolled ventilation air to flow outside of these areas both reduces the effectiveness of the system and potentially can dry existing timber supports, thereby increasing the risk of fires. Fire prevention will be addressed by building isolation walls and adding sensors to detect hazards and allow time to respond appropriately. A third hazard is water flow and sand flow due to water saturation. Planning for the diversion of water inflows is discussed in detail in Section 5.4.3.12, *Water Inflow Management*. Finally, deteriorating ground conditions are a hazard that will also need to be mitigated throughout the life of the Facility. The quality of the existing ground support as well as ground conditions varies throughout the OLR areas and rehabilitation including scaling and ground support installation will be required, depending on the frequency of use and the level of training and experience of the personnel requiring access.

### 5.9.2.3 Shaft Rehabilitation and Maintenance

Access to the OLR at the 4850L and above will be through the Ross Shaft and emergency egress through the Yates Shaft (Figure 5.9.5). Access to the OLR below the 4850L will be through the #6 Winze and emergency egress through the #8 Winze. All these shafts will be subjected to some level of either construction or rehabilitation. Access to the OLR during these shaft activities will require careful coordination with the Facility Management team such that safe access will always be provided. During shaft rehabilitation, there will be limited, phased access to OLR areas to allow for concurrent construction activities.

### 5.9.3 General Conditions at Re-entry

Homestake Mining Company (HMC) ceased mining operations in January 2002 and shut down the dewatering system in June 2003. The water level rose to a high point of 4,529 feet below the Ross Shaft collar in August 2008. Ground control systems that were submerged during this period included bolts and screening, with and without corrosion control, that experienced various degrees of deterioration. Homestake did not maintain ground control in extended areas after the ore was removed, and the ground control was not designed for extended life. For example, the ground control used in many areas is carbon steel with no galvanic coating to protect against corrosion. This type of ground control has a design life of 5-15 years, depending on regional conditions. Many stopes, or areas where ore was mined, were backfilled with sand to dispose of the waste material from mining and provide some ground support in these areas. In areas where stopes were submerged, the sand became saturated and in a few documented instances was able to flow out of the containment areas, which blocked some access.

As the water underground receded, the South Dakota Science and Technology Authority (SDSTA) staff thoroughly inspected the areas above the 5000L that were anticipated for scientific use. Their condition is now well known. Many of these levels have had some ground control repair and upgrades installed to provide safe access until additional ground control systems can be installed for DUSEL. Levels below the 5000L are underwater but those required for science or facility use will be made accessible after funds are available for #6 Winze rehabilitation. The ramp system from the 4850L to the 7400L has only been evaluated to the 5000L, but the conditions found in this section of the ramp indicated that the ramp would require considerable rehabilitation for safe entry in the future. As a result, the ramp system below the 5000L is not in the current plans for DUSEL construction or operations.

### 5.9.4 Fire/Life Safety

The strategy for fire/life safety at OLR varies depending on the level and its specific hazards. The fire/life safety approach is detailed in the level-specific discussions later in this section. Some general guidelines used for design include:

- When possible, two means of access/egress will be provided. If this is not possible, a strategy similar to that used for underground mining is employed. This could include refuge chambers, emergency respiration devices (oxygen supply or filtering self-rescuers), along with a trained DUSEL employee to escort scientists in these areas.
- Notification of an emergency event must be provided. The system for providing emergency notification will utilize the conventional “stench gas” system. This system incorporates the introduction of a chemical called ethyl mercaptan into the fresh air ventilation stream that has a strong odor, providing notification to personnel to evacuate.
- Timber will be removed or isolated whenever feasible to reduce fire hazards.
- All installations will limit flammable components to reduce fire risk.
- All individuals traveling outside the main campus areas on the 4850L and 7400L will be required to wear personal protective equipment (PPE) and carry self-rescuer devices as required by DUSEL Environment, Health, and Safety (EH&S) policies.
- Large rainfall events will require limiting or ceasing access to OLR areas until water levels are assessed and inspections are performed to ensure safe access.

### 5.9.5 Area-Specific Discussion for Major Component of OLRs

The electrical and communications infrastructure design to the substations at the 1700L and 4100L will be included as part of the DUSEL Project infrastructure. The development of services beyond these substations will be completed with non- Major Research Equipment and Facilities Construction (MREFC) resources through an operations and maintenance budget. Water supply for fire control at OLR will follow a similar philosophy, providing connections at each OLR shaft station that can be used as needed. Ground support will be upgraded through operations and maintenance as needed upon inspection as the experiments develop.

Excavation will have the most involvement with Final Design for OLR. Varying stress regimes and rock types dictate different excavation techniques. Removal of materials from these excavations will either require connection to the waste handling system or identification of unused areas to backfill. It is expected that the excavation design contractor selected for Final Design of the main Project will also include the design of OLR excavations as part of the deliverables. MREFC funding allows for services down the shafts to the OLR stations. Facility operations crews will provide ground support and utility installation to OLR experiment locations.

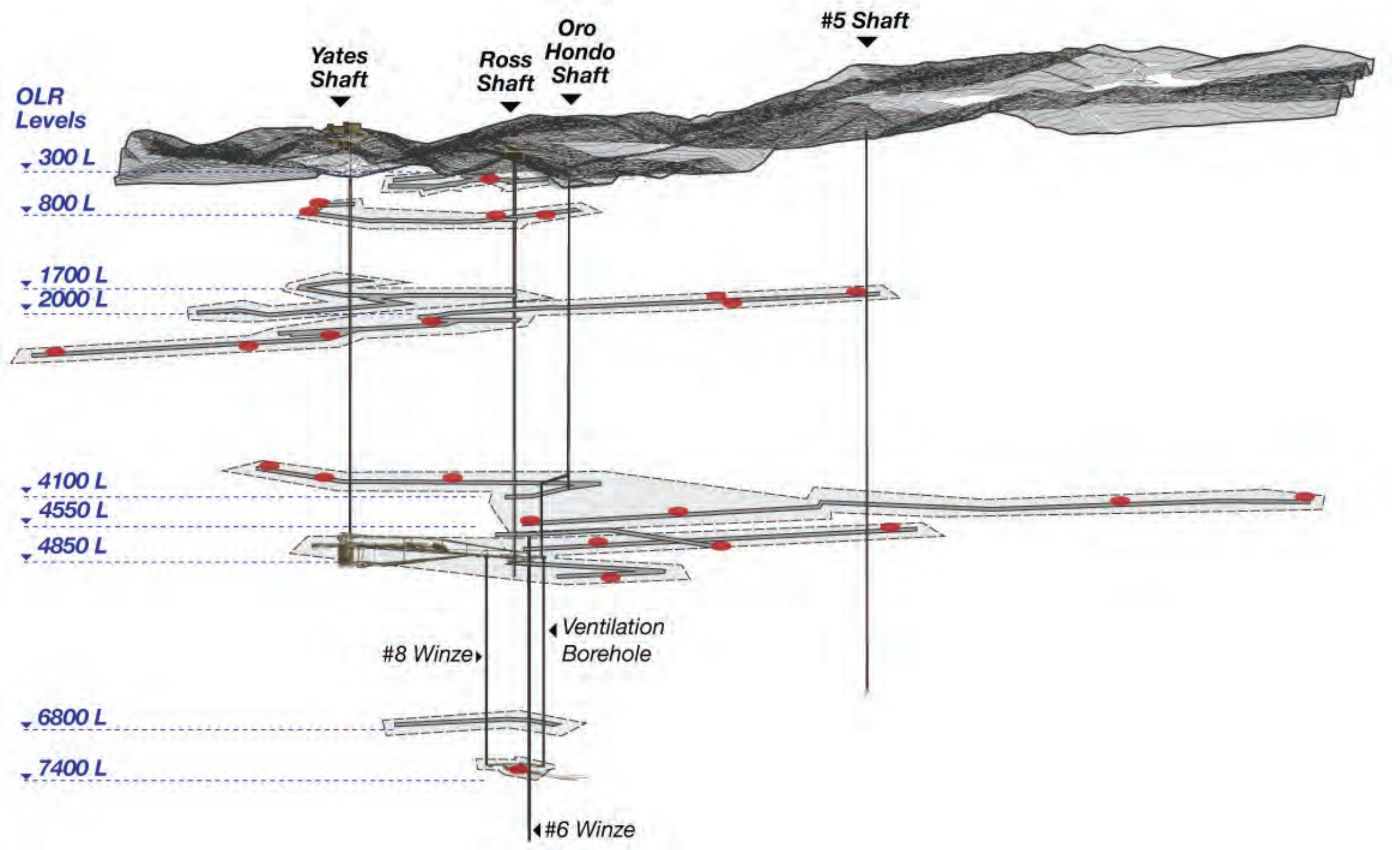

**Figure 5.9.5** Isometric of proposed DUSEL OLR experiments shown in red. [DKA]

Figure 5.9.5 shows an isometric view of the underground spaces with each level anticipated for use by OLR areas. Details are included in this section to explain the design for access and utilities to each level. For more information on the experiments, see Section 3.3.7 of this Preliminary Design Report.

### 5.9.5.1 300L

The 300L is one of the smaller areas in the OLR footprint but is unique because it has two portals to the surface; the Kirk Adit and Oro Hondo Adit (see Figure 5.9.5.1-1). The Kirk Adit breaks through to the surface near the site of the old Kirk ventilation fans (Ellison System) and the Oro Hondo Adit breaks through to the surface near the Oro Hondo Shaft. The Oro Hondo Adit contains a pipe that delivers water from Barrick's Grizzly Gulch tailing dam to the Sanford Laboratory Waste Water Treatment Plant. The 300L also connects with the Ross Shaft, which provides three means of egress. See Figure 5.9.5.1-2 for the 300L footprint and proposed experiment locations.

Ventilation through the 300L occurs naturally between the two adits. The level is isolated, with a door on the Ross Shaft to minimize air entering the shaft. In the winter, the air must be heated in the shafts to prevent ice damage to the shaft structure. Ice formation at the adits is also possible, but routine inspection and maintenance will be performed to ensure safe passage.

The 300L has been refurbished and upgraded as part of a project to replace the Grizzly Gulch pipe in 2010. Refer to Appendix 5.M (*Arup Preliminary Infrastructure Assessment Report*) for detailed long-term ground control recommendations. These ground control recommendations are more extensive than the work that has been completed to date, so periodic maintenance and spot bolting will be required throughout the life of the Project. No significant inflows are experienced at this level, as it is outside the Open Cut footprint. Water drainage on this level flows to the Ross Shaft and is discharged through the Ross Shaft dewatering system.

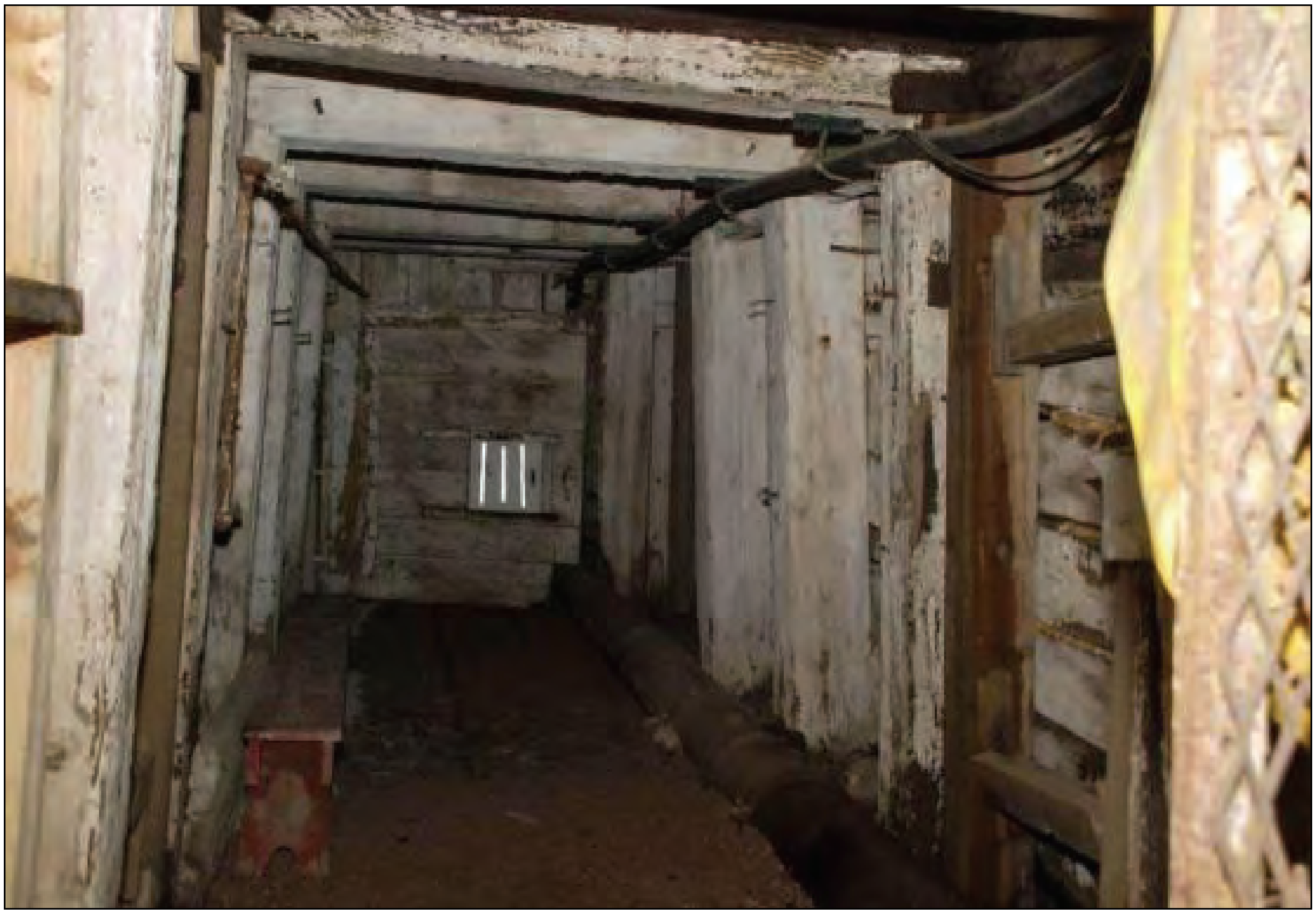

**Figure 5.9.5.1-1** Kirk Adit. [Bill Harlan, SDSTA]

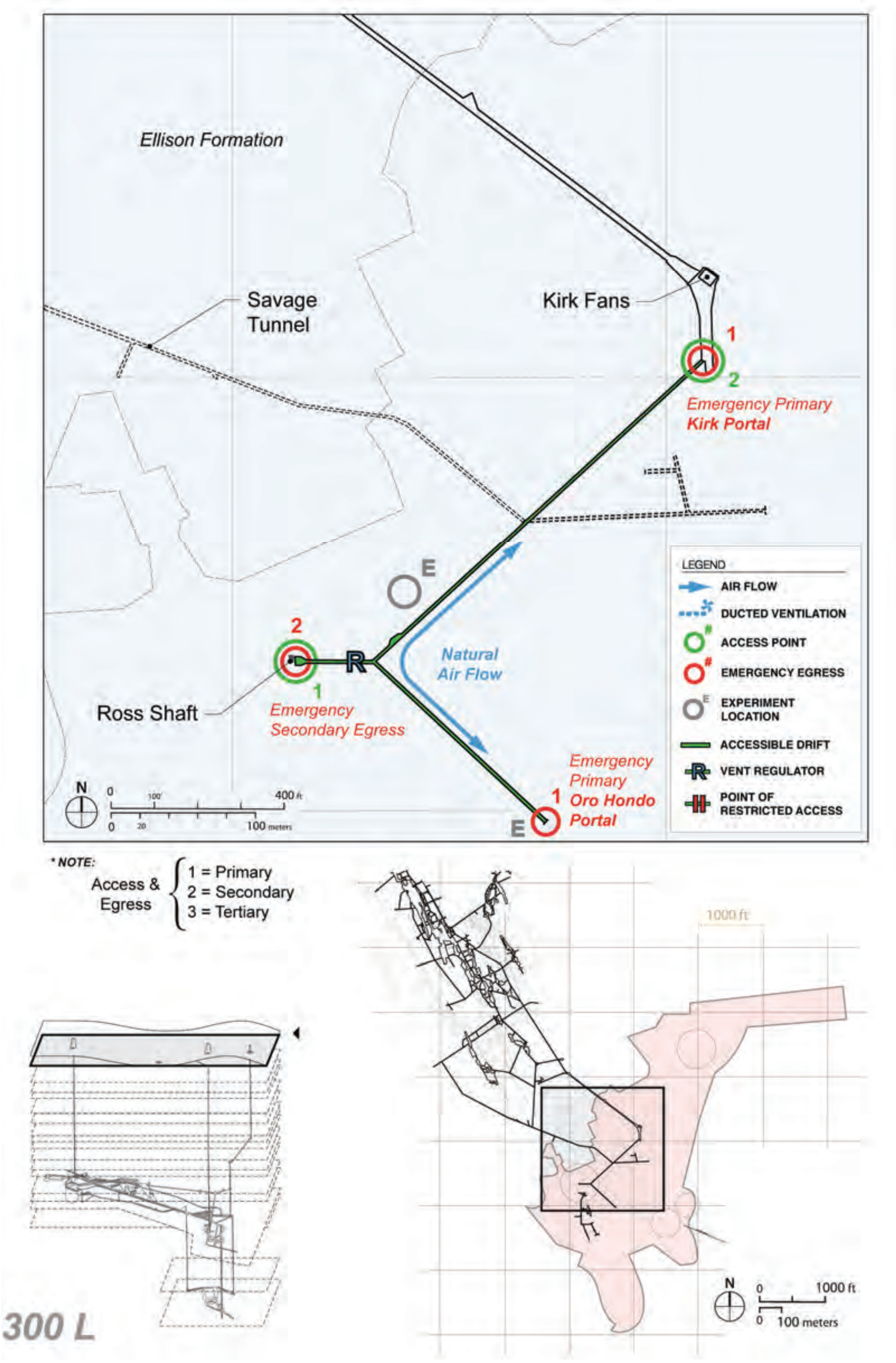

**Figure 5.9.5.1-2** 300L proposed experiment locations. [DKA]

### 5.9.5.2 800L

The 800L (Figure 5.9.5.2-1) connects to both the Yates and Ross Shafts, providing two means of egress for this level. An area previously used by HMC for blasting cap and explosives storage (Figure 5.9.5.2-2) can be used in the future for storage or laboratory space. Power and communications will be provided for this level.

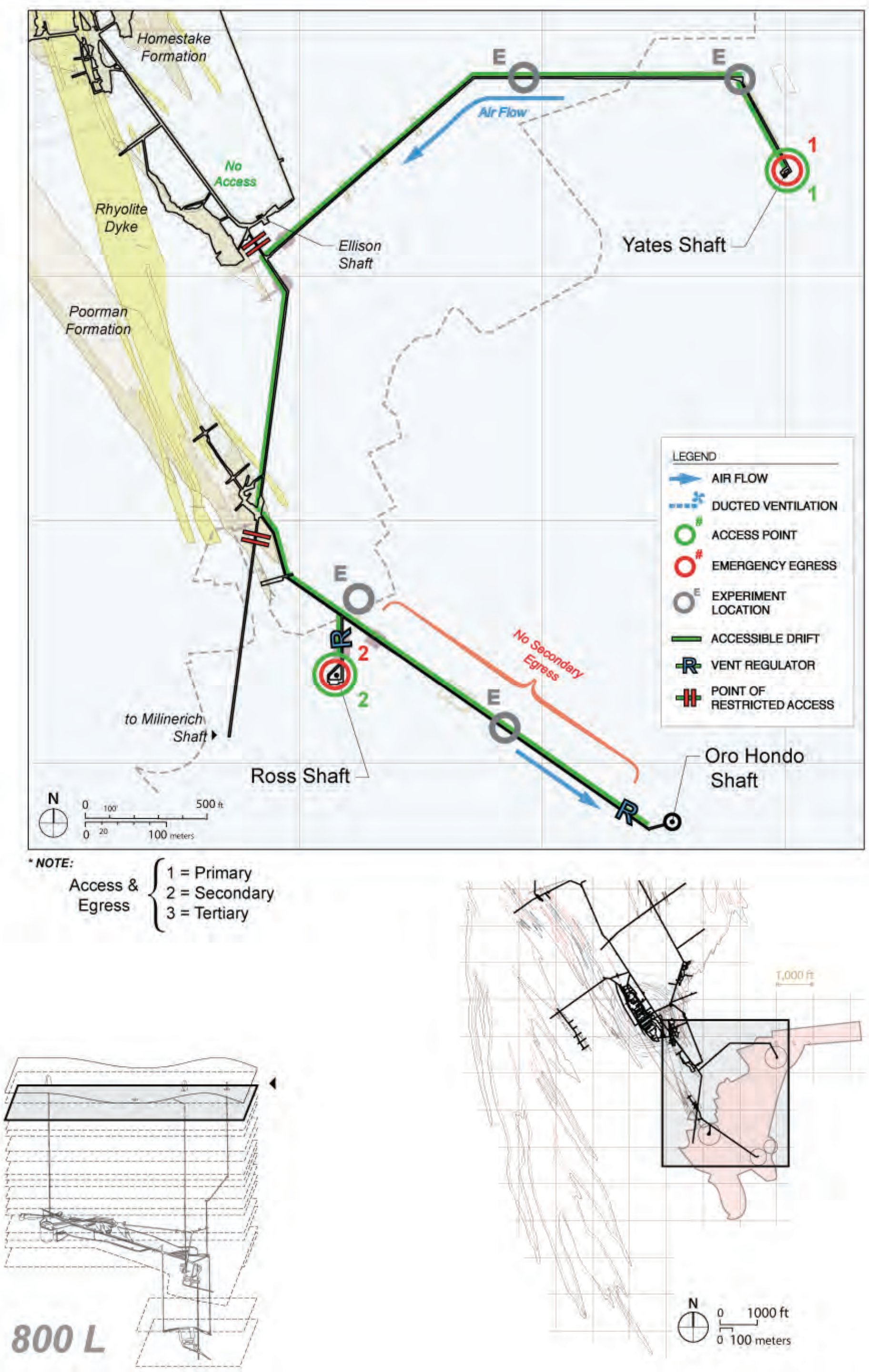

**Figure 5.9.5.2-1** 800L proposed experiment locations. [DKA]

Ventilation enters the level from the Yates Shaft, travels past the Ross, and exits into the Oro Hondo Shaft. Air from the Ross Shaft will be regulated through an air door near the station. Overall flow rate of air will be controlled by an existing regulator at the Oro Hondo Shaft. Two air doors near the Ellison Shaft prevent air from the Open Cut from entering the ventilation network.

Only minimal ground support was required at the 800L during the mining operations. The support that was installed is in satisfactory condition for early scientific needs, but spot bolting and pattern bolting will need to be performed for long-term use. Refer to Appendix 5.M (*Arup Preliminary Infrastructure Assessment Report*) for detailed long-term ground control recommendations.

Minor water inflows have been observed within the designed footprint of this level near a small area near the Milinerich Shaft drift. All water flowing on this level is directed toward the Ellison Shaft. Both of the shafts are outside the footprint intended for any future use, and therefore water flows will not impact science or facility needs.

Additional excavations are anticipated to support the experiments planned for this level. However, as of the completion of Preliminary Design, the volume of this excavation has not yet been defined. The 800L has been studied as a potential location for the liquid argon detector option for the Long Baseline Neutrino Experiment (LBNE), which is not included in the DUSEL MREFC-funded scope. A Conceptual Design for this experiment is being performed by the LBNE project.

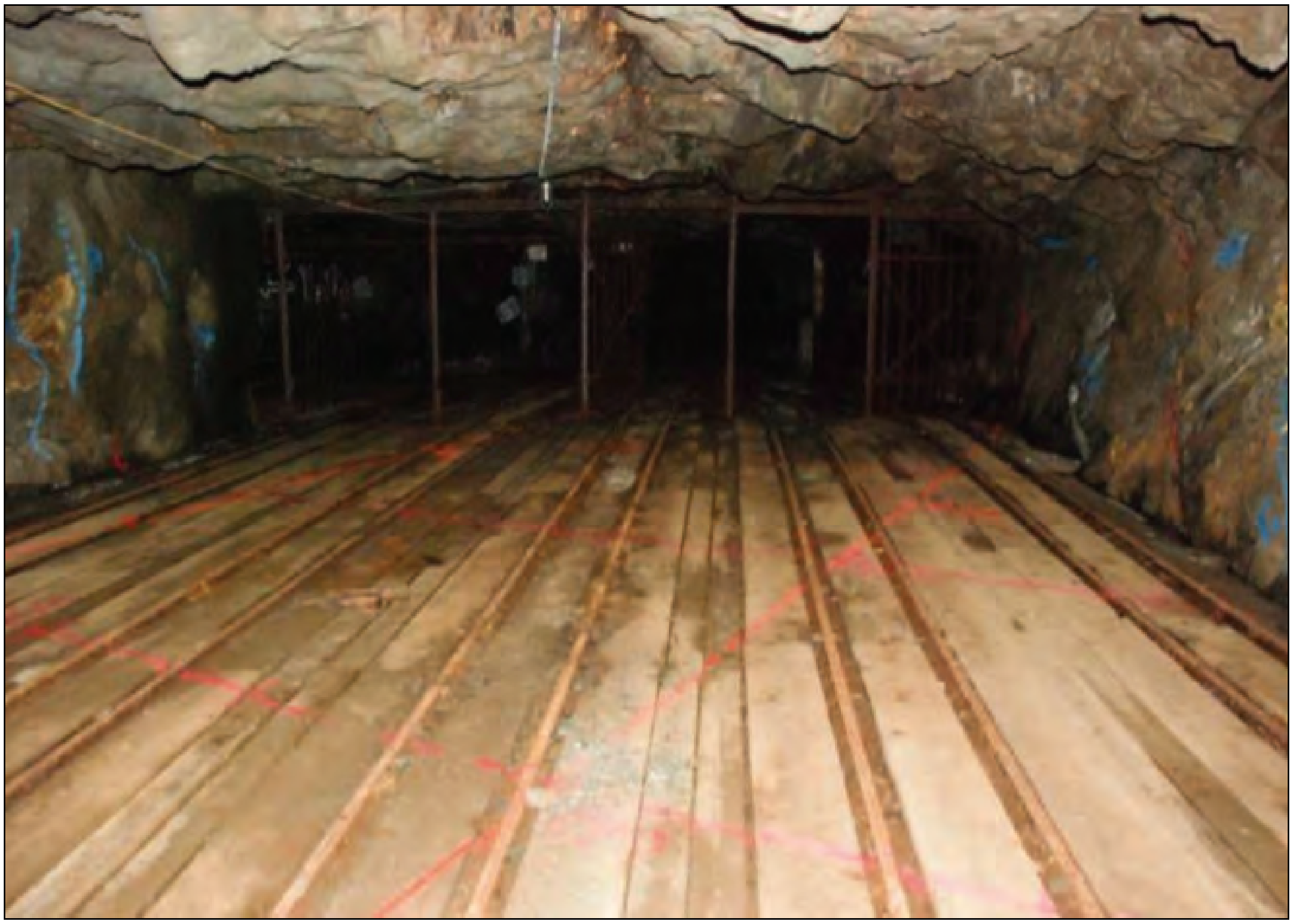

**Figure 5.9.5.2-2** 800L powder magazine. [Bill Harlan, SDSTA]

### 5.9.5.3 1700L

The 1700L (Figure 5.9.5.3-1) connects to both the Yates and Ross Shafts, providing two means of egress, but power and communications will not be provided on this level. The primary purpose of the 1700L is to provide secondary egress and ventilation from the 2000L using the 7 Ledge ramp system. A picture of the area near the 7 Ledge is shown in Figure 5.9.5.3-2.

Ventilation enters the 1700L from the Yates and Ross Shafts at regulated rates, flows across the 1700L, down the 7 Ledge ramp, and exhausts through an opening into the Oro Hondo Shaft on the 2000L. The overall flow rate will be controlled using an existing regulator at the Oro Hondo Shaft. The footprint of the 1700L has ventilation controls installed by SDSTA. See Figure 5.9.5.3-1 for the 1700L footprint.

Very little ground support was installed at this level during the HMC operations. The minimal support that was installed is in satisfactory condition for early scientific needs, but some spot bolting and pattern bolting will need to be installed for long-term use. Refer to Appendix 5.M (*Arup Preliminary Infrastructure Assessment Report*) for detailed long-term ground control recommendations.

Three water diversion walls and several ventilation control walls have been installed on this level. Ground support refurbishment consisting of spot bolting has been completed by the SDSTA. The 1700L is a critical link for water inflow control. Water flows from the existing B&M (Blackstone and McMaster) shafts to a borehole to the 1850L.

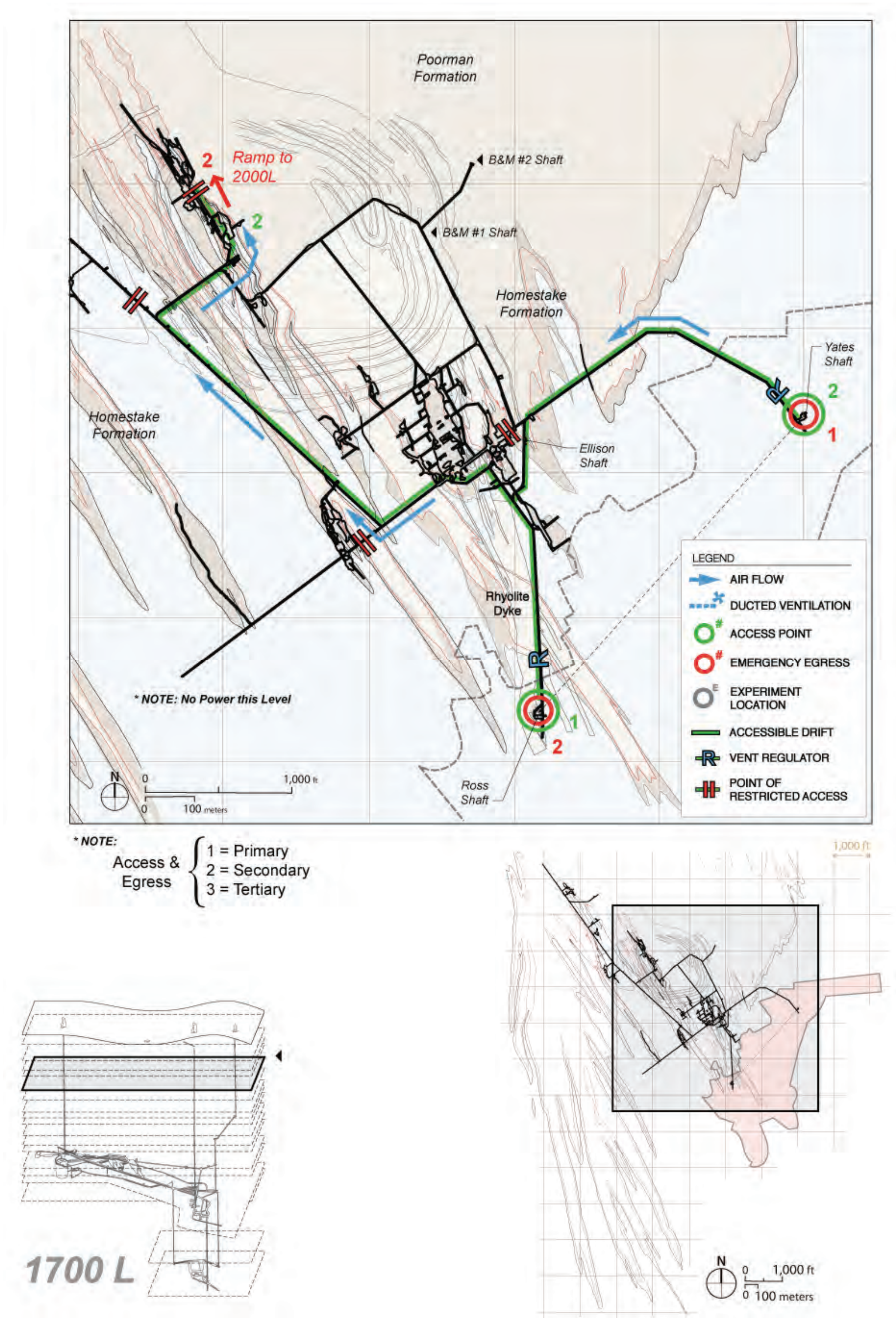

**Figure 5.9.5.3-1** 1700L egress route. [DKA]

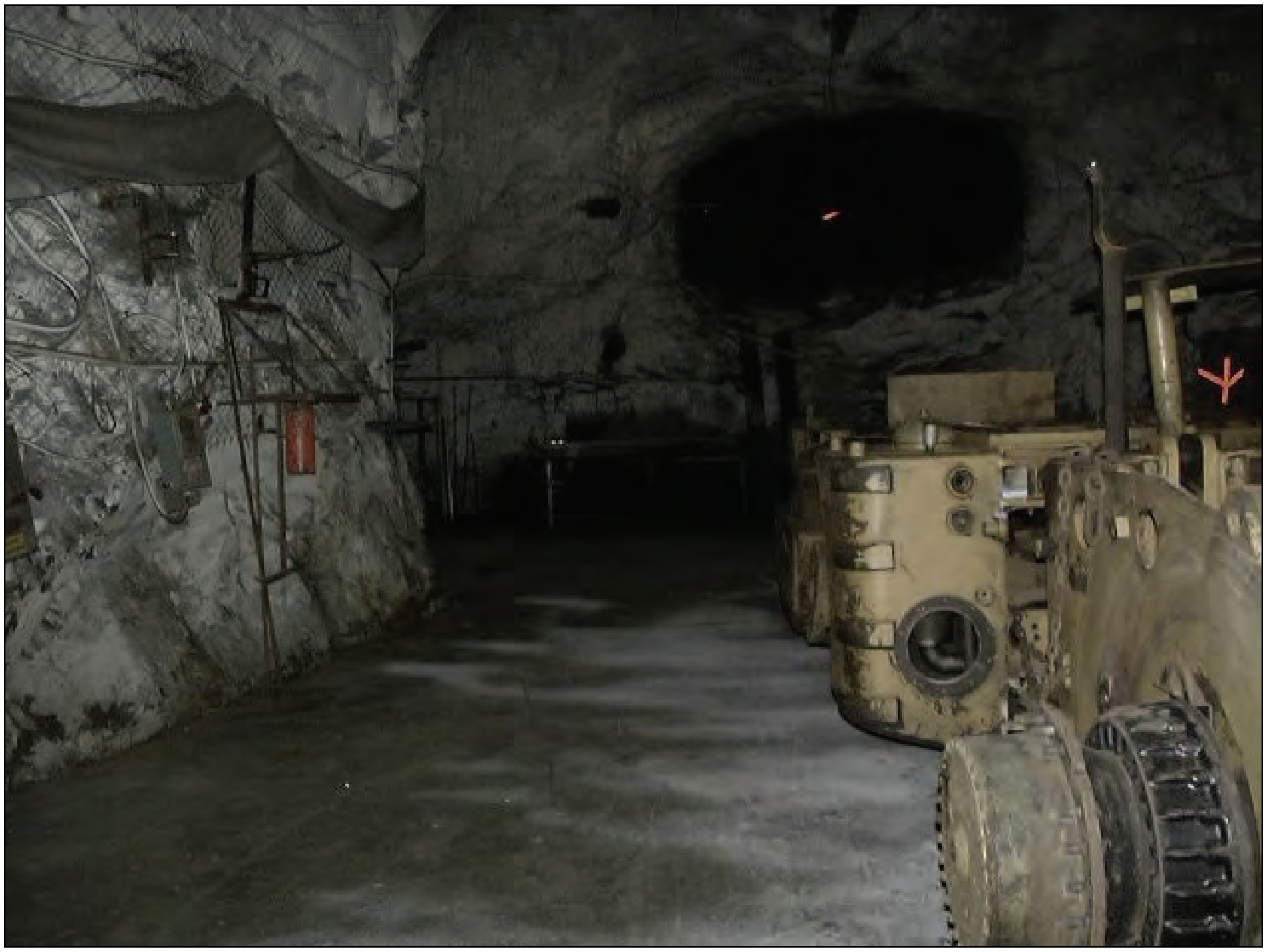

**Figure 5.9.5.3-2** 1700L, 7 Ledge motor barn at closure. [Tom Regan, SDSTA]

### 5.9.5.4 1700L to 2000L Ramp

The 1700L to 2000L ramp connects these two levels and intersects the 1850L (See Figure 5.9.5.4). The 1850L is isolated with ventilation stoppings in the ramp to restrict access and minimize any ventilation leakage. Numerous walled-off openings into old mine workings will need periodic inspection, although these walls appear to be in good condition with little evidence of water inflows. The inflows that do exist are directed to the 2000L. This ramp will be used for egress and ventilation purposes only. Power and communications will not be provided.

No ground support has been installed by the SDSTA beyond what was installed during HMC operations. Current ground conditions are in satisfactory condition for early scientific needs, but some spot bolting and pattern bolting will need to be performed for long-term use. Routine inspection of this system will be required through the life of the Project to ensure open, yet isolated, ventilation. Refer to Appendix 5.M (*Arup Preliminary Infrastructure Assessment Report*) for detailed long-term ground control recommendations.

There are two paths of egress on the ramp. One is up the ramp to the 1700L, then using either the Yates or Ross Shafts. The second is down the ramp to the 2000L, then using the Ross Shaft.

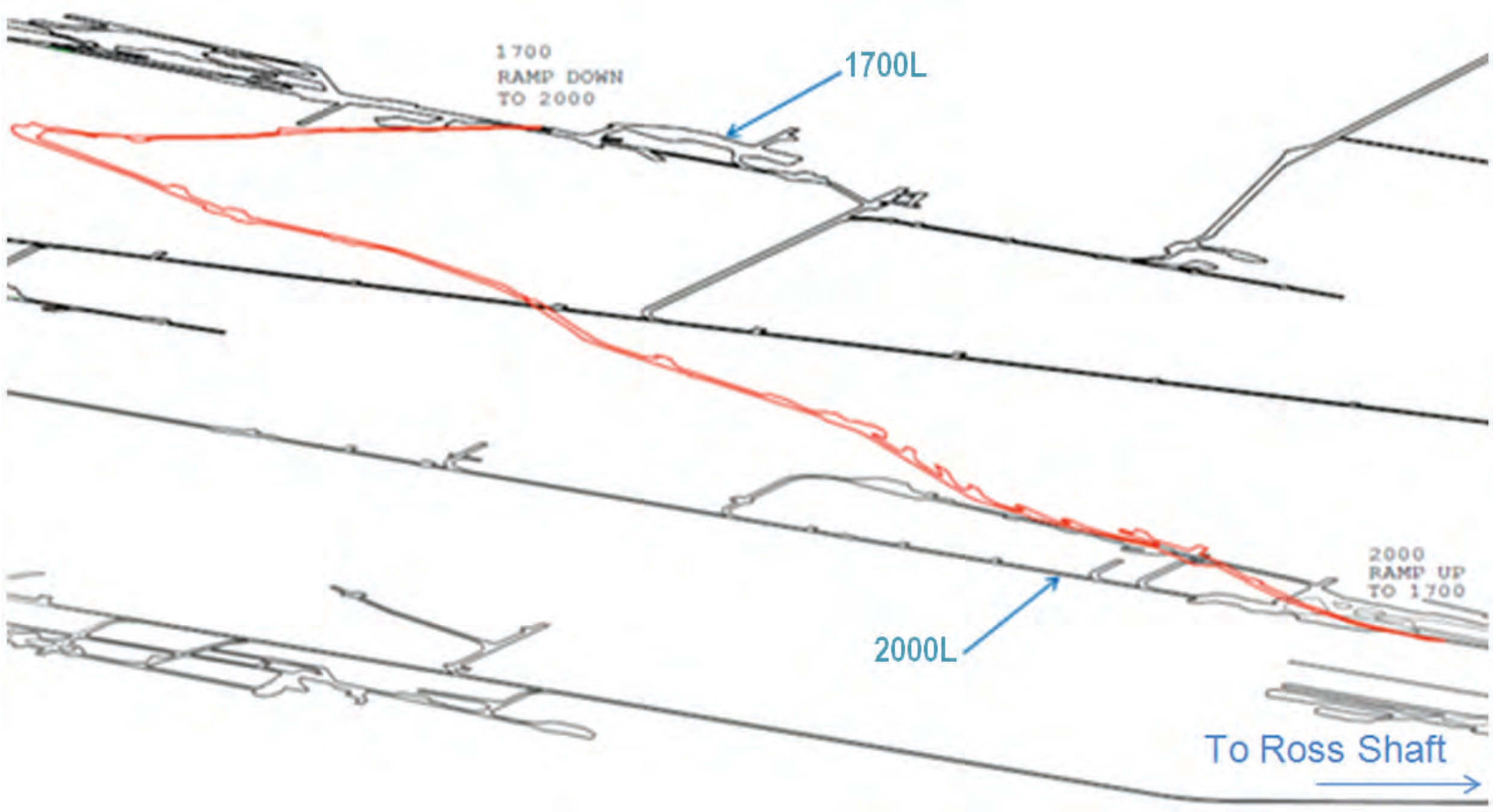

**Figure 5.9.5.4** 1700L to 2000L ramp (no currently proposed experiment locations). [Dave Plate, DUSEL]

### 5.9.5.5 2000L

The 2000L (Figure 5.9.5.5-1) only connects to the Ross Shaft, and secondary egress is provided through the ramp system described in Section 5.9.5.4. Power and communications will be provided to this level.

Fresh air ventilation enters from the 7 Ledge ramp and exhausts through the Oro Hondo Shaft. Flow rate will be controlled through an existing regulator at the Oro Hondo Shaft. The footprint of the 2000L has already been isolated with ventilation controls by the SDSTA.

Several areas of ponded water, or water sitting on the sill, exist at this level (Figure 5.9.5.5-2). Flow from these areas will be directed away from the experimental footprint toward the Ellison Shaft. This will be accomplished by installing two full concrete walls and one half-height diversion concrete wall to direct the water to the designed pathway for drainage.

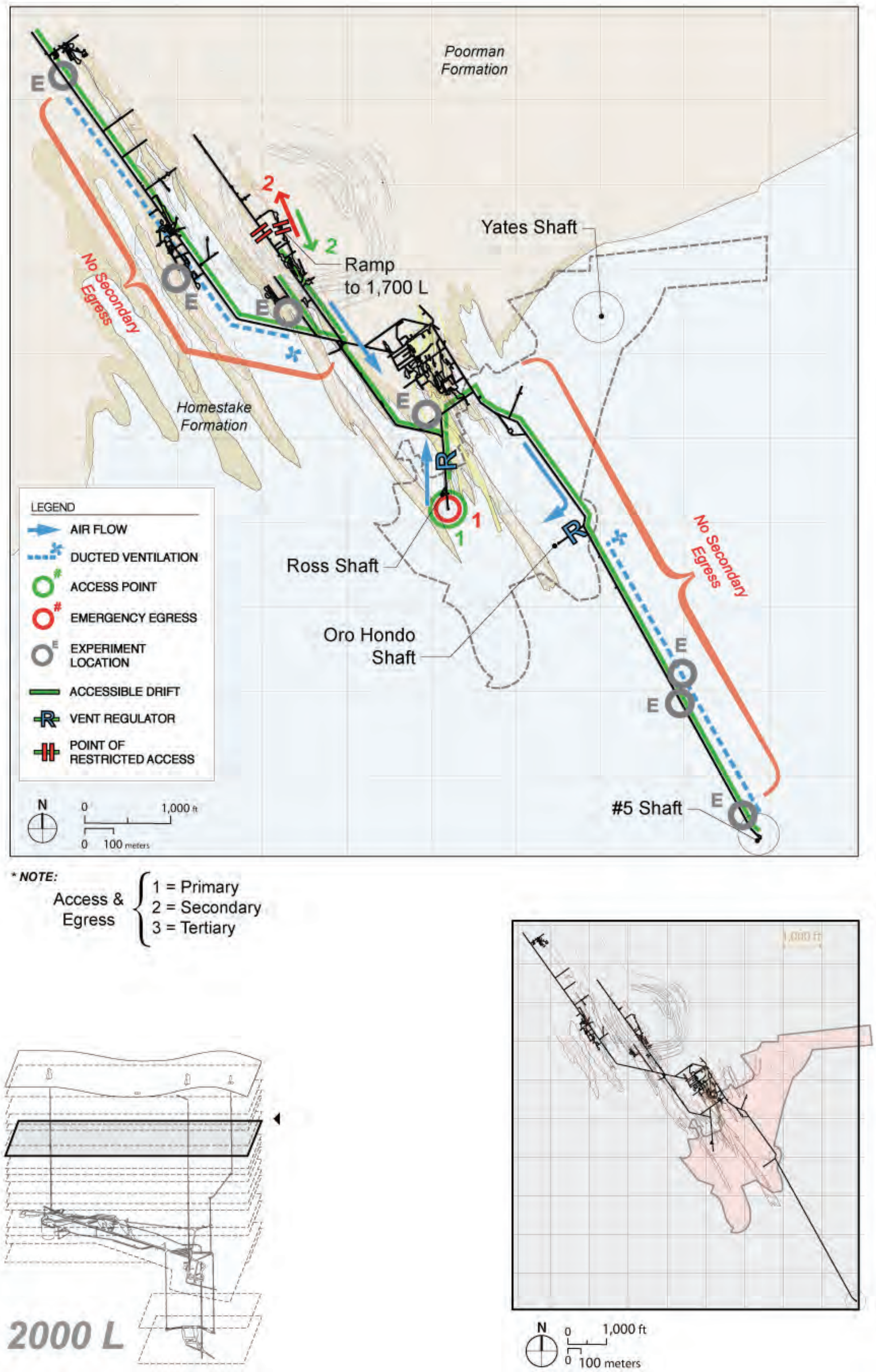

**Figure 5.9.5.5-1** 2000L proposed experiment locations. [DKA]

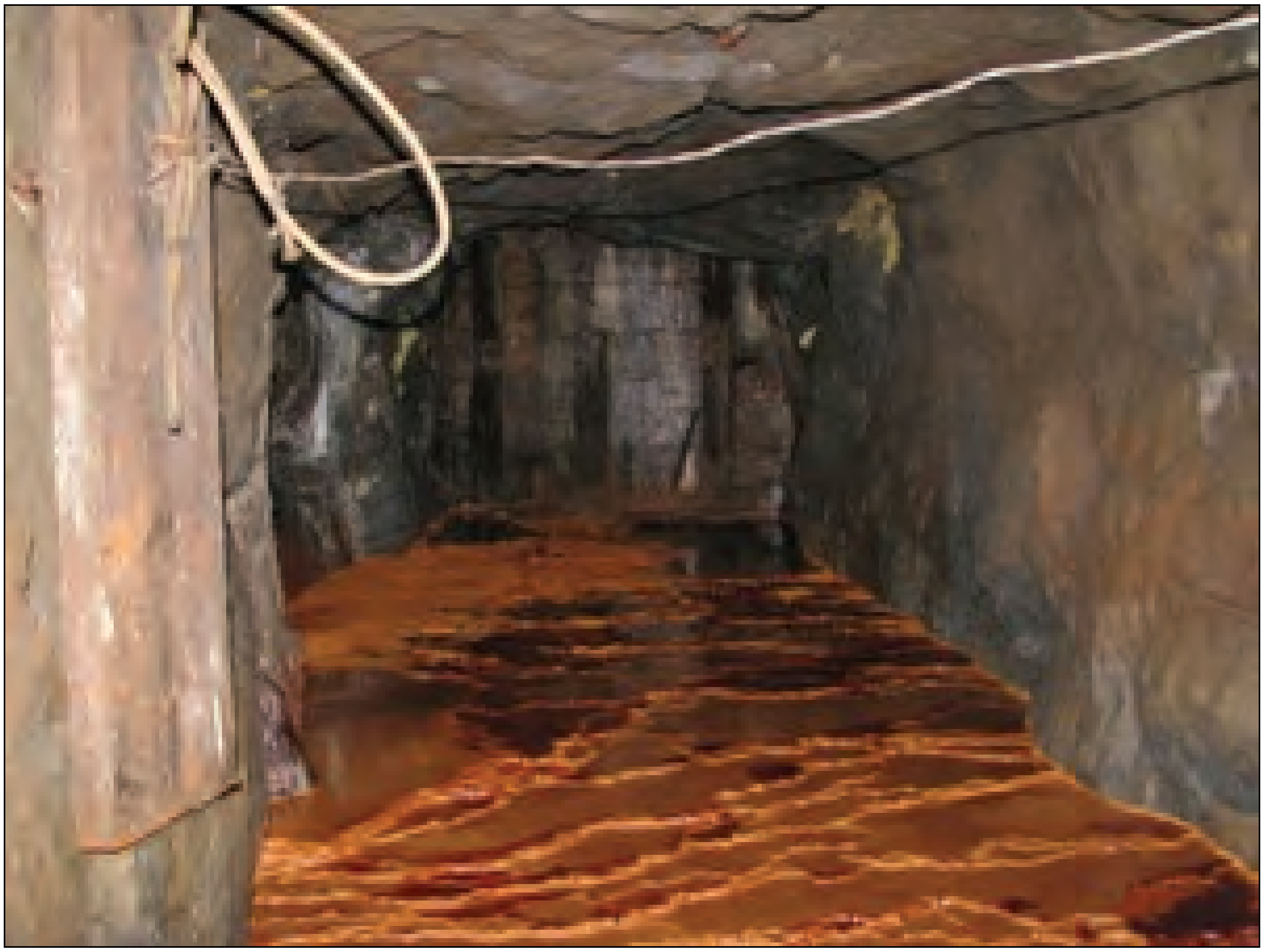

**Figure 5.9.5.5-2** Ponded water on the 2000L. [Tom Regan, SDSTA]

### 5.9.5.6 4100L

The 4100L has two distinct areas of interest. The first area extends between the Ross and Yates Shafts and has two means of access and egress. Power and communications to this area are provided from the substation located on this level. Fresh air ventilation for this area comes from both Ross and Yates Shafts and travels to the Oro Hondo system. Ventilation control is accomplished with air doors in the drifts near the Ross and Yates Shafts. A fan provides auxiliary ventilation to a dead-end drift north of the Yates Shaft.

The second area in the western portion of 4100L is accessed through the 17 Ledge system from either the 4550L or 4850L. This area will have restricted access since it is isolated by over 1.5 miles from the remainder of the underground campuses and covers over 2 miles of linear footage. Power and communication for this level is provided through the 17 Ledge ramp system. Fresh air will flow from the ramp system across the experimental area to the Oro Hondo exhaust system. Two dead-end drifts are included in the footprint. Ventilation for this area will be supplemented by ventilation fans and ductwork.

Following the installation of the sump at the 3500L (described in Section 5.4.3.12), significant water inflows are not anticipated on this level within the experimental footprint.

Ground support at this level has been refurbished by the SDSTA to provide safe access for early science and facility assessment activities. Three bulkheads were installed to isolate the footprint. Refer to Appendix 5.M (*Arup Preliminary Infrastructure Assessment Report*) for detailed long-term ground control recommendations. Figure 5.9.5.6-1 shows the 4100L ore dump feeding the Ross skipping system. Figure 5.9.5.6-2 shows the currently proposed experiment locations.

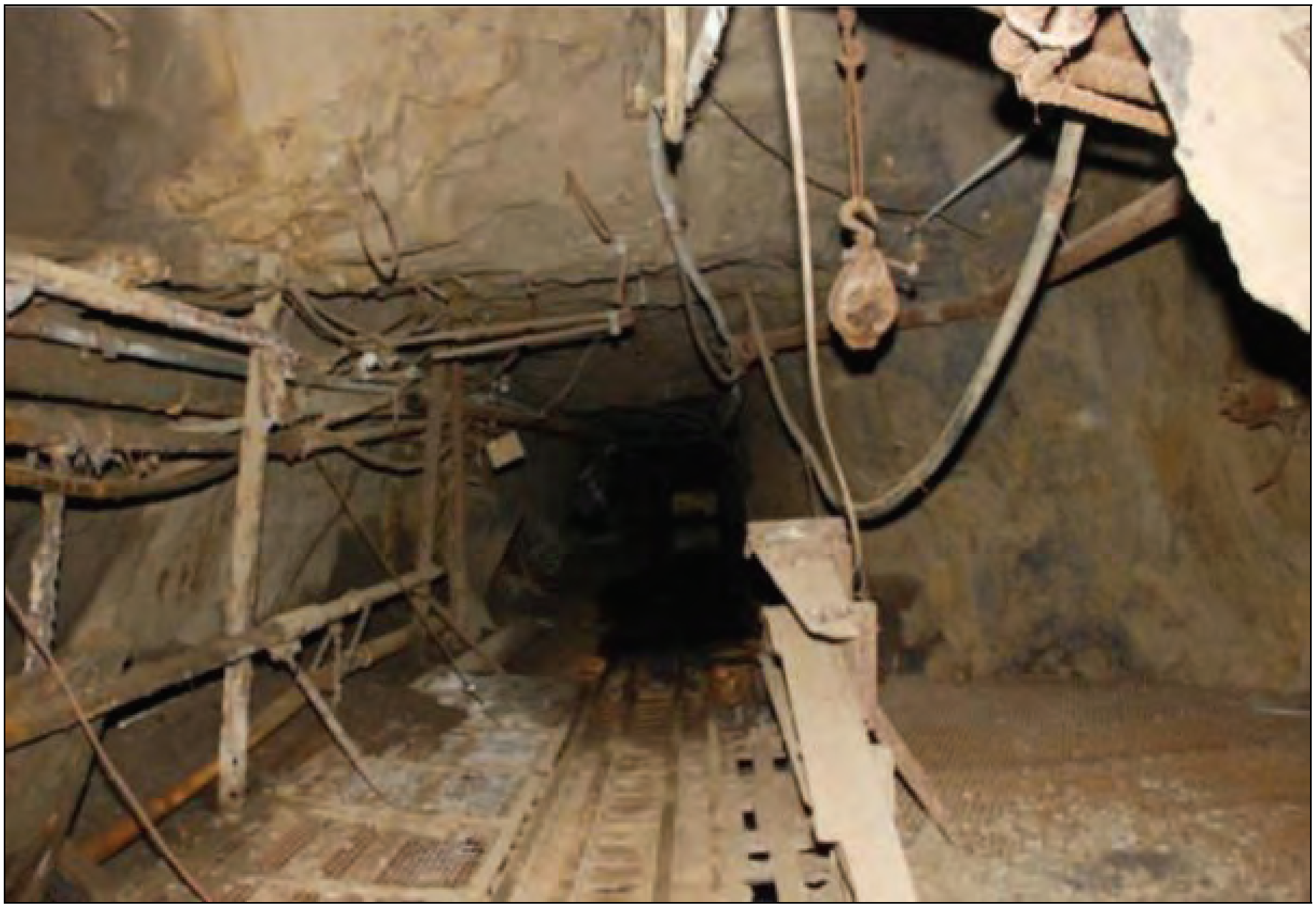

**Figure 5.9.5.6-1** 4100L Ross Station ore dump. [Bill Harlan, SDSTA]

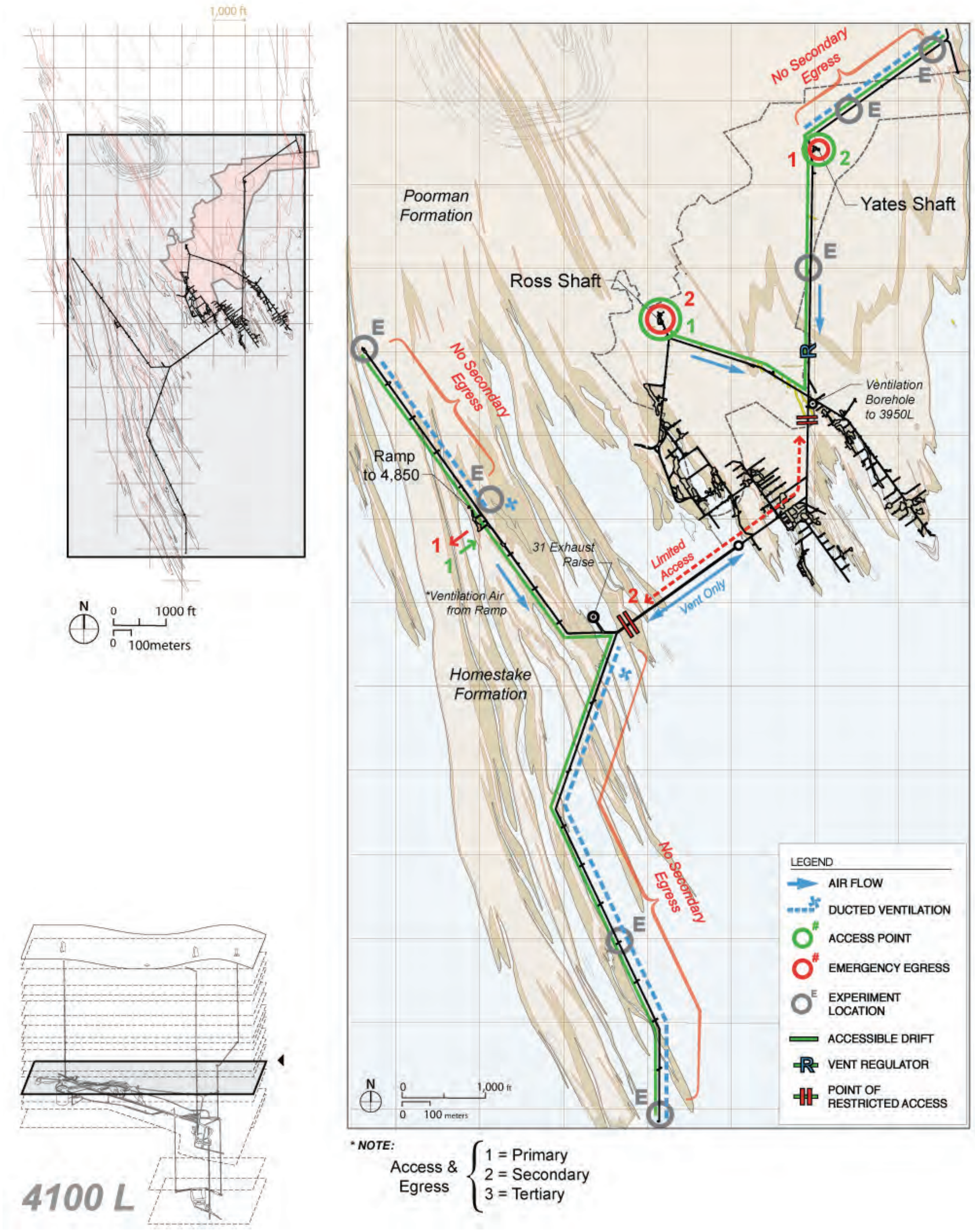

**Figure 5.9.5.6-2** 4100L proposed experiment locations. [DKA]

### 5.9.5.7 4550L

The 4550L can only be accessed though the Ross Shaft. Power and communications will be provided. The 4550L also provides access to the #6 Winze Hoistroom and shared use with the facility will be required.

Fresh air ventilation enters this level from the Ross Shaft and flows to the 17 Ledge ramp exhaust system. Flow rate of air overall will be controlled though an existing regulator on this level. The footprint of the 4550L has been isolated by SDSTA work completed to date.

There are two modes of egress on the 4550L, through either Ross Shaft or down the 17 Ledge ramp to the Yates Shaft on the 4850L.

Only minor water inflows exist on this level within the experimental footprint.

The 4550L is the highest level that was submerged by the water pool after the mine closed. Ground support deterioration was found at this level; however some ground support has been replaced by the SDSTA to provide access to the top of the #6 Winze. A significant amount of work remains to provide safe access through the desired experimental footprint on this level. Refer to Appendix 5.M (*Arup Preliminary Infrastructure Assessment Report*) for detailed long-term ground control recommendations. Figure 5.9.5.7-1 shows an area of the 4550L leading to the 17 Ledge. Figure 5.9.5.7-2 shows the proposed experiment locations.

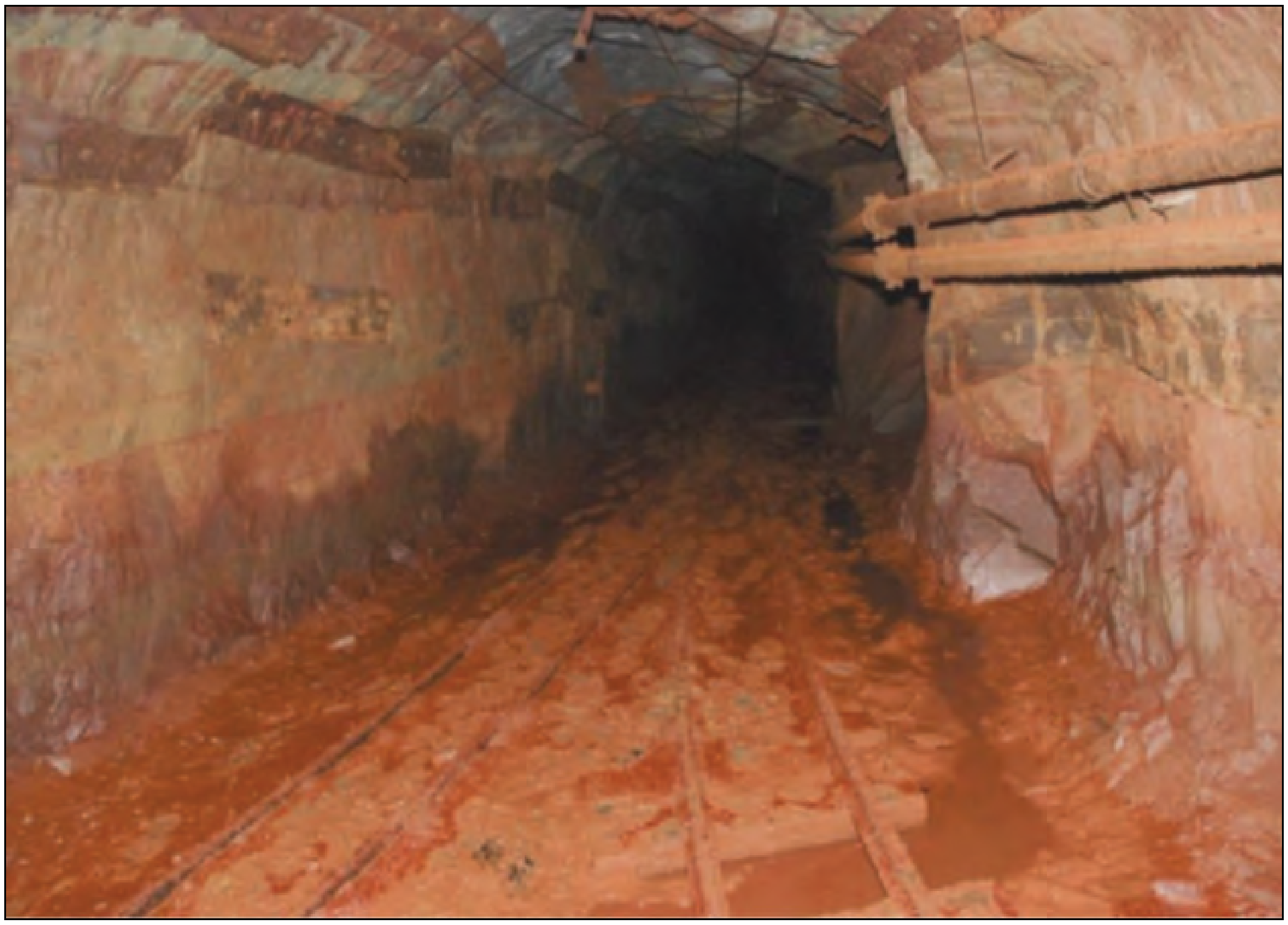

**Figure 5.9.5.7-1** 4550L Drift toward 17 Ledge. [Bill Harlan, SDSTA]

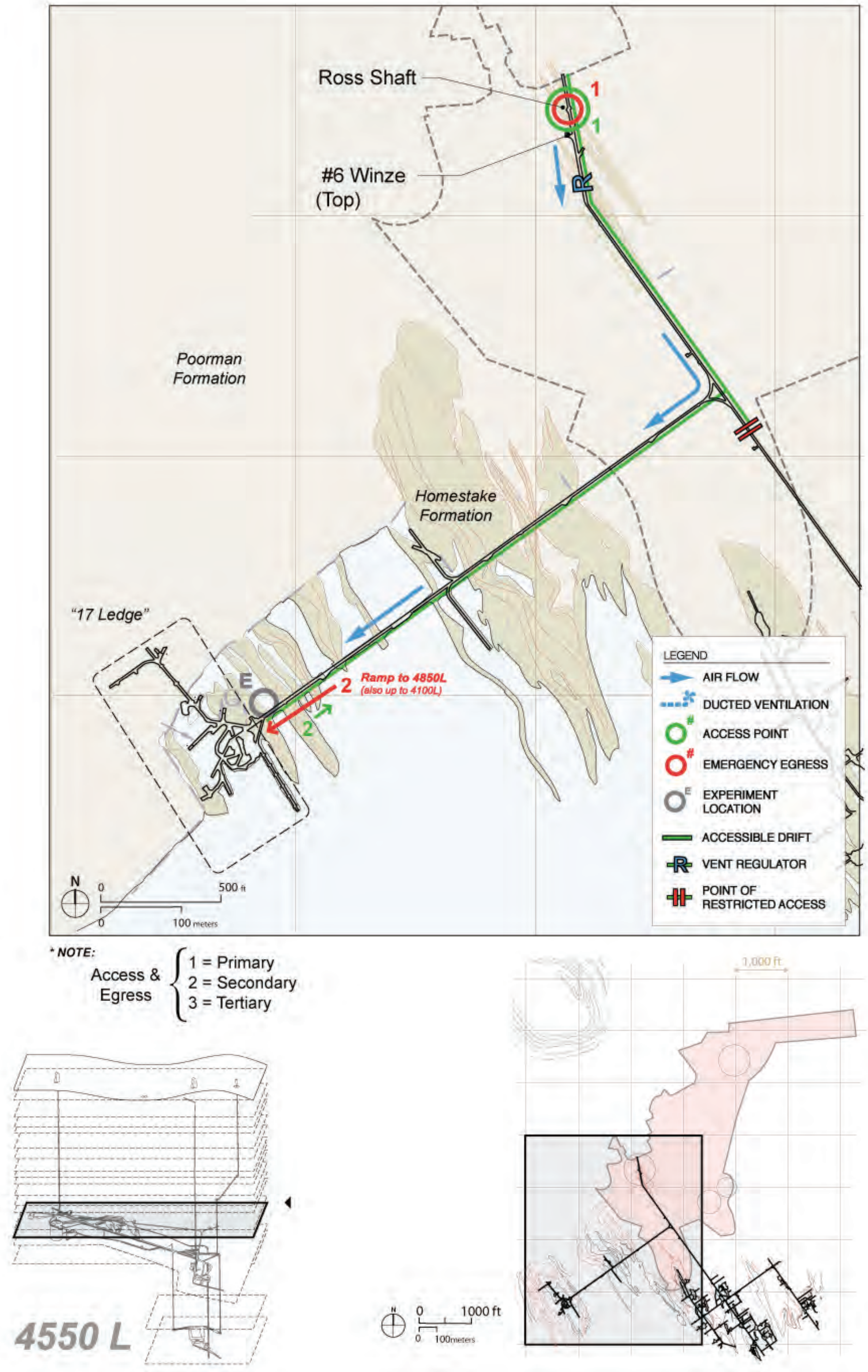

**Figure 5.9.5.7-2** 4550L proposed experiment locations. [DKA]

### 5.9.5.8 4100L to 4850L Ramp

The 4100L to 4850L (17 Ledge) ramp connects these two levels, intersecting the 4250L, 4400L, 4550L, and 4700L. The intersecting levels are isolated by ventilation controls in the ramp to restrict access and minimize ventilation leakage. There are numerous openings into the old mine workings that have been walled off, and these walls appear to be in good condition with little evidence of water seepage. Shared use between facility and OLR will take place in the ramp system. Power and communications will not be provided; however, supply wiring will pass through the ramp system to support the 4100L.

Fresh air ventilation enters the ramp from the Yates and Ross Shafts on the 4850L and 4550L at regulated rates and flows into the 4100L exhaust system. There are two modes of egress on the ramp: up the ramp to the 4100L, or down the ramp to the 4850L. Both levels have access to either the Ross or Yates Shafts.

The ground control on this ramp system is very similar to the 4550L (Section 5.9.5.7), with significant water damage due to submersion after mine closure. No upgrades to this ground support have been completed to date. Refer to Appendix 5.M (*Arup Preliminary Infrastructure Assessment Report*) for detailed long-term ground control recommendations.

Water flowing down this ramp should be very limited due to the water management system described in Section 5.4.3.12. Any water that does flow down the ramp system will be isolated from the main campus on the 4850L with a water diversion structure described in Section 5.4.3.12. Figure 5.9.5.8 shows a map of the ramp system described.

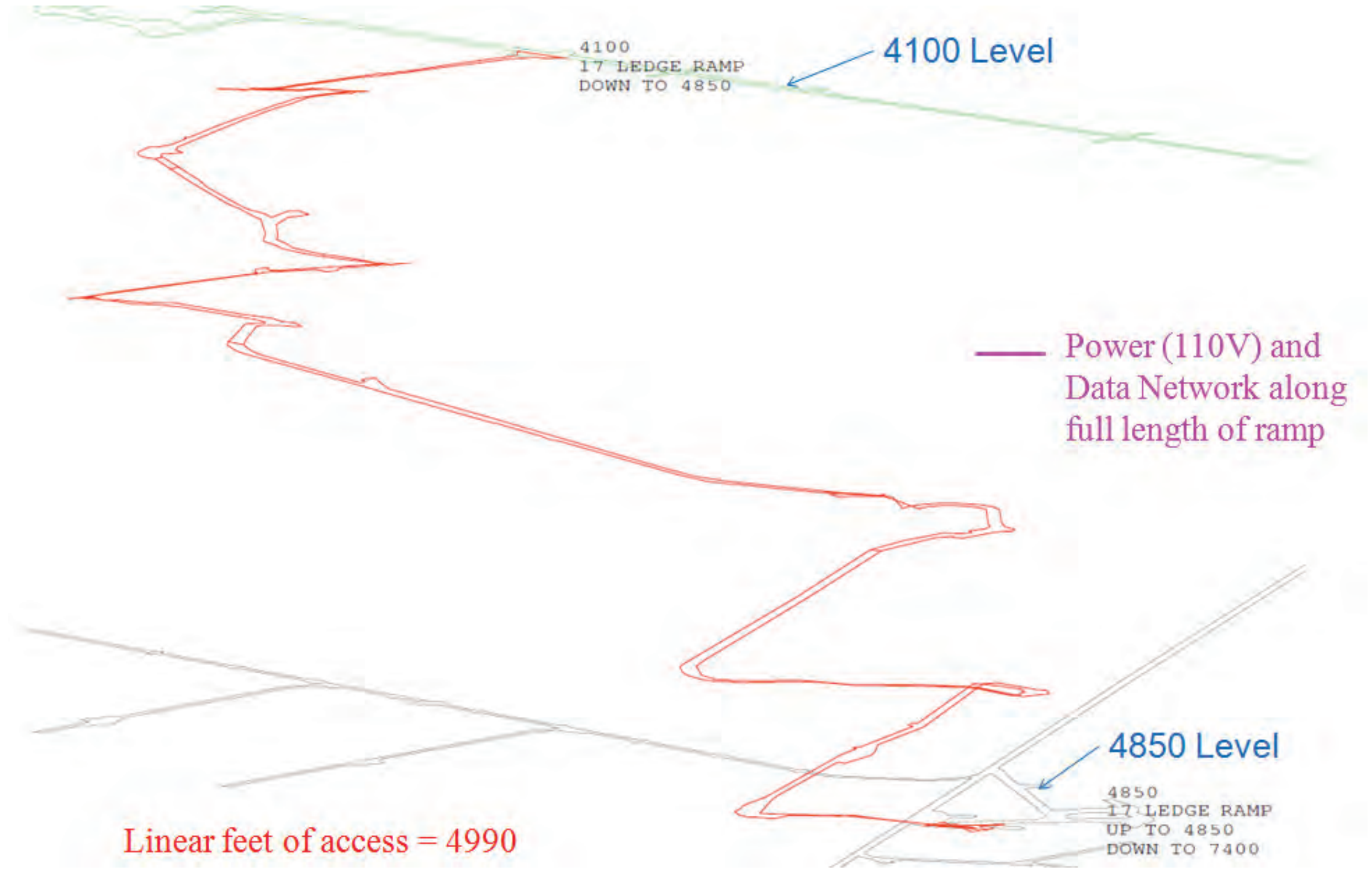

**Figure 5.9.5.8** 4100L to 4850L ramp (no proposed experiment locations). [Dave Plate, DUSEL]

### 5.9.5.9 4850L

The main campus and fire/life safety considerations of the 4850L design are described in detail in Chapter 5.4, *Underground Infrastructure Design*. Outside of the main campus footprint, another 6,030 feet of drifts are planned to be available for OLR experiments. One example of an area for OLR use is shown in Figure 5.9.5.9-1, which is an existing shop that can be repurposed for scientific needs. The footprint of the entire 4850L intended for science use (including physics experiments planned for OLR areas described in this Chapter, 5.9) is shown in Figure 5.9.5.9-2. Power and communications will be provided from the main campus mechanical/electrical rooms.

Fresh air ventilation to areas outside the main campus footprint on the 4850L will be provided from the Ross Shaft through the drifts to the 17 Ledge (4100L to 4850L ramp) or the #5 Shaft. A small drift extending from the drift toward the #5 Shaft will be provided with a fan and ductwork to boost flow through this area. Since the #5 Shaft is known to have failed in the past and will potentially fail again, it may be necessary to add a ventilation fan to support ventilation for experiments in this area as well. Little ground support has been rehabilitated in the areas outside the main campus footprint, but this will be required in advance of science access. Refer to Appendix 5.M (*Arup Preliminary Infrastructure Assessment Report*) for detailed long-term ground control recommendations.

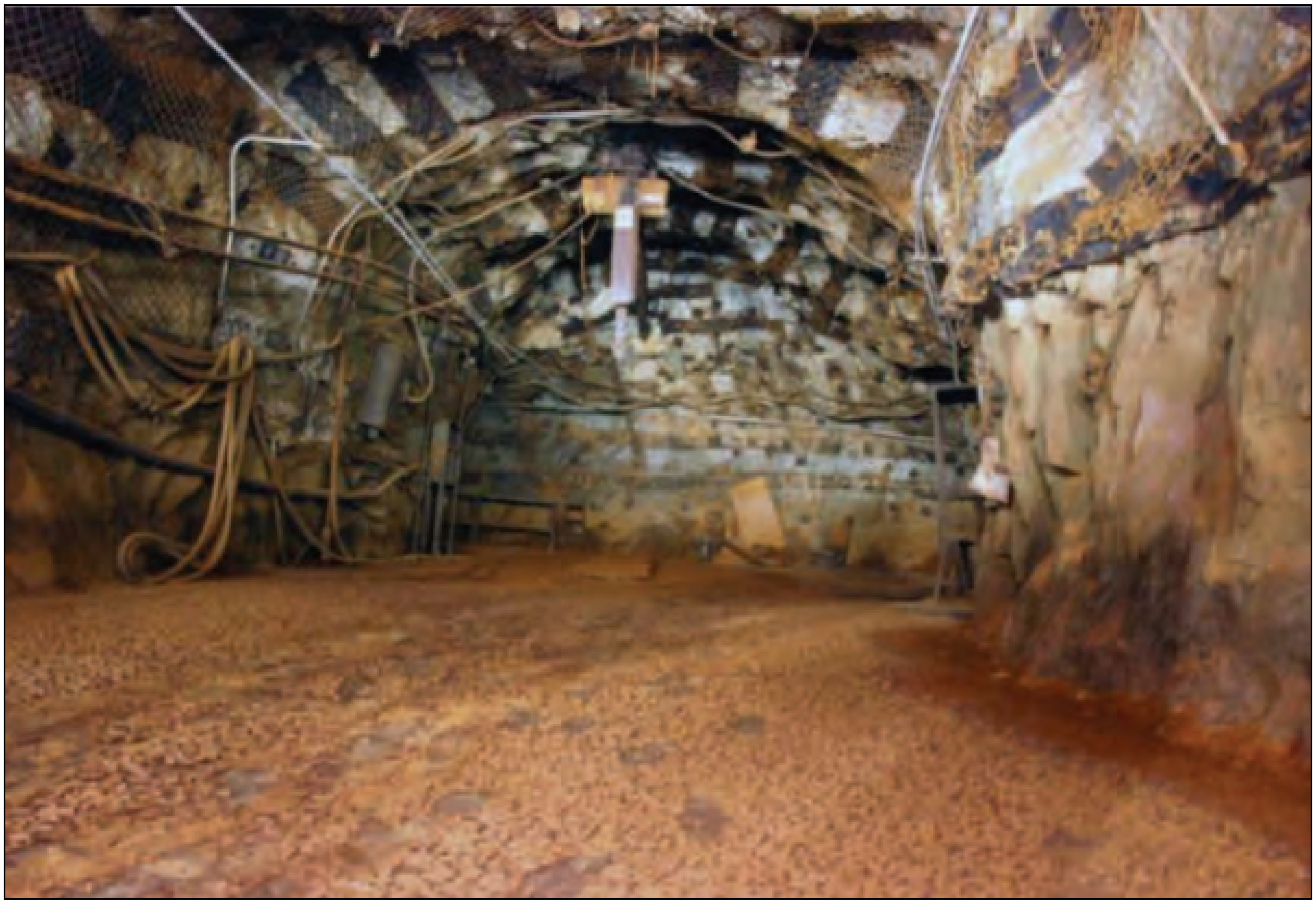

**Figure 5.9.5.9-1** 4850L 17 Ledge shop. [Bill Harlan, SDSTA]

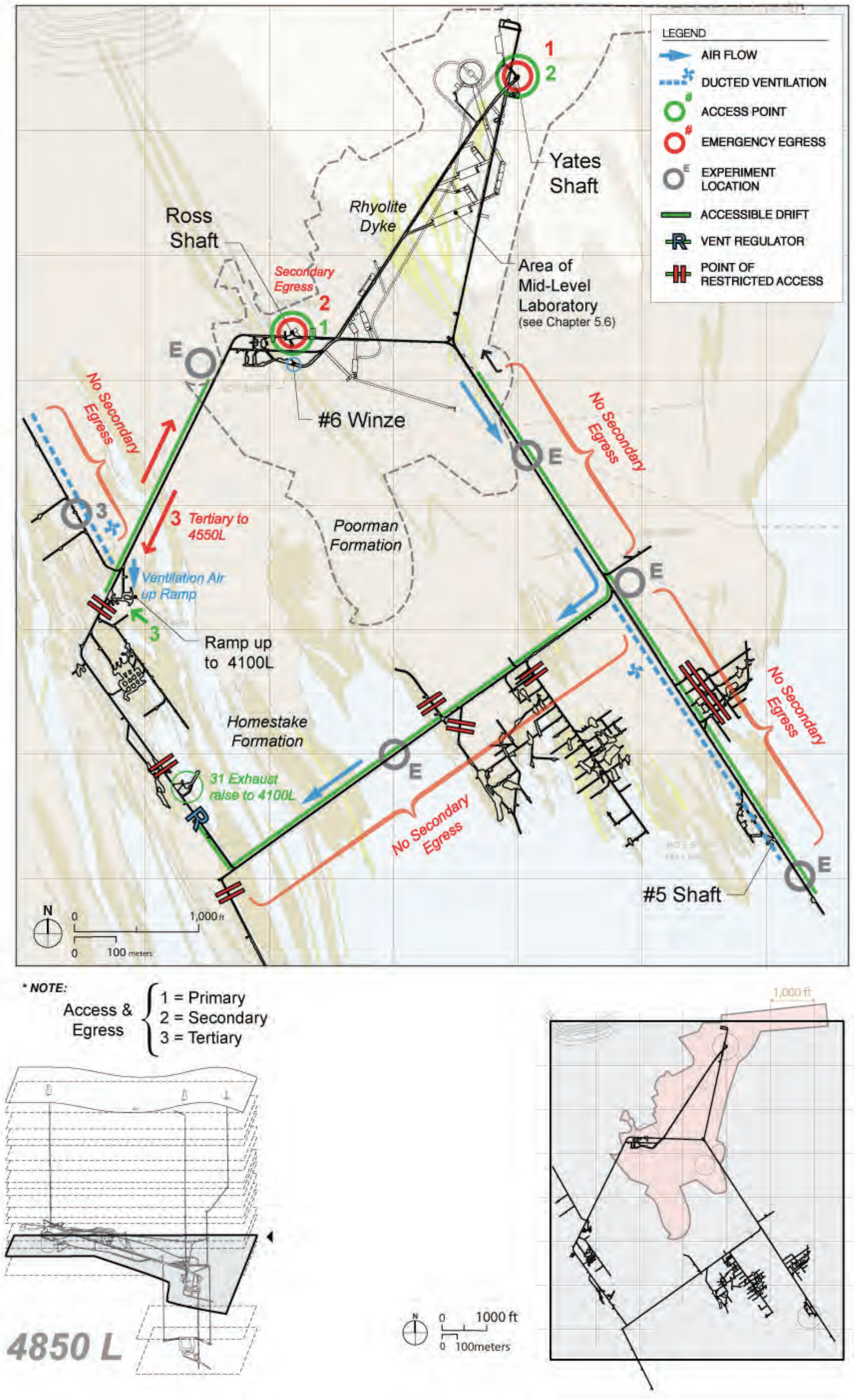

**Figure 5.9.5.9-2** 4850L proposed experiment locations. [DKA]

### 5.9.5.10 6800L

The only viable means of access to or egress from the 6800L is through the #6 Winze. The historic ramp system that previously provided secondary egress experienced severe deterioration following mine closure and is not planned for rehabilitation. Power and communications will be provided to this level. The 6800L will be used by DUSEL facility maintenance staff to inspect and monitor the North Drift plug. This plug was installed by Homestake when an exploration drift intersected a high-pressure water course. This level will also have a dewatering pump station with two pumps as part of the dewatering system.

Fresh air ventilation will be provided by a fan and ducting from the #6 Winze and will be exhausted back into the #6 Winze primary intake. Bulkheads will be installed to isolate this level from historic workings. This is important to isolate the laboratory footprint from water inflow hazards and to minimize ventilation leakage. A map of the intended footprint to be used on this level is shown in Figure 5.9.5.10.

Ground conditions at this level are unknown as of the completion of Preliminary Design. It is anticipated to require substantial rehabilitation, not only because it has been submerged in water, but also because of the high stress levels this far below grade. Since access has not been available to date, no ground control designs have been developed. An allowance has been included in the Project estimate for this work. Detailed geotechnical site investigations conducted upon safe re-entry will determine ground control requirements.

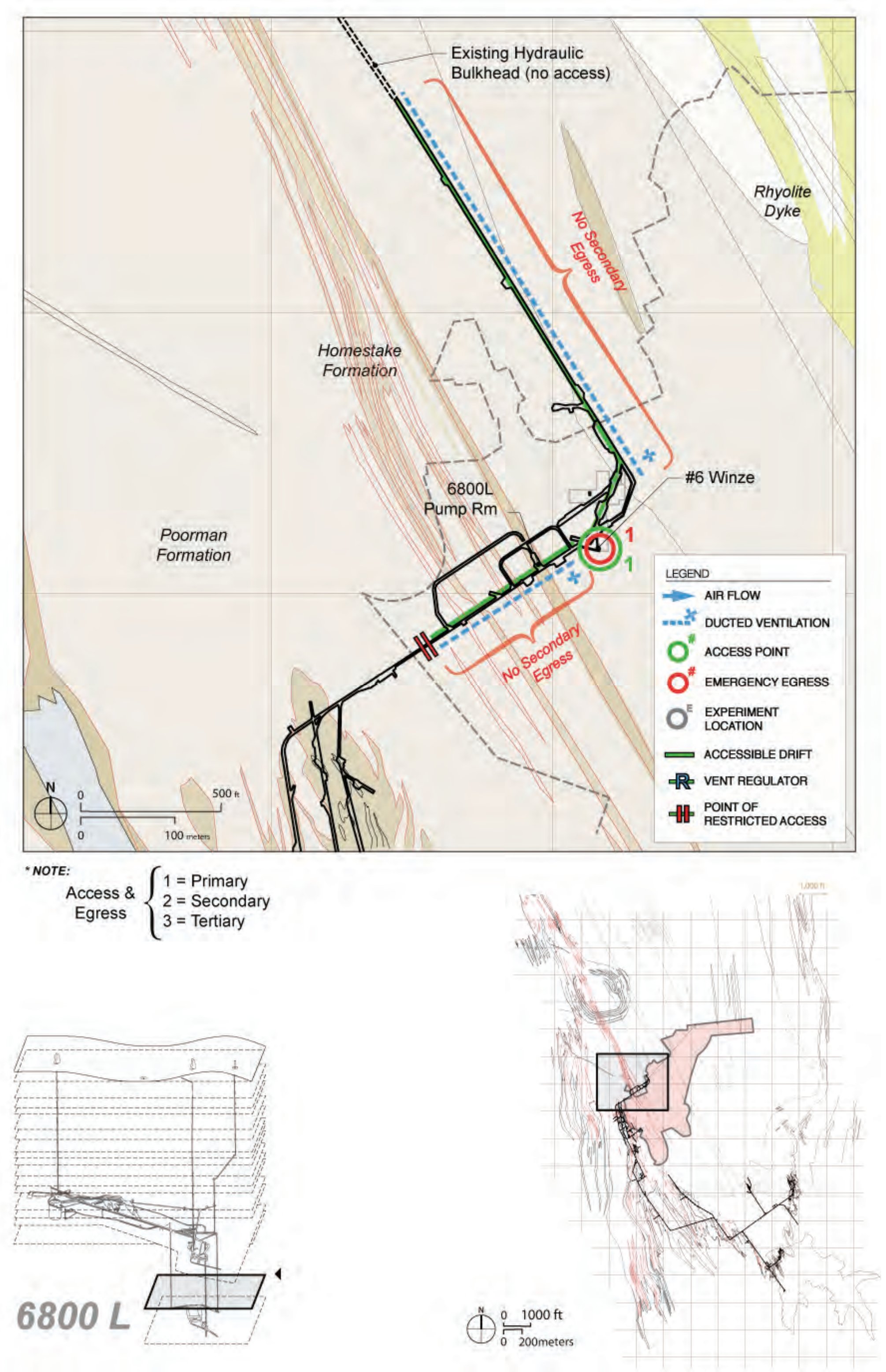

**Figure 5.9.5.10** 6800L proposed experiment locations. [DKA]

### 5.9.5.11 7400L

Details regarding design for the 7400L can be found in Chapter 5.8, *Deep-Level Laboratory Design at the 7400L (DLL)*. However, part of this campus includes a deep drill room, and is part of the OLR design scope. The deep drill room is intended to be used for drilling to extreme depths—well below where biological life has been studied previously.

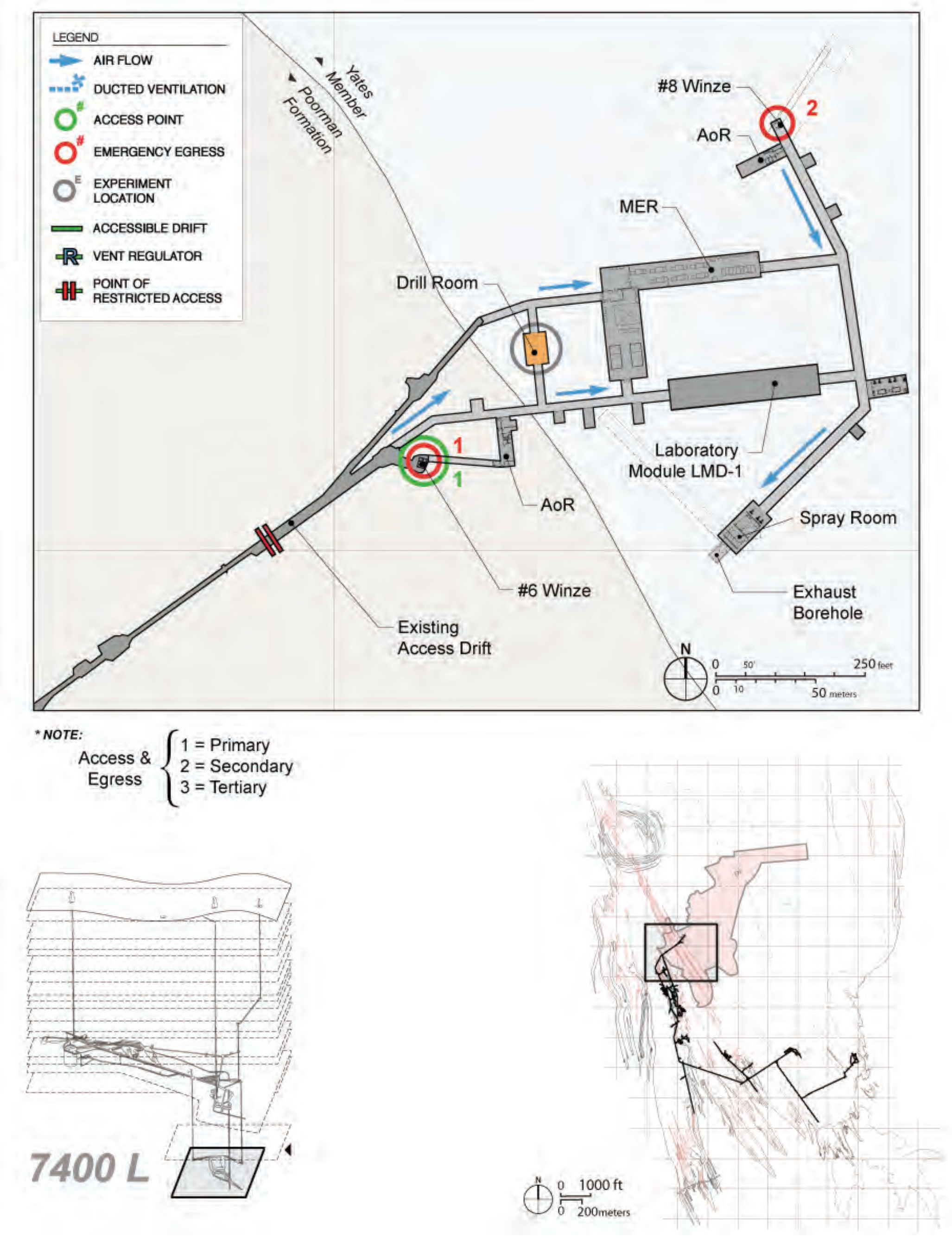

**Figure 5.9.5.11** 7400L proposed experiment locations. [DKA]

Similar to the 6800L, ground conditions at this level are unknown as of the completion of Preliminary Design. It is anticipated to require substantial rehabilitation, not only because it has been submerged in water, but also because of the high stress levels this far below grade. Since access has not been available to date, no ground control designs have been developed or proposed. Detailed geotechnical site investigations conducted upon safe re-entry will determine ground control requirements.

### 5.9.6 Proposed Experiments within OLR

Table 5.9.6 shows a list of potential experiments currently under consideration at all levels in the Project. Some examples of specific experiment installations can be found in Figures 5.9.6-1 and 5.9.6-2. Refer to Sections 3.3.7 and 3.8.3 for more details on proposed experiments within OLR.

| Level | Site | Experiment |
|---|---|---|
| 300 | 1 | $CO_2$ Sequestration |
| | 2 | EcoHydrology |
| | 2 | Transparent Earth (HPPP, MicroGravity, SQUID) |
| 800 | 3 | $CO_2$ Sequestration |
| | 1, 4 | EcoHydrology |
| | 2 | Transparent Earth (HPPP, MicroGravity, SQUID) |
| 2000 | 1, 3, 5 | EcoHydrology |
| | 3 | Fractured Processes |
| | Distributed | GEOX$^{TM}$ |
| | 1, 2, 5 | Transparent Earth (Broadband Seismic Array) |
| | 1, 2, 5 | Transparent Earth (Earth Passive EEM1) |
| | Distributed | Transparent Earth (Earth Electrical Array) |
| | 2, 4 | Transparent Earth (HPPP, MicroGravity, SQUID) |
| 4100 | 2, 4, 6 | EcoHydrology |
| | Distributed | GEOX$^{TM}$ |
| | 1, 3, 4, 6 | Transparent Earth (Broadband Seismic Array) |
| | 1, 3, 4, 6 | Transparent Earth (Earth Passive EEM1) |
| | 3 | Transparent Earth (Earth Passive EEM2) |
| | Distributed | Transparent Earth (Earth Electrical Array) |
| | 1, 3, 6 | Transparent Earth (Active Seismic Monitoring) |
| | 1, 3, 4, 5, 6 | Transparent Earth (HPPP, MicroGravity, SQUID) |
| 4550 | 1 | Transparent Earth (Broadband Seismic Array) |
| | 1 | Transparent Earth (Earth Passive EEM1) |
| | 1 | Transparent Earth (Active Seismic Monitoring) |
| | 1 | Transparent Earth (USGS Calibration Site) |

| Level | Site | Experiment |
|---|---|---|
| 4850 | 1 | Cavity Design |
| | 1 | Cavity Monitoring |
| | 3 | Coupled Processes |
| | 2, 6 | EcoHydrology |
| | 3 | Fractured Processes |
| | Distributed | GEOX™ |
| | 5, 6 | Transparent Earth (Broadband Seismic Array) |
| | 5, 6 | Transparent Earth (Earth Passive EEM1) |
| | 5, 6 | Transparent Earth (Active Seismic Monitoring) |
| | 4 | Transparent Earth (Earth Passive EEM2) |
| | 4 | Transparent Earth (Active Seismic Stress) |
| | Distributed | Transparent Earth (Earth Electrical Array) |
| 6800 | 4 | Transparent Earth (HPPP, MicroGravity, SQUID) |
| | Distributed | GEOX™ |
| 7400 | 1 | Transparent Earth (HPPP, MicroGravity, SQUID) |
| | 2 | EcoHydrology |
| | 2 | Fractured Processes |
| | Distributed | GEOX™ |
| | 3 | Transparent Earth (Broadband Seismic Array) |
| | 1 | Transparent Earth (HPPP, MicroGravity, SQUID) |

**Table 5.9.6** List of potential DUSEL experiments.

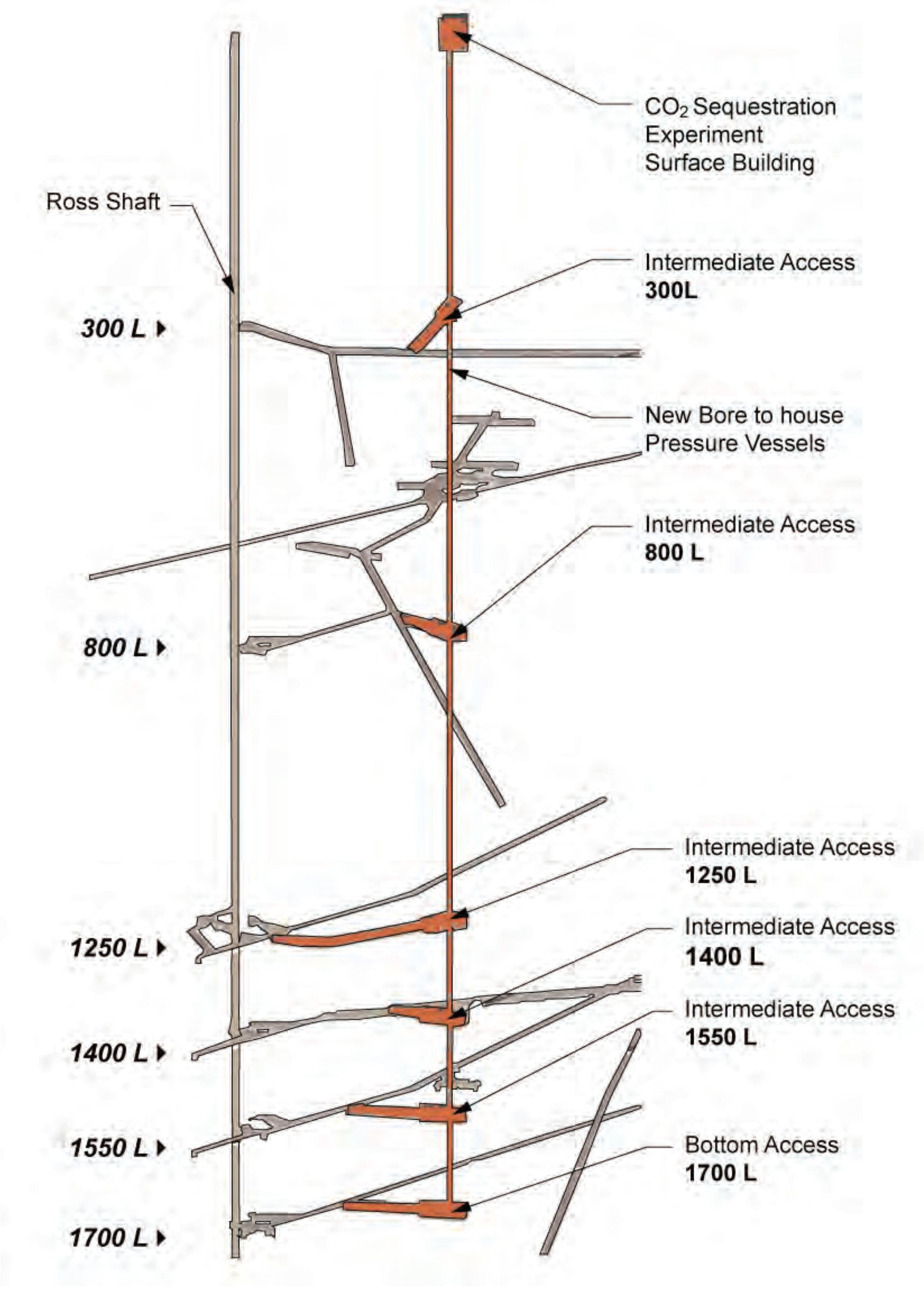

**Figure 5.9.6-1** $CO_2$ Sequestration experiment. [DKA]

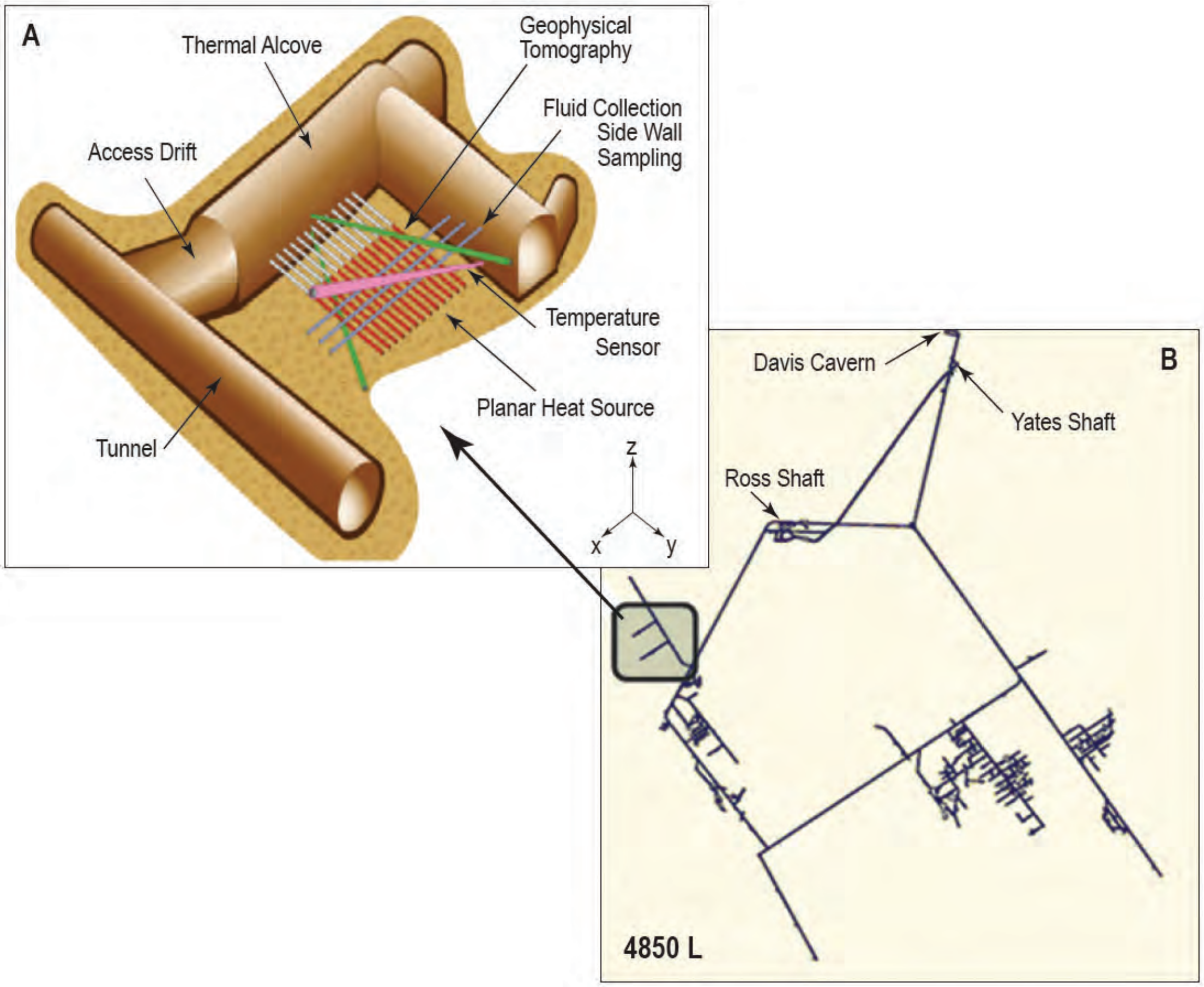

**Figure 5.9.6-2** Proposed Coupled Thermal-Hydrological-Mechanical-Chemical-Biological Processes Experiment (THMCB).

### 5.9.7 Scope Options and Scope Contingency

The only scope option for OLR is expansion of the footprint. This expansion has not been defined or requested by any experiment as of the PDR. As discussed at the beginning of this section, the nature of experiments planned for OLR is such that the scientists' desire is to have access to as much space as possible. The spaces have been limited in an effort to control costs of rehabilitation and maintenance. Some room for expansion will exist in the electrical infrastructure installed due to the code requirements and available sizes of transformers. Access to areas outside of the defined footprints will require careful inspection and review of the existing ground support, availability of power and communications, ventilation, infrastructure, and storm-water diversion flow patterns. If areas are identified that would benefit science and also use the existing infrastructure, expansion will be determined based on funding available for ground support refurbishment and other maintenance and operation considerations. No options are presented as part of the PDR for OLR.

The only contingencies carried for the OLR scope are those dictated by responsible engineering design and code requirements for electrical infrastructure. The ground support recommendations are conservative designs that are anticipated to provide for the life of DUSEL with routine maintenance.

## 5.10 Final Design and Construction Acquisition Plan

The facility designs developed during the Preliminary Design phase represent approximately 30% of the maturity expected for the final DUSEL construction-ready documents. This section discusses the work required to progress from the Preliminary Design phase through Final Design and Construction. The progression includes the development and execution of an acquisition plan to ensure an effective bidding process by creating bid packages, bidding the work, and careful selection of contractors. The foundation of this plan is the construction sequencing and schedule. In addition to the baseline scope outlined in this Preliminary Design Report (PDR), a defined set of scope options has been identified through the Preliminary Design phase and will be considered as options during Final Design and Construction. The following discussion addresses these various topics to provide a complete picture of the DUSEL approach to design completion, acquisition, and construction.

### 5.10.1 Summary of Deliverables for Final Design

Final Design activities will commence in February 2012, with appropriate review and approval by the National Science Board. Between the submission of the PDR and the commencement of Final Design, the Facility team will refine facility requirements and Trade Studies, and consider scope options in preparation for Final Design. As the Integrated Suite of Experiments (ISE) designs mature during this period, it is expected that science requirements will refine and the Project will continue to manage these requirements to inform the Final Design process. The ISE requirements-gathering and documentation process used during the Preliminary Design phase will be used during this transition period to ensure that the Facility design firms start Final Design with current requirement and interface definitions. This process is outlined in Volume 3, *Science and Engineering Research Program*, and Volume 9, *Systems Engineering*.

Design for the surface scope of work will advance during 2011 contingent upon adequate funding, which will allow all work scopes for the surface and underground to be developed concurrently during Final Design. The design for the Deep-Level Laboratory (DLL) campus will advance to support construction after the 7400L has been dewatered and safe access has been established to support detailed site investigations.

Similar to the approach used during the Preliminary Design phase, the deliverables for Final Design will include the following major deliverables, with approximate dates:

1. Detailed Basis of Design documents addressing all components of the surface and underground designs (March 2012)
2. 60% Final Design (August 2012)
3. 90% Final Design (January 2013)
4. 95% Final Design: Draft construction documents will be used to support the bidding process (April 2013)
5. 100% Final Design: Detailed construction drawings and specifications will accompany the signed construction contracts (September 2013)

Detailed cost estimates and schedules will be provided for overall project planning and coordination at each Final Design stage leading to the bidding process. Value exercises will be completed at each stage to verify a cost-effective design that is within the funding constraints.

During the Preliminary Design phase, a review of the Facility scope elements, the envisioned construction packaging approach, and the acquisition strategy for each package was performed to develop the project delivery approach for DUSEL. The Project, working in conjunction with the DUSEL design contractors and Construction Manager (CM), decided that a design-bid-build approach was most appropriate for the majority of the DUSEL work—particularly to address the complexities of the underground civil construction scopes. The intricacies of the expected interfaces between the science experiments and the facility require close coordination among the designers, DUSEL Project, experimenters, and the CM. The Project and supporting contractors regard the design-bid-build approach as most appropriate for the majority of the facility construction scopes. The early involvement of the CM during the preconstruction phases, including the Project's extensive use of independent estimates and constructability reviews by the CM, address the potential drawbacks of the design-bid-build approach.

In contrast, the Project has chosen a design-build approach for the Ross Surface Campus. Since the Ross Surface Campus primarily supports construction and facility maintenance efforts for the foreseeable future, a design-build approach provides the CM and its subcontractors more flexibility in designing a flexible and efficient construction support area that best meets their needs along with meeting the requirements of the DUSEL Operations Team for Facility maintenance.

### 5.10.2 Acquisition Plan

The acquisition plan defines the process of identifying qualified contractors to bid and safely perform the work within the cost and schedule targets established by the Project. The first component of the acquisition process, which takes place during Final Design, is performing market research to understand the current construction market, available qualified contractors, and the economic conditions that may impact the bidding process. Standardized market research forms are distributed by the CM to potential contractors to develop a bidder list that is used as bid packages are developed and distributed to industry. Competitive bidding is a requirement for the Project, and this process ensures that the groups competing for the work are qualified, have acceptable safety records, and are reputable. Efforts will be made to ensure minority, women-owned, and small or disadvantaged businesses are included in the market research and acquisition process.

A contractor prequalification process follows the market research analysis phase. Prequalification includes a more thorough evaluation of each potential contractor's safety and health records, financial history, and work history. Prequalification will be utilized where a two-step acquisition approach is appropriate for critical elements of construction such as underground related work that has specific safety requirements and requires an extensive evaluation of a contractor's safety performance record. A sample prequalification form is included in the McCarthy Kiewit deliverable included in Appendix 2.B.

Bid packages will be developed during the later stages of Final Design to support the bidding process. An initial set of planned bid packages was developed during Preliminary Design and will be refined up to the start of the bidding process. Bid packages divide the Project into subprojects based on scheduled starts, schedule for design completion, and phasing of work. Packages may include a variety of disciplines such as electrical, mechanical, and excavation. For example, one bid package requiring multiple disciplines or trades is the early Ross Shaft utility work, which includes both electrical cable installation and mechanical plumbing installation. This bidding structure allows for a diverse set of bid packages without creating the potential for conflicts in specific areas and times of the project. Packages are also formed to generate interest by the industry, resulting in a more competitive bidding process. Twenty packages are currently identified in the preliminary acquisition plan for the MREFC-funded Project. An additional 10 packages that were developed as part of the overall design but are operations- and deferred-maintenance-related activities will be performed using Research and Related Activities (R&RA) funding, as discussed in Volume 10. Each bid package will be reviewed as a separate project; however, multiple bid packages may be awarded to one contractor. It is also possible that every bid package could be awarded to separate contractors. The following list of bid packages is organized by facility level and design contract.

**Identified Bid Packages**

**4850L Mid-Level Laboratory (MLL) Campus and Upper Levels**

- Specialized Equipment
  - S1—Prepurchase Early Specialized Equipment
  - S2—Install Early Specialized Equipment
  - S3—Prepurchase Specialized Hoisting Equipment
  - S4—Install Specialized Hoisting Equipment
- Infrastructure
  - A1—Early Ross Shaft Mechanical, Electrical, and Plumbing (MEP) Work
  - A2—Main Infrastructure Work
  - A3—LM-1/LM-2 (Laboratory Modules) Work
  - A4—LC-1 (Large Cavity) LBNE work
- Excavation
  - G1—Main Excavation (includes all levels/areas except 7400L and Long Baseline Neutrino Experiment [LBNE] work)
  - G2—LBNE work
- Surface
  - H1—Procurement and Early Equipment Prepurchase
  - H2—Early Underground Infrastructure Work and Utilities
  - H3—New Assembly Building
  - H4a—Renovation of Existing Structures (Design/Build)
  - H4b—Renovation of Existing Structures (Design/Bid/Build)
  - H5—New Education and Public Outreach Visitor Center and Surface/Site Improvements

**7400L Deep-Level Laboratory (DLL) Campus**

- Specialized Equipment
  - S5—Prepurchase and Install Specialized Hoisting Equipment
- Infrastructure
  - A5—DLL Infrastructure
  - A6 —DLL Laboratory Construction
- Excavation
  - G5—Main Excavation (includes all levels below 4850L except LBNE-specific excavation)

Once the bid packages are assembled, they will be distributed in accordance with Project schedule and acquisition policies. These bid packages will include the following items:

- Invitation to Bid
- Work Category Breakdown
- Additional Subcontractor Conditions
- Bid Proposal Form
- Bid Bond Form
- Performance and Payment Bond Forms
- Customized Master Subcontract
- Purchase Order Documents
- Funding Source Flow-Down Requirements
- Insurance Requirements

To limit “risk of loss,” the University of California system will require DUSEL subcontractors to be bonded, and all subcontractor bonds must be provided by U.S. Treasury Approved Sureties. Submission deadlines will be established for all bids, after which the bids will be reviewed and evaluated using the specific selection criteria identified for the bid package, including face-to-face interviews as appropriate and past-performance reference checks. Additional acquisition plan details, including samples of the forms discussed in this section, can be found in section IV of the 100% Preliminary Design Phase McCarthy Kiewit Deliverables (Appendix 2.B).

### 5.10.3 Rehabilitation and Construction Schedule

The DUSEL Facility construction schedule has been developed to efficiently establish science access for the ISE while minimizing impacts of construction on the early science experiments. Working on levels at depths up to 7,400 feet below the surface, with schedules constrained by the physical limitations of conveyances to deliver personnel and materials, demands significant logistical requirements and coordination. The discussion below addresses key points in the overall construction sequencing to achieve science access to the underground laboratories; for a detailed schedule, see the 100% Preliminary Design Phase McCarthy Kiewit Deliverables (Appendix 2.B). Section 5.3.3.5.5, *General MLL Excavation Sequence,* provides a written description of the excavation sequence as well.

The early science programs located on the 4850L near the Davis Campus will be operational during DUSEL construction with periods of downtime during key construction periods. The Facility team and early science collaborations have worked together to develop an excavation sequence optimized for minimal interruptions in early science operations. Davis Campus access is currently provided via the Ross Shaft, with fresh air provided via the Yates Shaft. Power and communications are currently provided from the Ross Shaft, while potable and industrial water are provided through the Yates Shaft. Through work completed by Sanford Laboratory and non-MREFC funding sources, alternative paths for electrical, communication, and water services will be provided across the 4100L to new boreholes directly to the Davis Campus prior to MREFC-funded construction. A chilled-water system will be installed near the Davis Campus in the East Access Drift, with services crossing the area known as the Big X, where the East and West Access Drifts intersect (Figure 5.10.3-1).

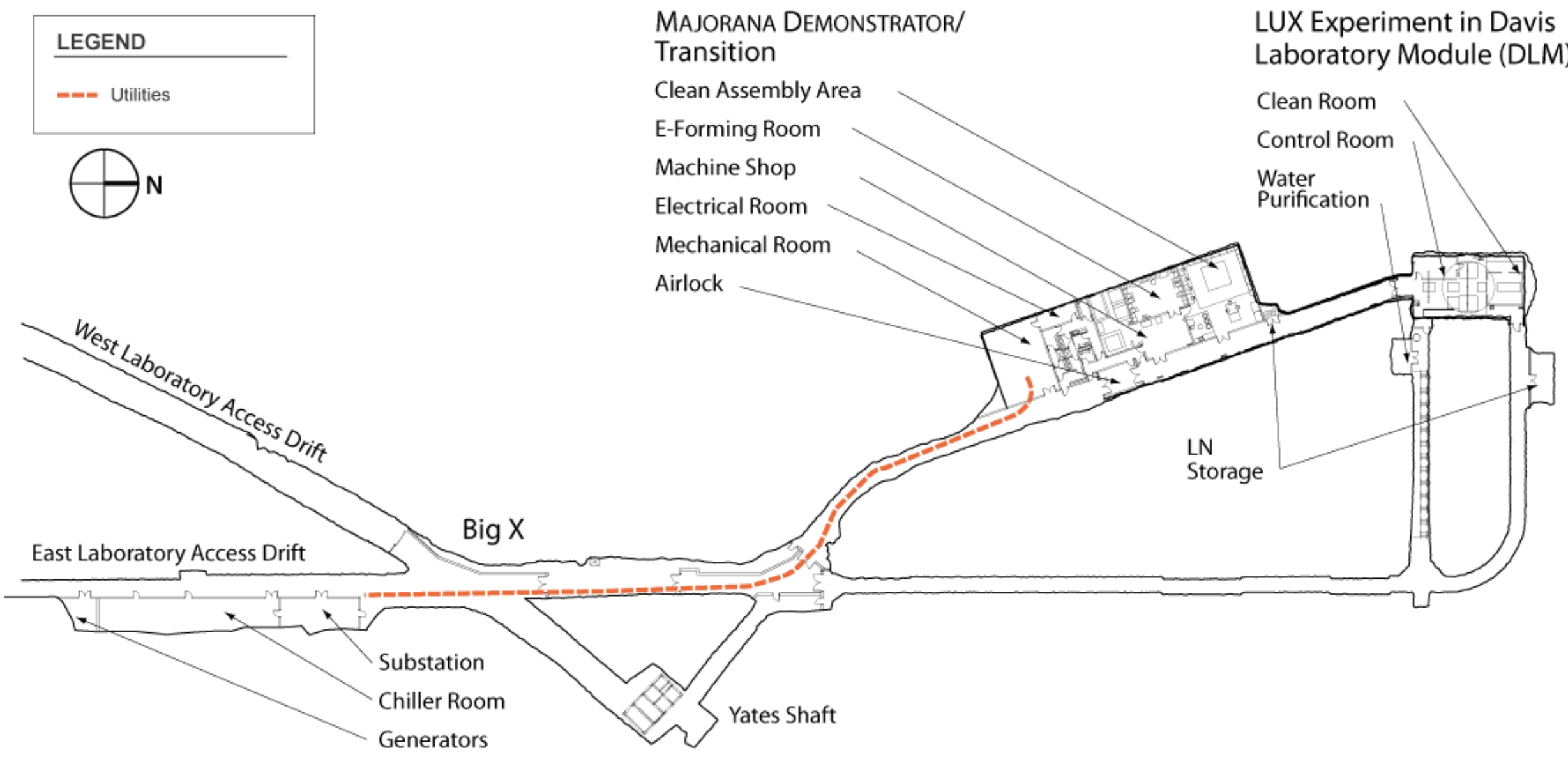

**Figure 5.10.3-1** Davis Campus and the Big X. [J. Willhite, W. Zawada, DUSEL; DKA]

DUSEL excavation will begin at the Ross Shaft and progress along the West Access Drift toward the Yates Shaft to provide a mucking passage. During excavation of the West Access Drift, science operations at the Davis Campus will be suspended because safe access cannot be provided through an active excavation area. To minimize the overall impact on the Davis Campus experiments, the excavation sequence will be conducted concurrently at a second location, the Big X. Once the Big X area and the access drifts to the new ventilation borehole are completed and utilities are restored, experiments can resume and access will be available from the Ross Shaft around the #4 Winze Wye, and along the East Access Drift as shown in Figure 5.10.3-2. This will continue to be the access path until the Yates Shaft rehabilitation is completed, at which point it becomes the primary access for the Davis Campus. Utilities will be connected from the MLL Campus to the Davis Campus at the experiment's convenience and has no impact on the construction schedule.

In addition to the Davis Campus, several biology, geology, and engineering (BGE) experiments will be in operation during the construction phase of the Project, on the MLL, DLL, and Other Levels and Ramps (OLR). The impact of construction on the BGE experiments will vary depending on the nature of the experiments and distance from construction activities; requirements to protect each BGE experiment are unique and will be reviewed with the DUSEL science liaison and collaboration to mitigate negative impacts as feasible. In some cases, the experiments may benefit from studying the effects of construction.

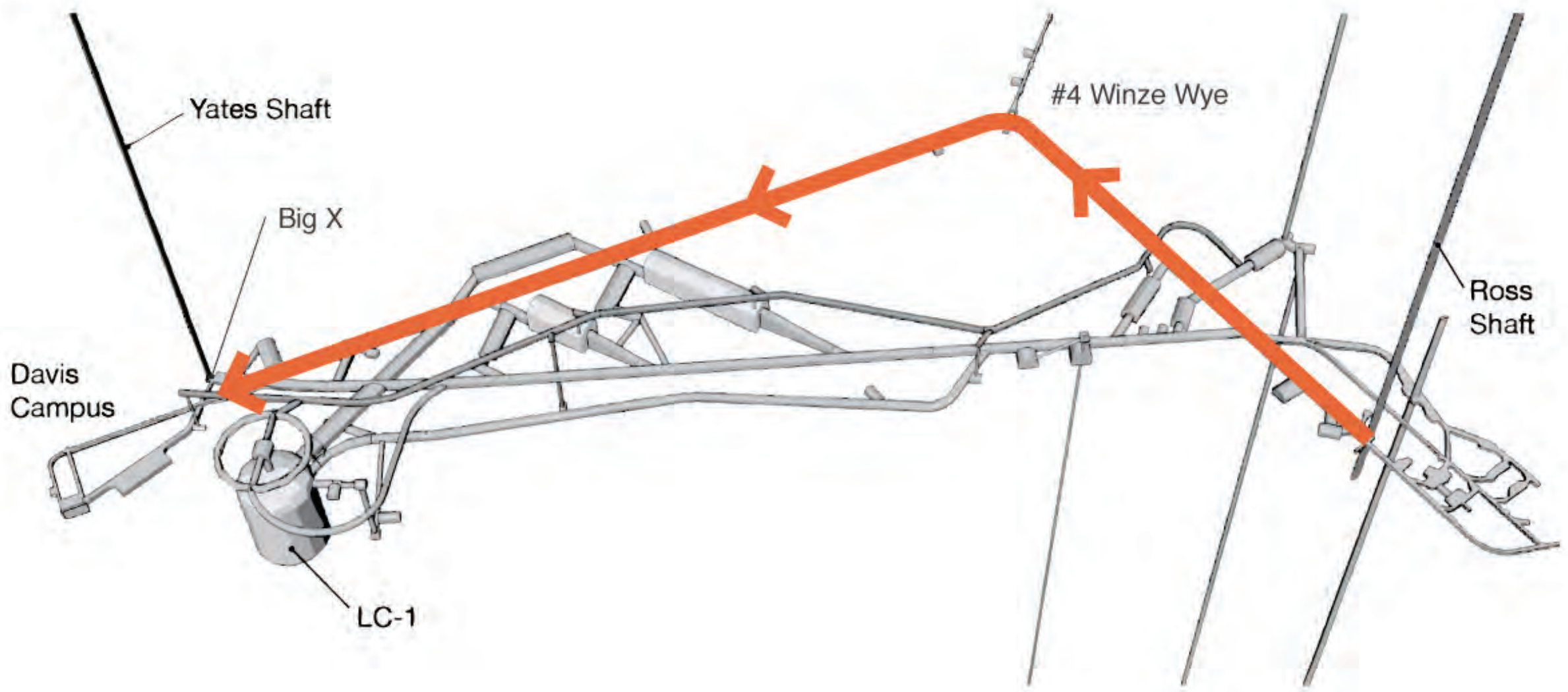

**Figure 5.10.3-2** Access path to Davis Campus prior to completion of Yates Shaft rehabilitation. [DKA]

After establishing safe access to the Davis Campus, the excavation sequence then focuses on maximizing excavation crew efficiency and productivity through multiple face developments and optimizing infrastructure development paths. This allows both excavation and infrastructure installation to occur simultaneously, reducing the overall construction duration. Details of the excavation sequence are discussed in Chapter 5.3, *Geotechnical Site Investigations and Analysis*, along with an overview of the sequence for each level. In summary, the construction sequence critical path progresses as follows:

### Construction Sequence

### Facility Rehabilitations Activities (R&RA-Funding)

- Prior to commencement of the MREFC-funded construction Project, several key operations and maintenance rehabilitation projects will be completed with R&RA funding.
  - Ross Shaft rehabilitation, including the waste rock skipping system
  - Yates Hoist rehabilitation
  - Waste rock handling system rehabilitation, including civil works
  - #6 Winze Hoist rehabilitation to allow access for 7400L geotechnical evaluations
  - Perform deferred maintenance on surface buildings such as roof repair and tuck pointing

### 4850L MLL Campus and Upper-Level Construction Activities (MREFC-Funding)

- Yates Shaft rehabilitation, including installation of plumbing, electrical, and communications
- Ross Shaft utility installation (plumbing, electrical, communications) is completed with MREFC funding prior to waste rock handling system commissioning, as this work cannot be done while skipping rock.

- Waste rock handling system installation of new equipment (tramway dump station and pipe conveyor)
- Excavation begins when the waste rock handling system is completed as described above to minimize early science impact.
- The ventilation path is established through the new borehole from the 4850L to the 3950L and associated excavations while the path for early science is established.
- Excavation proceeds through the West Drift from Ross to Yates.
  - Excavation of the 4700L ventilation drift and the 5060L large cavity access drift occurs concurrently with the 4850L West Drift excavation.
  - Once sufficient separation distance between excavation and locations where infrastructure can be installed is achieved, infrastructure installation will begin at the Yates Shaft and develop toward the Ross Shaft.
- As each laboratory module (LM) is reached, crews alternate between slashing the drift and excavating the LMs, blasting in one area and moving to the other while material is removed.
- LC-1 major excavation begins as soon as the 5060L ramp reaches the bottom of the large cavity excavation envelope.
- Once the West Drift is completed, the East Drift is slashed from the Davis Chiller area toward the 4850L LMs.
- 4850L utilities are installed as soon as the spaces become available.
- LM-1 and LM-2 are available for outfitting when commissioning of MEP systems begins.
- LC-1 is available for equipment installation as soon as excavation is completed.

**7400L DLL Campus Construction Activities (MREFC-Funding)**

- #6 Winze rehabilitation completes approximately when LC-1 excavation begins on the 4850L.
- 7400L development begins when the #6 Winze rehabilitation completes.
- 7400L excavation establishes a path to the egress and ventilation boreholes as the borehole pilot holes are drilled. The 7400L drift development should reach the pilot holes near the time that the pilot holes reach the 7400L.
- 7400L development continues by slashing out the LM from the excavation already made to establish ventilation, and adding the utility rooms.

**Surface Construction Activities (MREFC-Funding)**

- Establishing utilities and buildings to support the underground construction begins when MREFC funds are available.
- Development of the surface buildings for science use is concurrent with underground laboratory experiment installation to reduce potential damage during construction with high volumes of materials stored and transported through both shafts.

### 5.10.4 Facility Scope Options

The Preliminary Design for the DUSEL Facility has established a scope baseline from which cost estimates and construction schedules are developed. Within this baseline is a scope that meets the defined facility requirements. In the process of developing this baseline, many decisions were made based on analysis and Trade Studies. The design has been optimized through multiple rounds of Value Engineering to provide a design baseline that meets the science requirements at the lowest reasonable cost.

Facility scope options are additional capabilities that could be included in the facility baseline to meet science requirements but do not fit within the currently allocated project cost and schedule baseline. Many of these items were once included in the design, but were removed because they were not required to support science. These scope options would be considered, as funding is available. A summary of scope options for each main area in the Facility is described in Table 5.10.4. Many of these items are discussed throughout Volume 5 as noted in the cross-reference column.

| Scope Description | Estimated Cost (Costs include directs and indirects, but no Management Reserve) (in FY 2010 $M) | Section Cross Reference |
|---|---|---|
| **Ellison Hill Access Road and Parking**<br>A new Ellison Hill access road would provide safer access to the site, particularly during the winter season, and would reduce the Yates Campus vehicle traffic through residential areas along Mill Street and East Summit Street. Parking could also be provided along this road. | $7 | 5.2.8 |
| **Commons Building**<br>A Commons building, consisting of a 275-seat auditorium lecture space, meeting and breakout rooms, and full food-service cafeteria, could provide additional science, education, and outreach capabilities. | $4-5 | 5.2.8 |
| **Additional Administrative and Science Offices, Laboratory Space**<br>Additional administration and science offices as well as laboratory space would support a larger science community. (Note for review committee: FAC team will further develop this option to provide more specific information.) | $1-2 | 5.2.8 |
| **Expanded Sanford Center for Science Education (SCSE)**<br>Additional exhibition space in the SCSE would enhance the education and outreach capabilities. (Note for review committee: FAC team will further develop this option to provide more specific information.) | $1.5 | 5.2.8 |
| **Additional Underground Laboratory Modules**<br>Additional laboratory modules would provide additional science space at either the 4850L or 7400L. | $30<br>per module | 5.6 and 5.8 |
| **Increased Laboratory Module Dimensions**<br>Laboratory module dimensions would be increased to provide additional science space at either the 4850L or 7400L. | $5-20<br>per module depending on dimension changed | 5.6 and 5.8 |
| **Enhanced Laboratory Module Finishes**<br>Better laboratory module finishes, including a finish coat of smooth shotcrete and paint, would improve the cleanliness of the environment in the laboratory modules. | $1<br>per module | 5.6 |
| **Redundant Electrical Power Systems**<br>Redundant normal power routed through both the Yates and Ross Shafts for the 4850L and the #6 Winze and #8 Winze for the 7400L would provide single fault tolerance in the normal power system to ensure laboratory modules' power could be maintained with increased reliability. Note that the design for this is included in the 100% Preliminary Design, but the cost excludes it. | $5 | 5.4.5.9 |
| **Surface Generator to Support the 7400L**<br>A surface generator for the 7400L would remove the two generators currently included on the 7400L and replace them with surface-based generators. This option would eliminate the need for diesel storage on the 7400L, along with reducing the exhaust fumes that will be generated during emergencies. | $2 | 5.4.5.9 |

| **Scope Description** | **Estimated Cost** (Costs include directs and indirects, but no Management Reserve) (in FY 2010 $M) | **Section Cross Reference** |
|---|---|---|
| **Yates Shaft Auxiliary Hoist**<br>An auxiliary hoist in the Yates Shaft would provide a dedicated personnel conveyance while the Supercage is in use for material transportation, as well as additional emergency egress capability. Note that this is included in the 100% Preliminary Design, but excluded from the cost estimate. | $9 | 5.4.5.1 |
| **Modified #8 Winze Design to Allow for Primary Access Use**<br>The #8 Winze is designed for emergency egress only, as it lacks some capabilities required to classify it for primary access. Modifying the design to allow primary access would provide additional access to the 7400L while the #6 Winze would then be used for material delivery. The design could also be modified to provide a larger conveyance, enhancing both material delivery and emergency egress capability. This option was included in the initial Basis of Estimate. | $22 | 5.4.5.2 |
| **Monolithic Concrete Lining for the Yates Shaft**<br>The current Yates Shaft design includes a shotcrete lining; a full monolithic concrete lining would reduce future maintenance of this shaft. This option was removed after 60% Preliminary Design. | $11 | 5.4.5.1 |
| **New Lined Shaft to Replace the Oro Hondo Shaft**<br>The Oro Hondo Shaft provides primary ventilation but has continuously spalled since it was constructed, requiring routine material removal. A new concrete-lined borehole to replace the Oro Hondo Shaft would reduce the need for this ongoing material-removal process and would reduce future operations costs. Another option would rehabilitate the existing Oro Hondo Shaft and install a new lining. It is anticipated that both options are approximately the same cost. | $15 | 5.4.5.7 |
| **Architectural Finish Levels in Underground Campuses**<br>Improved architectural finishes throughout the underground facility in drifts, Areas of Refuge (AoRs), and ancillary spaces, including smoother shotcrete, paint, and other improvements. This would improve the overall appearance and reduce ongoing maintenance activities throughout the Facility. This option was removed after 90% Preliminary Design. | $2 | 5.4.5.5 |
| **Radon Control Measures in Underground Facilities**<br>Low radon control measures, including a supply duct and specialized concrete and shotcrete, would reduce background radiation, creating an enhanced environment for some of the anticipated science activities. The supply duct was removed after the 30% Preliminary Design. | $5 | 5.4.5.7 and 5.3 |
| **Underground Sewage Treatment**<br>Including underground sewage treatment in the Facility design would reduce the amount of sewage material that will need to be regularly removed by Operations staff after dewatering the solids underground. This scope was removed after 60% Preliminary Design. | $3 | 5.4.5.14 |
| **Expanded OLR Footprints**<br>An expanded OLR footprint would provide additional experimental space as well as improved diversity for geological study. (Note for review committee: FAC team will further develop this option to provide more specific information) | >$5 | 5.9 |

**Table 5.10.4** Facility Design Scope Options.

### 5.10.5 Future Value Engineering

At the Preliminary Design stage of the Project, a variety of options have been evaluated to optimize the design within the anticipated budget through the Value Engineering process. Additional options will be suggested as the Project progresses into the Final Design and will follow the same Value Engineering process described in Section 5.1.6.4, *Value Engineering*, and in Volume 9, *Systems Engineering*. Examples of Value Engineering opportunities that are in the process of evaluation as this PDR is being developed include the waste rock handling system location and the location of services for LC-1. The waste rock handling system is centered at the Ross Campus in the current design to provide separation between the science access and the waste rock handling system, allowing for future Facility development without impact to science access. Providing this functionality at the Yates Campus presents opportunity for significant savings over the current approach but may cause undesired interference between construction and science. It also may provide the opportunity to reduce the amount of rehabilitation work required for the Ross Shaft, which could represent a significant cost savings or at least allow for the rehabilitation and thus the cost of the Ross Shaft to be spread over time. It has not been fully evaluated. The LBNE has offered the opportunity to relocate some equipment into LC-1. If this proves viable, the Yates mechanical and electrical equipment could be relocated, reducing the size or possibly eliminating a large excavation. The results of the analysis of these and other options will continue to optimize the design throughout the Final Design and into Construction itself.

### 5.10.6 Conclusion

Volume 5 outlined the design approach and team structure, current site conditions at Homestake, and the Facility Preliminary Design developed in response to the Facility requirements developed through collaboration with the ISE. The Preliminary Design represents a 30% level of completion and provides a strong foundation for the Final Design, Bidding, and Construction phases to realize an efficient and safe facility to support science and operations at the DUSEL Facility. A sound construction sequence and schedule have been outlined with scope options to address additional science and facility operations requirements, as funding is available.

## Volume 5 References

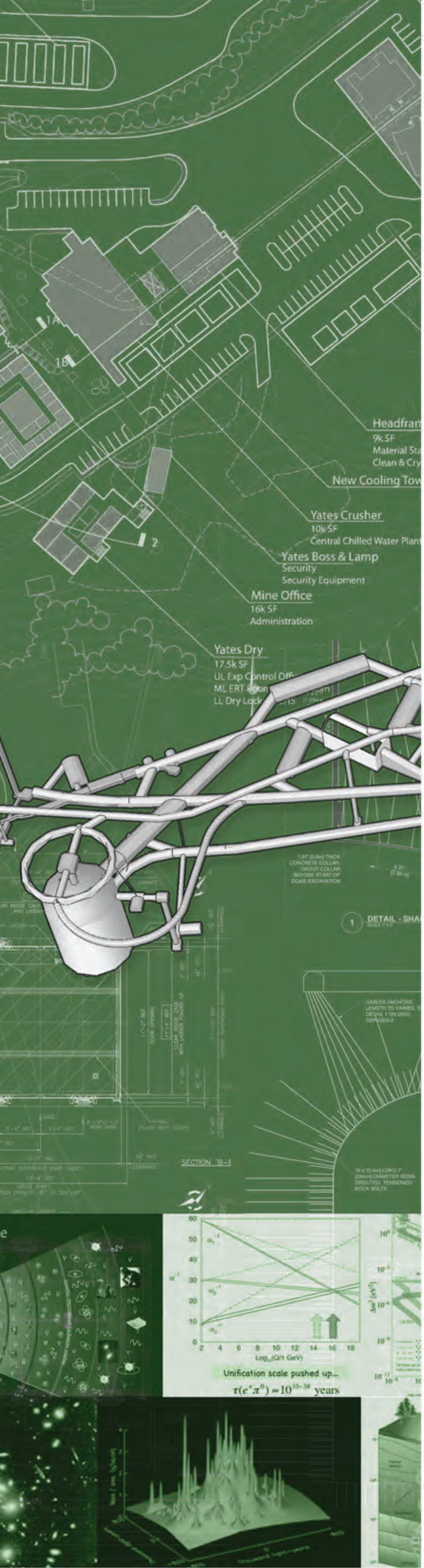

# Preliminary Design Report

May 2011

## Volume 6:

## Integrated Environment, Health, and Safety Management

DUSEL
Deep Underground Science and Engineering Laboratory

This page intentionally left blank

# Integrated Environment, Health, and Safety Management

## Volume 6

The South Dakota Science and Technology Authority (SDSTA), using state-controlled funds, is establishing the underground Sanford Laboratory at Homestake (Sanford Laboratory) in advance of the DUSEL Construction Project. To safely rehabilitate the site and to design, construct, and operate the Sanford Laboratory's modest early-scientific scope, an initial Integrated Safety Management (ISM) system and an Environment, Health, and Safety (EH&S) program have been developed. The development and implementation of these directly benefit the DUSEL design effort and are being used as the starting point for DUSEL's ISM system and EH&S program. The creation of a single EH&S organization to oversee the EH&S activities for both Sanford Laboratory and DUSEL has initiated the process of integrating their management structures and organizations, thereby preparing for DUSEL Final Design. The integration extends through all EH&S activities, including internal and external reviews. This approach permits the DUSEL Project to establish, implement, and improve many of the safety policies and procedures in advance of DUSEL's need for significantly more comprehensive requirements. The EH&S Policies and Procedures (P&P) are uniformly applied to SDSTA and DUSEL employees and contractors. The SDSTA staff, many of whom previously worked for Homestake Mining Company (HMC), has made very significant steps to embrace the safety culture appropriate for a modern research laboratory setting as well as to gain experience in working with and overseeing construction contractors.

This Volume presents Sanford Laboratory's ISM system and EH&S program and the Project's plans to expand them to meet DUSEL's requirements for Final Design, Construction, and Operations. The combined Sanford Laboratory and DUSEL EH&S approach builds upon Sanford Laboratory's substantial early efforts and will continue to evolve as the DUSEL ISM system and EH&S program. Both are essential elements in the development of the DUSEL Facility and experiment designs.

### 6.1 Integrated Environment, Health, and Safety Management System

The DUSEL ISM Project policy holds that everyone is responsible for conducting work and operations in a safe and environmentally sound manner. This expectation is applicable to employees, Facility users, visiting scientists, contractors, and their subcontractors.

Within the scope of this policy, it is the objective of the Project to systematically integrate excellence in EH&S into the management of work practices at all levels so that its mission is achieved while protecting the public, the workers, the environment, and physical and intellectual property, as detailed in Appendix 6.A, *Integrated Safety Management System*. This integration is accomplished by using the principles and core functions of the ISM system to ensure that the overall management of EH&S functions and activities is an integral part of work practices, and to seek improvement in management and performance at every opportunity. Within this policy, it is important to recognize that the use of the word "safety" refers to the reduction or elimination of all hazards, including hazards to health and environment.

An ISM system typically includes seven Guiding Principles and five Core Functions. The ISM system Guiding Principles are:

- **Line Management Responsibility for Safety.** Line Management is directly responsible for the protection of the public, the workers, and the environment. As a resource to line management, the EH&S Department advises, consults, audits, and provides independent feedback to the Project's senior management.
- **Clear Roles and Responsibilities.** Clear and unambiguous lines of authority and responsibility for ensuring safety shall be established and maintained at all organizational levels of the Project, its contractors, and experimental users.
- **Competence Commensurate with Responsibilities.** Personnel shall possess the experience, knowledge, skills, and abilities necessary to discharge their responsibilities.
- **Balanced Priorities.** Resources shall be effectively allocated to address safety, programmatic, and operational considerations. Protecting the public, the workers, and the environment shall be a priority whenever activities are planned and preformed.
- **Identification of Safety Standards and Requirements.** Before work is performed, the associated hazards shall be evaluated and an agreed-upon set of safety standards and requirements shall be established that, if properly implemented, will provide adequate assurance that the public, the workers, and the environment are protected from adverse consequences.
- **Operations Authorization.** The conditions and requirements to be satisfied for operations to be initiated and conducted shall be clearly established and agreed-upon.

The Project's ISM system Core Functions are:

- **Define the Scope of Work.** Missions are translated into work, expectations are set, tasks are identified and prioritized, and resources are allocated.
- **Analyze the Hazards.** Hazards associated with work are identified, analyzed, and categorized.
- **Develop and Implement Hazard Controls.** Applicable standards and requirements are identified and agreed-upon, controls to prevent/mitigate hazards are identified, the safety envelope is established, and controls are implemented.
- **Perform Work within Controls.** Readiness is confirmed and work is performed safely.
- **Provide Feedback and Continuous Improvement.** Feedback information on the adequacy of the controls is gathered, opportunities for improving the definition and planning of work are identified and implemented, line and independent oversight is conducted, and, if necessary, regulatory enforcement actions occur.

Figure 6.1 illustrates attributes for performance assurance provided by the Guiding Principles and the processes for integrating EH&S provided by the Core Functions.

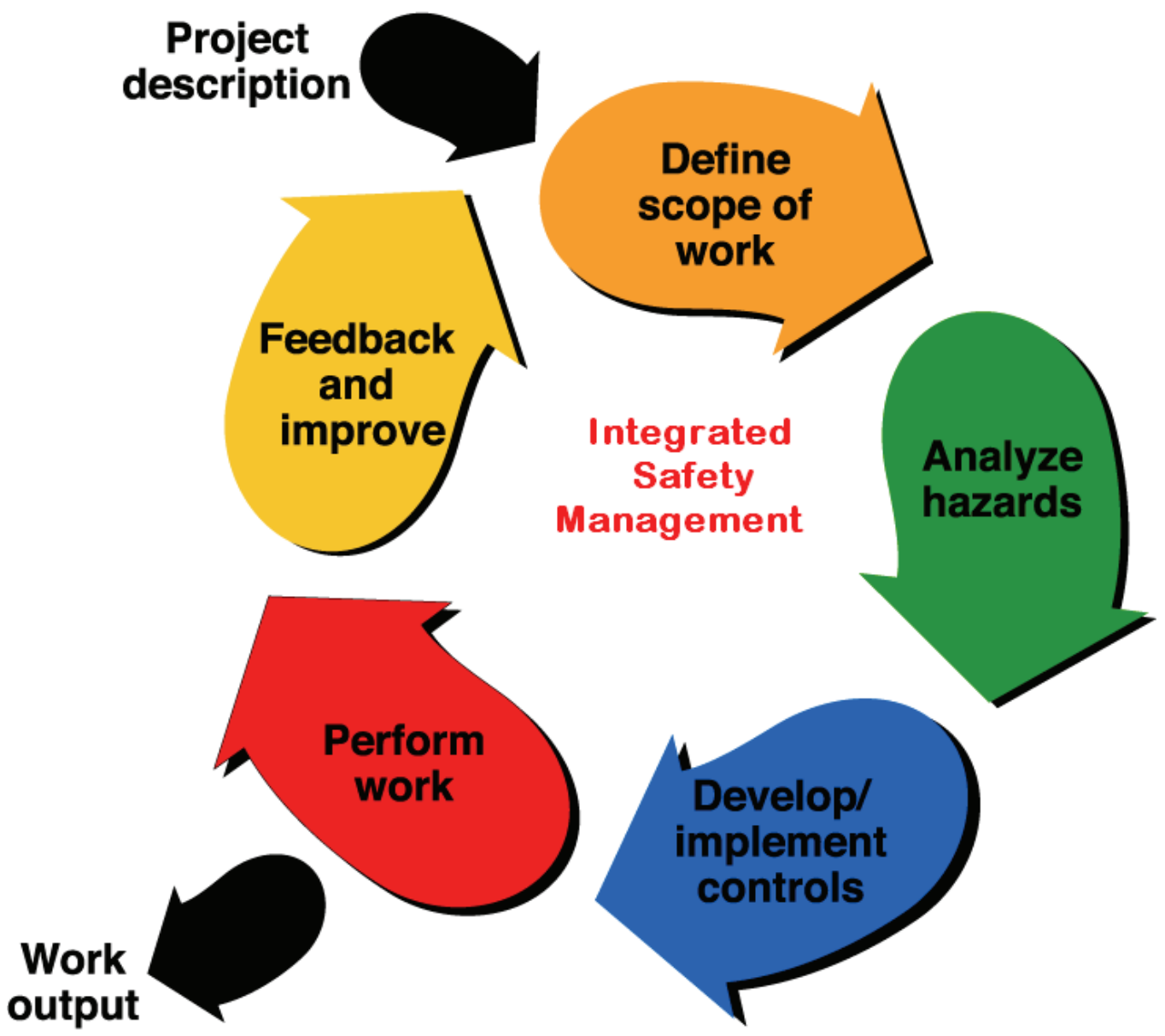

**Figure 6.1** ISM Core Functions describe the processes for integrating safety into all activities.

The *EH&S Manual* contains the set of EH&S P&P and serves as the mechanism for implementing the ISM system. The manual presently contains 31 P&P that provide administrative direction (e.g., contractor safety, job hazard analysis, incident reporting and investigation) or standard methods for controlling known hazards (e.g., oxygen deficiency hazards, confined spaces, compressed gases). See Appendix 6.B., *EH&S Manual Table of Contents*, with hotlinks to the entire EH&S Manual, available online.

Currently, all changes to the ISM P&P, including the creation of new ones, receive a Project-wide review to ensure that hazards are properly mitigated and hazard-control systems within a specific functional area do not conflict with controls established in other functional areas. This practice of rigorous P&P review and approval will continue as the ISM system matures. These reviews include subject-matter experts on the Laboratory Safety Committee drawn from the Science and Operations departments, the professional EH&S staff on site, and third-party subject-matter experts when appropriate. This review process allows experts from various disciplines and technical skills to provide input on the proposed EH&S P&P before they are changed or adopted.

The P&P provide a mechanism for the Project's organizations, or departments, to tailor the institutional EH&S procedures to meet the needs within organizations when necessary. There are variations in implementation at the department level or activity level only when required by the nature of the operations. To reduce implementation variations, the departments participate in the development of EH&S procedures and considerable effort is made to obtain universal buy-in as the procedures are developed. Uniformity is imposed where implementation by one organization may have a negative impact on another (e.g., training requirements and traffic enforcement are nearly universal).

The foundation of ISM is line responsibility; the line organization must have the authority, responsibility, and be held accountable for integrating EH&S into all work it does. Line responsibility for EH&S is woven into the organizational structure and all aspects of the EH&S program.

The Project's P&P identify the EH&S responsibilities of all employees. They further call out the roles and responsibilities of all parties (including management, various EH&S personnel, other staff with special

EH&S responsibilities, contractors, and visiting scientists) while performing work at or otherwise using the Sanford Laboratory facility.

A key to balancing priorities is to ensure that those who make the decisions are authorized to do so and that they have accurate information about the nature of the work, the hazards, and appropriate controls. In addition, resources must be effectively allocated to address EH&S, programmatic, and operational considerations.

Implementing a successful EH&S program requires both management leadership and employee engagement. The work will be conducted safely and with minimal environmental impact only if workers are involved in the process of planning the work and identifying potential hazards associated with the work. Mechanisms in place to provide for worker involvement include participation in activities such as: the Laboratory Safety Committee, with workers drawn from each department; identification of Project and task hazards and the controls necessary to mitigate those hazards; programmatic and organization self-assessments; review and comment on draft EH&S P&P; development of and teaching of EH&S training courses; and incident investigations and Lessons Learned activities.

A key tenet of ISM is that the people who perform the work also participate in the planning process, e.g., analyzing the hazards, determining the controls, and implementing the controls. This avoids disconnects between those planning the controls and those doing the work and implementing the controls. It also makes full use of worker knowledge of hazards and incorporation of Lessons Learned into the work planning process. This concept of worker involvement is applied during the design of the Facility, and the planning and conduct of work activities. The Project actively encourages open discussion about hazards and safety concerns at all levels. All employees and contractors are instructed that they all have the authority and obligation to stop unsafe work. Project management promulgates and enforces a strict policy that discussion of safety issues and concerns by staff will occur without fear of reprisal.

Before work is performed, hazards are identified and analyzed so that appropriate controls can be developed. Hazard analyses are performed at the facility level for major projects such as DUSEL, as well as at the activity level for individual science activities and operations tasks. The complexity and formality of the hazard identification process and subsequent development of work controls is tailored to the nature and scope of each work activity.

Engineering and administrative controls are put in place to prevent and mitigate EH&S hazards identified during the analyses, and controls applicable to routinely encountered hazards are captured in the *EH&S Manual.* The controls are tailored to the work being performed, and employee, Facility, Facility user, and contractor participation are extremely important in this area of work planning. Lessons Learned for incidents or earlier experiences with similar work activities are integrated into the hazard controls.

## 6.2 Hazard Analysis and Control

A broad range of hazards will be present during DUSEL construction and operation. These hazards will require close cooperation between the design/construction process and the controls identified by the hazard analysis. The evolving nature of the planned scientific activities and the resultant facility design demand an iterative hazard analysis process wherein each phase of the design process incorporates the necessary controls identified by the preceding phase of hazard analysis. Both the design and hazard control elements are informed by the requirements of the science programs, as defined at that phase. As the science requirements are refined and the hazard analysis and design processes move to subsequent phases, the level of detail and specificity for the hazard controls increases. This iterative process helps assure that ISM tenets are met in an efficient and effective manner, specifically that: a) facilities, systems, and components needed to meet mission requirements are designed, constructed, and operated in accordance with applicable regulatory requirements; and b) potential hazards to personnel and the environment presented by the Project are systematically identified and controlled as part of the planning and design process.

To that end, a hierarchy of documented EH&S hazard analysis and control processes has been established, consisting of: a *Preliminary Hazards Analysis* (PHA, see Appendix 6.C) completed in conjunction with the Conceptual and Preliminary Design phases of the DUSEL Project; a Hazard Analysis Report (HAR) and Environmental Impact Statement (EIS) to be prepared during the Final Design phase; and task-specific hazard analyses to be conducted as part of the Project Construction phase and on into the implementation of science activities. These hazard analysis and control processes are described in more detail in the following sections. In all instances, the results of the EH&S hazard analyses are coordinated with the DUSEL Risk Registry (see Appendix 9.AD) maintained by the Project, with any EH&S hazards that present a credible threat to the overall Project being placed on the Risk Registry for tracking and management. Hazards presenting a lesser level of risk that are not included on the Risk Registry are tracked via other mechanisms maintained by the EH&S Department or other Project divisions or departments.

The major hazard analysis and control processes are described in the following sections and are shown within the overall ISM framework for conducting work in Figure 6.2. Within that framework, the hazard analysis and control processes represent the second and third ISM core functions to Analyze the Hazards and Develop and Implement Hazard Controls.

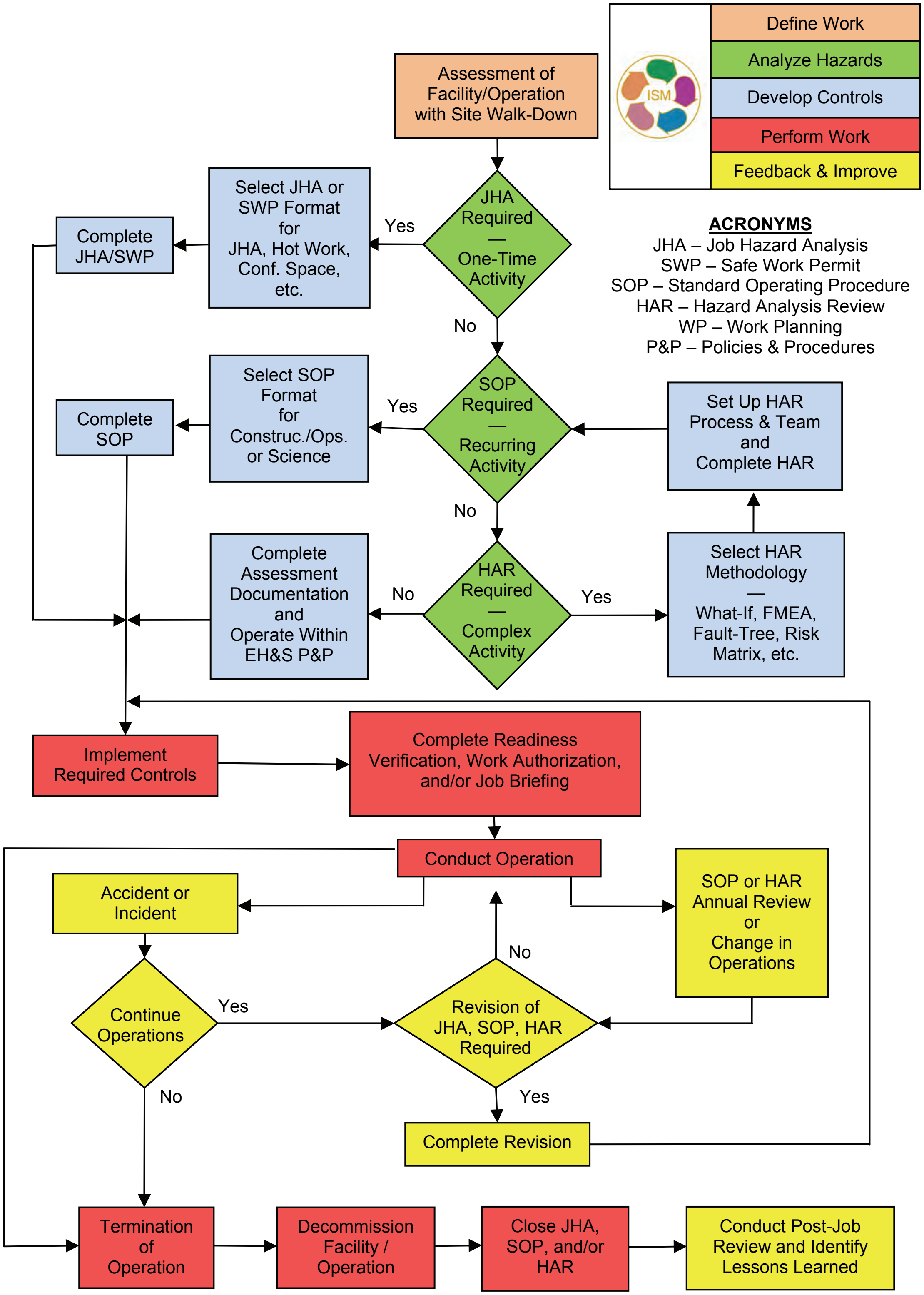

**Figure 6.2** The Project ISM process for planning and conducting work.

### 6.2.1 Preliminary Hazard Analysis

A principal component of an effective EH&S program is ensuring that hazards have been properly identified and controlled through facility design and administrative procedures. To facilitate the understanding of these issues at the Conceptual and Preliminary phases of the DUSEL Project, a PHA (see Appendix 6.C) was conducted and documented to identify the hazards that will be encountered during the Project's Construction and Operations stages.

The DUSEL Project PHA began concurrent with the Conceptual Design phase to help ensure that all significant EH&S hazards were identified and adequately addressed in the early design work. Each of the identified issues was further analyzed as the Project advanced to the Preliminary Design phase.

To support development of the PHA, a baseline hazards list was developed as a first step in identifying the potential hazards of the DUSEL Project. This list utilized best available information, drawing on data from the DUSEL Conceptual Design, existing safety-basis documentation, subject-matter expertise (from conventional facilities, the mining and science communities, engineering firms, and EH&S organizations), and Lessons Learned from similar deep underground science facilities. It also included a preliminary (pre-mitigation) risk assessment of the existing underground and support infrastructures as a subsystem of the entire Project. Design and operational controls were then applied to the list of hazards to mitigate the risks to an acceptable level, with close coordination with the design team to assure that the desired controls could be sustained by the resultant facility. The final list of major hazard categories addressed in the PHA consists of: Construction, Natural Phenomena, Environmental, Waste, Fire, Electrical, Noise and Vibration, Cryogenics (including oxygen-deficient atmospheres), Confined Spaces, Chemicals and Hazardous Materials, Material Handling, and Experimental Equipment and Operations.

Teams of subject-matter experts were formed where necessary to address significant or difficult hazards specific to the DUSEL Project, with these teams utilizing appropriate internal and third-party expertise. Examples include one team to address fire and life safety hazards unique to underground science facilities, and another to address oxygen deficiency hazards (ODHs) presented by the underground use of cryogens.

The PHA was added to the set of configuration-controlled documents for the Project to assure that all revisions are fully vetted with the Project design team. The PHA was also coordinated with all DUSEL design contractors to ensure that it was included in the Preliminary Design process.

### 6.2.2 Hazard Analysis Report

An HAR will be completed in conjunction with the Final Design phase of the Project. The HAR will be based on the results of the PHA and will incorporate detailed analysis (both qualitative and quantitative where practicable) of the final set of hazards identified for construction and operation of DUSEL. It will remain in effect for the life of the Facility and will establish the Facility operating basis from an EH&S standpoint. As such, the HAR will be a configuration-controlled document and will be subject to routine review and revision. It is anticipated that the review cycle will be annual, or when changes to the underground facility and/or operations are proposed that would significantly alter the hazard and risk profile of the DUSEL.

A HAR is typically developed in parallel with an EIS, as described in the following section, so that the full spectrum of EH&S hazards is adequately addressed. A Fire Hazard Analysis (FHA) may also be developed in parallel with the HAR, or it may be incorporated into the HAR. The exact approach will be determined as the Final Design phase progresses. As with the PHA, significant hazards identified by the HAR and/or FHA will be coordinated with the Risk Registry maintained for the DUSEL Project.

### 6.2.3 Environmental Impact Statement

The South Dakota-controlled funding for Sanford Laboratory has been used primarily for the dewatering of the mine, repair and replacement of site infrastructure, and modification of underground areas to support “early science” research. The nature of these activities and the source of funds used to conduct them exempt them from review under the National Environmental Policy Act (NEPA). NSF is funding design and planning work at the laboratory through a contract managed by the UC Berkeley. A determination has been made that a NEPA Categorical Exclusion applies to these NSF-funded design activities, conducted through UC Berkeley, and further NEPA actions are not required in advance of the design phases of the Project.

Prior to the Construction phase, the DUSEL Project must complete a NEPA review, as construction constitutes a major federal action with the potential to cause environmental impact. Therefore, an EIS has been deemed appropriate for construction activities and future operations as a federally funded laboratory. The duty to prepare NEPA documentation falls on NSF, as it is the federal agency proposing to construct and operate a new underground facility for science and engineering research at Homestake. NSF recognized this duty early in the planning process and contracted with Argonne National Laboratory (ANL) to prepare the necessary NEPA documentation in the form of an EIS. The EIS must be completed, published in the Federal Register, and a 30-day public comment period following publication completed prior to the expenditure of any federal funds by NSF on the proposed construction. Since major construction is not proposed until FY 2014, it is expected that the EIS process can be completed in sufficient time to allow NSF to make a decision on proceeding with the Project after considering the environmental impacts analyzed in the EIS.

Much progress has been made in gathering the information required for development of the EIS. This includes conducting public discussion through Sanford Laboratory to obtain external stakeholder input concerning the nature of the proposed construction and the research that will be conducted at the Facility during the operational phase. This process will continue via formal public scoping meetings that will seek input from all interested or affected stakeholders before publication of the EIS.

One of the major experiments proposed for the Facility during the Operational phase requires special construction not only at DUSEL but also at the Fermi National Accelerator Laboratory (Fermilab) near Chicago. This experiment is known as the Long Baseline Neutrino Experiment (LBNE). Since the expenditure of federal funds at both facilities is required for this experiment to occur, the construction necessary at Fermilab becomes a “connected action,” and it is the duty of the agency funding that construction at Fermilab to prepare necessary NEPA documentation. The agency funding Fermilab is the Department of Energy (DOE). As a consequence, DOE plans to support the preparation of the EIS by identifying itself as a cooperating agency with NSF and preparing a separate Environmental Assessment (EA, another form of NEPA documentation) for the portion of the work necessary at Fermilab. When finalized, it is expected these two NEPA documents will refer to each other and be mutually supportive in their analyses and conclusions. To ensure that this occurs, the EIS team has also been contracted to prepare the EA for the proposed modifications at Fermilab.

### 6.2.4 Task-Specific Hazard Analysis/Work Planning

In addition to the HAR and EIS, individual work tasks during DUSEL Construction will be further evaluated to identify and control EH&S hazards presented by each work task, in line with hazard analysis and work planning processes already in effect. Standardized hazard analysis and work planning processes will be utilized as described in the EH&S Manual, including Job Hazard Analyses (JHAs), Safe Work Permits (SWP), and Standard Operating Procedures (SOPs), with all of the processes incorporating ISM tenets such as worker involvement. Further details concerning the work planning processes are provided in a following section on EH&S Programs. In all instances, the task-specific hazard analysis and work planning processes will maintain the baseline EH&S risk levels specified for the DUSEL Project in the HAR and EIS.

## 6.3 EH&S Laws, Regulations, and Best Practice

### 6.3.1 Regulatory Standards and Applicability

The Homestake was once a world-class gold mine. Safety was governed exclusively by the Mine Safety and Health Administration (MSHA) and environmental issues were administered and enforced by the state of South Dakota and the Environmental Protection Agency (EPA). A change in EH&S governance took place to reflect new operations during the transition from a mine to a research facility. Operations at Sanford Laboratory are of three primary work types: surface operations, underground operations, and early science activities. The activities on both the surface and underground include construction, maintenance, and general operation activities, all focused on dewatering the underground and building the necessary infrastructure for the future underground laboratory. Early science activities are located both in the surface and underground laboratories but are small in scale compared with future plans included in the DUSEL Project.

The Project uses applicable standards and best practices primarily drawn from the Occupational Safety and Health Administration (OSHA), MSHA, the state of South Dakota, EPA, the International Fire Code and International Building Code (IFC/IBC), the National Fire Protection Association (NFPA), the Nuclear Regulatory Commission (NRC), the Food and Drug Administration (FDA), and national laboratories and international consortiums with similar operations and hazards. OSHA's 29 CFR 1926 (*Safety and Health Regulations for Construction*) and 29 CFR 1910 (*Occupational Safety and Health Standards*) are considered the most applicable of the available standards, especially for surface activities. MHSA's 30 CFR (*Mineral Resources*) standards are referenced for underground activities when the OSHA standards do not sufficiently address a given hazard. Early science activities use applicable OSHA, Nuclear Regulatory Commission (NRC), and FDA standards, but also rely heavily on best practices from national laboratories such as the Fermilab standards concerning the use of cryogenic gases.

In 2008 the MSHA determined that it no longer has safety jurisdiction over any of the operations at the Homestake site, as no mineral extraction is or will be involved in current or future activities. Subsequently, SDSTA and the state of South Dakota reached an agreement that establishes the South Dakota Office of Risk Management as having the responsibility to oversee and provide inspection and audit of the Sanford Laboratory, similar to its risk management responsibilities for other state institutions and facilities.

The authority to administer and enforce environmental rules and regulations has been delegated to the South Dakota Department of Environment and Natural Resources (SD DENR) by the EPA. The environmental programs, or regulations, overseen by the state are as strict as or stricter than federal rules. State programs that are absent or less strict than federal regulations defer to federal regulation. An example of this state deferral is the Safe Drinking Water Act and its associated underground injection rules.

Several government agencies may be the AHJs for radiation exposures, with the NRC and the FDA having authority for radioactive sources and radiation-generating machines, respectively. A limited number of small radioactive sources for early science activities are used on the Sanford Laboratory site under an NRC license issued to the University of South Dakota (USD), with Sanford Laboratory listed as a satellite site on the USD license. The EH&S department is initiating the process of identifying the necessary licenses for radioactive sources and radiation-generating equipment anticipated to be used at DUSEL.

### 6.3.2 Maintenance of EH&S Codes and Standards Set

The Project's set of codes and standards was originally drafted to be all encompassing. It was generated by merging standards sets from several national laboratories. As a result, the original document was very long, awkward to use, and hard to interpret, especially for design contractors and scientists. To make a more functional, user-friendly document, the set of codes and standards was edited to include only those elements applicable to hazards presented by current operations and design needs, and to maintain regulatory compliance.

This set is captured in tabular format in *EH&S Standards for DUSEL/Sanford Laboratory* (See Appendix 6.D.), a controlled document. As such the document can be revised as necessary to accommodate changing hazards, needs, and requirements. The set has been revised during the DUSEL Preliminary Design phase to incorporate additional codes and standards that have been identified as appropriate by the design team, the EH&S Department, and science collaborations. A vetting protocol similar to that found in the DOE Necessary and Sufficient Standards process is utilized to determine if a standard qualifies for adoption. This protocol helps ensure that changes to the set of standards and codes are necessary for the hazards present and sufficient to control the hazards, but are not unnecessary or burdensome.

## 6.4 EH&S Programs

The set of standards, codes, and best practices applicable to the Project's facility and operations is converted to user-friendly procedures and communicated to the workforce via a uniform set of P&P captured in the *EH&S Manual*. The policies establish the basic rule or principle for a specific EH&S topic to which workers are held accountable and the procedures provide the actions to be taken to achieve the outcome prescribed by the policy. Multiple procedures may be in effect for a single policy when discrete actions are required. These P&P form the basis of what is commonly known as the EH&S Programs. A program represents the multiple P&P and support functions (e.g., training, recordkeeping) that collectively respond to specific hazards or operational needs. A concerted effort was made during 2010 to ensure that the *EH&S Manual* contains the basic set of programs necessary for control of hazards presented by current operations and to support design of the DUSEL Facility and the science activities that will occupy the Facility. An overview of major EH&S Programs follow.

### 6.4.1 Work Planning

To assure that the necessary hazard controls (as defined by applicable standards and best practices) are applied to discrete work tasks, a work action planning procedure is included in the *EH&S Manual*. Within this process a JHA, SWP, or SOP is completed for planned work that will present EH&S hazards. A JHA is typically completed for tasks to be conducted a single time (e.g., construction, non-routine maintenance or repairs, short-duration science/research activity) and may be supplemented by a SWP if specific hazards are present (e.g., confined spaces, open flames), and an SOP is completed for ongoing tasks that will be performed multiple times (e.g., routine maintenance, regular work assignments, long-term science/research activities).

Completion of either a JHA or SOP includes a formal identification of the hazards presented by each step of the work task, development of engineering and administrative controls necessary to mitigate the hazard to an acceptable level, and development of step-wise procedures to be followed during completion of the task. The process and subsequent documentation may also serve as the authorization basis/approval for conduct of the work task. The work planning process is completed by the persons who will be performing the task, thereby assuring worker involvement per the ISM criteria, with support from subject-matter experts and EH&S staff when necessary. To ensure that hazards continue to be controlled, the process is repeated on a regular basis for SOPs, or whenever new tasks, personnel, and/or hazards are introduced for work covered by either a JHA or SOP.

The formats used for documenting the work planning process may vary, particularly between science and operations activities; however, certain minimum information must be included in all formats. The same work planning processes will be applied to DUSEL construction, providing the benefit that hazards presented by all underground activities (i.e., science, operations, construction) will be controlled by common EH&S processes.

### 6.4.2 Environmental Monitoring

The Environmental Monitoring Program is composed of various processes to ensure regulatory compliance and to monitor the site for potential impacts to the environment. The processes assuring compliance are for water quality (discharge permits), air quality (emissions inventory), and hazardous wastes (tracking spreadsheets). Regulatory requirements are listed for each plan. Proactive monitoring is currently practiced at the site. Solid waste, groundwater, and air emissions are routinely sampled and analyzed for their effect on human health and compliance with regulations. Any new project on site is reviewed for its impact to the environment as part of the Environmental Monitoring Program.

Environmental monitoring is also provided by regulatory agency and third-party audits. Within the past 36 months, these outside parties have examined Waste Water Treatment Plant operations, National Pollutant Discharge Elimination System (NPDES) permit compliance, storm-water permit compliance, Spill Prevention Control and Countermeasure (SPCC) compliance, Resource Conservation and Recovery Act (RCRA) compliance, and air quality compliance. In 2009 and 2010, SDSTA received the South Dakota Department of Environmental Compliance recognition award for full compliance with the Clean Water Act. This is a significant achievement, considering that over 2 billion gallons of water has been discharged since startup of the water treatment plant, and validates the effectiveness of environmental monitoring conducted by SDSTA.

### 6.4.3 Hazardous Material Management

A hazardous materials management process is in place for all hazardous materials brought onto and created at the site. Hazardous materials considered for use on site are requested through Purchasing and screened by the EH&S Department. The EH&S department must approve the use of the hazardous materials before on-site delivery and to ensure proper hazard communication training and on-site Material Safety Data Sheet (MSDS) availability All hazardous materials used at the site are tracked on spreadsheets that contain information on quantities purchased, usage, disposal, waste type, storage, training, composition, location, and responsible person/collaboration. This spreadsheet is reviewed regularly for mass balance and accuracy. Hazardous materials management plans will be established for the construction phases of the DUSEL Project in accordance with the previously described task-specific work planning processes.

### 6.4.4 Recycling and Waste Disposal

Waste minimization is an important objective of the Project. Recycling is practiced for steel, copper, aluminum, and cardboard. Other items such as paper and glass are expected to be included as the program grows. The waste disposal plan identifies on-site waste and how it is managed and contained. Waste items generated at the site are itemized and tracked using spreadsheets to determine monthly generator status and disposal method/disposition.

### 6.4.5 Energy and Resource Management

The Project has a societal responsibility to be energy efficient and a good steward of its resources. This responsibility extends to being energy efficient and conserving the quality of surrounding natural resources, and efforts are being designed to lower long-term Project costs and minimize impacts to the environment. These efforts focus on pursuit of Leadership in Energy and Environmental Design (LEED) criteria and the Sustainable Sites Initiative (SSI) for the Surface Facility, as detailed in Section 5.2.3 of the PDR.

The Project places a high regard on resource stewardship, as reflected in the operation of the Waste Water Treatment Plant (WWTP). The primary goal of the WWTP is permit compliance. This goal has been met since the start-up of the plant in 2008. The secondary goal of the WWTP is to dewater the underground facility in a timely manner. This goal has been met, as the water level in the underground facility, as of January 1, 2011, is at the 5,331-foot level. The Project has been efficient and innovative in modifying the Homestake WWTP to remove ferric iron in the groundwater, managing the resultant iron sludge (in dewatering tubes), and using biological treatment rather than chemical treatment to remove ammonia. This work has resulted in reduced costs and produced a product (an iron hydroxide filter cake) that is not a waste but a commercial product used in cement manufacture.

### 6.4.6 Occupational Health

A subset of the EH&S programs focuses on protecting the health and well-being of the workforce. This subset presently consists of programs for Hearing Conservation, Respiratory Protection, Personal Protective Equipment, and Lead Management. These programs are aligned with known hazards presented by the construction and maintenance of underground work areas, which are potentially noisy, dusty activities presenting significant health and bodily injury exposures.

The Occupational Health Programs are coordinated with industrial hygiene procedures designed to monitor and control health hazards presented to workers, and they incorporate appropriate health monitoring mechanisms (e.g., sound level monitoring, audiometric testing, air contaminant sampling) to validate the effectiveness of the exposure controls.

Further developments of the Occupational Health Program are being pursued in coordination with anticipated increases in underground construction activities, and may include provision of on-site medical staff such as an occupational health nurse (OHN) and, in some cases, exploiting the capabilities at the nearby Lead-Deadwood Hospital. The presence of an OHN would not only enhance the monitoring and control of health exposures, but also would also provide a level of on-site treatment for minor injuries and allow for effective medical case management when work-related injuries or illnesses occur. Emergency medical treatment would be coordinated with local medical facilities, as described in a later section on emergency preparedness. Case management has been demonstrated to effectively control the severity of injuries and illnesses (e.g., number of lost workdays, need for invasive treatments), thereby reducing the impact on the worker and the Facility. Such services can also be obtained through third-party providers, and the appropriate mixture of services, either in-house or third-party, will be incorporated into the Occupational Health Program.

### 6.4.7 Worker Safety

The largest subset of EH&S programs address worker safety, also commonly known as industrial safety or occupational safety. There are currently 13 procedures in this subset addressing the topics of Control of Hazardous Energy Sources (Lockout/Tagout), Confined Spaces, Hazard Communication, Electrical Safety, Industrial Hygiene, Cranes and Hoists, Slings and Rigging, Powered Industrial Trucks, Fall Protection, Hot Work, Compressed Gases, Surface Transportation, and Underground Transportation. These documents respond to the most severe hazards presented by work conducted at the Facility, as identified by the PHA previously described.

A significant set of guidance documents, forms, checklists, etc., are available to facilitate implementation of the worker safety programs by the persons performing the work tasks, with the appropriate materials being referenced in and linked to each worker safety P&P. Additional worker safety P&P are under development to respond to lower-level hazards presented by the operations.

### 6.4.8 Fire Prevention and Life Safety

Prevention and suppression of fires and management of products of combustion is critical to life safety and protection of physical assets in laboratory environments, particularly when the laboratory activities are located deep underground. A fire protection program is in effect that establishes a level of fire prevention and protection sufficient to minimize loss from fire and related hazards consistent with regulatory requirements, and consensus standards. Where practicable, Highly Protected Risk (HPR) criteria typical of the best protected class of industrial risks are applied as a best practice.

The program requires development of fire prevention practices and procedures, adequate design and quality construction, protection of facilities with fixed fire detection and suppression systems, procedures for testing and maintenance of fire protection systems and equipment, providing necessary fire fighting capabilities, providing adequate water supplies, and participation by all site personnel.

New construction or modification to the existing facility or fire protection systems is reviewed and approved by appropriate engineering staff, project managers, and EH&S staff. External review and approval is provided by the city of Lead as the AHJ for fire and building code matters. These reviews and approvals also address life safety elements that protect facility occupants, such as alarm and communication systems, means of egress, refuge facilities, and associated operational protocols. These reviews assure that a satisfactory level of protection is being provided, the applicable fire protection standards and HPR criteria are being met, the design and installation plan is satisfactory to the AHJ, and acceptance tests are adequate to assure proper operation. The organization responsible for the system installation or modification or for the experiment design is also responsible for documenting these reviews, with support from the EH&S Department.

A unique mixture of codes and standards are utilized to respond to fire and life safety hazards presented by both surface and underground operations, with a separate guidance document, the *Subterranean Design Criteria*, being included in the *EH&S Manual* for the application of these standards to the underground facility. As previously noted, third-party experts are used during the design of fire and life safety systems that respond to complex hazards, particularly in the underground areas.

### 6.4.9 Emergency Preparedness and Response

Credible emergencies are identified through the hazard analysis processes that all design, construction, and scientific and operational activities are subjected to, and emergency plans and response capabilities are developed commensurate with the identified credible emergencies. These processes address potential emergencies arising from fairly straightforward operations on the surface, including natural and man-made events, as well as more complicated and severe situations presented by underground operations. Response protocols are developed in accordance with applicable regulatory requirements and consensus standards and include: internal and external notification systems; surface and underground communication systems; worker tracking mechanisms that support accountability and rescue; supplies of response equipment and materials located on the surface and underground; a qualified, organized, and trained cadre of rescue personnel; and coordination with external emergency planning and response agencies.

The notification and communication protocols, response procedures, training and exercise requirements, and provision of support materials are detailed in the *Emergency Response Plan* (ERP) that is part of the *EH&S Manual*. The ERP is reviewed and revised on a regular basis to ensure that it continues to address the full set of credible emergencies identified and to incorporate Lessons Learned arising from training, exercises, responses, etc. The ERP is also referenced during work planning activities such as preparation of a JHA or SOP to ensure that necessary and appropriate emergency procedures are incorporated into the work planning documents. If the planned work presents a credible emergency scenario that is not addressed by the ERP, additional controls must be applied to the work plan to eliminate that emergency scenario or specialized emergency response procedures must be developed and implemented for the duration of the work activity.

The Emergency Response Team is composed of 28 members trained to respond to a wide variety of credible emergencies. The team has many skills: six members are on the Lead and Deadwood Fire Departments; one member is the Lawrence County Emergency Response Manager; four members are on Lawrence County Search and Rescue squad; eight members are 40-hour Hazardous Waste Operations (HAZWOPER) trained; four members are emergency medical technicians; and 25 members are trained in Incident Command (Levels 100 and 200). Twenty-seven members of the team have been active in mine rescue and collectively have a total of more than 300 years of experience. The team captain is an internationally recognized leader in mine rescue and mine safety. The team is in the process of organizing its talents, holding monthly eight-hour training sessions, and updating annual training and certifications (e.g., HAZWOPER, EMT, firefighter, Self-Contained Breathing Apparatus [SCBA]). All members of the team have primary jobs, many of which are with the Project or other emergency response disciplines. It is planned to expand the team to 60 members during the peak of DUSEL construction activities, in line with the hazards expected to be present at that time.

Response protocols are established with regional emergency planning and response organizations. These resources include the Lawrence County Emergency Management Commission, Lead and Deadwood Fire Departments, Lawrence County Sheriff's office (dispatch), Lawrence County Search and Rescue, the South Dakota National Guard located in Rapid City, Life Flight, and the Rapid City Hazardous Materials Response team. All of the Lawrence County response groups have a copy of the ERP and are trained on call-out procedures. Future training exercises will include participation by many of these outside organizations.

Medical first-aid capabilities are maintained on site via trained staff and provision of necessary supplies, and these capabilities may be expanded with the addition of the OHN position, as previously described. Off-site medical services are provided by the Lead-Deadwood Regional Hospital, including a 24-hour-a-day emergency department and emergency medical transport via ground ambulances. More advanced levels of emergency medical service are available at the Rapid City Regional Hospital, with the Lead-Deadwood and Rapid City Hospitals being part of a coordinated emergency medical system covering the Black Hills region. Trauma Level II facilities are available in Sioux Falls, South Dakota, and Cheyenne, Wyoming, those facilities being approximately 300 and 200 air miles from Lead, respectively. Trauma Level I facilities are available in Denver, Colorado, approximately 300 air miles away. Emergency medical air services are provided by Black Hills Life Flight based in Rapid City. Within the Black Hills coordinated emergency medical system, Life Flight operates both helicopter and fixed-wing aircraft, thereby assuring that necessary combinations of accessibility (i.e., helicopter) and transport speed (i.e., fixed wing) are available. To facilitate timely transport to the Rapid City facilities, a helicopter-landing zone has been established near the Sanford Laboratory Administration Building, and the layout of the zone and site access procedures have been coordinated with the appropriate external agencies, including Black Hills Life Flight.

Some restrictions are currently imposed on underground operations due to existing underground conditions that present a higher-than-desirable level of risk and/or potential for emergency situations. These controls include limits on the total number of persons allowed underground at one time and higher ratios of "guides" to "visitors." and have been determined through hazard analysis to maintain an acceptable level of risk. These access and operational restrictions will be relaxed and/or eliminated as the higher hazard conditions are remediated, such as reduction of fire hazards by elimination of the wooden structure in the Yates Shaft and provision of multiple means of egress.

In addition to the temporary underground access and operational restrictions, emergency response considerations are also being addressed for time periods when both DUSEL construction and early and intermediate science activities (e.g., LUX Experiment, MAJORANA DEMONSTRATOR Project) are being conducted underground.

### 6.4.10 Science Safety

While the set of EH&S programs described so far is applied to both science and operations activities as warranted by the hazards present, an additional set of programs is in place to specifically address hazards presented only by science activities. This includes a program for EH&S review of experiments that tailors work planning and equipment design review processes to the unique needs of Science activities. Programs also are established for Cryogenic System Reviews and control of ODHs due to the large quantities of cryogens planned for use in the underground laboratory areas. As previously noted, third-party experts are used during the design review for cryogenic systems in order to fully respond to the complex hazards presented, particularly in the underground areas.

Further EH&S programs will be added to this set as unique hazards are identified for future science and DUSEL activities, and additional third-party subject-matter experts will be utilized as necessary to evaluate these unique hazards.

As mentioned in Volume 5, *Facility Preliminary Design*, EH&S considerations are actively incorporated into the DUSEL design. Details of the impact on the DUSEL Facility and infrastructure design on systems such as ventilation, Areas of Refuge (AoRs), egress, and additional fire-life-safety systems are presented in that Volume.

### 6.4.11 Incident Reporting, Notification, and Investigation

To insure that incidents are properly reported and timely notification is provided to affected stakeholders, the *EH&S Manual* includes P&P for incident reporting and incident notification. The procedures describe which organizations, including external response agencies, receive immediate reports of incidents so that they can respond to the incident and provide support and assistance. In addition to the first-responder notifications, a further level of notification is provided for stakeholders such as SDSTA, DUSEL, and NSF. Timelines are provided for all notifications, with multiple time frames being provided, depending on the severity of the incident. Likewise, the stakeholders being notified will vary, dependent on the type and severity of incident, and may include external regulatory agencies at the state and federal levels. The sequence followed for stakeholder incident notification is shown in Figure 6.4.11.

Investigations are initiated in accordance with the P&P with the makeup of the investigation team being tailored to the type of incident. Third-party subject-matter experts are added to the investigation teams when warranted.

These reporting and notification mechanisms are continually updated as construction, operations, and science activities continue to evolve, and appropriate changes will be made to accommodate DUSEL construction.

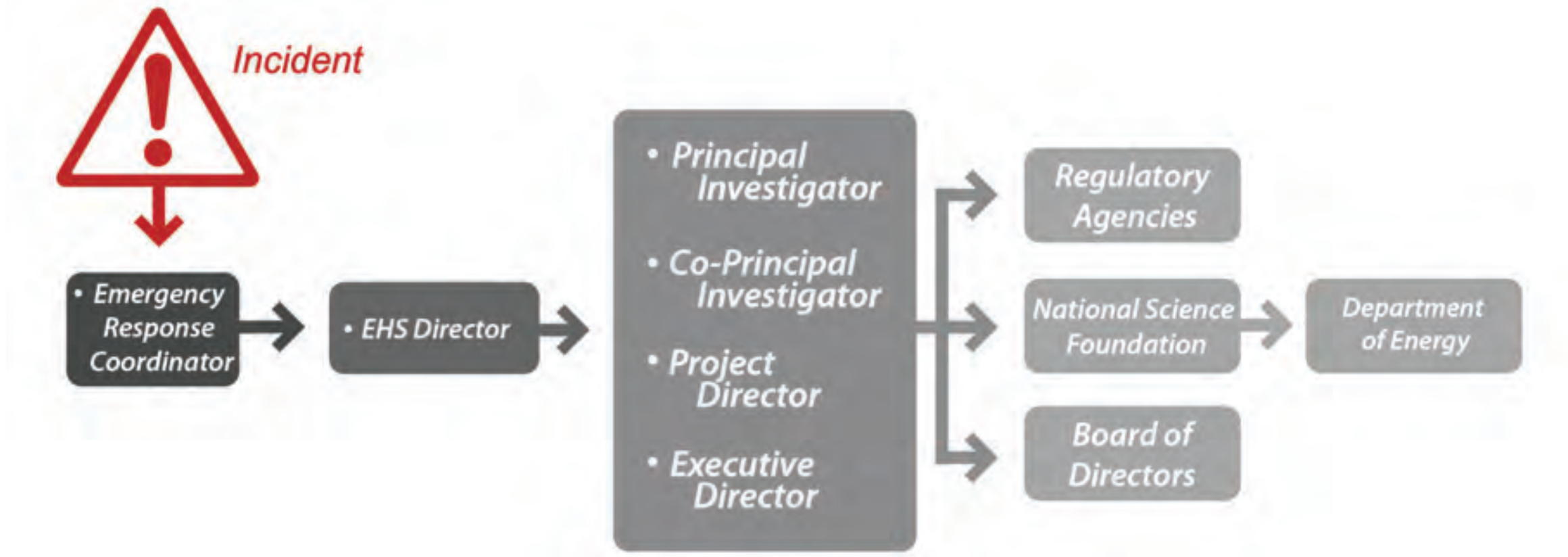

**Figure 6.4.11** Incident notification sequence for stakeholders.

### 6.4.12 Regional Communication and Public Information

SDSTA maintains several lines of communication with the local community and the surrounding region, including local, state, and federal government officials and agencies; local schools; South Dakota universities; and local business groups. Laboratory representatives provide in-person briefings once a month to the Lead and Deadwood City Commissions and the Lawrence County Commission. The ERP includes steps to notify local media by phone and by electronic mailing list. In addition, both the DUSEL Project and SDSTA maintain public-access Web sites. The Communications Department maintains an electronic mailing list for members of the public who request e-mail updates about the Laboratory.

All of these communication mechanisms are available and are used as necessary to provide EH&S-related information to the local and regional community. It is planned that the public communication mechanisms will be used extensively during the stakeholder input sessions conducted as part of the EIS preparation prior to DUSEL Construction.

## 6.5 EH&S Training

The Project is improving and expanding its existing EH&S Training Program. The initial Training Program targeted construction work and operations personnel, using OSHA requirements as its foundation. This approach was appropriate during early operations, when the vast majority of the workforce was involved in construction activities. However, as the workforce evolves, a more proactive approach is needed to provide training that is relevant to the work being done.

To facilitate and expedite the restructuring of the EH&S Training Program, the Laboratory enlisted the help of a subject-matter expert at Lawrence Berkeley National Laboratory (LBNL). This consultant visited Sanford Laboratory multiple times in 2010 and helped develop a strategy to restructure the current EH&S Training Program to meet future requirements. This resulted in a new EH&S Training Policy for the Laboratory.

The revised EH&S Training Program focuses on four key elements described in the following sections: Training Assessments, Training Content, Training Delivery, and Record Keeping.

### 6.5.1 Training Assessments

The Project has adopted an assessment methodology consisting of standardized functional groups for safety training, based on the work activities currently under way, as well as for planned future work (i.e., DUSEL construction, science and research). Hazard assessments of the activities performed by these functional groups are under way to determine the training requirements for each group. This approach will allow for the addition of individually required training not specifically addressed in the group requirements. The functional group requirements and those of individual workers will be reviewed annually.

### 6.5.2 Training Content

Over the past two years, the EH&S Department has assembled a training library of videos and PowerPoint presentations obtained through various vendors and focused primarily on OSHA training requirements. The Project plans to supplement this library with safety-training materials available from national laboratories, such as LBNL and Fermilab. DUSEL will rely heavily on the national laboratories as a source of training material for science-related EH&S issues.

### 6.5.3 Training Delivery

The current primary delivery method for safety training at the Laboratory is person-to-person in a classroom setting. This training is often supplemented with multimedia presentations from the current training library, but requires significant involvement from the trainer. During early operations, this was an effective and efficient method training method. However, as the Laboratory grows and training requirements change, including a need for standardized and consistent training messages, this approach may not be the most efficient way to provide safety training. In the future, the Laboratory will rely more on stand-alone multimedia (e.g., videos, interactive training materials). This will include the ability to conduct some required training online prior to arriving at the Laboratory. To facilitate this plan, the Laboratory recently hired a multimedia specialist to assist with the development of this program.

### 6.5.4 Record Keeping

The Laboratory recently purchased a commercial software package for documenting the Safety Training Program. The proposed Safety Program and its documentation packages have been structured with the software to optimize its current capabilities. For the future, adopting a more sophisticated system (similar to those used at the national laboratories) is under discussion.

## 6.6 EH&S Review Process—Inspection, Surveillance, and Oversight

The Project is steadfastly committed to achieving its scientific mission and goals with minimal risk to staff, users, visitors, the general public, and the environment. The *EH&S Manual* and its referenced standards are intended to provide clear and uniform guidance for accomplishing this mission in ways that satisfy Project obligations and concerns with respect to hazards at the Facility. The EH&S review process of inspections, surveillance, and oversight serves as quality assurance and a continuous improvement mechanisms to provide validation that a safe work environment is being provided and that safe work practices are being followed, and identifies methods to improve when such conditions do not exist.

The *EH&S Manual* provides guidelines for performing inspections, surveillance, and oversight of work environments and the execution of work. The conduct of EH&S reviews applies to all work and occupied work spaces at the Facility, to construction, to experimental work and maintenance, and to operations activities. The EH&S review process consists of systematic inspections, routine walkthrough inspections, and general safety-awareness observations expected from all staff members. These processes serve to inform the Project's management team of the status of implementation of EH&S program requirements.

Project management requires that all assigned spaces be inspected periodically, and the DUSEL Project will maintain a similar program of inspections. Activities specifically addressed will be construction, experiments, maintenance, and operations activities. Within this constraint, management determines the frequency and extent of EH&S-related inspections by considering the value provided by the inspection in the prevention and/or mitigation of the risks of the work or the results produced. Quarterly inspections are considered to be the basic minimum frequency.

All working groups conduct periodic inspections of selected locations to meet the established frequency commitment. Responsible managers may increase the priority or frequency of inspections. Managers cognizant of the risks created by the work under their purview should ensure that inspection activities are appropriate to those risks by reviewing inspection reports and accompanying inspectors on occasion.

Independent third-party oversight is provided through the conduct of MSHA Criteria Inspections of underground areas on a bimonthly basis. These inspections are conducted by qualified consultants and evaluate underground conditions against applicable MSHA requirements and standards. Third-party OSHA criteria inspections have also been completed at the Surface Facility via the Occupational Safety and Health Consultation Program conducted by South Dakota State University. Corrective action plans are developed, implemented, and tracked as necessary following each MSHA or OSHA criteria inspection.

These inspection and oversight processes will be extended to the DUSEL Construction phase. The EH&S Department is currently evaluating the level and type of inspection/oversight that will be provided by Project's employees, McCarthy Kiewit (the DUSEL construction management contractor), and third-party providers.

## 6.7 Independent External EH&S Reviews

The Project has adopted an evidence-based peer-review process to evaluate the effectiveness of its EH&S processes and program elements. The independent external review team members have been drawn from national laboratories, the federal government, universities, and industry. The focus is on high-risk areas of exposure such as:

- Life safety
- Emergency response
- Fire prevention
- Surface and underground construction
- Project Safety Program

To date, the Project has conducted three independent external safety reviews: 1) Project Safety Program, June 21-23, 2010; 2) Fire Protection Design Review, July 13- 14, 2010; and 3) EH&S Oversight Committee Review, August 24-27, 2010.

The EH&S Oversight Committee (EHSOC) reports to the Vice Chancellor for Research at UC Berkeley. The EHSOC advises the UC Berkeley Vice-Chancellor of Research and the SDSTA Board of Directors on all aspects of overall the DUSEL/Sanford Laboratory EH&S program and its implementation. The primary focus of the EHSOC is to review the DUSEL/Sanford Laboratory EH&S vision, mission, strategy, plans, and implementation progress as compared to best practices for a national laboratory; identify high-level gaps; and offer opportunities for improvements commensurate with a world-class user research facility. The EHSOC will conduct an external assessment at least twice each year to evaluate the state of EH&S programs, systems, and performance.

Other reviews in planning stages are:

- Emergency Response Capability
- Contractor Safety Review

In addition to external reviews of the process and program elements described above, a review of the Project design was commissioned through Hughes Associates, Inc., in August 2010. The intent of this review was to provide a third-party analysis of the safety elements included in the design, ensuring that the design appropriately addressed applicable codes and standards. The report generated from this independent review is included in Appendix 6.E, *Third Party Review Letter of the Fire and Life Safety (FLS) Design for DUSEL.* The recommendations of this report were incorporated in the Facility Preliminary Design.

## 6.8 EH&S Resources

Long-range planning is under way to help assure that necessary EH&S support resources are available during DUSEL design and construction, as well as during setup and long-term operation of underground science activities. The primary planning mechanism being used is the DUSEL Project Work Breakdown Structure (WBS) that tracks the levels and types of EH&S staffing through the year 2022. The data in the EH&S WBS is coordinated with a spreadsheet maintained by the EH&S Department for similar data through the year 2024, that being the year that DUSEL construction and setup has ended and ongoing science activities are in place. The EH&S Department spreadsheet also tracks the number of workers anticipated underground (both construction and science) as well as the types of activities being conducted (e.g., excavation, science setup, science operations) for each year, as this information will affect the number and type of EH&S support staff needed. Details are found in PDR Volume 10, *Operations Plans*, concerning the number and type of EH&S staff planned for each phase of the DUSEL Project.

To help assure that planned staff resources are appropriate, a baseline analysis is being conducted for the EH&S portion of the DUSEL Project WBS against a similar project completed at the Spallation Neutron Source, a DOE facility at Oak Ridge National Laboratory.

To supplement this planning process, information is being obtained from McCarthy Kiewit concerning the types of EH&S support services they will be providing during the Construction phase, to include their direct employees and/or third-party services. Third-party EH&S services may also be arranged directly by the DUSEL Project as the situation warrants, particularly in cases where the services and/or support staff will only be required for a discrete time period. EH&S support services for the DUSEL Project will then consist of the mixture of Project staff, McCarthy Kiewit provided staff, and third-party staff.

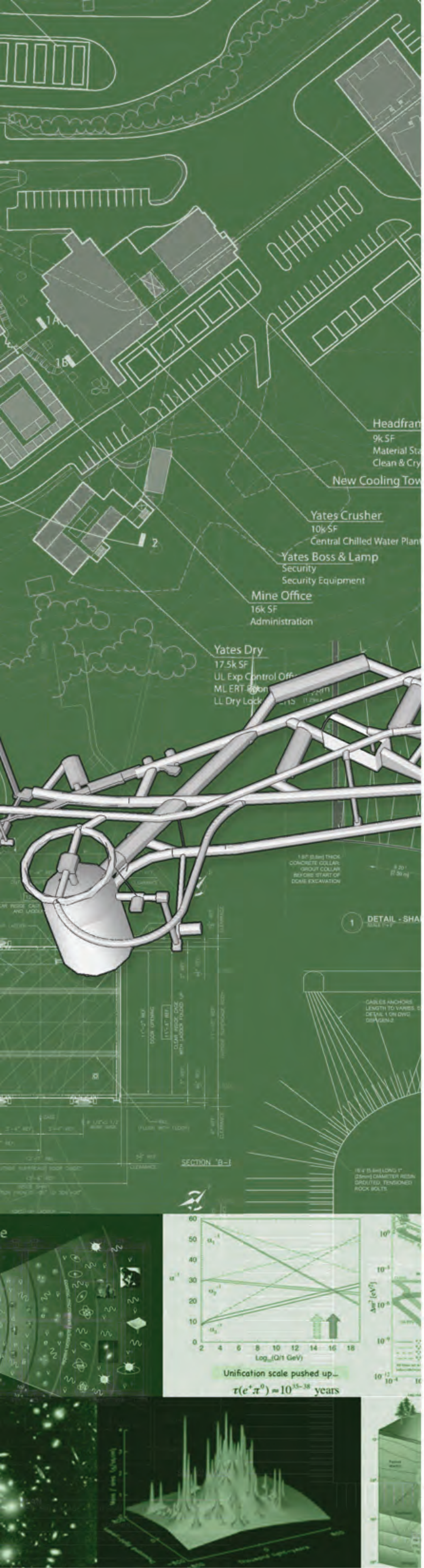

# Preliminary Design Report

May 2011

# Volume 7:
# Project Execution Plan

DUSEL
Deep Underground Science and Engineering Laboratory

# Project Execution Plan

Volume 7

## 7.0 Introduction

### 7.0.1 Purpose and Structure

The Deep Underground Science and Engineering Laboratory (DUSEL) Project Execution Plan (PEP) documents the planning, management, and oversight for the Design, Construction, and Operational phases of the DUSEL Project. This covers both funding received through the National Science Foundation (NSF) Major Research Equipment and Facility Construction (MREFC) and Research and Related Activities (R&RA) funding. This PEP provides information related to Project authority, approval, and funding; and provides overviews of management structure, organization, and Project baselines for cost, schedule, and technical scope. The PEP is an evolving document that matures as the Project scope, requirements, and cost estimates are refined. The NSF *Large Facilities Manual*[1] specifies the essential elements included in this PEP. Additional detailed information is contained in the appendices referenced in this Volume. Table 7.0.1 provides a cross-reference between the NSF *Large Facilities Manual* PEP requirements and references within this Preliminary Design Report (PDR) to specific PEP information that addresses each requirement.

In addition to being a volume within the PDR, the PEP will be issued as a standalone document. This document will be the primary agreement regarding Project planning and objectives between the Physics Division within the Directorate of Mathematical and Physical Sciences (MPS) of NSF and the University of California at Berkeley (UC Berkeley). The PEP will be reviewed and revised periodically to reflect Project maturity.

As used within this volume, the term “DUSEL” refers to all activities associated with DUSEL and its Design, Construction, and Operations regardless of funding source or phase. This PEP defines the baseline for the MREFC-funded construction project. The PDR discusses other funding channels related to the preparation and performance of the MREFC-funded construction project.

| NSF Large Facilities Manual, Appendix 3—PEP Requirements | |
|---|---|
| **Topic** | **Compliance (Volume, Chapter, or Appendix References)** |
| Research objectives underlying facility proposal<br>(Science Requirements) | 7.1.1 Science Objectives and Requirements<br>3.6 ISE Requirements Process<br>3.7 Generic Physics Requirements<br>3.8 Science-Driven Facility Infrastructure Requirements |
| Necessary infrastructure to support research | 7.1.3 Facility Infrastructure Necessary to Obtain Research Objectives<br>5 Facility Preliminary Design<br>9 Systems Engineering |
| Work breakdown structure (WBS)<br>WBS dictionary | 7.2 Work Breakdown Structure; Appendix 7.A WBS Dictionary<br>8 Project Management Control |
| Basis of estimate | 7.3 Project Budget<br>2 Cost, Schedule, and Staffing |
| Risk management approach and results | 7.4 Management Reserve<br>7.7 Project Risk Analysis and Management<br>2 Cost, Schedule, and Staffing |
| Contingency methods and budget | 7.4 Management Reserve<br>2 Cost, Schedule, and Staffing |
| Resource loaded project schedule | 7.5 Project Schedule<br>2 Cost, Schedule, and Staffing |
| Organizational structure | 7.8 Project Organization, Governance, Oversight, and Advisory Functions |
| Interagency and international partnerships | 7.9 Interagency Partnerships |
| Acquisition plans; Subcontracting strategy | 7.10 Project-Wide Acquisition Plans<br>5.10 Final Design and Acquisition Plans |
| Reporting and controls and PMCS functions | 7.11 Project Controls Systems<br>8 Project Management Control |
| Project governance | 7.8 Project Organization, Governance, Oversight, and Advisory Functions |
| Configuration control plans | 7.13.1 Configuration Management; Appendix 7.B CCB Charter<br>9 Systems Engineering; Appendix 9.D Configuration Management Plan |
| Contingency management | 7.4 Management Reserve<br>2 Cost, Schedule, and Staffing |
| Oversight plans and advisory functions | 7.8 Project Organization, Governance, Oversight, and Advisory Functions |
| Quality control and quality assurance plans | 7.15 Quality Assurance and Control; Appendix 7.C Quality Assurance Policy; Appendix 7.D Quality Assurance Surveillance Plan |
| Environmental plans, permitting, and assessment | 7.16.3 Environmental Plans, Permitting, and Assessment<br>6 Integrated EH&S Management (6.3, 6.4.2, 6.4.3) |
| Safety and health issues | 7.16 Environment, Health, and Safety (EH&S)<br>7.16.1 EH&S Hazards, Risk Assessment, and Mitigation Strategy<br>7.16.2 Integrated Safety Management (ISM) System<br>6 Integrated EH&S Management (6.1, 6.2) |
| Systems engineering requirements | 7.1.2 Level 1 Requirements and Key Performance Parameters<br>7.13 Systems Engineering<br>9 Systems Engineering; Appendix 9.E DUSEL Project Requirements |
| Systems integration and commissioning<br>Plans for transitioning to operational status | 7.14 Systems Verification |
| Estimates of operational cost for the facility | 7.3.2 Estimated Operations Costs (R&RA)<br>10 Operations Plans |

**Table 7.0.1** Cross-reference of NSF PEP requirements to DUSEL PEP and PDR contents.

### 7.0.2 Project Execution Plan Approval and Revisions

This version of the PEP, Volume 7 of the PDR, is the initial version, Version 1.0.

The standalone version of the PEP will be approved by the following:

- Physics Division (MPS-NSF): Cognizant Program Officer
- UC Berkeley: DUSEL Principal Investigator/Executive Director; DUSEL Project Director

## 7.1 Project Description

The DUSEL construction will in large part be funded from the MREFC Account. The Project documented in the PDR is an $875 million construction project (FY 2010 dollars) funded from the MREFC Account: $575 million is allocated for the Facility, including Management Reserve; $300 million is allocated for science experiment construction projects. The science funds are protected by a financial firewall from use for Facility construction. The MREFC-funded construction is planned for eight years, including approximately one year of Schedule Reserve.

The Facility design includes rehabilitation and deferred maintenance implemented under operations and maintenance activities through NSF R&RA funding. These R&RA-funded items are focused on Facility rehabilitation and maintenance to provide safe access to the Facility and to reduce risk. The implementation of these R&RA-funded items is addressed in Volume 10, *Operations Plans*. R&RA-funded elements of the design were developed along with the MREFC-funded elements to ensure that their implementation was cohesive and resulted in a well-integrated DUSEL Facility design—regardless of the funding source—and to provide NSF with adequate planning for post-construction Operations requirements.

The science objectives that will be met by the DUSEL Facility, the highest-level Project requirements, and the DUSEL Facility infrastructure to meet the Project requirements are summarized in the following sections.

### 7.1.1 Science Objectives and Requirements

The DUSEL science objectives are briefly stated in the following sections. The examples were informed through the NSF's S4 solicitation process. Additional candidates for final consideration may result from future solicitations and peer review. The requirements for the civil construction of the Facility needed to meet these objectives are guided by our expectations and goals for a generic Integrated Suite of Experiments (ISE). Although a specific set of experiments has not been identified, there are design aspects of proposed experiments that led to generic design requirements as described in Chapters 3.6, *ISE Requirements Process*; 3.7, *Generic Physics Requirements*; and 3.8, *Science-Driven Facility Infrastructure Requirements*, enabling the Project to create a Facility design capable of supporting the entire suite of candidate experiments.

#### 7.1.1.1 Long Baseline Neutrino Experiment

The Department of Energy's (DOE's) Long Baseline Neutrino Experiment (LBNE) will be included in the ISE. The selection among the options for implementation of this experiment will ultimately determine many of the Facility requirements. The current baseline design for the DUSEL Facility describes in detail the requirements and implementation of one Large Cavity and associated infrastructure for a water Cherenkov detector. Additional Facility options have been described and requirements for these options gathered, but at a less-detailed level at this time. These options include the addition of at least one more Large Cavity or a Large Cavity of greater dimensions, and facilities and infrastructure for a liquid argon detector (or detectors).

#### 7.1.1.2 Proton Decay Searches

The ability to search for proton decay with the detectors is likely to be an integral part of the LBNE and thus will be included in the ISE. Separate detectors aimed solely at proton decay searches are not currently included in the ISE. Specific requirements that enable the proton-decay search will be included in the Facility design.

#### 7.1.1.3 Detection of Astronomical Neutrinos

The detection of solar neutrinos and neutrinos from supernovae and the big bang are possible goals for LBNE's large detectors and thus likely part of the ISE. Separate detectors aimed solely at detection of these neutrinos are not currently included in the ISE. Specific requirements that enable detection are considered in the Facility design.

#### 7.1.1.4 Dark Matter Searches

At least one Generation Three (G3) dark-matter experiment will be included in the ISE, chosen from those proposed.

#### 7.1.1.5 Neutrinoless Double-Beta Decay Searches

At least one neutrinoless double-beta decay experiment will be included in the ISE, chosen from the proposed experiments.

#### 7.1.1.6 Nuclear Astrophysics Experiments

The Dakota Ion Accelerators for Nuclear Astrophysics (DIANA) is currently the only proposal for this area of science. The final Facility design at the Mid-Level Laboratory (MLL) Campus will allow DIANA to be implemented if it is selected to be one of the DUSEL experiments.

#### 7.1.1.7 Low Background Counting and Materials Assay Facility

Basic aspects of infrastructure for low-background counting and materials assay will be included in the design of the DUSEL Facility to support this program. More advanced R&D and a Facility in this area are represented currently by the Facility for Assay and Acquisition of Radiopure Materials (FAARM) proposal. The Facility design at the MLL Campus will allow aspects of FAARM to be implemented if it is selected to be one of the DUSEL experiments.

#### 7.1.1.8 Biology, Geology, and Engineering Experiments

The Facility design will accommodate among the ISE a subset of the S4 proposed biology, geology, and engineering (BGE) experiments (or experiments proposed in future). The number and type of experiments will be determined, based on reviews of scientific merit and resources.

#### 7.1.1.9 Support Activities and Staging Areas

As a part of the underground laboratory campuses, the Facility will provide areas for machining, low-activity materials fabrication, staging areas for assembly and installation, and other support activities for the experimental program. An initial scope for such areas is part of the baseline Facility design.

#### 7.1.1.10 Research and Development

Areas for R&D of new concepts for physics and BGE experiments are an important aspect of the experimental program. An initial scope for such areas, both above- and below-ground, will be part of the Facility design.

### 7.1.2 Level 1 Requirements and Key Performance Parameters

The Project highest-level requirements (Level 1) and key performance parameters (KPPs) are documented in the *DUSEL Project Requirements*, Appendix 9.E. These requirements guide the design of the Facility. The requirements in that document will be used during the verification process (See Chapter 7.15), along with the Level 2 and 3 requirements (See Volume 9, *Systems Engineering*) in order to provide pass/fail criteria to determine Facility acceptance. The Project will measure and track performance using the KPPs to determine satisfaction of Level 1 requirements to reduce and mitigate Project risk.

### 7.1.3 Facility Infrastructure Necessary to Obtain Research Objectives

The DUSEL Facility provides the necessary infrastructure to meet the Project Level 1 requirements for scientific research and support of daily operations and maintenance activities. This is detailed in Volume 5, *Facility Preliminary Design*. Working with the scientific collaborations to understand and document science requirements, the DUSEL Facility Preliminary Design has been crafted to support the Level 1 physics requirements and consists of four principal elements:

1. A Surface Campus supporting experimental efforts; support to the underground operations; and an education and public outreach center called the Sanford Center for Science Education (SCSE)
2. A research campus at the 4850L, the MLL Campus, that consists of two laboratory modules (LM-1 and LM-2), the Davis Laboratory Module (DLM), and Davis Transition Area (DTA)
3. A research campus at the 7400L, the Deep-Level Laboratory (DLL) Campus, for physics that consists of one laboratory module (LMD-1)
4. Facility infrastructure supporting activities at these campuses

The DUSEL Facility design scope to support the Project requirements for BGE experiments are:

1. Surface buildings, including a drill-core archive and surface support laboratories
2. A research campus at the 4850L, the MLL
3. A research campus at the 7400L, the DLL, that includes one drill room for ecohydrology research
4. Existing ramps, shafts, and levels (Other Levels and Ramps [OLR]) within the Facility, extending from the Surface to the 7400L, and extending across significant portions of the underground Facility footprint
5. Associated infrastructure, including utility stations to support OLR activities

To support the LBNE and proton decay, the Project has included in the design a Large Cavity (LC-1), accessed via the 4850L, and associated infrastructure for one 100 kT water Cherenkov detector (WCD).

**Surface Facility**

The DUSEL Project includes two distinct Surface Campuses: Ross and Yates. The Yates Campus will be developed as the primary campus for science, administration, and education and public outreach activities, including the SCSE. The Ross Campus will be developed with a primary mission of supporting construction, operations, and maintenance activities.

The Surface Campuses include a mix of adaptive reuse of many existing surface buildings and infrastructure and the addition of two new buildings on the Yates Campus—the SCSE to support the education and public outreach program and a new science experiment assembly building to support experiment surface assembly and fit checks. Existing surface buildings not required for future DUSEL operations will be removed.

**Underground Facility**

The MLL Campus at the 4850L includes two laboratory modules that are designed to house a suite of experiments and, therefore, have similar sectional sizes and configuration. Both are 24 m in height and 20 m in width, with LM-1 50 m in length and LM-2 100 m in length. LM-1 will nominally support two smaller or one larger physics experiments. LM-2 is designed to support three to four experiments. Several BGE experiments will also be located on the 4850L.

The DLL Campus at the 7400L includes one laboratory module (LMD-1) to house from one to two physics experiments in a 15 m high by 15 m wide by 75 m length space. Additionally, a room is located on the DLL to support deep-drilling research activities.

DUSEL BGE experiments will be located within the OLR. The OLR include approximately 19 miles (30 km) of existing excavated space and will be equipped with utilities such as power, water, and cyberinfrastructure to provide safe access and support for science operations.

The underground infrastructure that supports science laboratory and facility operation requirements includes the shafts, winzes, and hoists to provide primary and secondary access and egress. It also includes systems such as fire and life safety, ventilation, water inflow management and dewatering, electrical power, cyberinfrastructure and communications, transportation, waste rock handling, and plumbing. These systems provide the main connections between the underground and the surface facilities.

The detailed DUSEL Facility design is described in Volume 5, *Facility Preliminary Design.* The Facility requirements and the ISE interface requirements can be found in Volume 9, *Systems Engineering*. These requirements were allocated to each design contract and include applicable codes and standards such as the Occupational Safety and Health Administration (OSHA), Mine Safety and Health Administration (MSHA), National Environment Protection Act (NEPA), South Dakota State Historical Preservation Office (SHPO), International Building Code (IBC), and National Fire Protection Association (NFPA) code and regulation requirements called out by the Environmental Health and Safety organization (See Volume 6, *Integrated Environment, Health, and Safety Management*). To provide requirements verification during the Preliminary Design phase, formal requirements compliance matrices providing the design's conformance to the DUSEL requirements are discussed in Volume 9.

## 7.2 Work Breakdown Structure

The complete DUSEL Project and Operations work scope is captured within the DUSEL Work Breakdown Structure (WBS) and is used for planning, managing, and reporting. The WBS is divided into four major categories based on major systems and functions. The top-level (Level 1) of the WBS captures the entirety of the Project and Operations in all phases. Level 2 captures the scope of work in the major systems. The following is a description of the Level 2 elements within the WBS:

- **DUS.PRJ**. Project-Wide Systems. The scope of this WBS is to guide and unify the entire Project and to provide the organizational structure, performance requirements, standards, procedures, and methods to be followed in the delivery of all work scope contained within the Project. Included in this element of the WBS are the Project-wide crosscutting elements:
  - Project Management & Controls (DUS.PRJ.PMO)
  - Business Systems (DUS.PRJ.BUS)
  - Systems Engineering (DUS.PRJ.SYS)
  - Environmental Health and Safety (DUS.PRJ.EHS)
  - Information Technology (DUS.PRJ.ITS)
  - Education and Outreach (DUS.PRJ.EDO)
  - Quality Assurance and Control (DUS.PRJ.QAC)
- **DUS.FAC**. Facility. Included in this WBS element are the design, engineering, and construction of the Facility and supporting infrastructure, including outsourced contracts, management, and oversight to support those efforts. Included in this element of the WBS are the following major subsystems within the DUSEL Facility:
  - Management (DUS.FAC.MGT)
  - Surface (DUS.FAC.SUR)
  - Underground Infrastructure (DUS.FAC.UGI)
  - 4850L Mid-Level Laboratory (DUS.FAC.MLL)
  - 7400L Deep-Level Laboratory (DUS.FAC.DLL)
  - Other Levels and Ramps (DUS.FAC.OLR)
  - Large Cavity for LBNE (DUS.FAC.LGC)
- **DUS.OPS**. Operations. Included in this WBS element are the operation and maintenance of the Facility during all Project phases.
- **DUS.SCI**. Science Programs and Integrated Suite of Experiments. Included in this WBS element is the DUSEL effort to manage experiment development, experiment requirements definition, Facility and experiment interface definition, and experiment integration into the Facility design.

The Level 3 WBS elements are the major subsystems and crosscutting systems within the Project. Figure 7.2 shows the WBS structure down to the third level for each of the four major WBS elements. The DUSEL Project WBS Dictionary, under configuration control, defines the scope for each WBS element down to Level 4. The WBS Dictionary is in Appendix 7.A. Details on how the WBS is maintained and used within the Project are in Volume 8, *Project Management Control*.

The DUSEL WBS captures all scopes of work that are part of the Project and Operations, regardless of funding source. Funding source designations and phases allow the differentiation of the individual scopes of work associated with a particular funding type.

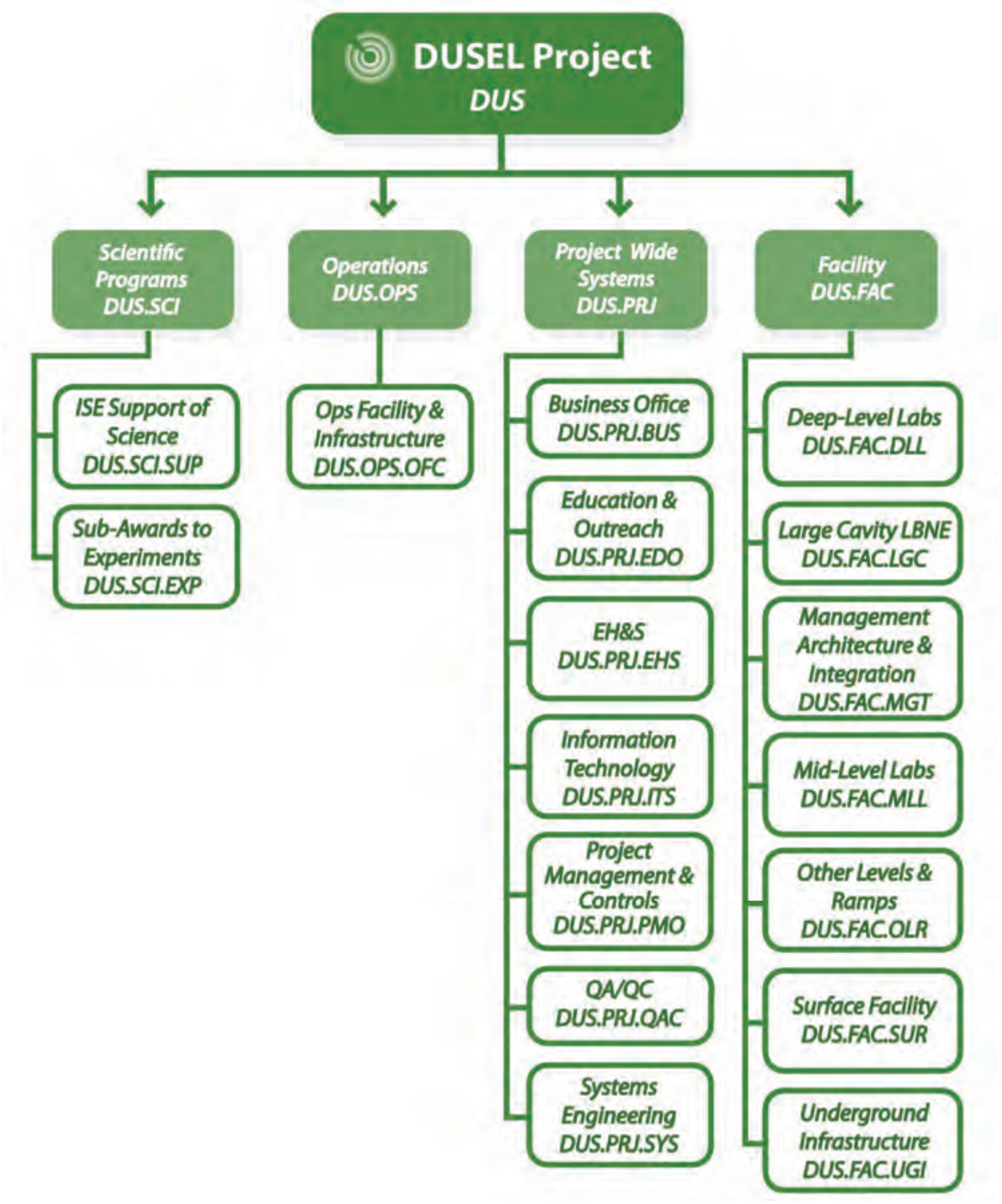

**Figure 7.2** DUSEL WBS structure. [DKA]

## 7.3 Project Budget

This chapter describes the NSF MREFC Account and R&RA funding supporting the DUSEL Project. Additional support is derived from SDSTA-controlled sources as described in Volume 2, *Cost, Schedule, and Staffing*.

### 7.3.1 MREFC-Funded Project Baseline Budget

Figures 7.3.1-1 and 7.3.1-2 present the DUSEL MREFC-funded Project baseline budget in unescalated FY 2010 dollars and escalated *then-year* dollars, respectively. Additional detail is provided in Volume 2, and the discussion of the technical facility can be found in Volume 5, *Facility Preliminary Design*.

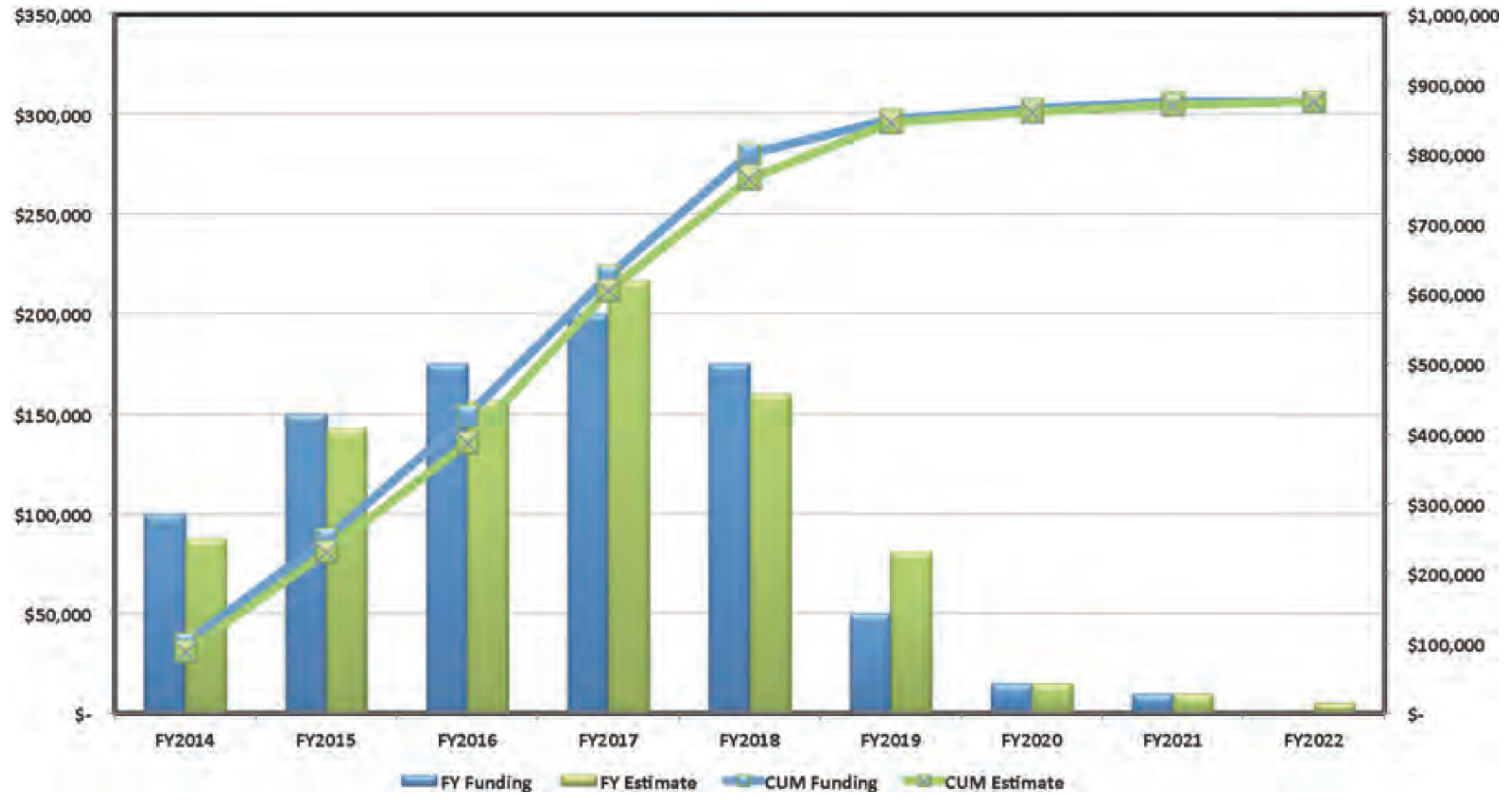

**Figure 7.3.1-1** DUSEL MREFC estimates in unescalated FY 2010 thousand dollars.

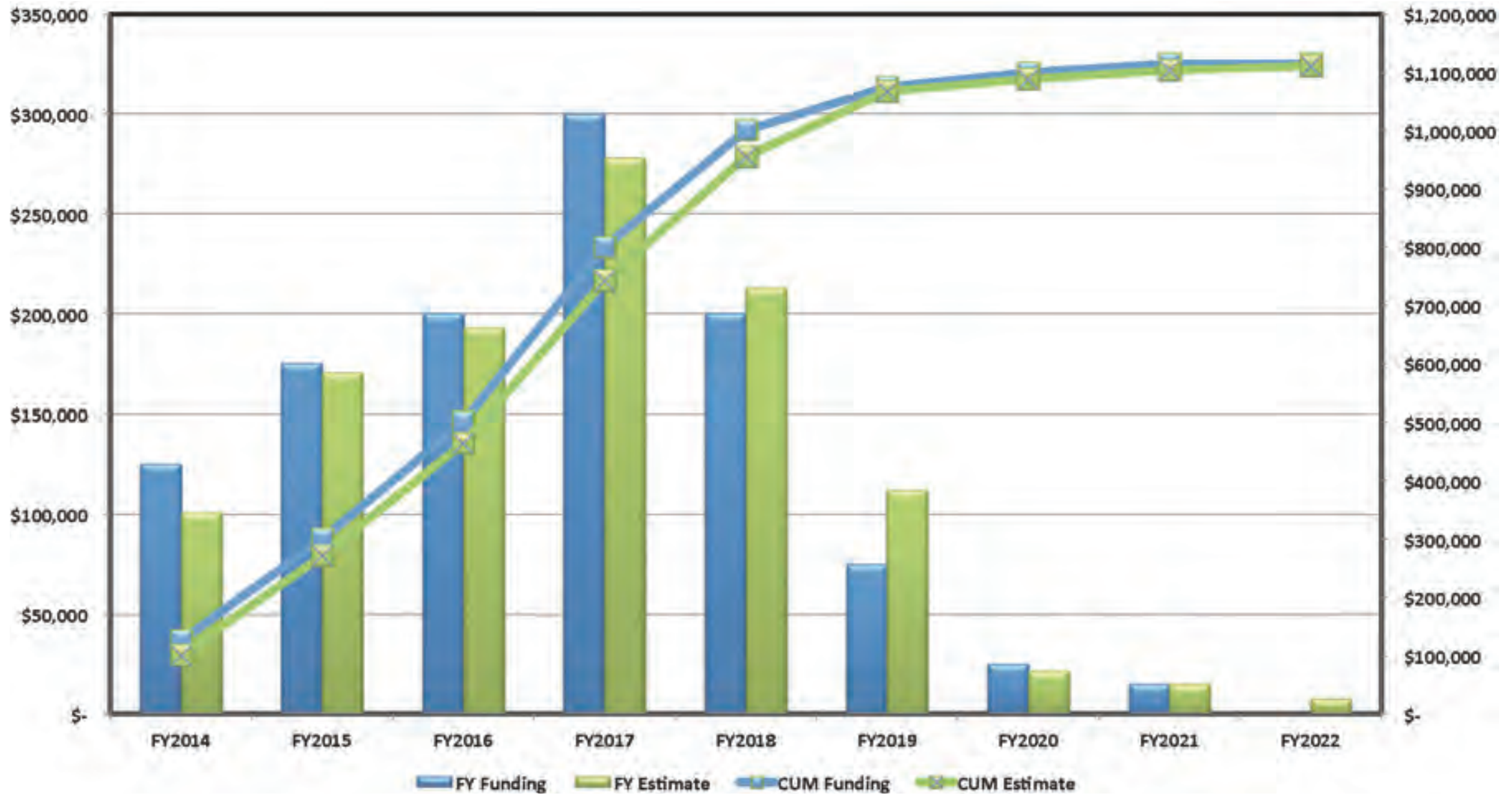

**Figure 7.3.1-2** DUSEL MREFC estimates in escalated then-year thousand dollars.

### 7.3.2 Estimated Operations Costs (R&RA)

Consistent with the NSF *Large Facilities Manual*, estimated costs of DUSEL activities (Design, Construction, and Operations) are applied to the appropriate NSF budget account—R&RA or MREFC. NSF policy states that the R&RA account will be used to fund concept, development, operations maintenance, renewal, or termination costs.

Operations costs estimated as part of DUSEL are consistent with these defined requirements for R&RA funding. These include the Final Design, maintaining safe access for design development and the initial scientific program, operations concurrent with the construction project, deferred-maintenance activities, and eventual operations. Figures 7.3.2-1 and 7.3.2-2 chart the annual R&RA expenditures by category for DUSEL in unescalated FY 2010 dollars and *then-year* dollars, respectively. More detailed R&RA cost information may be found in Volume 2, and scope of the Operations by time phase is described in Volume 10, *Operations Plans*.

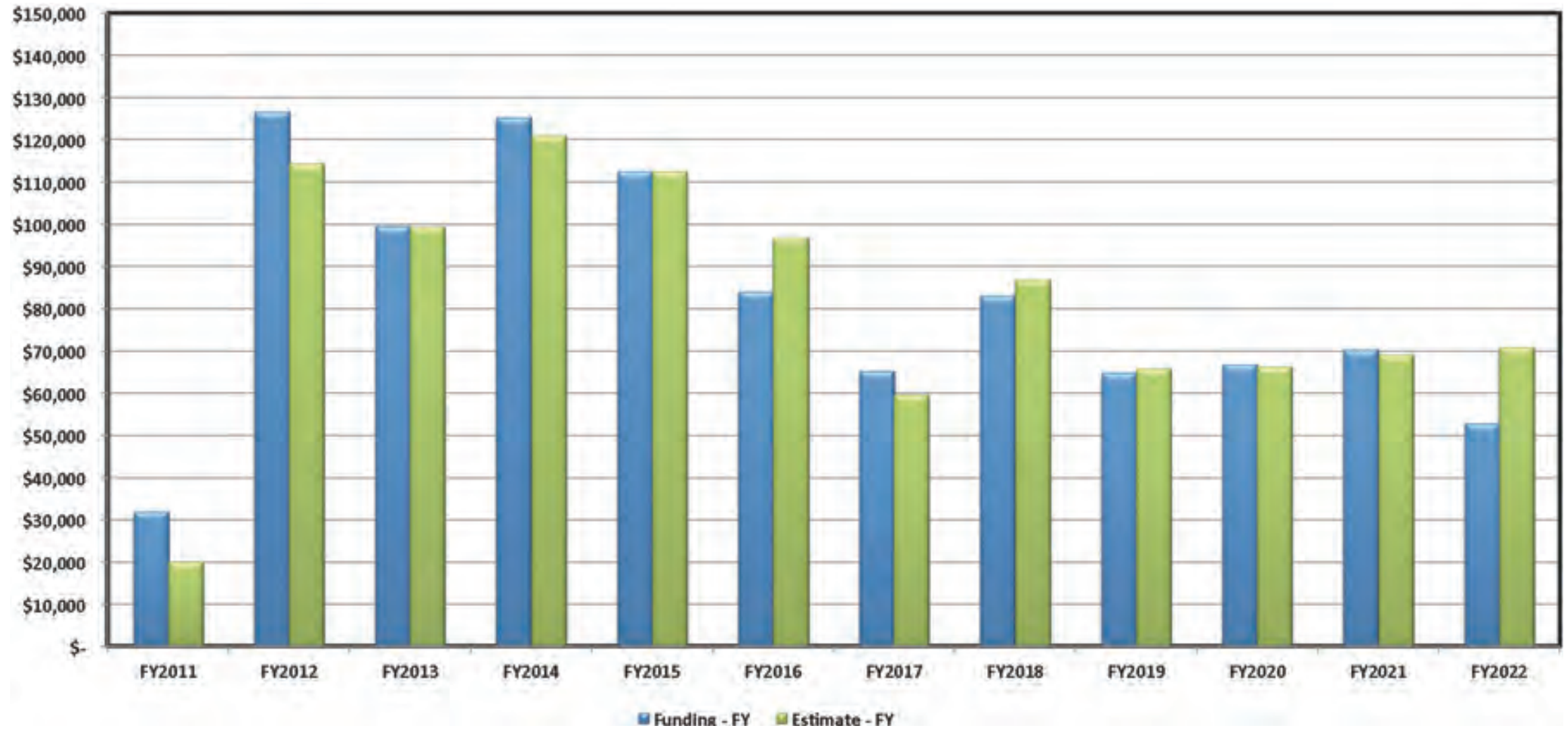

**Figure 7.3.2-1** R&RA projected costs by year and category in unescalated FY 2010 thousand dollars. Note that FY 2011 represents a six-month period of funding.

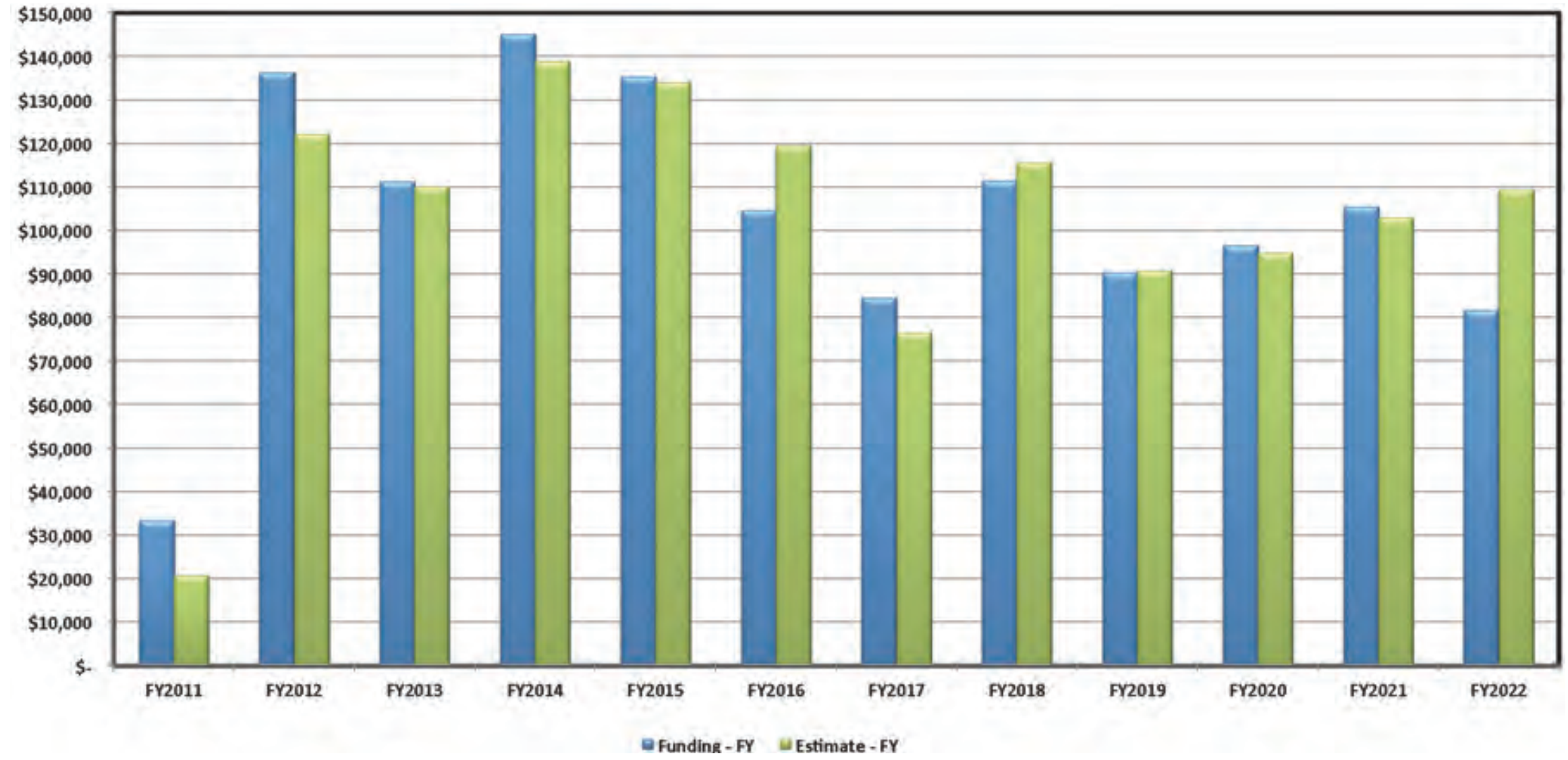

**Figure 7.3.2-2** R&RA projected costs by year and category in escalated then-year thousand dollars. Note that FY 2011 represents a six-month period of funding.

## 7.4 Management Reserve

As discussed in Volume 2, the DUSEL MREFC Account funding supports two primary scope elements: the Facility (DUS.FAC) and the science program (DUS.SCI). Within the Facility, a clear baseline is being established of $575 million including contingency, as detailed in Volume 2. The Management Reserve level was determined based on a bottom-up analysis and cross-checked with risk and uncertainty analysis. The schedule for development of the science program lags behind the Facility design development. Consequently, at this time it is not possible to establish commensurate baseline and control estimates for the scientific program. The $300 million allocated for the scientific program is a planning package that includes Management Reserve allowance to support the execution of the construction of the experiments. The configuration control process described in Chapter 7.13 governs the use of Management Reserve. Volume 2 describes in more detail the development and control of Management Reserve.

## 7.5 Project Schedule

Figure 7.5-1 is the summary schedule of the DUSEL Project. Figure 7.5-2 shows the critical path sequence and Table 7.5 lists the Level 1 and Level 2. These milestones constitute the baseline milestones that are subject to the configuration management processes and thresholds described in Section 7.13.1.1, *Configuration Management Thresholds*. Additional detail concerning the schedule and milestones are presented in Volume 2 and the discussion of the technical scope associated with the milestones is presented in Volume 5.

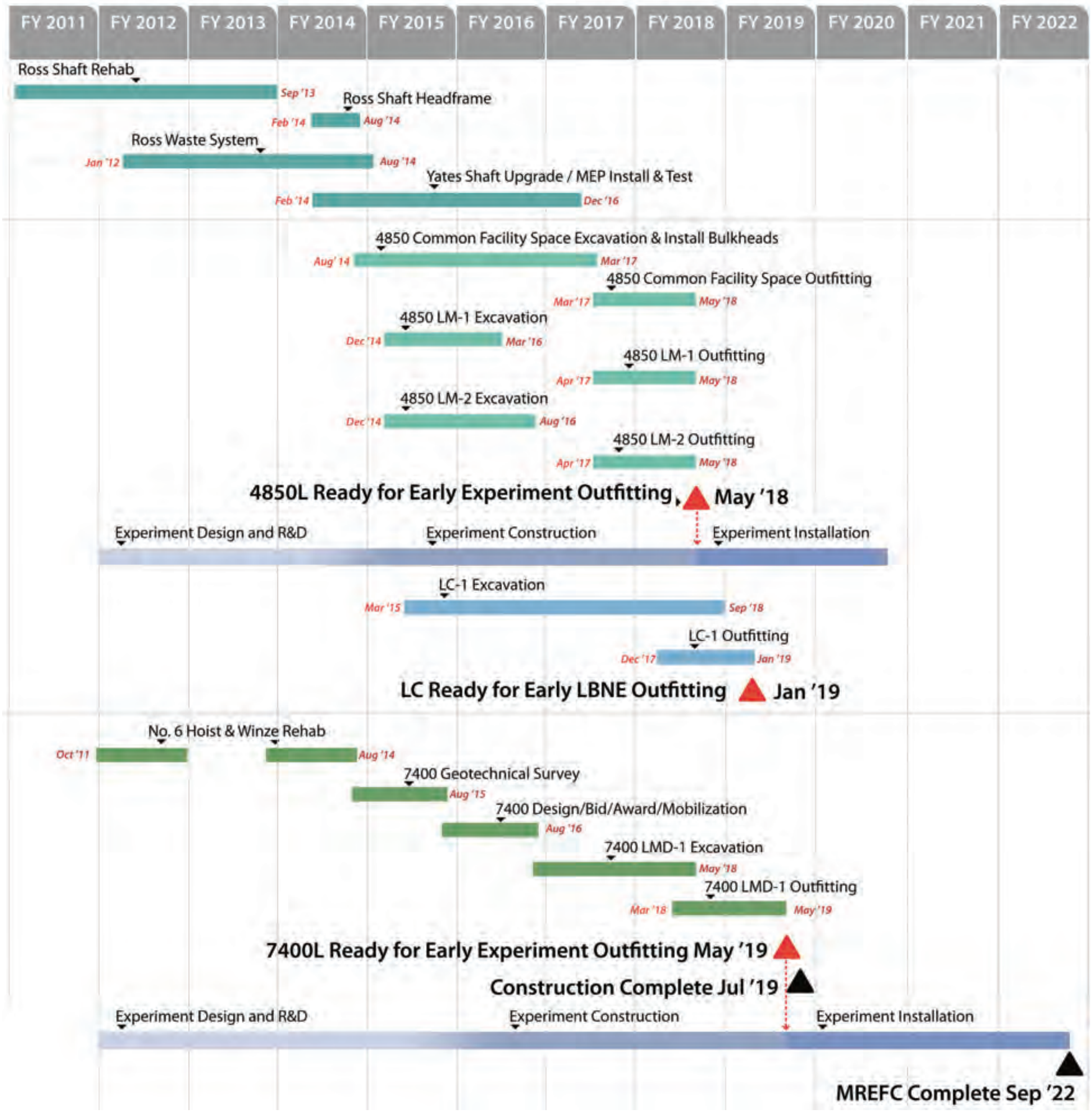

**Figure 7.5-1** DUSEL summary schedule. [DKA]

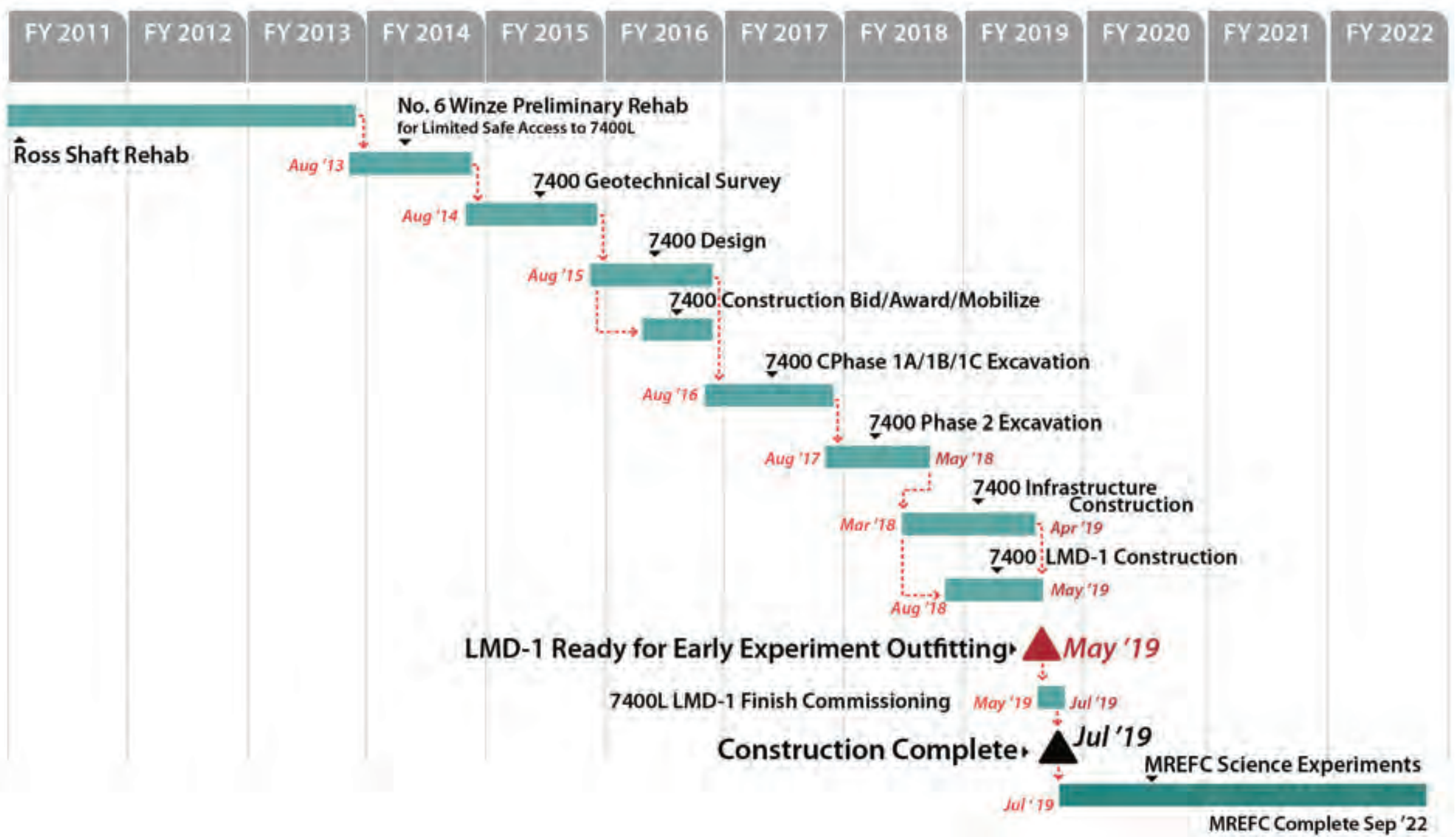

**Figure 7.5-2** DUSEL critical path sequence. [DKA]

| Level 1 External—Major Milestones | |
|---|---|
| **Milestone Description** | **Date** |
| S3 Award Complete | 15-Sep-08 |
| S4 Awards Announced | 4-Aug-09 |
| NSF Approval of CA2 Funds to Complete PDR Efforts | 24-Sep-09 |
| LBNE CD-0 DOE Approval | 8-Jan-10 |
| NSF Release of additional funds to complete PDR efforts | 12-Jan-10 |
| PDR 30% A&E Cost Estimates & Schedule Complete | 7-Apr-10 |
| PDR 60% A/E & CM Cost Estimates, Reconciled Cost Estimates & Schedules Complete | 4-Aug-10 |
| PDR 90% A/E & CM Cost Estimates, Reconciled Cost Estimates & Schedules Complete | 7-Oct-10 |
| PDR 100% A/E & CM Cost Estimates, Reconciled Cost Estimates & Schedules Complete | 22-Nov-10 |
| NSF Annual Review Complete: Spring 2011 | 18-Apr-11 |
| Preliminary Design Report (PDR) submitted to NSF | 29-Apr-11 |
| NSF Start PDR Baseline Review by National Science Board | 2-May-11 |
| NSF Approval of DUSEL Funding Proposal for R&RA-funded Final Design 2012-2013 | 16-Jun-11 |
| NSF Approval of DUSEL Funding Proposal for R&RA-funded Operations 2012-2013 | 16-Jun-11 |
| Final Design - Contract Award | 2-Feb-12 |
| NSF Annual Review Complete: Spring 2012 | 19-Apr-12 |
| LBNE CD-1 DOE Review Complete | 6-Jul-12 |
| LBNE CD-1 DOE Approval | 6-Sep-12 |
| Final Design - 60% | 17-Sep-12 |
| MREFC Construction Funding - NSF Authorized | 1-Oct-12 |
| Final Design - 90% | 14-Feb-13 |

| Level 1 External—Major Milestones | |
|---|---|
| **Milestone Description** | **Date** |
| Final Design Review (FDR) Complete | 1-Apr-13 |
| NSF Annual Review Complete: Spring 2013 | 18-Apr-13 |
| Construction Bid Package - Prepare | 14-May-13 |
| Final Design Review (FDR) Approval | 3-Jun-13 |
| NSF Approval of DUSEL Funding Proposal for R&RA-funded Operations 2014-2022 | 18-Jun-13 |
| Final Design - 95% | 23-Jul-13 |
| Construction Bid Package - Release | 13-Aug-13 |
| Construction Bid Package - Award Contract | 18-Nov-13 |
| Ross Shaft Rehabilitation Complete | 31-Jan-14 |
| MREFC Construction Funding - NSF Released - Start On Site Work | 3-Feb-14 |
| NSF Annual Review Complete: Spring 2014 | 18-Apr-14 |
| Ross Shaft and Waste Handling Available | 2-Aug-14 |
| NSF Annual Review Complete: Spring 2015 | 20-Apr-15 |
| DLL Final Design - 60% | 24-Dec-15 |
| DLL Final Design Review (FDR) Complete | 28-Jan-16 |
| Yates Shaft Full Ventilation Available | 4-Feb-16 |
| DLL Final Design - 90% | 1-Mar-16 |
| DLL Construction Bid Package - Prepare | 29-Mar-16 |
| DLL Final Design Review (FDR) Approval | 31-Mar-16 |
| NSF Annual Review Complete: Spring 2016 | 19-Apr-16 |
| Yates Shaft Rehabilitation Complete | 1-Jun-16 |
| DLL Final Design - 95% | 13-Jun-16 |
| DLL Construction Bid Package - Release | 13-Jun-16 |
| DLL Construction Bid Package - Award Contract | 15-Jul-16 |
| #6 Winze Rehabilitation Complete | 26-Oct-16 |
| NSF Annual Review Complete: Spring 2017 | 20-Apr-17 |
| NSF Annual Review Complete: Spring 2018 | 19-Apr-18 |
| MLL Lab Module 2 Construction Complete (ready for researcher fitout) | 15-Nov-18 |
| MLL Lab Module 1 Construction Complete (ready for researcher fitout) | 16-Nov-18 |
| NSF Annual Review Complete: Spring 2019 | 18-Apr-19 |
| LGC Large Cavity 1 Construction Complete (ready for LBNE fitout) | 31-Jul-19 |
| DLL Lab Module 1 Construction Complete (ready for researcher fitout) | 9-Dec-19 |
| NSF Annual Review Complete: Spring 2020 | 17-Apr-20 |
| MREFC-funded Facility Construction Complete | 27-Jul-20 |
| NSF Annual Review Complete: Spring 2021 | 20-Apr-21 |
| NSF Approval of DUSEL Funding Proposal for R&RA-funded Operations 2022- | 18-Jun-21 |
| NSF Annual Review Complete: Spring 2022 | 20-Apr-22 |
| MREFC Experiments Construction Complete | 29-Mar-24 |

**Table 7.5** DUSEL Level 1 milestones. Milestones reflect late finish dates that include Schedule Reserve.

## 7.6 Basis of Estimate

The DUSEL Facility Preliminary Design estimates are at a level necessary to establish a baseline, budget authorization, or control budget based on relevant consensus standards such as ASTM E 2516-06[2] or AACE 17R-97.[3] These standards classify the cost estimates, based on the method used to generate them, into two general categories: stochastic and deterministic. A proper budget authorization or control (a Class 3 Estimate), according to ASTM E 2516-06, is generally mixed, but primarily stochastic. The Owner's costs with the vetting and cross-referencing of the expert judgment approach reflect sound Class 3 estimates and represent 11% of the MREFC-funded Project estimate. All of the DUSEL estimates based on the completed design are primarily deterministic and represent 89% of the estimated DUSEL MREFC-funded construction costs and are therefore regarded as Class 2 estimates. Specific information on the detailed development of the basis and quality of the DUSEL Facility estimates are found in Volume 2, *Cost, Schedule, and Staffing*.

In addition to the estimating approaches outlined above, all facility estimates and schedules generated by the design contractors during Preliminary Design were independently verified by the construction manager and reconciled through a formal estimate and schedule review process described in Volume 5, *Facility Preliminary Design*.

## 7.7 Project Risk Analysis and Management

The DUSEL Project manages risk through a comprehensive strategy that emphasizes risk identification and effective risk management plans that avoid, mitigate, transfer, or possibly accept risks throughout the Project life cycle. The Principle Investigator/Executive Director (PI/ED) is assisted by a Risk Management Team (RMT) consisting of the Project Director (Chairperson); Facility Project Manager; Environment, Health, and Safety (EH&S) Director; and Systems Engineering and Integration Manager. The RMT meets regularly to review, monitor, and recommend action on identified risks. The Project's objective is to maintain contingency commensurate with Project risks through all phases in order to ensure that the full Project scope is completed within budget and on schedule. The Risk Management Plan (RMP) is documented in Appendix 9.C.

> **Project risk** is a measure of the potential inability to achieve Project objectives within defined scope, cost, schedule, and technical constraints.
>
> **Project risk management** is the continual process of identifying, quantifying, planning, responding to, and controlling risk events to maximize the potential for the success of and minimize the cost and schedule impacts of an activity.
>
> **Project risk events** are defined as individual occurrences or situations that may occur in the future that are determined to have potential negative or positive impacts on a project.
>
> A **Project Risk Registry** showing each identified risk, its relative ranking, and mitigation approach is approved and monitored by the RMT. The top Project risks in the Risk Registry are in Appendix 9.AD.

Project risks that are managed in the RMP do not include the detail of the EH&S or OSHA hazards, which are identified and managed by the EH&S department. The EH&S department conducts a separate comprehensive risk-based hazard analysis, and identifies avoidance and mitigation strategies in support of the effort to reduce Project risk. The assessment of these hazards provides input to the risk management process if they are deemed Project risks that may increase cost, cause schedule delays, or reduce the performance of the Project.

DUSEL risk management takes a comprehensive view of the DUSEL Project and Operations to identify and address specific project risks that require assessment, mitigation strategies, and tracking. While the initial risk assessment will be focused on the establishment of a valid baseline, risk assessment will be an ongoing process throughout the Project life cycle.

During the Preliminary Design phase, risks were identified and assessed according to their probability and impact to the success of DUSEL to focus the Project's risk management efforts. The risk exposure rankings include levels of Insignificant, Minor, Moderate, High, and Critical. The risks represented in Table 7.7 are the risks currently captured in the risk registry with exposure rankings of High and Critical.

| DUSEL Critical and High Risks | | |
|---|---|---|
| **ID / Level** | **Risk Statement and Description** | **Summary of Risk Mitigation Plan** |
| 222/<br>Critical | **IF selection of the LBNE detector design options and location is delayed, THEN project schedule delays and costs underestimates will occur.**<br>Because the PDR design has scheduled excavation and outfitting in order to maximize schedule performance between the Lab Modules and the Large Cavity, a delay in selection and design of the LBNE WCD at 4850L could cause interference with Lab Module design and construction. | LBNE to determine design decision across the various options before the start of the DUSEL Final Design phase. |
| 198/<br>High | **IF R&RA funding is delayed or not in sufficient amount to support accessing the Facility to the 7400L, THEN geotechnical data collection will be delayed, which may impact completion of final DLL Campus design and construction.**<br>Because the 7400L is only at conceptual level of design maturity, it is crucial to complete geotechnical studies of the level required to further the design to support the construction schedule outlined in the baseline. | Establish need date for funding, which is 15 months prior to DLL Final Design start.<br>• Build required funding into Management Reserve early to ensure funding availability<br>If delayed funding then:<br>• Combine Preliminary and Final Design investigations to one campaign to compress schedule prior to a Final Design start.<br>• Delay large excavations in DLL until geotechnical investigations are complete.<br>• Optimize the geotechnical site investigations to targeted high-risk areas early in support of DLL. |
| 226/<br>High | **IF graphitic shears and faults exist in DLL Campus area, THEN large rooms may need to be moved or—worst case—the entire campus moved to avoid graphitic shears and faults, resulting in higher costs and schedule delays.**<br>Because the 7400L is only at conceptual level of design maturity and the 7400L is currently underwater, it is not possible to understand the extent to which the 7400L contains these faults. As soon as the level is accessible and geotechnical data is collected, the outcome of this risk and more detailed mitigation steps will be created. | DUSEL has requested a proposal from Golder Associates to scope a proposal to address early geotechnical site investigation from the 4850L and as early as possible from the 7400L during the #6 Winze rehabilitation. |
| 81/<br>High | **IF rehabilitation of the Ross Shaft Hoist and Headframe takes longer or costs more than planned, THEN it will result in delays and cost increases during construction.**<br>The refurbishment of the shafts is part of the planned operations activity in advance of construction and may have an impact on MREFC-funded activities if schedule is impacted due to unforeseen issues with the current shaft or insufficient funding. | Rehabilitation of the Ross Shaft will be performed prior to the MREFC-funded construction. In this way, the impact of schedule delays will be minimized. During the Preliminary Design phase, detailed Non-Destructive Testing (NDT) will be performed to develop a thorough understanding of the shaft and supporting infrastructure's condition. |

**Table 7.7** DUSEL Critical and High Category Risks.

## 7.8 Project Organization, Governance, Oversight, and Advisory Functions

The DUSEL Project is organized to establish clear lines of authority and responsibility to perform the Project's scope of work. The roles are related to the scope of work at a given level and carry through the institutions involved in DUSEL. The DUSEL Project anticipates the formation of the DUSEL LLC during the Final Design phase to oversee the Project.

### 7.8.1 Institutional Organization Roles and Responsibility

#### 7.8.1.1 University of California

The Regents of the University of California (UC Regents) will be sole NSF awardee and the contractor for all aspects of the National Science Board (NSB)-approved construction Project and Operations. Consequently, the UC Regents are accountable to NSF for the DUSEL Project and Operations. UC Berkeley is responsible on behalf of the UC Regents for the management and execution of both the DUSEL Project and DUSEL Operations. UC Berkeley and the UC Regents have prime responsibility and are accountable to NSF for the Design, Construction, and Operations of the Facility as well as active participation in the scientific program.

Within UC Berkeley, DUSEL reports to the office of the Vice Chancellor for Research and institutionally relies on the Research Enterprise Services unit within the Vice Chancellor's office for support. The Principal Investigator/DUSEL Laboratory Director (PI/LD) shall be a member of the UC Berkeley staff. UC Berkeley will maintain a staff office reporting to the PI/LD.

In discharging its responsibility for DUSEL, UC Berkeley will have one principal subaward to the DUSEL Limited Liability Company (DUSEL LLC) for the Final Design, Construction, and Operations of the DUSEL Project at Homestake. In order to properly discharge its responsibility for the management and direction of the scientific program, UC Berkeley will likely have one or more subawards to institutions engaged in underground scientific programs, including activities within the Joint Institute for Underground Science discussed in Volume 3.

#### 7.8.1.2 DUSEL Limited Liability Company

On November 17, 2010, the UC Regents approved the participation in, and formation of, a limited liability company to operate and manage DUSEL.[4] The legal entities participating in the DUSEL LLC to manage and operate DUSEL are the UC Regents, the South Dakota Board of Regents (SD Regents), and the SDSTA. DUSEL LLC will be organized under the laws of South Dakota with the UC Regents having majority interest in order to ensure that UC Berkeley adequately executes its responsibility for DUSEL to the NSF (note that the legal entity of the cooperative agreement with the NSF is the UC Regents).

The responsibility of the DUSEL LLC is to handle the day-to-day management of the Final Design, Construction, and Operations of DUSEL. The DUSEL LLC is being formed to streamline both the management and subcontract oversight of DUSEL and to manage risk and financial liability. The formation of LLCs to manage the construction and operation of facilities such as DUSEL is common practice for the UC Regents. The DUSEL LLC will accomplish a number of goals:

- It provides insulation from upward legal and financial liability associated with DUSEL construction and operations to its participating legal entities.

- It centralizes and focuses effective management of on-site operations of DUSEL, with complete accountability to the participating legal institutions and, by extension, to NSF.
- It provides a centralization and standardization of subcontract and subaward management, making proper oversight by UC Berkeley and NSF simpler and more direct.
- By centralizing finance and accounting functions for DUSEL, it simplifies management and oversight.
- It provides for a proper flow-down of contractual requirements and accountability from NSF through UC Berkeley throughout all entities needed for the successful execution of Design, Construction, and Operations of DUSEL.

Clear institutional accountability is maintained while appropriate financial and management responsibility is delegated as shown in Figure 7.8.1.2.

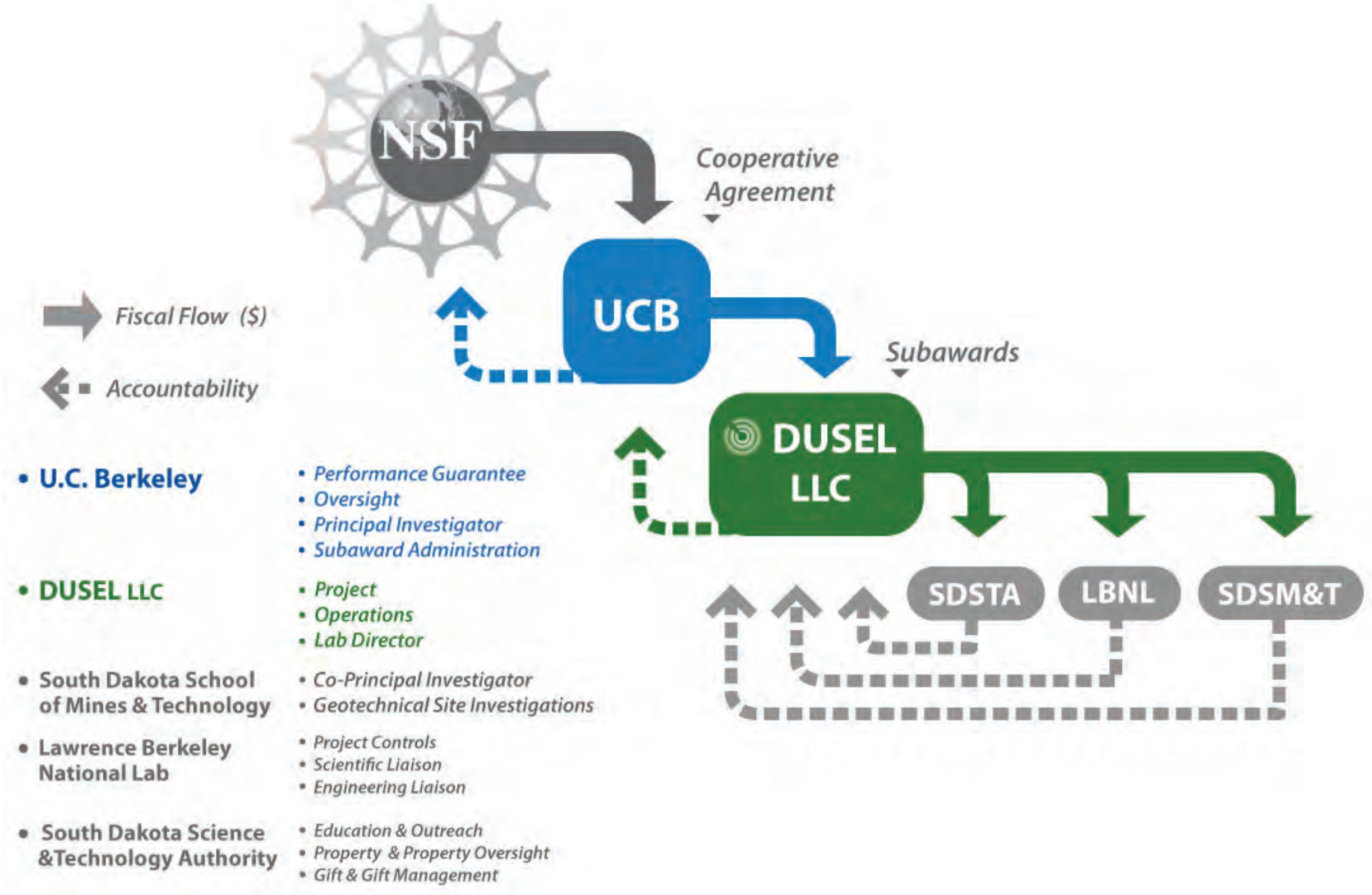

**Figure 7.8.1.2** DUSEL institutional accountability and fiscal flow.

During the NSB-approved Final Design phase, the DUSEL LLC will be responsible for the management of the Final Design, design subcontractors, construction management subcontract, and subawards to South Dakota School of Mines and Technology (SDSM&T), SDSTA, and Lawrence Berkeley National Laboratory (LBNL) in discharging their specific roles related to DUSEL. The DUSEL LLC will be responsible for maintaining safe access at the Homestake site, both underground and on the surface. During construction, the DUSEL LLC, in addition to managing concurrent operations and scientific access as appropriate, will be responsible for the management and direct oversight of the construction, including the general contractor and all subcontractors. During the Operations phase, the DUSEL LLC will be responsible for the day-to-day operations, maintenance, incremental improvements, and support of the science program at the Homestake site. Additionally, throughout all phases of the Project, the DUSEL

LLC will be responsible for ensuring compliance of operations and activities in accordance with the lease conditions to the SDSTA with relation to the Property Donation Agreement (PDA) governing the Homestake site.

#### 7.8.1.3 Institutional Governance—Sub-PI

The DUSEL Project and Operations activities are organized around the deliverables and scope of work defined in the WBS. The division of responsibilities between organizations is not rigorously aligned with the WBS. Nevertheless, to ensure clear governance, each institution's responsibilities are defined through subcontracts. Each institution has a Sub-Principal Investigator (Sub-PI) who is the point of contact and accountable to UC Berkeley or the DUSEL LLC (depending upon the subcontract) to ensure the execution of the scope of work delegated to the institution. The Sub-PI ensures that institutional means and measures provide adequate controls and administration of funds transferred to the institution to accomplish its respective scope of work. Finally, the Sub-PI provides both project and line supervision as appropriate for respective institution's staff working on DUSEL.

#### 7.8.1.4 South Dakota School of Mines and Technology

SDSM&T is the home institution of the Co-Principal Investigator (Co-PI) and the Facility Project Manager (Level 2 Manager). It is responsible for the technical direction of the Facility design development and an important seat for the development of the scientific program, especially for the BGE experiments. Within SDSM&T, the DUSEL Project reports to the Vice President of Research. SDSM&T is an active participant in the development of the design, trades, options, risk analysis and management, scientific program and liaison work, and day-to-day operations of DUSEL. SDSM&T will be under subcontract from the DUSEL LLC.

#### 7.8.1.5 South Dakota Science and Technology Authority

The SDSTA plays a key role as the legal owner of the Homestake property and facilities and one of the institutional entities that control the DUSEL LLC. During the Preliminary Design and Transition phases, the SDSTA has been responsible for day-to-day operations of Sanford Laboratory and the Homestake site and facilities. This has included dewatering the underground facility, the treatment of the groundwater, and its discharge in compliance with regulations and permits. The SDSTA has been responsible for maintaining safe access to the Homestake site and infrastructure both underground and at the surface, and is responsible for the execution of the projects designed to address deferred maintenance and necessary safety infrastructure.

The day-to-day operations and maintenance responsibilities of the site and the existing Sanford Laboratory will be transferred to the DUSEL LLC once it is established. After the transfer, the SDSTA's role is primarily as property owner, including ensuring that compliance with the PDA and the requirements of the T. Denny Sanford gifts are maintained. The SDSTA will be under direct subcontract from the DUSEL LLC for those activities and scope that it retains.

#### 7.8.1.6 Lawrence Berkeley National Laboratory

LBNL provides scientific and engineering liaison support in the development of the DUSEL scientific program. It also provides project management and a project controls support role to UC Berkeley. LBNL is also directly involved within a full range of the scientific programs and endeavors associated with DUSEL in multiple domains in physics and geosciences. During the Final Design and MREFC-funded construction phases, LBNL will be under subcontract to the DUSEL LLC.

#### 7.8.1.7 Black Hills State University

Black Hills State University (BHSU) is responsible for the development of the education and public outreach activities within DUSEL. It will be under subcontract to the DUSEL LLC. The DUSEL Education and Outreach Director (Level 3) is a BHSU faculty member. Organizationally, the BHSU DUSEL efforts report to the BHSU President.

### 7.8.2 Project Governance, Project Organization Roles, and Responsibilities

DUSEL has developed a project organization that focuses on the major responsibilities and work scope of the DUSEL Project and emphasizes clear roles and responsibilities and effective management. The Project is organized along the lines of the WBS and focused on advancing the Project deliverables. The description in this section is focused on the organization that will manage all phases of the DUSEL Project and Operations starting in Final Design and through to full steady-state Operations after the conclusion of the DUSEL MREFC-funded construction project.

The DUSEL Project is organized and governed through a structure focused and optimized for the execution of the Project and the day-to-day Facility operations. The management approach is closely aligned to the ultimate Facility development, Project deliverables, and Operations, defined by the DUSEL WBS (See Chapter 7.2). Roles and responsibilities are broken down along the WBS lines into the following major hierarchical classifications:

- Level 1 Managers: DUSEL Total Project and Operations-wide responsibilities
- Level 2 Managers: Major system responsibilities (e.g., DUS.FAC, DUS.OPS)
- Level 3 Crosscutting Managers: Major subsystems that cut across the major systems of the Project (e.g., DUS.PRJ.EHS and DUS.PRJ.SYS)
- Level 3 Managers: Subsystem responsibilities (e.g., DUS.FAC.UGI)
- Control Account Managers (CAM)

Figure 7.8.2 shows the DUSEL organizational structure and the key positions within the organization.

The strength of this project organization is that it reflects the division of work scope; it emphasizes interfaces and connections; and it defines clear roles and responsibilities to effectively execute the construction project and Facility operations.

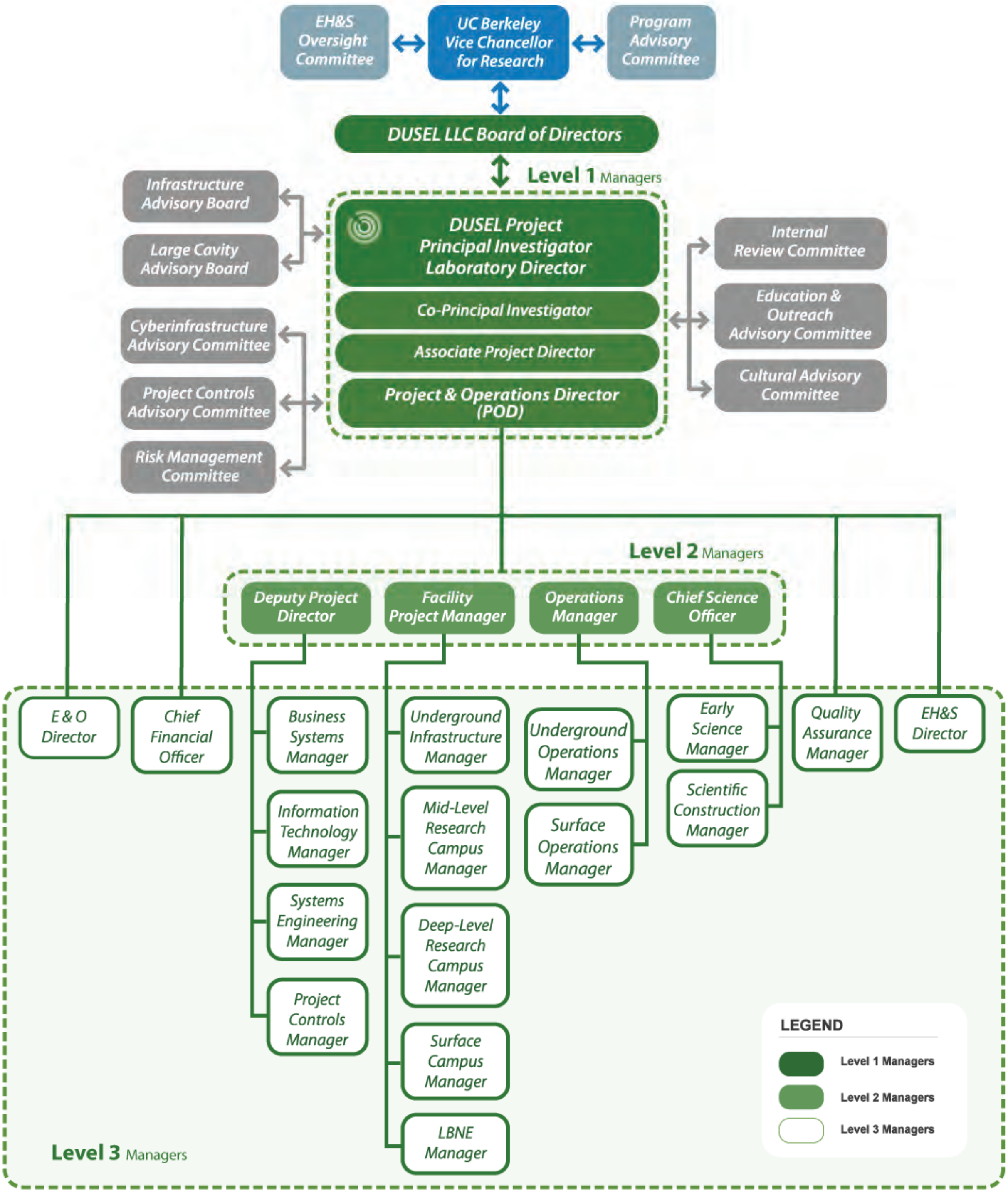

**Figure 7.8.2** DUSEL LLC organization chart. [DKA]

### 7.8.2.1 Level 1 Project-Wide Management

The Level 1 Project-Wide Management, also referred to as the Central Project Directorate, consists of the Principal Investigator/Laboratory Directory (PI/LD), the Co-PI, the Associate Director, and the Project and Operations Director (POD). The Level 1 Project Management directs and oversees all of the DUSEL Project.

#### 7.8.2.1.1 Principal Investigator / Laboratory Director.

The PI/LD is the key individual accountable for the Project success. He or she is the principal spokesperson for the Project, and the main point of contact for all matters related to the Project. The PI/LD is responsible for establishing the scientific reach and mission of the Project and ensuring that goals and objectives continue to be met as the design and construction and operations progress. The PI/LD is responsible for the ensuring the completion of the Project and through the operations and other DUSEL activities ensure the safe and successful delivery of the scientific program.

The PI/LD is the chair of the Configuration Control Board and thus has final approval and control over contingency and other configuration controlled items (see Section 7.13.1).

The PI/LD is a UC Berkeley employee who reports to the Vice Chancellor for Research. The PI/LD is also accountable to NSF for the successful execution of the DUSEL Project and Operations.

The Principal Investigator is also the DUSEL Laboratory Director and as such is the Chief Executive Officer of the DUSEL LLC, reports to the DUSEL LLC Board of Directors, and is accountable for all DUSEL activities at Homestake as well as the complete DUSEL construction project. In this role, the PI/LD also has responsibility for all subcontracts made by the DUSEL LLC and holds joint appointment with UC Berkeley and the DUSEL LLC.

Reporting to the PI/LD are the POD, the DUSEL LLC Chief Financial Officer, the DUSEL Chief Science Officer (CSO), the DUSEL EH&S Director, the DUSEL Education and Outreach Director, and the DUSEL Quality Assurance Manager.

#### 7.8.2.1.2 Co-Principal Investigator

The Co-PI is a secondary point of contact for DUSEL primarily for external Project matters. The Co-PI acts on behalf of the PI/LD in the PI/LD’s absence. The Co-PI is a voting member of the Configuration Control Board (CCB).

#### 7.8.2.1.3 Associate Laboratory Director

The Associate Laboratory Director (ALD) is a secondary point of contact for DUSEL primarily for internal Project matters. The ALD is a voting member of the CCB.

#### 7.8.2.1.4 Project and Operations Director

The DUSEL POD is responsible for the day-to-day management and execution of the DUSEL Project and Operations at the Homestake site. The POD is responsible for ensuring the adequacy and effective management of cost, schedule, risk, safety, and quality within the Project, and operations consistent with Project and Operations goals and objectives. The POD is a member of the CCB and the Risk Management Team (RMT) chair. Reporting to the POD are all Level 2 major system managers, with the exception of the Chief Science Officer. The POD ensures that proper interfaces, prioritization, and balance are maintained between the DUSEL construction project and the day-to-day physical plant operations.

### 7.8.2.2 Level 2 Major System Management

Level 2 major system managers are responsible and accountable for managing and ensuring proper execution within cost, schedule, and scope objectives of the DUSEL baseline. Level 1 and Level 2 management constitute the Senior Management of DUSEL. Level 2 managers are responsible for development of policy and key procedures within the major systems of DUSEL. There are four Level 2 major system managers: the CSO, the Facility Project Manager, the Operations Manager, and the Deputy Project Director. They are responsible for identifying risks, interfaces, and uncertainties that could impact the Project or Operations from their respective major systems. All Level 2 managers are members of the CCB.

#### 7.8.2.2.1 Chief Science Officer

The CSO is responsible for overseeing the scope and management of the DUSEL scientific program, including the management of the ISE included in the DUSEL science program (DUS.SCI). The CSO is responsible for ensuring the development of the proper scientific requirements and interfaces with the DUSEL Facility. All experimental scientific and engineering liaisons assigned to work with the various ISE collaborations report to the CSO. Likewise, the management and oversight of the early science program report to the CSO.

#### 7.8.2.2.2 Facility Project Manager

The Facility Project Manager oversees and manages by far the largest single major system (DUS.FAC) and is responsible for and manages the entire scope, cost, schedule, and quality of all of the underground and surface scope of the DUSEL Facility construction. All Facility Level 3 managers report to the Facility Project Manager.

#### 7.8.2.2.3 Operations Manager

The Operations Manager is responsible for maintaining safe access to both the surface and underground facilities at Homestake. The Operations Manager is responsible for maintaining the hoists, hoisting operations, pumps and water treatment, and all other aspects of day-to-day physical plant operations.

#### 7.8.2.2.4 Deputy Project Director

The Deputy Project Director is a primary point of contact for the project management approach, structure and design, and execution, including long-term policy and procedures. The Deputy Project Director is the Level 2 manager of all Level 3 crosscutting systems that do not directly report to the PI/LD. The Deputy Project Director is responsible for managing and developing policy and procedures that impact both the Operations and management of the Project.

### 7.8.2.3 Level 3 Subsystem Management

The Level 3 Subsystem Management can be divided into two categories: major scope subsystems and crosscutting system managers.

#### 7.8.2.3.1 Level 3 Subsystem Managers

The Level 3 subsystem managers are responsible for integrated scope, cost, and schedule of their respective subsystems. They are responsible for identifying risks, uncertainties, interfaces, and requirements that affect their subsystems. They have the authority to develop and implement procedures within the subsystem that do not span more than one Level 3 area. Often this position is the technical point of contact and liaison with external subcontractors. Examples of Level 3 major scope subsystem managers include the Underground Infrastructure Manager (DUS.FAC.UGI) and the Surface Facility Manager (DUS.FAC.SUR).

#### 7.8.2.3.2 Level 3 Crosscutting System Managers

The Level 3 crosscutting system managers are responsible for areas within DUSEL that have project-wide impact and implications. These areas must be coordinated, developed, and established to ensure the successful execution of the DUSEL Project and Operations. The crosscutting systems are generally characterized by not having physical scope deliverables, but instead having policy, procedure, or process deliverables. Level 3 crosscutting system managers either report to the POD or through the Deputy Project Director (DUS.PRJ.PMO, DUS.PRJ.BUS, DUS.PRJ.ITS, DUS.PRJ.SYS) or directly to the Laboratory Director (DUS.PRJ.EHS, DUS.PRJ.QAC, DUS.PRJ.EDO, and Chief Financial Officer [CFO]). The Level 3 crosscutting system managers are responsible for developing policy and procedures for crosscutting office and project-wide systems. They are responsible for the scope, budget, and schedule of their respective systems, and also for identifying risks and issues and presenting them to the Risk Management Team (RMT). They are responsible for developing corrective actions and mitigation plans for risks and issues within their respective systems.

##### 7.8.2.3.2.1 Environment, Health, and Safety Director

The EH&S department has a central and crucial role within the Project (see Volume 6). A fully functioning Integrated Safety Management (ISM) system is fundamental to the successful development of DUSEL. The EH&S effort is a single unified organization across the Project Design, Construction, and Operations. The EH&S effort is integral to the design and execution of the construction phases; day-to-day operations and maintenance of the site and Facility both aboveground and below; scientific program development; user support and interaction; education and public outreach; and is responsible for all visitors to DUSEL. The EH&S Director is responsible for developing policy and general procedures for the complete and global implementation of an integrated EH&S management system. The EH&S Director reports to the PI/LD in order to ensure independence and proper global focus on environmental stewardship and health and safety across the Project. The EH&S Director is responsible for ensuring that the proper knowledge of regulatory requirements is known within the organization to ensure compliance with any applicable regulatory requirements.

##### 7.8.2.3.2.2 Chief Financial Officer

The CFO holds the fiduciary responsibility for the DUSEL LLC. To retain independence and accountability to the DUSEL LLC Board of Directors, the CFO reports to the PI/LD. The CFO is responsible for ensuring that appropriate financial controls and independent audits are in place.

##### 7.8.2.3.2.3 Quality Assurance Manager

The Quality Assurance Manager (QAM) reports to the PI/LD and is responsible for ensuring that policy and general procedures are developed in a manner analogous to the ISM system. The goal of the Quality

Assurance Program is to ensure the DUSEL organization understands the role of quality for the development and execution of the Project, and day-to-day operations. The QAM is responsible for ensuring that the necessary controls, audits, and reviews are in place to verify the expected levels of quality.

#### 7.8.2.3.2.4 Business Systems Manager

The DUSEL Business Systems Manager is responsible for overseeing all business, contract procurement, human resource systems, and physical assets management. Subcontract development and management of a project and organization the size of DUSEL requires the implementation of advanced systems and best practices to ensure compliance with federal, state, parent institution, and funding agency requirements. The Business Systems Manager reports to the Deputy Project Director.

#### 7.8.2.3.2.5 Systems Engineering Manager

The Systems Engineering Manager is responsible for the overall systems engineering including requirements, interface, configuration, and risk management. The Systems Engineering Manager reports to the Deputy Project Director.

#### 7.8.2.3.2.6 Project Controls Manager

The Project Controls Manager (PCM) is responsible for developing and maintaining the project control systems necessary to manage the financial, schedule, and technical performance of the DUSEL construction project and Operations. These systems include the processes and mechanisms necessary to develop and manage a complete cost and schedule estimate over all phases of DUSEL. The PCM is responsible for developing and maintaining the necessary systems for a resource-loaded baseline schedule and for using Earned Value Management measures and analysis to quantify project performance. The PCM is a voting member of the CCB in order to provide data regarding the potential impact of any proposed changes on the Project baseline. The PCM reports to the Deputy Project Director.

#### 7.8.2.3.2.7 Information Technology Manager

The Information Technology Manager is responsible for the design architecture of the information and data systems within DUSEL and the implementation of the necessary computing and information management systems to support the scientific program and the required business, financial, quality, and safety computing systems. The Information Technology Manager reports to the Deputy Project Director.

#### 7.8.2.3.2.8 Education and Outreach Director

The Education and Outreach Director reports directly to the PI/LD and is responsible for developing and managing the education and public outreach—including cultural— programs of DUSEL.

### 7.8.2.4 Control Account Managers

In addition to the positions described above, additional control account managers (CAMs) are named—depending on the size and complexity of the scope—to ensure the proper level of management and control are in place during the execution of the Project. The CAM is responsible for integrated scope, cost, and schedule of his or her respective scope. The CAM is responsible for identifying risks, uncertainties, interfaces, and requirements that impact the respective scope. For the respective work scope, the CAM must understand and report cost and schedule variances.

### 7.8.3 Project Oversight and Advisory Bodies

Several oversight and advisory bodies have been organized within DUSEL to provide peer advice and review, thus ensuring that the execution of the Project and Operations are consistent with best practices and identifying areas for improvement. These oversight and advisory bodies are shown on the organization chart in Figure 7.8.2.

#### 7.8.3.1 Internal Review Committee

In addition to the annual reviews conducted by NSF, an Internal Review Committee composed of subject matter and management experts in the areas relevant to DUSEL has been organized. The Internal Review Committee conducts a review annually or as needed, ideally spaced approximately six months in advance of the NSF Annual Review. A formal report is issued to the Central Project Directorate and the addressing of recommendations is tracked as part of the management and corrective action tracking systems.

#### 7.8.3.2 Program Advisory Committee

The DUSEL Program Advisory Committee (PAC) advises the UC Berkeley Vice Chancellor for Research on all matters related to the DUSEL scientific program. The PAC provides review and advice about the nature, scope, and plans of the DUSEL scientific program. The PAC will review all expressions of interest, letters of intent, and proposals for scientific activities at the DUSEL Facility. The PAC is anticipated to meet annually, at a minimum, and may meet more frequently as needed.

#### 7.8.3.3 EH&S Oversight Committee

The Environment, Health, and Safety Oversight Committee (EHSOC) has been appointed by the UC Berkeley Vice Chancellor of Research and is made up of external experts in a number of EH&S disciplines relevant to DUSEL. The EHSOC helps to ensure that support organizations, administrative policies, processes, and systems—including strategic and operational plans—are adequate to enable DUSEL's mission, including early science, design, construction, and maintenance and operations. The EHSOC conducts an independent assessment of all aspects of ISM and environmental and industrial health within DUSEL. The EHSOC deliberations include evaluating the strategic goals, action plans, and organizational aspects of the support systems and identifying areas for improvement. The committee issues a formal report to the UC Berkeley Vice Chancellor for Research with findings and recommendations that are tracked within the management and corrective action tracking systems to ensure that they are addressed.

The EHSOC independent assessments are to:

- Conduct a critical oversight and examination of EH&S performance
- Provide feedback on proficiencies and performance gaps and recommend corrective actions to close gaps
- Provide advice on best management practices for DUSEL
- Advise on development and maintenance of continuous improvement, Lessons Learned, and operational experience improvement
- Respond to other evaluations and requests by the UC Berkeley Vice Chancellor for Research

#### 7.8.3.4 Infrastructure Advisory Board

Reporting to the PI/LD, the Infrastructure Advisory Board (IAB) is charged with providing technical expertise and advice to the DUSEL Central Project Directorate on issues of Facility infrastructure and support, and life-safety. The IAB is composed of senior subject matter experts selected to specifically assess and address underground facility infrastructure topics.

The IAB is charged with the review of design strategies, risk assessments, quality assurance plans, Facility Preliminary and Final Design plans, the infrastructure baseline design, and parametric cost comparisons for evaluating design options.

#### 7.8.3.5 Large Cavity Advisory Board

The Large Cavity Advisory Board (LCAB) is an independent advisory board responsible for reviewing and advising DUSEL on geotechnical investigations, ground support design, and excavation design of the Large Cavity in support of the LBNE. The LCAB reports to the DUSEL PI/LD and is composed of world-recognized specialists in underground civil construction and excavation.

#### 7.8.3.6 Cyberinfrastructure Advisory Committee

The purpose of the DUSEL Cyberinfrastructure Advisory Committee (CIAC) is to provide an expert panel of Information Technology (IT) professionals, DUSEL experimenters, and DUSEL Facility Operations staff to advise the DUSEL Project on cyberinfrastructure and communications requirements, designs, construction approaches, operations issues, and Lessons Learned as the Project leads the implementation and operation of cyberinfrastructure capabilities at DUSEL. The overall responsibility for the CIAC and its activities resides with the DUSEL Project and Operations Director. The DUSEL Cyberinfrastructure Chief Engineer serves as the lead day-to-day interface to the CIAC.

#### 7.8.3.7 Education Advisory Committee

The Education Advisory Committee (EAC) is a body of science and education professionals with expertise relevant to the development of education and public outreach components of the DUSEL Project. This committee serves in an advisory capacity to the Central Project Directorate and supports development of the education and public outreach programs at DUSEL, including the Sanford Center for Science Education (SCSE). The EAC is charged to:

- Review progress in planning the education and public outreach programs, including program development, stakeholder engagement, SCSE Facility design, and building institutional support for the SCSE
- Help prioritize opportunities and refine educational goals
- Foster partnerships regionally, nationally, and internationally and with special attention to culturally diverse audiences

#### 7.8.3.8 Cultural Advisory Committee

The Cultural Advisory Committee (CAC) advises the Central Project Directorate on policies and initiatives that support DUSEL's commitment to develop the Project in a manner that is consistent with South Dakota and regional cultures. The roles and responsibilities of the committee are to:

- Formulate and submit for consideration a cultural management strategy to recruit, retain, and mutually develop objectives for American Indians and other under-represented groups
- Promote and encourage DUSEL Senior Management to actively integrate regional culture into the Project's design process, procedures, and operations
- Review the program planning for DUSEL and advise on the effectiveness and prioritization of its cultural policies and initiatives

#### 7.8.3.9 Project Controls Advisory Committee

The Project Controls Advisory Committee reports to the Project and Operations Director and is composed of subject matter experts who will provide periodic surveillance of the DUSEL Project Management Control System (PMCS) to ensure continued compliance with the standards associated with Earned Value Management and other project control standards.

#### 7.8.3.10 Risk Management Advisory Committee

The Risk Management Advisory Committee reports to the Project and Operations Director and is composed of subject matter experts who will provide periodic surveillance of the DUSEL Project Risk Management activities to provide guidance on risk management approaches and independent advice on risk analyses and management plans.

## 7.9 Interagency Partnerships

A durable and adaptable agreement on roles and responsibilities between NSF and DOE will be established in a Memorandum of Understanding that defines the relationship between the two agencies and the responsibilities of each in the design and development of the DUSEL Facility and the proposed suite of experiments. The lines of authority and responsibilities for the management of the DUSEL physics experiments are shown in the organization chart of the Joint Oversight Group (JOG) in Figure 7.9.

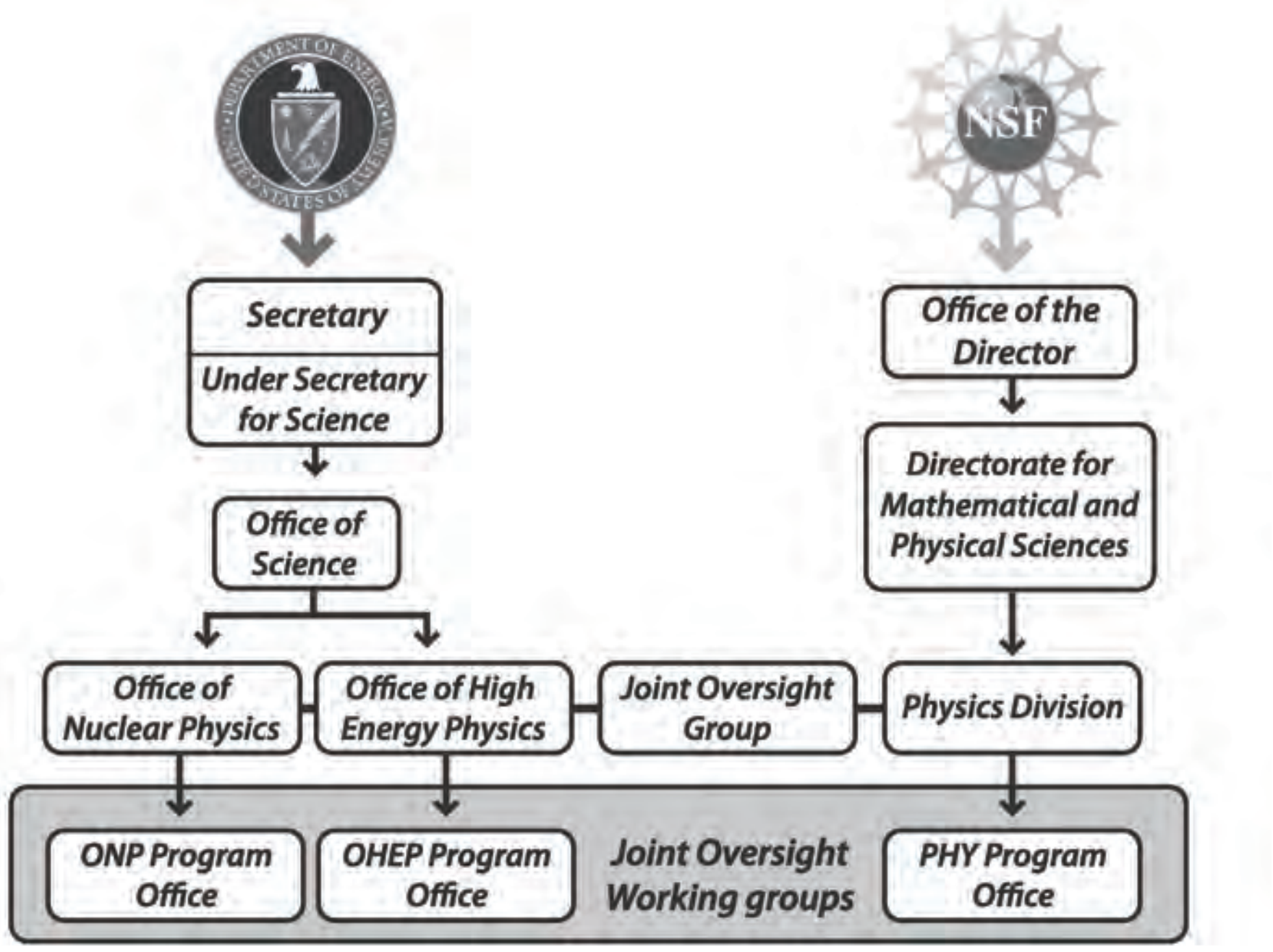

**Figure 7.9** Organization of NSF-DOE joint agency oversight of the DUSEL physics experiments.

The JOG is co-chaired by the Associate Director of DOE Nuclear Physics (NP), the Associate Director of DOE High Energy Physics (HEP), and the Associate Director of NSF Division of Physics (PHY). The associate directors at DOE report to the Deputy Director of the Office of Science (SC) at DOE, and the Director of NSF PHY reports to the Assistant Director for Mathematical and Physical Sciences (MPS).

The specific responsibilities of the JOG include:

- Oversight and review of the plans, budgets, schedules and milestones, and status reports of the DUSEL physics experiments and the DUSEL Facility
- Oversight and coordination of the integration and interfaces within the experimental program, and between the experiments and the Facility
- Oversight of the DUSEL physics experiments selection process
- Prior notification and consultation on the assignments of designated university staff or DOE national laboratory staff as managers of the DUSEL physics programs
- Ensuring that timely and effective technical, cost, schedule, and management reviews are conducted
- Assignment of DOE and NSF personnel to the JOG working groups
- Conducting joint reviews of the DUSEL physics experiments, as required

- Tracking and closing action items generated at JOG meetings

The JOG shall also engage in other activities it deems appropriate and within its programmatic responsibilities.

### 7.9.1 Joint Agency Management Model

Stewardship responsibilities apply to the design and planning activities of the DUSEL physics programs. The JOG co-chairs will oversee and approve stewardship roles and responsibilities as the DUSEL physics program evolves.

Design, early-generation experiments, technology development and prototyping, and demonstration projects for potential DUSEL experiments are supported by the programs within DOE HEP, DOE NP, and NSF PHY.

## 7.10 Project-Wide Acquisition Plans

DUSEL acquisition strategy covers preconstruction services to go from design completion to the commencement and completion of construction. The DUSEL construction acquisition plan (discussed in Chapter 5.10) is based on establishing policy, strategy, and procedures that help to ensure the most cost-effective and efficient approach to the realization of DUSEL. The general approach for the DUSEL MREFC-funded Project is to select a Construction Manager (CM), preferably chosen as a CM at risk that understands the technical challenges, risks, and has a demonstrated experience in the field of underground civil and large scientific facility construction. The acquisition approach is consistent with the applicable state and federal regulations and is based on competitive bidding and selecting contractors to provide the best value.

### 7.10.1 Project Acquisition

The DUSEL Project team will acquire services from a multitude of vendors with extensive experience building scientific laboratories and/or extensive experience building underground facilities. At the time of construction, the laboratory will be managed by the DUSEL LLC. The DUSEL LLC will be responsible for managing all subawards and subcontracts. The primary subawards of the DUSEL effort will be LBNL, SDSM&T, and SDSTA. Other subawards working with the DUSEL LLC may include other universities, such as BHSU, to address specific program areas such as the education and public outreach program. All entities will be accountable to the DUSEL LLC for successful completion of the efforts assigned to them.

During the Preliminary Design phase, a review of the Facility scope elements, the envisioned construction packaging approach, and the acquisition strategy for each package was performed to develop the project delivery approach for DUSEL. The Project, working in conjunction with the DUSEL design contractors and CM, decided that a design-bid-build approach was most appropriate for the majority of the DUSEL work—particularly to address the complexities of the underground civil construction scopes. The intricacies of the expected interfaces between the science experiments and the facility require close coordination among the designers, DUSEL Project, experimenters, and the CM. The Project and supporting contractors regard the design-bid-build approach to most appropriate for the majority of the facility construction scopes. The early involvement of the CM during the preconstruction phases, including the Project's extensive use of independent estimates and constructability reviews by the CM, address the potential drawbacks of the design-bid-build approach.

In contrast, the Project has chosen a design-build approach for the Ross Surface campus. Since the Ross Surface campus primarily supports construction and Facility maintenance efforts for the foreseeable future, a design-build approach provides the CM and subcontractors more flexibility in designing a flexible, efficient construction support area that best meets their needs along with meeting the requirements of the DUSEL Operations team for Facility maintenance.

The acquisition plan defines the process of identifying qualified contractors to bid and safely perform the work within the cost and schedule targets established by the Project. The first component of the acquisition process, which takes place during Final Design, is performing market research to understand the current construction market, available qualified contractors, and the economic conditions that may impact the bidding process. Standardized market research forms are distributed by the CM to potential contractors to develop a bidder's list that is used as bid packages are developed and distributed to industry. Competitive bidding is a requirement for the Project, and this process ensures that the groups competing for the work are qualified, have acceptable safety records, and are reputable. Efforts will be made to ensure minority, women-owned, and small or disadvantaged businesses are included in the market research and acquisition process.

A contractor prequalification process follows the market research analysis phase. Prequalification includes a more thorough evaluation of each potential contractor's safety and health records, financial history, and work history. Prequalification will be utilized where a two-step acquisition approach is appropriate for critical elements of construction such as underground related work that have specific safety requirements and require an extensive evaluation of a contractor's safety performance record.

Bid packages will be developed during the later stages of Final Design to support the bidding process. An initial set of planned bid packages was developed during Preliminary Design and will be refined up to the start of the bidding process. Bid packages divide the Project into subprojects based on scheduled starts, schedule for design completion, and phasing of work. Packages are also formed to generate interest by the industry, resulting in a more competitive bidding process.

The acquisition approach will include a cost commitment plan that will outline financial needs in relationship to the selected contracts. Financial needs will be assessed though the collection of cost estimates from contractors, as well as project cost estimates from CAMs. Estimates will be coordinated with the Project Controls team, which will integrate project cost and schedule information from the bidding process with the Project Management Control System (PMCS). Further discussion of the construction acquisition process is described in Chapter 5.10.

## 7.11 Project Controls Systems

Project Controls is responsible for cost-estimate management and scheduling of the DUSEL Project. It maintains the work breakdown structure (WBS) and organization of activities and functions within the top four levels of the WBS. The Project Controls team is responsible for establishing and maintaining the PMCS and generating reports on project status for Central Project Directorate to demonstrate how the Project is proceeding according to schedule and budget. During Final Design, Project Controls will establish an Earned Value Management System (EVMS) that will be operational prior to the start of MREFC-funded construction. Project Controls works closely with all other departments to ensure adherence to budgets and schedules. Changes to configured items, budgets, and schedules are reviewed and approved by the Configuration Control Board (CCB), an advisory board whose chair has the ability to veto recommendations of the CCB. They also work with the finance team to confirm that Management Reserve and all other budget line items are in order.

The PMCS is in place to schedule and optimize Project resources; determine Project status (a comparison of work accomplished and resources expended to the baseline plan); compute and track earned value; evaluate project risk with respect to cost and schedule; and manage the change process by evaluating the effect of changes to the cost and schedule baseline. The PMCS includes the software tools for development of Project cost and schedule databases as well as processes and procedures to organize and manage Project costs and schedules.

The Project Controls office work scope includes the maintenance of the PMCS used by the DUSEL Project and is described in Volume 8, *Project Management Control*. This includes project cost and schedule reporting and the EVMS requirements and processes.

## 7.12 Business Systems

The Business Office includes finance and accounting, human resources, contracts and procurement, asset management, administrative services, and communications during all Project phases.

During Preliminary Design, the Project has relied on the business and financial controls and systems of the collaborating institutions. As the DUSEL LLC is established during Final Design, the primary responsibility for these controls and systems will transfer to the DUSEL LLC with oversight maintained by UC Berkeley. A process to identify and select the most suitable systems and processes to support efficient DUSEL LLC operations was initiated during the Preliminary Design phase. The selection of systems and controls will be refined during Final Design and will guide the establishment of the Project-wide business systems framework.

The Finance and Accounting department maintains the overall DUSEL budget and provides financial data of the actual accounted expenditures to the Central Project Directorate and Project Controls to use in the measurement of project performance. This department is responsible for asset management and scheduling external audits, as well as preparing and processing invoices and managing contract obligations. The Finance and Accounting department, under the DUSEL LLC, remain accountable to UC Berkeley.

The Human Resources department within DUSEL is responsible for hiring and management of all full-time staff within DUSEL LLC; provide oversight of benefits providers, assisting personnel with benefits-related issues, and the oversight for effort-wide policy and procedure tracking; and assists with employee development and relations, as well as labor relations.

The Contract Management and Procurement department is responsible for establishing and maintaining all subawards and agreements related to the Project. The department administers all outsourced contracts, ranging from major design and construction efforts to small consulting agreements, as required for the Project execution. Procurement functions to support day-to-day operations during all Project phases.

Asset Management is handled by the Finance department within the Business Office. DUSEL LLC will follow generally accepted accounting principles (GAAP) in the organization of its finances and asset management, including the capitalization of all expenditures for assets with a life of greater than one year and a cost of $5,000 or more. All DUSEL assets acquired by its collaborating institutions in support of the DUSEL Project will be tracked by systems in place within the institution and verified to be compliant with the terms and conditions of the cooperative agreement between NSF and the UC Regents and with the subaward flow-down requirements. Delegation of authority concerning responsibilities is documented and verified, including the procurement process (purchase orders and contracts), cash management, salary and payroll procedures, fixed asset inventory policy, and surplus property disposal procedures.

The services of the Business Office and related functions are described by time-phase across the Project in Volume 10, *Operations Plans*.

## 7.13 Systems Engineering

DUSEL Systems Engineering is responsible for the technical integration of the Project, including management of project requirements and interfaces, Value Engineering and Trade Studies, risk management and assessment, configuration management, and systems verification. Systems Engineering provides the processes to facilitate technical integration of the Project and also provides the document structure for capturing technical information. Systems Engineering is responsible for supporting the technical integration and verification of the Facility and its interface to the science experiments for successful delivery of the DUSEL Project. A detailed discussion on the Systems Engineering efforts and activities is included in Volume 9, *Systems Engineering*.

### 7.13.1 Configuration Management

Configuration management is critical to protecting the integrity and consistency of the Project baseline. The configuration management activities address the following objectives:

- Configuration items are identified and controlled.
- Configuration control is established and enforced.
- DUSEL Project baseline is identified.
- Configuration items are only promoted to a baseline by meeting established criteria.
- Status history of the baselines and changes to it are maintained.
- The accuracy, completeness, and integrity of the baseline is protected and monitored.

The configuration management standards are based on ANSI/EIA-649 and the Software Engineering Institute (SEI) Capability Maturity Model Integrated (CMMI) configuration management principles. As part of the configuration management process, a Configuration Control Board (CCB) has been established and is the central review and decision-making body for evaluating and approving all changes to the baseline. The CCB charter is included in Appendix 7.B. Additionally, a Configuration Management Plan has been prepared to define the processes to manage and control configuration items and CCB activities. The Configuration Management Plan is included Appendix 9.D.

The Configuration Change Authority, the entity responsible for approving changes to items under formal configuration management, is the DUSEL PI/LD who chairs the CCB. The CCB is the advisory body to the CCB Chair and is responsible for evaluating all proposed changes to items under formal configuration management control.

The membership of the DUSEL CCB consists of:

**CCB Chairperson**

- DUSEL Principal Investigator

**CCB Members**

- Co-Principal Investigator
- Project Director
- Project Controls Manager
- Systems Engineering Manager
- Environmental, Health, and Safety (EH&S) Director
- Education and Public Outreach Director

- Scientific Programs and Integrated Suite of Experiments Project Manager
- Facility Project Manager
- Quality Assurance Manager
- CCB Administrator (ex officio)

#### 7.13.1.1 Configuration Management Thresholds

The configuration management thresholds define the approval levels within the organization for each class of configuration items. The Project CCB approval level threshold matrix, Table 7.13.1.1, documents the change management approval levels within the DUSEL Project.

| | **Approval Level** | | |
|---|---|---|---|
| | **Class 1**<br>**National Science Foundation** | **Class 2**<br>**Configuration Control Board** | **Class 3**<br>**Level 2 Manager** |
| **Technical** | Changes to the Project purpose or goals as described in the Key Performance Parameters (Level 1 Requirements)<br>*Appendix 9.E* | Changes to the Project's Level 2 Requirements<br>*Appendix 9.I* | Changes that impact a Level 3 Requirement<br>*Appendix 9.G and 9.H* |
| **Cost** | During MREFC-funded activities: Changes to the Total Project Cost or changes in cost allocation between Facility and Experiments<br><br>During Transition and Final Design: Any new or change in subcontract or subaward amount ≥$2 million<br><br>During Preliminary Design phase: Any new or change in subcontract or subaward amount ≥$250,000<br><br>*(Ref. NSF Cooperative Agreement PHY-0940801)* | During MREFC-funded activities: Changes to the WBS Level 2 Cost Baseline or any cumulative change at WBS Level 3 of >$5 million or 5% WBS, whichever is smaller<br><br>During Transition and Final Design: Any WBS Level 3 change of >$100,000 to the design baseline<br><br>During Preliminary Design phase: Any change ≥$25,000 | Changes to the cost baseline at WBS Level 3 or lower that are less than the thresholds of Class 2 |
| **Schedule** | Any change in an NSF milestone (Level 1)<br>*See Table 7.5* | Any change in a Level 2 milestone or Project Office milestone<br>*See Volume 2, Cost, Schedule, and Staffing* | Any change in a Level 3 milestone<br>*See Volume 2, Cost, Schedule, and Staffing* |
| **Policy, Plans, and Other Configuration Items** | Key Performance Parameters and Level 1 Requirements | Any general policy and any plan designated as key on the Configuration Items List. | All non-key plans and any procedures designated as key on the Configuration Items List |
| **Note:** The responsible Level 3 Manager may approve changes to configuration items that fall below the thresholds in this table. | | | |

**Table 7.13.1.1** DUSEL Configuration Management Threshold Matrix.

## 7.14 Systems Verification

The Systems Verification program addresses the testing and commissioning processes used in accepting the DUSEL Facility. As Construction nears completion, the Project will evaluate if the Facility functions as designed in compliance with the requirements, specifications, and interfaces used to inform the design process. All major portions of the Facility will be subjected to these evaluations before acceptance and transition to Operations. The Systems Engineering team leads the Systems Verification process, including the planning for verification during the design phases, which requires a dedicated Commissioning Agent contractor as part of the overall verification approach. This process provides a level of unbiased independence during testing and commissioning.

### 7.14.1 Systems Verification Plan

The DUSEL Systems Verification Plan documents the overall testing and commissioning approach and is lead by the Systems Engineering team. As the Systems Verification Plan addresses a broad range of acceptance activities for the entire Facility, and will provide the framework for the overall verification plan and then links to lower-level documentation, including testing plans, acceptance criteria, and detailed procedures. The Verification Plan will include and reference the following items:

- **Roles and responsibilities.** Responsibilities of DUSEL Project departments, design contractors, and Construction Manager as related to the verification program
- **Testing and commissioning plans.** Detailed plans addressing the end-to-end verification plan for the Facility, including quality assurance in relationship to verification
- **Construction and verification flow.** Sequence of integration and test processes, including test events and objectives, subsystem component delivery inputs, and review points
- **Verification matrices.** Tables addressing the traceability from the requirements and interface documentation to the testing plans and procedures to record and track the requirements verification

The Systems Verification Plan will be developed during Final Design to support bid packaging and address the NSF *Large Facilities Manual* requirements for commissioning planning.

### 7.14.2 Facility Verification and Commissioning Plans

Facility construction will be a phased process. As each major element nears completion, low-level Facility Verification and Commissioning Plans will be completed for each element in accordance with the overall DUSEL Systems Verification Plan. These Facility Verification and Commissioning Plans will address the Facility contractor testing and commissioning plans and procedures, including plans to be prepared by the independent commissioning agent. The following stakeholders will be involved during this process:

- **Independent commissioning agent.** The commissioning agent will be engaged during Final Design and will work with the DUSEL Systems Engineering and Facility teams to develop detailed verification and commissioning plans for each of the systems and subsystems.
- **DUSEL Systems Engineering.** The Systems Engineering team is responsible for the requirements and Systems Verification Plan and will work with the commissioning agent to ensure that mechanisms are in place for the commissioning of major Facility systems and subsystems before they are transitioned to Operations as well as ensuring compliance with the Systems Verification Plan.

- **DUSEL EH&S.** The commissioning agent will work closely with the EH&S department to ensure that proper verification, testing, and documentation of life-safety systems and other infrastructure systems are accepted before they are transitioned to Operations.
- **DUSEL Operations.** Operations will be involved in the acceptance activities, as the "operator" of the systems. The Operations team will ensure proper resources, training, and transfer of both knowledge and documentation are provided for successful operations and maintenance.
- **DUSEL Science.** The Science department will participate along with individual experiment collaborations in the experiment interface acceptance process.
- **DUSEL Facility.** The Facility team will oversee contractor participation as part of the verification process as the official owner's representative.
- **Facility Construction Contractors.** The contractors are responsible for delivering a facility that meets the requirements, specifications, and interfaces defined for the Facility. They participate in the testing and commissioning process. This includes development and execution of test and pre-operations start-up plans and procedures that are performed in advance of the overall DUSEL verification program.

### 7.14.3 Operations Readiness Review

Operations Readiness Reviews will be held in conjunction with the Systems Verification process. These reviews will be held to ensure that the verification and commissioning plans have been executed successfully before the acceptance of each element prior to science use. The DUSEL Project is responsible for developing the Operational Readiness Review approach, and it will consider the following items:

- Verification Plans
- Approach to acceptance
- Approach to entry of science into the Facility
- EH&S standards and process for verification of these standards
- Commencement of operations acceptance criteria
- Cybersecurity standards and the process for verification to these standards
- Maintenance approach
- Staff transitions and training
- Phasing of Construction and Operations
- Approach to use and occupancy permits
- Plans for operating the Facility
- Approach to decommissioning
- Evaluation and presentation of verification test and commissioning results

## 7.15 Quality Assurance and Control

The expected quality assurance and control outcomes and principles used to guide the Project will be defined in the DUSEL Global Quality Assurance Policy. In particular, the quality policy will outline the functions and associated actions that will be performed by DUSEL to ensure that process requirements are continuously and consistently performed in accordance with specified standards. Additionally, the policy will ensure that an appropriate level of quality control procedures are in place and are followed by contractors, service providers, and DUSEL personnel. A draft of this policy was initiated during Preliminary Design and will be placed under configuration control prior to Final Design.

The DUSEL Quality Assurance Policy, Appendix 7.C, will specify the processes and procedures used by the Project to achieve the desired outcomes, including the quality requirements for the Design, Construction, and Operations of the Facility. This plan will cover the following aspects:

- Quality Assurance Program
- Personnel training and qualification
- Quality improvement—Corrective Action and Preventive Action Program
- Work processes
- Quality management review
- Quality system audits

### 7.15.1 Quality Assurance Surveillance Plan

The *Quality Assurance Surveillance Plan* (QASP), Appendix 7.D, provides a systematic method to evaluate performance on the major design and construction contracts for DUSEL. Along with roles and responsibilities, the QASP specifies:

- Performance standards
- What will be monitored
- How monitoring will take place
- Who will conduct the monitoring
- How monitoring efforts and results will be documented

## 7.16 Environment, Health, and Safety (EH&S)

EH&S requirements are integrated into management and work practices at all levels to assure that the DUSEL Project mission is achieved while protecting the public, staff, visitors, contractors, subcontractors, and the environment. This is accomplished through the Integrated Safety Management (ISM) system, which requires that hazards are identified and mitigated; work is authorized only after EH&S analysis is complete; and oversight of work is conducted by management and staff. The ISM policies and procedures make it clear that everyone is responsible for conducting work and operations in a safe and environmentally sound manner. The Project approach to EH&S can be found in Volume 6, *Integrated Environment, Health, and Safety Management*.

### 7.16.1 EH&S Hazards, Risk Assessment, and Mitigation Strategy

The complexity and formality of the hazard identification process and subsequent development of work controls is tailored to the nature and scope of each work activity**.** The process for hazard analyses, risk assessment, and mitigation strategies is detailed in Chapter 6.2, *Hazard Analysis and Control*.

### 7.16.2 Integrated Safety Management (ISM) System

The EH&S department is responsible for setting the policies, plans, and procedures for the ISM system. The DUSEL ISM Project policy holds that everyone is responsible for conducting work and operations in a safe and environmentally sound manner. This expectation is applicable to employees, Facility users, visiting scientists, contractors, and their subcontractors.

Within the scope of this policy, it is the objective of the Project to systematically integrate excellence in EH&S into the management of work practices at all levels so that its mission is achieved while protecting the public, the workers, the environment, and physical and intellectual property, as detailed in Appendix 6.A, *Integrated Safety Management System Policy*. This is accomplished by using the principles and core functions of the ISM system to ensure that the overall management of EH&S functions and activities is an integral part of work practices, and to seek improvement in management and performance at every opportunity. Within this policy, it is important to recognize that the use of the word “safety” refers to the reduction or elimination of all hazards, including hazards to health and environment.

Chapters 6.1, *Integrated Environment, Health, and Safety Management System*, and 6.2, *Hazard Analysis and Control*, provide the details of the DUSEL implementation.

### 7.16.3 Environmental Plans, Permitting, and Assessment

The EH&S department is responsible for specifying the codes and standards applicable to the Project and regulatory compliance for water quality, air quality, and hazardous materials. Site monitoring shall be used to assess potential impacts to the environment. Chapters 6.3 and 6.4 provide the details of the DUSEL implementation.

The responsibility to prepare NEPA documentation falls upon NSF, as it is the federal agency proposing to construct and operate new underground facilities to support science and engineering research at the former Homestake Gold Mine. NSF recognized this duty early in the planning process and contracted with Argonne National Laboratory (ANL) to prepare the necessary NEPA documentation. Anticipating that construction activities associated with this Project would, without further analysis, have the potential to cause significant environmental impact, NSF decided to prepare an Environmental Impact Statement (EIS) rather than an Environmental Assessment (EA). As such, full disclosure and public participation is not only encouraged, but required.

## 7.17 Conclusion

As outlined in Section 7.0.1, this PEP is a living document and will be updated as the Project evolves during Final Design in preparation for the Construction phase as the Project scope, requirements, and cost estimates are refined. This initial PEP has been issued as part of the PDR and will be issued as a standalone document in subsequent versions. The PEP will be reviewed and revised periodically to reflect Project maturity under the control of the DUSEL CCB.

As the PEP and project planning evolve during Final Design, references to PDR sections used throughout this PEP will be replaced with references to specific baseline plans that describe the management processes and procedures that are used in the daily execution of the DUSEL Project.

# Volume 7 References

This page intentionally left blank

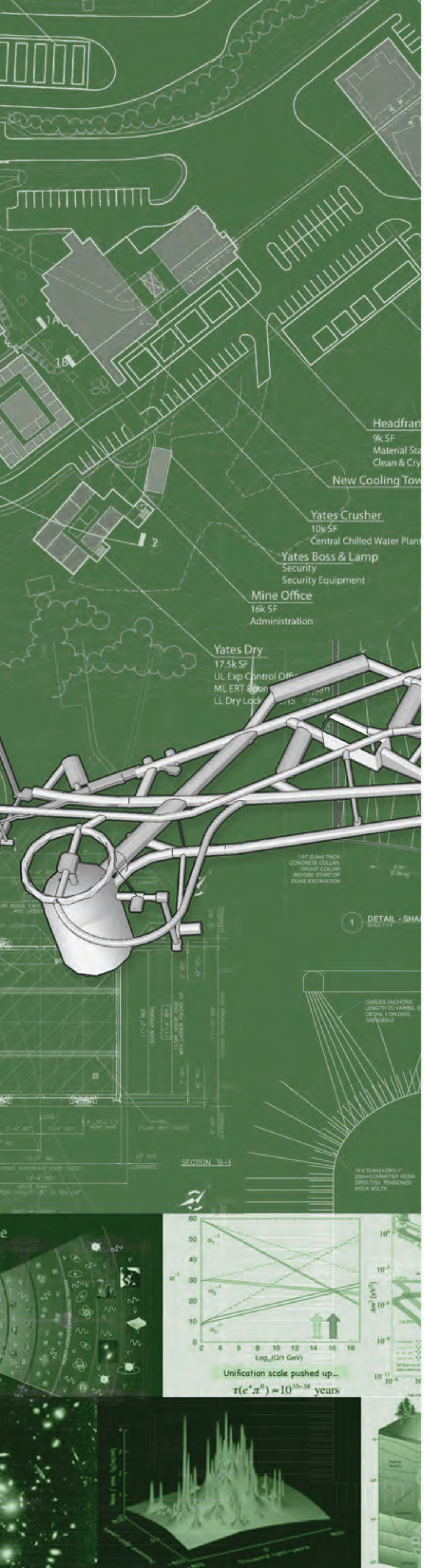

# Preliminary Design Report

May 2011

# Volume 8:
# Project Management Control

**DUSEL**
Deep Underground
Science and
Engineering Laboratory

This page intentionally left blank

# Project Management Control

## Volume 8

## 8.1 Executive Summary

### 8.1.1 Introduction

DUSEL Senior Management understands that a fully functioning and dynamic Project Management Control System (PMCS) is imperative for developing and managing the cost and schedule of the Project. The PMCS provides the DUSEL stakeholders, Central Project Directorate and Senior Management, and the Level 2 and 3 managers' visibility into the Project so that an accurate representation of the entire project is readily available detailing the work has been planned and performed, projected schedule completion dates, associated costs, and an estimate to complete. The PMCS will enhance the management of the budget and schedule by providing performance indicators designed to identify early-warning signals and support the Project in implementing corrective actions where necessary.

The National Science Foundation (NSF) requires that its large facility projects implement and use an Earned Value Management System (EVMS). The DUSEL Project is executed through a Cooperative Agreement between the NSF and the University of California at Berkeley (UC Berkeley) and this requirement to support the DUSEL Project's EVMS implementation and use during construction flows to the Project from UC Berkeley to its subawardees, and major construction contractors. This practice is consistent with Department of Energy (DOE) requirements as well. As a result, the DUSEL PMCS must include and address EVMS requirements as it is developed during the Preliminary and Final Design phases in preparation for a Construction start. While a full EVMS certification of the PMCS is not required, the Project will implement an advisory panel of subject matter experts to perform a review of the system prior to the start of the Construction phase, and the panel will perform periodic surveillance to insure continued compliance throughout the life of the Project.

### 8.1.2 Project Management Control System Overview

The DUSEL PMCS process and organization are designed to comply with the American National Standards Institute (ANSI)/Government Electronic Industries Alliance (GEIA) Standard 748-B-2007. The ANSI/GEIA-748 standard represents industry's best practices and is the official standard for EVMS. The DUSEL PMCS also fulfills the requirements of Office of Management and Budget (OMB) Circular No. A–11 (2008), Part 7, Section 300—Planning, Budgeting, Acquisition, and Management of Capital Assets. The DUSEL EVMS is a key component of the organization, methods, and procedures adopted by the DUSEL Project to ensure that its mission and functions are properly executed.

The DUSEL PMCS addresses the seven principles of EVMS as defined by the ANSI standard:

- Plan all work scope for the Project to completion.
- Break down the Project work scope into finite pieces that can be assigned to a responsible person or organization for control of the technical, schedule, and cost objectives.
- Integrate the Project work scope, schedule, and cost objectives into a Performance Measurement Baseline (PMB) against which accomplishments may be measured. Changes to the baseline are controlled.
- Use actual costs incurred and recorded in accomplishing the work performed.
- Objectively assess accomplishments at the work performance level.
- Analyze significant variances from the plans, forecast impacts, and prepare an estimate at completion based on performance to date and work to be performed.
- Use EVMS information in management processes.

These principles are integrated into a comprehensive system that develops and maintains the baseline; tracks Project cost, schedule, and scope; and provides for the generation of timely performance measurement data and reports. Performance measurement reports provide management with objective Project information critical to monitoring progress, identifying significant issues, and implementing corrective actions as needed.

The DUSEL PMCS is designed to provide project managers with a tool set to promote optimal planning, accurate reporting, and effective control through the standardization of processes used in Project scope, schedule, and budget management.

## 8.2 General Requirements

The foundation of the PMCS is the Work Breakdown Structure (WBS), which includes all of the scope defined for the DUSEL Project. All the key components of the EVMS identified in Section 8.1.2 will be linked directly to the WBS. It provides a common framework to organize, display, and define DUSEL systems and subsystems.

The balance of this section outlines the key requirements identified for each of the respective PMCS components.

### 8.2.1 Cost Estimate Requirements

Reflecting the Project's current technical baseline assumptions, the DUSEL cost estimate is mature and will serve as a starting point for the other key components within the PMCS. It contains a wealth of detailed information that is linked to the Integrated Project Schedule (IPS) and the PMB via activity IDs/work packages.

### 8.2.2 Integrated Project Schedule Requirements

The resource loaded activities within the IPS address the Project's work scope, have realistic durations, and are logically linked to enable the identification of a project critical path and identify near-critical path activities as well as provide an efficient mechanism for performing "what-if" scenarios. The IPS contains all milestones, differentiates between milestone levels (i.e. Level 1, 2, 3, etc.), interfaces milestones across WBS elements, and supports illustration of the Project's current schedule status. The IPS provides a project-summary schedule that is based on the bottom-up work package detail.

### 8.2.3 Earned Value Management System Requirements

The EVMS is used to establish and maintain the PMB, which is the time-phased budget baseline as defined by the IPS. The PMB is an approved course of action for the Project's work against which the ultimate project execution is compared. Variations from the original, baseline plan are measured to facilitate effective project management. The PMB summarizes the Project's base budget, and will incorporate actual accounting expenditures and schedule performance.

The time-phased budget is escalated based on defined escalation factors applied for each fiscal year. In addition to escalation rates, the EVMS will use approved labor and indirect rates as well as any other project-specific rates as defined by the participating institutions.

Among the many outputs and reports, a main output of the EVMS is a Cost Performance Report (CPR), the standard report used to outline the Project's Earned Value status, including a revised Estimate at Complete (EAC).

### 8.2.4 Reporting Requirements

The DUSEL PMCS will provide the required reports for management to assess, control and manage the Project, as well as provide the needed reporting to Project stakeholders. PMCS information is posted online through the Project's document management system, Xerox DocuShare Web Portal, to enhance communication and integration across the Project. Appropriate security permissions are implemented to protect the confidentiality of sensitive information.

### 8.2.5 PMCS Change Control Requirements

DUSEL has a Configuration Control Board (CCB) and a change control process, which is used to assess, control, and authorize programmatic and technical changes to the Project. The PMCS is a subsidiary component to the Project's CCB as it relates to changes to a project's Integrated Project Schedule (IPS) and/or Performance Measurement Baseline (PMB). Changes are controlled to maintain the integrity of the Project cost and schedule baselines.

## 8.3 System Design

### 8.3.1 Overview

The DUSEL PMCS includes the Detail Cost Estimate, the IPS, and the EVMS. The DUSEL WBS is the hierarchical structure used as the basis for organization in all three PMCS tools; the WBS Dictionary is included as Appendix 7.A. The first level of the WBS is aligned with the Project and subsequently broken down into major systems and subsystems. Each major system has a responsible Level 2 manager and each subsystem has a responsible Level 3 manager who is also designated as the Control Account Manager (CAM). The full DUSEL Organization Chart is provided in Appendix 1.A, listing the CAMs and management structures.

The Detail Cost Estimate is a database that incorporates all owner and subcontractor cost estimates for executing the Project. Estimates are collected from all CAMs for owner costs and from subcontractors for the construction costs. In addition, the Detail Cost Estimate includes the work scope as defined in the WBS Dictionary and estimate justification/basis of estimate for each WBS element. The detail estimates included in the Detail Cost Estimate document are subsequently summarized and used in the IPS and EVMS. Traceability is established between all three tools using the WBS.

The IPS, is a network-driven schedule that covers the entire work scope of DUSEL defined in the WBS and consists of milestones, resource loaded detail activities, and logical relationships. The activities in the IPS are assigned control account and work package numbers that are reflected in the EVMS tool. The combination of the control account and the work package numbers will serve as the key fields for linking the IPS to the EVMS. The start and finish dates of an IPS activity will define the EVMS work package. Additionally, each activity linked to a work package will define the resources required to complete the work scope as well as the time-phasing of the resources

The CAMs are required to provide status at least once a month on a pre-defined date typically on or near the last business day of the month. For discrete activities, revised dates and progress will subsequently be imported from the IPS tool into the EVMS tool. The combination of the resources assigned and the status will subsequently be used for the earned value calculation.

At the close of each accounting month, actual costs by resource category (labor, materials, or travel) are exported from the UC Berkeley accounting system for each control account and subsequently imported into the EVMS tool. Traditional earned value calculations, cost and schedule variances, performance indices and revised forecasts are available for management reporting and review. The DUSEL CCB defines the thresholds associated with the WBS that require a supporting variance explanation.

Once the IPS is baselined and the PMB has been established, the PMCS is managed via the DUSEL CCB. The CCB is described in Volume 9, *Systems Engineering*.

Figure 8.3.1 provides an illustration at a high level on the flow of project management data and how the respective databases are integrated with one another to form the PMCS.

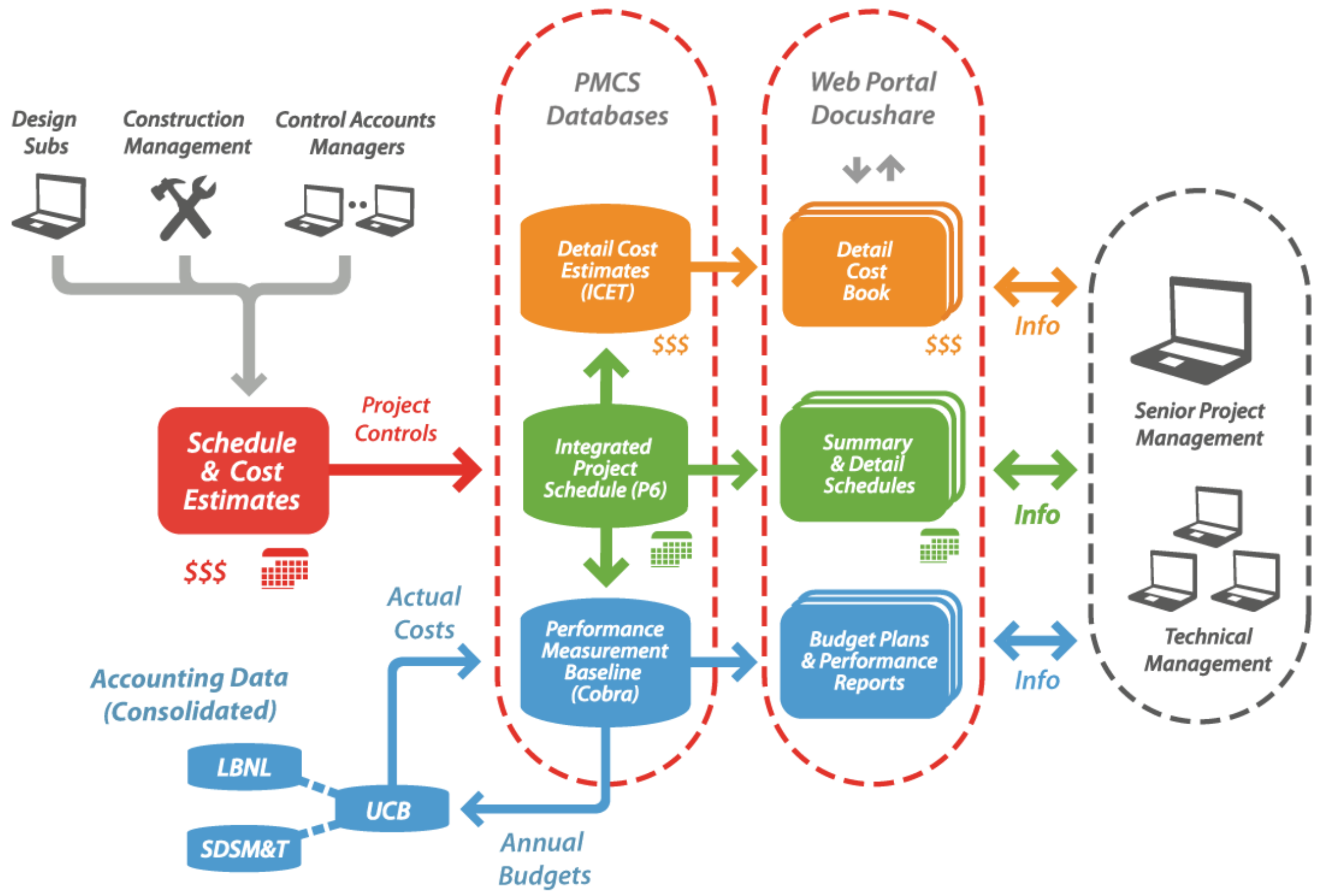

**Figure 8.3.1** DUSEL Project Management Control System overview. [DKA]

### 8.3.2 System Design Assumptions

The following assumptions are used as a basis for the design of the DUSEL PMCS:

1. All project schedule data and all project budget data can be shared freely within all organizations in the Project. Permissions may need to be assigned as a function of partnerships or other limited data-access requirements that are put in place in the future.
2. Summary-level actual cost data can be shared freely among all authorized persons.
3. Schedule status for the IPS is collected at the end of each calendar month.

### 8.3.3 Work Breakdown Structure

The foundation of the PMCS is a formal WBS that has been established to organize, display, and define the DUSEL systems and subsystems. The WBS is a product-oriented hierarchy that identifies all the scope elements of the DUSEL Project and their parent/child relationships. The WBS will closely mirror the organization and project deliverables as determined by the Central Project Directorate and will include elements to reflect the efforts required to manage, design, and integrate the system components. The scope of work for each WBS element is described thoroughly in the WBS Dictionary. Each fourth-level WBS element has been estimated, planned, and budgeted. Control accounts were developed at the fourth level for all funding types, including Conceptual Design, Preliminary Design, Final Design, Construction, and Operations.

The WBS and associated WBS Dictionary are maintained as supporting elements of the PMCS, both the structure and the Dictionary are under configuration control (see Appendix 9.D for the DUSEL Configuration Management Plan). A truncated WBS, to Level 3, is shown in Table 8.3.3. The entire DUSEL WBS is available to the Project team through DocuShare, and the WBS Dictionary is included in Appendix 7.A.

| WBS Code | WBS Name |
|---|---|
| **DUS** | DUSEL |
| **DUS.PRJ** | DUSEL Project-Wide Systems |
| **DUS.PRJ.PMO** | Project Management & Controls |
| **DUS.PRJ.BUS** | Business Office |
| **DUS.PRJ.SYS** | Systems Engineering |
| **DUS.PRJ.ITS** | Information Technology System |
| **DUS.PRJ.EHS** | Environment, Health, & Safety |
| **DUS.PRJ.EDO** | Education & Outreach |
| **DUS.PRJ.QAC** | Quality Assurance/Quality Control |
| **DUS.FAC** | DUSEL Facility |
| **DUS.FAC.MGT** | Facility Project Management, Architecture, and Integration |
| **DUS.FAC.SUR** | Surface Facility and Infrastructure |
| **DUS.FAC.UGI** | Underground Facility Infrastructure |
| **DUS.FAC.OLR** | Other Levels & Ramps |
| **DUS.FAC.MLL** | 4850L Mid-Level Laboratories |
| **DUS.FAC.DLL** | 7400L Deep-Level Laboratories |
| **DUS.FAC.LGC** | Large Cavity (LGC) for Long Baseline Neutrino Experiment (LBNE) |
| **DUS.SCI** | Science Scope Integrated Suite of Experiments |
| **DUS.SCI.SUP** | DUSEL Support of Science |
| **DUS.SCI.EXP** | DUSEL Sub Awards to Experiments |
| **DUS.OPS** | DUSEL Operations |
| **DUS.OPS.OFC** | OPS-Facility & Infrastructure Operations & Maintenance |

**Table 8.3.3** DUSEL WBS to Level 3.

### 8.3.4 Organizational Breakdown Structure

The Organizational Breakdown Structure (OBS) is a functionally oriented structure indicating organizational relationships and used as a framework for the assignment of work responsibilities. The organizational structure is progressively detailed downward toward the lower levels of management. DUSEL's OBS also includes various funding organizations and partners that can be used to generate reports for a respective organization. The OBS is presented in Figure 8.3.4.

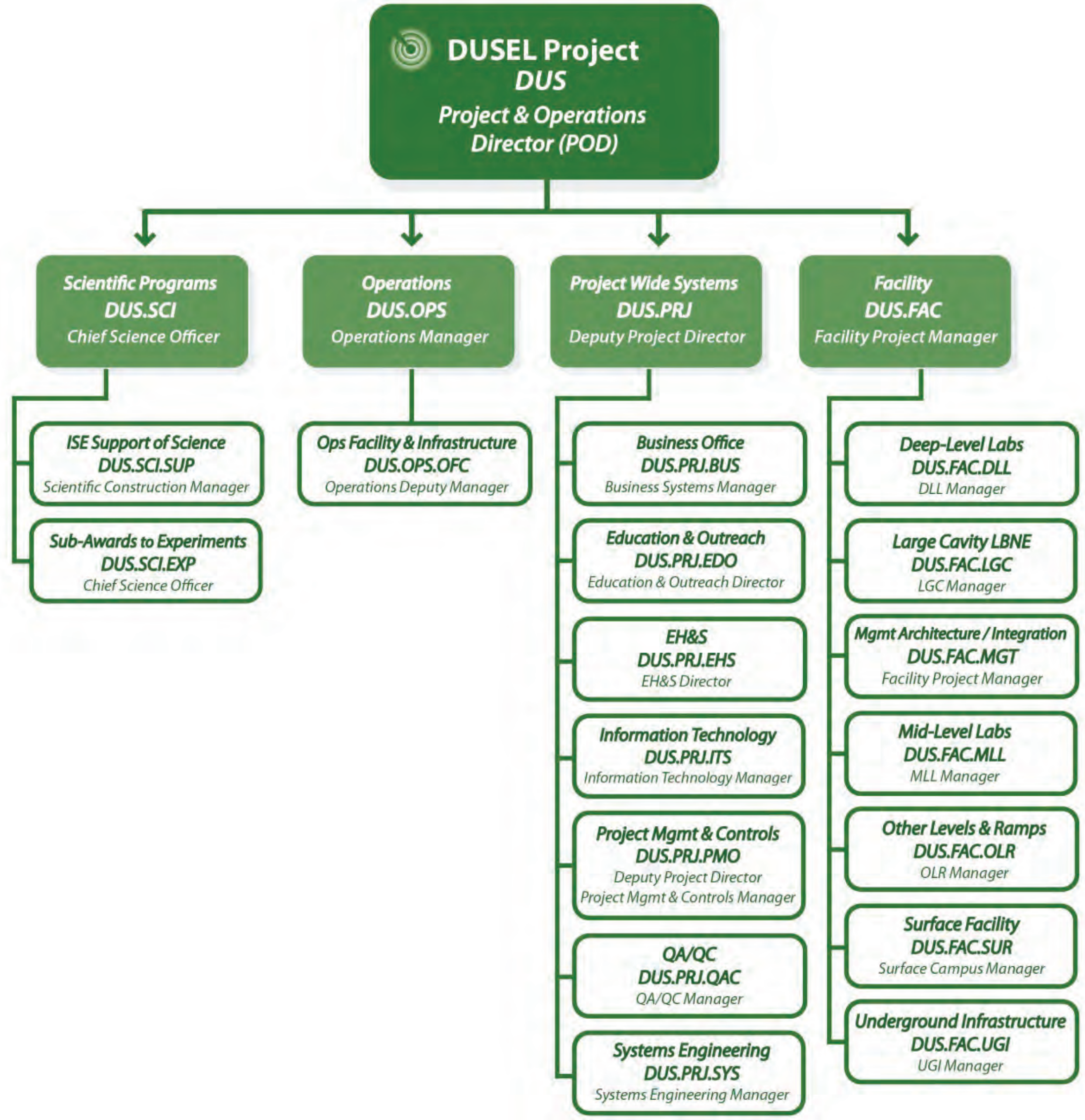

**Figure 8.3.4** DUSEL Organizational Breakdown Structure. [DKA]

### 8.3.5 PMCS Integrated Project Schedule System Design

The IPS represents all of the planned effort and the logical flow of how that work will be accomplished. One key objective of the IPS is to provide a tool to efficiently analyze and monitor the critical path and activities near the critical path for the entire Project. The IPS will be baselined as part of the PMB, allowing for a benchmark by which the current working schedule can be measured against the original plan.

The following schedule-related management objectives have been established for the DUSEL IPS:

1. Plan and schedule all activities required to meet the scientific, technical, and project objectives with minimum schedule risk, including key milestone completion dates.
2. Develop detailed networked schedules that encompass all WBS elements and activities with logical relationships and resource assignments.
3. Identify project-level major milestones (Level 1, 2, 3, and any lower level milestones).
4. Integrate planning and schedule information from all collaboration participants and contractors.
5. Identify critical and near-critical activities.
6. Create a project baseline schedule based on the above steps that has stakeholder approval.
7. Integrate activities in the IPS with work packages in the EVMS application. Assign resources for each activity/work package to create and manage the time-phased budget that is defined by the detailed schedule.
8. Develop the process for collecting, updating, and reporting the progress for active activities.
9. Serve as a key communication tool for the entire Project.

### 8.3.6 Characteristics of the IPS

The IPS will include the following characteristics:

1. Resource assignments at the activity level that can be time-phased to the allocated budget in the cost system work-package.
2. A minimum of imposed dates.
3. The ability to identify the critical path(s) and near-critical paths at the task level, system level, subsystem level or total project. Relative to the critical path, the IPS will identify which parts of the schedule have a significant amount of float (slack).
4. The most recent reported status as of an effective time-now date.
5. The ability to perform “what-if” scenarios to provide a method for modeling and identification of the impacts from new requirements.
6. Strategic planning of major procurement efforts with the ability to track individual contracts as determined by the Level 2 managers.
7. Extensive code fields that can be used for summarization, roll-ups, filtering, and sorting to satisfy management reporting needs.
8. Identification of all interface milestones—or handoffs—between the CAMs.
9. Three levels of milestones—Level 1 is controlled by the DUSEL agency level customer, Level 2 is controlled by the DUSEL Central Project Directorate, Level 3 is controlled by the Level 2 managers.

### 8.3.7 DUSEL Scheduling and Reporting Code Scheme

Various filtering and sorting capabilities from the IPS are available based on codes assigned to each activity. Activity codes are used to filter certain activities out of the IPS for a particular requirement or need. Code fields are required to be populated for proper integration into the DUSEL IPS and EVMS. A list of the activity codes to be used on the DUSEL Project is given in Table 8.3.7. The following code fields are available and are assigned to each activity in the IPS.

| Code | Description | Required |
|---|---|---|
| **WBS** | Lowest level of Work Breakdown Structure | Yes |
| **Control Account Manager** | Unique identifier for the CAM responsible for the activity | Yes |
| **Responsible Organization** | Identifies the institution performing the work | Yes |
| **Site** | Identifier that describes the physical location where the work is managed | Yes |
| **Phase Type** | Used to describe the particular type of work performed, i.e., design, procure, fabrication, test, install, shipping, etc. | No |
| **Milestone Levels** | Defines key milestones, milestone type, and the program ownership level | Yes |
| **Activity ID** | Unique identifier for each activity | Yes |
| **Funding Source** | Indicates Funding Source (MREFC or R&RA) of funds for the work package | Yes |
| **Contractor** | Project subcontractors responsible for performing work | No |
| **Control Account #** | Identifies the control account number | Yes |
| **EVM ID** | One of the key fields to integrate activities to the cost/EVMS application | Yes |
| **Funding Type** | Indicates Funding Type (separate MREFC and R&RA activities into sub-categories) | Yes |
| **CCR** | A way to link activities to PMCS Change Request | Only if part of a CCR |
| **Level** | This designates the geographic level (surface and below) of work scope. | No |
| **EVMT** | Earned Value Management Technique | For all work packages |

**Table 8.3.7** IPS coding structure.

### 8.3.8 Reporting Schedule Progress

CAMs are required to provide status for their schedules at least once a month. Status is currently collected in a manual process with interaction between the CAMs and their respective Project Controls support contact. The Project is planning to develop a capability within DocuShare to collect the inputs through a Web-accessible DocuShare interface. In this scenario, the CAMs would complete a progress update form online and submit the changes to the Project Controls support contact. The Project Controls contact would then review input into the IPS and iterate with the CAMs as needed. The status is collected on or near the last business day of each month, consistent with the accounting month-end close.

| Activity ID | Activity Name | Orig Dur | CAM | Planned Start | Planned Finish |
|---|---|---|---|---|---|
| **FAC.MGT.01.CA3** | **FAC Management & Integration** | **189** | | **4/1/11** | **12/30/11** |
| MGT_116300 | Facility Project Mgmt, Arch. & Integ. - SDSMT - FY11 | 128 | MH | 4/1/11 | 9/30/11 |
| MGT_116305 | Facility Project Mgmt, Arch. & Integ. - SDSMT - FY12 | 61 | MH | 10/3/11 | 12/30/11 |
| MGT_206000 | CA3 Planning Package for Design and CM contracts - FY11 | 128 | MH | 4/1/11 | 9/30/11 |
| MGT_206005 | CA3 Planning Package for Design and CM contracts - FY12 | 61 | MH | 10/3/11 | 12/30/11 |
| **FAC.MGT.01.FD1B** | **FAC Management & Integration** | **1194** | | **1/3/12** | **9/30/16** |
| MGT_117000 | Facility Project Mgmt, Arch. & Integ. - SDSMT - FY12 | 189 | MH | 1/3/12 | 9/28/12 |
| MGT_117100 | Facility Project Mgmt, Arch. & Integ. - SDSMT - FY13 | 251 | MH | 10/1/12 | 9/30/13 |
| MGT_117200 | Facility Project Mgmt, Arch. & Integ. - SDSMT - FY14 | 83 | MH | 10/1/13 | 1/31/14 |
| MGT_501010 | Final Design Construction Management Support - Main | 376 | MH | 1/3/12 | 6/28/13 |
| MGT_502020 | Final Design Construction Management Support - 7400 | 503 | MH | 10/1/14 | 9/30/16 |
| MGT_503090 | Final Design Commissioning Planning | 376 | MH | 1/3/12 | 6/28/13 |
| **FAC.MGT.01.FD1D** | **FAC Management & Integration** | **523** | | **1/3/12** | **1/31/14** |
| MGT_117300 | Facility Project Mgmt, Arch. & Integ. - SDSMT - FY12 | 189 | MH | 1/3/12 | 9/28/12 |
| MGT_117400 | Facility Project Mgmt, Arch. & Integ. - SDSMT - FY13 | 251 | MH | 10/1/12 | 9/30/13 |
| MGT_117500 | Facility Project Mgmt, Arch. & Integ. - SDSMT - FY14 | 83 | MH | 10/1/13 | 1/31/14 |
| MGT_503030 | Final Design Construction Management Support - LBNE | 376 | MH | 1/3/12 | 6/28/13 |
| **FAC.MGT.01.MREFC** | **FAC Management & Integration** | **1422** | | **2/3/14** | **9/30/19** |
| MGT_118900 | Facility Project Mgmt, Arch. & Integ. - SDSMT - FY14 | 168 | MH | 2/3/14 | 9/30/14 |
| MGT_119000 | Facility Project Mgmt, Arch. & Integ. - SDSMT - FY15 | 251 | MH | 10/1/14 | 9/30/15 |
| MGT_119100 | Facility Project Mgmt, Arch. & Integ. - SDSMT - FY16 | 252 | MH | 10/1/15 | 9/30/16 |
| MGT_119200 | Facility Project Mgmt, Arch. & Integ. - SDSMT - FY17 | 250 | MH | 10/3/16 | 9/29/17 |
| MGT_119300 | Facility Project Mgmt, Arch. & Integ. - SDSMT - FY18 | 250 | MH | 10/2/17 | 9/28/18 |
| MGT_119400 | Facility Project Mgmt, Arch. & Integ. - SDSMT - FY19 | 251 | MH | 10/1/18 | 9/30/19 |

**Table 8.3.8** Schedule status report.

### 8.3.9 Resource Breakdown Structure

The DUSEL Resource Breakdown Structure (RBS) reflects the common resource pool shared for the DUSEL Project that allows each CAM to assign resources to each detail activity. The requirements of the DUSEL RBS include the ability to assign labor resources by an assortment of labor grades as well as by functional discipline and location. In addition, travel and materials are available in the resource pool.

The RBS is the Budget Element file required within the EVMS application. The DUSEL RBS is presented in Appendix 8.A.

### 8.3.10 Responsibility Assignment Matrix

The DUSEL Responsibility Assignment Matrix (RAM) correlates the work defined in the WBS to the DUSEL functional organization defined in the OBS. The RAM lists the WBS control accounts on one axis and the assigned CAMs on the other. The intersections shown in the matrix illustrate which CAM is responsible for the management and execution of a given control account's scope. The RAM currently lists CAMs by position title and not the specific individual's name. The position titles will be replaced with specific name assignments prior to the start of Final Design.

The RAM is required within the EVMS and is presented in Appendix 8.B.

### 8.3.11 EVMS Cost System Design

The DUSEL EVMS contains time-phased budgets, reports of work performed/value earned, and actual costs. These are captured formally as the Budgeted Cost of Work Scheduled (BCWS), the Earned Value or Budgeted Cost of Work Performed (BCWP) for each work package, and the Actual Cost of Work Performed (ACWP) for each control account and is summarized to the each level of the WBS. The EVMS provides visibility on the planned budget and forecast as it is affected by schedule status. Using the commercial off-the-shelf cost-management software from the Deltek Corporation called Cobra®, the EVMS provides the following functionality:

1. Defines a time-phased budget for each control account and its work packages (labor, travel, materials, and supplies)
2. Allows for the importing/integrating of actual expenses from the accounting system at month-end
3. Calculates Earned Value for each work package using an assigned earned value method and progress imported from the schedule
4. Uses one common RBS
5. Incorporates Direct rates for labor resources
6. Incorporates the associated Indirect rates as defined by the participating institutions
7. Incorporates one or more escalation rates
8. Identifies and maintains a contingency log at the DUSEL project-level that tracks allocation of contingency to each control account to produce a revised PMB
9. Maintains multiple/historical PMBs
10. Supports a change management process to ensure that approved changes are captured and incorporated into the baseline in a controlled and timely manner
11. Monitors thresholds for variance reporting
12. Provides EVMS reporting by defined code tables
13. Calculates schedule variances (SVs) associated with work accomplished and planned budget through a designated accounting month-end

14. Calculates cost variances (CVs) associated with work accomplished and costs incurred through a designated accounting month-end
15. Generates tabular and graphic views of BCWS, BCWP, and ACWP. Calculates schedule and cost performance indexes (SPI, CPI) for each work package.
16. Allows for the creation of multiple cost types/classes, providing the ability to do what-if analysis and create an ETC without altering the current PMB profile
17. Accommodates fiscal-year and calendar-year requirements
18. Extensive cost-performance and earned-value reporting

### 8.3.12 Earned Value Methods

The Earned Value methods shown in Table 8.3.12 can be applied to a work package.

| EV Method | Description |
|---|---|
| **50-50 Rule** | 50% of the budget amount is earned at the activity's actual start and 50% at the actual finish date. |
| **0-100 Rule** | 100% of the budget is earned at the activity's actual finish date. |
| **100-0 Rule** | 100% of the budget is earned at the activity's actual start date. |
| **Level of Effort (LOE)** | Work that has no definable or easily measurable output. Typically general and supportive in nature. Examples are supervision, project administration, contract administration. |
| **Milestone** | Multiple milestones can be defined for each work package as a means of measuring progress. Milestones are typically used to identify points within a work package where costs are accrued. For example, a milestone may mark the completion of a phase of the work package.<br>The milestones in the Earned Value application typically are linked to activities in the schedule. The finish date of a milestone is the same as the finish date for the linked activity.<br>Unlike a scheduling application, the Earned Value application does allow for the percentage of a milestone to be recorded. |
| **Physical Assessment of Percent Complete** | This method uses a subjective percent complete as the measure of completed work. The Level 2 manager assigns a best assessment of the percent of work completed to date. The value that is earned represents the assessed progress and tangible achievement of the planned effort. This method is applied to "discrete" tasks that can be directly measured and have a specific end product or end result. |
| **Planning Package** | Used for planning tasks out in the future that are not ready to be broken down into detailed work plans. A total budget value may be assigned to the planning package, but the budget has not been distributed over time. |

**Table 8.3.12** Earned Value management techniques.

### 8.3.13 Cost and Schedule Integration

The activities in the IPS are assigned a work package/EVM ID number that is reflected in the EVMS tool as the work package number. The combination of the control account number and the work package number will serve as the key fields for linking the IPS to the EVMS. The start and finish dates of an IPS activity define the start and finish dates of the EVMS work package. The resources and time-phasing defined by the IPS activity define the resources and time-phasing of the EVMS work package.

When CAMs report progress each month, it will be imported from the IPS to the EVMS (Cobra) application.

### 8.3.14 Accounting Integration

At the close of each month-end accounting cycle, the actual project cost data from Lawrence Berkeley National Laboratory (LBNL), South Dakota School of Mines and Technology (SDSM&T), South Dakota Science and Technology Authority (SDSTA), and the University of California at Berkeley (UC Berkeley) is consolidated in the UC Berkeley accounting system for each control account and is broken down by each resource type: labor, non-labor, and travel. Actual costs are divided into Direct and Indirect.

Within the UC Berkeley accounting system the "chart string" number is the unique alphanumeric charge number used for collecting costs. The chart string is linked to the DUSEL WBS/control account numbers and the expenditure categories within the accounting system map to the resource types within a given control account. Table 8.3.14 represents an example of the flat file that is inserted into the PMCS.

| Control Account | Resource Type | Direct | Indirect |
|---|---|---:|---:|
| FAC.DLL.01.CA1 | Labor | $2,694.10 | $1,643.40 |
| FAC.LGC.01.CA1 | Non-Labor | $125,269.94 | $65,699.42 |
| FAC.LGC.01.CA1 | Labor | $22,272.73 | $13,586.37 |
| FAC.MGT.01.CA1 | Labor | $115,518.73 | $70,466.43 |
| FAC.MGT.01.CA1 | Non-Labor | $655,511.35 | $242,539.20 |
| FAC.MLL.01.CA1 | Labor | $165,879.21 | $101,186.32 |
| FAC.MLL.01.CA1 | Non-Labor | $3,153,253.40 | $1,166,703.70 |
| FAC.OLR.01.CA1 | Labor | $26,725.45 | $16,302.52 |
| FAC.SUR.01.CA1 | Labor | $10,714.28 | $6,535.71 |
| FAC.SUR.01.CA1 | Non-Labor | $302,200.00 | $111,814.00 |
| FAC.UGI.01.CA1 | Labor | $180,681.90 | $233,793.35 |
| FAC.UGI.01.CA1 | Non-Labor | $2,192,949.20 | $811,391.19 |
| OPS.MGT.01.CA1 | Labor | $42,670.00 | $54,747.00 |
| PRJ.BUS.01.CA1 | Labor | $72,181.17 | $58,863.18 |
| PRJ.EDO.01.CA1 | Labor | $190,576.91 | $79,140.94 |
| PRJ.EDO.01.CA1 | Non-Labor | $39,997.09 | $14,798.92 |
| PRJ.EDO.05.CA1 | Labor | $0.00 | $0.00 |
| PRJ.EHS.01.CA1 | Non-Labor | $19,726.03 | $10,553.43 |
| PRJ.EHS.01.CA1 | Labor | $71,317.21 | $47,816.44 |
| PRJ.PMO.01.CA1 | Labor | $733,481.59 | $651,647.97 |
| PRJ.PMO.01.CA1 | Non-Labor | $978,194.83 | $340,403.78 |
| PRJ.PMO.01.CA1 | Travel | $257,940.02 | $95,071.25 |
| PRJ.PMO.02.CA1 | Labor | $222,230.00 | $307,414.00 |
| PRJ.PMO.02.CA1 | Non-Labor | $177,264.08 | $94,836.28 |
| PRJ.PMO.03.CA1 | Labor | $5,394.00 | $5,714.00 |
| PRJ.PMO.03.CA1 | Non-Labor | $29,256.38 | $15,652.16 |
| PRJ.PMO.03.CA1 | Travel | $112,224.23 | $0.00 |
| PRJ.SYS.01.CA1 | Labor | $304,224.53 | $397,237.78 |
| SCI.BGE.01.CA1 | Labor | $292,435.00 | $467,156.00 |

| Control Account | Resource Type | Direct | Indirect |
|---|---|---|---|
| SCI.DBD.01.CA1 | Labor | $4,217.00 | $5,851.00 |
| SCI.DKM.01.CA1 | Labor | $25,255.00 | $42,327.00 |
| SCI.LBC.01.CA1 | Labor | $11,350.00 | $14,569.00 |
| SCI.LBN.01.CA1 | Labor | $50,597.00 | $81,542.00 |
| SCI.MGT.01.CA1 | Labor | $157,270.00 | $240,521.00 |
| SCI.NAS.01.CA1 | Labor | $59,231.00 | $81,784.00 |

**Table 8.3.14** UC Berkeley accounting system file that integrates with the PMCS.

### 8.3.15 Reporting

The ability to generate reports in a timely manner is key to the effectiveness of the DUSEL PMCS. The reporting functionality is intended to serve all levels of the DUSEL Project team as well as external partners and customers.

In addition to generating reports from each respective PMCS tool, a Web-based system (DocuShare) is in place to access and display cost and schedule data for all levels of the organization. Permissions are established for each user within the Project so that they are only able to view what they have been authorized to view. The reports are coded, filtered, and sorted in a manner consistent with the permissions granted for each user. Custom reports specific to a single user will also be available based on the requirements they provide.

Table 8.3.15-1 is representative of the common reports that are available output from the PMCS. Gantt charts will include dependencies to show critical and near-critical paths.

| Name | Fields | Filter Criteria | Comments/Features |
|---|---|---|---|
| **Schedule** | | | |
| **Activities in Progress** | Activity ID, Description., Duration, Early and Late dates, Physical % Complete, Total Float | Activities with Physical % Complete > 0, but <100% | Heading by WBS, sort by WBS, then Start.<br>Table + Gantt Chart |
| **Activities behind Schedule** | Activity ID, Description, Duration, Early Start, Early Finish, Baseline Start, Baseline Finish, Physical % Complete, Total Float | Activities with Early Finish > Baseline Finish | Heading by WBS, sort by WBS, then Start.<br>Table + Gantt Chart |
| **Critical Activities** | Activity ID, Description, Duration, Early Start, Early Finish, Total Float | Activities with Total Float < or = to 0 | Heading by WBS, sort by WBS, then Start.<br>Table + Gantt Chart |
| **Critical Path** | Activity ID, Description, Duration, Early Start, Early Finish, Total Float | The longest path within the schedule i.e. the path with least amount of float | Heading by WBS, sort by WBS, then Start.<br>Table + Gantt Chart |
| **Disconnects with EVMS** | WBS, Work Package, Description, Budget, Units, Early Start (Cobra), Early Start (Schedule App.), Early Finish (Cobra), Early Finish (Schedule App.) | Work packages that have either a start or finish date disconnect and are not complete. | Heading by WBS, Tabular report |

| Name | Fields | Filter Criteria | Comments/Features |
|---|---|---|---|
| **Gantt Chart** | Activity ID, Description, Duration, Early Start, Early Finish, Physical % Complete, Total Float | Use predefined filters | Heading by WBS, sort by WBS, then Start.<br>Gantt Chart with or without relationships<br>Drill-down capability<br>Adjustable date ribbon (Year/Qtr., Year/Month etc.)<br>Selection of one or more bars to compare current start/finish dates to baseline dates |
| **Milestones** | Activity ID, Description, Duration, Early Start, Early Finish, Baseline Start, Baseline Finish, Physical % Complete, Total Float | Activities with 0 duration<br>Completed vs. Not Completed<br>Select by level<br>Interface Milestones | Heading by WBS, sort by WBS, then Start.<br>Table + Gantt Chart<br>Various symbols associated with milestone level |
| **Schedule Updates** | Activity ID, Description, Responsible Person, Duration, Early Start, Early Finish, Actual Start,<br>Actual Finish, Expect Start, Expected Finish,<br>Current Physical % Complete, New Physical % Complete | Activities in progress + 60-day look-ahead | Heading by WBS, sort by WBS, then Start.<br>Table<br>The Actual Start,<br>Actual Finish, Expect Start, Expected Finish,<br>and New Physical % Complete are input fields. |
| **Cost Estimate** | | | |
| **Basis of Estimate** | WBS, Basis of Estimate (BOE) | | Paragraph Report<br>Export to Word |
| **Budget and Contingency** | Chart of budget and contingency by year<br>WBS<br>Budget Total<br>Contingency Total<br>Contingency %<br>Total Budget + Contingency | Filter by Base Year or Then Year | Dynamic chart by selecting WBS<br>Select by calendar by Fiscal Year or Calendar Year |
| **Cost Book** | WBS, Estimator(s), Total Labor, Total Non-Labor, Travel, Other (G&A, Fringe etc.) Contingency,<br>Contingency %, WBS Dictionary, BOE Labor, BOE Non-Labor, BOE Travel, Risk Tech Factor, Risk Tech Multiplier, Risk Cost Factor, Risk Cost Multiplier, Risk Schedule Factor, Risk Schedule Multiplier, Line Item Resource, Line Item Start, Line Item End,<br>Estimate Type, Unit Cost, Unit Quantity, Base Yr. Cost | All or WBS | See DUSEL example<br>Export to Word |
| **FTEs** | FTE Chart | Filter by Base Year or Then Year | Dynamic chart by selecting WBS<br>Select by calendar by Fiscal Year or Calendar Year<br>Line Chart<br>Tabular<br>Interactive by WBS |
| **Resource Type** | Pie chart for Labor, Non-Labor, Travel, Other, and Contingency | Filter by Base Year or Then Year | Dynamic chart by selecting WBS<br>Select by calendar by Fiscal Year or Calendar Year |

| Name | Fields | Filter Criteria | Comments/Features |
|---|---|---|---|
| **WBS Summary Report** | WBS, Description, Labor, Non-Labor, Travel, Other, Contingency, Contingency %, Total Cost | Filter by Base Year or Then Year | Tabular report and/or drill down<br>Export to Word |
| **WBS Dictionary** | WBS, WBS Description, Dictionary (Long Description), Comments, Scoping Options, User Defined Field | Filter by WBS | Tabular report and/or drill-down<br>Export to Word |
| **EVMS** | | | |
| **Bull's-Eye Chart** | SV %, CV% | Filter by WBS | Cobra generated |
| **Commitments (Obligations)** | | | Generate a Funding vs. Commitment report for any level of the project. |
| **Contingency Plan** | WBS, OBS, Title, Date Approved, Date Implemented, Amount Base Year Dollars, Amount Then Year Dollars, Balance | All, WBS, OBS | Link to Cobra<br>Tabular report |
| **CPR Reports** | | | |
| **Format 1 - WBS** | | Filter by WBS | |
| **Format 2 - OBS** | | Filter by WBS | |
| **Format 3 - Baseline** | | Filter by WBS | |
| **Format 4 - BAC** | | Filter by WBS | |
| **Format 5 - Explanations** | | Filter by WBS | |
| **Work Authorization Plan** | | | |
| **CSSR** | | Filter by WBS | |
| **Disconnects with IPS** | WBS, Work Package, Description, Budget, Units, Early Start (Cobra), Early Start (Schedule App.), Early Finish (Cobra), Early Finish (Schedule App.) | Work packages that have either a start of finish date disconnect and are not complete. | Heading by WBS, Tabular report |
| **Drill-down** | WBS, WBS Description, Early Start (Cobra), Early Finish (Cobra), Physical Percent Complete, Actuals_Cum (Cobra), EarnedValue_Cum (Cobra), Planned_Cum (Cobra), Cost Variance_Cum, CPI, SPI, BAC, EAC, Variance | All | Drill-down by WBS |
| **EVMS** | WBS, Planned_Cur, Earned_Cur, Actual_Cur, SchedVar, CostVar, SPI_Cur, CPI_Var, Planned_Cum, Earned_Cum, Actual_Cum, SchedVar_Cum, CostVar_Cum, SPI_Cum, CPI_Cum, BAC, EAC, Variance, Percent Complete | Filter by Base Year or Then Year | Charts and tables<br>Drill-down<br>Generate three curve EV graphs (PV, EV, AC) for the total project level, the Level 2 Manager level and the lowest-level WBS. |
| **SPI/CPI** | SPI, CPI by Month | Filter by WBS | Line Graph |
| **Variance Explanations** | Explanation_Cum, Impact_Cum, Corrective Action_Cum, Explanation_Cur, Impact_Cur, Corrective Action_Cur | Filter by WBS | Narrative |

| Name | Fields | Filter Criteria | Comments/Features |
|---|---|---|---|
| **Action Items** | | | |
| **Action Items** | Date Requested, Username (Requester), Description, WBS, Responsible Party, Target Finish, Actual Finish, Status | Filter/Sort by WBS, Requester, Responsible Party | Tabular report<br>Export to Word |
| **Weekly Activity Reports** | Entered On, Week Ending, Group (Project), Username, Accomplishments, Issues, Action Items,<br>Entry,<br>Week Ending Date, Approval Level 1, Approval Level 2 | Filter/Sort by WBS, Group, Username, Section (Accomplishment, Issues, Action Items) | Tabular report<br>Export to Word |

**Table 8.3.15-1** Typical reports available from the PMCS.

Table 8.3.15-2 provides the monthly reporting cycle schedule for reporting the status and updating the EVMS tools and generating monthly management reports.

| | | | | | | |
|---|---|---|---|---|---|---|
| Month End | 1 | 2 | 3 | 4 | 5 | 6 |
| | Collect and Input Month End Status | | | | | |
| | Final LBNL Month End Data | | | | | |
| 7 | 8 | 9 | 10 | 11 | 12 | 13 |
| | Review Updated Status Reports with CAMs and Iterate | | | | | |
| | Final SDSMT Month End Data | | Final UCB Month End Data | | | |
| 14 | 15 | 16 | 17 | 18 | 19 | 20 |
| | Develop Variance Analysis Reports | | | | | |
| 21 | 22 | 23 | 24 | 25 | 26 | 27 |
| | Review Month End Performance Reports with Project Management | | | | | |
| 28 | 29 | 30 | 31 | | | |
| | Distribute Schedule Reports for Statusing | | | | | |
| | | | Month End Report Ready | | | |

**Table 8.3.15-2** PCMS monthly reporting cycle.

## 8.3.16 Baseline Configuration Control

It is inevitable that changes will occur, and the configuration management process ensures these changes are captured and incorporated into the baseline in a controlled and timely manner. The configuration management process is initiated by the need to modify the IPS and/or the PMB.

The CCB will approve all changes and will be used specifically in the following situations:

- Budget moves from one Level 2 manager to another
- Budget moves from one Control account to another
- Any change to BCWS that is within the current month or next month

- Any transfer of work scope between funding organizations
- Increases in work scope that require the allocation of project contingency
- Customer-requested change
- An approved re-baseline
- A technical scope change/contract change (external or internal)
- Renegotiated subcontract activity
- Correction of errors in the baseline budget/schedule

The DUSEL configuration management process describes how the IPS and/or PMB is managed and controlled throughout the life of the project. The change process clearly traces the current version of the IPS/PMB back to the originals. The process ensures all changes to the Project are controlled, documented, and managed in a consistent manner. This will ensure both visibility and control of the baseline to ensure that timely and accurate management information is always available.

Figure 8.3.16 represents a flowchart of the current PMCS configuration control process.

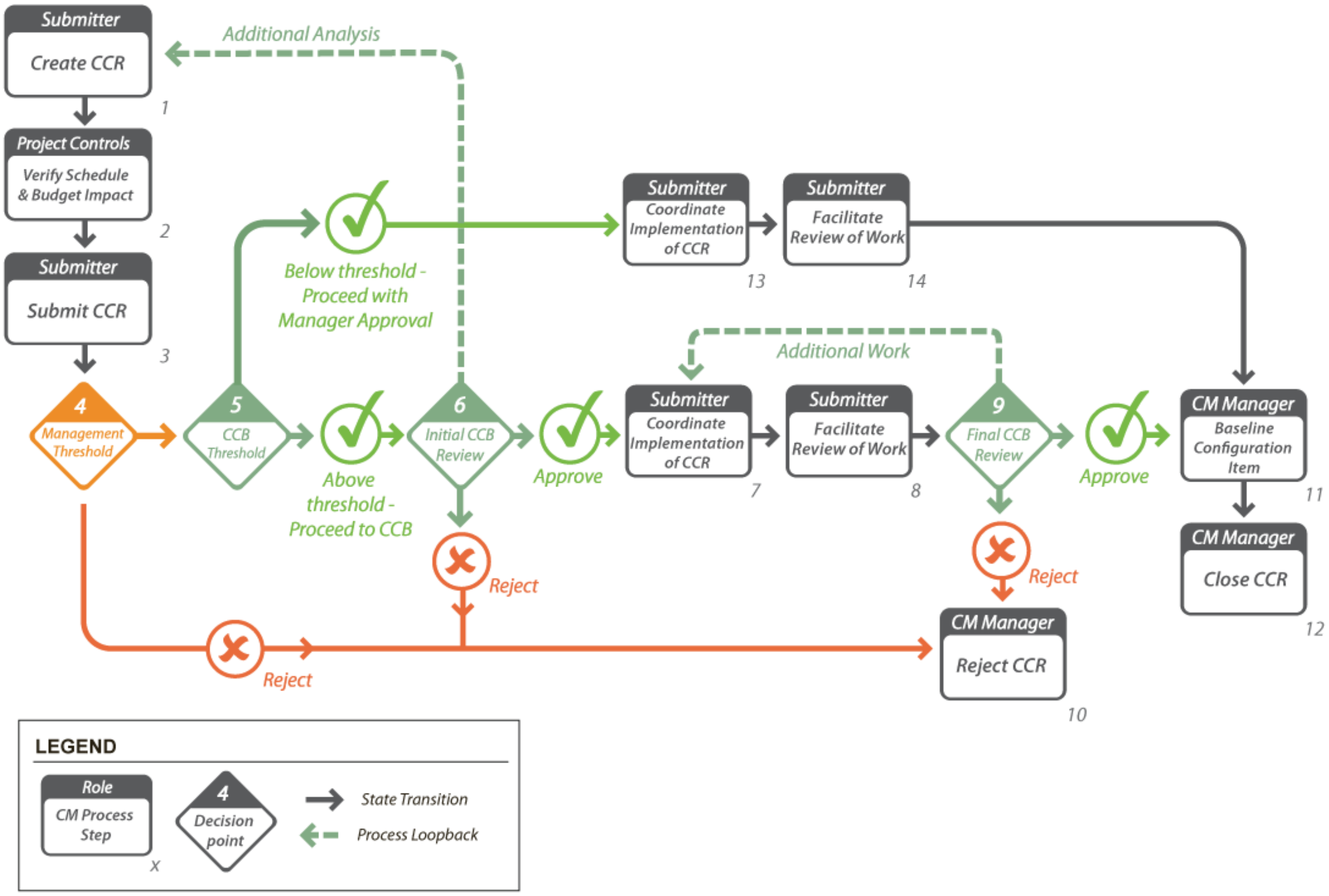

**Figure 8.3.16** Configuration Control Board flowchart. [DKA]

## 8.3.17 Security

Stakeholders, Central Project Directorate, Level 2 managers, and designated CAMs are granted the corresponding level of security permissions so that they can only read and/or update the relevant reports and pertinent data consistent with their authorization.

Security for a respective PMCS database is assigned to individual users and/or groups and controlled by the Project Controls team.

### 8.3.18 Glossary

**Actual Cost of Work Performed (ACWP), a.k.a. Actual Cost (AC).** The costs actually incurred and recorded in accomplishing the work performed within a given time period. ACWP is also referred to as Actual Cost (AC).

**Apportioned Effort.** Work scope that by itself is not readily divisible into short-span work packages but which is directly related in direct proportion to other measurable effort.

**Budget At Complete (BAC).** The total authorized budget for accomplishing the scope of work usually presented in the desired unit (hours or dollars) by cost element.

**Budgeted Cost of Work Performed (BCWP), a.k.a. Earned Value (EV).** The sum of the budgets for completed work packages and completed portions of open work packages, plus the applicable portion of the budgets for level of effort and apportioned effort within a given time period. BCWP is also referred to as Earned Value (EV).

**Budgeted Cost of Work Scheduled (BCWS), a.k.a. Planned Value (PV).** The sum of the budgets for all work packages and planning packages scheduled to be accomplished plus the applicable portion of the budgets for level of effort and apportioned effort within a given time period. A time-phased budget plan. BCWS is also referred to as Planned Value (PV).

**Constraint Date.** A time constraint or restriction applied to an activity's start or finish date. Example: Start No Earlier Than, Finish No Later Than.

**Control Account.** A significant subset of a project in which planned and actual costs are accumulated and compared to Earned Value for budget-management purposes. It represents the work assigned to one responsible organizational element and one work-breakdown structure element.

**Cost Element.** A unit of cost, typically in the form of direct labor, direct materials, other direct costs, and indirect or burdened costs.

**Cost Performance Index (CPI).** The ratio of Earned Value to actual cost expenditure for a specified time period. A value of 1.00 indicates efficiency as budgeted. Values greater than 1.00 indicate greater output than budgeted (under run) while values less than 1.00 indicate output less than budgeted (overrun). A value of .8 means that for every $1 spent, $0.80 value was accomplished.

**Cost Performance Report (CPR).** A contractually required report submitted to a customer that represents overall project cost and schedule performance in dollars.

**Cost Variance (CV).** The difference between Earned Value (BCWP or EV) and Actual Cost (ACWP or AC) (CV = BCWP – ACWP) at a specific point in time. A positive value indicates a favorable position and a negative value indicates an unfavorable condition. A positive or negative value that exceeds a designated threshold may also require a variance explanation.

**Critical Path.** The series of activities that determines the duration of the project. The critical path is usually defined as those activities with the least amount of float or less than or equal to a specified value, often zero. It is the longest continuous path through the project.

**Earned Value (EV).** An objective measure of work accomplished, based on its budgeted value within a specified time period. Using the Earned Value management process, management can readily compare how much work has actually been completed against the amount of work planned for accomplishment. Earned Value analysis requires project management to plan, budget, and schedule the authorized work scope in the time-phased plan. The time-phased plan is the incremental "planned value" culminating into a performance measurement baseline. As work is accomplished, it is "earned."

**Earned Value Management.** A management technique that relates resource planning to schedules and to technical cost and schedule requirements. All work is planned, budgeted, and scheduled in time-phased increments constituting a cost and schedule measurement baseline. Project performance is then measured relative to that baseline.

**Estimate At Completion (EAC).** The estimated total cost for the authorized work. It equals actual cost to date plus an estimate of the cost of the authorized work remaining (usually based on its authorized budget).

**Estimate To Complete (ETC).** A time-phased estimate of costs to complete all authorized but incomplete work from a specified time to completion. Typically forecast by cost element.

**Free Float.** The amount of time an activity can be delayed without affecting the earliest start of any of its succeeding activities.

**Integrated Project Schedule (IPS).** The IPS provides a summary view of all planned work, providing the ability to analyze and monitor the critical path for the entire project. The IPS is baselined, allowing for a benchmark by which the project's current working schedule can be measured against the original plan. The IPS database contains all cost and schedule data and includes the necessary coding to identify key elements in the project such as Work Breakdown Structure (WBS), Organizational Breakdown Structure (OBS), work packages, major project milestones, and handoff milestones.

**Level 2 Manager.** DUSEL uses an integrated product teaming approach for managing the DUSEL Project. These integrated product teams are an essential element in management process and are being used during all phases of the Project's life cycle. These teams consist of professionals representing diverse disciplines with the specific knowledge, skills, and abilities to support the Project Manager in successfully executing the project. The Level 2 managers for the DUSEL Project will consist of the department heads and group leaders.

**Interface/Handoff Milestone.** The term applied to the detail schedule milestone representing the agreed-to date for the delivery of product/data from one Level 2 manager to one or more recipient Level 2 managers. A milestone that identifies the point of integration from one Level 2 manager to another.

**Level of Effort (LOE).** Work that has no definable or easily measurable output. It is generally characterized by a uniform rate of activity over a specified period of time. Examples are supervision, project administration, and contract administration.

**Management Reserve (a.k.a. Contingency) (MR).** An amount of contract budget set aside by the Project Manager at the start of the project. It is reserved for those initially unknown tasks that, when identified in the future, are in scope to the contract but out of scope to a work package.

**Milestone.** A significant event in the life of a project, usually the start or completion of one or more major deliverables. Also known as an activity with zero duration.

**Organizational Breakdown Structure (OBS).** The organization structure, usually functional, that defines integrated product team (Level 2 manager) responsibilities for the management and performance of the project work scope down to the levels at which work is performed and managed.

**Performance Measurement Baseline (PMB).** The time-phased budget plan against which project performance is measured. The PMB is the sum of both the distributed budgets and the undistributed budget. The PMB is equal to the Contract Budget Baseline (CBB) less Management Reserve (MR) unless the over-target baseline has been established with prior customer notification. It is a time-framed summation of the planned work.

**Planning Package (PP).** Used for planning tasks out in the future that are not ready to be broken down into detailed work plans. A total budget value may be assigned to the PP, but the budget has not been distributed over time.

**Resource.** Any item (except time) required to accomplish an activity. Resources can be people, equipment, facilities, funding, or anything else needed to perform the work of a project. Can be planned/measured in labor hours or non-labor dollars.

**Schedule Performance Index (SPI).** The ratio of work accomplished to work planned for a specified time period. A value of 1.00 indicates productivity was exactly as planned. A value greater than 1.00 indicates more total work has been accomplished than scheduled; a value less than 1.00 indicates less total work has been accomplished then scheduled.

**Schedule Variance (SV).** The difference between Earned Value (BCWP or EV) and Planned Value (BCWS or PV) (SV = BCWP – BCWS) within a specified time. Also stated as the difference between what was accomplished and what was planned. A positive value is a favorable condition, while a negative value is unfavorable. A positive or negative value that exceeds a designated threshold may require a variance explanation.

**Statement of Work (SOW).** Contractual document that defines the work scope requirements for the project. It may include a list of specific deliverables or describe specific technical, cost, and scheduling targets. It is usually a narrative description of the work to be performed.

**Total Float.** The amount of time an activity can be delayed or expanded before it delays the completion date (or target completion date) of the project.

**Undistributed Budget.** The budget applicable to contract effort that has not yet been identified to the Level 2 managers.

**Undefinitized Work.** Authorized work for which a firm contract value has not been negotiated or otherwise determined.

**Variance at Completion.** The Budget At Completion (BAC) less the Estimate At Completion (EAC). It represents the amount of expected overrun or underrun. A positive value is an underrun, while a negative value is an overrun. A value that exceeds a designated threshold may require a variance explanation.

**Work Breakdown Structure (WBS)**. An indentured listing or graphic of all the products, components, work tasks, and services to be accomplished by the project at various nested levels, from the overall project to individual work packages. It organizes, displays, and defines the product to be developed and/or produced with other products.

**Work Package.** Short-span tasks or material items identified for accomplishing work required to complete the project or contract. A significant subset of a cost account, where planned and actual costs are captured and compared to EV for budget-management purposes.

This page intentionally left blank

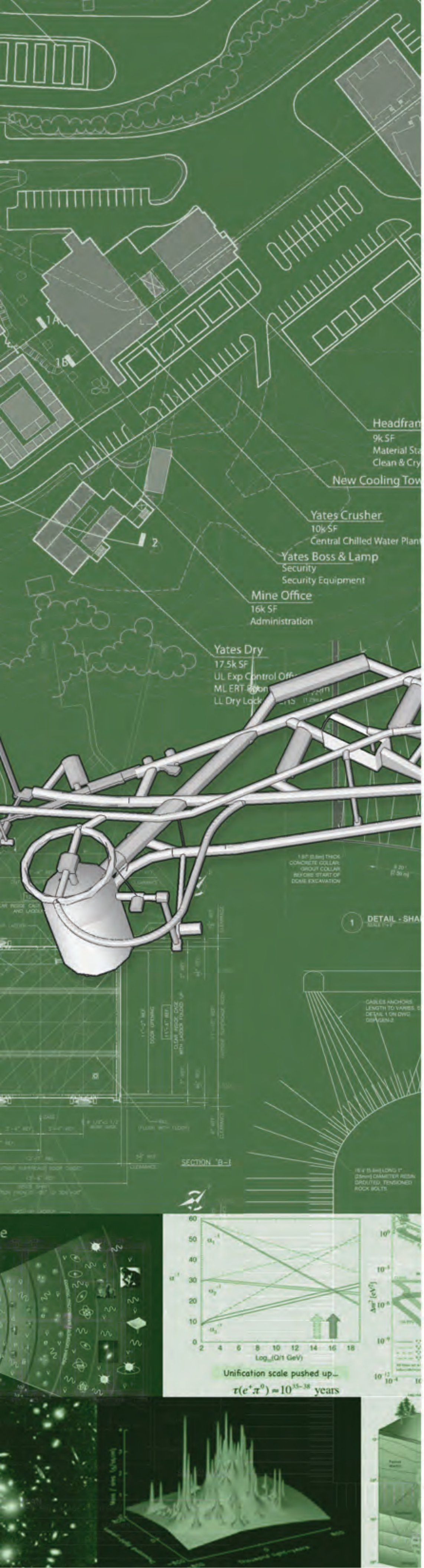

# Preliminary Design Report

May 2011

# Volume 9:
# Systems Engineering

DUSEL
Deep Underground
Science and
Engineering Laboratory

This page intentionally left blank

# Systems Engineering

## Volume 9

This section outlines the overall structure and staffing of the Systems Engineering department within the DUSEL Project, the main areas of focus in support of the Preliminary Design effort, and the major emphasis of the Systems Engineering department to support the start of Final Design. This section details the processes used by Systems Engineering to achieve Project integration with respect to requirements and interfaces, risk management, and design development. Volume 3, *Science and Engineering Research Program*, of this Preliminary Design Report (PDR) goes into more detail on the method for developing science-driven Facility requirements. Volume 5, *Facility Preliminary Design*, of this PDR provides additional detail on the key requirements and interfaces of the Facility design along with delineating the design in response to the stated requirements.

### 9.1 Systems Engineering Organization

The Systems Engineering effort is led by the Systems Engineering (SE) Manager. This position is a Level 3 Work Breakdown Structure (WBS) manager who is based in Lead, South Dakota, and who reports to the Deputy Project Director. The SE team is responsible for the following functions:

- Project Requirements
- Interface Control
- Value Engineering Management and Trade Studies
- Risk Management
- Configuration Management
- Systems Verification

Detailed explanations of the SE team's responsibilities can be found in Appendix 9.A, S*ystems Engineering Management Plan (SEMP)*. With the support of the other Level 2 and 3 managers (e.g., the Facility Project Manager), the SE team has integrated the Systems Engineering requirements and interface development process into the DUSEL Project. Through the end of the Construction phase, the Systems Engineering department will include the following staff:

- **Requirements and Interfaces Engineers.** These engineers are responsible for the development of requirements and interfaces that detail the needs of all stakeholders in the program and provide evaluation criteria for future verification efforts. They also participate in the configuration management processes, as well as take part in and sometimes lead Value Engineering and Trade Studies for the program. These engineers also identify and work on assigned risks in order to provide information to be used for risk assessment. During the Final Design phase, this effort will be staffed by two engineers, and then decreased to one engineer during Construction.
- **Configuration Management and Risk Management Engineer.** This engineer is involved in the planning and implementation of the Project Configuration Control and the Risk Management system as well as day-to-day operations of these processes.

- **Verification Engineer.** This engineer is responsible for the planning and oversight of all aspects of verification of requirements and interfaces. This engineer will work with commissioning agents; Facility; Environment, Health, and Safety (EH&S); and Science to ensure all aspects of the system are verified.

## 9.2 Systems Engineering Approach

The Systems Engineering efforts have focused on the following areas during the Preliminary Design phase:

- Development of the DUSEL Facility requirements (including the needs of the experimental program, EH&S, and Education and Outreach [E&O]) and the *Requirements Compliance Matrix*, Appendix 9.B
- Development of the interfaces between Science and the Facility as documented in an Interface Requirement Document (IRD) for each interface
- Maturation of the Project Risk Registry to support Project Controls' quantitative Project contingency recommendations
- Development and implementation of the DUSEL Configuration Management processes

The application of Systems Engineering practices has well positioned the DUSEL Project to move forward through design and construction while supporting Operations activities in parallel with the Facility development.

The Systems Engineering model tailored for DUSEL represents an adaptation of the engineering approach typically used in an aerospace system development environment. This model, as defined in the DUSEL SEMP and supporting plans, outlines a requirements development approach that first defines the highest-level DUSEL mission objectives. These fundamental mission needs are then decomposed and elaborated through successive, more detailed levels of requirements down to the facility functional and performance specifications or "build to" requirements that are provided to the architecture and engineering contractors to drive the Facility design development. Traceability exists between all requirements levels to ensure that all requirements have a tie to the mission objectives. The various levels requirements are used not only for design development but also serve as the basis for the end-to-end verification and commissioning program used to evaluate if the Facility meets the DUSEL science and operational needs. Facility acceptance is based upon this overall verification program during construction and is grounded in satisfaction of the defined requirements. Figure 9.2 illustrates the requirements decomposition approach, its link to the facility design, and the tie between the DUSEL requirements and verification plans that are generated and executed as the Facility is commissioned and accepted during the Construction phase.

The facility design compliance matrices were developed during Preliminary Design and demonstrate that the facility design satisfies the technical requirements derived from science needs. The technical requirements also allow the Project to control design scope and communicate baseline functions and performance levels. The Project risks have been documented, analyzed, and are being actively managed. The basis for Project contingency is supported through a defined quantitative risk analysis process, as defined in the Appendix 9.C, *Risk Management Plan*. The documented interface requirements create a framework to ensure that the design remains coordinated and integrated among all stakeholders. Figure 9.2 provides a graphic representation of the Systems Engineering process, demonstrating the relationship between each of the SE activities over the duration of the Project.

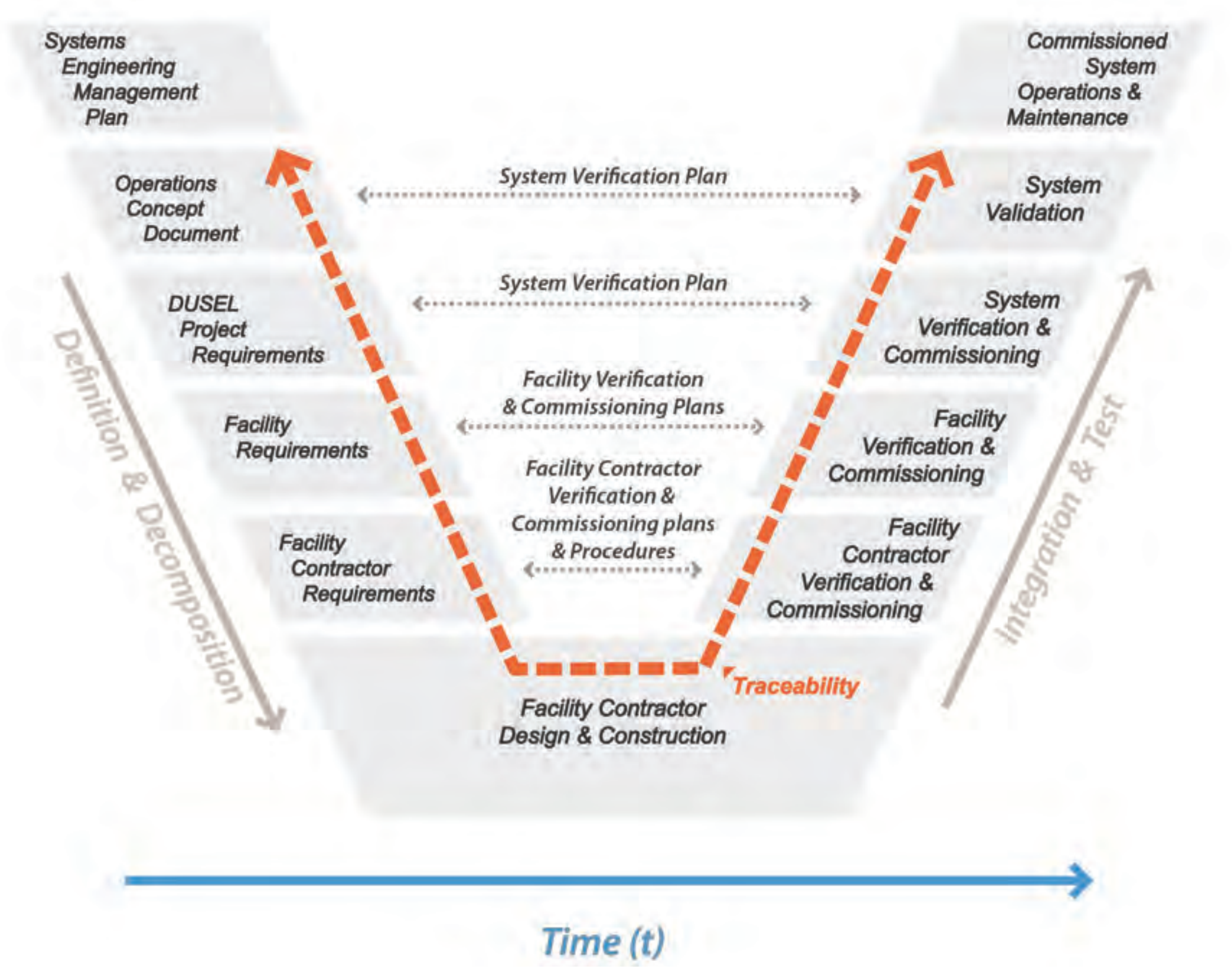

**Figure 9.2** DUSEL Systems Engineering model.

### 9.2.1 Requirements and Interface Development

The design of the DUSEL Facility, and thus the Facility requirements, are driven by the needs of the Science program, the Project's commitment to safety, and support for the E&O program. The processes established by Systems Engineering ensure that all stakeholders are involved in the requirements definition, derivation, and refinement process through interviews, working group meetings, and peer reviews. Details on the Systems Engineering processes can be found in Appendix 9.A *Systems Engineering Management Plan*. Systems Engineering established the peer-review process described in the Appendix 9.D *Configuration Management Plan* to ensure stakeholder feedback and involvement.

Systems Engineering has developed a requirements and interface document hierarchy tailored to the DUSEL Project that reflects Project organization, major interfaces, and functional aspects of the design. This structure also supports documentation of external constraints, such as limitations of utilities and variances to the requirements. Detailed explanations of the content of and relationships between these interface documents can be found in Appendix 9.A.

The highest-level DUSEL requirements and goals are documented in Appendix 9.E, *DUSEL Project Requirements Document*. This document includes the DUSEL Key Performance Parameters, which are fundamental metrics used in evaluating the Project's performance in satisfying the highest-level stakeholder goals for the DUSEL Project. This document is approved by the DUSEL Principal

Investigator, Co-Principal Investigator, Associate Project Director, and the Project and Operations Director.

Interface requirements between the Facility and the Integrated Suite of Experiments (ISE) are captured in Appendix 9.F, *Integrated Suite of Experiments Interface Requirements Documents (ISE IRD)*. The process for determining the general ISE requirements is detailed in Chapter 3.6, *ISE Requirements Process.* As the Systems Engineering and Configuration Management processes ensure adequate and proper stakeholder review of all released documents, DUSEL scientists and engineers participate in the requirements-development working group meetings and in the peer review of the two detailed lower-level Facility requirements documents, Appendix 9.G, *Facility Infrastructure Requirements Document;* and Appendix 9.H, *Facility Spaces Requirements Document*.

EH&S high-level requirements reside in Appendix 9.I, *DUSEL Facility Requirements Document*; interface requirements with Science and EH&S reside in the ISE IRD; and lower-level design detail requirements reside in *Facility Infrastructure Requirements Document* (Appendix 9.G), and *Facility Spaces Requirements Document* (Appendix 9.H). The EH&S requirements were developed based on interviews with EH&S personnel and fire/life/safety engineers; EH&S participation in requirements-development working group meetings; and the peer review of *ISE IRD*, *DUSEL Facility Requirements Document*, *Facility Infrastructure Requirements Document*, and *Facility Spaces Requirements Document*.

E&O high-level requirements are also contained in the *DUSEL Facility Requirements Document*. These requirements were developed from interviews with E&O personnel, from participation from E&O personnel during requirements-development working group meetings, and during the peer-review process. The Education and Outreach Director is a signatory of this document.

The primary requirements of concern for the Facility designers are the Level 3 subsystem requirements that reside in the *Facility Infrastructure Requirements Document* (Appendix 9.G) and the *Facility Spaces Requirements Document* (Appendix 9.H). These documents were reviewed by the Facility Integrated Product Teams (IPTs), as well as DUSEL Science, LBNE, and EH&S. These requirements are directly applicable to the design of the Facility and were allocated to the design contractors to drive the design and confirm design compliance with the requirements.

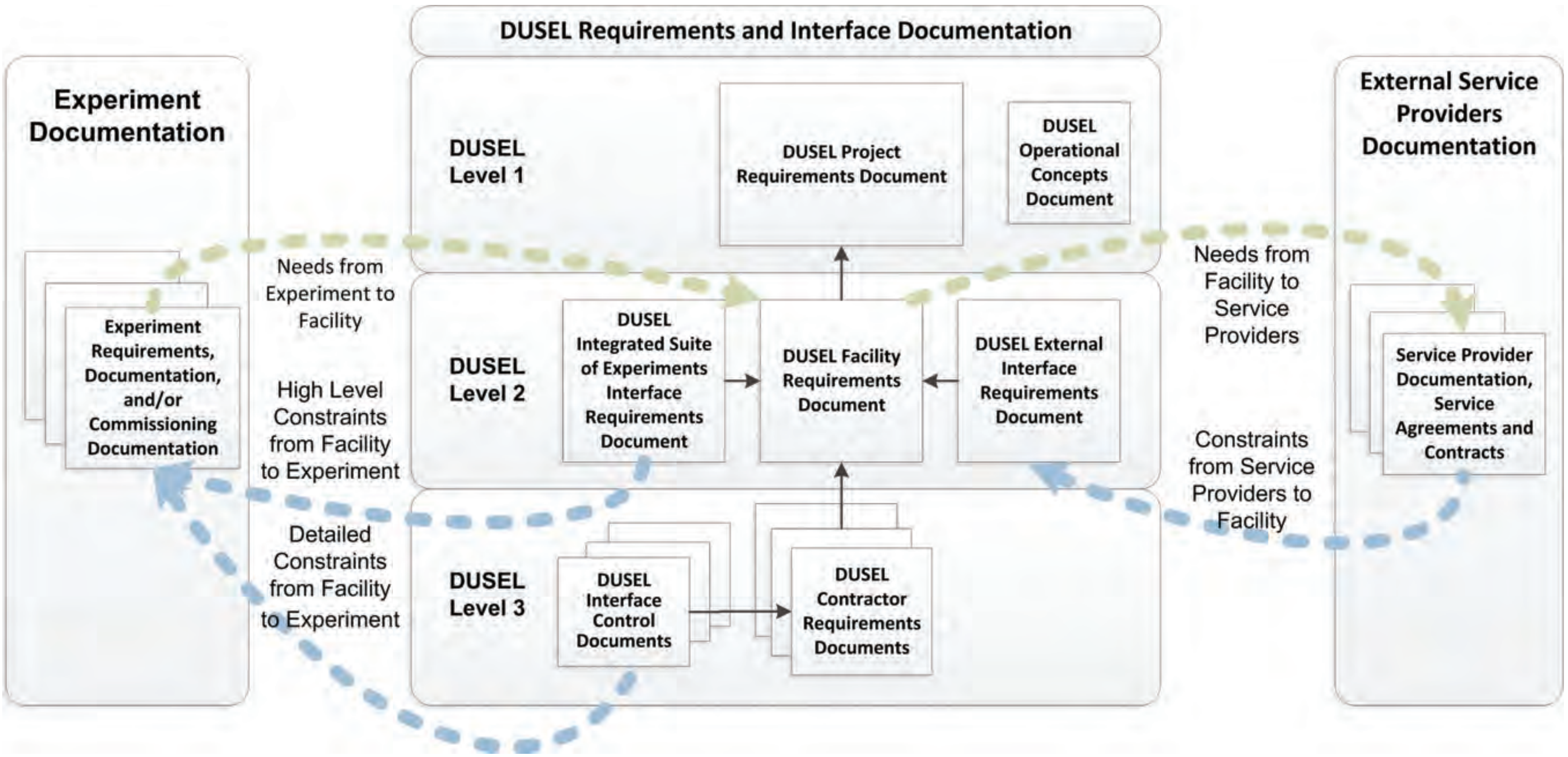

**Figure 9.2.1** Interface and requirements document structure. [Charing Townsend, DUSEL]

#### 9.2.1.1 Requirements Maturity

The requirements developed for the Preliminary Design effort are focused on the 4850L Mid-Level Laboratory (MLL) Campus, EH&S, and E&O. Requirements for the 7400L Deep-Level Laboratory (DLL) Campus and the Surface Campuses (Ross and Yates), the Davis Campus, and the Other Levels and Ramps (OLR) as well as Facility operational requirements will be further refined in the Transition phase to support the Final Design development.

To track the continual process of requirements maturity, each issued requirement document lists all open requirement issues in a summary table, which is commonly referred to as a TBX table. TBX is a generic designation that refers to either a "to be reviewed/refined/revised" (TBR) or a "to be determined" (TBD) requirement. As the Project matures, TBRs and TBDs will be removed as part of the requirements definition, derivation, and refinement process. The SE Manager also tracks forward work that needs to be performed to fill gaps in existing requirements, e.g., determination of reliability requirements, incorporation of construction and experiment assembly phases into requirements, induced environments, and the most effective way to specify codes and standards. All these items will be systematically addressed and closed out to support the start of Final Design. This closeout will be done in a collaborative effort led by Systems Engineering and involving participation of the Facility and Science liaison Project engineers. Systems Engineering will prioritize the TBX items based on a combination of criticality and maturity level of the item. The priorities will be periodically reviewed by the program and reprioritized to ensure efforts are in line with the Project priorities and areas of highest risk. It is typical to have closure of nearly all TBD values at the beginning of the Final Design phase with a small, but necessary, list of TBR values.

Systems Engineering leads the development of the *Operations Concept Document* (Appendix 9.J) to describe how the Facility will be used, i.e., a functional concept definition and rationale from the user's perspective. The document's purpose is to ensure that operational needs are clearly understood and incorporated into the requirements and Facility design.

#### 9.2.1.2 Interface Control

Physical, functional, and organizational interfaces occur at many levels across the DUSEL Project. At the highest level, the DUSEL Facility interfaces with its external environment. The external environment is defined as all things outside the scope of DUSEL and these interfaces are referred to as external constraints. Within the DUSEL Facility, interfaces exist among the various design contractor design scopes, as well as between various elements within a contractor scope. Examples of the latter are interfaces between different utility systems such as between the cyberinfrastructure subsystem and the power subsystem. Each type of interface is handled by the Project in a manner appropriate for the complexity, risk, and organizational aspects of the interface. Systems Engineering focuses on the integration and documentation of the technical aspects of the interface, which include clearly establishing the roles and responsibilities of each side of the interface.

**External Interfaces.** External interface requirements are captured in the *External Interface Requirements Document* (EIRD) (Appendix 9.K). In many cases, the interface definition will first be documented in contracts, inter-Project agreements, memoranda of understanding, and formal variance documentation. The goal of the EIRD is to consolidate those requirements and serve as a point of entry for traceability into the requirements management system. The EIRD is an ongoing development effort that will ramp up during the Transition phase with continued maintenance during the Final Design phase. Systems

Engineering will track variances recommended by design contractors in their design reports. This process ends when the Authority Having Jurisdiction (AHJ) has approved the variance.

**Science-to-Facility Interfaces.** Because the function of the DUSEL Facility is to host science experiments, a defined link between experiment requirements and the Facility requirements is essential. The *Integrated Suite of Experiments Interface Requirements Document* (ISE IRD) (Appendix 9.F) creates a closed-loop environment that establishes interface control and prohibits unilateral requirement changes. By showing traceability from the Facility requirements to the ISE IRD, compliance with the IRD is clearly identified.

**Internal Facility Interfaces.** Because all interfaces are not of equal technical risk or complexity, they do not all require documentation apart from the normal assembly and build-to drawings. Typically, the following cases of interfaces do warrant unique and controlled interface control documentation:

- Interfaces between contractors
- Interfaces with long lead procurements
- Interfaces with complex geometry, critical clearances
- Safety-critical interfaces
- Interfaces between items on the critical path

Interfaces between internal Facility elements are addressed by design contractors collaborating with one another or through communication through DUSEL Technical Representatives, as discussed in Volume 5, *Facility Preliminary Design*. Where applicable, these interfaces are documented in the design reports from each design contractor and reviewed by DUSEL Technical Representatives.

**Interface Control Documents.** As individual experiments are selected, the DUSEL Project will develop IRDs that provide detailed design information on the interface between the experiment and the Facility. These documents will contain drawings, figures, tables, and descriptive text critical to the smooth and successful integration of an experiment with the Facility.

### 9.2.1.3 Requirements Management

The DUSEL Project utilizes the IBM Rational DOORS software system to manage Facility and interface requirements. DOORS is a widely used requirements-management software tool that allows users to maintain requirements in a logical fashion while establishing traceability. Systems Engineering has developed the DUSEL-tailored architecture and schema to support the program from 100% Preliminary Design through Facility commissioning. The *Requirements Management Plan* (Appendix 9.L) establishes the process used in the maintenance of the DUSEL requirements database within the DOORS system.

While DOORS is used to manage the requirements database, the database contents are controlled through the configuration management process in the form of the requirements documents described above. These user-friendly documents support the review and approval of requirements and provide broader access to the information by the stakeholders.

## 9.2.2 Value Engineering Management and Trade Studies

DUSEL Project Value Management, explained in the *Systems Engineering Management Plan* (Appendix 9.A), consists of internal Project Trade Studies and Value Engineering (VE) as part of the design contractor effort. Trade Studies can be initiated anytime during the design process, are not tied to any design milestone, and are usually initiated within the DUSEL Project to evaluate alternatives that can result in a change to requirements and/or Facility designs.

The Facility Technical Representative or subsystem engineer initiates Trade Studies. Requirements for the study and the trade space are identified. Input from the design contractors may be requested to assist in defining the design options and evaluating alternatives. During Preliminary Design, the Facility Project Manager approved and the SE Manager concurred with the final report to formalize the decision. The reports are archived in DocuShare, the Project's document management system.

The Trade Studies completed and documented to support this PDR are listed in Table 9.2.2. The table describes the trade options, whether the design baseline was changed as a result of the trade, and the resolution that was implemented. Changes to the baseline as a result of a Trade Study recommendation were implemented through the VE process or by formal direction to the design contractors.

| **DUSEL Preliminary Design Trade Studies** | | |
|---|---|---|
| **ID / Status** | **Title / Description** | **Resolution** |
| 385 / Closed | **7400L *Emergency Escape, Secondary Egress* (Appendix 9.M)**<br>Trade Study of the 7400L evaluated the condition of the ramp system at the time of the Homestake Mining Company (HMC) closure vs. an alternative escape-way design. | A newly constructed raise bore for secondary egress was chosen over rehabilitating the existing ramp system from the 5000L to the 7400L and resulted in lower technical risk and a $2.75 million savings. |
| 418 / Closed | **31 *Exhaust Raise vs. New Raise Bore* (Appendix 9.N)**<br>Reorientation of the 7400L limited the usefulness of the existing #31 exhaust raise as a ventilation path. Large ductwork may be required to create an exhaust path from the new campus location, increasing excavation sizes and costs. The exhaust raise itself also has several costs that would not be required with a new raise bore, including adding doors at each intersecting level and creating ground support on the 7400L from the #6 Winze to the raise. | A newly constructed raise bore was chosen for ventilation exhaust, providing less Project risk and savings of approximately $12.2 million, as compared with rehabilitating the existing exhaust path. |
| 379 / Closed | ***Waste Rock Disposal* (Appendix 9.O)**<br>Excavated material for the development of the underground laboratory campuses will require a disposal system capable of meeting the excavation schedule and providing a final disposal location for as much as 3 million tons of material. This Trade Study determined the optimal system to transport this material to the Open Cut between two options—pipe conveyor vs. truck haulage via the Kirk Road. | It was accepted that the trucking option introduced significant public concern due to the traffic through residential areas and complications with trucking during inclement weather. The pipe conveyor option was chosen and proved to cost $3.6 million less than trucking and avoided the safety and environmental concerns outlined with the trucking option. |
| 422 / Closed | ***Shaft Pipe Connections* (Appendix 9.P)**<br>Trade Study was conducted of the cost of installation and parts as well as the installation schedule of four options for joining steel pipe ends. The options evaluated were: welded pipe, threaded pipe, flanged pipe, or grooved-end (Victualic) pipe. | Grooved-end joints were chosen based on cost, flexibility, ease of maintenance, and installation time. |
| 2 / Closed | ***Yates Headframe Reinforcement* (Appendix 9.Q)**<br>Both the Yates and Ross Headframes were built in the 1930s using standard engineering design practices for the time. Over the past 80 years, improvements in wire rope strengths and experience from mine accidents have prompted codes requiring headframes to be capable of withstanding "all potential loads." This Trade Study evaluated four options for headframe reinforcement to meet the structural requirements for the DUSEL Project. | An option to use detaching hooks on all conveyances was chosen. This option provides the highest level of safety with the lowest investment and least impact on the historic nature of the site. |

| DUSEL Preliminary Design Trade Studies | | |
|---|---|---|
| **ID / Status** | **Title / Description** | **Resolution** |
| 408 / Closed | ***LMD-1 (7400L Lab Module 1) Size and Configuration Options*** **(Appendix 9.R)**<br>The baseline design for LMD is 15 m wide, 15 m high, by 75 m long. This Trade Study developed the approximate costs for 20% increases in all dimensions. The intent of this study was not to change the baseline design for the PDR but to provide cost differentials for the various options. | The purpose of this Trade Study is to document the estimated cost impact of various size options for LMD. The baseline will remain unchanged for the PDR. The Trade Study report outlines the specific dollar impacts to various dimension changes. |
| 374 / On Hold | ***Evaluation of Alternative Shapes of Large Cavities*** **(Appendix 9.S)**<br>The study evaluated the stability of alternative sizes and shapes, including cylindrical as well as mailbox cavern geometries. The report includes estimates of incremental cost changes associated with the increase in cavern radius and estimates of incremental cost changes per meter of length in the mailbox cavern design. Six variants of large cavern sizes and shapes were evaluated. | This Trade Study titled *Evaluation of Alternative Shape of Large Cavities* was submitted to LBNE and outlines the cost deltas for various configuration changes. |
| 323 / Closed | ***Ventilation and Utilities Drift*** **(Appendix 9.T)**<br>This study examined the redesign of the 4850L emergency ventilation system from a design using ductwork located within the 4850L access drifts to one that replaces it with an overhead 4 m by 4 m exhaust ventilation drift approximately 100 feet above the 4850L. This would allow for reduction in the overall dimensions of the 4850L ventilation drift and a clear separation between areas for personnel and equipment access and exhaust airflow. | The overhead exhaust ventilation drift was chosen and resulted in a cost savings of $760,000, with improved safety. |
| 394 / Closed | ***Purified Water Routing to Experiments other than LBNE*** **(Appendix 9.U)**<br>The LBNE project is planning a water purification system that consists of a Level 1 purification plant at the surface, which does initial purification with stainless steel pipes and depressurization stations to bring water down the Yates Shaft to the Large Cavity at the 4850L to an underground Level 2 purification and recirculation system. Under consideration is whether or not to distribute Level 1 purified water to Lab Modules (LMs) for use by other experiments. | **4850L Experiments**<br>Provide Level 1 purified water to experiments at the 4850L. This provides a definite benefit at a very modest cost and satisfies the requirement that facility infrastructure serve multiple experiments.<br>**7400L Experiments**<br>Experiments at the 7400L requiring purified water will be supplied with industrial water and not purified water. The cost for providing purified water to the 7400L will be much higher with less definite benefit. |
| 380 / Closed | ***Central Utility Plant*** **(Appendix 9.V)**<br>This study examined methods to optimize the underground cooling system configuration for the lowest total cost of ownership. The baseline was a surface chiller installation with a 5,000 gpm closed-loop chilled-water recirculation to the 4850L and 7400L. Other options considered were: 1) separate chillers installed at the surface, 4850L, and 7400L; 2) an ice plant on the surface to reduce the volume flow through the shaft. | The conditions and volumes assumed at the 60% Preliminary Design milestone support heat removal underground. The surface ice plant option was considerably more expensive both in initial investment and ongoing Operations and Maintenance (O&M) cost. An underground plant was chosen as the most economical solution and lowered total cost over a surface installation by over $12 million—a 50% cost reduction. |
| 369 /Closed | ***Crane Configurations in Lab Modules*** **(Appendix 9.W)**<br>The design baseline as represented up to the 60% Preliminary Design reports included a 10 T bridge crane in each LM. The cranes were to be mounted directly to the LM walls. Science desired a higher-capacity bridge crane and also a monorail crane in each LM. This study examined the costs for this change. | Based on knowledge of experiment installations, the suggested baseline configuration would include both a 20 T bridge crane and a 40 T monorail. The cost differential of this change was marginal at $1.5 million, given its positive impact on Science installation. |

| DUSEL Preliminary Design Trade Studies | | |
|---|---|---|
| **ID / Status** | **Title / Description** | **Resolution** |
| 1 / Closed | ***Spacing of Yates Steel Sets*** **(Appendix 9.X)**<br>The original Yates Shaft steel set design was for a 16-foot set spacing (Option A). This Trade Study compared a 20-foot spacing (Option B). The evaluation included engineering constraints, material cost, and construction efficiencies. | Both Option A and B are structurally feasible.<br>The installation cycle time and overall advance productivity is expected to be similar between options. This is due to the increased time taken to remove the timber-based sets currently installed in the Yates Shaft<br>Material cost is approximately 20% lower for Option B. The total savings is approximately $7.3 million.<br>Option B (20-foot spacing) was selected for the 100% Preliminary Design and estimate. |
| 377 / Closed | ***UG Mobile Vehicle Type*** **(Appendix 9.Y)**<br>Underground mobile equipment selection for the 4850L and 7400L Campuses is important to the operation of the facility. Equipment will be needed for underground personnel transport, site servicing, and material handling. This Trade Study compared track and trackless guidance systems. | Due to the significant cost differential between the track and guided trackless systems, a trackless unit-load automatic guided vehicle (AGV) solution was selected. |
| 409 / On Hold | ***Lab Module 1 Modified for DIANA*** **(Appendix 9.Z)**<br>If the nuclear astrophysics accelerator proposal, DIANA, is part of the ISE, it will be installed as the sole experiment in LM-1. It is important that DIANA is sufficiently shielded from other experiments so that neutrons potentially produced are at or below the naturally occurring background. This will require some specialization in the form of an egress maze at both entries. This experiment, if approved, will likely operate through the duration of the laboratory's life. This study examined the cost changes associated with a DIANA-specific LM-1 design. | The purpose of this Trade Study is to document the estimated cost impact of customizing LM-1 for DIANA installation. It was determined that customizing LM-1 to the DIANA experiment will cost $1.73 million less than the baseline LM-1 but the change will be quite challenging from a constructability perspective. The baseline remained unchanged for the PDR. This Trade Study is on hold. |
| 368 / On Hold | ***Lab Module 2 Size and Configuration Options*** **(Appendix 9.AA)**<br>The baseline design for LM-2 is 20 m wide, 24 m high, and 100 m long. This Trade Study developed costs for 20% deviations in all dimensions. The intent was not to change the baseline design for the PDR but to provide cost differentials for the various options. | The purpose of this Trade Study was to document the estimated cost impact of various size options for LM-2. The Trade Study report outlines the various options and associated cost deltas. The baseline remained unchanged for the PDR. |

**Table 9.2.2** DUSEL Preliminary Design Trade Studies.

VE to support the PDR baseline design was performed jointly by the Project Facility, EH&S, Science, and Operations personnel; the design contractors; and the Construction Manager. The following items were evaluated during the VE process in order to align estimates provided at the end of the design phase with the Project budget:

- Change in quality (materials of construction, specification of product, etc.)
- Opening up the specification to alternative suppliers
- A constructability variation
- System alternatives
- Program or functional change
- Scope modification (deletion, shelled areas, postponement)
- Phasing/sequence or schedule variation

Additionally, broader design considerations were evaluated, such as:

- Overall system considerations/alternatives
- Functional area relationships and adjacencies
- Efficiency of space
- Overall level of quality
- General sequencing and phasing
- Changing Project requirements

The VE process was coupled with the Facility cost reconciliation efforts at the end of each interim design completion milestone during Preliminary Design and is detailed in the *Value Engineering Management Plan* (Appendix 9.AB). The VE goal was to fit the construction project within the allocated budget, resulting in a cost-effective design that meets the needs of Science. During the Preliminary Design phase, VE items presented to the Configuration Control Board (CCB) for approval were based on the Project thresholds, as stated in Volume 7, *Project Execution Plan* (PEP), and executed through the Configuration Management System. The list of facility-related VE items that were approved by the DUSEL CCB is described in Appendix 9.AC, *Preliminary Design Value Engineering Items*.

### 9.2.3 Design Compliance to Project Requirements

As part of the overall verification process, the status of the 100% Preliminary Design compliance to the Project-issued requirements is provided by each design contractor in the form of a Design Compliance Matrix. Any differences between the Project-issued requirements and the 100% Preliminary Design will be resolved prior to the start of Final Design.

A Design Compliance Matrix will also be required of each design contractor at every major Final Design milestone. Systems Engineering will track the requirements with which the design does not comply and actively work to resolve the noncompliance with Facility, Science, E&O, EH&S, and other departments as applicable.

### 9.2.4 Continuous Risk Management and Risk Assessment

Identification and management of risks has continued throughout the Preliminary Design phase. The risk-management process is a key management tool designed to identify threats to the DUSEL PEP and proactively mitigate the associated impacts through avoidance, reduction, transfer, or acceptance. Multiple risk workshops were held with the DUSEL Project team and design contractors to identify and evaluate the Project risk exposure, define mitigation actions, and avoid risks via VE and Trade Studies as the design and PEP have matured. The *Risk Management Plan,* Appendix 9.C and described in the PEP, integrates Project risk analysis with the Project controls cost and schedule quantitative analysis process. The qualitative risk analysis maintained within the Project Risk Registry directly supports the quantitative Project controls Monte Carlo cost and schedule analysis implemented via the Primavera Pertmaster tool.

Risk workshops and the inclusion of risk reviews within the monthly Facility Integration Workshops allowed total Project team participation in the risk-management process. This resulted in the identification of 244 risk items tracked within the Risk Registry during Preliminary Design with a summary provided in Appendix 9.AD, *Risk Registry PDR Summary*. The risk items are analyzed and categorized into active risks and watch-list risks. The watch-list risks are identified as either Project issues or actions within the required scope of work that are expected to be resolved through baseline plan execution. Active valid risks are identified as potential future unplanned events that may affect the baseline plan. Valid risks and associated mitigation strategies are actively monitored by the Risk Management Team (RMT). Project issues and requirement in scope-of-work items are placed on watch-list status within the registry. The watch-list items are monitored by the Risk Manager and Risk Owner to ensure that normal Project execution resolves the potential risk. The Risk Manager and Owner are responsible for bringing watch-list items of concern to the RMT, where each item may be recategorized as an active Valid Risk, or Project management may raise the item's priority within normal Project controls. The active risks were ranked and are tracked, and their status is provided at periodic reviews.

The risk-ranking distribution of the active risks is shown in Figure 9.2.4. The single "critical" risk is related to the selection of the LBNE detector configuration design, which has a significant impact on the DUSEL Facility design. The three "high" risks include 1) the need for Research and Related Activities (R&RA) funding in advance of DUSEL construction to support site rehabilitation activities, and 2) geotechnical graphitic shears or faults on the 7400L that can impact 7400L DLL Campus design and construction and 3) the possibility that the rehabilitation of the Ross Shaft may take longer and cost more than planned. All of the risks outlined in Figure 9.2.4 are outlined in the Risk Registry.

Project Safety risks are specific unique events that would cause a threat to health and safety and therefore major disruption to the Project plan. The Project-level risks summarize the detailed environmental, health and safety risks identified and managed by the DUSEL EH&S department. DUSEL EH&S conducts a separate comprehensive risk-based hazard analysis, and identifies avoidance and mitigation strategies in support of the overall effort to reduce Project risk. The assessment of these hazards may provide input to the Project risk-management process if they are deemed Project risks that may increase Project cost, cause Project schedule delays, or reduce the performance of the Project. The EH&S risk-based hazard analysis is discussed in Volume 6, *Integrated Environment, Health, and Safety Management.*

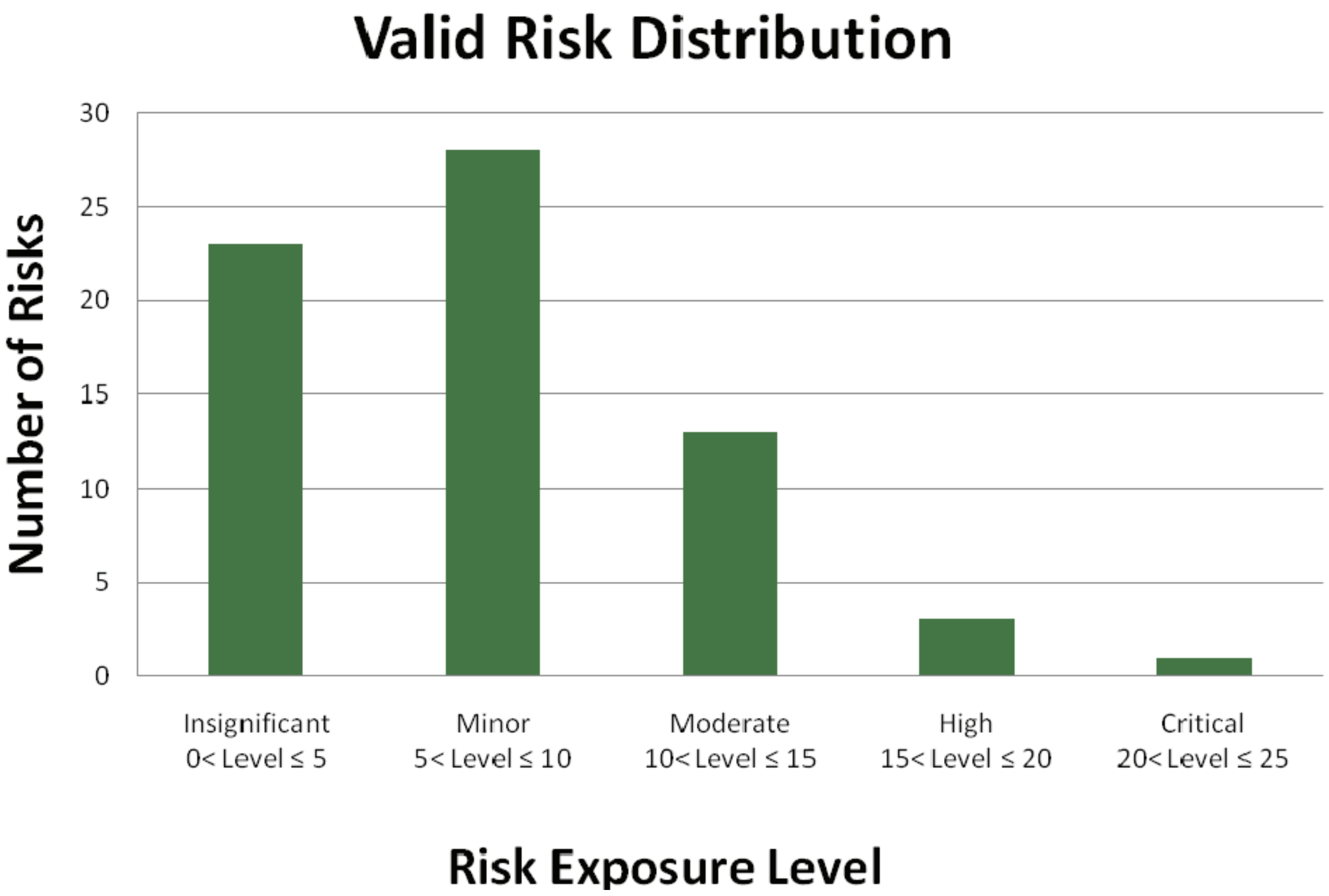

**Figure 9.2.4** Exposure distribution of active risks.

Risk exposures ranked higher than 10 have active mitigation plans that are reviewed and approved by Project management. These moderate, high, and critical risks are discussed in the related technical sections of this PDR. Refinement of these mitigation strategies will continue into the next phase of the Project. The Project Controls team performs a risk analysis and allocates adequate Management Reserve for the identified Project risks. Further refinement of mitigation planning will continue in the next Project phase to ensure management insight into the progress of risk retirement.

### 9.2.5 Configuration Management

The *Configuration Management Plan* (Appendix 9.D) establishes the DUSEL Configuration Management processes. The *Configuration Management Plan* is an extension of the *Systems Engineering Management Plan* and the PEP. The plan describes, in detail, the DUSEL processes to manage and control all Project configuration items. Following the processes laid out in the *Configuration Management Plan* should result in lowering Project risks that may arise from failure to identify, control, complete, and communicate changes effectively. The *Configuration Management Plan* is used in conjunction with other Project management plans and standards to ensure quality management of processes, process changes, and baseline and requirements change tracking.

The DUSEL Configuration Management System requires that changes to Project configuration items follow a formal change-control process specified in the Configuration Management Plan. The process requires a submitter to fill out a Configuration Change Request (CCR). The submitter provides a detailed description of the change and also provides accurate information concerning the technical, cost, and schedule implications. A Project Threshold Matrix has been established to determine the approval level required for a given change. The Project Threshold Matrix is defined in the PEP.

Approved changes enter an implementation phase, followed by a review process, baseline of configuration item(s), and finally closure. The Configuration Manager is responsible for overseeing all aspects of the Configuration Management process and providing technical guidance and assistance to staff when necessary.

The Configuration Management process is managed in DocuShare. DocuShare, the document-management system that provides electronic, Internet-accessible retention and retrieval of DUSEL documents, is administered and maintained by the Document Controls Manager. The Document Controls Manager is responsible for assigning accounts and permissions to authorized personnel. The Configuration Manager works closely with the Document Controls Manager to maintain strict control over configuration items. Tight controls ensure the integrity of the Project baseline.

A Configuration Management Wiki, accessible from DocuShare, allows easy Project access to the *Configuration Management Plan*, CCB thresholds, CCB agendas and minutes, the CCB Charter, DUSEL CCRs, and the Configuration Items List. The Configuration Items List includes all DUSEL-configured items and items planned to be placed under Configuration Control. The list order is based on the Work Breakdown Structure (WBS) of the responsible department. A link to the latest version of each configured item is included in the Configuration Items List to ensure access to the latest version. The Configuration Manager is responsible for implementing Configuration Control to ensure only authorized persons can modify the Configuration Items List after the document has been placed under configuration control.

The peer-review process is defined in the *Configuration Management Plan*. The peer-review process provides a formal process for document review during the implementation phase. It includes a tracking mechanism for reviewers' comments and decisions resulting from each peer review. The peer-review process provides a forum for Project stakeholder input into important documents to:

- Ensure the design is consistent with requirements
- Ensure technical agreement
- Provide consistency across the Project
- Uncover and resolve issues
- Identify and validate dependencies

Systems Engineering oversees the peer-review process. A peer-review area exists in DocuShare solely for the management and disposition of all peer reviews on the Project. As the Project gears up to produce Configuration Controlled documents, the Configuration Management team will provide Configuration Management training to Project personnel. They will also support the ongoing Configuration Management day-to-day operations. The *Configuration Management Plan* and supporting procedures will be updated as needed.

### 9.2.6 System Verification

The Systems Engineering effort includes the coordination of tasks to assure integration of verification activities into the construction process. Systems Engineering will work with the Facility Project Manager, Technical Representatives, Operations personnel, design contractors, the Construction Manager, and an independent commissioning agent to develop a verification plan based on the system requirements and interfaces. Systems Engineering will also assist in the technical acceptance criteria of the Facility.

In the construction industry, it is common for design contractors to specify subsystem functional testing and for a commissioning agent to specify functional and performance system tests. Test, verification and commissioning planning will all begin early in Final Design.

The Facility Verification Plan will incorporate component requirements, subsystem requirements, system requirements, internal interface and external interface verification of facility function, and performance. The Verification Plan will define specific test and commissioning events required to complete the DUSEL verification effort as well as the flow of appropriate progressive functionality tests throughout the construction period. The Verification Plan will be in place to support construction bid packages through commissioning. A Verification Matrix will be developed to track verification. The matrix will list each requirement, the level of integration at which the requirement is verified, and the method of verification. The Verification Plan will include, at a minimum, the following topics:

- **Roles and responsibilities.** Responsibilities of DUSEL Project departments, design contractors, and Construction Manager relating to requirements verification
- **Construction and test plans.** Facility and infrastructure, work authorization, and quality-assurance implementation; these may impart delivery or verification requirements on subsystems or reflect requirements for hardware protection (e.g., contamination control)
- **Construction flow.** Sequence of integration and test processes, including test events and objectives, subsystem component delivery inputs, and review points
- **Verification Matrix.** Table that records and tracks the verification of requirements

Manufacturer-recommended testing protocols will be followed when applicable. Performance tests will be required when installed equipment functions as part of a larger system or subsystem based on performance parameters provided by the design contractors. The independent commissioning agent will develop the commissioning plan.

Verification of the constructed Facility to the requirements will be part of the acceptance criteria for the Facility handover.

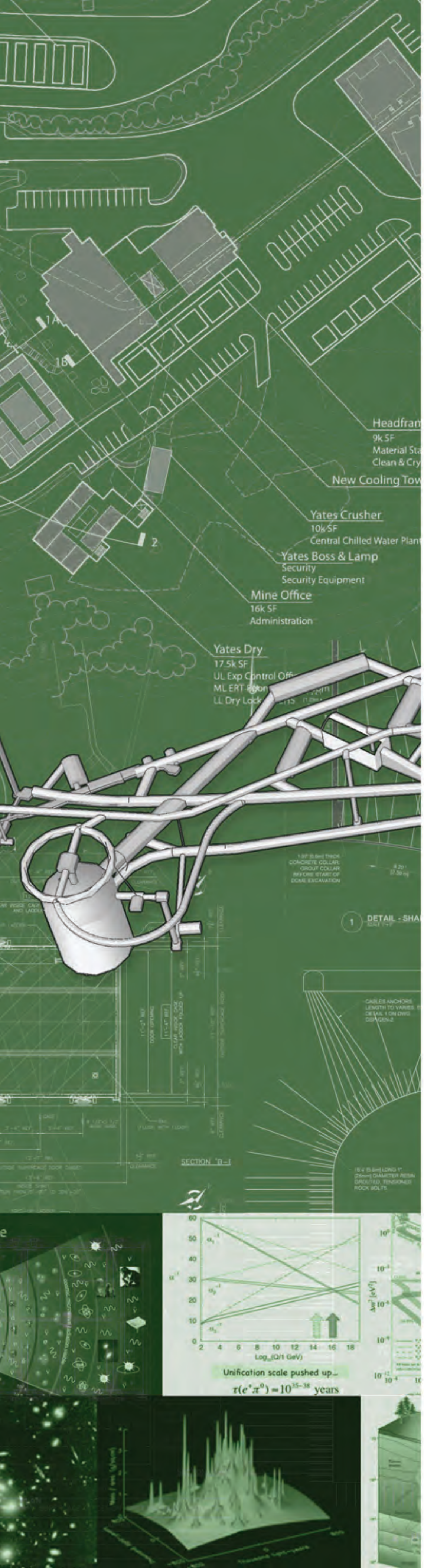

# Preliminary Design Report

May 2011

# Volume 10:
# Operations Plans

DUSEL
Deep Underground Science and Engineering Laboratory

This page intentionally left blank

# Operations Plans

## Volume 10

### 10.0 Introduction

Operations activities at the Sanford Laboratory began in 2006 as the state of South Dakota assumed ownership of the Homestake property from Barrick Gold Corporation (Barrick). The initial site reopening and then commencement of dewatering activities started an evolutionary process for Sanford Laboratory operations under the auspices of the South Dakota Science and Technology Authority (SDSTA) that began with fewer than five staff members in 2004 and has grown to an operational entity of over 100 staff in 2010. The SDSTA staff members cover a range of functions, including facility operations; early science; engineering; environment, health, and safety (EH&S); education and public outreach (E&O); and administration. The SDSTA provides a bridge from Barrick site ownership of the former operating mine to eventual DUSEL Construction, Facility Operations, and science Operations as a new major laboratory facility.

As the SDSTA commenced site operations in 2006, staff members were hired and site improvements initiated through financial support from the state of South Dakota, a generous donation from philanthropist T. Denny Sanford, and a grant from the U.S. Department of Housing and Urban Development. The central focus of the operations activity has been dewatering of the underground facility to counteract the impacts of groundwater inflows. Dewatering efforts successfully lowered the water level from a high of 4,529 feet below the surface in mid-2009 down to 5,331 feet in early 2011. In addition to dewatering, the SDSTA has devoted significant resources to establish and expand safe access to the Facility in support of early science activities on several underground levels. This work has included underground hazard assessment and mitigation activities such as initial shaft rehabilitation in the Ross and Yates Shafts; removal of old utilities; installation of water controls, airflow controls, and ground support; and preparations for early science users of Sanford Laboratory. The SDSTA's efforts have significantly assisted in assessment of the current site conditions to inform the DUSEL Preliminary Design development process from the surface to the 5000L.

In support of early science, the SDSTA reopened the 4850L in support of the Large Underground Xenon (LUX) and MAJORANA DEMONSTRATOR experiments, which included expansion of the 4850L-based Davis Campus. Beyond site rehabilitation and capital improvement activities, the SDSTA has established a well-functioning operations organization that includes a maturing EH&S program managed jointly by the SDSTA and DUSEL. The SDSTA support also includes early education and public outreach efforts and joint efforts with DUSEL to liaise with underrepresented communities of the region.

From 2007, when the National Science Foundation (NSF) selected Homestake to be the future DUSEL site, the DUSEL Project through NSF sponsorship has focused on development of design concepts and then Preliminary Designs to support an eventual Major Research Equipment and Facilities Construction (MREFC)-funded construction effort. The DUSEL Project is led by the University of California at Berkeley (UC Berkeley) and includes two subawardees: the South Dakota School of Mines and Technology (SDSM&T) and the Lawrence Berkeley National Laboratory (LBNL). Black Hills State

University (BHSU) supports the DUSEL Project through a subaward from SDSM&T, providing leadership for the development of education and public outreach programs.

In parallel with the SDSTA site operations, the DUSEL Project focused on Facility design development in support of this Preliminary Design Report (PDR), and science program definition and advancement, including development of experiment concept designs and requirements definition. DUSEL has also sponsored education and public outreach program development to support design efforts for a new Sanford Center for Science Education (SCSE). The DUSEL Project has worked in partnership with the SDSTA to develop a comprehensive EH&S program, business services, Project Controls, Systems Engineering, and quality assurance to support the Final Design and Construction phases. During the Preliminary Design development, the DUSEL Project formulated initial plans for site operations. With DUSEL-funded site operations starting in 2011, construction starting in 2014, and post-construction Operations starting in 2022, the DUSEL Operations Plans address a wide range of functions over this time span. The Operations Plans include:

- Current SDSTA activities
- Final Design activities, operations activities in support of Final Design, and preparation for Construction
- Operations support during construction
- Steady-state operations and maintenance after Construction is complete

In 2011 during the DUSEL Transition phase, the DUSEL Project through NSF funding is scheduled to assume funding responsibilities for Sanford Laboratory operations. As DUSEL accepts leadership responsibilities for Facility maintenance and operations, the SDSTA's role will transition from leading day-to-day site operations lead to becoming long-term site owner. The SDSTA will represent the state of South Dakota in future laboratory operations activities and maintain the long-term relationship with Barrick as required by the Property Donation Agreement (PDA) between the state of South Dakota and Barrick.

In advance of a Final Design start in early 2012, the SDSTA and DUSEL Project collaborators will form a single legal entity through a limited liability company (LLC) to be named the DUSEL LLC. This DUSEL LLC will be led by UC Berkeley with direct involvement from the state of South Dakota on the DUSEL LLC Board of Directors. The DUSEL LLC will oversee and manage daily site operations and also serve as the lead entity for the management of DUSEL Construction activities through a contract with a construction management firm.

In preparation of the PDR, the DUSEL Project developed a detailed operations budget plan, including staffing to delineate the resource requirements to support operations activities for construction and steady-state operations. Figure 10 outlines this staffing profile and depicts a newly combined SDSTA and DUSEL team under the DUSEL LLC starting in FY 2011as assumed in this PDR.

In advance of a DUSEL Construction start in 2014, Facility deferred-maintenance activities must be addressed to support safe access to the site and reduce risk. These items are also essential prerequisites to a MREFC-funded Construction start in 2014; the DUSEL Project planned cost and schedule baseline depend on the accomplishment of these activities prior to the start of construction. Operations activities that must occur in advance of a construction start include the rehabilitation of the Ross Shaft infrastructure, maintenance of the waste rock handling system, and rehabilitation of underground ventilation systems—all benefit from a start to rehabilitation activities in FY 2012. Required replacement of the Yates Shaft and #6 Winze infrastructure, both deferred-maintenance items, is planned to be addressed in parallel with construction and require R&RA funding as well to support the Construction schedule timeline. The specific R&RA funding levels required are essential to maintaining the Construction schedule and are outlined in Volume 2, *Cost, Schedule, and Staffing*. This issue has been documented as a "high" risk in the DUSEL Project Risk Register (Appendix 9.AD). Further technical information on these required deferred-maintenance items is provided in Volume 5, *Facility Preliminary Design*.

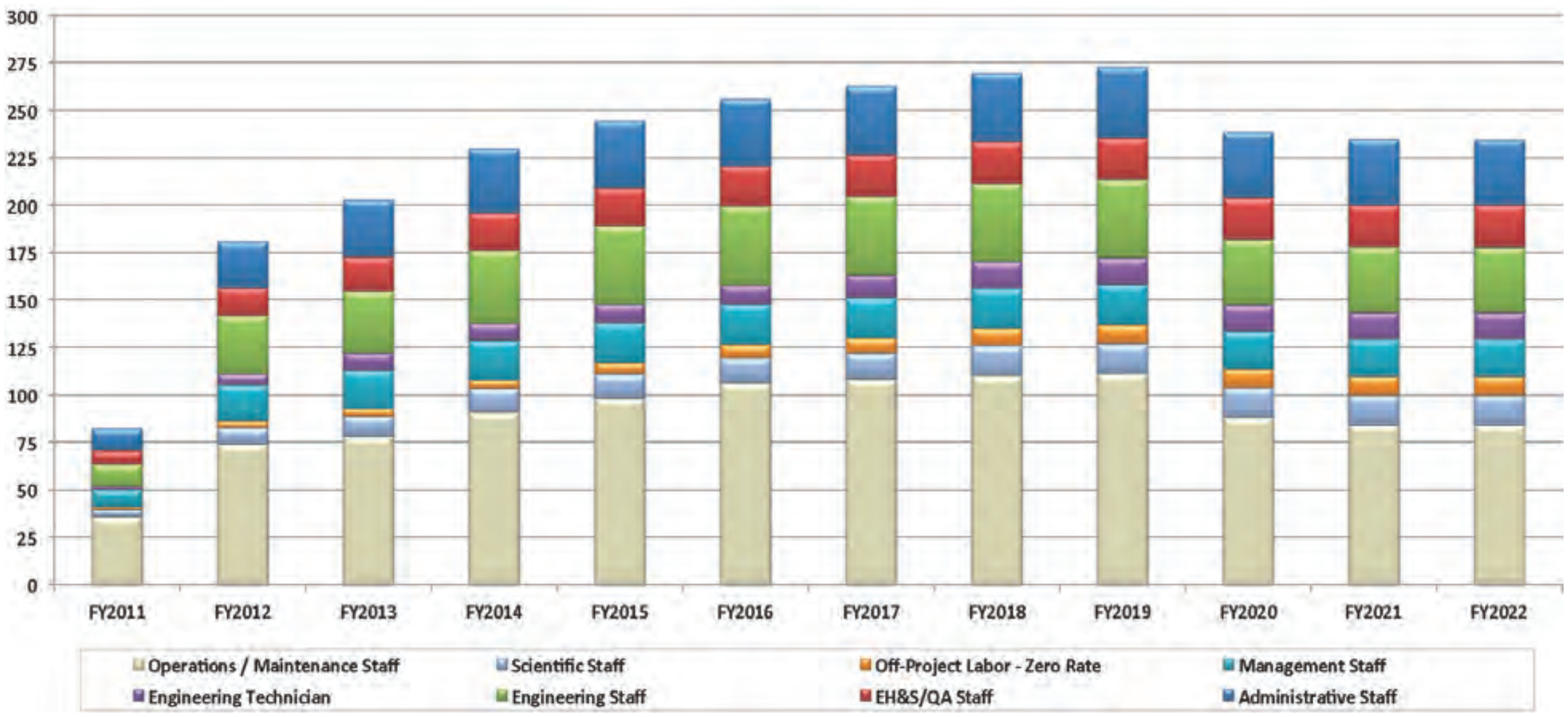

**Figure 10.0** Planned DUSEL operations staffing (FY 2011 represents six months' staffing to correspond with the planned start of the Transition phase in April 2011).

This volume provides a description of the operations activities outlined above, including an overview of the major functions and the expected change for each as the Project transitions from the current state through Final Design, Construction, and on to Steady-State Operations. While design and construction activities are a central focus during the preparation and execution of the Construction phase, Operations activities continue in parallel with them and play a major role in the development and maintenance of the site and in the execution of a successful DUSEL science program.

The DUSEL management structure is primarily addressed in Volume 7, *Project Execution Plan*, and an overview of the planned organization to support major operations and maintenance activities is outlined in this volume. Each chapter provides an overview of operations during each phase of the Project (current conditions, Final Design, Construction, and Steady-State Operations), including the management structure of the Project, primary departmental activities and approximate level of effort, major operations and maintenance activities to support safe access, rehabilitation and construction, science program development, and education and public outreach activities.

## 10.1 Current Operations

The South Dakota Legislature created the SDSTA in 2004 to foster science and technology in the state. Barrick Gold Corporation donated the Homestake property to the SDSTA in 2006. The SDSTA owns the Homestake campus, which includes 186 acres on the surface and 7,700 acres underground. In 2007, NSF concluded its comprehensive site selection process, which considered multiple sites and site attributes and selected Homestake as the preferred site for a deep underground science and engineering laboratory. The NSF solicitation process is described in Volume 1, *Project Overview*. The SDSTA's mission, funded by state and private sources, is to convert the former Homestake Mining Company's gold mine into an underground laboratory, now known as the Sanford Laboratory at Homestake. The underground campus is currently under construction at the 4850L by the SDSTA.

In 2007, Homestake was selected as the future DUSEL site with a collaboration led by UC Berkeley that included LBNL and the SDSM&T. UC Berkeley was awarded a cooperative agreement with NSF to advance design concepts and develop the Preliminary Design for a DUSEL Facility at Homestake. The DUSEL team is also leading the coordination with the Integrated Suite of Experiments (ISE) for the advancement of the scientific program for DUSEL. Together, these institutions, the SDSTA and collective DUSEL collaboration members, are working as a unified team to develop the full scope for DUSEL.

### 10.1.1 Organizational Management and Staffing Approach

The SDSTA and DUSEL staff will merge during 2011 into a single entity, which will be formed as a limited liability company—called the DUSEL LLC. Current organizational management structures for the SDSTA and DUSEL are described in the sections below. The DUSEL LLC will combine the efforts into a unified organization in anticipation of the Final Design phase. Although the two currently exist as separate legal entities, the overall SDSTA/DUSEL organization is represented as participating in a single, integrated project.

#### 10.1.1.1 SDSTA/Sanford Lab Organizational Management

The SDSTA created the organizational management configuration for Sanford Laboratory. The SDSTA Executive Director reports to the seven-member SDSTA Board of Directors. The Governor of South Dakota appoints board members to meet quarterly to approve annual budgets for Sanford Laboratory. The South Dakota Legislature’s Joint Appropriations Committee also reviews these budgets. The South Dakota Legislature appropriated state funds to reopen the former mine site and operate the Sanford Laboratory in preparation for DUSEL.

The SDSTA's annual budget includes funding for infrastructure maintenance, rehabilitation efforts to provide safe access to the Facility, and capital projects, including facilities to support early science experiments. A Configuration Control Board (CCB) that includes members of the SDSTA Board, SDSTA staff, and the DUSEL Project must approve capital budget changes.

The Internal Executive Committee of the SDSTA reviews major policies such as underground access. The committee includes the SDSTA’s Executive Director, Chief Financial Officer, Director of Engineering, and Operations Director. The SDSTA is organized by departments, including Operations (divided into Surface and Underground Operations), Administration, Engineering, Science, Communications, Environment Health and Safety (EH&S), and Education and Outreach (E&O). SDSTA Administration, directed by a Chief Financial Officer, includes accounting, procurement, contract management, information technology, and administrative services. Other departments include Engineering, Science Liaison, and Communications, which report to the Executive Director. Two departments at Sanford Laboratory (EH&S and E&O) were merged early with the DUSEL functions through UC Berkeley. An organizational chart for the full DUSEL Project, including SDSTA staff, is included in Appendix 1.A for reference.

#### 10.1.1.2 DUSEL Organizational Management

The DUSEL organizational management configuration reflects that of large complex scientific facility projects—e.g., the Spallation Neutron Source, the Linear Coherent Light Source, and the Facility for Rare Isotope Beams—and is characteristic of scientific projects of this size and complexity where multiple institutions are involved. UC Berkeley leads the institutional collaboration and is accountable to NSF for the design and development of DUSEL. Two other institutions are currently under direct subaward from UC Berkeley: SDSM&T, with primary responsibility for the DUSEL Facility infrastructure design; and LBNL, with primarily responsibility in development of the DUSEL science program. In support of DUSEL education and public outreach programs, Black Hills State University (BHSU) contributes to DUSEL's development through an SDSM&T subaward.

A Project organization has been developed that focuses on the major responsibilities and work scope of the DUSEL/Sanford Laboratory Project and emphasizes clear roles and responsibilities, controls, and management. The DUSEL Project efforts are organized along the Project's Work Breakdown Structure (WBS) and are focused on advancing the Project deliverables for design, safety, hardware, and access. For additional discussion on the WBS, see Volume 7, *Project Execution Plan*.

Within UC Berkeley, the DUSEL Project reports to the Office of the Vice Chancellor for Research. The Senior Project Management consists of the Principal Investigator/Executive Director, Co-Principal Investigator, Associate Director, and Project Director.

Reporting directly to the DUSEL Senior Management are all crosscutting Project and operations systems: Business Systems, Project Controls, Information Technology, EH&S, Systems Engineering, and E&O.

The remainder of the Project is divided into the major functions: the Facility design and construction and the scientific program and experimental integration.

In addition to line oversight of the Project, the UC Berkeley Vice Chancellor for Research has three oversight committees that provide EH&S oversight, project oversight and internal review, and program and scientific advice.

#### 10.1.1.3 Staffing Description

The SDSTA, as of Dec. 31, 2010, employs 102 full-time and 20 part-time personnel at Sanford Laboratory, headed by the SDSTA Executive Director, who reports to the SDSTA Board of Directors. The DUSEL Project, headed by UC Berkeley, employs 55 full-time staff for a total project effort of 157 full-time and 20 part-time personnel.

### 10.1.2 Project Operations

Several departments provide day-to-day operational functions for the advancement of DUSEL through SDSTA, UC Berkeley, LBNL, SDSM&T, and BHSU. These institutions all provide staffing to the various departments to support the unified DUSEL team. Additionally, SDSM&T and SDSTA share skilled staff on a reimbursable basis through a cooperative agreement. The skill sets shared include science liaison support, information technology, geology, environmental engineering, cultural liaison, project engineering and management, and administration. The following section outlines the departments that make up the SDSTA/DUSEL team.

#### 10.1.2.1 Business Services

Business services at SDSTA are directed by the chief financial officer and include contract administration, payroll, human resources, and procurement.

**Insurance**
Included in Business Services is risk financing to ensure that proper insurance coverages are in place to manage risk. The SDSTA insurance coverage includes general liability, environmental pollution, workers' compensation, auto, and directors' and officers' liability. In addition, the South Dakota Legislature set up a $10 million indemnification fund to protect Barrick Gold Corporation from liability for the SDSTA and a $1 million closure fund. Professional liability required for design activities in advance of MREFC funding has been secured through the designers for activities prior to Final Design.

**Contract Management and Procurement**
Contract Management and Procurement services during the initial Project start-up and Preliminary Design phases of DUSEL are provided through three separate entities, depending on the funding source and the responsible institution: UC Berkeley, SDSM&T, and SDSTA. Each of the contract and procurement processes has been coordinated to ensure that all insurance and procurement requirements have been met for all three entities. These functions are currently performed by two FTEs and an outsourced consulting agreement to support major contract modifications and new major contract procurements.

**Intellectual Property, Technology Transfer, and Copyright Policy**
For the experiment research and development activities taking place as a part of the DUSEL Project, the intellectual property, technology transfer, and copyright policies will follow the policies from the users' sponsoring institution on patents, copyrights, trademarks, and tangible research results. These are commonly referred to as intellectual property, as they relate to sponsored research agreements. In general, these policies will reflect the academic policies of the Project participants and the sponsoring funding agencies. The DUSEL LLC, when established, will consider and generate policy acceptable to the DUSEL stakeholders.

**Support Services**
Support Services is primarily an administrative function that includes staff to provide administrative support to the full Project team. Functions included in this group are the information management system (DocuShare system), administrative support services, and workshop and project review coordination. These functions are currently performed by approximately five FTEs.

**Human Resources**

Human Resources (HR) services are currently provided by each of the hiring institutions including UC Berkeley, LBNL, SDSM&T, BHSU, and SDSTA. Human Resources is responsible for the recruitment of all staff employees working for DUSEL as well as the coordination of all benefit plans. Working closely with senior management, they help develop job descriptions, post open positions, and assist in the hiring process. HR responsibilities at SDSTA are currently performed by one FTE. Human Resources from the other institutions is a shared resource with each institution in all cases and is not reflected in the Project staffing levels presented in this PDR.

**Communications**

The SDSTA communications department, at two FTEs, includes a communications director and a multimedia specialist. The communications director oversees media relations, including the distribution of press releases and other materials to state, regional, and national media. The communications department produces written materials, including a weekly lab newsletter, and provides content for the monthly DUSEL newsletter. The department also manages content of the Sanford Laboratory Web site, www.sanfordlab.org, and coordinates outreach to the general public, including the *Deep Science for Everyone* lecture series and the annual Neutrino Day science festival at the Sanford Laboratory.

#### 10.1.2.2 Information Technology

The Information Technology (IT) department is responsible for all centrally managed computer and network technologies and currently consists of four FTEs. These four FTEs are funded jointly by the SDSTA and the DUSEL Project. IT oversees the maintenance of nearly 250 desktop and laptop computers and servers. IT also manages an advanced gigabit fiber network to support both surface and underground networking requirements, including facility management and early science. This fiber network provides connections to commercial Internet providers and the Internet2 research network. The department hosts software applications to support Project Operations, including e-mail and typical office applications, project scheduling, and Earned Value Management Systems, document management, and technical requirements management.

#### 10.1.2.3 Finance

As the contracts are held by various Project partners, the finances are tracked by each institution. Invoices are paid by the institution that holds the applicable agreement or contract. Contract changes requiring allocation of Management Reserve are controlled through the CCB. The SDSTA finance department functions are performed by four FTEs, funded by the SDSTA. Finance functions in this phase are provided by each sponsoring institution and are not reflected in the staff count in the PDR until later Project phases.

#### 10.1.2.4 Project Controls

The DUSEL Project Controls department is a centralized team of four FTEs and is responsible for cost estimate management and scheduling of the DUSEL Project, including tracking all Project activities against the baseline budget and schedule. Project Controls generates reports on Project status to support management in tracking Project progress. They also work with the finance team to confirm the Management Reserve and all budget line items are in order.

#### 10.1.2.5 Operations and Engineering

The SDSTA Operations Department includes 77 FTEs, and Operations personnel are on duty 24 hours a day, seven days a week. They have provided ongoing maintenance for the entire facility, from surface grounds and buildings to underground facilities and infrastructure, and they have also refurbished existing infrastructure and constructed new facility (surface and underground). The SDSTA Operations staff manages a system of inspections and preventive maintenance for all Sanford Laboratory equipment, including:

- Shafts and hoists
- Electrical and fiber-optic cables
- Ventilation fans and water-management systems
- Heavy equipment used on the surface and underground
- Dewatering systems, including pumps and pump columns underground to the Waste Water Treatment Plant (WWTP) on the surface

Within the SDSTA, Operations staff also excavated new underground facility for early science on the 4850L. Their work to create safe access to the underground also will assist DUSEL staff and design contractors to better mitigate risk for DUSEL Construction.

The SDSTA Operations staff includes 31 infrastructure technicians and 13 facility technicians. Infrastructure technicians construct and maintain underground structures and systems, including maintaining and operating the Ross and Yates Shafts and Hoists and the movement of materials in and out of the underground facility. Facility technicians maintain equipment, mechanical systems, heating ventilation and air conditioning (HVAC) systems, electrical systems, plumbing, and surface infrastructure.

The current Operations staff also includes nine hoist operators, three industrial electricians, four state-certified WWTP operators, one rope technician, three technical support leads, one property maintenance technician, and two security guards.

The SDSTA Engineering Department, with four FTEs, provides engineering support to daily operations, including design and implementation of Facility capital projects, and includes an engineering director, a mechanical project engineer, an engineering technician, and a ventilation technician. Department engineers assist in the dewatering of the underground facility and in the design and operation of the WWTP and related systems. Engineers support the inspection, design, and repairs of the Ross and Yates Shafts. They also provide project management and engineering services for the excavation and outfitting of the Davis Campus on the 4850L in support of early science activities.

#### 10.1.2.6 Facility

Facility staffing levels are currently at 11 FTEs. The Facility team is responsible for the development of the DUSEL Facility design, including oversight and management of the major architecture and engineering design scopes and construction management services in support of the Preliminary Design development. The Facility team provides engineering leadership to each of the design development contract scopes to integrate the full scope of the DUSEL Facility design. Staff composition includes project management and engineering support such as civil/structural, electrical, geotechnical, mechanical systems, hydrology, and surface buildings and infrastructure systems.

#### 10.1.2.7 Systems Engineering

A Systems Engineering (SE) department of three FTEs is finalizing and implementing plans and processes that will be the basis of the program in the areas of risk and configuration management. The SE team is responsible for requirements and interface documentation to communicate the Project's needs. The SE team coordinates the Value Engineering (VE) process and assists with design Trade Studies on an as-needed basis.

#### 10.1.2.8 Environment, Health, and Safety

To safely rehabilitate the site and to design, construct, and operate Sanford Laboratory in support of early science development, initial Integrated Safety Management (ISM) and Environment, Health, and Safety (EH&S) programs have been developed. The DUSEL EH&S program is described in Volume 6, *Integrated Environment, Health, and Safety Management*.

The EH&S department includes nine FTEs who develop and implement the ISM system and EH&S programs necessary for the DUSEL Project at Sanford Laboratory. Within this organization, the EH&S Director is supported by an administrative assistant and three managers to oversee and guide staff and activities. The management positions consist of: an Environmental Manager to monitor compliance with applicable Environmental Protection Agency (EPA) regulations; a Safety and Health Manager to monitor compliance with all requirements, codes, and standards (Occupational Safety and Health Administration [OSHA], National Fire Protection Association [NFPA], International Building Code/International Fire Code [IBC/IFC], and others as applicable) and support the general safety, training, and health of operations, maintenance, construction, and science collaboration workers; and an Experimental Health and Safety Manager to support early science at Sanford Laboratory and DUSEL science projects through provision of expertise in laboratory and experimental safety.

Professional EH&S staff consist of an operations safety officer, a site safety specialist, an electrical safety engineer, a construction safety specialist, an industrial hygienist, and a technical assistant.

EH&S also sponsors a 30-person Emergency Response Team (ERT) drawn from the SDSTA staff and from regional emergency-response organizations. The EH&S department coordinates the ERT activities, including training and management of exercises and actual incidents.

Full-time EH&S staff are supplemented by third-party subject matter experts when necessary for specific areas of expertise (e.g., oxygen deficiency hazards, cryogen safety, risk assessment), with consultants drawn from sponsoring organizations or external consulting firms.

The primary EH&S functions supported and maintained include:

- Identification and risk-based ranking of hazards presented by operations and science activities for mitigation, and identification codes and standards necessary to maintain the residual risk
- Development and implementation of an EH&S Manual containing policies and procedures (P&P) that provide effective control mechanisms for identified hazards. This includes management of EH&S processes necessary for an effective ISM system, including safety committees and panels, self-assessment and inspection programs, and coordination of incident investigation, trending, and Lessons Learned processes. Also included is establishment of an EH&S training program.

- Initiation of a Facility-wide environmental assessment to be incorporated into the Environmental Impact Statement (EIS) that will be completed during the Final Design process through Argonne National Laboratory (ANL) (discussed in Chapter 10.2, *Operation Plans during Final Design*)
- Maintenance of permits, licenses, etc., in accordance with local, state, and federal regulations

#### 10.1.2.9 Quality Assurance

Quality assurance (QA) is the systematic monitoring and evaluation of various aspects of a project, service, or facility performed with the intention of maximizing the probability that established standards of quality will be attained throughout the Project's life cycle. DUSEL will establish, document, implement, and maintain a quality management system and continually improve its effectiveness for the duration of the Project. The quality management system and the supporting assurance requirements will be developed in accordance with the following standards and guidelines:

- DOE 414.1C—Quality Assurance
- ISO 9001—Quality Management Systems
- ISO 14001—Environmental Management Systems
- OHSAS 18001—Occupational Health & Safety Management System

The QA program is provided through two FTEs. The QA Program Manager is responsible for the development of a QA program and for the implementation activities that support and enhance the quality management system and the QA goals and objectives. A QA management review process will be implemented where senior management will review the organization's quality management system at regular intervals to ensure its continuing suitability, adequacy, and effectiveness. These reviews will include assessment opportunities for improvement and will identify the need for changes to the quality management system. Two QA management reviews are planned prior to the start of Final Design. A quality awareness training program and a corrective / preventative action program will also be implemented in advance of Final Design.

#### 10.1.2.10 Science

In support of SDSTA's early science program, three FTEs include a science liaison director who supervises two laboratory supervisors. In addition to the FTEs described above, two temporary science liaison specialists work with scientists conducting early science experiments at Sanford Laboratory.

The DUSEL team consists of senior and term scientists, and mechanical engineers. The DUSEL science and engineering staff consist of approximately seven FTEs.

#### 10.1.2.11 Education and Public Outreach

Education and Outreach (E&O) activities are accomplished through a partnership of SDSTA, DUSEL, Black Hills State University (BHSU), the South Dakota Department of Education, and volunteer scientists. Core staff members for E&O (four FTEs) include an Education and Outreach Director, Deputy Director of Education and Outreach, Cultural Coordinator, and Science Education Specialist. E&O programs rely on expertise from staff across the SDSTA and DUSEL Project.

### 10.1.3 Preliminary Design Activities to Prepare for Final Design

During current Operations, the Project team focused on activities to complete the Preliminary Design Report and prepare for both the Transition period activities and to position the Project to be ready for Final Design. To support these activities, staff members were hired to support increased design activities. The Transition phase activities during 2011 and early 2012 focus on site assessment and design work required to adjust design scope for reduced funding and schedule scenarios; support an efficient Final Design start with current requirements; advance understanding of experiment requirements; and maintain integrated, controlled requirements and design baseline.

The major scope elements that will be addressed during this Transition phase include:

- 4850L geotechnical site investigations for Final Design
- Phase 3 surface site assessment and advancing the surface design to complete 30% Design
- ISE and Facilities requirements refinement as ISE designs advance
- Targeted technical support from designers and construction manager to retain key staff, maintain design integration, and support efficient Final Design start

### 10.1.4 Maintenance to Support Safe Access and Facility Rehabilitation

Since opening Sanford Laboratory in 2007, the SDSTA has performed initial rehabilitation of the Ross and Yates Shafts, shown in Figure 10.1.4, to provide two means of safe access and egress to underground levels in support of design development and early science. Initial work on the Ross Shaft was completed to 5,000 feet underground and the initial work on the Yates Shaft was completed to 4,850 feet underground. SDSTA personnel and contractors conducted level-by-level inspections to identify and mitigate hazards.

SDSTA has constructed a campus for an interim laboratory at the 4850L, with facilities at two locations: the Davis Campus includes an expanded Davis Laboratory Module (DLM) that is being outfitted for installation of the LUX dark-matter experiment and a new Davis Transition Area (DTA), which provides space for both the MAJORANA DEMONSTRATOR and LUX experiments. The SDSTA constructed a temporary clean room for a copper electroforming facility to support the MAJORANA DEMONSTRATOR. SDSTA also created support infrastructure for about 20 smaller experiments at the 300L, 800L, 2000L, 4100L, and 4850L. Additional information on the current conditions and rehabilitation efforts for the infrastructure systems completed to date are described in Chapter 5.4, *Underground Infrastructure Design*. Current conditions of the surface facilities are described in Chapter 5.2, *Surface Facility and Infrastructure*.

In addition, SDSTA installed a system of pumps and pipe columns to dewater the underground facility. The system lifts water from 6,800 feet underground to a re-engineered WWTP (see Chapter 5.2, *Surface Facility and Infrastructure*, for a description of the system), which discharges in compliance with a permit from the South Dakota Department of Environment and Natural Resources. Since starting dewatering in August 2008 when the water level had reached 4,529 feet below the surface, the pumping system has dewatered the Facility to 5,331 feet underground as of January 1, 2011, and provided access for design and assessment to all levels down to and including the 5000L.

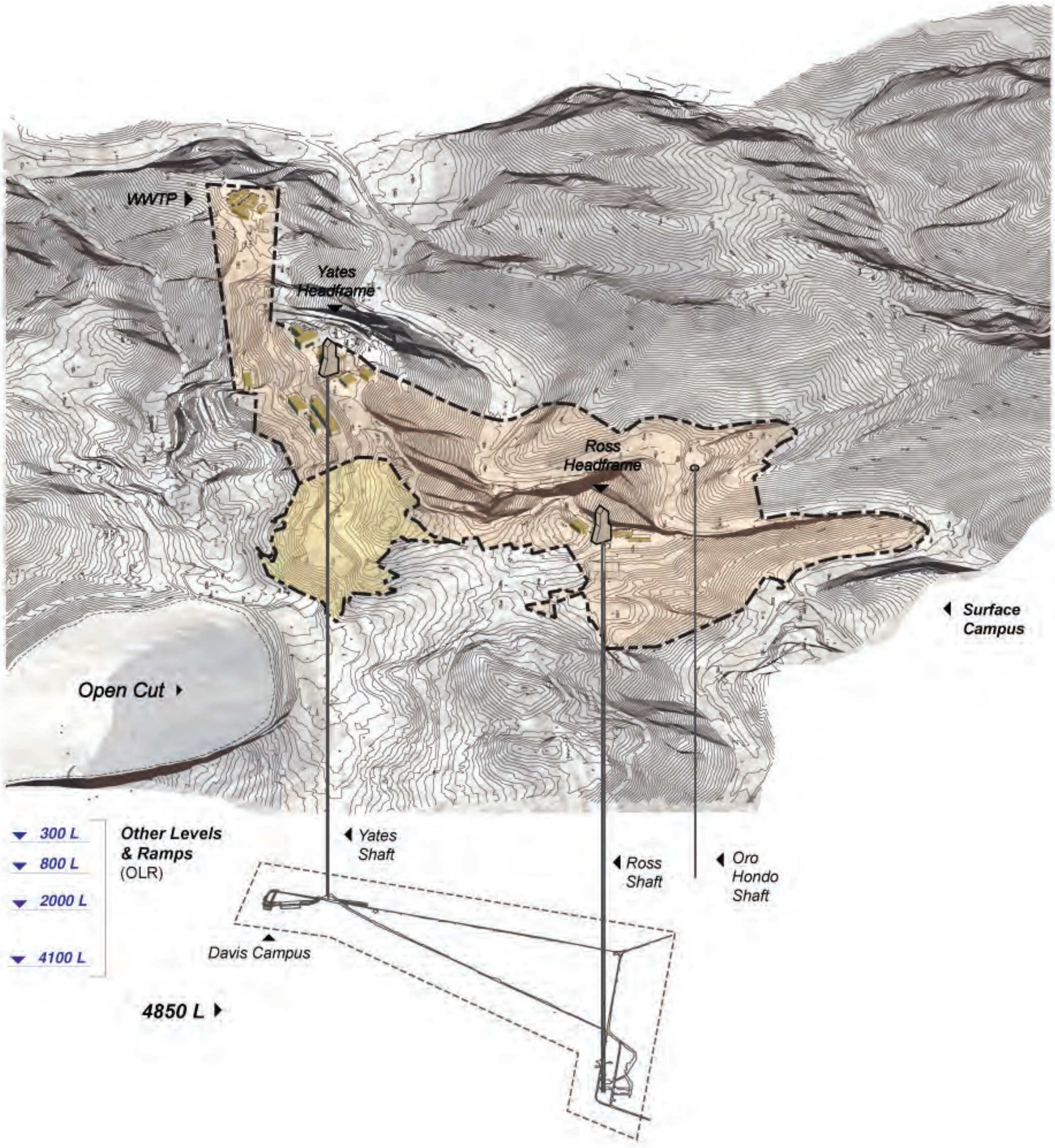

**Figure 10.1.4** Current design at Sanford Laboratory. Area shown in yellow is a potential future expansion of the SDSTA property. [DKA]

## 10.1.5 Operations and Maintenance

On the 186-acre Sanford Laboratory Surface Campus, Operations crews and contractors have remodeled a number of existing buildings for use as offices, shops, warehouses, an education facility, and early science, as well as removed hazardous materials, and prepared for the DUSEL Project.

The Yates Administration Building was remodeled to accommodate working space for approximately 60 staff, and a number of conference rooms.

The Yates Education Building was remodeled by SDSTA in late 2010 and has six offices, a small classroom, a large conference room for up to 100 occupants, and storage space.

The Ross Dry Building was remodeled in 2010; the old dry facilities to support former mining operations were removed and SDSTA installed lockers and showers for 70 operations staff, offices for six supervisors, a room for storage of personal protective equipment (PPE), and a safety training room where members of the ERT train and perform maintenance on specialized rescue equipment.

The former warehouse building, located on the Yates Surface Campus (see also Chapter 5.2) was remodeled and outfitted as a laboratory, including a clean room, and is currently used for early science by the LUX dark-matter experiment. The primary parking area for staff and scientists is located on the Yates Campus, and during 2010, SDSTA resurfaced the parking area to meet near-term needs and can now accommodate 110 vehicles.

#### 10.1.5.1 Dewatering System

Initially in 2008, a temporary system of small submersible pumps, working in concert with refurbished 700-horsepower pumps on upper levels and a re-engineered WWTP on the surface, began to lower the water level as described above.

That early dewatering system has since been upgraded to include large submersible pumps to continue the dewatering process to the 7700L to support construction of a skip loading pocket at the 7500L with additional space to capture large rainfall events and to control future inflows. The system, described more completely in Chapter 5.4, *Underground Infrastructure Design*, begins with a submersible pump 7,800 feet underground in the #6 Winze. The pump uses two 750-horsepower motors to lift up to 2,000 gallons per minute (gpm) to a sump located on the 5000L. From there, a series of 750-horsepower pumps lift water up the Ross Shaft and reverse cascade to sumps at the 3650L, the 2450L, the 1250L, and from there to a large surge tank on the surface called the Mill Reservoir. The Mill Reservoir has two chambers: Water from underground is stored in one chamber and, under a long-term agreement with Barrick, water from Grizzly Gulch tailings impoundment is fed to the other chamber. Water from both chambers of the Mill Reservoir is pumped or gravity fed to the WWTP, where the groundwater is processed through a series of sand filters and fabric bags. Then water from Grizzly Gulch is mixed with the filtered groundwater and directed through a series of rotating biological contactors that remove trace amounts of ammonia before the water is directed through a clarifying tank and a final set of polishing sand filters. Blending groundwater with water from Grizzly Gulch is beneficial for cooling the warm water from underground and to reduce the concentration of total dissolved solids (TDS). The WWTP also removes ammonia from Grizzly Gulch water before treated water is discharged into Gold Run Creek under the terms of the abovementioned water-discharge permit and approved by the EPA. The WWTP staff monitors water quality daily, weekly, monthly, and annually. Parameters monitored included TDS, total suspended solids, temperature, pH, and metals concentrations. SDSTA technicians and independent consultants also monitor aquatic life annually in Gold Run Creek and Whitewood Creek.

#### 10.1.5.2 Shaft Inspection and Rehabilitation for Safe Access

Siemag and Spencer Engineering recertified the Ross Hoist in 2007, and at that point, Dynatec Corp. began reentry of the shaft to inspect down to the 4550L. Working with Dynatec, SDSTA crews completed the visual inspection of the Ross Shaft to the 5000L, including individual assessments of 850 structural steel sets. Steel-set members were replaced where necessary to ensure the shaft is functional at reduced speeds and loads, which are adequate for the current needs to support early science and access for site investigations. Ground control systems were repaired or replaced where necessary, and SDSTA crews stripped all old utility lines out of the Ross Shaft from the surface to the 5000L. A new pump column was installed from the 5000L to the 3650L to support dewatering of the facility. SDSTA crews installed new communications fiber, 12 kV and 5 kV electrical cables in the Ross Shaft to the 5000L.

While the visual inspections performed by Dynatec and SDSTA were adequate to ensure safe access for reentry, additional non-destructive testing (NDT), comprising ultrasonic and direct measurement of steel members, is being performed. G.L. Tiley is using the information from the NDT to develop finite element models of each set and to determine where structural sets would be susceptible to damage in the event that the emergency brakes (dogs) are deployed to stop the conveyance at the loads and speeds required for DUSEL Construction. The results of this study will define the scope of repair work to allow this conveyance to operate at these loads and speeds required to support DUSEL development.

Under contract with SDSTA, Siemag and Spencer also recertified the Yates Hoist, and RCS Construction reentered the Yates Shaft in 2008, inspecting and assessing the shaft and reopening it to the 4850L, allowing SDSTA crews to conduct level and station inspections and complete the following activities:

- Old utilities were removed to the 1100L.
- Ground-control timbers (lacing) were removed, inspected, and reinstalled or replaced.
- Timber shaft guides were replaced.
- Cables and a 5 kV electrical cable were installed from the surface to the 4850L.
- Two water lines were installed in the Yates Shaft, including a 2-inch pipe carrying city water for fire control and a 4-inch pipe for industrial water for construction.

The SDSTA purchased a new cage for the Yates Shaft and an emergency cage-arrest system using dogging ropes is planned to be installed in calendar year 2011. The Yates Shaft is planned to be operational as the primary access to the 4850L for the LUX and MAJORANA DEMONSTRATOR experiments.

By April 2011, SDSTA technicians and contractors will complete excavations, ground control, shotcrete, and basic infrastructure to prepare for installation of the MAJORANA DEMONSTRATOR and LUX experiments at the Davis Campus. Figure 10.1.5.2 shows the 4850L and the associated spaces discussed in this section. SDSTA infrastructure crews enlarged the Davis Cavity, raising the back (ceiling) by 8 feet, to accommodate the LUX dark-matter detector. SDSTA technicians also excavated the DTA to 135 feet long by 50 feet wide by 17 feet tall. Both LUX and MAJORANA DEMONSTRATOR will use the Davis Campus for early science. By April 2011, extensometers will be installed in the Davis Campus, including the DTA, to monitor ground movement. Shotcrete will be applied by SDSTA contractors in the DTA, DLM, and in a 75-foot drift connecting the two spaces. In addition, SDSTA crews excavated cavities for chillers and electrical equipment, and expanded the Yates Shaft station at the 4850L to allow moving large equipment to the Davis Campus.

On the Ross Shaft side of the 4850L, SDSTA Operations crews rehabilitated a former electrical shop to create a temporary electro-forming laboratory for the MAJORANA DEMONSTRATOR experiment. Additional ground-control measures were installed, and a rock dump at the 4850L station was sealed with a steel wall. Contractors applied shotcrete at the Ross Station and down a drift, approximately 300 yards, to the electro-forming laboratory where shotcrete also was applied. A class-1000 clean room was installed for electro-forming baths, which will be in operation by April 2011.

SDSTA Operations staff also installed lighting on the 4850L from the Governor's Corner intersection through the Ross Station to the electroforming laboratory—a distance of about 500 yards.

Other equipment and infrastructure improvements performed by SDSTA at the 4850L include:

- Installation of an enclosed unisex bathroom next to the Ross Shaft Station
- First aid stations equipped with Automatic External Defibrillators (AED), stretchers, and other emergency supplies at both the Ross and Yates Shafts
- Carbon monoxide detectors at the Yates and Ross Shafts and at the #4 Winze, connected to warning systems underground and on the surface
- Fire-protection systems, including sprinklers at the electro-forming laboratory
- Hazardous-materials spill kits
- A magazine for storage of explosives and a magazine for blasting caps, constructed to standards set by the federal Bureau of Alcohol, Tobacco, and Firearms

The 4850L, by April 2011, will also be equipped with Voice over Internet Protocol (VoIP) telephones and Internet connections at the Governor's Corner and the electro-forming laboratory. As described in Chapter 5.5, *Cyberinfrastructure Systems Design*, Femco copper-wire telephones also service the Ross and Yates Shafts as well as various other underground locations. Leaky feeder radio systems provide communications throughout the Yates and Ross Shafts. Hoist operators also have access to Femco, radio, and telephone communications. Chapter 5.5 provides more details on current condition of underground communication systems.

SDSTA infrastructure technicians upgraded the narrow-gauge rail system on the 4850L, and they operate and maintain the electric locomotives. SDSTA technicians also operate and maintain the load-haul-dump (LHD) loaders, jumbo drills, jackleg drills, and other equipment used underground, including explosives, to perform excavation activities.

Two exhaust fans currently provide ventilation for the underground facility; as air is exhausted out of the Oro Hondo Shaft and the #5 Shaft, fresh air is pulled down through the Yates and Ross Shafts. Two 350-horsepower reversible fans are available—one located at the Oro Hondo Shaft and one at the #5 Shaft. Underground, more than 24 air doors control the air circulation provided by the exhaust fans on the surface. CAI Construction, under the supervision of SDSTA operations staff, installed nearly 30 Kennedy stoppings, or steel air doors, which are used to direct the flow of air underground. In addition, small booster fans move air through canvas vent bags to supply fresh air to work areas lacking sufficient flow-through ventilation. SDSTA technicians monitor and maintain this ventilation system to provide and maintain safe access for early science and to support design investigations. A discussion on the current conditions of the ventilation system is included in Section 5.4.3.6, *Drifts and Ramps Required for Access, Egress, and Ventilation*.

Two upgraded electrical substations provide power to hoists, fans, and experiments. Operations staff installed 12 kV and 5 kV cables in the Ross Shaft to the 5000L and 5 kV cables in the Yates Shaft to the 4850L.

A complete description of the existing infrastructure systems is discussed in Chapter 5.4, *Underground Infrastructure Design*.

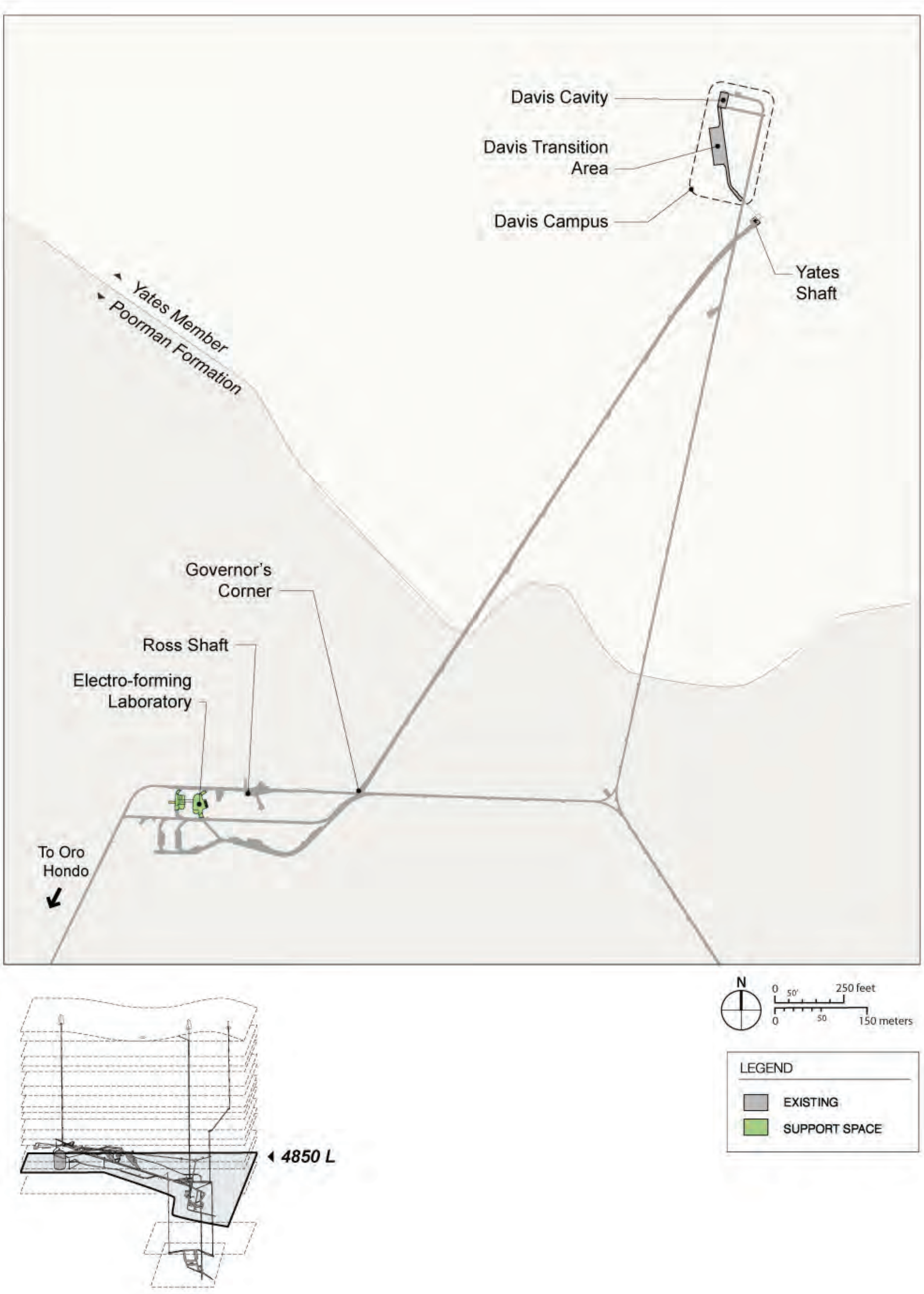

**Figure 10.1.5.2** 4850L and existing associated spaces. [DKA]

### 10.1.6 Early Science Activities

The science activities planned for 2011 until early 2012 are divided into two areas: 1) the continued development of the experimental program at Sanford Laboratory and 2) integration of potential DUSEL experiments into the design of the DUSEL Facility. An early science program, recommended by an external peer Program Advisory Committee (PAC), is under way at Sanford Laboratory (see Chapter 3.4 for a description of the experiments).

In 2011, electroforming for the MAJORANA DEMONSTRATOR will be under way in a clean room adjacent to the Ross Shaft at the 4850L and will continue into 2012 and be supported, in part, by SDSTA staff. The outfitting of the Davis Campus to house the LUX and MAJORANA DEMONSTRATOR experiments is planned to be completed by the end of 2011 and operation of the LUX experiment in the surface assembly building will start in early 2011 and continue until the underground area is ready for installation at the end of 2011. The MAJORANA DEMONSTRATOR experiment will begin to assemble a clean room in the DTAs by the end of 2011 and install equipment in the clean room once completed. SDSTA staff will also support these activities and in collaboration with EH&S personnel, will organize and take part in readiness reviews of LUX and MAJORANA DEMONSTRATOR (and other experiments) prior to beginning the underground installation.

Experiments in biology and geosciences were initiated at Sanford Laboratory during Preliminary Design. The science liaison team responsible for integration of proposed experiments and the resulting development of the Facility requirements will undertake design studies in collaboration with the prospective experimental community to refine the design of the 4850L and 7400L laboratory modules. The team will also work with the science community to better understand future computing needs at the DUSEL site. A key goal for 2011 is to update the Interface Requirements Document (IRD) with Systems Engineering that specifies experimental requirements to be ready for Final Design of the Facility in early 2012. In addition, the science liaison team will support the deliberations of the DUSEL PAC.

Researchers at Sanford Laboratory must follow an Experiment Implementation Policy (EIP) that has established rules, procedures, and guidelines for research, including policies on insurance, documentation, and hazard analysis. Researchers also must complete an Experimental Planning Statement (EPS) that includes a project summary, a list of equipment, a list of infrastructure needs, a list of potential hazards, access requirements, an experiment schedule, and a decommissioning plan. Approximately 24 scientific collaborations have started this process and are described in Volume 3.

### 10.1.7 Early Education & Outreach

Early Education and Outreach (E&O) efforts fall into two broad categories: 1) early implementation of education programs on behalf of Sanford Laboratory; and 2) planning the Sanford Center for Science Education (SCSE), which will house DUSEL's education and public outreach program. Existing Sanford Laboratory efforts include professional development opportunities for K-12 teachers, facility tours, hands-on science activities for K-12 school groups, public lectures on site and throughout South Dakota, research internships, the annual Neutrino Day science festival, summer programs for high school and college students, and development of curricular materials. All these activities represent prototype programs being tested, evaluated, and refined for integration into the DUSEL education and public outreach programs.

Concurrent with the implementation of early E&O programs, staff members support the design of the SCSE Facility, develop content and program plans, build relationships with a strong cultural emphasis, conduct focus groups and needs assessments, market research, and refine the SCSE business plan.

Integrated E&O activities including the implementation of prototype programs and the design of the SCSE are overseen by the Education Governing Board (EGB) as described in Volume 4. Additionally, an Education Advisory Committee (EAC) provides outside expertise and guidance.

Funding for E&O has come from a wide variety of sources, including the SDSTA, DUSEL for Preliminary Design planning, a dedicated NSF award for planning the SCSE programs, and NSF funding provided through South Dakota's EPSCoR Office (Experimental Program to Stimulate Competitive Research). Numerous smaller grants and contributions from the U.S. Department of Education, the South Dakota Department of Education, the South Dakota Department of Tourism and State Development, and 3M Corporation have also supported the program development.

Specific E&O projects taking place prior to the start of Final Design include the Davis-Bahcall Summer Scholars program, which involves study at Sanford Laboratory, Gran Sasso Laboratory, and Princeton University; the Neutrino Day science festival; Dave Bozied Summer Internships; the Deep Science for Everyone lecture series; the development and pilot testing of a modern physics course in Sioux Falls high schools; weeklong content workshops for K-12 teachers; and South Dakota GEAR UP (Gaining Early Awareness and Readiness for University Programs). During 2010 alone, approximately 5,000 educators, students, and members of the general public participated directly in early E&O activities conducted at or through Sanford Laboratory. In addition, Sanford Laboratory’s Web site attracted approximately 50,000 unique visitors in two years. Sanford Laboratory produced a 30-minute video aired on South Dakota Public Broadcasting that was distributed to high schools and universities throughout South Dakota talking about the science to take place at DUSEL.

Cultural outreach, led by DUSEL's cultural and diversity liaison coordinator and advised by DUSEL's Cultural Advisory Committee, is dedicated to creating programs and opportunities of regional cultural interest at DUSEL. The Cultural Advisory Committee includes 11 members, including members of three South Dakota American Indian tribes, as well as SDSTA and DUSEL staff. SDSTA and DUSEL staff have communicated with leaders from all nine tribes in South Dakota and with members of tribes in the region.

The E&O group also has begun a process to create a Virtual DUSEL (vDUSEL), a computer-generated, interactive model that will allow remote exploration of the underground laboratory. BHSU has developed a partnership with Dakota State University (DSU) to begin planning and design of vDUSEL.

## 10.2 Operations Plans during Final Design

The current schedule for the development of DUSEL anticipates the start of Final Design in February 2012. At this point, the Project will have received approval from NSF to complete the Facility design and proceed into the next funding phase. By the start of Final Design, the DUSEL LLC will have been fully formed and will be the management entity for the operations and activities at Sanford Laboratory under the leadership of UC Berkeley. The primary role of the SDSTA after the formation of the DUSEL LLC will be to retain the site ownership and maintain relationships with the state of South Dakota and Barrick Gold Corporation.

The focus of efforts during Final Design is fivefold:

- Basic Facility maintenance and operations
- Operations of the science program, including early science, refining the DUSEL scientific collaboration selection, and supporting the experiment design process
- Advancing the Facility design to prepare for DUSEL Construction
- Developing the education and public outreach efforts to build future capacity and increase awareness of the plans and scientific research at DUSEL
- Developing the supporting capacities required for Construction and then Operations of the DUSEL Facility and science program.

Final Design will complete in two years with the start of Construction in February 2014. With access to the 7400L deferred until 2013 due to dewatering progress, the Deep-Level Laboratory (DLL) Campus design activities will be completed by mid-2016. It is assumed that the activities described in this Chapter 10.2, *Operations during Final Design*, are built on the efforts described in Chapter 10.1, *Current Operations*, which serve as a foundation for the growth and development of the DUSEL Project.

### 10.2.1 Organizational Management and Staffing Approach

As the DUSEL LLC will be fully established by the start of Final Design, it will manage the full operations and staffing of the DUSEL Project. As previously discussed, during the early development and Preliminary Design of the DUSEL Project, major activities were performed through multiple entities, including UC Berkeley, SDSTA, SDSM&T, and BHSU. With the DUSEL LLC in place, UC Berkeley will maintain the primary DUSEL relationship with NSF through a Cooperative Agreement, and the DUSEL LLC will provide the overall day-to-day management direction for the Project and its development. The DUSEL LLC will manage design and engineering contracts and will subcontract to collaborating institutions for appropriate services and functions. Chapter 7.8, *Project Organization, Governance, Oversight, and Advisory Functions,* describes the organizational structure and key positions in greater detail. Additional information on the full staffing program is available in Volume 2, *Cost, Schedule, and Staffing*.

The Central Project Directorate (also referred to as the Level 1 Managers) will include the Principal Investigator/Laboratory Director (PI/LD), Co-Principal Investigator (Co-PI), Associate Laboratory Director (ALD), and Project and Operations Director (POD). The Central Project Directorate directs and oversees all DUSEL functions and activities.

The PI/LD is the key individual accountable for the Project success. He or she is the principal spokesperson for the Project, and the main point of contact for all matters related to the Project. The PI/LD is responsible for establishing the scientific reach and mission of the Project and ensuring that goals and objectives continue to be met as the Design, Construction, and Operations progress. The PI/LD is responsible Project completion, safe operations, and the successful delivery of the scientific program. The PI/LD reports to the DUSEL LLC Board of Directors and the UC Berkeley Vice Chancellors Office for Research.

The Co-PI is a secondary point of contact for DUSEL primarily for external Project matters. The Co-PI acts on behalf of the PI/LD in the PI/LD's absence.

The ALD is a secondary point of contact for DUSEL primarily for internal Project matters. The POD will manage the daily operations of DUSEL at Sanford Laboratory including procurement support; Facility Maintenance, Construction and Operations; Human Resources; Information Technology; Communications; Project Controls; and Systems Engineering.

Reporting to the Central Project Directorate are the four Level 2 major systems managers, including the Deputy Project Director, Facility Project Manager, Operations Manager, and Chief Science Officer (CSO).

The Deputy Project Director is a primary point of contact for the project management approach and is responsible for managing and developing policy and procedures that impact both the operations and management of the Project.

The Facility Project Manager oversees and manages the Facility development and is responsible for managing the entire scope, cost, schedule, and quality of all of the underground and surface scope of the DUSEL Facility construction.

The Operations Manager is responsible for maintaining safe access to both the surface and underground facility at Homestake, including all aspects of day-to-day physical plant operations.

The CSO will be responsible for managing relationships with and activities of the Integrated Suite of Experiments (ISE), including early science and the science construction program.

Other positions that report directly to the Central Project Directorate, but are Level 3 crosscutting system managers, include the Education and Outreach Director, Chief Financial Officer (CFO), the Quality Assurance Manager, and the Environment, Health, and Safety (EH&S) Director. Together, the Central Project Directorate, the Level 2 major systems managers, and the Level 3 crosscutting system managers form the DUSEL Senior Management as shown in Figure 10.2.1-1.

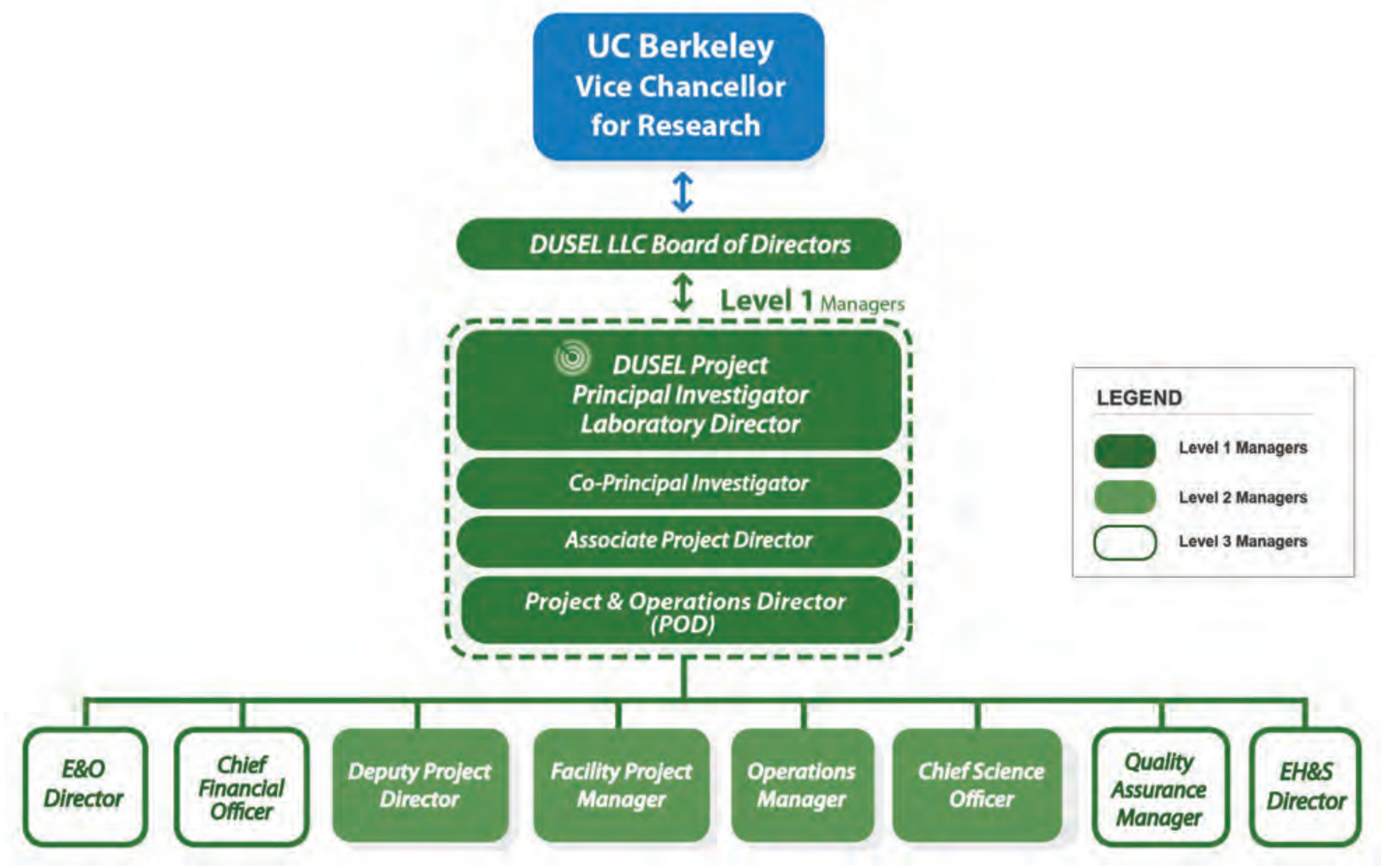

**Figure 10.2.1-1** Organizational structure of DUSEL LLC Senior Management. [DKA]

The E&O Director is responsible for leading the education and public outreach efforts for DUSEL, including the development of the program for the Sanford Center for Science Education (SCSE) and engaging the region through cultural outreach efforts.

The CFO is responsible for the financial accounting system, day-to-day financial transactions, asset management, and contract administration.

The Quality Assurance Manager reports directly to the Central Project Directorate to allow direct access to and communication with key Project stakeholders to quickly and appropriately manage all quality management and audit activities.

The EH&S Director reports directly to the Central Project Directorate because of the central and crucial role of EH&S across the Project to develop and maintain a functioning Integrated Safety Management (ISM) system.

The positions reporting to the POD include the Deputy Project Director, Facility Project Manager, and the Operations Manager. The Level 3 subsystem managers reporting to the Deputy Project Director are the Business Systems Manager, the Information Technology Manager, the Systems Engineering Manager, and the Project Controls Manager. Level 3 subsystem managers reporting to the Facility Project Manager include the Deputy Facility Project Manager, the Underground Infrastructure Manager, the Mid-Level Laboratory Campus Manager, the Deep-Level Laboratory Campus Manager, the Surface Campus Manager, and the Long Baseline Neutrino Experiment (LBNE) Manager. The Level 2 and Level 3 positions reporting to the POD are shown in Figure 10.2.1-2, below.

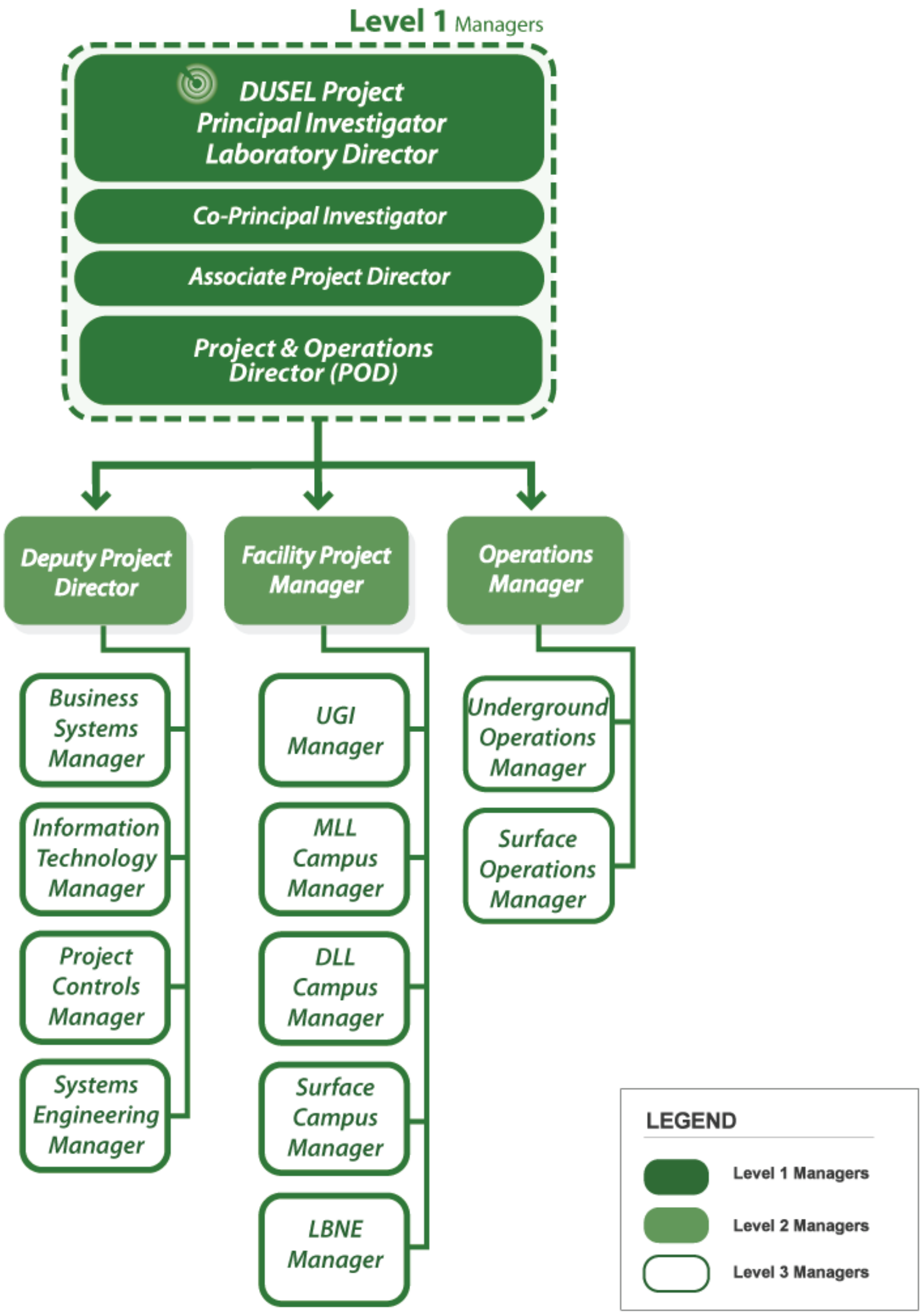

**Figure 10.2.1-2** Organizational structure of the POD and Level-3 Subsystem Managers. [DKA]

As the construction project is located in Lead, South Dakota, current plans assume that the center of activity and most staffing functions will be located at the Project site. UC Berkeley will remain accountable to NSF for the implementation and operations of the DUSEL Project, enabled by its participation in the DUSEL LLC Board of Directors and serving as the primary institution for the development of the Project.

The staffing levels for each phase of the Project described in this volume are general; specific details on staffing levels are included in Volume 2, *Cost, Schedule, and Staffing*. The descriptions outlined here are intended to describe on a time-phase basis the approximate level of effort supporting the development of the operations of the Project. The staffing levels Project-wide grew during Preliminary Design to provide a base capacity for the design development activities. To support the demands of Final Design as well as to prepare for construction activities, a small increase in procurement, contract management, and Facility staff is expected in the Final Design phase.

### 10.2.2 Project Operations

The following departments provide day-to-day operational functions for the DUSEL Project.

#### 10.2.2.1 Business Services

Reporting through the Deputy Project Director, Business Services will be responsible for all business administrative activities of the DUSEL LLC. The specific departments and an outline of each function are noted below.

**Insurance**

With the formation of the LLC in advance of a Final Design start, the insurances carried by the SDSTA during pre-DUSEL LLC operations will be combined with coverages required for pre-construction operations and ongoing design work to support Final Design activities. Commercial general liability, workers' compensation, environmental, directors and officers' liability, and auto, for example, will continue to be carried at limits appropriate for the preconstruction phase. Property insurance will be included to insure selected elements of the SDSTA property and equipment.

During the Final Design period, insurance coverages will migrate to policies coordinated through the University of California Regents (UC Regents) Risk Management Office to capitalize on the significant buying power of UC Regents in the insurance industry. In support of facility-design activities, professional liability insurance to address potential design errors and omissions will be purchased by the DUSEL LLC through a project-level policy to cover all design-related efforts to ease administration, enhance coverage specific to DUSEL, and lower overall insurance costs in comparison to relying on individual practice policies provided by each design firm.

**Procurement and Contract Management**

The Procurement department, working under the direction of the Business Systems Manager, is responsible for purchasing all materials and supplies required to support ongoing operations and maintenance, and for the management of programs and assets. Toward the conclusion of Final Design, the Procurement department will increase to prepare for the start of MREFC-funded construction activities. Adequately preparing for construction activities through early procurement of long-lead items and developing contracts to support the construction acquisition schedule will reduce the operational risks and required contingencies during construction. The Procurement department will establish purchase orders for all vendors and contractors working on the Project. Support for all of the procurement functions described above is anticipated to require approximately three FTEs.

The Contract Management department, also working under the direction of the Business Systems Manager, will manage and maintain relationships with all vendors supporting operations, construction, as well as science user support agreements. All outsourced service contracts, contractors, and any relationship that requires a contract with the DUSEL Project will be managed through this department. The Contract Management team will include two FTEs.

**Intellectual Property, Technology Transfer, and Copyright Policy**

For the experiment research and development activities taking place as a part of the DUSEL Project, the intellectual property, technology transfer, and copyright policies will follow the policies from the users' sponsoring institution on patents, copyrights, trademarks, and tangible research results. These are commonly referred to as intellectual property, as they relate to sponsored research agreements. In general, these policies will reflect the academic policies of the Project participants and the sponsoring funding agencies. The DUSEL LLC will consider and generate a policy acceptable to the DUSEL stakeholders.

**Support Services**

Support Services is primarily an administrative function that includes staff that provides administrative support to the Project team. This group is responsible for the information management system, administrative support services, workshops, and project review logistics. As currently understood, these functions will be performed by approximately six FTEs.

**Human Resources**

Human Resources is responsible for the recruitment of all staff employees working for DUSEL as well as the coordination of all benefit plans. Working closely with Senior Management, Human Resources helps develop job descriptions, post open positions, and assist in the hiring process. These activities will be performed by one FTE during Final Design.

**Communications**

The Communications department is responsible for public relations through a variety of media. These efforts include a newsletter, Web sites, and regular communications with the local community. The Communications efforts during Final Design are performed by three FTEs.

**User Support**

The User Support Office (USO) provides administrative management and support to individual researchers and collaborations working at DUSEL. The USO develops and supports processes to facilitate badging, coordinating safety training, office assignments, and assistance with business functions. USO staff serve as Facility user advocates by coordinating with other DUSEL resources that support researcher technical needs. The USO will also assist with scientific workshop and Project review meeting logistics. During Final Design, it is anticipated that the USO will include two FTEs.

### 10.2.2.2 Information Technology

The IT department will be responsible for all centrally managed computer and network technologies utilized in support of DUSEL including personal computers, printers, software, servers, data management, data archive and back-up, and other IT infrastructure as further described in Chapter 5.5, *Cyberinfrastructure Systems Design*. The team will support approximately 250 desktop and laptop computers, approximately 24 servers, and fiber and copper based backbone networks to support both surface and underground operations.

During this phase, the IT department will provide support to both staff and an increase in science activity. Added technologies will include Linux support services, enhanced network monitoring and administration, increased Voice over IP (VoIP) support services, and video teleconferencing systems. It is anticipated that the cybersecurity plan will be enhanced to address the addition of services and users during Final Design along with development of a Data Management Plan to address NSF data retention requirements. The IT staff will be increased from four to six FTEs.

### 10.2.2.3 Finance

The Finance and Accounting department is responsible for the reconciliation of all budgets and management of cash flow to support the full DUSEL Project. The department maintains the overall

DUSEL budget and provides financial detail to the Project Controls Team to prepare reports for NSF and the Central Project Directorate. This department is also responsible for asset management, coordination of audits, preparing and processing invoices, and managing contract obligations. Budgets are established by Senior Management. During Final Design, these functions will be supported by approximately four FTEs.

#### 10.2.2.4 Project Controls

Project Controls is responsible for cost-estimate management and scheduling of the DUSEL Project. It maintains the Work Breakdown Structure (WBS). The Project Controls team generates reports on Project status for the Central Project Directorate to demonstrate how the Project is proceeding according to the schedule and budget. During Final Design, Project Controls will establish an Earned Value Management System (EVMS) that will be operational prior to the start of MREFC funding. Project Controls works closely with all other departments to ensure adherence to budgets and schedules. Changes to configured items, budgets, and schedules are reviewed and approved by the Configuration Control Board (CCB). During Final Design, the activities of this department will require approximately five FTEs.

#### 10.2.2.5 Operations

During the Final Design phase, the Operations department, which reports to the Operations Manager, will include approximately 84 FTEs. This increases the level of effort by seven FTEs over the existing operations level to support safety, rehabilitation and refurbishment work, and materials management.

The Operations staff during Final Design includes management and technical staff to manage, plan, and execute operations and maintenance, including hoist operators, shaft maintenance technicians, security guards, WWTP operators, facility technicians, and warehouse and materials handling staff. This staffing level supports routine Facility operations and maintenance activities plus performance of activities to provide safe access to support design and construction activities, including ground support, water inflow management, pump system maintenance, and waste rock handling system rehabilitation projects.

#### 10.2.2.6 Facility

Facility staffing levels during the Final Design phase will increase over the Preliminary Design staffing levels of 11 FTEs. The Facility team will continue to be led by the Facility Project Manager and through a Deputy Facility Project Manager funded by the LBNE project. During Final Design, the Facility team will add 12 FTEs for a total of 23 FTEs. This increased staff level is essential to properly perform all engineering and project management functions that are needed to deliver a sound Final Design and to support the bidding process in preparation for construction. Included in this increase is staffing to address Facility engineering support to Operations. The Facility and Operations teams also support the science activities discussed in Section 10.2.7, *Science Activities*.

DUSEL Facility staffing provides one manager for each of the four major architecture and engineering design contracts, and engineering support to each contract scope to integrate design outputs into a cohesive Facility design. Staff increases for Final Design include a DLL Campus Manager, LBNE Facility Project Engineer funded by the LBNE project, a Project Engineer supporting the DLL and MLL integration, and additional engineering support for civil/structural, low voltage/electrical, geotechnical, mechanical systems, shaft/mechanical, Other Levels and Ramps (OLR)/hydrology, and Surface Facility infrastructure. During Final Design, Facility engineering support to daily operations, previously provided by the SDSTA, will transition to the DUSEL Facility team. This includes one electrical engineer and one mechanical engineer.

#### 10.2.2.7 Systems Engineering

During Final Design, the Systems Engineering (SE) department (six FTEs) will execute and refine plans and processes developed during Preliminary Design for risk, configuration, requirements, interface management, systems verification, Value Engineering, and Trade Studies.

Risk management is a continual process and has a central role in DUSEL Operations; periodic reviews will be conducted at which new risks will be identified and old risks retired as designs and program development are refined and mitigation plans put in place. The risk management process is described in the Risk Management Plan (Appendix 9.C).

The DUSEL Project requires a system in which an officially approved baseline is established. To assist in this process, the SE team is responsible for managing the control of the Project baseline and other configuration items, including policies, procedures, and plans. The Configuration Management Plan (Appendix 9.D) establishes the process for management of configuration-controlled documents. During Final Design, the SE team will manage the configuration management system, including the CCB as described in the Configuration Management Plan.

During Preliminary Design, the SE team, in collaboration with the Project stakeholders, developed a hierarchical requirement, interface, and design document structure that identifies how high-level stakeholder needs are translated into technical requirements, designs, and interfaces. The disciplined development and review process including control through the CCB ensures that all aspects of the Project are aligned and in agreement with the baseline. These processes will continue to be used throughout Final Design, along with Value Engineering and Trade Study processes to ensure that requirements and design changes are fully analyzed and approved prior to implementation.

#### 10.2.2.8 EH&S

Incremental increases in the extent and sophistication of EH&S functions and programs during Final Design, with a matching increase to EH&S department staffing levels to support these functions and programs, are anticipated. The increases will be driven by advancing science activities (including underground deployment of early science experiments) and increased surface and underground operations as the Project prepares for construction activities. EH&S staffing levels will increase during Final Design, including contracted consultants, will reach approximately 17 FTEs to address increasing EH&S support requirements.

For Final Design, the EH&S involvement will include participation in regular design reviews, Risk Management Team, CCB, and specialty groups such as the Fire and Life Safety Review Team, the Oxygen Deficiency Hazard (ODH) Panel, and coordination with the QA and SE departments. The frequency and duration of activities is expected to increase as the design matures, particularly as detailed designs for safety and emergency preparedness are matured.

Two specific EH&S tasks that will be completed during Final Design are the preparation of a Hazard Analysis Report (HAR) and support of the Environmental Impact Statement (EIS). The HAR will be based on the Preliminary Hazard Analysis (PHA) completed during the Preliminary Design phase and will use quantitative and qualitative risk assessment processes to identify hazards and appropriate controls to manage the risks from the identified hazards. The HAR development is an iterative process that identifies hazards and controls required to mitigate risks that will be incorporated into the Final Design.

The EIS is being prepared by Argonne National Laboratory (ANL), under contract to NSF. It identifies controls and alternatives for the DUSEL Project elements that present potentially unacceptable environmental impacts. Fulfilling National Environmental Policy Act (NEPA) requirements, the EIS must be completed before federal funds can be expended for DUSEL construction. While completion of the EIS is performed under funding direct to ANL, DUSEL is participating by providing information and review of draft and final documents. Environmental Assessments (EAs) are also being prepared by Fermilab for portions of the LBNE that will be conducted at Fermilab as all EA documentation, including an EIS, is site specific.

Simultaneously, the existing ISM system and EH&S programs will be enhanced to accommodate expanded surface activities, an increase in scientific collaboration presence on site, and support of DUSEL Operations and maintenance activities, particularly providing safe access during design and construction. This will include developing programs to accommodate collaborations, individual scientists, site visitors, and DUSEL contractors within ISM processes and protocols, including orientation, training, and oversight. During Preliminary Design, the SDSTA and DUSEL formed a joint EH&S department that proved effective and will be further developed under the DUSEL LLC.

#### 10.2.2.9 Quality Assurance

During Final Design, the Quality Assurance (QA) team (two FTEs) will continue to manage the quality system as described in Chapter 10.1. The efforts of the Quality Assurance Manager (reporting to the Central Project Directorate) and the Quality Assurance Engineer will support both construction and operations activities.

Quality Assurance Management Reviews will be conducted during Final Design to support the QA requirements. The following lists the additional QA activities that will take place during Final Design:

- QA plans and procedures will be implemented.
- The DUSEL Global QA Policy and supporting documentation will be developed.
- The Internal Quality Audit schedule will be implemented.
- Design and supplier evaluation and management will assist in preparing for construction.
- Internal quality audits will be performed to determine the status of the department, process, and assessed systems.
- Quality Training schedules will be established and QA will coordinate with Senior Management to analyze training programs and determine changes required to ensure that personnel, suppliers, and scientists receive appropriate training.
- A Corrective/Preventive Action Program tool will be implemented and fully integrated into the online training systems.

### 10.2.3 Final Design Activities to Prepare for MREFC-Funded Construction

The Preliminary Design phase established the baseline design and advanced the overall design efforts to about 30% total design. As discussed in Chapter 5.10, *Final Design and Construction Acquisition Plan*, during the Final Design phase, the Facility team, consisting of DUSEL Project staff and the outsourced contractors described in Chapter 5.1, *Facility Design Overview*, will advance the design to 100% design, referred to as the construction documents, and execute the bidding phase in preparation for construction activities. Advancing the design requires regular review of the design documents, cost estimates, construction schedules, risks, and developing a coordinated acquisition plan. The acquisition plan ensures an effective bidding process by carefully prequalifying contractors, creating bid packages, and bidding the Project.

As the structure of the design team solidified during Preliminary Design, a strong coordinated Preliminary Design package resulted; the structure of the outsourced design teams is anticipated to remain largely the same. Any changes to the design team organization for Final Design would be made based on information received from the design contractors in response to a Request for Proposal (RFP) for Final Design services that will be issued during the Transition phase. Similar to Preliminary Design, as the design teams are advancing the Facility design, the construction manager will provide independent estimates and constructability review, finalize the construction acquisition plan, and plan for construction activities.

The Final Design process will include milestones that track overall design progress and provide coordination and reconciliation points between design contractors and the Construction Manager across the full scope of work. At each of these major design milestones, cost estimates and schedules will be reconciled between the designers and Construction Manager. Value Engineering (VE) will be performed to optimize the design and interfaces will be clarified as required. These milestones and the current projected completion timeframe include detailed Basis of Design (BOD) Documents, 60%, 90%, 95%, and 100% Design. At each milestone, the cost estimates, schedules, design-to-cost targets, constructability, requirements, and design interfaces will be reviewed not just within the DUSEL Project team, but also with external Project stakeholders as appropriate. Through all of these milestones, the design team will be mindful of the scope options, identified either through the design-to-cost VE process or through design interface discussions, to provide the highest value design that best meets the Project requirements.

A major focus of activity during Final Design is preparing for construction through acquisition planning—including the process of identifying available resources within industries and the local region, understanding the construction schedule and sequence, dividing the design into bid packages, and beginning to share the acquisition plan within the industry to cultivate strong interest and thus competition for the DUSEL construction work.

The Facility design team will also support the operations and maintenance activities required to create and maintain safe access and maintain and rehabilitate existing systems at the Homestake site. These activities are described in Chapter 10.3, *Operations Plans during Construction.*

### 10.2.4 Maintenance Activities to Support Safe Access and Facility Rehabilitation

During Final Design, deferred maintenance activities will be addressed, as described in Chapter 5.4, *Underground Infrastructure Design*, to provide safe access to facilities and to support design and construction activities and maintain investments in infrastructure already in place, positioning the Project to be prepared to better address contingencies and risk during future phases.

During Final Design, safe access and a safe and healthy working environment underground are essential for the DUSEL design team to continue their efforts to characterize conditions and refine designs accordingly. The projects required to provide this safe access are listed below, and discussed in greater detail, including current conditions of existing systems, in Volume 5, *Facility Preliminary Design*.

- The Ross Shaft and Hoists require completion of deferred-maintenance activities to allow for safe access to and egress from the underground spaces. Design efforts during the Transition phase will outline the specific rehabilitation plans. The #6 Winze will be rehabilitated to the extent possible as water recedes in order to address hazards that may impact continued dewatering activities.
- Water flow in the underground spaces must be controlled to ensure safe access to all underground levels. This work will include ground support upgrades, diversion walls, and boreholes to direct water inflow away from spaces intended for facility occupancy. Work needs to be done to create safe access as well as maintenance of these elements throughout the life of the Facility.
- The waste rock handling system through the Ross Shaft and on the surface will be restored to a reliable operating condition through routine maintenance and replacement of some damaged or obsolete equipment. Safe access through the Tramway level is required for both this waste rock handling system and rehabilitation efforts, including ground support and repair to the Tramway rail system.

By addressing these work elements early, safe access can be provided to support current maintenance efforts, and deferred maintenance repairs can be accomplished without conflicting with future construction activities.

### 10.2.5 Facility Operations and Maintenance

During Final Design, the DUSEL Operations team will continue to perform dewatering activities to gain access to the deep levels of the underground facility, and will maintain continuous operation and maintenance of the pumping system and WWTP.

The Operations staff will continue operations, inspections, and maintenance for the Ross and Yates Shafts and Hoists. To minimize disruption of utility services to the early science experiments and ongoing maintenance activities, a secondary circuit currently mothballed in the Yates Shaft will be reconnected to provide a secondary power supply to the 4850L.

A coordinated hoisting schedule will be developed for the Yates and Ross Hoists to facilitate access for maintenance crews, design teams and their subcontractors, early science collaborations, ISE and S4 collaborations, and other authorized and escorted visitors. The Operations team will continue to maintain facilities and repair equipment as needed in addition to supporting early science activities. The parts and supplies warehouse and the on-site material delivery will continue with the support of the Operations team.

Work will continue into the Final Design period on several safety-related projects, including the central control monitoring, pump station/substation improvements, redundant alarm, personnel monitoring, and ventilation system improvements.

Operations staff will contribute to support the planning and design effort to develop commissioning plans, operations manuals, and training plans for new systems established during the Construction phase. This work will be coordinated with Systems Engineering and other DUSEL departments to ensure a coordinated process.

The small fleet of equipment owned by SDSTA will continue to be used and will refuel on site. Roads and grounds-maintenance activities, including tree trimming and snow removal, will continue. Site security will be coordinated with early construction activities and is anticipated to continue on a full-time basis.

### 10.2.6 Science Activities

Science activities from early 2012 to early 2014 will consist of two primary components: 1) continued support of the experiments at the Sanford Laboratory site and 2) integration of experimental requirements into the Final Design of the Facility. Enhanced support of the Program Advisory Committee (PAC) will be required as specific experiments are reviewed and selected for DUSEL. More background and a timetable of milestones for these activities are provided in Chapter 3.10, *Organization and Management of the Research Program.*

#### 10.2.6.1 Science Integration and Support

By early 2012, LUX will be in full operation. The MAJORANA DEMONSTRATOR will outfit and occupy the Davis Transition Area (DTA) in the Davis Campus, install detector elements, and be in operation by 2013. Biology and geosciences experiments will continue, and new experiments in this scientific area are likely. The Center for Ultralow-Background Experiments at DUSEL (CUBED) experiment is expected to be ready for underground operation by 2013. These experiments will require additional technical support from the DUSEL staff.

The DUSEL science integration team will be responsible for supporting the integration of experiments into the Facility Final Design. The science integration team will be responsible for the interface of experiments to the Facility, for establishing and maintaining a safe working environment (in collaboration with EH&S), and for coordination and implementation of common aspects to support multiple experiments. An increase in integration activities, and therefore staff, will be needed during the Final Design period.

The science integration team will continue to consist of scientific, engineering, and technical staff to support the early science research at Sanford Laboratory and to support planning for integration of experiments into the DUSEL Facility. Functions will be added when required to support experiments. These consist of additional engineering functions, technical support, and shops (such as machining, electrical, biology/geology, and low-background counting). Each major element is described below.

Scientific staff is essential to lead the integration and support effort. It will consist of senior personnel with management and coordination responsibilities; permanent staff positions with well-identified responsibilities for major activities; and term scientists with specific responsibilities. Scientific staff will support the early science activities started at Sanford Laboratory, interface with the Facility design team, support design integration with the experimental community, and assist the DUSEL PAC in the review of potential experiments.

Engineering staff covering the mechanical, electrical, and computer science disciplines will be required for integration and support of collaborations. Mechanical engineers will support the integration of future experiments at DUSEL and support early science at Sanford Laboratory during Final Design.

A modest growth in technical staff is anticipated during Final Design to support the operation of early science experiments (see Volume 3.4 for a description of the experiments).

Planning for implementation of mechanical, electrical, and support laboratories (e.g., chemistry and geology) will take place during early stages of Final Design, to be installed during MREFC-funded Construction.

The low-background counting and materials assay capabilities are critical to the success of the early science and DUSEL physics program. Some capability in this area will be established in the Davis Campus before Final Design, using equipment procured through Project partners and anticipated DUSEL physics experiments. Planning for procurement of this system will be completed in Final Design.

#### 10.2.6.2 Experiment Design and Research and Development

The experiment design and research development at DUSEL includes Preliminary and Final Design activities in support of the DUSEL experiments (MREFC-funded) and related Research and Development (R&D) activities. NSF-supported experiment design and R&D for the Integrated Suite of Experiments (ISE) at DUSEL through early Final Design is planned to continue as part of the three-year funding of the S4-supported groups for physics and biology, geology, and engineering (BGE) experiments. The design of the ISE and their related R&D will, in general, not be complete until later in Final Design. Additional funds will be required during Final Design to complete the required design and R&D and the funding source for these activities has not yet been identified. Funding may be provided by NSF, through DUSEL subawards to research groups, or a combination of the two. Planning efforts have assumed that a full implementation of the NSF-DOE stewardship model for the experiments is in place and operational by mid-Final Design. Thus, design and R&D funds for a particular experiment are anticipated to predominantly be supplied by the stewarding funding agency. Background discussion on this stewardship model is included in Chapter 3.10.

Experiment design and engineering were started during Preliminary Design through the S4 support of experimental proposals by NSF, through DOE support, through institutional, and non-U.S. support. The design of prospective DUSEL experiments will not be completed by the time the S4 funding is currently expected to cease. Thus, funds to support engineering design and for subcontracts to specialized design firms (e.g., familiar with underground laboratories, clean rooms, etc.) will be required during the Final Design phase. Planning assumes the NSF funds required for this purpose are available for a continuation of the S4 activities. DOE or other funds will also be required but are not included in the estimates. It is anticipated that FY 2013 and FY 2014 will be peak years for experiment design activity, primarily Final Design, related to the ISE. As noted above, the mechanism for disbursement of design funds remains to be determined.

R&D activities, including advanced prototypes, are required to complete the design and begin construction of the ISE deployed at DUSEL supported by MREFC funding. The R&D estimate for the ISE is based on extrapolations of ongoing S4-funded groups (through FY 2012) and assumes full implementation of the stewardship model for experiments is in place by FY 2013.

Experimental research includes all aspects of support provided to SDSTA/DUSEL for scientific research related to Sanford Laboratory and DUSEL experiments, including early science experiments and later the initial deployment of experiments at DUSEL. Materials, supplies, and minor equipment, supported by MREFC funding, are included. Support of DUSEL scientific staff performing research is also included. Funding for material, supplies, and services from the state of South Dakota for early science experiments is planned to continue through FY 2012. NSF support will be required starting in FY 2013. Support from NSF for research time of DUSEL scientists is limited until mid-Final Design (until FY 2013), with partial support of a few DUSEL personnel engaged in LUX, MAJORANA DEMONSTRATOR, and BGE experiments.

### 10.2.7 Early Education and Public Outreach Activities

Education and Outreach (E&O) efforts during Final Design include continued planning of the Sanford Center for Science Education (SCSE) as well as continued implementation of programs started during Preliminary Design. E&O staff will work in collaboration with architects and engineers to complete the design of the SCSE Facility and continue to develop the institution, including organizational design, program and exhibit planning, and fundraising. A full business plan and a Conceptual Design of the exhibit space are scheduled for completion during Final Design.

Concurrent with the planning and design efforts, prototype education programs will continue to be developed, implemented, evaluated, and refined. An increasing number of K-12 school groups will come on site for hands-on activities, interact with project scientists and engineers, and take site tours of the Surface Campus as safe access can be provided. Increasing numbers of teachers will participate in graduate-level classes that focus on underground science content, both on site and regionally. Student interns will continue visits to Sanford Laboratory for summer placements across a wide array of disciplines, not only within science and engineering, but also in areas such as mass communications and science education. Project staff will expand programs off site, and continue to build relationships, partnerships, and cultural engagement. Off-site programs will include visits by Project staff to schools and communities throughout the surrounding region, testing of educational content and activities, and deployment of prototype hands-on exhibits in existing regional science and community centers.

Staffing for E&O will increase during Final Design from four to eight FTEs with half the positions to be supported through DUSEL preconstruction R&RA funding and the other half supported by resources outside the R&RA funding profile. Management and senior professional staff will include a Director of Education and Outreach, a Deputy Director of Education and Outreach, a Cultural and Diversity Liaison and Coordinator, an Education Manager, an Education Researcher, and a Scientist Liaison Manager.

The planning efforts with the early implementation of E&O programs as well as the allocation of resources—fiscal, physical, and human—will continue to be reviewed by the Education Governing Board. During Final Design, the SCSE Foundation will be established and will take fiduciary responsibility for the SCSE Endowment. Through this process, the Project’s Education Advisory Committee will continue to provide review and input on the institution and program development.

Cultural outreach and engagement efforts, led by the Cultural Coordinator with support from other E&O personnel, will continue to expand during Final Design. The Cultural Coordinator will convene and obtain input from DUSEL's Cultural Advisory Committee, engage in regional community activities, facilitate E&O programs with strong cultural elements, explore existing cultural center models, and provide input while the SCSE is being completed.

Increased emphasis will be placed on distance education and developing online resources, including virtual tours, computer simulations, and webcams. Part of this emphasis will be on developing a robust road map for moving forward with the cyberinfrastructure and distance education planning.

Further prioritization and refinement of program delivery mechanisms will also happen during this time. Key to the early success of the SCSE will be a focus on developing programs in just a few highly strategic areas. Focusing exercises will be conducted during Final Design with input from national experts about priorities and will be funded, in part, by a dedicated planning grant from NSF for SCSE program and institutional development.

## 10.3 Operations Plans during Construction

The current schedule for the development of DUSEL anticipates the start of construction in February 2014. By this point, the Project will have received approval from NSF to start with MREFC-funded Construction with appropriate spending authority. The DUSEL LLC will be fully in place and will be the management entity for the Operations and maintenance activities at Sanford Laboratory, with leadership from UC Berkeley. The primary role of the SDSTA after the formation of the DUSEL LLC will be to retain the site ownership and maintain relationships with the state of South Dakota and Barrick Gold Corporation.

The focus of efforts during MREFC-funded Construction is fivefold:

- Continued facility maintenance and operations, including engineering support to operations
- Operations of the early science program, finalizing the DUSEL scientific collaboration selection, and supporting the experiment design process
- Executing the MREFC-funded Construction within the approved baseline
- Continuing the development of the education and public outreach efforts and preparing for the opening of the SCSE
- Developing the supporting capacities and organizational management required to support the operations of the DUSEL facility and program

As Facility construction comes to a conclusion in 2018 for the 4850L and in 2019 for the 7400L, the installation of scientific experiments will begin as safe access can be provided to the scientific collaborations. With access to the 7400L deferred until 2013 due to dewatering progress, the Deep-Level Laboratory (DLL) Campus design activities will be completed in 2016. It is assumed that the activities described in this Chapter 10.3, *Operations during Construction*, are built on the efforts described in Chapter 10.2, *Operations Plans during Final Design*, which serve as a foundation for the growth and development of the DUSEL Project and its scientific program.

### 10.3.1 Organizational Management and Staffing Approach

The organizational management structure of the DUSEL LLC as described in Section 10.2.1, *Organizational Management and Staffing Approach*, is expected to remain constant into the Construction phase of the Project although the staffing numbers are expected to increase (Section 10.3.2, *Project Operations*).

### 10.3.2 Project Operations

The following departments provide day-to-day operational functions for the DUSEL Project during Construction.

#### 10.3.2.1 Business Services

Increases in Business Services staffing to support construction activities will take place during the end of Final Design and early in the Construction period. Staff increases will include procurement and contract management to support the acquisition of required materials and contractors to efficiently carry out the Project-wide acquisition plans. Business Services during Construction will be provided by approximately 23 FTEs.

#### Insurance

Included in Business Services is risk financing to ensure that proper insurance coverages are in place to manage risk. During the Construction phase, insurances required to support ongoing Operations will remain in place, in parallel with insurances specific to execution of the MREFC-funded Construction. Commercial general liability, workers' compensation, environmental, directors and officers liability, property, and auto will continue to be carried at limits appropriate for Operations during the Construction phase. Professional liability will continue to cover errors and omissions related to Facility designs as they are implemented through Construction.

Additional coverages will be purchased through a "wrap-up" or "project" policy approach to address construction activities. These project policies will be administered similar to the professional liability insurance program established during Final Design but will specifically address construction activities. These insurances will include an Owner's Controlled Insurance Program (OCIP) that addresses commercial general liability and workers' compensation for contractor staff and activities, builder's risk insurance, and contractor's pollution and environmental liability insurance. These project policies are purchased by the DUSEL LLC, and the contractors are included as named insured on the policies so that their DUSEL-related construction work is covered appropriately. As with professional liability, these project insurances lower the overall insurance cost, simplify policy administration and potential claims processing, and provide coverage specific to DUSEL when compared with the alternative approach of each contractor providing his or her own insurance. Project policies have shown to reduce overall insurance costs on large construction projects by 0.5% to 2% of total construction costs.

#### Procurement and Contract Management

Continuing the functions described in Chapter 10.2, the Procurement and Contract Management functions will grow to support the MREFC-funded Construction efforts, requiring an additional contract manager. These departments will continue to support the ongoing operations of the DUSEL Project, the MREFC-funded construction activities, and the operations of the SCSE.

#### Intellectual Property, Technology Transfer, and Copyright Policy

For the experiment R&D activities taking place as a part of the DUSEL Project, the intellectual property, technology transfer, and copyright policies will follow the policies from the users' sponsoring institution on patents, copyrights, trademarks, and tangible research results. These are commonly referred to as intellectual property, as they relate to sponsored research agreements. In general, these policies will reflect the academic policies of the Project participants and the sponsoring funding agencies. The DUSEL LLC will consider and generate a policy acceptable to the DUSEL stakeholders.

#### Support Services

Support Services will continue to serve the full DUSEL Project as described in Chapter 10.2. As the Facility construction is under way, it is anticipated that the coordination and support for workshops and reviews focused on the scientific program and the Integrated ISE will increase.

#### Human Resources

Human Resources efforts are not anticipated to increase over the level of effort described in Chapter 10.2. This department will continue to provide support services to the Project, but will shift its focus from increasing staff in order to support design or construction, to focusing on identifying staff for positions required for the steady-state Operations phase of DUSEL.

**Communications**

During Construction, it is anticipated that significant efforts in communications and public relations will be required to keep the local community, the interested general public, and the scientific community apprised of the construction progress and activities. Regular updates to the DUSEL Web site and other social media sites will be used to communicate information on DUSEL progress. In addition, members of the Project team will participate in community events to engage the local community. Communications efforts will also be closely coordinated with the E&O efforts through the SCSE, which will serve as a primary interface with the interested public. During this time, DUSEL will also communicate news from the developing scientific program.

**User Support Office**

As the scientific program develops and an increased number of researchers are on site, the activities of the User Support Office (USO) to facilitate administrative processes for researchers working at DUSEL (badging, coordinating safety training, office assignments, and assistance with business functions) will also increase. In concert with other business services functions, the USO will support reviews and workshops for the science collaborations and assist with the process of integrating facility users into the DUSEL community.

### 10.3.2.2 Information Technology

The IT department is responsible for all centrally managed computer and network technologies. The IT team will be supporting approximately 300 desktop and laptop machines, approximately 24 servers, and a gigabit Ethernet fiber network for both surface and underground operations.

During construction, there will be an increase in support staff and the need for increased capacity for storage and backups. As many systems put in place during earlier phases of the Project will reach their end of life cycles and current technology increases, some software applications and hardware will require replacement or upgrades. Increased support for E&O and SCSE operations, as well as increased support for science technologies as laboratory spaces become available, will result in increased network activity. The IT staff will increase from six to 18 FTEs.

### 10.3.2.3 Finance

The role of the Finance department as described in Chapter 10.2 will continue during MREFC-funded Construction but will experience a significant increase in activity to support the construction efforts. This department will be responsible for maintaining all budgets and managing cash flow to support both Construction and ongoing operations, requiring close coordination with all departments, but particularly Facilities, Project Controls, and Operations. The department will continue to maintain the overall DUSEL budget and provide financial detail to the Project Controls team to prepare reports for NSF and the Central Project Directorate. As this department is also responsible for asset management, scheduling of audits, preparing and processing invoices, and managing contract obligations, the staff to support these functions will increase from four FTEs during Final Design to five FTEs during Construction.

#### 10.3.2.4 Project Controls

Project Controls is responsible for cost estimate management and scheduling of the DUSEL Project. They maintain the Work Breakdown Structure (WBS). The Project Controls team generates reports on Project status for Senior Management to demonstrate how the Project is proceeding according to the overall project schedule and timeline. During Construction, Project Controls will implement and manage an Earned Value Management System (EVMS). Project Controls works closely with all other departments to ensure adherence to baseline budgets and schedules. They also work with the Finance team to confirm the Management Reserve and all other budget line items are in order. During Construction, the activities of this department will require approximately five FTEs.

#### 10.3.2.5 Operations

The MREFC-funded Construction period begins in 2014 with the construction of the waste rock conveyor system. Completion of this system enables excavation at the 4850L to begin, which subsequently ramps up to an around-the-clock underground construction operation. Some activities during excavation are best suited to working multiple-shift days, whereas aboveground movement of waste rock will primarily be operated during daytime hours. In addition, the volume of materials arriving on site, and the need for an around-the-clock presence by Operations staff, will increase and remain at increased levels during the excavation period. As systems and shops are constructed and begin to be turned over for regular operations, and scientific laboratory fit-out work begins, the volume and diversity of materials to be transported underground will increase, and more scientific collaboration groups and their contractors will need logistics support. Additional materials handling and rigging staff are needed, continuing through the experiment installation period (2020). The added staff members are in warehouse and riggers (growing from one to seven between 2011 and 2016); shop technicians (an additional three FTEs); and central operations center (an additional five FTEs), which operates continuously. Peak staffing is reached with 134 total FTEs, dropping back down to 104 once the waste rock removal activities taper off and most of the ground support is installed for other levels and ramps by 2020.

#### 10.3.2.6 Facility and Engineering

During the Construction phase, the Facility staffing will increase by one FTE over Final Design levels to 24 FTEs. This increase will result from the addition of one mechanical engineer to lead Yates Shaft rehabilitation activities. Facility staffing during the Construction phase will include approximately eight FTEs for Project Management-related support and approximately 16 FTEs for engineering support. Two FTEs of engineering support, as in Final Design, will be dedicated to support operations during the Construction phase. As during Final Design, the Facility team will be led by the Facility Project Manager, and the Deputy Project Manager will be funded through the LBNE project as a Fermi National Accelerator Laboratory (Fermilab) employee. The Level 3 LBNE Facility Project Engineer will also continue into the Construction phase, funded by the LBNE as a Fermilab employee. All other Facility department employees (22 FTEs) will be funded by DUSEL, including Facility engineering support to Operations.

#### 10.3.2.7 Systems Engineering

During the Construction phase, the Systems Engineering (SE) team (six FTEs) will continue its role in the execution plans and processes in the areas of risk management and configuration management. During Construction, SE will focus heavily on the verification of Facility requirements and interfaces with the Facility and science experiments. The SE team will work closely with the Facility team to execute a defined Verification Plan, which includes the involvement of an independent commissioning agent.

Risk management will continue during Construction just as it did during Final Design, in order for the members of the Facility, SE, E&O, EH&S, and Science teams to identify, classify, and score Project risks. The risks and the information associated with them are provided to Project Controls and Finance as tools for decision-making and effective program management. Mitigation steps and risk-identification will use configuration management and quality processes.

SE team duties regarding configuration management will continue through Construction, as it is key to have this process established and to verify that all changes are communicated to all stakeholders.

The requirements, interface, and design document set developed during Final Design will be used to communicate changes during Construction, when the SE team will assist with communicating these changes to stakeholders and coordinating changes to requirements, interfaces, and design documents (using the configuration management process). The interfaces with the Science teams will also be captured to confirm that all requirements are met.

The Verification Plans initiated during Final Design will continue during the Construction phase and will include roles and responsibilities, Facility and science commissioning, construction flow, verification matrices, and developing operational manuals and documentation.

#### 10.3.2.8 EH&S

EH&S functions and activities funded by R&RA and other non-MREFC accounts will continue incremental increases during DUSEL Construction. These increases will be driven by increased EH&S support required for early science projects and expanded administrative functions such as training and recordkeeping for increased numbers of staff on site.

MREFC-funded EH&S activities related to DUSEL Construction are expected to increase significantly during this period, primarily for direct EH&S support of MREFC-funded Construction performed by contractors. These construction activities increase physical hazards that require comparable increases in EH&S construction specialist oversight and industrial hygiene monitoring. Alternative methods to hiring full-time staff for these positions—including third-party consulting firms or having services provided by the construction management firm—will be investigated, as the positions will be phased-out following DUSEL Construction completion. EH&S oversight responsibility cannot be completely delegated to a third party, as DUSEL will need to maintain control over the process.

The total EH&S staff will increase from 17 to 29 FTEs (both R&RA and MREFC funded) at the peak of construction and will decline slightly to about 25 FTEs toward the end of Construction.

Increases in the number of dedicated Emergency Response Team (ERT) members are also planned for this period to maintain a core ERT on site. These positions will also conduct training, perform safety inspections, practice emergency response procedures, and work collaboratively with the DUSEL Project. Although DUSEL contractors will maintain emergency response capabilities specific to their work areas, these efforts will not be counted as part of the core ERT that DUSEL will maintain. However, capabilities and procedures maintained by the contractors will be closely coordinated with and monitored by the ERT to ensure seamless communication and response mechanisms across the DUSEL Project.

#### 10.3.2.9 Quality Assurance

Once the Quality Assurance (QA) system is successfully implemented, changes will be based on audit results, customer satisfaction data, results of management reviews, and data and trend analysis from the management of goals and objectives.

The QA program will continue to be provided by two FTEs. The efforts of the Quality Assurance Manager and the Quality Assurance Engineer will support both Construction and Operations activities.

The Internal Quality Audit schedule will be implemented. Construction and supplier evaluation and management will be the main focus during 2014 and 2015. The internal audits will be expanded to include more field assessments and supplier management and control over second- and third-tier subcontractors.

#### 10.3.2.10 Science

Scientific staff will continue to lead the integration and support effort. DUSEL scientific staff will be involved in all of the major initial experiments to support the integration of specific experiments into the DUSEL Facility, including installation and commissioning. This model has worked well at existing science user-facilities such as accelerators and light sources.

The scientific staff at the start of Construction will still be located both in Lead, South Dakota, and in Berkeley, California, to support scientific development at the Homestake site with increased activity for the early science experiments. By the end of Construction, the scientific staff will include 15 FTEs (not including research time) in support of the DUSEL experiments.

Mechanical and Electrical Engineering staff will increase during the Construction to support experiments. Direct computing support of experiments is anticipated to be limited (one FTE) and come from IT resources described above. Total technical staff in support of experiments will reach 12-14 FTEs during MREFC-funded construction. Administrative support of the science integration will be provided by one FTE for total scientific staff, reaching 30 FTEs by the end of Construction.

#### 10.3.2.11 Education and Outreach

Overall staffing for E&O is scheduled to increase within the early construction period from eight to 17 FTEs across all funding sources (DUSEL Project funds [seven FTEs] and external funds [10 FTEs]). Management and senior professional staff at the time the SCSE Facility opens will include the Education and Outreach Director, Cultural Liaison Coordinator, Cultural Outreach Manager, two Scientist liaisons (one to focus on physics and the other to focus on BGE), Education Manager, Education Researcher, SCSE Operations Manager, and Development Director (externally funded).

### 10.3.3 Facility Support of Construction Activities

The Facility team will lead construction activities working with the construction manager (CM), who will hold the individual subcontracts in support of Construction execution. DUSEL will also coordinate the activities and involvement of the design firms in providing Construction administration support, including quality assurance activities, interpretation of the construction documents, and developing document updates as required, including support for change order processing. During Final Design, commissioning plans will be developed by the SE team with involvement from the Facility team. During Construction, these plans will be executed as part of an overall commissioning and acceptance of the Facility to the requirements and specifications developed as part of the design.

Site logistics will involve management of the site, staff, and materials coming onto the site during the Construction period—both Construction- and Operations-related activities including science. The management of these workflows and planning of activities will be led primarily by the CM, with involvement from DUSEL staff, including the Facility, Operations, and EH&S teams. To minimize interference with Construction, the CM will maintain the schedule in coordination with all involved stakeholders. Central logistics and warehousing functions to deal with materials flowing to the site will be managed by DUSEL for Operations and Construction.

### 10.3.4 Maintenance to Support Safe Access and Facility Rehabilitation

During Construction, continued maintenance activities will be conducted to maintain safe access to the Facility and supporting ongoing Construction activities. Continued maintenance to sustain investments and existing infrastructure are necessary to reduce risk during Construction and to prepare fully for steady-state Operations. Safe access and a safe working environment are essential for the MREFC-funded Construction activities to proceed as scheduled, budgeted, and planned. Similar to the items discussed in Section 10.2.4, *Maintenance Activities to Support Safe Access and Facility Rehabilitation,* the items discussed below will be completed during the Construction phase and will be coordinated with the Construction schedule to synchronize and sequence the work most efficiently.

- The Yates Shaft will be rehabilitated, including modifications to the Yates Hoist electrical and mechanical systems and hoisting-related headframe modifications.
- Many air doors and water control structures will be required outside of the primary physics campuses at the 4850L and 7400L. Ground control is required for safe access to many areas on Other Levels and Ramps (OLR).
- Generators will provide standby power for many areas of the Facility. Some examples of standby loads include hoists, lighting, breathing air compressors, and critical water pumps.
- Mobile equipment all-terrain vehicles (ATVs) for underground rescue and maintenance work and related underground vehicles to support Facility maintenance are supplied.

### 10.3.5 Facility Operations and Maintenance

The DUSEL Operations team will continue to perform similar operations and maintenance services during Construction as it did during the Final Design phase. By the start of Construction, the Facility will be dewatered to below the 7400L; however, maintenance of the dewatering system will continue. The WWTP will continue to operate to support removal of groundwater inflow into the Facility.

The Operations team will participate in commissioning of the newly constructed systems in a process managed by the SE and Facility teams, and executed using a combination of a third-party commissioning agent, contractors, and DUSEL staff. Operations staff will participate in acceptance testing and will be trained on operation and maintenance of the new systems. The systems will be tested in accordance with the Systems Verification Plan (see Volume 9) developed during Final Design. Preventative maintenance activities, managed by the Operations staff, will begin as systems are commissioned. This commissioning process will occur repeatedly throughout the Construction period as new systems are completed.

As the site will have an increased level of activity during Construction, the advance planning and scheduling of periodic maintenance and coordination with on-site contractors will be critical.

Operations will continue to support the needs of early science, providing maintenance and repair services, as well as assisting individual collaborations to safely install experiment equipment. Ground support, water inflow management, and air door installation activities will continue.

Concurrent with the Facility construction, research collaborations will finalize their experiment designs and planning. The Operations team will assist with scheduling utility connections and material movement to the experiment locations. The Operations team will review experiment installation plans to confirm compatibility with Facility design.

As the underground lab modules (LMs) are completed, the scientific collaborations will be granted limited and controlled access to the LMs in support of final experiment layout activities.

The DUSEL Operations department expenses include the provision of site-wide utilities, general service contracts (site maintenance, janitorial), consumable items required to support maintenance, diesel fuel, and oil and grease for vehicles. To streamline site logistics, DUSEL will have a supply contract to provide all fuel and petroleum products through the Operations budgets through appropriate funding sources, including R&RA funds.

The Surface Campus, including both Ross and Yates, require ongoing maintenance, including interior and exterior building surfaces; resurfacing roads and parking lots; replacing or adding curbs, sidewalks, and retaining walls; landscaping; and general Facility maintenance to maintain the condition of the Facility.

### 10.3.6 Science Integration and Support

A significant growth in the number of science personnel related to science integration and support will be required during MREFC-funded Construction compared with the level in place at the end of 2013. The design of the initial DUSEL experiments and the related Research and Development (R&D) will be completed from 2014 onward, primarily during 2014-2016. Design for future experiments and related R&D, after completion of the MREFC-funded experiments, at DUSEL will occur toward the end of the Facility Construction period.

To support the experiment installation during the Construction phase, mechanical and electrical shops with modest capabilities will be located on the Surface Campus. A shop with modest capability will also exist underground to support experiments. Support for the BGE experiments, including chemical analysis to store and analyze biological, chemical, and rock samples, will be provided.

Basic capability in low-background counting will be implemented by 2014 and will consist of a minimum of two instruments located in the Davis Campus. Further details on how this will be accomplished will be developed prior to the installation of the ISE.

The engineering design of the ISE will be largely complete during the first few years of the Facility Construction phase, including all large-scale physics and BGE experiments. Some smaller BGE or physics experiments may require minimal design support later in the Facility Construction period.

Funding to support the identification of new experiments and minimal associated R&D has been allocated to allow for the suite of experiments to be expanded beyond the initial set. Additionally, research efforts for DUSEL scientists ramp up substantially in the first few years of the Facility Construction period. These efforts include data analysis for early science experiments and simulation work for the ISE. Once DUSEL experiments are operational toward the end of the MREFC-funded Construction period, approximately 40% to 50% of DUSEL scientist effort will be spent on research.

### 10.3.7 Education & Outreach

E&O activities during early Construction will be similar to activities during Final Design, but will be increased with added logistical dimensions. In addition to supporting the development of the SCSE Facility, programs, exhibits, and the institution, cultural outreach and engagement will be sustained and expanded, and refinement and evaluation of prototype programs will continue.

As an organizational unit within DUSEL, the Education and Public Outreach department will rely heavily upon the other divisions of DUSEL, including Facilities, Operations, Information Technology, EH&S, Human Resources, and Business Services.

The SCSE Facility is scheduled for completion in 2018. Prior to beneficial occupancy of the new Facility, the interim home base for E&O, the Yates Education Building, will be renovated and repurposed according to the final configuration of the Surface Campus design. During the period when the Yates Education Building is under renovation, a temporary off-site home for E&O will be established. During the peak Construction period, E&O will continue to engage with students, teachers, and the general public through an off-site location in Lead. The E&O programs will also utilize facilities in nearby universities and schools. Once the Facility construction is complete, the installation of exhibits and fit-out of programmatic spaces will occur to support the opening of the SCSE to the public as soon as safe access to the site can be established. The focus of the E&O programs will continue to build awareness and knowledge of the scientific program and build cultural and community relationships supporting the continued growth of DUSEL.

## 10.4 Ongoing DUSEL Scientific and Facility Operations

The current schedule for the development of DUSEL anticipates construction activities to be complete in 2022, when the Facility and organization will enter steady-state Operations. At this point, the Construction will be complete, the initial science installations will be complete, and the Facility will operate under an NSF-approved budget through the DUSEL LLC. As described in Chapter10.3, the primary role of the SDSTA after the formation of the DUSEL LLC will be to retain the site ownership and maintain relationships with the state of South Dakota and Barrick Gold Corporation.

The focus of efforts during steady-state Operations is fourfold:

- Continuing Facility maintenance and operations
- Continuing science operations
- Continuing development of the education and public outreach efforts through the SCSE
- Continuing the supporting capacities and organizational management required to support the operations of the DUSEL Facility and program

It is assumed that the activities described in this Chapter 10.4 are built on the efforts described in Chapters 10.1 through 10.3, which serve as a foundation for the growth and development of the DUSEL Project.

### 10.4.1 Organizational Management and Staffing Approach

The organizational management structure of the DUSEL LLC as described in Section 10.2.1, *Organizational Management and Staffing Approach*, is expected to remain in place for the operation of DUSEL. New departments may be created as needed to support operations, but it is not anticipated that the senior management organizational structure will change.

The staffing requirements to support Operations, as opposed to Design and Construction activities, will change and staff will be repurposed as appropriate to meet these needs before additional staff are hired. Staffing in the Facility team will reduce to address the change in engineering support for Construction to support for Operations.

### 10.4.2 Project Operations

The following departments provide day-to-day operational functions for DUSEL.

#### 10.4.2.1 Business Services

The Business Services department will retain the functions t necessary to support laboratory operations, and they will be provided by approximately 23 FTEs during steady-state Operations.

**Insurance**

Included in Business Services is risk financing to ensure that proper insurance coverages are in place to manage risk. As Construction activities conclude and DUSEL enters the steady-state Operations phase, the construction-related insurances supporting the MREFC-funded activities will cease. Operations-related coverages will continue forward and include commercial general liability, workers' compensation, environmental, directors and officers liability, property, and auto, for example. As experiments enter research operations, it is anticipated that the DUSEL LLC will need to provide insurance coverage for experiment activities and research personnel. Specific requirements of the experiment-related liability and pollution coverages will have been developed during Final Design. These insurances will be made available to experiment collaborations in an attempt to simplify policy acquisition, lower costs, and simplify administration.

**Procurement and Contract Management**

As the Project transitions from Construction and into steady-state Operations, the procurement and contract management activities will be reduced to reflect ongoing operational needs. Procurement contracts for outsourced services will include Facility maintenance support, supplier agreements, commissioning of systems as needed, and support of operational activities to maintain the site. The services of this department will also support the SCSE for traveling exhibits and other services as required to support an active science outreach and education center.

**Support Services**

The administrative support services function includes staff that provide administrative support to the full Project. The focus during Operations will be in support of the ongoing science users, including support for workshops and reviews. The Support Services group will also interface with all DUSEL visitors, support visitor identification through EH&S activities, and coordinate with departments Project-wide to assist in the smooth day-to-day Operations of the laboratory.

**Intellectual Property, Technology Transfer, and Copyright Policy**

For the experiment R&D activities taking place as a part of the DUSEL Project, the intellectual property, technology transfer, and copyright policies will follow the policies from the users' sponsoring institution on patents, copyrights, trademarks, and tangible research results. These are commonly referred to as intellectual property, as they relate to sponsored research agreements. In general, these policies will reflect the academic policies of the Project participants and the sponsoring funding agencies. The DUSEL LLC will consider and generate a policy acceptable to the DUSEL stakeholders.

**Human Resources**

The functions of the Human Resources department will be similar to those described in Chapter 10.2, *Operations Plans during Final Design*, and the department will primarily focus on maintaining staff resources required to operate the Facility, reducing employee turnover, and general human resource functions.

**Communications**

Communications efforts of the operational DUSEL Facility will be closely coordinated with the SCSE programs, which will serve as a primary interface with the public. DUSEL will also communicate news from the developing scientific program, and support regular updates to Web sites, newsletters, and relationship building with various constituencies.

**User Support**

The User Support Office (USO) facilitates support to all scientific groups resident at DUSEL, and those that wish to conduct research on the site. Researchers wishing to conduct experiments at DUSEL will contact the USO, which will assist the scientists in establishing relationships within science support and other departments as necessary. Once a research project has been accepted, the USO works with the interested party to develop agreements with the sponsoring institution. The USO will also support the resident research groups by assisting with workshop planning.

#### 10.4.2.2 Information Technology

The IT department is responsible for all centrally managed computer and network technologies utilized at DUSEL. After Construction, support will include not only business, facility, and operations functions, but also science collaborations and the SCSE. To maintain this support, the IT staff will have approximately 18 FTEs.

#### 10.4.2.3 Finance

The role of the Finance department will decrease during steady-state Operations and will function in a manner closer to what is described in Chapters 10.1 and 10.2. This department will be responsible for maintaining all budgets and managing cash flow to support steady-state Operations requiring close coordination with all departments, but particularly Facilities, Project Controls, and Operations. The department will continue to maintain the overall DUSEL budget and provide financial detail to the Project Controls team to prepare reports for the NSF and Senior Management. This department is also responsible for asset management, scheduling of audits, preparing and processing invoices, and managing contract obligations. The staff to support these functions will include five FTEs.

#### 10.4.2.4 Project Controls

Project Controls is responsible for cost estimate management and scheduling of the DUSEL Project. The department maintains the Work Breakdown Structure (WBS) and generates reports on Project status for Senior Management to demonstrate how the Project is proceeding according to the Project baseline. Project Controls works closely with all other departments to ensure adherence to budgets and schedules. It also works with the Finance team to confirm that all budget line items are in order. During steady-state Operations, the activities of this department will decrease significantly, to support only small projects, and therefore the staffing level will require approximately five FTEs during this phase.

#### 10.4.2.5 Operations

Once Construction—and more specifically the waste rock removal and ground support installation—activities are complete, the Operations department staff will be reduced to a steady-state Operations staffing of approximately 100 FTEs. The Operations staff during this steady-state phase will support building and equipment maintenance and science equipment integration support. They will also monitor ground support systems, and maintain safe access to the Facility.

Most Operations staff will work during regular business hours; however, the hoists will be operated 24 hours a day to provide continuous underground access for research and maintenance personnel. The WWTP will also continue to operate continuously. The Command and Control Center (CCC) will be staffed continuously to monitor the status of life safety, Facility monitoring, and other infrastructure systems.

#### 10.4.2.6 Facility and Engineering

Facility staffing levels during the steady-state Operations phase will decrease to reflect staffing requirements for ongoing laboratory operations and maintenance activities, including Operations-related projects and laboratory support to experiment installation and checkout. The Facility staffing level will be reduced from 24 FTEs during Construction to 17 FTEs during the steady-state Operations phase. Operations support staffing will consist of three FTEs of project management support including the Facility Project Manager and 14 FTEs of engineering support staffing. The LBNE Level-3 Facility Project Engineer will remain on staff and be funded by the LBNE project. DUSEL engineering support to Operations staff remaining onboard include geological and geotechnical engineering, DLL and MLL integration engineering, OLR integration and ventilation engineering support, electrical and mechanical engineering, safety and transportation systems engineering, and surface buildings and infrastructure engineering support.

#### 10.4.2.7 Systems Engineering

During the Operations phase, the SE team will have two FTEs, but will continue its role in the execution and improvement of plans and processes for changes to the Facility due to science experiment installation, technology upgrades, or Facility changes.

Risk management will be continued during Operations at a lower level than during Final Design and Construction to identify, classify, and score Project risks. The Risk Registry will continue as a management tool for decision-making and effective program management.

SE team duties regarding configuration management will continue during Operations per the previously described plans.

The requirements interface and design document set discussed previously will be used as the baseline for any changes to the system required during Operations. Changes to requirements documents due to new scientific users or upgrades to the Facility will be the major scope of work during this time frame.

#### 10.4.2.8 EH&S

EH&S functions and activities funded by R&RA and other non-MREFC accounts are expected to evolve following completion of DUSEL Construction transitioning to steady-state Operations. These changes in focus will be driven by increased EH&S support for ongoing science collaboration presence on site and expanded administrative responsibilities for increased number of people on site. As Construction activities conclude, the EH&S staff and support functions will be completely transitioned to R&RA and other funding sources.

MREFC-funded EH&S positions that were solely in support of Construction will be eliminated. However, staff with construction-related EH&S expertise will be retained within the DUSEL EH&S program to support minor construction and ongoing maintenance activities.

The overall level of EH&S staffing is expected to decrease from the staffing level during Construction as the steady-state Operations phase is entered. Slight increases in science and general administrative support staff will be matched or exceeded by reductions in construction support staff. The anticipated EH&S staffing level will be approximately 25 FTEs.

#### 10.4.2.9 Quality Assurance (QA)

Once the Quality Assurance (QA) system is successfully implemented, system changes will be based on audit results, customer satisfaction data, results of management reviews, and data and trend analysis toward meeting DUSEL's goals and objectives. The QA level of effort will be maintained through the steady-state Operations phase, requiring two FTEs.

The QA staff will focus quality audits on new experiments as they are commissioned and decommissioned. Internal quality audit reports will continue to be reviewed and reported to Senior Management to support decision-making. The DUSEL Global Quality Assurance Policy and supporting documentation will be reviewed to ensure continued suitability. Annual quality reviews will be performed, and quality training will continue in order to promote a quality culture with staff, suppliers, and scientists. Continuous improvement goals, objectives, and targets will be analyzed to ensure that monitoring and measurement activities and corrective action tools are effective.

#### 10.4.2.10 Science

Scientific staff will continue to lead the integration and support effort, with the support of the USO and other DUSEL departments. DUSEL scientific staff will be involved in all major initial experiments to support the integration of specific experiments into the Facility, including installation and commissioning. After installation of the experiments is complete, DUSEL scientific staff will focus primarily on research activities, developing and supporting new research, working with the scientific community at DUSEL, and supporting the education and public outreach efforts to communicate DUSEL science to the general public. The staffing level to support science activities will remain at approximately 40 FTEs through a variety of funding sources.

#### 10.4.2.11 Education and Public Outreach

The staffing level of the SCSE at opening day, as described in Chapter 10.3, is sufficient to deliver the initial set of SCSE programs designed to accomplish DUSEL's goals but will grow over time as new partnerships emerge, programs are developed, and funding is available through additional grants.

The organizational structure for Education and Public Outreach will remain consistent throughout steady-state Operations. The E&O Director will continue to report within the DUSEL organizational structure, and the SCSE endowment will continue to support new initiatives and complement NSF funding for steady-state Operations. Total staffing to support E&O activities will be maintained at 17 FTEs through a variety of funding sources.

### 10.4.3 Facility Support of Ongoing Operations

The Facility team will provide engineering and project management support to Operations and maintenance activities, as well as work with the USO and science liaisons during experiment installation to confirm that the Facility infrastructure will support the experiment design. As outlined in Section 10.4.2.6, Facility team staffing will be significantly reduced from Design and Construction levels, reflecting the change in activity level on site. Engineering disciplines within the Facility group will be targeted to address ongoing operational needs and to be a resource to the Operations team. The Facility team will support regular assessments of the infrastructure, Facility, and systems; will verify continued safe access and operations for the entire Facility; and will engage in the process to plan, oversee, and manage capital upgrades and related operational support and maintenance projects.

### 10.4.4 Operations and Maintenance

The primary function of the Operations team is to operate and maintain the DUSEL Complex. A full preventative maintenance program will be implemented, in which equipment and systems will undergo routine maintenance in accordance with best practices, manufacturer requirements, and maintenance manuals developed specifically for the DUSEL Project. The shafts and hoists, WWTP, and CCC will operate continuously to monitor the status of infrastructure systems, facility monitoring systems, and life safety systems. The CCC will be staffed by trained and qualified technicians at all times to monitor trouble indicators and alarms. Operations staff will support scientific collaborations during installation and experiment operations for Facility and material handling needs. Warehouse, fleet, mechanical and electrical shops, and energy management services will be provided.

Visitors to the site will be directed to the SCSE or the Yates Administration Building, where they will check in. Signage will be installed to direct visitors to the correct location. Visiting scientists will be registered through Support Services in the Administration Building. Site access beyond these two areas will be controlled for the safety of the visitors, facility, and staff.

Security staff will regularly monitor the Facility both surface and underground. All staff, visiting researchers, volunteers, and visitors will be required to display an identification badge while on site (visitors to the SCSE may be supplied with a sticker or similar system). Badges for those who are on site regularly will serve as photo identification and as key cards, facilitating efficient access-control procedures based on security clearance, training, and other required authorizations. As described in Chapter 5.5, *Cyberinfrastructure Systems Design*, a radio frequency identification (RFID) system will be put in place to track high-value assets, high-risk substances, and personnel.

### 10.4.5 Planning for Future Science Activities

With DUSEL Construction completed and in steady-state Operations, the future vitality of the science program will depend upon the Central Project Directorate's close attention and access to informed advice concerning discoveries and developments, not only from DUSEL's science experiments but also those worldwide in a variety of science areas. New directions and new opportunities for enhancing the Project's scientific reach can be expected to occur; therefore, planning for future scientific opportunities will be an important aspect of the responsibilities of the entire DUSEL scientific and technical staff. Upgrades to the ISE, new experiments, and, very likely, completely new concepts will arise. The planning process will be the same as the process for selection of the ISE at DUSEL. Expressions of interest, letters of intent, technical proposals or some combination will be required for all experiments. These will be reviewed by the DUSEL staff and by the DUSEL Program Advisory Committee (PAC) or subcommittees of the PAC.

### 10.4.6 Education and Outreach

The SCSE will grow, mature, and evolve throughout DUSEL steady-state Operations. The SCSE will serve regional students, educators, and members of the general public throughout the year and will gradually expand and refine on-site, off-site, and distance education programs.

Attendance projections for the SCSE vary seasonally and will likely change through different phases of DUSEL's life cycle. Logistics and programming will be carefully orchestrated to accommodate large visitor audiences in summer and to make optimal use of the educational spaces to reach and serve expanded audiences during the academic year.

A potential draw for audiences wishing to visit DUSEL and the SCSE is an interest in experiencing authentic science in action. All will be able to experience the underground campuses remotely through webcams and simulations. Select audiences may be escorted underground to see the installed and operating scientific experiments.

Important to the SCSE's long-term success will be engaging exhibits and programs, mirroring and reflecting the dynamic science and engineering research at DUSEL. While the SCSE is expected to have a handful of signature exhibits and programs that remain stable over time, ongoing content development, exhibit design, and program development will continue. Evaluation, educational research, and dissemination will increase the impact of the SCSE beyond the specific science and engineering disciplines of DUSEL.

## 10.5 Summary

In this Volume we present the Operations requirements as the Project transitions from Preliminary Design (Chapter 10.1) to Final Design (Chapter 10.2), through Construction (Chapter 10.3) and into steady-state Operations (Chapter 10.4). The operation and staff planning efforts described throughout this Volume have been developed to support the overall development, budget, and schedule planning described in this Preliminary Design Report (PDR). The Project anticipates that these plans will be revisited as the Project plans are refined during Final Design and Construction. Opportunities will exist for efficiencies to be gained and processes to be refined, either through collaboration with other institutions or through engaging new technologies.

The Operations Plans outlined in this Volume have been benchmarked and compared to existing laboratories with similar components, as well as through extrapolation of the current maintenance and operations efforts by the SDSTA (see Appendix 10.A, *Comparative Analysis of DUSEL Operations Requirements to Other Research Facilities*). While no single facility matches precisely or completely the full scope of the DUSEL Project, the Operations planning team, scientific staff, and design teams have examined a variety of facilities with comparable components to aid in the understanding of the requirements and effort required to design, construct, and operate DUSEL. The Operations planning has included investigations to evaluate the size of underground and surface campuses, required staff levels, budgets, scientific program and requirements, and public outreach and education programs at other institutions. These institutions include SNOLab in Sudbury, Ontario; Gran Sasso (LNGS) in L'Aquila, Italy; Soudan Underground Laboratory in Soudan, Minnesota; Laser Interferometer Gravitational-wave Observatory (LIGO) Visitors Center in Livingston, Louisiana; Spallation Neutron Source in Oak Ridge, Tennessee; and Fermilab, including the Lederman Science Center in Batavia, Illinois. Additionally, project-specific data from the current SDSTA Operations have informed the DUSEL Operations Plans development. The benchmarking research has validated the current estimate on a program-to-program comparison basis with components of these other institutions. The current activities at Sanford Laboratory are valuable to the planning process.

DUSEL Project staff have been involved in the development of the science program, Facility design, Operations planning, E&O, and project organization and business systems with other projects, thereby providing an expertise and insight to shape the DUSEL development and planning process. Many key facility operations and maintenance staff at Sanford Laboratory bring experience from working at the former Homestake Mining Company and are familiar with the condition, maintenance requirements, and operation requirements for the systems and infrastructure required for underground access. Additionally, as described in Volume 7, *Project Execution Plan*, the Project has engaged a number of advisory committees to review DUSEL plans and provide advice and outside (off-project) expertise.

The operations, scientific, and research support requirements briefly described above and completed during the Preliminary Design phase form the basis for the Operations Plans as described in this Volume. Work to refine these plans, including Operations budgets, are described in Volume 2, *Cost, Schedule, and Staffing*. These plans will continually be refined during Final Design and Construction to develop an efficient and successful organization to manage the operations of DUSEL.

The competition-sensitive nature of the estimates contained in Volume 2 of the Preliminary Design Report resulted in this Volume being redacted from this distribution of the Report.

The size, format and complexity of the Report Appendices prevent their inclusion with the main report. Requests for the appendices should be sent to the DUSEL Project Principal Investigator.

*This work was supported by the National Science Foundation under Cooperative Agreements PHY-0717003 and PHY-0940801 with the University of California, Berkeley. Any opinions, findings, and conclusions or recommendations expressed in this material are those of the author(s) and do not necessarily reflect the views of the National Science Foundation.*

For additional information concerning the report please contact:

Dr. Kevin T. Lesko
DUSEL Principal Investigator
UC Berkeley
ktlesko@berkeley.edu
www.DUSEL.org

The competition-sensitive nature of the estimates contained in Volume 2 of the Preliminary Design Report resulted in this Volume being redacted from this distribution of the Report.

The size, format and complexity of the Report Appendices prevent their inclusion with the main report. Requests for the appendices should be sent to the DUSEL Project Principal Investigator.

*This work was supported by the National Science Foundation under Cooperative Agreements PHY-0717003 and PHY-0940801 with the University of California, Berkeley. Any opinions, findings, and conclusions or recommendations expressed in this material are those of the author(s) and do not necessarily reflect the views of the National Science Foundation.*

For additional information concerning the report please contact:

Dr. Kevin T. Lesko
DUSEL Principal Investigator
UC Berkeley
ktlesko@berkeley.edu
www.DUSEL.org